\documentclass[12pt,british,german,english]{extreport}
\usepackage[T1]{fontenc}
\usepackage[latin9]{inputenc}
\usepackage{geometry}
\usepackage{color}
\usepackage{babel}
\usepackage{float}
\usepackage{mathtools}
\usepackage{enumitem}
\usepackage{amsmath}
\usepackage{amssymb}
\usepackage{cancel}
\usepackage{stmaryrd}
\usepackage{stackrel}
\usepackage{graphicx}
\usepackage{setspace}
\usepackage{wasysym}
\usepackage{esint}
\usepackage[unicode=true,pdfusetitle,
 bookmarks=true,bookmarksnumbered=false,bookmarksopen=false,
 breaklinks=false,pdfborder={0 0 0},pdfborderstyle={},backref=false,colorlinks=true]
 {hyperref}

\makeatletter

\providecommand{\tabularnewline}{\\}

\@ifundefined{date}{}{\date{}}
\usepackage{lmodern}
\usepackage[T1]{fontenc}
\usepackage{caption}
\numberwithin{equation}{section}
\numberwithin{figure}{section}
\numberwithin{table}{section}
\usepackage{upgreek}
\usepackage{titlesec}
\titlespacing*{\section}
{0pt}{5.5ex plus 1ex minus .2ex}{4.3ex plus .2ex}
\titlespacing*{\subsection}
{0pt}{5.5ex plus 1ex minus .2ex}{4.3ex plus .2ex}
\titlespacing*{\subsubsection}
{0pt}{5.5ex plus 1ex minus .2ex}{4.3ex plus .2ex}
\titlespacing*{\paragraph}
{0pt}{5.5ex plus 1ex minus .2ex}{4.3ex plus .2ex}

\renewcommand\chapter{\par
                    \thispagestyle{plain}%
                    \global\@topnum\z@
                    \@afterindentfalse
                    \secdef\@chapter\@schapter}

\@ifundefined{showcaptionsetup}{}{%
 \PassOptionsToPackage{caption=false}{subfig}}
\usepackage{subfig}
\makeatother

\begin{document}
\begin{doublespace}
\begin{center}
\textbf{\LARGE{}$ $}\\
\textbf{\LARGE{}}\\
\textbf{\LARGE{}}\\
\textbf{\LARGE{}}\\
{\huge{}Magnetic Compression of Compact Tori\vspace{0.5cm}
 Experiment and Simulation}\\
{\large{}\vspace{1.5cm}
}{\large\par}
\par\end{center}
\end{doublespace}

\begin{center}
{\large{}A Thesis Submitted to the }{\large\par}
\par\end{center}

\begin{center}
{\large{}College of Graduate and Postdoctoral Studies }{\large\par}
\par\end{center}

\begin{center}
{\large{}in Partial Fulfillment of the Requirements }{\large\par}
\par\end{center}

\begin{center}
{\large{}for the degree of Doctor of Philosophy }{\large\par}
\par\end{center}

\begin{center}
{\large{}in the Department of Physics and Engineering Physics}{\large\par}
\par\end{center}

\begin{center}
{\large{}University of Saskatchewan }{\large\par}
\par\end{center}

\begin{center}
{\large{}Saskatoon}{\large\par}
\par\end{center}

\begin{center}
{\large{}Canada}\\
{\large{}\vspace{2cm}
By}{\large\par}
\par\end{center}

\begin{center}
{\large{}Carl Dunlea\vspace{1cm}
}\\
$\varcopyright$ Carl Dunlea, August 2019. All rights reserved.
\par\end{center}

\thispagestyle{empty} 

\newpage\pagenumbering{roman} \setcounter{page}{1}\cleardoublepage \phantomsection \addcontentsline{toc}{chapter}{Permission to use}

\chapter*{Permission to use}

In presenting this thesis in partial fulfillment of the requirements
for a Postgraduate degree from the University of Saskatchewan, I agree
that the Libraries of this University may make it freely available
for inspection. I further agree that permission for copying of this
thesis in any manner, in whole or in part, for scholarly purposes
may be granted by the professor or professors who supervised my thesis
work or, in their absence, by the Head of the Department or the Dean
of the College in which my thesis work was done. It is understood
that any copying or publication or use of this thesis or parts thereof
for financial gain shall not be allowed without my written permission.
It is also understood that due recognition shall be given to me and
to the University of Saskatchewan in any scholarly use which may be
made of any material in my thesis. Requests for permission to copy
or to make other use of material in this thesis in whole or part should
be addressed to: \\
\\
Head of the Department of Physics and Engineering Physics, Room 163,
116 Science Place, University of Saskatchewan, Saskatoon, Saskatchewan,
Canada S7N 5E2, \\
\\
or: \\
\\
Dean, College of Graduate and Postdoctoral Studies, University of
Saskatchewan, 116 Thorvaldson Building, 110 Science Place Saskatoon,
Saskatchewan, Canada S7N 5C9.

\newpage\cleardoublepage \phantomsection \addcontentsline{toc}{chapter}{Abstract}
\begin{abstract}
{\footnotesize{}The magnetic compression experiment at General Fusion
was a repetitive non-destructive test to study plasma physics applicable
to magnetic target fusion compression. A compact torus (CT) is formed
with a co-axial gun into a containment region with an hour-glass shaped
inner flux conserver, and an insulating outer wall. External coil
currents keep the CT off the outer wall (radial levitation) and then
rapidly compress it inwards. The optimal external coil configuration
greatly improved both the levitated CT lifetime and the recurrence
rate of shots with good compressional flux conservation. As confirmed
by spectrometer data, the improved levitation field profile reduced
plasma impurity levels by suppressing the interaction between plasma
and the insulating outer wall during the formation process. Significant
increases in magnetic field, electron density, and ion temperature
were routinely observed at magnetic compression in the final external
coil configuration tested, despite the prevalence of an instability,
thought be an external kink mode, at compression. Matching the decay
rate of the levitation currents to that of the CT currents resulted
in a reduced level of MHD activity associated with unintentional compression
by the levitation field, and a higher probability of long-lived CTs.
The DELiTE (Differential Equations on Linear Triangular Elements)
framework was developed for spatial discretisation of partial differential
equations on an unstructured triangular grid in axisymmetric geometry.
The framework is based on discrete differential operators in matrix
form, which are derived using linear finite elements and mimic some
of the properties of their continuous counterparts. A single-fluid
two-temperature MHD model is implemented in this framework. The inherent
properties of the operators are used in the code to ensure global
conservation of energy, particle count, toroidal flux, and angular
momentum. The development of the discrete forms of the equations solved
is presented. The code was applied to study the magnetic compression
experiment. The numerical models developed to simulate CT formation,
levitation, and magnetic compression are reported. A model for anisotropic
thermal diffusion has been formulated and implemented to the code.
A method for determining the $q$ profile of the CT was established
- simulated CT $q$ profiles indicate that the magnetic compression
experiment could be improved by modifying the $q$ profile to regimes
with increased stability against kink modes. Comparisons between simulated
and experimental diagnostics are presented. A model of plasma/neutral
fluid interaction was developed and included in the framework. The
source rates of species momentum and energy due to ionization and
recombination were derived using a simple method that enables determination
of the volumetric rate of thermal energy transfer from electrons to
photons and neutral particles due to radiative recombination, which
has been neglected in other studies. The implementation of the model
has enabled clarification of the mechanisms behind the increases in
CT electron density that are routinely observed on the SPECTOR plasma
injector well after CT formation. This understanding helps account
for the exceptionally significant increase in electron temperature,
and markedly reduced electron density, observed during the electrode
edge biasing experiment conducted on SPECTOR. It is thought that edge
fueling impediment, a consequence of a biasing-induced transport barrier,
is largely responsible for the observed modifications to temperature
and density.\thispagestyle{plain}}{\normalsize{}\pagenumbering{roman} \setcounter{page}{2}}{\normalsize\par}
\end{abstract}
\cleardoublepage \phantomsection \addcontentsline{toc}{chapter}{Acknowledgments}

\subsection*{{\huge{}Acknowledgments}}

I would like to thank my thesis supervisors Professors Akira Hirose
and Chijin Xiao, especially for guidance and assistance with the CHI
(Co-axial Helicity Injection) on the STOR-M tokamak over the second
year of the PhD program. That was a great introduction to tokamak
hardware and operation, and to laboratory work. Thanks to Dave McColl
for his advice on various high power circuit modifications. A description
of the CHI experiment is presented in \cite{CHI_STORM}. Thanks to
Professor Xiao in particular for helping arrange formalities over
the final stages of the program, to Professor Andrei Smolyakov for
the initial contact that led to this research work, and to my academic
committee for general support and advice with the thesis formatting.
Thanks to Debbie and Marj for always being friendly and willing to
offer practical advice. The couple of winters I spent in Saskatoon
would not have been as enjoyable without various wonderful people
who I was lucky to meet there, especially Yanli, Winston, Lili, and
Joel. At General Fusion (GF), many people were supportive and helped
with the development of the work presented here. In particular, Wade
Zawalski, Stephen Howard, and Kelly Epp were involved with the experiment
and I learned a lot from them. Stephen helped guide the development
of an equilibrium solver to model magnetic compression, and was very
helpful with the experiment and with data analysis. Many of the improvements
and developments that are presented here were achieved towards the
end of the experimental phase of the project. After the experiment
was decommissioned, Ivan Khalzov, in his spare time, supported the
development of the brand new code framework that is described in detail
in part two of this thesis; I am extremely grateful for his guidance.
The work Stephen and I did on the equilibrium solver during a large
part of 2015 is not presented here, but some lessons were learned
from it and applied to the MHD code. Thanks to Meritt Reynolds and
Charlson Kim at GF for discussions and insight, and to Masayoshi Nagata
for useful feedback. Thanks also to Blake Rablah, Curtis Gutjahr and
James Wilkie for assistance and advice relating to circuits, diagnostics,
and hardware during the experiment (particularly towards the end),
to Alex Mossman, Peter O' Shea, Mike Donaldson, Michael Delage, and
Michel Laberge for general support at various times, and to many of
the great people at GF for being friendly and assisting with different
aspects of the project. Thanks to the University of Saskatchewan ICT
Research Computing Facility for advice and computing time. I would
like to acknowledge the funding provided in part by General Fusion
Inc., Mitacs, University of Saskatchewan, and NSERC. Thanks to my
parents and family for their continual encouragement. Finally, I would
like to thank my lovely Yanli for her support and patience throughout
this process.\pagenumbering{roman} \setcounter{page}{3}

\newpage\cleardoublepage \phantomsection \addcontentsline{toc}{chapter}{Table of contents}

\tableofcontents{}

\newpage\cleardoublepage \phantomsection \addcontentsline{toc}{chapter}{List of tables}

\listoftables
\newpage\cleardoublepage \phantomsection \addcontentsline{toc}{chapter}{List of figures}

\listoffigures

\newpage{}

\chapter*{List of Symbols}

\addcontentsline{toc}{chapter}{List of symbols}

\begin{doublespace}
\begin{flushleft}
\begin{tabular}{lll}
$\alpha$ &  & subscripts $\alpha=i,\,e,\,n$ denote ion, electron and neutral particles/fluids\tabularnewline
$\beta$ &  & subscripts $\beta=r,\,\phi,\,z$ denote radial, toroidal and axial
vector components \tabularnewline
$\gamma$ &  & adiabatic gas constant\tabularnewline
$\Gamma$ &  & boundary of system volume or boundary of computational domain\tabularnewline
$\epsilon_{0}$ &  & vacuum permittivity (8.85 $\times10^{-12}$ {[}F/m{]})\tabularnewline
$\zeta$ &  & density diffusion coefficient {[}m$^{2}$/s{]}\tabularnewline
$\eta'$ &  & resistivity {[}$\Omega$-m{]}\tabularnewline
$\eta$ &  & resistive diffusion coefficient {[}m$^{2}$/s{]}\tabularnewline
$\theta$ &  & subscript, poloidal vector component\tabularnewline
$\kappa$ &  & thermal conductivity {[}(m-s)$^{-1}${]}\tabularnewline
$\lambda_{mfp}$ &  & mean free path {[}m{]}\tabularnewline
$\mu_{0}$ &  & vacuum permeability (4$\pi$$\times10^{-7}$ {[}H/m{]})\tabularnewline
$\mu$ &  & dynamic viscosity {[}kg m$^{-1}$ s$^{-1}${]}\tabularnewline
$\nu$ &  & viscous diffusion coefficient {[}m$^{2}$/s{]}\tabularnewline
$\boldsymbol{\pi}$ &  & viscosity tensor {[}J/m$^{3}${]}\tabularnewline
$\rho$ &  & mass density {[}kg/m$^{3}${]}\tabularnewline
$\rho_{c}$ &  & charge density {[}C/m$^{3}${]}\tabularnewline
$\tau_{ii}$ &  & ion-ion collision time {[}s{]}\tabularnewline
$\tau_{ie}$ &  & electron-ion collision time {[}s{]}\tabularnewline
$\phi$ &  & toroidal flux {[}Wb{]}\tabularnewline
$\chi$ &  & thermal diffusion coefficient {[}m$^{2}$/s{]}\tabularnewline
$\Psi$ &  & poloidal flux {[}Wb{]}\tabularnewline
\end{tabular}
\par\end{flushleft}
\end{doublespace}

\newpage{}

\begin{doublespace}
\begin{flushleft}
\begin{tabular}{lll}
$\psi$ &  & poloidal flux per radian {[}Wb/radian{]}\tabularnewline
$\omega$ &  & angular speed {[}s$^{-1}${]}\tabularnewline
$\omega_{c}$ &  & cyclotron frequency {[}radians/s{]}\tabularnewline
$\mathbf{B}$ &  & magnetic field {[}T{]}\tabularnewline
$d\Gamma$ &  & elemental area {[}m$^{2}${]}\tabularnewline
$dS$ &  & elemental area {[}m$^{2}${]}\tabularnewline
$dV$ &  & elemental volume {[}m$^{3}${]}\tabularnewline
$d\mathbf{V}$ &  & elemental volume in velocity space {[}m$^{3}$ s$^{-3}${]}\tabularnewline
$e$ &  & electron charge {[}C{]}\tabularnewline
$\mathbf{E}$ &  & electric field {[}V/m{]}\tabularnewline
$h$ &  & Planck's constant (6.63$\times10^{-34}$ {[}J-s{]})\tabularnewline
$I_{comp}$ &  & compression coil current signal measured for a particular shot {[}kA{]}\tabularnewline
$I_{form}$ &  & formation current signal measured for a particular shot {[}kA{]}\tabularnewline
$I_{lev}$ &  & levitation coil current signal measured for a particular shot {[}kA{]}\tabularnewline
$I_{main}$ &  & main coil current setting for a particular shot {[}A{]}\tabularnewline
$I_{shaft}$ &  & shaft current signal measured for a particular shot {[}kA{]}\tabularnewline
$\mathbf{J}$ &  & current density {[}A/m$^{2}${]}\tabularnewline
$m$ &  & particle mass {[}kg{]}\tabularnewline
$n$ &  & number density {[}m$^{-3}${]}\tabularnewline
$p$ &  & pressure {[}Pa{]}\tabularnewline
$q$ &  & particle charge {[}C{]} (or safety factor)\tabularnewline
$\mathbf{q}$ &  & heat flux density {[}J m$^{-2}$ s$^{-1}${]}\tabularnewline
$Q_{ie}$ &  & volumetric rate of ion-electron heat exchange {[}J m$^{-3}$ s$^{-1}${]}\tabularnewline
$\mathbf{R}$ &  & collisional friction force density {[}N/m$^{3}${]}\tabularnewline
$T$ &  & temperature {[}J{]} or {[}eV{]}\tabularnewline
$t_{comp}$ &  & delay between firing formation capacitor banks and compression capacitor
banks {[}$\upmu$s{]} \tabularnewline
$|t_{lev}$| &  & delay between firing levitation capacitor banks and formation capacitor
banks {[}$\upmu$s{]} \tabularnewline
\end{tabular}\\
\begin{tabular}{lll}
$t_{gas}$ &  & delay between firing gas puff valves and formation capacitor banks
{[}$\upmu$s{]} \tabularnewline
$U_{K}$ &  & kinetic energy {[}J{]}\tabularnewline
$\dot{U}_{K\pi}$ &  & rate of decrease of $U_{K}$ due to viscous dissipation {[}J/s{]}\tabularnewline
$U_{Th}$ &  & thermal energy {[}J{]}\tabularnewline
$\dot{U}_{Th\pi}$ &  & rate of increase of $U_{Th}$ due to viscosity {[}J/s{]} \tabularnewline
$\dot{U}_{Thideal}$ &  & resistivity-independent rate of change of $U_{Th}$ {[}J/s{]} \tabularnewline
$\dot{U}_{Th\theta\eta}$ &  & rate of increase of $U_{Th}$ due to ohmic heating associated with
poloidal currents {[}J/s{]} \tabularnewline
$\dot{U}_{Th\phi\eta}$ &  & rate of increase of $U_{Th}$ due to ohmic heating associated with
toroidal currents {[}J/s{]} \tabularnewline
$U_{M}$ &  & magnetic energy {[}J{]}\tabularnewline
$U_{M\theta}$ &  & magnetic energy associated with poloidal field {[}J{]}\tabularnewline
$U_{M\phi}$ &  & magnetic energy associated with toroidal field {[}J{]}\tabularnewline
$\dot{U}_{M\theta ideal}$ &  & resistivity-independent rate of change of $U_{M\theta}$ {[}J/s{]} \tabularnewline
$\dot{U}_{M\phi ideal}$ &  & resistivity-independent rate of change of $U_{M\phi}$ {[}J/s{]} \tabularnewline
$\dot{U}_{M\theta\eta}$ &  & rate of decrease of $U_{M\theta}$ due to resistive decay of poloidal
magnetic field {[}J/s{]} \tabularnewline
$\dot{U}_{M\phi\eta}$ &  & rate of decrease of $U_{M\phi}$ due to resistive decay of toroidal
magnetic field {[}J/s{]} \tabularnewline
$\mathbf{v}$ &  & fluid velocity {[}m/s{]}\tabularnewline
$\mathbf{V}$ &  & particle velocity {[}m/s{]}\tabularnewline
$V_{comp}$ &  & voltage to which  compression capacitor banks are charged for a particular
shot {[}kV{]}\tabularnewline
$V_{form}$ &  & voltage to which  formation capacitor banks are charged for a particular
shot {[}kV{]}\tabularnewline
$V_{gun}$ &  & voltage signal measured between machine electrodes {[}kV{]}\tabularnewline
$V_{lev}$ &  & voltage to which  levitation capacitor banks are charged for a particular
shot {[}kV{]}\tabularnewline
$V_{th}$ &  & particle thermal speed {[}m/s{]}\tabularnewline
$Z_{eff}$ &  & volume averaged ion charge\tabularnewline
 &  & \tabularnewline
$*$ &  & matrix multiplication\tabularnewline
$\circ$ &  & element-wise matrix multiplication\tabularnewline
$\oslash$ &  & element-wise matrix division\tabularnewline
\end{tabular}
\par\end{flushleft}
\end{doublespace}

\newpage{}

\chapter*{List of Abbreviations}

\addcontentsline{toc}{chapter}{List of abbreviations}

\begin{doublespace}
\begin{flushleft}
\begin{tabular}{lll}
ATC &  & Adiabatic Toroidal Compressor\tabularnewline
CT &  & Compact Torus\tabularnewline
CHI &  & Coaxial Helicity Injection \tabularnewline
DAQ &  & Data AcQuisition \tabularnewline
DELiTE &  & Differential Equations on Linear Triangular Elements\tabularnewline
FEMM &  & Finite Element Method Magnetics\tabularnewline
FIR &  & Far Infra Red\tabularnewline
FRC &  & Field Reversed Configuration\tabularnewline
GF &  & General Fusion Inc. \tabularnewline
H-mode &  & High Confinement Mode\tabularnewline
ICF &  & Inertial Confinement Fusion\tabularnewline
LCFS &  & Last Closed Flux Surface\tabularnewline
MCF &  & Magnetic Confinement Fusion\tabularnewline
MHD &  & Magneto Hydro Dynamics \tabularnewline
MRT &  & Magnetised Ring Test \tabularnewline
MTF &  & Magnetic Target Fusion\tabularnewline
PCS &  & Plasma Compression, Small \tabularnewline
PI3 &  & Plasma Injector 3\tabularnewline
SMRT &  & Super Magnetised Ring Test \tabularnewline
SPECTOR &  & Spherical Compact Toroid\tabularnewline
TS &  & Thomson Scattering\tabularnewline
ULQ &  & Ultra Low Q\tabularnewline
\end{tabular}
\par\end{flushleft}
\end{doublespace}

\newpage{}

\pagenumbering{arabic} \setcounter{page}{1}

\chapter{Introduction\label{chap:Introduction}}

\section{Background - plasma and nuclear fusion}

Plasma is a collection of ions, electrons and neutral particles that
exhibits long range collective \foreignlanguage{british}{behaviour}
due to the electromagnetic forces between the charged particles.
\begin{figure}[H]
\centering{}\includegraphics[scale=0.6]{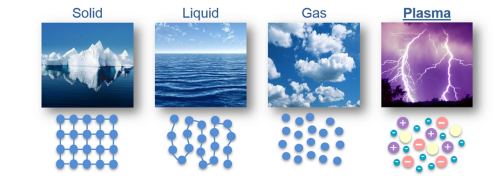}\caption{\label{fig:4thstate}$\,\,\,\,$Fourth state of matter {\footnotesize{}(image
credit: http://www.grinp.com/plasma/physics.html) }}
\end{figure}

When a solid, such as ice (see figure \ref{fig:4thstate}), is continuously
heated, it changes phase to become a liquid, then a gas, and then
a plasma. Plasma is known as the fourth state of matter - most of
the visible universe is in the plasma state. 
\begin{figure}[H]
\centering{}\subfloat[Plasma Properties ({\scriptsize{}Image credit: Contempory Physics
Education Project})]{\includegraphics[scale=0.45]{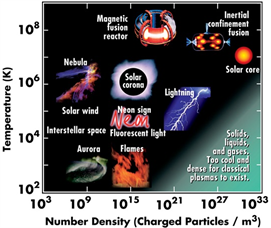}}\hfill{}\subfloat[Rosette nebula\protect \\
({\scriptsize{}Image credit: http://cs.astronomy.com/asy/m/nebulae/466016.aspx})]{\includegraphics[scale=0.3]{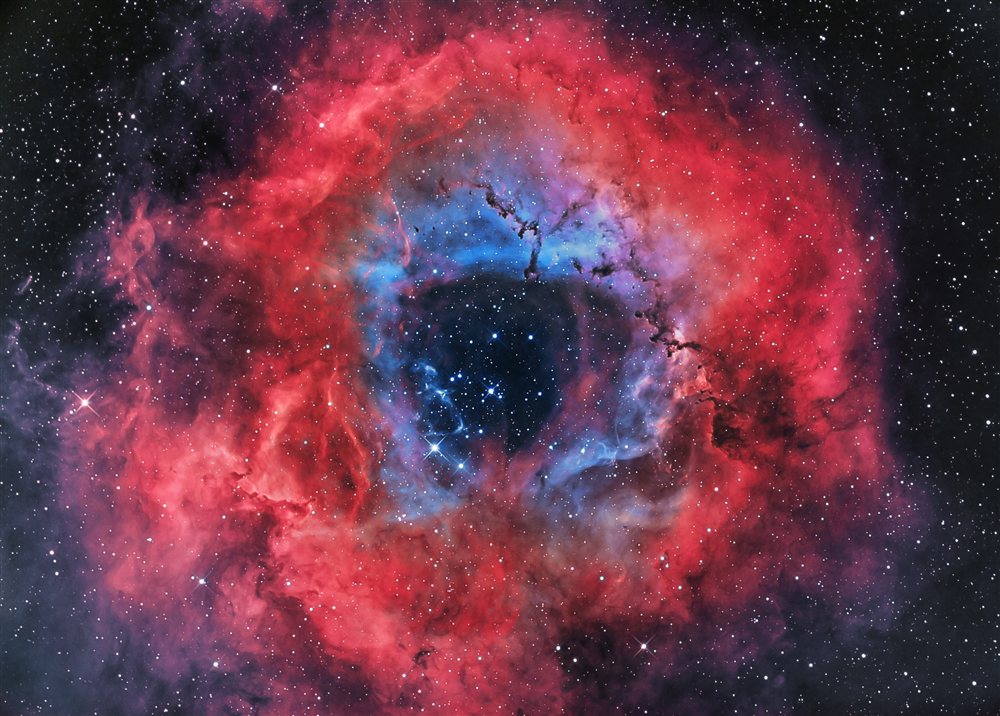}}\caption{$\,\,\,\,$Depiction of plasma properties and Rosette nebula\label{fig:Rossete}}
\end{figure}
 Figure \ref{fig:Rossete}(a) indicates the number densities and temperatures
of various plasmas - for comparison, note that the number density
of air is $\sim10^{25}\mbox{ m}^{-3}$. Terrestrial plasmas include
the aurora, lightning, flames, and the plasmas found in neon and fluorescent
light tubes. Solar nebulae ($e.g.$, figure \ref{fig:Rossete}(b)),
where stars are formed, are predominantly composed of very tenuous
plasmas, while the extreme density of plasmas characteristic of inertial
confinement fusion devices approaches that of the solar core.

When two positively charged ions are forced so close together that
the mutually attractive nuclear force overcomes the mutually repulsive
Coulomb force, nuclear fusion will occur - the ions fuse together
in a process that releases energy. 
\begin{figure}[H]
\subfloat[Sun -{\scriptsize{} Image credit: https://www.nationalgeographic.com }]{\centering{}\includegraphics[scale=0.12]{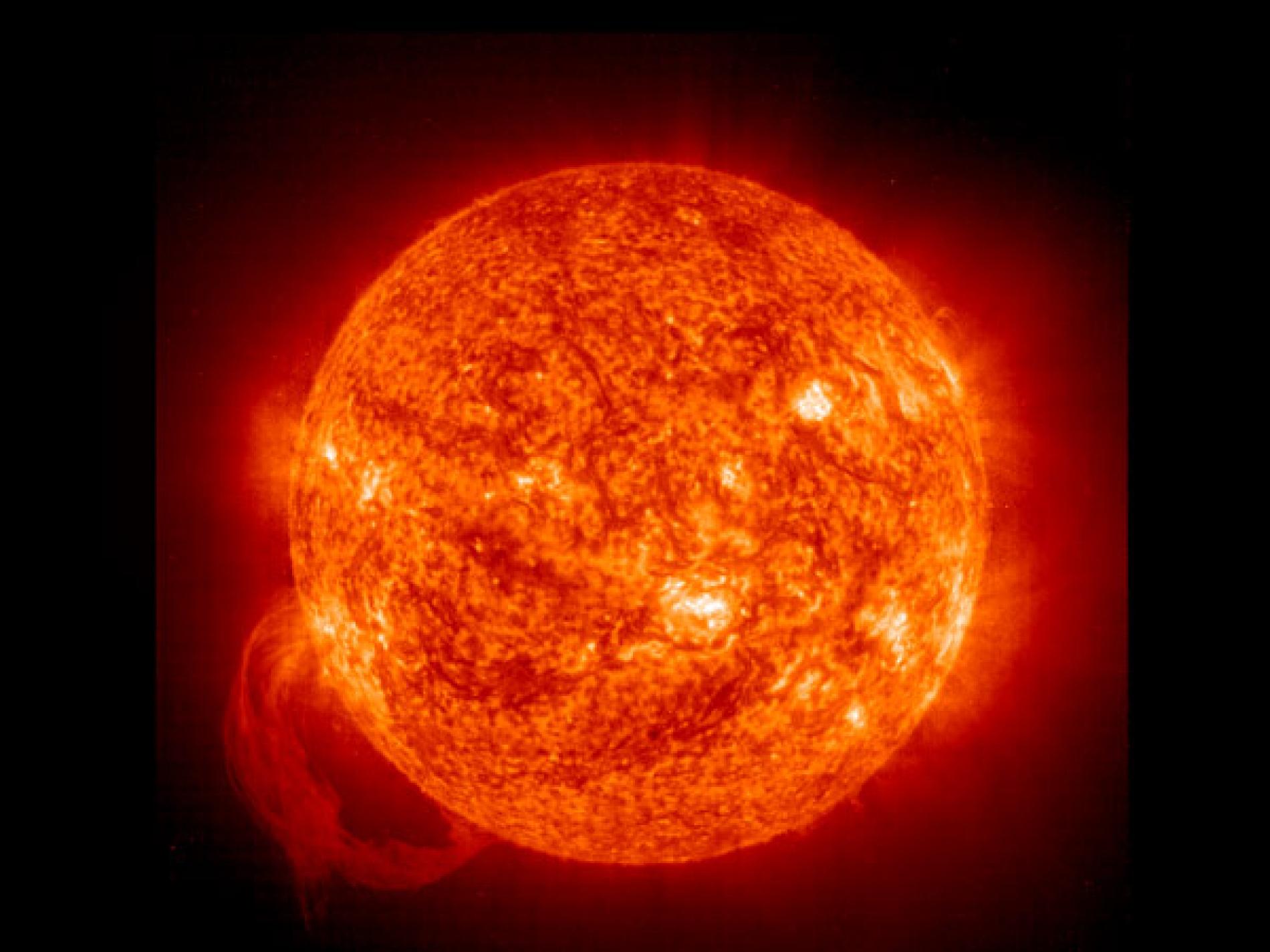}}\hfill{}
\subfloat[D-T fusion -{\scriptsize{} Image credit: http://www.fz-juelich.de }]{\centering{}\includegraphics[scale=0.7]{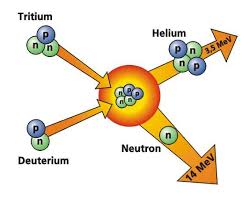}}

\caption{$\,\,\,\,$The Sun and schematic of D-T fusion reaction\label{fig:Sun_DT}}

\end{figure}
Fusion powers the stars - in solar plasmas (figure \ref{fig:Sun_DT}(a)),
the ions are forced close together by immense gravitational forces.
Stars are powered by fusion of hydrogen and its isotopes for most
of their lives, and, if they are hot enough, fuse helium and then
heavier elements as the hydrogen and then the helium supplies are
exhausted. 

Unlike nuclear fission, fusion is not associated with long-term radioactivity,
and the fuel source for the most easily attainable forms of fusion
are abundant on earth. The fusion reaction which requires one of the
lowest temperatures to initiate is the deuterium-tritium (D-T) reaction
($\ensuremath{\mbox{D}+\mbox{T}\rightarrow\mbox{He}^{4}\,(3.5\,\mbox{MeV})+\mbox{n}\,{\rm (14.1\,MeV)}}$),
depicted in figure \ref{fig:Sun_DT}(b). Deuterium and tritium are
isotopes of hydrogen. Deuterium can be extracted from sea-water and
tritium can be obtained from neutron activation of lithium-6. For
the D-T reaction, when the ions are heated to $\sim10\mbox{ keV}\,=$
$\sim10^{8}$ Kelvin, they have enough energy to overcome the Coulomb
barrier and fuse, releasing a high energy neutron which escapes the
plasma. In this type of $neutronic$ ($i.e.,$ neutron-producing reactions)
fusion, the energetic neutron is used to heat an external water supply,
producing steam which is harnessed to generate electricity with turbines
in the usual way. The Lawson criterion (triple product form), given
by the product of the plasma number density, the energy confinement
time and the plasma (ion) temperature, specifies the minimum value
of that product required for fusion ignition (the condition where
sufficient energy is produced by fusion reactions to self-sustain
further reactions): $n\tau_{E}T_{i}\geq3.1\times10^{21}$ keV-s-m$^{-3}$. 

\begin{figure}[H]
\centering{}\subfloat[Tokamak schematic{\footnotesize{}}\protect \\
{\footnotesize{}Image credit: modified from https://www.sciencealert.com }]{\includegraphics[height=5.5cm]{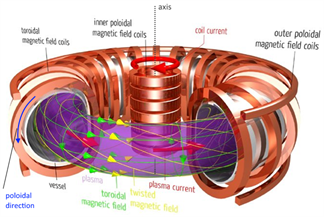}}\hfill{}\subfloat[Schematic of ICF process{\footnotesize{}}\protect \\
{\footnotesize{}Image credit: https://scitechdaily.com}]{\includegraphics[height=4cm]{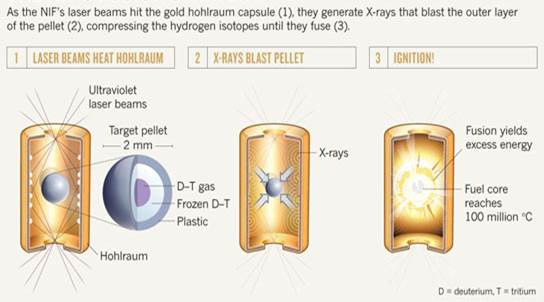}}\caption{$\,\,\,\,$MCF and ICF fusion\label{fig:Tok_icf}}
\end{figure}
The two most prevalent and well-researched potential methods towards
achieving a fusion power generation industry are magnetic confinement
fusion (MCF) and inertial confinement fusion (ICF). The tokamak depicted
in figure \ref{fig:Tok_icf}(a) is based on magnetic confinement.
Ions and electrons, the charged particles that constitute the bulk
of the plasma, gyrate around and stream along magnetic field lines.
A toroidal magnetic field is produced by poloidally-directed currents
in the toroidal field coils that surround the vacuum vessel. Note
that the poloidal and toroidal directions are indicated by the blue
and green arrows respectively in figure \ref{fig:Tok_icf}(a). Neutral
gas ($e.g.,$ a mix of deuterium and tritium) is injected into the
vacuum vessel and is broken down to form a plasma by the action of
a toroidal electric field that is induced by transformer action as
a result of the poloidal magnetic field due to toroidal currents in
the inner poloidal field coils (the standard vertical ferromagnetic
iron core shaft inside the inner poloidal coils, which concentrates
the magnetic field within it, is not shown in the figure). The same
toroidal electric field drives a toroidal plasma current - this current
produces a poloidal magnetic field. The combination of the toroidal
and poloidal magnetic fields result in a twisted magnetic field. The
charged plasma particles gyrate around, and stream along these twisted
helical field lines, resulting in confinement of the particles within
the vacuum vessel. 

The twisted helical field prevents the particle escape that would
occur due to the combination of curvature, $\nabla B$, and $\mathbf{E}\times\mathbf{B}$
particle drifts if the confinement was by toroidal field alone. Particle
drift for species $\alpha$ ($i.e.,$ ion or electron) due to magnetic
field curvature, which results in a field-perpendicular centrifugal
force on the particle, is given by 
\[
\mathbf{V}_{\alpha\,c}=\frac{V_{\alpha\parallel}^{2}}{\omega_{c\alpha}B^{3}}\mathbf{B}\times(\mathbf{B}\cdot\nabla)\mathbf{B}
\]
Here, $V_{\alpha\parallel}$ is the particle speed in the direction
parallel to the magnetic field, $\omega_{c\alpha}=q_{\alpha}B/m_{\alpha}$
is the cyclotron frequency, where $m_{\alpha}$ is the particle mass
and $q_{\alpha}$ is the particle charge. The $\nabla B$ drift is
given by $\mathbf{V}_{\alpha\,\nabla B}=\frac{V_{\alpha\perp}^{2}}{2\omega_{c\alpha}B^{2}}\mathbf{B}\times\nabla B$.
Here, $V_{\alpha\perp}$ is the particle speed in the direction perpendicular
to the magnetic field. Note that $\nabla B$, resulting from having
closer spacing between the toroidal field coils at the inboard side
of the machine, is directed horizontally towards the machine axis.
With the assumption that the magnetic field is due to currents outside
the volume of interest so that $\nabla\times\mathbf{B}=0\Rightarrow$
$(\mathbf{B}\cdot\nabla)\mathbf{B}=B\nabla B$ , the expression for
the combined curvature and $\nabla B$ drifts is:
\[
\mathbf{V}_{\alpha\,c}+\mathbf{V}_{\alpha\,\nabla B}=\frac{1}{\omega_{c\alpha}B^{2}}(V_{\alpha\parallel}^{2}+\frac{1}{2}V_{\alpha\perp}^{2})\mathbf{B}\times\nabla B
\]
This combined drift is vertically directed and is in opposite directions
for electrons and ions because of its dependence on the sign of $q_{\alpha}$.
Left unchecked, the combined drift leads to charge separation and
a vertically directed electric field is established. This field would
then result in $\mathbf{E}\times\mathbf{B}$ particle drifts in the
horizontal direction. The $\mathbf{E}\times\mathbf{B}$ drifts are
much faster than the combined curvature and $\nabla B$ drifts, so
the particles would ultimately reach the outboard vessel walls due
to the $\mathbf{E}\times\mathbf{B}$ drifts, which have been established
by the action of the curvature and $\nabla B$ drifts. The superimposition
of a poloidal magnetic field on the toroidal field prevents particle
loss by this drift mechanism - electrons, which are much lighter and
faster than ions, stream along the twisted helical field lines and
neutralise the polarization that would arise, in the absence of poloidal
field, due to the curvature and $\nabla B$ drifts.

In principle, in magnetic confinement devices, the confined plasma
would be heated to fusion conditions by ohmic heating due to the plasma
current, with possible supplementary heating by external sources such
as ion or electron cyclotron resonance heating, neutral beam heating,
Bernstein waves etc. Overall, magnetic confinement devices work by
confining the plasma at low densities $(\sim10^{20}\,[\mbox{m}^{-3}])$,
and the long term goal is for continuous steady state operation.

Inertial confinement fusion devices are pulsed systems and use inertial
forces to confine a plasma for very short durations, during which
the plasma is heated very rapidly to fusion conditions. Typical densities
are $\sim10^{32}\,[\mbox{m}^{-3}]$, with confinement times of less
than a nanosecond. A schematic of the inertial confinement process
used at the National Ignition Facility (NIF) is illustrated in figure
\ref{fig:Tok_icf}(b). X-rays generated by the action of powerful
laser beams directed on the inner surfaces of the gold \textquotedbl hohlraum\textquotedbl ,
cause the explosion of the outer plastic layer of a $\sim2\mbox{\mbox{\mbox{ mm}}}$
diameter capsule. The resultant implosion of the fuel at the center
of the capsule compresses and momentarily confines the fuel while
heating it to plasma state and then to fusion conditions. In principle,
the implosion must be precisely spherically-symmetric so as to avoid
instabilities that would lead to confinement loss. 

Magnetic target fusion (MTF) occupies the middle ground between the
methods of magnetic and inertial confinement. Like ICF, it is a pulsed
system, but with much longer confinement times $(\sim1\mbox{\,ms})$.
In principle, plasma with low density $(\sim10^{20}\,[\mbox{m}^{-3}])$
and temperature $(\sim100\mbox{ eV})$ is magnetically confined, and
then compressed, by the action of inertial forces on a movable external
flux conserver, to fusion conditions, characterised by moderate density
$(\sim10^{23}\,[\mbox{m}^{-3}])$ and high temperature $(\sim10\mbox{\,keV})$,
such that the Lawson criterion is met. It is expected that the major
obstacle to MCF, which is the long-term maintenance of an instability-free
magnetically confined plasma, may be overcome by reducing the requirement
to that of preserving stability for a short time. At the same time,
it is hoped that the major obstacle to ICF - the difficulty in achieving
an extremely rapid, nearly perfectly symmetric compression - can be
surmounted by compressing the plasma, which is magnetically confined,
over a much longer timescale - the aim is for inertial compression
over $\sim100\,\upmu$s compared with $\sim200$ ps in the ICF regime.

\section{General Fusion - overview\label{sec:General-Fusion--}}

\begin{figure}[H]
\centering{}\subfloat[Reactor overview]{\includegraphics[height=4cm]{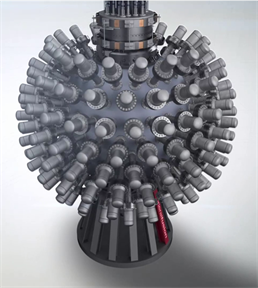}}\hfill{}\subfloat[CT formation into cavity]{\includegraphics[height=4cm]{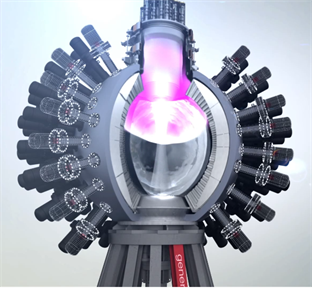}}\hfill{}\subfloat[CT relaxation]{\includegraphics[height=4cm]{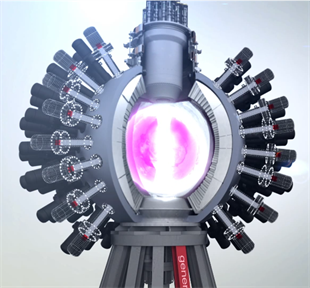}}\hfill{}\subfloat[Piston action $\rightarrow$ cavity collapse $\rightarrow$ CT compression]{\includegraphics[height=4cm]{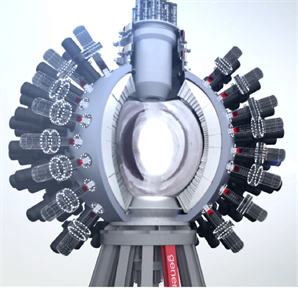}}\caption{\label{fig:GFreact}$\,\,\,\,$Current GF reactor concept {\footnotesize{}(image
credit: General Fusion) }}
\end{figure}
General Fusion (GF), founded in 2002, is a private, investor-funded
company. GF is developing a MTF power plant, based on the concept
of compressing a compact torus (CT) plasma to fusion conditions by
the action of external pistons on a liquid lead-lithium shell surrounding
the CT\cite{Laberge}. The most recent design for the proposed prototype
GF reactor, with outer radius $\sim10$ m, is depicted in figure \ref{fig:GFreact}(a).
A plasma injector, based on a magnetized Marshall gun, sits on top
of a hollow metal sphere that is partially filled with liquid metal
(lead-lithium mixture). The sphere is surrounded by many pistons that
are attached to its surface. Due to the action of jets inside the
sphere, the liquid metal is rotating toroidally inside the hollow
metal sphere. Centrifugal forces due to the toroidal rotation causes
an approximately spherical vacuum cavity to open up inside the spherical
shell of rotating liquid metal. As indicated in figure \ref{fig:GFreact}(b),
the plasma injector forms a compact torus (CT) into the cavity. A
central shaft, which carries vertically-directed current to provide
additional CT-stabilising toroidal field can be seen in the figure.
In this test-prototype design, the solid metal shaft would be damaged
by the immense inboard pressures during compression, and would have
to be replaced after each compression shot. Future designs may use
liquid metal jets as a renewable shaft, or an alternative stabilisation
method may be found - the ultimate goal is for a reactor design that
can accomplish a compression shot every second. The CT is allowed
to self-organise and stabilise - figure \ref{fig:GFreact}(c). The
inner surface of the cavity in the liquid metal is the outer magnetic
flux conserver for the CT. As depicted in figure \ref{fig:GFreact}(d),
the pistons attached to the outer surface of the hollow sphere are
triggered simultaneously, resulting in the addition to the sphere
of more liquid metal that is directed with spherical symmetry radially
inwards through a honeycomb-like mesh located at the inner surface
of the hollow sphere. The additional liquid metal causes the cavity
to collapse over $\sim100\,\upmu$s. In principle, this causes the
CT plasma to be heated to fusion conditions.

\begin{figure}[H]
\centering{}\subfloat[MRT]{\includegraphics[height=6cm]{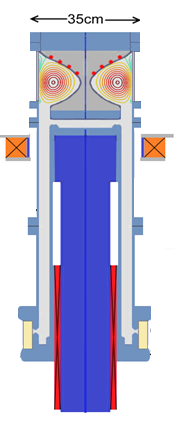}}\hfill{}\subfloat[SPECTOR]{\includegraphics[height=6cm]{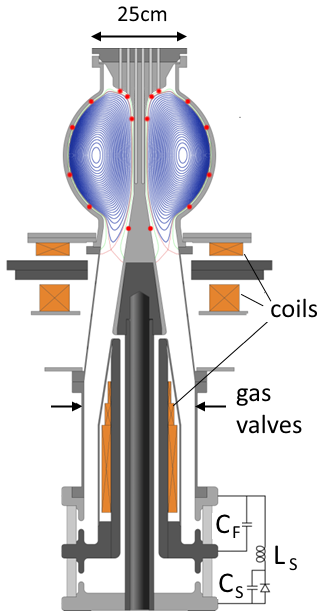}}\hfill{}\subfloat[PI3]{\includegraphics[height=6cm]{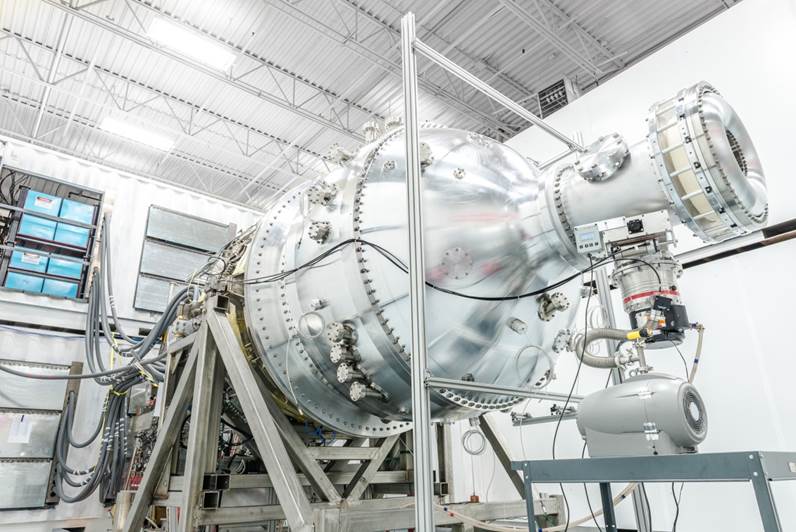}}\caption{\label{fig:GFinjectors}$\,\,\,\,$GF plasma injectors{\footnotesize{} }}
\end{figure}
Figure \ref{fig:GFinjectors} shows three of the four most recent
plasma injectors at GF. Not shown here, the \textquotedbl tokamak\textquotedbl{}
machine was a short-term experiment based on the MRT (Magnetised Ring
Test) design, but with a separate circuit, in addition to the principal
formation circuit, to drive axial central shaft current, thereby producing
CTs with additional stabilising toroidal field, and having an addition
control mechanism for CT $q$ profiles. It was found that maintaining
long-duration shaft current led to longer-lived, more stable CTs.
That experiment paved the way for the SPECTOR (Spherical Compact Toroid)
and PI3 (Plasma Injector 3) designs which also have separate shaft
current circuits. All four injectors operate based on the principle
of the magnetised Marshall gun \cite{RACE_Hammer,RACE Molvik}. The
operation is described in section \ref{subsec:Machine-operation-principles}
below. The MRT and SPECTOR machines are both $\sim2$m in height,
while PI3 is $\sim5$m in length. The MRT machines were in operation
until 2016, when it was replaced by SPECTOR, which is still operating.
PI3 was built in 2017, and has been undergoing testing until now.

To study the plasma physics of compressed CTs, GF has conducted several
PCS (Plasma Compression, Small) tests. In a PCS test, which takes
place outdoors in a remote location, a CT is compressed by symmetrically
collapsing the outer aluminum flux conserver with the use of chemical
explosives. PCS tests are destructive, and therefore do not employ
the full array of diagnostics used in CT formation and characterization
experiments in the laboratory, and can only be executed every few
months, the time needed to fabricate and commission a new injector.
There have been sixteen PCS tests to date - PCS tests $1\rightarrow11$
have been conducted with MRT machines, test $12$ was on the \textquotedbl tokamak\textquotedbl{}
machine, and tests $13\rightarrow16$ have been on SPECTOR machines.
The PCS program goal of significant compressional ion heating has
not yet been achieved. Significant progress, in particular after the
transition to the SPECTOR design, has been made in improving the stability
of compressed CTs.

The magnetic compression experiment at GF was designed as a repetitive
non-destructive test to study plasma physics applicable to MTF compression.
Although not intended to compress a CT to the same degree as a PCS
test, a principal goal of the magnetic compression experiment was
to gain insights to the performance and stability of compressed CTs,
and thereby advance the PCS program.\\

\section{Magnetic geometry of compact tori\label{subsec:Magnetic-geometry-of}}

Spheromaks are self-organizing, force-free plasma configurations with
comparable toroidal and poloidal magnetic fluxes that are self-sustained
by internal currents. A spheromak is an example of a solution to MHD
equilibrium. In equilibrium, the MHD equation of motion reduces to
$\nabla p=\mathbf{J}\times\mathbf{B}$. Almost all laboratory plasmas
have $\beta=2\mu_{0}\,p/B^{2}\ll1$, so it is a good approximation
to set $\nabla p\sim0$, so that $\mathbf{J}\parallel\mathbf{B}$
or, using Ampere's law, $\nabla\times\mathbf{B}\parallel\mathbf{B}$.
Equivalently, we can write $\nabla\times\mathbf{B}=\lambda\mathbf{B}$.
This is the equation of force-free equilibrium. Magnetic helicity
is defined as $K=\int\mathbf{A\cdot B}\,dV$, where $\mathbf{A}$
is the vector potential and $dV$ is an elemental volume. Examples
of fields that contain helicity include those that contain twisted,
knotted, kinked or linked flux tubes, and force-free fields \cite{Bellan_Spheromaks,Bellan_conf,Bellan2}. 

Magnetic helicity can be well described as follows \cite{Berger}:
If we approximate the magnetic field in a closed volume, with the
boundary condition $\mathbf{B\cdot}\hat{\mathbf{n}}$=0 ($i.e.,$
perfectly conducting walls), by a collection of $N$ closed flux tubes,
the quantity 
\[
K=\Sigma_{i=1}^{N}\Sigma_{j=1}^{N}L_{ij}\,\phi_{i}\,\phi_{j}
\]
describes the total mutual flux linkage contained in the volume. Here,
\[
L_{12}=-\frac{1}{4\pi}\ointop_{1}\ointop_{2}\frac{d\mathbf{x}}{d\sigma}\cdot\frac{\mathbf{r}}{r^{3}}\times\frac{d\mathbf{y}}{d\tau}\,d\tau\,d\sigma
\]
is the Gauss linking number ($cf.$ mutual inductance for two closed
conducting loops). The linking number quantifies the linking of two
closed loops, which are labeled here as loops 1 and 2, parametrized
by $\sigma$ and $\tau$ respectively (points on loop 1 are labeled
as $\mathbf{x}(\sigma)$, points on loop 2 are labeled as $\mathbf{y}(\tau$)),
with $\mathbf{r}=\mathbf{y-x}$. Letting $N\rightarrow\infty$ (so
that $\phi_{i}\rightarrow0$), then 
\[
K=-\frac{1}{4\pi}\int\int\mathbf{B}(\mathbf{x})\cdot\left(\frac{\mathbf{r}}{r^{3}}\times\mathbf{B}(\mathbf{y})\right)d^{3}x\,d^{3}y
\]
Use of the Coulomb gauge ($\nabla\cdot\mathbf{A}=0$) implies that
$\mathbf{A}(\mathbf{x})=-\frac{1}{4\pi}\int\frac{\mathbf{r}}{r^{3}}\times\mathbf{B}(\mathbf{y})\,d^{3}y$,
leading to the standard definition of helicity as $K=\int\mathbf{A\cdot B}\,dV$. 

Taylor \cite{Taylor} found that for any magnetized plasma with finite
helicity, there is a unique minimum energy state, characterised by
force free currents ($\mathbf{J}\parallel\mathbf{B}$) and a uniform
inverse length $\lambda$. Spheromak magnetic configurations occupy
a state of minimum energy subject to a given magnetic helicity; the
minimum energy final state of the self-organizing process is defined
by Woltjer-Taylor theory \cite{Taylor,Woltjer}. It can be shown \cite{Berger}
that for all quasi-ideal MHD dynamics, that total magnetic helicity
is conserved; plasmas reach a state of minimal energy equilibrium
while maintaining a constant level of helicity. Magnetic helicity
is conserved even in some processes that require finite plasma resistivity,
such as magnetic reconnection. As shown in \cite{Taylor,Woltjer},
helicity is approximately conserved at low resistivity and exact helicity
conservation is regained as $\eta'\rightarrow0$ (not only at the
limit $\eta'$= 0), where $\eta'\,[\Omega-\mbox{m}]$ is the plasma
resistivity. The minimization of plasma energy subject to the constant
helicity constraint can be done using the method of Lagrange multipliers
and the force-free equation 
\begin{equation}
\nabla\times\mathbf{B}=\lambda\mathbf{B}\label{eq:0.01}
\end{equation}
with constant eigenvalue $\lambda$ {[}m$^{-1}${]}, is the result.
With fixed $\lambda$ and the boundary condition being that $\mathbf{B}$
is parallel to the surface in a closed cylindrical volume, the lowest
order solution to the force-free equation gives the magnetic field
components of a spheromak in that geometry \cite{Schaffer}. 
\begin{figure}[H]
\begin{raggedright}
\subfloat[Tokamak $B$ profiles]{\centering{}\includegraphics[width=6cm,height=6cm]{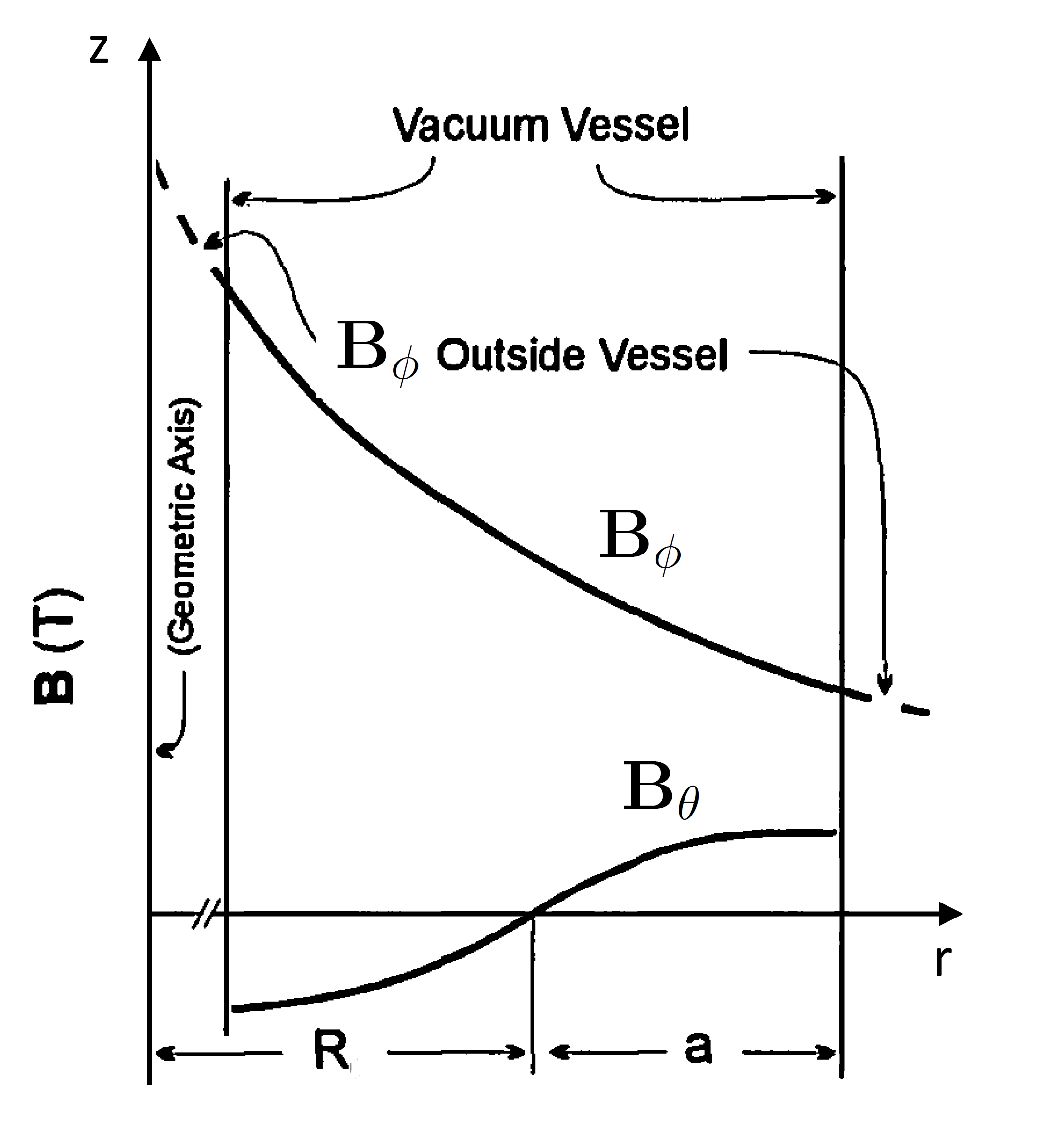}}\hfill{}\subfloat[Spheromak $B$ profiles]{\centering{}\includegraphics[width=6cm,height=6cm]{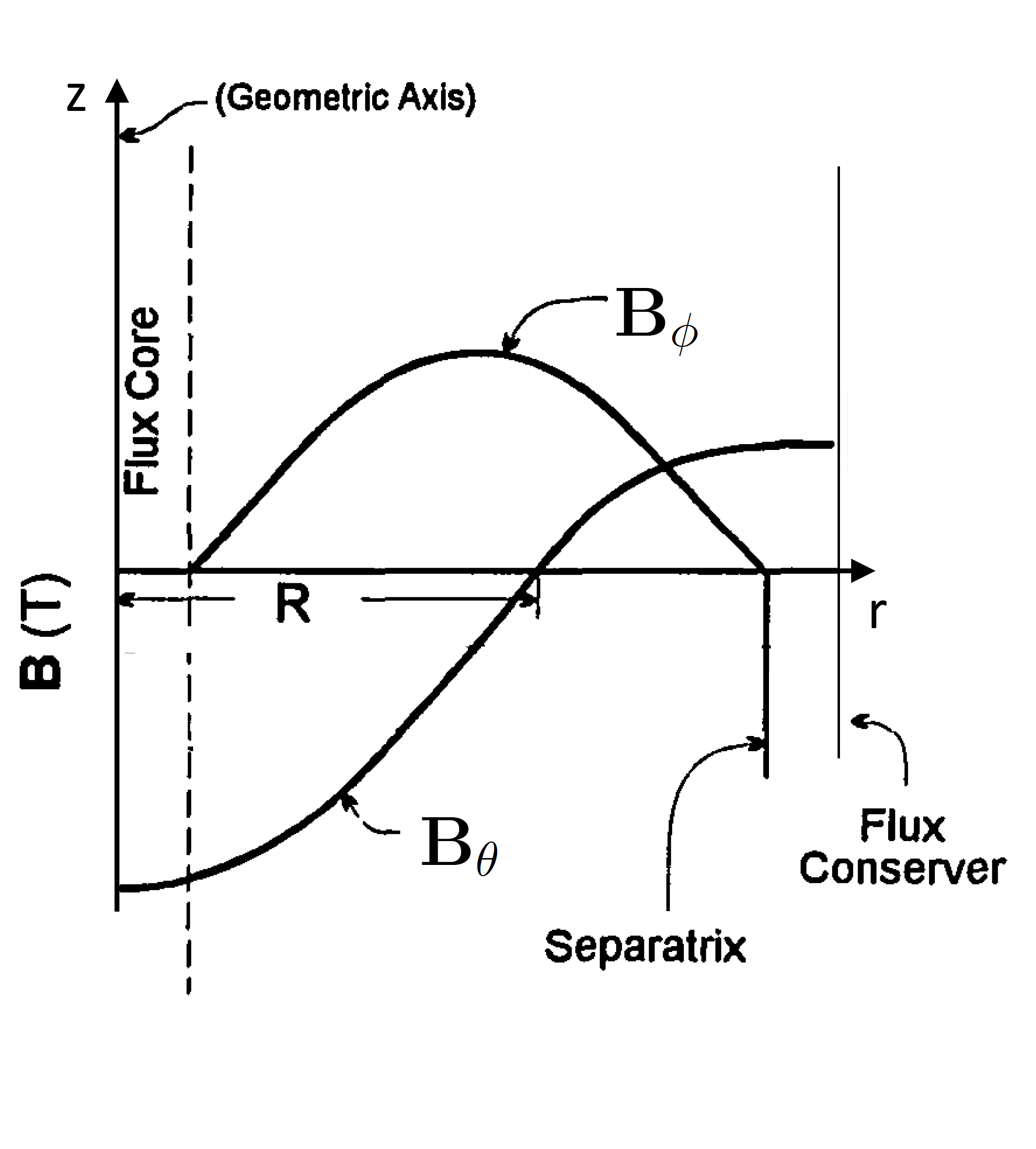}} 
\par\end{raggedright}
\centering{}\caption{\label{fig:Bprof_toka_sphero}$\,\,\,\,$Tokamak and spheromak magnetic
field profiles}
\end{figure}
Figure \ref{fig:Bprof_toka_sphero}(a) indicates the profiles for
the toroidal and poloidal components of the magnetic field for the
case of a tokamak. Note that $B_{\phi}$, the toroidal field, which
is produced by currents in external coils (as depicted in figure \ref{fig:Tok_icf}(a)),
is much larger than $B_{\theta}$, the poloidal field, which is produced
by internal plasma toroidal current, and that it increases towards
the geometric axis where the external coils are closer together. By
contrast, in a spheromak, the field components are of the same order
(figure \ref{fig:Bprof_toka_sphero}(b)), and both field components
are due to internal plasma currents. In the flux core and outside
the separatrix, $B_{\theta}$ is finite due to toroidal currents in
the spheromak. On the other hand, just as there is no field outside
a long solenoid whose ends have been joined to form a torus, $B_{\phi}=0$
at the spheromak edge.

Injection of spheromaks with significant helicity-content into a tokamak
plasma has been proposed as a means for current drive. In helicity
injection current drive, helicity conserving magnetic relaxation processes
incorporate helicity added to the system by dissipating any excess
free energy, increasing parallel currents and keeping the plasma close
to the Taylor state \cite{REdd}. Helicity is injected by driving
current on injector magnetic flux, $\Psi_{main}$. The helicity injection
rate is given by $2V_{gun}\Psi_{main}$ \cite{Brown_Bellan,JArboe_review_spheromak}.
$V_{gun}$, the injection voltage, is the voltage across the helicity
injection device electrodes, and $\Psi_{main}$ is the injector bias
flux. The spheromak's field reconnects (relaxes) to that of the tokamak
to add helicity. It is because helicity is conserved even in the presence
of turbulent tearing that current can be expected to be driven as
a result of helicity injection \cite{4brown}. A CHI (Co-axial Helicity
Injection) device was devised to inject current to STOR-M. I worked
with the CHI device on the STOR-M tokamak in the plasma physics laboratory
at the University of Saskatchewan for about a year before starting
on the magnetic compression project at General Fusion. Various circuit
modifications were made to the CHI circuits, including adding capacitor
bank extensions, and experimenting with high current switch and triggering
configurations. Spheromak injection into STOR-M usually resulted in
disruption of the tokamak discharge. After modifying the CHI device
to operate at increased power, it looked like tokamak current was
increased by a few kiloamps just prior to disruption, but careful
testing proved that the signals indicating a current drive were actually
spurious, caused by inductive pickup. The CHI device was attached
to a portable vacuum chamber, to characterise the CTs produced. Magnetic
probes were constructed, using the principles outlined later in section
\ref{subsec:Magnetic-field-measurements}, to measure poloidal and
toroidal field near the CT edge. Langmuir probes were made and returned
reasonable estimates for edge CT density and temperature. Further
details of the helicity injection experiment are presented in \cite{CHI_STORM}.

\section{Scaling laws for magnetic compression\label{subsec:Scaling-laws-for}}

If a magnetically confined plasma is compressed on a time-scale less
than the resistive magnetic decay time of the plasma, the magnetic
fluxes are conserved. In cylindrical coordinates with azimuthal symmetry,
poloidal flux at the magnetic axis can be expressed, assuming that
the plasma torus has a circular poloidal cross-section, as 
\begin{equation}
\Psi_{axis}=\int_{R}^{R+a}\mathbf{B}_{\theta}\cdot d\mathbf{s}\sim2\pi Ra\,<B_{\theta}>\label{eq:0.011}
\end{equation}
where $d\mathbf{s}=ds\widehat{\mathbf{z}}$, and $ds=2\pi r\,dr$
is an annular elemental area of radius $r$, with its center at $r=0$,
and with the same axial coordinate as the magnetic axis, $a$ {[}m{]}
and $R$ {[}m{]} are the minor and major radii of the plasma torus,
and $<B_{\theta}>$ is the average poloidal field between $r=R$ and
$r=R+a$. Note that $\psi=\frac{1}{2\pi}\Psi$ is the poloidal flux
per radian. The axisymmetric MHD code that was developed in support
of this work (part two of the thesis) evolves $\psi,$ rather than
$\Psi$, and the various other fields that are evolved are expressed
in terms of $\psi$. Both $\Psi$ and $\psi$ will be referred to
in the thesis. 
\begin{figure}[H]
\begin{centering}
\includegraphics[scale=0.4]{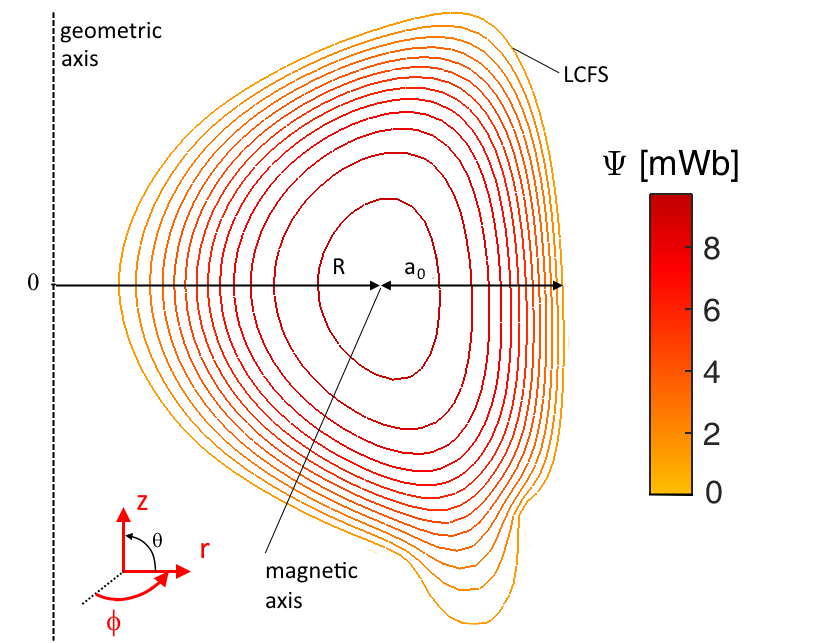}\caption{\label{fig:CT--contours}$\,\,\,\,$CT $\Psi$ contours for magnetic
compression geometry, with geometric definitions}
\par\end{centering}
\end{figure}
For irregular axisymmetric toroidal geometries with non-circular poloidal
cross-sections, as associated with the magnetic compression experiment,
the relationship expressed in equation \ref{eq:0.011} also approximately
holds, where, as indicated in figure \ref{fig:CT--contours}, $a\rightarrow a_{0}=a(\theta=0)$
is the distance from the magnetic axis to the last closed poloidal
flux surface (LCFS) at poloidal angle $\theta=0$. The magnetic axis,
LCFS, and scaling of $\Psi$ are also indicated in the figure. Note
that the poloidal field should be approximately perpendicular to the
line denoted as $a_{0}$ in the figure for the relationship to hold.
With poloidal flux conservation, equation \ref{eq:0.011} yields the
compression scaling laws $\frac{B_{\theta2}}{B_{\theta1}}=\frac{a_{1}R_{1}}{a_{2}R_{2}}$,
where the subscripts $2$ and $1$ indicate post-compression and pre-compression
parameters respectively. Introducing the notation 
\[
\widetilde{x}=\frac{x_{2}}{x_{1}}
\]
the compression scaling for $B_{\theta}$ can be equivalently expressed
as $\widetilde{B}_{\theta}=\widetilde{a}^{-1}\widetilde{R}^{-1}$
or $B_{\theta}\rightarrow a^{-1}R^{-1}$, $ie$., the post compression
poloidal field magnitude scales inversely with the product of the
post compression minor and major radii. For simplicity, the subscript
$2$ will be dropped for the remainder of this section. Toroidal flux
can be expressed to the first order as $\phi=\int\mathbf{B}_{\phi}\cdot d\mathbf{s}_{\phi}\sim S\,<B_{\phi}>$,
where $S$ is the cross-sectional area of the plasma torus in the
poloidal plane, and $<B_{\phi}>$ is the average toroidal field in
the cross-section, yielding the compression scaling for the toroidal
field as $B_{\phi}\rightarrow S^{-1}$. An assumption of adiabatic
compression implies that there is no energy lost due to loss of particles
and their associated energy. If the compression time is short compared
with the particle confinement time, the compression scaling for the
particle number density is $n\rightarrow V^{-1}$, where $V$ is the
volume of the plasma torus. If the compression time is also short
compared with the thermal confinement time, adiabaticity implies constant
$p\rho^{-\gamma}=nT\,\left(nm_{i}\right)^{-\frac{5}{3}}$ so that,
using the scaling for $n$, $T\rightarrow n^{\frac{2}{3}}\rightarrow V^{-\frac{2}{3}}$.\\
\\
Here, $\gamma=\frac{c_{p}}{c_{v}}$ is the adiabatic gas constant,
where $c_{p}\,${[}J$\,$kg$^{-1}$K$^{-1}${]} is the specific heat
capacity at constant pressure, and $c_{v}\,${[}J$\,$kg$^{-1}$K$^{-1}${]}
is the specific heat capacity at constant volume. $\gamma=\frac{N+2}{N}$,
where $N$ is the number  of degrees of freedom of the bulk species.
For a monatomic gas, $N=3,$ so that $\gamma=\frac{5}{3}.$ For a
given material, the availability of translational, rotational and
intra-molecular exchanges characterise the value of $N$. Stokes theorem
and the integral of Ampere's law in the poloidal plane yields $B_{\theta}L=\mu_{0}I_{p},$
where $L$ is the path length along the LCFS, so, using the scaling
for $B_{\theta}$, $I_{p}\rightarrow L\,a^{-1}R^{-1}$. Toroidal beta
(standard beta) is the ratio of the plasma pressure to the toroidal
field contribution to magnetic pressure: $\beta_{\phi}=\frac{p}{p_{M}}=\frac{2\mu_{0}nT}{B_{\phi}^{2}}$.
Using the scalings for $n,\,T$, and $B_{\phi}$, this gives the scaling
for toroidal beta as $\beta_{\phi}\rightarrow V^{-\frac{5}{3}}S^{2}$.
In the same manner, the compression scaling for poloidal beta is found
as $\beta_{\theta}\rightarrow V^{-\frac{5}{3}}a^{2}R^{2}$. For geometries
with circular poloidal-plane cross-sections, the volume of the plasma
torus can be approximated as $V\sim2\pi^{2}Ra^{2}$, the cross-sectional
area is $S=\pi a^{2}$, and the LCFS path length is $L=2\pi a$, so
the scalings can be expressed in terms of $a$ and $R$. 
\begin{table}[H]
\centering{}%
\begin{tabular}{|c|c|c|c|c|c|c|c|}
\hline 
parameter & $B_{\theta}$ & $B_{\phi}$ & $n$ & $T$ & $I_{p}$ & $\beta_{\phi}$ & $\beta_{\theta}$\tabularnewline
\hline 
scaling &  &  &  &  &  &  & \tabularnewline
(general case) & $a_{0}^{-1}R^{-1}$ & $S^{-1}$ & $V^{-1}$ & $V^{-\frac{2}{3}}$ & $L\,a_{0}^{-1}R^{-1}$ & $V^{-\frac{5}{3}}S^{2}$ & $V^{-\frac{5}{3}}a_{0}^{2}\,R^{2}$\tabularnewline
\hline 
scaling  &  &  &  &  &  &  & \tabularnewline
(circular $\times$ section) & $a^{-1}R^{-1}$ & $a^{-2}$ & $a^{-2}R^{-1}$ & $a^{-\frac{4}{3}}R^{-\frac{2}{3}}$ & $R^{-1}$ & $a^{\frac{2}{3}}R^{-\frac{5}{3}}$ & $a^{-\frac{4}{3}}R^{\frac{1}{3}}$\tabularnewline
\hline 
\end{tabular}\caption{\label{tab:Scaling-parameters-for}$\,\,\,\,$Parameter scalings for
adiabatic compression }
\end{table}
The principal adiabatic compression scalings for toroidal geometry
with circular and irregular poloidal cross-sections are collected
in table \ref{tab:Scaling-parameters-for}. In chapter \ref{chap:Simulation-results},
the adiabatic compression scalings for the general case will be compared
with MHD simulation results.

The compressional evolution of plasma temperature is determined by
non-adiabatic as well as adiabatic processes. Adiabatic processes
include the compressional work associated with the external magnetic
field and the loss of electron energy due to ionization of neutral
particles \cite{Dendy,ZeroD Adiabatic Comp Scaling} (note that electron
thermal energy is converted to potential energy in the ionization
process). Impurity radiation, radiative recombination losses in an
optically thin plasma, losses associated with finite particle confinement
times, and ohmic heating are examples of non-adiabatic processes \cite{ZeroD Adiabatic Comp Scaling}.
Ohmic heating involves resistive heating by plasma currents, and can
be viewed as the conversion of magnetic to thermal energy, part of
which is lost from the system by heat diffusion. Classical ohmic heating
power assumes that plasma resistivity is determined by the Spitzer
formula: 
\begin{equation}
\eta'[\Omega-\mbox{m}]=\frac{\pi Z_{eff}\,e^{2}\sqrt{m_{e}}\,\varLambda}{\left(4\pi\epsilon_{0}\right){}^{2}\left(T_{e}\,[\mbox{J}]\right)^{1.5}}\label{eq:0.1}
\end{equation}
Here, $Z_{eff}$ is the effective (volume averaged) ion charge, $e$
is the electron charge, $m_{e}$ is the electron mass, $\varLambda\sim10$
is the Coulomb logarithm, $\epsilon_{0}$ is vacuum permittivity,
and $T_{e}$ is the ion temperature in Joules. However, anomalous
resistivity can occur when the ratio of the electron fluid drift velocity
to the thermal or Alfven velocities is too high \cite{ZeroD Adiabatic Comp Scaling}.
Both ratios are expected to decrease with adiabatic compression, so
it seems unlikely, but remains unconfirmed, that anomalous resistivity
will increase during compression \cite{ZeroD Adiabatic Comp Scaling,Furth}.\\
\\
It is interesting how ohmic heating power density scales with adiabatic
compression: ohmic heating power density is $P_{\Omega}=\eta'J_{\phi}^{2}$
{[}W/m$^{3}${]}. The Spitzer resistivity formula gives the compression
scaling for resistivity: $\eta'\propto T_{e}^{-\frac{3}{2}}\Rightarrow\eta'\rightarrow V$.
Toroidal current density is $J_{\phi}=I_{p}/S$. With the approximation
$V\sim RS$, this implies that $\Rightarrow J_{\phi}\rightarrow L\,a_{0}^{-1}V^{-1}$,
so that $P_{\Omega}\rightarrow L^{2}\,a_{0}^{-2}V^{-1}$. Power density
lost from the plasma due to line radiation is $P_{rad}\propto n_{e}\,n_{imp}$,
where $n_{imp}$ {[}m$^{-3}${]} is the impurity number density, so
that the associated compression scaling is $P_{rad}\rightarrow V^{-2}$. 

In cases with circular poloidal cross-section, where $P_{\Omega}\rightarrow V^{-1}$,
the relation $P_{rad}\rightarrow\left(P_{\Omega}\right)^{2}$ implies
that there may be no compressional heatings of electrons for a plasma
where there is an initial balance between ohmic heating and radiative
losses \cite{ZeroD Adiabatic Comp Scaling}. Mitigating this is the
effect that increased $n_{e}$ leads to an increased rate of ionization,
so that the impurities can evolve to a less radiative charge state.
The condition in which this effect allows ohmic compressional heating
to be effective in radiation dominated plasmas is explored in \cite{Carolan}.
In that work, it was shown that when the pre-compression electron
temperature is greater than 25 eV in plasmas where oxygen is the dominant
impurity, or greater than 20 eV when carbon is the dominant impurity,
and if $n_{e}\tau_{p}\gtrsim1\times10^{15}$ s/m$^{3}$ ($\tau_{p}$
is the particle confinement time), then compression can raise electron
temperature significantly even for radiation dominated plasmas. In
\cite{ZeroD Adiabatic Comp Scaling}, the compression of a low temperature
spheromak was simulated with a time dependent power balance model.
It was found that, since the input energy due to ohmic heating is
much greater than the work done in compressing the ions and electrons,
that the relationship between $\tau_{p}$ and the time over which
compression is executed is not important as long as both are much
shorter than the magnetic decay time. The importance of the effect
of identical adiabatic compression scalings for toroidal current density
and electron density, in cases with circular poloidal cross-section,
was highlighted. This allows high temperature states to be achieved
through the effect of increasing $n_{e}$ leading to an increased
ionization rate and less radiative impurity ions. An increased ionization
rate leads to higher $Z_{eff}$, which in turn leads to a further
increase in ohmic heating power, as $\eta'\propto Z_{eff}$. It was
found that the most significant compressional electron temperature
increases are for cases where the radiation power loss dominates the
power loss caused by particle recycling \cite{ZeroD Adiabatic Comp Scaling}.\\
\\
Compression in minor radius, while keeping major radius constant,
can be achieved by ramping up the toroidal field in a tokamak. In
this compression scenario, $C>1$, the geometric compression ratio,
is defined as $C=\widetilde{a}^{-1}=\frac{a_{1}}{a}$, or $a\rightarrow C^{-1}$.
The vertical field due to toroidal currents in external poloidal field
coils (see figure \ref{fig:Tok_icf}(a)) is also increased so as to
maintain constant $R$. This type of compression was done on the TUMAN
series of tokamaks \cite{Tuman-2Aconf}. In this scenario, with circular
poloidal cross-section, poloidal beta increases as 
\[
\beta_{\theta}\rightarrow a^{-\frac{4}{3}}R^{\frac{1}{3}}\Rightarrow\frac{\beta_{\theta}}{\beta_{\theta1}}=\left(\frac{a}{a_{1}}\right)^{-\frac{4}{3}}\left(\frac{R}{R_{1}}\right)^{\frac{1}{3}}=C^{+\frac{4}{3}}
\]
while toroidal beta decreases as $\beta_{\phi}\rightarrow C^{-\frac{2}{3}}$.
In a tokamak, where $B_{\phi}\gg B_{\theta}$, the magnetic energy
in the plasma can be approximated as $U_{M}=\frac{1}{2\mu_{0}}\int B_{\phi}^{2}\,dV$.
The volume of integration doesn't change with compression, since the
toroidal field is externally imposed, so that magnetic energy scales
as $U_{M}\rightarrow S^{-2}$ (general-case scalings). The plasma
thermal energy is $U_{Th}=\frac{1}{\gamma-1}\int p\,dV$, so that
$U_{Th}\rightarrow nT\,V\rightarrow V^{-\frac{2}{3}}$. For constant
major radius compression in cases with circular poloidal cross-section,
the total magnetic energy of the machine ($U_{M}\rightarrow a^{-4}$)
increases much faster than the thermal energy of the plasma ($U_{Th}\rightarrow a^{-\frac{4}{3}}$),
so that this type of compression is very inefficient in terms of magnetic
energy input to thermal energy gain.

Compression that maintains constant aspect ratio $R/a$ , with circular
poloidal cross-section, was designated as $Type\,A$ compression in
an early paper \cite{Furth} which has been referenced in many subsequent
papers dealing with magnetic compression. $Type\,A$ compression is
achieved by sharply ramping up external $B_{z}$, the vertical field
due to toroidal currents in external poloidal field coils at the same
time as external $B_{\phi}$ is increased. In this case, both $R$
and $a$ scale in proportion to $C^{-1}$. Since $R$ is not constant,
the $1/R$ scaling of $B_{\phi}$ has to be accounted for when estimating
the total machine magnetic energy, which is again approximately $U_{M}=\frac{1}{2\mu_{0}}\int B_{\phi}^{2}\,dV$.
The pre-compression toroidal field is $B_{\phi1}(R)=\frac{B_{\phi11}R_{1}}{R}$,
where $B_{\phi11}$ is the pre-compression toroidal field at pre-compression
major radius $R=R_{1}$. The post-compression toroidal field is $B_{\phi2}(R)=\frac{B_{\phi22}R_{2}}{R}$,
where $B_{\phi22}$ is the post-compression toroidal field at post-compression
major radius $R=R_{2}$. This leads to a compression scaling for toroidal
field at a fixed radius $R$: $\frac{B_{\phi2}(R)}{B_{\phi1}(R)}=\frac{B_{\phi22}R_{2}}{B_{\phi11}R_{1}}=\left(\frac{a}{a_{1}}\right)^{-2}\frac{R}{R_{1}}=C^{2}C^{-1}=C$,
so that $B_{\phi}(R)\rightarrow C$. The total machine magnetic energy
adiabatic compression scaling for $Type\:A$ compression is then $U_{M}\rightarrow C^{2}$.
The expression for plasma thermal energy scaling remains as $U_{Th}\rightarrow a^{-\frac{4}{3}}R^{-\frac{2}{3}}$.
For $Type\,A$ compression, this is $U_{Th}\rightarrow C^{2}$, same
as the scaling for magnetic energy. Thus, $Type\,A$ compression is
far more energy-efficient than the constant $R$ method of compression.
Toroidal and poloidal beta scale equally in proportion to $C$. 

External toroidal field coil current is unmodified in $Type\,B$ compression
\cite{Furth}, which was also defined for the case with circular poloidal
cross-section. Increasing external $B_{z}$ pushes the plasma inwards
towards the machine axis, where the toroidal field is larger. In this
compression scenario the geometric compression ratio is defined as
$C=\frac{R_{1}}{R}$, or $R\rightarrow C^{-1}$. Minor radius scales
as $a\rightarrow C^{-\frac{1}{2}}$, and toroidal beta increases more
than poloidal beta at compression. The expression for scaling of toroidal
field at a fixed radius $R$ is the same as that for $Type\,A$ compression:
$\frac{B_{\phi22}(R)}{B_{\phi11}(R)}=\left(\frac{a}{a_{1}}\right)^{-2}\frac{R}{R_{1}}.$
For $Type\,B$ compression this leads to $B_{\phi}(R)\rightarrow constant$,
so that the total machine magnetic energy scales with the magnetic
energy associated with $B_{z}$, which is orders of magnitude less
than the total machine magnetic energy, and is comparable to the thermal
energy, which scales as $U_{Th}\rightarrow C^{\frac{4}{3}}$.

\section{Prior magnetic compression experiments\label{subsec:Prior-magnetic-compression}}

The experiment on which this work is based represents the first time
that magnetic compression has been attempted on plasmas produced by
a magnetised Marshall gun. In the past, mostly in the 1970's and 1980's,
there were several studies looking at magnetic compression of conventional
tokamak plasmas. Magnetic compression of spheromaks was the focus
of the S-1 experiment, and there have also been a few studies of magnetic
compression of reversed field pinches (RFPs) and field reversed configurations
(FRCs).

\subsection{S-1 spheromak }

\begin{figure}[H]
\begin{raggedright}
\subfloat[]{\centering{}\includegraphics[width=8cm,height=8cm]{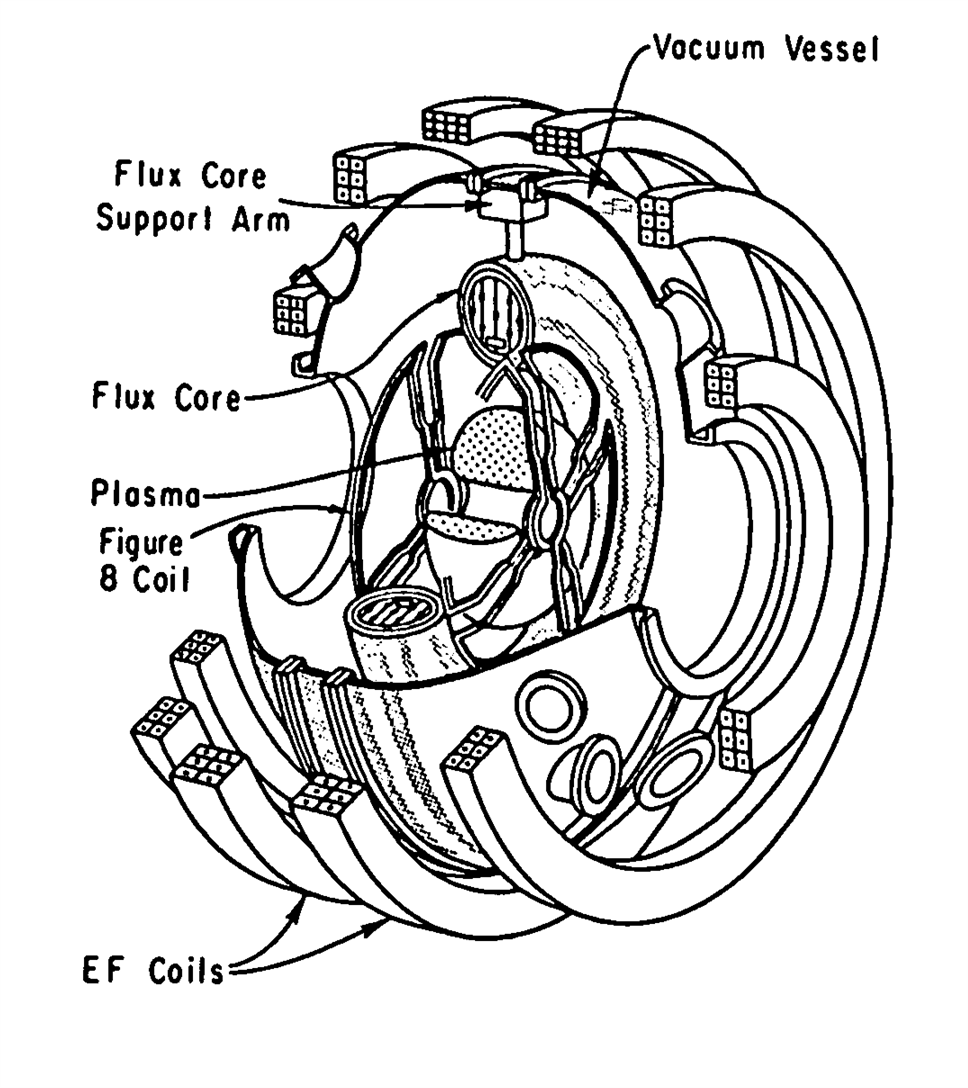}}\hfill{}\subfloat[]{\centering{}\includegraphics[width=8cm,height=8cm]{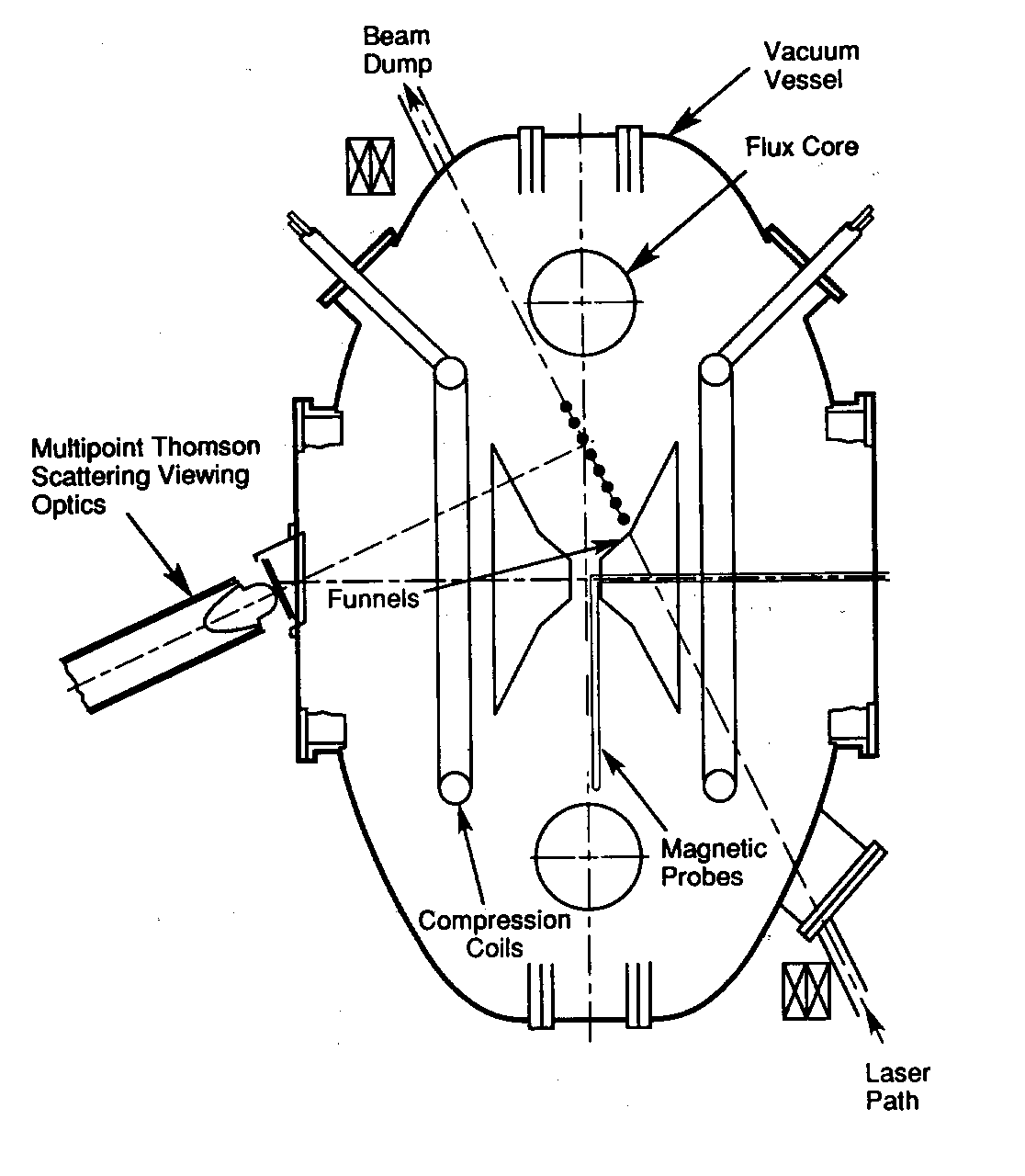}} 
\par\end{raggedright}
\centering{}\caption{\label{fig:S1}$\,\,\,\,$Schematics of S-1 device {\footnotesize{}(image
credit \cite{S1spheromak,S1_compression})}}
\end{figure}
The S-1 device ($R=50$ cm, $a=25$ cm, pre-compression) \cite{S1spheromak,S1_compression}
used a toroidally shaped flux core to generate spheromaks by inductive
means. The flux core indicated in figures \ref{fig:S1}(a) and (b)
contains both poloidal and toroidal windings. A quasi steady-state
equilibrium axial magnetic field is produced by an additional two
toroidal turns in the flux core and by toroidal currents in the external
equilibrium field (EF) coils. To prevent outgassing, while maintaining
rapid flux penetration, the flux core coils are covered with a very
thin layer (<0.5 mm) of resistive inconel. Despite its complexity,
inductive formation has the advantage, over the more conventional
formation technique using magnetized Marshall guns (\cite{RACE_Hammer,RACE Molvik},
see also section \ref{subsec:Machine-operation-principles} for the
operating principles), that there are no large formation currents
flowing between electrodes, thereby eliminating the source of impurities
to the plasma that arises at the electrode surfaces. 

Stabilization of the tilt instability (which can be envisaged by seeing
the spheromak as a magnetic dipole directed against the background
axial field) and shift instabilities, during formation and compression,
was achieved on S-1 with the use of passive conducting stabilisers.
The \textquotedbl figure 8\textquotedbl{} stabilizers used initially
(see figure \ref{fig:S1}(a)), before the installation of the in-vessel
magnetic compression coils, required that more than 10\% of spheromak
poloidal flux remained linked to the flux core \cite{S1spheromak,S1_tilt}
for MHD stability. The funnel-shaped stabilisers used later removed
the requirement for flux linking. These are depicted, along with the
in-vessel compression coils, in figure \ref{fig:S1}(b). Note that
compression coils located inside the vacuum vessel were also part
of the recent design proposal, presented in \cite{Woodruff_AdiabaticComp},
for a spheromak magnetic compressor. Current with a rise time of 100
to 200$\,\upmu$s was driven in the S-1 compression coils with a 1
MJ capacitor bank. The plasma-internal magnetic probe array and the
sampling locations for the ten-point Thomson scattering system used
to find electron temperature and density are indicated in figure \ref{fig:S1}(b).

Pre-compression S-1 spheromaks had toroidal plasma current $I_{p}\sim$200
kA, corresponding to peak poloidal field $B_{\theta}\sim$0.15 T.
If a plasma is compressed self-similarly, so that the minor radius
decreases in the same proportion as the major radius, toroidal and
poloidal fluxes are expected to be conserved, so that the Taylor equilibrium
state is maintained. A geometrical compression factor of $\frac{R_{1}}{R}\sim\frac{a_{1}}{a}\sim C\sim$1.6
(recall that $R_{1}$ and $a_{1}$ denote the pre-compression major
and minor radii, while $R$ and $a$ denote the post compression radii)
was achieved on the S-1 device \cite{S1_compression}, leading the
researchers to classify the regime as $Type\,A$ compression (see
section \ref{subsec:Scaling-laws-for}). The deviation of major radius
compression scaling from that of minor radius compression scaling
was found to be less than 10\%. The availability of spheromak-internal
temperature, density, and magnetic field point diagnostics allowed
the researchers to produce shot-averaged poloidal flux contours over
the compression cycle, and enabled determination of the compression
scalings for density, temperature and magnetic field, and comparison
with the predicted scalings. 
\begin{figure}[H]
\begin{centering}
\subfloat{\centering{}\includegraphics[width=15cm,height=8cm]{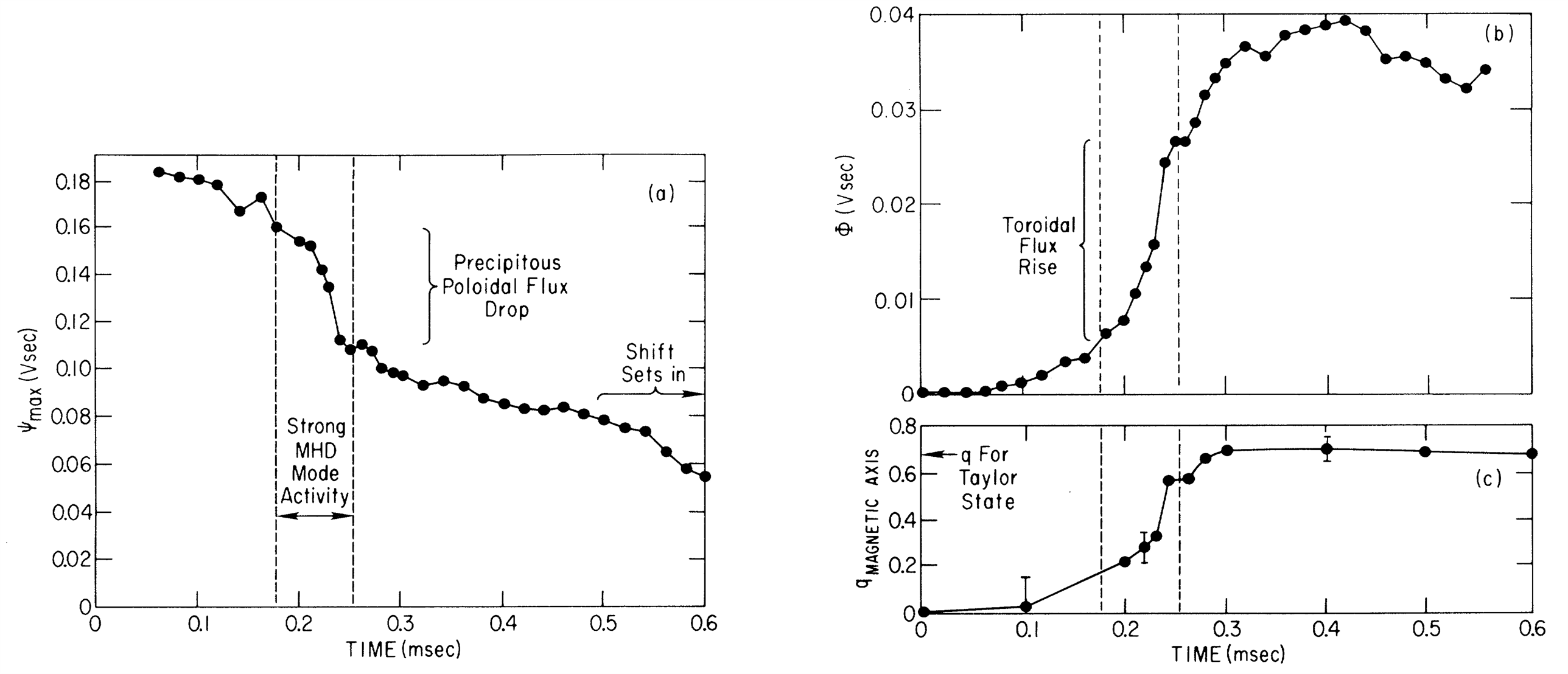}} 
\par\end{centering}
\centering{}\caption{\label{fig: S1relax}$\,\,\,\,$S-1 device; evolution of fluxes and
$q_{axis}$ {\footnotesize{}(image credit: modified from \cite{Janos1})}}
\end{figure}
Fine spatial resolution of the magnetic field measurements allowed
for experimental confirmation of several aspects of basic theory.
Figure \ref{fig: S1relax} indicates the experimentally determined
evolution of the poloidal and toroidal magnetic fluxes and of $q$
at the magnetic axis. The relaxation of the magnetic configuration
towards the minimum energy (Taylor) state is characterised by the
rapid conversion of poloidal to toroidal flux from 180$\,\upmu$s
to 250$\,\upmu$s \cite{Taylor,Janos,Janos1}. Good spatial resolution
of field measurements also enabled experimental confirmation of the
approximate conservation of the individual fluxes, and of the absence
of flux conversion, during compression. The level of conservation
of the fluxes was improved by increasing the compression field - this
may have been related to the increased electron temperature (reduced
resistivity) seen with more compression. The total spheromak magnetic
energy was measured directly and shown to increase with compression.
It was shown experimentally that the local electron beta remained
constant during compression ($i.e.,\,n_{e0}T_{e0}\sim B_{\phi0}^{2}$,
where the subscript $0$ indicates peak values around the magnetic
axis), and that the compression was non-adiabatic. Density scaled
as $\frac{n_{e}}{n_{e1}}\propto C$, compared with the prediction
of $\frac{n_{e}}{n_{e1}}\sim C^{3}$ for the case of particle conservation.
Particle losses were also observed for uncompressed S-1 spheromaks,
but $\tau_{p},$ the particle confinement time, was found to decrease
significantly with compression. This was thought to have been due
partially to the effect of geometric shrinking, and largely due to
some extra particle loss mechanism, that was attributed to the enhanced
fluctuation level that was associated with increased drift velocities. 

Electron current density ($J_{\phi}\sim I_{p}/\pi a^{2}$) was seen
to increase significantly at compression, with approximately the predicted
scaling of $\frac{J_{\phi}}{J_{\phi1}}\sim\frac{I_{p}}{I_{p1}}\left(\frac{a_{1}}{a}\right)^{2}\sim C\,C^{2}=C^{3}$.
Ion temperatures were measured based on Doppler broadening of impurity
line radiation. The contribution of Stark broadening to the observed
temperature broadening was determined to be small for the observed
transitions at the electron densities measured. Electron and ion temperatures
were observed to increase at compression, but not with the adiabatic
scaling of $\frac{T}{T_{1}}\sim C^{2}$. Peak $T_{e}$ rose from \textasciitilde 40
eV to \textasciitilde 100 eV with compression, and ion temperatures
of up to 500 eV were measured at compression. Prior to compression,
$T_{i}$ was generally greater than $T_{e}$ by a factor of two to
four, and was up to a factor of five greater than $T_{e}$ at peak
compression. In general, it was found that $T_{i}$ correlated well
with $T_{e}$, and $T_{i}$ also correlated with the level of the
fluctuations recorded in $B.$ It was concluded that $T_{e}$ increased
at compression largely due to increased ohmic heating as $I_{p}$
increased, and that ion temperature increases were largely due to
anomalous non-collisional heating mechanisms (microinstabilities such
as drift waves and ion cyclotron waves) that were excited by the large
values of $J_{\phi}/n_{e}$ at compression. These microinstabilities
were thought to be related to fluctuations observed in magnetic field
measurements, which were enhanced at compression \cite{S1_compression}. 

\subsection{ATC tokamak}

\begin{figure}[H]
\begin{centering}
\subfloat{\centering{}\includegraphics[width=10cm,height=11cm,keepaspectratio]{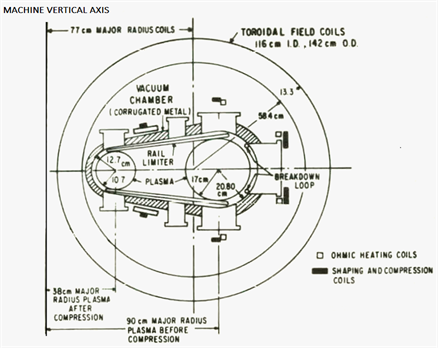}} 
\par\end{centering}
\centering{}\caption{\label{fig:ATC}$\,\,\,\,$Poloidal cross-section of ATC device {\footnotesize{}(image
credit: \cite{ATCconf})}}
\end{figure}
The Adiabatic Toroidal Compressor (ATC) experiment ($R=90$ cm, $a=17$
cm, before compression), which was operated in the 1970's, employed
$Type\,B$ compression. As in standard tokamaks, the vacuum vessel
in enclosed by the toroidal field coils, but in the ATC, molybdenum
rail limiters guide the plasma inwards at compression, which is activated
by increasing toroidal current in the compression coils depicted in
figure \ref{fig:ATC} from 2 to 10 kA over 2 ms. To allow for rapid
penetration of compression field, the vacuum vessel shell was quite
electrically resistive, made from 0.7 mm thick stainless steel. The
principal obstacle to successful compression was that radial field
errors would cause the plasma column to shift vertically, leading
to excessive interaction between the plasma and the rail limiters
\cite{ATCpaper}, and ultimately a disruptive instability. The radial
field errors were thought to be due to the effect of eddy currents
induced in the vacuum vessel. A solution allowing completion of the
full compression cycle was found by imposing an optimised radial correction
field, and installing passive stabiliser coils that opposed any remaining
vertical shift \cite{ATCpaper}. A Thomson scattering (TS) system
was used to obtain measurements of $T_{e}$ and $n_{e}$ at points
inside the plasma column. TS electron densities were calibrated using
microwave interferometry measurements - vertical interferometry chords
were located at the pre-compression and post-compression major radii
of 90 cm and 38 cm. Charge exchange spectroscopy was used to measure
ion temperature. 

Compression in $a$, $R$, $B_{\phi}$, $I_{p}$, $n_{e}$, and $T_{i}$
was observed with scalings consistent with predictions for $Type\,B$
compression.
\begin{table}[H]
\centering{}%
\begin{tabular}{|c|c|c|}
\hline 
parameter & Pre-compression & Post-compression\tabularnewline
\hline 
\hline 
$R$ & 90 cm & 38 cm\tabularnewline
\hline 
$a$ & 17 cm & 10 cm\tabularnewline
\hline 
$B_{\phi}$ & 1.5 T & 4.6 T\tabularnewline
\hline 
$I_{p}$ & 60 kA & 150 kA\tabularnewline
\hline 
$T_{i}$ & 200 eV & 600 eV\tabularnewline
\hline 
$T_{e}$ & 1 keV & 2 keV\tabularnewline
\hline 
$<n_{e}>$ & $1-2\times10^{19}$ m$^{-3}$ & $10^{20}$ m$^{-3}$\tabularnewline
\hline 
\end{tabular}\caption{\label{tab:ATCparams}$\,\,\,\,$Parameter compression scalings on
ATC device }
\end{table}
Variation of ATC parameters for typical compressed discharges are
indicated in table \ref{tab:ATCparams}. While $T_{i}$ scaled as
$T_{i}\rightarrow C^{\frac{4}{3}},$ where $C=90/38=2.37$, $T_{e}$
was found to scale in proportion to $C$, and the explanation given
\cite{ATCpaper} was that $\tau_{Ei}>\tau_{comp}\sim\tau_{Ee}$, where
$\tau_{Ei}$ and $\tau_{Ee}$ are the ion and electron energy confinement
times, and $\tau_{comp}\sim2$ ms is the time over which the compression
field is increased to its maximum, so that $T_{e}$ could not be expected
to follow the adiabatic scaling law. Since, $T_{e}\rightarrow C$
in the ATC experiment, while $n_{e}$ and $B_{\theta}$ scale with
the $Type\,B$ predictions, it could be expected that $\beta_{\theta e}=2\mu_{0}n_{e}T_{e}/B_{\theta}^{2}$
should remain constant at compression, and this was verified, at least
in the case with relatively low 40 kA pre-compression plasma current. 

The ATC compression mechanism is similar to the \textquotedbl radial
magnetic pumping\textquotedbl{} scheme proposed in 1969 \cite{Artsimovich},
in which it was suggested that a tokamak plasma would be maintained
over several $B_{z}$ compression cycles, and that ions could be heated
further at each compression. However, each ATC discharge terminated
in a disruption when the plasma column was pushed on to the inner
limiter that protects the vacuum vessel. In \cite{ATCpaper}, it was
recommended that high frequency magnetic compression on ATC would
be technically difficult, but that low frequency compression on a
device, with dimensions increased to five to ten times those of ATC,
might lead to fusion initiation. 

\subsection{Ultra Low q device }

The scenario of attaining ohmic ignition of fusion through the combination
of an ultra-low-$q$ (ULQ) discharge and adiabatic magnetic compression
was explored in \cite{Inoue}. The safety factor in the cylindrical
approximation with $R\gg a$ is $q(r)=\frac{rB_{\phi}}{RB_{\theta}}$.
With this definition, the current density can be expressed, using
Ampere's law, as $J_{\phi}(r)=\frac{2B_{\theta}}{\mu_{0}r}=\frac{2B_{\phi}}{\mu_{0}qR}$.
The ULQ regime explored on the REPUTE-1 device ($R=82$ cm, $a=22$
cm), has safety factor $q$ less than one, which, compared with traditional
tokamak regimes, enables higher toroidal current density for the same
toroidal field, so that ohmic heating is increased, reducing the need
for supplementary heating. Application of minor radius magnetic compression
leads to a further increase in ohmic heating. 

Low $q$ operation is also associated with high beta ($i.e.,$ low
toroidal field), which is advantageous for the design of compact,
relatively low cost reactors \cite{Kamada}. The energy confinement
time, which is key to meeting the Lawson criterion, is generally higher
for high $q$ - for ohmically heated plasmas the energy confinement
time is found experimentally to scale as 
\begin{equation}
\tau_{E}\sim0.07\,\bar{n}_{e}a\,R^{2}q_{edge}\label{eq:0.2}
\end{equation}
where $\bar{n}_{e}$ is the average electron density in units of $10^{20}$
m$^{-3}$, and $q_{edge}$ is the edge safety factor \cite{WessonTokamaks}.
In general, tokamaks have high $q$ and low beta, while reversed field
pinches (RFPs) have low $q$ and high beta. 
\begin{figure}[H]
\begin{centering}
\subfloat{\centering{}\includegraphics[width=7cm,height=8cm]{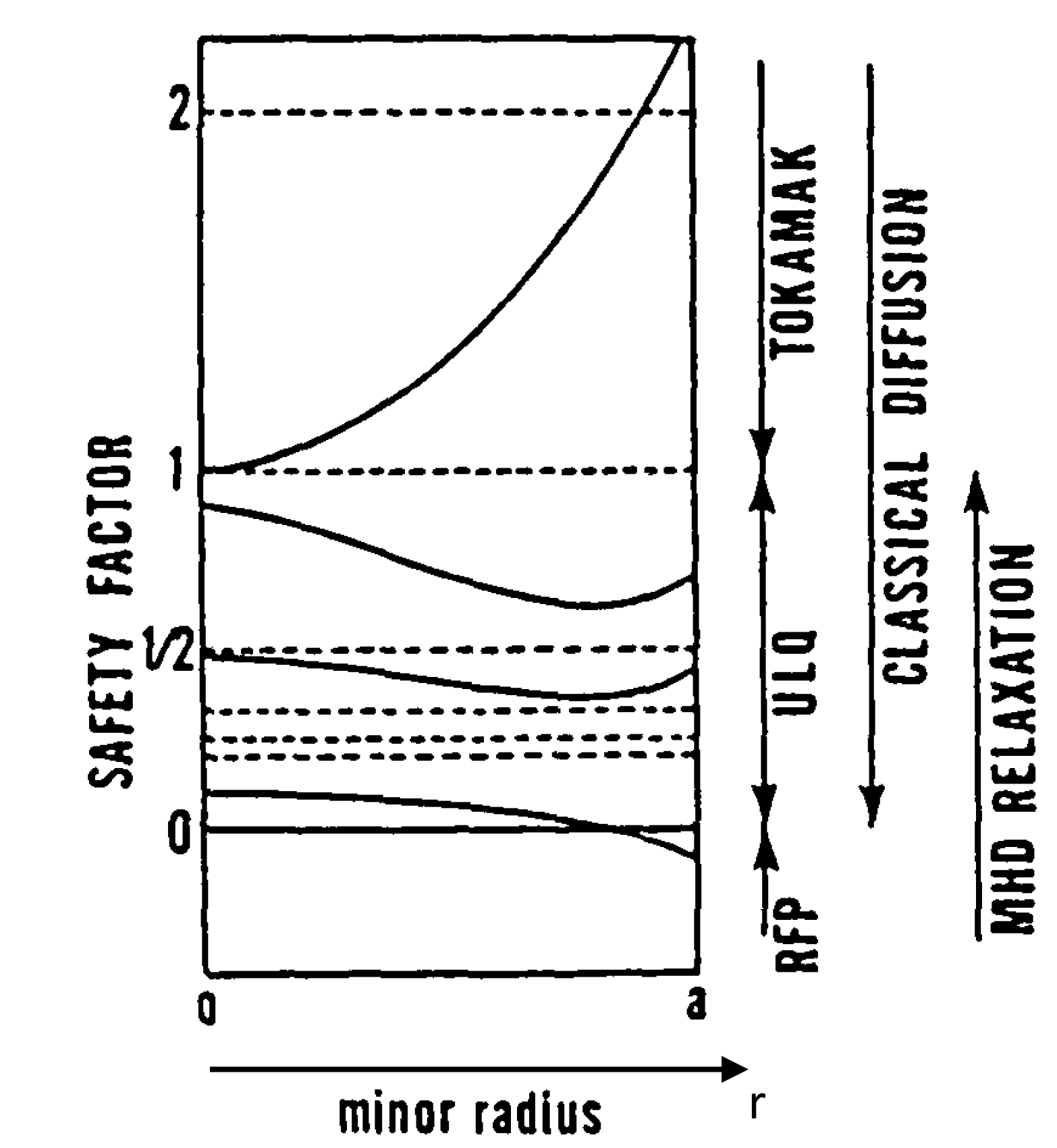}} 
\par\end{centering}
\centering{}\caption{\label{fig:qULQ}$\,\,\,\,$$q$ profiles (ULQ device){\footnotesize{}
(image credit: modified from \cite{Kamada})}}
\end{figure}
Radial $q$ profiles for various regimes are indicated in figure \ref{fig:qULQ}.
The dissipative processes which determine the $q$ profile of an equilibrium,
independent of the initial state of the plasma, vary from regime to
regime. As shown in \cite{yoshida}, tokamaks and RFPs are on different
dissipative branches. In general, two distinct dissipative processes
can be identified - resistive magnetic field diffusion, which leads
to $\frac{dq}{dr}>0$ and peaked current density profiles in tokamaks,
and MHD relaxation associated with m (poloidal mode number) = 1 kink
activity \cite{Taylor,Taylor0}, which leads to $\frac{dq}{dr}<0$
and flat current density profiles in RFPs \cite{Kamada}. The ULQ
regime $q$ profile is a result of competition between the two processes.
When $q<1$, such as in ULQ, spheromak, and RFP regimes, the m=1 helical
kink mode is enhanced and stabilised through MHD relaxation \cite{WatanabePhd}.
MHD analysis indicated that, with sufficient magnetic shear and moderate
magnetic Reynolds numbers, the ULQ regime was marginally stable against
both ideal and resistive modes \cite{Kamada}. Note that $R_{m}$,
the magnetic Reynolds number, indicates the relative magnitude of
the advective and diffusive contributions to the rate of change of
magnetic field. The MHD equation for the rate of change of magnetic
field is given by $\frac{\partial\mathbf{B}}{\partial t}=\nabla\times\mathbf{v\times B}+\nabla\times\left(\eta\,\nabla\times\mathbf{B}\right)$,
yielding $R_{m}=\frac{vL}{\eta}$, where $v$ is the plasma fluid
velocity, $L$ is the system's characteristic length scale, and $\eta$
{[}m$^{2}$/s{]}$=\frac{\eta'\,[\Omega-\mbox{m}]}{\mu_{0}}$ is the
plasma resistive diffusion coefficient.

Power density balance in an ohmically heated fusion plasma is given
by \cite{Inoue,Hutch_OhmIG} 
\begin{align*}
P\,\mbox{[W/m}^{3}] & =\frac{du_{tot}}{dt}=P_{\Omega}+P_{\alpha}-P_{L}-P_{B}\\
 & =\eta'J_{\phi}^{2}+\frac{n^{2}}{4}Q_{\alpha}<\sigma V>-\frac{3nT}{\tau_{E}}-C_{B}\,n^{2}\sqrt{T}
\end{align*}
where $u_{tot}$ {[}J/m$^{3}${]} is the total plasma energy density,
$\eta'\,[\Omega-\mbox{m}]\propto T_{e}^{-1.5}$ is the plasma resistivity,
and $P_{\Omega},\,P_{\alpha},\,P_{L}$ and $P_{B}$ are the contributions
to power density balance from ohmic heating, alpha particle heating,
transport losses, and Bremsstrahlung radiative losses respectively.
$Q_{\alpha}$ {[}J{]} is the (3.5 MeV) alpha particle energy for D-T
fusion, $<\sigma V>$ {[}m$^{3}$/s{]} is the fusion reaction rate,
$\tau_{E}$ {[}s{]} is the energy confinement time, and $C_{B}\propto Z_{eff}$
is a coefficient that scales the losses associated with Bremsstrahlung
losses. This model assumes that Bremsstrahlung radiative losses are
much greater than losses due to impurity line radiation \cite{Hutch_OhmIG},
which is a reasonable assumption at high temperatures when ions have
been stripped of all their electrons. In addition, cyclotron radiative
losses are neglected, without clear justification. Following the widespread
observation that energy confinement time in magnetic confinement devices
generally scales with density (equation \ref{eq:0.2}), the assumption
is made that $\tau_{E}=Kn$, where $K$ is some constant. The expression
for power density balance can then be re-expressed as 
\begin{equation}
P=C_{\Omega}J_{\phi}^{2}T^{-\frac{3}{2}}+C_{\alpha}n^{2}T^{\frac{7}{2}}-C_{L}T-C_{B}n^{2}T^{\frac{1}{2}}\label{eq:009}
\end{equation}
where $C_{\Omega}=C_{\Omega}(Z_{eff}),\,C_{\alpha}=C_{\alpha}(Z_{imp},\,Z_{eff}),\,C_{L}\propto\frac{1}{K}$,
and $C_{B}\propto Z_{eff}$ are known approximations from empirical
scalings \cite{Inoue,Hutch_OhmIG}. The steady-state equilibrium power
balance condition is $P=0$, and the stability of the power-balance
equilibrium is determined by the signs of $\frac{\partial P}{\partial n}$
and $\frac{\partial P}{\partial T}$. For example, if $\frac{\partial P}{\partial T}<0$,
a perturbation to the equilibrium balance leading to a temperature
increase would result in a loss of energy, while a temperature reduction
would lead to an increase of energy. The ignition point of interest
is the \textquotedbl thermal runaway\textquotedbl{} point, where
a small perturbation in $n$ or $T$ moves the plasma into the regime
where the alpha heating is dominant \cite{Inoue,Hutch_OhmIG}. The
ignition saddle point values for $T$ and $n$ that satisfy $P=\frac{\partial P}{\partial n}$$=\frac{\partial P}{\partial T}=0$
can be calculated. For example given inputs $\tau_{E}=2.5$ s, $Z_{eff}=1.5$,
and $Z_{imp}=8$ to enable assessment of $C_{\Omega},\,C_{\alpha},\,C_{L},\mbox{ and }C_{B}$,
it is found that $T=5.3$ keV and $n=5\times10^{20}$ m$^{-3}$ are
required for ignition.\\
\\
If magnetic compression, with increasing external toroidal field only,
keeping $R$ fixed, is carried out, an updated expression for the
power density balance applies. The adiabatic compression scalings
in table \ref{tab:Scaling-parameters-for} can be used, with $a^{-1}\rightarrow C$,
to modify equation \ref{eq:009}:

\begin{equation}
P=C_{\Omega}C^{2}J_{\phi1}^{2}T_{1}^{-\frac{3}{2}}+C_{\alpha}C^{\frac{26}{3}}n_{1}^{2}T_{1}^{\frac{7}{2}}-C_{L}C^{\frac{4}{3}}T_{1}-C_{B}C^{\frac{14}{3}}n_{1}^{2}T_{1}^{\frac{1}{2}}\label{eq:010}
\end{equation}
Again, parameters with subscript $1$ denote pre-compression values.
Adiabatic compression enhances the ohmic heating input ($P_{\Omega}\rightarrow C^{2}$)
more than the transport losses $(P_{L}\rightarrow C^{\frac{4}{3}})$,
and enhances the alpha particle heating $(P_{\alpha}\rightarrow C^{\frac{26}{3}})$
more than the Bremsstrahlung losses $(P_{B}\rightarrow C^{\frac{14}{3}})$,
so that ohmic ignition can be achieved more easily at reduced $\tau_{E}$
and $J_{\phi}$. Recalling that $\tau_{E}=Kn$, and noting that $C_{L}$
is given as $C_{L}=\frac{0.048}{K}\,[\frac{\mbox{MW}}{\mbox{m}^{3}}\mbox{s\ensuremath{\,}keV}^{-1}]$
\cite{Inoue}, equation \ref{eq:010}, with $P=0$, can be used to
find an expression for the energy confinement time as 
\begin{equation}
\tau_{E}=\frac{0.048\,n_{1}C^{\frac{10}{3}}T_{1}}{C_{\Omega}C^{2}J_{\phi1}^{2}T_{1}^{-\frac{3}{2}}+C_{\alpha}C^{\frac{26}{3}}n_{1}^{2}T_{1}^{\frac{7}{2}}-C_{B}C^{\frac{14}{3}}n_{1}^{2}T_{1}^{\frac{1}{2}}}\label{eq:011}
\end{equation}
With $T_{1}=600$ eV and $n_{1}=2\times10^{19}$ m$^{-3}$, the ignition
parameters $T=5.3$ keV and $n=5\times10^{20}$ m$^{-3}$ can, in
principle, be achieved with $C=5$. These pre-compression values can
be used in equation \ref{eq:011}, to find the energy confinement
time as $\tau_{E}=0.15$ s, which is around 5\% of the 2.5 s energy
confinement time that would be required for ignition in the case without
compression, as a result of the intensification of the ohmic and alpha
particle heating power with adiabatic compression. 

It was envisaged that the external $B_{\phi}$ would be maintained
after compression, and that the compressed regime with an unprecedented
level of ohmic heating would naturally expand in minor radius, as
determined by the magnetic diffusion timescale of the plasma, attaining
the more stable tokamak configuration. In \cite{Inoue}, no particular
emphasis was placed on assessing the stability of the low $q$ plasma
during compression. The main results of the magnetic compression experiment
on REPUTE-1 were reported in \cite{WatanabePhd,WatanabeREPUTE1}.
Post-compression transition to the tokamak regime was not reported,
but $q_{a}$ increased from around 0.3 to 0.65 at compression, before
settling, post-compression at around 0.5. Reduction in minor radius
at compression, and post-compression expansion into the increased
toroidal field, was observed. A significant reduction of visible light
emission was observed during magnetic compression, as a result of
reduced plasma-wall interaction as the plasma column is pushed away
from the vacuum vessel walls. The observed increase in plasma current
after magnetic compression was attributed to reduced resistivity due
to plasma-wall interaction reduction. The post-compression reduction
in $q_{a}$ was attributed to the increase in plasma current. As $q_{a}$
variation led to rational surfaces at the edge, increased MHD activity
was observed, and consequent plasma-wall interaction was seen to increase
again, but to levels lower than those before compression. 

\subsection{Merging-compression formation of tokamak plasmas}

Merging-compression is a spherical tokamak (ST) plasma formation method
that involves the merging and magnetic reconnection of two plasma
rings, followed by inward radial magnetic compression of the resultant
single torus to form a spherical tokamak plasma configuration. The
initial tori are formed inductively around coils internal to the vacuum
vessel, and the compression coils are also internal, an approach with
some similarities to that developed on the S-1 device. This ST plasma
formation method has the advantage of eliminating the need for a traditional
central solenoid - in an ST, space is limited in the central post
and is inadequate for solenoids capable of inducing toroidal plasma
currents in the MA range \cite{Gryaznevich}. The merging phase leads
to efficient transformation of magnetic to kinetic, then thermal energy
(up to 15 MW of ion heating power was recorded on on MAST), and also
leads to a rapid increase of plasma current \cite{Gryaznevich}. The
merging compression ST formation method was first used on START \cite{Gryaznevich1,Sykes}
in 1991, and then in MAST \cite{Sykes1,Gryaznevich2}, and is currently
employed on the compact high field spherical tokamak ST40 at Tokamak
Energy Ltd. \cite{Gryaznevich}.

\subsection{Magnetic compression of FRCs}

In the experiment described in \cite{Rej}, FRC (field reversed configuration)
ion temperatures of up to 2 keV were achieved with magnetic compression.
FRCs can be formed by merging two spheromaks with opposite helicity
and are self-confined by almost purely poloidal fields. Helion Energy
Inc. has compressed FRCs, attaining ion temperatures of around 5 keV;
a set of independently triggered formation and acceleration coils
are used to form and merge two oppositely directed supersonic FRCs
\cite{Slough}.

\section{Thesis outline\label{sec:Thesis-outline}}

Part one of this work presents an overview of the magnetic compression
experiment, the principal tests undertaken, and the main findings.
A paper \cite{exppaper} based on the material in part one has been
published in the arXiv electronic journal, and has been submitted
to a peer-reviewed journal. Most of the results presented in the paper
were obtained over the time after GF officially stopped the project
and allowed me several weeks additional time to work alone with the
machine and obtain data for the PhD thesis. In that period, various
modifications to the external coil configuration were implemented,
leading to significantly improved performance of levitated and magnetically
compressed CTs. Useful discussions and technical assistance from the
co-authors was helpful in achieving the results obtained. In particular,
Stephen Howard helped guide the development of the CT separatrix measurement
technique. Experimental data included in the paper (and in this thesis)
was written to file using GF IGOR-PRO data analysis software and processed
using a MATLAB code that I wrote for data analysis and presentation.
Four conference posters \cite{APSposter2016,APSposter2017,EPSposter2018,ICPPposter2018}
have been presented, partially based on the experimental results. 

Chapter \ref{Chap:Exp_Overview} is a brief description of the key
features of the magnetic compression experiment. The machine and its
working principles, and main operational and plasma diagnostics are
presented. 

In chapter \ref{Chap:MagLev}, the principal external coil configurations
explored to hold plasma off the insulating wall surrounding the CT
containment region by the action of a levitation field formed by toroidal
currents in external coils are described. Interaction between plasma
and the insulating wall during the CT formation process was a major
obstacle to progress with the experiment, as it led to plasma impurities
and radiative cooling. The main results from the various configurations
tested are presented, and modifications to the experiment that lead
to improved performance of levitated CTs are described. 

The principal observations and findings relating to CT behaviour at
magnetic compression constitute chapter \ref{Chap:Magnetic-Compression}.
Once again, various configurations were experimented with, and progress
was made towards successful compression in the most recent configuration,
leading to regular achievement of significant increases in CT magnetic
field, density, and ion temperature at compression. The signs of an
MHD instability that was prevalent at compression are described, along
with a possible mechanism behind the observations.\\
\\
\\
Part two of the thesis, and appendices \ref{chap:Kinetic,MhD,GS},
\ref{sec:Formulation-of-discretized} and \ref{chap:Gen code setup},
focus on the development of the new DELiTE (Differential Equations
on Linear Triangular Elements) framework that was developed for spatial
discretisation of partial differential equations on a triangular grid
in axisymmetric geometry. Two papers \cite{SIMpaper,Neut_paper},
based on the material presented in part two have been submitted to
arXiv and to peer-reviewed journals. Ivan Khalzov, the co-author on
these papers, helped guide the development of the conservative finite
element code framework. With his guidance, I modified his original
isothermal version of the MHD code to include a model for energy evolution
while conserving net system energy. My contribution also included
implementation of various simulated diagnostics including the $q$
profile solver, development of models for CT formation, levitation
and compression, implementation of the resulting sets of equations
to code, running the simulations, and performing code verification
and validation checks. I developed and implemented, with some advice
from Dr. Khalzov, a model for interaction between plasma and neutral
fluids, and models for maintenance of toroidal flux conservation with
the inclusion of a partially electrically insulating computational
domain. I implemented the methods for data saving and plotting, and
timestep adjustment. Three conference posters \cite{APSposter2017,EPSposter2018,ICPPposter2018},
partially based on the material in part two, have been presented.

Chapter \ref{chap:Core-Code-Development} presents the development
of discrete differential operators in matrix form, which are derived
using linear finite elements to mimic some of the properties of their
continuous counterparts, and constitute the core of the framework.
A single-fluid two-temperature MHD model is implemented in the framework
in order to study the magnetic compression experiment. Note that the
continuous set of equations that describe the MHD model and various
forms of the Grad-Shafranov equation that describe MHD equilibrium
in the axisymmetric case, are derived in appendix \ref{chap:Kinetic,MhD,GS}.
The full set of continuous equations that describe the axisymmetric
MHD model are presented in chapter \ref{chap:Core-Code-Development},
and various conservation properties of the system are demonstrated.
As demonstrated in appendix \ref{sec:Formulation-of-discretized},
the discrete differential operators and the discrete forms of the
mass and energy conservation equations are used to derive a discrete
form of the momentum equation which, as a consequence of the inherent
properties of the operators, ensures that total energy of the system
described by the complete set of discretised MHD equations is conserved
with appropriate boundary conditions. In chapter \ref{chap:Core-Code-Development},
it is demonstrated how global conservation of energy, particle count,
toroidal flux, and angular momentum for the discrete system of equations
is ensured due to operator properties.

The various methods constructed to simulate CT formation into a levitation
field, and magnetic compression, are described in chapter \ref{chap:Implementation-of-models}.
The boundary conditions applied to the various fields are discussed,
and the techniques established to couple the vacuum field solution
in insulating regions to the full MHD solutions in the remainder of
the domain are presented. The coupling of the solutions leads to a
more physical representation of the processes involved in the interaction
between plasma and the insulating wall during CT formation; special
care is taken to maintain toroidal flux conservation. The development
and implementation of a range of simulated diagnostics that have experimental
counterparts, as well as additional simulated diagnostics including
evaluation of CT $q$ profile, and other CT-internal simulated measurements,
are reported.

Chapter \ref{chap:Simulation-results} introduces some of the primary
code inputs. Simulation results, including time-evolution of the various
fields solved for, and comparisons between experimentally-obtained
and simulated diagnostics are presented. The experiment lacked adequate
CT-internal plasma diagnostics for evaluation of how closely the adiabatic
compression scalings were followed, but the scalings are assessed
from simulations for which the available experimental diagnostics
closely match their simulated counterparts. 

The development and implementation to code of the discrete forms of
the equations that describe interaction between plasma and neutral
fluids, when ionization, recombination, and charge exchange processes
are included in the model, is outlined in chapter \ref{chap:Neutral-models}.
Simulation and experimental results indicating interaction between
plasma and neutral fluids are compared. 

Main conclusions, and suggestions for further improvements to the
experiment and its simulation are presented in chapter \ref{chap:Concluding-remarks}.

As mentioned above, appendix \ref{chap:Kinetic,MhD,GS} consists of
a derivation of the MHD equations, using principles of basic kinetic
theory, which are solved for in the code. This lays the groundwork
for chapter \ref{chap:Neutral-models}, in which expressions for the
terms in the coupled sets of plasma fluid and neutral fluid equations,
that pertain to reactive collisions, are developed. Derivations of
various equilibrium models are also presented in appendix \ref{chap:Kinetic,MhD,GS}.

A discrete form of the MHD equations which ensures (with appropriate
boundary conditions) total energy conservation, is derived in appendix
\ref{sec:Formulation-of-discretized}. Appendix \ref{chap:Gen code setup}
describes general code aspects, such as the computational grid, timestepping
methods, a description of the numerical solution for the Grad-Shafranov
equation, a general discussion of diffusion coefficients and their
implementation, and outcomes of code validation and verification. 

A description of an edge-biasing experiment conducted on the SPECTOR
plasma injector, and initial results, are presented in appendix \ref{chap:First-results-from}.
A paper \cite{Spector_biasing} based on the material presented in
appendix \ref{chap:First-results-from} has been submitted to the
arXiv electronic journal, and to a peer-reviewed journal. The results
obtained in that work were obtained over a period of two weeks. I
was allowed to try the biasing experiment on SPECTOR at GF and based
the report for the University comprehensive exam on the work. My contribution
to the material in the paper was the design and construction of the
electrode assembly and biasing circuit, and all data and circuit analysis.
Discussions and technical assistance from the GF team was helpful
in accomplishing the results. The insertion of a disc-shaped molybdenum
electrode, biased at up to $+100$V, into the edge of the CT, resulted
in up to 1 kA radial current being drawn. Electron temperature, as
measured with a Thomson-scattering diagnostic, was found to increase
by a factor of up to 2.4 in the optimal configuration. $\mbox{H}_{\alpha}$
intensity was observed to decrease, and CT lifetimes increased by
a factor of up to 2.3. A significant reduction in electron density
was observed; this is thought to be due to the effect of a transport
barrier, established by edge biasing, impeding CT fueling, where the
fueling source is neutral gas that remains concentrated around the
machine gas valves after CT formation, an effect that was verified
by simulations that are presented in chapter \ref{chap:Neutral-models}.
Particularly because of the issues encountered with plasma impurities
in the magnetic compression experiment, special care was taken to
choose a plasma-compatible materials for the biasing electrode assembly.
\\
\newpage{}

\part*{PART 1: EXPERIMENT\label{part:Experiment}\addcontentsline{toc}{part}{PART 1: EXPERIMENT} }

\chapter{Experiment overview\label{Chap:Exp_Overview}}

The SMRT (Super Magnetised Ring Test) magnetic compression experiment,
which was operated at General Fusion from 2013 to 2016, was designed
as a repetitive non-destructive test to study plasma physics applicable
to magnetic target fusion compression. SMRT was identical to the MRT
machines (section \ref{sec:General-Fusion--}), with the exception
that the aluminum flux conserver surrounding the CT confinement region
was replaced with an insulating outer wall. Also, in the SMRT design,
currents in external coils surrounding the containment region produce
a magnetic field which applies a radial force on the plasma that \textquotedbl levitates\textquotedbl{}
it off the outer wall during CT formation and relaxation, and then
rapidly compresses it inwards. GF had hoped to be able to produce
and magnetically compress levitated CTs with the same characteristics
as the non-levitated CTs produced in standard MRT machines. With extensive
lab-based diagnostics, this would give insight to the behavior of
CTs that were compressed in the PCS experiments (section \ref{sec:General-Fusion--}).
By 2016, GF had shifted the focus of the PCS experiments to producing
and explosively compressing spherical tokamak plasmas which had far
better performance than their spheromaks. Partly because of the investment
that would be required in order to redesign an experiment that would
magnetically compress spherical tokamaks, and partly because of the
problems with impurities and possibly other factors associated with
the existing magnetic compression design, GF decided to move resources
to other projects that are part of their scientific objectives.

This chapter begins with an overview of the magnetic compression device,
with a description of the operating principles in section \ref{subsec:Machine-operation-principles}.
The principal experimental diagnostics used are presented, along with
details of their functionality, in section \ref{subsec:Diagnostics}.

\section{Machine operation principles\label{subsec:Machine-operation-principles}}

\begin{figure}[H]
\centering{}\includegraphics[scale=0.6]{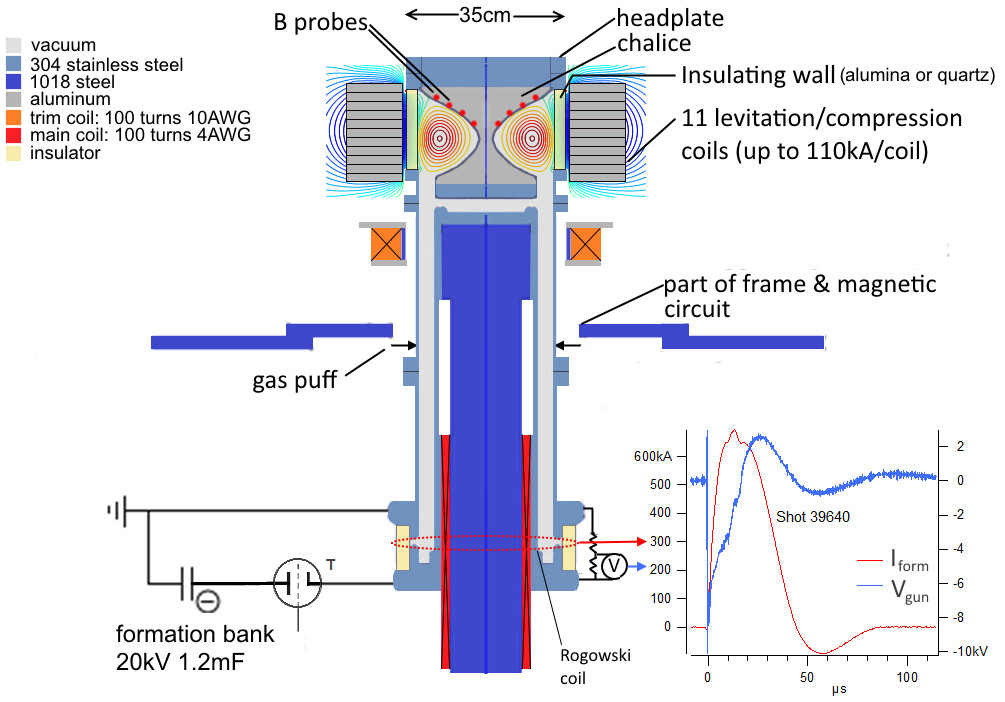}\caption{\label{fig:Machine-Schematic-1}$\,\,\,\,$Machine schematic}
\end{figure}

Figure \ref{fig:Machine-Schematic-1} shows a schematic of the magnetic
compression device, with spheromak and levitation $\psi$ contours
from an equilibrium model superimposed. Measurements of formation
current $(I_{form}(t))$, and voltage across the formation electrodes
$(V_{gun}(t))$ are also indicated. Note that the principal materials
used in the machine construction, and some key components, are indicated
by the color-key at the top left of the figure. The sequence of machine
operation is as follows:\\

\vspace{1cm}
\begin{table}[H]
\begin{centering}
\begin{tabular}[t]{lll}
(i) & $t\sim-3$ s  & Main coil is energised with steady state ($\sim4$s duration) \tabularnewline
 &  & current ($I_{main}$)\tabularnewline
(ii) & $t=t_{gas}\sim-400\,\upmu$s  & Gas is injected into vacuum \tabularnewline
(iii) & $t=t_{lev}\sim-400\,\upmu\mbox{s}\rightarrow-40\,\upmu\mbox{s}$  & Levitation banks, charged to voltage $V_{lev}$, are fired\tabularnewline
(iv) & $t=0\mbox{s}$  & Formation banks, charged to voltage $V_{form}$, are fired\tabularnewline
(v) & $t=t_{comp}\sim40\,\upmu\mbox{s}\rightarrow150\,\upmu\mbox{s}$  & Compression banks, charged to voltage $V_{comp}$, are fired\tabularnewline
\end{tabular}
\par\end{centering}
\caption{\label{tab:Sequence-of-machine}$\,\,\,\,$Sequence of machine operation }

\end{table}
\vspace{1cm}
\begin{figure}[H]
\subfloat{\raggedright{}\includegraphics[scale=0.5]{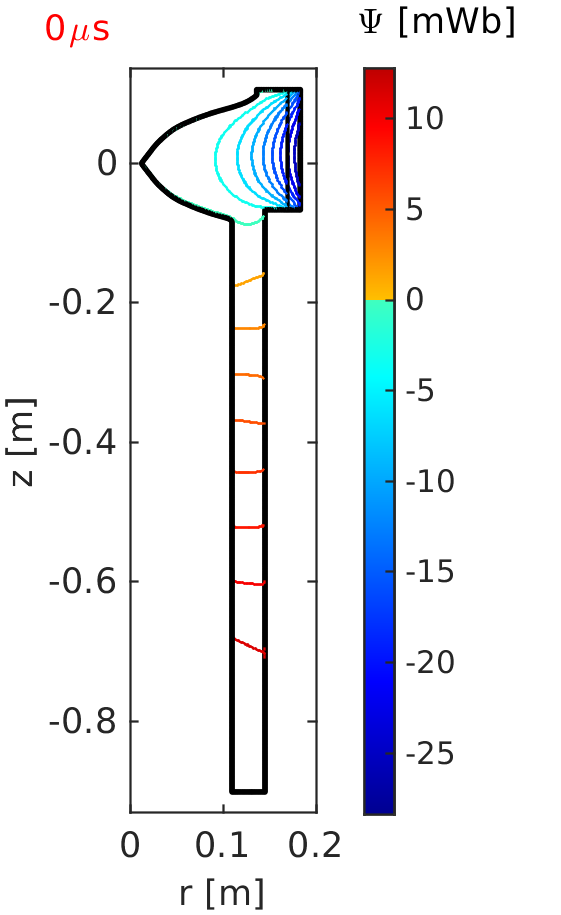}}\hfill{}\subfloat{\raggedright{}\includegraphics[scale=0.5]{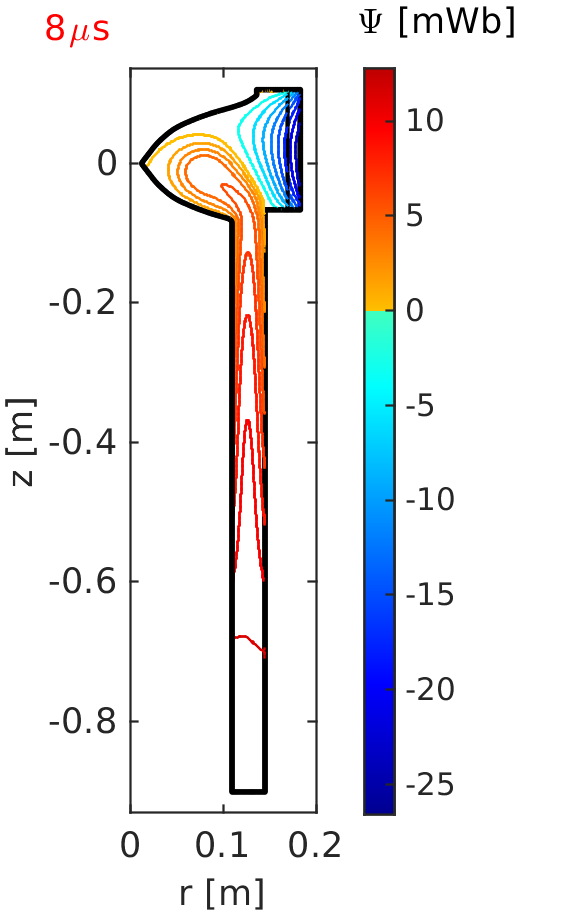}}\hfill{}\subfloat{\raggedright{}\includegraphics[scale=0.5]{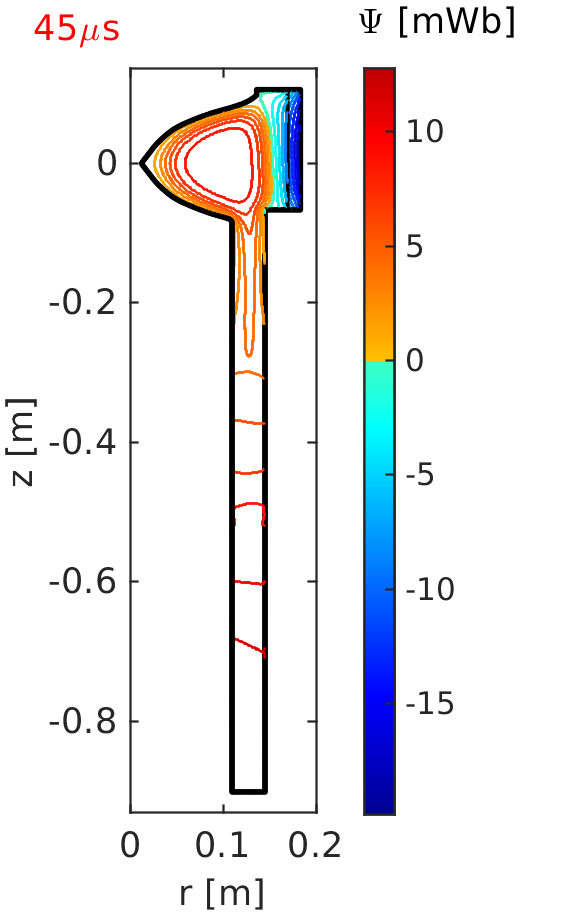}}

\subfloat{\raggedright{}\includegraphics[scale=0.5]{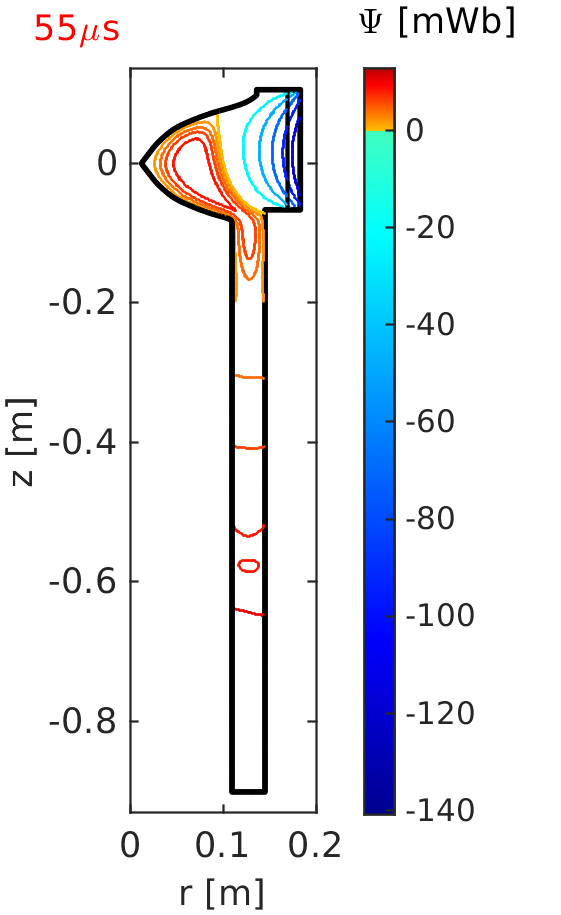}}\hfill{}\subfloat{\raggedright{}\includegraphics[scale=0.5]{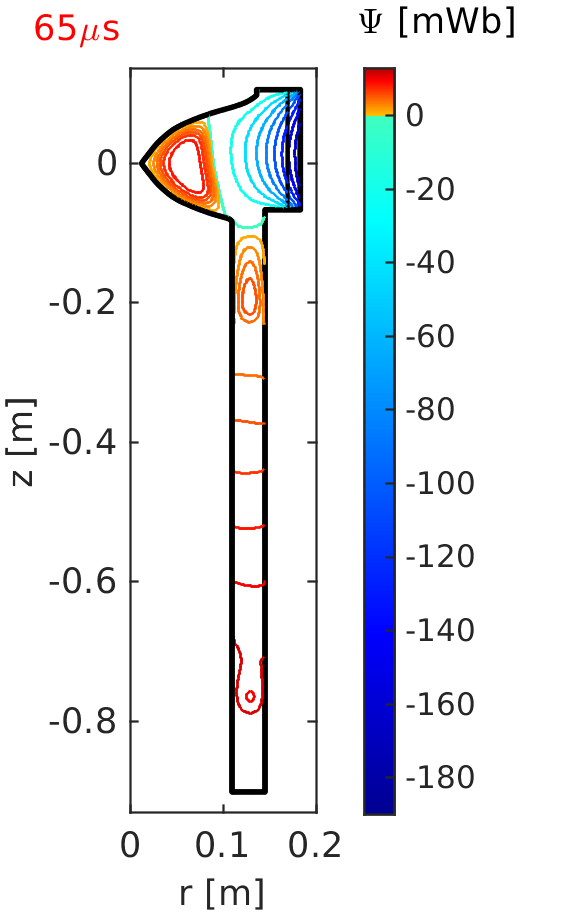}}\hfill{}\subfloat{\raggedright{}\includegraphics[scale=0.5]{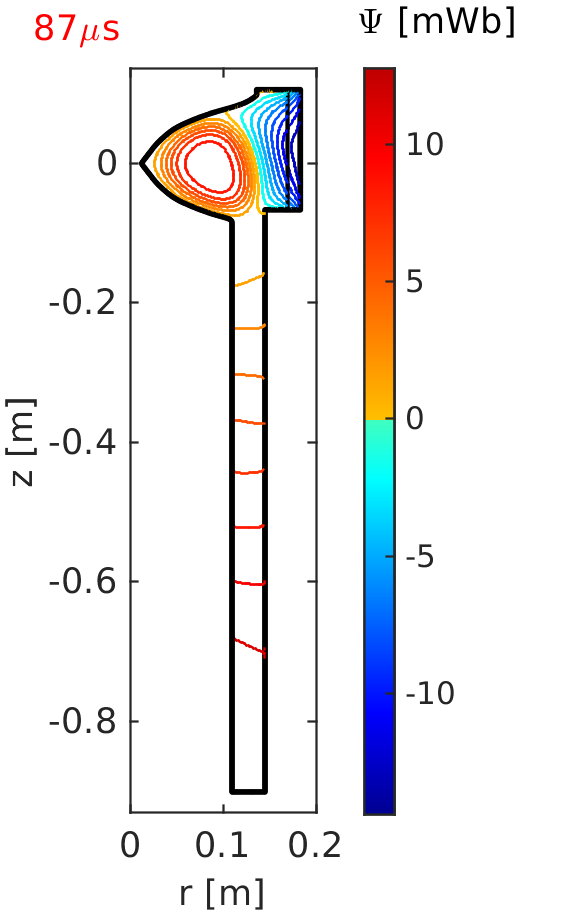}}

\caption{\label{fig:MHDform-1}$\,\,\,\,$$\psi$ contours from MHD simulation,
with CT formation and compression}
\end{figure}
To better illustrate the sequence of operation, $\psi$ contours from
an MHD simulation (see part two of the thesis) of the magnetic compression
experiment are shown in figure \ref{fig:MHDform-1}. Note that $\psi$
contours represent poloidal field lines, and that the vertical black
line at the top-right of the figures at $r\sim17\mbox{ cm}$ represents
the inner radius of the insulating wall. Vacuum field only is solved
for to the right of the line, and the plasma dynamics are solved for
in the remaining solution domain to the left of the line. The inner
radius of the stack of eleven levitation/compression coils (which
are not depicted here) is located at the outer edge of the solution
domain, at $r\sim18\mbox{ cm}$. Simulation times are notated in red
at the top left of the figures. Note that the colorbar scaling changes
over time; max$(\psi)$ decreases slowly over time as the CT decays,
while min$(\psi)$ increases as the levitation current in the external
coils decays, and then drops off rapidly as the compression current
in the external coils is increased, starting at $t_{comp}=45\,\upmu\mbox{s}$
in this simulation. At time $t=0,$ the stuffing field ($\psi>0$)
due to currents in the main coil fills the vacuum below the containment
region, and has soaked well into all materials around the gun, while
the levitation field fills the containment region. Simulated CT formation
is initiated with the addition of toroidal flux below the gas puff
valves located at $z=-0.43$ m; initial intra-electrode radial formation
current is assumed to flow at the z-coordinate of the valves. As described
in detail in chapter \ref{chap:Implementation-of-models}, toroidal
flux addition is scaled over time in proportion to $\int_{0}^{t}V_{gun}(t')\,dt'$.
Open field lines that are resistively pinned to the electrodes, and
partially frozen into the conducting plasma, have been advected by
the $\mathbf{J}_{r}\times\mathbf{B}_{\phi}$ force into the containment
region by $t=8\,\upmu\mbox{s}$ ($\mathbf{J}_{r}$ is the radial formation
current density across the plasma between the electrodes, and $\mathbf{B}_{\phi}$
is the toroidal field due to the axial formation current in the electrodes).
By $45\,\upmu\mbox{s}$, open field lines have reconnected at the
entrance to the containment region to form closed CT flux surfaces.
The presence of a pressure gradient directed towards the CT center
allows formation of a toroidal diamagnetic current which in turn helps
sustain the CT poloidal field. At these early times, open field lines
remain in place surrounding the CT. Compression starts at $45\,\upmu\mbox{s}$
and peak compression is at $65\,\upmu\mbox{s}$. The CT expands again
between $65\,\upmu\mbox{s}$ and $87\,\upmu\mbox{s}$ as the compression
current in the external levitation/compression coils reduces. Note
that at $55\,\upmu$s, magnetic compression causes closed CT poloidal
field lines that extend down the gun to be pinched off at the gun
entrance, where they reconnect to form a second smaller CT. Field
lines that remain open surrounding the main CT are then also reconnectively
pinched off, forming additional closed field lines around the main
CT, while the newly reconnected open field lines below the main CT
act like a slingshot that advects the smaller CT down the gun, as
can be seen at $65\,\upmu\mbox{s}$.\\

A pulse-width modulation system was used for current control in the
main coil circuit. The working gas was typically $\mbox{He},\,\mbox{H}_{2},$
or $\mbox{D}_{2}$, with valve plenum pressure $\sim30$ psi (gauge),
and optimal vacuum pressure $\sim1\times10^{-8}$ Torr. The formation
capacitor bank ($240\mbox{ kJ bank consisting of }24\times50\,\upmu\mbox{F},\,20\mbox{ kV}$
capacitors in parallel) drives up to 1 MA of current, with a half
period of $50\,\upmu$s (see figure \ref{fig:Machine-Schematic-1}).
The original configuration had six levitation/compression coils, with
each coil having its own levitation and compression circuit. The $120$
kJ levitation bank consisted of $2\times50\,\upmu\mbox{F},\,20\mbox{ kV}$
capacitors in parallel for each coil, and there were four of these
capacitors in parallel for each coil for the $240$ kJ compression
bank. 
\begin{figure}[H]
\centering{}\includegraphics[scale=0.32]{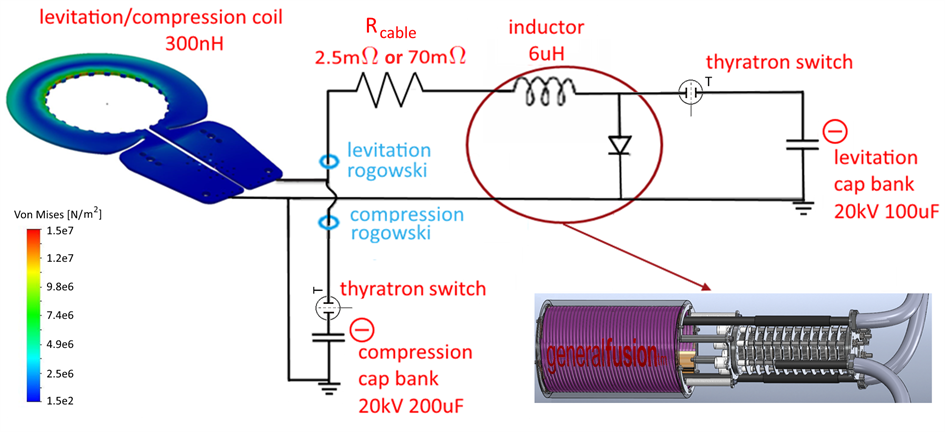}\caption{\label{fig:Levitation-and-compression-1}$\,\,\,\,$Levitation and
compression circuit }
\end{figure}
Figure \ref{fig:Levitation-and-compression-1} illustrates the circuit
for one of the single-turn levitation and compression coils. Each
coil (or coil-pair in the case of the configuration with eleven coils)
had a separate identical circuit. Unlike the crowbarred levitation
currents, the compression currents are allowed to ring with the capacitor
discharge. 
\begin{figure}[H]
\centering{}\includegraphics[scale=0.5]{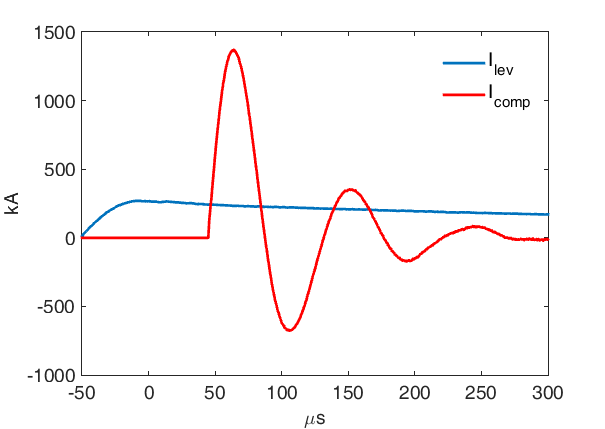}\caption{\label{fig:Ilev_comp}$\,\,\,\,$Total levitation and compression
currents }
\end{figure}
Levitation and compression current profiles are shown in figure \ref{fig:Ilev_comp},
for a case with $V_{lev}=12$ kV, $V_{comp}=18$ kV, $t_{lev}=-50\,\upmu$s,
and $t_{comp}=45\,\upmu$s. Note that the currents indicated are divided
(approximately equally) between the coils. With 2.5 m$\Omega$ resistance
cables (denoted as $R_{cable}$ in figure \ref{fig:Levitation-and-compression-1})
in place between the holding inductors and coils, the crowbarred levitation
current drops to zero at around 3 ms.

\section{Diagnostics\label{subsec:Diagnostics}}

The experiment used a variety of measurements to characterise machine
operation and diagnose plasma behaviour. In this section an overview
of the measurements employed will be given, followed by brief descriptions
of the operating principles for each diagnostic in sections \ref{subsec:Magnetic-field-measurements}
to \ref{subsec:Optical-diagnostics}.

Formation, levitation and compression currents were measured using
Rogowski coils, while formation voltage was measured with a voltage
divider arrangement. The locations of the current and voltage measurement
devices are indicated in figures \ref{fig:Machine-Schematic-1} and
\ref{fig:Levitation-and-compression-1}.

Plasma diagnostics included magnetic field measurements at magnetic
probes embedded in the chalice (the inner flux conserver depicted
in figure \ref{fig:Machine-Schematic-1}), and interferometers to
record line averaged electron density. Ion temperature was estimated
using an ion Doppler diagnostic. Visible light emission is recorded
by two survey spectrometers which have variable exposure durations,
and by six fiber-coupled photodiodes that record time-histories of
total optical emission. 

The absence of magnetic field, and spatially resolved density and
temperature measurements internal to the CT, means that there is a
high level of uncertainty about internal plasma dynamics. Internal
plasma diagnostics are only useful if the degree of plasma perturbation
due to the diagnostic is tolerable. Relative to more conventional
magnetically confined plasmas, the small size and high density of
the CTs studied in this project impose a lower tolerance on what can
acceptably be inserted into the plasma. Non-invasive diagnostics,
that can provide spatially resolved information, include Thomson scattering
for electron density and temperature, and polarimetry for magnetic
field measurements. Although Thomson scattering is employed on other
projects at GF, it was never implemented on the magnetic compression
experiment. A polarimetry diagnostic is under development, but again,
was not used on this experiment.
\begin{figure}[H]
\begin{centering}
\includegraphics[scale=0.5]{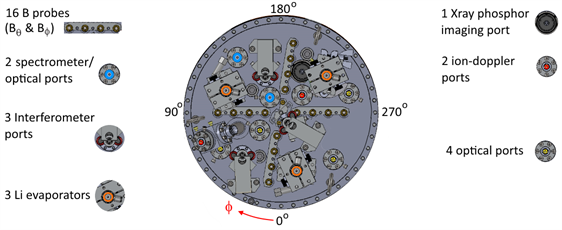}
\par\end{centering}
\centering{}\caption{\label{fig:Machine-headplate-schematic-1}$\,\,\,\,$Machine headplate
schematic }
\end{figure}
Figure \ref{fig:Machine-headplate-schematic-1} shows the schematic
of the machine headplate indicating main diagnostics and lithium gettering
ports. 
\begin{figure}[H]
\begin{centering}
\includegraphics[scale=0.4]{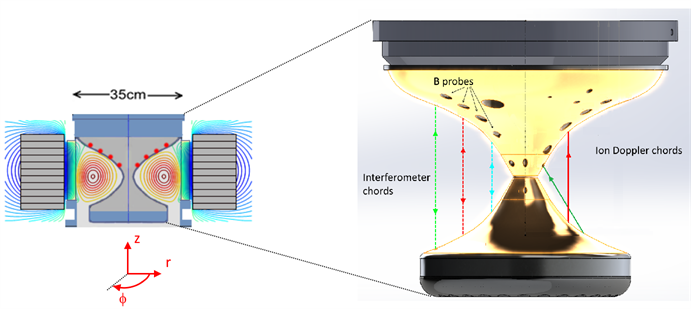}
\par\end{centering}
\centering{}\caption{\label{fig:Chalice} $\,\,\,\,$Chalice schematic }
\end{figure}
Figure \ref{fig:Chalice} depicts the tungsten-coated aluminum chalice
which is the inner flux-conserver, indicating the locations of some
of magnetic probe ports and the lines-of-sight for the ion-Doppler
and interferometer diagnostics. For ease of depiction, the ion-Doppler/interferometer
chords are shown to be located on the same poloidal plane. Line-averaged
electron density was obtained along chords at $r=35\mbox{\mbox{ mm}},\,r=65\mbox{\mbox{ mm}}$
and $r=95\mbox{\mbox{ mm}}$ using dual $1310\mbox{ nm}$ and $1550\mbox{ nm}$
He-Ne laser interferometers. Dual wavelength interferometers were
used to enable compensation for errors due to machine vibration during
a shot. Each beam is split in two - one of the split beams travels
through the plasma and is reflected off a retroreflector in the base
of the chalice and then passes through the plasma a second time before
being transmitted away from the injector by optic fiber and merging
with the other split beam (which has travelled an equivalent path
but with the distance traveled by the first beam through the plasma
being replaced by an equal distance through air), for interference
analysis leading to line-averaged $n_{e}$. An indication of ion temperature,
along the vertical chord at $r=45\,\mbox{\mbox{mm}}$ and the diagonal
chord with its upper point at $r\sim20\mbox{\,\mbox{mm}}$, was found
from Doppler broadening of line radiation from singly ionized helium. 

Each of the sixteen magnetic probes was located at the closed end
of a thin-walled stainless steel tube embedded in axially directed
holes in the chalice. The $r,\,\phi$ coordinates of the magnetic
probes, where $r=0$ is defined as being at the machine axis, are:\hfill{}
\begin{table}[H]
\centering{}{\footnotesize{}}%
\begin{tabular}{|c|c|c|c|c|c|c|c|c|c|c|c|c|c|c|c|c|}
\hline 
{\footnotesize{}$\mathbf{r}\,[\mbox{\mbox{mm}}]$} & {\footnotesize{}$26$} & {\footnotesize{}$26$} & {\footnotesize{}$39$} & {\footnotesize{}$39$} & {\footnotesize{}$52$} & {\footnotesize{}$52$} & {\footnotesize{}$64$} & {\footnotesize{}$64$} & {\footnotesize{}$77$} & {\footnotesize{}$77$} & {\footnotesize{}$90$} & {\footnotesize{}$90$} & {\footnotesize{}$103$} & {\footnotesize{}$103$} & {\footnotesize{}$116$} & {\footnotesize{}$116$}\tabularnewline
\hline 
{\footnotesize{}$\phi\,${[}deg.{]}} & {\footnotesize{}$90$} & {\footnotesize{}$270$} & {\footnotesize{}$10$} & {\footnotesize{}$190$} & {\footnotesize{}$90$} & {\footnotesize{}$270$} & {\footnotesize{}$10$} & {\footnotesize{}$190$} & {\footnotesize{}$90$} & {\footnotesize{}$270$} & {\footnotesize{}$10$} & {\footnotesize{}$190$} & {\footnotesize{}$90$} & {\footnotesize{}$270$} & {\footnotesize{}$10$} & {\footnotesize{}$190$}\tabularnewline
\hline 
\end{tabular}\caption{\label{tab: coordinates-ofBprobes}$\,\,\,\,$$r,\,\phi$ coordinates
of magnetic probes (original configuration)}
\end{table}

\subsection{Magnetic field measurements - B probes\label{subsec:Magnetic-field-measurements}}

A standard magnetic probe is one way to measure magnetic field. The
voltage signal induced in a coil of wire due to a changing field can
be integrated to give the field magnitude over time. Faraday's law
of induction is: $\nabla\times\mathbf{E}=-\dot{\mathbf{B}}$. Integrating
over the area of the probe's coil, using Stoke's theorem ($\int\nabla\times\mathbf{a}\cdot d\mathbf{S}=\int\mathbf{a}\cdot d\mathbf{l}$),
and noting that $V=\int\mathbf{E}\cdot d\mathbf{l}$, a formula for
the voltage induced in a single loop of wire with loop-area $A$ is
arrived at: $V=-A\dot{B}$. Here, it's assumed that the magnetic field
is uniform within the loop. Only the component of the magnetic field
that is perpendicular to the plane of the loop contributes to the
induced voltage. For a coil with $N$ turns, 
\begin{equation}
B(t)=-\frac{1}{NA}\int_{0}^{t}V(t')\,dt'\label{eq:770}
\end{equation}
The measured voltage signal can be time-integrated using an RC integrator,
or can be integrated numerically. Numerical integration has the advantage
of not attenuating the incoming signal. The voltage signal is proportional
to $\dot{B},\,N$ and $A$, so for a given (estimated) value of $\dot{B}$,
the product $NA$ must be sufficiently large that the voltage signal
can be picked up by the data acquisition (DAQ) system and also not
be dominated by background noise. On the other hand keeping the probe
as small as possible is good so that the perturbation of the plasma
due to the presence of the probe is minimised. Equation \ref{eq:770}
can be rearranged as 
\[
NA\sim\frac{V_{min}\tau_{est}}{B_{est}}
\]
to give an estimate for the minimum required value of $NA$. Here,
$V_{min}$ is the minimum induced probe voltage that will allow an
adequate signal to noise ratio while meeting the minimum DAQ sensitivity,
and $B_{est}$ and $\tau_{est}$ are estimates for the field magnitude
that will be measured, and its characteristic rise time. 

Another criterion that must be considered when designing a magnetic
probe is the probe response time, $i.e.,$ the minimum time period
over which changes in the magnetic field can be captured by the probe.
This minimum time period is given by $\tau_{LR}=\frac{L}{R}$ where
$L$ is the coil inductance, and $R$ is the resistance of the combined
coil and coil-to-DAQ system circuit. The coil inductance can be calculated
from one of the two following standard formulae: If $w=\frac{h_{c}}{r_{c}}<1$,
where $r_{c}$ is the radius of the circular coil turns, and $h_{c}$
is the height of the coil (for a coil with non-overlapping turns which
are stacked one on top of the other, $h_{c}=N\,d_{w}$, where $d_{w}$
is the diameter of the coil wire), then for an air-cored coil 
\[
L=\frac{\mu_{0}N^{2}\pi r_{c}^{2}}{h_{c}}\left(1-\frac{8w}{3\pi}+\frac{w^{2}}{2}-\frac{w^{4}}{4}+\mathcal{O}w^{6}\right)\mbox{[H]}
\]
If $w>1$, then 
\[
L=\mu_{0}N^{2}\pi r_{c}^{2}[\mbox{H}]
\]
In either case, the number of turns should be minimised, in order
to reduce $L$ and maintain good high frequency response, while still
having the product $NA$ sufficient for adequate signal strength.
Additional considerations include that the coil winding must be protected
from heat, and that the diffusion time of the magnetic field through
the heat shielding material must be small enough that the probe can
pick up the required information. The B probe coils used on the magnetic
compression experiment each had ten turns of 34 AWG magnet wire (0.86
$\Omega$/m), with cross-sectional area 0.02 mm$^{2}$. Each of the
probes located at the coordinates indicated in table \ref{tab: coordinates-ofBprobes}
had two ten-turn windings, which were individually oriented so as
to measure poloidal and toroidal field components. The coils were
wound on Delrin forms which were inserted in hollow thin-walled stainless
steel tubes with one closed end. The outer surfaces of the closed
ends of the tubes, and the inner surface of the chalice, were tungsten
coated. Tungsten coating of plasma facing surfaces reduces the level
of sputtering of impurity ions into the plasma, which in turn reduces
the level of plasma cooling due to line radiation. Each tube was inserted
in vertical holes through the chalice, as depicted in figure \ref{fig:Chalice},
with the closed end of the tubes flush with the chalice walls. The
skin diffusion time for the magnetic field through the closed ends
of the tubes can be estimated as $\tau_{diff}\sim\frac{t^{2}\mu}{\eta'}$,
where $t=0.5$ mm is the tube wall thickness, $\mu=\mu_{0}\mu_{r}$
is the material permeability (relative permeability $\mu_{r}\sim1$
for stainless steel, so $\mu=\mu_{0}$), and $\eta'\,[\Omega-\mbox{m}]$
is the material resistivity (\textasciitilde 1.16$\times10^{-6}\,\Omega-\mbox{m}$
for stainless steel), so that $\tau_{diff}\sim0.3\,\upmu$s. With
a coil design such that $\tau_{LR}<\tau_{diff}$, this implies that
changes in field magnitude over times $t\gtrsim0.3\,\upmu$s are captured.

The usual practice is to calibrate magnetic probes using a known field
so that a calibration factor can be found for converting the integrated
voltage signal to a field amplitude. From equation \ref{eq:770},
the calibration factor is $1/NA$. $N$ may be known accurately but
$A$ can only be approximated, particularly for coils with more than
a single layer of windings. 

Particular care was taken to reduce, as much as possible, the noise
pickup on the electrical signals. The magnet wire coil feedthroughs
were arranged in twisted pairs as far as the machine headplate. From
there, shielded cables transmitted the signals a short distance to
electrically shielded boxes, where the electrical signals are electronically
integrated, using a passive low-pass filter type integrator, and the
integrated signals were converted to optical signals. Standard optic
cables transmitted these signals a distance of several meters to an
electrically shielded area, where they were converted back to electrical
signals, digitised ($i.e.$, converted to digital data suitable for
computer processing), and stored for analysis.

\subsection{Current measurements - Rogowski coils}

\begin{figure}[H]
\centering{}\includegraphics[width=7cm,height=5cm]{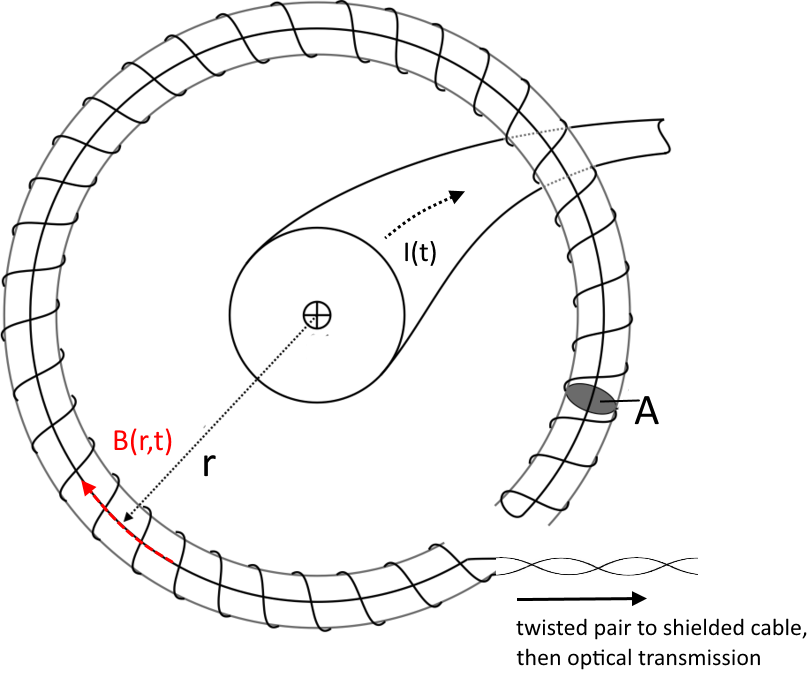}\caption{\label{fig:Rogowski-coil-schematic}$\,\,\,\,$Rogowski coil schematic}
\end{figure}
The principles behind magnetic probe operation also apply to Rogowski
coils, which are used to measure time varying currents. Figure \ref{fig:Rogowski-coil-schematic}
is a schematic of a Rogowski coil. $N$ turns of wire, with each turn
having cross-sectional area $A$, are wrapped around around an internal
torus-shaped form. The wire is returned along the toroidal path inside
the poloidal windings, so that there is no net toroidal turn. Using
Stoke's theorem, Ampere's law defines the magnetic field produced
by a current as $B(r,t)=\frac{\mu_{0}I(t)}{2\pi r}$, where $r$ is
the distance from the current path in the direction perpendicular
to the current flow. In combination with equation \ref{eq:770}, this
implies that the amplitude of a time varying current can be determined
from the time-integral of the measured induced voltage in a coil as
\begin{equation}
I(t)=-\frac{2\pi r}{NA\mu_{0}}\int_{0}^{t}V(t')\,dt'\label{eq:770.5}
\end{equation}
A Rogowski coil, encircling the cables that connected each levitation/compression
coil to its associated levitation inductor and capacitor bank, was
used to measure the levitation current for each of the levitation/compression
coils. Similarly, Rogowski coils were used to record the compression
current for each of the levitation/compression coils. As depicted
in figure \ref{fig:Machine-Schematic-1}, a single Rogowski coil encircling
the lower machine insulator was used to record formation current.

\subsection{Density measurements - interferometry}

Line averaged electron density can be measured through determination
of the phase shift of an electromagnetic wave traversing the plasma.
The amount of phase shift caused by crossing through the plasma depends
on the line-averaged refractive index of the plasma which, in turn,
depends on the line averaged electron density. A standard interferometric
technique is based on splitting a laser beam into two beams which
are directed along two paths of equal length, with a segment of one
of the paths being through the plasma. The beams are then merged and
the relative phase shift is assessed. 
\begin{figure}[H]
\centering{}\includegraphics[width=7cm,height=7cm]{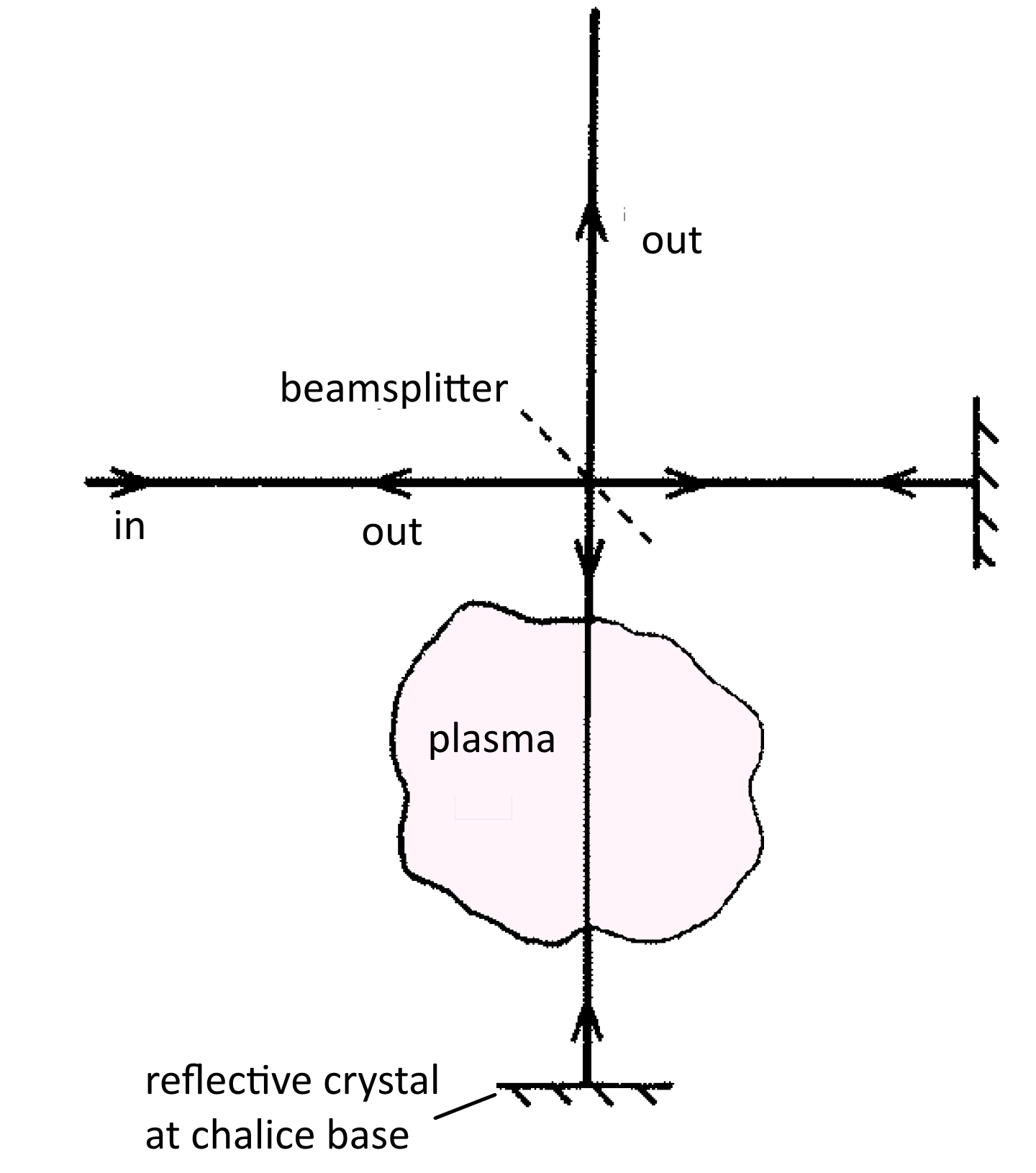}\caption{\label{fig:Michelson-interferometer-schemat}$\,\,\,\,$Michelson
interferometer schematic }
\end{figure}
The schematic is indicated in figure \ref{fig:Michelson-interferometer-schemat}
for the case of a Michelson-type interferometer, which was used in
the magnetic compression experiment - note that the plasma-traversing
beam crosses through the plasma twice. Retroreflectors positioned
at the base of the chalice (figure \ref{fig:Chalice}) reflect the
beam back up through the plasma. The reference beam is directed along
a path of equal length in ambient air.

After traversing their respective paths, the electric fields associated
with the beams are $Ee^{i\omega t}$ and $Ee^{i(\omega t+\phi)}$,
with phase difference $\phi$ between them, and the total electric
field of the merged beams is $E_{t}=Ee^{i\omega t}(1+e^{i\phi})$.
The merged beam power is proportional to $|E_{t}|^{2}$. Using basic
trigonometric identities, it can be shown that $|E_{t}|^{2}=2E^{2}(1+\mbox{cos}(\phi))$,
so that the phase shift can be found by measuring the merged beam
power.

Geometric optics defines that the phase lag applied to a wave due
to travelling a distance $L$ in a medium of refractive index $N=\frac{c}{v_{phase}}=\frac{ck}{\omega}$
is 
\[
\phi=\int_{0}^{L}N\frac{\omega}{c}dl=\int_{0}^{L}k\,dl
\]
Here, $c=3\times10^{8}$ m/s is the speed of light in a vacuum, $v_{phase}\,[\mbox{m/s}]$
is the phase speed of the wave, $k\,[\mbox{m}^{-1}]=\frac{2\pi}{\lambda}$
is the wavenumber, where $\lambda\,${[}m{]} is the wavelength, and
$\omega\,\mbox{[radian/s}]=kv_{phase}$ is the angular frequency of
the wave. The wavenumber $k$ for the reference beam is $k=k_{0}=\frac{\omega}{c}$.
The plasma-traversing beam also has wavenumber $k=k_{0}$ for the
plasma-external segment of its path, and $k=k_{plasma}$ for the segment
through the plasma. Hence, after merging, the phase difference between
the two beams will be 
\begin{equation}
\Delta\phi=\int_{0}^{L_{p}}(k_{plasma}-k_{0})\,dl\label{eq:770.6}
\end{equation}
where $L_{p}$ is the length of the path through the plasma. In order
to find an expression relating the phase difference to electron density,
the dispersion relation describing how the wavenumber depends on the
wave frequency must be defined. Systems of linear differential equations
can be studied using Fourier analysis - if any one quantity in the
equations varies sinusoidally with a particular frequency, then all
other quantities in the equations must vary at the same frequency
\cite{Goldston}. The plasma fluid equations are not linear, so there
is nonlinear coupling between frequencies. The equations can be linearised
if the perturbation amplitudes are considered to be small relative
to the steady state quantities, $i.e.,\,\widetilde{x}\ll x$ for all
quantities $x$ and perturbations $\widetilde{x}$. The process of
linearisation involves the first order expansion of the equations,
where second and higher order terms are neglected. All oscillating
quantities are represented with exponential notation, for example
the plasma number density is represented as $n=n_{0}+\widetilde{n}$,
where $\widetilde{n}\sim e^{i(\mathbf{k\cdot r}-\omega t)}$. Gradients
and time derivatives of perturbed quantities can be represented as
$\nabla\rightarrow i\mathbf{k}$ and $\frac{\partial}{\partial t}\rightarrow-i\omega$.
\\
\\
Taking the curl of Faraday's law, and using Ampere's law gives 
\begin{align}
\nabla\times\nabla\times\mathbf{E} & =-\nabla\times\dot{\mathbf{B}}\nonumber \\
\Rightarrow-\nabla^{2}\mathbf{E}+\nabla(\nabla\cdot\mathbf{E}) & =-\frac{\partial}{\partial t}\left(\nabla\times\mathbf{B}\right)\nonumber \\
 & =-\frac{\partial}{\partial t}\left(\mu_{0}\mathbf{J}+\mu_{0}\epsilon_{0}\dot{\mathbf{E}}\right)\label{eq:770.61}
\end{align}
From Gauss's law, $\nabla\cdot\mathbf{E}=\frac{\rho_{c}}{\epsilon_{0}}$,
where $\rho_{c}$ {[}Coulomb/m$^{3}${]} is the charge density. In
a vacuum, $\rho_{c}=0$ and $\mathbf{J}=\mathbf{0}$, so the vacuum
wave equation is $\nabla^{2}\widetilde{\mathbf{E}}=\frac{1}{c^{2}}\frac{\partial\mathbf{^{2}\widetilde{\mathbf{E}}}}{\partial t^{2}}$,
where the relation $c^{2}=\frac{1}{\mu_{0}\epsilon_{0}}$ has been
used. The perturbation to the electric field scales as $\widetilde{\mathbf{E}}\sim e^{i(\mathbf{k\cdot r}-\omega t)}$,
so that dispersion relation for electromagnetic waves in a vacuum
can be expressed as 
\[
\omega^{2}=c^{2}\,k^{2}
\]
In a plasma, quasineutrality holds over distances greater than the
Debye length, so $\nabla\cdot\mathbf{E}=\frac{\rho_{c}}{\epsilon_{0}}=0$.
Hence, in terms of the field perturbations, 
\begin{equation}
\nabla^{2}\widetilde{\mathbf{E}}=\frac{\partial}{\partial t}\left(\mu_{0}\widetilde{\mathbf{J}}+\frac{1}{c^{2}}\dot{\widetilde{\mathbf{E}}}\right)\label{eq:770.62}
\end{equation}
In plasma, $\widetilde{\mathbf{J}}\neq\mathbf{0}$ so an expression
for $\mathbf{\widetilde{J}}$ must be found in order to find the dispersion
relation. The electron momentum equation in an unmagnetised cold plasma
is $m_{e}n_{e}\left(\mathbf{\dot{v}}_{e}+\left(\mathbf{v}_{e}\cdot\nabla\right)\mathbf{v}_{e}\right)=-en_{e}\mathbf{E}$,
where $\mathbf{v}_{e}$ is the electron fluid velocity. This can be
linearised about an equilibrium with $\mathbf{v}_{e0}=\mathbf{0}$,
leading to $i\omega m_{e}\mathbf{\widetilde{v}}_{e}=e\widetilde{\mathbf{E}}$.
Over the short timescales associated with high frequency phenomena,
the ions can be considered to be fixed in space, so that $\widetilde{\mathbf{J}}=-n_{0}e\,\mathbf{\widetilde{v}}_{e}$.
Combination of these last two expressions leads to $\widetilde{\mathbf{J}}=\frac{i\,n_{0}e^{2}\widetilde{\mathbf{E}}}{\omega m_{e}}$,
so that, referring to equation \ref{eq:770.62}, $\frac{\partial}{\partial t}\left(\mu_{0}\widetilde{\mathbf{J}}\right)\rightarrow\mu_{0}\frac{n_{0}e^{2}\widetilde{\mathbf{E}}}{m_{e}}=\frac{\omega_{pe}^{2}}{c^{2}}\,\widetilde{\mathbf{E}}$,
where $\omega_{pe}=\sqrt{\frac{e^{2}n_{e}}{\epsilon_{0}m_{e}}}$ is
the electron plasma frequency. Thus, the dispersion relation for high
frequency waves in an unmagnetised plasma is 
\begin{equation}
\omega^{2}=c^{2}k_{plasma}^{2}+\omega_{pe}^{2}\label{eq:770.7}
\end{equation}
This dispersion relation also holds for magnetized plasmas in which
$\omega_{ce}\gg\omega$, where $\omega_{ce}=\frac{eB}{m_{e}}$ is
the electron cyclotron frequency \cite{Goldston}. Note that if the
wave frequency $\omega$ is less than the plasma frequency, then $k$
and $N$ are imaginary and the wave is evanescent. For wave propagation,
$\omega>\omega_{pe}\Rightarrow n_{e}<\frac{\omega^{2}\epsilon_{0}m_{e}}{e^{2}}$.
The wave can propagate through the plasma only if the plasma density
$n_{e}$ is less than some critical density: 
\begin{equation}
n_{c}=\frac{\omega^{2}\epsilon_{0}m_{e}}{e^{2}}\label{eq:771}
\end{equation}
Equation \ref{eq:770.7} implies that the beam wavenumber for the
path segment through the plasma can be expressed as $k_{plasma}=\frac{1}{c}\sqrt{\omega^{2}-\omega_{pe}^{2}}$.
Hence, with $k_{0}=\frac{\omega}{c}$, the phase difference (equation
\ref{eq:770.6}) can be re-expressed as 
\[
\Delta\phi=\int_{0}^{L_{p}}\left(\frac{\omega}{c}\left(1-\left(\frac{\omega_{pe}}{\omega}\right)^{2}\right)^{\frac{1}{2}}-\frac{\omega}{c}\right)\,dl
\]
In the limit $\omega\gg\omega_{pe}$, $\left(1-\left(\frac{\omega_{pe}}{\omega}\right)^{2}\right)^{\frac{1}{2}}\approx1-\frac{1}{2}\left(\frac{\omega_{pe}}{\omega}\right)^{2}$,
so this can be reduced to 
\[
\Delta\phi=-\frac{1}{2}\frac{\omega}{c}\int_{0}^{L_{p}}\left(\frac{\omega_{pe}}{\omega}\right)^{2}\,dl
\]
Using the definitions for the plasma frequency and the critical density,
this implies that 
\[
\Delta\phi=-\frac{\omega}{2c\,n_{c}}\int_{0}^{L_{p}}n_{e}\,dl
\]
Therefore, the line averaged electron density $\frac{1}{L_{p}}\int_{0}^{L_{p}}n_{e}\,dl$
can be assessed through determination of the phase shift, which is
a function of the power of the merged beams.

. 

\subsection{Temperature measurements - ion-Doppler spectroscopy\label{subsec:Temperature-measurements--}}

When bound electrons in atoms or partially ionized ions undergo transitions
from upper to lower discrete energy levels ($i.e.,$ quantum states),
photons are emitted. The photon energy is given by $E_{p}=E_{i}-E_{j}=h\nu=\frac{hc}{\lambda}$,
where $E_{i}$ is the energy of the upper level which the electron
originally occupied, and $E_{j}$ is the energy of the lower level
which the electron occupies after the transition, $h=6.63\times10^{-34}\,[\mbox{J-s}]$
is Planck's constant, $c$ is the speed of light, while $\lambda$
{[}m{]} and $\nu\,[\mbox{s}^{-1}]$ are the wavelength and frequency
respectively of the radiation associated with the photon. When the
radiation emitted by a group of particles is measured, radiation at
frequencies/wavelengths corresponding to dominant transitions appear
as relatively bright spectral lines in an otherwise uniform and continuous
radiation spectrum. When prior knowledge of the dominant transitions
for different types of particles is available, spectral lines can
be used to identify the particles whose radiation spectrum is being
measured. Due to various spectral line broadening mechanisms, in particular
natural broadening, pressure broadening, and Doppler broadening, the
radiation emitted as a result of transition between any two particular
energy levels is spread over a range of frequencies. 

Natural broadening arises because, as a consequence of the uncertainty
principle, the quantum states of an atom or ion have a small spread
in energy \cite{Hutchinson}. The lifetime of the atom or ion in a
particular state $k$ is finite due to spontaneous transitions to
lower quantum states. The effective spread in energy is given by $\Delta E\approx\frac{h}{2\pi\tau_{k}}$,
where the lifetime in state $k$ is given by $\tau_{k}=\frac{2}{\Sigma_{j}A_{kj}}$,
where $A_{kj}\,[\mbox{s}^{-1}]$ is the probability per unit time
for spontaneous transition from state $k$ to state $j$. The corresponding
broadening of the range of wavelengths for radiation emitted due to
transitions from state $k$ is given by $\Delta\lambda=\frac{hc}{\Delta E}\approx2\pi c\tau_{k}$
\cite{Hutchinson}. 

Density broadening (also known as pressure or collisional broadening)
arises from the influence of nearby particles on the emitting atom
or ion. The collisional approach to calculating density broadening
considers that collisions between the atom (or ion) and electrons
cause broadening by interrupting the emission of the wave train. An
uncertainty in the frequency of the emission is introduced, analogous
to the uncertainty in the upper quantum state energy level that leads
to natural broadening \cite{Tallents}. The quasistatic approach to
calculating density broadening looks at the effect of the electric
fields due to nearby particles, which perturb the atomic energy levels
in the emitting atom or ion. Shifts in quantum state energy levels
are known as Stark shifts, so this mechanism is known as Stark broadening
\cite{Hutchinson}. The measurements of line widths of Stark-broadened
lines can be used to determine ion densities in plasmas where density
is high enough that Stark broadening is the dominant broadening effect.

Due to the Doppler effect, if a moving object emits a wave with a
frequency $\nu_{0}$ in its own reference frame, the frequency measured
in the frame of a stationary observer is modified by the effect of
the object's motion. If the object is moving at velocity $V$ towards
or away from the observer, where $V$ is positive if the object is
moving towards the observer, or negative if moving away, then the
observed frequency is given by $\nu=\nu_{0}(1+\frac{V}{c})$. In a
plasma, individual ions (and atoms) have a distribution of thermal
speeds, towards and away from the point of observation, and the net
effect is a spread in observed wavelengths and frequencies, $i.e.,$
broadening of the observed spectral line. When thermal motion causes
a particle to move towards the observer, the emitted radiation will
be shifted to a higher frequency. Likewise, when the emitter moves
away, the frequency will be lowered. If $F(V_{L})$ is the particle
distribution function for the velocity component along the line of
sight between the particles and the point of observation, ($i.e$.,
$F(V_{L})\,dV_{L}$ is the fraction of particles with velocity between
$V_{L}$ and $V_{L}+dV_{L}$ along the line of sight), then the corresponding
frequency distribution is $F(\nu)\,d\nu=F\left(\left(\frac{\nu}{\nu_{0}}-1\right)c\right)\frac{dV_{L}}{d\nu}\,d\nu$.
If the distribution for the velocity component along the line of sight
is given by a Maxwellian:
\[
F(V_{L})\,dV_{L}=\sqrt{\frac{m_{i}}{2\pi T_{i}}}\,\mbox{exp}\left(\frac{-m_{i}V_{L}^{2}}{2T_{i}}\right)\,dV_{L}
\]
where $m_{i}$ and $T_{i}$ are the mass and temperature (in Joules)
of the atom or ion, then the corresponding frequency distribution
is
\[
F(\nu)\,d\nu=\frac{c}{\nu_{0}}\sqrt{\frac{m_{i}}{2\pi T_{i}}}\,\mbox{exp}\left(\frac{-m_{i}c^{2}(\nu-\nu_{0})^{2}}{2T_{i}\,\nu_{0}^{2}}\right)\,d\nu
\]
which is a Gaussian profile ($i.e.$, of the form $g(x)=a\,\mbox{exp}\left(\frac{-(x-b)^{2}}{2\sigma^{2}}\right)$,
where $b$ is the expected (mean) value, and $\sigma$ is the standard
deviation), with variance $\sigma^{2}=\frac{T_{i}\nu_{0}^{2}}{m_{i}c^{2}}=\frac{V_{thi}^{2}\nu_{0}^{2}}{2c^{2}}$,
where $V_{thi}=\sqrt{\frac{2T_{i}}{m_{i}}}$ is the ion thermal speed.
The full-width-half-maximum for a Gaussian profile $g(x)$ is given
by $g_{FWHM}=2\sigma\sqrt{2\,\mbox{ln}(2)}$, so for frequency broadening
due to the ion Doppler effect, the full-width-half-maximum is $\nu_{FWHM}\sim1.7\,\frac{\nu_{0}}{c}V_{thi}$.
When Doppler broadening is the dominant broadening mechanism, as is
the case for many applications, determination of $\nu_{FWHM}$ enables
an estimation of the ion temperature along the line of sight of the
diagnostic. Doppler broadening can be measured for either the majority
ion species, or any prevalent impurity ion. For hydrogen plasmas,
where the first ionisation energy is $13.6$ eV, Doppler broadening
of the hydrogen line is only useful for diagnosing cooler edge plasma.
Therefore, measurement of the temperature of impurity ions with higher
ionization energies is more practical in many cases. When the thermal
equilibration time between the majority ions and the impurity ion
is short, as is often the case for plasmas with moderate density \cite{Hutchinson},
this measurement gives a good estimate of the majority ion temperature.
In the magnetic compression experiment, where working gas was generally
helium, the ion-Doppler diagnostic was focused on the He II line at
468.5nm. 

While Doppler broadening of the observed spectral line leads to a
frequency distribution with a Gaussian profile, natural broadening
and density broadening result in frequency distributions with a Lorentzian
profiles. Generally, temperature ($i.e.,$ Doppler) and density broadening
are the dominant broadening mechanisms - natural broadening is rarely
observed directly, except in nebular environments. The combination
of temperature broadening and density broadening leads to a Voigt
profile for the frequency distribution. This is the convolution of
Gaussian and Lorentzian profiles, and has no simple analytic form.
Ion-Doppler measurements over magnetic compression, presented in chapter
\ref{Chap:Magnetic-Compression} were carefully analysed to assess
the relative contributions of temperature and density broadening to
the observed broadening of the He II line. In some cases, by assessing
the quality of the fits of photon count against photon frequency profiles
to Lorentzian and Gaussian profiles, line broadening that at first
glance seemed to indicate a particularly significant temperature increase
at compression, was deemed to be due to density broadening. With reference
to data presented in \cite{Kunze,Pittman}, an error in the temperature
measurement (He II line at 468.5 nm) due to density broadening has
been evaluated as $\sim$17 eV for a density of 1.6$\times10^{22}$
m$^{-3}$, and the error falls off in proportion to $n_{e}^{0.83}$.
Density along the ion-Doppler chords is not directly evaluated, but
from this information, along with observations of the variation of
line averaged density with radius, sensible estimates to the contribution
of density broadening to the indicated temperature increases can be
evaluated on a shot-to-shot basis. 

\subsection{Optical diagnostics\label{subsec:Optical-diagnostics}}

The intensity of optical emission along vertical chords at the locations
of the optical ports indicated in figure \ref{fig:Machine-headplate-schematic-1}
was recorded over time. Optic fibres carry the signals to an electrically
screened area where they are converted to electrical signals using
photodiodes, then the electrical signals are digitised and stored
for analysis. 

\newpage{}

\chapter{Magnetic Levitation\label{Chap:MagLev}}

This chapter includes a description, with experimental results, of
the main configurations tested to form a CT into a levitation field,
which applies a radial force on the plasma that \textquotedbl levitates\textquotedbl{}
it off the outer insulating wall. Configurations are presented in
the chronological order in which they were tested. The original configuration,
comprising six coils around a ceramic insulating tube, is the focus
of section \ref{subsec:6-coils-config_ceramic}. Section \ref{subsec:6-coilsconfig, quartz}
presents results from the configuration with six coils around a quartz
insulating tube that had increased internal radius. Section \ref{subsec:Toroidal-MHD-modes}
discusses the fluctuations that were routinely observed on poloidal
field signals associated with CTs produced in standard MRT machines,
but were absent for magnetically levitated CTs. This discrepancy led
to the theory that mode-locking might be impeding coherent CT toroidal
rotation, and was a consideration that led to the design of the 25-turn
levitation coil configuration that is the focus of section \ref{subsec:Multi-turn-coil}.
Along with improved toroidal symmetry of the levitation field, the
25-turn coil had a modified poloidal field profile, and closed the
gaps located above and below the stack of six discrete coils in the
original 6-coil configuration. This led to reduced plasma-wall interaction
during CT formation, and significantly improved levitated CT performance.
Section \ref{subsec:semipermshell} outlines the semi-permeable shell
concept, and the experiment that was conducted to test if it was the
action of the levitation field itself, rather than plasma impurity
considerations, that was responsible for the absence of fluctuations
on levitated CT poloidal field signals, and for the relatively poor
performance of levitated CTs compared with CTs produced in MRT machines.
Shortly after that test, the magnetic compression experiment was scheduled
to be decommissioned. Section \ref{subsec:11-coil-configuration}
focuses on levitated CT performance in the final configuration tested
during the extra time allocated to me to get data for this thesis.
That configuration, comprising eleven coils around the quartz insulating
tube had a levitation field profile similar to that of the multi-turn
coil, and also led to significantly improved levitated CT performance.
The strategy of matching the decay rates of levitation current and
CT currents was developed in that configuration, and led to an understanding
of the compressional instability that is discussed in chapter \ref{Chap:Magnetic-Compression}.
The method developed to experimentally measure the outboard equatorial
CT separatrix is described in section \ref{subsec:Using-side-probe-data}.
A comparison of total spectral power and levitated CT lifetime for
the principal configurations tested is presented in section \ref{subsec:Comparison-of-total}.
This chapter concludes with a summary in section \ref{subsec:SummaryLev}.

\section{\label{subsec:6-coils-config_ceramic}6-coil configuration, ceramic
outer insulating wall}

\begin{figure}[H]
\begin{raggedright}
\subfloat[]{\centering{}\includegraphics[scale=0.5]{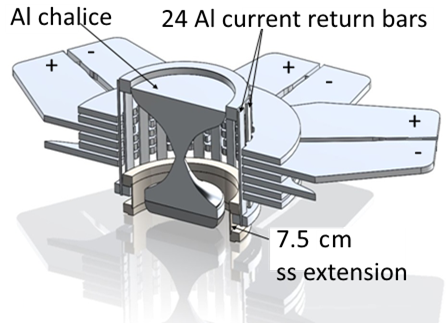}}\hfill{}\subfloat[]{\centering{}\includegraphics[scale=0.3]{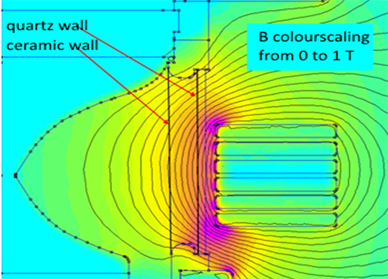}} 
\par\end{raggedright}
\centering{}\caption{\label{fig:Schematic-of-6}$\,\,\,\,$6-coil configuration schematic
and FEMM solution for levitation field }
\end{figure}
Figure \ref{fig:Schematic-of-6}(a) indicates, for the original configuration
with six coils, the coils, chalice, stainless steel extension, and
aluminum return current bars that carry axial current outside the
insulating wall. The inner radii of the original ceramic (alumina
- Al$_{2}$O$_{3}$) wall and the quartz (silica - $\mbox{SiO}_{2}$)
wall that was tested later are shown in figure \ref{fig:Schematic-of-6}(b).
This is an output plot from the open-source FEMM (Finite Element Method
Magnetics) program \cite{FEMM}, with 30 kA per levitation coil and
an input current frequency of 800 Hz. Contours of $\psi_{lev}$, poloidal
levitation flux, are shown, with the plot colour-scaling being proportional
to $|B|.$ FEMM models alternating currents as sinusoidal waveforms
in time, so we chose $|t_{lev}|=300\,\upmu$s to be the quarter-period,
giving a frequency of $800$ Hz. In the 6-coil configuration, it was
found that CT lifetimes could be increased by $\sim10\%$ by firing
the levitation capacitors at $t_{lev}\sim-300\,\upmu$s, well before
firing the formation capacitors. This allows the levitation field
to soak into the stainless steel above and below the wall, resulting
in line-tying (field-pinning); magnetic field that is allowed to soak
into the steel can only be displaced on the resistive timescale of
the metal, which is longer than the time it takes for the CT to bubble-in
to the containment region. Note that the principal materials used
in the construction of the magnetic compression machine are indicated
in figure \ref{fig:Machine-Schematic-1}. As confirmed by MHD simulations
(see section \ref{subsec:Simulated-plasma-wall-interactio}), this
line-tying effect is thought to have reduced plasma-wall interaction
and CT impurity inventory by making it a little harder for magnetised
plasma entering the confinement region to push aside the levitation
field. As described in section \ref{subsec:Boundary-conditions},
FEMM models were used to produce boundary conditions for $\psi$,
pertaining to the peak values of toroidal currents in the main, levitation,
and compression coils at the relevant frequencies, for MHD and equilibrium
simulations. For MHD simulations, boundary conditions for $\psi_{lev}(\mathbf{r},t)$
and $\psi_{comp}(\mathbf{r},t)$ are scaled over time according to
the experimentally measured waveforms for $I_{lev}(t)$ and $I_{comp}(t)$. 

The $7.5$$\mbox{ cm}$ high stainless steel extension indicated in
figure \ref{fig:Schematic-of-6}(a) was an addition to the original
configuration that also helped reduce the problem of plasma-wall interaction
- in the original design without the extension, the ceramic insulating
outer wall extended down an additional 7.5 cm. With the original levitation
field profile from six coils, and without the extension, CTs were
short-lived, up to $\sim100\,\upmu$s as determined from the poloidal
B-probes embedded in the aluminum inner flux conserver at $r=52\mbox{ mm}$
(see figures \ref{fig:Machine-Schematic-1}, \ref{fig:Machine-headplate-schematic-1},
and \ref{fig:Chalice}), compared with over $300\,\upmu$s on MRT
injectors with an aluminum outer flux conserver. CT lifetime was increased,
up to $\sim170\,\upmu$s, with the addition of the steel extension.
The extension mitigated the problems of sputtering of steel at the
lower alumina/steel interface, and of plasma interaction with the
insulating wall during the formation process. \\
\\
\begin{figure}[H]
\begin{raggedright}
\subfloat[]{\centering{}\includegraphics[width=8cm,height=5cm]{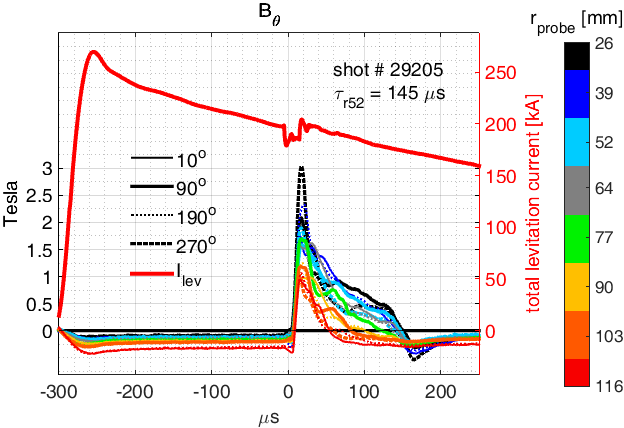}}\hfill{}\subfloat[]{\centering{}\includegraphics[width=8cm,height=5cm]{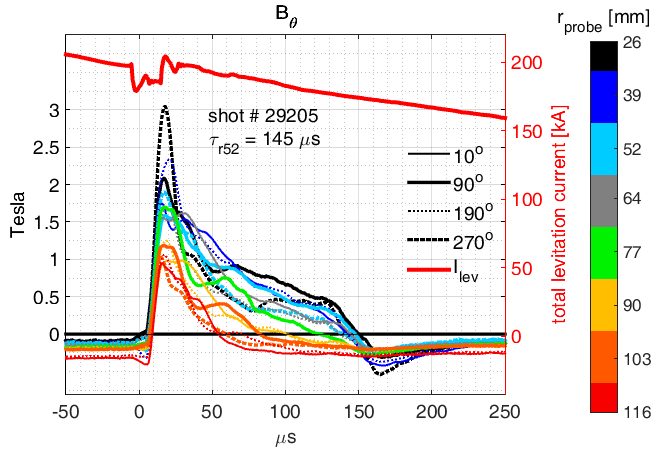}}
\par\end{raggedright}
\centering{}\caption{\label{fig:Bp_29205lev}$\,\,\,\,$$B_{\theta}$ for for shot  29205,
six coils, ceramic insulating wall}
\end{figure}
Figures \ref{fig:Bp_29205lev}(a) and (b) (close-up view) shows $B_{\theta}(r,\,t),$
the poloidal field traces measured at the magnetic probes located
in the chalice wall, for shot  29205. With optimisation of the machine
settings ($V_{form},\,V_{lev},\,I_{main}$, $t_{lev}$ and $t_{gas}$
- see table \ref{tab:Sequence-of-machine}) for longest lived CTs,
this was a typical shot in the configuration with six coils and a
ceramic (alumina) outer insulating wall, with the steel extension
in place. As indicated in table \ref{tab: coordinates-ofBprobes},
there are sixteen probes located in the chalice - eight of these are
located at four different radii ($r=39,\,64,\,90$, and 116 mm) at
toroidal angle $\phi=10^{o}$ and $\phi=190^{o}$, and there are an
additional eight probes at ($r=26,\,52,\,77$, and 103 mm) at $\phi=90^{o}$
and $\phi=270^{o}$. Magnetic probe signals are colored by the r-coordinates
of the probe locations, with toroidal coordinates of the probe locations
denoted by linestyle, as denoted in the plot legends. Note that $B_{\theta}$
is the field component parallel to the chalice surface in the poloidal
plane. Total levitation current, measured with Rogowski coils, is
also indicated (thick red lines, right axes). CT lifetime is gauged
using the $\tau_{r52}$ metric (indicated in figures \ref{fig:Bp_29205lev}(a)
and (b)), which is the time at which the average of the poloidal field
measured at the two probes at $r=$52 mm crosses zero. 

For shot  29205, $t_{lev}=-300\,\upmu$s, so with a current rise time
of $\sim40\,\upmu$s in the levitation coils, the poloidal levitation
field measured at the probes reaches its maximum negative value at
$t\sim-260\,\upmu$s, when the total levitation current is around
$260$ kA (\textasciitilde 43 kA per coil). After its peak, the poloidal
levitation field decays at the same rate as the levitation current.
Formation capacitors are fired at $t=0\,\upmu$s and, referring to
figure \ref{fig:MHDform-1}, it takes $\sim10-20\,\upmu$s for the
gun (stuffing) flux to be advected up to the probe locations. The
total levitation current has dropped to around $200$ kA by the time
the magnetised plasma bubbles into the confinement region. Until this
time ($\sim10\,\upmu$s), the poloidal field measured at the probes
is purely the levitation field. When the magnetised plasma bubbles
into the confinement region, it displaces the levitation field, and
the polarity of the measured field reverses - the stuffing field has
opposite polarity to the levitation field, $i.e.,$ the toroidal currents
in the main solenoidal coil and the levitation coils are in opposite
directions. Over the next several tens of $\,\upmu$s, during and
after reconnection of poloidal field to form closed flux surfaces,
the CT undergoes Taylor relaxation during which part of the poloidal
flux is converted to toroidal flux. The CT shrinks and is displaced
inwards by the relatively constant levitation field as the CT currents
and fields decay resistively. Starting at the outer probes and progressing
inwards towards the inner probes, the CT field measured at the probes
is replaced by the levitation field. After $\sim150\,\upmu$s, the
poloidal field measured at all the B probes is the levitation field
once again. Note that, when levitation field is being measured at
the probes, that $|B_{\theta}|$ is larger at the outer probes, due
to the $1/(r_{coil}-r_{probe})$ scaling of levitation field with
levitation current in the external coils. On the other hand, when
CT field is being measured at the probes, $B_{\theta}$ is larger
at the inner probes, due to the $1/r_{probe}^{2}$ scaling of CT field
with CT flux - poloidal field lines are bunched together progressively
more at smaller radii.

\begin{figure}[H]
\begin{centering}
\subfloat[]{\centering{}\includegraphics[width=8cm,height=5cm]{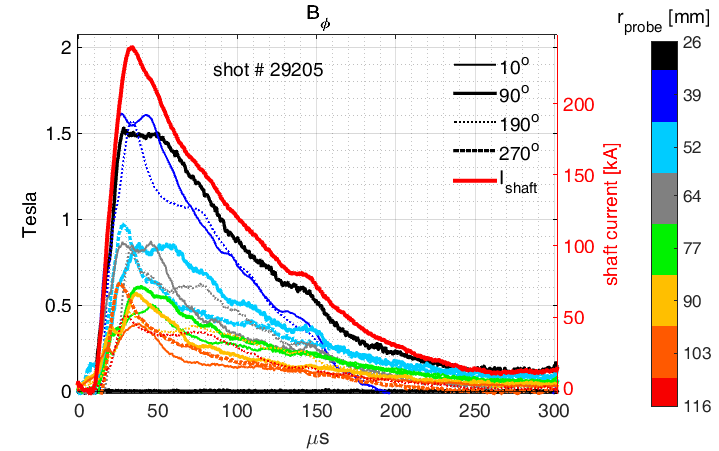}}\hfill{}\subfloat[]{\centering{}\includegraphics[width=8cm,height=5cm]{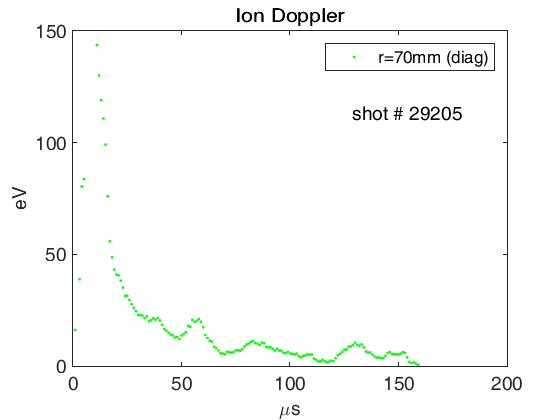}}
\par\end{centering}
\centering{}\caption{\label{fig:Bt_Ti_29205}$\,\,\,\,$$B_{\phi}$ and $T_{i}$ for shot
 29205, six coils, ceramic insulating wall}
\end{figure}
Figure \ref{fig:Bt_Ti_29205}(a) shows the toroidal field, $B_{\phi}(r,\,t)$,
measured at the chalice probes, for shot  29205. The toroidal field
is due to poloidal shaft current in the aluminum chalice, and not
due to poloidal CT current - recall that the field outside a toroidal
solenoid is zero - in this analogy the plasma's poloidal current constitutes
the solenoid's current. Shaft current is induced to flow in the metal
surrounding the CT confinement area as the system tries to conserve
the toroidal flux introduced at CT formation, and continues to decay
away resistively for several tens of microseconds after the CT currents
have resistively decayed. $I_{shaft}$ (thick red traces in figure
\ref{fig:Bt_Ti_29205}(a), right axis) is calculated from $B_{\phi}(r,\,t)$
using Ampere's law. As the CT shrinks due to resistive decay of CT
currents, compounded by mild magnetic compression due to being pushed
on by the relatively constant levitation field, which decays much
more slowly than the CT currents, increasing proportions of poloidal
shaft current can divert from the initial paths in the aluminum bars
shown in figure \ref{fig:Schematic-of-6}(a), and flow through the
ambient plasma surrounding the CT. This will be clarified in section
\ref{subsec:Compressional-Instability}. Shaft current increases when
it flows along the reduced inductance path through the ambient plasma.
There is evidence in \ref{fig:Bt_Ti_29205}(a) of mild magnetic compression
by the levitation field starting at around $130\,\upmu$s on shot
29205. This is evident from the overall rise in shaft current at $130\,\upmu$s,
and from the increase in $B_{\phi}$ at the probes, particularly at
the inner probes located at $(r,\,\phi)=$ $(26\mbox{ mm},\:90^{\circ})$,
$(39\mbox{ mm},\:10^{o})$, and $(52\mbox{ mm},\:90^{\circ})$. As
outlined in section \ref{subsec:Levitation-field-decay}, this unintentional
compression event was eliminated in the 11-coil configuration with
the implementation of a levitation circuit modification.

Figure \ref{fig:Bt_Ti_29205}(b) shows an ion Doppler measurement
for the same shot. Only the ion Doppler diagnostic looking along the
green colored diagonal chord with its lower point $r=70$ mm, indicated
in figure \ref{fig:Chalice}, was functioning for this shot. A peak
ion temperature measurement in that region, of around $150$ eV, is
recorded at around $15\,\upmu$s, when plasma enters the confinement
region. Viscous heating is the predominant ion heating mechanism during
CT formation, as ions are rapidly advected up the gun into the confinement
region. Further ion acceleration occurs at early times around the
confinement region entrance as open field lines reconnect to form
closed flux surfaces. The jets associated with this reconnection process
leads to additional ion viscous heating. The subsequent ion cooling
due to thermal transport to the vessel walls has an offset due to
collisional heat exchange with electrons that are being heated ohmically.
The slight increase in ion temperature seen here at $\sim130\,\upmu$s
is likely due to compressional heating associated with (unintentional)
magnetic compression of the CT by the levitation field at around the
same time. Interferometer electron density measurements are not available
for shot 29205, so the error in measured ion temperature associated
with density broadening can not be assessed for this shot. This error
will be discussed in section \ref{subsec:Levitation-field-decay}
when presenting electron density and ion temperature measurements
for shots for which both diagnostics are available. 
\begin{figure}[H]
\begin{centering}
\subfloat[]{\centering{}\includegraphics[width=8cm,height=5cm]{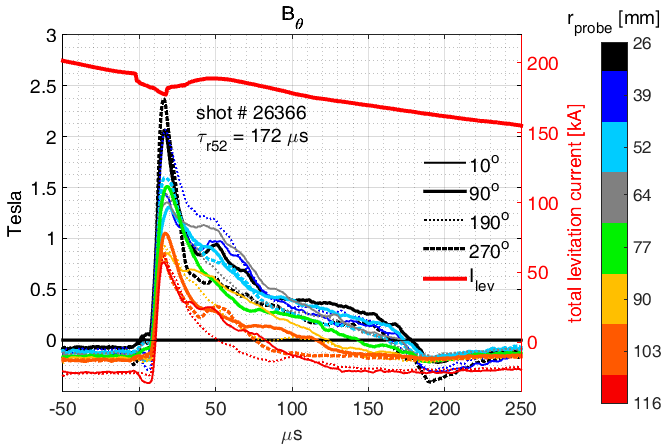}}\hfill{}\subfloat[]{\includegraphics[width=8cm,height=5cm]{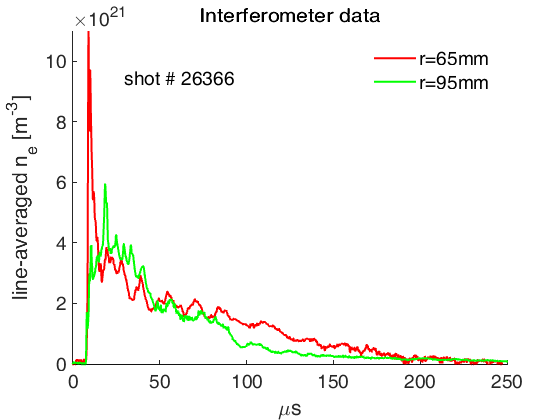}

}
\par\end{centering}
\centering{}\caption{\label{fig:Bp_ne_26366}$\,\,\,\,$$B_{\theta}$ and $n_{e}$ for
shot  26366, six coils, ceramic insulating wall}
\end{figure}
Figure \ref{fig:Bp_ne_26366}(a) shows the measured poloidal field
for shot  26366, which, with $\tau_{r52}=170\,\upmu$s, was the longest-lived
shot taken in the 6-coil configuration with the ceramic outer insulating
wall. Repeatability of long lived shots was poor, until the implementation
of the levitation circuit modification described in section \ref{subsec:Levitation-field-decay}.
Figure \ref{fig:Bp_ne_26366}(b) indicates the line averaged density
recorded for this shot. There are initial spikes in density at $\sim15\,\upmu$s,
when plasma enters the confinement region at high speed and is compressed
against the upper chalice walls. MHD simulations indicate that closed
CT flux surfaces have formed by around $30\,\upmu$s. The CT decays
resistively and shrinks over time as it is pushed inwards by the relatively
constant levitation field. The inward movement of the CT magnetic
axis, around which density is expected to be concentrated, may explain
the fact that the line-averaged density along the inner chord at $r=65$
mm approaches and sometime surpasses (as it does in this shot) the
density at $r=95$ mm. The innermost interferometer chord at $r=35$
mm was generally not operational. 

\section{\label{subsec:6-coilsconfig, quartz}6 coil configuration, quartz
outer insulating wall}

An insulating wall with increased internal radius was tested (original
alumina tube with $r_{in}=144$ mm was replaced with a quartz tube
with $r_{in}=170$ mm). The resistive part of $\dot{\psi}$ is $\dot{\psi}_{\eta}=\eta\Delta^{^{*}}\psi$
(see section \ref{par:Expressionfor_psidot}), where $\Delta^{^{*}}$
is the elliptic Laplacian-type operator used in the Grad-Shafranov
equation, and $\eta\,[\mbox{m}^{2}/\mbox{s}]$ is the magnetic diffusivity,
so CT lifetime should scale with $l^{2}$, where $l$ is some characteristic
length scale associated with the CT. The radius of the inboard wall
of the chalice at $z=0$ is $r_{w}\sim20\mbox{ mm}$. The minor CT
radius for a given $r_{in}$ would be approximately $a\sim\frac{r_{ins}-r_{w}}{2}$,
so assuming that $l\sim a$, we have $\frac{l_{quartz}^{2}}{l_{ceramic}^{2}}\sim1.5$.
\begin{figure}[H]
\begin{raggedright}
\subfloat[]{\centering{}\includegraphics[width=8cm,height=5cm]{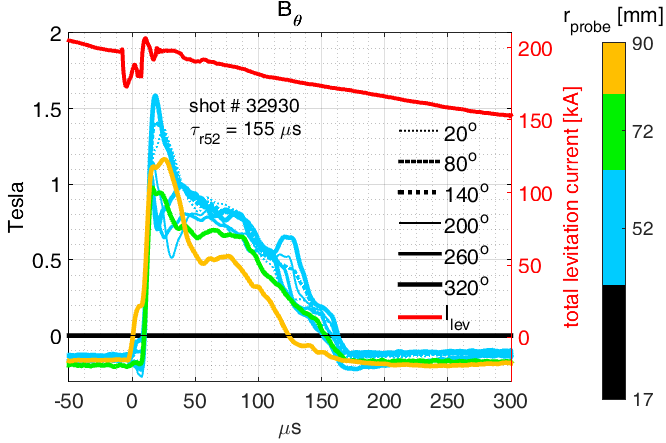}}\hfill{}\subfloat[]{\centering{}\includegraphics[width=8cm,height=5cm]{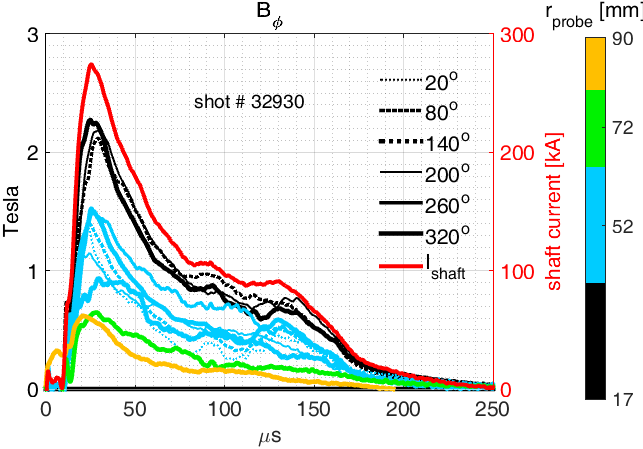}}
\par\end{raggedright}
\centering{}\caption{\label{fig:Bp_Bt_32930}$\,\,\,\,$$B_{\phi}$ and $B_{\theta}$ for
shot  32930, six coils, quartz insulating wall}
\end{figure}
From this rough estimate, an increase from $\sim170\,\upmu$s to $\sim260\,\upmu$s
would be expected with the transition to a larger radius insulating
tube, if the tube material was not changed. Figure \ref{fig:Bp_Bt_32930}(a)
indicates how, contrary to expectations, CT lifetime decreased noticeably
($\sim170\,\upmu$s to $\sim150\,\upmu$s) with the transition, so
that in terms of CT lifetime, quartz was almost $twice$ as bad as
ceramic - shot  32930 produced one of the longest-lived CTs in the
configuration with six coils around the quartz wall. The quartz wall
led to more plasma impurities, and consequent further radiative cooling,
and therefore an increased rate of resistive decay. Note that, for
the configuration with six coils and the quartz wall, the chalice
was replaced with a different model, which had the same geometric
shape, but had fewer probes at different locations, which are indicated
in the following table:\hfill{} \\
\begin{table}[H]
\centering{}{\footnotesize{}}%
\begin{tabular}{|c|c|c|c|c|c|c|c|c|c|c|c|}
\hline 
{\footnotesize{}$\mathbf{r}\,[\mbox{\mbox{mm}}]$} & {\footnotesize{}$17$} & {\footnotesize{}$17$} & {\footnotesize{}$17$} & {\footnotesize{}$52$} & {\footnotesize{}$52$} & {\footnotesize{}$52$} & {\footnotesize{}$52$} & {\footnotesize{}$52$} & {\footnotesize{}$52$} & {\footnotesize{}$72$} & {\footnotesize{}$90$}\tabularnewline
\hline 
{\footnotesize{}$\phi\,${[}deg.{]}} & {\footnotesize{}$80$} & {\footnotesize{}$200$} & {\footnotesize{}$320$} & {\footnotesize{}$20$} & {\footnotesize{}$80$} & {\footnotesize{}$140$} & {\footnotesize{}$200$} & {\footnotesize{}$260$} & {\footnotesize{}$320$} & {\footnotesize{}$320$} & {\footnotesize{}$320$}\tabularnewline
\hline 
\end{tabular}\caption{\label{tab: coordinates-ofBprobes-1}$\,\,\,\,$$r,\,\phi$ coordinates
of magnetic probes (intermediate configuration)}
\end{table}

With six probes located toroidally every $60^{o}$ at $r=52$ mm,
the updated chalice design was intended to give more information on
MHD activity associated with toroidal modes. The primary motivation
for the re-design was for implementation on other GF plasma injectors,
which routinely indicated magnetic fluctuations with toroidal mode
number $n=2$ on the measured $B_{\theta}$ signals, as determined
by examining the $B_{\theta}$ signals from probes located at the
same radius $180^{o}$ apart toroidally. Six probes toroidally spaced
at the same radius could, in principle, enable identification of modes
up to $n=5$. The two inner probes on the original chalice at $r=26$
mm were replaced with three probes at $r=17$ mm. This change was
implemented because it was considered that during the PCS experiments,
the CT may be compressed to radii smaller than 26 mm, in which case
the new inboard probes would give additional information. It turned
out that the poloidal field measurements at the three inner probes
was never trustworthy, so these measurements are not shown in figure
\ref{fig:Bp_Bt_32930}(a). $t_{lev}$ was set to $-300\,\upmu$s for
shot 32930. Looking at the rise of $B_{\theta}$ at the probe at $(r,\,\phi)=(52\mbox{ mm},\,320^{o})$
at $t\sim120\,\upmu$s, there is evidence of mild (unintentional)
magnetic compression associated with the mismatch between the decay
rates of CT and external levitation currents.

Figure \ref{fig:Bp_Bt_32930}(b) shows the measured toroidal field,
and the crowbarred shaft current, for shot  39230. Unlike the poloidal
field measurements at the $r=17$ mm probes, toroidal field measurements
at those probes were acceptable. Again, the rise in $I_{shaft}$ at
$t\sim120\,\upmu$s, provides evidence of mild, unintended, magnetic
compression. 
\begin{figure}[H]
\begin{centering}
\subfloat{\centering{}\includegraphics[width=8cm,height=5cm]{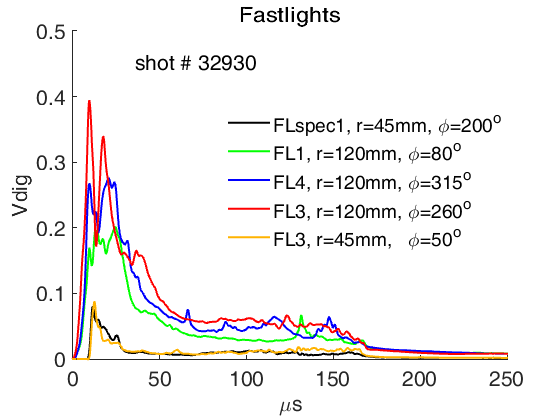}}
\par\end{centering}
\centering{}\caption{\label{fig:FL32930}$\,\,\,\,$Optical intensity measurements for
shot  32930, six coils, quartz insulating wall}
\end{figure}
Figure \ref{fig:FL32930} indicates, for the same shot, the light
intensity, presented in units of voltage induced in the photodiodes,
measured along the vertical chords at the optical port locations for
this configuration. The ports located at $r=120$ mm have a view down
the gun barrel, so the plasma breakdown is observed. Toroidal symmetry
of the signal magnitude indicates an even breakdown. While the level
of symmetry varies from shot to shot, repeated shots with asymmetric
optical signals early in time can be an indication of a some electrode
imperfection that would require correction. Optical intensity increases
in regions of increased density where line radiation intensity, which
scales as $n_{e}^{2}/Z_{eff}$, is high. Bremsstrahlung radiation
can also contribute to optical radiation emission, but is generally
negligible compared with optical emission due to line radiation at
the temperatures associated with this experiment. Optical intensity
at the inner chords at $r=45$ mm rises from zero later than at the
outer 120 mm chords due to a delay in the plasma reaching the inner
regions after it enters the CT containment area. Light intensity remains
lower at the inner chords partly because the optical emission there
is from the CT only, with no source from down the gun, where intra-electrode
currents may continue to flow for extended times, and also due to
the extended lengths of the parts of the outer chords that pass through
the CT. 

\section{CT rotation and toroidal MHD modes\label{subsec:Toroidal-MHD-modes}}

Before introducing the next configuration used to study levitated
CTs, it is worth discussing some details relating to the lack of observation,
during the magnetic compression experiment, of magnetic fluctuations
with toroidal mode number $n=2$, that were routinely observed on
the measured poloidal field signals of CTs produced with standard
MRT machines. The toroidal mode number is obtained by phase analysis
of signals from probes located at different toroidal angles. 
\begin{figure}[H]
\subfloat[]{\includegraphics[width=8cm,height=5cm]{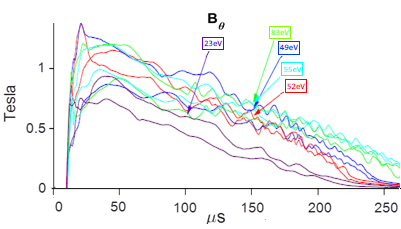}}\hfill{}\subfloat[]{\includegraphics[width=8cm,height=5cm]{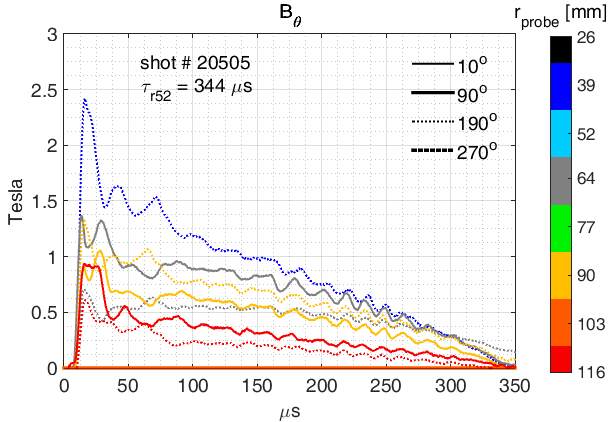}}\caption{\label{fig:n2-(magic-shots)}$\,\,\,\,$$B_{\theta}$ fluctuations
with toroidal mode $n=2$}
\end{figure}
Figure \ref{fig:n2-(magic-shots)}(a) shows $B_{\theta}$ traces at
a single probe from a selection of shots taken with MRT-type injectors,
from 2013, around the time when titanium gettering was first used
on GF injectors. The burgundy traces are from two shots prior to titanium
gettering, and do not have the $n=2$ fluctuations exhibited by the
shots after gettering. Electron temperature, as measured using Thomson-scattering,
indicates that electron temperature increased by a factor of between
two and four as a result of gettering. Ions and electrons cool and
recombine if they reach the walls of the vacuum vessel. Gettering
materials such as titanium and lithium act as a reservoir that absorbs
these cold particles and reduces the rate at which they are \textquotedbl recycled\textquotedbl{}
back into the plasma \cite{Causey}. Figure \ref{fig:n2-(magic-shots)}(b)
shows $B_{\theta}$ traces, also with $n=2$ fluctuations, from the
best shot taken on the magnetic compression experiment during the
period when the insulating outer wall was replaced with an aluminum
flux conserver. This configuration was implemented in order to confirm
(which it did) that the initial poor performance of levitated CTs
(with six coils and before the stainless steel extension shown in
figure \ref{fig:Schematic-of-6}(a)) was not due to an inherent machine
defect ($e.g.,$ gas injection, electrode conditioning etc.) and was
indeed due the unknowns associated with CT formation into a levitation
field. Note that only some of the magnetic probes were functioning
for this shot. Since the probes at $r=52$ mm were also not working,
the lifetime parameter $\tau_{r52}$ is estimated for shot 20505 by
interpolation using the field traces at $r=39$ and $r=64$ mm. 

These $n=2$ fluctuations are evidence of coherent CT toroidal rotation,
and were absent on shots taken on the magnetic compression device
with the outer insulating wall and levitation coils in place. There
was concern that rotation could be impeded by mode-locking caused
by toroidal asymmetry in the levitation field, introduced by the gaps
in toroidal levitation current associated with single-turn coils.
A set of six new coils (coil outline is depicted in figure \ref{fig:Levitation-and-compression-1}),
which reduced the field error, was manufactured. Analysis with the
MAGNET code, an add-on to the Solidworks program, indicated that field
asymmetry would be improved significantly, by a factor of $\sim10$,
with the new coils. Also, a 25-turn, high inductance ($160\,\upmu$H)
coil was experimented with - this reduced the original field error
by a factor of $\sim100$ - see section \ref{subsec:Multi-turn-coil}.
It turned out that levitation field asymmetry associated with the
original single-turn coils was not likely a problematic issue at the
level of performance achieved. At the settings for low-flux CTs, no
improvement in CT lifetime or symmetry was seen with either the new
set of discrete coils or the 25-turn coil, and there was no additional
evidence of CT rotation, or that a mode-locking issue had been alleviated.
X-ray phosphor videos indicated the likelihood of, but couldn't confirm
CT rotation. In retrospect, ion Doppler or Mach probe diagnostics
would ideally have been used to systematically check for rotation.
The $n=2$ fluctuations observed on $B_{\theta}$ signals with standard
$\mbox{MRT}$ machines may be connected with internal reconnection
events that occurred upon exceeding a threshold in CT temperature
that was not attained on the magnetic compression experiment due to
radiative cooling associated with impurities from the insulating wall.
As described in section \ref{subsec:Stainless-steel-shell-test},
$n=2$ fluctuations were also observed when the levitation field was
allowed to soak through a relatively resistive stainless steel outer
flux conserver. In that case, the levitation field is present but
the impurity cooling problem is alleviated. Coherent CT rotation was
confirmed later in the experiment; $n=1$ fluctuations regularly appeared
on the $B_{\theta}$ traces when CT toroidal field was increased with
the use of $80$ kA crowbarred formation current. This experiment
presented in section \ref{subsec:Additional-tests-on}.\\

\section{25-turn coil\label{subsec:Multi-turn-coil}}

A 25-turn single-conductor levitation coil was built to test if levitated
CT performance could be improved by:
\begin{enumerate}
\item Reducing the field errors associated with the 6-coil configuration 
\item Reducing plasma-wall interaction during the formation process
\end{enumerate}
To test the first hypothesis, the coils was designed with a single
length of conductor (insulated AWG-4 copper wire). To test the second
hypothesis, the coil was made with a height that filled the gaps that
were present above and below the 6-coil stack, outboard of the insulating
wall. Filling the gaps resulted in a large concentration of poloidal
flux and resultant magnetic pressure at the top and bottom of the
coil - this proved not to be an issue with sufficient clamping force.

\subsection{Coil design parameters}

\begin{table}[H]
\centering{}\includegraphics[scale=0.6]{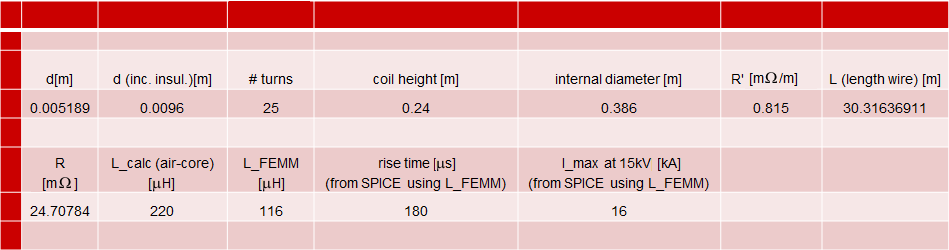}\caption{\label{tab:Multi-turn-coil-parameters}$\,\,\,\,$25-turn coil parameters}
\end{table}
Table \ref{tab:Multi-turn-coil-parameters} shows the parameters that
were chosen for the 25-turn coil. A lower bound on the coil inductance
was imposed by the number of turns required to cover the entire height
of the quartz wall. The upper bound on inductance was set by the maximum
desired current rise time, and the minimum desired levitation field,
which depends on the coil current, which is constrained by the available
capacitor voltage and capacitance. The value of $220\,\upmu$H for
the inductance of the air-cored coil with these parameters were obtained
using inductance calculators\cite{inductance calc1,inductancecalc2},
which take the coil-wire's insulation thickness into account. This
value was confirmed by an actual measurement when the coil was made.
FEMM was used to calculate the value of $116\,\upmu$H for the inductance
with the aluminum chalice present inside the coil. 
\begin{figure}[H]
\centering{}\includegraphics[scale=0.4]{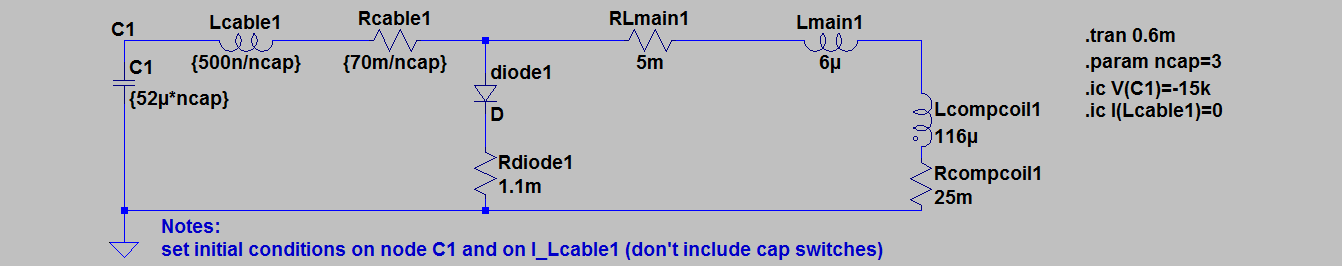}\caption{\label{fig:Spice-model-schematic}$\,\,\,\,$LT-SPICE model schematic
for the 25-turn coil}
\end{figure}
 Figure \ref{fig:Spice-model-schematic} shows the LT-SPICE model
used to simulate the coil currents and rise times. The main (holding)
$6\,\upmu$H inductor, and diode (stack) in the model, are depicted
at the bottom-right of figure \ref{fig:Levitation-and-compression-1}.
\begin{figure}[H]
\centering{}\includegraphics[scale=0.5]{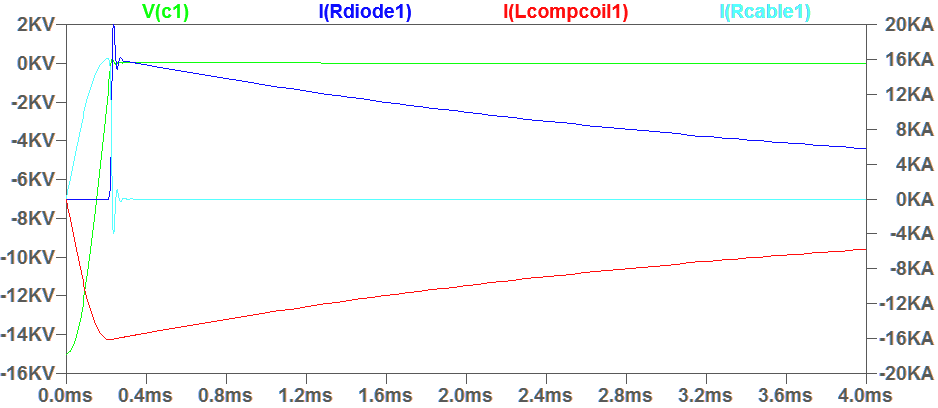}\caption{\label{fig:Spice-model-outputs}$\,\,\,\,$Circuit model outputs for
25-turn coil - no series resistance}
\end{figure}
 Figure \ref{fig:Spice-model-outputs} shows the traces of $V_{capacitor},\,I_{diode},\,I_{coil},$
and $I_{cable}$ at $V_{lev}=15$ kV  obtained from the model. Note
that the main inductor-driven crowbarred current continues to flow
in the diode stack $I(Rdiode1)$ and coil $I(Lcompcoil1)$ after the
capacitor voltage, and the current in $Rcable1$, change polarity
and drop to zero (refer to figure \ref{fig:Spice-model-schematic}).
The current in the coil ($\sim15$ kA with $V_{lev}=15$ kV  ) at
$t=t_{rise}=\sim180\,\upmu$s was used in FEMM along to evaluate the
resultant levitation field, with $t_{rise}$ (the time it takes for
the levitation current in the coil to attain its maximum value) determining
the frequency input to FEMM to be $f=\frac{1}{4(t_{rise})}=\sim1.4$
kHz. 
\begin{figure}[H]
\centering{}\includegraphics[scale=0.6]{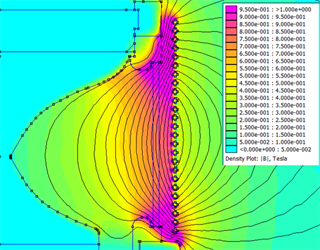}\caption{\label{fig:FEMM-output-for}$\,\,\,\,$FEMM output for the 25-turn
coil}
\end{figure}
The FEMM output for the 25-turn coil is shown in figure \ref{fig:FEMM-output-for}.
Compared with levitation field profile for the 6-coil configuration,
depicted in figure \ref{fig:Schematic-of-6}(b), it can be seen that
the field profile with the 25-turn coil should fix the problem associated
with the gaps above and below the 6-coil stack. Levitation field can
easily be displaced through those gaps as magnetised plasma enters
the pot at high speed - ions are the free to impact the wall and sputter
high $Z$ impurities into the plasma, resulting in performance degradation
due to radiative cooling. It seems reasonable that, on the 6-coil
setup, a levitation field that is strong enough to keep the CT off
the insulating wall above and below the coil stack may be too strong
to allow successful plasma entry to the pot. When the levitation field
is too strong, the entry to the CT containment region can be partially
blocked, especially when line tying effects impede entry further when
the field is allowed to soak well into the steel outboard of the entry
to the containment region. As shown in section \ref{subsec:Effect-of-levitation},
the MHD simulation code that we developed confirms that a field profile
like that associated with the 25-turn coil reduces plasma-wall interaction
during the formation process. The major drawback with the 25-turn
coil was that it wouldn't allow fast compression because of its extended
current-rise time. $t_{rise}$ $\sim180\,\upmu$s $cf.$ $\sim20\,\upmu$s
for the discrete coils in the compression circuit; in any case it
would probably break if placed under the high $\mathbf{J\times B}$
forces associated with large compression currents.

\subsection{Coil manufacture}

To make the coil, a $30$ m length AWG-4 wire was wound on a dense
foam cylindrical core that had been machined to size (and wrapped
in Mylar sheet for subsequent removal of the coil), and the winding
was wrapped in eight layers of heavy fiberglass cloth. That assembly
was then placed in a mould that was made by gluing the base of a hollow
cardboard tube ($\sim42\mbox{\mbox{ cm}}$ inner diameter, designed
for making concrete pillars) to a thick cardboard base. The top of
the tube was sealed off with plastic sheeting and a moderate vacuum
($\sim0.5\mbox{ psi})$ was established in the assembly by attaching
a plastic tube ($1\mbox{\mbox{ cm}}$ inner diameter) leading from
the vacuum pump to a sealed feedthrough made at the bottom of the
cardboard tube. Another sealed feedthrough was made at the top of
the cardboard tube through which activated epoxy resin was pumped
at up to $30$ psi. This arrangement had the desired result of reducing
the number and size of structurally-detrimental air bubbles in the
fiberglass as it set. Upon setting, the coil was removed from its
mould for finishing.

\subsection{\label{subsec:25turn-Levitated-CT-pefomance}Levitated CT performance
with the 25-turn coil}

\begin{figure}[H]
\centering{}\includegraphics[scale=0.5]{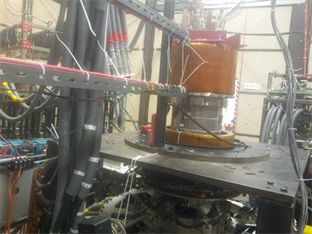}\caption{\label{fig:photo25-turn-coil-on}$\,\,\,\,$25-turn coil on machine}
\end{figure}
Figure \ref{fig:photo25-turn-coil-on} shows the 25-turn coil on the
machine with a $2$ m long feedthrough intended to reduce field error.
In the 6-coil configuration, the longest-lived CTs were achieved at
generally low settings for $V_{form}$, $I_{main}$, and $V_{lev}$
(note that for optimal CT performance, these machine settings scale
with one another), resulting in low-flux CTs. For example, $V_{form}\mbox{ and }I_{main}$
would typically have been $12$ kV  and $45$ A, compared with $16$
kV  and $70$ A for best performance on MRT machines. Note that $V_{form}=16$
kV  correspond to a peak formation current of $I_{form}\sim700$  kA,
while $I_{main}=70$ A corresponds to a gun flux of around 12 mWb.
Increasing these parameters on the magnetic compression injector in
the 6-coil configuration led to increased impurity levels and degraded
lifetime further. Initial testing on the 25 turn coil configuration
was done at typical optimal low flux settings of $V_{form}\sim12$
kV, $V_{lev}\sim12$ kV, and $I_{main}\sim45\mbox{ A}$. At these
setting, there was no real improvement in CT lifetime, or any indication
that the CTs produced was behaving any differently than previously.
This indicated that the improved toroidal symmetry of the field produced
with the coil did not benefit CT performance. However, one of the
advantages of the single continuous coil winding is that it facilitated
the addition of resistance to the circuit. The decay time constant
of the crowbarred levitation current is determined by $\tau_{decay}\sim L/R$,
and can be reduced by adding resistance to the circuit in series with
the cable between the coil and the main inductor. Note that the cable
resistance is included in the resistance denoted as $RLmain1$ in
the LT-SPICE model depicted in figure \ref{fig:Spice-model-schematic}.
\begin{figure}[H]
\centering{}\includegraphics[scale=0.55]{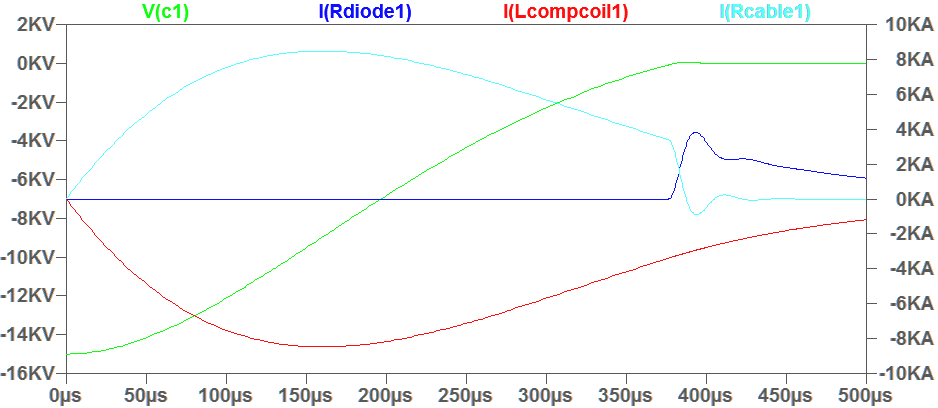}\caption{\label{fig:25turn_1Ohm_Circuit-model-outputs}$\,\,\,\,$Circuit model
outputs for 25 turn coil - $1\,\Omega$ series resistance}
\end{figure}
As shown in figure \ref{fig:25turn_1Ohm_Circuit-model-outputs}, with
$1\,\Omega$ resistance in series, the current in the coil decays
to zero from its peak at $\sim160\,\upmu$s to $<2\mbox{ kA}$ over
$\sim300\,\upmu$s, which is around the total time over which the
CTs decay completely after bubble in. With the original circuit design,
the levitation field would decay very slowly relative to the CT (see
figure \ref{fig:Ilev_comp}), so the CT is being effectively compressed
by the levitation field. In fact, as discussed later in section \ref{subsec:Levitation-field-decay},
due to this effect, the asymmetry or instability that was apparent
during practically every instance of magnetic compression was also
evident in most levitation-only shots. By matching the levitation
field decay time to that of the CT, the CT is allowed to retain the
size that it would have if it was being held in place and stabilised
by eddy currents induced in an outboard flux-conserver instead of
an outboard levitation field. This will be discussed further, and
it will be shown in section \ref{sec:rsep} that CTs retained more
volume when the levitation field decay time was optimised.
\begin{figure}[H]
\centering{}\includegraphics[scale=0.4]{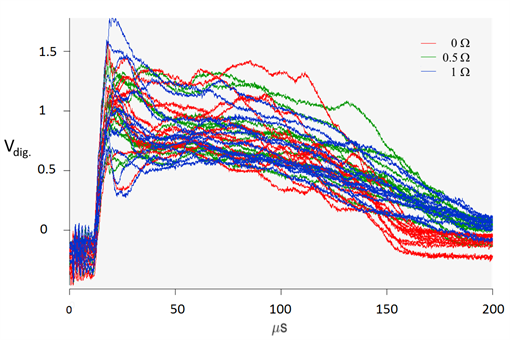}\caption{\label{fig:Performance-improvement-with}$\,\,\,\,$Performance improvement
with series resistance (25-turn coil)}
\end{figure}

When resistance was added to the circuit, CT lifetimes in the 25-turn
coil configuration increased by up to $20\%$, as indicated in figure
\ref{fig:Performance-improvement-with}, which shows traces of measured
$B_{\theta}$ at a single magnetic probe at $r=52\mbox{\mbox{ mm}}$
for shots with different series resistances (one trace per shot).
It can be seen how, with the addition of series resistance, the $B_{\theta}$
signals also indicate a smoother decay that is consistent with a reduction
in fluctuations that are associated with MHD activity due to unintentional
magnetic compression.

\begin{figure}[H]
\centering{}\subfloat{\centering{}\includegraphics[width=12cm,height=7cm]{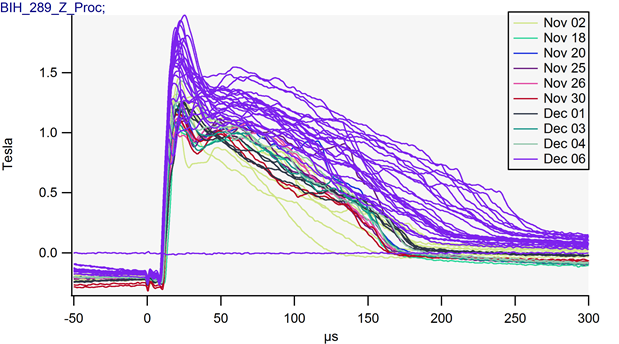}}\caption{\label{fig:25turn-Improved-performance-with-1}$\,\,\,\,$Improved
performance with the 25-turn coil at high flux settings}
\end{figure}
When the coil was tested at higher formation settings such as $V_{form}=16$
kV, $V_{lev}=16$ kV, and $I_{main}\sim60\mbox{ A}$, CT lifetimes
improved dramatically. Figure \ref{fig:25turn-Improved-performance-with-1}
shows the performance improvement with the 25-turn coil at the higher
formation settings. Traces are of measured $B_{\theta}$ at a single
magnetic probe at $r=52$ mm, for a selection of shots (one signal
per shot). The traces from the shots taken with the 25-turn coil in
place at the higher formation settings (on December 6$^{th}$ 2015)
are in dark purple. The remaining traces in figure \ref{fig:25turn-Improved-performance-with-1}
are from shots taken in the 6-coil configuration with the quartz wall,
at low-flux settings (recall that higher formation settings led to
performance degradation in the 6-coil configurations). 
\begin{figure}[H]
\centering{}\subfloat{\includegraphics[width=8cm,height=5cm]{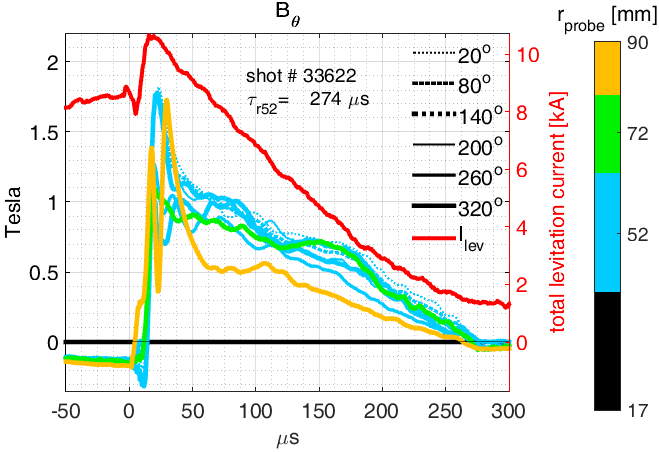}}\hfill{}\subfloat{\includegraphics[width=8cm,height=5cm]{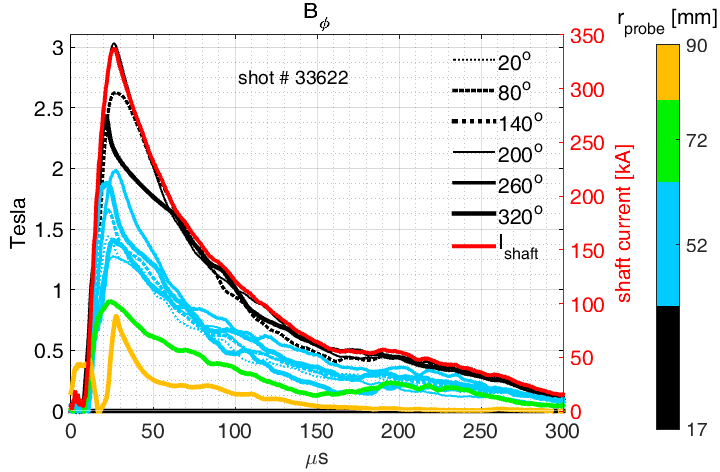}}\caption{\label{fig:Bp_Bt_33622}$\,\,\,\,$$B_{\theta}$ and $B_{\phi}$ for
shot 33622, 25-turn coil, quartz insulating wall}
\end{figure}
Figures \ref{fig:Bp_Bt_33622}(a) and (b) show traces of measured
$B_{\theta}$ and $B_{\phi}$ at the chalice magnetic probes for shot
 33622 in the 25-turn configuration. As indicated in figure \ref{fig:Bp_Bt_33622}(a)
(right axis), with $1\,\Omega$ series-resistance in the levitation
circuit, the coil current peak is $\sim10\mbox{\mbox{ kA}}$ and decays
at around the same rate as the CT currents. The single levitation
capacitor was fired at $t_{lev}=-150\,\upmu$s, which is the rise
time of the levitation current for this configuration. Comparing figures
\ref{fig:Bp_Bt_32930}(a) and \ref{fig:Bp_Bt_33622}(a), it can be
seen that the CT lifetime increased substantially, by almost 80\%
from $\tau_{r52}=155\,\upmu$s to $\tau_{r52}=274\,\upmu$s. It can
also be seen how, for the 25-turn coil configuration, the levitation
current increases by a much greater proportion when the plasma enters
the confinement region at around $15\,\upmu$s. As discussed below,
this is due to the increased inductance of the levitation coil in
the 25-turn coil configuration.

\begin{figure}[H]
\centering{}\includegraphics[scale=0.7]{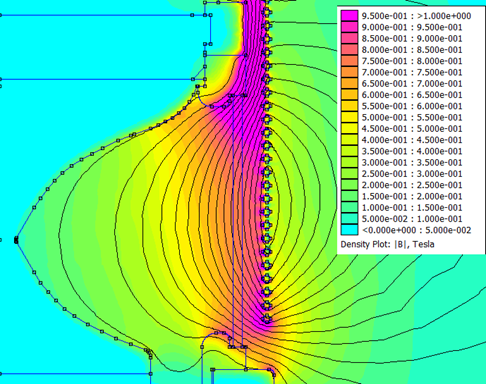}\caption{\label{fig:26turn_1inchup}$\,\,\,\,$FEMM model, 25-turn coil moved
up 25 mm, 10 kA at 1250 Hz }
\end{figure}
 The 25-turn coil was tested after being shifted up 25 mm \textendash{}
from FEMM (see figure \ref{fig:26turn_1inchup}), the field at the
top of the quartz wall is unchanged while the field at the bottom
was then comparable to that with the standard six coils at $30\mbox{\mbox{ kA}}$/coil.
On the same day (approximately the same machine conditions), 40 shots
with the coil down, then another 40 with it shifted up 25 mm, were
taken. Over 10 of the shots with the coil down had lifetimes >150$\,\upmu$s,
compared with just one shot with the coil up. This indicated that
having a coil outboard of the bottom of the quartz wall was partly
responsible for the improved performance. Due to machine geometry
constraints, a similar test with the coil moved down could not be
done.

There was some discussion about the cause of the benefit seen with
the 25-turn coil. One candidate for the improvement was the increase
in the ratio, between the coil inductance and the main circuit holding
inductor, that was associated with the 25-turn coil. When magnetised
conducting plasma enters the pot (confinement region) it reduces the
inductance of the part of the levitation circuit that includes the
levitation/compression coil and the material that the coil encompasses.
The levitation current increases when the coil inductance is reduced
when plasma enters the pot. If the percentage rise of the levitation
current is maximised, it means that levitation current prior to plasma
entry to the pot can be minimised. This reduction in levitation field
reduces the likelihood that the levitation field will be strong enough
to partially block plasma entry to the pot, while still allowing the
field that is present when the plasma does enter to be strong enough
to levitate the plasma away from the insulating wall. 
\begin{figure}[H]
\centering{}\includegraphics[width=16cm,height=6cm]{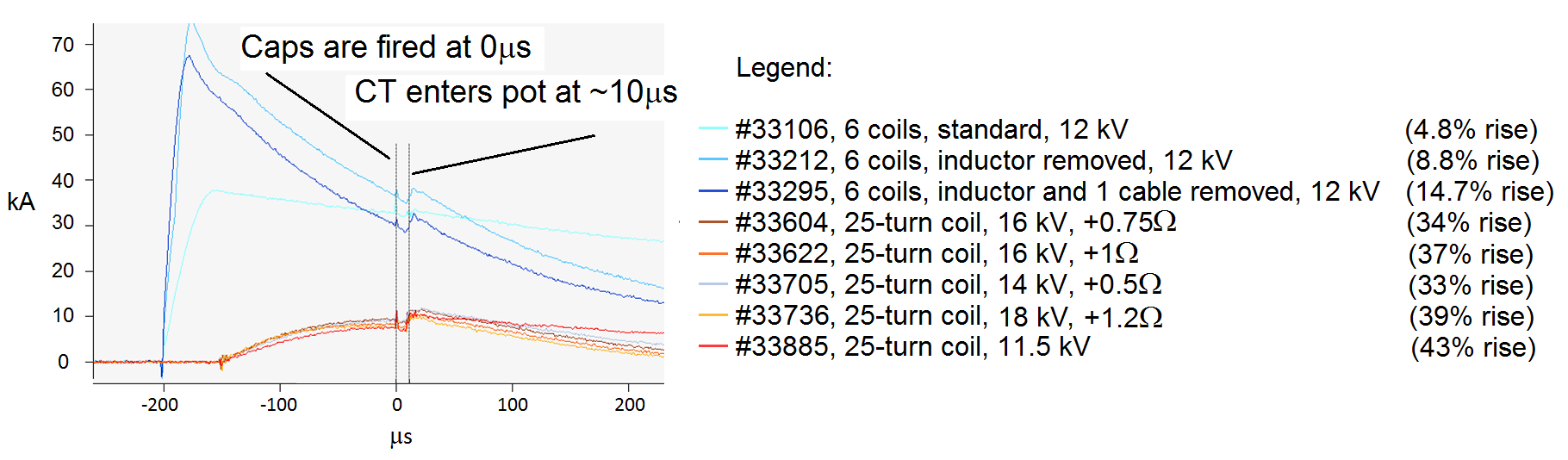}\caption{\label{fig:Percentage-current-rise}$\,\,\,\,$Percentage current
rise for various levitation circuit configurations }
\end{figure}
The currents indicated in figure \ref{fig:Percentage-current-rise}
are for one of the six coils (blue traces), and for the 25-turn coil
(red $\rightarrow$ orange traces). Comparing the data from shots
 33016 and 33885, the percentage rise in levitation current when plasma
enters the pot is up to nine times higher with the multiturn coil.
For the discrete coils, $L_{coil}/L_{main}=600\mbox{ nH}/6\,\upmu\mbox{H}=0.1$.
For the multiturn coil, $L_{coil}/L{}_{main}=116\,\upmu\mbox{H}/6\,\upmu\mbox{H}\,\sim\,20$.
The two darker blue traces in figure \ref{fig:Percentage-current-rise}
are from shots 33212 and 33295, taken when the main inductors were
removed from the levitation circuit, which has the effect of increasing
the peak levitation current at the same $V_{lev}$, and increasing
the decay rate of levitation current as $\tau_{LR}\sim L/R$ is reduced.
This configuration was experimented with on 6-coil configuration,
but no benefit to CT lifetime was observed. Shot  33295 was taken
after also removing one of the two $5\mbox{ m}\Omega$, $500\mbox{ nH}$
coaxial cables that connect the main inductor to the coil for each
of the six levitation coil circuits. Note that levitation capacitors
were prefired $200\,\upmu$s before formation for the shots indicated
here with the 6-coil setup (blue traces), in order to achieve the
line-tying benefit that is discussed in section \ref{subsec:6-coils-config_ceramic}.
On the 25-turn configuration coil (red $\rightarrow$ orange traces),
the levitation capacitors were prefired as late as it is possible
to (at $|t_{lev}|\sim t_{rise}\sim150\,\upmu$s before formation)
while maintaining maximum field at the time of plasma entry to the
CT confinement region. As discussed in section \ref{subsec:11-coil-configuration},
the line-tying strategy was not useful when the field profile was
optimised to reduce plasma-wall interaction. 

The 25-turn coil proved to be beneficial because it reduced plasma-wall
interaction at high formation settings. At low formation settings,
without addition levitation circuit series resistance, CT lifetimes
showed no improvement over those seen with the 6-coil configuration.
This indicates that neither improved toroidal symmetry of the field
produced, nor the increase of the ratio between the coil inductance
and the main circuit holding inductor, were wholly responsible for
the improvement seen, although they may have enhanced performance
at the settings for high-flux CT production. That the inductance ratio
increase was not wholly responsible for the improvement is further
confirmed by the fact that no benefit was seen when the main inductors
were removed from the levitation circuit on the 6-coil setup, and
supported again by the performance improvement observed with the 11-coil
configuration (which had a levitation field profile similar to that
of the 25-turn configuration) at high formation settings with approximately
the same coil inductance, see section \ref{subsec:11-coil-configuration}.
The benefit seen with series resistance in the 25-turn coil levitation
circuit was a completely separate effect that was obtained by optimizing
the decay rate of levitation current to match the decay rate of CT
currents.

\section{Semi-permeable shell\label{subsec:semipermshell}}

One of the proposals for a configuration that would allow repetitive
magnetic compression while retaining the possibly stabilising benefits
of a metal wall, eliminating the need for a levitation field and the
problems associated with plasma impurity contamination due to the
insulating outer wall, was the semi-permeable shell system. This would
involve a thin-walled metal tube (shell) in place of the insulating
wall. A thin inconel shell with a thickness of $\sim0.5\mbox{\mbox{ mm}}$
was the most likely candidate for success. Inconel was chosen because
of its relatively high resistivity and good structural properties.
The classical skin depth formula is 
\begin{equation}
\delta_{skin}=\sqrt{2\eta'/(\mu_{0}\,\omega)}\label{eq:101}
\end{equation}
The angular frequency of the compression field can be calculated by
approximating the compression waveform as a sine wave. The rise time
of the compression current is around $20\,\upmu\mbox{s}\Rightarrow\omega\sim2\pi/(4\times20\times10^{-6})\sim7.8\times10^{4}$
rad/s. Using the values $\eta'_{inconel}=1.03\times10^{-6}\,[\Omega-\mbox{m}]$
and vacuum permeability $\mu_{0}=4\pi\times10^{-7}\,[\mbox{H/m}]$,
this gives $\delta_{skin}\sim5$ mm. In principle the amplitude of
a field would be attenuated by a factor of $1/e\sim0.37$ by a conductor
with thickness $\delta_{skin}$, so the semi-permeable shell concept
initially seemed feasible. 

An analysis was undertaken using FEMM to help calculate the heating
of the inconel shell. The compression current waveform was Fourier
decomposed using MATLAB and its component frequencies and amplitudes
were successively input to FEMM in a loop via a LUA script. For each
FEMM analysis based on a single amplitude and frequency, the current
density in the shell was calculated to find the corresponding temperature
rise in the shell using the shell-material's resistivity and specific
heat capacity, and the duration of the compression current. The temperature
increments due to each Fourier harmonic were then added to find the
final estimated maximum temperature. Using this process, and also
the more straightforward method using FEMM with a single frequency
and amplitude based on the approximation of the compression waveform
as a sine wave, it was found that the central region of the inconel
shell's cross-section would reach a temperature of $\sim900^{\circ}$C
after just one shot (melting point of inconel is $1400^{\circ}$C),
although that estimate didn't account for thermal diffusion. The maximum
pressure due to $\mathbf{J}\times\mathbf{B}$ forces at magnetic compression
on the shell was calculated to be $30$ MPa, based on FEMM outputs.
With inconel as the semi-permeable shell material, the concept appeared
worth a try - compressional heating of the shell was deemed acceptable,
and it was envisaged that the thin shell could be structurally supported
against compressional forces by welding a hundred or so steel pins
(diameter \textasciitilde{} 2 mm, length \textasciitilde{} 1  cm)
pointing radially outwards from the outer tube surface and embedding
it in an epoxy tube.

A similar analysis was used to assess the feasibility of using a $30\,\upmu$m
(porous) coating of tungsten on the existing alumina $(\mbox{A\ensuremath{l_{2}O_{3}})}$
wall. The classical skin-depth for porous tungsten with an estimated
resistivity of $\eta'=\eta'_{W}/2=1.1\times10^{-7}\,[\Omega-\mbox{m}]$,
($\eta'_{W}$ is the resistivity of pure tungsten) at $\omega=8.16\times10^{4}$
rad/s, is $\delta_{skin}=1.6\,\mbox{\mbox{mm}}$. The method outlined
above indicated that the field would be attenuated by just $17\%$,
but that part of the porous coating would reach a peak temperature
of $\sim18000^{\circ}$C (melting point of pure tungsten is $6800^{\circ}$C).
Another major issue was that the bonding strength of the coating to
the alumina was unknown, and the coating would have to withstand huge
pressure forces. Maximum pressure on the shell was calculated to be
$20$ MPa. Overall, the concept of using a thin coating of tungsten
on an alumina shell appeared out of reach due excessive heating and
material bonding issues.

The same procedure indicated that a $30\,\upmu$m coating of lithium
would melt and be probably be ionized over a single magnetic compression
shot. A peak temperature of $\sim12000^{\circ}$C at compression would
be attained (melting point of lithium is $180^{\circ}$C), and $780$
kJ of energy per mole of lithium would be imparted - the $1^{st}$
and $2^{nd}$ ionization energies of lithium are $520$ and $7300$
kJ/mol. This raised the concern that plasma contamination with lithium
impurities might be an issue when lithium was used as a gettering
agent during magnetic compression. Time-resolved spectroscopic analysis
could have helped to clarify this, but was not available during the
experiment.

\subsection{Stainless-steel shell test\label{subsec:Stainless-steel-shell-test}}

A test with a stainless steel outer flux conserver was conducted largely
with the goal of observing the effect of allowing levitation field
to interact with a CT that was supported with a conducting wall. This
investigation was largely driven by concern over the absence, as discussed
in section \ref{subsec:Toroidal-MHD-modes}, of the $n=2$ fluctuations,
commonly observed with MRT injector-produced CTs, on levitated CT
$B_{\theta}$ signals. With a stainless shell with wall thickness
3.8 mm, wrapped in three layers of $1.5\mbox{\mbox{ mm}}$ diameter
copper wire, replacing the insulating wall, the maximum CT lifetime
was only $\sim150\,\upmu$s even after around 600 cleaning shots $cf.$
$\sim200\,\upmu$s with the aluminum shell (standard MRT machine)
after around 600 cleaning shots. Compared with the aluminum flux conserver,
the resistivity of the stainless steel flux conserver is increased
by a factor of ten, leading to more magnetic field soakage, and consequent
impurity sputtering, radiative cooling, and reduced CT lifetimes.
After lithium coating, lifetime increased to $\sim250\,\upmu$s with
hydrogen plasmas, and $\sim200\,\upmu$s with helium plasmas. 
\begin{figure}[H]
\centering{}\includegraphics[width=9cm,height=5cm]{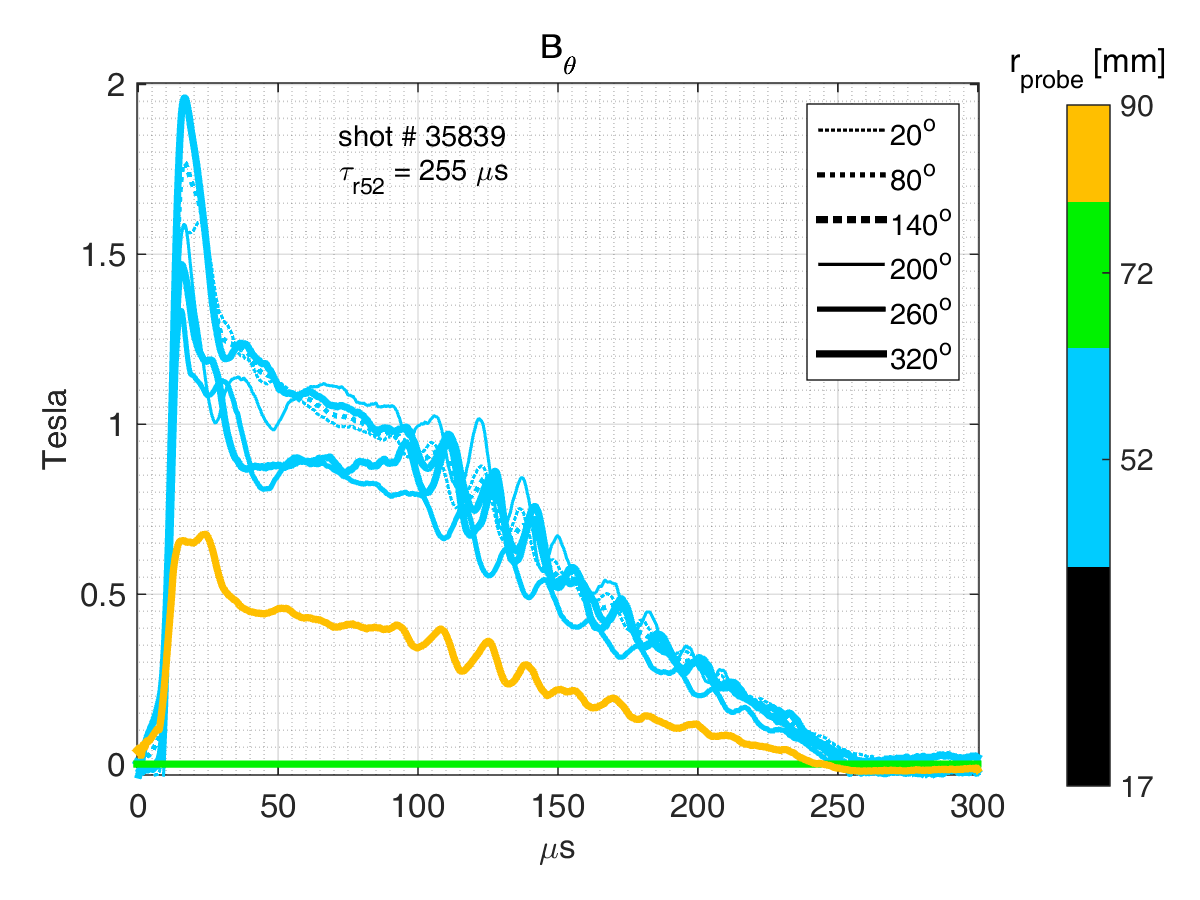}\caption{\label{Bp-ssflyer}$\,\,\,\,$$B_{\theta}$ for CT with the stainless
steel flux-conserver}
\end{figure}
 With hydrogen, CTs often exhibited $n=2$ fluctuations on the $B_{\theta}$
signals, as indicated in figure \ref{Bp-ssflyer}. Note that not all
of the magnetic probes were functioning - the signals relevant to
the probes at $r=17$ mm and $r=77$ mm have been zeroed out in figure
\ref{Bp-ssflyer}. Removing the copper wire resulted in a reduction
of CT lifetime to $\lesssim160\,\upmu$s for hydrogen and $\lesssim190\,\upmu$s
for helium, both without apparent $n=2$ activity. Placing the 25-turn
coil around the bare steel shell resulted, with a short delay $(|t_{lev}|$)
between firing of levitation and formation capacitors, and low $B_{lev}$,
in CT lifetimes of $\sim210\,\upmu$s with hydrogen, $\sim250\,\upmu$s
with helium, with $n=2$ behaviour being observed in both cases. With
$|t_{lev}|>300\,\upmu$s, and low $B_{lev}$, the CT life was maximized,
$\sim270\,\upmu$s with $\mbox{He}$, with occasional weak $n=2$.
With high $B_{lev}$, especially at long $|t_{lev}|$, the CT lifetimes
were less consistent, with behaviour that is characteristic of unintentional
low-level compression - note that these tests were done without series
resistance in place in the levitation circuit. With high $B_{lev}$,
$n=2$ fluctuations were present only with short $|t_{lev}|$.

This set of results confirms that a high, relatively constant levitation
field is detrimental to CT performance - as shown in sections \ref{subsec:25turn-Levitated-CT-pefomance}
and \ref{subsec:Levitation-field-decay}, performance is better when
the levitation field decays at the same rate as $\psi_{CT}.$ The
fact that CT lifetimes with the steel shell were lower than the CT
typical lifetimes observed an aluminum outer flux-conserving shell,
but were increased when the 25-turn coil was placed around the steel
shell suggests that stainless steel is not a good material with regard
to impurities and that levitation field can reduce impurity levels
by keeping the CT off the wall. The improvement seen with the levitation
field was enhanced by increasing $|t_{lev}|$ through the effect of
line-tying, as outlined in section \ref{subsec:6-coils-config_ceramic}.
When the levitation field was increased, the unintentional low-level
compression is more effective and starts earlier in time resulting
in diminished performance. With long $|t_{lev}|$ and high $B_{lev}$,
the effect of blocking the entry of magnetized plasma to the CT containment
region also become problematic, especially because the levitation
field is being tied in the entire resistive shell.

It has been confirmed that $n=2$ fluctuations are a sign of internal
MHD activity associated with increased electron temperature, as discussed
in section \ref{subsec:Toroidal-MHD-modes}. It was thought that this
correlation, and the absence of the fluctuations on levitated CTs,
was a sign that levitated CTs were colder than flux-conserved CTs,
and the problems encountered with plasma wall interaction in the levitation
configurations made that scenario more likely. However, the longest-lived
CTs produced with the 25-turn configuration endured for up to 10\%
longer than, and may therefore be assumed to be hotter than the CTs
produced in the configuration with the stainless steel flux conserver.
It may be that the levitation field acts to damp out helically propagating
magnetic fluctuations at the outboard CT edge and that internal MHD
activity is relatively unchanged. The $n=1$ magnetic fluctuations
(see section \ref{subsec:Additional-tests-on}), observed when 80
kA additional crowbarred shaft current was applied to the machine
in the eleven coil configuration, partially confirmed coherent toroidal
rotation of levitated CTs, and may have been a result of more vigorous
MHD activity that remained apparent despite damping. 

\subsection{Effective skin depth for thin shells}

As outlined in section \ref{subsec:semipermshell}, the concept of
using an inconel semi-permeable shell configuration seemed feasible.
However, towards the end of the series of tests with the stainless
steel shell, it was realised that when $d\sim\delta_{skin}$ ($d$
is the shell thickness and $\delta_{skin}$ is the classical skin
depth - equation \ref{eq:101}), the interior of a shell with inner
radius $R_{in}$ is effectively screened from an externally applied
field if:
\begin{equation}
d>\delta_{eff}=\frac{\delta_{skin}^{2}}{R_{in}}\label{eq:102}
\end{equation}
This relationship \cite{EM screening by metals,EMscreeningMITweb},
which effectively scuppered the semi-permeable shell concept, can
be derived as follows: 
\begin{figure}[H]
\begin{raggedright}
\subfloat[Schematic of thin conducting shell with external solenoid ]{\raggedleft{}\includegraphics[scale=0.6]{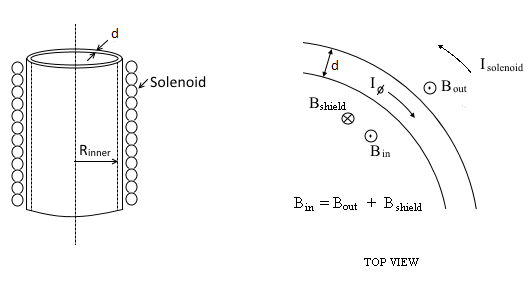}}\hfill{}\subfloat[Shielding of external $B_{out}$ versus $d$ in units of $\delta_{eff}$]{\includegraphics[scale=0.5]{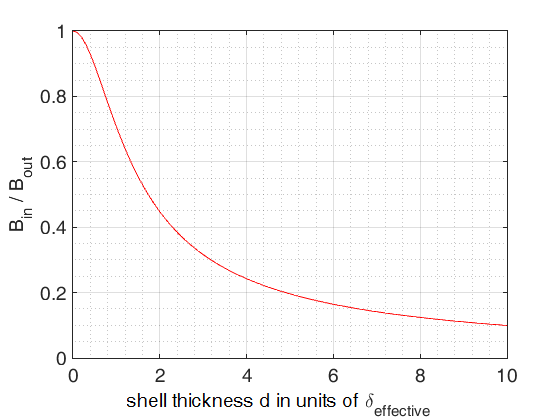}} 
\par\end{raggedright}
\centering{}\caption{$\,\,\,\,$Electro-magnetic shielding in a thin conducting shell\label{fig:thin Shell}}
\end{figure}
With reference to figure \ref{fig:thin Shell}(a), when current flows
in a solenoid surrounding an electrically conducting shell, the field
inside the shell is $B_{in}=B_{out}+B_{shield}$, where $B_{out}$
is the field applied external to the shell, and $B_{shield}$ (with
orientation opposite to that of $B_{out})$ is the field produced
by the current driven by the electric field induced along the shell
circumference. The induced electric field is found using Faraday's
law, taking the integral over the area enclosed by the shell, and
using Stokes' theorem:
\begin{equation}
E_{\phi}=-\frac{R_{in}}{2}\dot{B}_{in}
\end{equation}
Assuming the time dependence $B_{in}\sim B_{in0}\:e^{-i\omega t}$,
this leads to: $E_{\phi}=\frac{i}{2}\omega R_{in}B_{in}$. By Ohm's
law, $\mathbf{J}=\sigma\mathbf{E}$, so that the current per unit
shell height is $I_{\phi}=\frac{i}{2}\omega\sigma R_{in}B_{in}d$.
Here, $\sigma\,[\mbox{S/m}]=\frac{1}{\eta'}$ is the electrical conductivity
of the shell material in Siemens/meter. Ampere's law defines the relation
between $I_{\phi}$ and $B_{shield}$: $\nabla\times\mathbf{B=}\mu_{0}\mathbf{J}\Rightarrow\int\mathbf{B}_{shield}\cdot d\mathbf{l}=\mu_{0}\,I_{\phi}\Rightarrow$
\[
B_{shield}=\frac{i}{2}\mu_{0}\,\omega\sigma R_{in}B_{in}d
\]
Here, Stokes theorem has been used again, and the integral was taken
over a rectangular cross-sectional area of unit shell height in a
plane of constant azimuthal angle. In terms of conductivity, $\delta_{skin}^{2}=2/(\mu_{0}\,\omega\sigma)$
, so that 
\[
B_{shield}=iR_{in}B_{in}d/\delta_{skin}^{2}
\]
Since $B_{in}=B_{out}+B_{shield}$, $\frac{B_{in}}{B_{out}}=1+\frac{B_{shield}}{B_{out}}=1+\frac{B_{shield}}{B_{in}}\frac{B_{in}}{B_{out}}=1+\frac{iR_{in}d}{\delta_{skin}^{2}}\frac{B_{in}}{B_{out}}\Rightarrow\frac{B_{in}}{B_{out}}\left(1-\frac{iR_{in}d}{\delta_{skin}^{2}}\right)=1$.
Substituting $b=\frac{R_{in}d}{\delta_{skin}^{2}}$, this leads to

\begin{align*}
|\frac{B_{in}}{B_{out}}| & =|\frac{1}{1-ib}|=|\frac{1+ib}{1+b^{2}}|=\sqrt{\left(\frac{1}{1+b^{2}}\right)^{2}+\left(\frac{b}{1+b^{2}}\right)^{2}}=\frac{1}{1+b^{2}}\sqrt{1+b^{2}}\\
 & =\frac{1}{\sqrt{1+\left(\frac{R_{in}d}{\delta_{skin}^{2}}\right)^{2}}}=\frac{1}{\sqrt{1+(d/\delta_{eff})^{2}}}
\end{align*}
where $\delta_{eff}=\frac{\delta_{skin}^{2}}{R_{in}}$. Figure \ref{fig:thin Shell}(b)
is a plot of $|\frac{B_{in}}{B_{out}}|$ against shell thickness $d$
in units of $\delta_{eff}$, the effective skin depth. This plot holds
for any material at any field frequency when $d\sim\delta_{skin}$.
For $R_{in}=169\mbox{\mbox{ mm}}$ (the inner radius of the quartz
wall for the magnetic compression experiment), and $\delta_{skin}=4.5\mbox{\mbox{ mm}}$
(classical skin depth for high resistivity inconel, $\sigma\sim1\times10^{6}\,$S/m,
at 12.5 kHz, the frequency of the magnetic compression fields), $\delta_{eff}=0.12\mbox{\mbox{ mm}}$.
From figure \ref{fig:thin Shell}(b), it can be seen that field strength
inside the shell would be attenuated by \textasciitilde 25\% for
$d\sim0.1\mbox{\mbox{ mm}}$, and attenuated by 80\% for $d\sim0.5\mbox{\mbox{ mm}}$.

This information seemed to make the semi-permeable shell concept impossible,
as the minimum shell thickness considered to be structurally adequate
(with additional reinforcement from horizontal steel pins on the outer
wall of the shell that would be embedded in an external fiberglass
form) was around $0.5$ mm, and the corresponding 80\% attenuation
of compression field would be unacceptable. 

\section{\label{subsec:11-coil-configuration}11-coil configuration, quartz
outer insulating wall}

The 11-coil configuration was tested after GF had decided to stop
the magnetic compression experiment in February 2016, after the last
few months of testing with a stainless steel flux conserver. I was
given a few additional weeks (three, extended to seven) to try and
get additional data for the PhD. The 11-coil configuration consisted
of five coil pairs and one single coil, and approximately reproduced
the levitation field profile of the 25-turn coil, allowing for formation
and compression of higher-flux CTs with correspondingly increased
lifetimes (CT lifetime scales with $\psi_{CT}$). Each coil-pair was
assembled using one of the original coils and one of the newer coils
that were designed to increase toroidal symmetry in the levitation/compression
field. The remaining newer coil was included on its own, positioned
3rd from the bottom of the 11-coil stack, to further increase the
field at the bottom of the wall. In contrast to the 25-turn coil configuration,
magnetic compression was possible with the 11-coil setup. As shown
in chapter \ref{Chap:Magnetic-Compression}, magnetic compression
performance was greatly improved over that with the 6-coil configuration,
due to the modified compression (and/or levitation) field profile.

\begin{figure}[H]
\begin{raggedright}
\subfloat[11 coils on machine]{\raggedleft{}\includegraphics[scale=0.5]{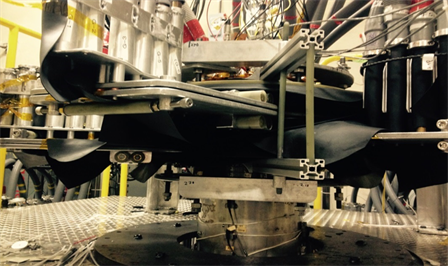}}\hfill{}\subfloat[FEMM model - 11 coils]{\includegraphics[scale=0.5]{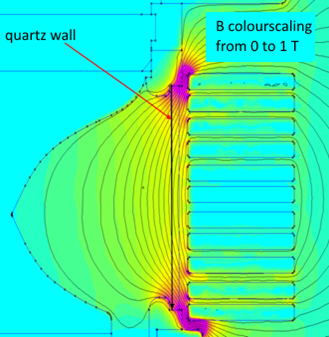}} 
\par\end{raggedright}
\centering{}\caption{\label{fig:11-coil-configuration}$\,\,\,\,$11-coil configuration}
\end{figure}
Figure \ref{fig:11-coil-configuration}(a) shows the 11-coil stack
installed on the machine - the single coil is on the right. Each coil/coil-pair
is connected to its own levitation circuit via the two outer co-axial
cables in the cable connecting bracket attached to the coil/coil-pair.
One of the six brackets can be seen in the upper left foreground.
Each of the inner four co-axial cables in each bracket links individually
to a single $52\,\upmu\mbox{F},\,20\mbox{ kV}$ compression capacitor
and thyratron switch. Figure \ref{fig:11-coil-configuration}(b) shows
a FEMM output plot for the 11-coil setup with $16\mbox{\mbox{ kA}}$
per coil and a frequency of $4\mbox{ kHz}$, corresponding to the
experimentally determined optimal delay of $|t_{lev}|=50\,\upmu$s
between the firing of the levitation and formation capacitor banks.
The 11-coil configuration was an entirely new build, with new parts
that had to be baked and vacuum conditioned before testing. Baking
involved wrapping the machine in insulating material and maintaining
it under vacuum at around 150$^{o}$C for three to four days to remove
water. An inner flux conserving  \textquotedbl chalice\textquotedbl{}
with magnetic probe locations listed in table \ref{tab: coordinates-ofBprobes}
was used, and a new quartz wall was installed. A different material
(alumina or boron-nitride) would have been far better for CT performance,
but there wasn't time to wait an additional few weeks to manufacture
an insulating tube from a more plasma-compatible material. After baking
the assembly, the first CT with a lifetime of greater than $150\,\upmu$s
was achieved within four days of cleaning shots, compared with over
thirty days for the same milestone on the configuration with six coils
around the quartz insulating wall. This may have been largely due
to the improved levitation field profile, but also may be because,
with limited time available, care was taken to pre-bake the chalice
and quartz tube at 200$^{o}$C for two days in a low pressure oven,
and to avoid contact of the parts with (water-containing) air during
installation on the machine. 

\subsection{Levitation field decay rate\label{subsec:Levitation-field-decay}}

The strategy of adding resistance to the levitation circuits, to match
the decay rates of external levitation and CT currents, as initially
tested on the 25-turn coil, was used again on the 11-coil configuration.
\begin{figure}[H]
\subfloat[$B_{\theta}$, $2.5\mbox{ m}\Omega$ cables]{\raggedright{}\includegraphics[width=8cm,height=5cm]{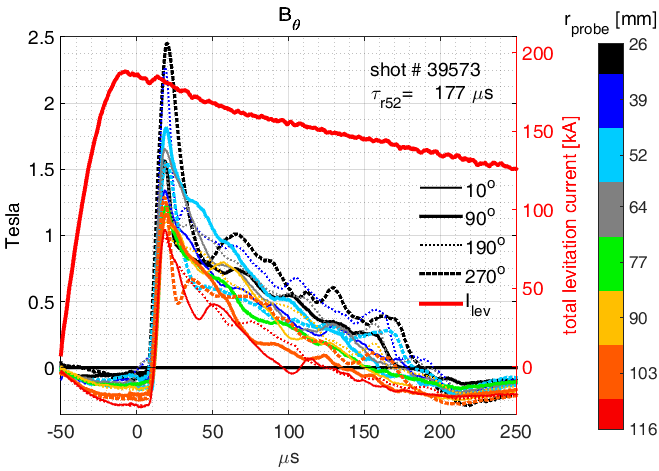}}\hfill{}\subfloat[$B_{\theta}$, $70\mbox{ m}\Omega$ cables]{\raggedleft{}\includegraphics[width=8cm,height=5cm]{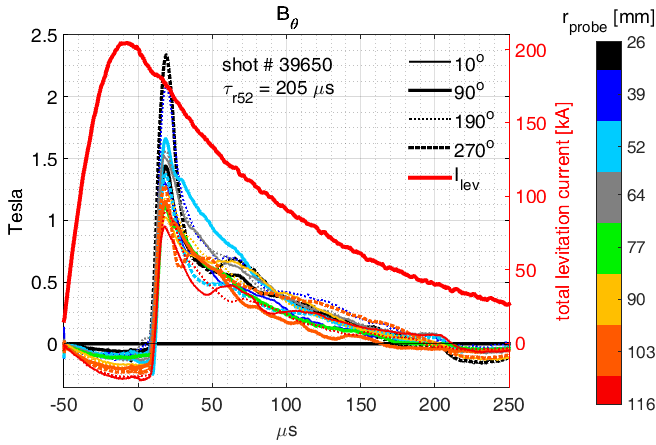}}

\subfloat[$B_{\phi}$, $2.5\mbox{ m}\Omega$ cables ]{\raggedright{}\includegraphics[width=8cm,height=5cm]{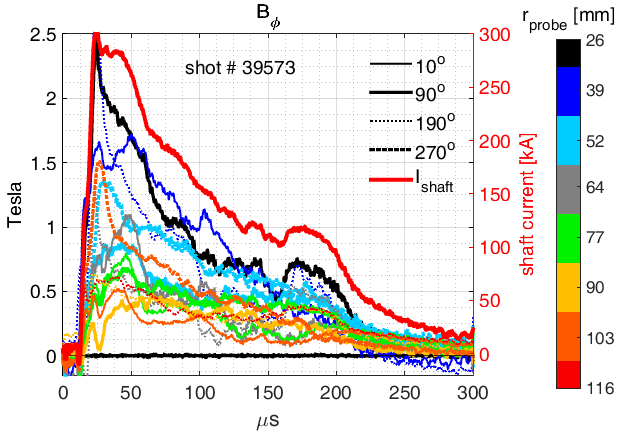}}\hfill{}\subfloat[$B_{\phi}$, $70\mbox{ m}\Omega$ cables]{\raggedleft{}\includegraphics[width=8cm,height=5cm]{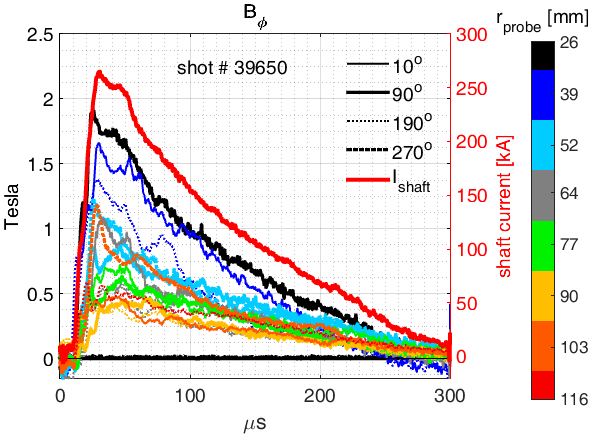}}

\caption{\label{fig:Poloidal-field-for}$\,\,\,\,$$B_{\theta}$ and $B_{\phi}$
for levitated CTs (11-coil configuration) }
\end{figure}
Figure \ref{fig:Poloidal-field-for} shows $B_{\theta}$ and $B_{\phi}$
 for two shots with different cable resistances for the 11-coil configuration.
These two shots were both at $V_{form}=16$ kV  - additional settings
for shot  39573 included $V_{lev}=13.8$ kV  and $I_{main}=73\mbox{ A}$
(corresponding to \textasciitilde 12 mWb gun flux), and for shot
 39650, $V_{lev}=16$ kV  and $I_{main}=70\mbox{ A}$. To achieve
the same levitation current with the additional cable resistance,
$V_{lev}$ was increased for shot  39650. It can be seen how the decay
rate of $I_{lev}$ approximately matches that of the CT toroidal current
(as determined by positive $B_{\theta}$) with a $70\mbox{ m}\Omega$
cable replacing the original pair of $5\mbox{ m}\Omega$ cables in
parallel ($i.e.$, total $2.5\mbox{ m}\Omega$) between the main holding
inductor and coil in each levitation circuit. A much higher rate of
\textquotedbl good\textquotedbl{} shots, smoother decays of $B_{\theta}$
and $B_{\phi}$ (less apparent MHD activity), and a lifetime increase
generally of around $15-20\%$, was observed with the $70\mbox{ m}\Omega$
cables in place. There is evidence of mild magnetic compression starting
at around $150\,\upmu$s on shot 39573 (figure \ref{fig:Poloidal-field-for}(c)),
even though the compression capacitors were not fired. In this shot,
this is particularly evident from the rise $B_{\phi}$ (especially
at the $(26\mbox{\mbox{ mm}},\:90^{\circ})$ and $(52\mbox{\mbox{ mm}},\:190^{\circ})$
probes) and from the overall rise in shaft current at $150\,\upmu$s
- see discussion in section \ref{subsec:6-coils-config_ceramic}.
The CT is decaying far faster than the levitation field with the low
resistance cables in place, so that it is being compressed more and
more as $\psi_{CT}$ decreases.

\begin{figure}[H]
\subfloat[$n_{e}$, $2.5\mbox{ m}\Omega$ cables]{\raggedright{}\includegraphics[width=8cm,height=5cm]{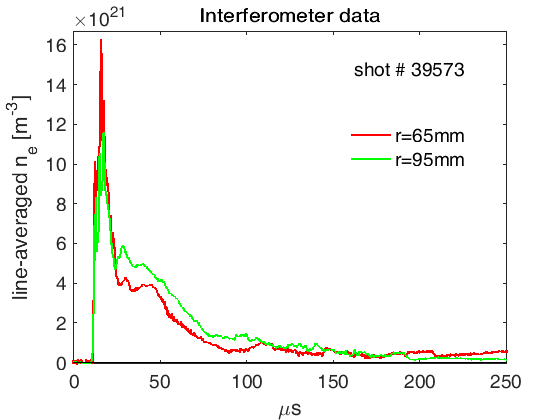}}\hfill{}\subfloat[$n_{e}$, $70\mbox{ m}\Omega$ cables]{\raggedleft{}\includegraphics[width=8cm,height=5cm]{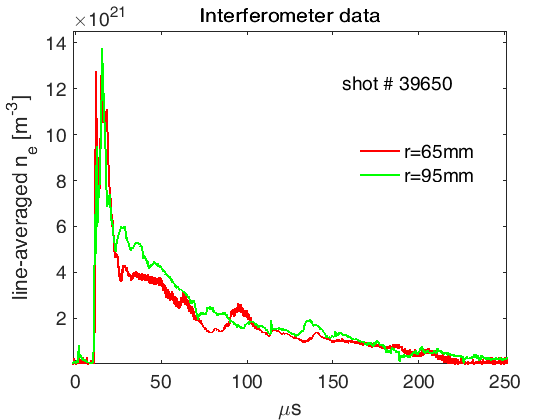}}

\subfloat[Optical intensity, $2.5\mbox{ m}\Omega$ cables ]{\raggedright{}\includegraphics[width=8cm,height=5cm]{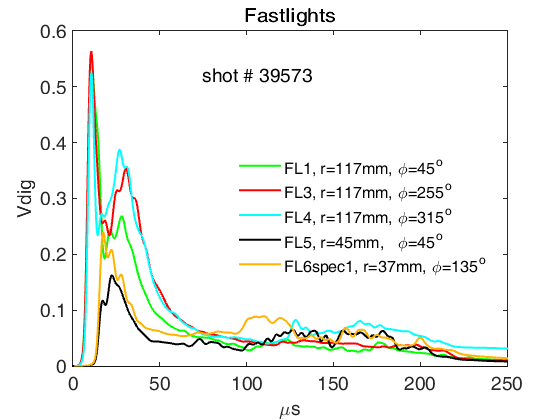}}\hfill{}\subfloat[Optical intensity, $70\mbox{ m}\Omega$ cables]{\raggedleft{}\includegraphics[width=8cm,height=5cm]{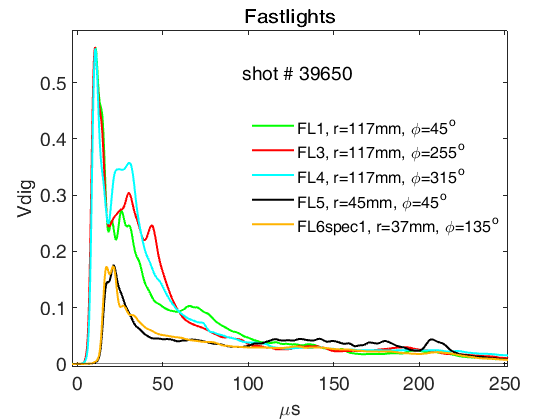}}

\subfloat[$T_{i}$, $2.5\mbox{ m}\Omega$ cables ]{\raggedright{}\includegraphics[width=8cm,height=5cm]{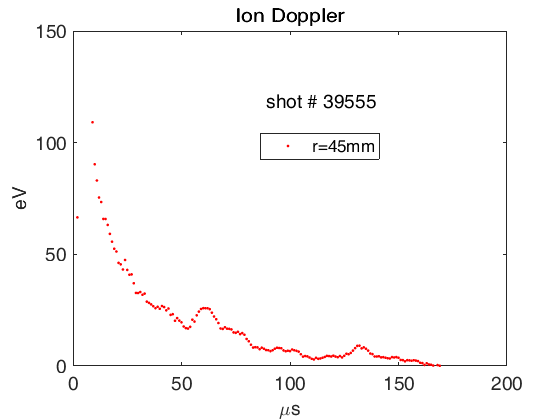}}\hfill{}\subfloat[$T_{i}$, $70\mbox{ m}\Omega$ cables ]{\raggedright{}\includegraphics[width=8cm,height=5cm]{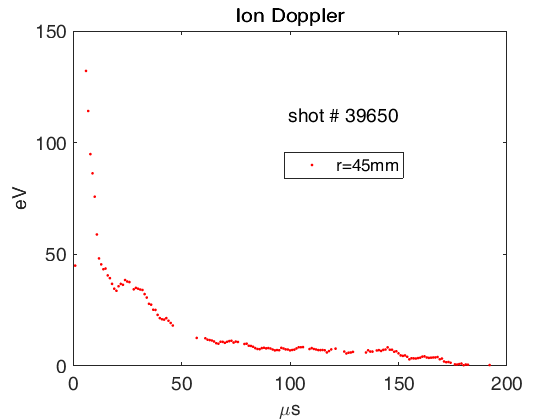}}

\caption{\label{fig:Ti_n_FL_11coils}$\,\,\,\,$$n_{e}$, optical intensity,
and $T_{i}$ for levitated CTs (11-coil configuration) }
\end{figure}
 Figures \ref{fig:Ti_n_FL_11coils}(a) and (b) compare the line averaged
density measurements for shots  39573 and  39650. Corresponding to
the longer CT lifetime and larger CT size associated with shot  39650,
the density falls more gradually compared with the density decay rate
for shot  39573. MHD activity may lead to enhanced particle loss.
In general, shots with implementation of decay rate matching exhibited
less evidence of strong internal MHD activity. Associated with this,
as can be seen by comparing figures \ref{fig:Ti_n_FL_11coils}(c)
and (d), optical intensity measurements are relatively fluctuation
free in shot 39650, especially around 150$\,\upmu$s when the CT in
shot 39573 is extinguished. Optical intensity increases in regions
of increased density, where line radiation intensity, which scales
as $n_{e}^{2}/Z_{eff}$, is high. Ion Doppler data was not well resolved
for shot 39573, and only the diagnostic looking along the vertical
chord at $r=45$ mm (see figure \ref{fig:Chalice}) was working for
shot 39650. Shot 39555 was similar to shot 39573 and was also taken
in the configuration with low resistance cables. Figures \ref{fig:Ti_n_FL_11coils}(e)
and (f) indicate how fluctuations in measured ion temperature are
more apparent with the low resistance cables in place. In both cases,
the maximum ion temperature is at $\sim$10$\,\upmu$s when plasma
enters the CT confinement region, due to high velocities achieved
by the plasma through the action of formation forces, which leads
to high levels of ion viscous heating. With reference to data presented
in \cite{Kunze,Pittman}, an error in the temperature measurement
(He II line at 468.5nm) due to density broadening has been evaluated
as $\sim$17 eV for a density of 1.6$\times10^{22}$ m$^{-3}$, and
the error falls off in proportion to $n_{e}^{0.83}$. Density along
the ion-Doppler chords is not directly evaluated, but from this information,
along with observations of the variation of line averaged density
with radius, the error in the $T_{i}$ measurement along the $r=45$
mm chord for shot 39650 can be roughly estimated as being $\sim$12
eV at $20\,\upmu$s and $\sim$2 eV at $100\,\upmu$s .

\subsection{Additional tests on the 11-coil configuration\label{subsec:Additional-tests-on}}
\begin{itemize}
\item Around ten coil configurations, including configurations with passive
and open-circuited coils (up to four configurations per day) were
tested. Recall that the experiment with eleven coils was conducted
in a quite a rush over a few weeks after the experiment had been scheduled
for decommissioning. There was no obviously outstandingly beneficial
configuration apart from the setup with all coils on with $70\mbox{ m}\Omega$
cables. It would have been worth testing with the single coil located
near the bottom of the stack moved towards the top of the stack, to
see if increased levitation field at the top of the wall was more
beneficial than increased field at the bottom, but again there wasn't
time to do this.
\end{itemize}
\begin{figure}[H]
\subfloat[$B_{\theta}$ ]{\raggedright{}\includegraphics[width=8cm,height=5cm]{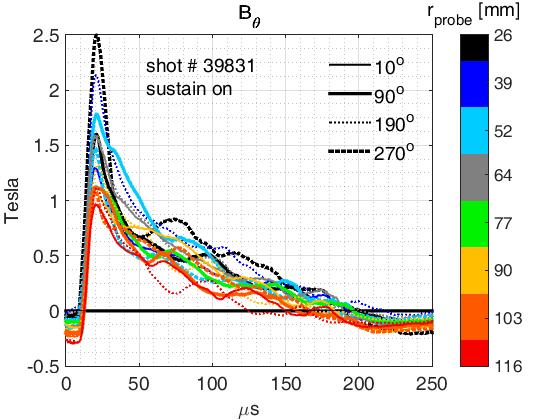}}\hfill{}\subfloat[$T_{i}$]{\raggedright{}\includegraphics[width=8cm,height=5cm]{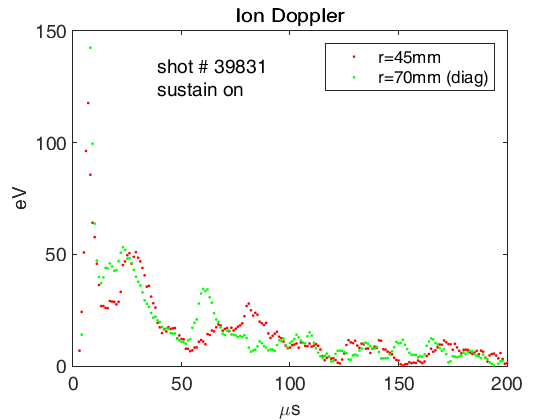}}

\subfloat[$B_{z}$ on side probes]{\raggedleft{}\includegraphics[width=8cm,height=5cm]{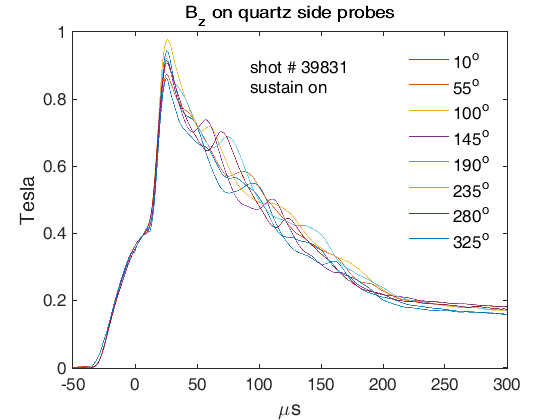}}\hfill{}\subfloat[Optical data ]{\raggedleft{}\includegraphics[width=8cm,height=5cm]{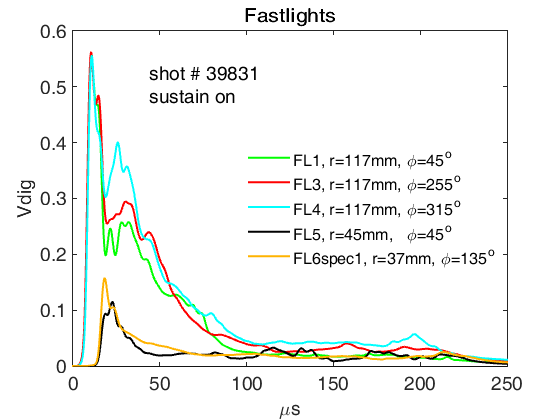}}

\caption{$\,\,\,\,$$B_{\theta}$, $T_{i}$, $B_{z}$ and optical data with
sustained $I_{shaft}$\label{shot39831} }
\end{figure}

\begin{itemize}
\item A few shots were taken with up to $80\mbox{\mbox{ kA}}$ sustained
shaft current. This involved connecting one of the levitation circuits,
including its two capacitors and main holding inductor, in parallel
with the formation banks across the formation electrodes. In that
configuration, the lower coil-pair was left open-circuited. Sustainment
at far higher currents significantly extends CT lifetimes in standard
MRT machines. It wasn't possible to try and add more capacitors because
no more levitation circuits could be borrowed without degrading levitation
to unacceptable levels. In fact, the reduction in levitation current
had to be compensated for by replacing the 70 m$\Omega$ cables in
two of the five remaining the levitation circuits with 2.5 m$\Omega$
cables. There wasn't time in the available period to set up new circuits.
Oscillations with toroidal mode number $n=1$ were observed on the
$B_{\theta}$ signals on some shots with $80\mbox{\mbox{ kA}}$ sustained
current. The fluctuations recorded on measured $B_{\theta}$, $T_{i}$,
$B_{z}$ and optical signals for shot 39831 are indicated in figure
\ref{shot39831}, where the $B_{z}$ signals were recorded at probes
located on the outside of the quartz wall, as described below in section
\ref{sec:rsep}. In general, fluctuations in these signals are enhanced
with the unintentional magnetic compression scenario associated with
lower resistance cables in place in the levitation circuits, but seem
to be further enhanced in this case. This suggests enhanced MHD activity
with additional shaft current, which may be related to the effect
of the compressional instability in the presence of increased CT toroidal
field. It would have been interesting to observe the effect of sustained
$I_{shaft}$ with 70 m$\Omega$ cables in place on all the levitation
circuits. The $n=1$ mode on $B_{\theta}$ signals confirmed, for
the first time on this experiment, coherent CT rotation and the existence
of closed flux surfaces. Apart from that, there were no other noticeable
effects associated with the additional driven shaft current.
\end{itemize}

\section{\label{sec:rsep}Using side-probe data to find CT outboard separatrix
radius over time\label{subsec:Using-side-probe-data}}

A set of eight magnetic probes with windings to measure $B_{r},\,B_{\phi}$,
and $B_{z}$, were attached to the outside of the insulating wall
at $\phi=10^{\circ},\,55{}^{\circ},\,100^{\circ},\,145{}^{\circ},\,190^{\circ},\,235{}^{\circ},\,280^{\circ},\,\mbox{ and }325{}^{\circ}$.
The probes were installed at $z=0\,\mbox{\mbox{mm}}$ ($i.e.,$ the
chalice waist) on the earlier configuration with six coils around
the ceramic (alumina) wall (with stainless steel extension in place),
and again at $z=6\,\mbox{\mbox{mm}}$ on the configuration with eleven
coils around the quartz wall. The probes measured the levitation field
which is compressed when the plasma enters the pot. A bigger CT will
displace a greater proportion of the levitation flux, so that $B_{\theta}(\phi,\,t)$$=B_{\theta}(r_{s}(\phi,\,t))$,
where $B_{\theta}(\phi,\,t)$ is the poloidal field measured at the
side probe, and $r_{s}(\phi,\,t)$ is the radius of the CT's separatrix
at the z-coordinate of the probe. By definition of the separatrix,
$\psi_{CT}(\phi,\,t)+\psi_{lev}(\phi,\,t)=0$ at $r_{s}(\phi,\,t),$
where $\psi_{CT}$ and $\psi_{lev}$ are the contributions to $\psi$
that arise due to CT currents and external coil currents respectively.
We expect that $\psi_{lev}(\phi,\,t)\approx\psi_{lev}(t),$ with any
deviation from toroidal symmetry being due either to coil misalignment,
or asymmetry associated with discrete coils. However, the CT poloidal
flux, and therefore $r_{s},$ can vary with toroidal angle, depending
on MHD activity in the CT. The $r$ component of the experimentally
measured field at the probes proved to be negligible, so we made the
approximation $B_{z}\approx B_{\theta}.$ We used a set of FEMM models
to estimate the value of $B_{z}$ that would be measured at the probes
for varying $r_{s}$.
\begin{figure}[H]
\subfloat[$r_{sF}=168\mbox{\mbox{ mm}}$.]{\raggedright{}\includegraphics[width=7cm,height=4cm]{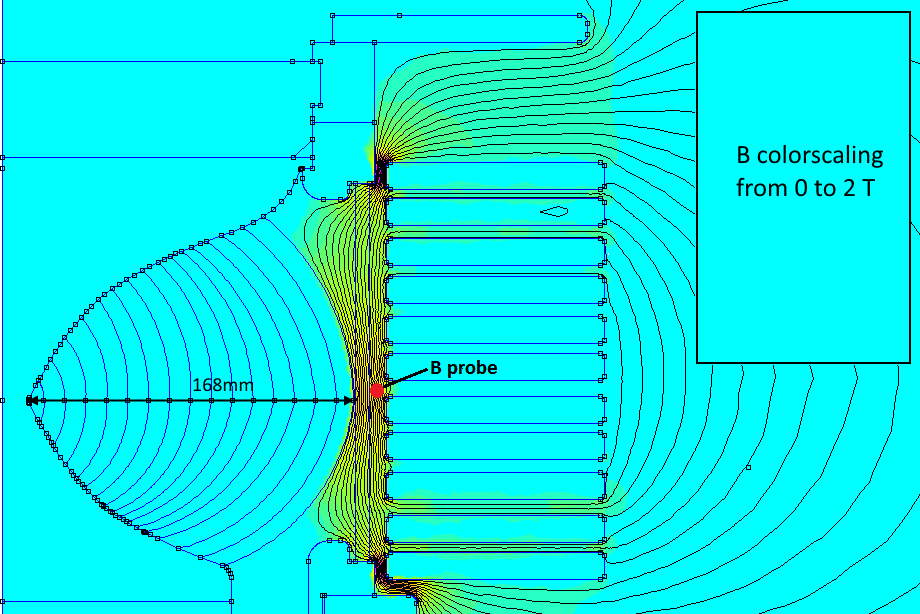}}\hfill{}\subfloat[$r_{sF}=60\mbox{\mbox{ mm}}$.]{\raggedleft{}\includegraphics[width=7cm,height=4cm]{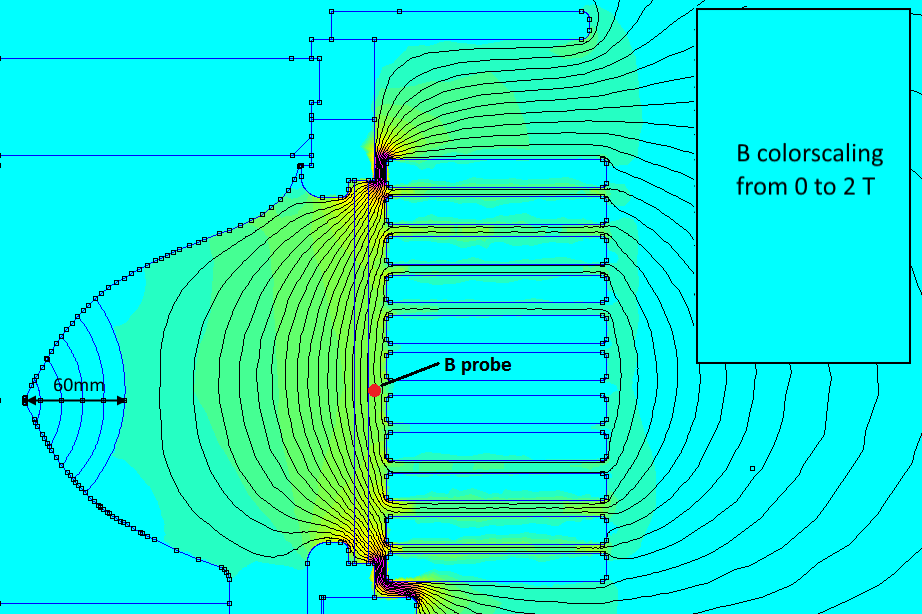}}

\caption{\label{rsepFEMM-1}$\,\,\,\,$FEMM models for finding $r_{s}(\phi,t)$}
\end{figure}
Figure \ref{rsepFEMM-1} shows two of the seventeen FEMM models used
to find $r_{s}$ for the 11-coil configuration. A material with artificially
high conductivity ($\sigma=10^{12}$ S/m) was assigned to the areas
representing the nested \textquotedbl plasmas\textquotedbl{} in FEMM.
The seventeen models are identical except that, starting with the
model  with $r_{sF}=168\mbox{\mbox{ mm}}$ (note $r_{quartz}=170\mbox{\mbox{ mm}})$,
the outermost of the set of nested shells representing \textquotedbl plasma\textquotedbl{}
material is removed for each successive model. Figure \ref{rsepFEMM-1}(a)
shows the FEMM solution (contours of $\psi,$ coloured by $|B|$)
for the model with $r_{sF}=168\mbox{\mbox{ mm}}$, and figure \ref{rsepFEMM-1}(b)
shows the solution for the case with $r_{sF}=60\mbox{\mbox{ mm}}$,
where $r_{sF}$ is the radius, at the same z coordinate as the probes,
of the outermost layer of \textquotedbl plasma material\textquotedbl{}
in the FEMM model.

With the levitation coil currents in the models determined from experimental
measurements, and the coil current frequencies set to a high value
($\sim1$ MHz), FEMM was run for each model. The high conductivity
of the material representing plasma, and the high current frequency,
ensure minimal penetration of levitation field into the \textquotedbl plasma\textquotedbl{}
region, so that the true separatrix radius is modelled. A LUA script
was written so that each of the models can be loaded successively
and run automatically through FEMM, and the required data for each
solution can be written to file for processing. The required data
consists, for each model, of $r_{sF}$ and $B_{zF}(r_{sF})$, where
$B_{zF}(r_{sF})$ is the FEMM solution for $B_{z}$ at the probe location
$((r,\,z)=(177\mbox{\mbox{ mm}},\,6\mbox{\mbox{ mm})}$ for the 11-coil
configuration) for a given $r_{sF}$. The process was repeated for
another set of FEMM models based on the 6-coil configuration with
the reduced insulator inner radius ($r_{ceramic}=144\mbox{\mbox{ mm}})$,
with the probe location being $(r,\,z)=(161\mbox{\mbox{ mm}},\,0\mbox{\mbox{ mm})}$. 

\subsection{Levitation only shots with 70$\mbox{ m}\Omega$ cables\label{subsec:Rsep_lev_70mOhm}}

\begin{figure}[H]
\subfloat[$B_{z}$ on side probes, shot 39650]{\raggedright{}\includegraphics[width=8cm,height=5cm]{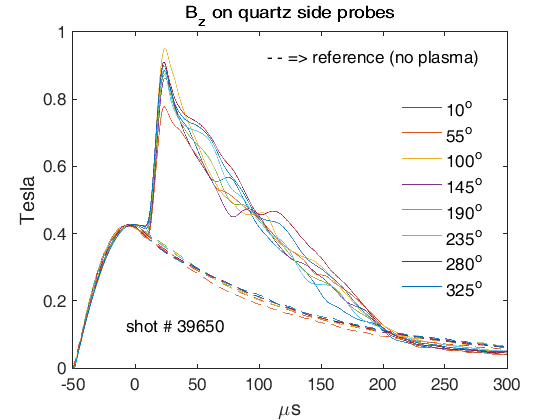}}\hfill{}\subfloat[FEMM outputs and functional fit ]{\raggedleft{}\includegraphics[width=8cm,height=5cm]{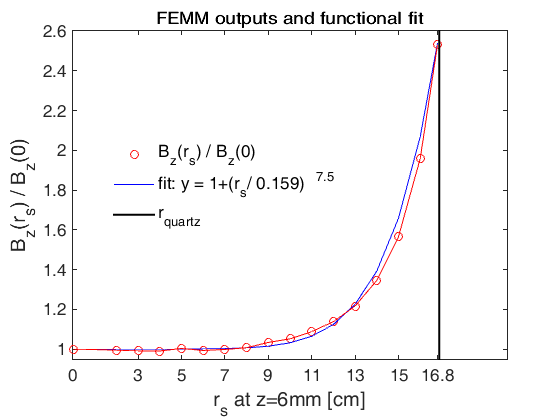}}

\caption{$\,\,\,\,$Experimental data and functional fit to FEMM data (70 m$\Omega$
cables)\label{Bz_rsep39650_1} }
\end{figure}
 Figure \ref{Bz_rsep39650_1}(a) shows the $B_{z}(\phi,\,t)$ signals
measured at the eight probes for shot $39650$, along with $B_{zref}(\phi,\,t)$,
the reference signals which are the averages of the $B_{zref}(\phi,\,t)$
signals from three levitation-only shots taken without charging or
firing the formation banks. The $B_{z}(\phi,\,t)$ and $B_{zref}(\phi,\,t)$
signals were calibrated using $B_{zF}(0)$ ($i.e.,$ no superconducting
\textquotedbl plasma\textquotedbl{} material in the FEMM model) to
determine the peak field amplitude at $\sim0\,\upmu$s (before CT
entry in the case of $B_{z}(\phi,\,t)$). Shot $39650$ was taken
in the 11-coil configuration with $70\mbox{ m}\Omega$ cables in place
between each main levitation inductor and coil-pair/coil, with $|t_{lev}|=50\,\upmu$s.
Figure \ref{Bz_rsep39650_1}(b) shows $B_{zF}(r_{sF})/B_{zF}(0)$
plotted against $r_{sF}$ using data from the set of FEMM models for
the 11-coil configuration. A function of the form $y=1+(r_{sF}/0.159)^{7.5}$
was found to be a good fit to the data. Using the data from each of
the eight probes, this functional fit is inverted to find $r_{s}(\phi,\,t)$
at each of the toroidal angles associated with the probes. At each
probe, we have recorded $B_{zref}(\phi,\,t)$, and $B_{z}(\phi,\,t)$,
so $r_{s}(\phi,\,t)$ can be found using the formula 
\begin{equation}
r_{s}(\phi,\,t)=0.159\,(B_{z}(\phi,\,t)\,/\,B_{zref}(\phi,\,t)\,-1)^{\frac{1}{7.5}}\label{eq:103-1}
\end{equation}
Note that $r_{s}(\phi,\,t)$ becomes complex-valued if $B_{zref}(\phi,\,t)>B_{z}(\phi,\,t)$
- care has to be taken to ensure that probe signals are properly calibrated,
and signals from any probes that have unusual responses must be ignored,
in order for the method to work. Note also that for $r_{s}\lesssim9$
cm, the slope of the function fit in \ref{Bz_rsep39650_1}(b) is too
flat to be successfully inverted with good accuracy. 
\begin{figure}[H]
\begin{centering}
\subfloat[]{\raggedright{}\includegraphics[scale=0.5]{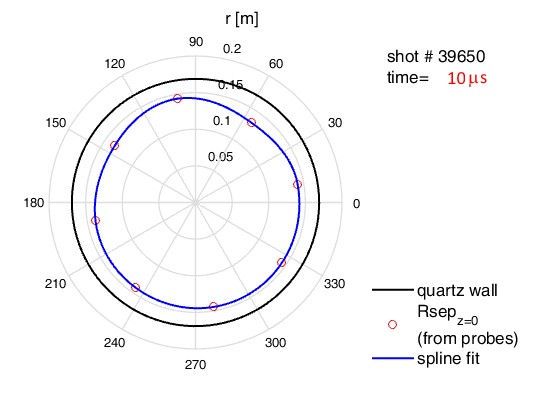}}\hspace{1.5cm}\subfloat[]{\raggedleft{}\includegraphics[scale=0.5]{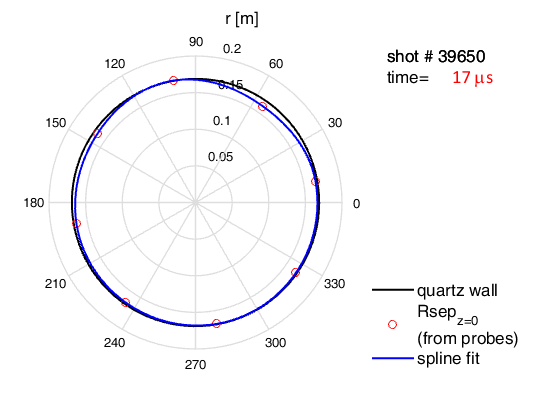}}
\par\end{centering}
\begin{centering}
\subfloat[]{\raggedright{}\includegraphics[scale=0.5]{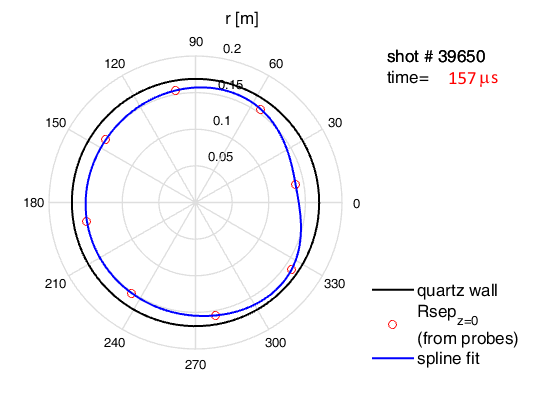}}\hspace{1.5cm}\subfloat[]{\raggedleft{}\includegraphics[scale=0.5]{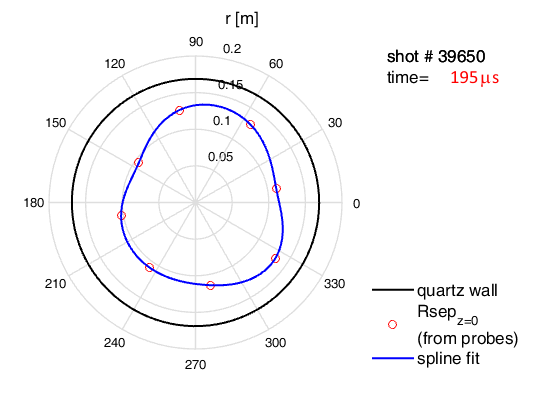}}
\par\end{centering}
\caption{\label{fig:rsep39650VID}$\,\,\,\,$Time-sequence of $r_{s}(\phi,t)$
for shot $39650\:(70\mbox{ m}\Omega\:\mbox{cables})$ }
\end{figure}
Figure \ref{fig:rsep39650VID} shows images from a video that is the
output of a MATLAB code that finds $r_{s}(\phi,\,t)$ based on $B_{z}(\phi,\,t)$
recorded at the side probes during shot $39650$, and the functional
fit in equation \ref{eq:103-1} that was obtained using the FEMM models.
It can be seen that the plasma enters the confinement region at $t=10\,\upmu$s,
and that at $t=17\,\upmu$s the CT fills the space right up to the
inner radius of the quartz wall, at $z=6\,\mbox{\mbox{mm}}$ (z coordinate
of the side probes). It remains at around this size and then starts
to shrink at around $157\,\upmu$s. At this time it looks like it
is being pushed in more at around $\phi=10^{\circ}$. At $195\,\upmu$s,
there are signs of an $n=3$ mode - the CT is being pushed in more
at around $\phi=10^{\circ}$ and $150^{\circ}$, and is reacting by
starting to bulge outwards at $\phi=80{}^{\circ},210^{\circ}\,\mbox{and}\:330^{\circ}$.
The CT is gone shortly after $195\,\upmu$s.
\begin{figure}[H]
\begin{centering}
\subfloat{\raggedright{}\includegraphics[width=8cm,height=5cm]{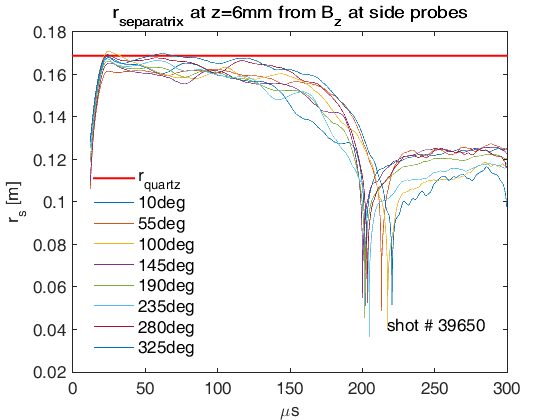}}
\par\end{centering}
\caption{\label{fig:rsep39650_2}$\,\,\,\,$$r_{s}(\phi,t)$ for shot $39650\:(70\mbox{ m}\Omega$
cables) }
\end{figure}
Figure \ref{fig:rsep39650_2} is a plot of the modelled $r_{s}(\phi,\,t)$
against time for shot 39650. As also indicated in figure \ref{fig:rsep39650VID}(c),
the CT starts to shrink in from the inner radius of the wall at around
$150\,\upmu$s. As mentioned, the evaluation for $r_{s}(\phi,\,t)$
is not valid when $B_{zref}(\phi,\,t)>B_{z}(\phi,\,t)$. It can be
seen in figure \ref{Bz_rsep39650_1}(a) that (due to inaccuracies
in probe responses etc.) $B_{zref}(\phi,\,t)>B_{z}(\phi,\,t)$ after
around $200\,\upmu$s. 

\subsection{Levitation only shots with 2.5$\mbox{ m}\Omega$ cables}

\begin{figure}[H]
\subfloat[$B_{z}$ on side probes, shot 29205]{\raggedright{}\includegraphics[width=8cm,height=5cm]{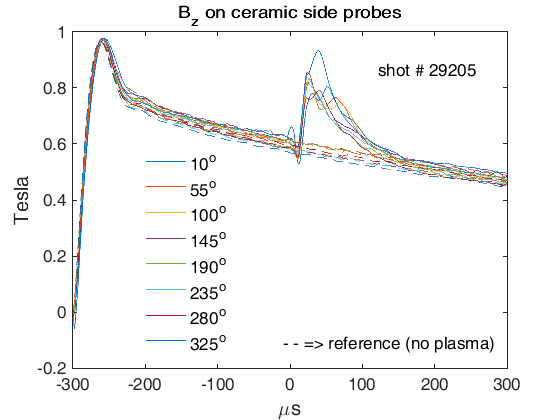}}\hfill{}\subfloat[FEMM output and functional fit ]{\raggedleft{}\includegraphics[width=8cm,height=5cm]{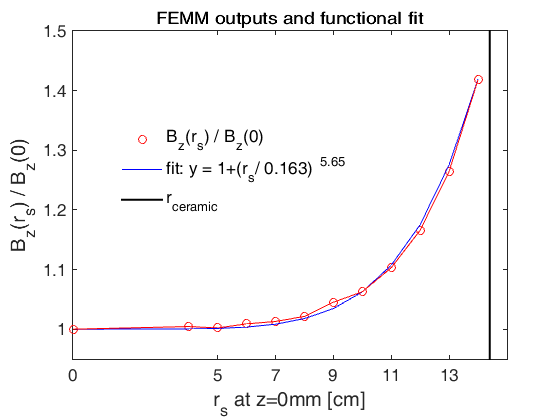}}

\caption{$\,\,\,\,$Experimental data and functional fit to FEMM data (2.5
m$\Omega$ cables)\label{Bz_rsep29205_1} }
\end{figure}
Figure \ref{Bz_rsep29205_1}(a) shows the $B_{\theta}(\phi,\,t)$
and $B_{zref}(\phi,\,t)$ signals for shot $29205$. Shot $29205$
was taken in the configuration with six coils surrounding the shortened
alumina insulating wall, with $2\times5\mbox{ m}\Omega$ cables in
parallel between the main levitation inductors and the coils, and
with $|t_{lev}|=300\,\upmu$s to allow for enhanced field line pinning
and reduced plasma-wall interaction in the 6-coil configuration. Figure
\ref{Bz_rsep29205_1}(b) shows $B_{zF}(r_{sF})/B_{zF}(0)$ plotted
against $r_{sF}$ for the 6-coil configuration. A function of the
form $y=1+(r_{s}/0.163)^{5.65}$ was found to be a good fit to the
FEMM data and the procedure followed to get $r_{s}(\phi,\,t)$ from
the experiment data is as described above.
\begin{figure}[H]
\begin{centering}
\subfloat[]{\raggedright{}\includegraphics[scale=0.5]{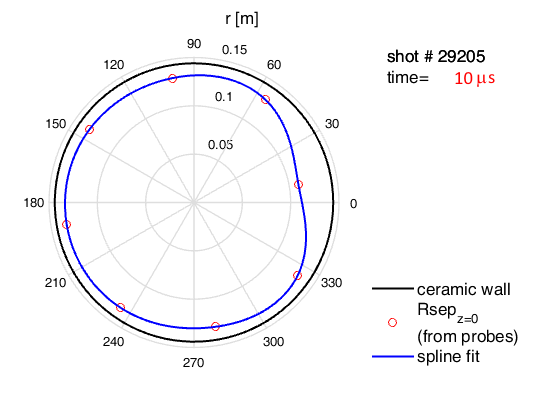}}\hspace{1.5cm}\subfloat[]{\raggedleft{}\includegraphics[scale=0.5]{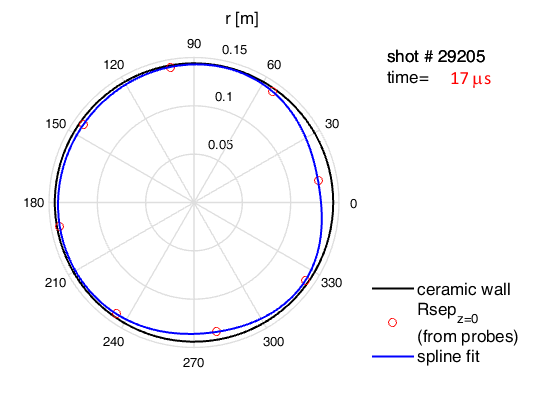}}
\par\end{centering}
\begin{centering}
\subfloat[]{\raggedright{}\includegraphics[scale=0.5]{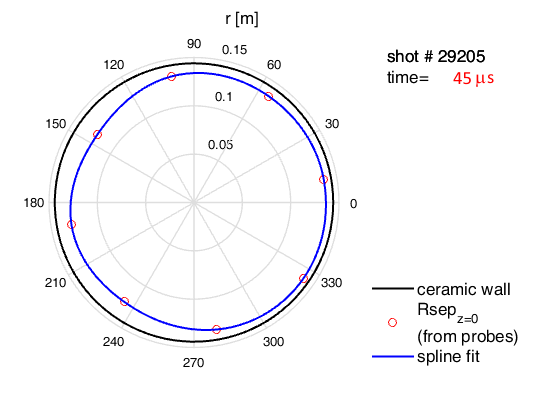}}\hspace{1.5cm}\subfloat[]{\raggedleft{}\includegraphics[scale=0.5]{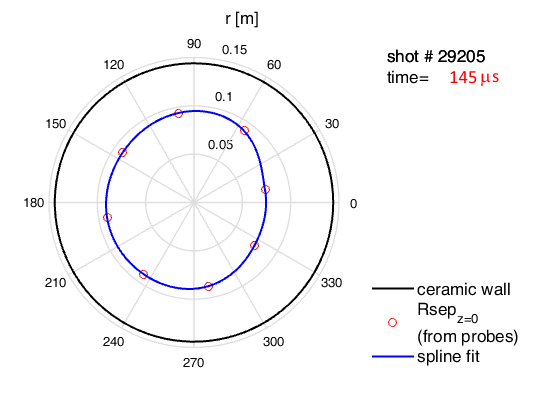}}
\par\end{centering}
\caption{\label{resp29205_2}$\,\,\,\,$Time-sequence of $r_{s}(\phi,t)$ for
shot $29205\ (2.5\mbox{ m}\Omega$ cables) }
\end{figure}
Figure \ref{resp29205_2} shows images at four times, indicating $r_{s}(\phi,\,t)$
based on the FEMM model outputs. As in figure \ref{fig:rsep39650VID},
it can be seen that the plasma enters the confinement region at $t=10\,\upmu$s,
and that at $t=17\,\upmu$s the CT fills the space right up to the
inner radius of the insulating wall. With the low resistance cables,
for a shot on the 6-coils with ceramic wall configuration, the levitation
field is constantly compressing the CT. It can be seen (figure \ref{resp29205_2}(c))
how the CT has already started to shrink at $45\,\upmu$s, whereas
the CT retains its maximum volume up until around $157\,\upmu$s when
the levitation field decay rate is optimized (figure \ref{fig:rsep39650VID}(c)).
The CT continues to be pushed inwards rapidly and is extinguished
shortly after $145\,\upmu$s.
\begin{figure}[H]
\subfloat[$r_{s}(\phi,\,t)$ for shot $29205\:(2.5\mbox{ m}\Omega$ cables) ]{\raggedright{}\includegraphics[width=8cm,height=5cm]{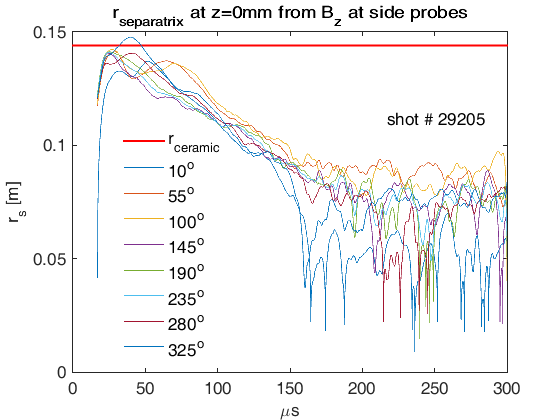}}\hfill{}\subfloat[$r_{s}(\phi,\,t)$ for shot $39573\:(2.5\mbox{ m}\Omega$ cables) ]{\raggedleft{}\includegraphics[width=8cm,height=5cm]{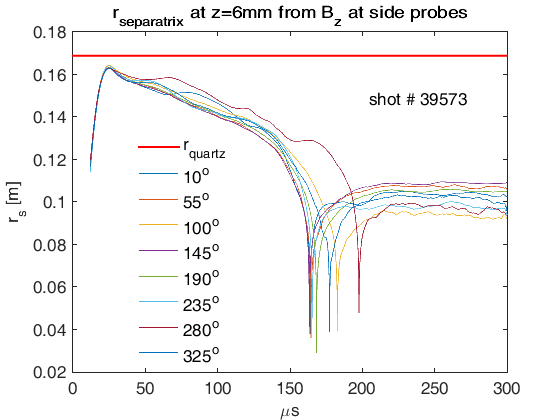}}

\caption{$\,\,\,\,$$r_{s}(\phi,\,t)$ for shots $29205\:\mbox{and 39573 }(2.5\mbox{ m}\Omega$
cables) \label{fig:rsep29205_39573}}
\end{figure}
Figure \ref{fig:rsep29205_39573}(a) is a plot of the modelled $r_{s}(\phi,\,t)$
for the same shot. As also indicated in figure \ref{resp29205_2}(c),
the CT starts to shrink in from the inner radius of the wall at around
$50\,\upmu$s. Calculated $r_{s}$ is not valid after around $150\,\upmu$s,
when $B_{z}(\phi,\,t)\,\leq\,B_{zref}(\phi,\,t)$ (see figure \ref{Bz_rsep29205_1}(a)).
Figure \ref{fig:rsep29205_39573}(b) is a plot of the modelled $r_{s}(\phi,\,t)$
for shot 39573, which also had the original levitation circuits with
$2\times5\mbox{ m}\Omega$ cables in parallel between the main levitation
inductors and the coils, but was taken on the 11-coil configuration,
and therefore the functional fit indicated in figure \ref{Bz_rsep39650_1}(b)
was used to extract $r_{s}(\phi,\,t)$. The CT in shot 39573 ($V_{form}=16\mbox{ kV}$),
lives longer than that in shot 29205 ($V_{form}=12\mbox{ kV}$). However,
the CTs in shots 29205 and 39573 are similar in that they both shrink
rapidly in comparison with shot 39650 (figure \ref{fig:rsep39650_2}
), in which the decay-rate matching strategy was used.

\section{Comparison of total spectral power measurements and levitated CT
lifetimes for various configurations\label{subsec:Comparison-of-total} }

\begin{figure}[H]
\begin{centering}
\subfloat[]{\raggedright{}\includegraphics[width=8.6cm,height=5cm]{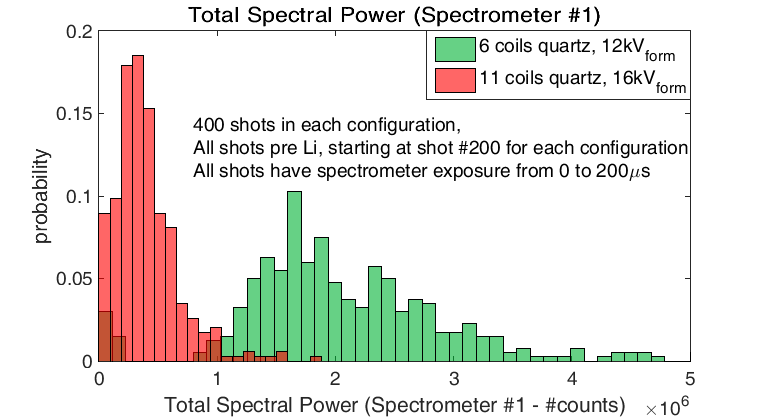}}\hfill{}\subfloat[]{\raggedleft{}\includegraphics[width=7.5cm,height=5cm]{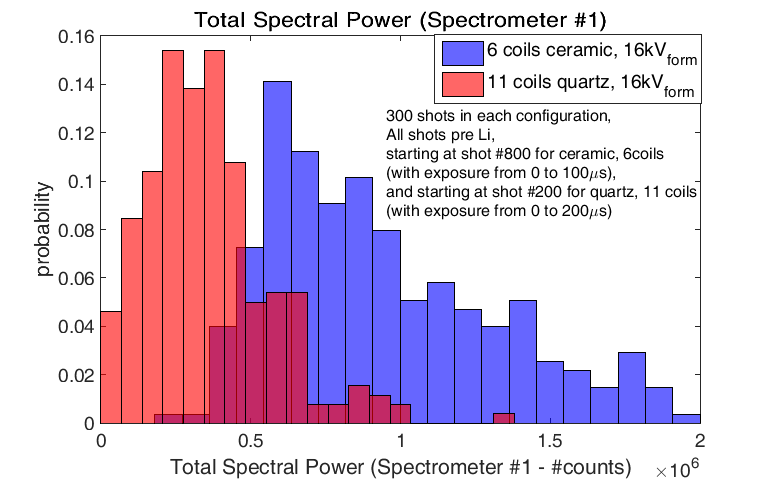}}
\par\end{centering}
\centering{}\caption{\label{fig:11coils_spectrometerdata}$\,\,\,\,$Spectrometer data}
\end{figure}
Figure \ref{fig:11coils_spectrometerdata}(a) shows normalised histograms
comparing total spectral power recorded with spectrometer \#1 for
shots with the quartz wall in place, for the 6-coil and 11-coil configurations.
Data from 400 pre-lithium shots for each configuration, all with spectrometer
exposure from $0$ to $200\,\upmu$s, is included. Each shot in the
selection was taken after at least 200 post-baking cleaning shots
in the relevant configuration. Having the same spectrometer exposure
time and the same number of cleaning shots prior to the shots being
selected for comparison in each configuration is important for a valid
comparison. The validity of the data was verified by comparing the
total spectral power recorded with the measured intensity of plasma
optical emission at around the same location as the spectrometer,
and finding a good correlation. The locations of spectrometer and
optical ports on the machine headplate are indicated in figure \ref{fig:Machine-headplate-schematic-1}.
Spectrometer \#1 is the inner of the two spectrometers, and both spectrometers
had vertical lines of sight. Even at increased formation voltage,
total spectral power is around four times lower with eleven coils.
This is particularly unusual because on a given configuration, it's
expected that higher formation current ($i.e.,$ $V_{form}$) leads
to increased ablation of electrode material and consequently increased
impurity levels and total spectral power. 

Due to variations in spectrometer exposure settings throughout the
experiment, the available data does not allow for direct comparison
of recorded spectral power between the ceramic and quartz wall on
the 6-coil configuration. As discussed previously, CT lifetimes decreased
significantly with the transition from the ceramic to quartz wall,
so it is expected the spectral power would be far higher with the
quartz wall. Figure \ref{fig:11coils_spectrometerdata}(b) shows data
comparing total spectral power recorded with spectrometer \#1 for
the 6-coil (ceramic wall) and 11-coil (quartz wall) configurations.
Spectrometer exposure was $0$ to $100\,\upmu$s for the selection
of shots taken in the 6-coil configuration, and $0$ to $200\,\upmu$s
for those taken in the 11-coil configuration. Naturally, spectral
power increases with increase exposure time. There were four times
as many prior cleaning shots, which tend to reduce impurity and spectral
power levels, for the configuration with the ceramic wall and six
coils. Despite the different exposure times and number of cleaning
shots, and despite the quartz wall, total spectral power is several
times lower with the 11-coil configuration. Comparing the spectral
powers for shots in the 6-coil configuration with the quartz wall
(figure \ref{fig:11coils_spectrometerdata}(a)) and ceramic wall (figure
\ref{fig:11coils_spectrometerdata}(b)), it can be seen how the spectral
power is indeed significantly higher in the case with the quartz wall,
even though the shots with the ceramic wall were taken at higher formation
voltage. There were, however, four times as many prior cleaning shots,
and spectrometer exposure was reduced, for the configuration with
the ceramic wall, so this last comparison is not conclusive. 

It is thought that the 11-coil setup reduced impurities and the associated
energy losses due to line radiation because it reduced the level of
interaction between plasma and the outer insulating wall during the
bubble-in process. The data presented in figure \ref{fig:11coils_spectrometerdata}
confirms that the benefit from the reduction of plasma-insulating
wall interaction appears more significant than any impurity increase
caused by either the transition from the ceramic to the quartz wall,
or plasma-electrode interaction ($i.e.,$ increased $V_{form}$).

\begin{figure}[H]
\centering{}\includegraphics[height=4.5cm]{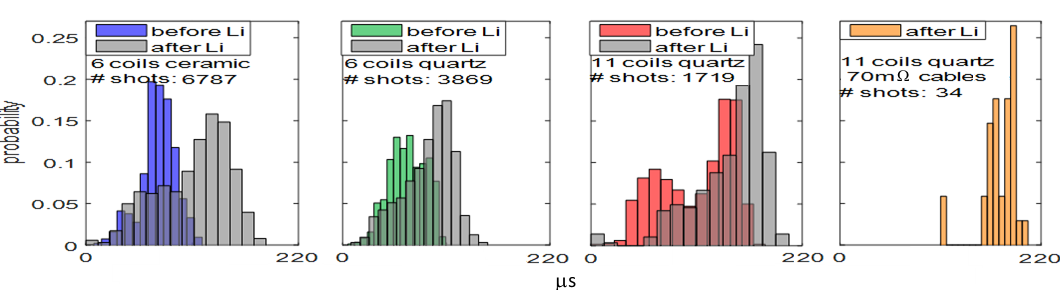}\caption{\label{fig:Effect-of-lithium}$\,\,\,\,$Comparison, across configurations,
of CT lifetimes and effect of lithium gettering.}
\end{figure}
As indicated in the normalised histograms in figure \ref{fig:Effect-of-lithium},
pre-lithium CT lifetimes were longer with the ceramic wall, despite
the smaller volume. Lithium gettering was very effective on the ceramic
wall ($\sim70\%$ lifetime increase), but not so effective on quartz
($\sim30\%$ lifetime increase). Lifetime increased significantly
with the 11-coil configuration. The \textquotedbl double-Gaussian\textquotedbl{}
shape of the (before Li) distribution for eleven coils may be due
to the $\sim35$\% of shots taken in that configuration in suboptimal
machine-parameter space ($i.e.$ values of $V_{form},\,V_{lev},\:I_{main}$,
and $t_{gas}$) that were rapidly explored in the last days of the
experiment in new configurations such as without levitation inductors,
with additional crowbarred sustain current, and with passive or open-circuited
levitation/compression coils. Note that of the $10,000+$ shots from
which data is taken for this levitated CT lifetime comparison, and
the \textasciitilde 40,000 shots taken in total over the three year
duration of the experiment, only 34 shots in the best of the configurations
tested - eleven coils with $70\mbox{ m}\Omega$ cables - are shown
because the 11-coil configuration was explored rapidly in the days
before the experiment was decommissioned. It can be seen that the
repeatability of good shots was significantly improved in that configuration.

\section{Summary\label{subsec:SummaryLev}}

The principal obstacle to the progress of the magnetic compression
experiment was the problem associated with radiative cooling that
arose due to interaction at CT formation between plasma entering the
containment region and the insulating outer wall that was required
in the design to allow operation of the compression field. In the
original six-coil configurations, plasma being rapidly advected into
the containment region during the formation process was able to displace
the levitation field into the large gaps above the coil stack, and
come into contact with the insulating wall. This issue was only identified
during the experiment with the 25-turn coil configuration, shortly
before the experiment was scheduled for decommissioning, and was only
partly resolved with the eleven coil configuration over the last few
weeks of extended operation. 
\begin{figure}[H]
\begin{raggedright}
\subfloat[six coils, ceramic wall]{\centering{}\includegraphics[width=8cm,height=5cm]{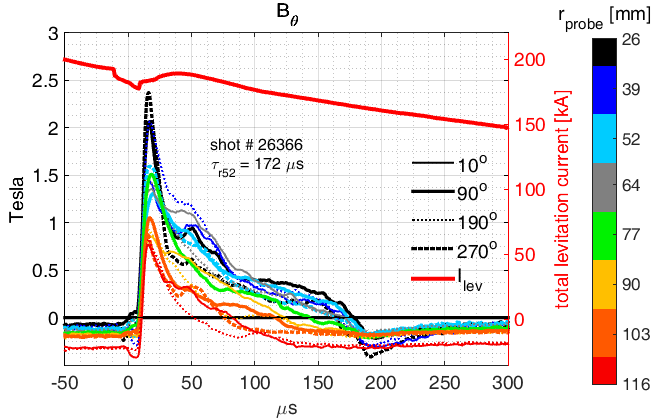}}\hfill{}\subfloat[six coils, quartz wall]{\centering{}\includegraphics[width=8cm,height=5cm]{fig_44.png}}
\par\end{raggedright}
\begin{raggedright}
\subfloat[25 turn coil, quartz wall]{\centering{}\includegraphics[width=8cm,height=5cm]{fig_57.png}}\hfill{}\subfloat[11 coils, quartz wall]{\centering{}\includegraphics[width=8cm,height=5cm]{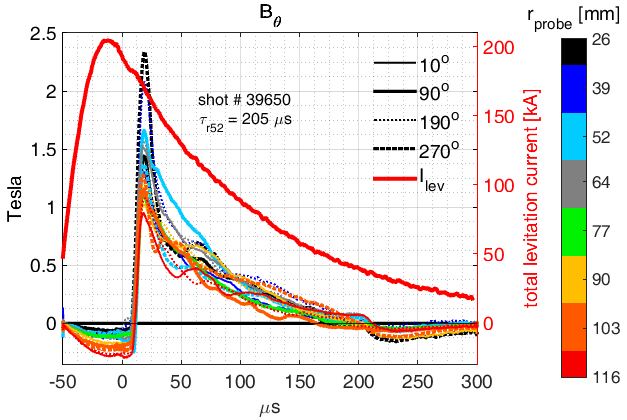}}
\par\end{raggedright}
\begin{raggedright}
\subfloat[no coils, stainless steel wall]{\centering{}\includegraphics[width=8cm,height=5cm]{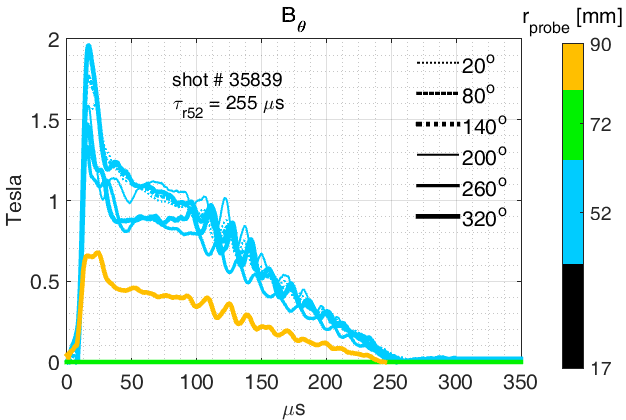}}\hfill{}\subfloat[no coils, aluminum wall]{\centering{}\includegraphics[width=8cm,height=5cm]{fig_48.png}}
\par\end{raggedright}
\centering{}\caption{\label{fig:-for-four4configs}$\,\,\,\,$$B_{\theta}$ for six configurations}
\end{figure}
Figure \ref{fig:-for-four4configs} shows poloidal field traces for
the six principal configurations tested. Comparing figures \ref{fig:-for-four4configs}(a)
and (b), and noting, as outlined in section \ref{subsec:6-coilsconfig, quartz},
that a 50\% increase in CT lifetime was expected with the switch to
the larger internal radius insulating tube, it can be seen how quartz
was significantly worse than ceramic as a plasma-facing material.
For these two shots, $|t_{lev}|$ was $300\,\upmu$s - as mentioned
in section \ref{subsec:6-coils-config_ceramic}, and confirmed by
MHD simulations (see section \ref{subsec:Effect-of-increasing}),
the strategy of allowing the levitation field more time to soak into
the steel above and below the insulating wall led to a reduced level
of plasma-wall interaction and slightly increased CT lifetimes in
the 6-coil configurations.

Apparently, plasma/material interaction during formation was reduced
with the modified levitation field profiles of the 25-turn coil and
eleven-coil configurations, in which current carrying coils extended
along the entirety of the outer surface of the insulating wall. Spectrometer
data and observations of CT lifetime confirm that the improved design
led to reduced levels of plasma impurities and radiative cooling.
Consistent with this explanation for the improvement, at the same
initial CT poloidal flux, as determined by the voltage on the formation
capacitors and the current in the main coil, CT lifetimes were around
the same for the six-coil, 25-turn coil, and eleven-coil configurations.
However the 25-turn coil and eleven-coil configurations allowed for
the successful formation of higher flux, physically larger, CTs -
formation voltage could be increased from $12$ kV  to $16\mbox{ kV}$
and main coil current could be increased from 45A to 70A, corresponding
to an increase in $\Psi_{gun}$ from around 8 mWb to 12 mWb. In contrast,
the benefit of increased initial CT flux was outweighed by the performance
degradation due to increased wall interaction in the six coil setup.
With CT lifetimes of up to $274\,\upmu$s, the longest-lived levitated
CTs were produced with the 25-turn coil configuration (figure \ref{fig:-for-four4configs}(c)),
despite the presence of the quartz wall. This was almost double the
maximum $\sim150\,\upmu$s lifetimes seen with six coils around the
quartz wall, but still less than the $\sim340\,\upmu$s lifetimes
observed without levitation with an aluminum flux conserver (figure
\ref{fig:-for-four4configs}(f)). The eleven coil configuration, with
a field profile similar to that of the 25-turn coil setup, also enabled
the production of relatively high-flux CTs with correspondingly increased
lifetimes (figure \ref{fig:-for-four4configs}(d)). Levitation circuit
modification to match the decay rates of the levitation and plasma
currents led to more stable, larger, longer lived plasmas, and a greatly
increased rate of good shots, by avoiding unintentional magnetic compression
during CT levitation. This strategy was implemented with the 25-turn
coil and eleven-coil configurations. The experimental technique developed
to measure the CT outboard separatrix confirmed that optimisation
of levitation field damping led to CTs that remained physically larger
over extended times. In general, the recurrence rate of good shots
in the 25-turn coil configuration was poor compared with that in the
11-coil configuration. However, it remains unclear why the longest-lived
CTs produced with the 25-turn configuration outlived those produced
in the 11-coil configuration. One possible explanation is that the
toroidal symmetry of the levitation field played a role at high formation
settings. A second explanation may be the effect of the exceptionally
thick lithium coating that was applied during the test with the 25-turn
coil configuration. 

A third possible explanation is that the ratio of the coil inductance
to the levitation circuit holding inductance was increased from $L_{coil}/L_{main}=600\mbox{ nH}/6\,\upmu\mbox{H}=0.1$
for the 11-coil configuration to $L_{coil}/L{}_{main}=116\,\upmu\mbox{H}/6\,\upmu\mbox{H}\,\sim\,20$
for the 25-turn coil. When conductive plasma enters the pot (confinement
region) it reduces the inductance of the part of the levitation circuit
that includes the levitation/compression coil and the material that
the coil encompasses. The levitation current increases when the inductance
is reduced as plasma enters the pot. If the percentage rise of the
levitation current is increased, by increasing the ratio $L_{coil}/L{}_{main}$,
it means that levitation current prior to plasma bubble-in can be
minimised. This reduction in $I_{lev}$ reduces the likelihood that
the levitation field will be strong enough to partially block plasma
entry to the pot, while still allowing the field that is present,
when the plasma does enter, to be strong enough to levitate the plasma
away from the insulating wall. Comparing figures \ref{fig:-for-four4configs}(c)
and (d), it can be seen how the levitation current increases significantly
at bubble-in for the 25-turn coil only. FEMM models indicate that
the levitation fluxes found to be optimal at moderately high formation
settings for the 25-turn and 11-coil configurations were approximately
the same prior to plasma entry to the containment region. It may be
that the increased levitation flux at CT entry in the 25-turn configuration
was more efficient at keeping plasma off the wall. The optimal settings
for $|t_{lev}|$ in the two configurations were limited by $t_{rise}$,
the rise time of the levitation current for the particular configuration.
While the strategy of increasing $|t_{lev}|$ to allow the levitation
field more time to soak into the steel above and below the insulating
wall led to slightly increased CT lifetimes on the 6-coil configurations,
it was found that $|t_{lev}|$ should be reduced to as low a value
as possible on the 25-turn coil and 11-coil configurations for best
performance. Reducing $|t_{lev}|$ reduces the likelihood that the
levitation field will impede, through the line tying effect, plasma
entry to the containment region at formation. The benefit of slightly
reducing plasma-wall interaction by \emph{increasing} $|t_{lev}|$,
and the line-tying effect, outweighed the detrimental effect of pot-entry
blocking in the 6-coil configuration only. With the high inductance
25-turn coil, optimal $|t_{lev}|$ was equal to $t_{rise}\sim150\,\upmu$s,
while for the 11-coil configuration, optimal $|t_{lev}|$ was set
to $t_{rise}\sim50\,\upmu$s. It may be that allowing the level of
levitation flux that was present in the containment region upon plasma
entry in the 25-turn configuration to soak into the steel above and
below the wall, even for 50$\,\upmu$s in the 11-coil configuration,
degraded performance by impeding plasma entry to the containment region.
The requirement for increased $|t_{lev}|$, and consequent pot-entry
blocking may have been the cause of the poor repeatability of good
shots in the 25-turn configuration. Future designs should optimise
between the ideals of minimal coil inductances while maximising the
coil to levitation circuit inductance ratios.

The 25-turn coil extended farther above and below the insulating wall
than the stack of eleven coils - a fourth possible explanation for
the (occasional) improved performance of the 25-turn coil over the
11-coil configuration is that the increased levitation field, relative
to that for the 11-coil configuration, at the top and bottom of the
insulating wall, played a key role. At low formation settings, without
addition levitation circuit series resistance, levitated CT lifetimes
in the 25-turn configuration were comparable to those in both the
11-coil and 6-coil configurations. It is clear that the feature shared
by the 25-turn and 11-coil configurations, of closing the gaps that
remained above and below the coil stack in the 6-coil configurations,
was responsible for enabling the formation of high flux CTs with correspondingly
increased lifetimes, and that the unconfirmed mechanism that enabled
(occasional) even better performance in the 25-turn configuration
was also effective only at high formation settings.

Compared with the aluminum flux conserver, a stainless steel wall
led to more impurities and shorter CT lifetimes (figures \ref{fig:-for-four4configs}(e)
and (f)), likely due to more CT field-diffusion into the material,
leading to enhanced sputtering. Toroidal $n=2$ magnetic perturbations
were observed on CTs produced with both stainless steel and aluminum
outer flux conservers, and remained even when a moderate levitation
field was allowed to soak through the stainless steel wall, but were
absent in all configurations tested in which a CT was held off an
outer insulating wall by a levitation field. It is known that $n=2$
fluctuations are a sign of internal MHD activity associated with increased
electron temperature. However, the longest-lived CTs produced with
the 25-turn coil configuration endured for up to 10\% longer than,
and may therefore be reasonably assumed to be hotter than the CTs
produced with the stainless steel outer flux conserver. It is possible
that the levitation field acts to damp out helically propagating magnetic
fluctuations at the outboard CT edge and that internal MHD activity
is relatively unchanged. The $n=1$ magnetic fluctuations observed
when 80 kA additional crowbarred shaft current was applied to the
machine in the eleven-coil configuration confirmed coherent toroidal
CT rotation, and may have been a result of more vigorous MHD activity
that remained apparent despite damping.

As presented in \cite{FowlerComments}, the e-folding time characterising
the decay rate of spheromak magnetic field is:
\begin{equation}
\tau_{mag}=\frac{1}{2Z_{eff}}R^{2}T_{e0}\sqrt{T_{eE}}
\end{equation}
where $R\,[\mbox{m}]$ is an estimate for the outer CT separatrix
($\sim0.15\,$m for the magnetic compression experiment), $T_{e0}\,[\mbox{keV}]$
is the peak electron temperature and $T_{eE}\,[\mbox{keV}]$ is the
edge electron temperature (expected to be $<10\,\mbox{eV}$). The
fact that the best levitated CT lifetimes were $\sim20\%$ less than
lifetimes seen on the machine when the insulating outer wall was replaced
with the standard MRT aluminum flux conserver may be accounted for
by the higher impurity-associated $Z_{eff}$ of levitated CTs. It
is clear that the switch from the alumina wall to the quartz wall
led to an increased impurity level. The 11-coil configuration reduced
the impurity problem but a further reduction in impurities and $Z_{eff}$
would have been achieved in a configuration with eleven coils on an
alumina insulating wall. It seems clear that CT lifetimes comparable
with those achieved with the aluminum flux conserver could be attained
in such a configuration. An obvious further improvement would be to
implement an even more compatible plasma facing material. It is well
known that boron nitride (BN) is a good material for minimising impurity
levels in cases when an insulating material comes in contact with
plasma, and is often used as an insulating component on probes that
are inserted into a plasma. Compared with alumina, boron has lower
$Z$ than aluminum, and nitrogen is less reactive than oxygen, and
is thus less likely to combine with other impurities to form high
$Z$ molecules that would lead to further radiative cooling. While
standard hot-pressed BN, which has a hexagonal crystal structure,
contains up to 5\% oxygen and other impurities, pyrolytic boron nitride
(PBN), which has a diamond-like crystal structure, is an extremely
pure (>99.999\%) form of BN that is manufactured using a chemical
vapor deposition (CVD) process. Due to the manufacturing process,
PBN is generally available only in thin sheets or thin-walled tubes.
An eleven coil configuration including a PBN tube with 2 mm wall thickness
located with a close fit inside an alumina tube would be a natural
next step toward improving levitated CT performance. The alumina tube
would be able to support vacuum and should ideally have an inner radius
at least equal to that of the quartz tube which it would replace.
It may be, as discussed, that the 25-turn coil had some additional
advantages over the eleven-coil configuration, that could be reproduced
in a new build. 

A redesign of the experiment involving levitation and compression
coils located inside the vacuum vessel, as was implemented on the
S-1 magnetic compression \cite{S1_compression}, device would involve
a lot of work but would avoid the problem of impurity contamination
associated with the external insulating wall. The design of vacuum-tight
power transmission to the internal coils would be a challenge. It
would be worth attempting a minor change with a new insulating wall
material on the eleven coil configuration, and assess the results,
before attempting such a major modification. \newpage{}

\chapter{Magnetic Compression\label{Chap:Magnetic-Compression} }

The focus of this chapter is on the results obtained when CTs were
magnetically compressed. Section \ref{subsec:OverviewMag} presents
an overview of the principal observations from representative compression
shots taken in the main configurations that were introduced in chapter
\ref{Chap:MagLev}. It will be shown that, particularly with the eleven
coil configuration, significant increases in magnetic field, density,
and ion temperature were routinely observed at compression. Various
metrics for magnetic compression performance were developed and are
presented here. The compressional instability that was observed on
most compression shots is discussed in section \ref{subsec:Compressional-Instability}.
A comparison of compressional performance between configurations is
presented in section \ref{subsec:Comparison-of-compression}. The
cause of the high frequency fluctuations observed on the field measurements
for many compression shots taken in the six coil configuration is
discussed. The method developed to experimentally evaluate the CT
outboard equatorial separatrix, descried in chapter \ref{Chap:MagLev},
is applied to a compressed shot, and indicates a significant geometric
compression factor of around two. Indications, from scintillator data,
of fusion neutron production during CT formation, is presented in
section \ref{subsec:Scintillator-data}. The chapter concludes with
a summary in section \ref{sec:SummaryCOMP}. 

\section{Overview of results\label{subsec:OverviewMag}}

\begin{figure}[H]
\centering{}\includegraphics[width=16cm,height=10cm]{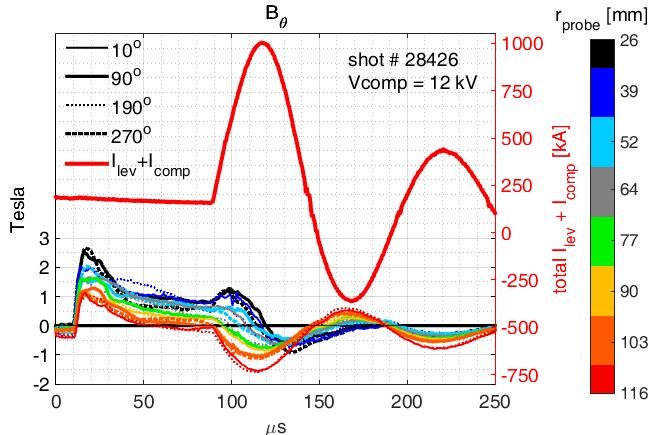}\caption{\label{fig:Bp_28426}$\,\,\,\,$$B_{\theta}$ and $I_{lev}+I_{comp}$
for shot  28426 (6-coil configuration)}
\end{figure}
Figure \ref{fig:Bp_28426} shows $B_{\theta}$ traces and total levitation
plus compression current (right axis) for shot  28426, which was taken
in the 6-coil configuration, with $V_{comp}=12$ kV  and $t_{comp}=90\,\upmu$s.
Shot 28426 is a representative compression shot for the 6-coil configuration.
The levitation capacitors were fired at $t_{lev}=-300\,\upmu$s (the
original low resistance 2.5 m$\Omega$ cables were in place), and
by the time the plasma has entered the confinement region at around
$15\,\upmu$s, the total levitation current (divided approximately
equally between the six coils) has dropped to $\sim200$ kA. By $t=t_{comp}=90\,\upmu$s,
the total levitation current has dropped further, to around $150$
kA. At this time, the compression capacitors are fired, and the total
compression current rises over $t_{rise}\sim25\,\upmu$s to its peak,
for $V_{comp}=12$ kV, of around 850 kA, so that the total combined
levitation and compression current is around $1$ MA at the time peak
compression, around $115\,\upmu$s. In this shot, the CT is compressed
inwards beyond the probes at $r=77\mbox{ mm}$, so $B_{\theta}$ at
the probes located at $r\geq77\mbox{ mm}$ ($i.e.,$ green, light
orange, dark orange, and red traces) is a measurement of the compression
field after $\sim90\,\upmu$s, while, in general, the CT poloidal
field is measured at $r<77\mbox{ mm}$, until $\sim120\,\upmu$s.
Note that the combined levitation and compression current changes
polarity at around 150$\,\upmu$s and again at $\sim190\,\upmu$s,
because the compression current is allowed to ring with the capacitor
discharge.
\begin{figure}[H]
\centering{}\includegraphics[width=16cm,height=10cm]{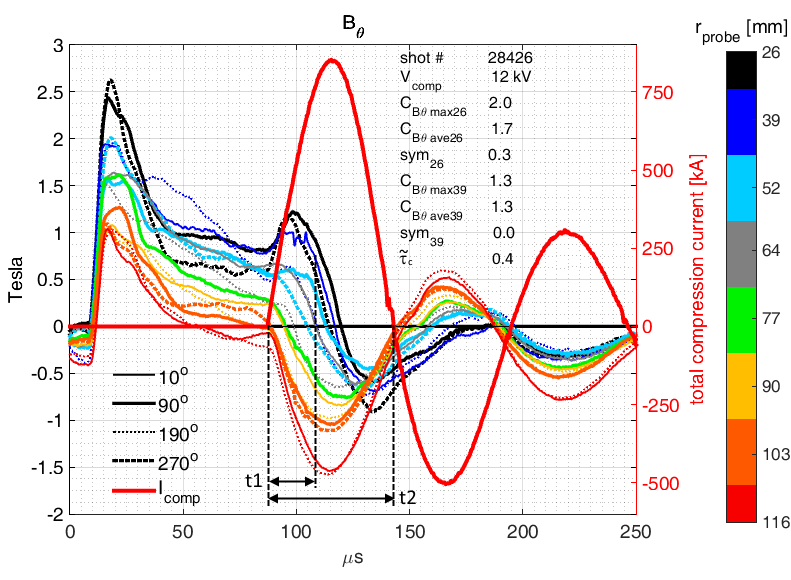}\caption{\label{fig:Bp_28426-1}$\,\,\,\,$$B_{\theta}$ and $I_{comp}$ for
shot  28426 (6-coil configuration)}
\end{figure}
The $B_{\theta}$ traces for shot  28426 are produced again in figure
\ref{fig:Bp_28426-1}, where, for clarity, the compression current,
this time without the superimposed levitation current, is also shown.
Experimental data at GF is usually analysed and visualised using IGOR-PRO
software, but for the magnetic levitation and compression data, some
MATLAB scripts were written so that data from individual shots could
be visualised quickly without waiting one to two minutes for the IGOR
program to load the data for each shot. In figure \ref{fig:Bp_28426-1},
some of the Matlab-calculated parameters that were considered good
indications of the quality of the compression shot are displayed.
$C_{B\theta max}(r)$ and $C_{B\theta ave}(r)$ are the maximum and
average of the two poloidal magnetic compression ratios, obtained
at the two probes located at radius $r$ mm, $180^{o}$ apart toroidally,
$e.g.,$ $C_{B\theta max26}=max(C_{B\theta_{90^{\circ}r26}},\,C_{B\theta_{270^{\circ}r26}})$,
where, for example, $C_{B\theta_{270^{\circ}r26}}=B_{\theta CTpeak}/B_{\theta CTpre}$,
where $B_{\theta CTpeak}$ and $B_{\theta CTpre}$ are the values
of $B_{\theta},$ measured with the probe at $r=26\mbox{\mbox{ mm}},\,\phi=270^{\circ}$,
at the peak of compression and just before compression respectively.
The values of $C_{B\theta max26}=1.7\,\mbox{and }C{}_{B\theta ave26}=1.3$
obtained for this shot are quite low, for the reasons discussed below.
Note that when $|B_{\theta CTpre}|\,<\,|B_{\theta levpre}|$, where
$|B_{\theta levpre}|\,(\mbox{note \,}B_{\theta levpre}<0)$ is the
absolute value of the poloidal levitation field that would be measured
at the probe just before compression if no CT was present, a \textquotedbl levitation
offset\textquotedbl{} should be included when calculating these compression
ratios. In those cases, the compression ratio would be $C_{B\theta}=(B_{\theta CTpeak}-B_{\theta levpeak})/(B_{\theta CTpre}-B_{\theta levpre})$,
where $B_{\theta levpeak}$ is the poloidal levitation field that
would be measured at the probe at peak compression if no CT was present.
However, these cases (anyway, only one shot which is being ignored)
are not interesting because there\textquoteright s not much left to
compress if $|B_{\theta CTpre}|\,<\,|B_{\theta levpre}|$. This is
especially true for shots with the decay rate of the levitation field
matching the CT's decay rate, when $|B_{\theta levpre}|$ is very
low - for example $B_{\theta levpre}\sim-0.01$ T for shot  39735
(figure \ref{fig:39735Bp}), while $B_{CTpre}\sim0.3$ T. 

Parameters $sym_{r}$ give an indication of the toroidal asymmetry
of the magnetic compression at the probes located at radius $r\,\mbox{mm}$.
Shots with $sym_{r}$ close to zero have toroidally symmetric compression
at radius $r\,\mbox{mm}$. With parameters $sym_{26}=0.3$, and $sym_{39}=0$,
shot  28426 had quite symmetric compression at $r=26$ mm, and very
symmetric compression at $r=39$ mm. Asymmetry of the magnetic field
at compression was typical for both the 6-coil and 11-coil configurations,
with only a slight improvement with the 11-coil configuration - this
shot has exceptionally symmetric compression of poloidal field.

The parameter $\widetilde{\tau}_{c}$ indicates the level of compressional
flux conservation, and is calculated as $\widetilde{\tau}_{c}=t1/t2$,
where $t1$ and $t2$ are indicated in figure \ref{fig:Bp_28426-1}.
$t2\sim50\,\upmu$s is the half-period of the compression current,
and $t1$ is the time from $t_{comp}$ to the average of the times
when $B_{\theta}$ at the two $r=26\mbox{ mm}$ probes fall to their
pre-compression values (at $t=t_{comp}$). If the CT doesn't lose
flux during compression, due to, for example, enhanced resistive flux
loss, or some instability that disrupts flux surfaces, the measured
$B_{\theta}$ at the inner probes rises and falls approximately in
proportion to the compression current, and $t1\sim t2$. Shots for
which flux is not lost due to any mechanism besides the usual resistive
loss are characterised by $\widetilde{\tau}_{c}\sim1$. This characterisation
method assumes that the CT is not being compressed to a radius less
than $26\mbox{\mbox{ mm}}$. If that did happen, the indication of
$B_{\theta}$ increase at $26\mbox{\mbox{ mm}}$ should disappear
early ($\widetilde{\tau}_{c}\ll1$), and then there would be no data
whatsoever available to assess the compression beyond $26\mbox{\mbox{ mm}}$.
If the CT is being compressed beyond $26\mbox{\mbox{ mm}}$, and stays
stable, it may expand back to $r>26\mbox{\mbox{ mm}}$ after the peak
in compression field, but there are no examples of that occurrence
in the data.

Shots taken in the 6-coil configuration generally became \textquotedbl unstable\textquotedbl ,
and lose poloidal flux, during magnetic compression - the low value
of parameter $\widetilde{\tau}_{c}$ of around 0.4 for shot  28426
is typical for shots taken in the 6-coil configuration. Compression
ratios are low for shots that don't conserve their poloidal flux during
compression, so the low value of $C_{B\theta ave26}$=1.7 for shot
28426 is also typical for shots taken in the 6-coil configuration.
The proportion of poloidal flux conserving shots with high magnetic
compression ratios increased dramatically with the transition to the
11-coil configuration, as indicated later in section \ref{subsec:Comparison-of-compression}. 

If the CT remains stable during compression, it expands to its pre-compression
state (apart from resistive flux losses and thermal losses) between
$t\sim t_{comp}+25\,\upmu$s and $t\sim t_{comp}+50\,\upmu$s. When
the compression current changes direction at $t\sim t_{comp}+50\,\upmu$s,
the CT poloidal field reconnects with the compression field, and a
new CT with polarity opposite to that of the previous CT is induced
in the containment region, compressed, and then allowed to expand.
The process repeats itself at each change in polarity of the compression
current until either the plasma loses too much heat, or the compression
current is sufficiently damped. As shown in chapter \ref{chap:Simulation-results},
MHD simulations model this effect while closely reproducing experimental
measurements for $B_{\theta}$, line-averaged $n_{e}$, and $T_{i}$
(from the ion-Doppler diagnostic), and Xray-phosphor imaging indicates
the compressional heating of up to three distinct plasmoids on many
compression shots. Note also that, as shown in section \ref{subsec:Simulation-of-comp_flux_loss},
MHD simulations support the idea that loss of CT poloidal flux at
compression leads to the collapse in poloidal field that is characterised
by having parameter $\widetilde{\tau}_{c}$ less than one. 

\begin{figure}[H]
\centering{}\includegraphics[scale=0.8]{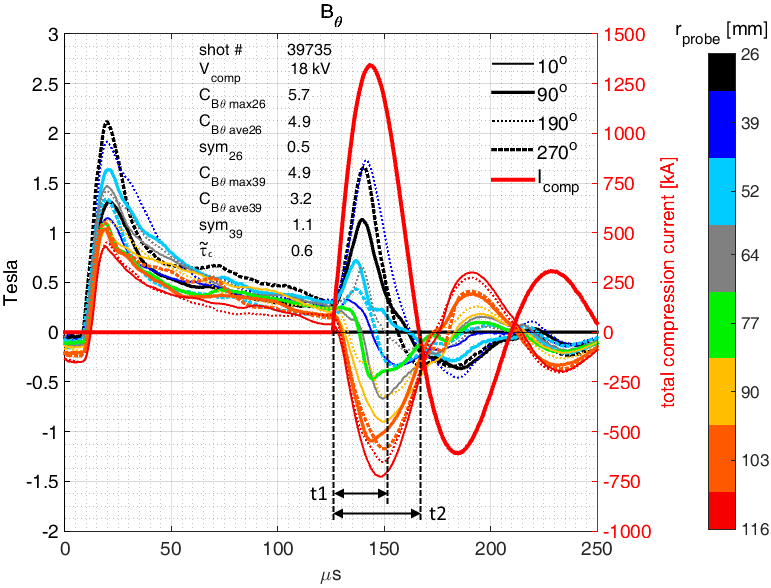}\caption{\label{fig:39735Bp}$\,\,\,\,$$B_{\theta}$ and $I_{comp}$ for shot
 39735 (11-coil configuration)}
\end{figure}
Figure \ref{fig:39735Bp} shows $B_{\theta}$ traces and total compression
current (without levitation current) for shot  39735, in the 11-coil
configuration. This is one of the few shots that were compressed late
in time, at $130\,\upmu$s. The voltage on the compression capacitors
was close to their maximum setting, at $18\mbox{ kV}$, resulting
in $\sim1.3\mbox{ MA}$ of total compression current. The values of
$C_{B\theta max26}=5.7\,\mbox{and }C{}_{B\theta ave26}=4.9$  obtained
for this shot are particularly high, partly because the CT was compressed
late when most of its poloidal flux had resistively decayed away.
Shot  39735, has parameter $\widetilde{\tau}_{c}=0.6$, $sym_{26}=0.5$
and $sym_{39}=1.1$, which classify it as a shot with quite asymmetric
compression (especially, as indicated by $sym_{39},$ in the poloidal
plane defined by $\phi=10^{\circ}--190^{\circ}$), that lost a significant
proportion of its flux. 
\begin{figure}[H]
\centering{}\subfloat[]{\includegraphics[width=8.5cm,height=5cm]{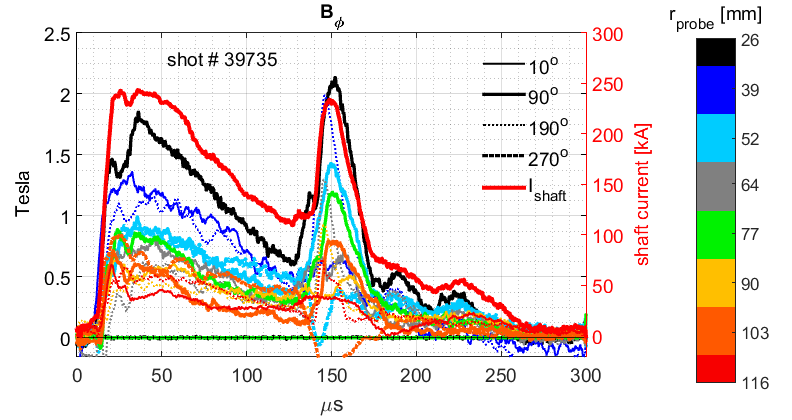}}\hfill{}\subfloat[]{\includegraphics[width=6.7cm,height=5cm]{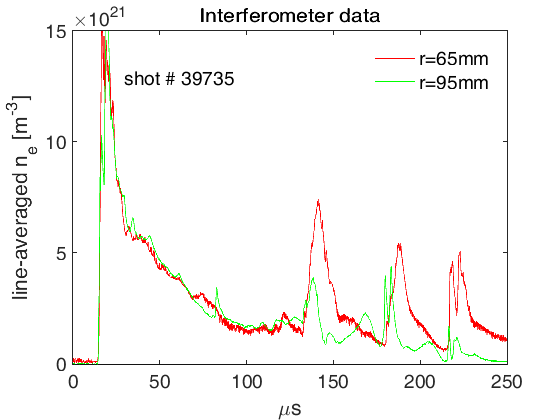}
\raggedleft{}}\caption{\label{fig:Bpand--traces}$\,\,\,\,$$B_{\phi}$ and $n_{e}$ traces
for shot  39735 (11-coil configuration)}
\end{figure}
Figure \ref{fig:Bpand--traces} shows measured toroidal field and
line-averaged electron density for shot  39735. As discussed in sections
\ref{subsec:6-coils-config_ceramic} and \ref{subsec:Compressional-Instability},
$B_{\phi}$ rises at compression as shaft current increases when it
is able to divert from the aluminum bars outside the insulating wall
to a lower inductance path through ambient plasma outside the CT.
For the 1st, 2nd and 3rd compressions, this is particularly evident
from the rise of the $r=26\mbox{ mm}$ probe signal. An obvious exception
is during the 1st compression at $\sim150\,\upmu$s, when the toroidal
field at $\phi=270^{\circ}$ drops off - this is an indication of
the compressional instability that is discussed later in section \ref{subsec:Compressional-Instability}.
The measured electron densities shown in figure \ref{fig:Bpand--traces}(b)
are line-averaged quantities obtained with He-Ne laser interferometers
looking down the vertical chords at $r=65\mbox{\mbox{ mm}}$ and $r=95\mbox{\mbox{ mm}}$
that are indicated in figure \ref{fig:Chalice}. The three distinct
density peaks correspond to the three CT compressions. Line-averaged
electron density at $r=65$ mm increased by a factor of around five
at the main compression cycle. From the time difference between the
peaks at compression of the two $n_{e}$ signals, the electron density
front at the main compression is found to move inwards at $\sim10\mbox{km/s}$.
\begin{figure}[H]
\centering{}\subfloat{\centering{}\includegraphics[width=8cm,height=5cm]{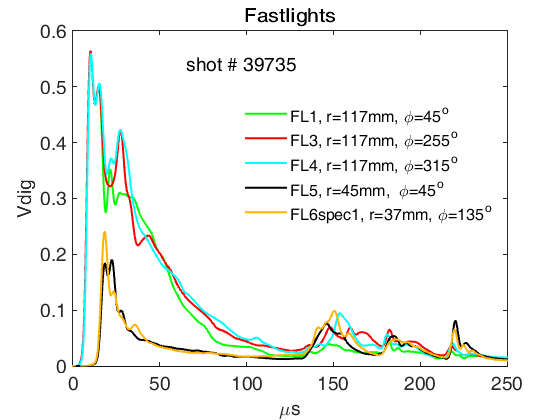}}\caption{\label{fig:FL39735}$\,\,\,\,$Optical intensity measurements for
shot  39735 (11-coil configuration)}
\end{figure}
Figure \ref{fig:FL39735} shows measured optical intensity for shot
 39735. Plasma breakdown is indicated as being toroidally symmetric.
Light intensity increases at each of the three compressions when density
and associated line radiation intensity increases. At $t_{comp}=130\,\upmu$s,
intensity at $r=117$ mm, $\phi=45^{o}$(green trace) does not rise
to the same level as at the other 117 mm chords, indicating toroidally
asymmetric compression. Ion-Doppler data is not available for this
shot. 
\begin{figure}[H]
\centering{}\includegraphics[scale=0.6]{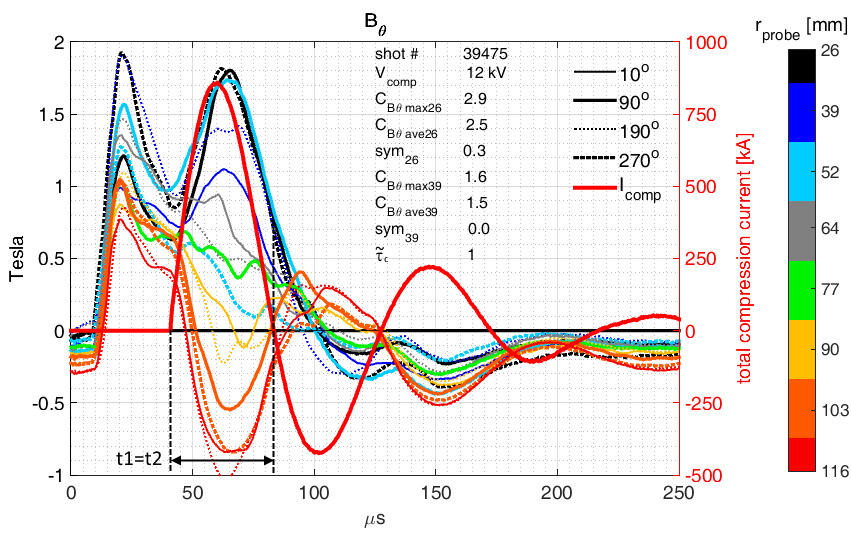}\caption{\label{fig:Bp39475}$\,\,\,\,$$B_{\theta}$ for shot  39475 (11-coil
configuration)}
\end{figure}
Figure \ref{fig:Bp39475} shows the measured $B_{\theta}$ for shot
$39475$. This shot had the more usual choice of $t_{comp}=45\,\upmu\mbox{s}$
at a moderate formation voltage of $12\mbox{ kV}$. The magnetic compression
ratios calculated are also moderate, at $C_{B\theta max26}=2.9$ and
$C_{B\theta ave26}=2.5$. With the calculated parameters $\widetilde{\tau}_{c}=1$,
$sym_{26}=0.3$ and $sym_{39}=0$, this shot conserved poloidal flux
at compression, and the compression was symmetric. As indicated in
section \ref{subsec:Comparison-of-compression}, flux conserving shots
were common only in the 11-coil configuration. Note that $B_{\theta270^{\circ}r26}>B_{\theta90^{\circ}r26}$,
$B_{\theta270^{\circ}r52}<B_{\theta90^{\circ}r52}$ , $B_{\theta190^{\circ}r39}>>B_{\theta10^{\circ}r39}$
, while $B_{\theta190^{\circ}r64}\sim B_{\theta10^{\circ}r64}$, from
formation until the start of compression, just as in shot  39735 (figure
\ref{fig:39735Bp}) - it's likely that magnetic probe calibrations
are largely responsible for this.
\begin{figure}[H]
\subfloat[]{\raggedright{}\includegraphics[width=8.5cm,height=5cm]{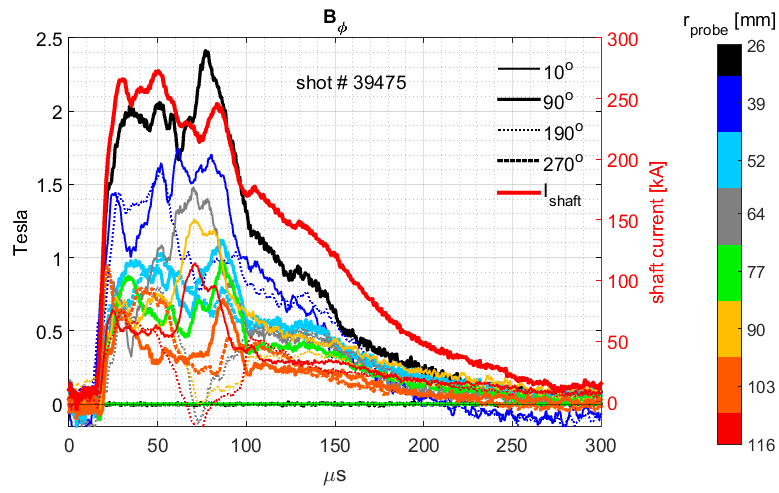}}\hfill{}\subfloat[]{\raggedleft{}\includegraphics[width=6.7cm,height=5cm]{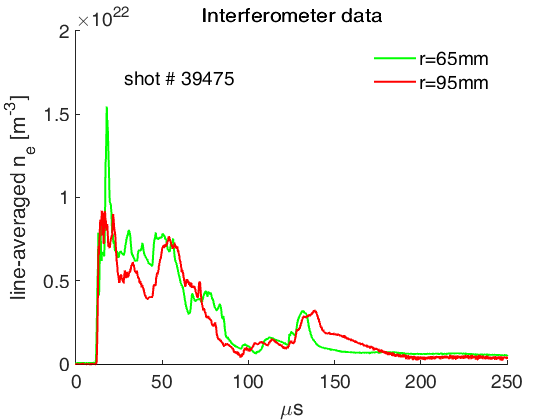}}

\caption{\label{fig:Bt_n_39475}$\,\,\,\,$$B_{\phi}$ and $n_{e}$ for shot
 39475 (11-coil configuration)}
\end{figure}
Figure \ref{fig:Bt_n_39475} shows the measured $B_{\phi}$ and line-averaged
electron density traces for shot $39475.$ While not as distinct as
for shot  39735 (figure \ref{fig:Bpand--traces}(b)), there are again
three density peaks corresponding to the three compressions. The $B_{\phi}$
signals for this shot are an excellent exemplification of the instability
that was observed on most compression shots. It can be seen how $B_{\phi}$
at all four probes at $190^{\circ}$ drops during compression, while
the field at the other toroidal angles rises. The angle at which the
signal drops varies, apparently randomly, from shot to shot, but shots
were generally quite consistent in displaying this behaviour.
\begin{figure}[H]
\subfloat[]{\raggedright{}\includegraphics[width=8.5cm,height=5cm]{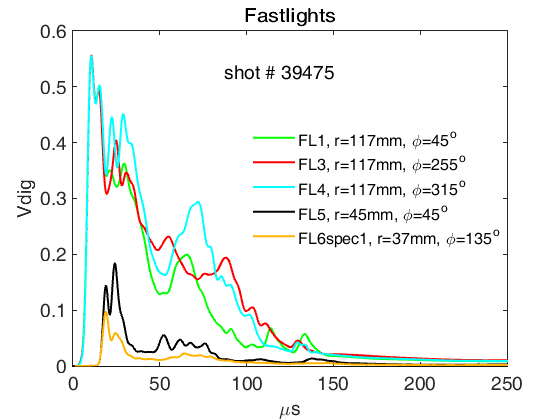}}\hfill{}\subfloat[]{\raggedleft{}\includegraphics[width=6.7cm,height=5cm]{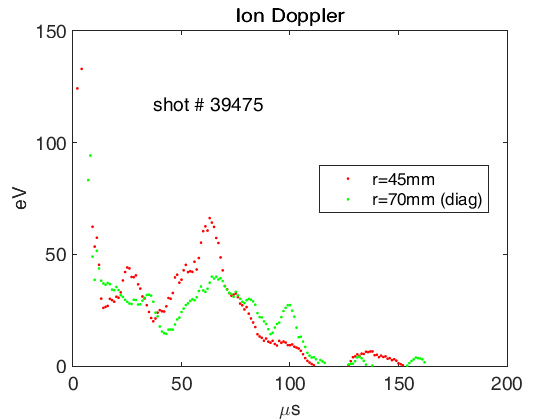}}

\caption{\label{fig:FL_ID_39475}$\,\,\,\,$Optical intensity and $T_{i}$
measurements for shot 39475 (11-coil configuration)}
\end{figure}
Figure \ref{fig:FL_ID_39475}(a) shows measured optical intensity
for shot  39475. Plasma breakdown is again indicated as being toroidally
symmetric. The light intensity increases at compression when density
and associated line radiation intensity increases. At $t_{comp}=45\,\upmu$s,
intensity at $r=117$ mm, $\phi=255^{o}$ (red trace) does not rise
to the same level as at the other 117 mm chords, indicating toroidally
asymmetric compression. Compared with data for shot 39735 (figure
\ref{fig:FL39735}), it can be seen how the factors by which optical
intensity increases at compression are relatively minor for this shot,
especially at the inner chords at $r=37$ mm and $r=45$ mm. This
is likely related to the reduced value of $V_{comp}$ for this shot.
Figure \ref{fig:FL_ID_39475}(b) shows the ion-Doppler temperature
measurement for shot 39475, where the diagnostic was focused along
the two chords indicated in figure \ref{fig:Chalice}. At moderate
$V_{comp}=12$ kV, the ion temperature increase at compression is
still significant, indicated as rising from around 20 eV to 70 eV
at the $r=45$ mm chord. 
\begin{figure}[H]
\subfloat[]{\raggedright{}\includegraphics[width=8.5cm,height=5cm]{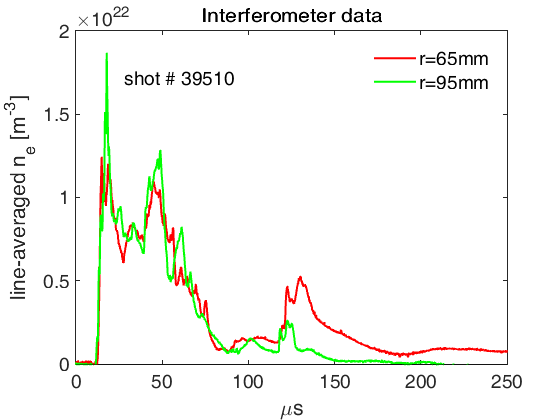}}\hfill{}\subfloat[]{\raggedleft{}\includegraphics[width=6.7cm,height=5cm]{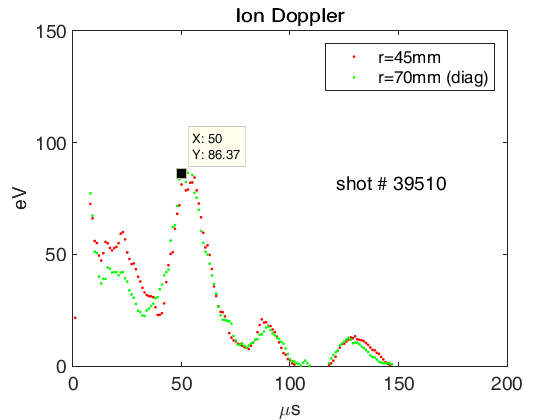}}

\subfloat[]{\raggedright{}\includegraphics[width=8.5cm,height=5cm]{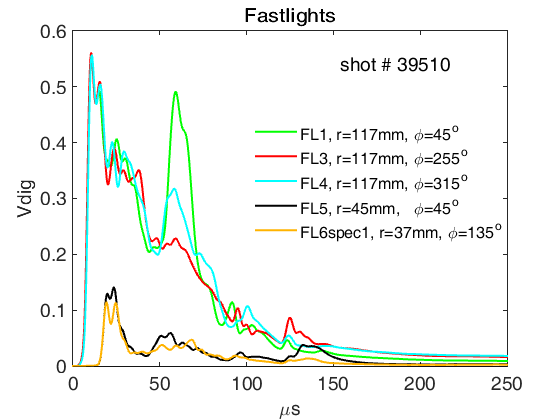}}\hfill{}\subfloat[]{\raggedleft{}\includegraphics[width=6.7cm,height=5cm]{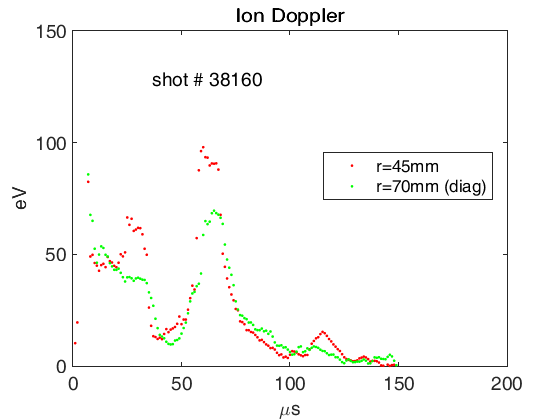}}

\caption{\label{fig:ne_ID_39510-_ID38160}$\,\,\,\,$$n_{e}$, $T_{i}$ and
optical intensity measurements for shot 39510, $T_{i}$ for shot 38160
$\,\,\,$(11-coil configuration)}
\end{figure}
Figures \ref{fig:ne_ID_39510-_ID38160}(a) and (b) show the density
and ion-Doppler temperature measurements for shot  39510, which was
also a flux conserving shot, but with $t_{comp}=40\,\upmu$s, and
increased compressional energy, with $V_{comp}=18$ kV. With increased
compressional energy, the indicated increases in both density and
ion temperature are more significant than in shot 39475. For this
shot, an increase in ion temperature by a factor of around four, from
$\sim25$ eV to $\sim100$ eV, is measured in the region of the ion
Doppler chords. A maximum error in the temperature measurement (He
II line at 468.5nm) due to density broadening has been evaluated as
$\sim$12 eV for density levels associated with shot 39510 at peak
compression \cite{Kunze,Pittman}. Figure \ref{fig:ne_ID_39510-_ID38160}(c)
shows measured optical intensity for shot  39510. Plasma breakdown
is indicated as being toroidally symmetric, but this diagnostic indicates
that compression may have been quite asymmetric. The light intensity
increases at compression when density and associated line radiation
intensity increases, particularly at $r=117$ mm, $\phi=45^{o}$ (green
trace). Figure \ref{fig:ne_ID_39510-_ID38160}(d) shows the ion-Doppler
measurement for shot  38160, which had $t_{comp}=45\,\upmu$s and
$V_{comp}=14$ kV. The density measurement for shot 38160, which indicated
typical density values, was not considered to be trustworthy - extreme
increases in optical emission at the time of compression indicated
that the density may have been underestimated. Although the ion temperature
rise at compression looks significant for this shot, careful analysis
indicated that, around peak compression, Lorentzian profiles are better
fits than Gaussian profiles to plots of photon count against photon
frequency, indicating the likelihood that density broadening rather
than temperature broadening was the dominant broadening mechanism
for shot 38160 (see discussion in section \ref{subsec:Temperature-measurements--}).
Gaussian profiles were better fits to the data for shots 39475 and
39510, so it is believed that the ion-Doppler measurements for those
shots are true indications of temperature rise at compression. Note
that ion-Doppler temperature increases at compression were significant
on the 11-coil configuration only.

\begin{figure}[H]
\subfloat[$B_{\theta}$]{

\includegraphics[width=7.5cm,height=5cm]{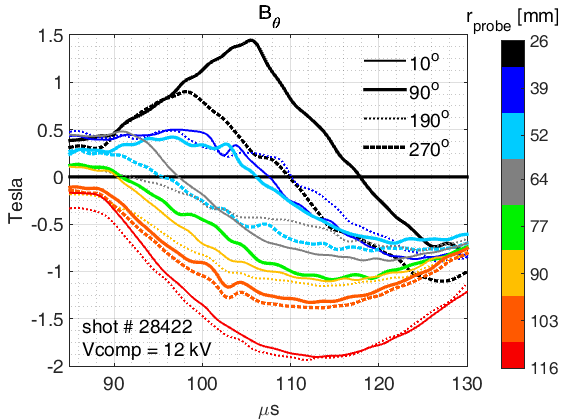}}\hfill{}\subfloat[$B_{\phi}$]{\raggedleft{}\includegraphics[width=7.5cm,height=5cm]{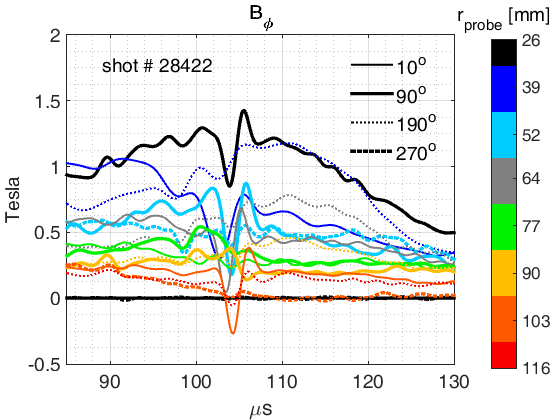}}\caption{\label{figfasctfluct28422}$\,\,\,\,$$B_{\theta}$ and $B_{\phi}$
for shot 28422 (6-coil configuration), $t_{comp}=90\,\upmu$s}
\end{figure}
Many compression shots on the 6-coils-ceramic build, some on the 6-coils-quartz
build, but apparently none on the 11-coils configuration, have high
frequency $B_{\phi}$ spikes - fluctuations up to one Tesla, with
half-periods of $\sim3$$\,\upmu\mbox{s}$, see, for example, figure
\ref{figfasctfluct28422}. These events are evident primarily on the
$B_{\phi}$ probes but are often evident on the $B_{\theta}$ probes
too. The fact that the insulation on the last build with eleven coils
was more robust to account for the proximity of the top and bottom
coil-pairs to the steel vessel, points to the likelihood that a coil-to-coil
or coil-to-aluminum-bar electrical short was involved in the events
on earlier configurations. Confirming this, the few shots for which
there is a clear indication of a coil short on the X-ray-phosphor
images do display this type of behaviour on their $B_{\phi}$ traces.
An external poloidal arc would induce a $B_{\phi}$ fluctuation, and
if there was a toroidal component to the arc, then also a fluctuation
in $B_{\theta}$. On the other hand, we expect the plasma to adjust
itself to preserve its pressure-balanced structure, so toroidal magnetic
perturbations should result in poloidal ones too.

\section{Compressional instability\label{subsec:Compressional-Instability}}

\begin{figure}[H]
\centering{}\includegraphics[scale=0.6]{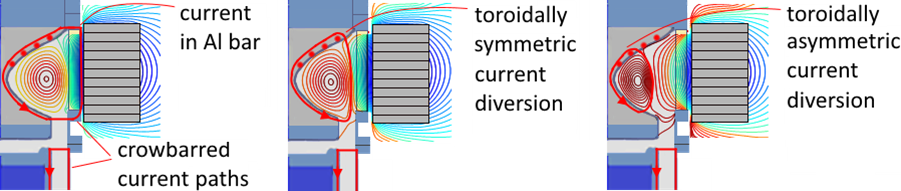}\caption{\label{fig:divertedcurrent}$\,\,\,\,$Asymmetric current diversion}
\end{figure}
Figure \ref{fig:divertedcurrent} shows a possible explanation for
the compressional instability that was routinely observed. After the
$50\,\upmu$s formation capacitor-driven pulse, toroidal-flux conserving
crowbarred current continues to flow, primarily along two separate
paths as indicated. In addition, it is likely that there is a third
current path, consisting of the merger of the two paths indicated
here. Referring to the upper path, initially most of the outboard
part of this current is in the aluminum bars depicted in figure \ref{fig:Schematic-of-6}(a).
Shaft current, and $B_{\phi}$ at probes, rise at compression as the
current path shifts symmetrically to a lower inductance path (central
subfigure); now the outboard part of the current loop travels through
the ambient plasma outboard of the CT. The asymmetric current diversion
depicted in the right subfigure will be discussed after outlining
how the symmetric shifting current path mechanism is reproduced in
MHD simulations: 
\begin{figure}[H]
\subfloat[]{\raggedright{}\includegraphics[width=6.7cm,height=5cm]{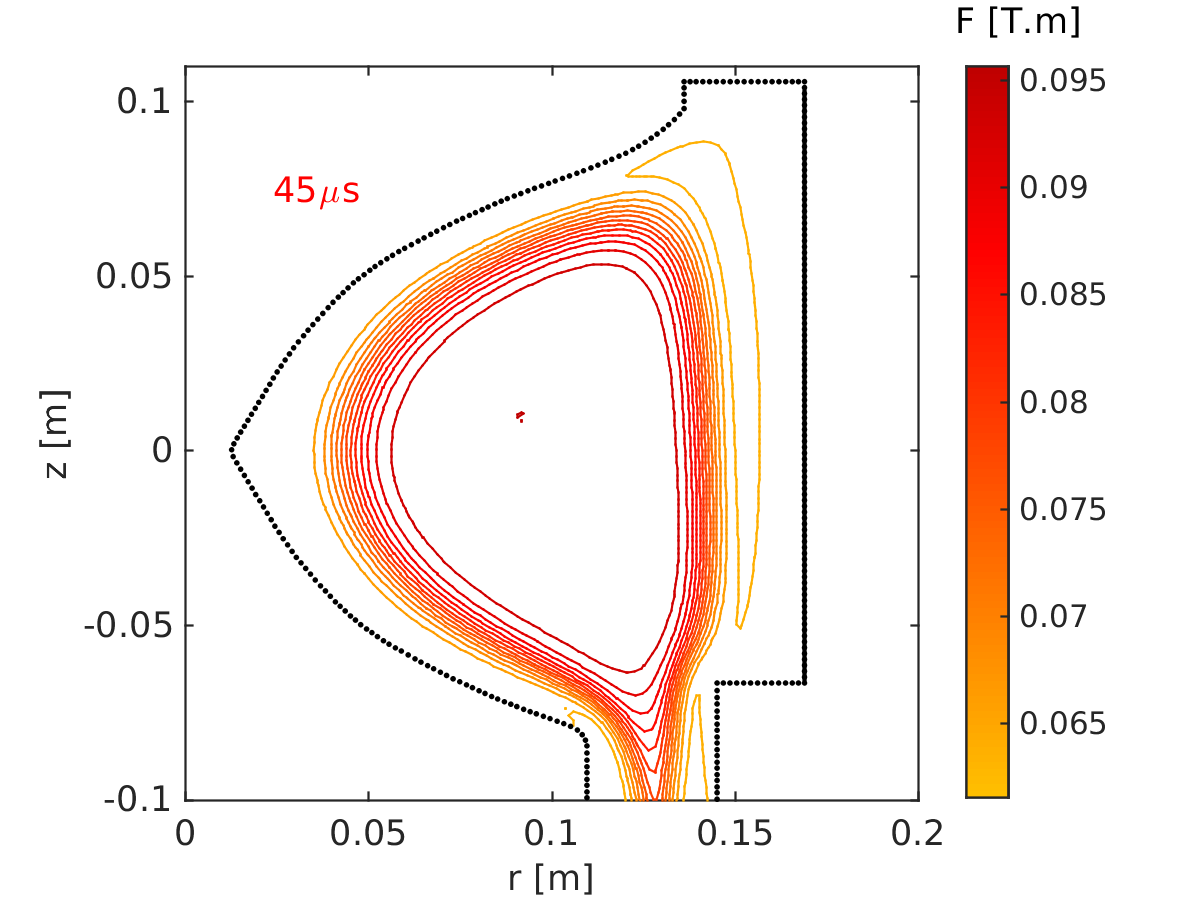}}\hfill{}\subfloat[]{\raggedleft{}\includegraphics[width=6.7cm,height=5cm]{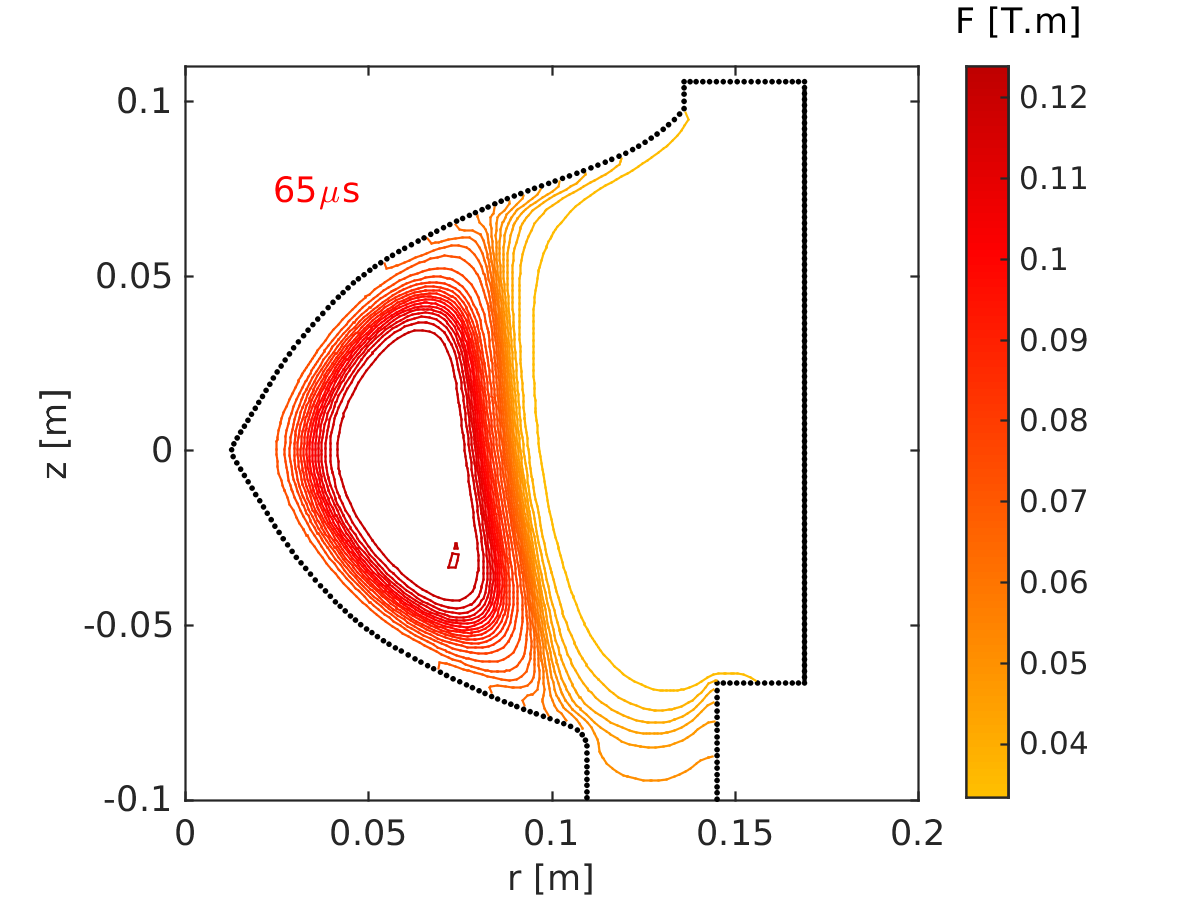}}

\caption{\label{fig:CompInstab_F-1}$\,\,\,\,$$f$ contours at magnetic compression,
simulation  1968}
\end{figure}
Figure \ref{fig:CompInstab_F-1} shows contours of $f$ at $45\,\upmu$s
just prior to magnetic compression, and at $65\,\upmu$s, at peak
magnetic compression. Contours of $f=rB_{\phi}$ represent paths of
poloidal current. Closely spaced contours indicate regions of high
gradients of $f$, which in turn are regions of high currents. The
MHD equations implemented to the code are formulated such that the
code has various conservation properties (see section \ref{subsec:Conservation-properties}),
including conservation of toroidal flux. It can be seen in figure
\ref{fig:CompInstab_F-1}(b) how the imposition of toroidal flux conservation
leads to the induction, at magnetic compression, of poloidal currents
flowing from wall-to-wall through the ambient plasma just external
to the outboard boundary of the CT.

If some mechanism causes the CT to be compressed more at a particular
toroidal angle (an effect which the axisymmetric MHD code cannot reproduce),
the inductance of the current path at that angle will be reduced further
and more current will flow there (right subfigure in figure \ref{fig:divertedcurrent}),
enhancing the instability. This is analogous to the mechanisms behind
external kink and toroidal sausage type instabilities. As the current
path moves inwards past the probes at a particular toroidal angle,
$B_{\phi}$ at the probes will change polarity at that toroidal angle,
as is observed on most compression shots ($e.g.,$ figures \ref{fig:Bpand--traces}(a)
and \ref{fig:Bt_n_39475}(a)). As the CT decompresses as $I_{comp}$
decreases, the current path returns towards its pre-compression path.
It is noteworthy that although the magnetically compressed CTs generally
exhibit this instability, there is a noticeable correlation in that
the compression shots that have a high value of $\widetilde{\tau}_{c}$
($i.e.,$ apparent flux conservation during compression) seem to exhibit
the clearest manifestation of the instability, through the behaviour
of the $B_{\phi}$ signals - shot $39475$ (figure \ref{fig:Bt_n_39475}(a))
is a good example of this. As mentioned earlier, even levitated, but
non-compressed shots, exhibited this behaviour to some degree ($e.g.,$
figure \ref{fig:Poloidal-field-for}(c)), in cases where the levitation
currents were not optimised to decay at near the rate of the CT currents. 

\section{Comparison of compression parameters between configurations\label{subsec:Comparison-of-compression}}

The impedances (effective resistance to alternating current, due to
combined effects of ohmic resistance and reactance) of the coils arrays
are slightly different for the 6-coil and 11-coil configurations,
leading to a few percent variation in peak compression current at
the same $V_{comp}$. At $V_{comp}=18$ kV, measured peak $I_{comp}$
was $\sim200\mbox{\mbox{ kA}}$ per coil in the 6-coil configuration,
compared with $\sim210\mbox{\mbox{ kA}}$ per coil-pair (and $\sim210\mbox{\mbox{ kA}}$
in the single coil 3rd from the bottom of the stack) in the 11-coil
configuration. At the same $V_{comp}$, compressional flux, as estimated
from $\mbox{FEMM}$ model outputs, is around $1.7$ times higher in
the 6-coil configuration, relative to the 11-coil configuration, due
to the large gaps (see figure \ref{fig:Schematic-of-6}(b) $cf.$
figure \ref{fig:11-coil-configuration}(b)), above and below the coil
stack, that ease entry of compressional flux into the containment
region for the 6-coil configuration. 

In general, as external compressional flux is increased, the level
of CT flux conservation at compression is reduced, while, for the
same level of flux conservation, magnetic compression ratios are increased.
A fair comparison of compression performance metrics across builds
can be obtained by comparing the metrics for shots with the same level
of external compressional flux. For around the same external compressional
flux, we would ideally compare shots with $V_{comp}=$14 kV  in the
11-coil configuration against shots with $V_{comp}=$8.2 kV  (which
were not taken) in the 6-coil configuration, but a reasonable comparison
of compression parameter trending can be made looking at shots with
$V_{comp}=14\mbox{ kV}$ for the 11-coil configuration, and $V_{comp}=7\mbox{ kV}\rightarrow9\mbox{ kV}$
in the 6-coil configuration. 
\begin{figure}[H]
\centering{}\includegraphics[scale=0.35]{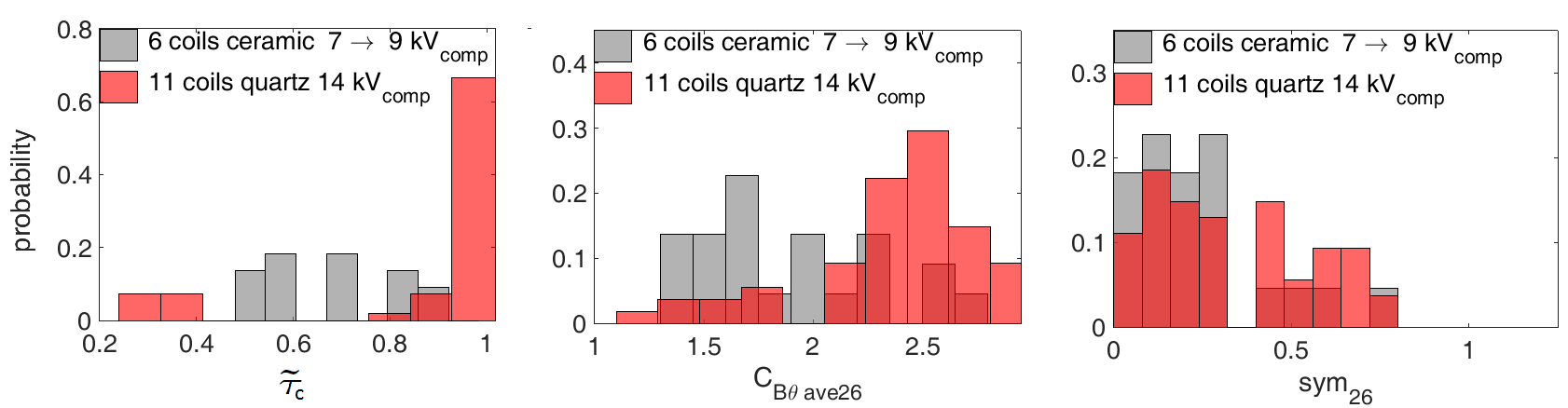}\caption{\label{fig:Comp_comparsion}$\,\,\,\,$Comparison of compression parameters}
\end{figure}
Figure \ref{fig:Comp_comparsion} shows normalised histograms of key
compression parameters that were defined below figure \ref{fig:Bp_28426-1}.
The recurrence rate of shots that conserved CT flux at compression
was significantly improved in the 11-coil configuration. Around $70\%$
of shots had good CT flux conservation ($i.e.,\,\widetilde{\tau}_{c}\sim1$)
in the 11-coil configuration, while only $\sim10\%$ of shots conserved
$\sim80\%$ of flux ($i.e.,\,\widetilde{\tau}_{c}\sim0.8$) in the
6-coil configuration. Poloidal magnetic compression ratios (characterised
by $C_{B\theta ave26}$) would be expected to be low when CT flux
is lost, and it can be seen how the ratios are nearly doubled on average
in the 11-coil configuration. Compression asymmetry (characterised
by $sym_{26}$) remains poor in both configurations. While reduced
plasma wall interaction at formation and consequent reduced impurity
radiation cooling in the 11-coil configuration was certainly behind
the huge improvement in lifetimes of levitated CTs (figures \ref{fig:Effect-of-lithium}
and \ref{fig:-for-four4configs}), it seems likely, but can't be confirmed
without further experiment or 3D simulation, that a different mechanism
was responsible for the orders of magnitude improvement in the rate
of shots with good CT flux conservation at compression. Supporting
this, shots taken in the 11-coil configuration with compression fired
late when plasma has had time for significant diffusive cooling ($e.g.,$
figure \ref{fig:39735Bp}) generally conserved \emph{more} flux than
those fired early in time in the 6-coil configuration ($e.g.,$ figure
\ref{fig:Bp_28426-1}). The improvement is likely to be largely due
to the compression field profile itself, which led to more uniform
outboard compression, as opposed to largely equatorial outboard compression
with the six coil configuration. Equatorially-focused outboard compression
may have caused the CT to bulge outwards and upwards/downwards above
and below the equator, leading to poloidal field reconnection, CT
depressurisation and possible disruption as a consequence.

\section{Experimentally measured CT outboard separatrix radius for compression
shots\label{subsec:Rsep_comp}}

Using the method outlined in section \ref{sec:rsep}, it is possible
to determine the CT separatrix at the equator ($z\sim0\mbox{\mbox{ mm}}$)
for compression shots. 
\begin{figure}[H]
\subfloat[averaged $B_{z}$ (from side probes), shot 39738]{\raggedright{}\includegraphics[width=8cm,height=5cm]{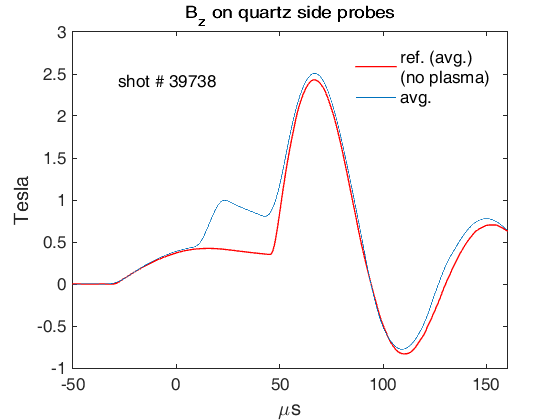}}\hfill{}\subfloat[Measured $r_{s}$ for shot 39738 ]{\raggedleft{}\includegraphics[width=8cm,height=5cm]{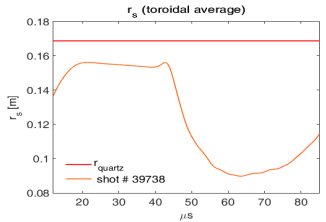}}

\caption{$\,\,\,\,$Measured CT separatrix radius for compression shot 39738\label{fig:rsepCOMP39738} }
\end{figure}
Figure \ref{fig:rsepCOMP39738}(a) shows the averages of the eight
$B_{\theta}(\phi,\,t)$ and $B_{zref}(\phi,\,t)$ signals for shot
$39738$, which had $V_{comp}=18$ kV, $|t_{lev}|=30\,\upmu$s, $t_{comp}=45\,\upmu$s,
and was taken in the 11-coil configuration with 70 m$\Omega$ cables
in place in the levitation circuits, so that the functional fit indicated
in figure \ref{Bz_rsep39650_1}(b) was used to find $r_{s}(t)$, which
is shown in figure \ref{fig:rsepCOMP39738}(b). For compression shots,
it is convenient to find the toroidally averaged $r_{s}$ using toroidally
averaged probe data. As seen in figure \ref{fig:rsepCOMP39738}(a),
at compression, the reference $B_{z}$ is very close to $B_{z}$ when
the CT is present, so that errors in probe signal response can lead
to instances when $B_{zref}(\phi,\,t)>B_{z}(\phi,\,t)$, and consequent
complex-valued $r_{s}(\phi,\,t)$ solutions (see section \ref{subsec:Using-side-probe-data}).
Using the toroidally averaged signals reduces the likelihood of this
error. For shot 39738, a radial compression factor, in terms of equatorial
outboard CT separatrix, of $C_{s}=1.7$ is indicated. Note that $r_{s}\sim9$
cm at peak compression. As outlined in section \ref{subsec:Rsep_lev_70mOhm},
when $r_{s}\lesssim9$ cm, the slope of the functional fit in \ref{Bz_rsep39650_1}(b)
is too flat to be successfully inverted with good accuracy. For this
reason, $C_{s}$ cannot be evaluated if the CT is compressed more
than in shot  39378. An example of a shot in which compression is
too strong for successful evaluation of $C_{s}$ is shot  39735 (figure
\ref{fig:39735Bp}) which also has $V_{comp}=18$ kV, but is compressed
later ($t_{comp}=130\,\upmu$s), when pre-compression CT flux has
decayed to lower levels and therefore compression is more extreme.

\section{Scintillator data\label{subsec:Scintillator-data}}

\begin{figure}[H]
\centering{}\includegraphics[scale=0.6]{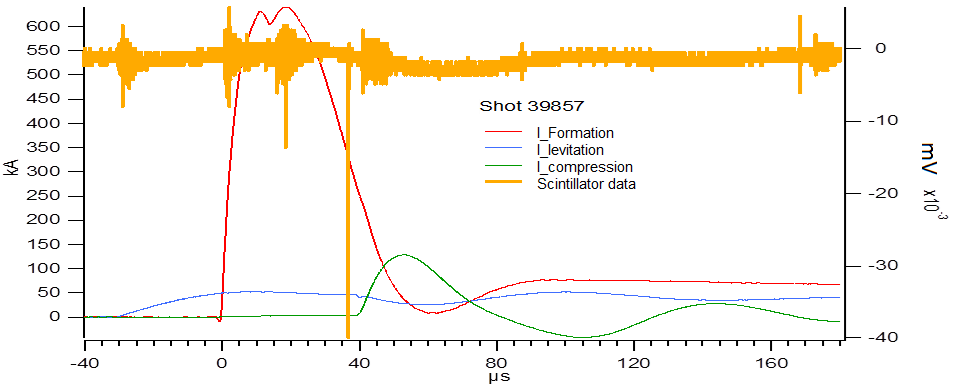}\caption{\label{fig:Scintillator-data-for}$\,\,\,\,$Scintillator data for
shot 39857}
\end{figure}
\begin{figure}[H]
\subfloat[]{\raggedright{}\includegraphics[scale=0.5]{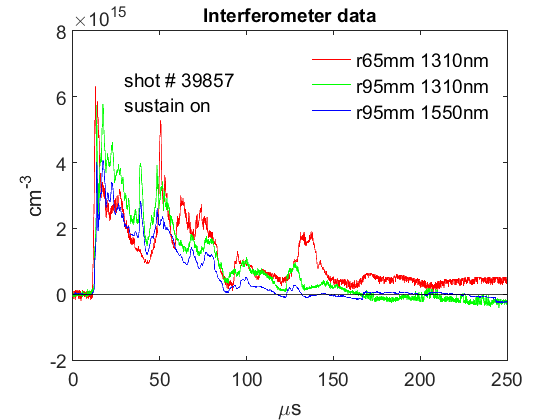}}\hfill{}\subfloat[]{\raggedleft{}\includegraphics[scale=0.5]{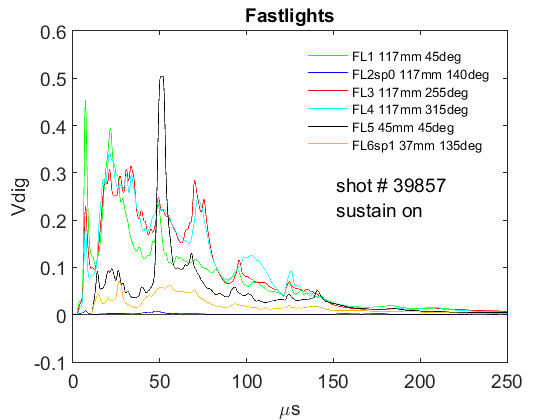}}\caption{\label{n_FL_39857}$\,\,\,\,$$n_{e}$ and optical data for shot 39857}
\end{figure}

\begin{itemize}
\item Scintillator spikes were observed at the very beginning of magnetic
compression on three of the twenty shots taken with deuterium on the
last night of the period during which the 11-coil configuration was
tested. The spikes coincide with the initial rises towards elevated
peaks observed on the density and optical emission signals. Note scintillator
data should be shifted $3\,\upmu$s forward in figure \ref{fig:Scintillator-data-for}.
Shot 39857 was one of the three shots on which this behaviour was
observed. 
\item Scintillator spikes were seen on two of the twenty deuterium shots,
at the time when plasma entered the CT containment region. This is
most likely related to viscous heating of ions during CT formation.
\end{itemize}

\section{Summary\label{sec:SummaryCOMP}}

The recurrence rate of shots in which the CT poloidal flux was conserved
during magnetic compression is an indication of resilience against
a disruption-inducing instability during compression, and was increased
from around 10\% to 70\% with the transition to the levitation/compression
field profile of the eleven-coil configuration. The improvement is
likely to be largely due to the compression field profile itself,
which led to more uniform outboard compression, as opposed to the
largely equatorial outboard compression associated with the six coil
configuration. The effect of having a reduced impurity concentration
and increased CT plasma temperature prior to compression initiation,
as a consequence of the improved levitation field profile, may also
have played a role. Due to improved flux conservation at compression,
magnetic compression ratios increased significantly with the eleven
coil configuration. Magnetic compression usually did not exhibit good
toroidal symmetry. 

In the eleven coil configuration, poloidal field at the CT edge, at
fixed $r=26$ mm, increased by a factor of up to six at compression,
while line averaged electron density at fixed $r=65$ mm was observed
to increase by 400\%, with the electron density front moving inwards
at up to 10 km/s. Ion Doppler measurements, at fixed $r=45$ mm indicated
ion temperature increases at magnetic compression by a factor of up
to four. Increases in poloidal field, density, and ion temperature
at compression were significant only in the eleven coil configuration.
Experimentally measured CT outer equatorial separatrix indicated a
radial compression factor of up to $1.7$, which is close to the maximum
compression factor that can be resolved by the diagnostic. It may
be that larger compression factors, in terms of CT outer equatorial
separatrix, which cannot be evaluated, were achieved with shots when
compression was fired later in time. Ideally it would be good to have
diagnostics to get information about internal CT characteristics.
Internal magnetic probes would be difficult to implement without interference
and plasma contamination and due to the small size of the CT containment
region, but non-invasive diagnostics such as a Thomson scattering
system to assess internal electron temperature and density profiles
would be useful and practical. 

Indications of an instability, thought to be an external kink, occurred
very frequently during magnetic compression and during under-damped
magnetic levitation. Levitation circuit modification to match the
decay rates of the levitation and plasma currents led to more stable,
longer lived plasmas, and a greatly increased rate of good shots,
by avoiding unintentional magnetic compression during CT levitation.
An obvious improvement to the experiment design would be to drive
additional shaft current and raise the $q$ profile to MHD stable
regimes. 

\newpage{}

\part*{PART 2: SIMULATIONS\label{part:Simulations}\addcontentsline{toc}{part}{PART 2: SIMULATIONS} }

\chapter{Core code development and MHD model\label{chap:Core-Code-Development}}

This chapter is focused on the core aspects of the DELiTE (Differential
Equations on Linear Triangular Elements) framework that was developed
for spatial discretisation of partial differential equations on an
unstructured triangular grid in axisymmetric geometry, and applied
to study the magnetic compression experiment. The framework, which
is implemented in MATLAB, is based on discrete differential operators
in matrix form, which are derived using linear finite elements and
mimic some of the properties of their continuous counterparts. A single-fluid
two-temperature MHD code is implemented in this framework. The inherent
properties of the operators are used to ensure global conservation
of energy, particle count, toroidal flux, and angular momentum. 

The concept of conservation schemes and the importance of conservation
properties in numerical methods is the focus of section \ref{sec:Conservation-schemes}.
An overview of the finite element method used is presented in section
\ref{subsec:Finite-element-method}. Definitions of key notations
used, and descriptions of mappings between elemental and nodal quantities,
are presented in section \ref{sec:Key-notations-and}. In section
\ref{sec:Differential-operators}, the finite element method is applied
to derive various first and second order derivative matrix operators
that mimic some of the properties of their continuous counterparts. 

In section \ref{sec:Axisymmetric-two-temperature-MHD}, expressions
for $\dot{\psi}$ and $\dot{f}$ are developed. The particular form
of the expression for $\dot{f}$ leads, together with properties of
the discrete operators, to conservation of toroidal flux for the discretised
system. The continuous forms of the equations defining the axisymmetric
two-temperature MHD model are presented and used to demonstrate various
conservation properties that the discrete model will mimic. 

Note that in appendix \ref{sec:Formulation-of-discretized}, the derivative
matrix operators, discrete forms of the mass and energy conservation
equations, and the expressions for $\dot{\psi}$ and $\dot{f}$, are
used to develop a discrete form of the momentum equation which, using
the special properties of the differential operators, ensures that
total energy of the system described by the complete set of discretised
MHD equations is conserved with appropriate boundary conditions. In
section \ref{sec:Discrete-form_cons_props}, the full set of discretised
MHD equations is presented, along with the closure models, including
the model for anisotropic thermal diffusion. Global conservation of
mass, toroidal flux, angular momentum, and energy for the system of
equations in discrete form is demonstrated. A summary in section \ref{sec:SummaryCoreCodeDev}
concludes the chapter.

\section{Conservation schemes\label{sec:Conservation-schemes}}

Ideally, a numerical discretisation method should, as well as accurately
representing the continuous form of the mathematical equations that
describe a particular physical system, reproduce the physical properties
of the system being modelled. In practice, such properties are often
expressed as conservation laws, and their maintenance can be just
as important as standard numerical method assessment gauges like convergence,
stability, accuracy, and range of applicability. Numerical solutions
that contradict basic physical principles by, for example, destroying
mass or energy are inherently unreliable when applied to novel physical
regimes. On the other hand, many numerical methods do not have strong
conservation properties, or conserve only some naturally conserved
quantities, but can still regularly produce informative results that
have a resemblance to real-world observations. However, as more complicated
physical systems are modelled, physically incorrect numerical solutions
can go unnoticed. This is especially true in simulating complex physical
models, such as fluid dynamics or magnetohydrodynamics (MHD), where
the maintenance of conservation laws puts more constraints on numerical
schemes and, thus, helps to avoid spurious solutions. It is best to
deal with the physical fidelity of the model at the numerical method
design level \cite{Perot2}, and try to replicate, in the discretised
form, as many of the conservation laws inherent to the original physical
system as possible. 

Numerical methods with discrete conservation properties are well known
in computational fluid dynamics (for example, \cite{Perot2,Caramana,Margolin,Zhang}),
and in computational MHD \cite{Derigs,Liu,Franck}. In this chapter,
a novel numerical scheme for a two-dimensional (axisymmetric) compressible
MHD system is presented. The scheme is based on a continuous Galerkin
linear finite element method on an unstructured triangular mesh and
by construction has global (for the whole domain) conservation of
mass, energy, toroidal flux and angular momentum. A novelty of the
code is that all discrete spatial differential operators are represented
as matrices, and the discretized MHD equations are obtained by simply
replacing the original continuous differentiations with the corresponding
matrix operators. 

Note that by global conservation of a quantity in our numerical method,
we imply that there is a discretised analogue of the continuity equation
for that quantity, and, when integrated over the volume, its fluxes
are completely cancelled in the interior of the domain, even though
the explicit form of these fluxes are not always given. As shown in
\cite{Hughes} and \cite{Perot2}, global conservation for a method
with local support ($i.e.,$ local stencil) also implies local conservation.
To enable the development of a numerical formulation with the aforementioned
conservation properties, the discrete differential operators must
obey a property equivalent, for a scalar field $u(\mathbf{r})$ and
a vector field $\mathbf{p}(\mathbf{r})$, to 
\begin{equation}
\int u\nabla\cdot\mathbf{p}\,dV+\int\mathbf{p}\cdot\nabla u\,dV=\int u\mathbf{p}\cdot d\boldsymbol{\Gamma}\label{eq:500.0}
\end{equation}
so that the discrete forms of the differential product rule and divergence
theorem are satisfied \cite{Perot}. This is the essence of mimetic
schemes, also known as support operator methods, to which this work
is closely related \cite{Perot2,Shashkov}.

\section{Finite element method overview\label{subsec:Finite-element-method}}

Physical laws are usually expressed in terms of partial differential
equations, for which there is often no analytical solution. Various
discretization methods solve for an approximation of the equations.
In the finite element method, the solution domain is divided into
finite element subdomains, and an approximate solution is developed
over each element. The element-specific solutions are assembled to
obtain the approximate overall domain solution. Continuous functions
$u(\mathbf{r})$ are approximated by piecewise continuous functions
$U(\mathbf{r})$ using a combination of basis functions, as 
\begin{equation}
u(\mathbf{r})\approx U(\mathbf{r})=\overset{N}{\underset{n=1}{\Sigma}}U_{n}\,\phi_{n}(\mathbf{r})\label{eq:500}
\end{equation}
where $\phi_{n}(\mathbf{r})$ is the basis function associated with
node $n$, and $U_{n}$ constitute a set of $N$ scalar coefficients.
Each finite element is associated with a number of nodes, located
at discrete points on the solution domain, at which the solution to
$U(\mathbf{r})$ will be evaluated. The basis functions in the finite
element method depend on the dimensionality of the problem, and the
order of the approximation sought. A principle of the finite element
method is to require the approximate solution $U(\mathbf{r})$ to
satisfy the differential equation in a weighted-integral sense. For
example, in one dimension, if the partial differential equation is
$-\frac{\partial}{\partial x}\left(x\frac{\partial u(\mathbf{r})}{\partial x}\right)+u(\mathbf{r})=0$,
the requirement for the approximate solution can be written as 
\begin{eqnarray}
\int_{0}^{L}wR\,dx & = & 0\label{eq:501}
\end{eqnarray}
where the integral is over the solution domain spanning from $x=L$
to $x=0$, $w$ is a weight function, and the residual is $R=-\frac{\partial}{\partial x}\left(x\frac{\partial U(\mathbf{r})}{\partial x}\right)+U(\mathbf{r})$.
A linearly independent equation is obtained for each independent function
$w$ - in this way, sufficient equations can be generated in order
to solve for the unknowns coefficients $U_{n}$. The method of using
the basis functions as weight functions is known as the Galerkin method.
In the example here, the basis function must be twice differentiable
because the partial differential equation is second order. To weaken
the basis function continuity requirements, integration by parts is
used to remove the second differentiation on $U(\mathbf{r})$ in equation
\ref{eq:501}, and apply it to $w$. This method reduces equation
\ref{eq:501} to the so-called \textquotedbl weak formulation\textquotedbl :
\\
\begin{eqnarray}
\int_{0}^{L}\left(\frac{\partial w}{\partial x}\left(x\,\frac{\partial U(\mathbf{r})}{\partial x}\right)+w\,U(\mathbf{r})\right)\,dx-\left[w\,x\,\frac{\partial U(\mathbf{r})}{\partial x}\right]_{0}^{L} & = & 0\label{eq:502}
\end{eqnarray}
In this work, we develop linear approximations to $U(\mathbf{r})$.
To develop the finite element discretisation, we drew inspiration
from material presented in \cite{PICwebsite}, which in turn, is based
on material in \cite{Strang}, in which a finite element method is
used to solve Laplace and Poisson equations in two dimensions. The
two-dimensional computational domain, with azimuthal symmetry in cylindrical
coordinates, is represented by an unstructured triangular mesh. The
freely-available mesh generator DISTMESH \cite{MeshPersson1,MeshPersson2,MeshPersson3}
was adapted to provide the computational grid - see appendix \ref{subsec:Computational-grid}
for details on the mesh generation. Nodes are located at triangle
vertices. In the linear finite element method, any continuous function
$u(\mathbf{r})$ is approximated as a piecewise continuous function
$U(\mathbf{r})$ that is linear across each triangular element:
\begin{equation}
u(\mathbf{r})\approx U(\mathbf{r})=\overset{N_{e}}{\underset{e=1}{\Sigma}}\,U^{e}(\mathbf{r})\label{eq:500.01}
\end{equation}
where $N_{e}$ is the number of elements, and $U^{e}(\mathbf{r})$
represents $U(\mathbf{r})$ within element $e$: 
\begin{equation}
U^{e}(\textbf{\ensuremath{\mathbf{r}}})=A^{e}+B^{e}r+C^{e}z\label{eq:500.02}
\end{equation}
Here, $A^{e},\,B^{e}$ and $C^{e}$ are constants that are specific
to element $e$. These coefficients will be derived in section \ref{subsec:Dre Dze}.
Equivalent to equation \ref{eq:500.01}, and analogous to equation
\ref{eq:500} for the linear approximation method, $U(\mathbf{r})$
may be defined using a combination of basis functions as 
\begin{equation}
u(\mathbf{r})\approx U(\mathbf{r})=\overset{N_{n}}{\underset{n=1}{\Sigma}}\,U_{n}\,\phi_{n}(\mathbf{r})\label{eq:500.03}
\end{equation}
where $N_{n}$ is the number of nodes, $U_{n}$ represents the values
of $U(\mathbf{r})$ at node $n$, and $\phi_{n}(\mathbf{r})\equiv\phi_{n}(r,\,z)$
is the basis function associated with node $n$. 
\begin{figure}[H]
\centering{}\includegraphics[width=9.2cm,height=7cm]{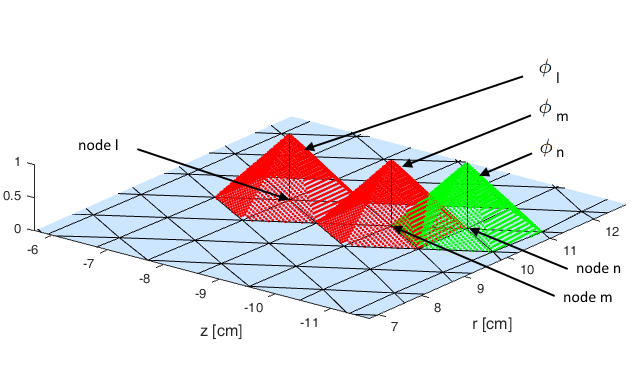}\caption{\label{fig:Linear-basis-function}$\,\,\,\,$Linear basis function
depiction for triangular elements}
\end{figure}
The linear basis functions have the forms of pyramids, as depicted
in figure \ref{fig:Linear-basis-function}, which indicates a portion
of a computation mesh, with depictions of the basis functions for
example nodes $l,\,m$ and $n$. Each node $n$ is associated with
a pyramid function $\phi_{n}$, which has an elevation of one above
the node, and falls linearly to zero at the immediately surrounding
nodes. Each pyramid function $\phi_{n}$ has $e_{n}$ sides, where
$e_{n}$ is the node-specific number of triangular elements surrounding
node $n$. Note that the pyramid shapes associated with basis functions
for nodes on the domain boundary have, in addition, one or more vertical
sides at the boundaries. The remaining individual pyramid-side functions
are defined by sections of planes that are tilted relative to the
$r-z$ plane. Thus, each node is associated with $e_{n}$ tilted planes.
In total, there are $K$ tilted planes defined in the solution domain,
where $K=\overset{N_{n}}{\underset{n=1}{\Sigma}}e_{n}=3N_{e}$. Each
triangular element is associated with three tilted planes. The tilted
planes are truncated to constitute pyramid sides by defining the functions
representing the pyramid sides as 
\begin{equation}
\psi_{n}^{e}(\textbf{\ensuremath{\mathbf{r}}})=a_{n}^{e}+b_{n}^{e}r+c_{n}^{e}z\label{eq:502.2}
\end{equation}
with the additional truncating property that $\psi_{n}^{e}(\textbf{\ensuremath{\mathbf{r}}})=0$
for all points located at $\mathbf{r}$ which lie outside the triangular
element associated with $\psi_{n}^{e}$. The notation $\psi_{n}^{e}$
indicates that each pyramid side function is associated with a particular
node $n$, \emph{and} with a particular triangular element $e$. The
coefficients $a_{n}^{e},\,b_{n}^{e}$, and $c_{n}^{e}$ are also associated
with a particular node and element, and are such that $\psi_{n}^{e}=1$
at its associated node $n$, and $\psi_{n}^{e}=0$ at the other two
nodes in the triangular element $e$. In summary, the basis functions
$\phi_{n}(\textbf{\ensuremath{\mathbf{r}}})$, which define pyramid
shapes with a peak elevation of one at node $n$, can be expressed
as 
\begin{equation}
\phi_{n}(\textbf{\ensuremath{\mathbf{r}}})=\overset{e_{n}}{\underset{}{\Sigma}}\,\psi_{n}^{e}(\textbf{\ensuremath{\mathbf{r}}})\label{eq:502.3}
\end{equation}
where the summation is over the pyramid side functions associated
with node $n$, each of which is non-zero only over its associated
triangular element. The basis functions have the property that $\phi_{n}(r_{j},\,z_{j})=\delta_{nj}$,
where $\delta_{nj}$ is the Kronecker delta, and $(r_{j},\,z_{j})$
are the coordinates of node $j$. Noting that the volume of a pyramid
is given by 
\begin{equation}
V_{pyramid}=sH/3\label{eq:502.32}
\end{equation}
where $s$ is the pyramid base area and $H$ is the pyramid height,
this leads naturally, for any continuous function $U(\mathbf{r})$
(including the piecewise linear approximation), to the property 
\begin{equation}
\int\phi_{n}(\mathbf{r})\,U(\mathbf{r})\,dr\,dz=(U_{n}+\mathcal{O}(h_{e}))\,\int\phi_{n}(\mathbf{r})\,dr\,dz\approx\frac{U_{n}\,s_{n}}{3}\label{eq:502.4}
\end{equation}
where $U_{n}=U(r_{n},z_{n})$, $s_{n}$ is the support area of node
$n$ (area of the base of the pyramid function defined by $\phi_{n}(\mathbf{r})$),
pyramid height $H=1,$ and $h_{e}$ is the typical element size. This
identity is analogous to the integral property of the Dirac delta
function. In deriving the property, it is assumed that the function
$U(\mathbf{r})$ is sufficiently smooth that it is approximately constant
(to order $h_{e}$) in the support area of node $n$. We neglect the
term of order $h_{e}$ because our numerical scheme has overall the
first order accuracy, as defined by the use of linear basis functions. 

Another important property of the basis functions is related to partition
of unity - the sum of all the basis functions in the domain, at any
point in the domain ($i.e.,$ at non-nodal locations, as well as at
nodal locations), is equal to one. This property also hold for the
pyramid side functions, $i.e.,$ $\overset{N_{n}}{\underset{n=1}{\Sigma}}\phi_{n}(\textbf{\ensuremath{\mathbf{r}}})=\overset{N_{n}}{\underset{n=1}{\Sigma}}\psi_{n}(\textbf{\ensuremath{\mathbf{r}}})=1$.

\section{Key notations and node-to-element mapping\label{sec:Key-notations-and} }

In this chapter, the notation $\underline{X}$ (or $\underline{x})$
will be used to denote vectors of dimensions $[N_{n}\times1]$ that
contain node-associated quantities, while $\underline{X^{e}}\,[N_{e}\times1]$
denotes vectors of element-associated quantities. The notation $\underline{\underline{X^{e}}}$
will be used to denote matrices of dimensions $[N_{e}\times N_{n}]$
that operate on vectors of nodal quantities $\underline{X}$, to produce
vectors of elemental quantities $\underline{Y^{e}}$, while $\underline{\underline{X_{n}}}\,[N_{n}\times N_{e}]$
operates on vectors of elemental quantities $\underline{X^{e}}$ to
produce vectors of nodal quantities $\underline{Y}$. The notations
$\underline{\underline{X}}$ and $\widehat{\underline{\underline{X}}}$
will be used to denote square matrices with dimensions $[N_{n}\times N_{n}]$
and $[N_{e}\times N_{e}]$ respectively. Notations defining the various
matrix dimensions are collected in table \ref{tab:Notations-for-matrices}.
\begin{table}[H]
\centering{}\includegraphics[scale=0.15]{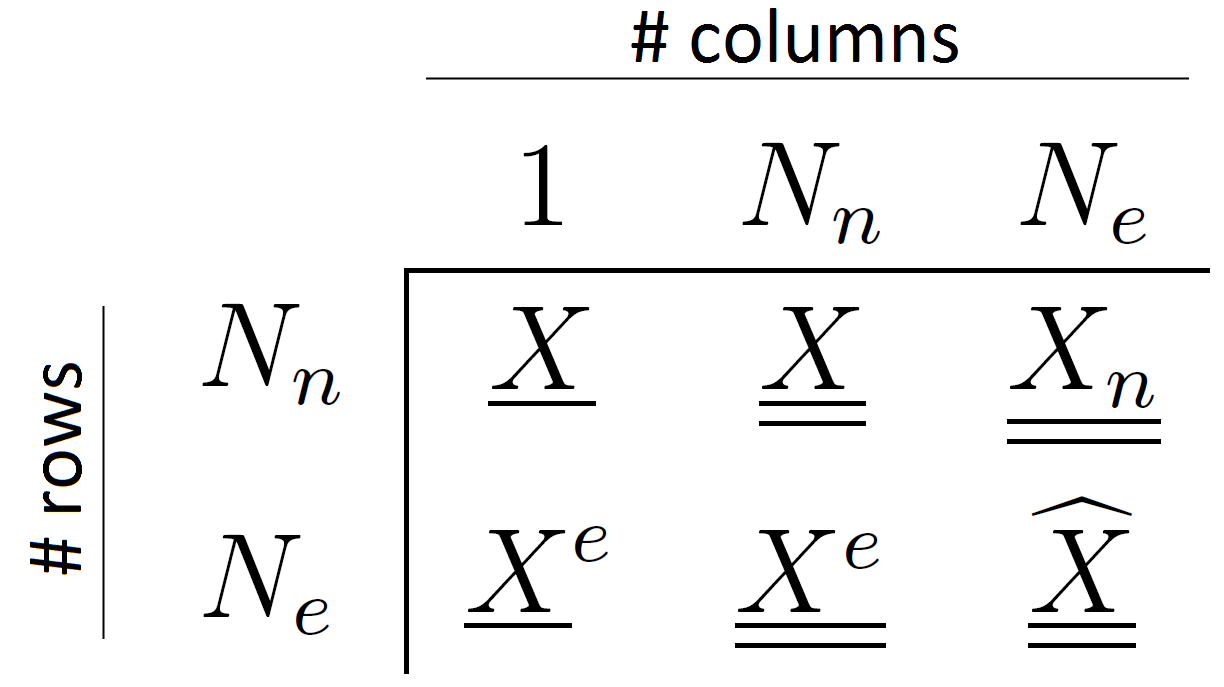}\caption{\label{tab:Notations-for-matrices}$\,\,\,\,$Notations for matrices
of various dimensions}
\end{table}

We introduce node-to-element averaging as
\begin{equation}
<\underline{X}>^{e}=\frac{1}{3}\underline{\underline{M^{e}}}*\underline{X}\label{eq:502.31}
\end{equation}
Here, the connectivity matrix $\underline{\underline{M^{e}}}$ has
dimensions $[N_{e}\times N_{n}]$. Each row of $\underline{\underline{M^{e}}}$
corresponds to a particular triangular element, and has only three
non-zero entries. $\underline{\underline{M^{e}}}(e,\,n)=1$ for the
column indexes $n$ corresponding to the indexes $n$ of the nodes
located at the vertices of the triangle with index $e.$ The symbol
$*$ represents regular matrix multiplication. The radial coordinates
of element centroids are defined with $\underline{r^{e}}=(r_{1}^{e},\,r_{2}^{e},...r_{N_{e}}^{e})^{T}=<\underline{r}>^{e}$,
where $\underline{r}=(r_{1},r_{2},...r_{N_{n}})^{T}$ contains the
$r$ coordinates of the nodes. The superscript $T$ implies the transpose
operation. Similarly, the axial centroid coordinates are defined as
$\underline{z^{e}}=<\underline{z}>^{e}$. The vector of nodal support
areas $\underline{s}$, containing the areas of the bases of the pyramid
functions associated with the nodes, is related to $\underline{s^{e}}$,
the vector of elemental areas, as $\underline{s}=\underline{\underline{M^{e}}}^{T}*\underline{s^{e}}$. 

Volume integrals over the computational domain can be approximated
in two ways, corresponding to nodal or elemental representations of
the integrand function 
\begin{align*}
\int u(\mathbf{r})\,dV & \approx\underline{dV}^{T}*\underline{U}\,\,\,\mbox{ or }\,\,\,\int u(\mathbf{r})\,dV\approx\underline{dV^{e}}^{T}*\underline{U^{e}}
\end{align*}
where $\underline{dV}=\frac{2\pi}{3}\,\underline{s}\circ\underline{r}$
contains the volumes associated with each node, which are found by
integrating the node-associated areas from $0$ to $2\pi$ in the
toroidal direction, and $\underline{dV^{e}}=2\pi\,\underline{s^{e}}\circ\underline{r^{e}}$
contains the elemental volumes. The factor of three in the former
expression arises because each elemental area is shared by three nodes.
Note that these two approximations do not give identical results for
vectors related by equation \ref{eq:502.31}, unless the original
integrand function is constant. The symbol $\circ$ represents the
Hadamard product, piecewise element-by element multiplication ($i.e.,$
$(\underline{a}\circ\underline{b})_{i}=a_{i}b_{i}$), and the symbol
$\oslash$ represents Hadamard division, piecewise element-by element
division ($i.e.,$ $(\underline{a}\oslash\underline{b})_{i}=a_{i}/b_{i}$).

Defining $\underline{\underline{R}}$, $\underline{\underline{S}}$,
$\widehat{\underline{\underline{R}}}$ and $\widehat{\underline{\underline{S}}}$
as the diagonal arrays constructed from $\underline{r}$, $\underline{s}$,
$\underline{r^{e}}$, and $\underline{s^{e}}$, we define a volume-averaging
operator 
\begin{equation}
\underline{\underline{W_{n}}}=\underline{\underline{R}}^{-1}*\underline{\underline{S}}^{-1}*\underline{\underline{M^{e}}}^{T}*\widehat{\underline{\underline{S}}}*\widehat{\underline{\underline{R}}}\label{eq:515}
\end{equation}
 that is used to map element-based quantities to node-based quantities,
as
\begin{equation}
<\underline{U^{e}}>=\underline{\underline{W_{n}}}*\underline{U^{e}}\label{eq:515.1}
\end{equation}
This operator satisfies the following identity: 
\begin{equation}
\underline{dV}^{T}*\left(\underline{Q}\circ\left(\underline{\underline{W_{n}}}*\underline{U^{e}}\right)\right)=\underline{dV^{e}}^{T}*\left(\underline{Q^{e}}\circ\underline{U^{e}}\right)\label{eq:516}
\end{equation}
where $\underline{U^{e}}$ and $\underline{Q}^{e}$ are the discrete
representations, defined at the element centroids, of the approximations
to the continuous functions $u(\mathbf{r})$ and $q(\mathbf{r})$,
and $\underline{Q}$ is the discrete representation, defined at the
nodes, of $q(\mathbf{r})$, and is related to $\underline{Q^{e}}$
by equation \ref{eq:502.31}. A proof of this identity follows:

\begin{align*}
\underline{dV}^{T}*\left(\underline{Q}\circ\left(\underline{\underline{W_{n}}}*\underline{U^{e}}\right)\right) & =\frac{2\pi}{3}\left(\underline{s}\circ\underline{r}\right)^{T}*\left(\underline{Q}\circ\left(\underline{\underline{R}}^{-1}*\underline{\underline{S}}^{-1}*\underline{\underline{M^{e}}}^{T}*\widehat{\underline{\underline{S}}}*\widehat{\underline{\underline{R}}}*\underline{U^{e}}\right)\right)\\
 & =\frac{2\pi}{3}\underline{Q}^{T}*\left(\underline{\underline{M^{e}}}^{T}*\widehat{\underline{\underline{S}}}*\widehat{\underline{\underline{R}}}*\underline{U^{e}}\right)\\
 & =\frac{2\pi}{3}\left(\widehat{\underline{\underline{S}}}*\widehat{\underline{\underline{R}}}*\underline{U^{e}}\right)^{T}*\left(\underline{Q}^{T}*\underline{\underline{M^{e}}}^{T}\right)^{T} & \mbox{(transpose)}\\
 & =2\pi\left(\widehat{\underline{\underline{S}}}*\widehat{\underline{\underline{R}}}*\underline{U^{e}}\right)^{T}*\left(\frac{1}{3}\underline{\underline{M^{e}}}*\underline{Q}\right)\\
 & =\underline{dV^{e}}^{T}*\left(\underline{Q^{e}}\circ\underline{U^{e}}\right) & \mbox{(use eqn. \ref{eq:502.31})}
\end{align*}
Note that the matrix transpose relation

\begin{equation}
\left(\mathbb{A}*\mathbb{B}\right){}^{T}=\mathbb{B}{}^{T}*\mathbb{A}{}^{T}\label{eq:516.01}
\end{equation}
for matrices $\mathbb{A}$ and $\mathbb{B}$, is used to transpose
the scalar on the right side of the equation in the third last step
of the derivation above. In the particular case with $\underline{Q}=\underline{1}$,
the identity becomes 
\begin{equation}
\underline{dV}^{T}*\left(\underline{\underline{W_{n}}}*\underline{U^{e}}\right)=\underline{dV^{e}}^{T}*\underline{U^{e}}\label{eq:516.1}
\end{equation}

\section{Differential operators\label{sec:Differential-operators}}

\subsection{First order node-to-element differential operators\label{subsec:Dre Dze}}

\begin{figure}[H]
\centering{}\includegraphics[width=8cm,height=2.5cm]{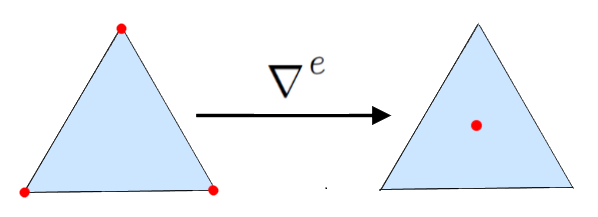}\caption{\label{fig:Drc}$\,\,\,\,$Node-to-element differential operator mechanism }
\end{figure}
The node-to-element derivative operator matrices are defined such
that 
\begin{alignat*}{1}
\underline{U_{r}^{'e}} & =\underline{\underline{Dr^{e}}}*\underline{U}\\
\underline{U_{z}^{'e}} & =\underline{\underline{Dz^{e}}}*\underline{U}
\end{alignat*}
Referring for example to the radial derivative operator, application
of $\underline{\underline{Dr^{e}}}$, a matrix with dimensions $[N_{e}\times N_{n}]$,
to the vector $\underline{U}$, returns $\underline{U_{r}^{'e}}\,[N_{e}\times1],$
containing the values for the radial derivatives of $U(\mathbf{r})$
inside the triangular elements. The node-to-element gradient and divergence
operations, for cylindrical coordinates with azimuthal symmetry are
\\
 
\begin{align}
\underline{\underline{\nabla^{e}}}\,\,\underline{U} & =\left(\underline{\underline{Dr^{e}}}*\underline{U}\right)\hat{\mathbf{r}}+\left(\underline{\underline{Dz^{e}}}*\underline{U}\right)\hat{\mathbf{z}}\label{eq:505.01}
\end{align}
and 
\begin{align}
\underline{\underline{\nabla^{e}}}\cdot\underline{\mathbf{P}} & =\left(\left(\underline{\underline{Dr^{e}}}*\left(\underline{r}\circ\underline{P_{r}}\right)\right)+\left(\underline{\underline{Dz^{e}}}*\left(\underline{r}\circ\underline{P_{z}}\right)\right)\right)\oslash\underline{r^{e}}\label{eq:505.02}
\end{align}
where $\underline{P_{r}}$ and $\underline{P_{z}}$ are the nodal
representations of the $r$ and $z$ components the continuous vector
field $\mathbf{p}(\mathbf{r})$. A schematic of the node-to-element
gradient operation mechanism is indicated in figure \ref{fig:Drc}. 

To derive the $\underline{\underline{Dr^{e}}}$ and $\underline{\underline{Dz^{e}}}$
operators we use the elemental representation of $U(\mathbf{r})$,
equations \ref{eq:500.01} and \ref{eq:500.02}. Equation \ref{eq:500.02}
defines the values of the approximation $U(\mathbf{r})$ at the nodes
as 
\begin{equation}
U_{\varepsilon}=A^{e}+B^{e}r_{\varepsilon}+C^{e}z_{\varepsilon}\label{eq:503-1}
\end{equation}
where $\varepsilon=i,\,j,\,k$ denotes the indexes of the nodes at
the vertices of triangular element $e$. In the following, the element-specific
node indexes $i,\,j,\,k$ will be replaced with the indexes $1,\:2$
and $3$ for simplicity. From equation \ref{eq:500.02}, the radial
and axial first spatial derivatives of $U^{e}(\mathbf{r})$ are constants
across the element $e$, and are given by $\frac{\partial U^{e}}{\partial r}=B^{e}$,
and $\frac{\partial U^{e}}{\partial z}=C^{e}$. Expressions for $B^{e}$
and $C^{e}$ are found using equation \ref{eq:503-1}: 
\begin{align*}
U_{1}-U_{3} & =B^{e}(r_{1}-r_{3})+C^{e}(z_{1}-z_{3})\\
U_{2}-U_{3} & =B^{e}(r_{2}-r_{3})+C^{e}(z_{2}-z_{3})
\end{align*}

\begin{alignat}{1}
\Rightarrow B^{e} & =\frac{\partial U^{e}}{\partial r}=\frac{U_{1}(z_{2}-z_{3})+U_{2}(z_{3}-z_{1})+U_{3}(z_{1}-z_{2})}{2s^{e}}\nonumber \\
C^{e} & =\frac{\partial U^{e}}{\partial z}=\frac{U_{1}(r_{2}-r_{3})+U_{2}(r_{3}-r_{1})+U_{3}(r_{1}-r_{2})}{-2s^{e}}\label{eq:503.2}
\end{alignat}
The triangle area $s^{e}$ is introduced here, assuming that vertices
are number counterclockwise, noting that $2s^{e}=|\mathbb{R}^{e}|$,
where $|\mathbb{R}^{e}|$ is the determinant of array $\mathbb{R}^{e}=\left(\begin{array}{ccc}
1 & r_{1} & z_{1}\\
1 & r_{2} & z_{2}\\
1 & r_{3} & z_{3}
\end{array}\right)$, so that 
\[
2s^{e}=(r_{1}-r_{3})(z_{2}-z_{3})-(r_{2}-r_{3})(z_{1}-z_{3})
\]
The element-specific spatial derivatives can also be expressed in
terms of the coefficients in the element-specific pyramid-side functions
$\psi_{1}^{e},\,\psi_{2}^{e}$, and $\psi_{3}^{e}$. As noted earlier,
each pyramid-side function is specific to a particular node, and to
a particular element, and is defined as $\psi_{n}^{e}(\textbf{\ensuremath{\mathbf{r}}})=a_{n}^{e}+b_{n}^{e}r+c_{n}^{e}z$,
and has the property $\psi_{n}^{e}(r_{j},\,z_{j})=\delta_{nj}$. This
yields, for each element $e$, the identity $\mathbb{R}^{e}*\mathbb{C}^{e}=\mathbb{I}$,
or 
\[
\left(\begin{array}{ccc}
1 & r_{1} & z_{1}\\
1 & r_{2} & z_{2}\\
1 & r_{3} & z_{3}
\end{array}\right)*\left(\begin{array}{ccc}
a_{1}^{e} & a_{2}^{e} & a_{3}^{e}\\
b_{1}^{e} & b_{2}^{e} & b_{3}^{e}\\
c_{1}^{e} & c_{2}^{e} & c_{3}^{e}
\end{array}\right)=\left(\begin{array}{ccc}
1 & 0 & 0\\
0 & 1 & 0\\
0 & 0 & 1
\end{array}\right)
\]
 
\begin{align}
\Rightarrow\mathbb{C}^{e} & =\left(\mathbb{R}^{e}\right)^{-1}=\frac{1}{2s^{e}}\left(\begin{array}{ccc}
\left(r_{2}z_{3}-r_{3}z_{2}\right) & \left(r_{3}z_{1}-r_{1}z_{3}\right) & \left(r_{1}z_{2}-r_{2}z_{1}\right)\\
\left(z_{2}-z_{3}\right) & \left(z_{3}-z_{1}\right) & \left(z_{1}-z_{2}\right)\\
\left(r_{3}-r_{2}\right) & \left(r_{1}-r_{3}\right) & \left(r_{2}-r_{1}\right)
\end{array}\right)\label{eq:504.1}
\end{align}
Comparing equations \ref{eq:503.2} with equation \ref{eq:504.1},
it is evident that 
\begin{align}
\frac{\partial U^{e}}{\partial r} & =U_{1}b_{1}^{e}+U_{2}b_{2}^{e}+U_{3}b_{3}^{e}=\left(\begin{array}{ccc}
b_{1}^{e} & b_{2}^{e} & b_{3}^{e}\end{array}\right)*\left(\begin{array}{c}
U_{1}\\
U_{2}\\
U_{3}
\end{array}\right)\nonumber \\
\frac{\partial U^{e}}{\partial z} & =U_{1}c_{1}^{e}+U_{2}c_{2}^{e}+U_{3}c_{3}^{e}=\left(\begin{array}{ccc}
c_{1}^{e} & c_{2}^{e} & c_{3}^{e}\end{array}\right)*\left(\begin{array}{c}
U_{1}\\
U_{2}\\
U_{3}
\end{array}\right)\label{eq:505}
\end{align}
Given the values of the approximation for the piecewise linear function
$U(\mathbf{r})$ on the triangle vertices, and the element-specific
array $\mathbb{C}^{e}$ evaluated using equation \ref{eq:504.1},
equations \ref{eq:505} can be used to determine the values of the
spatial derivatives of $U(\mathbf{r})$ at the interior of each element.
$\underline{\underline{Dr^{e}}}$ and $\underline{\underline{Dz^{e}}}$
are initially defined as sparse all-zero arrays. Reverting to node-specific
notation, where $i,\,j,\,k$ are the indexes of the nodes at the vertices
of element $e$, the values $b_{i}^{e},\,b_{j}^{e}$ and $b_{k}^{e}$
are inserted in row $e$ of $\underline{\underline{Dr^{e}}}$ with
placements at the column indexes $i,\,j,\,k$. Similarly, the values
$c_{i}^{e},\,c_{j}^{e}$ and $c_{k}^{e}$, for each element, are used
to assemble $\underline{\underline{Dz^{e}}}$. The resulting derivative
operators produce exact derivatives for nodal functions with linear
dependence on $r$ and $z$, and have first order accuracy ($i.e.,\,\mathcal{O}(h_{e})$)
when applied to nonlinear functions. The operators introduced in the
following sections are all based on these node-to-element derivative
operators, and so they all have the same accuracy. \\
\\
Here, we will demonstrate a property of the node-to-element operators
that will be used to demonstrate angular momentum conservation later
in section \ref{subsec:Angular-momentum-conservation}. For any continuous
scalar functions $f$ and $g$, the identity (Stoke's theorem) holds
that 
\begin{equation}
\int\left(\nabla f\times\nabla g\right)\cdot d\mathbf{S}=\int\nabla\times\left(f\,\nabla g\right)\cdot d\mathbf{S}=\int f\,\nabla g\cdot d\mathbf{l}\label{eq:505.001}
\end{equation}
With azimuthal symmetry, $\int\left(\nabla f\times\nabla g\right)\cdot d\mathbf{S}=\int\left(f_{z}'\,g_{r}'-f_{r}'\,g_{z}'\right)\widehat{\boldsymbol{\phi}}\cdot d\mathbf{S}=\int\left(f_{z}'\,g_{r}'-f_{r}'\,g_{z}'\right)ds_{\phi}$,
where $ds_{\phi}$ is an elemental area in the poloidal ($r-z$) plane.
Hence, the discrete form of $\int\left(\nabla f\times\nabla g\right)\cdot d\mathbf{S}$,
using the node-to-element differential operators, is 
\begin{alignat}{1}
\left(\int\left(\nabla f\times\nabla g\right)\cdot d\mathbf{S}\right)_{disc.} & =\underline{s^{e}}^{T}*\left(\left(\underline{\underline{Dz^{e}}}*\underline{f}\right)\circ\left(\underline{\underline{Dr^{e}}}*\underline{g}\right)-\left(\underline{\underline{Dr^{e}}}*\underline{f}\right)\circ\left(\underline{\underline{Dz^{e}}}*\underline{g}\right)\right)\nonumber \\
 & =\left(\underline{\underline{Dz^{e}}}*\underline{f}\right)^{T}*\underline{\underline{\widehat{S}}}*\left(\underline{\underline{Dr^{e}}}*\underline{g}\right)-\left(\underline{\underline{Dr^{e}}}*\underline{f}\right)^{T}*\underline{\underline{\widehat{S}}}*\left(\underline{\underline{Dz^{e}}}*\underline{g}\right)\nonumber \\
 & =\underline{f}^{T}*\underline{\underline{Dz^{e}}}^{T}*\underline{\underline{\widehat{S}}}*\underline{\underline{Dr^{e}}}*\underline{g}-\underline{f}^{T}*\underline{\underline{Dr^{e}}^{T}}*\underline{\underline{\widehat{S}}}*\underline{\underline{Dz^{e}}}*\underline{g}\nonumber \\
 & =\underline{f}^{T}*\left(\underline{\underline{Dz^{e}}}^{T}*\underline{\underline{\widehat{S}}}*\underline{\underline{Dr^{e}}}-\underline{\underline{Dr^{e}}}^{T}*\underline{\underline{\widehat{S}}}*\underline{\underline{Dz^{e}}}\right)*\underline{g}\nonumber \\
 & =\underline{f}^{T}*\left(\underline{\underline{B}}^{T}-\underline{\underline{B}}\right)*\underline{g}\nonumber \\
 & =\underline{f}^{T}*\underline{\underline{A}}*\underline{g}\label{eq:505.0011}
\end{alignat}
where $\underline{\underline{B}}=\underline{\underline{Dr^{e}}}^{T}*\underline{\underline{\widehat{S}}}*\underline{\underline{Dz^{e}}}$.
Setting $\underline{f}=\underline{\phi_{m}}$ and $\underline{g}=\underline{\phi_{n}}$,
where $\underline{\phi_{m}}$ and $\underline{\phi_{n}}$ are the
vectors defining the nodal values of basis functions $\phi_{m}$ and
$\phi_{n}$ respectively ($i.e.,$ $\underline{\phi_{m}}(k)=\delta_{mk}$
and $\underline{\phi_{n}}(k)=\delta_{nk}$), equations \ref{eq:505.0011}
and \ref{eq:505.001} imply that 
\begin{equation}
A_{mn}\equiv A(m,n)=\underline{\phi_{m}}^{T}*\underline{\underline{A}}*\underline{\phi_{n}}=\int\phi_{m}\,\nabla\phi_{n}\cdot d\mathbf{l}\label{eq:505.002}
\end{equation}
We will show that each element of $\underline{\underline{A}}$ is
zero at internal nodes, $i.e.,$ $\left(\underline{\underline{A}}\right)_{int}=0$.
There are four cases to consider: (1) $m\neq n$ and nodes $m$ and
$n$ are not adjacent, (2) $m=n$, (3) $m$ and $n$ are the indexes
of adjacent internal nodes, and (4) $m$ and $n$ are the indexes
of adjacent nodes located on the boundary of the computational domain.
In case (1), $A_{mn}$ is obviously zero. In case (2), $A_{mn}=\underline{\phi_{m}}^{T}*\left(\underline{\underline{B}}^{T}-\underline{\underline{B}}\right)*\underline{\phi_{m}}=\underline{\phi_{m}}^{T}*\underline{\underline{B}}^{T}*\underline{\phi_{m}}-\underline{\phi_{m}}^{T}*\underline{\underline{B}}*\underline{\phi_{m}}$
. The first term scalar term here can be transposed, so that $A_{mn}=0$
in case (2). 
\begin{figure}[H]
\centering{}\includegraphics[width=9.2cm,height=7cm]{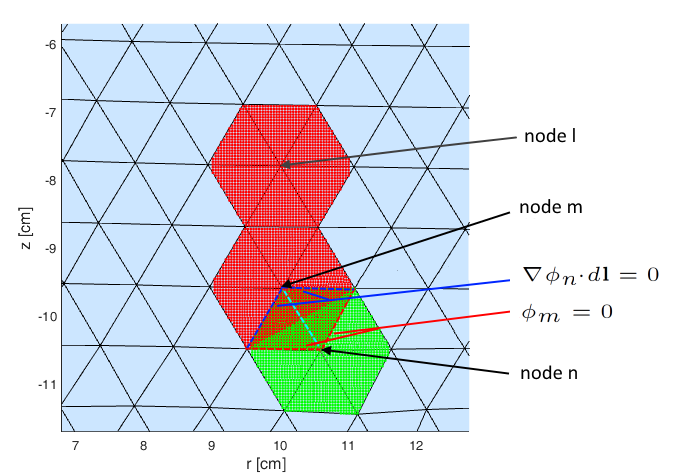}\caption{\label{fig:Linear-basis-function_topview}$\,\,\,\,$Linear basis
function depiction for triangular elements (top view (1)). }
\end{figure}
Figure \ref{fig:Linear-basis-function_topview} is a top-view of figure
\ref{fig:Linear-basis-function}. Note the dark blue dashed lines
represent the part of the boundary of basis function $\phi_{n}$ that
overlaps with $\phi_{m}$, and the red dashed lines represent the
part of the boundary of $\phi_{m}$ that overlaps with $\phi_{n}$.
Referring to equation \ref{eq:505.002} and to figure \ref{fig:Linear-basis-function_topview},
the only contribution to $A_{mn}$ in case (3) is from the contour
integral $\int\phi_{m}\,\nabla\phi_{n}\cdot d\mathbf{l}$ along the
boundary (depicted with dark blue and red dashed lines) of the diamond-shape
representing the overlapping area of the basis functions $\phi_{m}$
and $\phi_{n}$. It can be seen that $\phi_{m}=0$ along part of the
contour, and $\nabla\phi_{n}\cdot d\mathbf{l}=0$ along the remaining
parts, so that $A_{mn}=0$ in case (3). In case (4), when $m$ and
$n$ are the indexes of adjacent nodes located on the boundary of
the computational domain, the integral $\int\phi_{m}\,\nabla\phi_{n}\cdot d\mathbf{l}$
would be finite along the light blue dashed line connecting nodes
$m$ and $n$ (in case (4), the light blue dashed line would represent
part of the computational domain boundary). Therefore, $A_{mn}$ is
finite only when $m$ and $n$ are the indexes of adjacent nodes located
on the boundary of the computational domain, leading to the identity
\begin{equation}
\left(\underline{\underline{Dz^{e}}}^{T}*\underline{\underline{\widehat{S}}}*\underline{\underline{Dr^{e}}}-\underline{\underline{Dr^{e}}}^{T}*\underline{\underline{\widehat{S}}}*\underline{\underline{Dz^{e}}}\right)_{int}=0\label{eq:505.003}
\end{equation}

\subsection{First order element-to-node differential operators\label{subsec:Drn}}

\begin{figure}[H]
\centering{}\includegraphics[width=6cm,height=2.5cm]{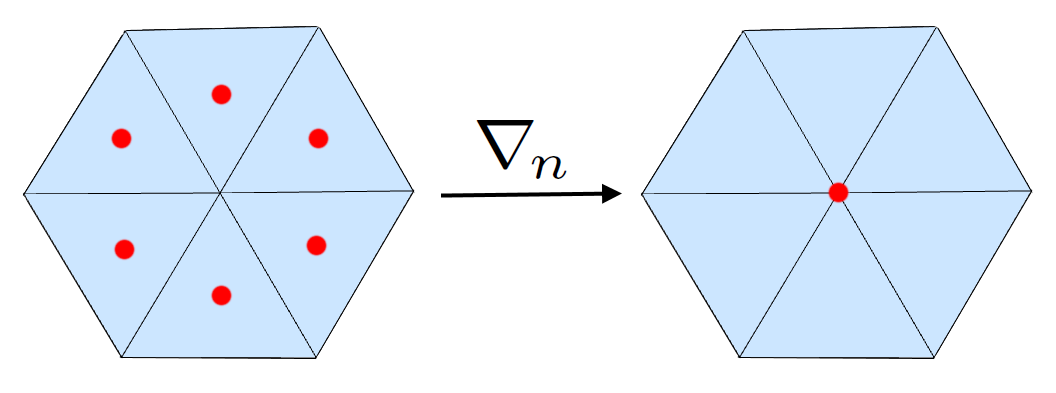}\caption{\label{fig:Drn}$\,\,\,\,$Element-to-node differential operator mechanism}
\end{figure}
The element-to-node derivative operator matrices are defined such
that 
\begin{align*}
\underline{U_{r}'} & =\underline{\underline{Dr_{n}}}*\underline{U^{e}}\\
\underline{U_{z}'} & =\underline{\underline{Dz_{n}}}*\underline{U^{e}}
\end{align*}
Referring for example to the radial derivative operator, application
of the $\underline{\underline{Dr_{n}}}$ operator, which is a matrix
with dimensions $[N_{n}\times N_{e}]$, to the vector $\underline{U^{e}}\,[N_{e}\times1]$,
which contains the values of $U(\mathbf{r})$ at the element centroids,
returns $\underline{U_{r}'}\,[N_{n}\times1],$ the values for the
radial derivatives of $U(\mathbf{r})$ at the nodes. The element-to-node
gradient and divergence operations, for cylindrical coordinates with
azimuthal symmetry are 
\begin{align}
\underline{\underline{\nabla{}_{n}}}\,\,\underline{U^{e}} & =\left(\underline{\underline{Dr_{n}}}*\underline{U^{e}}\right)\hat{\mathbf{r}}+\left(\underline{\underline{Dz_{n}}}*\underline{U^{e}}\right)\hat{\mathbf{z}}\label{eq:515.01}
\end{align}
and 
\begin{align}
\underline{\underline{\nabla_{n}}}\cdot\underline{\mathbf{P}^{e}} & =\left(\left(\underline{\underline{Dr_{n}}}*\left(\underline{r^{e}}\circ\underline{P_{r}^{e}}\right)\right)+\left(\underline{\underline{Dz_{n}}}*\left(\underline{r^{e}}\circ\underline{P_{z}^{e}}\right)\right)\right)\oslash\underline{r}\label{eq:515.02}
\end{align}
A schematic of the element-to-node gradient operation mechanism is
indicated in figure \ref{fig:Drn}. The element-to-node differential
operators are derived in relation to the node-to-element operators
by requiring the discrete form of equation \ref{eq:500.0} to hold:
\begin{equation}
\underline{dV}{}^{T}*\left[\underline{\mathbf{P}}\cdot\left(\underline{\underline{\nabla_{n}}}\,\,\underline{U^{e}}\right)\right]+\underline{dV^{e}}{}^{T}*\left[\underline{U^{e}}\circ\left(\underline{\underline{\nabla^{e}}}\cdot\underline{\mathbf{P}}\right)\right]=0\label{eq:515.03}
\end{equation}
or 
\begin{equation}
\underline{dV}{}^{T}*\left[\underline{U}\circ\left(\underline{\underline{\nabla_{n}}}\cdot\underline{\mathbf{P}^{e}}\right)\right]+\underline{dV^{e}}{}^{T}*\left[\underline{\mathbf{P}^{e}}\cdot\left(\underline{\underline{\nabla^{e}}}\,\,\underline{U}\right)\right]=0\label{eq:515.031}
\end{equation}
In this derivation, it is assumed that the boundary term $\int u\mathbf{p}\cdot d\boldsymbol{\Gamma}=0$,
$i.e.,$ there is no flux of the continuous vector field $u\mathbf{p}$
through the boundary. For arbitrary discrete nodal representations
$\underline{\mathbf{P}}$ (equation \ref{eq:515.03}) and $\underline{U}$
(equation \ref{eq:515.031}), a consequence of this assumption is
that the element-to-node gradient operator produces valid results
at the boundary nodes only if the original continuous function $u$
is zero at the boundary $(i.e.,\,u|_{\Gamma}=0)$. Similarly, the
element-to-node divergence operator produces valid results at the
boundary nodes only if the original continuous function $\mathbf{p}$
has no component perpendicular to the boundary $(i.e.,\,\mathbf{p_{\perp}}|_{\Gamma}=0)$.
In the following, we will refer to these conditions as the natural
boundary conditions. For the terms involving radial derivatives, equation
\ref{eq:515.031} implies that

\begin{align*}
 & \underline{dV}^{T}*\left(\underline{U}\circ\left(\left(\underline{\underline{Dr_{n}}}*\left(\underline{r^{e}}\circ\underline{P_{r}^{e}}\right)\right)\oslash\underline{r}\right)\right) &  & =-\underline{dV^{e}}^{T}*\left(\underline{P_{r}^{e}}\circ\left(\underline{\underline{Dr^{e}}}*\underline{U}\right)\right)\\
 & \Rightarrow\frac{2\pi}{3}\,\underline{s}^{T}*\left(\underline{U}\circ\left(\underline{\underline{Dr_{n}}}*\left(\underline{r^{e}}\circ\underline{P_{r}^{e}}\right)\right)\right) &  & =-2\pi\,\underline{s^{e}}^{T}*\left(\underline{r^{e}}\circ\underline{P_{r}^{e}}\circ\left(\underline{\underline{Dr^{e}}}*\underline{U}\right)\right)\\
 & \Rightarrow\frac{1}{3}\underline{U}^{T}*\left(\underline{\underline{S}}*\left(\underline{\underline{Dr_{n}}}*\left(\underline{r^{e}}\circ\underline{P_{r}^{e}}\right)\right)\right) &  & =-\left(\underline{r^{e}}\circ\underline{P_{r}^{e}}\right)^{T}*\left(\widehat{\underline{\underline{S}}}*\underline{\underline{Dr^{e}}}*\underline{U}\right)\\
 & \Rightarrow\frac{1}{3}\underline{U}^{T}*\left(\underline{\underline{S}}*\left(\underline{\underline{Dr_{n}}}*\left(\underline{r^{e}}\circ\underline{P_{r}^{e}}\right)\right)\right) &  & =-\underline{U}^{T}*\left(\underline{\underline{Dr^{e}}}^{T}*\widehat{\underline{\underline{S}}}*\left(\underline{r^{e}}\circ\underline{P_{r}^{e}}\right)\right) & \mbox{(transpose)}
\end{align*}
Hence, the element-to-node derivative operators are 
\begin{align}
\underline{\underline{Dr_{n}}} & =-3\underline{\underline{S}}^{-1}*\left(\underline{\underline{Dr^{e}}}^{T}*\widehat{\underline{\underline{S}}}\right)\nonumber \\
\underline{\underline{Dz_{n}}} & =-3\underline{\underline{S}}^{-1}*\left(\underline{\underline{Dz^{e}}}^{T}*\widehat{\underline{\underline{S}}}\right)\label{eq:515.0}
\end{align}
These definitions can alternatively be obtained using equation \ref{eq:515.03}.
Noting the particular case when $\underline{U}=\underline{1}$, then
equation \ref{eq:515.031} leads to the identity 
\begin{align}
\underline{dV}{}^{T}*\left[\underline{\underline{\nabla_{n}}}\cdot\underline{\mathbf{P}^{e}}\right] & =0\label{eq:515.04}
\end{align}

\subsection{Second order node-to-node differential operators\label{subsec:Lapl_delstar}}

\begin{figure}[H]
\centering{}\includegraphics[width=6cm,height=2.5cm]{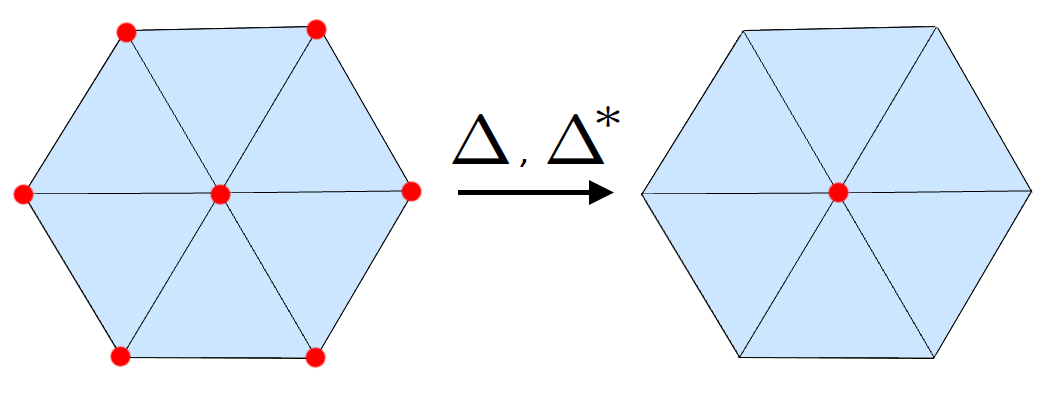}\caption{\label{fig:LaplDelstar}$\,\,\,\,$Node-to-node second order differential
operator mechanism}
\end{figure}
Discrete forms of node-to-node second order differential operators,
with dimensions $[N_{n}\times N_{n}]$, can be constructed using combinations
of the first order node-to-element and element-to-node operators.
For example, the discrete form of the Laplacian operator in cylindrical
coordinates, with azimuthal symmetry, is 
\begin{equation}
\underline{\underline{\Delta}}\,\,\underline{U}=\underline{\underline{\nabla_{n}}}\cdot\underline{\underline{\nabla^{e}}}\,\,\underline{U}=\left(\left(\underline{\underline{Dr_{n}}}*\left(\underline{r^{e}}\circ\left(\underline{\underline{Dr^{e}}}*\underline{U}\right)\right)\right)+\left(\underline{\underline{Dz_{n}}}*\left(\underline{r^{e}}\circ\left(\underline{\underline{Dz^{e}}}*\underline{U}\right)\right)\right)\right)\oslash\underline{r}\label{eq:514.7}
\end{equation}
Note that $\underline{\underline{\Delta}}\,\,\underline{U}$ may be
expressed as $\underline{\underline{\nabla_{n}}}\cdot\underline{\mathbf{P}^{e}}$,
where $\underline{\mathbf{P}^{e}}=\underline{\underline{\nabla^{e}}}\,\,\underline{U}$.
Therefore, equation \ref{eq:515.04} implies that
\begin{align}
\underline{dV}{}^{T}*\left[\underline{\underline{\Delta}}\,\,\underline{U}\right] & =0\label{eq:515.71}
\end{align}
The condition $\mathbf{p_{\perp}}|_{\Gamma}=0$, required for correct
evaluation of $\underline{\underline{\nabla_{n}}}\cdot\underline{\mathbf{P}^{e}}$
at the boundary nodes, is equivalent to having the normal component
of the gradient of the continuous function $u$ equal to zero at the
boundary. Thus, the operation $\underline{\underline{\Delta}}*\underline{U}$
produces correct results at the boundary nodes only if the natural
boundary condition is satisfied, which in this case is $\left(\nabla_{\perp}u\right)|_{\Gamma}=0.$ 

The discrete form of the elliptic operator used in the Grad-Shafranov
equation, in cylindrical coordinates with azimuthal symmetry is (see
equation \ref{eq:19})

\begin{equation}
\underline{\underline{\Delta^{^{*}}}}\,\,\underline{U}=\underline{r}^{2}\circ\left(\underline{\underline{\nabla_{n}}}\cdot\left(\left(\underline{\underline{\nabla^{e}}}\,\,\underline{U}\right)\oslash\underline{r^{e}}^{2}\right)\right)=\underline{r}\circ\left(\underline{\underline{Dr_{n}}}*\left(\left(\underline{\underline{Dr^{e}}}*\underline{U}\right)\oslash\underline{r^{e}}\right)+\underline{\underline{Dz_{n}}}*\left(\left(\underline{\underline{Dz^{e}}}*\underline{U}\right)\oslash\underline{r^{e}}\right)\right)\label{eq:515.72}
\end{equation}
The $\underline{\underline{\Delta}}$ and $\underline{\underline{\Delta^{^{*}}}}$
operator mechanisms are indicated in figure \ref{fig:LaplDelstar}.
Note that equation \ref{eq:515.04} implies that 
\begin{align}
\underline{dV}{}^{T}*\left[\left(\underline{\underline{\Delta^{^{*}}}}\,\,\underline{U}\right)\oslash\underline{r}^{2}\right] & =0\label{eq:515.71-1-1}
\end{align}
Again, the operation $\underline{\underline{\Delta^{^{*}}}}\,\,\underline{U}$
produces correct results at the boundary nodes only if the natural
boundary condition is satisfied, which in this case is $\left(\left(\nabla_{\perp}u\right)/r^{2}\right)|_{\Gamma}=0.$

\subsubsection{Alternative derivation of the second order node-to-node differential
operators}

It is illustrative to show how the second order node-to-node differential
operators can be derived directly using the fundamental finite element
methods, outlined in section \ref{subsec:Finite-element-method},
of constructing a weighted integral and obtaining the weak formulation
expression. This method also leads to the definitions of the first
order element-to-node differential operators, without employing the
procedure, outlined in section \ref{subsec:Drn}, of invoking the
discrete form of equation \ref{eq:500.0}.

The Laplacian operator is defined as $\Delta u(\mathbf{r})=\nabla\cdot\nabla u(\mathbf{r})$.
With the Galerkin method, the basis functions $\phi_{j}(\mathbf{r})$,
where $j=1:N_{n}$, are used as weighting functions. Defining 
\[
L(\mathbf{r})=\Delta U(\mathbf{r})=\nabla\cdot\nabla U(\mathbf{r})
\]
where $U(\mathbf{r})$ and $L(\mathbf{r})$ are the piecewise linear
approximations to the continuous function $u(\mathbf{r})$ and $\Delta u(\mathbf{r}),$
the relevant weighted integral equality is
\begin{align}
\int\phi_{j}(\mathbf{r})\,L(\mathbf{r})\,dr\,dz & =\int\phi_{j}(\mathbf{r})\nabla\cdot\nabla U(\mathbf{r})\,dr\,dz\nonumber \\
 & =\int\nabla\cdot\left(\phi_{j}\nabla U\right)\,dr\,dz-\int\nabla\phi_{j}\cdot\nabla U\,dr\,dz & \mbox{\mbox{(integration by parts) }}\nonumber \\
 & =\varoint\left(\phi_{j}\nabla U\right)\cdot d\boldsymbol{\mathcal{L}}-\int\nabla\phi_{j}\cdot\nabla U\,dr\,dz & \mbox{(using Green's theorem) }\label{eq:511.1}
\end{align}
Here, Greens theorem is used to reduce the integral over area to an
integral evaluated on the boundary $\mathcal{L}$. Note $d\boldsymbol{\mathcal{L}}=\widehat{\mathbf{n}}d\mathcal{L}$,
where $\widehat{\mathbf{n}}(\mathbf{r})$ is the unit outward normal
to the boundary. If $\left(\nabla_{\perp}U(\mathbf{r})\right)|_{\Gamma}=0$,
$i.e.,$ the component of $\nabla U$ perpendicular to the boundary,
is zero at the boundary, then the boundary integral can be neglected.
This will be discussed further below. Dropping the boundary term and
applying the linear basis function expansion (equation \ref{eq:500.03})
to $U(\mathbf{r})$, the resulting expression is 
\begin{align}
\int\phi_{j}\,L(\mathbf{r})\,dr\,dz & =-\overset{N_{n}}{\underset{i=1}{\Sigma}}U_{n}\int\nabla\phi_{j}\cdot\nabla\phi_{n}\,dr\,dz\label{eq:511.11}\\
\nonumber 
\end{align}
Equation \ref{eq:502.4} is used to expand the left side of this equation
as $L_{j}\,s_{j}/3$. On the right side, matrix $\underline{\underline{La}}$
of size $N_{n}\times N_{n}$ is introduced as 
\begin{equation}
\underline{\underline{La}}(j,\,n)=-\int\nabla\phi_{j}\cdot\nabla\phi_{n}\,dr\,dz\label{eq:508.1-1}
\end{equation}
Hence, equation \ref{eq:511.11} can be re-expressed in matrix form
as 

\begin{align}
\frac{\underline{L}(j)\,\underline{s}(j)}{3} & =\overset{N_{n}}{\underset{n=1}{\Sigma}}\left(\underline{\underline{La}}(j,\,n)\,\underline{U}(n)\right)\nonumber \\
\Rightarrow\underline{L}=\underline{\underline{\Delta}}*\underline{U} & =3\,\underline{\underline{S}}^{-1}*\underline{\underline{La}}*\underline{U}\nonumber \\
\Rightarrow\underline{\underline{\Delta}} & =3\,\underline{\underline{S}}^{-1}*\underline{\underline{La}}\label{eq:508.11}
\end{align}
Expressing the pyramid functions in terms of the pyramid side functions
according to equation \ref{eq:502.3}, then referring to equations
\ref{eq:508.1-1} and \ref{eq:508.11}, $\underline{\underline{\Delta}}$
can be expressed as 

\begin{align*}
\underline{\underline{\Delta}}(j,\,n) & =-\frac{3}{s_{j}}\,\int\left(\overset{e_{j}}{\underset{}{\Sigma}}\,\frac{\partial\psi_{j}^{e}(\mathbf{r})}{\partial r}\,\overset{e_{n}}{\underset{}{\Sigma}}\,\frac{\partial\psi_{n}^{e}(\mathbf{r})}{\partial r}+\overset{e_{j}}{\underset{}{\Sigma}}\,\frac{\partial\psi_{j}^{e}(\mathbf{r})}{\partial z}\,\overset{e_{n}}{\underset{}{\Sigma}}\,\frac{\partial\psi_{n}^{e}(\mathbf{r})}{\partial z}\right)\,dr\,dz\\
 & =-\frac{3}{s_{j}}\,\int\left(\overset{e_{j}\cap e_{n}}{\underset{}{\Sigma}}\,b_{j}^{e}\,b_{n}^{e}+\overset{e_{j}\cap e_{n}}{\underset{}{\Sigma}}\,c_{j}^{e}\,c_{n}^{e}\right)\,dr\,dz\\
 & =-\frac{3}{s_{j}}\,\left(\overset{e_{j}\cap e_{n}}{\underset{}{\Sigma}}\,b_{j}^{e}\,b_{n}^{e}+\overset{e_{j}\cap e_{n}}{\underset{}{\Sigma}}\,c_{j}^{e}\,c_{n}^{e}\right)s^{e}
\end{align*}
Here, $\overset{e_{j}\cap e_{n}}{\underset{}{\Sigma}}$ implies summation
over the set of elements surrounding node $n$ that overlap with the
elements surrounding node $j$ (see the example in section \ref{subsec:Alternative-derivations-N2N}
for clarification). Referring to equations \ref{eq:505}, it can be
seen how this implies that the discrete Laplacian operator, in matrix
form, is \\
\begin{equation}
\underline{\underline{\Delta}}=-\left(3\underline{\underline{S}}^{-1}\right)*\left(\underline{\underline{Dr^{e}}}^{T}*\widehat{\underline{\underline{S}}}*\underline{\underline{Dr^{e}}}+\underline{\underline{Dz^{e}}}^{T}*\widehat{\underline{\underline{S}}}*\underline{\underline{Dz^{e}}}\right)\label{eq:514}
\end{equation}
With the definition $\underline{\underline{\Delta}}\,\,\underline{U}=\underline{\underline{\nabla_{n}}}\cdot\underline{\underline{\nabla^{e}}}\,\,\underline{U}$,
this leads naturally to the definitions of the element-to-node operators
(equation \ref{eq:515.0}) - in this case, the properties of the element-to-node
operators outlined in section \ref{subsec:Drn} arise because the
boundary term in equation \ref{eq:511.1} was dropped. The method
outlined in this section results again in the expression for the Laplacian
and $\Delta^{^{*}}$ operations in cylindrical coordinates, as defined
by equations \ref{eq:514.7} and \ref{eq:515.72}, where the operator
properties \ref{eq:515.71} and \ref{eq:515.71-1-1} follow naturally.\\

\subsection{First order node-to-node differential operators \label{subsec:Dr_Dz}}

\begin{figure}[H]
\centering{}\includegraphics[width=6cm,height=2.5cm]{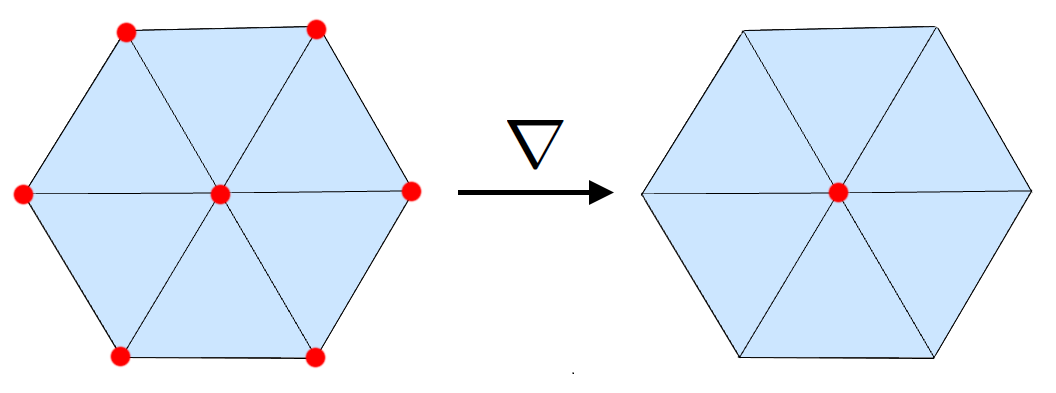}\caption{\label{fig:Dr}$\,\,\,\,$Node-to-node differential operator mechanism}
\end{figure}
The node-to-node derivative operator matrices $\underline{\underline{Dr}}$
and $\underline{\underline{Dz}}$, with dimensions $[N_{n}\times N_{n}]$,
are used to evaluate, at the nodes, the radial and axial derivatives
of any function $U(\mathbf{r})$ that is defined at the nodes, and
are defined as 
\begin{align*}
\underline{U_{r}'} & =\underline{\underline{Dr}}*\underline{U}\\
\underline{U_{z}'} & =\underline{\underline{Dz}}*\underline{U}
\end{align*}
The node-to-node gradient and divergence operations, for cylindrical
coordinates with azimuthal symmetry are 
\begin{align}
\underline{\underline{\nabla}}\,\,\underline{U} & =\left(\underline{\underline{Dr}}*\underline{U}\right)\hat{\mathbf{r}}+\left(\underline{\underline{Dz}}*\underline{U}\right)\hat{\mathbf{z}}\label{eq:510.21-1}
\end{align}
and 
\begin{align}
\underline{\underline{\nabla}}\cdot\underline{\mathbf{P}} & =\left(\left(\underline{\underline{Dr}}*\left(\underline{r}\circ\underline{P_{r}}\right)\right)+\left(\underline{\underline{Dz}}*\left(\underline{r}\circ\underline{P_{z}}\right)\right)\right)\oslash\underline{r}\label{eq:510.22-1}
\end{align}
A schematic of the node-to-node gradient operation mechanism is illustrated
in figure \ref{fig:Dr}. Here, we will look at the derivation of $\underline{\underline{Dr}}$,
the derivation of $\underline{\underline{Dz}}$ is analogous. The
node-to-node operators can be expressed in terms of the node-to-element
operators by considering the expansion of $U(\mathbf{r})$ in terms
of the piecewise-linear functions $U^{e}(\mathbf{r})$ (equation \ref{eq:500.01}):

\begin{align*}
U(\mathbf{r}) & =\overset{}{\underset{e}{\Sigma}}\,U^{e}(\textbf{\ensuremath{\mathbf{r}}})
\end{align*}
This implies that
\[
U_{r}'(\mathbf{r})=\overset{}{\underset{e}{\Sigma}}\,U_{r}^{e\,'}(\mathbf{r})
\]
Applying the Galerkin method, we obtain 
\begin{equation}
\int\phi{}_{j}(\mathbf{r})\,U_{r}'(\mathbf{r})\,dr\,dz=\int\phi{}_{j}(\textbf{\ensuremath{\mathbf{r}}})\,\overset{}{\underset{e}{\Sigma}}U_{r}^{e\,'}(\mathbf{r})\,dr\,dz\label{eq:507.01}
\end{equation}
where integral is taken over the support area of the basis function
$\phi_{j}(\mathbf{r})$. Note that $U_{r}^{e\,'}=B^{e}$ (equation
\ref{eq:500.02}). Expanding the right side of equation \ref{eq:507.01}
in terms of pyramid side functions according to equation \ref{eq:502.3},
and using equation \ref{eq:502.4} to expand the left side, this implies
that 
\begin{align*}
\frac{U_{rj}'s_{j}}{3} & =\overset{e_{j}}{\underset{}{\Sigma}}\int\psi_{j}^{e}(\textbf{\ensuremath{\mathbf{r}}})\,B^{e}\,dr\,dz\\
 & =\overset{e_{j}}{\underset{}{\Sigma}}\,B^{e}\int\psi_{j}^{e}(\textbf{\ensuremath{\mathbf{r}}})\,dr\,dz\\
 & =\overset{e_{j}}{\underset{}{\Sigma}}\,B^{e}\frac{s^{e}}{3}
\end{align*}
Therefore, in our method the derivative at a node is defined as the
area-weighted average of the elemental derivatives in the support
of that node. With use of connectivity matrix $\underline{\underline{M^{e}}}$,
this definition can be written as:

\begin{align}
\underline{U_{r}'} & =\left(\underline{\underline{S}}^{-1}*\left(\underline{\underline{M^{e}}}^{T}*\widehat{\underline{\underline{S}}}*\underline{\underline{Dr^{e}}}\right)\right)*\underline{U}\nonumber \\
\Rightarrow\underline{\underline{Dr}} & =\underline{\underline{S}}^{-1}*\left(\underline{\underline{M^{e}}}^{T}*\widehat{\underline{\underline{S}}}*\underline{\underline{Dr^{e}}}\right)\label{eq:510.1-1-1}
\end{align}
The derivation of $\underline{\underline{Dz}}$ follows from that
of $\underline{\underline{Dr}}$, with $\underline{\underline{Dz}}=\left(\underline{\underline{S}}^{-1}*\left(\underline{\underline{M^{e}}}^{T}*\widehat{\underline{\underline{S}}}*\underline{\underline{Dz^{e}}}\right)\right)$. 

Now we introduce some special properties of the node-to-node differential
operators. $U_{r}'(\mathbf{r})$, the radial spatial derivative of
the approximate solution $U(\mathbf{r})$, can alternatively be expressed,
using the basis function expansion (equation \ref{eq:500.03}), as
\begin{equation}
U_{r}'(\mathbf{r})=\overset{N_{n}}{\underset{n=1}{\Sigma}}U{}_{n}\,\phi_{rn}'(\mathbf{r})\label{eq:506}
\end{equation}
Applying the Galerkin method, we obtain 
\begin{equation}
\int\phi{}_{j}\,U_{r}'(\mathbf{r})\,dr\,dz=\overset{N_{n}}{\underset{n=1}{\Sigma}}U{}_{n}\int\phi{}_{j}\,\phi_{rn}'\,dr\,dz\label{eq:507}
\end{equation}
Again, from equation \ref{eq:502.4}, the left side of this equation
is $U_{rj}'\,s_{j}/3$. On the right side, matrix $\underline{\underline{Lr}}$
of size $N_{n}\times N_{n}$ is introduced as 
\begin{equation}
\frac{1}{3}\underline{\underline{Lr}}(j,\,n)=\int\phi{}_{j}\,\phi_{rn}'\,dr\,dz\label{eq:508.1}
\end{equation}
Hence, equation \ref{eq:507} can be re-expressed as 

\begin{equation}
\underline{U_{r}'}(j)\,\underline{s}(j)=\overset{N_{n}}{\underset{n=1}{\Sigma}}\,\left(\underline{\underline{Lr}}(j,\,n)\,\underline{U}(n)\right)\label{eq:509}
\end{equation}
or in matrix form: $\underline{\underline{S}}*\underline{\underline{Dr}}*\underline{U}=\underline{\underline{Lr}}*\underline{U}$
\begin{equation}
\Rightarrow\underline{\underline{Dr}}=\underline{\underline{S}}^{-1}*\underline{\underline{Lr}}\label{eq:509.1}
\end{equation}
With integration by parts and Green's theorem, equation \ref{eq:508.1}
implies that 
\begin{align*}
\underline{\underline{Lr}}(j,\,n) & =3\int\phi{}_{j}\,\phi_{rn}'\,dr\,dz\\
 & =3\int\left(-\phi_{rj}'\phi{}_{n}\,+\left(\phi{}_{j}\,\phi{}_{n}\right)_{r}'\right)\,dr\,dz\\
 & =3\left(-\int\phi{}_{n}\,\phi_{rj}'\,dr\,dz+\varoint\phi{}_{j}\,\phi{}_{n}\,dz\right)
\end{align*}
The boundary term vanishes at internal nodes, so that $\underline{\underline{Lr}}(j,\,n)=-\underline{\underline{Lr}}(n,\,j)$
at the internal nodes. This implies that all elements of the matrix
$\underline{\underline{Lr}}+\underline{\underline{Lr}}^{T}$ are equal
to zero, except for the elements corresponding to boundary nodes.
Hence, using equation \ref{eq:509.1}, and noting that the transpose
of a diagonal matrix is the same matrix, we obtain 
\begin{equation}
\left(\underline{\underline{S}}*\underline{\underline{Dr}}+\underline{\underline{Dr}}^{T}*\underline{\underline{S}}\right)_{int}=0\label{eq:510.23}
\end{equation}
where the subscript $int$ denotes all elements of the matrix that
correspond to interior (non-boundary) nodes. There is an analogous
identity for the $\underline{\underline{Dz}}$ operator:
\begin{equation}
\left(\underline{\underline{S}}*\underline{\underline{Dz}}+\underline{\underline{Dz}}^{T}*\underline{\underline{S}}\right)_{int}=0\label{eq:510.24}
\end{equation}
As a consequence of these identities, the node-to-node differential
operators have some important properties that mimic properties of
the corresponding continuous operators. In particular, the discrete
expression for the volume integral in equation \ref{eq:500.0} is
\begin{align*}
 & \underline{dV}{}^{T}*\left[\underline{U}\circ\left(\underline{\underline{\nabla}}\cdot\underline{\mathbf{P}}\right)+\underline{\mathbf{P}}\cdot\left(\underline{\underline{\nabla}}\,\,\underline{U}\right)\right]\\
 & =\underline{dV}{}^{T}*\left[\underline{U}\circ\left(\underline{\underline{Dr}}*\left(\underline{r}\circ\underline{P_{r}}\right)+\underline{\underline{Dz}}*\left(\underline{r}\circ\underline{P_{z}}\right)\right)\oslash\underline{r}+\underline{P_{r}}\circ\left(\underline{\underline{Dr}}*\underline{U}\right)+\underline{P_{z}}\circ\left(\underline{\underline{Dz}}*\underline{U}\right)\right]\\
 & =\frac{2\pi}{3}\left[\underline{U}^{T}*\underline{\underline{S}}*\left(\underline{\underline{Dr}}*\underline{\underline{R}}*\underline{P_{r}}+\underline{\underline{Dz}}*\underline{\underline{R}}*\underline{P_{z}}\right)+\underline{P_{r}}{}^{T}*\underline{\underline{S}}*\underline{\underline{R}}*\underline{\underline{Dr}}*\underline{U}+\underline{P_{z}}{}^{T}*\underline{\underline{S}}*\underline{\underline{R}}*\underline{\underline{Dz}}*\underline{U}\right]\\
 & =\frac{2\pi}{3}\,\underline{U}{}^{T}*\left[\left(\underline{\underline{S}}*\underline{\underline{Dr}}+\underline{\underline{Dr}}{}^{T}*\underline{\underline{S}}\right)*\underline{\underline{R}}*\underline{P_{r}}+\left(\underline{\underline{S}}*\underline{\underline{Dz}}+\underline{\underline{Dz}}{}^{T}*\underline{\underline{S}}\right)*\underline{\underline{R}}*\underline{P_{z}}\right]
\end{align*}
Note that in the last step, transpose operations on the scalar expressions
have been performed. Hence, due to equations \ref{eq:510.23} and
\ref{eq:510.24}
\begin{align}
\underline{dV}{}^{T}*\left[\left(\underline{U}\circ\left(\underline{\underline{\nabla}}\cdot\underline{\mathbf{P}}\right)+\left(\underline{\underline{\nabla}}\,\,\underline{U}\right)\cdot\underline{\mathbf{P}}\right)\right] & =0 & \left(\mbox{if }\ensuremath{\ensuremath{\underline{U}\,|_{\Gamma}}=0}\mbox{ or }\ensuremath{\underline{\mathbf{P}}_{\perp}|_{\Gamma}=0}\right)\label{eq:511.04}
\end{align}
For the particular case of $\underline{U}=\underline{1}$ , this implies
that 
\begin{align}
\underline{dV}{}^{T}*\left[\underline{\underline{\nabla}}\cdot\underline{\mathbf{P}}\right] & =\underline{0} & \left(\mbox{if }\ensuremath{\underline{\mathbf{P}}_{\perp}|_{\Gamma}=0}\right)\label{eq:511.041}
\end{align}

\subsubsection{Alternative derivations of the first order node-to-node differential
operators\label{subsec:Alternative-derivations-N2N}}

Note that while the expansion of $U(\mathbf{r})$ in terms of the
piecewise-linear functions $U^{e}(\mathbf{r})$ (equation \ref{eq:500.01})
was used to define the node-to-node operators, the basis function
expansion (equation \ref{eq:500.03}) was required to demonstrate
the operator properties in equations \ref{eq:510.23} to \ref{eq:511.041}.
It is illustrative to note that the operators can be defined equivalently
using the latter expansion. The procedure for deriving the radial
node-to-node derivative operator is demonstrated as follows. After
doing the expansion and applying the Galerkin method, the array of
integrals $\underline{\underline{Lr}}$, defined in equation \ref{eq:508.1},
can be expressed, after expressing the basis functions in terms of
the pyramid side functions according to equation \ref{eq:502.3},
as 

\[
\underline{\underline{Lr}}(j,\,n)=3\int\left(\overset{e_{j}}{\underset{}{\Sigma}}\,\psi_{j}^{e}(\textbf{\ensuremath{\mathbf{r}}})\,\overset{e_{n}}{\underset{}{\Sigma}}\,\psi_{rn}^{e'}(\textbf{\ensuremath{\mathbf{r}}})\right)\,dr\,dz
\]
Here, $\overset{e_{n}}{\underset{}{\Sigma}}\,\psi_{rn}^{e'}(\textbf{\ensuremath{\mathbf{r}}})$
implies summation of the radial derivatives of the pyramid side functions
$\psi_{n}^{e}(\textbf{\ensuremath{\mathbf{r}}})$ that are non-zero
over the triangular elements that have a vertex at node $n$. Similarly
$\overset{e_{j}}{\underset{}{\Sigma}}\,\psi_{j}^{e}(\textbf{\ensuremath{\mathbf{r}}})$
implies summation over the pyramid side functions $\psi_{j}^{e}$
defined over the elements immediately surrounding node $j$. Equation
\ref{eq:502.2} defines that 
\begin{align*}
\psi_{rn}^{e'}(\textbf{\ensuremath{\mathbf{r}}}) & =b_{n}^{e} &  & \mbox{(if \ensuremath{(r,\,z)} is in the triangular element associated with \ensuremath{\psi_{n}^{e})}}\\
 & \mbox{}\\
\psi_{rn}^{e'}(\textbf{\ensuremath{\mathbf{r}}}) & =0 &  & \mbox{(otherwise)}
\end{align*}
In either case, the constant can be taken outside the integral, and
$\underline{\underline{Lr}}(j,\,n)$ can be re-expressed as $\underline{\underline{Lr}}(j,\,n)=3\,\overset{e_{n}}{\underset{}{\Sigma}}\psi_{rn}^{e'}(\textbf{\ensuremath{\mathbf{r}}})\overset{e_{j}}{\underset{}{\Sigma}}\int\psi_{j}^{e}(\textbf{\ensuremath{\mathbf{r}}})\,dr\,dz$.
Using again the expression for the volume of a pyramid (equation \ref{eq:502.32}),
the remaining integral in $\underline{\underline{Lr}}(j,\,n)$ is
$\int\psi_{j}^{e}(\textbf{\ensuremath{\mathbf{r}}})\,dr\,dz=\frac{s^{e}}{3}$,
where $s^{e}$ is the area of the triangular element associated with
the pyramid side function $\psi_{j}^{e}.$ Thus
\begin{align}
\underline{\underline{Lr}}(j,\,n) & =\overset{e_{j}\cap e_{n}}{\underset{}{\Sigma}}\,b_{n}^{e}\,s^{e}\label{eq:509.3}
\end{align}
where $\overset{e_{j}\cap e_{n}}{\underset{}{\Sigma}}$ implies summation
over the set of elements surrounding node $n$ that overlap with the
elements surrounding node $j$. This can be clarified with an example.
\begin{figure}[H]
\centering{}\includegraphics[width=9.2cm,height=7cm]{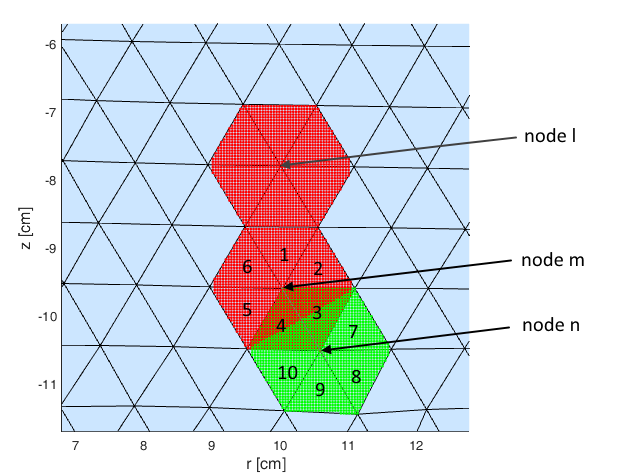}\caption{\label{fig:Linear-basis-function-1}$\,\,\,\,$Linear basis function
depiction for triangular elements (top view (2))}
\end{figure}
As shown in figure \ref{fig:Linear-basis-function-1} above, which
is a top view of figure \ref{fig:Linear-basis-function}, nodes $n$
and $m$ are adjacent, and are each surrounded by six triangular elements
in this example. Only elements 3 and 4 are in the overlapping regions
defined by bases of the pyramid functions $\phi_{n}$ and $\phi_{m}$.
Each element is associated with three pyramid-side functions and three
sets of coefficients $a_{n}^{e},\,b_{n}^{e}$, and $c_{n}^{e}$. In
this example, $\underline{\underline{Lr}}(n,\,m)=b_{m}^{3}s^{3}+b_{m}^{4}s^{4}$,
where, for example, $b_{m}^{3}$ is one of the three coefficients
defining the pyramid-side function that is non-zero only over element
3, and that has an elevation of one at node $m$, while $s^{3}$ is
the area of triangular element 3. Meanwhile, $\underline{\underline{Lr}}(m,\,n)=b_{n}^{3}s^{3}+b_{n}^{4}s^{4}$,
$\underline{\underline{Lr}}(m,\,m)=\overset{6}{\underset{i=1}{\Sigma}}b_{m}^{i}s^{i}$,
$\underline{\underline{Lr}}(n,\,n)=\overset{4}{\underset{i=3}{\Sigma}}b_{n}^{i}s^{i}+\overset{10}{\underset{i=7}{\Sigma}}b_{n}^{i}s^{i}$,
and since node $l$ is not adjacent to node $m$ nor to node $n$,
$\underline{\underline{Lr}}(l,\,m)=\underline{\underline{Lr}}(m,\,l)=\underline{\underline{Lr}}(l,\,n)=\underline{\underline{Lr}}(n,\,l)=0$.
Having developed an analytical expression for $\underline{\underline{Lr}}$
in terms of the pyramid side functions, $\underline{\underline{Dr}}$
can be evaluated using equation \ref{eq:509.1}. The procedure for
deriving $\underline{\underline{Dz}}$ using the basis function expansion
method is analogous.\\

The node-to-node differential operators can alternatively be defined
directly in terms of the node-to-element differential operators simply
by requiring the discrete form of equation \ref{eq:500.0} to hold.
The identity implies that when the continuous functions $\mathbf{p}$
and $u$ are such that $\mathbf{p}_{\perp}|_{\Gamma}=0$ or $u|_{\Gamma}=0$,
that $\int u\nabla\cdot\mathbf{p}\,dV=-\int\mathbf{p}\nabla u\,dV$.
Thus, with the boundary conditions $\mbox{ }\ensuremath{\ensuremath{\underline{U}\,|_{\Gamma}}=0}\mbox{ or }\ensuremath{\underline{\mathbf{P}}_{\perp}|_{\Gamma}=0}$:
\begin{align*}
\underline{dV}^{T}*\left\{ \underline{U}\circ\left(\left(\underline{\underline{Dr_{n}}}*\left(\underline{r}^{e}\circ\underline{P_{r}^{e}}\right)\right)\oslash\underline{r}\right)\right\}  & =-\underline{dV}^{T}*\left\{ \underline{P_{r}}\circ\left(\underline{\underline{Dr}}*\underline{U}\right)\right\} \\
\Rightarrow\underline{s}^{T}*\left(\underline{U}\circ\left(\underline{\underline{Dr_{n}}}*\left(\underline{r}^{e}\circ\underline{P_{r}^{e}}\right)\right)\right) & =-\underline{s}^{T}*\left(\underline{r}\circ\underline{P_{r}}\circ\left(\underline{\underline{Dr}}*\underline{U}\right)\right)\\
\Rightarrow\underline{U}^{T}*\left(\underline{\underline{S}}*\left(\underline{\underline{Dr_{n}}}*\left(\frac{1}{3}\underline{\underline{M^{e}}}*\left(\underline{r}\circ\underline{P_{r}}\right)\right)\right)\right) & =-\left(\underline{r}\circ\underline{P_{r}}\right)^{T}*\left(\underline{\underline{S}}*\underline{\underline{Dr}}*\underline{U}\right)\\
\Rightarrow\left(\frac{1}{3}\underline{\underline{M^{e}}}*\left(\underline{r}\circ\underline{P_{r}}\right)\right)^{T}*\left(\underline{\underline{Dr_{n}}}^{T}*\underline{\underline{S}}*\underline{U}\right) & =-\left(\underline{r}\circ\underline{P_{r}}\right)^{T}*\left(\underline{\underline{S}}*\underline{\underline{Dr}}*\underline{U}\right) & \mbox{(transpose)}\\
\Rightarrow\left(\underline{r}\circ\underline{P_{r}}\right)^{T}*\left(\frac{1}{3}\underline{\underline{M^{e}}}^{T}*\underline{\underline{Dr_{n}}}^{T}*\underline{\underline{S}}*\underline{U}\right) & =-\left(\underline{r}\circ\underline{P_{r}}\right)^{T}*\left(\underline{\underline{S}}*\underline{\underline{Dr}}*\underline{U}\right)\\
\Rightarrow\underline{\underline{Dr}} & =-\frac{1}{3}\underline{\underline{S}}^{-1}*\underline{\underline{M^{e}}}^{T}*\underline{\underline{Dr_{n}}}^{T}*\underline{\underline{S}}\\
 & =\underline{\underline{S}}^{-1}*\underline{\underline{M^{e}}}^{T}*\widehat{\underline{\underline{S}}}*\underline{\underline{Dr^{e}}} & \mbox{(use eqn.}\mbox{\ref{eq:515.0})}
\end{align*}
Again, the procedure for deriving $\underline{\underline{Dz}}$ using
the basis function expansion method is analogous. While this derivation
method is short and simple, demonstration of the special operator
properties requires, as outlined previously, the expansion of the
derivatives of $U(\mathbf{r})$ in terms of the basis functions.

\section{Axisymmetric two-temperature MHD model\label{sec:Axisymmetric-two-temperature-MHD}}

A two-temperature plasma MHD model, in which the electrons and ions
are allowed to have different temperatures, is a compromise between
the single-fluid and two-fluid MHD models. It has the relative computational
simplicity of the single-fluid model, but from the energy point of
view it treats electrons and ions as two separate fluids, and allows
implementation of different thermal diffusion coefficients for the
electrons and ions. In the following, SI units will be used, but temperatures
will be expressed in Joules unless otherwise indicated. As shown in
appendix \ref{chap:Kinetic,MhD,GS}, the MHD equations can be derived
from basic kinetic theory. The Grad-Shafranov equation, derived in
appendix \ref{subsec:Grad-Shafranov-equation}, describes MHD equilibrium
in the case with toroidal symmetry:
\[
\Delta^{*}\psi=-\left(\mu_{0}r^{2}\frac{dp}{d\psi}+f\frac{df}{d\psi}\right)
\]
Here, $\psi$ is the poloidal flux per radian, and $f(\psi)=rB_{\phi}$.
The numerical method implemented to code to solve the Grad-Shafranov
equation, using the discrete form of the $\Delta^{*}$ operator (section
\ref{subsec:Lapl_delstar}), is presented in appendix \ref{sec:Numerical-solution-of}.
In equilibrium, when $\mathbf{J}$ is parallel to $\mathbf{B}$, the
magnetic field topology is described by equation \ref{eq:0.01}: $\nabla\times\mathbf{B=}\lambda\mathbf{B}$.
As shown in appendix \ref{subsec:GS_linear_lambda}, this implies
that $\lambda=\lambda(\psi)$, with the consequence that general axisymmetric
force free states must have $\lambda$ constant on a flux surface.
Unlike the Taylor state, in which fields satisfy $\nabla\times\mathbf{B=}\lambda\mathbf{B}$
with constant $\lambda$, $\lambda$ can vary across flux surfaces
in general axisymmetric force free states. A simple model assuming
a linear dependence of $\lambda$ on $\psi$ \cite{Bellan_Spheromaks,Knox}
is presented in appendix \ref{subsec:GS_linear_lambda}, and is implemented
as an option when solving the Grad-Shafranov equation. Example solutions
are presented in appendices \ref{sec:Numerical-solution-of} and \ref{subsec:Equilibrium-solution-comparison}.

In addition to the continuous forms of the two-temperature MHD equations,
from appendix summary \ref{sec:SummaryKin_MHD_EQ}, we require expressions
for $\dot{\psi}$ and $\dot{f}$ to complete the axisymmetric model:\\

\subsection{\label{par:Expressionfor_psidot}Expression for $\dot{\psi}$}

Referring to Ampere's law, the reduced form of Ohm's law (equation
\ref{eq:476.4}) implies that $\mathbf{E}+\mathbf{v\times}\mathbf{B}=\eta\nabla\times\mathbf{B}$,
so that $E_{\phi}=\eta(\nabla\times\mathbf{B})_{\phi}-(\mathbf{v\times}\mathbf{B})_{\phi}$.
Here $\eta\,\mbox{[m}^{2}/\mbox{s}]=\frac{\eta'}{\mu_{0}}$ is the
magnetic diffusivity. The electric field is defined as $\mathbf{E=-\nabla}\Phi_{E}-\mathbf{\dot{\mathbf{A}}}$,
where $\mathbf{A}$ is the magnetic vector potential defined by $\mathbf{B=\nabla\times A}$,
and $\Phi_{E}$ is the electric potential, so that:
\begin{equation}
-\cancel{\nabla_{\phi}\Phi_{E}}-\dot{\mathbf{A}}_{\phi}=\eta\left(\frac{\partial B_{r}}{\partial z}-\frac{\partial B_{z}}{\partial r}\right)\widehat{\boldsymbol{\phi}}-\left(v_{z}B_{r}-v_{r}B_{z}\right)\widehat{\boldsymbol{\phi}}\label{eq:207-1}
\end{equation}
With toroidal symmetry, the first term goes to zero. Noting that $\psi=rA_{\phi}$
(equation \ref{eq:9.001}), and referring to equation \ref{eq:20.2}
for the definitions of the components of the axisymmetric magnetic
field, we can re-express this as 
\begin{equation}
\dot{\psi}=\eta\left(r\frac{\partial}{\partial r}\left(\frac{1}{r}\frac{\partial\psi}{\partial r}\right)+\frac{\partial^{2}\psi}{\partial z^{2}}\right)-v_{z}\frac{\partial\psi}{\partial z}-v_{r}\frac{\partial\psi}{\partial r}=\eta\Delta^{^{*}}\psi-(\mathbf{v}\cdot\nabla)\psi\label{eq:209-1}
\end{equation}
\begin{equation}
\Rightarrow\dot{\psi}=\dot{\psi}_{ideal}+\dot{\psi}_{\eta}=-(\mathbf{v}\cdot\nabla)\psi+\eta\Delta^{^{*}}\psi\label{eq:211}
\end{equation}
Here, $\dot{\psi}_{ideal}$ is the part of $\dot{\psi}$ that would
occur even in a perfectly conductor, for example in an ideal plasma,
and $\dot{\psi}_{\eta}$ is the part of $\dot{\psi}$ that is due
to diffusive effects associated with electrical resistivity.

\subsection{\label{subsec:expressionfor_fdot}Expressions for $\dot{f}$ }

\subsubsection*{General case}

Taking the curl of Ohm's law, we obtain: $\nabla\times\mathbf{E}=\nabla\times(\eta'\mathbf{J}-\mathbf{v\times}\mathbf{B})$.
Using Faraday's law $\left(\nabla\times\mathbf{E}=-\dot{\mathbf{B}}\right)$,
and Ampere's law, this implies that
\begin{equation}
\dot{\mathbf{B}}=\nabla\times(\mathbf{v}\times\mathbf{B}-\eta'\mathbf{J})=\nabla\times\left(\mathbf{v}\times\mathbf{B}\right)-\nabla\times(\eta\nabla\times\mathbf{B})
\end{equation}
To find $\dot{f}=r\dot{B}_{\phi}$, we want an expression for 
\begin{equation}
\dot{f}=r\left(\left(\nabla\times\left(\mathbf{v}\times\mathbf{B}\right)\right){}_{\phi}-\left(\nabla\times\left(\eta\nabla\times\mathbf{B}\right)\right)_{\phi}\right)\label{eq:212}
\end{equation}
As was with done with $\dot{\psi}$, $\dot{f}$ can be split into
an ideal part and a resistive part: 
\begin{equation}
\dot{f}=\dot{f}_{ideal}+\dot{f}_{\eta}=r\left(\nabla\times\left(\mathbf{v}\times\mathbf{B}\right)\right){}_{\phi}+r\left(-\nabla\times\left(\eta\nabla\times\mathbf{B}\right)\right)_{\phi}\label{eq:213.1}
\end{equation}
This can be alternatively expressed as a full divergence of terms.
Along with inherent properties of the differential operators, the
discrete form of the full-divergence expression will ensure conservation
of toroidal flux. To derive the full divergence expression, it is
convenient to expand $\dot{f}_{ideal}$ and $\dot{f}_{\eta}$ separately.
The following relationships will be used to expand $\dot{f}_{ideal}$:
\begin{align}
\nabla\left(\frac{v_{\phi}}{r}\right) & =\frac{1}{r}\nabla v_{\phi}+v_{\phi}\nabla\left(\frac{1}{r}\right)\nonumber \\
\Rightarrow\nabla v_{\phi} & =r\nabla\omega+\omega\widehat{\mathbf{r}}\label{eq:213.2}
\end{align}
Here, $\omega=v_{\phi}/r$ is the plasma angular speed. Replacing
$v_{\phi}$ in this expression with $\frac{f}{r}$:

\begin{align}
\nabla\left(\frac{f}{r}\right) & =r\nabla\left(\frac{f}{r^{2}}\right)+\left(\frac{f}{r^{2}}\right)\widehat{\mathbf{r}}\label{eq:213.3}
\end{align}

\begin{align*}
\dot{f}_{ideal} & =r(\nabla\times\mathbf{v}\times\mathbf{B})_{\phi}\\
 & =r\left(\mathbf{\cancel{v\left(\nabla\cdot\mathbf{B}\right)}-\mathbf{B\left(\nabla\cdot\mathbf{v}\right)}+\mathbf{\left(B\cdot\nabla\right)\mathbf{v}}-\mathbf{\left(v\cdot\nabla\right)B}}\right)_{\phi}\\
 & =r\left(\mathbf{-}B_{\phi}\mathbf{\mathbf{\nabla\cdot\mathbf{v}}+\mathbf{B\cdot\nabla}}v_{\phi}+\frac{B_{\phi}v_{r}}{r}-\mathbf{v}\cdot\nabla B_{\phi}-\frac{v_{\phi}B_{r}}{r}\right)\\
 & =r^{2}\left(\mathbf{-}\frac{f}{r^{2}}\nabla\cdot\mathbf{v}+\frac{1}{r}\mathbf{\mathbf{B\cdot}}\nabla v_{\phi}+\frac{f}{r^{3}}v_{r}-\frac{1}{r}\mathbf{v}\cdot\nabla\left(\frac{f}{r}\right)-\frac{\omega B_{r}}{r}\right)\\
 & =r^{2}\biggl(\mathbf{-}\frac{f}{r^{2}}\nabla\cdot\mathbf{v}+\frac{1}{r}\mathbf{\mathbf{B\cdot}}\left(r\nabla\omega+\omega\widehat{\mathbf{r}}\right)+\frac{f}{r^{3}}v_{r} & \mbox{}\\
 & \,\,-\frac{1}{r}\mathbf{v}\cdot\left(r\nabla\left(\frac{f}{r^{2}}\right)+\left(\frac{f}{r^{2}}\right)\widehat{\mathbf{r}}\right)-\frac{\omega B_{r}}{r}\biggr) & (\mbox{use eqns. \ref{eq:213.2}, \ref{eq:213.3}})\\
 & =r^{2}\left(\mathbf{-}\frac{f}{r^{2}}\nabla\cdot\mathbf{v}+\mathbf{\mathbf{B\cdot}}\nabla\omega+\cancel{\frac{\omega B_{r}}{r}}\cancel{+\frac{f}{r^{3}}v_{r}}-\mathbf{v}\cdot\nabla\left(\frac{f}{r^{2}}\right)-\cancel{\frac{f}{r^{3}}v_{r}}\cancel{-\frac{\omega B_{r}}{r}}\right)\\
\Rightarrow\dot{f}_{ideal} & =r^{2}\nabla\cdot\left(-\left(\frac{f}{r^{2}}\mathbf{v}\right)+\omega\mathbf{B}\right)
\end{align*}
To expand $\dot{f}_{\eta}$, an additional relation needs to be established.
Taking the curl of the poloidal component of the axisymmetric magnetic
field defined in equation \ref{eq:9.1} leads to
\begin{align*}
\nabla\times\mathbf{B}_{\theta} & =\nabla\times\left(\nabla\psi\times\nabla\phi\right)\\
 & =\mathbf{\nabla\psi\left(\cancel{\nabla\cdot\nabla\phi}\right)-\mathbf{\nabla\phi\left(\nabla\cdot\nabla\psi\right)}+\mathbf{\left(\nabla\phi\cdot\nabla\right)\nabla\psi}-\mathbf{\left(\nabla\psi\cdot\nabla\right)\nabla\phi}}\\
 & =-\left(\mathbf{\nabla\cdot\nabla\psi}-\frac{2}{r}\frac{\partial\psi}{\partial r}\right)\nabla\phi
\end{align*}
Note that, using equation \ref{eq:19}, $\Delta^{*}\psi$ can be written
as $\Delta^{*}\psi=r^{2}\nabla\cdot\left(\frac{\nabla\psi}{r^{2}}\right)=\nabla\cdot\nabla\psi-\frac{2}{r}\frac{\partial\psi}{\partial r}$,
leading to the relation 
\begin{equation}
\nabla\times\left(\nabla\psi\times\nabla\phi\right)=-\Delta^{*}\psi\,\nabla\phi\label{eq:213.7}
\end{equation}
for any function $\psi(r,\,z)$ defined in geometry with azimuthal
symmetry. Since, with axisymmetry, the curl of a poloidal vector is
toroidal, and vice-versa, we can write 
\begin{align*}
\dot{f}_{\eta}\,\widehat{\boldsymbol{\phi}} & =-r\left(\nabla\times(\eta\nabla\times\mathbf{B})\right){}_{\phi}\,\widehat{\boldsymbol{\phi}}\\
 & =-r\,\nabla\times\left(\eta\nabla\times\mathbf{B}_{\phi}\right)\\
 & =-r\,\nabla\times\eta\left(\nabla f\times\nabla\phi\right)\\
 & =-r\left(-\eta\,\Delta^{*}f\,\nabla\phi+\nabla\eta\times\left(\nabla f\times\nabla\phi\right)\right) & (\mbox{use eqn. \ref{eq:213.7}})\\
 & =-r\left(-\eta\,\Delta^{*}f\,\nabla\phi+\nabla f\,\left(\cancel{\nabla\eta\cdot\nabla\phi}\right)-\nabla\phi\,\left(\nabla\eta\cdot\nabla f\right)\right)\\
 & =r\left(\eta\,r^{2}\nabla\cdot\left(\frac{\nabla f}{r^{2}}\right)+\left(\nabla\eta\cdot\nabla f\right)\right)\frac{\widehat{\boldsymbol{\phi}}}{r}\\
\Rightarrow\dot{f}_{\eta} & =r^{2}\left(\eta\,\nabla\cdot\left(\frac{\nabla f}{r^{2}}\right)+\left(\nabla\eta\cdot\left(\frac{\nabla f}{r^{2}}\right)\right)\right)\\
= & r^{2}\nabla\cdot\left(\frac{\eta}{r^{2}}\nabla f\right)
\end{align*}
Hence, 
\begin{equation}
\dot{f}=\dot{f}_{ideal}+\dot{f}_{\eta}=r^{2}\,\nabla\cdot\left(-\left(\frac{f}{r^{2}}\mathbf{v}\right)+\omega\mathbf{B}+\frac{\eta}{r^{2}}\nabla f\right)\label{eq:214}
\end{equation}

\subsubsection*{Isothermal case}

The more general and useful case described above is that $\eta$ has
spatial dependence: $\eta=\eta(T_{e}(r,z))$. The isothermal version
of the code (in which energy is not conserved) assumes constant temperature,
with spatially constant $\eta$, so that 
\begin{equation}
\dot{f}_{iso}=\dot{f}_{ideal}+\dot{f}_{\eta iso}=r(\nabla\times\mathbf{v}\times\mathbf{B})_{\phi}+r\eta(\nabla^{2}\mathbf{B})_{\phi}\label{eq:213}
\end{equation}
Here,
\[
(\nabla^{2}\mathbf{B})_{\phi}=\nabla^{2}B{}_{\phi}-\frac{B_{\phi}}{r^{2}}=\frac{1}{r}\frac{\partial}{\partial r}\left(r\frac{\partial B_{\phi}}{\partial r}\right)+\frac{\partial^{2}B_{\phi}}{\partial z^{2}}-\frac{B_{\phi}}{r^{2}}=\frac{\partial^{2}B_{\phi}}{\partial r^{2}}+\frac{1}{r}\frac{\partial B_{\phi}}{\partial r}-\frac{B_{\phi}}{r^{2}}+\frac{\partial^{2}B_{\phi}}{\partial z^{2}}
\]
By the definition of the $\Delta^{^{*}}$ operator (equation \ref{eq:19}),
\begin{align*}
\Delta^{^{*}}(rB_{\phi}) & =r\frac{\partial}{\partial r}\left(\frac{1}{r}\frac{\partial(rB_{\phi})}{\partial r}\right)+\frac{\partial^{2}(rB_{\phi})}{\partial z^{2}}=r\frac{\partial}{\partial r}\left(\frac{1}{r}\left(r\frac{\partial B_{\phi}}{\partial r}+B_{\phi}\right)\right)+r\frac{\partial^{2}B_{\phi}}{\partial z^{2}}\\
 & =r\left(\frac{\partial^{2}B_{\phi}}{\partial r^{2}}+\frac{1}{r}\frac{\partial B_{\phi}}{\partial r}-\frac{B_{\phi}}{r^{2}}+\frac{\partial^{2}B_{\phi}}{\partial z^{2}}\right)
\end{align*}
so that 
\begin{equation}
\dot{f}_{\eta iso}=r\eta(\nabla^{2}\mathbf{B})_{\phi}=\eta\Delta^{^{*}}(rB_{\phi})=\eta\Delta^{^{*}}f\label{eq:215}
\end{equation}

\subsection{Continuous form of MHD model equations and demonstration of conservation
properties\label{subsec:Continuous_eqns_conservation_props} }

Combining these expressions for $\dot{\psi}$ and $\dot{f}$ (equations
\ref{eq:211} and \ref{eq:214}) with the continuous forms of the
two-temperature MHD equations (from appendix summary \ref{sec:SummaryKin_MHD_EQ}),
the full set of equations describing the axisymmetric MHD model is:

\begin{equation}
\dot{n}=-\nabla\cdot(n\mathbf{v})\label{eq: 479.01}
\end{equation}

\begin{equation}
\dot{\mathbf{v}}=-\mathbf{v}\cdot\nabla\mathbf{v}+\frac{1}{\rho}\left(-\nabla p-\nabla\cdot\overline{\boldsymbol{\pi}}+\mathbf{J\times}\mathbf{B}\right)\label{eq:479.02}
\end{equation}

\begin{equation}
\dot{p}_{i}=-\mathbf{v}\cdot\nabla p_{i}-\gamma p_{i}\,\nabla\cdot\mathbf{v}+(\gamma-1)\left(-\nabla\cdot\mathbf{q}_{i}+Q_{ie}-\overline{\boldsymbol{\pi}}:\nabla\mathbf{v}\right)\label{eq:479.1}
\end{equation}

\begin{equation}
\dot{p}_{e}=-\mathbf{v}\cdot\nabla p_{e}-\gamma p_{e}\,\nabla\cdot\mathbf{v}+(\gamma-1)\left(-\nabla\cdot\mathbf{q}_{e}-Q_{ie}+\eta'\mathbf{J}^{2}\right)\label{eq:479.2}
\end{equation}

\begin{eqnarray}
\dot{\psi} & = & -\mathbf{v}\cdot\nabla\psi+\eta\Delta^{*}\psi\label{eq:479.3}\\
\dot{f} & = & r^{2}\,\nabla\cdot\left(-\left(\frac{f}{r^{2}}\mathbf{v}\right)+\omega\mathbf{B}+\frac{\eta}{r^{2}}\nabla f\right)\label{eq:479.4}
\end{eqnarray}
where $\omega=v_{\phi}/r$ is the plasma angular speed, and $\eta$
{[}m$^{2}$/s{]}$=\eta'\,[\Omega-\mbox{m}]/\mu_{0}$ is the plasma
resistive diffusion coefficient. Note that $n$ is the ion number
density, so that the electron number density is $n_{e}=Z_{eff}n$,
where $Z_{eff}$ is the volume-averaged ion charge. We assume ideal
gas laws for ions and electrons: 
\begin{eqnarray}
p_{i} & = & nT_{i}\nonumber \\
p_{e} & = & Z_{eff}nT_{e}\label{eq:479.5}
\end{eqnarray}
Total plasma fluid pressure is $p=p_{i}+p_{e}$. The boundary conditions
and closure for this model (namely, definitions of thermal fluxes
$\mathbf{q}_{i}$ and $\mathbf{q}_{e}$, viscous stress tensor $\overline{\boldsymbol{\pi}}$
and ion-electron heat exchange rate $Q_{ie}$) will be discussed in
section \ref{subsec:Discretised-MHD-model}. 

With axisymmetry, the magnetic field can be represented in divergence-free
form according to equation \ref{eq:9.1}: 
\begin{equation}
\mathbf{B=\nabla\psi\times\nabla\phi+}f\nabla\phi\label{eq:480}
\end{equation}
Referring to equation \ref{eq:213.7}, the current density in this
case is
\begin{equation}
\mathbf{J}=\frac{1}{\mu_{0}}\nabla\times\mathbf{B}=\frac{1}{\mu_{0}}\left(-\Delta^{*}\psi\,\nabla\phi+\nabla f\times\nabla\phi\right)\label{eq:480.1}
\end{equation}
and the Lorentz force is 
\begin{equation}
\mathbf{J}\times\mathbf{B}=-\frac{1}{\mu_{0}r^{2}}\bigg(\Delta^{*}\psi\,\nabla\psi+f\,\nabla f\bigg)+\frac{\mathbf{B}\cdot\nabla f}{\mu_{0}r}\widehat{\boldsymbol{\phi}}\label{eq:480.2}
\end{equation}
The system described by equations \ref{eq: 479.01} - \ref{eq:479.4}
has a number of exact conservation laws expressed by corresponding
continuity equations. The obvious one is conservation of particles,
expressed by equation \ref{eq: 479.01}, with the boundary condition
$\mathbf{v}_{\perp}|_{\Gamma}=0$, corresponding to impermeable walls,
where the subscript $\perp$ implies the component perpendicular to
the boundary. Another conserved quantity is toroidal flux, defined
as 
\[
\Phi=\int B_{\phi}\,dr\,dz=\int\frac{f}{r}\,dr\,dz=\frac{1}{2\pi}\int\frac{f}{r^{2}}2\pi r\,dr\,dz=\frac{1}{2\pi}\int\frac{f}{r^{2}}\,dV
\]
Equation \ref{eq:479.4} implies that the rate of change of system
toroidal flux is 
\begin{equation}
\dot{\Phi}=\frac{1}{2\pi}\int\nabla\cdot\left(-\frac{f}{r^{2}}\mathbf{v}+\omega\mathbf{B}+\frac{\eta}{r^{2}}\nabla f\right)\,dV=\frac{1}{2\pi}\int\left(-\frac{f}{r^{2}}\mathbf{v}+\omega\mathbf{B}+\frac{\eta}{r^{2}}\nabla f\right)\cdot d\boldsymbol{\Gamma}\label{eq:480.5}
\end{equation}
With appropriate boundary conditions, for example $\mathbf{v}|_{\Gamma}=\mathbf{0}$
and $\left(\nabla_{\perp}f\right)|_{\Gamma}=0$, or $\mathbf{v}_{\perp}|_{\Gamma}=\mathbf{0}$
and $\left(B_{\theta\perp}\,v_{\phi}+\eta\left(\frac{\nabla_{\perp}\,f}{r}\right)\right)|_{\Gamma}=0$,
toroidal flux is conserved. Note that the electric field at the system
boundary is 
\begin{align*}
\mathbf{E}|_{\Gamma} & =\left(-\mathbf{v\times}\mathbf{B}+\eta\nabla\times\mathbf{B}\right)|_{\Gamma}\\
 & =\left(-\mathbf{v_{\phi}\times}\mathbf{B}_{\theta}-\mathbf{v_{\theta}\times}\mathbf{B}_{\phi}-\mathbf{v_{\theta}\times}\mathbf{B}_{\theta}+\eta\left(-\Delta^{*}\psi\,\nabla\phi+\nabla f\times\nabla\phi\right)\right)|_{\Gamma} & \mbox{\mbox{}}\\
 & =\left(-\mathbf{v_{\phi}\times}\left(\mathbf{B}_{\theta\parallel}+\mathbf{B}_{\theta\perp}\right)-B_{\phi}\,\left(\mathbf{v_{\theta\perp}}+\mathbf{v_{\theta\parallel}}\right)\times\widehat{\boldsymbol{\phi}}+(\mathbf{v}\cdot\nabla\psi)\nabla\phi+\eta\left(-\Delta^{*}\psi\,\nabla\phi+\frac{\nabla f}{r}\mathbf{\times}\widehat{\boldsymbol{\phi}}\right)\right)|_{\Gamma}
\end{align*}
Here, the subscript $\parallel$ implies the component parallel to
the boundary. Hence, the poloidal component of the electric field
parallel to the boundary is 
\begin{equation}
E_{\theta\parallel}|_{\Gamma}=\left(B_{\theta\perp}\,v_{\phi}-B_{\phi}\,v_{\theta\perp}+\eta\frac{\nabla_{\perp}f}{r}\right)|_{\Gamma}\label{eq:480.51}
\end{equation}
Therefore, either of the sets of boundary conditions listed above
for toroidal flux conservation correspond to having the poloidal component
of the electric field perpendicular to the boundary, the condition,
in the case of azimuthal symmetry, for perfectly electrically conducting
walls.

Conservation of angular momentum is established after noting that
in the axisymmetric case 
\[
(\mathbf{v}\cdot\nabla\mathbf{v})_{\phi}=v_{r}\frac{\partial v_{\phi}}{\partial r}+\frac{v_{r}v_{\phi}}{r}+v_{z}\frac{\partial v_{\phi}}{\partial z}=\frac{v_{r}}{r}\frac{\partial(rv_{\phi})}{\partial r}+\frac{v_{z}}{r}\frac{\partial(rv_{\phi})}{\partial z}=\frac{1}{r}\mathbf{v}\cdot\nabla(rv_{\phi})
\]
and that the $\phi$ coordinate of the divergence of the viscous stress
tensor, which is transpose symmetric (see equation \ref{eq:238}),
is 
\[
(\nabla\cdot\overline{\boldsymbol{\pi}})_{\phi}=\frac{1}{r}\frac{\partial}{\partial r}(r\pi_{r\phi})+\frac{\partial}{\partial z}(\pi_{z\phi})+\frac{\pi_{r\phi}}{r}=\frac{1}{r^{2}}\frac{\partial}{\partial r}(r^{2}\pi_{r\phi})+\frac{1}{r^{2}}\frac{\partial}{\partial z}(r^{2}\pi_{z\phi})=\frac{1}{r}\nabla\cdot(r\pi_{r\phi}\hat{\mathbf{r}}+r\pi_{z\phi}\hat{\mathbf{z}})
\]
The continuity equation for angular momentum density is then 
\begin{eqnarray}
 &  & \frac{\partial}{\partial t}(\rho v_{\phi}r)=m_{i}\dot{n}v_{\phi}r+\rho\dot{v}_{\phi}r\nonumber \\
 & = & -\biggl[m_{i}v_{\phi}r\nabla\cdot(n\mathbf{v})+\rho\mathbf{v}\cdot\nabla(rv_{\phi})\biggr]-\nabla\cdot(r\pi_{r\phi}\hat{\mathbf{r}}+r\pi_{z\phi}\hat{\mathbf{z}})+\frac{1}{\mu_{0}}\mathbf{B}\cdot\nabla f\label{eq:480.6}\\
 & = & -\nabla\cdot\bigg(\rho v_{\phi}r\mathbf{v}+r\pi_{r\phi}\hat{\mathbf{r}}+r\pi_{z\phi}\hat{\mathbf{z}}-\frac{f}{\mu_{0}}\mathbf{B}\bigg)\nonumber 
\end{eqnarray}
Hence, the rate of change of total system angular momentum is 
\[
\dot{P}_{\phi}=\frac{\partial}{\partial t}\left(\int\left(\rho rv_{\phi}\right)dV\right)=-\int\left(\rho v_{\phi}r\mathbf{v}+r\pi_{r\phi}\hat{\mathbf{r}}+r\pi_{z\phi}\hat{\mathbf{z}}-\frac{f}{\mu_{0}}\mathbf{B}\right)\cdot d\boldsymbol{\Gamma}
\]
With appropriate boundary conditions, total system angular momentum
is conserved. One set of appropriate boundary conditions is as follows.
The first term here vanishes with the boundary conditions for impermeable
walls, $\mathbf{v}_{\perp}|_{\Gamma}=\mathbf{0}$. Referring to equation
\ref{eq:238}, it can be seen that with the boundary condition $\left(\nabla_{\perp}\omega\right)|_{\Gamma}=0$,
the second and third terms vanish too. The boundary condition $\left(\nabla_{\perp}\omega\right)|_{\Gamma}=0$
is physical only in unusual cases; for example, there may be no viscosity
at the wall due to rotation of the wall. The last term is zero with
the boundary condition $\left(\nabla_{\parallel}\psi\right)|_{\Gamma}=0$
($i.e.,$ $\psi|_{\Gamma}=C$ (constant)), corresponding to having
the magnetic field parallel to the boundary.

The rate of change of the system's total energy is the sum of the
rates of change of kinetic, thermal, and magnetic energies: {\small{}
\begin{align*}
\dot{U}_{Total} & =\dot{U}_{K}+\dot{U}_{Th}+\dot{U}_{M}=\int\frac{\partial}{\partial t}\left[\frac{1}{2}\rho v^{2}+\frac{p}{\gamma-1}+\frac{B^{2}}{2\mu{}_{0}}\right]\:dV
\end{align*}
}Using the vector identity 
\begin{equation}
(\mathbf{A}\times\mathbf{B})\cdot(\mathbf{C}\times\mathbf{D})=(\mathbf{A}\cdot\mathbf{C})(\mathbf{B}\cdot\mathbf{D})-(\mathbf{A}\cdot\mathbf{D})(\mathbf{B}\cdot\mathbf{C})\label{eq:480.71}
\end{equation}
$B^{2}=(\nabla\psi\times\nabla\phi)\cdot(\nabla\psi\times\nabla\phi)=(\nabla\psi\cdot\nabla\psi)(\nabla\phi\cdot\nabla\phi)-(\nabla\psi\cdot\nabla\phi)(\nabla\phi\cdot\nabla\psi)+\left(\frac{f}{r}\right)^{2}$.
With azimuthal symmetry, $\nabla\psi\cdot\nabla\phi=0$, so that $B^{2}=\left(\frac{\nabla\psi}{r}\right)^{2}+\left(\frac{f}{r}\right)^{2}$.

{\small{}
\begin{align}
\Rightarrow\dot{U}_{Total} & =\dot{U}_{K}+\dot{U}_{Th}+\dot{U}_{M}=\int\left[\left(\frac{1}{2}\dot{\rho}v^{2}+\rho\mathbf{v}\cdot\dot{\mathbf{v}}\right)+\frac{\dot{p}}{\gamma-1}+\frac{1}{2\mu{}_{0}}\left(\frac{\partial}{\partial t}\left(\left(\frac{\nabla\psi}{r}\right)^{2}+\left(\frac{f}{r}\right)^{2}\right)\right)\right]\:dV\nonumber \\
\nonumber \\
\label{eq:480.61}
\end{align}
}Using the continuous MHD equations, this may be evaluated as
\begin{align}
\dot{U}_{Total} & =\int\nabla\cdot(\Sigma_{\beta}(\mathbf{v}_{\beta}Y_{\beta})+\mathbf{S}+\mathbf{q})\,dV\label{eq:480.7}
\end{align}
Here, the subscript $\beta$ refers to the spatial coordinates, $Y_{\beta}$
are complicated expressions involving various physical parameters
$(\eta,\,\gamma,\,\mu_{0}$), and fields $\rho,\,v_{\beta},\,...$
etc., $\mathbf{S}\,[\mbox{W/m}^{2}]\mathbf{=E\times}\mathbf{H}$ is
the Poynting vector, and $\mathbf{q}\mbox{\,[W/m}^{2}]$ is the heat
flux density vector. When the system is surrounded by solid walls,
$\mathbf{v}_{\perp}|_{\Gamma}=\mathbf{0}$. Applying the divergence
theorem,

\begin{align*}
\dot{U}_{Total} & =\int(\mathbf{S}+\mathbf{q})\,\cdot d\boldsymbol{\Gamma}
\end{align*}
Thus, system energy is conserved, apart from the fluxes of electromagnetic
and thermal energy lost through the system boundary. If additional
boundary conditions correspond to perfectly electrically conducting
and thermally insulating walls, then there will be no energy fluxes
out of the system, and total energy will be conserved. System energy
conservation with appropriate boundary conditions can be proved for
the continuous MHD equations as follows. Note that $-\mathbf{v}\cdot\nabla\mathbf{v}=-\nabla\left(v^{2}/2\right)+\mathbf{v}\times(\nabla\times\mathbf{v})$.
Referring to equation \ref{eq:480.1}, and using equation \ref{eq:480.71}
again, note also that $J^{2}=\frac{1}{\mu_{0}^{2}}\left(\left(\frac{\Delta^{*}\psi}{r}\right)^{2}+\left(\frac{\nabla f}{r}\right)^{2}\right)$.
Expressing the total system energy as the sum of kinetic, thermal
and magnetic energy, and referring to equations \ref{eq: 479.01}
- \ref{eq:479.4}, the continuity equation for energy density is 
\begin{eqnarray}
\dot{u}_{Total} & = & \dot{u}_{K}+\dot{u}_{Th}+\dot{u}_{M}\nonumber \\
 & = & \frac{\partial}{\partial t}\left(\frac{\rho\mathbf{v}^{2}}{2}+\frac{p}{\gamma-1}+\frac{1}{2\mu_{0}r^{2}}\left(\left(\nabla\psi\right){}^{2}+f^{2}\right)\right)\nonumber \\
 & = & \frac{m_{i}\dot{n}\mathbf{v}^{2}}{2}+\rho\mathbf{v}\cdot\dot{\mathbf{v}}+\frac{\dot{p}}{\gamma-1}+\frac{1}{\mu_{0}r^{2}}\left(\nabla\psi\cdot\left(\nabla\dot{\psi}\right)+f\dot{f}\right)\nonumber \\
 & = & -\left[\frac{m_{i}\mathbf{v}^{2}}{2}\nabla\cdot(n\mathbf{v})+\rho\mathbf{v}\cdot\nabla\frac{\mathbf{v}^{2}}{2}\right]\nonumber \\
 &  & -\left[\mathbf{v}\cdot\nabla p+\frac{\mathbf{v}\cdot\nabla p+\gamma p\nabla\cdot\mathbf{v}}{\gamma-1}\right]\nonumber \\
 &  & -\left[\frac{1}{\mu_{0}r^{2}}\Delta^{*}\psi(\mathbf{v}\cdot\nabla\psi)+\frac{1}{\mu_{0}r^{2}}\nabla\psi\cdot\nabla(\mathbf{v}\cdot\nabla\psi)\right]\nonumber \\
 &  & -\left[\frac{1}{\mu_{0}}\frac{f}{r^{2}}(\mathbf{v}\cdot\nabla f)+\frac{1}{\mu_{0}}f\,\nabla\cdot\bigg(\frac{f}{r^{2}}\mathbf{v}\bigg)\right]\nonumber \\
 &  & +\left[\frac{1}{\mu_{0}}\omega\mathbf{B}\cdot\nabla f+\frac{1}{\mu_{0}}f\nabla\cdot\bigg(\omega\mathbf{B}\bigg)\right]\nonumber \\
 &  & +\left[\frac{\eta(\Delta^{*}\psi)^{2}}{\mu_{0}r^{2}}+\frac{1}{\mu_{0}r^{2}}\nabla\psi\cdot\nabla(\eta\Delta^{*}\psi)\right]\nonumber \\
 &  & +\left[\frac{\eta(\nabla f)^{2}}{\mu_{0}r^{2}}+\frac{f}{\mu_{0}}\nabla\cdot\bigg(\frac{\eta}{r^{2}}\nabla f\bigg)\right]\label{eq:481.1-1}\\
 &  & -\biggl[\nabla\cdot(\mathbf{q}_{i}+\mathbf{q}_{e})\biggr]\nonumber \\
 &  & -\biggl[\mathbf{v}\cdot(\nabla\cdot\overline{\boldsymbol{\pi}})+\overline{\boldsymbol{\pi}}:\nabla\mathbf{v}\biggr]\nonumber \\
 & = & -\nabla\cdot\biggl(\frac{m_{i}n\mathbf{v}^{2}}{2}\mathbf{v}+\frac{\gamma p}{\gamma-1}\mathbf{v}+\frac{(\mathbf{v}\cdot\nabla\psi)}{\mu_{0}r^{2}}\nabla\psi\nonumber \\
 &  & +\frac{f^{2}}{\mu_{0}r^{2}}\mathbf{v}-\frac{1}{\mu_{0}}\omega f\,\mathbf{B}-\frac{\eta\Delta^{*}\psi}{\mu_{0}r^{2}}\nabla\psi-\frac{\eta f}{\mu_{0}r^{2}}\nabla f+\mathbf{q}_{i}+\mathbf{q}_{e}+\overline{\boldsymbol{\pi}}\cdot\mathbf{v}\biggr)\nonumber 
\end{eqnarray}
Note that the five magnetic terms involving $\mu_{0}$ in the final
full-divergence expression constitute the Poynting flux. Taking the
integral over the final expression for $\dot{u}_{total}$ over the
system volume, and applying Gauss's theorem, it can be seen how total
system energy is conserved with appropriate boundary conditions, for
example $\mathbf{v}|_{\Gamma}=\mathbf{0},\,\mathbf{q}_{i\perp}|_{\Gamma}=\mathbf{q}_{e\perp}|_{\Gamma}=\mathbf{0},\,(\nabla_{\perp}\psi)|_{\Gamma}=0\mbox{ and }(\nabla_{\perp}f)|_{\Gamma}=0$,
or $\mathbf{v}_{\perp}|_{\Gamma}=\mathbf{0},\,\mathbf{q}_{i\perp}|_{\Gamma}=\mathbf{q}_{e\perp}|_{\Gamma}=\mathbf{0},\,(\nabla_{\perp}\psi)|_{\Gamma}=0$
and\\
$\left(B_{\theta\perp}\,v_{\phi}+\eta\left(\frac{\nabla_{\perp}\,f}{r}\right)\right)|_{\Gamma}=0$.
Referring to equation \ref{eq:480.51}, note that either of these
sets of boundary conditions also eliminates energy loss associated
with Poynting flux. Note that the terms $+\frac{(\mathbf{v}\cdot\nabla\psi)}{\mu_{0}r^{2}}\nabla\psi$
and $-\frac{\eta\Delta^{*}\psi}{\mu_{0}r^{2}}\nabla\psi$ in the final
full-divergence expression in equation \ref{eq:481.1-1} constitute
the part of the Poynting flux that arises due to the toroidal component
of the electric field at the boundary. The boundary condition $(\nabla_{\perp}\psi)|_{\Gamma}=0$
or $(\Delta^{*}\psi)|_{\Gamma}=0$ is required to eliminate this contribution.
Referring to equation \ref{eq:479.3}, it can be seen that the combination
of boundary conditions $\mathbf{v}_{\perp}|_{\Gamma}=\mathbf{0}$
and $\psi|_{\Gamma}=0$ automatically leads to the boundary condition
$(\Delta^{*}\psi)|_{\Gamma}=0$. Therefore, the boundary condition
$(\nabla_{\perp}\psi)|_{\Gamma}=0$ included above in the lists of
requirements for maintenance of system energy conservation may be
replaced with the requirement $\psi|_{\Gamma}=0$. 

\section{Discrete form of MHD equations and conservation properties\label{sec:Discrete-form_cons_props}}

In this section, the full set of discretised equations for a two-temperature
MHD model is presented. The model has been constructed so as to preserve
the global conservation laws inherent to the original continuous system
of equations. Referring to the properties of the differential operators,
it will be demonstrated how the conservation characteristics associated
with the continuous form of the MHD equations (see section \ref{sec:Axisymmetric-two-temperature-MHD})
are replicated by the discrete form. In order to preserve the conservation
laws of the system in discrete form, the pairs of terms which constitute
full divergences, denoted with square brackets in equations \ref{eq:480.6}
and \ref{eq:481.1-1}, have been discretised in compatible way, $i.e.,$
the appropriate corresponding discrete operators are used in these
terms, and with particular boundary conditions, the operator properties
lead to the cancellation of each pair.

\subsection{Discretised MHD model\label{subsec:Discretised-MHD-model}}

The discretised equations for the axisymmetric MHD model, in a form
that ensures conservation of total system energy, are presented in
appendix summary \ref{sec:SummaryB}. The discrete form of the single
fluid energy equation (equation \ref{eq:127.112}) is partitioned
into parts pertaining to the ions and electrons, with the inclusion
of discrete forms of the species thermal diffusion and heat exchange
terms. With the inclusion of artificial density diffusion, the resulting
set of discretised equations is:\\
 
\begin{align}
\underline{\dot{n}} & =-\underline{\underline{\nabla}}\cdot(\underline{n}\circ\underline{\mathbf{v}})+\underline{\underline{\nabla_{n}}}\cdot\left(\underline{\zeta}\circ\underline{\underline{\nabla^{e}}}\,\,\underline{n}\right)\nonumber \\
\underline{\dot{v}_{r}} & =\underbrace{-\underline{\underline{Dr}}*\left(\frac{\underline{v}^{2}}{2}\right)-\underline{v_{z}}\circ\left(\underline{\underline{Dz}}*\underline{v_{r}}-\underline{\underline{Dr}}*\underline{v_{z}}\right)+\underline{v_{\phi}}\circ\left(\underline{\underline{Dr}}*\left(\underline{r}\circ\underline{v_{\phi}}\right)\right)\oslash\underline{r}}_{\mathclap{{-(\mathbf{v\cdot\nabla}\mathbf{v})_{r}=\left(-\nabla(v^{2}/2)+\mathbf{v}\times(\nabla\times\mathbf{v})\right)_{r}}}}\,\,\underbrace{-\left(\underline{\underline{Dr}}*\underline{p}\right)\oslash\underline{\rho}}_{{-\frac{1}{\rho}(\nabla p)_{r}}}\nonumber \\
 & \,\,\,\,\,\,\,\underbrace{-\underline{\varPi_{r}}\oslash\underline{\rho}}_{-\frac{1}{\rho}(\nabla\cdot\overline{\boldsymbol{\pi}})_{r}}\,\,+\underbrace{\left(\underset{}{-(\underline{\underline{Dr}}*\underbar{\ensuremath{\psi}})}\circ\left(\underline{\underline{\Delta^{^{*}}}}\,\,\underline{\psi}\right)-\left(\underline{f}\circ\left(\underline{\underline{Dr}}*\underline{f}\right)\right)\right)\oslash\left(\mu_{0}\,\underline{r}{}^{2}\circ\underline{\rho}\right)}_{\frac{1}{\rho}(\mathbf{J\times}\mathbf{B})_{r}}+\underline{f_{\zeta r}}\oslash\underline{\rho}\nonumber \\
\underline{\dot{v}_{\phi}} & =\underbrace{-\underline{\mathbf{v}}\cdot\left(\underline{\underline{\nabla}}\,\,\left(\underline{r}\circ\underline{v_{\phi}}\right)\right)\oslash\underline{r}}_{\mathclap{{-(\mathbf{v\cdot\nabla}\mathbf{v})_{\phi}=\left(-\nabla(v^{2}/2)+\mathbf{v}\times(\nabla\times\mathbf{v})\right)_{\phi}}}}\,\,\,\,\,\,\,\,\,\,\underbrace{-\underline{\varPi_{\phi}}\oslash\underline{\rho}}_{-\frac{1}{\rho}(\nabla\cdot\overline{\boldsymbol{\pi}})_{\phi}}+\underbrace{\left(\underline{\underline{W_{n}}}*\left(\underline{\mathbf{B}_{\theta}^{e}}\cdot\left(\underline{\underline{\nabla^{e}}}\,\,\underline{f}\right)\right)\right)\oslash\left(\mu_{0}\,\underline{r}\circ\underline{\rho}\right)}_{\frac{1}{\rho}(\mathbf{J\times}\mathbf{B})_{\phi}}+\underline{f_{\zeta\phi}}\oslash\underline{\rho}\nonumber \\
\underline{\dot{v}_{z}} & =\underbrace{-\underline{\underline{Dz}}*\left(\frac{\underline{v}^{2}}{2}\right)+\underline{v_{r}}\circ\left(\underline{\underline{Dz}}*\underline{v_{r}}-\underline{\underline{Dr}}*\underline{v_{z}}\right)+\underline{v_{\phi}}\circ\left(\underline{\underline{Dz}}*\left(\underline{r}\circ\underline{v_{\phi}}\right)\right)\oslash\underline{r}}_{\mathclap{{-(\mathbf{v\cdot\nabla}\mathbf{v})_{z}=\left(-\nabla(v^{2}/2)+\mathbf{v}\times(\nabla\times\mathbf{v})\right)_{z}}}}\,\,\underbrace{-\left(\underline{\underline{Dz}}*\underline{p}\right)\oslash\underline{\rho}}_{{-\frac{1}{\rho}(\nabla p)_{z}}}\,\,\nonumber \\
 & \,\,\,\,\,\,\,\underbrace{-\underline{\varPi_{z}}\oslash\underline{\rho}}_{-\frac{1}{\rho}(\nabla\cdot\overline{\boldsymbol{\pi}})_{z}}\,\,+\underbrace{\left(-\left(\underline{\underline{Dz}}*\underbar{\ensuremath{\psi}}\right)\circ\left(\underline{\underline{\Delta^{^{*}}}}\,\,\underline{\psi}\right)-\underline{f}\circ\left(\underline{\underline{Dz}}*\underline{f}\right)\right)\oslash\left(\mu_{0}\,\underline{r}{}^{2}\circ\underline{\rho}\right)}_{\frac{1}{\rho}(\mathbf{J\times}\mathbf{B})_{z}}+\underline{f_{\zeta z}}\oslash\underline{\rho}\nonumber \\
\underline{\dot{p}_{i}} & =-\underline{\mathbf{v}}\cdot\left(\underline{\underline{\nabla}}\,\,\underline{p_{i}}\right)-\gamma\,\underline{p_{i}}\circ\left(\underline{\underline{\nabla}}\cdot\underline{\mathbf{v}}\right)+(\gamma-1)\,\left[-\underline{\underline{\nabla_{n}}}\cdot\underline{\mathbf{q}_{i}^{e}}+\underline{Q_{ie}}+\underbrace{\underline{Q_{\pi}}}_{-\overline{\boldsymbol{\pi}}:\nabla\mathbf{v}}+\underline{Q_{\zeta}}\right]\nonumber \\
\underline{\dot{p}_{e}} & =-\underline{\mathbf{v}}\cdot\left(\underline{\underline{\nabla}}\,\,\underline{p_{e}}\right)-\gamma\,\underline{p_{e}}\circ\left(\underline{\underline{\nabla}}\cdot\underline{\mathbf{v}}\right)+(\gamma-1)\,\biggl[-\underline{\underline{\nabla_{n}}}\cdot\underline{\mathbf{q}_{e}^{e}}-\underline{Q_{ie}}\nonumber \\
 & \,\,\,\,\,\,\,+\underset{\eta'J_{\phi}^{2}}{\underbrace{\left(\underline{\eta}/\mu_{0}\right)\circ\left(\left(\underline{\underline{\Delta^{^{*}}}}\,\,\underline{\psi}\right)\oslash\underline{r}\right)^{2}}}+\underset{\eta'J_{\theta}^{2}}{\underbrace{\underline{\underline{W_{n}}}*\left(\left(\underline{\eta^{e}}/\mu_{0}\right)\circ\left(\left(\underline{\underline{\nabla^{e}}}\,\,\underline{f}\right)\oslash\underline{r^{e}}\right)^{2}\right)}}\biggr]\nonumber \\
\underline{\dot{\psi}} & =-\underline{\mathbf{v}}\cdot\left(\underline{\underline{\nabla}}\,\,\underline{\psi}\right)+\underline{\eta}\circ\left(\underline{\underline{\Delta^{^{*}}}}\,\,\underline{\psi}\right)\nonumber \\
\underline{\dot{f}} & =\underline{r^{2}}\circ\left[-\underline{\underline{\nabla}}\cdot\left(\underline{f}\circ\underline{\mathbf{v}}\oslash\underline{r}^{2}\right)+\underline{\underline{\nabla_{n}}}\cdot\left(\underline{\mathbf{B}_{\theta}^{e}}\circ\underline{\omega^{e}}\right)+\underline{\underline{\nabla_{n}}}\cdot\left(\underline{\eta^{e}}\circ\left(\underline{\underline{\nabla^{e}}}\,\,\underline{f}\right)\oslash\underline{r^{e}}^{2}\right)\right]\label{eq:517.3}
\end{align}
Here, $\underline{\eta^{e}}=<\underline{\eta}>^{e}$, and $\underline{\omega^{e}}=<\underline{v_{\phi}}\oslash\underline{r}>^{e}$
(equation \ref{eq:502.31}). $\underline{\underline{W_{n}}}$ is the
volume-averaging operator (equations \ref{eq:515} and \ref{eq:515.1}),
and $\underline{\mathbf{B}_{\theta}^{e}}=\left(-\left(\underline{\underline{Dz^{e}}}*\underline{\psi}\right)\hat{\mathbf{r}}+\left(\underline{\underline{Dr^{e}}}*\underline{\psi}\right)\hat{\mathbf{z}}\right)\oslash\underline{r^{e}}$.
These discretised equations are written in a form which is a direct
analogue of their continuous representations with inclusion of several
extra terms. Namely, the continuity equation has an artificial density
diffusion term, which is required for density smoothing and avoiding
negative density regions. The density diffusion coefficient $\zeta\,[\mbox{m}^{2}/\mbox{s}]$
can be spatially held constant or can be increased where density gradients
are high or density values approach zero. The components of force
per volume vector $\underline{\mathbf{f}_{\zeta}}=(\underline{f_{\zeta r}},\,\underline{f_{\zeta\phi}},\,\underline{f_{\zeta z}})^{T}$
in the velocity equations and the heating term $\underline{Q_{\zeta}}$
in the ion pressure equation are included to cancel the effect of
artificial density diffusion on the total system momentum and energy. 

To close the system of equations, we need to specify the forms of
the viscous terms $\underline{\boldsymbol{\varPi}}$ and $\underline{Q_{\pi}}$,
the species heat exchange term $\underline{Q_{ie}}$, the resistive
diffusion coefficient $\underline{\eta}$, and the heat flux density
terms $\underline{\mathbf{q}_{i}^{e}}$ and $\underline{\mathbf{q}_{e}^{e}}$.
As mentioned in appendix \ref{sec:Formulation-of-discretized}, we
have implemented, for simplicity, the unmagnetised version of the
viscous stress tensor. The components of $\underline{\boldsymbol{\varPi}}$
represent the discrete forms of the components of $\nabla\cdot\boldsymbol{\underline{\pi}}$,
and are defined by equation \ref{eq:127.1111}: 
\begin{align}
\underline{\varPi_{r}} & =\biggl[\underset{}{-2\underline{\underline{Dr_{n}}}*\left(\underline{\mu^{e}}\circ\underline{r^{e}}\circ\left(\underline{\underline{Dr^{e}}}*\underline{v_{r}}\right)\right)}-\underset{}{\underline{\underline{Dz_{n}}}*\left(\underline{\mu^{e}}\circ\underline{r^{e}}\circ\left(\underline{\underline{Dr^{e}}}*\underline{v_{z}}+\underline{\underline{Dz^{e}}}*\underline{v_{r}}\right)\right)}\biggr]\oslash\underline{r}\nonumber \\
 & \underset{}{+\frac{2}{3}\left(\underline{\underline{Dr_{n}}}*\left(\underline{\mu^{e}}\circ\left(\underline{\underline{\nabla^{e}}}\cdot\underline{\mathbf{v}}\right)\right)\right)}+\underset{}{2\,\underline{\mu}\circ\underline{v_{r}}\oslash\underline{r}^{2}}\nonumber \\
\underline{\varPi_{\phi}} & =-\left(\underline{\underline{\nabla_{n}}}\cdot\left(\underline{\mu^{e}}\circ\underline{r^{e}}^{2}\circ\left(\underline{\underline{\nabla^{e}}}\,\,\underline{\omega}\right)\right)\right)\oslash\underline{r}\nonumber \\
\underline{\varPi_{z}} & =\biggl[\underset{}{-2\underline{\underline{Dz_{n}}}*\left(\underline{\mu^{e}}\circ\underline{r^{e}}\circ\left(\underline{\underline{Dz^{e}}}*\underline{v_{z}}\right)\right)}-\underset{}{\underline{\underline{Dr_{n}}}*\left(\underline{\mu^{e}}\circ\underline{r^{e}}\circ\left(\underline{\underline{Dr^{e}}}*\underline{v_{z}}+\underline{\underline{Dz^{e}}}*\underline{v_{r}}\right)\right)}\biggr]\oslash\underline{r}\nonumber \\
 & \underset{}{+\frac{2}{3}\left(\underline{\underline{Dz_{n}}}*\left(\underline{\mu^{e}}\circ\left(\underline{\underline{\nabla^{e}}}\cdot\underline{\mathbf{v}}\right)\right)\right)}\label{eq:517.4}
\end{align}
Here, $\underline{\mu^{e}}=<\underline{\mu}>^{e}$ (equation \ref{eq:502.31}).
The discrete form of $-\overline{\boldsymbol{\pi}}:\nabla\mathbf{v}$,
which appears in the expression for $\underline{\dot{p}_{i}}$, is
given by equation \ref{eq:127.113}: 
\begin{align}
\underline{Q_{\pi}} & =\underline{\underline{W_{n}}}*\biggl[\underline{\mu^{e}}\circ\biggl\{2\left(\underline{\underline{Dr^{e}}}*\underline{v_{r}}\right)^{2}+2\left(\underline{\underline{Dz^{e}}}*\underline{v_{z}}\right)^{2}+\left(\underline{r^{e}}\circ\left(\underline{\underline{\nabla^{e}}}\,\,\underline{\omega}\right)\right)^{2}\nonumber \\
 & +\left(\underline{\underline{Dr^{e}}}*\underline{v_{z}}+\underline{\underline{Dz^{e}}}*\underline{v_{r}}\right)^{2}-\frac{2}{3}\left(\underline{\underline{\nabla^{e}}}\cdot\underline{\mathbf{v}}\right)^{2}\biggr\}\biggr]+2\,\underline{\mu}\circ\left(\underline{v_{r}}\oslash\underline{r}\right)^{2}\label{eq:517.5}
\end{align}
The representations in equations \ref{eq:517.4} and \ref{eq:517.5}
are consistent with the discrete form of energy conservation, as will
be explicitly demonstrated in section \ref{subsec:Energy-conservation}.
$\underline{\mu^{e}}\,[\mbox{kg}\,\mbox{m}^{-1}\mbox{s}{}^{-1}]=\underline{\rho^{e}}\circ\underline{\nu^{e}}$
specifies dynamic viscosity at element centroids, where $\underline{\nu^{e}}$
{[}m$^{2}\mbox{/s}${]} specifies kinematic viscosity. As discussed
in more detail in appendix \ref{subsec:Diffcoefs}, a certain minimum
level of artificial viscosity, which depends on the time-step, simulation
type, and mesh resolution, is required for numerical stability. Simulations
that involve CT formation and consequent extreme plasma acceleration
and steep gradients in the velocity fields are run with dynamic viscosity
set to constant $\mu=\rho_{0}\nu_{0}$, for typical code input $\nu_{0}\sim700$
{[}m$^{2}\mbox{/s}${]}, where $\rho_{0}\sim6\times10^{-6}$ {[}kg/m$^{3}${]}
is a typical representative mass density. 

The resistive diffusion coefficient may be held spatially constant,
or is based on the Spitzer formula

\begin{equation}
\underline{\eta}=\frac{m_{e}}{1.96\,e^{2}\,\mu_{0}}\oslash\left(Z_{eff}\,\underline{n}\circ\underline{\tau_{ei}}\right)=\left(418\,Z_{eff}\left(\underline{T_{e}}\,[\mbox{eV}]\right)^{-\frac{3}{2}}\right)\,[\mbox{m}^{2}\mbox{/s}]\label{eq:517.6}
\end{equation}
where the electron-ion collision time ($cf.$ equation \ref{eq:472.38})
is 
\[
\underline{\tau_{ei}}=\left(\frac{6\sqrt{2}\pi^{1.5}\epsilon_{0}^{2}\sqrt{m_{e}}}{\varLambda\,e^{4}Z_{eff}^{2}}\right)\,\left(\underline{T_{e}}\,[\mbox{J}]\right){}^{\frac{3}{2}}\oslash\underline{n}=\left(3.45\times10^{10}\,\frac{\left(\underline{T_{e}}\,[\mbox{eV}]\right)^{\frac{3}{2}}\oslash\underline{n}\,[\mbox{m}^{-3}]}{Z_{eff}^{2}}\right)\,[\mbox{s}]
\]
The heat exchange term $\underline{Q_{ie}}$ ($cf.$ equation \ref{eq:481.5})
gives the rate at which energy is imparted from the electrons to the
ions due to collisions between ion and electron fluids: 
\begin{align}
\underline{Q_{ie}} & =3(m_{e}/m_{i})\,Z_{eff}\,\underline{n}\circ\left(\underline{T_{e}}[\mbox{J}]-\underline{T_{i}}[\mbox{J}]\right)\oslash\underline{\tau_{ei}}\nonumber \\
 & =\left(7.6\times10^{-33}\,Z_{eff}^{3}\,\left(\underline{T_{e}}[\mbox{eV}]-\underline{T_{i}}[\mbox{eV}]\right)\circ\left(\underline{T_{e}}\,[\mbox{eV}]\right)^{-\frac{3}{2}}\circ\left(\underline{n}\,[\mbox{m}^{-3}]\right)^{2}/\mu_{i}\right)\,[\mbox{W/\ensuremath{\mbox{m}^{3}}}]\label{eq:517.61}
\end{align}
Here, $\mu_{i}$ is the ion mass in units of proton mass.

Gyrorotation of plasma particles along magnetic field lines implies
that cross-field thermal diffusion is impeded, while temperature equalizes
quickly along the field lines. To include the effect of significantly
enhanced thermal diffusion parallel to the magnetic field, we include
anisotropy in the model for thermal diffusion. The species heat flux
density can be expressed as (equation \ref{eq:481.41}) 
\begin{align*}
\mathbf{q}_{\alpha} & =-\left(\kappa_{\parallel\alpha}\nabla_{\parallel}T_{\alpha}+\kappa{}_{\perp\alpha}\nabla_{\perp}T_{\alpha}\right)\\
 & =-\left(\kappa_{\parallel\alpha}\nabla_{\parallel}T_{\alpha}-\kappa{}_{\perp\alpha}\nabla_{\parallel}T_{\alpha}+\kappa{}_{\alpha\perp}\nabla_{\parallel}T_{\alpha}+\kappa{}_{\perp\alpha}\nabla_{\perp}T_{\alpha}\right)\\
 & =-\left(\left(\kappa_{\parallel\alpha}-\kappa_{\perp\alpha}\right)\nabla_{\parallel}T_{\alpha}+\kappa_{\perp\alpha}\nabla T_{\alpha}\right)
\end{align*}
where $\kappa_{\parallel\alpha}$ and $\kappa_{\perp\alpha}$ {[}(m-s)$^{-1}${]}
are the thermal conductivities for species $\alpha$, pertaining to
thermal diffusion parallel and perpendicular to the magnetic field.
With azimuthal symmetry, the toroidal component of $\nabla_{\parallel}T_{\alpha}$
can be dropped as it will not make a finite contribution to $\nabla\cdot\mathbf{q}_{\alpha}$,
so that the discrete forms of $\mathbf{q}_{\alpha}$ may be expressed
as 
\begin{align}
\underline{\mathbf{q}_{\alpha}^{e}} & =-\left\{ \left(\underline{\kappa_{\parallel\alpha}^{e}}-\underline{\kappa_{\perp\alpha}^{e}}\right)\circ\left(\underline{\mathbf{B}_{\theta}^{e}}\circ\left(\underline{\mathbf{B}_{\theta}^{e}}\cdot\left(\underline{\underline{\nabla^{e}}}\,\,\underline{T_{\alpha}}\right)\right)\oslash\underline{B^{e}}^{2}\right)+\underline{\kappa_{\perp\alpha}^{e}}\circ\left(\underline{\underline{\nabla^{e}}}\,\,\underline{T_{\alpha}}\right)\right\} \label{eq:517.7}
\end{align}
Physically, the thermal conductivities vary with local conditions;\\
$e.g.,$ $\kappa_{\parallel\alpha}(\mathbf{r},t)=n_{\alpha}(\mathbf{r},t)\,\chi_{\parallel\alpha}(\mathbf{r},t)$.
As discussed in more detail in appendix \ref{subsec:Diffcoefs}, simulations
of the magnetic compression experiment were usually run with constant
conductivities, of the order $\kappa_{\parallel\alpha}=n_{0}\chi_{\parallel\alpha}$
and $\kappa_{\perp\alpha}=n_{0}\chi_{\perp\alpha}$, where $n_{0}\sim1\times10^{21}$
{[}m$^{-3}${]} is a typical representative number density, and the
thermal diffusion coefficients are held constant, for example $\chi_{\parallel e}\sim16000,\,\chi_{\parallel i}\sim5000,\,\chi_{\perp e}\sim240$
and $\chi_{\perp i}\sim120$ $[\mbox{m}^{2}\mbox{/s}]$. The high
parallel coefficients represent rapid equilibration of temperature
along field lines, while the perpendicular coefficients are chosen
to match the experimentally observed resistive decay rate of CT currents.
Enhanced thermal diffusion acts as a proxy for anomalous energy sink
mechanisms, including radiative losses. 

\subsection{Conservation properties\label{subsec:Conservation-properties} }

Here, we demonstrate that the discretised system of equations \ref{eq:517.3}
has the same global conservation laws as the original continuous system. 

\subsubsection{Particle count conservation \label{subsec:Particle-count-conservation}}

Integrating the discrete form of the continuity equation over volume,
we obtain the rate of change of the total number of particles in the
system:

\begin{align*}
\dot{N} & =\underline{dV}^{T}*\underline{\dot{n}}\\
 & =\underline{dV}^{T}*\left\{ -\underline{\underline{\nabla}}\cdot\left(\underline{n}\circ\underline{\mathbf{v}}\right)+\underline{\underline{\nabla_{n}}}\cdot\left(\underline{\zeta^{e}}\circ\left(\underline{\underline{\nabla^{e}}}\,\,\underline{n}\right)\right)\right\} 
\end{align*}
With boundary conditions $\underline{\mathbf{v}_{\perp}}|_{\Gamma}=\mathbf{0}$,
or $\underline{\mathbf{v}}|_{\Gamma}=\mathbf{0}$, identity \ref{eq:511.041}
determines that the first term is zero. The second term is always
zero, due to property \ref{eq:515.04}. Hence, total particle count
is conserved. Note that if no boundary conditions are explicitly applied
to density, $\underline{n}$ will automatically evolve to satisfy
the natural boundary condition $\left(\zeta\,\nabla_{\perp}n\right)|_{\Gamma}=0$,
as a consequence of the properties of the element-to-node divergence
operation (section \ref{subsec:Drn}).

\subsubsection{Toroidal flux conservation\label{subsec:Toroidal-flux-conservation}}

Analogous to equation \ref{eq:480.5}, the discrete expression for
the rate of change of system toroidal flux follows from the discrete
expression for $\dot{\underline{f}}$ in equation \ref{eq:517.3}
\begin{align*}
\dot{\Phi} & =\frac{1}{2\pi}\underline{dV}^{T}*\left\{ \dot{\underline{f}}\oslash\underline{r}^{2}\right\} \\
 & =\frac{1}{2\pi}\underline{dV}^{T}*\left\{ -\underline{\underline{\nabla}}\cdot\left(\underline{f}\circ\underline{\mathbf{v}}\oslash\underline{r}^{2}\right)+\underline{\underline{\nabla_{n}}}\cdot\left(\underline{\mathbf{B}_{\theta}^{e}}\circ\underline{\omega^{e}}\right)+\underline{\underline{\nabla_{n}}}\cdot\left(\underline{\eta^{e}}\circ\left(\underline{\underline{\nabla^{e}}}\,\,\underline{f}\right)\oslash\underline{r^{e}}^{2}\right)\right\} 
\end{align*}
Once again, with boundary conditions $\underline{\mathbf{v}_{\perp}}|_{\Gamma}=\mathbf{0}$,
identity \ref{eq:511.041} determines that the first term is zero.
The second and third terms are zero, due to property \ref{eq:515.04}.
Hence, system toroidal flux is conserved. Note that in this case $\underline{f}$
will automatically evolve to satisfy the natural boundary condition
$\left(\mathbf{B}_{\theta\perp}\omega+\eta\left(\nabla_{\perp}\,f\,\right)/r^{2}\right)|_{\Gamma}=0$
if no boundary conditions are explicitly imposed on $\underline{f}$
. In combination with the boundary conditions $\underline{\mathbf{v}_{\perp}}|_{\Gamma}=\mathbf{0}$,
this corresponds to having the poloidal component of the electric
field perpendicular to the boundary, the condition, in the case of
azimuthal symmetry, for a perfectly electrically conducting boundary
(see equation \ref{eq:480.51}). 

Simulations that include CT formation and compression are run with
the boundary conditions $\underline{v_{\beta}}|_{\Gamma}=0$ applied
explicitly to each velocity component, so that the automatically imposed
boundary condition for $f$, corresponding to the physical case of
a perfectly electrically conducting boundary, becomes $\left(\nabla_{\perp}\,f\,\right)|_{\Gamma}=0$.

As described in section \ref{sec:PHIconservation-with}, when part
of the computational boundary is modelled as an electrical insulator,
special care must be taken to define the explicitly applied boundary
condition for $f$ along the insulator boundary, in order to maintain
global toroidal flux conservation. In such cases, only the conducting
boundary regions are allowed to retain the naturally imposed boundary
conditions for $f$. 

\subsubsection{Angular momentum conservation\label{subsec:Angular-momentum-conservation} }

The rate of change of system angular momentum is 
\[
\dot{P}_{\phi}=\frac{\partial}{\partial t}\left(\int\left(\rho rv_{\phi}\right)dV\right)=m_{i}\int\left(rv_{\phi}\dot{n}+nr\dot{v}_{\phi}\right)dV
\]
Here, it will be shown that system angular momentum is conserved for
the discrete model when the terms corresponding to density diffusion
in the discrete expressions for $\dot{n}$ and $\dot{v}_{\phi}$ are
neglected. In section \ref{subsec:Maintenance-of-momentum}, it will
be shown that angular momentum conservation can be maintained even
with the inclusion of these terms. In discrete form, the rate of change
of angular momentum is
\begin{align}
\dot{P}_{\phi} & =m_{i}\,\underline{dV}^{T}*\biggl\{-\left(\underline{r}\circ\underline{v_{\phi}}\right)\circ\left(\underline{\underline{\nabla}}\cdot(\underline{n}\circ\underline{\mathbf{v}})\right)\nonumber \\
 & \,\,\,\,\,\,\,+\left(\underline{n}\circ\underline{r}\right)\circ\biggl\{-\underline{\mathbf{v}}\cdot\left(\underline{\underline{\nabla}}\,\,(\underline{r}\circ\underline{v_{\phi}})\right)\oslash\underline{r}-\underline{\varPi_{\phi}}\oslash\underline{\rho}\nonumber \\
 & \,\,\,\,\,\,\,+\underline{\underline{W_{n}}}*\left(\underline{\mathbf{B}_{\theta}^{e}}\cdot\left(\underline{\underline{\nabla^{e}}}\,\,\underline{f}\right)\right)\oslash\left(\mu_{0}\,\underline{r}\circ\underline{\rho}\right)\biggr\}\biggr\}\label{eq:517.8}
\end{align}
The first two terms here can be simplified to\\
$-m_{i}\,\underline{dV}^{T}*\left\{ \left(\underline{r}\circ\underline{v_{\phi}}\right)\circ\left(\underline{\underline{\nabla}}\cdot(\underline{n}\circ\underline{\mathbf{v}})\right)+\left(\underline{n}\circ\underline{\mathbf{v}}\right)\cdot\left(\underline{\underline{\nabla}}\,\,(\underline{r}\circ\underline{v_{\phi}})\right)\right\} $.
With boundary conditions\\
 $\underline{\mathbf{v}_{\perp}}|_{\Gamma}=\mathbf{0}$, this combination
vanishes due to identity \ref{eq:511.04}. With reference to the definition
for $\underline{\varPi_{\phi}}$ (equation \ref{eq:517.4}), if no
boundary conditions are explicitly applied to $\underline{v_{\phi}}$,
the properties of the element-to-node divergence operation will automatically
impose the natural boundary conditions $\left(\nabla_{\perp}\,\omega\,\right)|_{\Gamma}=0$,
and the third term in equation \ref{eq:517.8} will vanish due to
identity \ref{eq:515.04}. With reference to equation \ref{eq:516.1},
the fourth term in equation \ref{eq:517.8} can be expressed as 
\begin{align*}
 & \frac{1}{\mu_{0}}\,\underline{dV^{e}}^{T}*\left\{ \underline{\mathbf{B}_{\theta}^{e}}\cdot\left(\underline{\underline{\nabla^{e}}}\,\,\underline{f}\right)\right\} \\
 & =\frac{2\pi}{\mu_{0}}\left(\underline{s^{e}}\circ\underline{r^{e}}\right)^{T}*\left\{ \left(-\left(\underline{\underline{Dz^{e}}}*\underline{\psi}\right)\circ\left(\underline{\underline{Dr^{e}}}*\underline{f}\right)+\left(\underline{\underline{Dr^{e}}}*\underline{\psi}\right)\circ\left(\underline{\underline{Dz^{e}}}*\underline{f}\right)\right)\oslash\underline{r^{e}}\right\} \\
 & =\frac{2\pi}{\mu_{0}}\,\underline{s^{e}}^{T}*\left\{ -\left(\underline{\underline{Dz^{e}}}*\underline{\psi}\right)\circ\left(\underline{\underline{Dr^{e}}}*\underline{f}\right)+\left(\underline{\underline{Dr^{e}}}*\underline{\psi}\right)\circ\left(\underline{\underline{Dz^{e}}}*\underline{f}\right)\right\} \\
 & =\frac{2\pi}{\mu_{0}}\left(-\left(\underline{\underline{Dr^{e}}}*\underline{f}\right)^{T}*\left(\underline{\underline{\widehat{S}}}*\left(\underline{\underline{Dz^{e}}}*\underline{\psi}\right)\right)+\left(\underline{\underline{Dz^{e}}}*\underline{f}\right)^{T}*\left(\underline{\underline{\widehat{S}}}*\left(\underline{\underline{Dr^{e}}}*\underline{\psi}\right)\right)\right)\\
 & \frac{2\pi}{\mu_{0}}\,\underline{f}^{T}*\left(\left(-\underline{\underline{Dr^{e}}}^{T}*\underline{\underline{\widehat{S}}}*\underline{\underline{Dz^{e}}}+\underline{\underline{Dz^{e}}}^{T}*\underline{\underline{\widehat{S}}}*\underline{\underline{Dr^{e}}}\right)*\underline{\psi}\right)
\end{align*}
and vanishes due to identity \ref{eq:505.003}, when the boundary
condition $\psi|_{\Gamma}=0$ is applied. Thus for angular momentum
conservation, no boundary conditions are applied on $v_{\phi}$, and
$\psi|_{\Gamma}$ must be set to zero.\\
\begin{figure}[H]
\centering{}\subfloat{\includegraphics[width=13cm,height=7cm]{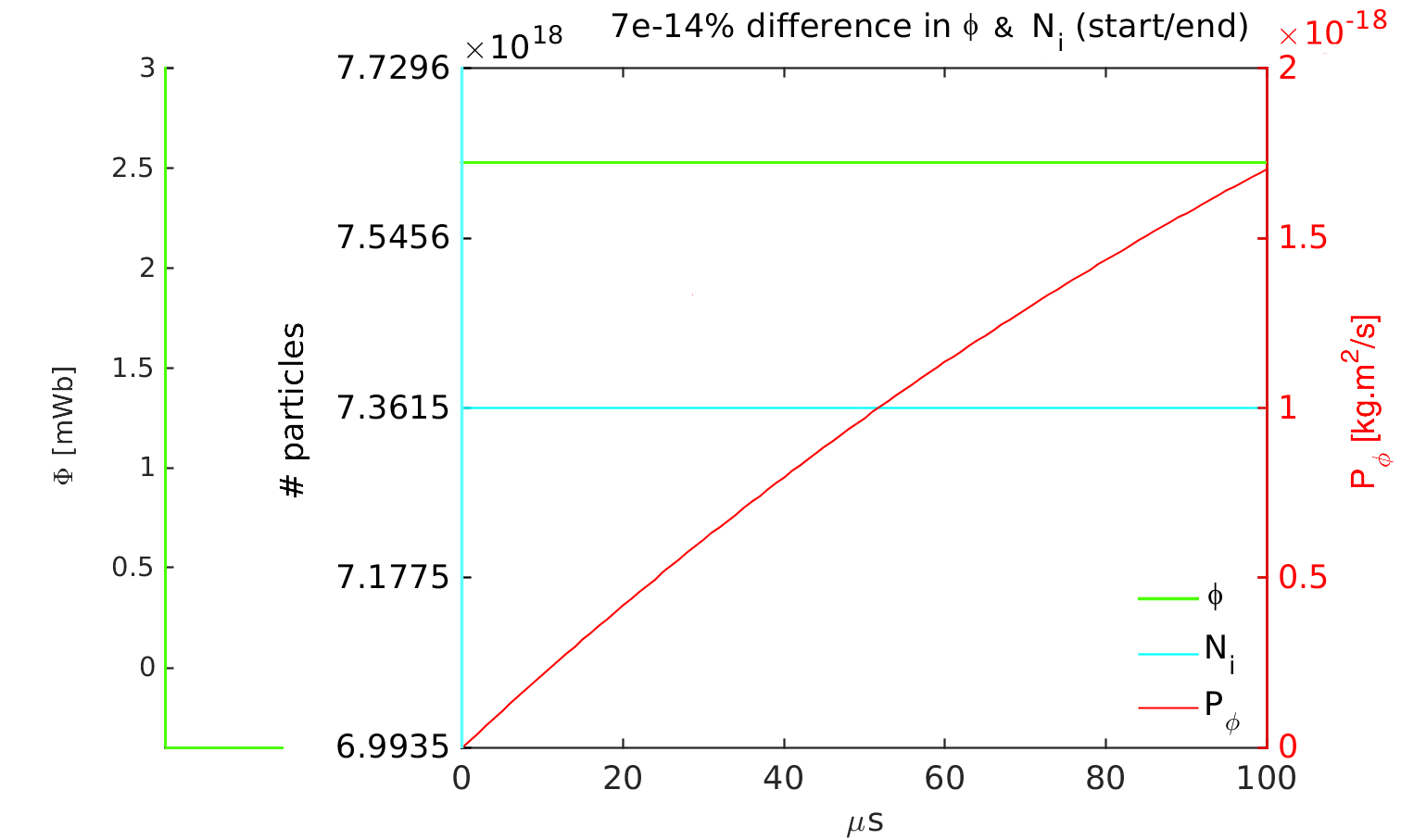}}\caption{\label{fig:Ni_Psi_Pp cons}$\,\,\,\,$Illustration of particle, toroidal
flux, and angular momentum conservation }
\end{figure}
Figure \ref{fig:Ni_Psi_Pp cons} shows how the total number of particles
($i.e.,$ \# ions, $N_{i})$, toroidal flux, and angular momentum,
calculated as the integrals over the computational domain, is conserved
to numerical precision for a $100\,\upmu$s simulation that started
from a Grad-Shafranov equilibrium, with explicitly applied boundary
conditions $v_{r}|_{\Gamma}=v_{z}|_{\Gamma}=0$, and $\psi|_{\Gamma}=0$.

\subsubsection{Energy conservation\label{subsec:Energy-conservation} }

Analogous to the continuous expression \ref{eq:481.1-1}, the discrete
expression for total system energy is

\begin{align}
\dot{U}_{Total} & =\dot{U}_{K}+\dot{U}_{Th}+\dot{U}_{M}\nonumber \\
 & =\frac{\partial}{\partial t}\,\left(\underline{dV}^{T}*\left\{ \frac{\underline{\rho}\circ\underline{\mathbf{v}}^{2}}{2}+\frac{\underline{p}}{\gamma-1}+\frac{1}{2\mu_{0}}\,\underline{f}{}^{2}\oslash\underline{r}{}^{2}\right\} +\underline{dV^{e}}^{T}*\left\{ \frac{1}{2\mu_{0}}\left(\underline{\underline{\nabla^{e}}}\,\,\underline{\psi}\right)^{2}\oslash\underline{r^{e}}{}^{2}\right\} \right)\nonumber \\
 & =\underline{dV}^{T}*\left\{ \frac{m_{i}\,\underline{\dot{n}}\circ\underline{\mathbf{v}}^{2}}{2}+\underline{\rho}\circ\underline{\mathbf{v}}\cdot\dot{\mathbf{\underline{v}}}+\frac{\dot{\underline{p}}}{\gamma-1}+\frac{1}{\mu_{0}}\,\underline{f}\circ\underline{\dot{f}}\oslash\underline{r}{}^{2}\right\} \nonumber \\
 & \,\,\,+\underline{dV^{e}}^{T}*\left\{ \frac{1}{\mu_{0}}\,\left(\underline{\underline{\nabla^{e}}}\,\,\underline{\psi}\right)\cdot\left(\underline{\underline{\nabla^{e}}}\,\,\underline{\dot{\psi}}\right)\oslash\underline{r^{e}}{}^{2}\right\} \nonumber \\
 & =-\left[\underline{dV}^{T}*\left\{ \frac{1}{2}\underline{v}^{2}\circ\left(\underline{\underline{\nabla}}\cdot(\underline{\rho}\circ\underline{\mathbf{v}})\right)+\frac{1}{2}(\underline{\rho}\circ\underline{\mathbf{v}})\cdot\left(\underline{\underline{\nabla}}\,\,\underline{v}^{2}\right)\right\} \right]\nonumber \\
 & \,\,\,-\left[\underline{dV}^{T}*\left\{ \underline{\mathbf{v}}\cdot\left(\underline{\underline{\nabla}}\,\,\underline{p}\right)+\frac{1}{\gamma-1}\left(\underline{\mathbf{v}}\cdot\left(\underline{\underline{\nabla}}\,\,\underline{p}\right)+\gamma\,\underline{p}\circ\left(\underline{\underline{\nabla}}\cdot\underline{\mathbf{v}}\right)\right)\right\} \right]\nonumber \\
 & \,\,\,-\left[\frac{1}{\mu_{0}}\,\underline{dV}^{T}*\left\{ \left(\underline{\mathbf{v}}\cdot\left(\underline{\underline{\nabla}}\,\,\underline{\psi}\right)\right)\circ\left(\underline{\underline{\Delta^{^{*}}}}\,\,\underline{\psi}\right)\oslash\underline{r}{}^{2}\right\} +\underline{dV^{e}}^{T}*\left\{ \left(\underline{\underline{\nabla^{e}}}\,\,\underline{\psi}\right)\cdot\left(\underline{\underline{\nabla^{e}}}\,\,\left(\underline{\mathbf{v}}\cdot\left(\underline{\underline{\nabla}}\,\,\underline{\psi}\right)\right)\right)\oslash\underline{r^{e}}{}^{2}\right\} \right]\nonumber \\
 & \,\,\,-\left[\frac{1}{\mu_{0}}\,\underline{dV}^{T}*\left\{ \left(\underline{f}\circ\underline{\mathbf{v}}\oslash\underline{r}^{2}\right)\cdot\left(\underline{\underline{\nabla}}\,\,\underline{f}\right)+\underline{f}\circ\left(\underline{\underline{\nabla}}\cdot\left(\underline{f}\circ\underline{\mathbf{v}}\oslash\underline{r}^{2}\right)\right)\right\} \right]\nonumber \\
 & \,\,\,+\left[\frac{1}{\mu_{0}}\,\underline{dV}^{T}*\left\{ \underline{\omega}\circ\left(\underline{\underline{W_{n}}}*\left(\underline{\mathbf{B}_{\theta}^{e}}\cdot\left(\underline{\underline{\nabla^{e}}}\,\,\underline{f}\right)\right)\right)+\underline{f}\circ\left(\underline{\underline{\nabla_{n}}}\cdot\left(\underline{\mathbf{B}_{\theta}^{e}}\circ\underline{\omega^{e}}\right)\right)\right\} \right]\nonumber \\
 & \,\,\,+\left[\frac{1}{\mu_{0}}\,\underline{dV}^{T}*\left\{ \underline{\eta}\circ\left(\left(\underline{\underline{\Delta^{^{*}}}}\,\,\underline{\psi}\right)\oslash\underline{r}\right)^{2}\right\} +\underline{dV^{e}}^{T}*\left\{ \left(\underline{\underline{\nabla^{e}}}\,\,\underline{\psi}\right)\cdot\left(\underline{\underline{\nabla^{e}}}\,\,\left(\underline{\eta}\circ\left(\underline{\underline{\Delta^{^{*}}}}\,\,\underline{\psi}\right)\right)\right)\oslash\underline{r^{e}}{}^{2}\right\} \right]\nonumber \\
 & \,\,\,+\left[\frac{1}{\mu_{0}}\,\underline{dV}^{T}*\left\{ \underline{\underline{W_{n}}}*\left(\left(\underline{\eta^{e}}\circ\left(\underline{\underline{\nabla^{e}}}\,\,\underline{f}\right)\oslash\underline{r^{e}}^{2}\right)\cdot\left(\underline{\underline{\nabla^{e}}}\,\,\underline{f}\right)\right)+\underline{f}\circ\left(\underline{\underline{\nabla_{n}}}\cdot\left(\underline{\eta^{e}}\circ\left(\underline{\underline{\nabla^{e}}}\,\,\underline{f}\right)\oslash\underline{r^{e}}^{2}\right)\right)\right\} \right]\nonumber \\
 & \,\,\,-\left[\underline{dV}^{T}*\left\{ \underline{\underline{\nabla_{n}}}\cdot\left(\underline{\mathbf{q}_{i}^{e}}+\underline{\mathbf{q}_{e}^{e}}\right)\right\} \right]\nonumber \\
 & \,\,\,-\left[\underline{dV}^{T}*\left\{ \underline{\mathbf{v}}\cdot\underline{\boldsymbol{\varPi}}-\underline{Q_{\pi}}\right\} \right]\label{eq:518}
\end{align}
Note that each set of square brackets has a counterpart in equation
\ref{eq:481.1-1}, the continuous form of $\dot{u}_{Total}$. The
poloidal magnetic energy is expressed in terms of the element-centered
gradient of $\underline{\psi}$, as required for consistency with
the definition of the second order $\underline{\underline{\Delta^{^{*}}}}$
differential operator. Here, it will be shown how, using the various
mimetic properties of the differential operators, that the terms in
each set of square brackets cancel when appropriate boundary conditions
are applied, leading to total system energy conservation.

The terms in the first set of square brackets here represents the
contribution to $\dot{U}_{K}$ due to advection, and vanish, with
boundary conditions $\mathbf{v_{\perp}}|_{\Gamma}=\mathbf{0}$, due
to equation \ref{eq:511.04} (where $\underline{\mathbf{P}}=\underline{\rho}\circ\underline{\mathbf{v}}$
and $\underline{U}=\underline{v}^{2}$). The terms in the second of
square brackets represents the contribution to $\dot{U}_{Th}$ from
compressional heating and, with the same boundary conditions, also
vanish due to equation \ref{eq:511.04} (with $\underline{\mathbf{P}}=\underline{\mathbf{v}}$
and $\underline{U}=\underline{p}$). 

The terms in the third, fourth and fifth sets of square brackets in
equation \ref{eq:518} represent the contribution to $\dot{U}_{K}$
that arises from the discrete forms of the components of $(\mathbf{J}\times\mathbf{B})$
in combination with the ideal (non-resistive) part of $\dot{U}_{M}$.
Using identity \ref{eq:515.72} to expand the $\underline{\underline{\Delta^{^{*}}}}$
operator, the terms in the third set of square brackets can be expressed
as 

{\small{}
\begin{align*}
-\frac{1}{\mu_{0}}\biggl[\underline{dV}^{T}*\left\{ \left(\underline{\mathbf{v}}\cdot\left(\underline{\underline{\nabla}}\,\,\underline{\psi}\right)\right)\circ\left(\underline{\underline{\nabla_{n}}}\cdot\left(\left(\underline{\underline{\nabla^{e}}}\,\,\underline{\psi}\right)\oslash\underline{r^{e}}^{2}\right)\right)\right\} +\underline{dV^{e}}^{T}*\left\{ \left(\left(\underline{\underline{\nabla^{e}}}\,\,\underline{\psi}\right)\oslash\underline{r^{e}}^{2}\right)\cdot\left(\underline{\underline{\nabla^{e}}}\,\,\left(\underline{\mathbf{v}}\cdot\left(\underline{\underline{\nabla}}\,\,\underline{\psi}\right)\right)\right)\right\} \biggr]
\end{align*}
}These terms cancel due to identity \ref{eq:515.031}, where $\underline{\mathbf{P}^{e}}=\left(\left(\underline{\underline{\nabla^{e}}}\,\,\underline{\psi}\right)\oslash\underline{r^{e}}^{2}\right)$
and $\underline{U}=\underline{\mathbf{v}}\cdot\left(\underline{\underline{\nabla}}\,\,\underline{\psi}\right)$. 

With boundary condition $\mathbf{v\perp}|_{\Gamma}=\mathbf{0}$, the
terms in the fourth set of square brackets cancel due to identity
\ref{eq:511.04}. Using equation \ref{eq:516}, where $\underline{Q}=\underline{\omega}=\underline{v_{\phi}}\oslash\underline{r}$,
the terms in the fifth set of square brackets can be expressed as 

\begin{align*}
\frac{1}{\mu_{0}}\,\left[\underline{dV^{e}}^{T}*\left\{ \left(\underline{\mathbf{B}_{\theta}^{e}}\circ\underline{\omega^{e}}\right)\cdot\left(\underline{\underline{\nabla^{e}}}\,\,\underline{f}\right)\right\} +\underline{dV}^{T}*\left\{ \underline{f}\circ\left(\underline{\underline{\nabla_{n}}}\cdot\left(\underline{\mathbf{B}_{\theta}^{e}}\circ\underline{\omega^{e}}\right)\right)\right\} \right]
\end{align*}
This combination cancels due to identity \ref{eq:515.031}, where
$\underline{\mathbf{P}^{e}}=\left(\underline{\mathbf{B}_{\theta}^{e}}\circ\underline{\omega^{e}}\right)$
and $\underline{U}=\underline{f}$. 

The terms in the sixth set of square brackets represent the rate of
increase of thermal energy due to ohmic heating from toroidal currents
(first term), in combination with the rate of decrease of magnetic
energy associated with poloidal field, due to resistive decay of the
toroidal currents (second term). As is true in the continuous case,
these terms are balanced in the discrete case. Using identity \ref{eq:515.72},
the combination may be expressed as 

{\footnotesize{}
\begin{align*}
 & \frac{1}{\mu_{0}}\left[\underline{dV}^{T}*\left\{ \left(\underline{\eta}\circ\left(\underline{\underline{\Delta^{^{*}}}}\,\,\underline{\psi}\right)\right)\circ\left(\underline{\underline{\nabla_{n}}}\cdot\left(\left(\underline{\underline{\nabla^{e}}}\,\,\underline{\psi}\right)\oslash\underline{r^{e}}{}^{2}\right)\right)\right\} +\underline{dV^{e}}^{T}*\left\{ \left(\left(\underline{\underline{\nabla^{e}}}\,\,\underline{\psi}\right)\oslash\underline{r^{e}}{}^{2}\right)\cdot\left(\underline{\underline{\nabla^{e}}}\,\,\left(\underline{\eta}\circ\left(\underline{\underline{\Delta^{^{*}}}}\,\,\underline{\psi}\right)\right)\right)\right\} \right]
\end{align*}
}and vanishes due to identity \ref{eq:515.031}, where $\underline{\mathbf{P}^{e}}=\left(\left(\underline{\underline{\nabla^{e}}}\,\,\underline{\psi}\right)\oslash\underline{r^{e}}{}^{2}\right)$
and $\underline{U}=\underline{\eta}\circ\left(\underline{\underline{\Delta^{^{*}}}}\,\,\underline{\psi}\right)$. 

The terms in the seventh set of square brackets in equation \ref{eq:518}
represent the rate of increase of thermal energy due to ohmic heating
from poloidal currents (first term), in combination with the rate
of decrease of magnetic energy associated with toroidal magnetic field,
due to resistive decay of the poloidal currents (second term). Again,
these terms cancel, which is physically representative. Referring
to equation \ref{eq:516.1}, the combination vanishes due to identity
\ref{eq:515.031}, where $\underline{\mathbf{P}^{e}}=\left(\underline{\eta^{e}}\circ\left(\underline{\underline{\nabla^{e}}}\,\,\underline{f}\right)\oslash\underline{r^{e}}^{2}\right)$
and $\underline{U}=\underline{f}$. 

With reference to the definitions of the discrete forms for the thermal
flux $\underline{\underline{\nabla_{n}}}\cdot\underline{\mathbf{q}_{\alpha}^{e}}$
, where $\underline{\mathbf{q}_{\alpha}^{e}}$ is defined in equation
\ref{eq:517.7}, it can be seen how the terms in the eighth set of
square brackets in equation \ref{eq:518} vanish due to equation \ref{eq:515.04}.
Boundary conditions that are explicitly applied to the pressure fields
break computational-domain-energy conservation, and enable thermal
losses in accordance with the thermal conduction model and the values
explicitly applied to $n_{0},\,\left(\chi_{\parallel\alpha}\right)|_{\Gamma}$
and $\left(\chi_{\perp\alpha}\right)|_{\Gamma}$, but thermal energy
fluxes through the boundary may be systematically accounted for. 

To deal with the viscosity related terms in the final set of square
brackets in equation \ref{eq:518}, the substitutions $\underline{\mathbf{P}_{1}}=\left(\underline{v_{r}}\oslash\underline{r}\right)\widehat{\mathbf{r}}$
, $\underline{\mathbf{P}_{2}}=\left(\underline{v_{z}}\oslash\underline{r}\right)\widehat{\mathbf{z}}$
and $\underline{\mathbf{P}_{3}}=\left(\underline{v_{z}}\oslash\underline{r}\right)\widehat{\mathbf{r}}+\left(\underline{v_{r}}\oslash\underline{r}\right)\widehat{\mathbf{z}}$
are made, and the expansion of terms, using equations \ref{eq:517.4},
\ref{eq:517.5}, and \ref{eq:516.1} is

{\small{}
\begin{align*}
 & \underline{dV}^{T}*\left\{ -\underline{\mathbf{v}}\cdot\underline{\boldsymbol{\varPi}}+\underline{Q_{\pi}}\right\} \\
 & =2\left[\underline{dV}^{T}*\left\{ \underline{\mathbf{P}_{1}}\cdot\left(\underline{\underline{\nabla_{n}}}\,\,\left(\underline{\mu^{e}}\circ\underline{r^{e}}^{2}\circ\left(\underline{\underline{\nabla^{e}}}\cdot\underline{\mathbf{P}_{1}}\right)\right)\right)\right\} +\underline{dV^{e}}^{T}*\left\{ \underline{\mu^{e}}\circ\underline{r^{e}}^{2}\circ\left(\underline{\underline{\nabla^{e}}}\cdot\underline{\mathbf{P}_{1}}\right)^{2}\right\} \right]\\
 & +2\left[\underline{dV}^{T}*\left\{ \underline{\mathbf{P}_{2}}\cdot\left(\underline{\underline{\nabla_{n}}}\,\,\left(\underline{\mu^{e}}\circ\underline{r^{e}}^{2}\circ\left(\underline{\underline{\nabla^{e}}}\cdot\underline{\mathbf{P}_{2}}\right)\right)\right)\right\} +\underline{dV^{e}}^{T}*\left\{ \underline{\mu^{e}}\circ\underline{r^{e}}^{2}\circ\left(\underline{\underline{\nabla^{e}}}\cdot\underline{\mathbf{P}_{2}}\right)^{2}\right\} \right]\\
 & +\left[\underline{dV}^{T}*\left\{ \underline{\mathbf{P}_{3}}\cdot\left(\underline{\underline{\nabla_{n}}}\,\,\left(\underline{\mu^{e}}\circ\underline{r^{e}}^{2}\circ\left(\underline{\underline{\nabla^{e}}}\cdot\underline{\mathbf{P}_{3}}\right)\right)\right)\right\} +\underline{dV^{e}}^{T}*\left\{ \underline{\mu^{e}}\circ\underline{r^{e}}^{2}\circ\left(\underline{\underline{\nabla^{e}}}\cdot\underline{\mathbf{P}_{3}}\right)^{2}\right\} \right]\\
 & -\frac{2}{3}\left[\underline{dV}^{T}*\left\{ \underline{\mathbf{v}}\cdot\left(\underline{\underline{\nabla_{n}}}\,\,\left(\underline{\mu^{e}}\circ\left(\underline{\underline{\nabla^{e}}}\cdot\underline{\mathbf{v}}\right)\right)\right)\right\} +\underline{dV^{e}}^{T}*\left\{ \underline{\mu^{e}}\circ\left(\underline{\underline{\nabla^{e}}}\cdot\underline{\mathbf{v}}\right)^{2}\right\} \right]\\
 & +\left[\underline{dV}^{T}*\left\{ \underline{\mathbf{\omega}}\circ\left(\underline{\underline{\nabla_{n}}}\cdot\left(\underline{\mu^{e}}\circ\underline{r^{e}}^{2}\circ\left(\underline{\underline{\nabla^{e}}}\,\,\underline{\omega}\right)\right)\right)\right\} +\underline{dV^{e}}^{T}*\left\{ \underline{\mu^{e}}\circ\underline{r^{e}}^{2}\circ\left(\underline{\underline{\nabla^{e}}}\,\,\underline{\omega}\right)^{2}\right\} \right]\\
 & -2\left[\underline{dV}^{T}*\left\{ \underline{\mu}\circ\left(\underline{v_{r}}\oslash\underline{r}\right)^{2}-\underline{\mu}\circ\left(\underline{v_{r}}\oslash\underline{r}\right)^{2}\right\} \right]
\end{align*}
}The terms in the first to fourth sets of square brackets here vanish
due to equation \ref{eq:515.03}, where, for example, for the first
set of square brackets, $\underline{\mathbf{P}}=\underline{\mathbf{P}_{1}}$
and $\underline{U^{e}}=\underline{\mu^{e}}\circ\underline{r^{e}}^{2}\circ\left(\underline{\underline{\nabla^{e}}}\cdot\underline{\mathbf{P}_{1}}\right)$.
The terms in the fifth set of square brackets vanish due to equation
\ref{eq:515.031} where $\underline{U}=\underline{\omega}$ and $\underline{\mathbf{P}^{e}}=\underline{\mu^{e}}\circ\underline{r^{e}}^{2}\circ\left(\underline{\underline{\nabla^{e}}}\,\,\underline{\omega}\right)$,
and the terms in the final set obviously cancel. Thus, the volume-integrated
rate of increase of thermal energy due to viscous heating is balanced
by the volume-integrated rate of decrease of kinetic energy due to
viscous dissipation, which is also a property of the physical system
(see appendix \ref{subsec:Viscous-heating}). 
\begin{figure}[H]
\centering{}\includegraphics[width=15cm,height=8cm]{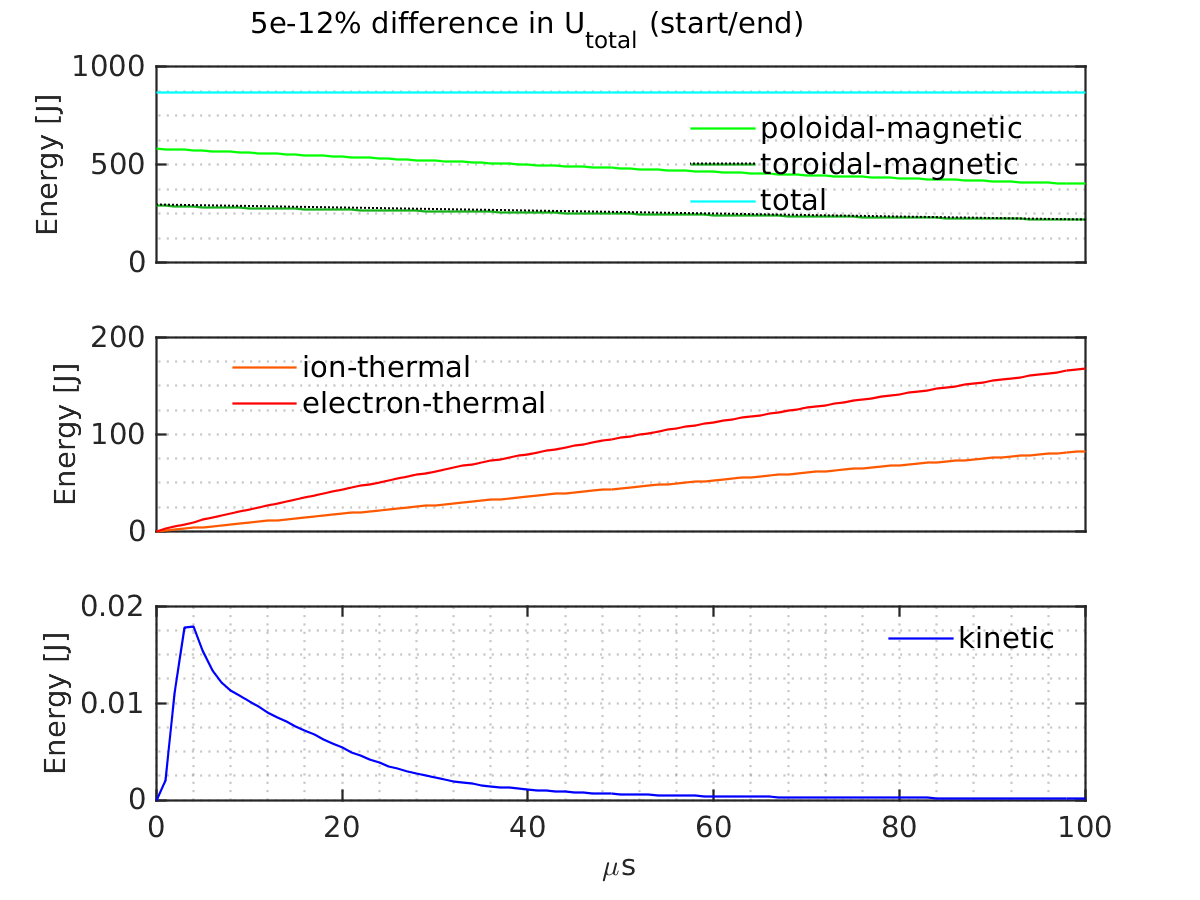}\caption{\label{fig:Energy-partition-of-}$\,\,\,\,$System energy evolution
for an MHD simulation starting with a Grad-Shafranov equilibrium }
\end{figure}
Figure \ref{fig:Energy-partition-of-} shows the partition of energy,
and how total energy is conserved approximately to machine precision
for a $100\,\upmu$s simulation with moderate time-step and mesh resolution,
using Runge-Kutta second order time-stepping, that started from a
Grad-Shafranov equilibrium. It can be seen that the magnetic energy
is largely due to poloidal field in this case, and that the decrease
in magnetic energy due to resistive decay of plasma currents is approximately
canceled (with some offset due to changes in kinetic energy) by the
increase thermal energy. The simulation started with cold plasma.
Thermal energy is imparted directly to the electron fluid by ohmic
heating and is collisionally transferred from the electrons to the
ions (term $Q_{ie}$ in equations \ref{eq:479.1} and \ref{eq:479.2}).
In summary, as the plasma currents decay resistively, the magnetic
energy decreases, while the thermal energy increases by the same amount
due to ohmic heating. For simulations starting with a Grad-Shafranov
equilibrium, as in this case, the plasma is in equilibrium and kinetic
energy is practically negligible. The only explicitly applied boundary
conditions are $v_{r}|_{\Gamma}=v_{z}|_{\Gamma}=0\mbox{, and }\psi|_{\Gamma}=0$.
In contrast, simulations presented in chapter \ref{chap:Simulation-results}
that include CT formation and magnetic levitation and magnetic compression,
are run with explicitly applied boundary conditions for $\psi$ that
are determined by experiment. This enables inward electromagnetic
energy fueling and outward electromagnetic losses (Poynting flux).
In those simulations, boundary conditions are explicitly applied to
the pressure fields to enable thermal losses in accordance with the
thermal conduction model, and boundary conditions for $f$ are explicitly
applied to the parts of the domain boundary that represent insulating
regions. 

\subsection{Maintenance of energy and momentum conservation with artificial density
diffusion\label{subsec:Maintenance-of-momentum}}

As mentioned earlier, the components of force per volume vector $\underline{\mathbf{f}_{\zeta}}=(\underline{f_{\zeta r}},\,\underline{f_{\zeta\phi}},\,\underline{f_{\zeta z}})^{T}$
in the velocity equations, and the heating term $\underline{Q_{\zeta}}$
in the ion pressure equation (equation \ref{eq:517.3}) are included
to cancel the effect of artificial density diffusion on the total
system momentum and energy. There are two models for the correction
terms. The first model is straightforward and maintains conservation
of energy and angular momentum when density diffusion is included
in the mass continuity equation. It is evaluated simply by treating
the density diffusion as a particle source and assessing the local
($i.e.,$ nodal) correction terms as the local effects of the source
on momentum and energy. However this model can lead locally to negative
ion pressure if the density gradients are extreme, so it is not suitable
for inclusion in simulations of CT formation and compression. The
model works satisfactorily for other simulation types, for example
the resistive decay of a magnetised plasma described by an equilibrium
model. The second model is more complicated - it maintains conservation
of volume integrated energy and can not cause negative ion pressure,
but angular momentum conservation is not maintained. Both models for
the correction terms compensate for additional $r$ and $z$ directed
momentum introduced due to density diffusion - for the first model
the compensation is local, while for the second model the compensation
is in the volume integrated sense. The simulations presented in chapter
\ref{chap:Simulation-results} use the second model. 

The correction terms for the first model are derived as follows. For
convenience, we use the notation $\underline{\zeta_{n}}=\underline{\underline{\nabla_{n}}}\cdot\left(\underline{\zeta}\circ\underline{\underline{\nabla^{e}}}\,\,\underline{n}\right)$.
Referring to equation \ref{eq:517.3}, note that $\underline{\dot{n}}|_{\zeta}=\underline{\zeta_{n}}$,
$\underline{\dot{v}_{\beta}}|_{\zeta}=\underline{f_{\zeta\beta}}\oslash\underline{\rho}$,
and $\underline{\dot{p}_{i}}|_{\zeta}=\underline{Q_{\zeta}}$ are
the parts $\underline{\dot{n}}$, $\underline{\dot{v}_{\beta}}$,
and $\underline{\dot{p}_{i}}$ respectively that are associated with
density diffusion. The local rate of change of the $\beta$ component
(where, here, $\beta=r,\,z$) of momentum per unit volume due to the
local particle source/sink terms arising due to density diffusion
is set to zero by design, resulting in 
\begin{align*}
\left(\frac{\partial}{\partial t}\left(m_{i}\,\underline{n}\circ\underline{v_{\beta}}\right)\right)_{\zeta} & =m_{i}\,\left(\underline{n}\circ\underline{f_{\zeta\beta}}\oslash\underline{\rho}+\underline{\zeta_{n}}\circ\underline{v_{\beta}}\right)=0\\
 & \Rightarrow\underline{f_{\zeta\beta}}=-m_{i}\,\underline{v_{\beta}}\circ\underline{\zeta_{n}}
\end{align*}
Naturally, for $\beta=\phi$, this expression for $\underline{f_{\zeta\beta}}$
also locally cancels additional angular momentum per unit volume introduced
by density diffusion. The rate of change of energy per unit volume
due to local particle source/sink terms associated with density diffusion
is also set to zero: 
\begin{align*}
\left(\frac{\partial}{\partial t}\left(\frac{1}{2}m_{i}\,\underline{n}\circ\underline{v}^{2}+\frac{1}{\gamma-1}\,\underline{\dot{p_{i}}}\right)\right)_{\zeta} & =\frac{1}{2}m_{i}\,\underline{\zeta_{n}}\circ\underline{v}^{2}+\underset{\beta}{\Sigma}\left(m_{i}\,\underline{n}\circ\underline{v_{\beta}}\circ\underline{f_{\zeta\beta}}\oslash\underline{\rho}\right)+\underline{Q_{\zeta}}=0\\
 & \Rightarrow\underline{Q_{\zeta}}=\frac{1}{2}m_{i}\,\underline{v^{2}}\circ\underline{\zeta_{n}}
\end{align*}
Hence, the contributions to $\dot{U}_{K}$ and $\dot{U}_{Th}$ arising
from the artificial density diffusion and relevant correction terms
cancel by design as:

{\small{}
\begin{align*}
\left(\dot{U}_{K}+\dot{U}_{Th}\right)_{\zeta} & =\underline{dV}^{T}*\left\{ m_{i}\left(\frac{1}{2}-1+\frac{1}{2}\right)\,\underline{v}^{2}\circ\underline{\zeta_{n}}\right\} =0
\end{align*}
}However, note that if density gradients are extreme, $\underline{\zeta_{n}}$
and hence $\underline{Q_{\zeta}}$ will have some very large negative
values, so that $\underline{p_{i}}$ can become negative.\\

For the second model of correction terms, $\underline{Q_{\zeta}}=\underline{0}$,
and a suitable choice of $\underline{f_{\zeta\beta}}$ that ensures
maintenance, in a volume integrated sense, of conservation of total
energy, is
\begin{equation}
\underline{f_{\zeta\beta}}=\frac{1}{2}m_{i}\,\zeta\left[\underline{\underline{W_{n}}*}\left(\left(\underline{\underline{\nabla^{e}}}\,\,\underline{n}\right)\cdot\left(\underline{\underline{\nabla^{e}}}\,\,\underline{v_{\beta}}\right)\right)+\underline{\underline{\nabla_{n}}}\cdot\left(\underline{v_{\beta}^{e}}\circ\left(\underline{\underline{\nabla^{e}}}\,\,\underline{n}\right)\right)-\underline{v_{\beta}}\circ\left(\underline{\underline{\Delta}}\,\,\underline{n}\right)\right]\label{eq:518-1}
\end{equation}
Note that while this term also cancels modifications to the $r$ and
$z$ directed momentum resulting from density diffusion, modification
to angular momentum is not compensated for. This method requires that
$\zeta$ is spatially constant so that, in this case $\underline{\zeta_{n}}=\zeta\,\underline{\underline{\Delta}}\,\,\underline{n}$.
Expressions for $\underline{f_{\zeta\beta}}$ that also conserve angular
momentum can be derived if $\zeta$ is made to be spatially dependent,
but this additional complication has been omitted for now.

Using expression in equation \ref{eq:518-1} for $\underline{f_{\zeta\beta}}$,
the total contribution to the volume integral of the $\beta$ component
of momentum per unit volume (where $\beta=r,\,z$) from terms associated
with density diffusion is 
\begin{align*}
\dot{P_{\beta}}|_{\zeta} & =\underline{dV}^{T}*\left\{ \left(\frac{\partial}{\partial t}\left(m_{i}\,\underline{n}\circ\underline{v_{\beta}}\right)\right)_{\zeta}\right\} \\
 & =\underline{dV}^{T}*\left\{ m_{i}\,\underline{\zeta_{n}}\circ\underline{v_{\beta}}+\underline{f_{\zeta\beta}}\right\} \\
 & =\underline{dV}^{T}*\biggl\{ m_{i}\,\zeta\,\left(\underline{\underline{\Delta}}\,\,\underline{n}\right)\circ\underline{v_{\beta}}+\frac{1}{2}m_{i}\,\zeta\biggl[\underline{\underline{W_{n}}*}\left(\left(\underline{\underline{\nabla^{e}}}\,\,\underline{n}\right)\cdot\left(\underline{\underline{\nabla^{e}}}\,\,\underline{v_{\beta}}\right)\right) & \mbox{}\\
 & \,\,\,\,\,\,\,+\cancel{\underline{\underline{\nabla_{n}}}\cdot\left(\underline{v_{\beta}^{e}}\circ\left(\underline{\underline{\nabla^{e}}}\,\,\underline{n}\right)\right)}-\underline{v_{\beta}}\circ\left(\underline{\underline{\Delta}}\,\,\underline{n}\right)\biggr]\biggr\} & \mbox{(use eqn. \ref{eq:515.04})}\\
 & =\frac{1}{2}m_{i}\,\zeta\,\underline{dV}^{T}*\left\{ \underline{\underline{W_{n}}*}\left(\left(\underline{\underline{\nabla^{e}}}\,\,\underline{n}\right)\cdot\left(\underline{\underline{\nabla^{e}}}\,\,\underline{v_{\beta}}\right)\right)+\underline{v_{\beta}}\circ\left(\underline{\underline{\Delta}}\,\,\underline{n}\right)\right\} \\
 & =\frac{1}{2}m_{i}\,\zeta\,\left[\underline{dV^{e}}^{T}*\left\{ \left(\underline{\underline{\nabla^{e}}}\,\,\underline{n}\right)\cdot\left(\underline{\underline{\nabla^{e}}}\,\,\underline{v_{\beta}}\right)\right\} +\underline{dV}^{T}*\left\{ \underline{v_{\beta}}\circ\left(\underline{\underline{\nabla_{n}}}\cdot\left(\underline{\underline{\nabla^{e}}}\,\,\underline{n}\right)\right)\right\} \right] & \mbox{(use eqn. \ref{eq:516.1})}\\
\Rightarrow\dot{P}_{\beta}|_{\zeta} & =0 & (\mbox{use eqn. }\ref{eq:515.031})
\end{align*}
Note that in the last step, identity \ref{eq:515.031} has been used,
with $\underline{\mathbf{P}^{e}}=\underline{\underline{\nabla^{e}}}\,\,\underline{n}$
and $\underline{U}=\underline{v_{\beta}}$. 

The total contribution to energy associated with $v_{\beta}$ (where
$\beta=r,\,\phi,\,z$) from terms associated with density diffusion
is {\small{}
\begin{align*}
\dot{U_{\beta}}|_{\zeta} & =\underline{dV}^{T}*\left\{ \left(\frac{\partial}{\partial t}\left(\frac{1}{2}\,m_{i}\,\underline{n}\circ\underline{v_{\beta}}^{2}\right)\right)_{\zeta}\right\} \\
 & =\underline{dV}^{T}*\left\{ \frac{1}{2}m_{i}\,\underline{\zeta_{n}}\circ\underline{v_{\beta}}^{2}+\underline{v_{\beta}}\circ\underline{f_{\zeta\beta}}\right\} \\
 & =\underline{dV}^{T}*\biggl\{\cancel{\frac{1}{2}m_{i}\,\zeta\,\left(\underline{\underline{\Delta}}\,\,\underline{n}\right)\circ\underline{v_{\beta}}^{2}}\\
 & \,\,\,\,+\frac{1}{2}m_{i}\,\zeta\,\underline{v_{\beta}}\circ\left[\underline{\underline{W_{n}}*}\left(\left(\underline{\underline{\nabla^{e}}}\,\,\underline{n}\right)\cdot\left(\underline{\underline{\nabla^{e}}}\,\,\underline{v_{\beta}}\right)\right)+\underline{\underline{\nabla_{n}}}\cdot\left(\underline{v_{\beta}^{e}}\circ\left(\underline{\underline{\nabla^{e}}}\,\,\underline{n}\right)\right)\cancel{-\underline{v_{\beta}}\circ\left(\underline{\underline{\Delta}}\,\,\underline{n}\right)}\right]\biggr\}\\
 & =\frac{1}{2}m_{i}\,\zeta\,\underline{dV}^{T}*\left\{ \underline{v_{\beta}}\circ\left[\underline{\underline{W_{n}}*}\left(\left(\underline{\underline{\nabla^{e}}}\,\,\underline{n}\right)\cdot\left(\underline{\underline{\nabla^{e}}}\,\,\underline{v_{\beta}}\right)\right)+\underline{\underline{\nabla_{n}}}\cdot\left(\underline{v_{\beta}^{e}}\circ\left(\underline{\underline{\nabla^{e}}}\,\,\underline{n}\right)\right)\right]\right\} \\
 & =\frac{1}{2}m_{i}\,\zeta\biggl[\underline{dV^{e}}^{T}*\left\{ \underline{v_{\beta}^{e}}\circ\left(\left(\underline{\underline{\nabla^{e}}}\,\,\underline{n}\right)\cdot\left(\underline{\underline{\nabla^{e}}}\,\,\underline{v_{\beta}}\right)\right)\right\} \\
 & \,\,\,\,\,\,\,+\underline{dV}^{T}*\left\{ \underline{v_{\beta}}\circ\left(\underline{\underline{\nabla_{n}}}\cdot\left(\underline{v_{\beta}^{e}}\circ\left(\underline{\underline{\nabla^{e}}}\,\,\underline{n}\right)\right)\right)\right\} \biggr] & \mbox{\mbox{(use eqn. \ref{eq:516})}}\\
\Rightarrow\dot{U}_{\beta}|_{\zeta} & =0 & (\mbox{use eqn. }\ref{eq:515.031})
\end{align*}
}In the last step, identity \ref{eq:515.031} has been used again,
now with $\underline{\mathbf{P}^{e}}=\underline{v_{\beta}^{e}}\circ\left(\underline{\underline{\nabla^{e}}}\,\,\underline{n}\right)$
and $\underline{U}=\underline{v_{\beta}}$. 

Hence, the modification to total system energy due to density diffusion
is canceled with the inclusion of the correction terms using either
correction model, thereby ensuring maintenance of global energy conservation.
Conservation of angular momentum is maintained using the first model
only, but that model is suitable only for simulations that don't involve
steep density gradients. Both models compensate for the modifications
to radially and axially directed momentum associated with density
diffusion. 

\section{Summary\label{sec:SummaryCoreCodeDev}}

It has been shown how the finite element method has been used to develop
first and second order differential matrix operators that mimic some
of the properties of their continuous counterparts. Namely, the operators
satisfy the discrete forms of the differential product rule and the
divergence theorem. Conservation of mass, toroidal flux, angular momentum,
and energy for the continuous system of MHD equations has been demonstrated.
A discrete counterpart set of equations has been developed, which
also satisfy the discrete forms of these conservation laws. A model
for anisotropic thermal diffusion has been formulated and implemented
to the code. In order to maintain energy conservation, and angular
momentum conservation in some scenarios, in the presence of artificial
density diffusion, correction terms have been developed and included
in the discrete momentum and energy equations.

\newpage{}

\chapter{Models for CT formation, levitation and magnetic compression, and
implementation of simulated diagnostics \label{chap:Implementation-of-models}}

In this chapter, the various models developed to simulate CT formation,
levitation and magnetic compression, and their implementation to the
code, are described. In section \ref{subsec:Boundary-conditions},
the boundary conditions used for the different fields that are evolved
in time are outlined. Magnetic levitation and compression are modelled
through the implementation of boundary conditions for $\psi$. Appropriate
boundary conditions on $f$ lead to toroidal flux conservation, which
is a key requirement to successfully reproducing physical properties
associated with the experiment. In section \ref{sec:Vacuum-field-in},
the method developed to model an insulating region, representing the
insulating tube located between the external levitation/compression
coils and the CT containment region, is described. The vacuum field
solution in the insulating region is coupled to the full MHD solutions
in the remainder of the computational domain. In section \ref{sec:External-sources-of},
the techniques established to simulate addition of toroidal flux to
the domain, representative of either the CT formation process, or,
for simulations starting with a Grad-Shafranov equilibrium, the flow
of shaft current around the CT containment region, are specified.
The strategy developed to maintain toroidal flux conservation when
an insulating region is included in the domain is outlined in section
\ref{sec:PHIconservation-with}. Section \ref{subsec:Simulated-diagnostics}
consists of an overview of the simulated diagnostics that were implemented
to the code. In addition to representations of the experiment's diagnostics,
these include methods developed to evaluate, based on various simulated
field distributions, the time evolutions of the $q(\psi)$ profile,
CT magnetic axis location, volume-averaged $\beta$, CT volume and
magnetic fluxes, system energy components, and maximum ion and electron
temperatures. The chapter concludes with a summary in section \ref{sec:SummaryImplemnationFformetc}.

\section{Boundary conditions\label{subsec:Boundary-conditions}}

. 

\subsection{Boundary conditions for density\label{subsec:bcs density} }

No boundary condition is explicitly applied to density. As described
in section \ref{subsec:Particle-count-conservation}, the natural
boundary condition 
\[
\left(\zeta\,\nabla_{\perp}n\right)|_{\Gamma}=0,
\]
is automatically imposed when density diffusion is included in the
model. This leads to $\left(\nabla_{\perp}n\right)|_{\Gamma}=0,$
if $\zeta$ is spatially constant.

\subsection{Boundary conditions for velocity\label{subsec:rbcs Vel} }

For simulations including CT formation and/or magnetic compression,
the plasma fluid velocity components are set to zero on the boundary
$i.e.,$ the no-slip condition for an impermeable boundary: 
\[
\mathbf{v}|_{\Gamma}=\mathbf{0}
\]
Since the plasma viscosity scales with $T^{5/2}$ (Braginskii formulae
- equations \ref{eq:472.47}), it might be expected that a $slip$
condition would be more suitable because $T\rightarrow0$ at the boundary,
so that there should be negligible viscosity at the wall, and a boundary
layer, associated with neutral gas flow at a wall, should be absent.
A $slip$ condition would involve setting wall-normal velocity components
to zero at the wall, and, to ensure unique solutions for the velocity
fields, setting the normal derivatives of the wall-tangential components
to zero at the wall. On the other hand, physically, friction with
cold newly recombined neutral particles at the wall should impede
wall-tangential plasma motion at the boundary. In addition, the stuffing
field of the Marshall gun is resistively pinned to the vertical electrode
walls, and plasma at the walls that is being advected up the gun is
partially frozen-into and retarded by the field at the electrode walls,
so that the $no-slip$ condition, as implemented, may be the more
physical choice for the plasma fluid velocity components. For the
neutral fluid, the $no\ slip$ condition is also implemented - the
suitability of this condition is less ambiguous for the neutral fluid.

As described in sections \ref{subsec:Angular-momentum-conservation}
and \ref{subsec:Maintenance-of-momentum}, conservation of system
angular momentum requires that no boundary conditions are explicitly
applied to $v_{\phi}$, in which case the natural boundary condition
$\left(\nabla_{\perp}\,\omega\,\right)|_{\Gamma}=0$ is automatically
imposed. However, angular momentum is conserved only in simulations
for which either no density diffusion is included in the model, or
the first of the two strategies described in section \ref{subsec:Maintenance-of-momentum}
for the corrective terms is implemented. These scenarios are suitable
for simulations looking at resistive evolution of CTs initially described
by a Grad-Shafranov equilibrium, but not for simulations that include
CT formation or magnetic compression.

\subsection{Boundary conditions for temperature\label{subsec:bcs T} }

Boundary conditions for pressures follow from equations \ref{eq:479.5},
with temperatures $T_{i}$ and $T_{e}$ set to approximately zero
at the boundary: 
\[
T_{i}|_{\Gamma}=T_{e}|_{\Gamma}=0.02\mbox{ eV}\approx20^{o}\mbox{C}
\]
If a neutral fluid is included in the simulation, $T_{n}$ is also
set to 0.02 eV at the boundary. Physically, plasma particles are expected
to enter and rattle around in microscopic wall cavities, where they
cool and recombine, and are then recycled back into the plasma region
as cold neutral particles which can then undergo further ionization. 

\subsection{Boundary conditions for $\psi$ \label{subsec:bcs psi}}

Toroidal currents in the main, and levitation/compression coils, constitute
sources of poloidal flux, which can be included in the model with
the application of appropriate boundary conditions for $\psi$. The
locations of the boundary points, that determine the solution domain
representing the vacuum region of the magnetic compression machine,
along with the insulating areas (air) around the levitation/compression
coils, and the ceramic (or quartz) insulating wall surrounding the
CT confinement region, were defined in a Matlab code which writes
the coordinates of the boundary points to file. For each machine configuration,
a $\mbox{LUA}$ script is an output of the code. After assigning the
material properties, solution frequencies, and peak coil currents
pertaining to either the main stuffing field, the levitation field
or the compression field, and running a $\mbox{FEMM}$ \cite{FEMM}
model (see example outputs in figures \ref{fig:Schematic-of-6}(b)
and \ref{fig:11-coil-configuration}(b)) for the relevant configuration,
the $\mbox{LUA}$ script can be loaded to the various $\mbox{FEMM}$
solutions, and the values of $\psi_{main},\,\psi_{lev},\,\mbox{and }\psi_{comp}$
at the boundary points listed in the $\mbox{LUA}$ script are written
to file. Field diffusion times in the most electrically conducting
parts of the machine are much longer than the timescales associated
with the levitation and compression fields. Magnetic boundary conditions
are applied at all boundary nodes, but levitation boundary values
are significant only on the levitation/compression coil boundaries
and on the boundaries of the regions representing the stainless steel
above and below the insulating wall, while compression boundary values
are significant only on the coil boundaries, and are approximately
zero at other boundary regions. 
\begin{figure}[H]
\subfloat[$\widetilde{I}_{lev}(t)$]{\begin{centering}
\includegraphics[width=7cm,height=5cm]{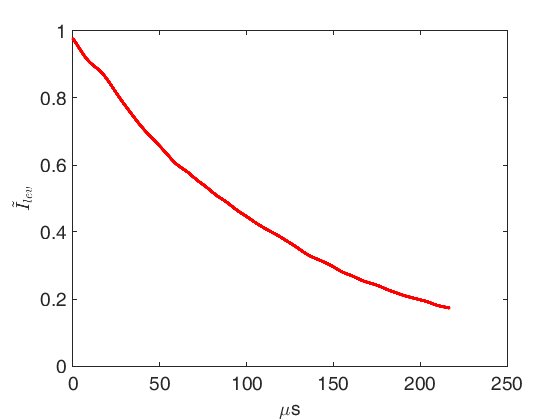}
\par\end{centering}
}\hfill{}\subfloat[$\widetilde{I}_{comp}(t)$]{\centering{}\includegraphics[width=7cm,height=5cm]{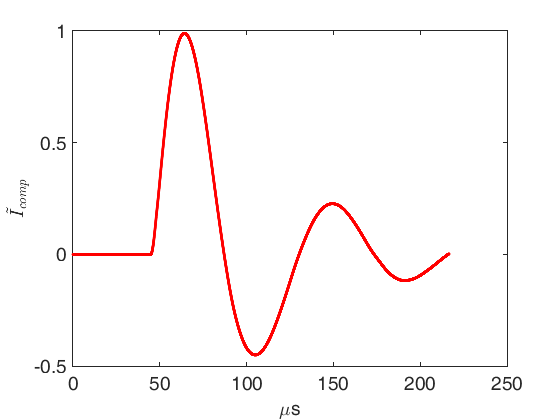}}\caption{$\,\,\,\,$Normalised levitation and compression current signals\label{fig:Ilev_comp_sim}}
\end{figure}
The boundary conditions for $\psi_{main}(\mathbf{r})$ are held constant
over time ($I_{main}$ is approximately constant over $\sim3$s before
firing the formation banks), while boundary values obtained for $\psi_{lev}(\mathbf{r},\,t)$
and $\psi_{comp}(\mathbf{r},\,t)$, which pertain to the peak levitation/compression
currents, are scaled over time according to the experimentally measured
waveforms for $I_{lev}(t)$ and $I_{comp}$(t):
\begin{equation}
\psi(\mathbf{r},\,t)|_{\Gamma}=\psi_{main}(\mathbf{r})|_{\Gamma}+\psi_{lev}(\mathbf{r},\,t)|_{\Gamma}+\psi_{comp}(\mathbf{r},\,t)|_{\Gamma}\label{eq:519}
\end{equation}
Figure \ref{fig:Ilev_comp_sim}(a) shows $\widetilde{I}_{lev}(t)$,
the normalised levitation current signal, as measured with Rogowski
coils during the experiment; similarly, figure \ref{fig:Ilev_comp_sim}(b)
shows $\widetilde{I}_{comp}(t)$. For this simulation, $t_{comp}=45\,\upmu$s,
and the simulation was run until around $220\,\upmu$s. Simulation
input parameter $Rcable$ was set to $1$ for this simulation - $i.e.,$
a levitation current profile pertaining to the levitation circuits
with the $70\mbox{ m}\Omega$ cables in place of the original $2.5\mbox{ m}\Omega$
cables was used to control $\psi_{lev}$ (see section \ref{subsec:Levitation-field-decay}).
Note that $\widetilde{I}_{lev}<1$ at $t=0$, because $t_{lev}=-50\,\upmu$s
- $i.e.,$ in the experiment, the levitation capacitor banks were
fired $50\,\upmu$s before the formation banks. With a rise time of
$\sim40\,\upmu$s, the levitation current peaks at $t\sim-10\,\upmu$s,
when $\widetilde{I}_{lev}=1$. The values of $\widetilde{I}_{lev}(t)$
and $\widetilde{I}_{comp}(t)$ are defined at the times defined by
the digitiser sampling time, $dt_{exp}=50$ ns. For evaluation at
the simulation times (typically $dt_{sim}\sim0.01$ ns), Matlab function
handles are defined in the MHD code, so that $\widetilde{I}_{lev}(t)$
and $\widetilde{I}_{comp}(t)$ can be obtained at the required simulation
times, either by interpolation, or by evaluating a high-order polynomial
fit to the experimental signal. 

The MHD code also has the option to set $\psi=0$ on the entire boundary,
representing perfectly conducting walls.

\subsection{Boundary conditions for $f$ \label{subsec:bcs F}}

In general, no boundary conditions are explicitly applied to $f$
on boundary regions representing electrical conductors, and the natural
boundary condition
\[
\left(\mathbf{B}_{\theta\perp}\omega+\eta\left(\nabla_{\perp}\,f\,\right)/r^{2}\right)|_{\Gamma}=0
\]
is automatically imposed. In combination with the boundary condition
$\underline{\mathbf{v}_{\perp}}|_{\Gamma}=\mathbf{0}$, this corresponds
to having the poloidal component of the electric field perpendicular
to the boundary, the condition, in the case of azimuthal symmetry,
for a perfectly electrically conducting boundary (see equation \ref{eq:480.51}).
Simulations that include CT formation and compression are run with
the boundary conditions $\underline{v_{\beta}}|_{\Gamma}=0$ applied
explicitly to each velocity component, so that the automatically imposed
boundary condition for $f$, corresponding to the physical case of
a perfectly electrically conducting boundary, becomes $\left(\nabla_{\perp}\,f\,\right)|_{\Gamma}=0$.
Since $f=rB_{\phi}=\mu_{0}I_{\theta}/2\pi$ (from Ampere's law when
$\frac{\partial}{\partial\phi}=0$), the boundary condition $\left(\nabla_{\perp}f\right)|_{\Gamma}=0$
implies, when $\mathbf{v}|_{\Gamma}=\mathbf{0},$ that any poloidal
currents that flow into or out of the wall ($e.g.,$ radial intra-electrode
formation current or crowbarred shaft current during magnetic compression)
flow perpendicular to the wall, $i.e.,$ $\mathbf{E}$ has no component
parallel to the boundary. Since no currents can flow in insulating
regions, $f$ must be spatially constant on the part of the boundary
representing the insulating wall surrounding the CT confinement region
(see section \ref{sec:Vacuum-field-in} below). As detailed in section
\ref{sec:PHIconservation-with}, the value of that constant, consistent
with conservation of toroidal flux in the combined domains, is calculated
at each timestep and explicitly applied as a boundary condition for
$f$ on the interface shared by the plasma and insulating domains. 

For formation simulations, sources of toroidal flux include formation
current driven through the plasma, and so CT formation can not be
modelled by setting boundary conditions. Similarly, during magnetic
compression, crowbarred shaft current is diverted through the plasma
at the CT edge, as schematically indicated in figure \ref{fig:divertedcurrent}.
This implies that for simulations that include magnetic compression,
the source of toroidal flux due to crowbarred shaft current can also
not be accurately modelled by setting boundary conditions. The methods
developed for modelling toroidal flux addition are described in section
\ref{sec:External-sources-of}. 

For simulations starting with a Grad-Shafranov equilibrium, the method
of applying boundary conditions on $f$, in order to model the reduction
of toroidal flux over time that results from decaying crowbarred shaft
current, produces reasonable results that compare well with experiment,
but only in cases where magnetic compression is not applied. This
method can be implemented by setting code input parameter $Shaftbc=1$.
In this case, as outlined in section \ref{subsec:Simulations-starting-withGS11},
experimentally-determined shaft current, taken from a reference non-compression
shot, is used to determine boundary conditions for $f$. The flux
conserving natural boundary condition of $\left(\nabla_{\perp}f\right)|_{\Gamma}=0$
is overwritten, so that the system does not conserve toroidal flux.
The simulated diagnostic for $B_{\phi}$ at the chalice magnetic probe
locations follows the prescribed crowbarred shaft current profile
exactly. When compression is simulated by modifying the boundary conditions
for $\psi$, simulated $B_{\phi}$ does not deviate from the prescribed
profile, so this method is generally not suitable for modelling compression.

The code also has the option to set $f|_{\Gamma}=0$, representing
an electrically insulating boundary through which toroidal flux can
leave the system.

\section{Vacuum field in insulating region\label{sec:Vacuum-field-in}}

\begin{figure}[H]
\subfloat[Full mesh]{\includegraphics[scale=0.5]{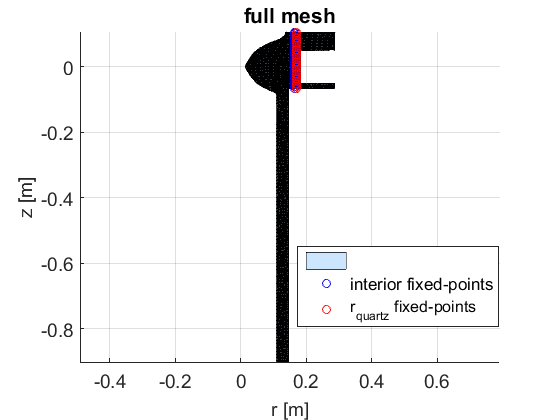}}
\hfill{}\subfloat[Full mesh (close up)]{\includegraphics[scale=0.5]{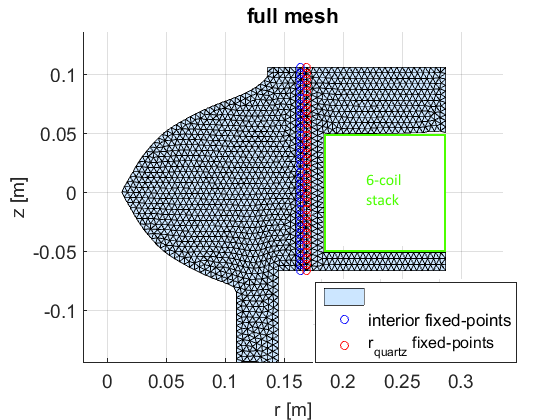}} 

\subfloat[Plasma mesh]{\includegraphics[scale=0.5]{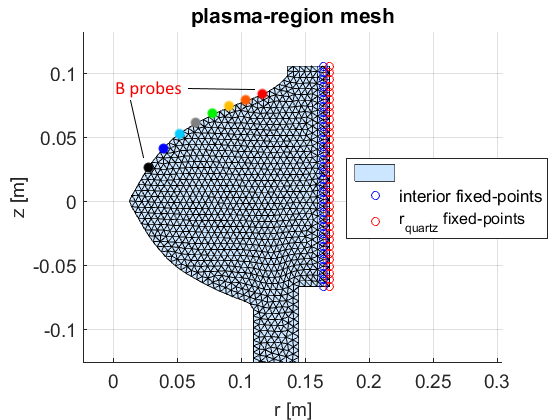}

} \hfill{}\subfloat[Insulator mesh]{\includegraphics[scale=0.5]{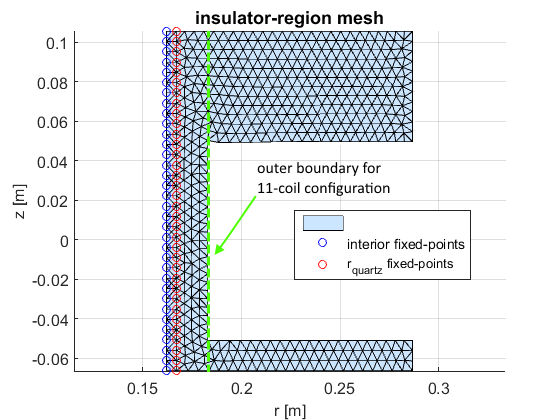}}

\caption{\label{fig:Grid-arrangement-with}$\,\,\,\,$Computational mesh with
insulating region, 6-coil configuration}
\end{figure}
In order to model the interaction of plasma with the insulating wall
during the CT formation process, a vacuum field should be solved for
in the insulating area between the inner radius of the insulating
outer wall and the levitation/compression coils. The insulating area
includes the wall itself as well as the air between the wall and the
coil-stack, and the air above and below the coil-stack. If this area
is included in the domain in which the plasma fields are solved, then
unphysical currents will be allowed to flow in insulating regions.
To solve for a vacuum field in the insulating region and couple it
to the plasma fields, the computational grid for the combined plasma
and insulating domain is split in two, as shown in figure \ref{fig:Grid-arrangement-with},
in which the original six levitation/compression coil configuration
is being modelled. The six coil stack representation is specifically
indicated in figure \ref{fig:Grid-arrangement-with}(b). For the 11-coil
configuration, in which the stack of coils extends along the entire
height of the insulating wall, the main part of the mesh (figure \ref{fig:Grid-arrangement-with}(c))
is unchanged, while the parts of the insulator mesh located above
and below the 6-coil stack are deleted, as indicated in figure \ref{fig:Grid-arrangement-with}(d).
Note that the magnetic probe locations are indicated in figure \ref{fig:Grid-arrangement-with}(c).
To couple between the plasma and vacuum solutions at each timestep,
we use the two vertical rows of shared fixed mesh points indicated
in the figure to exchange boundary values along the plasma/insulating
interface. 

No currents can flow in a vacuum, so 
\begin{equation}
\Delta^{^{*}}\psi_{v}=0\label{eq:111.1}
\end{equation}
where $\psi_{v}$ represents $\psi$ in a vacuum (or insulator). Defining
$\underline{\psi_{v}}=\underline{\psi_{v\Gamma}}+\underline{\psi_{vi}}$,
where $\underline{\psi_{v\Gamma}}$ and $\underline{\psi_{vi}}$ have
the values of $\psi$ set to zero at the nodes corresponding to internal
/ boundary nodes respectively, the discrete form of equation \ref{eq:111.1}
leads to $\underline{\underline{\Delta^{^{*}}}}\,\,\underline{\psi_{v}}=\underline{\underline{\Delta^{^{*}}}}\,\,\underline{\psi_{v\Gamma}}+\underline{\underline{\Delta_{0}^{^{*}}}}\,\,\underline{\psi_{vi}}=\underline{0}$
, so that: 
\begin{equation}
\underline{\psi_{vi}}=-\left(\underline{\underline{\Delta_{0}^{^{*}}}}\right)^{-1}*\left(\underline{\underline{\Delta^{^{*}}}}\,\,\underline{\psi_{v\Gamma}}\right)\label{eq:112}
\end{equation}
The discrete operator $\underline{\underline{\Delta_{0}^{^{*}}}}$
is defined in appendix \ref{sec:Numerical-solution-of}. $\underline{\underline{\Delta^{^{*}}}}$
and $\underline{\underline{\Delta_{0}^{^{*}}}}$ operators based on
the geometries of the plasma domain and the insulating domain were
constructed. With zero initial simulated plasma density everywhere,
$\psi$ in the entire domain is initially a vacuum field, so additional
$\underline{\underline{\Delta^{^{*}}}}$ and $\underline{\underline{\Delta_{0}^{^{*}}}}$
operators based on the geometry of the combined domain are required
in order to construct the initial vacuum $\psi$ solution. For formation
simulations, initially the time-relevant boundary values for $\psi$
are applied at the boundary points of the combined mesh, according
to equation \ref{eq:519}, and the vacuum field in the combined domain
is obtained using equation \ref{eq:112}. The sequence of subsequent
steps followed at each timestep to couple the solutions is as follows:
\begin{enumerate}
\item After evolving $\psi_{plasma}$ at the nodes in the plasma domain
according to equations \ref{eq:517.3}, boundary conditions $\psi(\mathbf{r},\,t)|_{\Gamma}$
are applied separately to the boundary points of the plasma and insulator
meshes.
\item The values at the nodes along the left boundary (\textquotedbl interior
points\textquotedbl{} - blue circles in figure \ref{fig:Grid-arrangement-with}(d))
of the insulator mesh are overwritten using the equivalent values
from $\psi_{plasma}$. 
\item Equation \ref{eq:112} is used to calculate $\psi_{v}$ in the insulating
area. Note that the operation $\underline{\underline{\Delta^{^{*}}}}\,\,\underline{\psi_{v\Gamma}}$
produces inaccurate results at the boundary points, because $\nabla_{\perp}\psi\neq0$
at the boundary, while the $\underline{\underline{\Delta^{^{*}}}}$
operator requires $\nabla_{\perp}\psi=0$ at the boundary for accurate
results at the boundary points. Therefore, for plotting purposes only,
the $\psi_{FEMM}$ boundary values, and then again the values at the
interior points from the plasma-grid solution, are reapplied to the
boundary points of the insulator grid.
\item Finally, the values at the outer right boundary of the plasma-grid
(red circles in figure \ref{fig:Grid-arrangement-with}(c)) are overwritten
using the equivalent values from the insulating region solution $\psi_{v}.$
\end{enumerate}
\newpage{}

\section{External sources of toroidal flux\label{sec:External-sources-of} }

The toroidal flux added to the system as a result of externally-driven
poloidal currents must be included in the model. Toroidal flux is
added due to formation current and due to crowbarred shaft current.
The crowbarred shaft current paths are schematically indicated in
figure \ref{fig:divertedcurrent}. Formation current or diverted crowbarred
shaft current driven through the plasma can not be modelled by setting
boundary conditions. When external sources of toroidal flux are included,
the continuous form of the expression for the time-rate of change
of $f$ is, referring to equation \ref{eq:214}, modified to: 
\begin{equation}
\dot{f}(\mathbf{r},\,t)=r^{2}\,\nabla\cdot\left(-\left(\frac{f}{r^{2}}\mathbf{v}\right)+\omega\mathbf{B}+\frac{\eta}{r^{2}}\nabla f\right)+\dot{f}{}_{external}(t)\label{eq:144.2}
\end{equation}
In practice, the expression for $f_{external}$ is added to the existing
value for $f$ at the beginning of each timestep, so that the natural
toroidal flux conserving boundary condition ($(\nabla_{\perp}f)|_{\Gamma}=0$)
is maintained on electrically conducting boundaries. Conservation
of the system's intrinsic toroidal flux implies that the total system
flux is equal to the initial flux plus the flux added due to $f_{external}(t)$,
at each time. For simulations which start with a static plasma and
include the formation process, $f_{external}(t)=f_{form}(z,\,t)$,
as defined in section \ref{subsec:Formation-simulations11}. Simulations
that start with a Grad-Shafranov equilibrium use $f_{external}(t)=F_{shaft}(t)$,
where the expression for $F_{shaft}(t)$ is derived in section \ref{subsec:Simulations-starting-withGS11}. 

\subsection{External sources of toroidal flux for formation simulations\label{subsec:Formation-simulations11}}

Physically, CT formation in a magnetized Marshall gun is achieved
as a result of $\mathbf{J}_{r}\times\mathbf{B}_{\phi}$ forces acting
on plasma, where $\mathbf{J}_{r}$ is the radial formation current
density across the plasma between the electrodes, and $\mathbf{B}_{\phi}$
is the toroidal field due to the axial formation current in the electrodes.
Open stuffing magnetic field lines that are resistively pinned to
the electrodes, and frozen into the conducting plasma, are advected
with the plasma by the $\mathbf{J}_{r}\times\mathbf{B}_{\phi}$ force,
into the containment region, where they reconnect to form CT closed
flux surfaces. Simulated formation is initiated with the addition
of toroidal flux below the physical locations of the gas puff valves
- radial formation current is assumed to occur at the z-coordinate
of the valves, where the gas density is initially highest.

Integrating Faraday's law in the poloidal plane over the area defining
the formation current path (up the outer gun electrode as far as the
region with the initially highest gas concentration at the z-coordinate
of the valves, through the plasma across the intra-electrode gap,
and down the inner electrode) we have: 
\begin{align}
 & \int\,\nabla\times\mathbf{E}_{\theta}(\mathbf{r},t)\cdot d\mathbf{S}=-\int\,\dot{\mathbf{B}}_{\phi}(\mathbf{r},t)\cdot d\mathbf{S}\nonumber \\
 & \Rightarrow\int\,\mathbf{E}_{\theta}(\mathbf{r},t)\cdot d\mathbf{l}=V(t)=-\dot{\Phi}_{form}(t)\nonumber \\
 & \Rightarrow\Phi_{form}(t)=-\intop_{0}^{t}V(t')\:dt'\label{eq:145}
\end{align}
Here, 
\begin{equation}
V(t)=V_{gun}(t)+I_{form}(t)\,R(t)\label{eq:145.001}
\end{equation}
where $V_{gun}(t)$ is the voltage measured across the formation electrodes
and $I_{form}(t)\,R(t)$ is the resistive voltage drop along the formation
current path. $\mathbf{E}_{\theta}(\mathbf{r},t)$ is the poloidal
electric field that is established through the application of $V_{gun}(t)$,
and $\dot{\mathbf{B}}_{\phi}(\mathbf{r},t)$ and $\dot{\Phi}_{form}(t)$
are the time-rates of change of the toroidal magnetic field and toroidal
flux induced in the area bounded by the path along which formation
current driven by $\mathbf{E}_{\theta}(\mathbf{r},t)$ flows. 

For the simplifying assumption of a fixed formation current path of
constant inductance, $-V(t)=\dot{\Phi}_{form}(t)=L\dot{I}_{form}(t)$,
where $I_{form}(t)$ is the formation current and $L$ is the inductance
of the formation current path. $R(t)$ in equation \ref{eq:145.001}
is a resistance that depends on the resistivity of the metal along
the formation current path, and also on the resistivity of the plasma
between the gun electrodes. In addition, it is expected that when
plasma, which is advected upwards during the formation process, occupies
the gap between the chalice and the inner gun electrode (see figure
\ref{fig:Machine-Schematic-1}), that part of the formation current
will flow on a path up the outer electrode, through the aluminum bars
located outside the insulating wall (see figure \ref{fig:Schematic-of-6}(a)),
down the chalice, across the plasma in the gap, and down the inner
electrode. Physically, $R(t)$ includes the contribution of the resistance
of the plasma in the gap shortly after initiation of the formation
process. Note that the chalice ($i.e.,$ the inner flux conserver
indicated in figure \ref{fig:Chalice}) and inner electrode are modelled
as a continuous conductor - we do not have sufficient information,
for example the voltage measured across the gap, to properly model
the effect of the gap. For simplicity, we assume here that $R$ is
constant in time. Thus the expression for the voltage measured across
the formation electrodes is

\begin{align}
-V_{gun}(t) & =L\dot{I}_{form}(t)+I_{form}(t)R\nonumber \\
 & =\dot{\Phi}_{form}(t)+\frac{\Phi_{form}(t)}{\tau_{LR}}\label{eq:150.1}
\end{align}
where $\tau_{LR}=L/R$ is the $LR$ time determining the formation
current decay rate. This time constant can be estimated from the e-folding
time of the toroidal field that is experimentally measured at the
probes embedded in the chalice. The toroidal field at the probes is
a measure of the crow-barred shaft current that flows poloidally around
the machine after formation. The value of $\tau_{LR}$ varies depending
on the formation current path and resistances of the intra-electrode
plasma and the plasma in the gap between the chalice and the gun inner
electrode, so it can vary from shot to shot depending on plasma conditions.
For relatively long-lived magnetically levitated CTs, $\tau_{LR}\sim90\,\upmu$s.
Reduced values of $\tau_{LR}$ can be chosen as a simulation input
when shorter-lived CTs ($e.g.,$ for simulations with relatively high
thermal diffusion coefficients, or low flux CTs associated with reduced
$V_{form}$ and $I_{main}$) are being modelled. Assuming that $\Phi_{form}(t)$
can be expressed as $\Phi_{form}(t)\sim C(t)e^{-\frac{t}{\tau_{LR}}}$,
then with the initial condition $\Phi_{form}(0)=0=C(0)$, equation
\ref{eq:150.1} yields $C(t)=-\intop_{0}^{t}V_{gun}(t')\,e^{\frac{t'}{\tau_{LR}}}\:dt'$,
so that 
\begin{equation}
\Phi_{form}(t)=-e^{-\frac{t}{\tau_{LR}}}\intop_{0}^{t}V_{gun}(t')\,e^{\frac{t'}{\tau_{LR}}}\:dt'\label{eq:145.01}
\end{equation}
By definition,
\begin{equation}
\Phi_{form}(t)=\int\frac{f_{form}(z,t)}{r}dr\;dz\label{eq:146}
\end{equation}
 $f_{form}(z,t)$ may be expressed as
\begin{equation}
f_{form}(z,t)=A_{form}(t)\,g_{form}(z,t)\label{eq:147}
\end{equation}
where $A_{form}(t)$ determines the amplitude of $f_{form}(z,t)$,
and $g_{form}(z,t)$ is a geometric profile that determines where
plasma formation current flows. The smooth logistic profile
\begin{equation}
g_{form}(z,t)=\frac{e^{m\,z_{I}(t)}}{e^{m\,z_{I}(t)}+e^{m\,z}}\label{eq:145-1}
\end{equation}
defines $g_{form}(z,t)$, where $m\sim40$ defines the profile slope,
and $z_{I}(t)$ is the $z$ coordinate at which the greatest concentration
of radial formation current flows between the machine electrodes.
For simplicity, we neglect the time dependence in $g_{form}$ by replacing
$z_{I}(t)$ with $z_{gp}=-0.43\mbox{ m}$, the $z$ coordinate of
the machine's gas puff valves, around which the greatest concentration
of radial formation current is expected to flow. Note, $z=0$ is defined
to be at the chalice waist. 
\begin{figure}[H]
\subfloat[$g_{form}(z)$ profile for formation flux input]{\begin{centering}
\includegraphics[width=7cm,height=5cm]{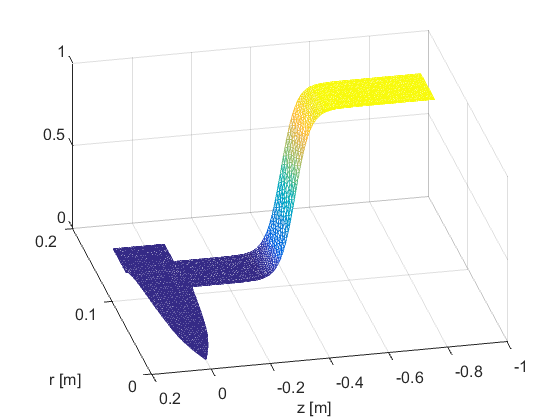}
\par\end{centering}
}\hfill{}\subfloat[Measured $V_{gun}(t)$ and calculated $\Phi_{form}(t)$]{\centering{}\includegraphics[width=7cm,height=5cm]{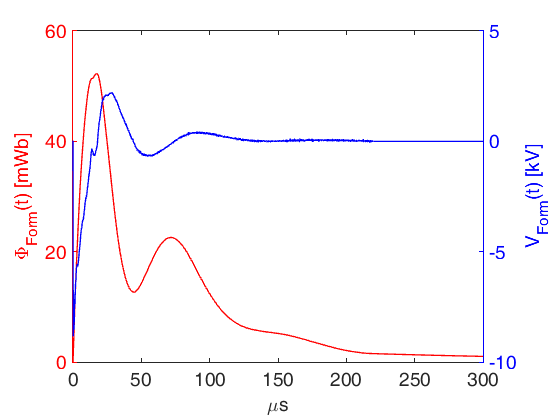}}\caption{\label{fig:gform_Phiform_Vform}$\,\,\,\,$Geometric profile for formation
flux input and $\Phi_{form}(t)$ scaling}
\end{figure}
Figure \ref{fig:gform_Phiform_Vform}(a) indicates the logistic function
formation profile. As outlined in section \ref{sec:Time-dependent-formation-profile},
it is possible to reproduce the measured formation current over the
bulk of the simulated formation process, when the time dependence
of $g_{form}$ is included. Figure \ref{fig:gform_Phiform_Vform}(b)
shows the experimentally measured formation voltage signal (right
axis) which is numerically integrated over time to evaluate $\Phi_{form}(t)$
(left axis) as defined by equation \ref{eq:145.01}.

Combining equations \ref{eq:146} and \ref{eq:147}, we have:
\begin{equation}
\Phi_{form}(t)=A_{form}(t)\,\int\frac{g_{form}(z)}{r}dr\;dz\label{eq:148}
\end{equation}
Together with equation \ref{eq:145.01}, this gives: 
\begin{equation}
A_{form}(t)=\frac{-e^{-\frac{t}{\tau_{LR}}}\intop_{0}^{t}V_{gun}(t')\,e^{\frac{t'}{\tau_{LR}}}\:dt'}{\int\frac{g_{form}(z)}{r}dr\;dz}\label{eq:149}
\end{equation}
Combining equations \ref{eq:147} and \ref{eq:149} yields the expression
for $f_{form}(z,t)$, which is used in equation \ref{eq:144.2} to
simulate the formation process: 
\begin{equation}
f_{form}(z,t)=\Phi_{form}(t)\,\kappa_{F}(z)\label{eq:150}
\end{equation}
where 
\begin{equation}
\Phi_{form}(t)=-e^{-\frac{t}{\tau_{LR}}}\intop_{0}^{t}V_{gun}(t')\,e^{\frac{t'}{\tau_{LR}}}\:dt'\label{eq:150-1}
\end{equation}
and

\begin{equation}
\kappa_{F}(z)=\frac{g_{form}(z)}{\int\frac{g_{form}(z')}{r}dr\;dz'}\label{eq:150-2}
\end{equation}
As outlined in section \ref{subsec:bcs psi}, the values of experimental
measurements are defined at the times defined by the digitiser sampling
time, $dt_{exp}=50\mbox{ ns}$. Matlab function handles in the MHD
code enable assessment of $f_{form}(t)$ at the required simulation
times ($dt_{sim}\sim0.01\mbox{ ns}$), either by interpolation, or
by evaluation of a high-order polynomial fit to the experimental signal
for $V_{gun}(t)$. 

\subsubsection{Time-dependent formation profile\label{sec:Time-dependent-formation-profile}}

\begin{figure}[H]
\subfloat[Geometry definitions]{\centering{}\includegraphics[width=6cm,height=5cm]{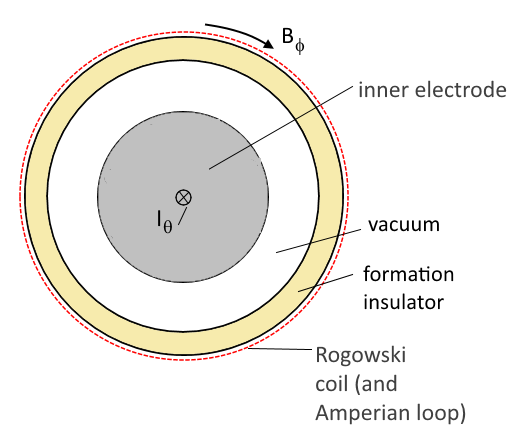}}\hfill{}\subfloat[Measured vs. simulated $I_{form}(t)$ (time-dependent $z_{I}$, sim.
2225)]{\centering{}\includegraphics[width=8cm,height=6cm]{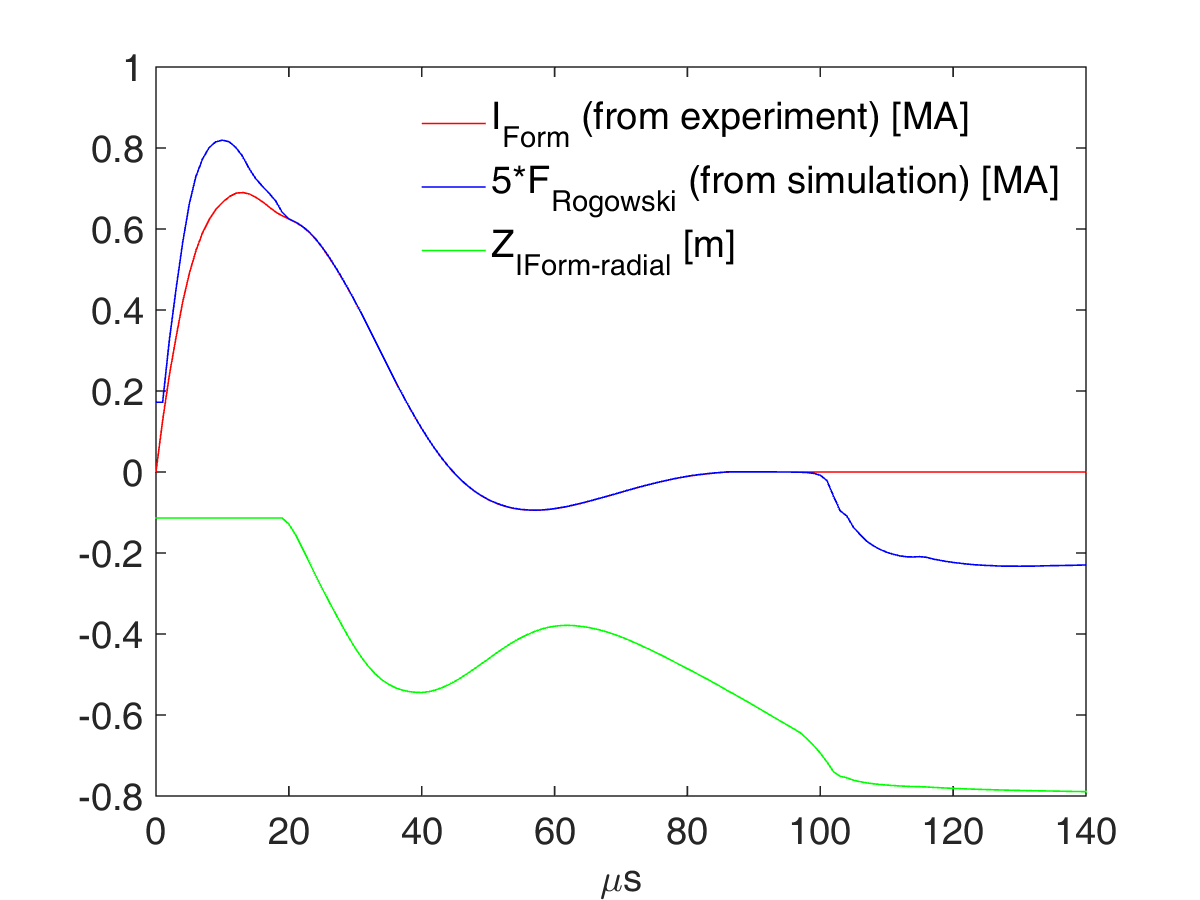}}\caption{\label{fig:Simulated-vs-measured}$\,\,\,\,$Geometry definitions
and measured vs. simulated $I_{form}(t)$ }

\end{figure}

As indicated in figure \ref{fig:Machine-Schematic-1}, formation current
and voltage measurements were recorded. $V_{gun}(t)$ across the electrodes
was measured using a voltage divider setup, and $I_{form}(t)$ was
measured using a Rogowski coil wrapped around the formation insulator
at the bottom of the machine. Prior to firing the formation capacitors,
gas is puffed into the vacuum vessel through the eight gas valves
located at $z=-0.43$ m. Gas (usually helium) is generally puffed
from a gas bottle at $\sim30$ psi - the gas valves, which have an
opening time of $\sim1$ ms, are generally opened $\sim400\,\upmu$s
before firing the formation banks. The gas can diffuse away from the
gas valve locations, but the highest density during the puff is always
located around the gas valves. We don't know the axial coordinate
($z_{I}(t)$) at which radial formation current, $I_{Form-r}(t)$
flows across the gun electrodes and through the plasma, but it is
expected to be concentrated where the gas cloud is densest. However
the formation current path through the plasma is also determined by
energy considerations - the system adjusts so as to have the lowest
possible energy, where the inductive contribution to the magnetic
energy is proportional to the inductance of the formation current
path ($U=\frac{1}{2}LI^{2}$). To accurately simulate the formation
process, it would be convenient to be able to calculate $z_{I}(t)$.
As shown in this section, it is possible to do that using a combination
of the $V_{gun}(t)$ and $I_{form}(t)$ measurements, and some simulated
parameters, at some, but not all, times during the formation process.
Integrating over the region enclosing axially-directed formation current
flowing down the inner electrode, in a plane of constant $z$, at
the $z$ coordinate of the Rogowski coil around the lower formation
insulator (see figure \ref{fig:Simulated-vs-measured}(a)), and using
Ampere's law with toroidal symmetry, we have $\int\nabla\times\mathbf{B}_{\phi}(r,t)\cdot d\mathbf{S}=\mu_{0}\int\mathbf{J}_{\theta}(r,t)\cdot d\mathbf{S}\,\Rightarrow rB_{\phi}=F_{Rog}=\frac{4\pi\times10{}^{-7}}{2\pi}I_{form}\,[\mbox{A}]$,
so that:
\begin{equation}
F_{Rog}=0.2\,I_{form}\,[\mbox{MA}]\label{eq:151}
\end{equation}
Here, the formation current is in mega-amps, and $F_{Rog}$ is the
total value of $rB_{\phi}$ at a fixed location $(r_{Rog},\,z_{Rog})$
just inboard of the Rogowski coil (physically, the Rogowski coil is
located just outboard of the simulation domain). We can define:
\begin{equation}
F_{Rog}=0.2\,I_{form}[\mbox{MA}]=F_{Form-Rog}+f_{plasma-Rog}\label{eq:154}
\end{equation}
$F_{Form-Rog}$ is the part of $F_{Rog}$ that is present due to the
voltage applied between the electrodes, and $f_{plasma-Rog}$ is the
part of $F_{Rog}$ that is present due to (poloidal) plasma currents.
Since $F_{Form-Rog}$ is the value of $f_{form}(z,t)$ (as defined
in equation \ref{eq:150}) at $(r_{Rog},\,z_{Rog})$, we have: 
\begin{equation}
f_{form}(z_{Rog},t)=\Phi_{form}(t)\,\kappa_{F}(z_{Rog},t)=0.2\,I_{form}[\mbox{MA}]-f_{plasma-Rog}
\end{equation}
Here, 
\[
\kappa_{F}(z_{Rog},t)=\frac{g_{form}(z_{Rog},t)}{\int\frac{g_{form}(z',t)}{r}dr\,dz'}
\]
and referring to equation \ref{eq:145-1}, 
\[
g_{form}(z_{Rog},t)=\frac{e^{m\,z_{I}(t)}}{e^{m\,z_{I}(t)}+e^{m\,z_{Rog}}}
\]
Choosing $z_{I}(t)<-0.1$ m (the upper coordinate of the straight
section of the Marshall gun barrel is at $z\sim-0.085$ m), $\int\frac{g_{Form}(z',t)}{r}dr\,dz'=h_{F}\,\mbox{ln}(r_{out}/r_{in})$,
where $r_{in}$ and $r_{out}$ are the inner and outer radii of the
gun electrodes, and $h_{F}=-z_{min}+z_{I}=0.9+z_{I}$ is the height
of the simulated formation current loop. Here, $z_{min}=-0.9$ m is
the $z$ coordinate of the bottom of the barrel of the magnetized
Marshall gun. Combining these expressions, we can write:
\begin{equation}
\Phi_{form}(t)\,g_{form}(z_{Rog},t)=(0.2\,I_{form}(t)[\mbox{MA}]-f_{plasma-Rog}(t))(0.9+z_{I}(t))\,\mbox{ln}(r_{out}/r_{in})
\end{equation}
Noting $z_{Rog}\sim-0.8$ m, so that $g_{Form}(z_{Rog},t)=1$ (see
figure \ref{fig:gform_Phiform_Vform}(a)), we arrive at the expression:
\begin{equation}
z_{I}(t)=\frac{-e^{-\frac{t}{\tau_{LR}}}\intop_{0}^{t}V_{gun}(t')\,e^{\frac{t'}{\tau_{LR}}}\:dt'}{(0.2\,I_{form}(t)[\mbox{MA]}-f_{plasma-Rog}(t))\,\mbox{ln}(r_{out}/r_{in})}-0.9\label{eq:155}
\end{equation}
By recording $f_{plasma-Rog}(t)$ (the part of $rB_{\phi}$ at $(r_{Rog},\,z_{Rog})$
that is present due to (poloidal) plasma currents) after each timestep
(or half/quarter timestep if using Runge-Kutta two/four timestepping),
we can use equation \ref{eq:155} to find the value of $z_{I}(t)$.
In practice, it's necessary to limit $z_{I}(t)$ to lie between some
upper and lower z coordinates. If $z_{I}(t)$ is too high up the gun,
the level of poloidal stuffing flux due to the main coil that will
be advected upwards with the accelerating plasma during the formation
process will be minimised, resulting in a low-flux CT. On the other
hand, $z_{I}(t)$ can't be too low because the gun has a finite length.
To test the effect of varying $z_{I}(t)$, we imposed upper/lower
limits on $z_{I}(t)$ of $-0.11$ m and $-0.8$ m. At each timestep,
updated values for $\kappa_{F}(z_{I}(t))$ are included in the expression
for $f_{external}(t)=f_{form}(t)$ (from equation \ref{eq:150}) in
the code's main time loop, with any calculated values of $z_{I}(t)$
that lie outside the imposed limits being replaced with the nearest
limit. By recording $F_{Rog}(t)$ (the total value of $rB_{\phi}$
at $(r_{Rog},\,z_{Rog})$), after each timestep, we can use equation
\ref{eq:151} to find the simulated formation current as $I_{form}\,[\mbox{MA}]=5F_{Rog}$.
Figure \ref{fig:Simulated-vs-measured}(b) shows a comparison of simulated
and measured $I_{form}(t)$, for a case with $z_{I}(t)$ limited to
$-0.11\mbox{ m}>z_{I}(t)>-0.8\mbox{ m}$. For the first $20\,\upmu$s
of simulation time, $z_{I}(t)$ was limited to the upper bound $max(z_{I})=-0.11$m.
The measured and simulated $I_{form}(t)$ match at times when $z_{I}(t)$
was free to vary, between $20\,\upmu$s and $\sim100\,\upmu$s. This
simulation includes the approximation for the resistive contribution
to $\Phi_{form}(t)$ (equation \ref{eq:150-1}) , with $\tau_{RC}=90\,\upmu$s.
When the resistive contribution is neglected ($i.e.,$ $\Phi_{form}(t)\rightarrow-\intop_{0}^{t}V_{gun}(t')\:dt'$),
the experimental $I_{form}(t)$ is matched only until $\sim50\,\upmu$s.
The high values of $z_{I}(t)$ over the first $20\,\upmu$s, when
a large proportion of the total formation current is driven, leads
to formation of a CT with lower poloidal flux (\textasciitilde 6
mWb $cf.$ \textasciitilde 10 mWb for simulations where $z_{I}$
is fixed at $z_{gp}=-0.43$ m), and consequentially low simulated
$B_{\theta}$ (maximum \textasciitilde 1.2 T at bubble-in, recorded
at the location of the magnetic probe embedded in the chalice at $r=26$
mm, $cf.$ $\sim2$ T for cases with fixed $z_{I}$). 

If more information, including a voltage measurement across the gap
between the chalice and the inner gun electrode, was available to
determine the distribution of formation currents within the machine,
an optimised expression for $\Phi_{form}(t)$ may lead to a better
match between simulated and measured $I_{form}(t)$. In principle,
it should be possible to get a match at all simulated times with a
complete model of the formation process, and if physically correct
plasma parameters were implemented. For example, classical (Spitzer)
resistivity scales with $T_{e}^{-1.5}$, and $T_{e}$ is in turn is
a function of $T_{i},\,\nu,\,n,\,\chi_{\parallel i},\,\chi_{\parallel e},\,\chi_{\perp i},\,\chi_{\perp e}$.
To match experimental $I_{form}$ at all times, we would likely need
to include an extended model for transport processes (neoclassical,
anomalous resistivity etc). Energy losses due to atomic processes
such as impurity line radiation, charge exchange, impact ionization
and recombination should also be included. Ultimately, the simulation
is 2D and neglects inherently 3D dissipation associated with turbulent
transport, this factor is likely another major barrier against seeing
perfect matches between measured and simulated quantities, including
the measured and simulated formation current. 

\subsection{External sources of toroidal flux for simulations starting with a
Grad-Shafranov equilibrium\label{subsec:Simulations-starting-withGS11}}

For the case where the plasma dynamics are evolved starting from an
initial Grad-Shafranov equilibrium, we include experimentally measured
crowbarred shaft current, taken from a reference non-compression shot,
as an external source for toroidal flux. The decrease in system toroidal
flux due to shaft current reduction can be modelled, while conserving
the system's intrinsic toroidal flux. Along with the explicitly applied
boundary conditions $\mathbf{v}|_{\Gamma}=\mathbf{0}$, the toroidal
flux conserving boundary condition $\left(\nabla_{\perp}f\right)|_{\Gamma}=0$
is automatically imposed, and the total system flux at each time is
equal to the initial flux plus the flux added due to the shaft current
source. 

In the experiment, there are at least two crowbarred shaft current
paths, as schematically indicated in figure \ref{fig:divertedcurrent}.
The magnitude of the crowbarred current in the upper path around the
CT confinement region can be calculated using Ampere's law and the
toroidal field measured at the magnetic probes embedded in the chalice.
There were also a total of twelve probes on the gun shaft, at four
toroidal locations, with $z$ coordinates $-42\mbox{\mbox{ cm}},\,-32\mbox{\mbox{ cm}}$
and $-22\mbox{\mbox{ cm}}$ (where $z=0$ is the coordinate of the
chalice waist), but those probes were not working reliably at any
time, and so the crowbarred current on the lower path can't be determined.
In any case, for simulations with an initial Grad-Shafranov equilibrium,
the computational mesh used is a shorter one that doesn't include
the gun region, so that the lower crowbarred current amplitude is
not required.

The measured values of $B_{\phi}(t)$ at the chalice probes, taken
from a reference non-compression shot, are averaged over toroidal
angle to find $F_{shaft}(t)$. Ampere's law can be integrated over
an Amperian circular loop with $r=r_{probe}$ in the plane at constant
$z=z_{probe}$, where (for the original chalice configuration) $r_{probe}$
is any of the eight chalice probe radii listed in table \ref{tab: coordinates-ofBprobes},
and $z_{probe}$ is the corresponding $z$ coordinate of the relevant
probe pair: 
\begin{align}
 & \int\nabla\times\mathbf{B_{\phi}}\cdot d\mathbf{S}=\mu_{0}\int\mathbf{J}_{\theta}\cdot d\mathbf{S}\nonumber \\
 & \Rightarrow2\pi rB_{\phi}(r,t)=\mu_{0}I_{shaft}\label{eq:145.3}
\end{align}
Note that $B_{\phi}(r,t)$ scales with $\frac{1}{r}$, so that $I_{shaft}$
and $F_{shaft}$ don't have dependence on $r$:

\begin{align}
F_{shaft}(t) & =rB_{\phi}(r,t)=\frac{\mu_{0}I_{shaft}(t)\mbox{[A]}}{2\pi}\nonumber \\
\Rightarrow F_{shaft}(t) & =0.2\;I_{shaft}(t)\mbox{[MA]}\label{eq:145.34}
\end{align}
For equilibrium-evolved simulations, $f_{external}(t)=F_{shaft}(t)$
is generally included as an external source in equation \ref{eq:144.2},
which describes the evaluation of $\dot{f}$ in the code's main time
loop. As outlined in section \ref{subsec:bcs F}, when code input
parameter $Shaftbc=1$, $F_{shaft}(t)$ is used to determine boundary
conditions for $f$, and $f_{external}(t)$ is set to zero. However
by default, due to the unsuitability of the boundary condition method
for modelling magnetic compression, $Shaftbc=0$ and no boundary conditions
on $f$ are applied except on the interface between the plasma and
insulating regions. 

As shown in figures \ref{fig:Poloidal-field-for}(c) and (d), the
experimentally measured crowbarred shaft current typically has a peak
of around $300\mbox{ kA}$ and decays to zero at a rate slower than
the decay rate of the plasma current, as indicated by the poloidal
field decay rate. 
\begin{figure}[H]
\subfloat[Normalised shaft current signal]{\begin{centering}
\includegraphics[width=7cm,height=5cm]{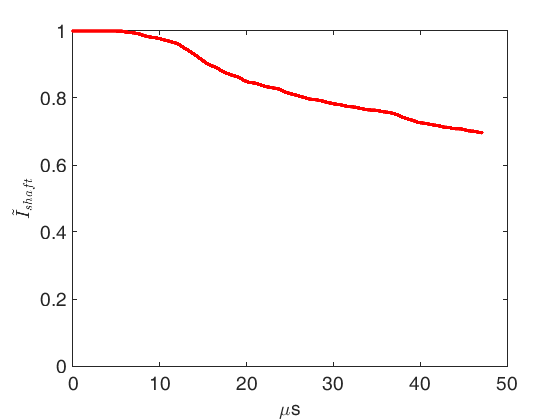}
\par\end{centering}
}\hfill{}\subfloat[$B_{\phi}$ for a simulation starting with a Grad-Shafranov equilibrium
(no compression)]{\begin{centering}
\includegraphics[width=7cm,height=5cm]{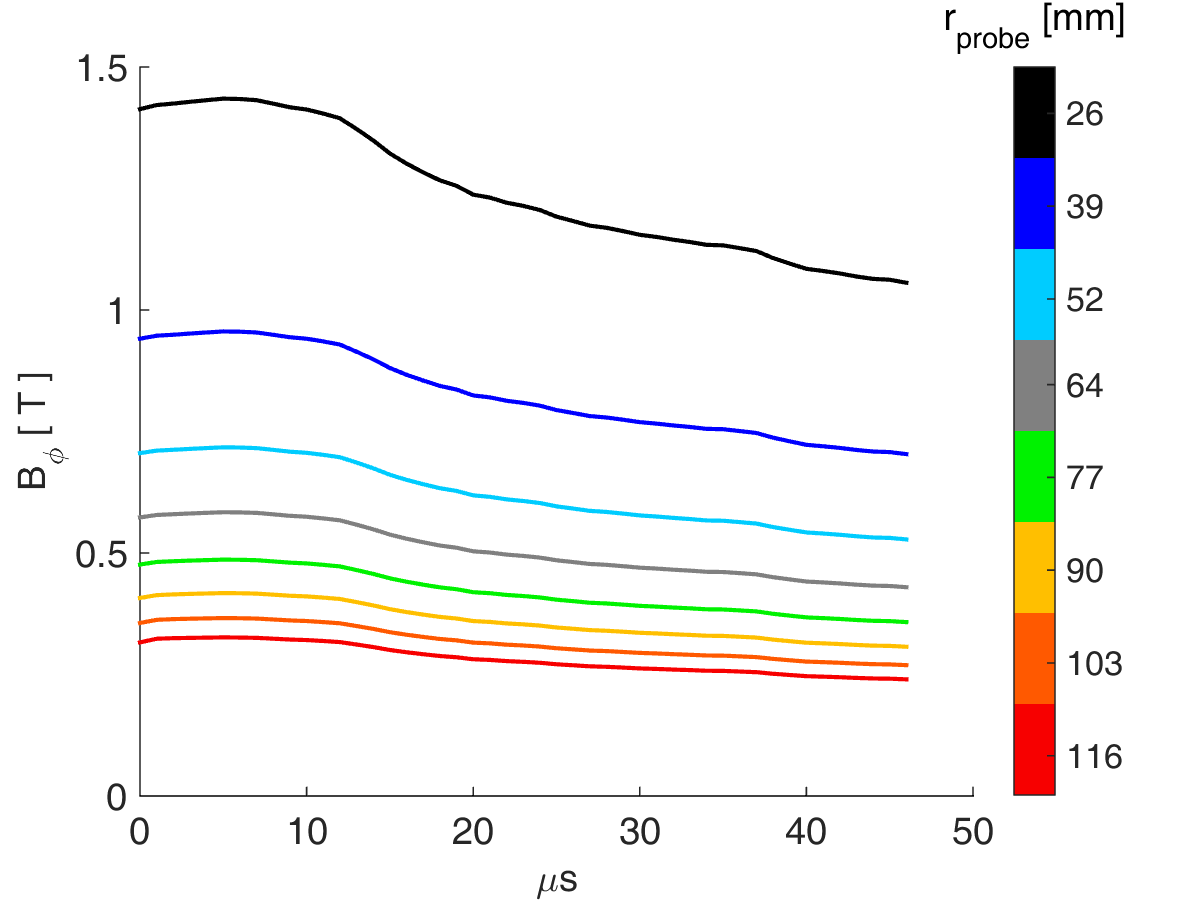}
\par\end{centering}
}

\subfloat[$B_{\phi}$ at compression for a simulation starting with a Grad-Shafranov
equilibrium]{\begin{centering}
\includegraphics[width=7cm,height=5cm]{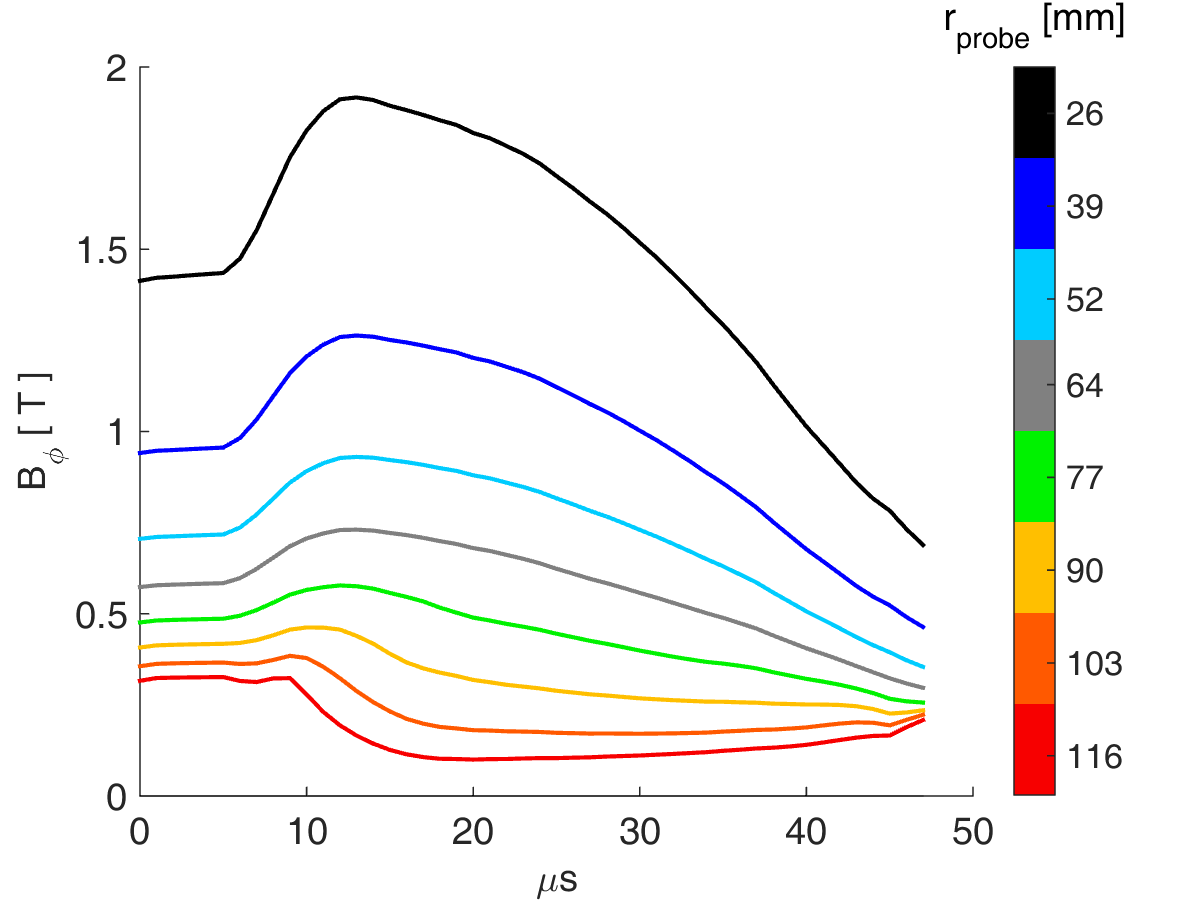}
\par\end{centering}
}\hfill{}\subfloat[$B_{\phi}$ from shot  39862 with $t_{comp}=90\,\upmu$s ]{\begin{centering}
\includegraphics[width=7cm,height=5cm]{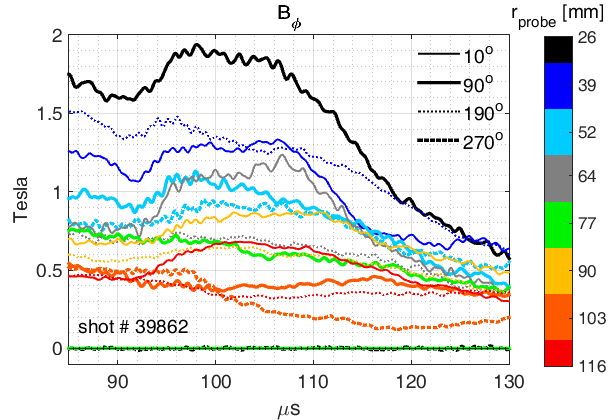}
\par\end{centering}
}\caption{$\,\,\,\,$Comparison of measured and simulated $B_{\phi}$ (Grad-Shafranov
equilibrium)\label{fig:Ishaft_sim_Btor_GS}}
\end{figure}
By default, simulations starting with a Grad-Shafranov equilibrium
use input profiles for $\widetilde{I}_{shaft}(t)$ and $\widetilde{I}_{lev}(t)$
that have their starting values set to the recorded values at $t=t_{comp}$,
where $t_{comp}$ is the code input parameter determining the time
of magnetic compression initiation. Figure \ref{fig:Ishaft_sim_Btor_GS}(a)
shows $\widetilde{I}_{shaft}(t)$, the crowbarred shaft current, calculated
using equation \ref{eq:145.3}, and normalised with its value at $t=t_{comp}$.
This profile, with values at simulation-time interpolated from the
digitised $B_{\phi}$ signals, which have sampling interval $dt_{exp}=50\mbox{ ns}$,
is used along with the value of $I_{shaft}$ at $t=t_{comp}$, to
find $f_{external}(t)=F_{shaft}(t)$ at each timestep. Note that the
current is held constant at its maximum value over the first $5\,\upmu$s
(this simulation example has code input parameter $t_{rise}=5\,\upmu$s).
For equilibrium-evolved simulations, the signal that determines the
scaling of the compression $\psi$ boundary conditions (see section
\ref{subsec:bcs psi}) is also held constant (at zero) over the first
$5\,\upmu$s, and the levitation-relevant signal is ramped up to its
starting value over the first $5\,\upmu$s. The Grad-Shafranov equilibrium
is calculated with the boundary condition $\psi|_{\Gamma}=0$ (as
described in appendix \ref{sec:Numerical-solution-of}), and it is
required to slowly ramp up the boundary conditions for $\psi_{lev}$
to the starting value, corresponding to $\psi_{lev}$ at $t=t_{comp}$,
in order to avoid sharp gradients in pressure that can lead to numerical
instability. 

For simulations with $Shaftbc=0$, the simulated diagnostic for $B_{\phi}$
at the chalice magnetic probe locations \emph{approximately} follows
the experimentally-determined shaft current and toroidal field profile,
except at compression. Figure \ref{fig:Ishaft_sim_Btor_GS}(b) shows
$B_{\phi}$ at the chalice probe locations from a simulation starting
with a Grad-Shafranov equilibrium, without compression and with $Shaftbc=0$.
It can be seen how, for example, individual $B_{\phi}$ traces are
not exactly constant over the first five microseconds, while the shaft
current signal is, as discussed, held constant over that period. Consistent
with the naturally imposed boundary condition $\left(\nabla_{\perp}f\right)|_{\Gamma}=0$
(when boundary conditions $\mathbf{v}|_{\Gamma}=\mathbf{0}$ are applied
explicitly), this is due to wall-perpendicular currents that act to
conserve system toroidal flux. If $Shaftbc$ is set to one, the $B_{\phi}$
traces precisely follow the prescribed profile. For compression simulations,
with $Shaftbc=0$, simulated $B_{\phi}$ qualitatively follows the
experimental measurement signals at compression as shaft current is
diverted through the plasma at the CT edge (see section \ref{subsec:Compressional-Instability}).
Figure \ref{fig:Ishaft_sim_Btor_GS}(c) shows $B_{\phi}$ at the chalice
probe locations from a simulation starting with a Grad-Shafranov equilibrium,
with magnetic compression. Compression starts at $t_{rise}=5\,\upmu$s.
Figure \ref{fig:Ishaft_sim_Btor_GS}(d) shows measured $B_{\phi}$
for representative shot  39862, which had $t_{comp}=90\,\upmu$s (note
that the $B_{\phi}$ probes at $(r,\,\phi)=(26$ mm, $270^{o}$) and
$(r,\,\phi)=(77$ mm, $270^{o}$) were not functioning for this shot).
The observed toroidal asymmetry is associated with the compressional
instability discussed in section \ref{subsec:Compressional-Instability},
and can't be captured in a 2D simulation. Qualitatively, toroidal
field increases more at compression at the inner probes, and as also
reproduced in the simulation, toroidal field decreases at the outer
probes as shaft current is diverted to a path inside the locations
of the outer probes.

\section{Toroidal flux conservation with insulating region\label{sec:PHIconservation-with}}

\subsection{\label{subsec:PHIcons_form}Formation simulations}

Integrating toroidal field over the area of the total domain (combined
plasma and insulating wall regions) in the poloidal plane, we have:

\begin{equation}
\Phi_{tot}(t)=\overset{\underbrace{\int_{\Omega_{P}}\frac{f_{P}(\mathbf{r},t)}{r}dr\,dz+\int_{\Omega_{I}}\frac{f_{I}(t)}{r}dr\,dz}}{\Phi_{PI}(t)}+\overset{\underbrace{\int_{\Omega_{P}+\Omega_{I}}\frac{f_{form}(z,t)}{r}dr\,dz}}{\Phi_{form}(t)}\label{eq:176}
\end{equation}
The subscripts $P$ and $I$ refer to plasma and insulating wall regions.
Note that the insulating wall area includes the insulating wall only,
and not the insulating (air) regions above and below the coil stack
as depicted in figure \ref{fig:Grid-arrangement-with}. Poloidal currents
in the aluminum bars located outboard of the insulating wall (depicted
in figure \ref{fig:Schematic-of-6}(a)) add toroidal flux to the system
only in the insulating wall and plasma regions. Since no currents
can flow in insulators, $f$ must be spatially constant in the insulating
wall inside aluminum bars. The flux-conserving natural boundary condition
that is imposed on $f_{P}$ (which is evolved according to equation
\ref{eq:144.2}) must be overwritten on the interface between the
insulating wall and the plasma domain. The second term in equation
\ref{eq:176} can be re-expressed as $\int_{\Omega_{I}}\frac{f_{I}(t)}{r}dr\;dz=f_{I}(t)\,h_{I}\,\mbox{ln}(r_{out}/r_{in})$,
where $h_{I}$ is the height of the rectangular cross-section of the
insulating wall, and $r_{out}$ and $r_{in}$ are the outer and inner
radii of the wall. Initial system toroidal flux is zero, and $\Phi_{PI}(t)$
is to be conserved by design, so the system's only source of toroidal
flux is $\Phi_{form}(z,t)$:
\begin{align}
 & \Phi_{tot}(t)=\Phi_{form}(t)\Rightarrow\Phi_{PI}(t)=0\nonumber \\
 & \Rightarrow\int_{\Omega_{P}}\frac{f_{P}(\mathbf{r},t)}{r}dr\;dz+\int_{\Omega_{I}}\frac{f_{I}(t)}{r}dr\;dz=0\nonumber \\
 & \Rightarrow\stackrel[i=1]{N_{n}}{\Sigma}\left(\frac{f_{P_{i}}(t)\,s_{i}}{3r_{i}}\right)+f_{I}(t)\left(h_{I}\,\mbox{ln}(r_{out}/r_{in})\right)=0\label{eq:140}
\end{align}
Here, $N_{n}$ is the number of nodes in the non-insulating part of
the domain ($i.e.,$ the plasma domain) in which the MHD equations
are solved, $s_{i}/3$ is the area, and $r_{i}$ is the radial coordinate,
associated with node $i$. The summation is over all nodes in the
plasma domain, which includes the fixed-point nodes along the inner
radius of the insulating wall, at which we want to evaluate $f_{I}$.
Equation \ref{eq:140} can be solved for the constant $f_{I}$ if
$f_{P_{i}}$ is temporarily set to zero at the fixed-point nodes along
the inner wall of the insulator (see figure \ref{fig:Grid-arrangement-with}),
so that $f_{P}\rightarrow f_{P0}$. Equation \ref{eq:140} is modified
as:
\begin{equation}
\stackrel[i=1]{N_{n}}{\Sigma}\left(\frac{f_{P0_{i}}(t)\,s_{i}}{3r_{i}}\right)+\left(\stackrel[j=1]{N_{int}}{\Sigma}\left(\frac{s_{j}}{3r_{j}}\right)+h_{I}\,\mbox{ln}(r_{out}/r_{in})\right)f_{I}(t)=0\label{eq:141}
\end{equation}
Here, $\stackrel[j=1]{N_{int}}{\Sigma}$ implies summation over the
interface fixed-point nodes along the inner insulating wall. In equation
\ref{eq:141}, $\tilde{L}_{ins}=h_{I}\,\mbox{ln}(r_{out}/r_{in})$
is related to the inductance of the insulating wall's cross-sectional
area (recall that the inductance of a co-axial cable is $L_{coaxial}=\frac{\mu_{0}}{2\pi}l\,\mbox{ln}(r_{out}/r_{in})$,
where $l$ is the length of the coaxial cable and $r_{out}/r_{in}$
are the cable's outer and inner radii), and $\tilde{L}_{int\Delta}=\stackrel[j=1]{N_{int}}{\Sigma}\left(\frac{s_{j}}{3r_{j}}\right)$
is related to the inductance of the area of the parts of the triangular
elements in the plasma domain that are associated with the interface
fixed point nodes. The resulting expression for $f_{I}$ is:
\begin{equation}
f_{I}(t)=\:\frac{-1}{\tilde{L}_{ins}+\tilde{L}{}_{int\Delta}}\stackrel[i=1]{N_{n}}{\Sigma}\left(\frac{f_{P0_{i}}(t)\,s_{i}}{3r_{i}}\right)\label{eq:142}
\end{equation}
The constant $f_{I}$ is calculated at each timestep and is applied
as a boundary condition for $f$ on the interface shared by the plasma
domain and the insulating domain, resulting in conservation of total
toroidal flux in the combined domains. Now, with the natural boundary
condition $\left(\nabla_{\perp}f\right)|_{\Gamma}=0$ imposed at all
boundary points in the plasma domain (in combination with the explicitly
applied boundary conditions $\underline{\mathbf{v}}|_{\Gamma}=\mathbf{0}$),
and with equation \ref{eq:142} used to overwrite the values for $f_{P}$
on the interface, intrinsic toroidal flux conservation is ensured
in the combined computational domain. 

\subsection{Simulations starting with a Grad-Shafranov equilibrium}

For the case where the plasma dynamics are evolved starting from an
initial Grad-Shafranov equilibrium, the expression for $\Phi_{tot}(t)$
($cf.$ equation \ref{eq:176}) is: 

\begin{equation}
\Phi_{tot}(t)=\overset{\underbrace{\int_{\Omega_{P}}\frac{f_{P}(\mathbf{r},t)}{r}dr\,dz+\int_{\Omega_{I}}\frac{f_{I}(t)}{r}dr\,dz}}{\Phi_{PI}}+\overset{\underbrace{\int_{\Omega_{P}+\Omega_{I}}\frac{F_{shaft}(t)}{r}dr\,dz}}{\Phi_{shaft}}\label{eq:184}
\end{equation}
$\Phi_{shaft}$ is the toroidal flux added to the combined plasma
and insulator domains due to crowbarred shaft current that flows in
the chalice wall, in the external aluminum bars, and through the ambient
plasma surrounding the CT, as depicted in figure \ref{fig:divertedcurrent}.
For the equilibrium-evolved simulations, the system's only source
of toroidal flux is $\Phi_{shaft}(t)$, and the initial toroidal flux
is $\Phi_{0}=\int_{\Omega_{p}}\frac{f_{gs}(\mathbf{r})}{r}dr\,dz$,
where $f_{gs}(\mathbf{r})$ is the $rB_{\phi}(\mathbf{r})$ field
associated with the Grad-Shafranov equilibrium. Using equation \ref{eq:184},
this implies that
\begin{align*}
\Phi_{tot}(t) & =\int_{\Omega_{P}}\frac{f_{gs}(\mathbf{r})}{r}dr\,dz+\cancel{\int_{\Omega_{P}+\Omega_{I}}\frac{F_{shaft}(t)}{r}dr\,dz}\\
 & =\int_{\Omega_{P}}\frac{f_{P}(\mathbf{r},t)}{r}dr\,dz+\int_{\Omega_{I}}\frac{f_{I}(t)}{r}dr\,dz+\cancel{\int_{\Omega_{P}+\Omega_{I}}\frac{F_{shaft}(t)}{r}dr\,dz}\\
 & \Rightarrow\int_{\Omega_{I}}\frac{f_{I}(t)}{r}dr\,dz=\int_{\Omega_{P}}\frac{1}{r}\left(f_{gs}(\mathbf{r})-f_{P}(\mathbf{r},t)\right)dr\,dz
\end{align*}
Following from the procedure described in subsection \ref{subsec:PHIcons_form},
the constant $f_{I}$ can be evaluated if $f_{P}$ and $f_{gs}$ are
temporarily set to zero at the fixed-point nodes along the inner wall
of the insulator, leading to the expression: 
\begin{equation}
f_{I}(t)=\:\frac{1}{\tilde{L}_{ins}+\tilde{L}{}_{int\Delta}}\stackrel[i=1]{N_{n}}{\Sigma}\left(\frac{\left(f_{gs0_{i}}-\;f_{P0_{i}}(t)\right)s_{i}}{3r_{i}}\right)\label{eq:139-1}
\end{equation}

\subsection{Illustration of toroidal flux conservation}

The total flux in the system is calculated at each simulation output
time using the formula 
\begin{equation}
\Phi_{tot}(t)=\stackrel[i=1]{N_{n}}{\Sigma}\left(\frac{f_{i}(t)s_{i}}{3r_{i}}\right)+\left(h_{I}\,\mbox{ln}(r_{out}/r_{in})\right)f_{I}(t)\label{eq:143}
\end{equation}
Note that in equation \ref{eq:143}, $\Phi_{tot}(t)$ refers to the
net toroidal flux in the total domain due to (i) the combined effects
of internal plasma poloidal currents in the plasma domain, and (ii)
flux addition related to $f_{external}(t)$. $f_{i}$ denotes the
value of $rB_{\phi}$ at node $i$ due to the combined effects of
flux addition due to externally-driven currents as well as internal
plasma poloidal currents. In contrast, $f_{P_{i}}$ in equations \ref{eq:140}
denotes the value of $rB_{\phi}$ at node $i$ due to internal plasma
poloidal currents only. $\Phi_{input}$ at each output time is calculated
using the formula: 
\begin{equation}
\Phi_{input}(t)=\stackrel[i=1]{N_{n}}{\Sigma}\frac{s_{i}}{3r_{i}}\left(f_{form}(z,t)+F_{shaft}(t)\right)+F_{shaft}(t)\left(h_{I}\,\mbox{ln}(r_{out}/r_{in})\right)\label{eq:144}
\end{equation}
Note that for formation simulations, $F_{shaft}(t)$ is automatically
set to zero, while for non-formation simulations, $f_{form}(z,t)$
is is automatically set to zero. Note that $f_{form}(z,t)\propto g_{Form}(z)$
(equation \ref{eq:150}), and $g_{Form}(z)$ is not defined in the
insulating region, so that $f_{form}(z,t)$ contributes directly to
$\Phi_{input}(t)$ in the plasma domain only. The initial toroidal
flux in the total domain is 
\[
\Phi_{initial}=\stackrel[i=1]{N_{n}}{\Sigma}\frac{s_{i}}{3r_{i}}\,f_{i\,(t=0)}
\]
Note that $f_{(t=0)}$ is zero at all nodes for formation simulations,
while $f_{(t=0)}=f_{GS}$ for simulations starting with a Grad-Shafranov
equilibrium.
\begin{figure}[H]
\subfloat[Simulated $\phi(t)$ (formation)]{\begin{centering}
\includegraphics[width=8cm,height=5cm]{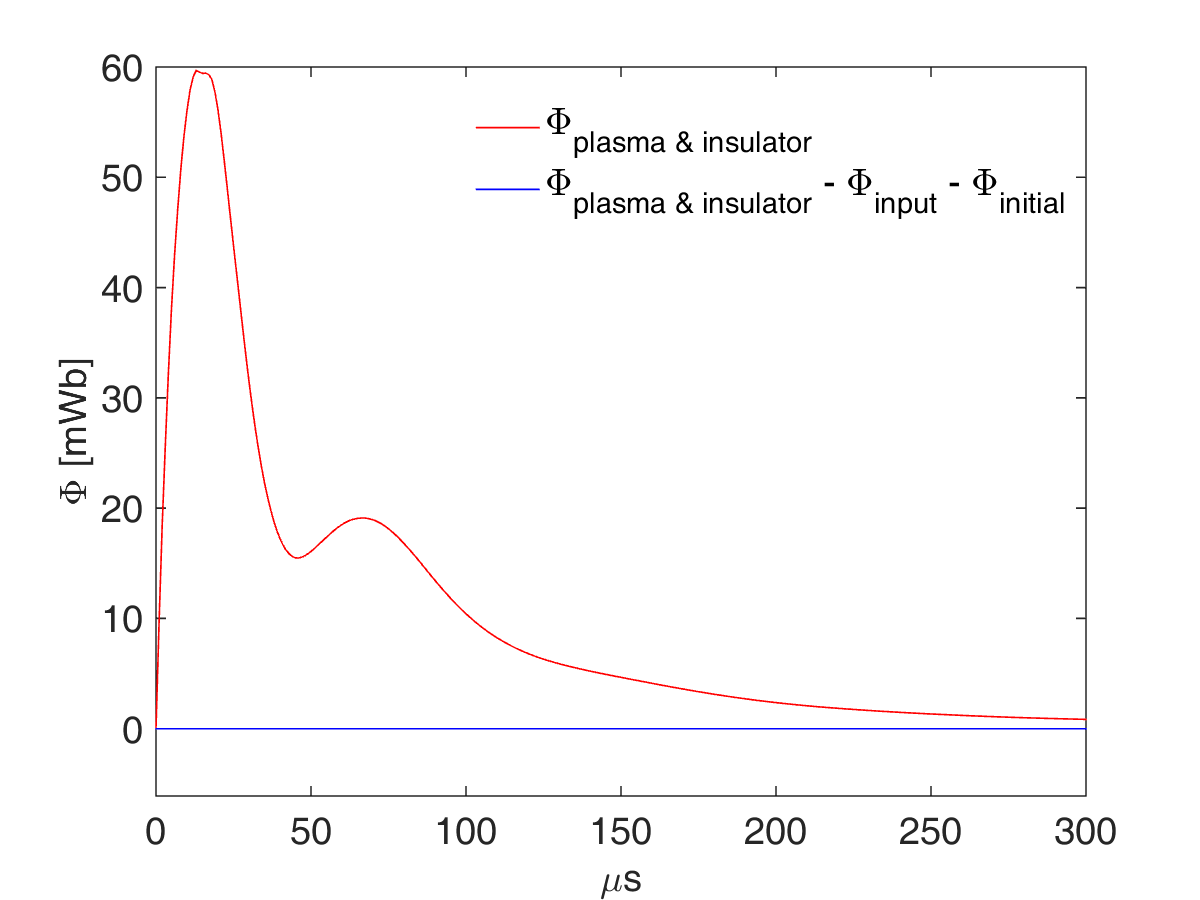}
\par\end{centering}
}\hfill{}\subfloat[Simulated $\phi(t)$ (initial Grad-Shafranov equilibrium)]{\centering{}\includegraphics[width=8cm,height=5cm]{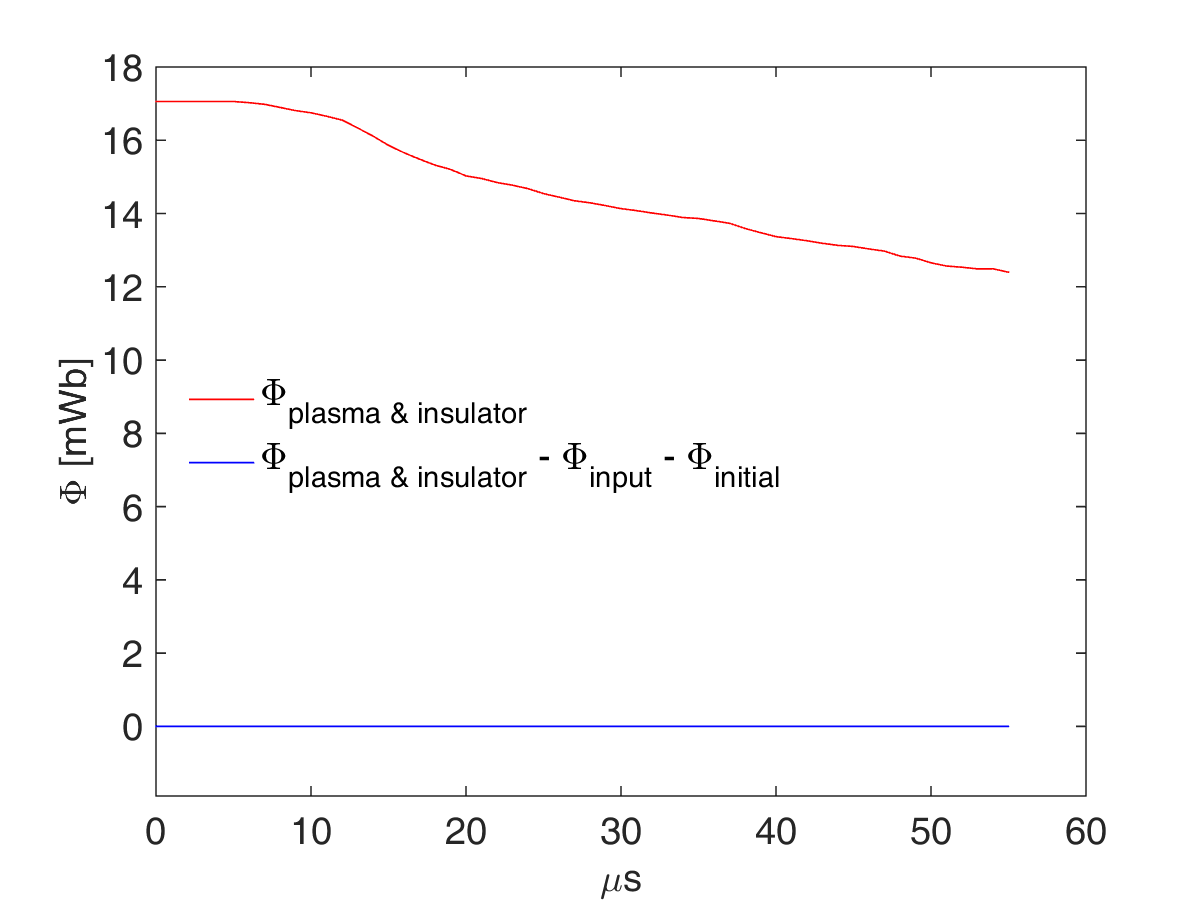}}\caption{$\,\,\,\,$Illustration of simulated toroidal flux conservation\label{fig:phi_cons}}
\end{figure}
Now, with the condition $\nabla_{\perp}f=0$ imposed on all boundary
points of the plasma grid (in combination with the explicitly applied
boundary conditions $\underline{\mathbf{v}}|_{\Gamma}=\mathbf{0}$),
and with equation \ref{eq:142} (formation simulations) or \ref{eq:139-1}
(non-formation simulations) used to overwrite the values for $f_{plasma}$
on the interface, we ensure intrinsic toroidal flux conservation in
the computational domain. Figure \ref{fig:phi_cons}(a) and (b) indicate
toroidal flux conservation over time for formation and non-formation
simulations respectively.

\section{Simulated diagnostics\label{subsec:Simulated-diagnostics}}

\subsection{Simulated diagnostics corresponding to experimental measurements\label{sec:Simulated-diagnostics_exp}}

The experimental diagnostics that can be modelled in simulations are
poloidal and toroidal magnetic field at the magnetic probes located
in the chalice wall, CT outer separatrix radius (see section \ref{subsec:Using-side-probe-data}),
line averaged electron density along the interferometer chords, and
ion temperature along the ion-Doppler chords. The poloidal field experimentally
measured at the probe locations is the boundary-tangential component
of the field at the probes in the $r-z$ plane. Simulated poloidal
field is calculated at the boundary nodes by calculating $t_{r}$
and $t_{z}$, the $r$ and $z$ components of the unit tangents to
the computational domain at the boundary nodes, and using the equation
\begin{equation}
\left(B_{\theta}\right)_{bn}=\left(\hat{\mathbf{t}}\cdot\mathbf{B}_{\theta}\right)_{bn}=\left(t_{r}B_{r}+t_{z}B_{z}\right)_{bn}\label{eq:35.14}
\end{equation}
Here, the subscript $bn$ denotes a particular boundary node. The
method using unit tangents allows for consistent evaluation of the
sign of the poloidal field, which varies across the separatrix between
the CT and levitation/compression field. $B_{r}$ and $B_{z}$ are
calculated at the boundary nodes using the discrete forms of equation
\ref{eq:9.01}: $\underline{B_{r}}|_{\Gamma}=\left(-\left(\underline{\underline{Dz}}*\underline{\psi}\right)\oslash\underline{r}\right)|_{\Gamma}$
and $\underline{B_{z}}|_{\Gamma}=\left(\left(\underline{\underline{Dr}}*\underline{\psi}\right)\oslash\underline{r}\right)|_{\Gamma}$.
Using equation \ref{eq:35.14}, the values calculated at the boundary
nodes are used to find, by interpolation, the wall-tangential poloidal
field at the probe locations. The simulated toroidal field at the
magnetic probe locations is also interpolated from the values at the
boundary nodes, where $\underline{B_{\phi}}|_{\Gamma}=\left(\underline{f}\oslash\underline{r}\right)|_{\Gamma}$.

Simulated CT outer equatorial separatrix radius ($r_{s}(t)$) is evaluated
by finding, at each code output time, the $r$ coordinate, at $z=0$,
of the intersection (where $B_{\theta}=0$) between the CT poloidal
field and the levitation field.

The simulated diagnostics for $n_{e}$ and $T_{i}$ are crude models,
simply the line-averaged quantities along the chords indicated in
figure \ref{fig:Chalice}.

\subsection{Additional simulated diagnostics}

In addition to the simulated diagnostics that are models of experimental
diagnostics, there are several other informative simulated diagnostics
that can't be reproduced experimentally due to the absence of experimental
measurements in the CT interior. These include time evolutions of
the magnetic field across the CT, the CT $q$ profile, CT magnetic
axis location, total poloidal flux of the CT, CT volume, and volume-averaged
CT beta. Also, the evolution of the system energy components, and
maximum ion and electron temperatures in the computational domain
can be tracked. The methods for evaluating the $q$ profile, and related
CT volume and volume-averaged beta are relatively involved, and will
be presented here.

\subsubsection{q profile\label{subsec:q-profile}}

The safety factor, $q(\psi)=\frac{\partial\phi(\psi)}{\partial\psi}=\frac{2\pi}{\iota}$
is defined as the number of toroidal transits a field line makes as
it completes one poloidal transit on a closed flux surface. The safety
factor is important because magnetically confined plasmas are magnetohydrodynamically
unstable if $q$\ensuremath{\le}2 at the last closed flux surface
\cite{WessonTokamaks}. Here, $\iota$ (iota), the rotational transform
(or field line pitch angle in radians) is $2\pi$ times the number
of poloidal transits a field line makes as it completes one toroidal
transit on a closed flux surface. To calculate $q$, use was made
of the contour matrix $cout$ that is returned with the MATLAB command
$[cout,\,hout]=tricontour1(rz,\,tri,\,field,\,edges,\,eINt,\,e2t,\,ContourLevels)$.
Here, $rz\,[N_{n}\times2]$ is a two column vector with $N_{n}$ rows,
the first and second columns contain the $r$ and $z$ coordinates
of the computational nodes. $tri\,[N_{e}\times3]$ is a three column
vector with $N_{e}$ rows that contains the indexes of the rows in
$rz$ that refer to the three nodes that define each triangular element.
$field\,[N_{n}\times1]=\underline{\psi}$, $edges,\;eINt,\mbox{ and }e2t$
define mesh connectivity data structures, and $ContourLevels$ is
a vector defining the values of $\psi$ at which contours are to be
drawn. $cout$ is an array with two rows that contains data that defines
the contours of $\psi$. Each $2\times n$ section (variable $n$)
of $cout$ contains data for a contour defined by $n-1$ spatial points.
The value in the first row of the first column of each section of
$cout$ contains the value of $\psi$ on the contour defined by the
particular section. The value in the second row of the first column
of each section of $cout$ is always an integer that indicates the
number of points that define the contour referenced in that section
of $cout$. For example, if the first section of $cout$ has dimensions
$2\times800$, then $cout(1,1)$ has the value of $\psi$ for the
contour defined by the section, $cout(2,1)=799$ indicates that $799$
points define the contour referenced in the section, while $cout(1,\,2:800)$
and $cout(2,\,2:800)$ contain the $r\mbox{ and }z$ coordinates of
the points that define the contour referenced in the section. By locating
the integer values in $cout$, the indexes of columns in $cout$ that
are the first columns of the individual sections of $cout$ are found.

$q$ is generally defined only on closed contours. The second and
last columns of each section of $cout$ that refers to a \emph{closed}
contour are identical and this relationship is used to find the indexes
of columns in $cout$ that contain starting point indexes for data
defining individual closed contours. For simulations that include
the computational domain of the gun, there can be closed $\psi$ contours
down the gun (for example, if part of the CT is pinched off during
magnetic compression - see figure \ref{fig:MHDform-1}), and we are
not interested in $q$ there. The indexes of columns in $cout$ that
contain starting point indexes for data for individual \emph{closed}
contours in the \emph{CT} \emph{confinement region} are found by placing
the constraint of a minimum $z$ coordinate on the spatial points
that define closed contours. 

For each closed poloidal flux contour (index $j$) in the confinement
region, $\underline{I^{e}}_{j}\,[N_{e}\times1]$, a logical vector
of ones and zeroes defining the rows in $\underline{r^{e}}\,[N_{e}\times1]$
and $\underline{z^{e}}\,[N_{e}\times1]$, that refer to triangle centroids
lying within the contour, are found. The average value of toroidal
magnetic flux $\phi=\int\mathbf{B_{\phi}}\cdot\mathbf{dS}$ within
a particular contour with index $j$ can be found as $\phi_{j}=S{}_{in_{j}}<\underline{f^{e}}_{in_{j}}\varoslash\underline{r^{e}}_{in_{j}}>$.
Here $S_{in_{j}}$ is the area enclosed by contour $j$, and $<\underline{f^{e}}_{in_{j}}\varoslash\underline{r^{e}}_{in_{j}}>$
is the average of $B_{\phi}=f/r$ within contour $j$, where $\underline{f^{e}}_{in_{j}}=\underline{f^{e}}\left(\underline{I^{e}}_{j}\right)$
and $\underline{r^{e}}_{in_{j}}=\underline{r^{e}}\left(\underline{I^{e}}_{j}\right)$
contain the values of $\underline{f^{e}}$ and $\underline{r^{e}}$
at triangular element centroids that lie within contour $j$. Note
that this averaging method is implemented because it produces smoother
results, for moderate mesh resolution, than the approximately equivalent
definition $\phi_{j}=\underline{s^{e}}_{in_{j}}^{T}*\left(\underline{f^{e}}_{in_{j}}\varoslash\underline{r^{e}}_{in_{j}}\right)$.
Triangle-centered quantities are used instead of nodal quantities
because there are more triangles than nodes so averaging is more accurate.
With $\psi_{j}$ and $\phi_{j}$ defined for each $\psi$ contour,
it is then straightforward to find $q_{j}=\frac{\partial\phi_{j}}{\partial\psi_{j}}$
using the Matlab gradient function. As detailed in section \ref{subsec:Equilibrium-solution-comparison},
this method yields $q(\psi)$ profiles that are an excellent match
to those produced by a well-established equilibrium code. 

\subsubsection{Evaluation of CT volume and volume-averaged $\boldsymbol{\beta}$}

At each code output time, the data associated with the $\psi$ contour
that defines the last closed flux surface (LCFS) can be used to evaluate
$V_{CT},$ the CT volume. Defining $j=0$ as the index of the poloidal
flux contour that defines the LCFS, 
\begin{equation}
V_{CT}=2\pi\underline{r^{e}}_{in_{0}}^{T}*\underline{s^{e}}_{in_{0}}\label{eq:300}
\end{equation}
where $\underline{r^{e}}_{in_{0}}^{T}=\left(\underline{r^{e}}\left(\underline{I^{e}}_{0}\right)\right)^{T}$
is the transpose of the vector containing the $r$ coordinates of
the triangular element centroids that lie within the LCFS, and vector
$\underline{s^{e}}_{in_{0}}=\underline{s^{e}}\left(\underline{I^{e}}_{0}\right)$
contains the areas of the triangular elements whose centroids lie
within the LCFS. 

Similarly, at each output time the volume-averaged total, poloidal,
and toroidal betas (ratios of fluid pressure to magnetic pressure)
can be calculated as 
\begin{eqnarray}
<\beta> & = & 2\mu_{0}\left(\underline{V}_{in_{0}}^{e}{}^{T}*\left(\underline{p_{i}^{e}}_{in_{0}}+\underline{p_{e}^{e}}_{in_{0}}\right)\right)\varoslash\left(\underline{V}_{in_{0}}^{e}{}^{T}*\left(\left(\underline{B_{r}^{e}}_{in_{0}}\right)^{2}+\left(\underline{B_{\phi}^{e}}_{in_{0}}\right)^{2}+\left(\underline{B_{z}^{e}}_{in_{0}}\right)^{2}\right)\right)\nonumber \\
<\beta_{\theta}> & = & 2\mu_{0}\left(\underline{V}_{in_{0}}^{e}{}^{T}*\left(\underline{p_{i}^{e}}_{in_{0}}+\underline{p_{e}^{e}}_{in_{0}}\right)\right)\varoslash\left(\underline{V}_{in_{0}}^{e}{}^{T}*\left(\left(\underline{B_{r}^{e}}_{in_{0}}\right)^{2}+\left(\underline{B_{z}^{e}}_{in_{0}}\right)^{2}\right)\right)\label{eq:301}\\
<\beta_{\phi}> & = & 2\mu_{0}\left(\underline{V}_{in_{0}}^{e}{}^{T}*\left(\underline{p_{i}^{e}}_{in_{0}}+\underline{p_{e}^{e}}_{in_{0}}\right)\right)\varoslash\left(\underline{V}_{in_{0}}^{e}{}^{T}*\left(\underline{B_{\phi}^{e}}_{in_{0}}\right)^{2}\right)\nonumber 
\end{eqnarray}
Here, $\underline{V}_{in_{0}}^{e}{}^{T}=2\pi\underline{r^{e}}_{in_{0}}\circ\underline{s^{e}}_{in_{0}}$
is the transpose of the vector containing the volumes associated with
the triangular elements whose centroids lie within the LCFS, and $\underline{p_{i}^{e}}_{in_{0}},\,\underline{p_{e}^{e}}_{in_{0}},\,\underline{B_{r}^{e}}_{in_{0}},\,\underline{B_{\phi}^{e}}_{in_{0}}$
and $\underline{B_{z}^{e}}_{in_{0}}$ contain the values of ion pressure,
electron pressure, and the $r,\,\phi$, and $z$ components of the
magnetic field at the triangle centroids inside the LCFS.

\section{Summary\label{sec:SummaryImplemnationFformetc}}

It has been shown how boundary conditions for $\psi$ are evaluated,
applied, and varied over time to model the levitation and compression
fields, while boundary conditions for $\psi$ corresponding to the
stuffing field are held constant over time. In combination with the
boundary conditions $\mathbf{v}|_{\Gamma}=\mathbf{0}$, the natural
imposition of the boundary condition $\left(\nabla f_{\perp}\right)|_{\Gamma}=0$
is required for conservation of the part of the toroidal flux associated
with $f_{\eta}$, which enables simulations to reproduce poloidal
currents that, in the experiment, are induced to flow primarily in
machine walls and across various regions of ambient plasma external
to the CT, in order to conserve toroidal flux.

The ability to model part of the domain as an insulating plasma-free
material is useful because it expands the code's range of applicability
as a problem-solving tool. As shown in chapter \ref{chap:Simulation-results},
this feature has been used to verify that penetration into the insulating
wall of poloidal field being advected into the containment region
during CT formation was an issue associated with the 6-coil configuration
that was alleviated in the 11-coil configuration. It has been shown
how particular effort was made to ensure maintenance of toroidal flux
conservation when an insulating region is included in the model.

The methods developed to simulate the CT formation process, by adding
toroidal flux associated with the measured formation voltage to the
domain, have been presented. Even with the simplifying assumption
that radial formation current between the electrodes flows at a fixed
axial location, the model is able to reproduce experimental measurements
to a very acceptable level, as will be shown in chapter \ref{chap:Simulation-results}.
Tests to allow variation of the axial location of intra-electrode
formation current, with upper and lower bounds, were able to precisely
reproduce the experimentally measured formation current signals over
the times when the axial location was unrestricted. 

Various simulated diagnostics have been implemented to the code. Some
of these are counterparts to the available experimental diagnostics.
Several other simulated diagnostics have been developed including
the time evolutions of the $q(\psi)$ profile, magnetic axis location,
$\beta$ profile, CT volume and magnetic fluxes, system energy components,
and maximum ion and electron temperatures. These additional simulated
diagnostics are based on various simulated field distributions.

\newpage{}

\chapter{Simulation input parameters and results\label{chap:Simulation-results}}

This chapter will begin with descriptions, in section \ref{sec:Code-input-parameters,},
of some of the key code input parameters. Some of the principal input
parameters for simulation  2353, which includes CT formation into
a levitation field, and magnetic compression, will be tabulated. Next,
in section \ref{subsec:Simulated-plasma-wall-interactio}, simulation
results which confirm that the level of plasma-wall interaction was
reduced by the line-tying effect of increased $|t_{lev}|$, and more
significantly, with the switch from the 6-coil to the 11-coil configuration,
will be presented. Contour plots depicting the evolution of various
fields for simulation  2353 will be shown in section \ref{subsec:Evolution-of-the}.
Presentation of the evolution of a selection of simulated fields for
the case of an extended simulation that includes the effect of the
ringing compression field will follow in section \ref{subsec:ring_psi_f}.
Proceeding this, in section \ref{subsec:Simulated-diag_CFexp}, a
selection of simulated diagnostics will be presented and compared
with the experimental counterparts, for levitation-only shots in the
cases with low and high $R_{cable},$ and for a selection of compression
shots. Presentation and discussion of further simulated diagnostics
that don't have experimental counterparts, such as simulated $q$
profile, compression scalings, internal magnetic fields, CT volume,
$\beta$, and magnetic axis location, will follow in section \ref{subsec:Results-from-additional}.
The chapter will conclude with a summary in section \ref{sec:SummarySIMresults}.

\section{Simulation input parameters\label{sec:Code-input-parameters,}}

\begin{table}[H]
\centering{}{\footnotesize{}}%
\begin{tabular}{|ccccccc|}
\hline 
 &  &  &  &  &  & \tabularnewline
\textbf{\footnotesize{}simulation \# } & {\footnotesize{}$\mathbf{Plato}$} & {\footnotesize{}$\mathbf{do_{formation}}$} & {\footnotesize{}$\mathbf{geometry}$} & {\footnotesize{}$\mathbf{meshtype}$} & \textbf{\footnotesize{}$\mathbf{h}_{\mathbf{e}}${[}mm}{\footnotesize{}{]}} & {\footnotesize{}$\mathbf{do_{insulator}}$}\tabularnewline
{\footnotesize{}2353} & {\footnotesize{}1} & {\footnotesize{}1} & {\footnotesize{}2} & {\footnotesize{}0} & {\footnotesize{}2} & {\footnotesize{}1}\tabularnewline
\hline 
 &  &  &  &  &  & \tabularnewline
{\footnotesize{}$\mathbf{vary_{H}}$} & \textbf{\footnotesize{}$\mathbf{\boldsymbol{\tau}_{\boldsymbol{LR}}}$
$[\boldsymbol{\upmu}$s{]}} & {\footnotesize{}$\mathbf{do_{2temp}}$} & {\footnotesize{}$\mathbf{T}_{\mathbf{0}}$}\textbf{\footnotesize{}
{[}eV{]}} & {\footnotesize{}$\mathbf{T}_{\boldsymbol{\Gamma}}$}\textbf{\footnotesize{}
{[}eV{]}} &  & \tabularnewline
{\footnotesize{}0} & {\footnotesize{}70} & {\footnotesize{}1} & {\footnotesize{}0.02} & {\footnotesize{}0.02} &  & \tabularnewline
\hline 
 &  &  &  &  &  & \tabularnewline
{\footnotesize{}$\mathbf{RK}$} & \textbf{\footnotesize{}$\mathbf{dt_{0}}$ {[}s{]}} & \textbf{\footnotesize{}$\mathbf{t_{out}}$ $[\boldsymbol{\upmu}$s{]}} & {\footnotesize{}$\mathbf{tsrf}$} & \textbf{\footnotesize{}$\mathbf{dt_{min}}$ {[}s{]}} & \textbf{\footnotesize{}$\mathbf{t_{sim}}$ $[\boldsymbol{\upmu}$s{]}} & \tabularnewline
{\footnotesize{}1} & {\footnotesize{}$8\times10^{-11}$} & {\footnotesize{}1} & {\footnotesize{}2} & {\footnotesize{}$1\times10^{-12}$} & {\footnotesize{}87} & \tabularnewline
\hline 
 &  &  &  &  &  & \tabularnewline
{\footnotesize{}$\mathbf{V}_{\mathbf{form}}$ {[}}\textbf{\footnotesize{}kV}{\footnotesize{}{]}} & {\footnotesize{}$\mathbf{V}_{\mathbf{lev}}$ {[}}\textbf{\footnotesize{}kV}{\footnotesize{}{]}} & {\footnotesize{}$\mathbf{V}_{\mathbf{comp}}$ {[}}\textbf{\footnotesize{}kV}{\footnotesize{}{]}} & {\footnotesize{}$\mathbf{I}_{\mathbf{main}}$ {[}}\textbf{\footnotesize{}A}{\footnotesize{}{]}} & \textbf{\footnotesize{}|$\mathbf{t_{lev}}$| $[\boldsymbol{\upmu}$s{]}} & {\footnotesize{}$\mathbf{R}_{\mathbf{cable}}$} & \textbf{\footnotesize{}$\mathbf{t_{comp}}$ $[\boldsymbol{\upmu}$s{]}}\tabularnewline
{\footnotesize{}16} & {\footnotesize{}16} & {\footnotesize{}18} & {\footnotesize{}70} & {\footnotesize{}60} & {\footnotesize{}1} & {\footnotesize{}45}\tabularnewline
\hline 
 &  &  &  &  &  & \tabularnewline
\textbf{\footnotesize{}$\mathbf{m}_{\mathbf{0}}$} & \textbf{\footnotesize{}$\mathbf{Z_{eff}}$} & {\footnotesize{}$\mathbf{n_{o}\,[\mbox{\textbf{m}}^{-3}]}$} & \textbf{\footnotesize{}$\mathbf{\boldsymbol{\sigma}_{n}^{2}}\,[\mbox{\textbf{m}}^{2}]$} & \textbf{\footnotesize{}$\mathbf{\boldsymbol{\zeta}\,[\mbox{\textbf{m\ensuremath{\mathbf{^{\mathbf{2}}}}/s}}]}$} & {\footnotesize{}$\mathbf{\boldsymbol{\nu}}_{\mathbf{num}}\mathbf{\,[\mbox{\textbf{m\ensuremath{\mathbf{^{\mathbf{2}}}}/s}}]}$} & {\footnotesize{}$\mathbf{\boldsymbol{\nu}}_{\mathbf{phys}}\mathbf{\,[\mbox{\textbf{m\ensuremath{\mathbf{^{\mathbf{2}}}}/s}}]}$}\tabularnewline
{\footnotesize{}4} & {\footnotesize{}1.3} & {\footnotesize{}$9\times10^{20}$} & {\footnotesize{}0.005} & {\footnotesize{}50} & {\footnotesize{}700} & {\footnotesize{}410}\tabularnewline
\hline 
 &  &  &  &  &  & \tabularnewline
{\footnotesize{}$\mathbf{vary_{\boldsymbol{\eta}}}$} & {\footnotesize{}$\boldsymbol{\eta}_{\mathbf{max}}\,[\mbox{\textbf{m\ensuremath{\mathbf{^{\mathbf{2}}}}/s}}]$} & {\footnotesize{}$\mathbf{vary_{\boldsymbol{\chi}}}$} & {\footnotesize{}$\mathbf{\boldsymbol{\chi}_{\parallel i}\,[\mbox{\textbf{m\ensuremath{\mathbf{^{\mathbf{2}}}}/s}}]}$} & {\footnotesize{}$\mathbf{\boldsymbol{\chi}_{\parallel e}\,[\mbox{\textbf{m\ensuremath{\mathbf{^{\mathbf{2}}}}/s}}]}$} & {\footnotesize{}$\mathbf{\boldsymbol{\chi}_{\perp i}\,[\mbox{\textbf{m\ensuremath{\mathbf{^{\mathbf{2}}}}/s}}]}$} & {\footnotesize{}$\mathbf{\boldsymbol{\chi}_{\perp e}\,[\mbox{\textbf{m\ensuremath{\mathbf{^{\mathbf{2}}}}/s}}]}$}\tabularnewline
{\footnotesize{}1} & {\footnotesize{}5000} & {\footnotesize{}0} & {\footnotesize{}5000} & {\footnotesize{}16000} & {\footnotesize{}120} & {\footnotesize{}240}\tabularnewline
\hline 
\end{tabular}\caption{\label{tab:Sim parameters}$\,\,\,\,$Selection of code input parameters
for simulation  2353 }
\end{table}
In total, there are over one hundred code input parameters. Some of
the principal code input parameters for example simulation  2353 results
are presented in table \ref{tab:Sim parameters}. $Plato=1$ implies
that the simulation was run remotely, using a VPN service, on the
high performance computing cluster ($Plato$) at the University of
Saskatchewan ($Plato=0$ for cases where the simulation is run directly
on a local computer). $do_{formation}=1$ implies that the simulation
started with CT formation, as opposed to starting with a Grad-Shafranov
equilibrium (in that case $do_{formation}$ would be set equal to
zero). $geometry=2$ implies that the simulation was run in a computational
domain representing the geometry for the 11-coil configuration in
the full Marshall gun. $meshtype=0$ and $h_{e}=2$ mm implies that
a computational grid with uniform triangular elements of size 2 mm
was used (see appendix \ref{subsec:Computational-grid}). $do_{insulator}=1$
implies that an insulating region was included in the simulation (see
section \ref{sec:Vacuum-field-in}). $vary_{H}=0$ implies that the
z coordinate of the radial component of formation current between
the gun electrodes was held constant at $z=z_{gp}=-0.43$ m, the location
of the gas puff valves (see section \ref{sec:Time-dependent-formation-profile}).
For a time-dependent formation current profile, $vary_{H}$ would
be set equal to one. $\tau_{LR}=70\,\upmu$s defines the timescale
associated with the resistive contribution to $\Phi_{form}$ (see
section \ref{subsec:Formation-simulations11}). $do_{2temp}=1$ implies
that the single fluid energy equation is split into ion and electron
components (equations \ref{eq:479.1} and \ref{eq:479.2}). $T_{0}$
and $T_{\Gamma}$ define the initial and boundary values for the ion
and electron temperatures - see section \ref{subsec:bcs T}.\\

The code inputs in the third row of table \ref{tab:Sim parameters}
pertain to timestepping, see appendix \ref{sec:Timestepping-methods}.
$RK=1$ implies that forward Euler timestepping was used. $dt_{0}$
defines the initial timestep, $t_{out}$ defines the interval between
the instances when the simulation data is saved to file, along with
plots of analysed field data. $tsrf$ defines the factor by which
the timestep is reduced before continuing the simulation from field
data saved to file at the last instance of data output, in the event
of a crash caused by having a timestep that is too large. $dt_{min}$
defines the minimum timestep that can be reached before the simulation
is terminated. $t_{sim}=87\,\upmu$s defines the duration of the simulation
- this value is chosen here because when compression is started at
$t_{comp}=45\,\upmu$s, the peak compression is at around $65\,\upmu$s,
and compression current falls to zero by around $87\,\upmu$s. In
this simulation the effect of compression current polarity change
(see figure \ref{fig:Ilev_comp_sim}(b)) is not examined. Results
from a simulation in which the compression current is allowed to ring
are presented in sections \ref{subsec:ring_psi_f} and \ref{par:Shot-=00002339735}.
\\

The code inputs in the fourth row of table \ref{tab:Sim parameters}
are related to the external magnetic conditions. For formation simulations,
the formation voltage waveform used to calculate $\Phi_{form}(t)$
in the expression for $f_{external}(z,t)=f_{form}(z,t)$ (section
\ref{subsec:Formation-simulations11}) is taken from an experimental
measurement for a typical shot with $V_{form}=16$ kV. This waveform
can be linearly scaled according to the formation voltage required
as an input parameter for the simulation. As described in section
\ref{subsec:bcs psi}, the boundary conditions for $\psi$, pertaining
to $\psi_{main},\,\psi_{lev},\,\mbox{and }\psi_{comp}$ are obtained
using FEMM models. For the FEMM model used to find $\psi_{main}$,
the dc current in the main coil was set to $I_{main}=$70 A, and the
boundary values for $\psi_{main}$ can be linearly scaled according
to the value of $I_{main}$ required for the simulation. Similarly,
the currents in the levitation and compression coils in the FEMM models,
used to obtain boundary conditions for $\psi_{lev}$ and $\psi_{comp}$,
were set to the experimentally measured values corresponding to experimentally
recorded $V_{lev}=16$ kV  and $V_{comp}=18$ kV  (voltages to which
the levitation and compression capacitors were charged) respectively.
Again, the boundary values for $\psi_{lev}$ and $\psi_{comp}$ can
be scaled linearly according to the values of $V_{lev}$ and $V_{comp}$
required for the simulation. $|t_{lev}|=60\,\upmu$s implies that
the boundary conditions for $\psi_{lev}$ are taken from a FEMM model
with the input current frequency set to a value (4170 Hz) corresponding
to having the quarter-period of the levitation current equal to $60\,\upmu$s
(see sections \ref{subsec:6-coils-config_ceramic} and \ref{subsec:Simulated-plasma-wall-interactio}).
$R_{cable}=1$ implies that the levitation current waveform corresponding
to inclusion of $70\mbox{ m}\Omega$ cables between each levitation
coil (or coil-pair) and levitation inductor was used to scale the
levitation component of the boundary conditions for $\psi$ over time
(see section \ref{subsec:Levitation-field-decay}, note that $R_{cable}=0$
corresponds to 2.5 m$\Omega$ cables). $t_{comp}=45\,\upmu$s means
that magnetic compression ($i.e.,$ superimposition of the $\psi_{comp}$
boundary conditions, scaled by experimentally recorded $\widetilde{I}_{comp}(t)$,
on the $\psi_{main}$ and $\psi_{lev}$ boundary conditions, see section
\ref{subsec:bcs psi}) is started $45\,\upmu$s into the simulation.
\\

The inputs in the fifth row of table \ref{tab:Sim parameters} define
some basic plasma parameters and the diffusion coefficients for density
and viscosity. $m_{0}=4$ implies that the ion mass is four proton
masses ($m_{p}=1.67\times10^{-27}$ kg), which is relevant for modeling
helium plasmas. $Z_{eff}=1.3$ defines the estimate for the (volume
averaged) ion charge, and determines the ratio of electron to ion
number density ($n_{e}=Z_{eff}\,n_{i})$. Since the plasma is being
modeled as a single fluid, while the energy equation is split into
ion and electron components, $Z_{eff}$ determines the ratio between
the ion and electron pressures ($p=p_{i}+p_{e}=n\,(T_{i}\,[\mbox{J}]+Z_{eff}\,T_{e}\,[\mbox{J}])$,
where $n\iff n_{i}$ ). $Z_{eff}$ also enters into the determination
of plasma Spitzer resistivity (equation \ref{eq:517.6}), the ion-electron
heat exchange term $Q_{ie}$ (equation \ref{eq:517.61}) and, for
neutral fluid dynamics (chapter \ref{chap:Neutral-models}), the rate
coefficients for ionization and recombination. Fully ionized helium
plasma would have $Z_{eff}=2$, here we assume that a proportion of
the helium atoms are only singly ionized. $Z_{eff}$ can be increased
to include some of the effects of the inclusion of high $Z$ impurities
in the plasma. $n_{0}\,[\mbox{m}^{2}/\mbox{s}]$ determines the plasma
number density - note that normalised density $\widetilde{n}(\mathbf{r},t)=\frac{n(\mathbf{r},t)}{n_{0}}$
is evolved over time. Along with $n_{0}$, $\sigma_{n}^{2}$ {[}m$^{2}${]}
defines the variance of the Gaussian function that determines the
initial plasma density distribution, which is centered around $z_{gp}=-0.43$
m, the $z$ coordinate of the physical location of the machine gas
puff valves. The initial density profile is given by 
\begin{equation}
n_{t0}(z)=n_{0}\,\left((n_{high}-n_{low})\,\widetilde{g}(z)+n_{low}\right)\label{eq:900}
\end{equation}
where $\widetilde{g}(z)=g(z)/\mbox{max}(g(z))$. Here, $g(z)=\frac{1}{\sqrt{2\pi\,\sigma_{n}^{2}}}\,\mbox{exp}\left(\frac{-\left(z-z_{gp}\right)^{2}}{2\sigma_{n}^{2}}\right)$.
$n_{high}$ and $n_{low}$ are typically set to 10 and 0.1 respectively.
Note that $n_{low}$ must be finite - density is not allowed to approach
too close to zero anywhere in the computational domain. See figure
\ref{fig: n_11coils}(a) for an example initial density profile. The
width of the base of the Gaussian is approximately 0.7 m, 0.45 m,
0.4 m, or 0.2 m for $\sigma_{n}^{2}$= 0.01, 0.005, 0.003, or 0.001
m$^{2}$ respectively. The sound speed of a neutral gas is given by
$V_{s}=\sqrt{\frac{\gamma p}{\rho}}$. At room temperature (0.02 eV),
for helium gas, this is $V_{s}\approx\sqrt{\frac{\frac{5}{3}(1.6\times10^{-19})(0.02)}{4(1.67\times10^{-27})}}\approx900$
m/s. In a typical shot, gas is puffed into the vacuum vessel through
the eight gas valves spaced toroidally on the outer electrode, around
$400\,\upmu$s before firing the formation capacitors. A simple estimate
of $d_{n}$, the maximum spread of the neutral gas around the valves
can be estimated using this time and the sound speed, as $d_{n}=(400\times10^{-6})(900)=0.36$
m, so that choosing $\sigma_{n}^{2}=0.005$ m$^{2}$, corresponding
to a Gaussian base width of around 0.45 m, is a reasonable estimate
for the initial spread. Simulations that don't include a model for
interaction between plasma and neutral fluids start with the crude
approximation that all the plasma fluid particles are ionized. The
density diffusion coefficient $\zeta$ is chosen as 50 m$^{2}$/s,
which, as mentioned in appendix \ref{subsec:Diffcoefs}, is close
to the minimum required value for sufficient density field smoothing
for formation simulations. Similarly, at 700 m$^{2}$/s, $\nu_{num}$
is close to the minimum value required for sufficient velocity field
smoothing, while $\nu_{phys}=410$ m$^{2}$/s, is chosen because it
leads to simulated ion temperature close to the levels indicated by
the ion-Doppler system (see appendix \ref{subsec:Viscous-diffusion}).\\

The inputs in the sixth row of table \ref{tab:Sim parameters} define
the resistive and thermal diffusion coefficients. For this simulation,
$vary_{\eta}=1$ so that the Spitzer formula for plasma resistivity
is used to calculate the (isotropic) resistive diffusion coefficient
at each timestep according to equation \ref{eq:517.6}. An upper limit
of $\eta_{max}$$=5000\,[\mbox{m}^{2}\mbox{/s}]$ is imposed to prevent
extremely high values of $\eta$ in regions where $T_{e}$ is low,
as this would lead to small timesteps, as outlined in appendix \ref{subsec:Diffcoefs}.
Input parameter $vary_{\chi}=0$ for this simulation, so that thermal
diffusion is determined by fixed coefficients $\chi_{\parallel i},\,\chi_{\parallel e},\,\chi_{\perp i}$
and $\chi_{\perp e}$. The constant thermal diffusion coefficients
used in this simulation are $\chi_{\parallel i}=5000,\,\,\,\chi_{\parallel e}=16000,\,\,\,\chi_{\perp i}=120\mbox{ and }\chi_{\perp e}=240\,\,[\mbox{m}^{2}/\mbox{s}]$.
The high parallel coefficients represent the physical case where particles
are free to stream unimpeded along field lines. At early simulation
times, all magnetic field lines are open ($i.e.,$ field lines do
not close on themselves), and connect to the walls of the vacuum vessel,
so that parallel thermal diffusion to the walls, where the boundary
conditions $T_{\Gamma}=0.02$ eV are applied, represents the most
significant loss of thermal energy. After CT formation, heat arising
from residual formation current down the gun below the CT continues
to be lost to the walls through parallel diffusion along open field
lines, but perpendicular thermal diffusion is the dominant heat loss
mechanism for the plasma constituting the CT itself, because CT field
lines are closed so that heat contained within the CT is not lost
by parallel diffusion. In the absence of magnetic compression, the
constant perpendicular thermal diffusion coefficients chosen for this
simulation lead to a CT lifetime that is comparable to experimentally
observed lifetimes of levitated CTs in the 11-coil configuration.
Note that an explicit model for cooling due to impurity line radiation
is not included in the simulation. The relatively high perpendicular
thermal diffusion coefficients act as a proxy to include the effects
of this cooling mechanism.\\
\\
Some additional input parameters for simulation 2353 pertain to dynamics
of neutral fluid. These parameters, and a selection of simulation
results relating to the neutral fluid dynamics, will be presented,
along with the model for interaction between the plasma and neutral
fluids, in chapter \ref{chap:Neutral-models}.

\section{Simulated plasma-wall interaction\label{subsec:Simulated-plasma-wall-interactio}}

\subsection{Effect of increasing levitation field soak-in\label{subsec:Effect-of-increasing}}

\begin{figure}[H]
\hfill{}\subfloat[$|t_{lev}|=60\,\upmu$s]{\raggedright{}\includegraphics[scale=0.5]{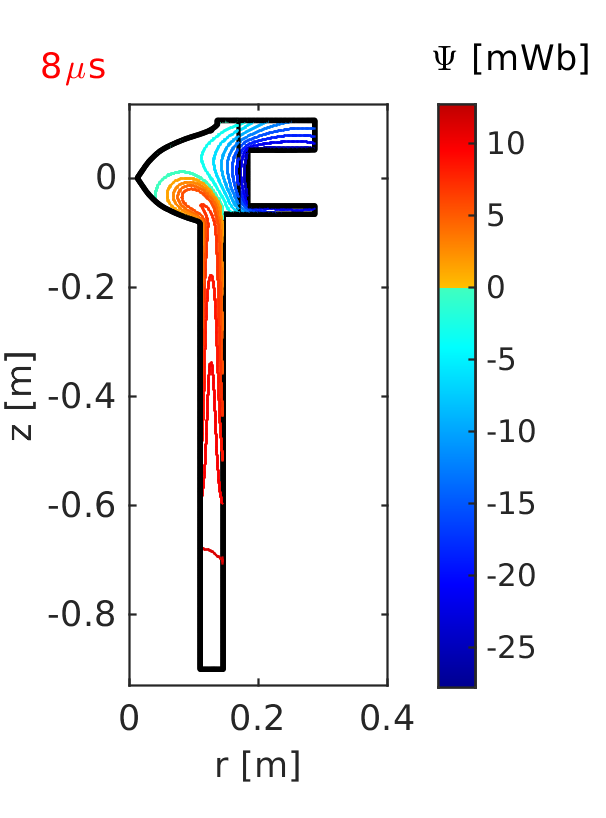}}\hfill{}\subfloat[$|t_{lev}|=60\,\upmu$s]{\includegraphics[scale=0.5]{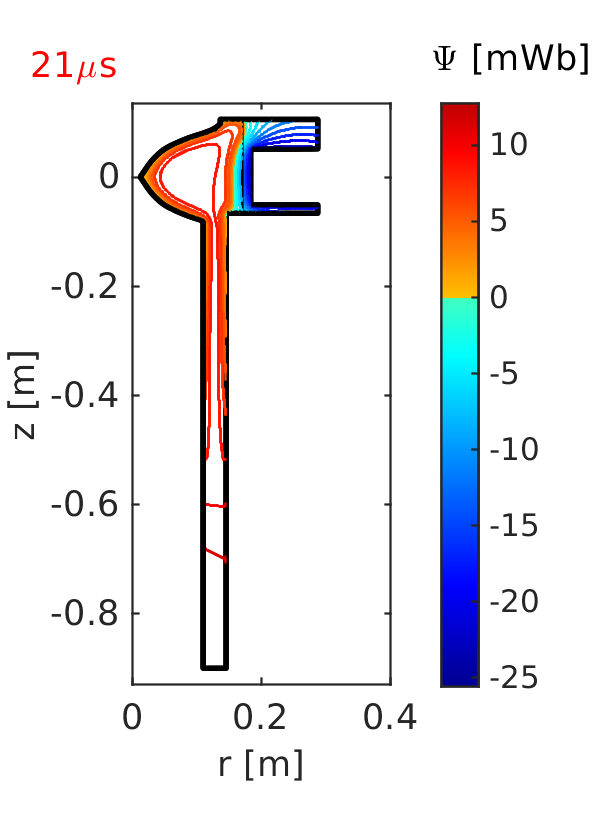}}\hfill{}

\subfloat[$|t_{lev}|=60\,\upmu$s]{\raggedright{}\includegraphics[width=7cm,height=5cm]{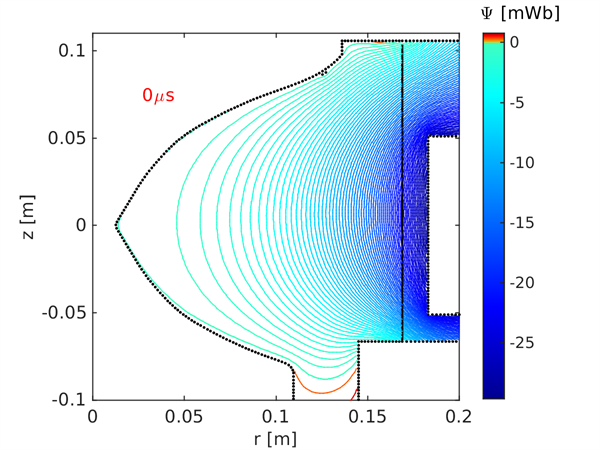}}\hfill{}\subfloat[$|t_{lev}|=60\,\upmu$s]{\raggedright{}\includegraphics[width=7cm,height=5cm]{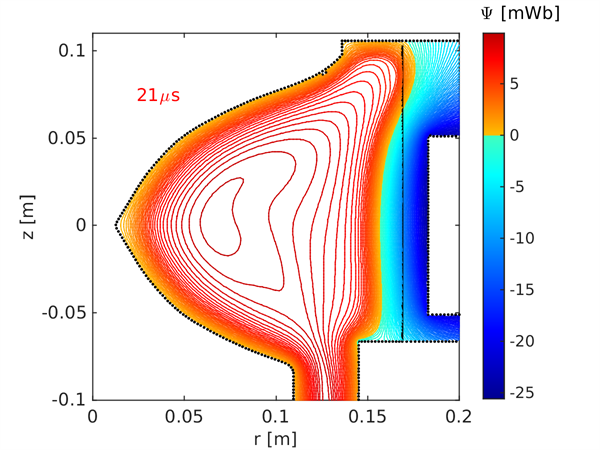}}

\subfloat[$|t_{lev}|=300\,\upmu$s]{\raggedright{}\includegraphics[width=7cm,height=5cm]{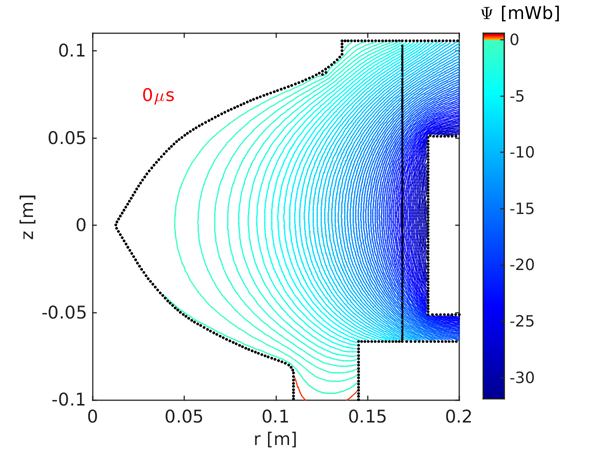}}\hfill{}\subfloat[$|t_{lev}|=300\,\upmu$s]{\raggedright{}\includegraphics[width=7cm,height=5cm]{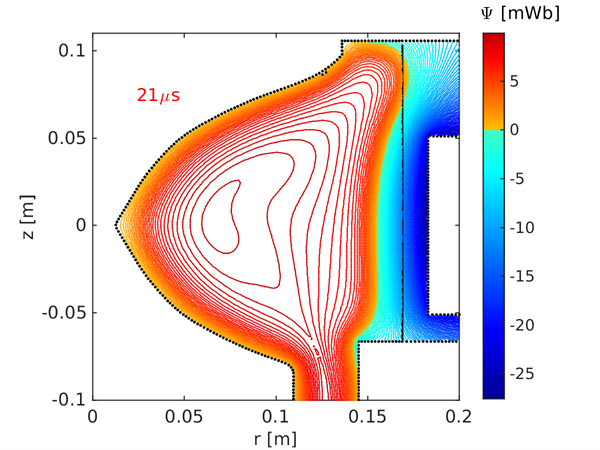}}

\caption{\label{fig: psi6_tsoak}$\,\,\,\,$Poloidal flux contours for 6-coil
configuration, effect of $t_{lev}$ }
\end{figure}
Figure \ref{fig: psi6_tsoak} shows how, for the 6-coil configuration,
MHD simulations confirm the reduction of plasma-wall interaction at
formation for the cases when the levitation field is allowed to soak
into the stainless steel above and below the insulating wall outboard
of the CT confinement region. As outlined in section \ref{subsec:6-coils-config_ceramic},
magnetic field that is allowed to soak into the steel can only be
displaced on the resistive timescale of the steel, which is longer
than the time it takes for the CT to bubble-in to the containment
region. The resulting line-tying effect is thought to have reduced
the level of CT poloidal field that penetrates the insulating wall
during the formation process.

Figures \ref{fig: psi6_tsoak}(a) and (b) show simulated $\psi$ contours
in the entire computational domain during the formation process at
$t=8\,\upmu$s and $t=21\,\upmu$s for the 6-coil configuration. The
simulation input parameters were approximately the same as for simulation
 2353, except that parameter $geometry$, which defines the machine
configuration, was set to three for this simulation. As depicted explicitly
in figure \ref{fig:Grid-arrangement-with}(b), the stack of six coils
is partly located in the blank rectangle at the top-right of figures
\ref{fig: psi6_tsoak}(a) and (b), centered around $z=0\mbox{ mm}$,
and extends off further to the right (not shown). The region above,
below, and just to the left of the coil-stack represents the air around
the stack. The vertical black line at $r=17\mbox{ cm}$ represents
the inner radius of the insulating wall, and the outer radius of the
insulating wall at $r=17.7\mbox{ cm}$ is not indicated. As described
in section \ref{sec:Vacuum-field-in}, only $\psi$, which determines
the vacuum poloidal field, is evaluated in the insulating region to
the right of the inner radius of the insulating wall. The solution
for $\psi$ is coupled to the full MHD solution in the remainder of
the domain. To maintain toroidal flux conservation, boundary conditions
for $f$, which has a finite constant value in the insulating wall
and is zero outside the current-carrying aluminum bars depicted in
figure \ref{fig:Schematic-of-6}(a), are evaluated for, and applied
to, the part of the boundary representing the inner radius of the
insulating wall, as described in section \ref{sec:PHIconservation-with}.
In this case (figures \ref{fig: psi6_tsoak}(a) and (b)), the boundary
conditions for $\psi$ are from a FEMM solution for a case with a
current frequency of 4170 Hz, corresponding to a quarter-period of
$|t_{lev}|=60\,\upmu$s. Figures \ref{fig: psi6_tsoak}(c) and (d)
show close ups of the CT confinement region for the same simulation,
at $t=0\,\upmu$s, and $t=21\,\upmu$s. It can be seen how, at $t=21\,\upmu$s,
the poloidal field at the leading edge of the forming CT intersects
the insulating wall above the coil stack. The simulations indicate
penetration of CT poloidal field into the insulating wall between
$t=13\,\upmu$s, and $t=33\,\upmu$s, with peak penetration at around
$t=21\,\upmu$s. Ions gyrorotating around and streaming along poloidal
field lines can intersect the wall in the 6-coil configuration and
sputter impurities into the plasma, leading to increased energy losses
due to line radiation.

For comparison, figures \ref{fig: psi6_tsoak}(e) and (f) show the
equivalent solutions for a case corresponding to $|t_{lev}|=300\,\upmu$s,
with corresponding boundary conditions for $\psi$ from a FEMM solution
for a case with a current frequency of 800 Hz. All other code input
parameters were identical to those used for the simulation presented
in figures \ref{fig: psi6_tsoak}(a)$\rightarrow$(d). Comparing figures
\ref{fig: psi6_tsoak}(c) and \ref{fig: psi6_tsoak}(e), it can be
seen how, at $t=0\,\upmu$s, the levitation field has soaked further
into the steel, particularly below the insulating wall, in the case
with increased $|t_{lev}|$. As a consequence of the additional energy
required to bend the soaked-in levitation field, poloidal field that
has been advected up the gun penetrates less deeply into the insulating
wall during formation in the case with $|t_{lev}|=300\,\upmu$s (figure
\ref{fig: psi6_tsoak}(d) $cf.$ figure \ref{fig: psi6_tsoak}(f)).
The simulations also indicate that the duration of interaction is
reduced with increased $|t_{lev}|$; the initial penetration is delayed
from $t=13\,\upmu$s to $t=15\,\upmu$s, and CT poloidal field stops
intersecting the wall sooner, at around $t=30\,\upmu$s ($cf.\,t=33\,\upmu$s).
\begin{figure}[H]
\subfloat[$|t_{lev}|=60\,\upmu$s]{\raggedright{}\includegraphics[width=7cm,height=5cm]{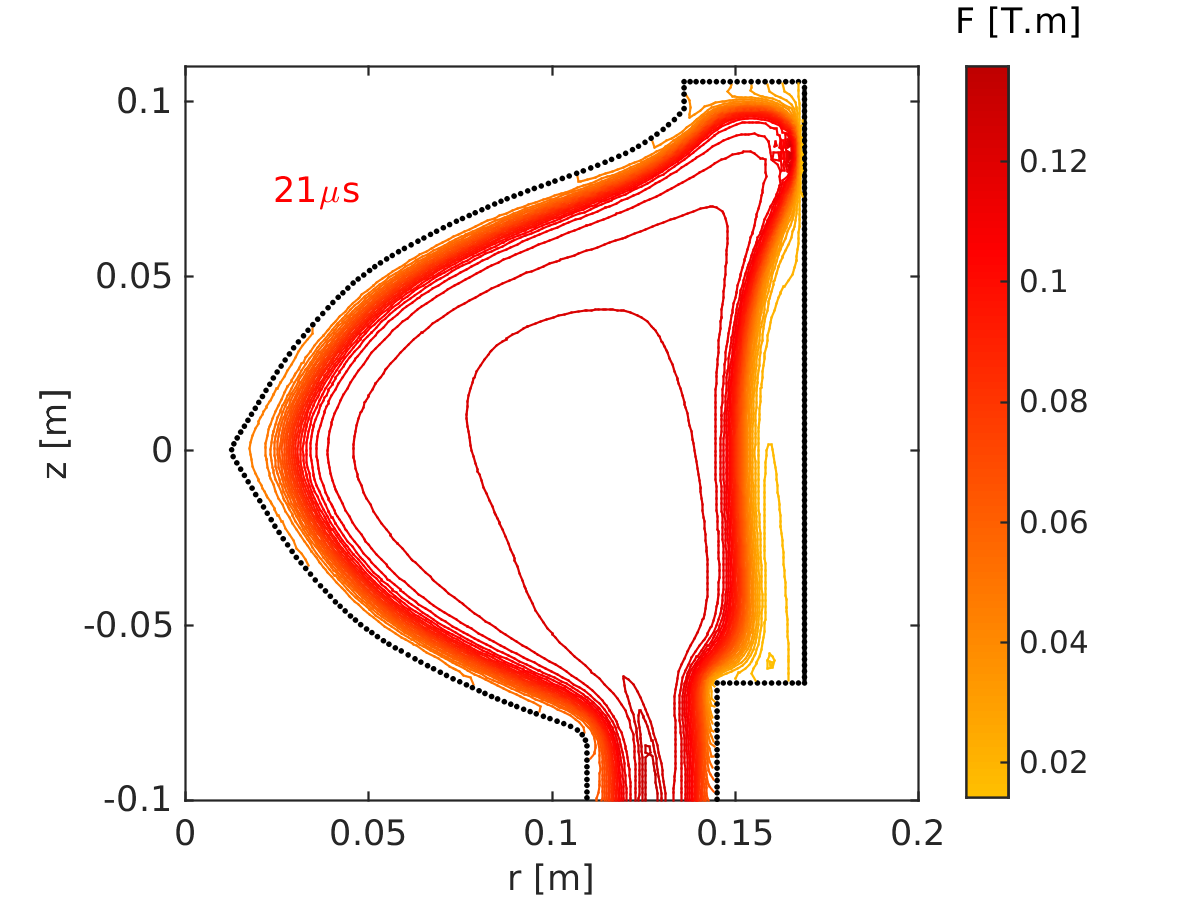}}\hfill{}\subfloat[$|t_{lev}|=300\,\upmu$s]{\raggedright{}\includegraphics[width=7cm,height=5cm]{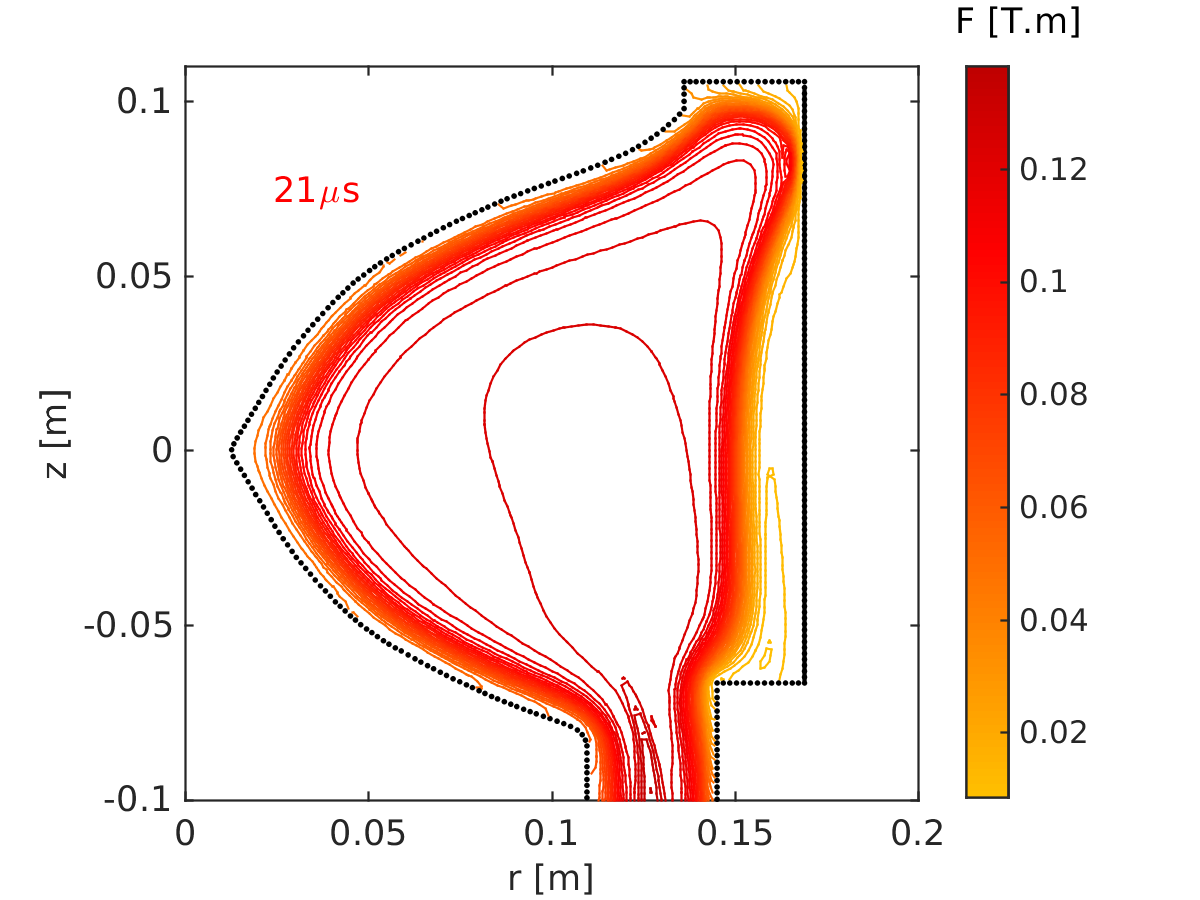}}

\caption{\label{fig: f6_tsoak-1}$\,\,\,\,$$f$ contours for 6-coil configuration,
effect of $t_{lev}$ }
\end{figure}
Figure \ref{fig: f6_tsoak-1} shows a comparison of contours of $f$
from the same simulations at $t=21\,\upmu$s, the time of peak field
penetration. Contours of $f$ represent lines of poloidal current
- it can be seen how the contours of $f$ are compressed up against
the top of the insulating wall, with less compression in the case
with $|t_{lev}|=300\,\upmu$s, indicating a reduced level of poloidal
current scraping against the wall. 

\subsection{Effect of levitation field profile (11-coil \emph{cf.} 6-coil configuration)\label{subsec:Effect-of-levitation}}

\begin{figure}[H]
\subfloat[]{\raggedright{}\includegraphics[width=7cm,height=5cm]{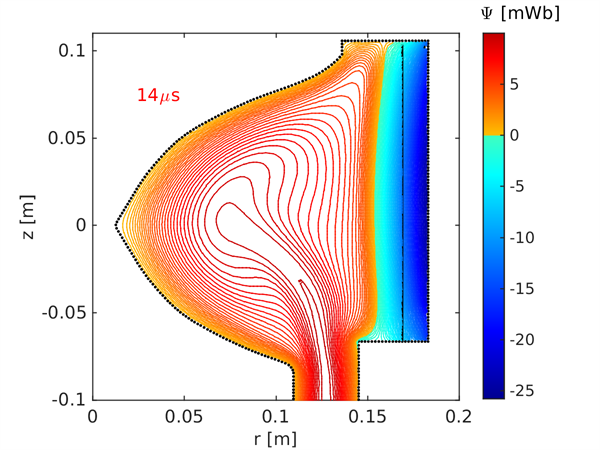}}\hfill{}\subfloat[]{\raggedright{}\includegraphics[width=7cm,height=5cm]{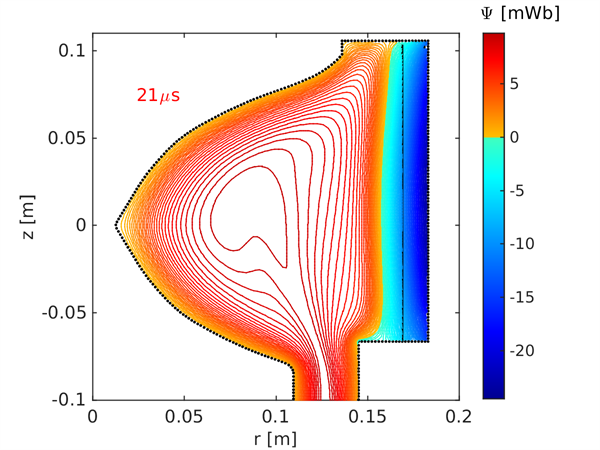}}

\subfloat[]{\raggedright{}\includegraphics[width=7cm,height=5cm]{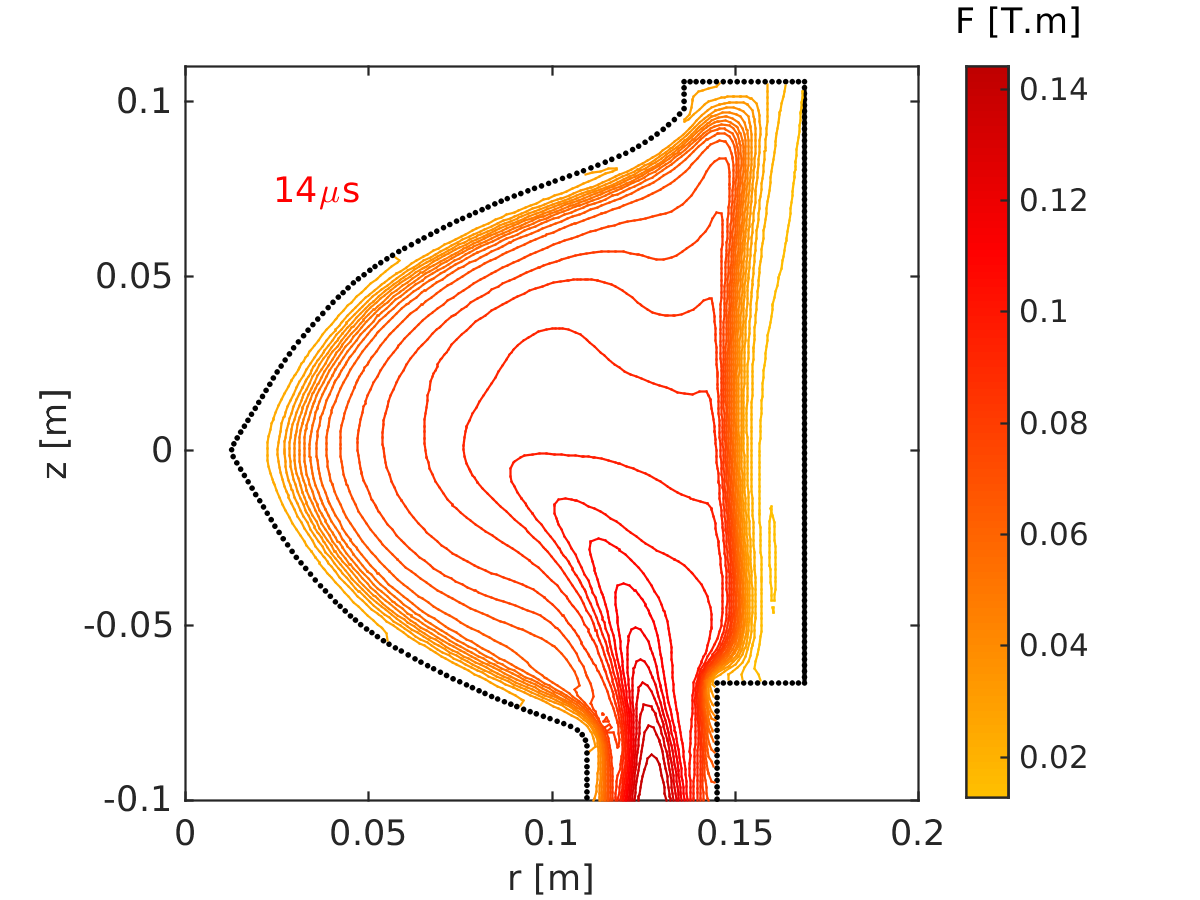}}\hfill{}\subfloat[]{\raggedright{}\includegraphics[width=7cm,height=5cm]{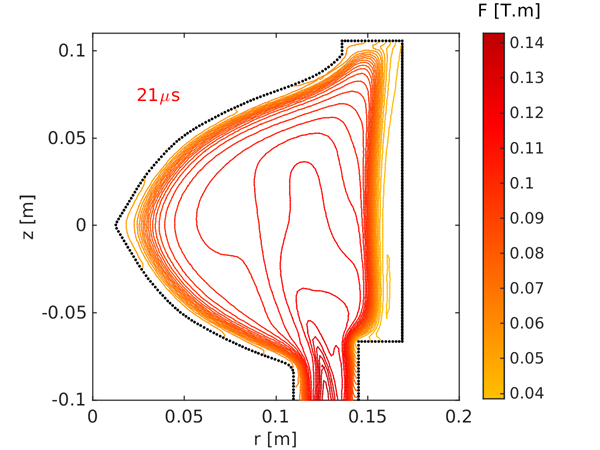}}

\caption{\label{fig: psi_f_11coils}$\,\,\,\,$$\psi$ and $f$ contours for
11-coil configurations ($t_{lev}=60\,\upmu$s) }
\end{figure}
Figure \ref{fig: psi_f_11coils} shows contours of $\psi$ and $f$
at $14$ and $21\,\upmu$s, during the simulated formation process
in the 11-coil configuration, from simulation  2353. In contrast with
the case for six coils (figures \ref{fig: psi6_tsoak} and \ref{fig: f6_tsoak-1}),
it can be seen how the leading edge of the forming CT is held off
the wall by the levitation field all along the height of the wall
for the 11-coil configuration. Note that the simulations presented
in figures \ref{fig: psi6_tsoak}, \ref{fig: f6_tsoak-1}, and \ref{fig: psi_f_11coils}
all have boundary conditions for $\psi_{main}$ and $\psi_{lev}$
from FEMM models, pertaining to $I_{main}=70$A, and with the total
levitation current such that $\psi_{lev}$ is approximately the same
for each configuration. Formation toroidal flux has been added to
the system in each case according to the experimentally measured $V_{gun}(t)$
signal for $V_{form}=16\mbox{ kV}$. The strategy of increasing $|t_{lev}|$
was found to be ineffective for the 25 turn coil and 11-coil configurations.
The optimal settings were $|t_{lev}|\sim150\,\upmu$s and $50\,\upmu$s
for the 25 turn coil and 11-coil configurations, approximately the
respective rise times of the levitation current (see discussion in
section \ref{subsec:SummaryLev}). Increasing $|t_{lev}|$ beyond
these values led to degraded CT performance in the 25 turn coil and
11-coil configurations. This is thought to be due to the effect of
partially blocking plasma entry to the CT containment region, and
reduced CT size, with increasing $|t_{lev}|$. Simulations confirm
that, as $|t_{lev}|$ is increased, the poloidal field of the CT becomes
more distorted during bubble-in, as it is forced to bend around the
levitation field that is resistively pinned to the steel at the outboard
side of the entrance to the confinement region. The poor recurrence
rate of good shots in the 6-coil and 25-turn configurations relative
to that in the 11-coil configuration is thought to be related to the
high values of $|t_{lev}|$ that were necessarily used with the 25
turn coil configuration, and used for optimal ($i.e.,$ occasionally
good) performance in the 6-coil configuration. In the 6-coil configuration,
the performance degradation associated with increased $|t_{lev}|$
was outweighed by the benefit of reduced plasma-wall interaction associated
with increased line-tying, however increasing $|t_{lev}|$ to more
than around $300\,\upmu$s also led to degradation of CT lifetime
with the 6-coil configuration. 

It is thought that the moderate improvement in CT lifetime seen with
increased $|t_{lev}|$ in the 6-coil configuration, and the more significant
improvement found with the switch to the levitation field profile
associated with the 25 turn coil and 11-coil configurations, are due
to the effect of reducing the level of plasma-wall interaction during
the CT formation process. Less interaction implies less ion bombardment
on the insulating wall, and consequent reductions in impurity concentration
and in the level of cooling associated with impurity line radiation.
From equation \ref{eq:211}, $\dot{\psi}_{\eta}\propto\eta$. Since
$\eta\propto T_{e}^{-1.5}$, reduced cooling leads to increased CT
lifetimes.

\section{Evolution of the various fields, simulation  2353\label{subsec:Evolution-of-the}}

In this section, contours of the evolved fields from simulation  2353,
as well as contours of current density, calculated using equation
\ref{eq:20.2}, will be presented at a selection of simulation times.
The simulations start with CT formation in the magnetized Marshall
gun. Plasma starts to bubble-into the containment region at around
8$\,\upmu$s, and closed CT flux surfaces start to form around 25$\,\upmu$s.
Magnetic compression is started at $t_{comp}=45\,\upmu$s, with peak
compression ($i.e.,$ peak compression coil current) occurring at
65$\,\upmu$s. The simulations continues until 87$\,\upmu$s, by which
time the primary magnetic compression cycle has completed and compression
coil current has dropped back to zero. Full field depictions as well
as close-up containment region views will be presented at various
output times for $\psi$ and $f$, while the remaining fields will
be presented at various output times in either of the two views. 

\begin{figure}[H]
\subfloat[]{\raggedright{}\includegraphics[scale=0.5]{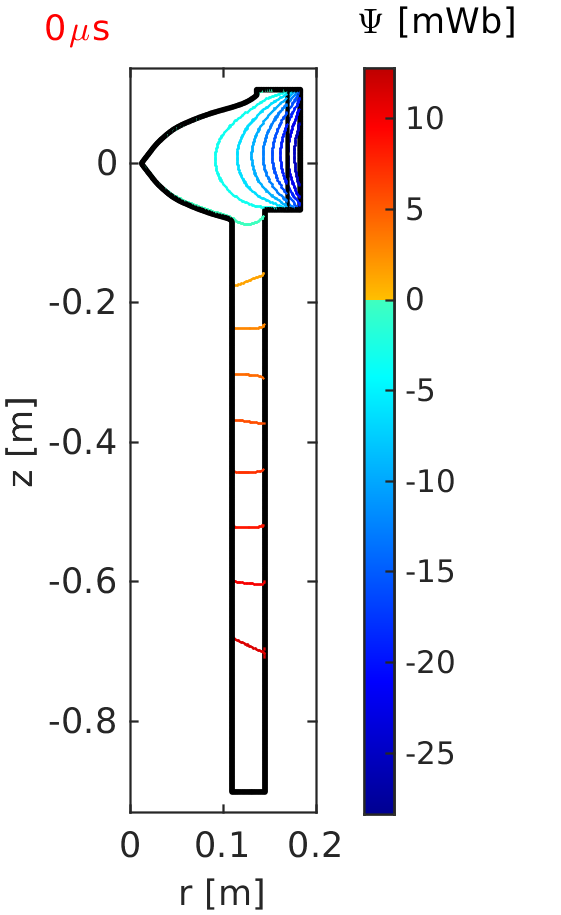}}\hfill{}\subfloat[]{\raggedright{}\includegraphics[scale=0.5]{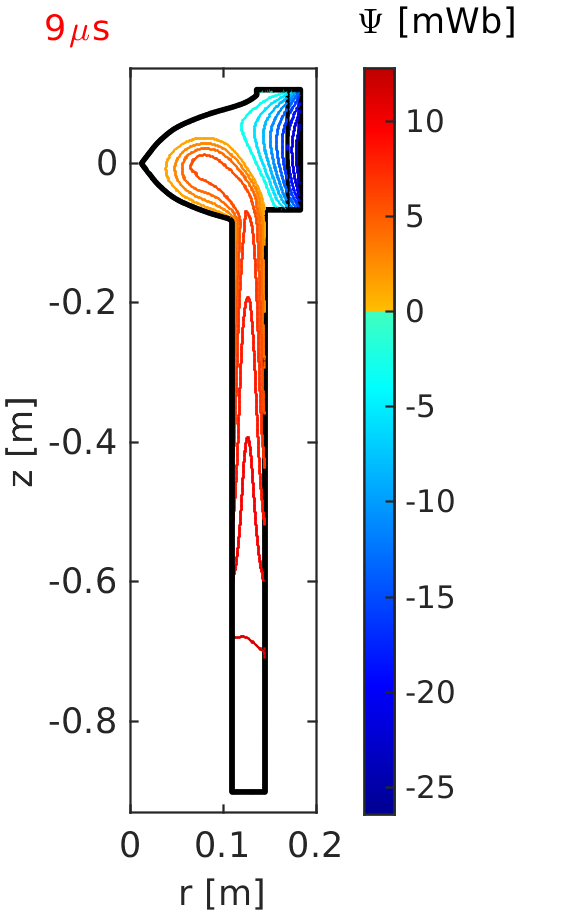}}\hfill{}\subfloat[]{\raggedright{}\includegraphics[scale=0.5]{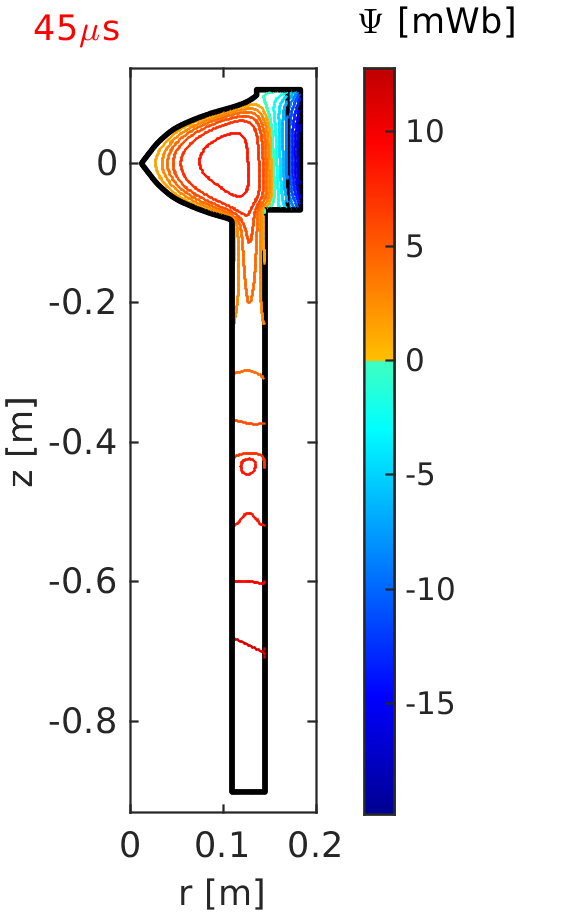}}

\subfloat[]{\raggedright{}\includegraphics[scale=0.5]{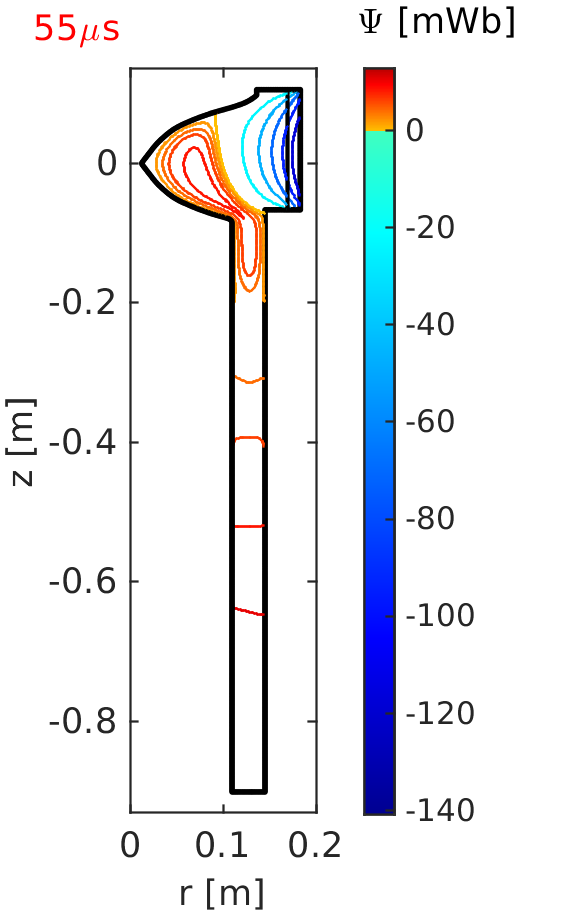}}\hfill{}\subfloat[]{\raggedright{}\includegraphics[scale=0.5]{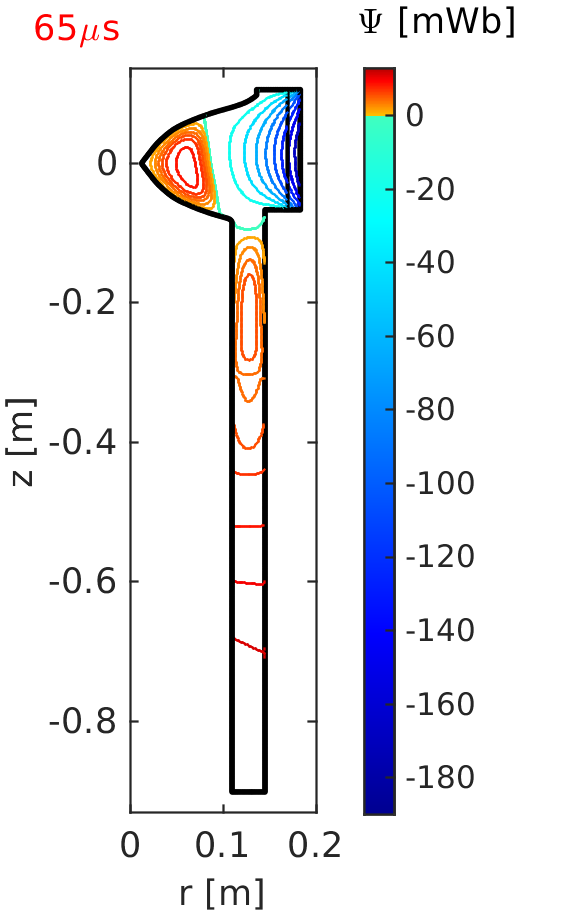}}\hfill{}\subfloat[]{\raggedright{}\includegraphics[scale=0.5]{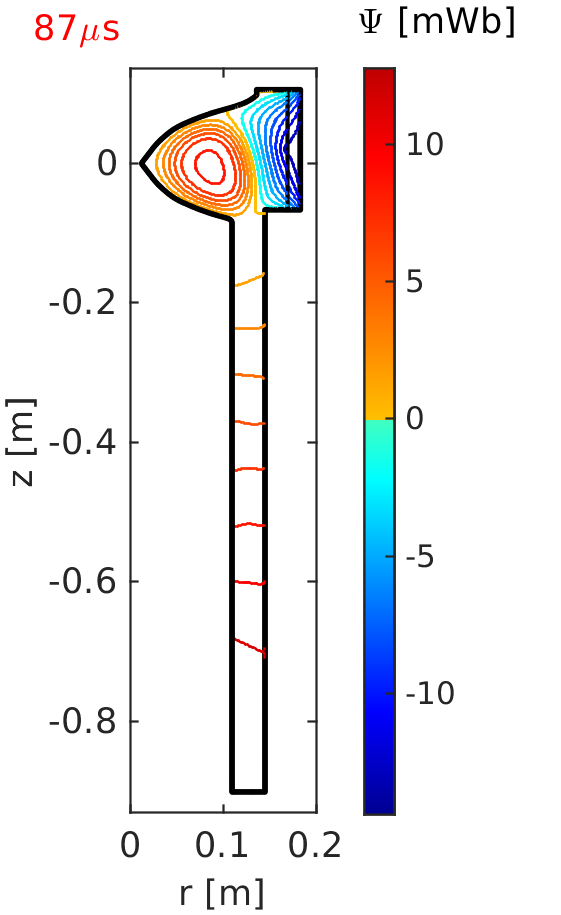}}

\caption{\label{fig: psi_11coils}$\,\,\,\,$Poloidal flux contours }
\end{figure}
Figure \ref{fig: psi_11coils} shows $\psi$ contours at various times.
Simulation times are notated in red at the top left of the figures.
Note that the colorbar scaling changes over time; max$(\psi)$ decreases
slowly over time as the CT decays, while min$(\psi)$ increases as
the levitation current in the external coils decays, and then drops
off rapidly as the compression current in the external coils is increased,
starting at $t_{comp}=45\,\upmu\mbox{s}$ in this simulation. At time
$t=0,$ the stuffing field due to currents in the main coil ($\psi>0$)
fills the vacuum below the containment region, and has soaked well
into all materials around the gun, while the levitation field fills
the containment region. Open field lines that are resistively pinned
to the electrodes, and frozen into the conducting plasma, have been
advected by the $\mathbf{J}_{r}\times\mathbf{B}_{\phi}$ force, into
the containment region by $t=9\,\upmu$s. By $45\,\upmu$s, open field
lines have reconnected at the entrance to the containment region,
and closed flux surfaces have formed in the containment region. The
primary compression cycle starts at $45\,\upmu$s and continues until
$65\,\upmu$s. Note that at $55\,\upmu$s (figure \ref{fig: psi_11coils}(d)),
magnetic compression causes closed CT poloidal field lines that extend
down the gun to be pinched off at the gun entrance, where they reconnect
to form a second smaller CT. Field lines that remain open surrounding
the main CT are then also reconnectively pinched off, forming additional
closed field lines around the main CT, while the newly reconnected
open field lines below the main CT act like a slingshot that advects
the smaller CT down the gun, as can be seen at $65\,\upmu\mbox{s}$.
Peak CT compression is at $65\,\upmu$s, and the CT re-expands as
$I_{comp}$ in the compression coils reduces, reaching maximum expansion
at around $87\,\upmu$s. 
\begin{figure}[H]
\subfloat[]{\raggedright{}\includegraphics[width=7cm,height=5cm]{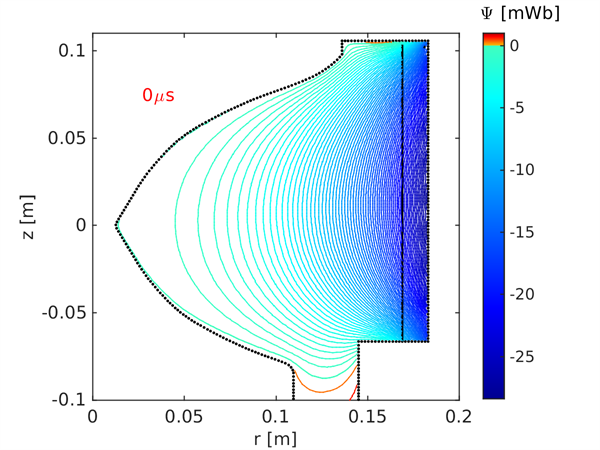}}\hfill{}\subfloat[]{\raggedright{}\includegraphics[width=7cm,height=5cm]{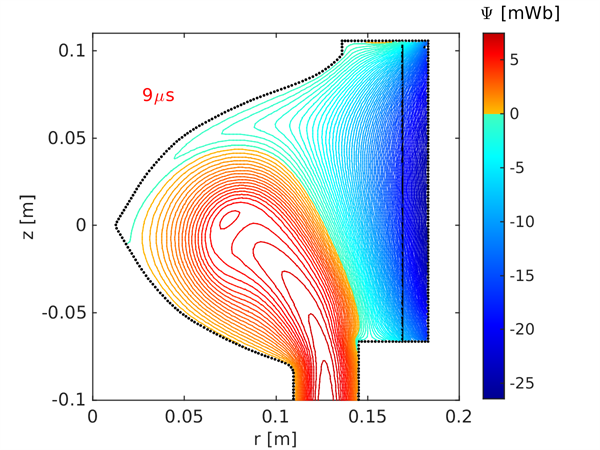}}

\subfloat[]{\raggedright{}\includegraphics[width=7cm,height=5cm]{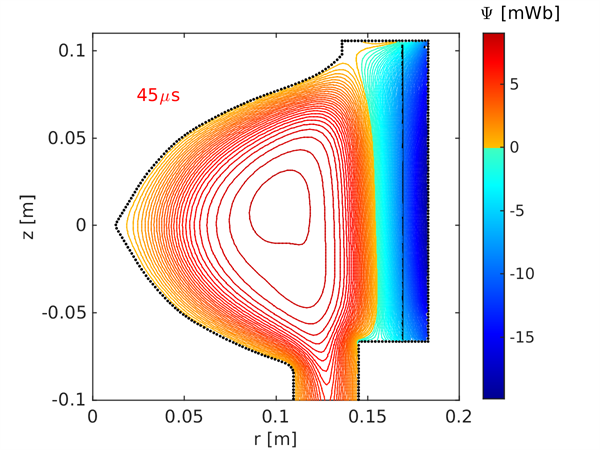}}\hfill{}\subfloat[]{\raggedright{}\includegraphics[width=7cm,height=5cm]{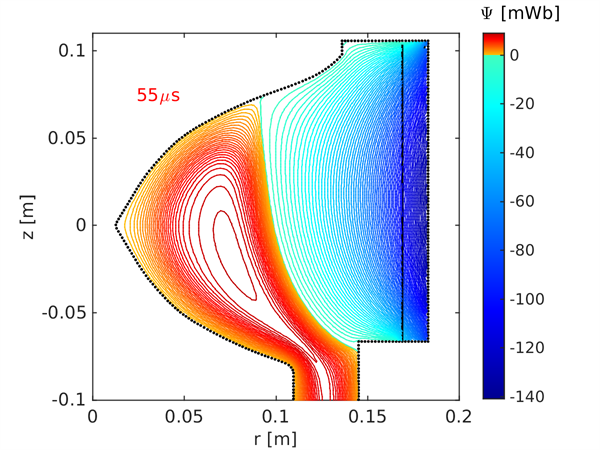}}

\subfloat[]{\raggedright{}\includegraphics[width=7cm,height=5cm]{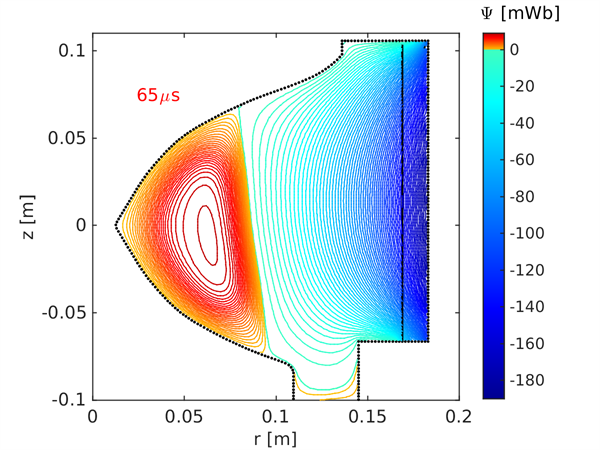}}\hfill{}\subfloat[]{\raggedright{}\includegraphics[width=7cm,height=5cm]{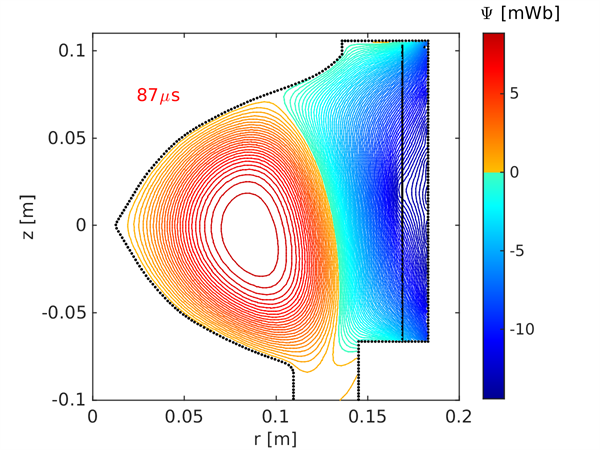}}

\caption{\label{fig: psiC_11coils}$\,\,\,\,$Poloidal flux contours (confinement
region) }
\end{figure}
Figure \ref{fig: psiC_11coils} shows close-up views, focused on the
CT containment region, of the $\psi$ contours presented in figure
\ref{fig: psi_11coils}.

\begin{figure}[H]
\subfloat[]{\raggedright{}\includegraphics[scale=0.5]{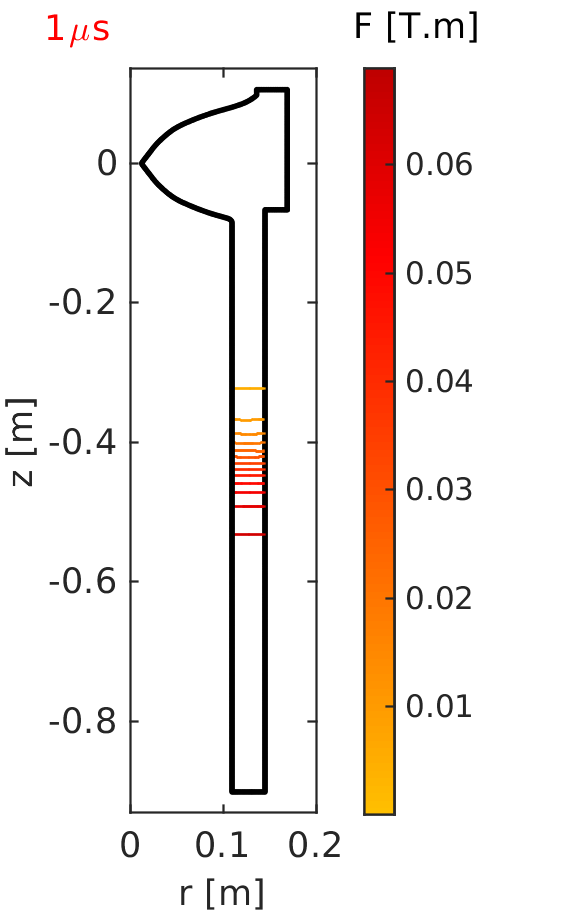}}\hfill{}\subfloat[]{\raggedright{}\includegraphics[scale=0.5]{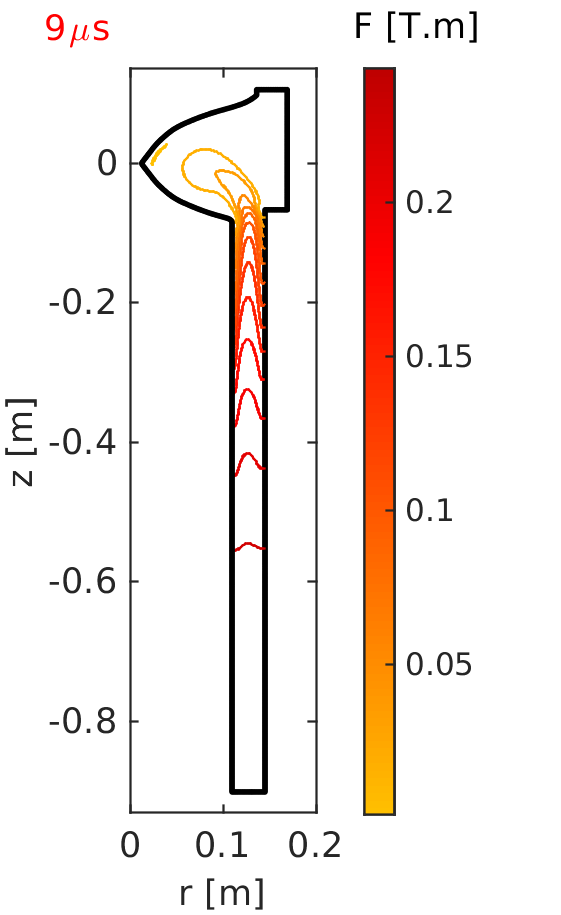}}\hfill{}\subfloat[]{\raggedright{}\includegraphics[scale=0.5]{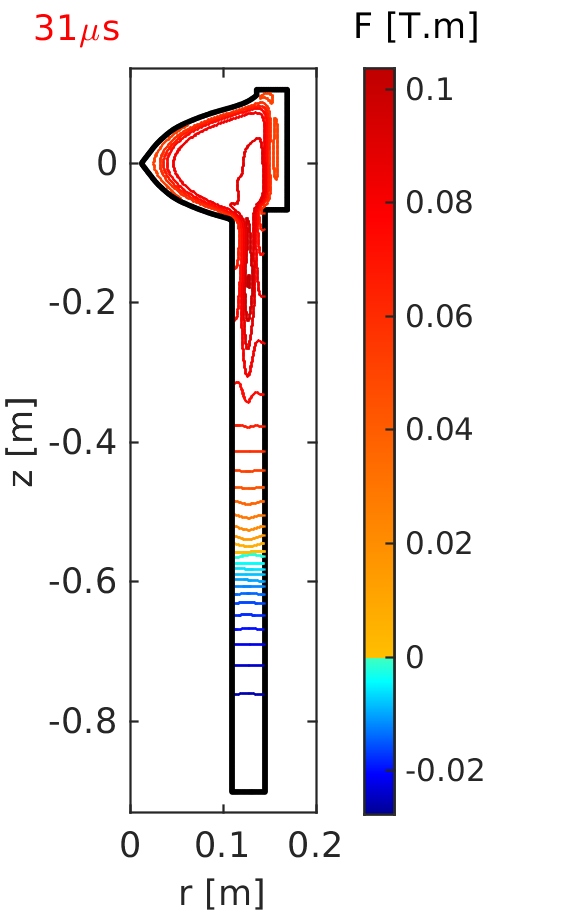}}

\subfloat[]{\raggedright{}\includegraphics[scale=0.5]{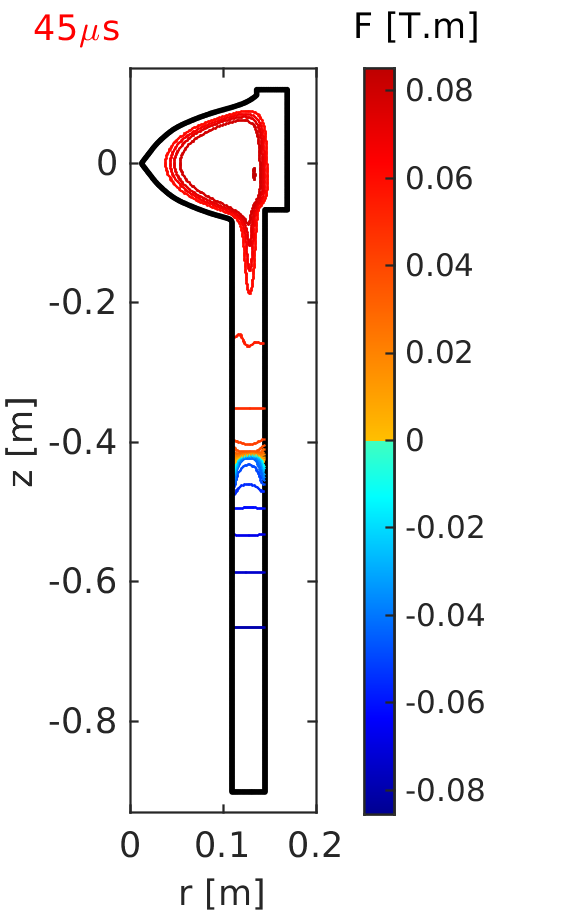}}\hfill{}\subfloat[]{\raggedright{}\includegraphics[scale=0.5]{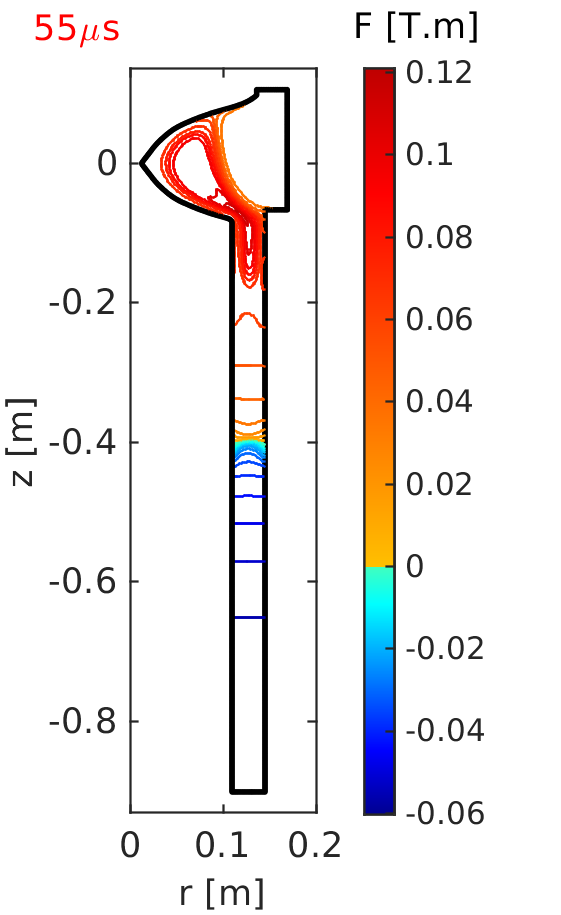}}\hfill{}\subfloat[]{\raggedright{}\includegraphics[scale=0.5]{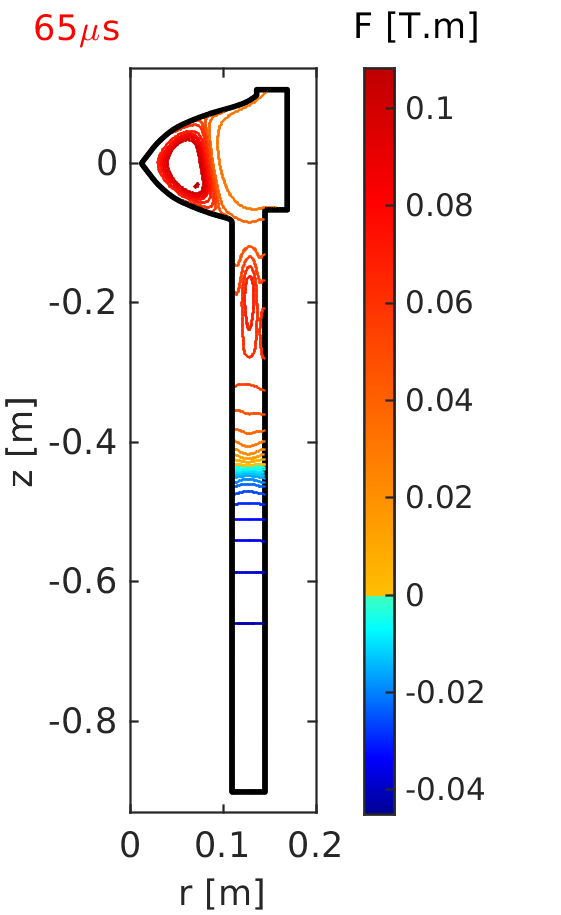}}

\caption{\label{fig: F_11coils}$\,\,\,\,$Lines of poloidal current ($f$
contours) }
\end{figure}
Figure \ref{fig: F_11coils} shows contours of $f$ at various times
from simulation  2353. Recall that contours of $f=rB_{\phi}$ represent
lines of poloidal current. Initially, $f$ is zero at all nodes. At
$1\,\upmu$s, it can seen how radial formation current between the
gun electrodes is concentrated around the gas valve location, at $z$$=-0.43$
m. Referring to equation \ref{eq:20.2}, it can be seen how radial
current corresponds to axial gradients in $f$. Closely spaced contours
indicate regions of high gradients and high poloidal currents. To
simulate formation, toroidal flux is being added to the domain according
to the geometric profile shown in figure \ref{fig:gform_Phiform_Vform}(a).
At $1\,\upmu$s, $f$ is constant, at its highest value in the domain,
below the lowermost contour at $z\sim-0.55$ m representing $f\sim0.07$
T-m. By $9\,\upmu$s, plasma has been advected upwards, and poloidal
current is flowing along open poloidal magnetic field lines that remain
resistively pinned to the electrodes down the gun. By $31\,\upmu$s
(figure \ref{fig: F_11coils}(c)) poloidal current is flowing along
closed and open poloidal magnetic field lines, and most of the toroidal
flux in the domain has been advected with the plasma into the containment
region. At this time, as the level of toroidal flux added to the domain
below the gas valve locations, which is equal to the total toroidal
flux in the domain, falls off (see figure \ref{fig:phi_cons}(a)),
radial currents between the electrodes further down the gun reverse
direction, and $f$ becomes negative, as consequences of toroidal
flux conservation and upward advection of toroidal flux. As shown
in figure \ref{fig: F_11coils}(d), (e) and (f), poloidal current
flows around the closed poloidal field lines of the main CT as it
is being compressed, and around the closed poloidal field lines of
the smaller pinched off CT at 65$\,\upmu$s. 

\begin{figure}[H]
\subfloat[]{\raggedright{}\includegraphics[width=7cm,height=5cm]{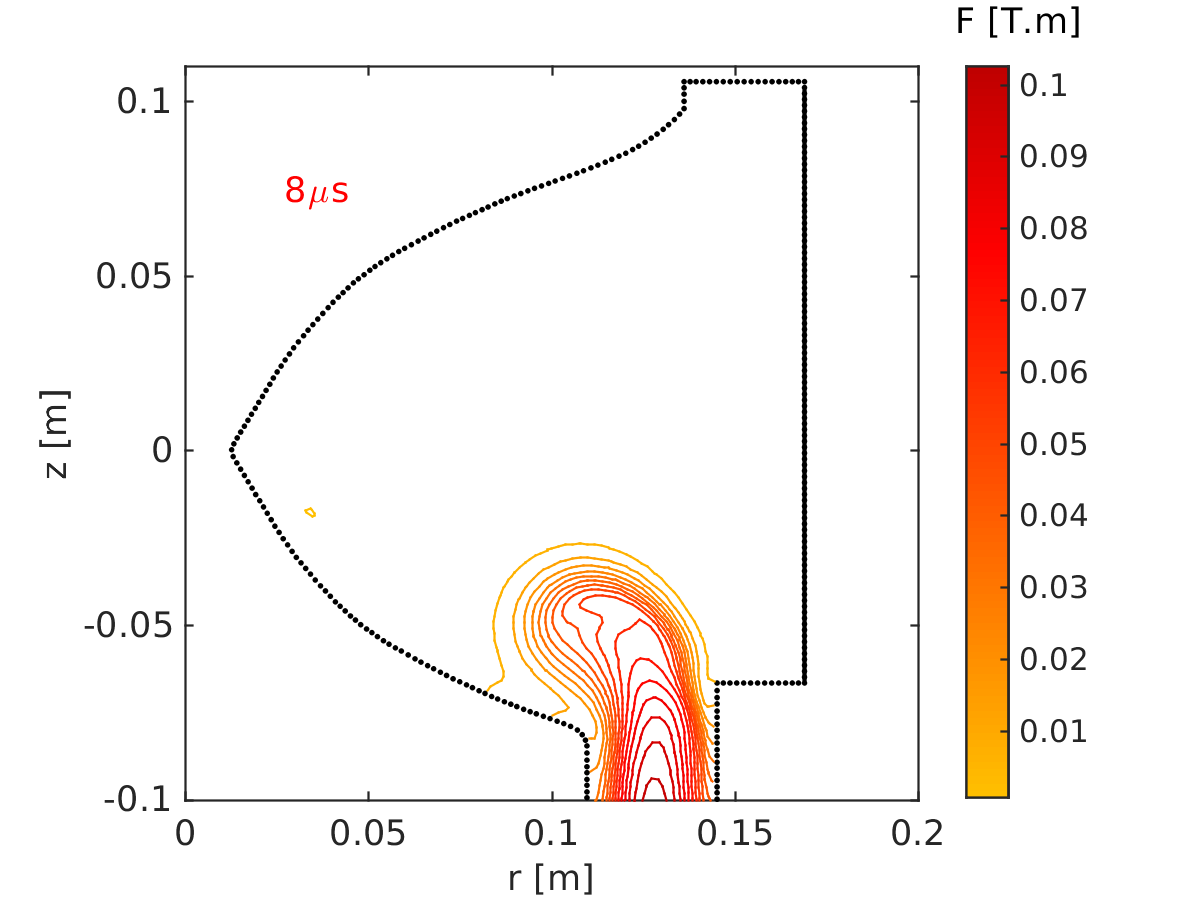}}\hfill{}\subfloat[]{\raggedright{}\includegraphics[width=7cm,height=5cm]{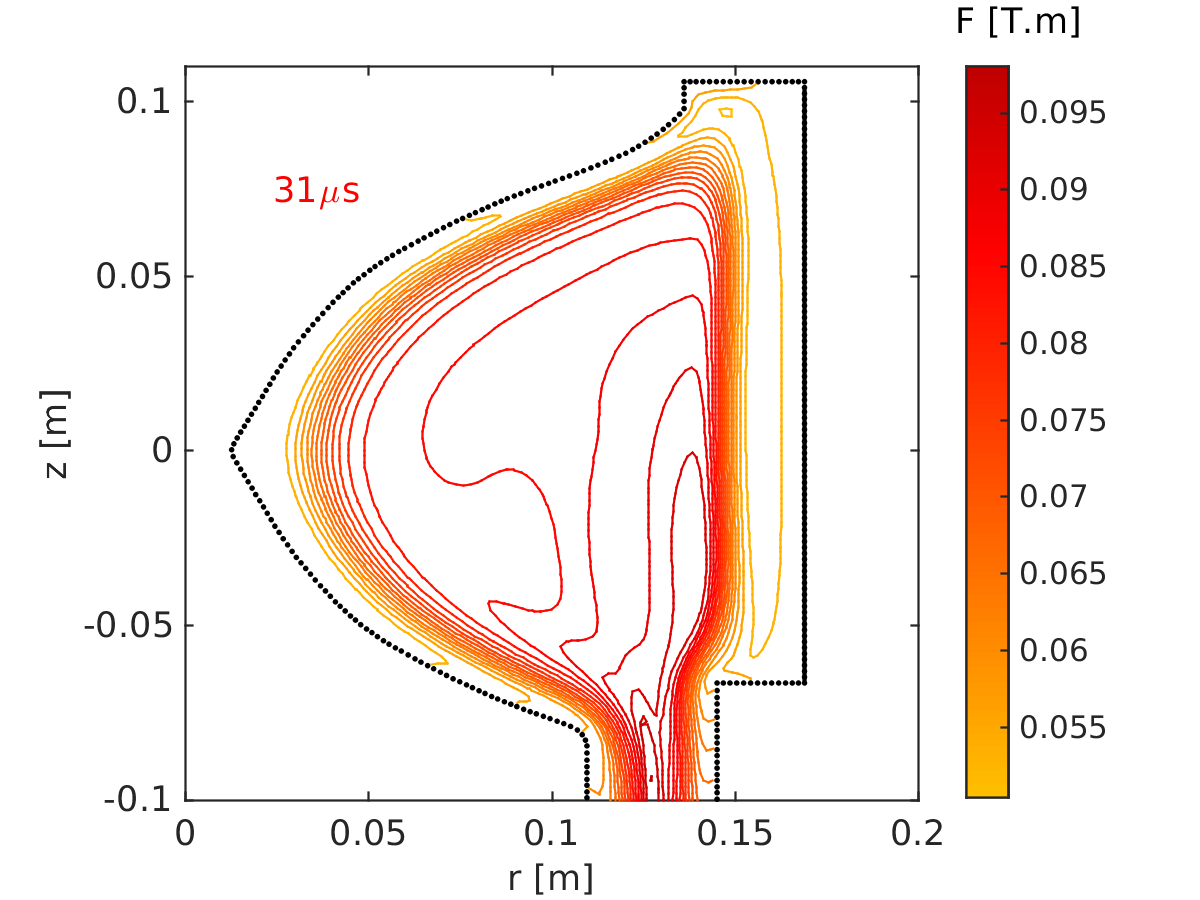}}

\subfloat[]{\raggedright{}\includegraphics[width=7cm,height=5cm]{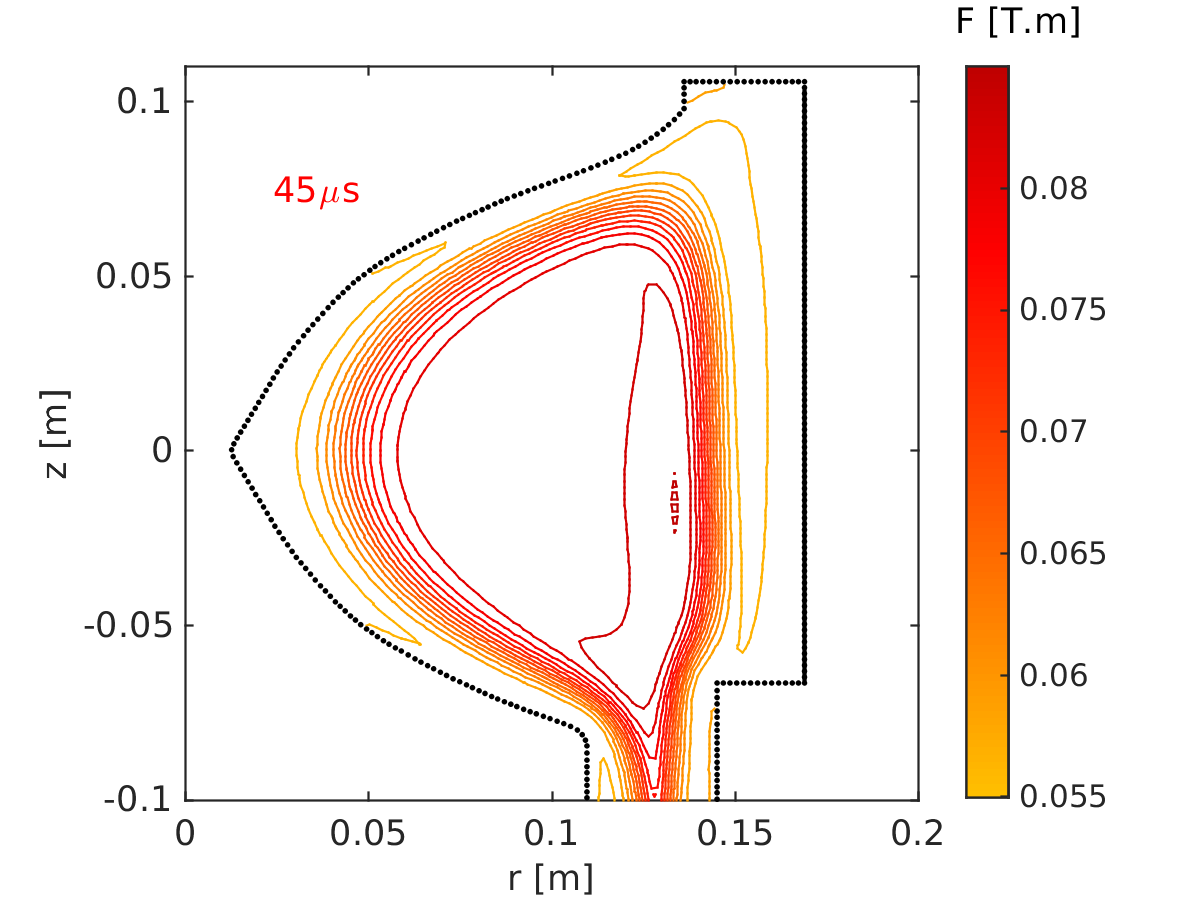}}\hfill{}\subfloat[]{\raggedright{}\includegraphics[width=7cm,height=5cm]{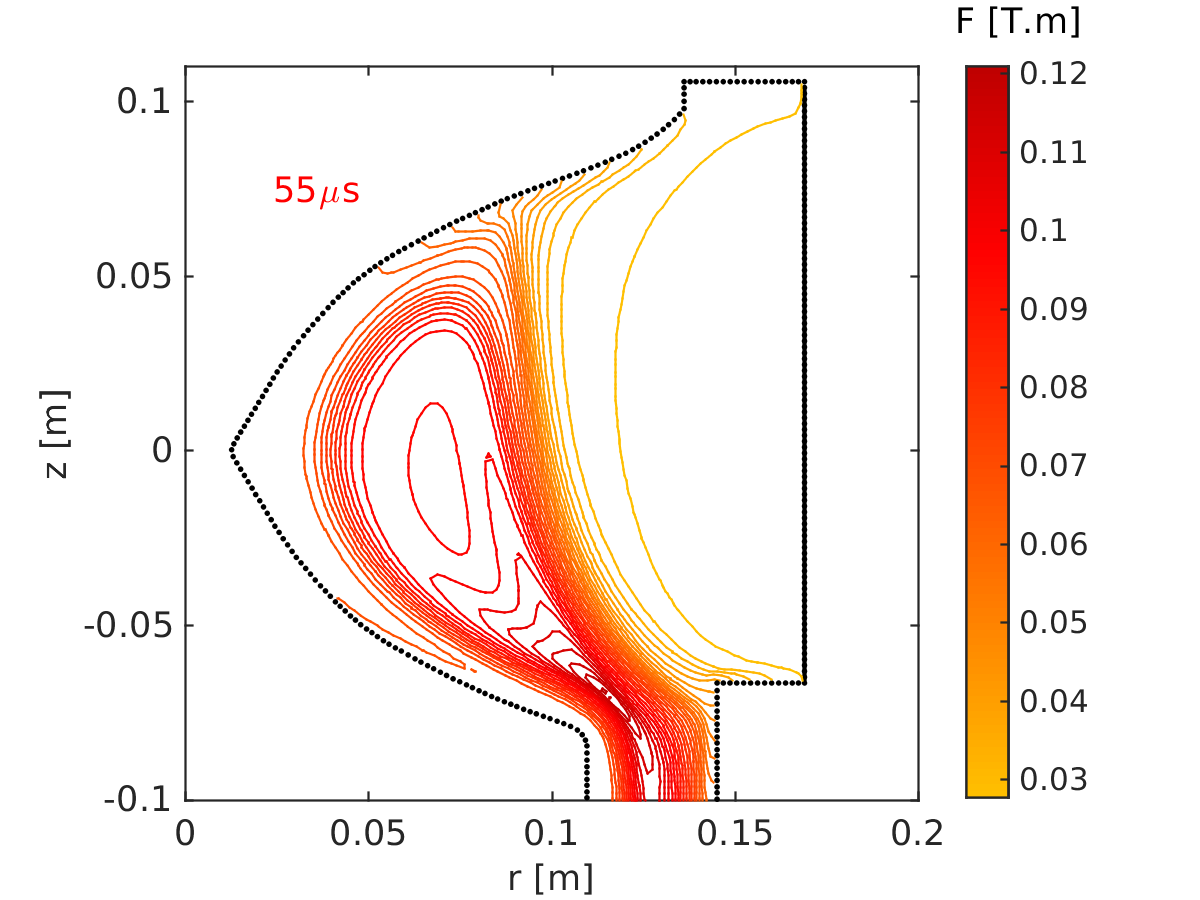}}

\subfloat[]{\raggedright{}\includegraphics[width=7cm,height=5cm]{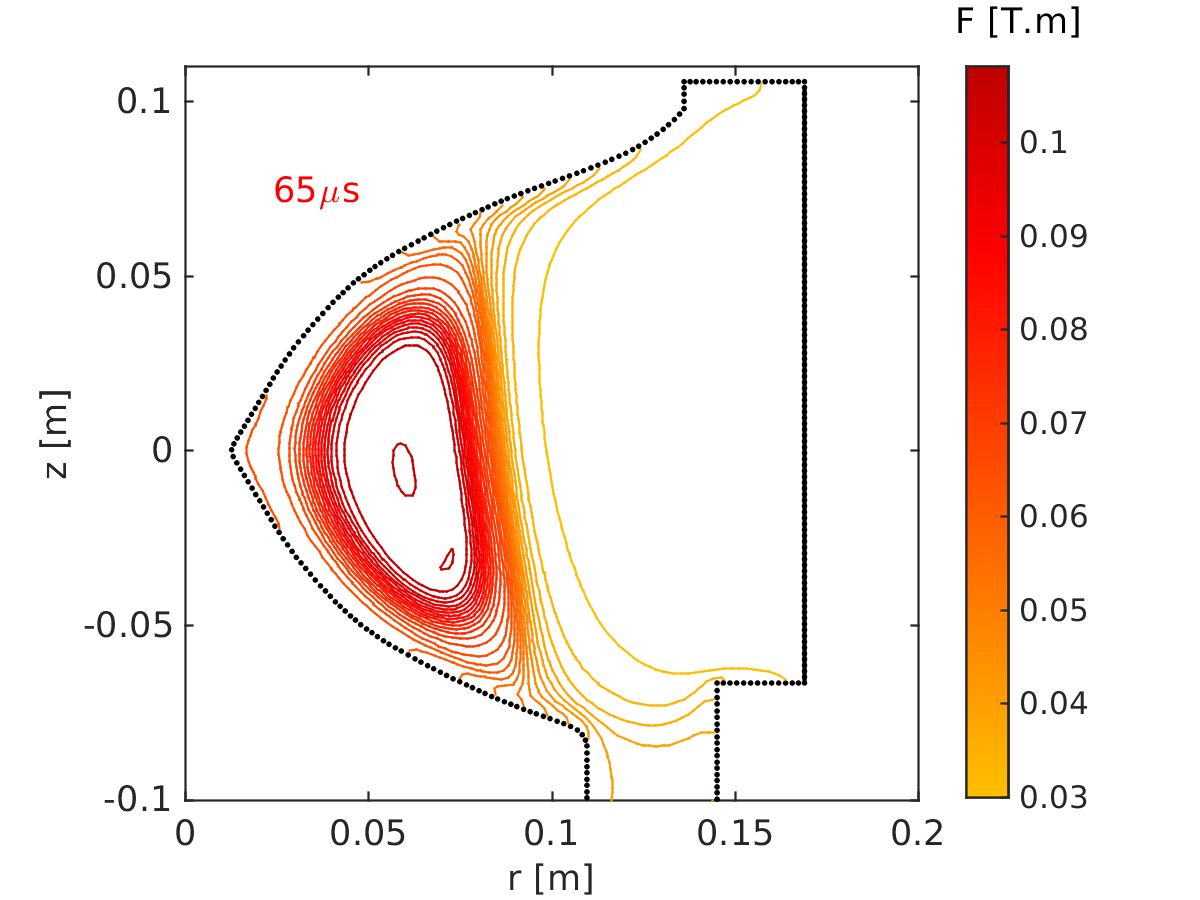}}\hfill{}\subfloat[]{\raggedright{}\includegraphics[width=7cm,height=5cm]{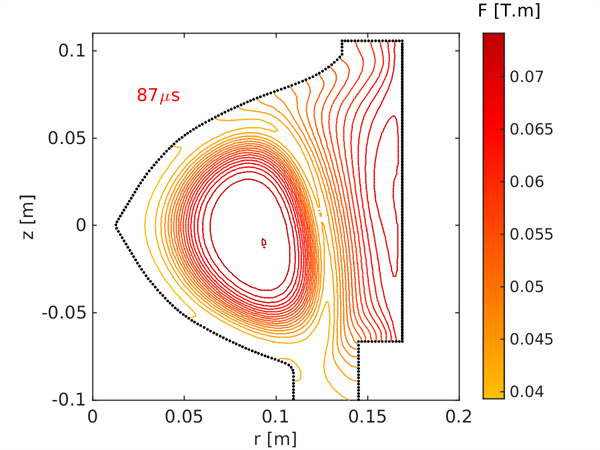}}

\caption{\label{fig: FC_11coils}$\,\,\,\,$Lines of poloidal current ($f$
contours), CT confinement region }
\end{figure}
Figure \ref{fig: FC_11coils} shows close-up views of $f$ contours
at various times in the CT containment region. It can be seen how
the imposition of toroidal flux conservation leads, at compression,
to the induction of poloidal currents flowing from wall-to-wall just
external to the outboard boundary of the CT (figures \ref{fig: FC_11coils}(d)
to (f)). This is a partial illustration of the mechanism behind the
experimentally observed compressional instability discussed in section
\ref{subsec:Compressional-Instability}. Note that 2D simulations,
which neglect inherently three dimensional turbulent transport and
flux conversion, generally overestimate the level of hollowness of
the field profiles.

\begin{figure}[H]
\subfloat[]{\raggedright{}\includegraphics[scale=0.5]{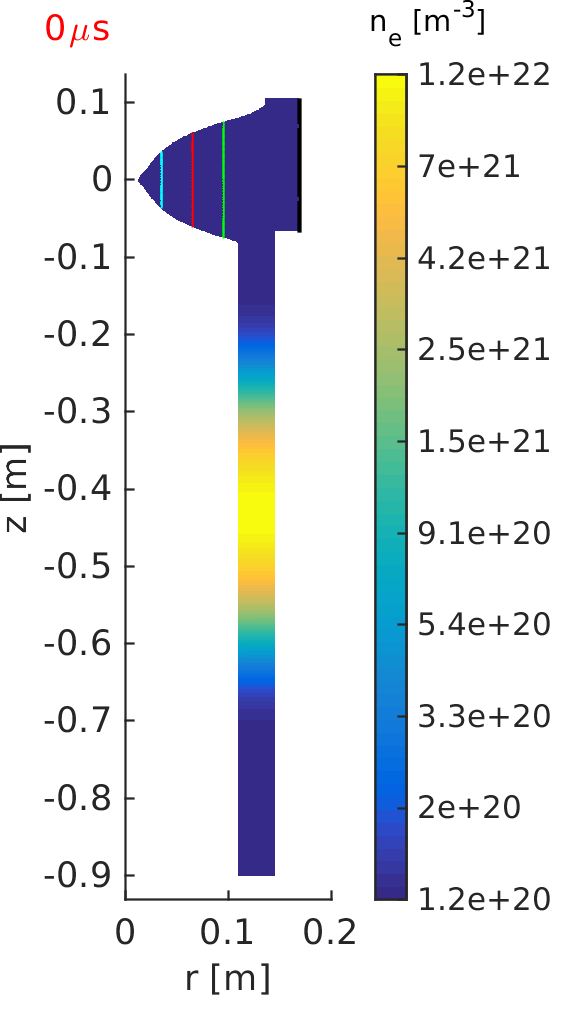}}\hfill{}\subfloat[]{\raggedright{}\includegraphics[scale=0.5]{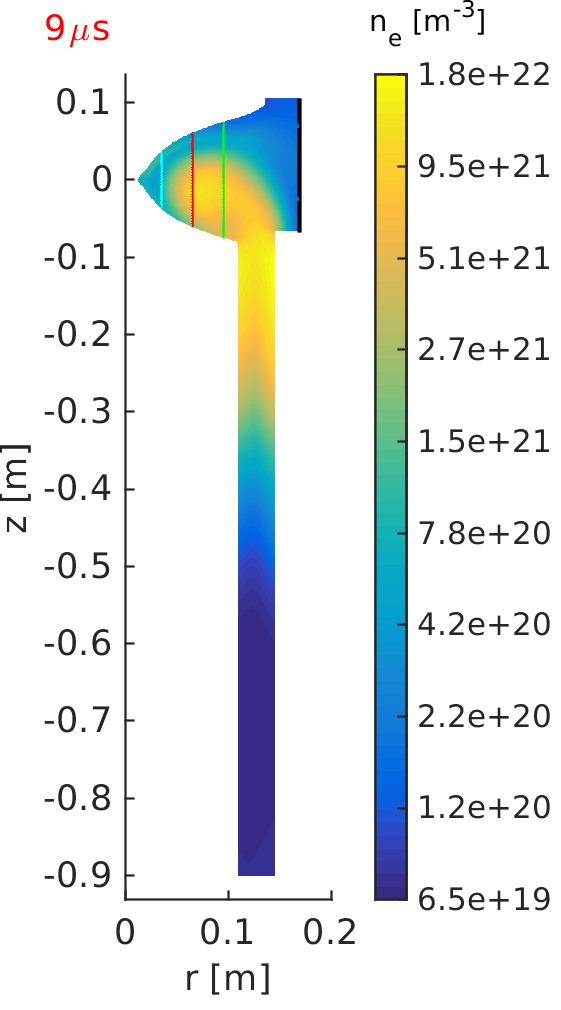}}\hfill{}\subfloat[]{\raggedright{}\includegraphics[scale=0.5]{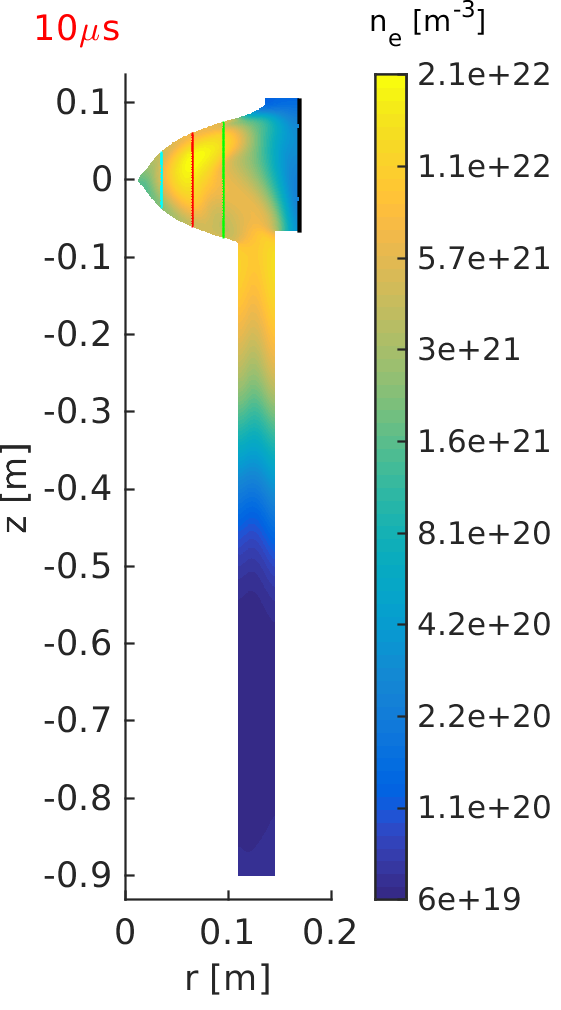}}

\subfloat[]{\raggedright{}\includegraphics[width=7cm,height=5cm]{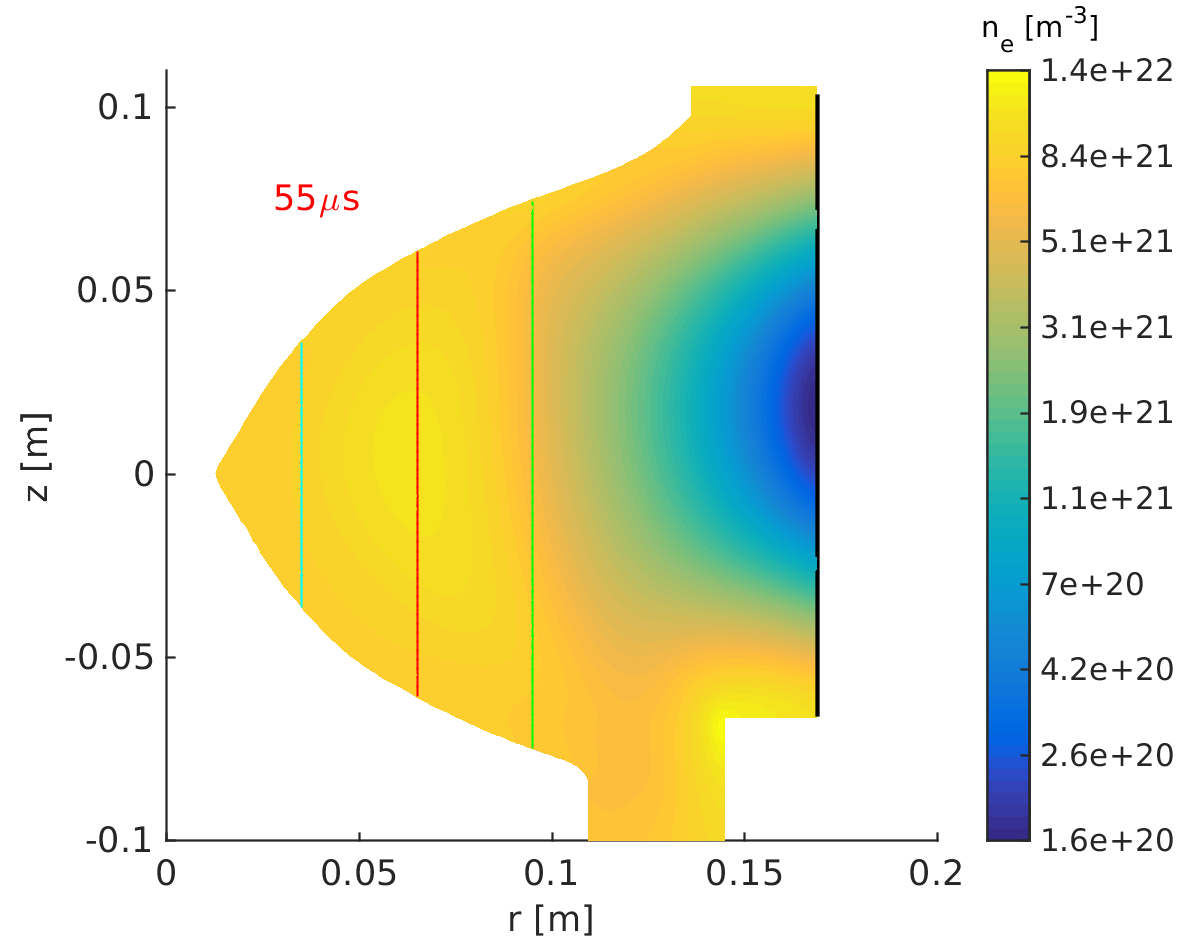}}\hfill{}\subfloat[]{\raggedright{}\includegraphics[width=7cm,height=5cm]{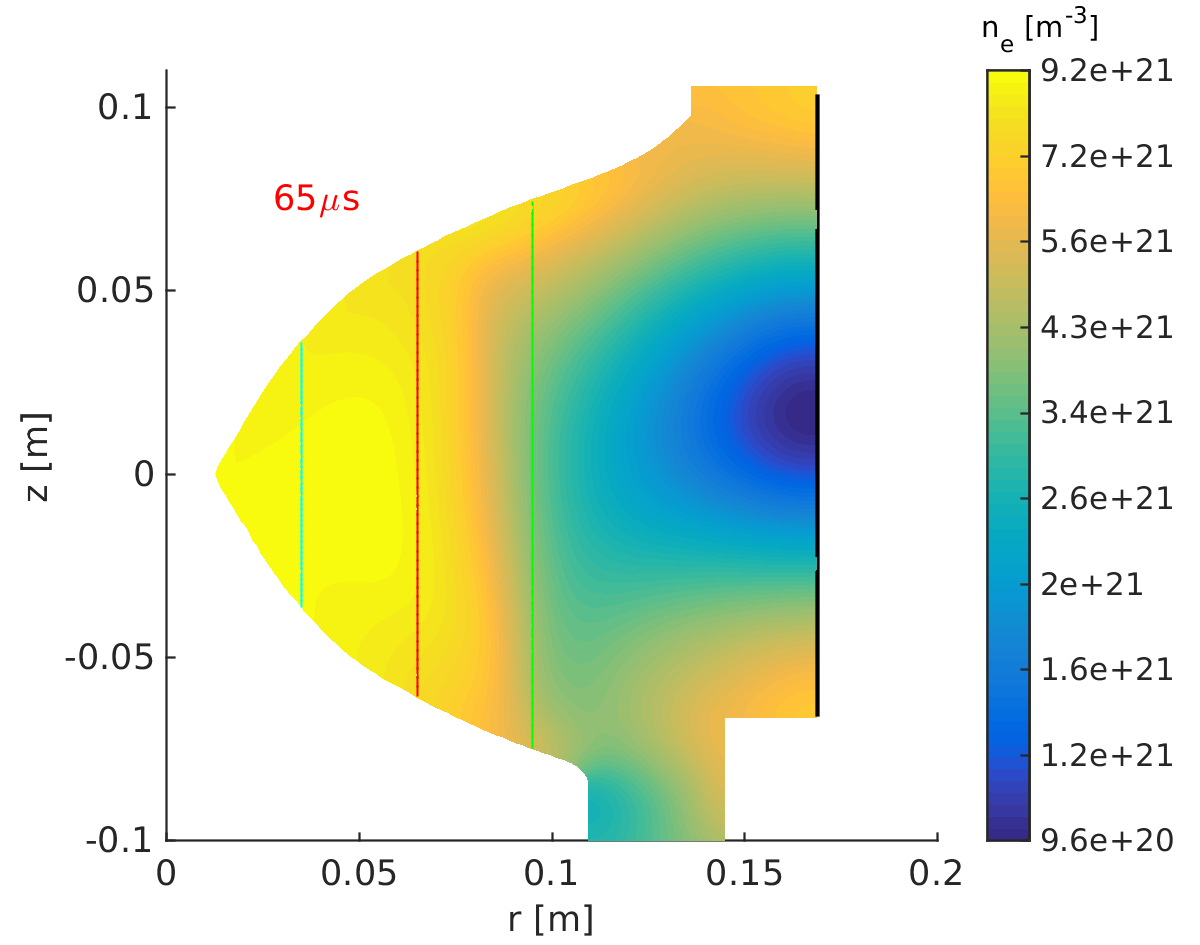}}

\caption{\label{fig: n_11coils}$\,\,\,\,$Electron density profiles}
\end{figure}
Figure \ref{fig: n_11coils}(a) shows the initial profile for the
electron density. As described with equation \ref{eq:900}, the initial
density profile is defined with a Gaussian profile centered around
the axial coordinate of the locations of the gas puff valves. Plasma
is advected up the gun during CT formation, and bubbles-in to the
containment region, as shown in figures \ref{fig: n_11coils}(b) and
(c). In figures \ref{fig: n_11coils}(d) and (e), it can be seen how
density rises over compression around the CT core, and that a region
of low density remains outboard of the compressed CT. The vertical
blue, red and green chords in the CT containment region in figures
\ref{fig: n_11coils}(a) to (e) represent the lines of sight of the
interferometer measurements ($cf.$ figure \ref{fig:Chalice}), along
which simulated line averaged-electron density is evaluated for comparison
with experimental data. 

\begin{figure}[H]
\subfloat[]{\raggedright{}\includegraphics[scale=0.5]{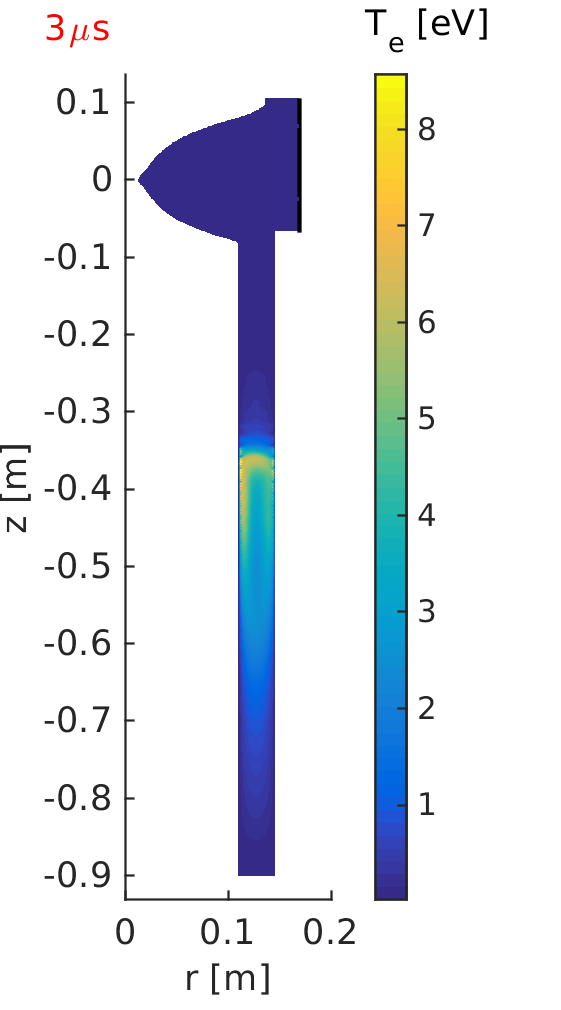}}\hfill{}\subfloat[]{\raggedright{}\includegraphics[scale=0.5]{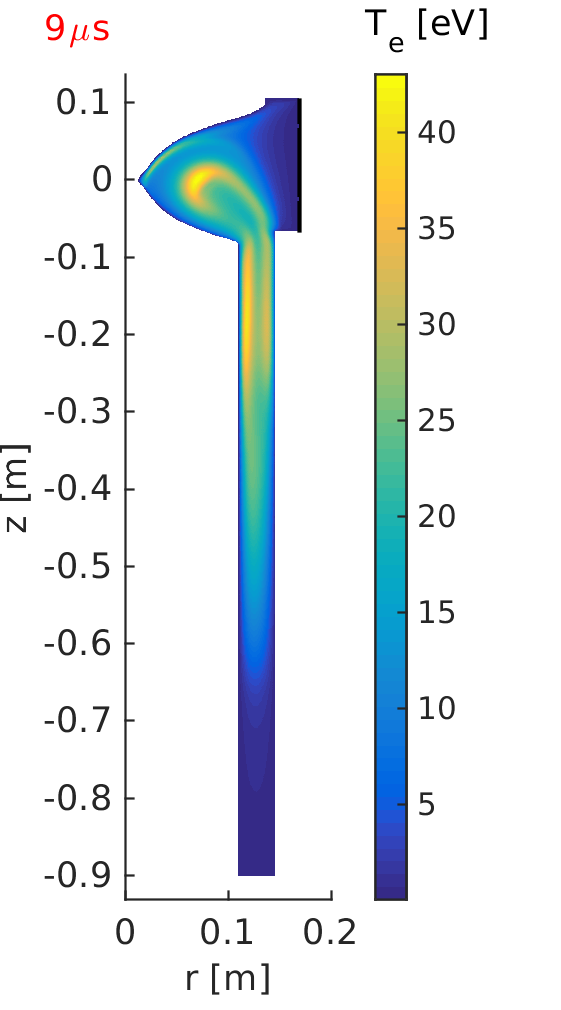}}\hfill{}\subfloat[]{\raggedright{}\includegraphics[scale=0.5]{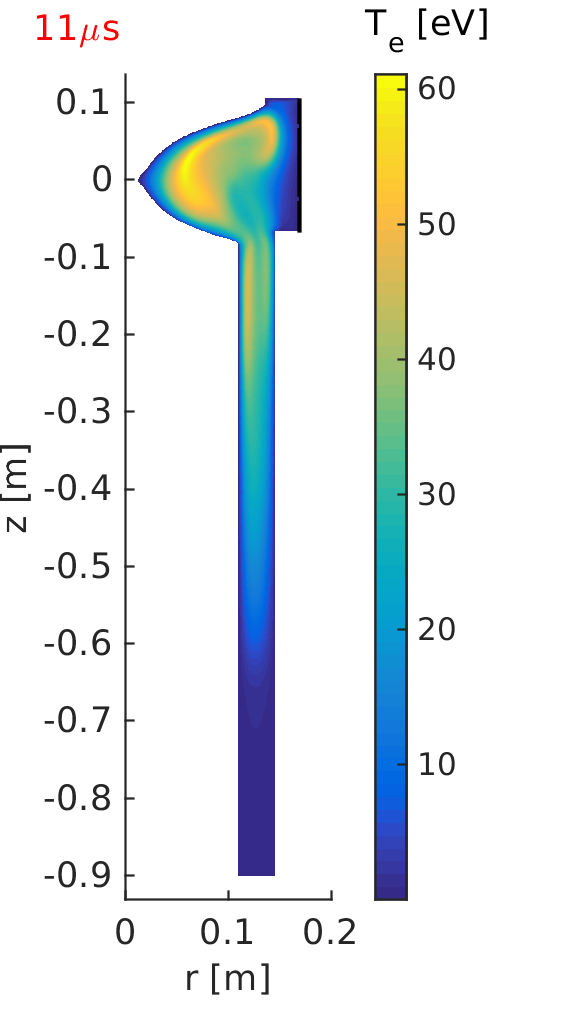}}

\subfloat[]{\raggedright{}\includegraphics[width=7cm,height=5cm]{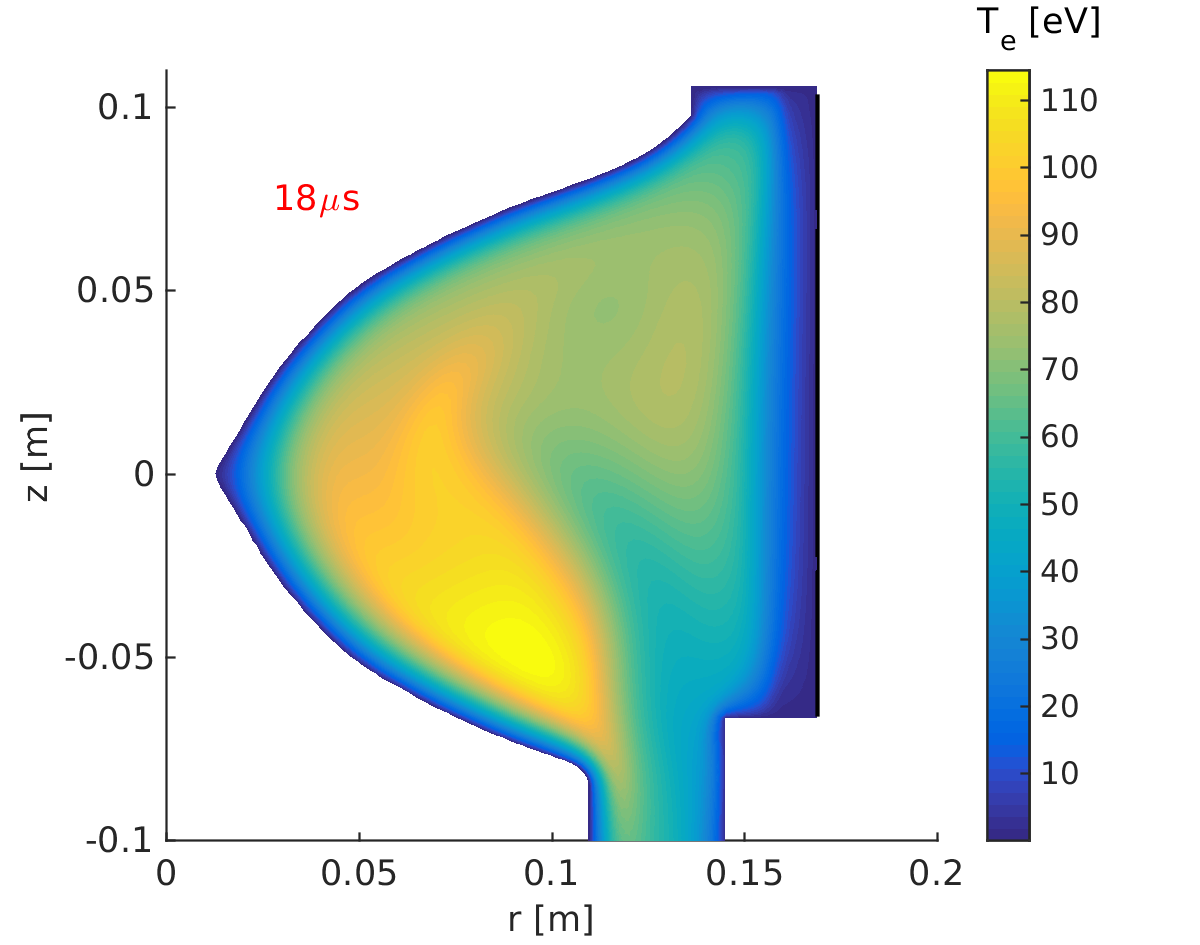}}\hfill{}\subfloat[]{\raggedright{}\includegraphics[width=7cm,height=5cm]{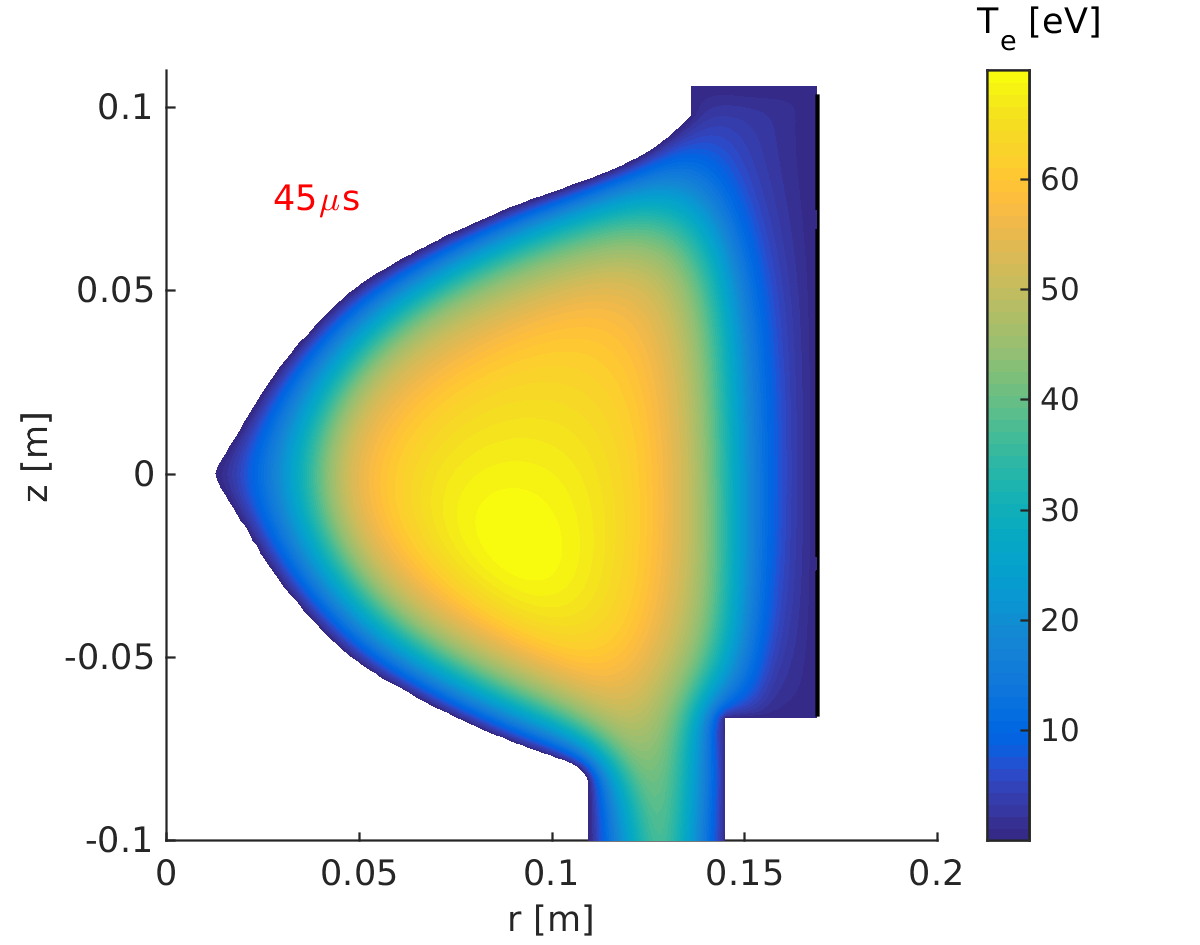}}

\subfloat[]{\raggedright{}\includegraphics[width=7cm,height=5cm]{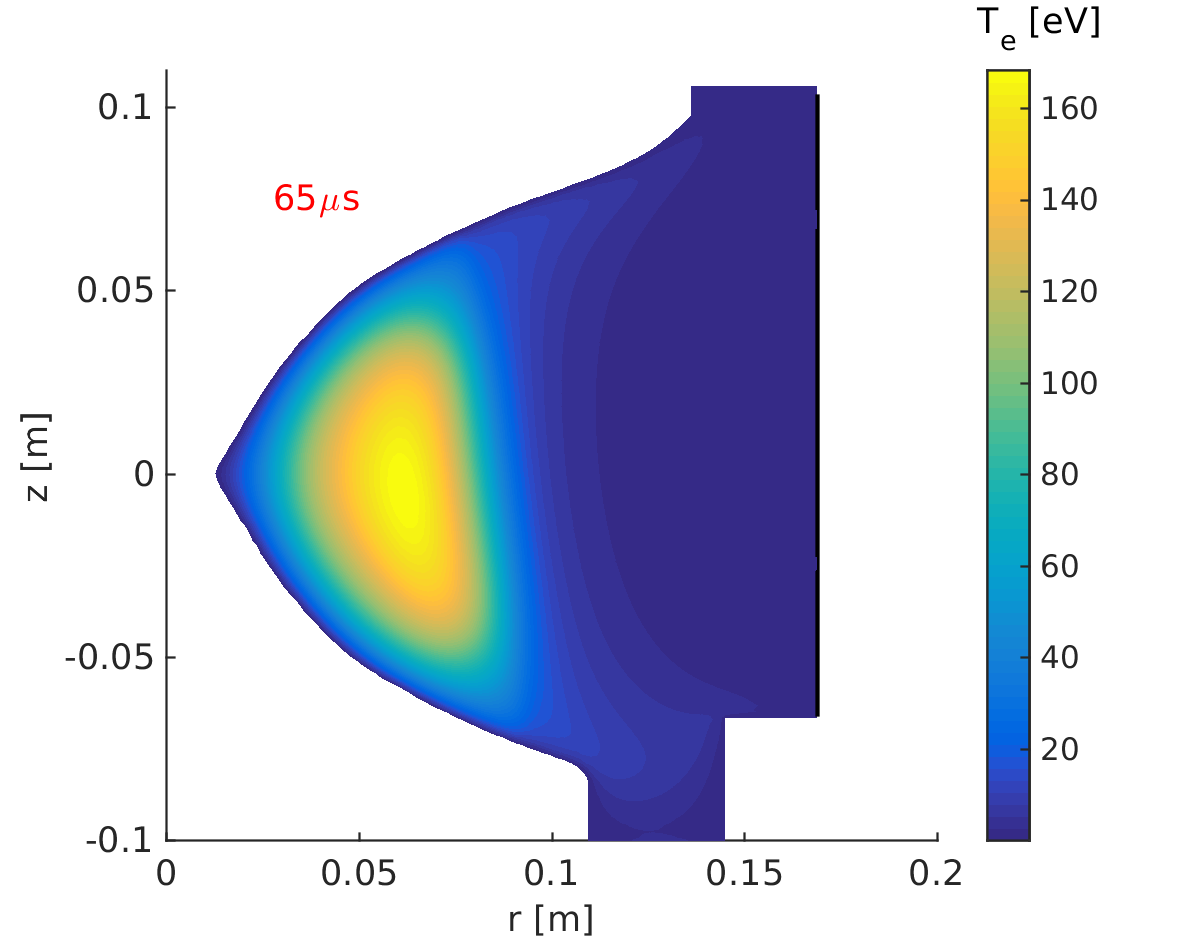}}\hfill{}\subfloat[]{\raggedright{}\includegraphics[width=7cm,height=5cm]{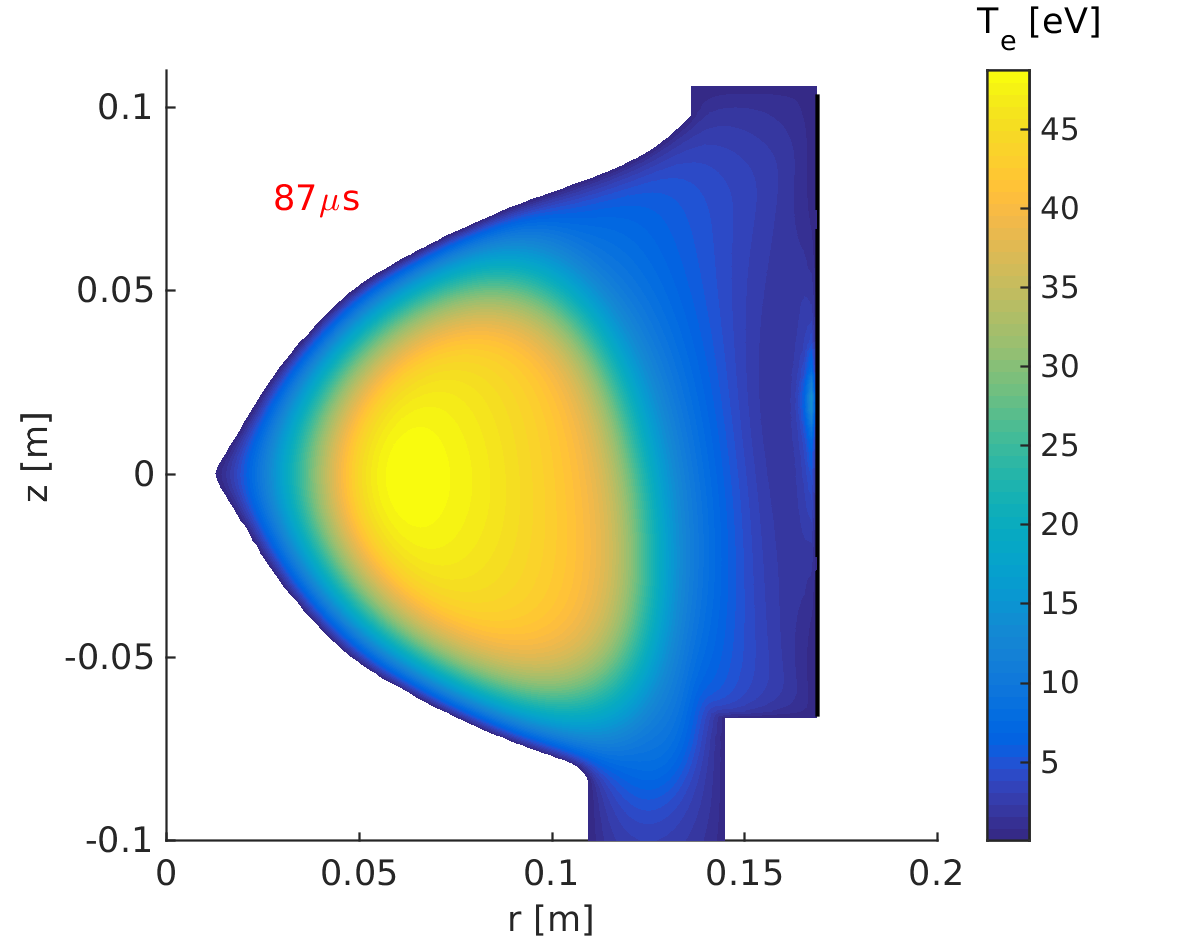}}

\caption{\label{fig: Te_11coils}$\,\,\,\,$Electron temperature profiles }
\end{figure}
\begin{figure}[H]
\subfloat[]{\raggedright{}\includegraphics[scale=0.5]{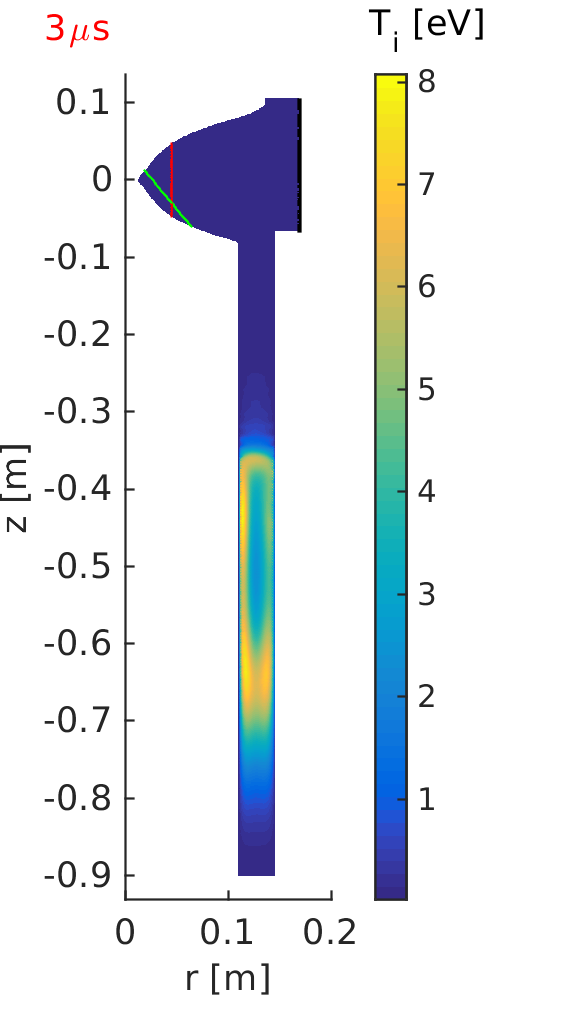}}\hfill{}\subfloat[]{\raggedright{}\includegraphics[scale=0.5]{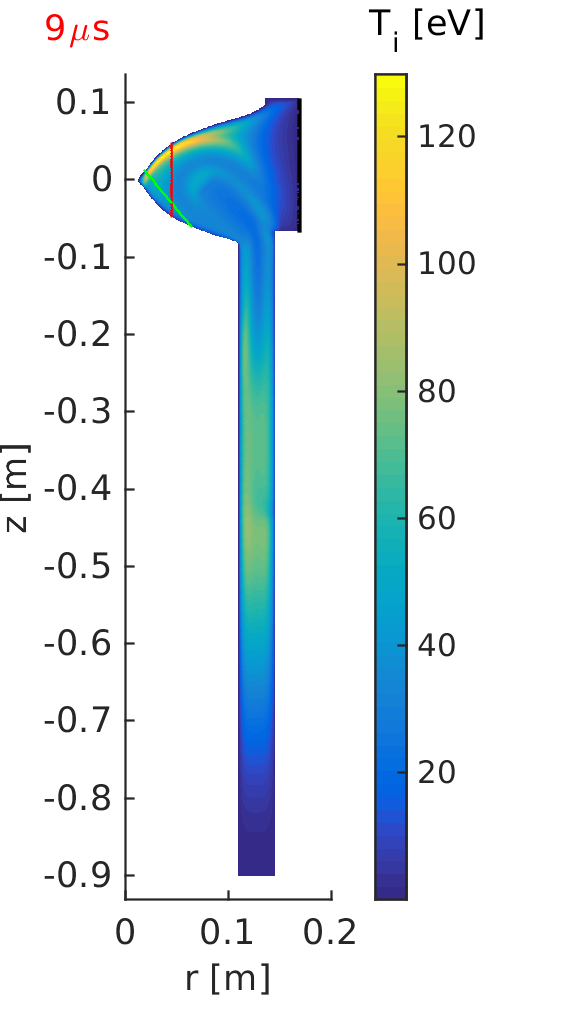}}\hfill{}\subfloat[]{\raggedright{}\includegraphics[scale=0.5]{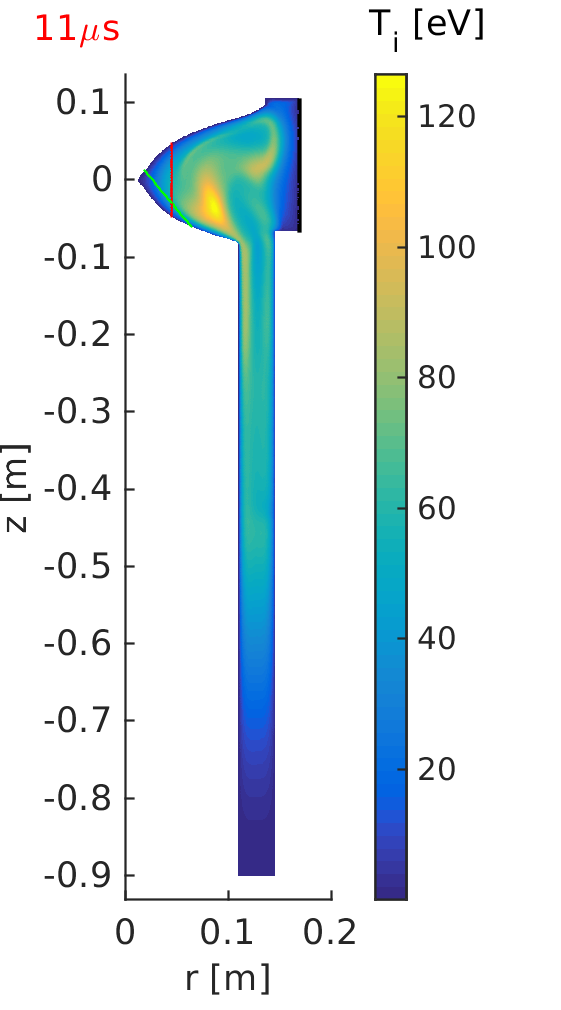}}

\subfloat[]{\raggedright{}\includegraphics[width=7cm,height=5cm]{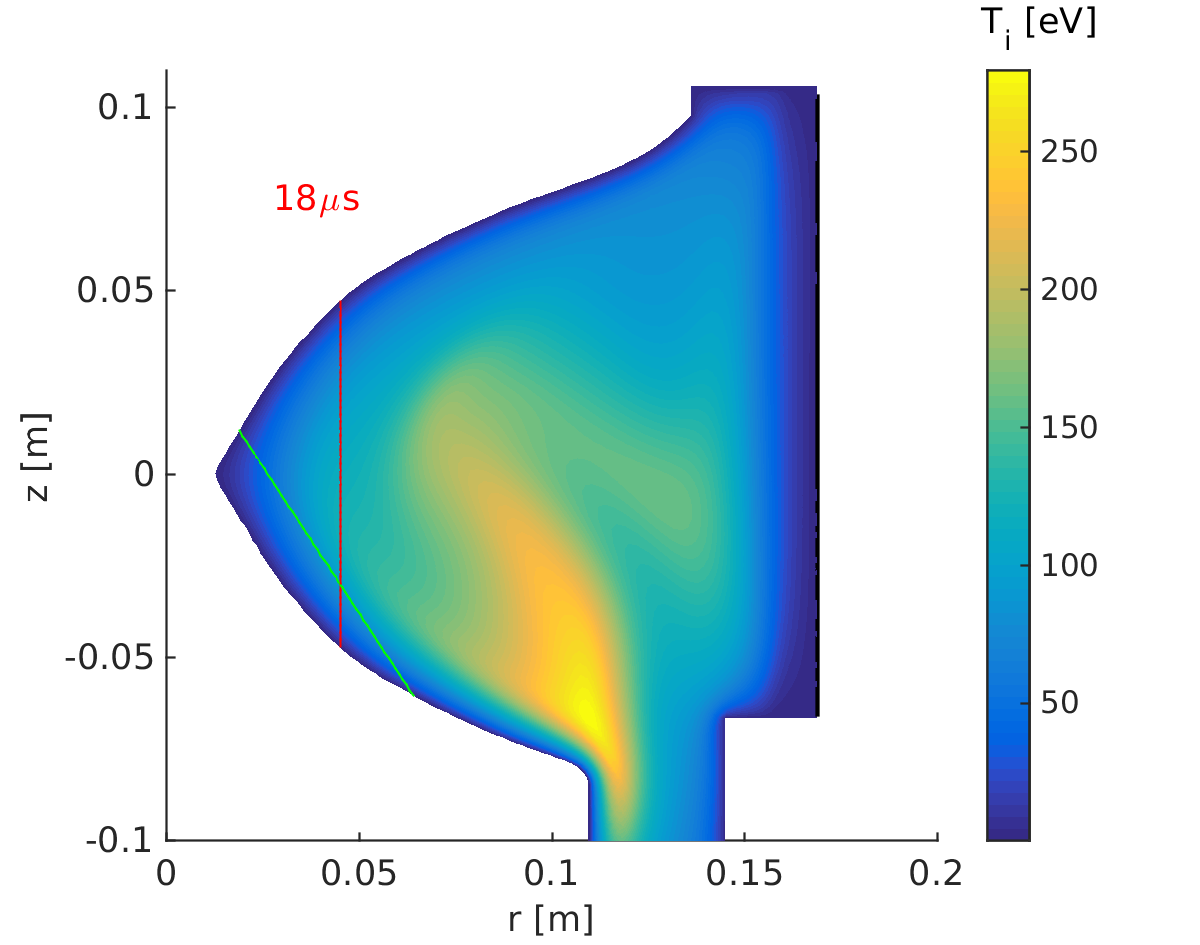}}\hfill{}\subfloat[]{\raggedright{}\includegraphics[width=7cm,height=5cm]{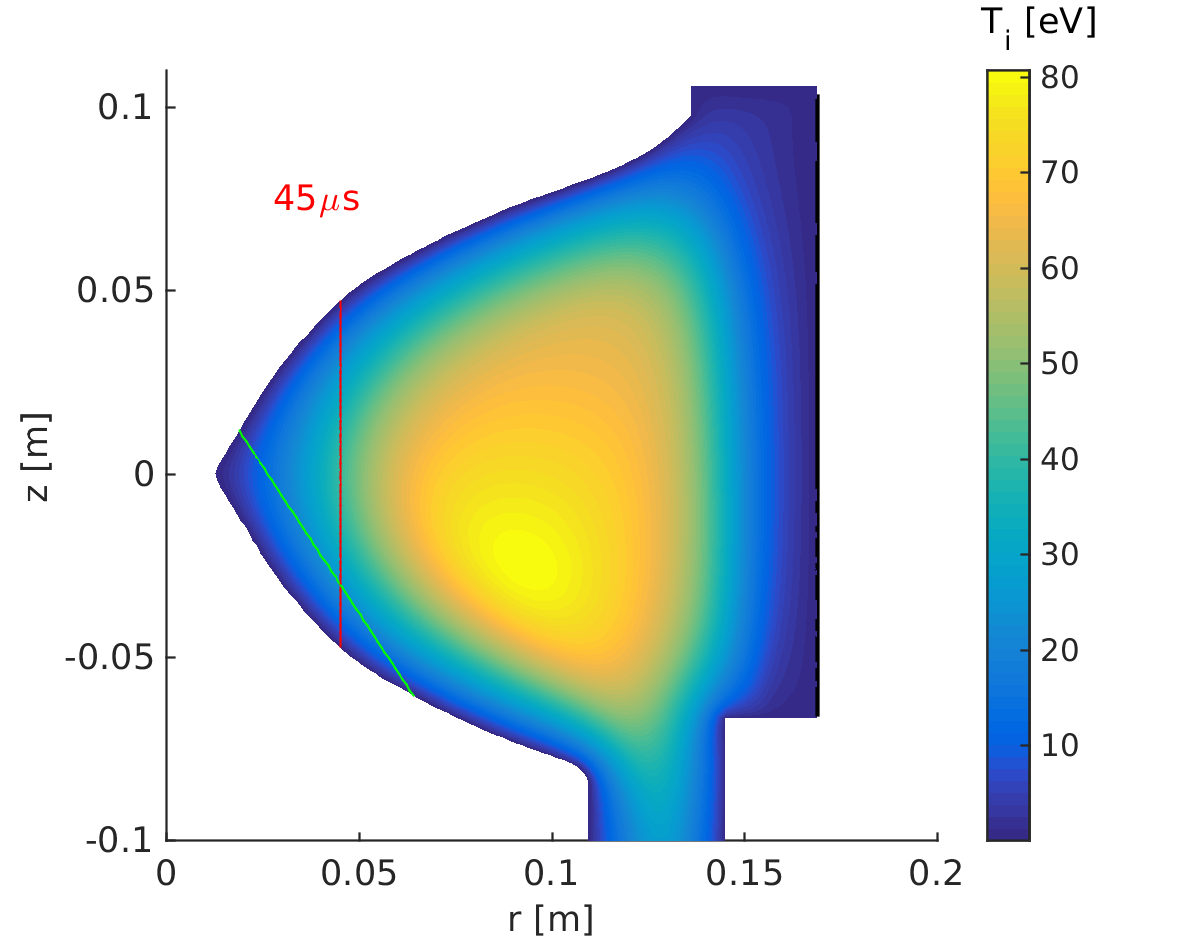}}

\subfloat[]{\raggedright{}\includegraphics[width=7cm,height=5cm]{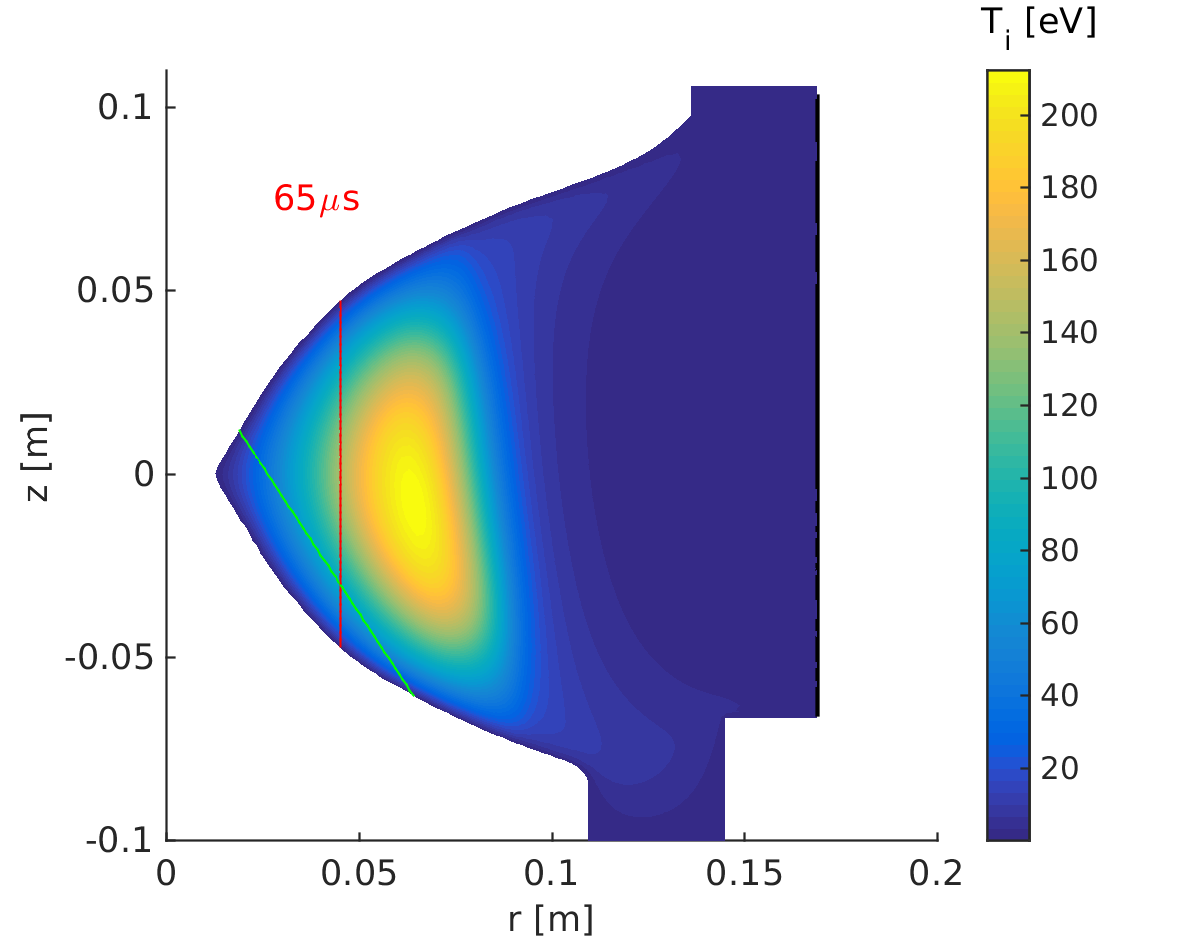}}\hfill{}\subfloat[]{\raggedright{}\includegraphics[width=7cm,height=5cm]{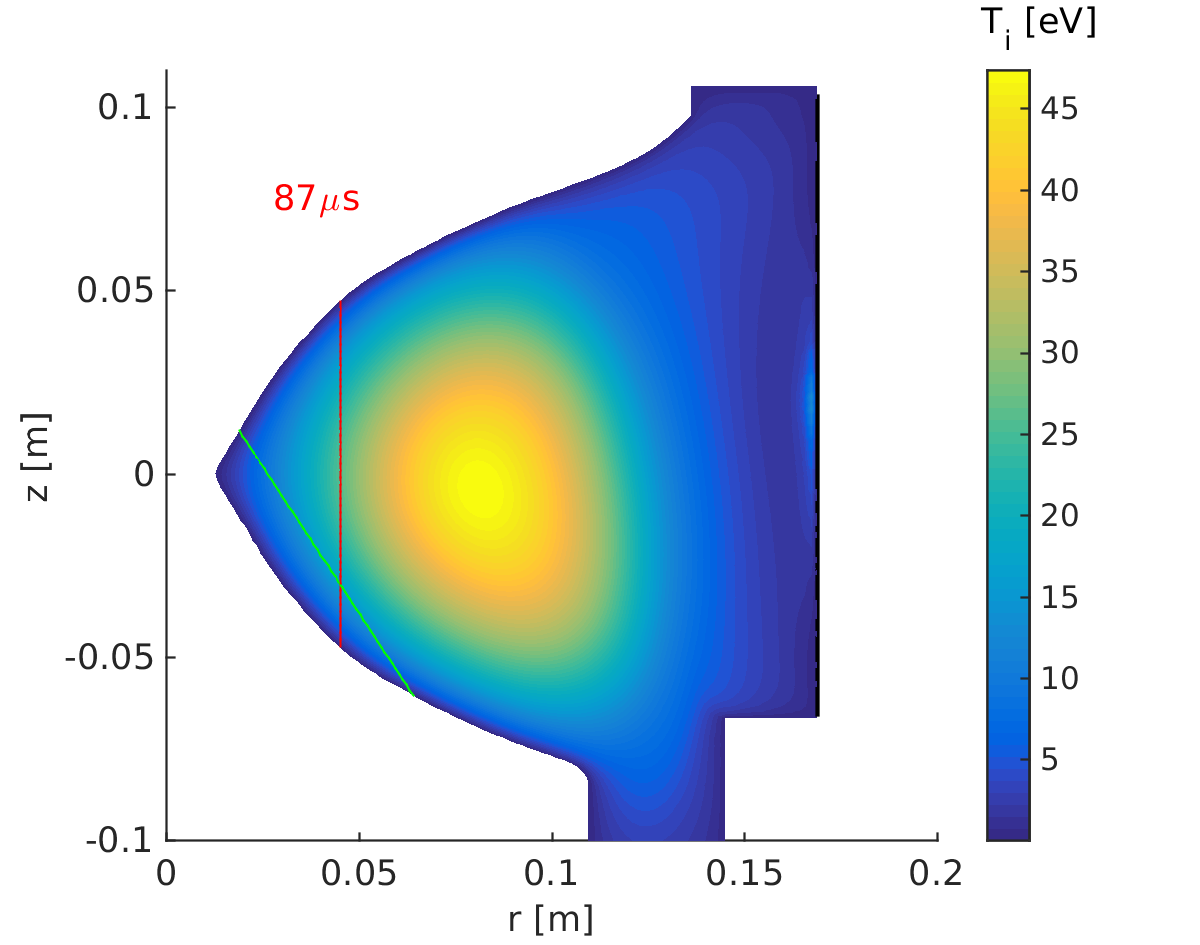}}

\caption{\label{fig: Ti_11coils}$\,\,\,\,$Ion temperature profiles }
\end{figure}
\begin{figure}[H]
\subfloat[]{\raggedright{}\includegraphics[scale=0.5]{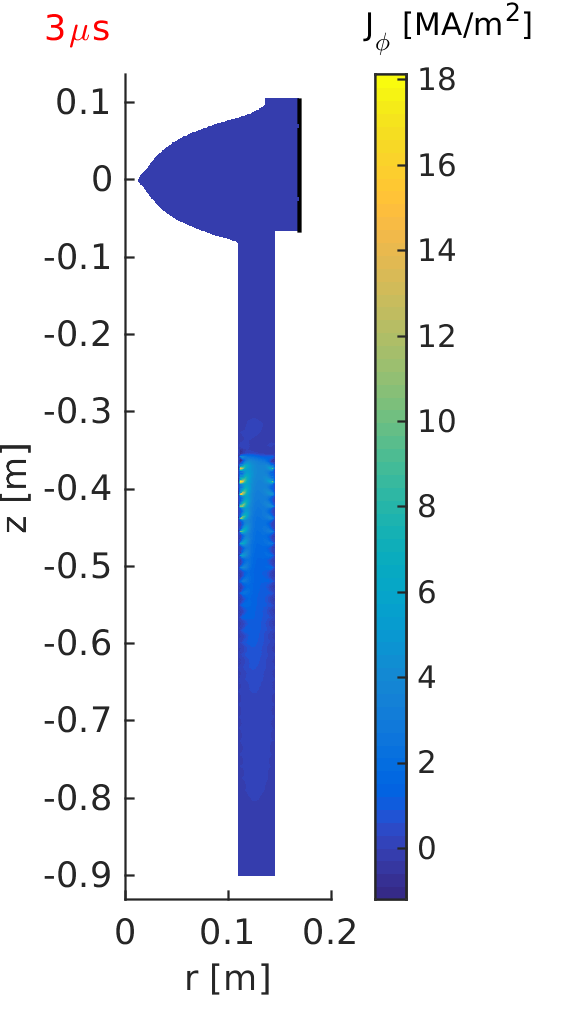}}\hfill{}\subfloat[]{\raggedright{}\includegraphics[scale=0.5]{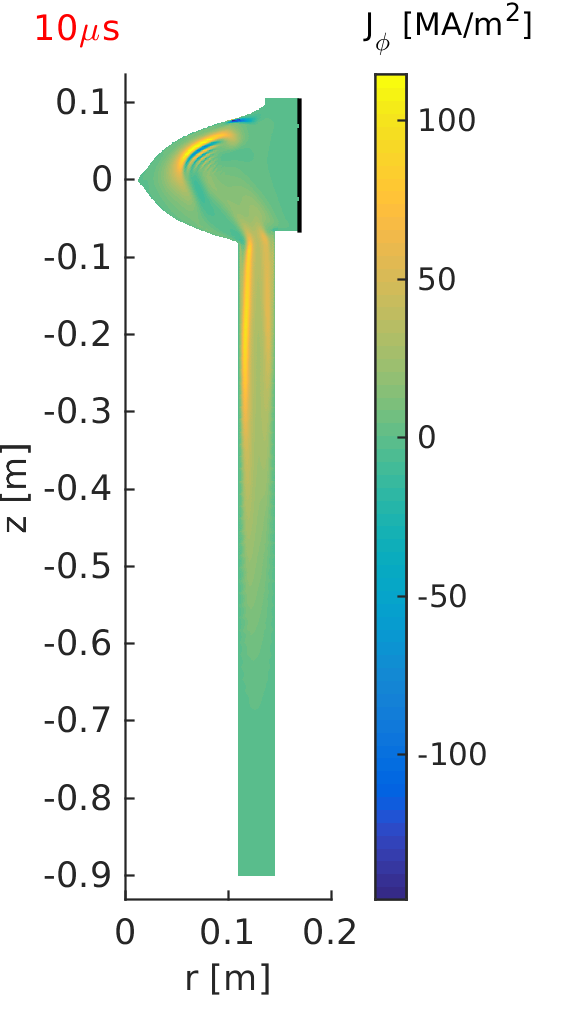}}\hfill{}\subfloat[]{\raggedright{}\includegraphics[scale=0.5]{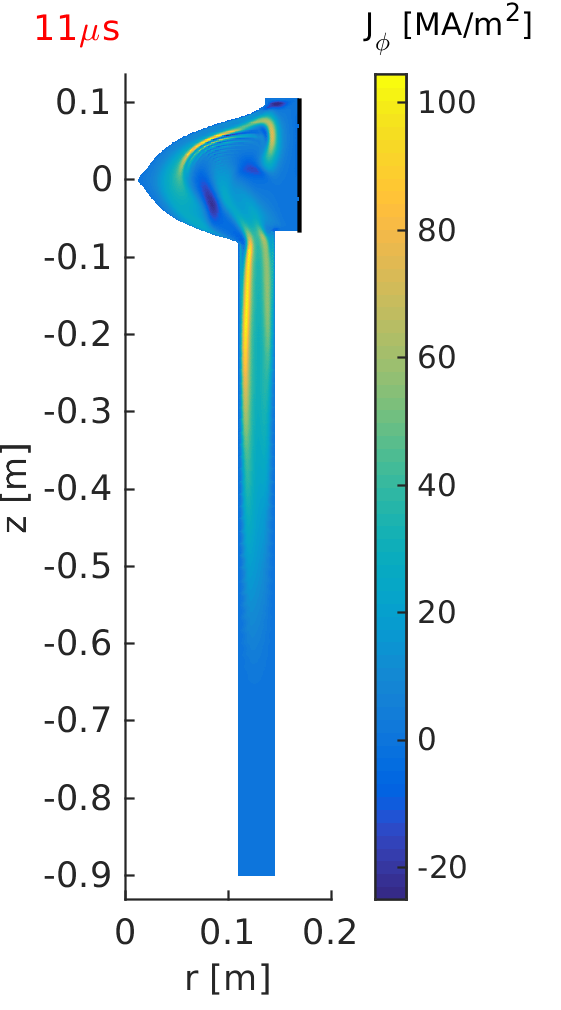}}

\subfloat[]{\raggedright{}\includegraphics[width=7cm,height=5cm]{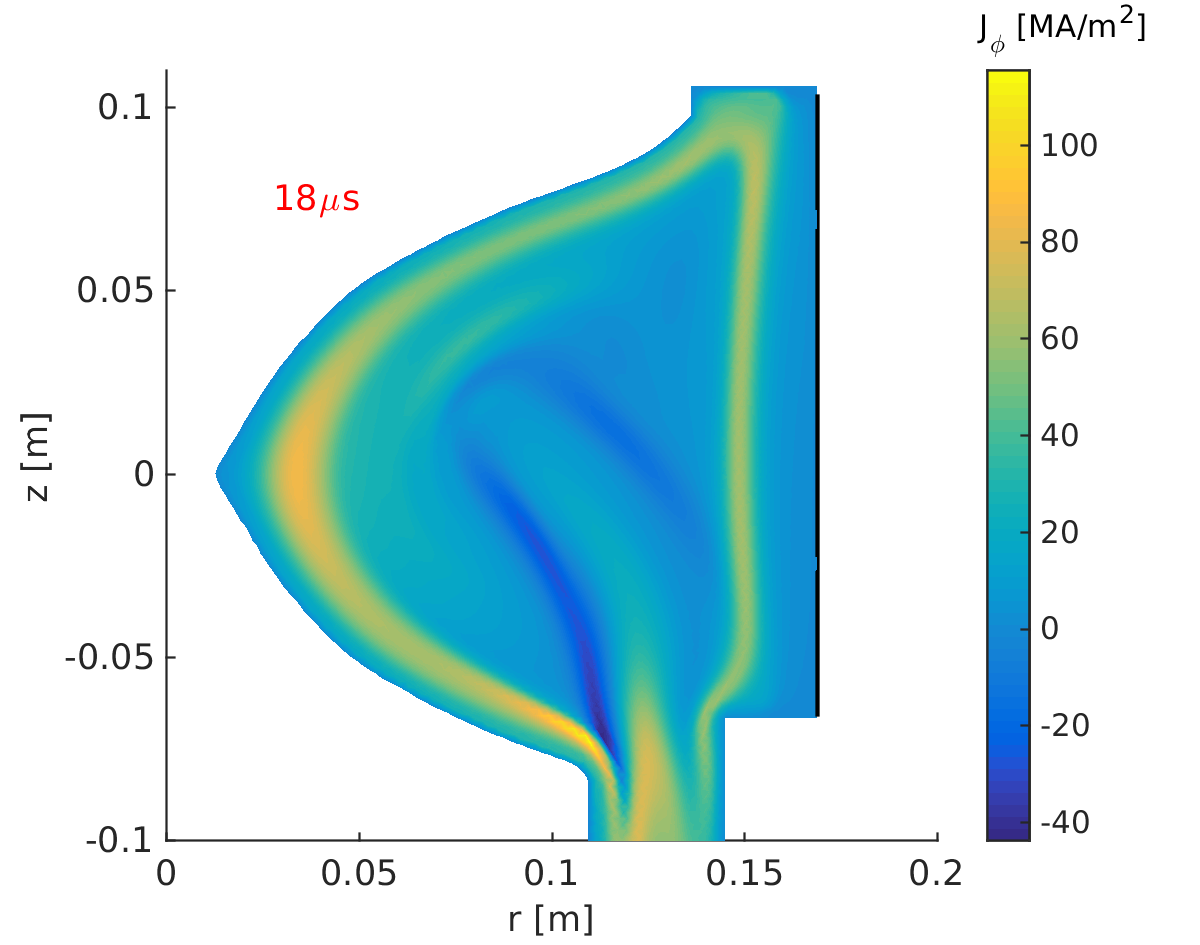}}\hfill{}\subfloat[]{\raggedright{}\includegraphics[width=7cm,height=5cm]{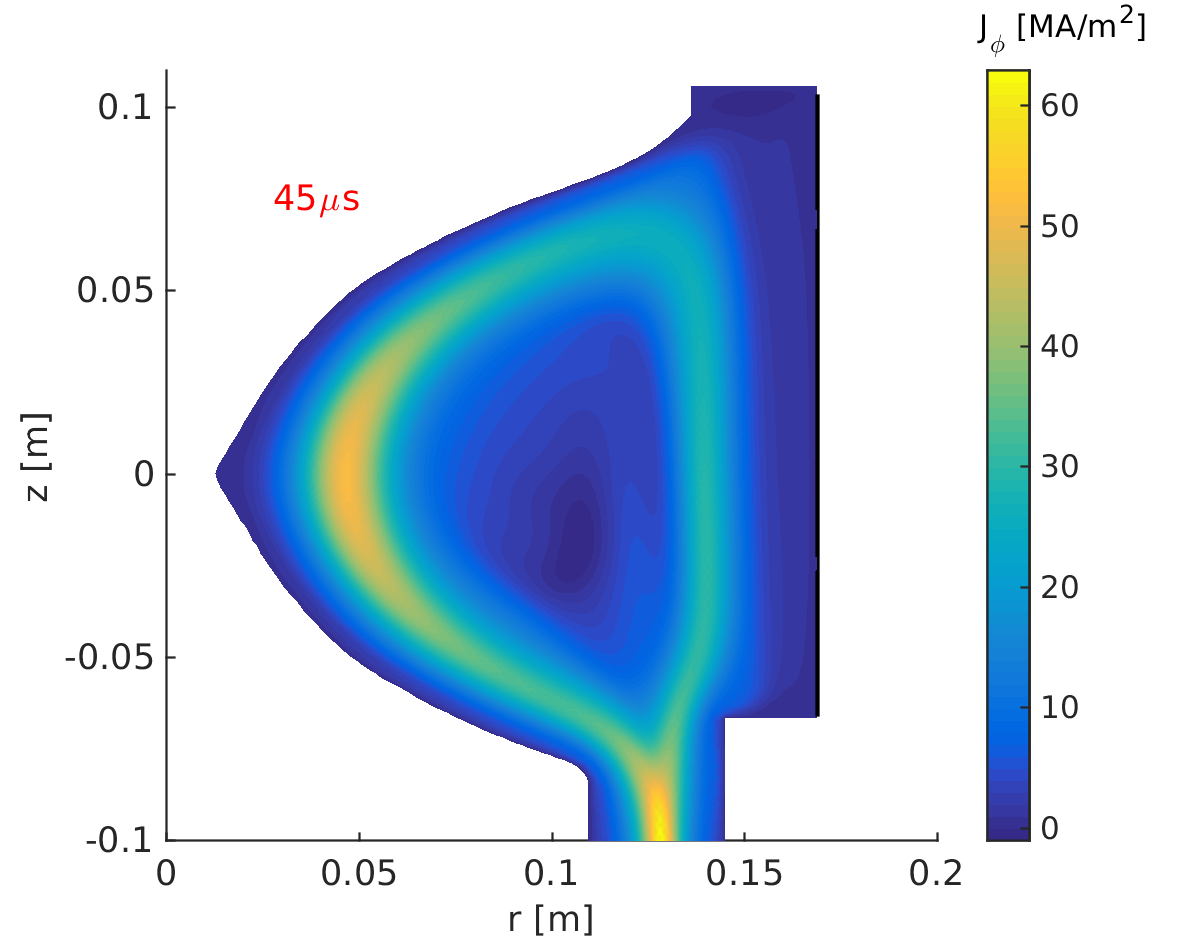}}

\subfloat[]{\raggedright{}\includegraphics[width=7cm,height=5cm]{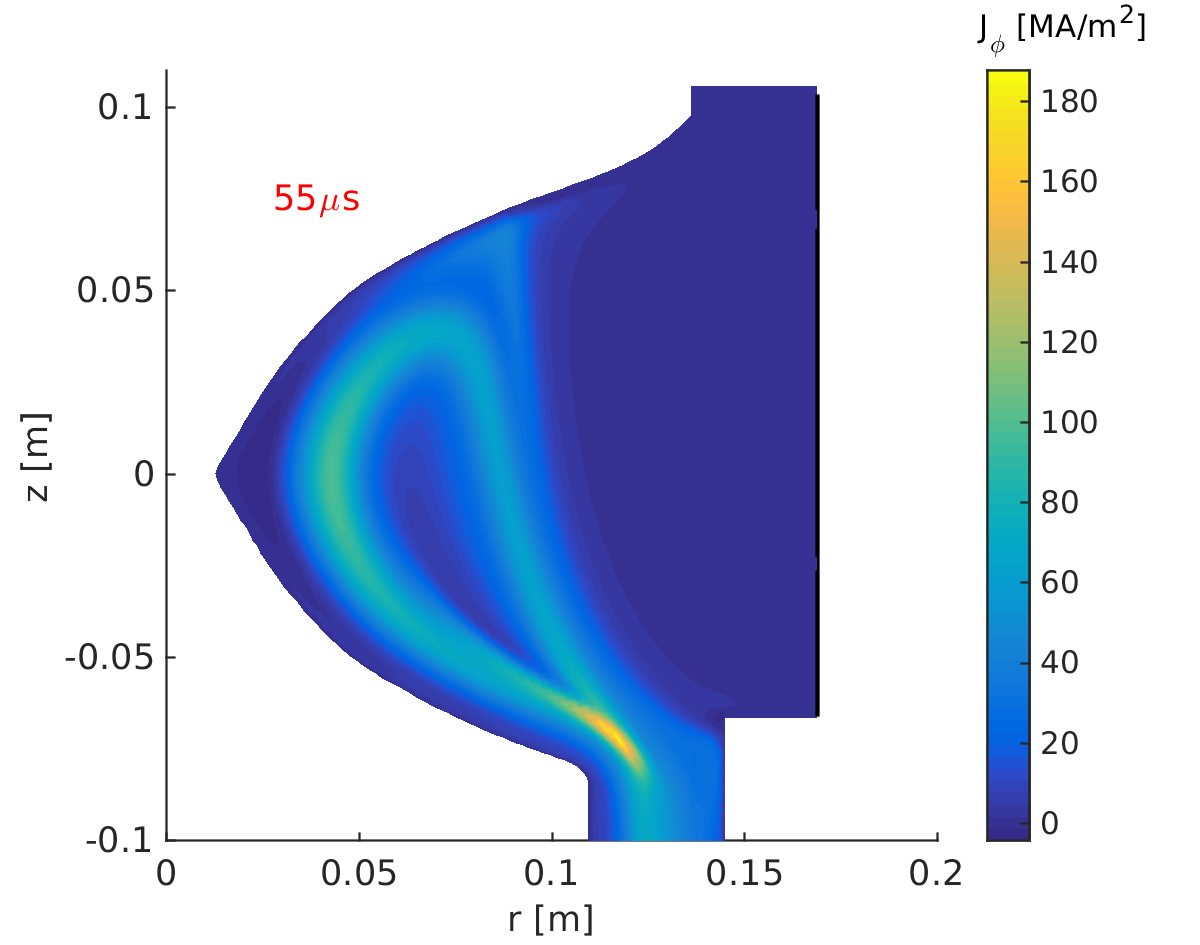}}\hfill{}\subfloat[]{\raggedright{}\includegraphics[width=7cm,height=5cm]{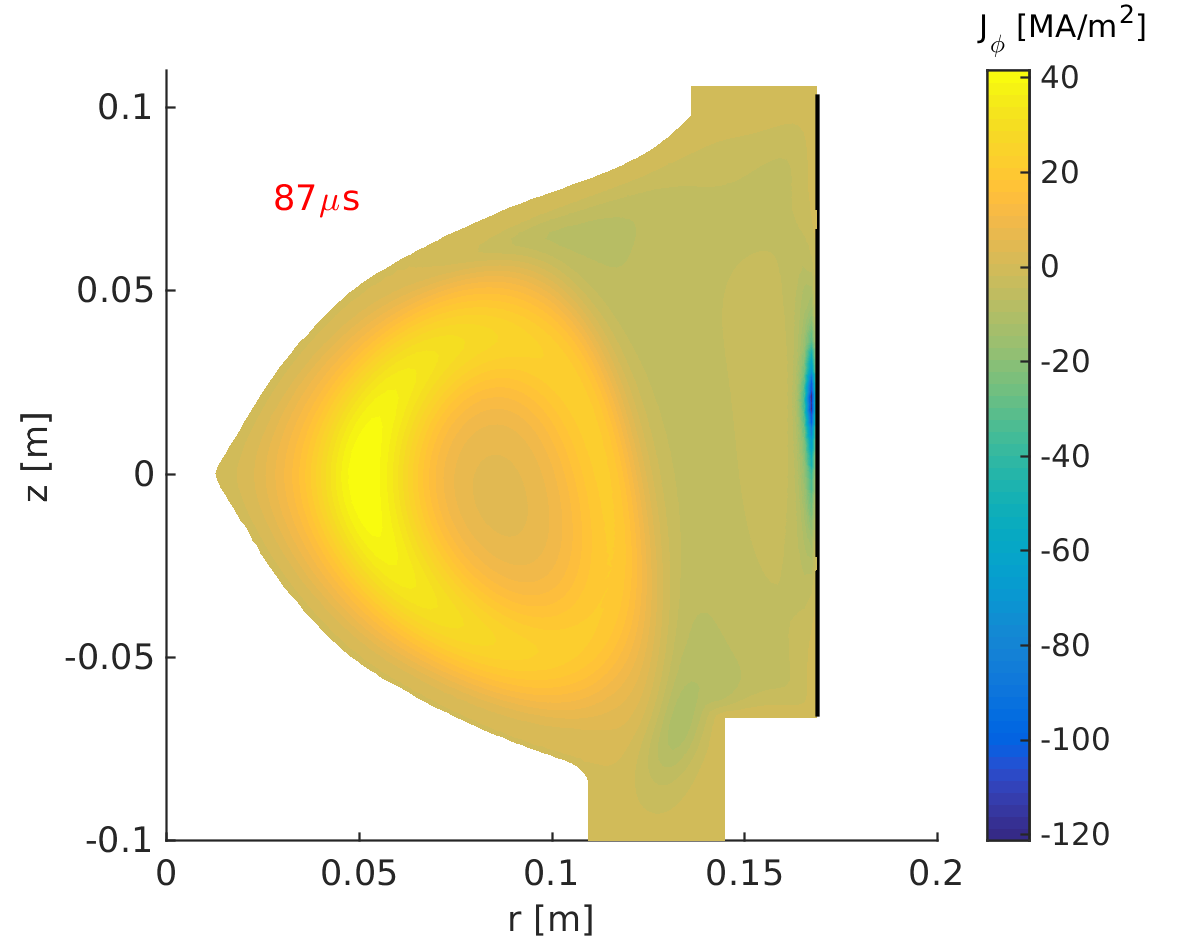}}

\caption{\label{fig: Jp_11coils}$\,\,\,\,$Toroidal current density profiles}
\end{figure}
\begin{figure}[H]
\subfloat[]{\raggedright{}\includegraphics[scale=0.5]{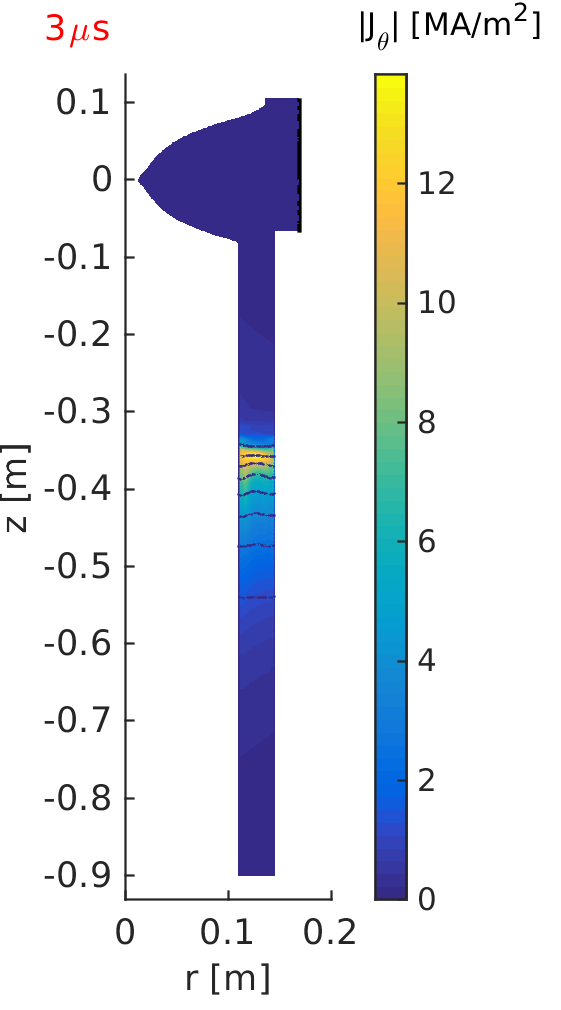}}\hfill{}\subfloat[]{\raggedright{}\includegraphics[scale=0.5]{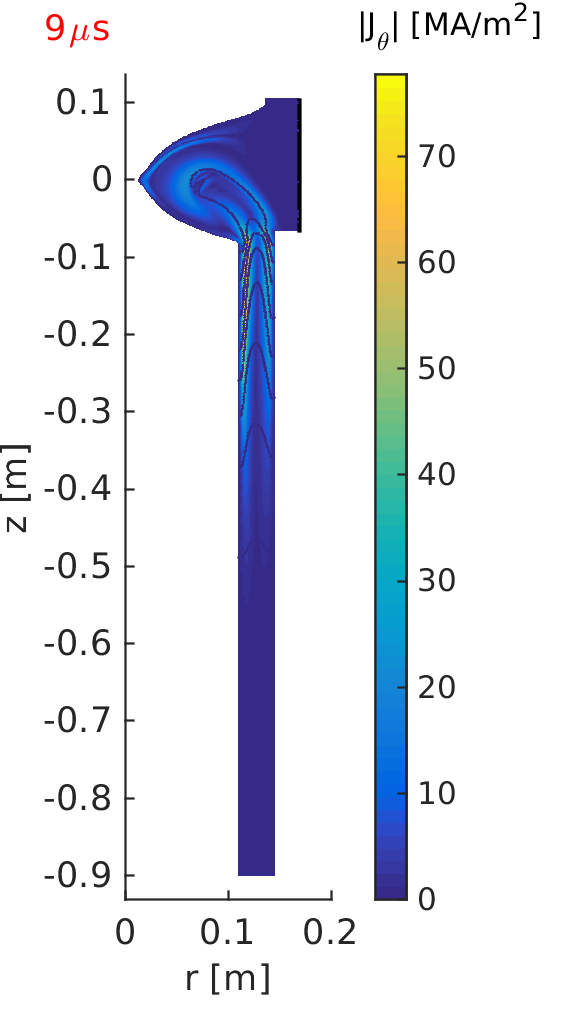}}\hfill{}\subfloat[]{\raggedright{}\includegraphics[scale=0.5]{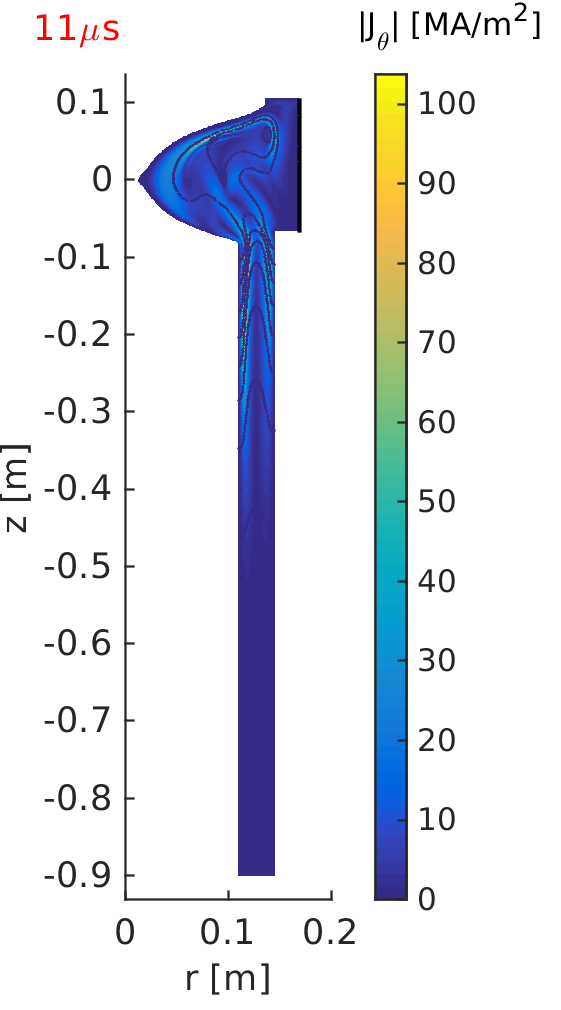}}

\subfloat[]{\raggedright{}\includegraphics[width=7cm,height=5cm]{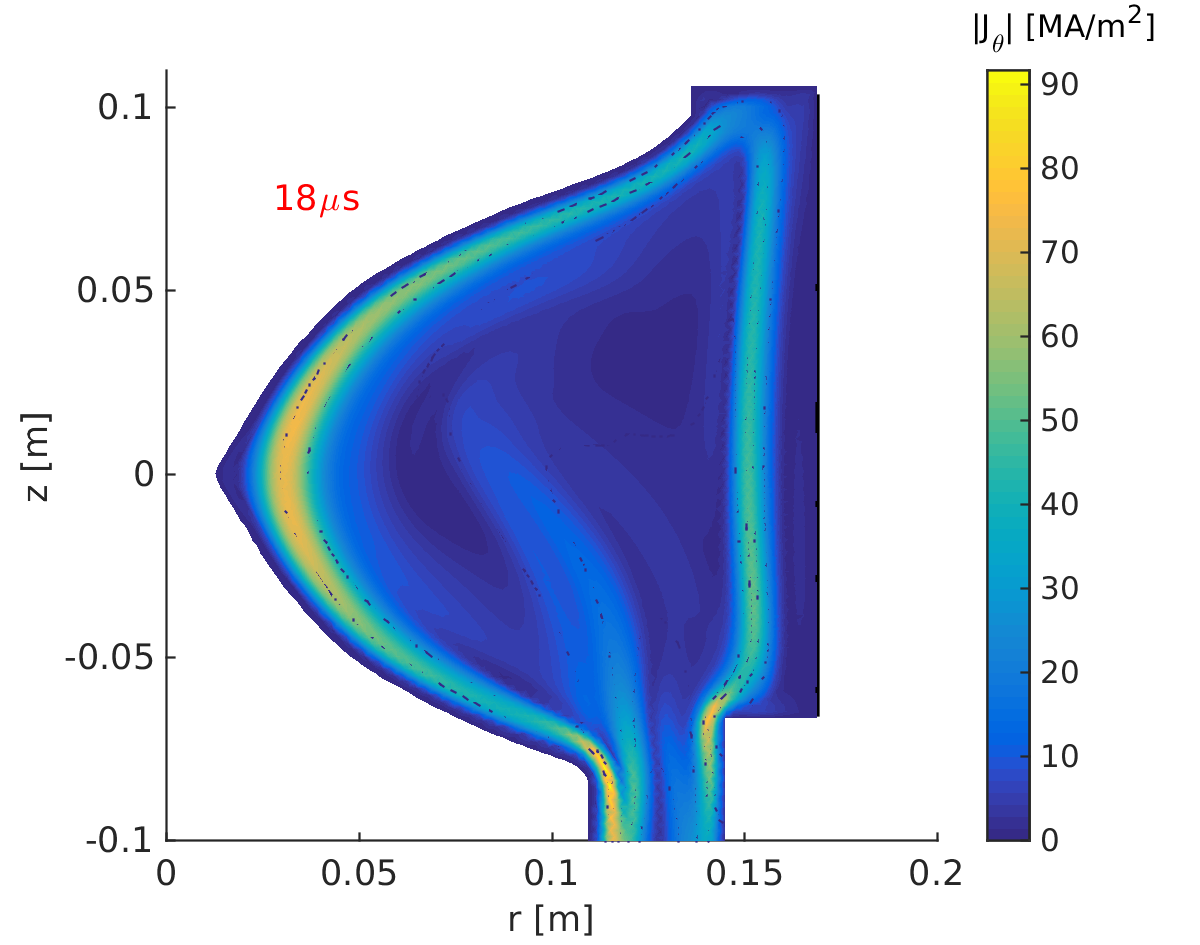}}\hfill{}\subfloat[]{\raggedright{}\includegraphics[width=7cm,height=5cm]{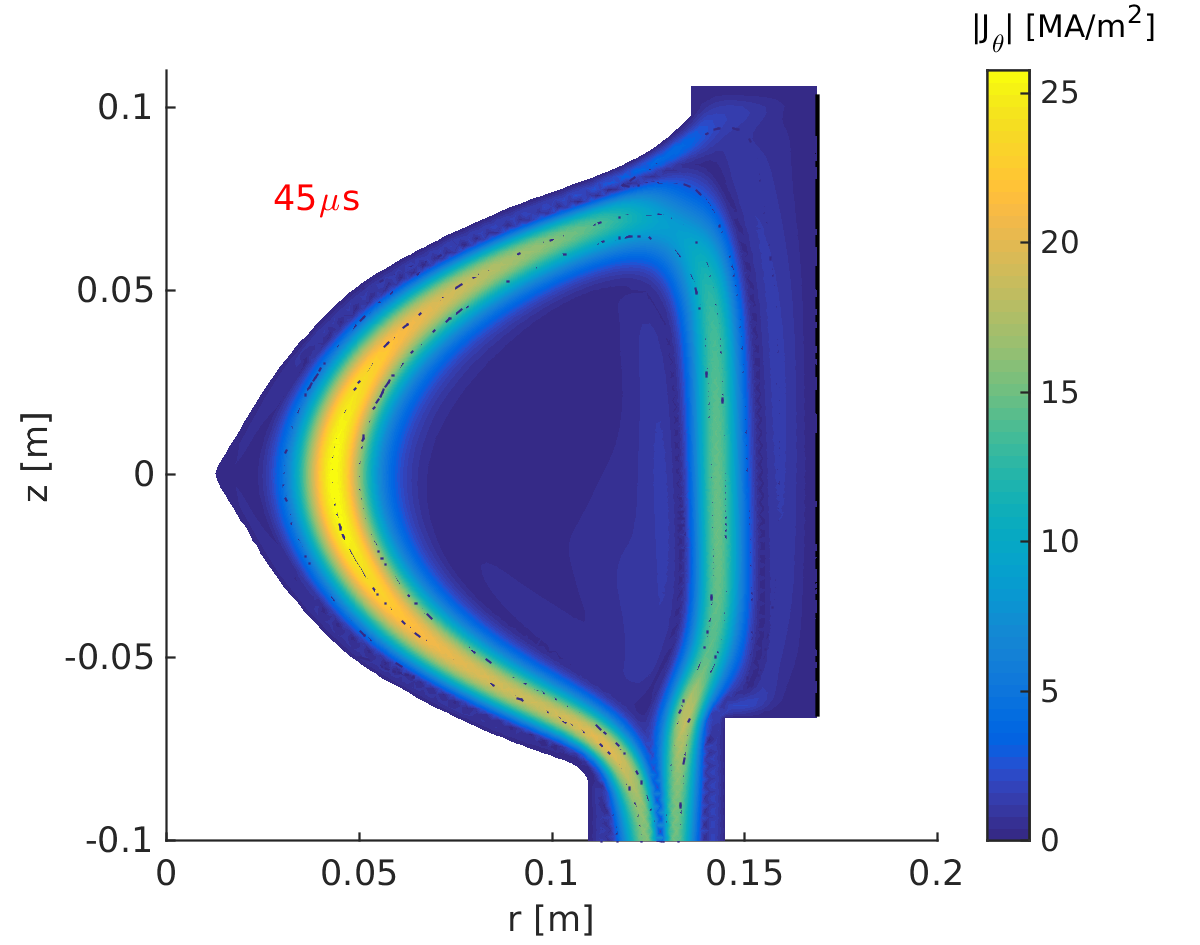}}

\subfloat[]{\raggedright{}\includegraphics[width=7cm,height=5cm]{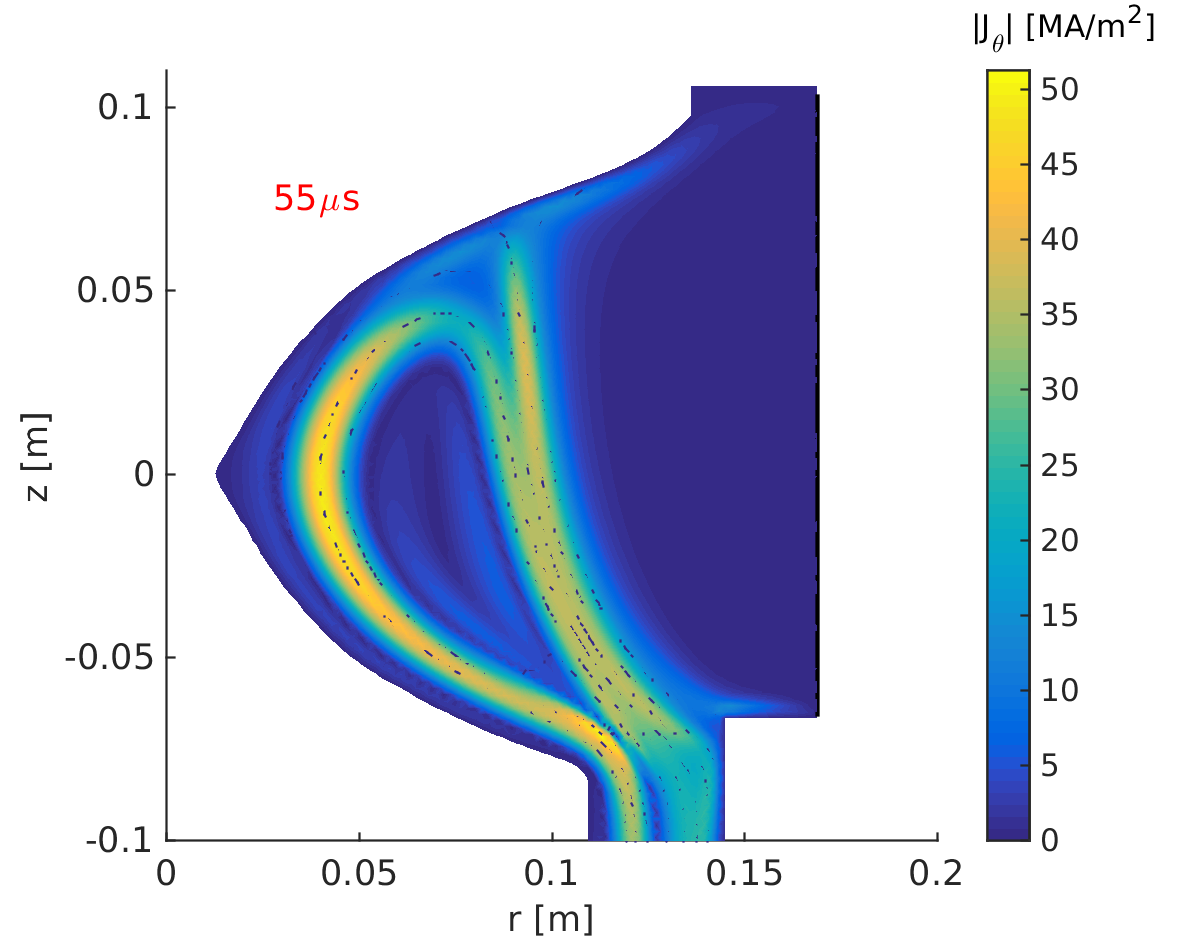}}\hfill{}\subfloat[]{\raggedright{}\includegraphics[width=7cm,height=5cm]{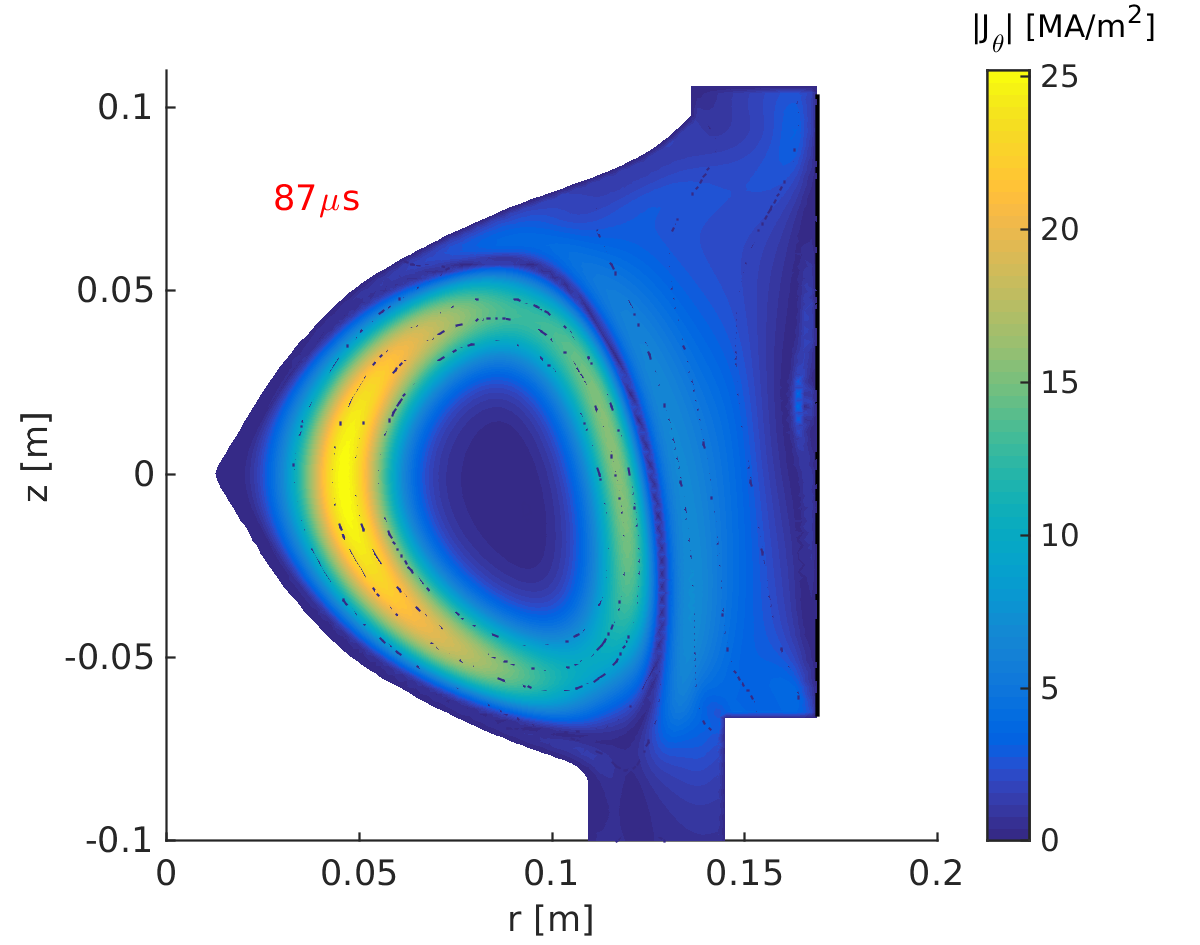}}

\caption{\label{fig: Jpol_11coils}$\,\,\,\,$Poloidal current density profiles }
\end{figure}

Formation current ohmically heats initially cold electrons as plasma
is pushed up the gun into the containment region (figures \ref{fig: Te_11coils}(a)
to (c)). The temperature of $115$ eV attained by electrons near the
containment region entrance at $18\,\upmu$s (figure \ref{fig: Te_11coils}(d))
is partially due to heat exchange with ions, which, as indicated in
figure \ref{fig: Ti_11coils}(d)), have been heated to around 280
eV near the same area at this time due to viscous heating. Thermal
diffusion through the boundary causes the electron temperature around
the CT core to be reduced to around 70 eV just prior to magnetic compression
at 45$\,\upmu$s. Electron temperature is more than doubled, to 165
eV, at peak magnetic compression at 65$\,\upmu$s. Comparing with
figures \ref{fig: Jp_11coils} and \ref{fig: Jpol_11coils}, it can
be seen how ohmic heating is a principal electron heating mechanism.
Compressional heating is the main heating mechanism during magnetic
compression, and is supplemented by enhanced ohmic heating. When compression
time is short compared with the plasma resistive diffusion time, the
adiabatic compression relationship of constant $pV^{\gamma}$ determines
compressional heating, and temperature should scale approximately
as $V^{-\frac{2}{3}}$, as in table \ref{tab:Scaling-parameters-for}.
As discussed in section \ref{subsec:SimDiagCompression-scalings},
electron and ion temperatures at compression increase more, while
density increases less, than the predicted increases based on the
adiabatic scalings, due to the inclusion of artificial density diffusion.
Compression coil current falls to zero by around 87$\,\upmu$s, by
which time the CT has re-expanded, and core electron temperature has
fallen to around 50 eV.

Figure \ref{fig: Ti_11coils} indicates simulated ion temperatures
at the same times referred to in figure \ref{fig: Te_11coils} for
electron temperatures. Over the early stages of the simulation, ion
temperature is significantly higher than electron temperature, due
to ion viscous heating during the formation process. Later, ohmic
heating of electrons is the main heating source and electron temperature
approaches ion temperature. Note that, as a result of compressional
heating in combination with heat exchange with the electrons, which
are heated ohmically by enhanced currents during compression, ion
temperature is more than doubled at peak compression, increasing from
around 80 to 210 eV from 45$\,\upmu$s to 65$\,\upmu$s. The vertical
red chord and diagonal green chord in the CT containment region in
figures \ref{fig: Ti_11coils}(a) to (g) represent the lines of sight
of the ion-Doppler measurements ($cf.$ figure \ref{fig:Chalice}),
along which simulated line-averaged ion temperature is evaluated for
comparison with experimental data. 
Figures \ref{fig: Jp_11coils} and \ref{fig: Jpol_11coils} shows
simulated toroidal and poloidal current density at various times.
The absolute value of poloidal current density, calculated as $|J_{\theta}|=\sqrt{J_{r}^{2}+J_{z}^{2}}$,
where $J_{r}$ and $J_{z}$ are found in terms of gradients of $f$
according to equation \ref{eq:20.2}, is plotted under contours of
$f$ in figure \ref{fig: Jpol_11coils}. Poloidal current is directed
from right to left in the gun barrel, and anticlockwise in the CT
containment region. Comparing with figure \ref{fig: Te_11coils},
it can be seen how ohmic heating is a principal electron heating mechanism.
Note the high concentration of $J_{\phi}$ at 55$\,\upmu$s (figure
\ref{fig: Jp_11coils}(f)), relating to the toroidally directed current
sheet present between oppositely directed poloidal field lines during
the magnetic reconnection process that occurs as closed CT field lines
that extend down the gun, and then open field lines surrounding the
CT, are pinched off during magnetic compression. A reconnection-related
current sheet is also evident when open poloidal field lines reconnect
to form closed CT flux surfaces, for example at 18$\,\upmu$s (figure
\ref{fig: Jp_11coils}(d)). It can be seen how the current density
profiles are remarkably hollow; as mentioned previously, 2D simulations
 generally overestimate the level of hollowness of the field profiles.
\begin{figure}[H]
\subfloat[]{\raggedright{}\includegraphics[scale=0.5]{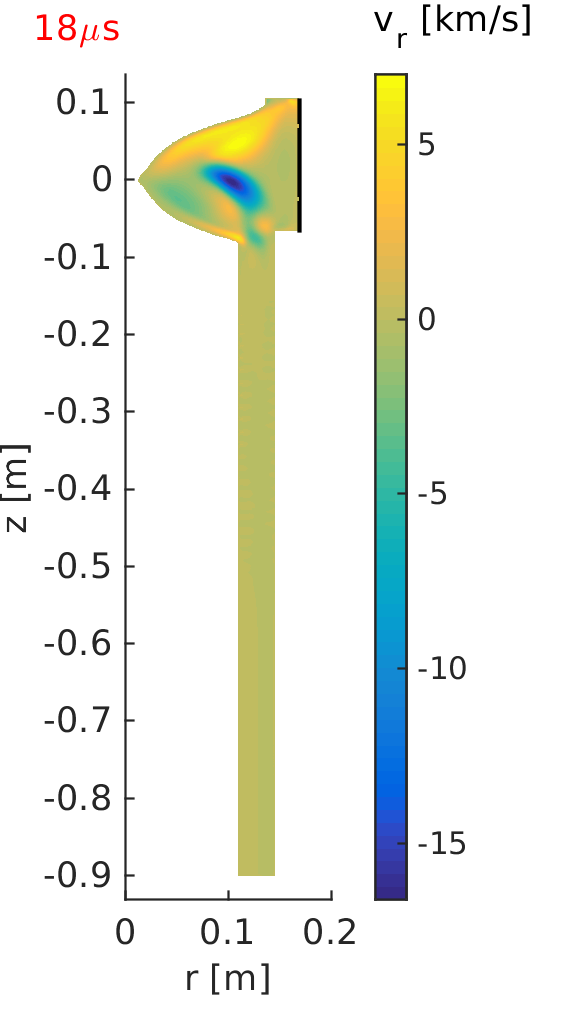}}\hfill{}\subfloat[]{\raggedright{}\includegraphics[scale=0.5]{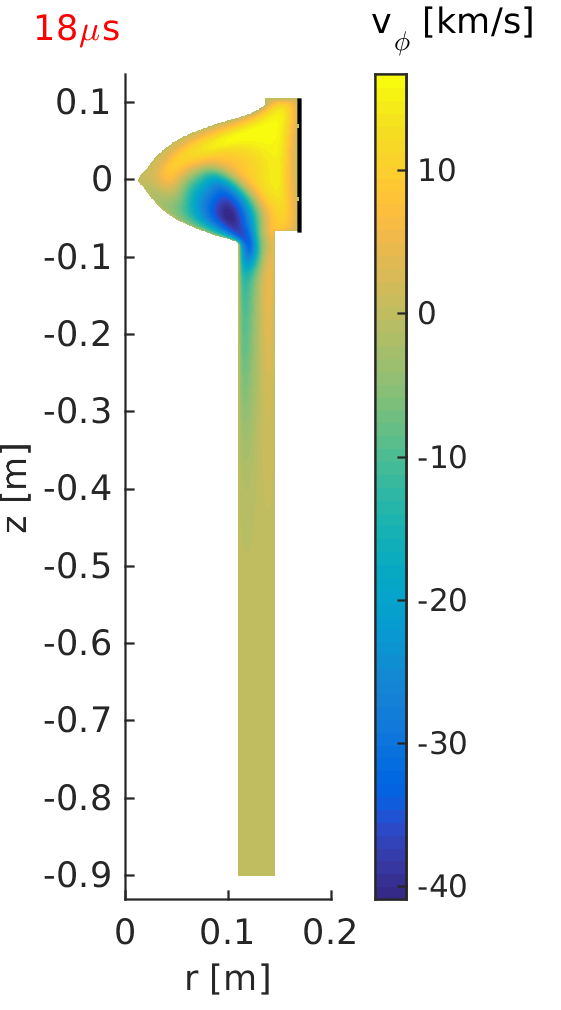}}\hfill{}\subfloat[]{\raggedright{}\includegraphics[scale=0.5]{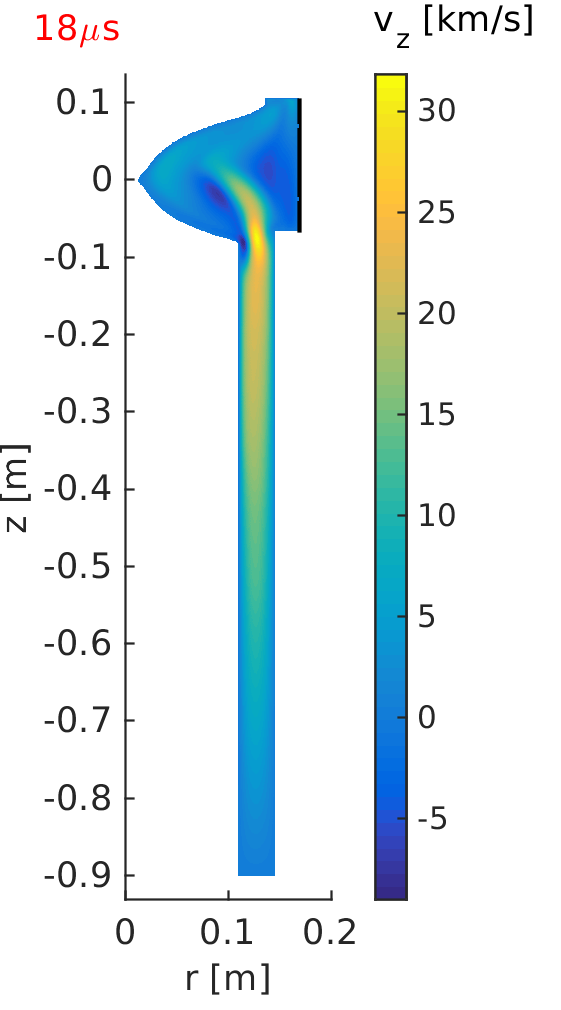}}

\subfloat[]{\raggedright{}\includegraphics[width=7cm,height=5cm]{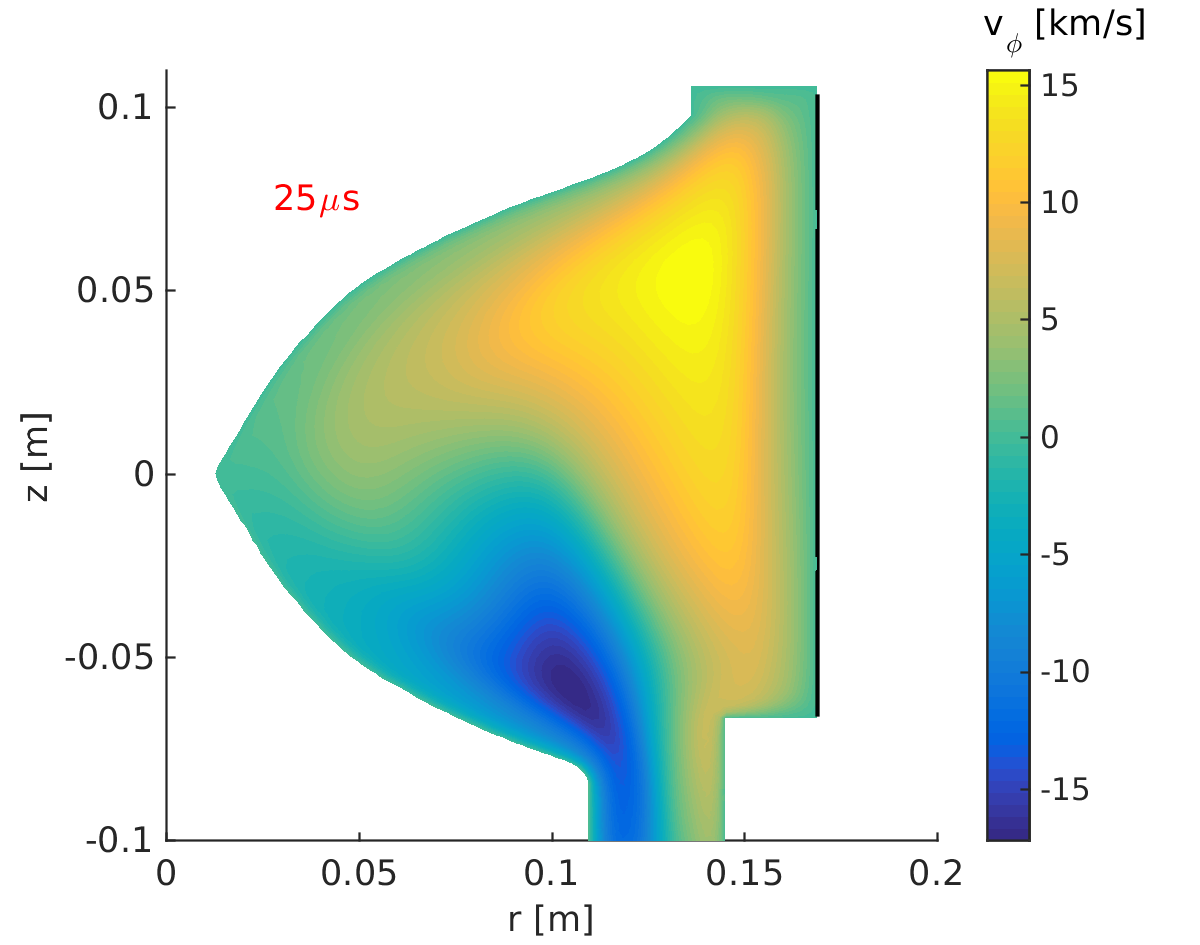}}\hfill{}\subfloat[]{\raggedright{}\includegraphics[width=7cm,height=5cm]{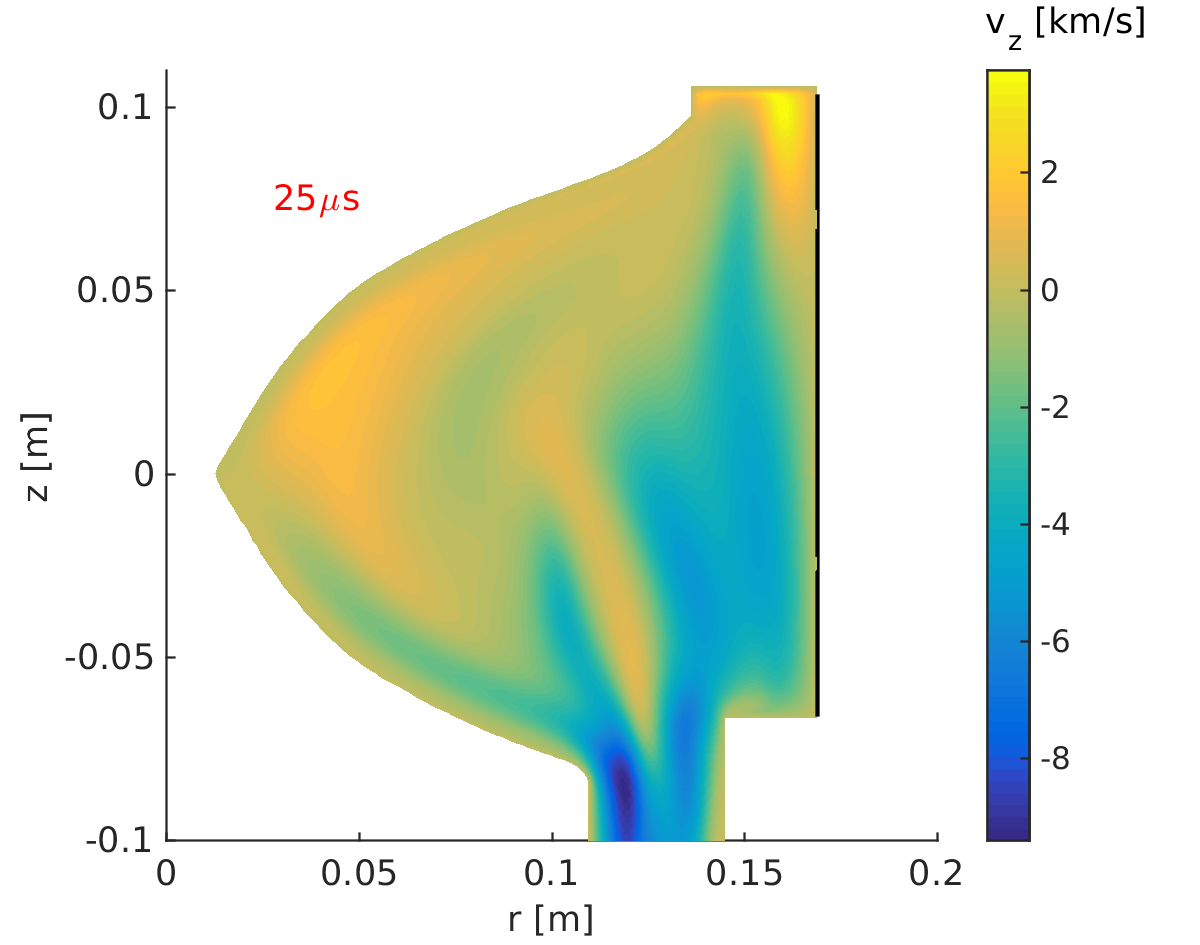}}

\subfloat[]{\raggedright{}\includegraphics[width=7cm,height=5cm]{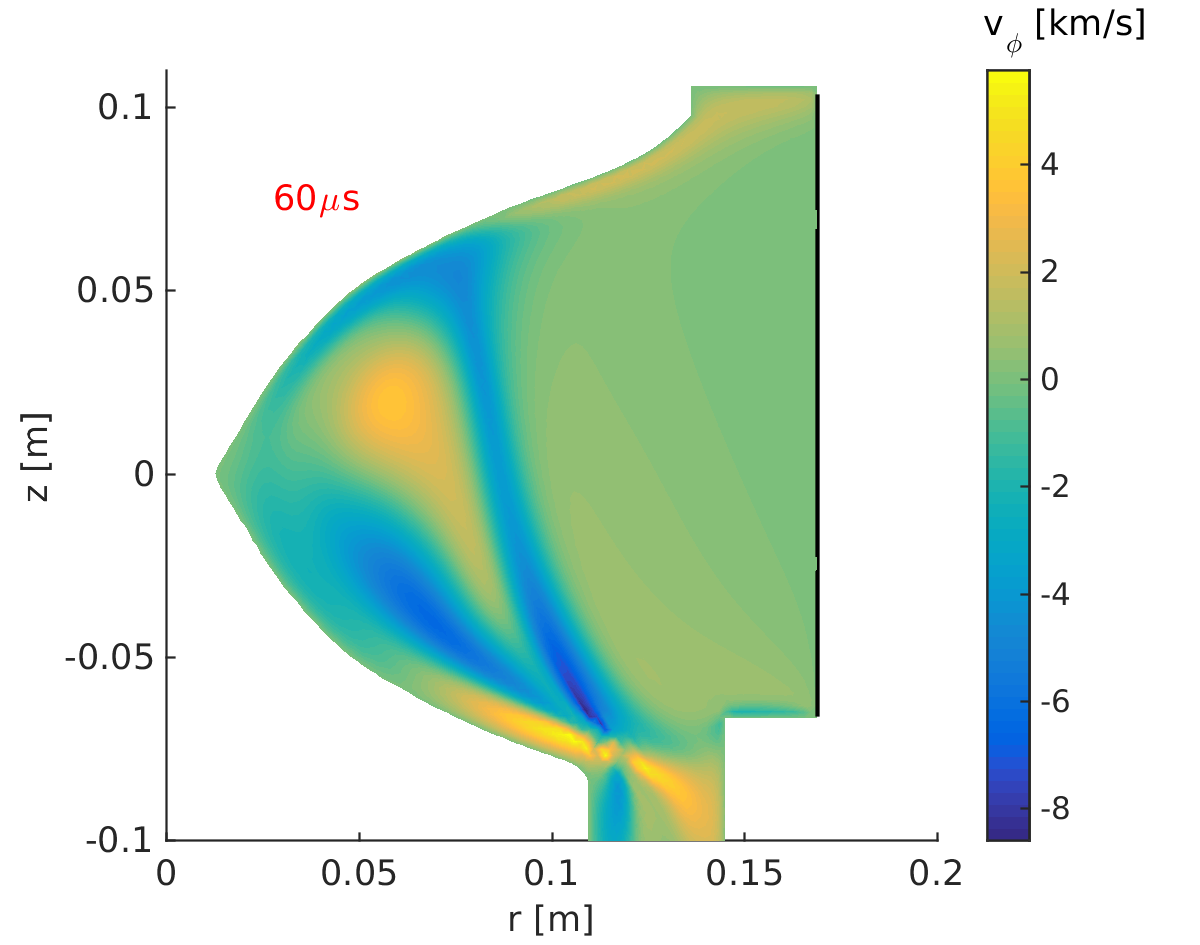}}\hfill{}\subfloat[]{\raggedright{}\includegraphics[width=7cm,height=5cm]{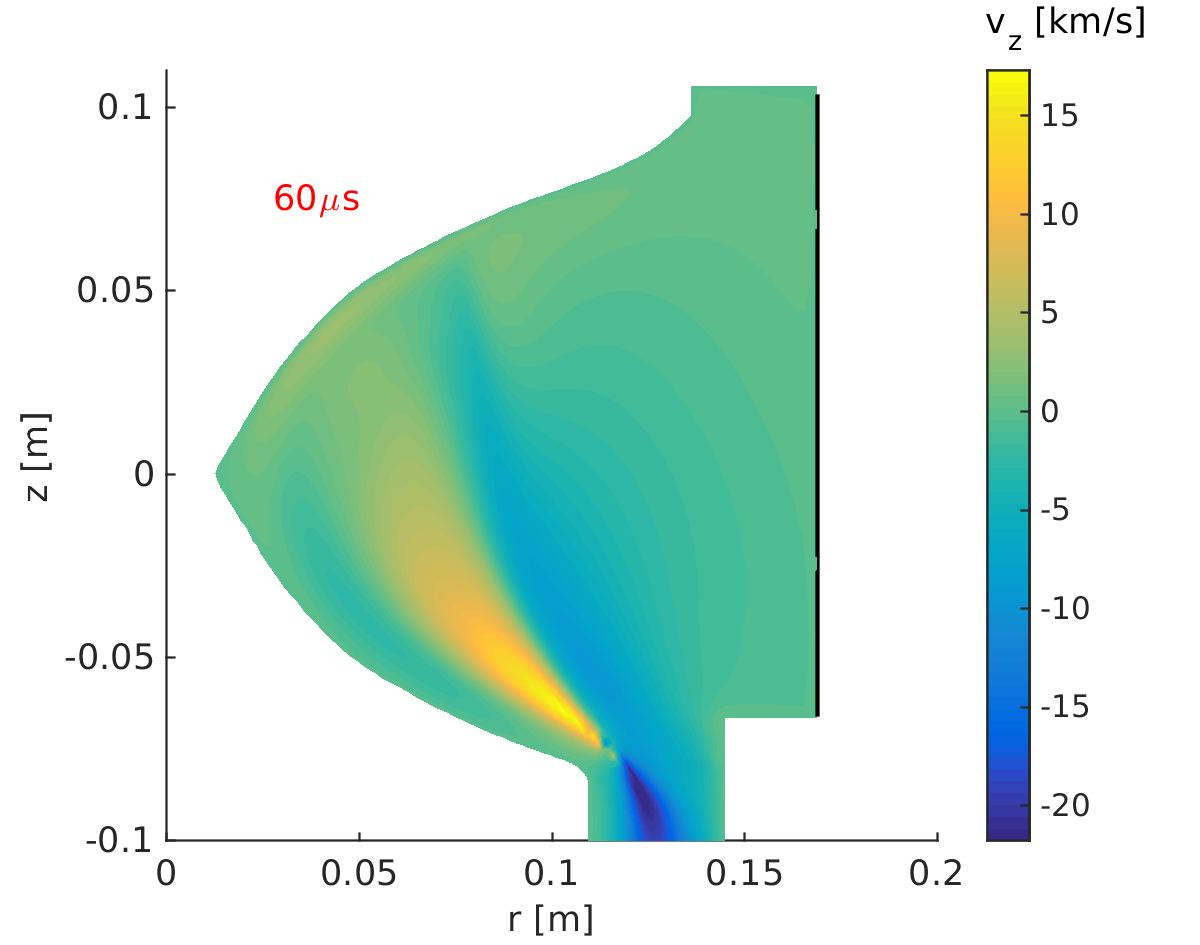}}

\caption{\label{fig: v_11coils}$\,\,\,\,$Profiles of various velocity components }
\end{figure}
Figure \ref{fig: v_11coils}(a) to (c) indicates profiles of $v_{r}$,
$v_{\phi}$ and $v_{z}$ from simulation 2353 at $18\,\upmu$s. The
dynamics are complicated, but some key features can be noted. Velocities,
in particular $v_{z}$, are especially high around this time during
the CT formation process, and lead to enhanced ion viscous heating
(see figure \ref{fig: Ti_11coils}(c)). Pre-compression profiles of
$v_{\phi}$ and $v_{z}$ are shown at $25\,\upmu$s in figures \ref{fig: v_11coils}(d)
and (e). As discussed in \cite{Morales}, internal flows are naturally
associated with plasma in toroidal geometry if the externally imposed
magnetic and electric fields are curl-free. When externally imposed
magnetic and electric fields are curl-free, the curl of the Lorentz
force is finite, so it cannot be balanced by the pressure gradient.
When the curl of the Lorentz force is finite, it acts as a source
of vorticity. Toroidal vorticity leads to poloidal rotation which
interacts with toroidal magnetic field leading to poloidal current
density. Poloidal current density and poloidal field lead to toroidal
velocities, which are directed according to the relative orientation
of the poloidal current density and poloidal magnetic field. The toroidal
velocities shown in figure \ref{fig: v_11coils}(d), when the CT is
approaching an equilibrium state, may result partially from this effect.
Simulations run starting with a Grad-Shafranov equilibrium exhibit
qualitatively similar toroidal velocity profiles, generally with either
a dipolar structure or a quadrupolar structure in $v_{\phi}$. The
particularly high downward-directed axial velocity near the entrance
to the CT containment region at $25\,\upmu$s in figure \ref{fig: v_11coils}(e)
is related to jets of plasma fluid associated with magnetic reconnection
during the formation of closed CT flux surfaces. Velocities are gradually
dissipated by viscous effects until $t_{comp}=45\,\upmu$s, when magnetic
compression is initiated and velocities rise sharply in reaction to
the various dynamics associated with compression. The structures in
the profiles of $v_{\phi}$ and $v_{z}$ near the entrance to the
CT containment region at 60$\,\upmu$s (figures \ref{fig: v_11coils}(f)
and (g)), just after halfway through the main compression cycle, are
related to jets of plasma fluid associated with magnetic reconnection
when open poloidal field lines surrounding the CT are pinched off.

\section{Evolution of fields when $\boldsymbol{I_{comp}}$ changes polarity,
simulation  2287\label{subsec:ring_psi_f}}

As described in chapter \ref{Chap:Magnetic-Compression}, when the
compression current in the coils changes direction (see figure \ref{fig:Ilev_comp}),
the CT poloidal field magnetically reconnects with the compression
field, and a new CT with polarity opposite to that of the previous
CT is induced in the containment region, compressed, and then allowed
to expand. The process repeats itself at each change in polarity of
the compression current until either the plasma loses too much heat,
or the compression current is sufficiently damped.
\begin{figure}[H]
\subfloat[]{\raggedright{}\includegraphics[width=7cm,height=5cm]{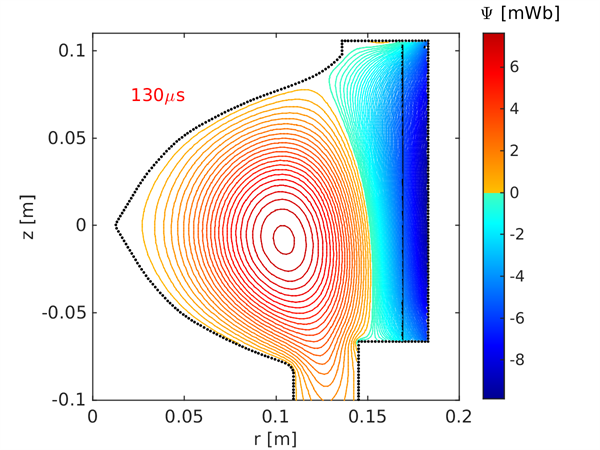}}\hfill{}\subfloat[]{\raggedright{}\includegraphics[width=7cm,height=5cm]{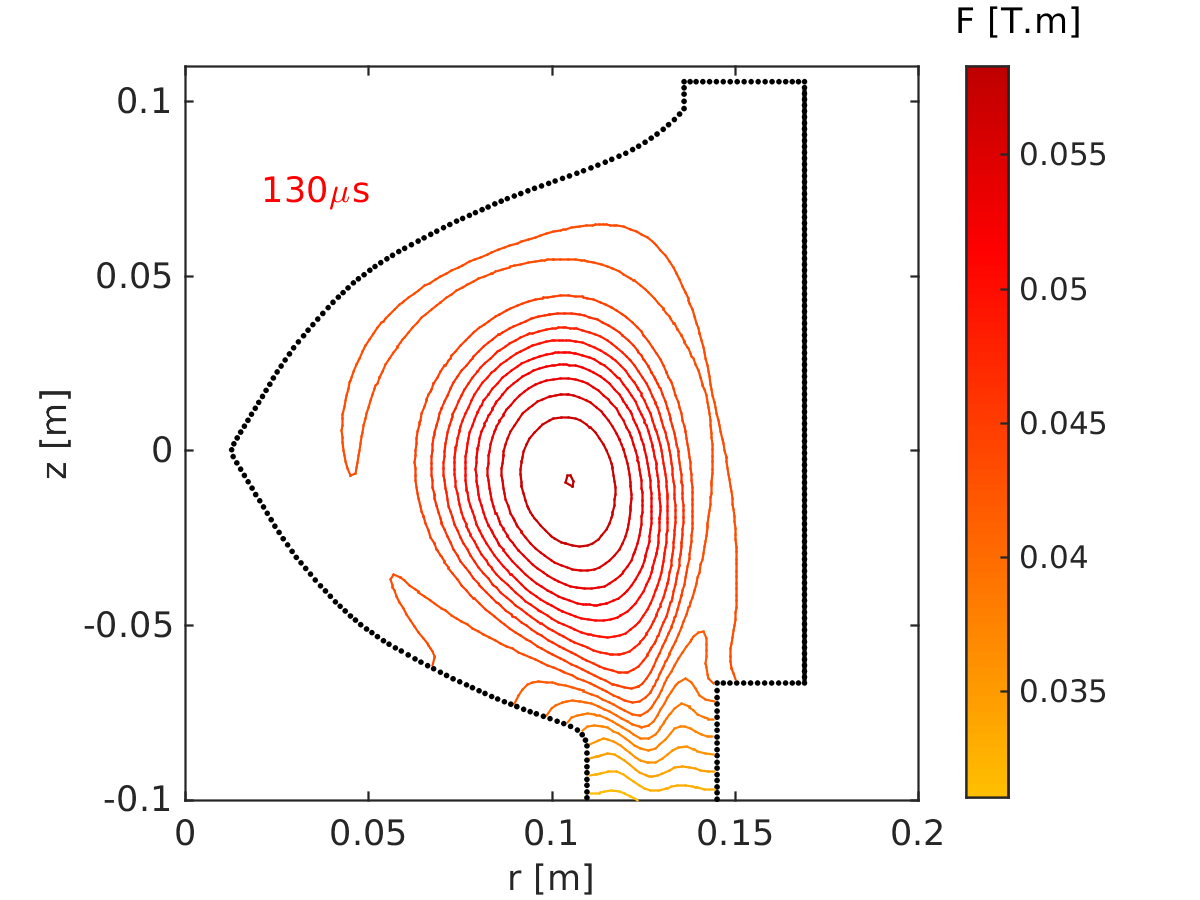}}

\subfloat[]{\raggedright{}\includegraphics[width=7cm,height=5cm]{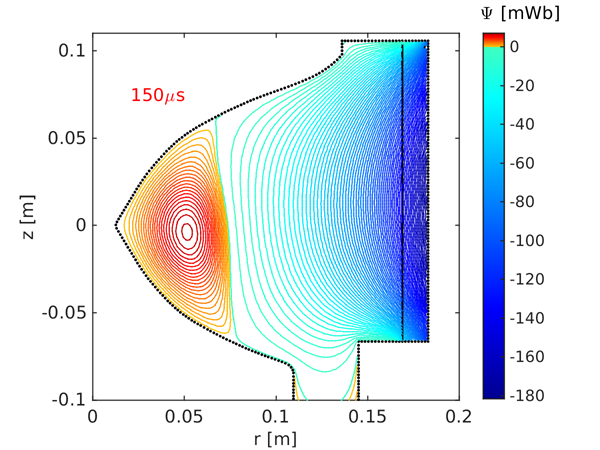}}\hfill{}\subfloat[]{\raggedright{}\includegraphics[width=7cm,height=5cm]{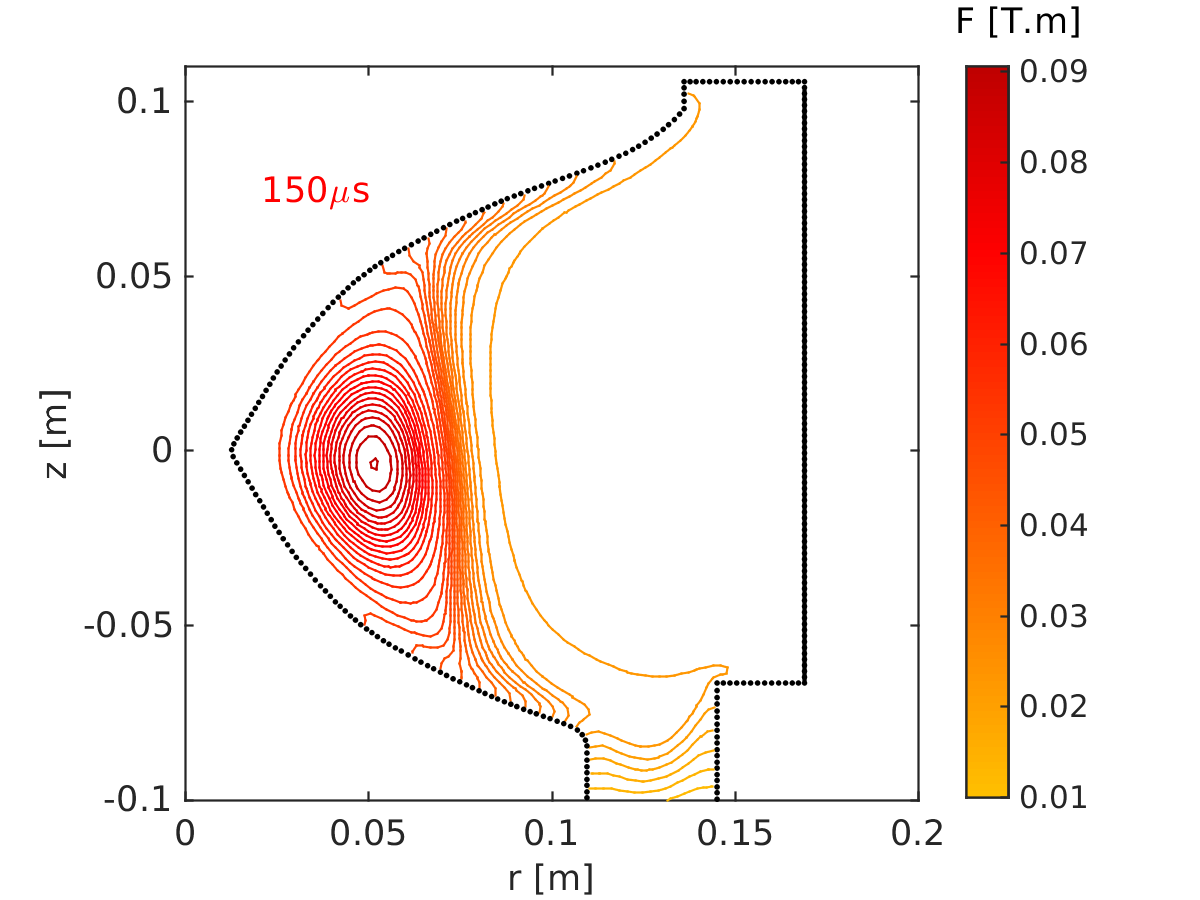}}

\subfloat[]{\raggedright{}\includegraphics[width=7cm,height=5cm]{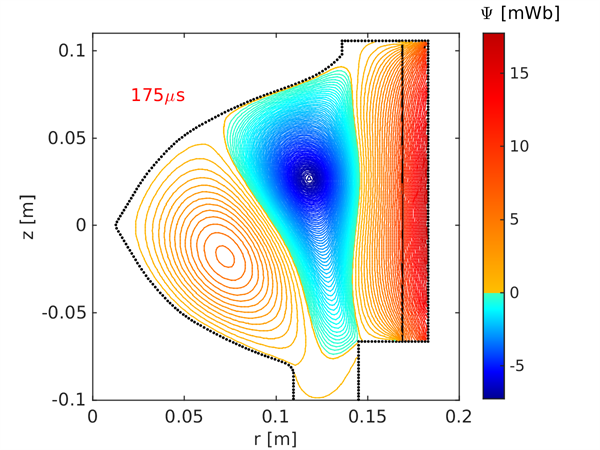}}\hfill{}\subfloat[]{\raggedright{}\includegraphics[width=7cm,height=5cm]{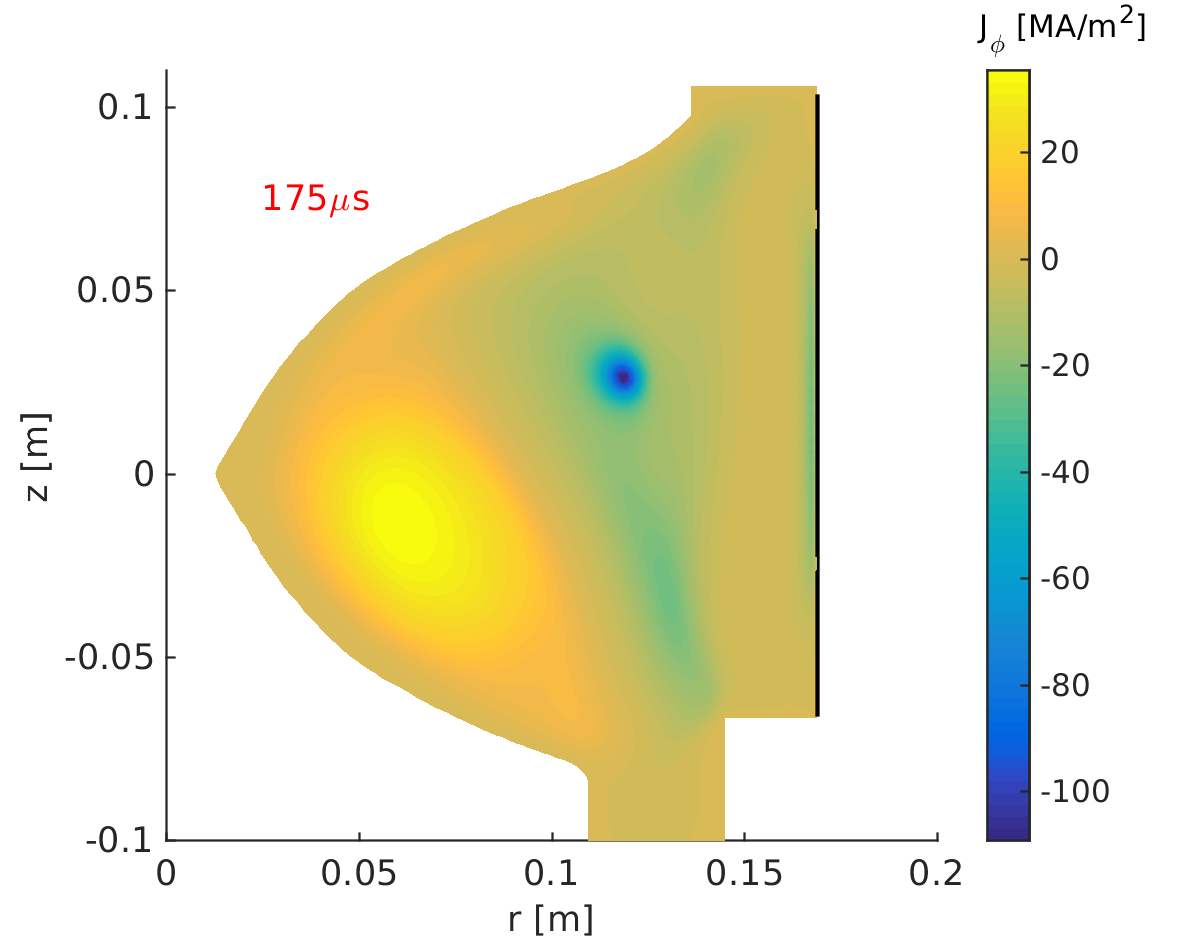}}

\subfloat[]{\raggedright{}\includegraphics[width=7cm,height=5cm]{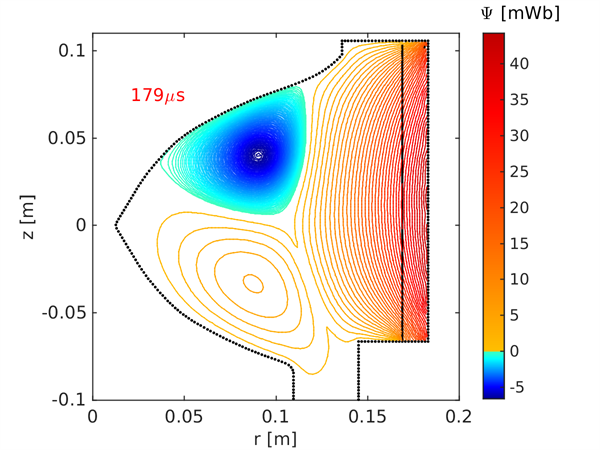}}\hfill{}\subfloat[]{\raggedright{}\includegraphics[width=7cm,height=5cm]{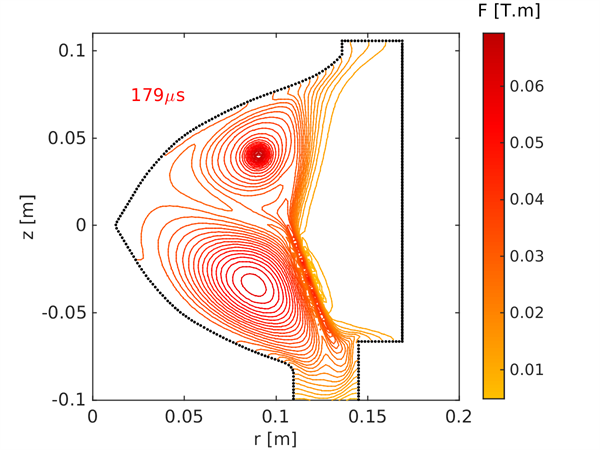}}

\caption{\label{fig: ring_psi_f}$\,\,\,\,$$\psi$ and $f$ contours, and
$J_{\phi}$ profiles, for a simulation with compression current reversal}
\end{figure}
\begin{figure}[H]
\subfloat[]{\raggedright{}\includegraphics[width=7cm,height=5cm]{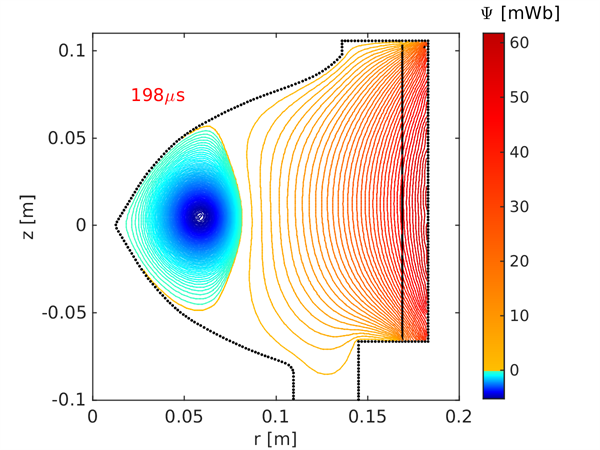}}\hfill{}\subfloat[]{\raggedright{}\includegraphics[width=7cm,height=5cm]{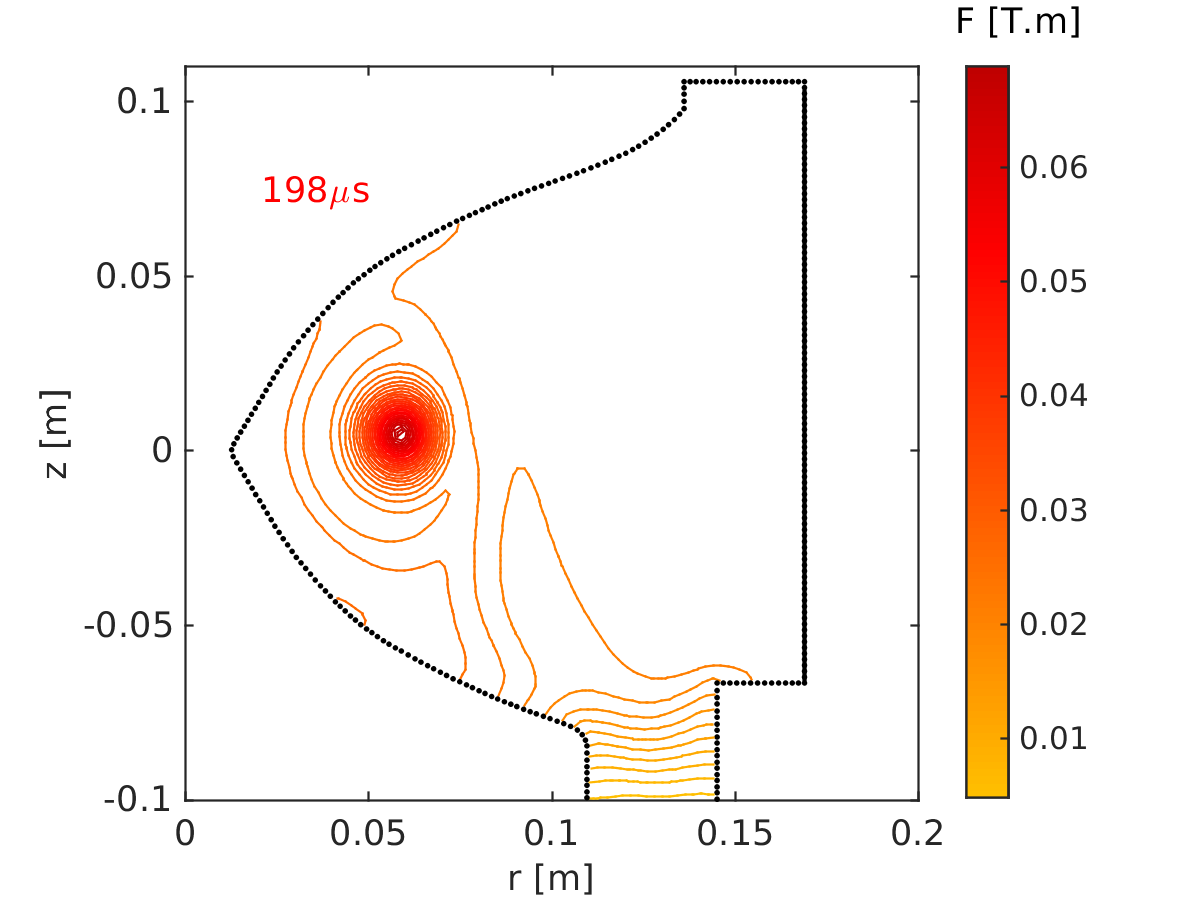}}

\subfloat[]{\raggedright{}\includegraphics[width=7cm,height=5cm]{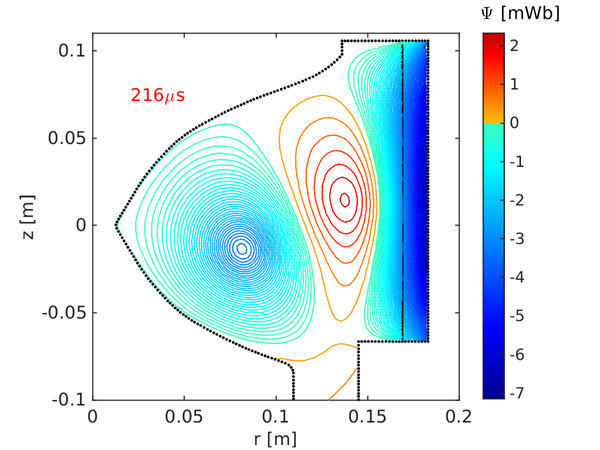}}\hfill{}\subfloat[]{\raggedright{}\includegraphics[width=7cm,height=5cm]{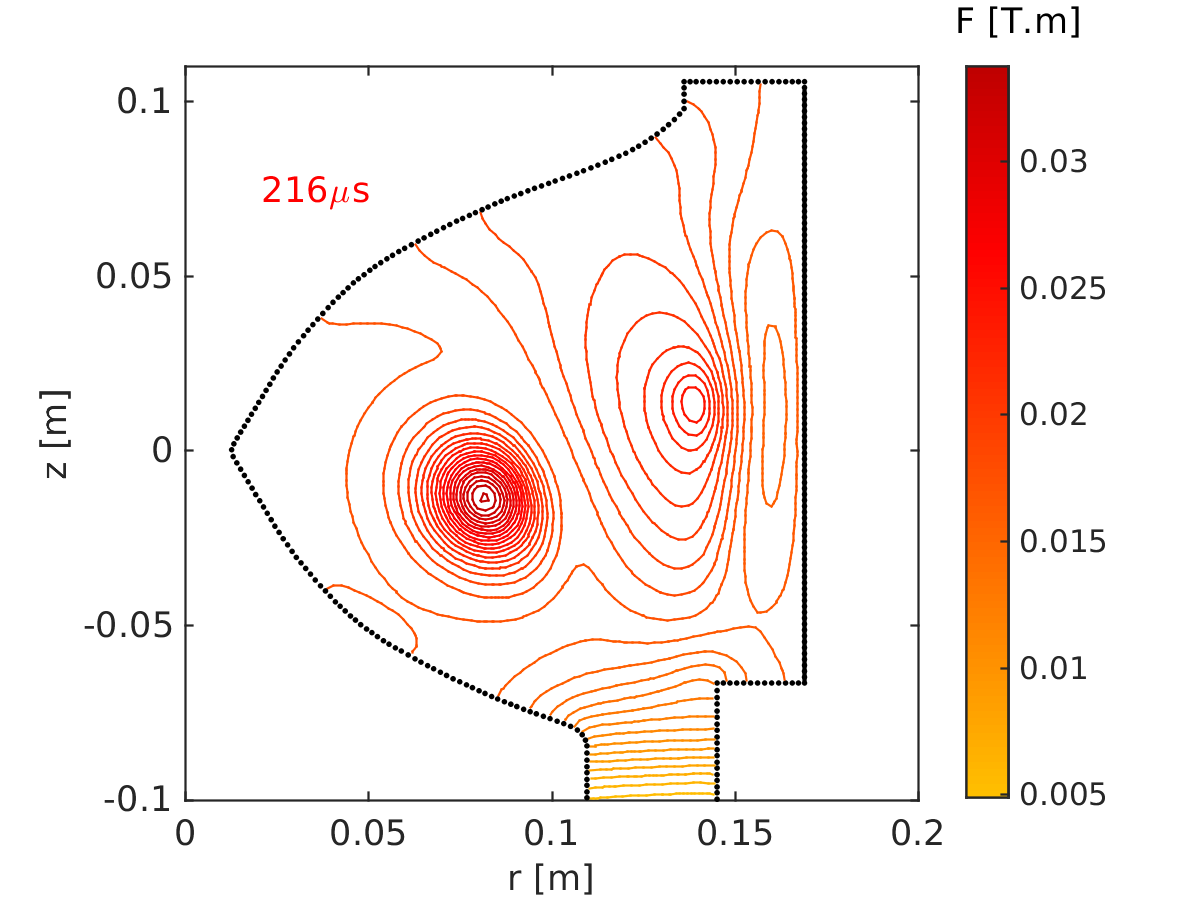}}

\subfloat[]{\raggedright{}\includegraphics[width=7cm,height=5cm]{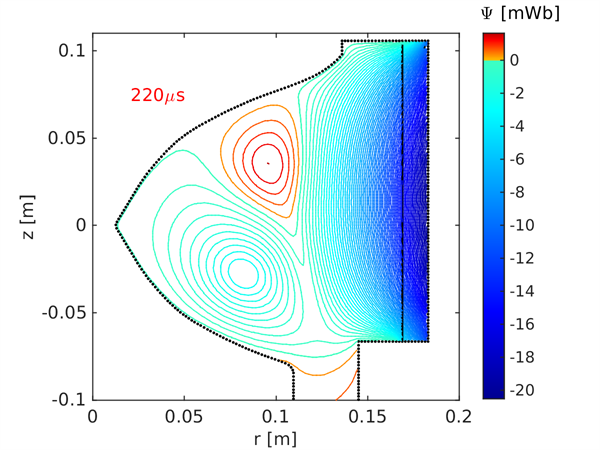}}\hfill{}\subfloat[]{\raggedright{}\includegraphics[width=7cm,height=5cm]{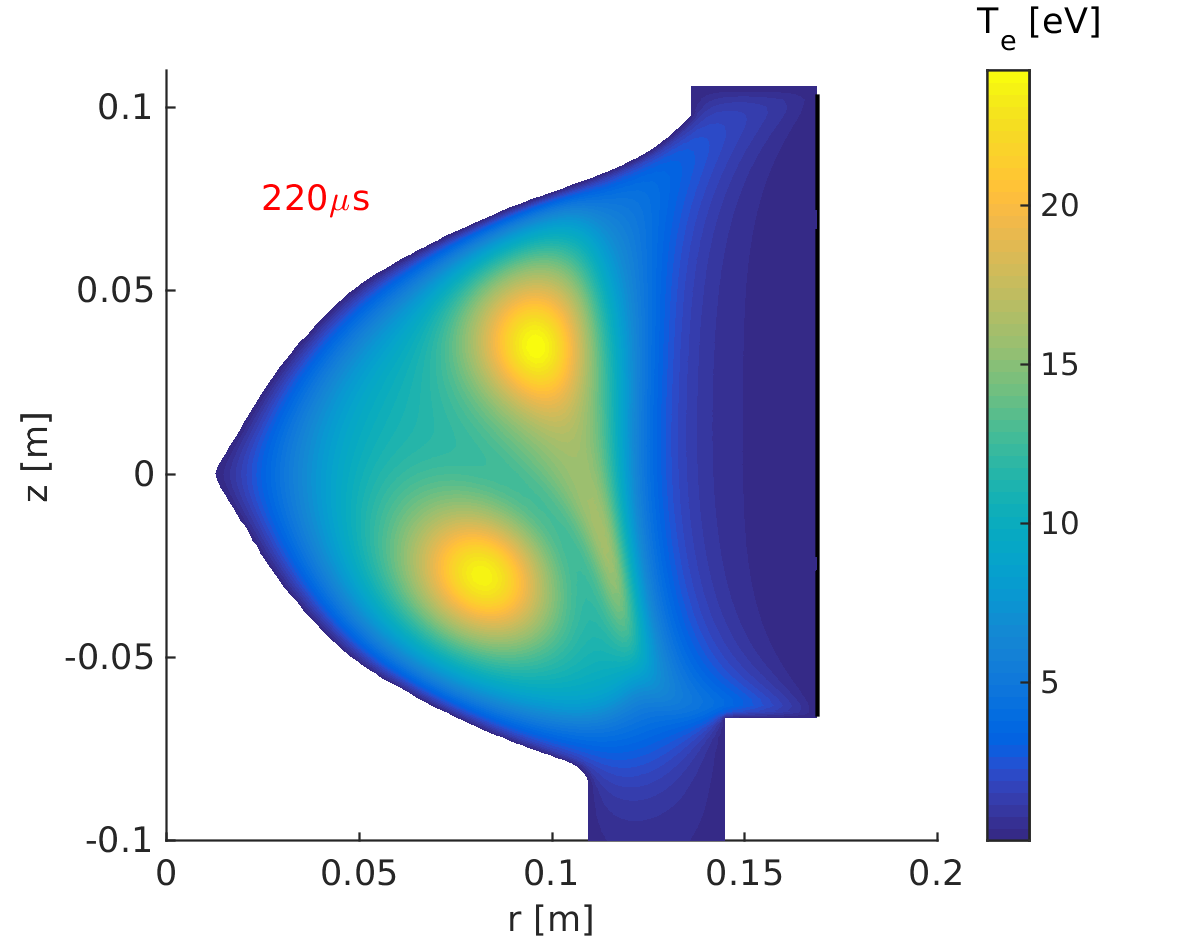}}

\subfloat[]{\raggedright{}\includegraphics[width=7cm,height=5cm]{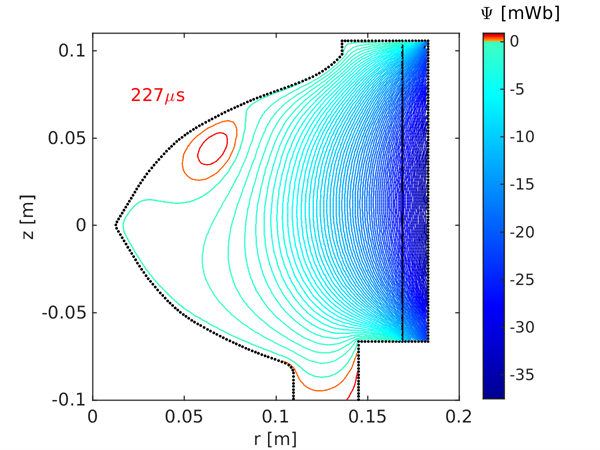}}\hfill{}\subfloat[]{\raggedright{}\includegraphics[width=7cm,height=5cm]{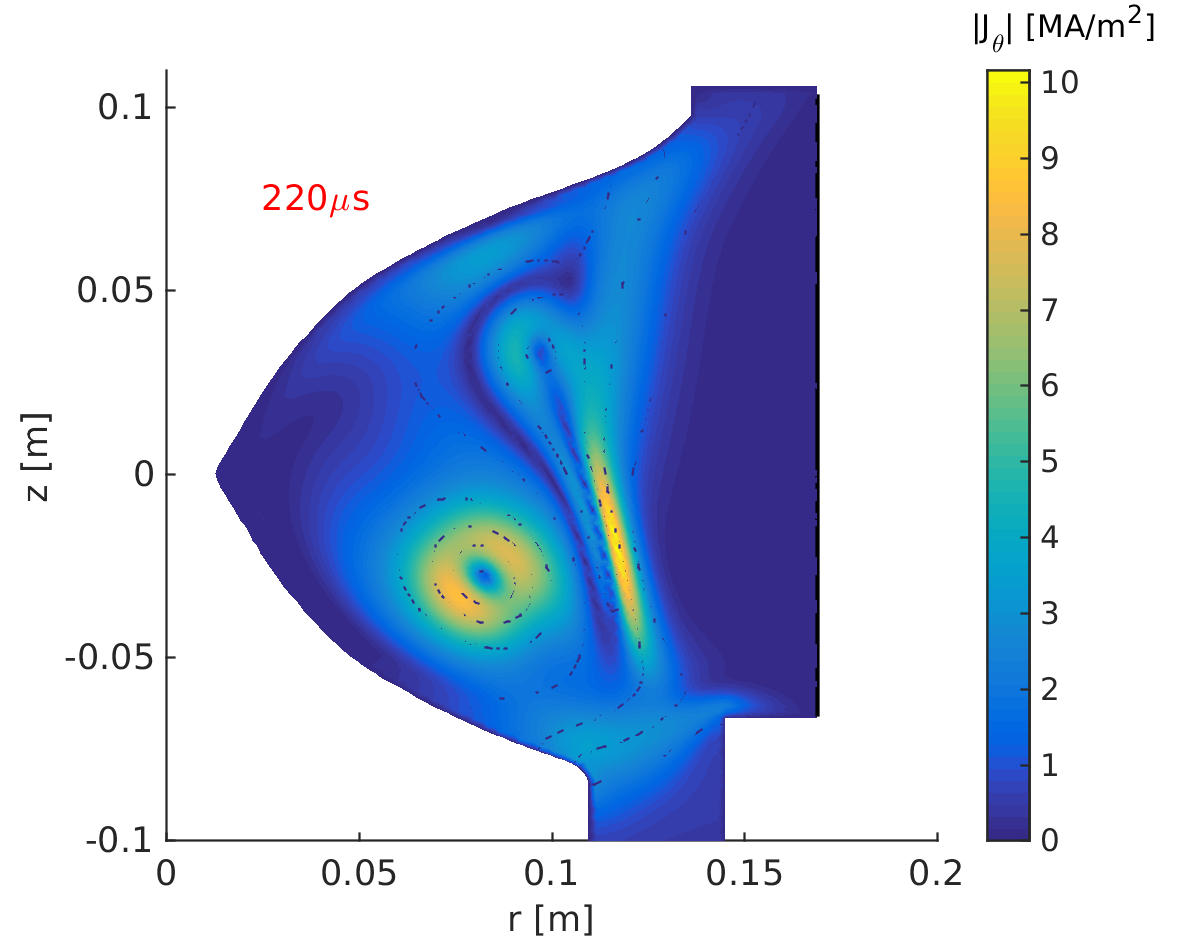}}

\caption{\label{fig: ring_psi_f-1}$\,\,\,\,$(sequence continued) $\,\,\,\,$$\psi$
and $f$ contours, and $T_{e}$ and $J_{\theta}$ profiles, for a
simulation with compression current reversal}
\end{figure}
Figures \ref{fig: ring_psi_f} and \ref{fig: ring_psi_f-1} show contours
of $\psi$ and $f$, and profiles of $J_{\phi}$, $T_{e}$ and $J_{\theta}$
from simulation  2287 at various times. Magnetic compression begins
at $t=t_{comp}=130\,\upmu$s, and peak compression is at $150\,\upmu$s.
By $175\,\upmu$s, the external compression field has changed polarity.
Toroidal currents are induced to flow in the ambient plasma initially
located outboard of the original CT, enabling the formation of a new
CT (blue closed contours) with polarity opposite to that of the original
CT (figures \ref{fig: ring_psi_f}(e) and (f)). The external compression
field starts to reconnect with the poloidal field of the original
CT at $179\,\upmu$s. The new induced CT is magnetically compressed
inwards by the increasing reversed polarity compression field, with
peak compression at around $198\,\upmu$s (figures \ref{fig: ring_psi_f-1}(a)
and (b)). The compression field polarity rings back to its original
state by $216\,\upmu$s, and a third CT is induced, with the same
polarity as the original CT (figures \ref{fig: ring_psi_f-1}(c) and
(d)). The poloidal field of the second CT starts to reconnect with
the compression field at around $220\,\upmu$s. Electron temperature
and poloidal current density at $220\,\upmu$s for the two co-existing
CTs is presented in figures \ref{fig: ring_psi_f-1}(f) and (h). By
$227\,\upmu$s, the poloidal flux of the third CT, which is being
compressed inwards during the third compression cycle, has almost
decayed away (figure \ref{fig: ring_psi_f-1}(g)). Comparison of simulated
diagnostics outputs for simulation  2287 with experimental measurements
from the relevant shot will be presented in section \ref{par:Shot-=00002339735}.

\section{Simulated diagnostics, and comparison with experimental results\label{subsec:Simulated-diag_CFexp}}

The experimental diagnostics that can be modelled in simulations are
poloidal and toroidal magnetic field measured at the magnetic probes
locations, the CT outer separatrix radius (see section \ref{subsec:Using-side-probe-data}),
line averaged electron density along the interferometer chords, and
ion temperature along the ion-Doppler chords. The methods developed
to implement the simulated diagnostics measurements to the code have
been described in section \ref{subsec:Simulated-diagnostics}. 

\subsection{Non-compression shots/simulations}

\begin{figure}[H]
\subfloat[$B_{\theta}$, $2.5\mbox{ m}\Omega$ cables]{\raggedright{}\includegraphics[width=8cm,height=5cm]{fig_66.png}}\hfill{}\subfloat[$B_{\theta}$, $70\mbox{ m}\Omega$ cables]{\raggedleft{}\includegraphics[width=8cm,height=5cm]{fig_67.png}}

\subfloat[$2.5\mbox{ m}\Omega$ cables (simulation  1343)]{\raggedright{}\includegraphics[width=7cm,height=5cm]{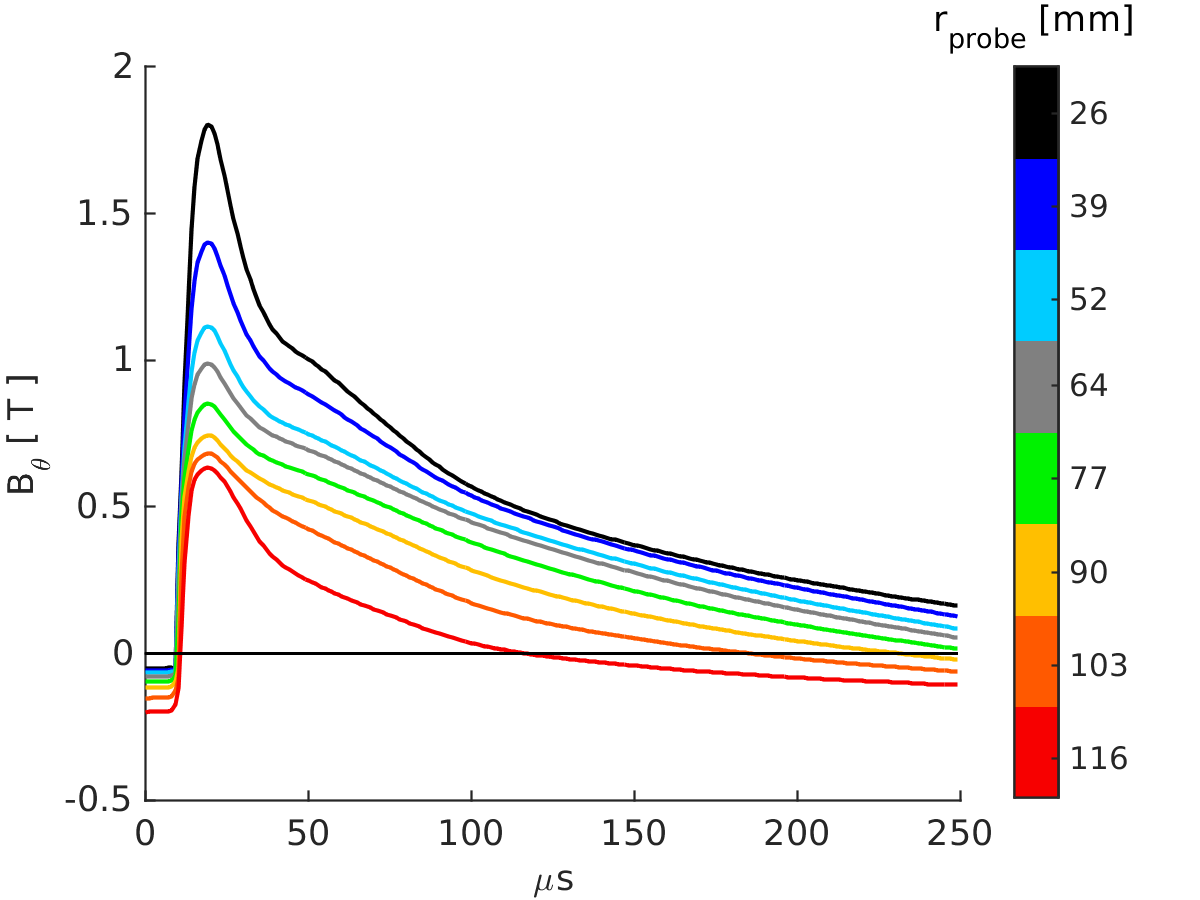}}\hfill{}\subfloat[$70\mbox{ m}\Omega$ cables (simulation  2174)]{\raggedleft{}\includegraphics[width=7cm,height=5cm]{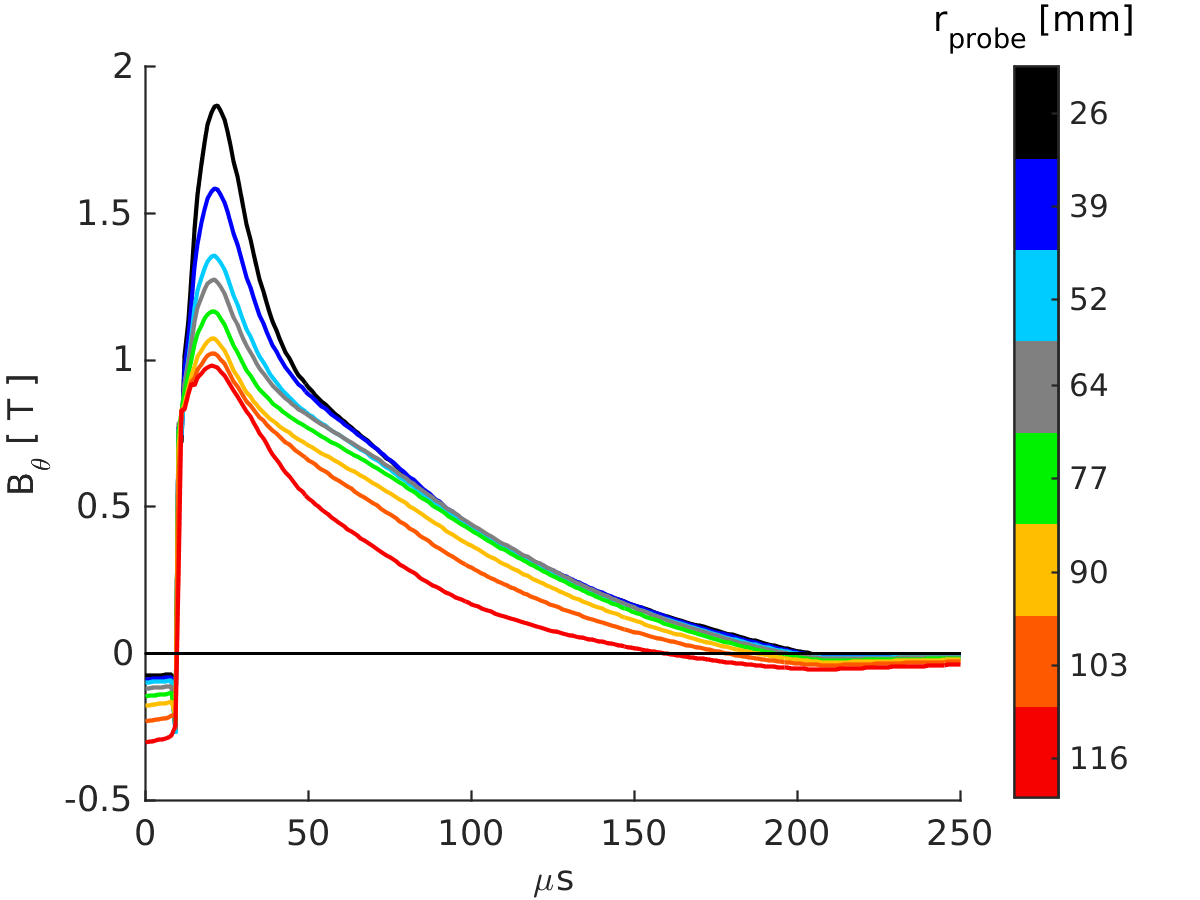}}

\caption{\label{fig:Bpol_meas_cf_sim}$\,\,\,\,$Comparison of measured and
simulated $B_{\theta}$ (levitation - eleven coils) }
\end{figure}
Figures \ref{fig:Bpol_meas_cf_sim}(a) and (b) show experimentally
measured $B_{\theta}$ for two shots with different resistances ($R_{cable}$)
of the cables between the levitation coil (or coil-pair) and main
levitation inductor in each of the six levitation circuits (see figure
\ref{fig:Levitation-and-compression-1}), for the 11-coil configuration
(for convenience, figures \ref{fig:Poloidal-field-for}(a) and (b)
have been reproduced here). For comparison, figures \ref{fig:Bpol_meas_cf_sim}(c)
and (d) show $B_{\theta}$ recorded at the probe locations for MHD
simulations in which the boundary conditions pertaining to the levitation
field were evolved over time according to the two different experimentally
measured waveforms for $I_{lev}$, which depend on $R_{cable}$. The
two $I_{lev}(R_{cable},\,t)$ waveforms are indicated in figure \ref{fig:Bpol_meas_cf_sim}(a)
and (b), right axes. The $B_{\theta}$ traces in figure \ref{fig:Bpol_meas_cf_sim}(c)
are from a simulation which had code input parameter $R_{cable}$
set to zero (see section \ref{sec:Code-input-parameters,}), while
the $B_{\theta}$ traces in figure \ref{fig:Bpol_meas_cf_sim}(d)
are from a simulation which had input $R_{cable}$ set to one. Additional
code input parameters for these two simulations were approximately
the same as for simulation 2353 (table \ref{tab:Sim parameters}),
but with $V_{comp}=0$ kV.

Comparing the $B_{\theta}$ traces in figures \ref{fig:Bpol_meas_cf_sim}(a)
and \ref{fig:Bpol_meas_cf_sim}(c), it can be seen how the comparison
is qualitatively good up until around $170\,\upmu$s, when the compressional
instability (discussed in section \ref{subsec:Compressional-Instability}),
which is not captured by the 2D MHD dynamics, causes the CT to be
extinguished rapidly. The comparison in figures \ref{fig:Bpol_meas_cf_sim}(b)
and \ref{fig:Bpol_meas_cf_sim}(d) remains good at all times, as the
compressional instability did not arise in the case with decay rate
matching. 

\begin{figure}[H]
\subfloat[$70\mbox{ m}\Omega$ cables (experiment) ]{\raggedright{}\includegraphics[width=8.6cm,height=5cm]{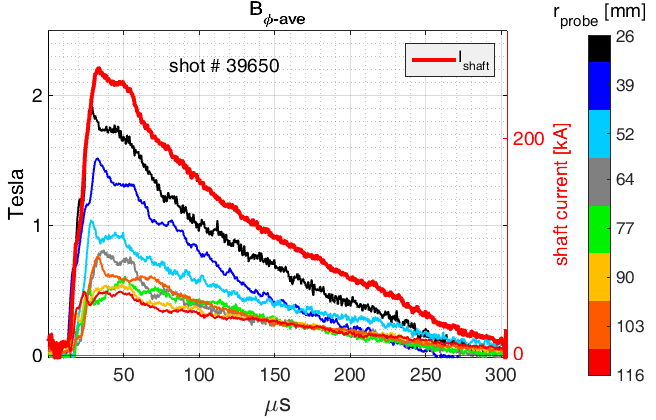}}\hfill{}\subfloat[$70\mbox{ m}\Omega$ cables (simulation  2174)]{\raggedleft{}\includegraphics[width=7.5cm,height=5cm]{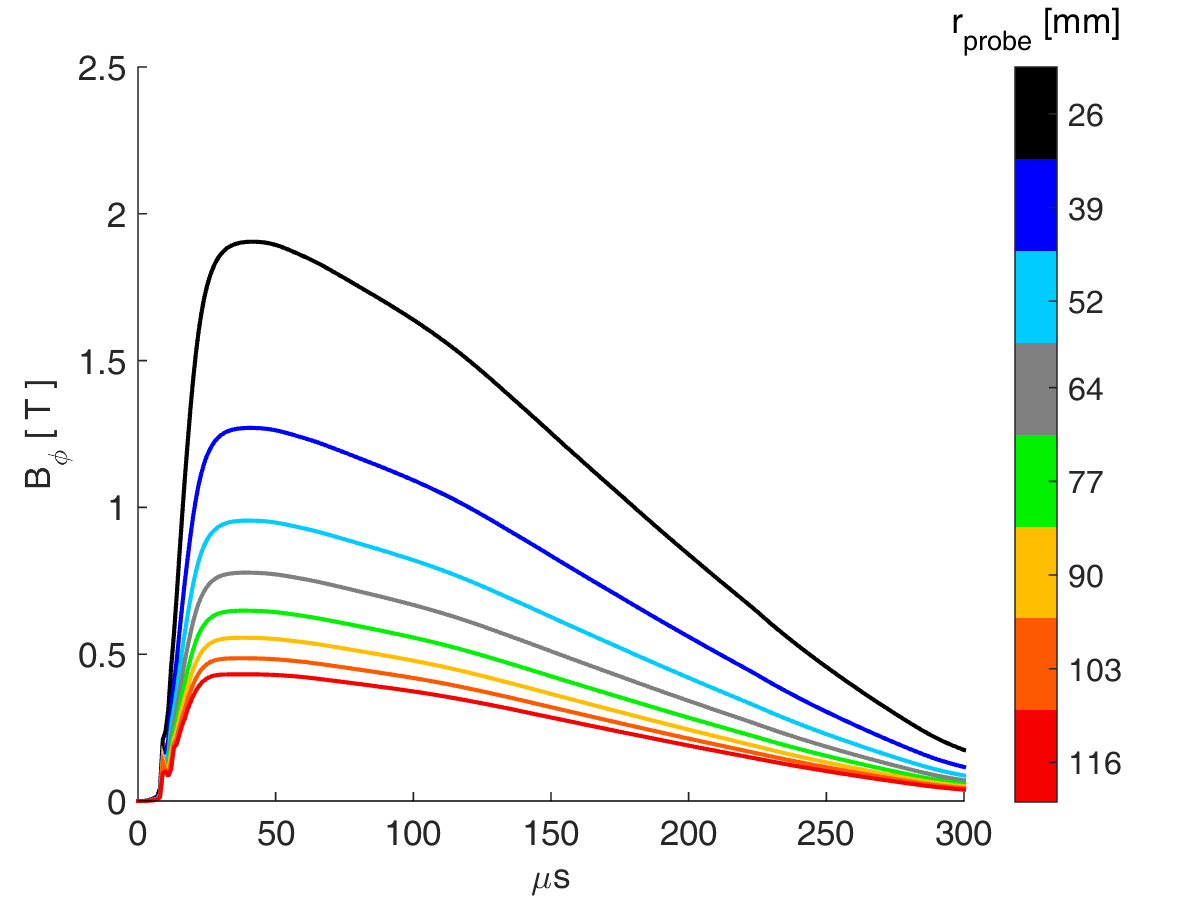}}

\caption{\label{fig:Bphi_exp_sim_comparison}$\,\,\,\,$Comparison of measured
and simulated $B_{\phi}$ (levitation - eleven coils) }
\end{figure}
Figure \ref{fig:Bphi_exp_sim_comparison}(a) shows experimentally
measured $B_{\phi}$ for a shot with $70\mbox{ m}\Omega$ cable resistance
taken with the 11-coil configuration. For ease of comparison, the
toroidal averages of the toroidal field traces measured at the two
probes 180$^{o}$ apart at each of the eight radii at which the chalice
magnetic probes are located (see table \ref{tab: coordinates-ofBprobes})
are shown here. With $70\mbox{ m}\Omega$ cable resistance, the compressional
instability that was routinely observed on levitation-only shots with
the $2.5\mbox{ m}\Omega$ cable resistance ($e.g.,$ see figure \ref{fig:Poloidal-field-for}(c))
is not manifested on the toroidal field measurements, so the simulated
toroidal field is a good match (2D simulations cannot reproduce the
compressional instability). Note that this simulation includes the
resistive contribution (section \ref{subsec:Formation-simulations11})
to $\Phi_{form}$; simulated $B_{\phi}$ decreases much more gradually
over time after formation, and doesn't match the experimental measurement,
if the resistive contribution is neglected.

\begin{figure}[H]
\begin{centering}
\subfloat[shot $39573\:(2.5\mbox{ m}\Omega\:\mbox{cables})$ ]{\raggedleft{}\includegraphics[width=7cm,height=5cm]{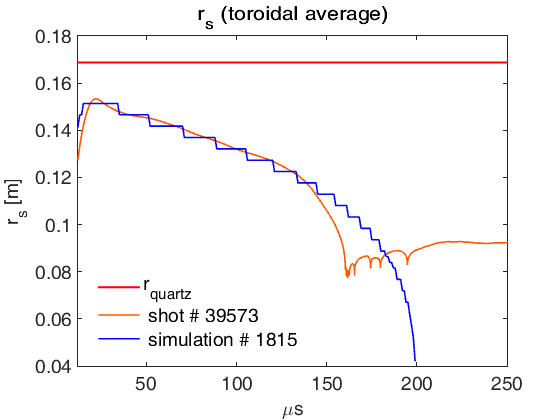}}\hfill{}\subfloat[shot $39650\:(70\mbox{ m}\Omega\:\mbox{cables})$ ]{\raggedleft{}\includegraphics[width=7cm,height=5cm]{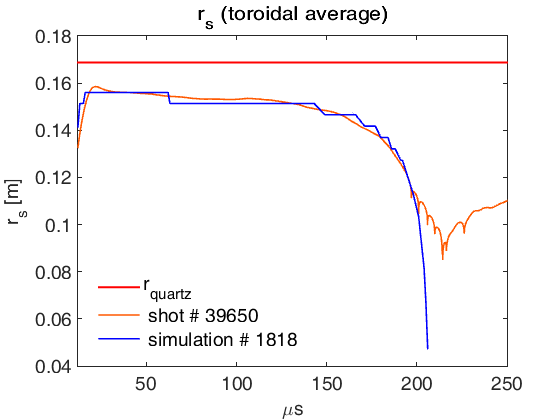}}
\par\end{centering}
\caption{\label{fig:Rsep_lev_comparison}$\,\,\,\,$Comparison of experimentally
measured and simulated CT separatrix radius }
\end{figure}
Figure \ref{fig:Rsep_lev_comparison}(a) shows the close match obtained
between the toroidally-averaged, experimentally inferred separatrix
radius, $r_{s}(t)$ (from figure \ref{fig:rsep29205_39573}(b), see
section \ref{subsec:Using-side-probe-data}), and the separatrix radius
determined by MHD simulation. Shot  39573 was taken with the 2.5 m$\Omega$
cables in place, and, for comparison, simulation  1815 had code input
parameter $R_{cable}=0$. 

Similarly, figure \ref{fig:Rsep_lev_comparison}(b) shows the match
obtained between the toroidally-averaged, experimentally inferred
separatrix radius (from figure \ref{fig:rsep39650_2}), and the simulated
separatrix radius. Shot  39650 was taken with the 70 m$\Omega$ cables
in place, and simulation  1818 was run with $R_{cable}=1$. 
\begin{figure}[H]
\subfloat[]{\raggedleft{}\includegraphics[width=7cm,height=5cm]{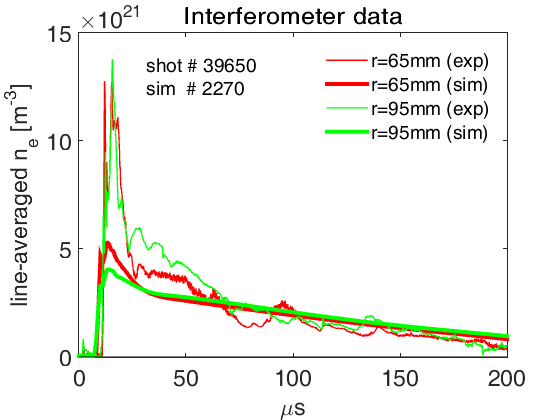}}\hfill{}\subfloat[]{\raggedright{}\includegraphics[width=7cm,height=5cm]{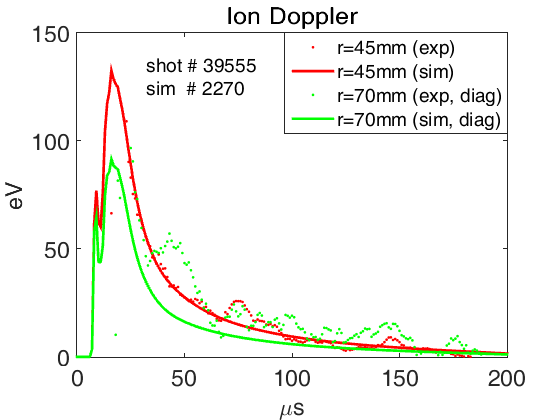}}

\caption{$\,\,\,\,$Comparison of measured and simulated $n_{e}$ and $T_{i}$
(levitation only)\label{fig:ne_IDlev}}
\end{figure}
Figures \ref{fig:ne_IDlev}(a) and \ref{fig:ne_IDlev}(b) indicate
the approximate agreement between experimentally measured and simulated
electron density and ion temperature in the case without magnetic
compression. The simulated diagnostics for $n_{e}$ and $T_{i}$ are
simply the line-averaged quantities along the chords indicated in
figures \ref{fig:Chalice}, \ref{fig: n_11coils} and \ref{fig: Ti_11coils}.
Note that the interferometer looking along the vertical chord at $r=35$
mm was generally not functioning during the experiment, and so the
experimentally measured and simulated line averaged electron density
along that chord hasn't been included in figure \ref{fig:ne_IDlev}(a).
In general, compared with the simulated line averaged electron density,
the experimentally measured density is higher initially and decays
much faster over the first $70\,\upmu$s or so. This may be related
to the rapid recombination of high $Z$ impurity ions that are initially
injected to the plasma during the formation process, through the action
of sputtering on the electrodes and the insulating outer wall of the
CT confinement region. This mechanism would not be captured by the
simulation. Another possible explanation is the action of artificial
density diffusion, which effectively relocates particles from high
density regions to low density regions. This effect is pronounced
when the density gradients are more extreme, such as during the formation
process. 

Code input parameters $\nu_{phys}$ and $\nu_{num}$ (see section
\ref{sec:Code-input-parameters,}) were set to $450$m$^{2}/$s and
$700$m$^{2}/$s respectively for simulation  2270, so that ion viscous
heating is reduced to an acceptable level, leading to a close match
to the experimentally measured ion temperature.

\subsection{Compression shots/simulations}

\subsubsection{Shots 39475, 39738, and 39510}

\begin{figure}[H]
\subfloat[$B_{\theta}$ for shot  39475]{\raggedright{}\includegraphics[width=7.5cm,height=5.5cm]{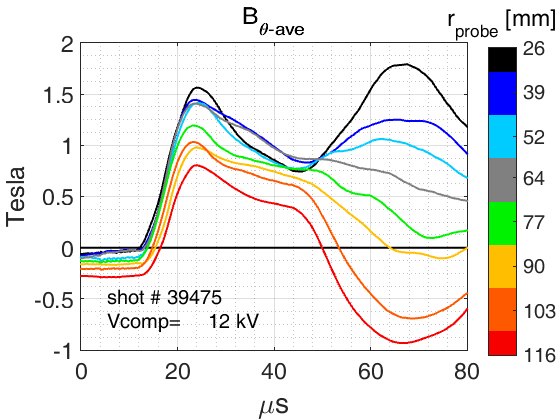}}\hfill{}\subfloat[$B_{\theta}$ for simulation  2350]{\raggedleft{}\includegraphics[width=8cm,height=5.5cm]{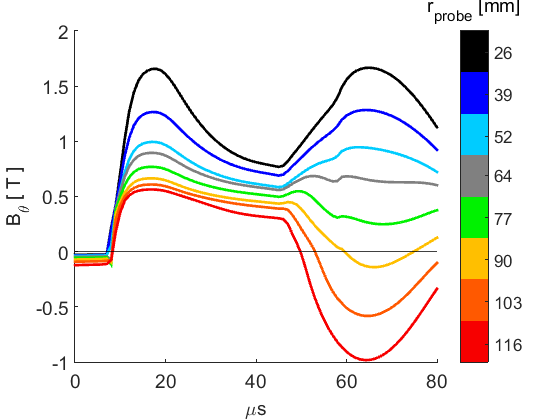}}

\caption{\label{fig:Bpol_meas_cf_sim39475}$\,\,\,\,$Comparison of measured
and simulated $B_{\theta}$ (compression - eleven coils) }
\end{figure}
Figure \ref{fig:Bpol_meas_cf_sim39475} indicates the comparison between
experimentally measured and simulated poloidal field at the magnetic
probe locations. For shot 39475 and simulation 2350, $V_{comp}=12$
kV  and $t_{comp}=45\,\upmu$s (code input parameters for simulation
2350 were the same as for simulation 2353 (table \ref{tab:Sim parameters}),
but with $V_{comp}=12$ kV). For ease of comparison, the toroidal
averages of the poloidal field traces measured at the two probes 180$^{o}$
apart at each of the eight radii at which the chalice magnetic probes
are located (see table \ref{tab: coordinates-ofBprobes}) are shown
in figure \ref{fig:Bpol_meas_cf_sim39475}(a).  These axisymmetric
MHD simulations allow for only resistive loss of flux and do not capture
inherently three-dimensional plasma instabilities that can lead to
poloidal flux loss. Shot $39475$ was a flux-conserving shot, and
therefore a good match between experimentally measured and MHD-simulated
poloidal field is obtained. 
\begin{figure}[H]
\subfloat[]{\raggedleft{}\includegraphics[width=7cm,height=5cm]{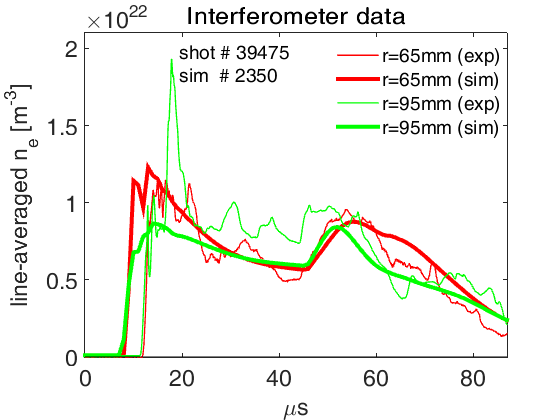}}\hfill{}\subfloat[]{\raggedright{}\includegraphics[width=7cm,height=5cm]{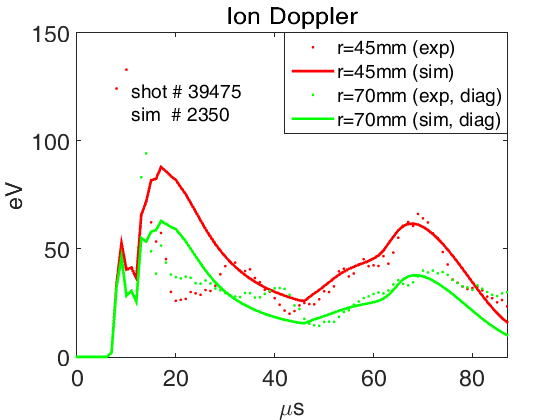}}
\begin{centering}
\subfloat[]{\raggedright{}\includegraphics[width=7cm,height=5cm]{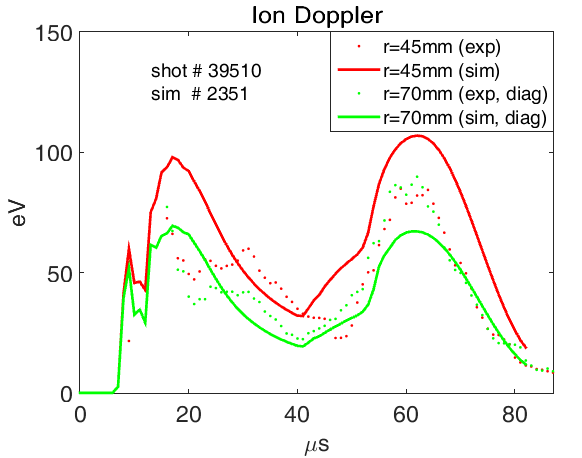}}
\par\end{centering}
\caption{\label{fig:ne_IDcomp-1}$\,\,\,\,$Comparison of measured and simulated
$n_{e}$ and $T_{i}$ (compression shots) }
\end{figure}
Figures \ref{fig:ne_IDcomp-1}(a) and \ref{fig:ne_IDcomp-1}(b) indicate
the agreements between experimentally measured and simulated electron
density and ion temperature in the case with magnetic compression
for shot 39475 and simulation 2350. The simulated line averaged electron
density along the interferometer chord at $r=35$ mm hasn't been included
in figure \ref{fig:ne_IDcomp-1}(a) because the experimental data
for that chord is not available. It can be seen how the time it takes
for plasma to be advected up the gun and enter the CT containment
region is underestimated by $\sim5\,\upmu$s in this simulation. The
electron density and ion temperature magnitudes during CT decay and
magnetic compression are approximately reproduced by the simulation.
Figure \ref{fig:ne_IDcomp-1}(c) shows the comparison between the
experimental and simulated ion-Doppler measurements for shot  39510
and simulation 2351. Like shot 39475, shot 39510 was also a flux conserving
shot, but with $t_{comp}=40\,\upmu$s, and increased compressional
energy, with $V_{comp}=18$ kV. Code input parameters for simulation
2351 were approximately the same as for simulation 2353 (table \ref{tab:Sim parameters}),
but with $t_{comp}=40\,\upmu$s. 
\begin{figure}[H]
\subfloat[]{\raggedright{}\includegraphics[width=7cm,height=5cm]{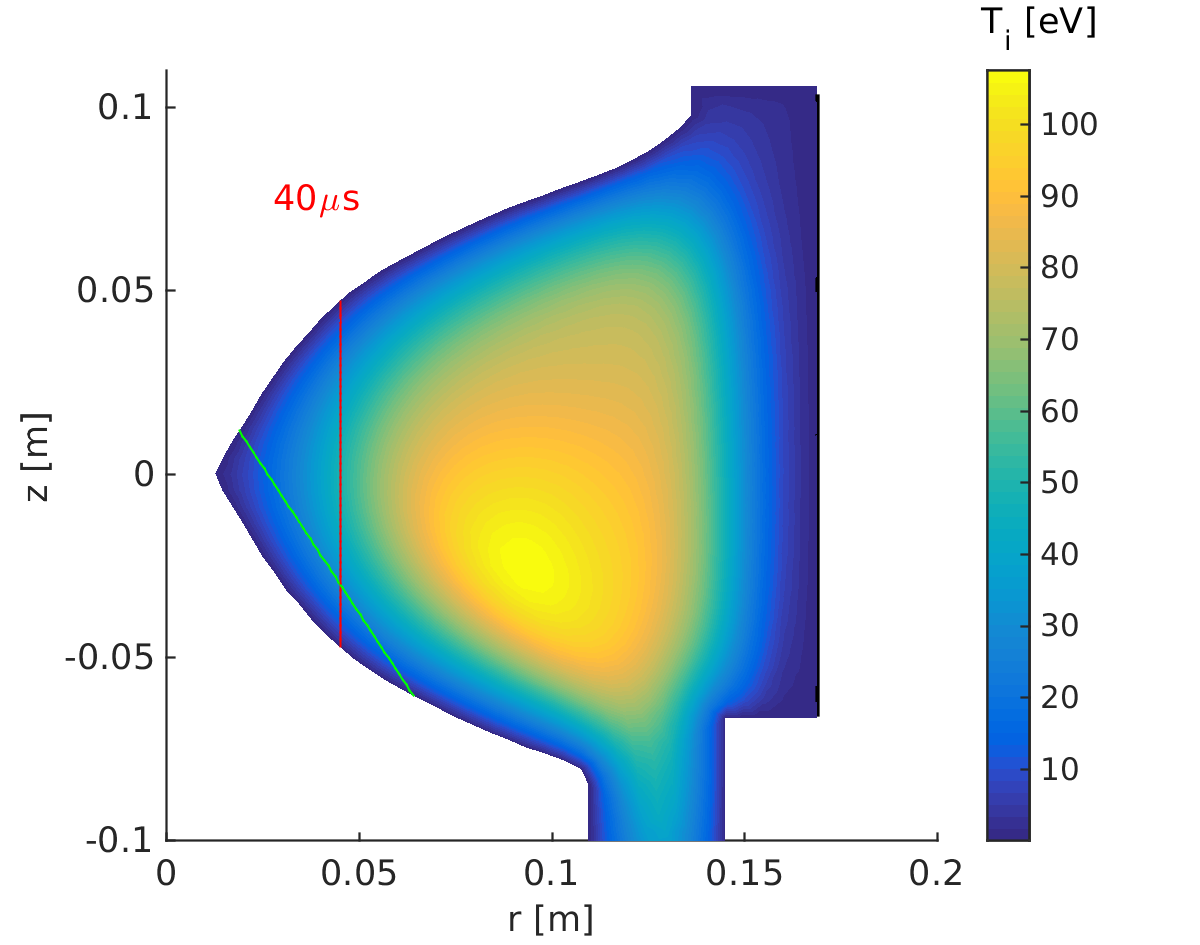}}\hfill{}\subfloat[]{\raggedright{}\includegraphics[width=7cm,height=5cm]{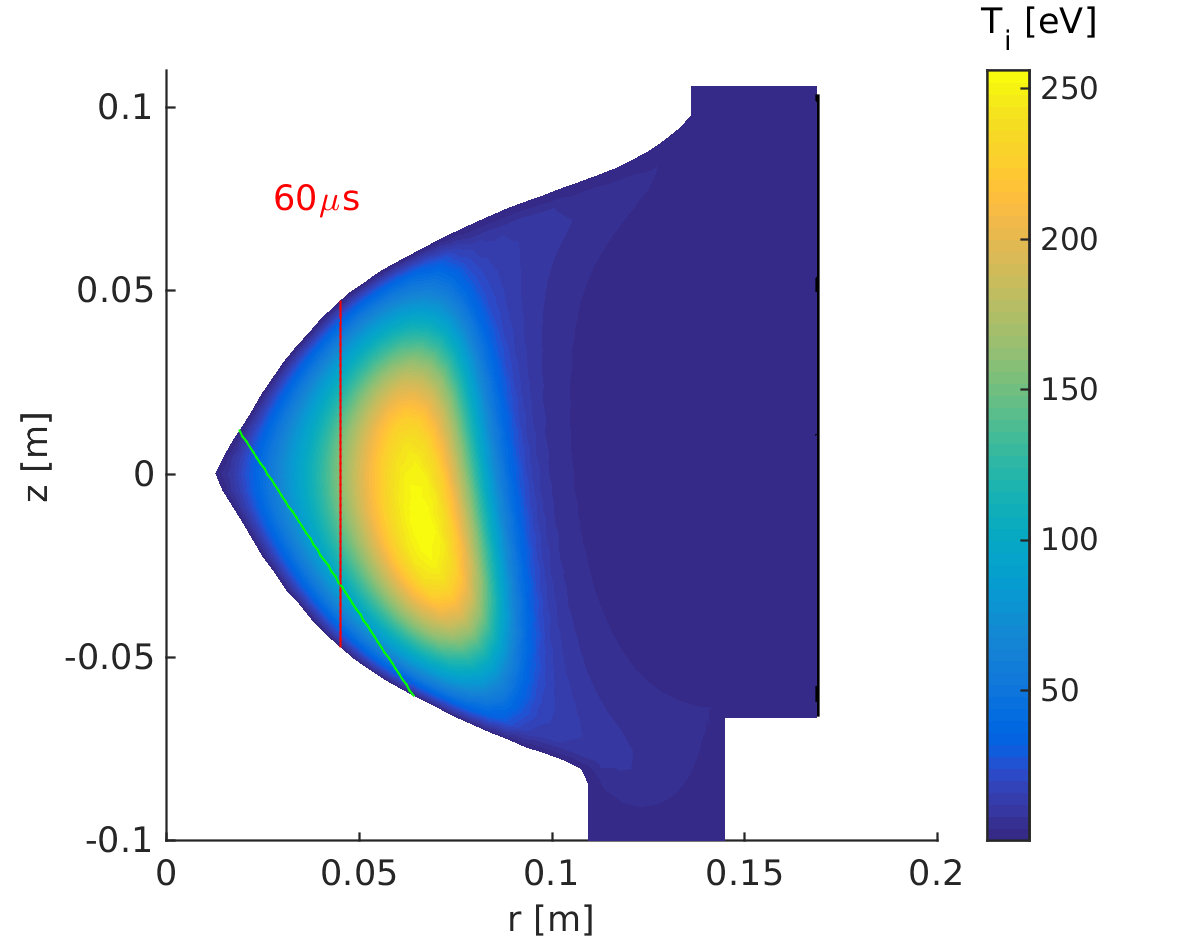}}

\caption{\label{fig:Ti}$\,\,\,\,$$T_{i}$ contours, simulation 2351 }
\end{figure}
When the experimental ion-Doppler measurement is matched by simulations,
simulated core ion temperature increases by a factor of around 2.5
over the main compression cycle, as indicated in figure \ref{fig:Ti},
in which contours of ion temperature, for a simulation of shot 39510,
are shown just prior to magnetic compression ($t=40\,\upmu$s) and
at around peak compression ($t=60\,\upmu$s). As seen from figure
\ref{fig:Chalice}, the ion-Doppler chords are located well away from
the CT core. 
\begin{figure}[H]
\begin{centering}
\subfloat{\raggedleft{}\includegraphics[width=10cm,height=6cm]{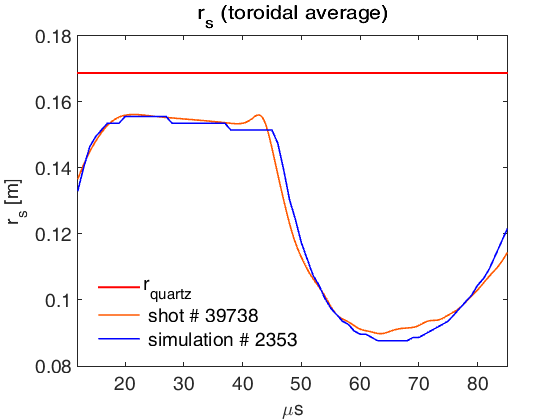}}
\par\end{centering}
\caption{\label{fig:Rsep_comp_comparison}$\,\,\,\,$Comparison of measured
and simulated $r_{s}$ with magnetic compression }
\end{figure}
Figure \ref{fig:Rsep_comp_comparison} shows the comparison obtained
between the toroidally-averaged, experimentally inferred separatrix
radius, $r_{s}(t)$ (from figure \ref{fig:rsepCOMP39738}(b)), and
the separatrix radius evaluated in MHD simulation 2353. Like shot
 39475 (for which side probe data, and hence experimental determination
of $r_{s}(t)$ was not available), shot $39738$ was a flux-conserving
shot, and a good match is found between experimentally determined
and simulated $r_{s}$. Note that $V_{comp}=18$ kV  for shot 39738
and simulation 2353.

\subsubsection{Shot  39735\label{par:Shot-=00002339735}}

\begin{figure}[H]
\begin{centering}
\subfloat[$B_{\theta}$, shot 39735 ]{\raggedright{}\includegraphics[width=8cm,height=5cm]{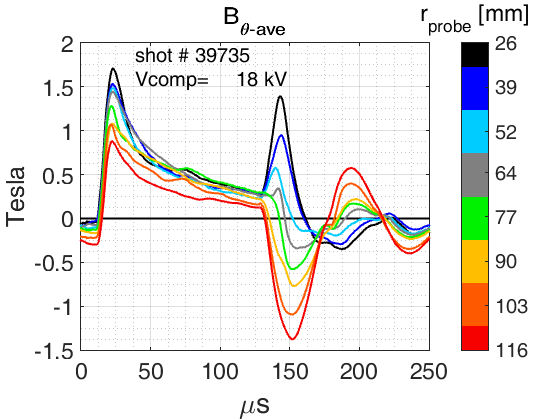}}\hfill{}\subfloat[$B_{\theta}$, simulation 2287]{\raggedleft{}\includegraphics[width=8cm,height=4.8cm]{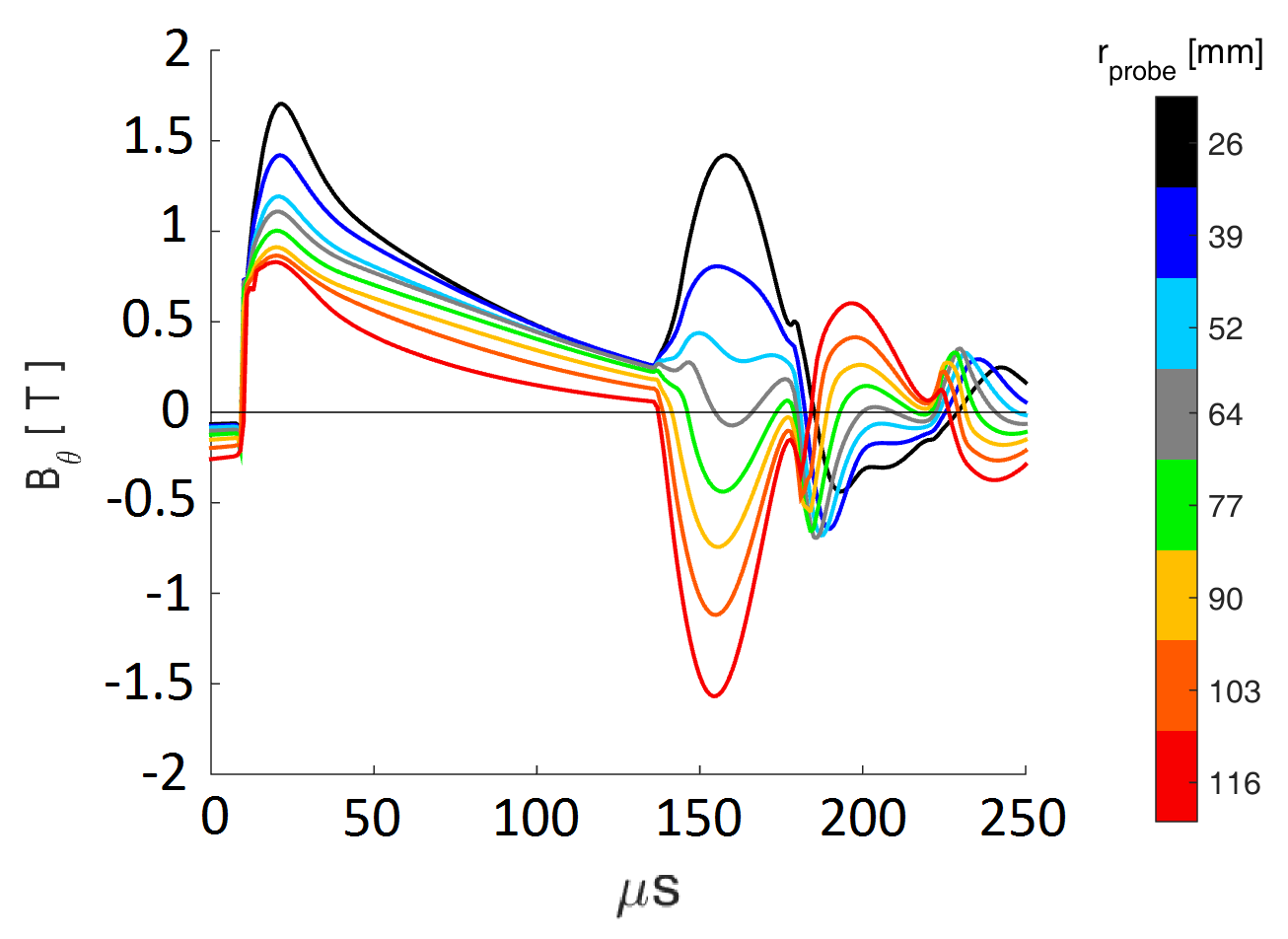}}
\par\end{centering}
\begin{centering}
\subfloat[$B_{\phi}$, shot 39735 ]{\raggedright{}\includegraphics[width=8cm,height=5cm]{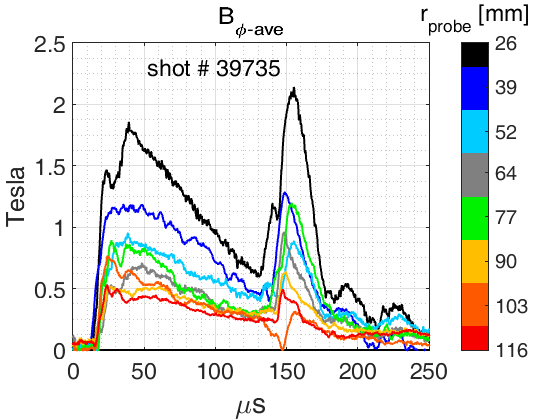}}\hfill{}\subfloat[$B_{\phi}$, simulation 2287]{\raggedleft{}\includegraphics[width=7.8cm,height=4.8cm]{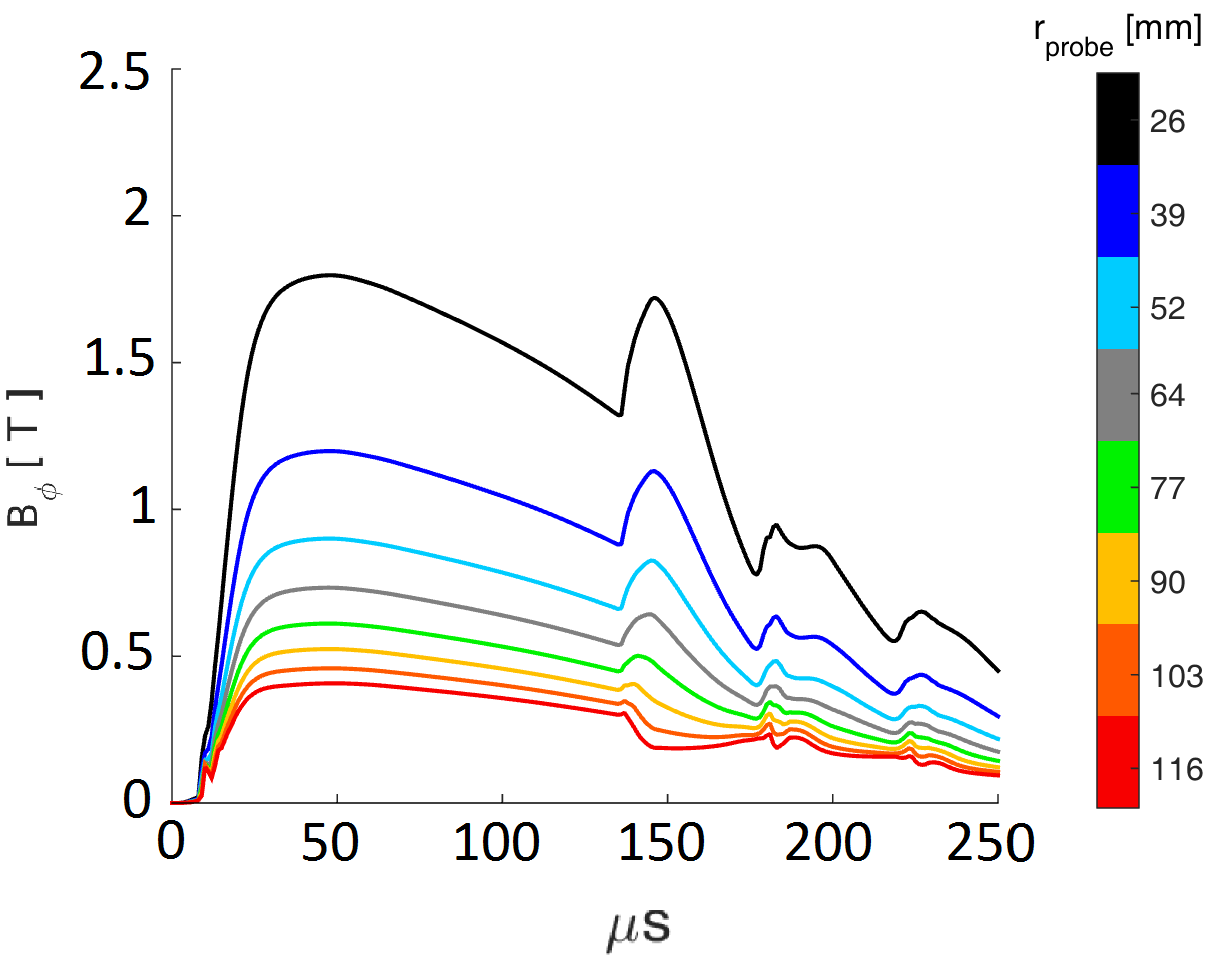}}
\par\end{centering}
\caption{$\,\,\,\,$Comparison of measured and simulated poloidal and toroidal
magnetic field at magnetic probe locations ($V_{comp}=18$ kV, 11-coil
configuration)\label{fig:Bpol_Btor_meas_cf_sim39735_2287} }
\end{figure}
Figure \ref{fig:Bpol_Btor_meas_cf_sim39735_2287} shows the comparison
between experimentally measured and simulated poloidal and toroidal
field, where the experimental measurements have been toroidally averaged
for clarity. Shot  39735 presented here (see also figures \ref{fig:39735Bp}
and \ref{fig:Bpand--traces}(a)) had $V_{comp}=18$ kV, and $t_{comp}=130\,\upmu$s.
The simulation was run until $t_{sim}=250\,\upmu$s, and includes
the time when the current in the compression coils changes polarity.
Poloidal flux and $f$ contours for this simulation were presented
in section \ref{subsec:ring_psi_f}. In shot  39735, the poloidal
field measured at the inner probes collapses at $\sim145\,\upmu$s,
while the compression coil current peaks at $\sim150\,\upmu$s. Because
of this, as outlined in section \ref{Chap:Magnetic-Compression},
shot  39735 had parameter $\widetilde{\tau}_{c}=0.6$, implying that
poloidal flux was not well conserved during compression. Apart from
resistive losses, the simulation conserves poloidal flux, so the poloidal
field at the inner probes (\ref{fig:Bpol_Btor_meas_cf_sim39735_2287}(b))
continues to rise until the compression coil current peaks. 

The compressional instability leads to toroidal field measurements
that are toroidally very asymmetric, and the axisymmetric code cannot
reproduce this. Comparison of figures \ref{fig:Bpol_Btor_meas_cf_sim39735_2287}(c)
and \ref{fig:Bpol_Btor_meas_cf_sim39735_2287}(d) shows how the simulated
$B_{\phi}$ does, in general, rise at the magnetic probes as crowbarred
shaft current increases when it is diverted to a lower inductance
path (as described in section \ref{subsec:Compressional-Instability}).
There is at least a qualitative agreement between the simulated $B_{\phi}$
and the toroidal-averages of the measured $B_{\phi}.$

\begin{figure}[H]
\begin{centering}
\subfloat{\raggedleft{}\includegraphics[width=10cm,height=6cm]{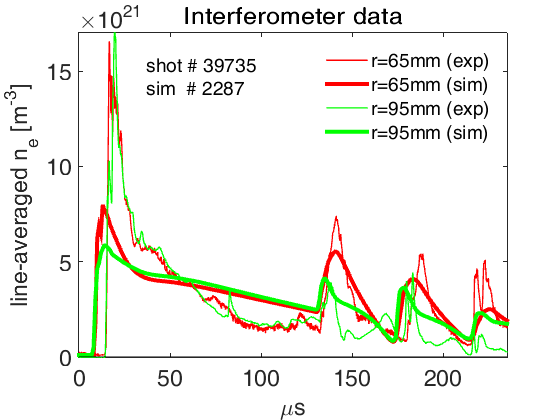}}
\par\end{centering}
\caption{$\,\,\,\,$Comparison of measured and simulated $n_{e}$ (compression
shot  39735)\label{fig:n_39735_2287}}
\end{figure}
Figure \ref{fig:n_39735_2287} indicates the qualitative agreement
between experimentally measured and simulated electron density. Note
that ion-Doppler and CT outer separatrix measurements were not available
for shot  39735. 

\subsubsection{Simulation of compressional flux loss\label{subsec:Simulation-of-comp_flux_loss}}

\begin{figure}[H]
\subfloat[$B_{\theta}$ for shot  29089 ]{\raggedright{}\includegraphics[width=7cm,height=5cm]{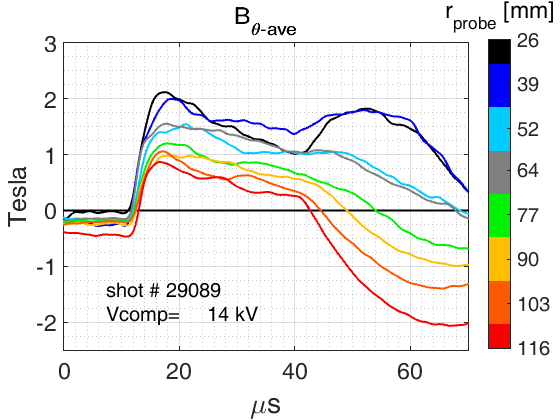}}\hfill{}\subfloat[$B_{\theta}$ for simulation  1720]{\raggedleft{}\includegraphics[width=9cm,height=5cm,keepaspectratio]{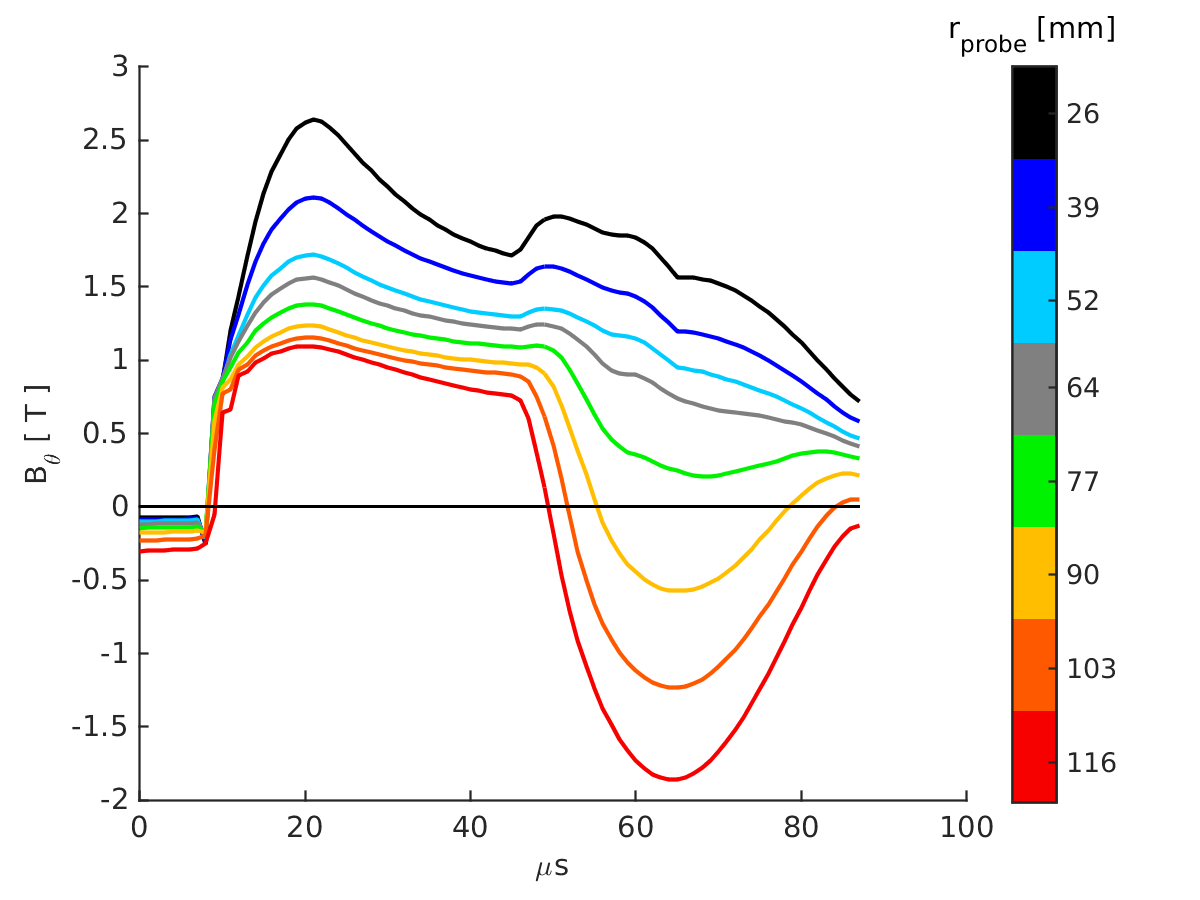}}

\caption{\label{fig:Bpol_meas_cf_sim29089_1720}$\,\,\,\,$Measured $cf.$
simulated $B_{\theta}$ (compression - 6 coils, with flux loss) }
\end{figure}
As discussed in section \ref{subsec:Comparison-of-compression}, it
was usual, for shots taken in the 6-coil configuration, for the poloidal
field measured at the inner probes to collapse during magnetic compression,
leading to low values of the experimentally determined compressional
parameter $\widetilde{\tau}_{c}$. It was assumed that low values
of $\widetilde{\tau}_{c}$ implied poor compressional flux conservation.
To try and verify this assumption, simulations were run in which CT
poloidal flux loss is imposed. Figure \ref{fig:Bpol_meas_cf_sim29089_1720}(a)
shows a typical example of measured poloidal field for a compression
shot taken in the 6-coil configuration. Poloidal field at the inner
probes initially rises at compression, but then rapidly collapses.
For comparison, poloidal CT flux is forced to decrease by 70\% over
the 20$\,\upmu$s between $t_{comp}=45\,\upmu$s and peak compression
at $\sim65\,\upmu$s in simulation  1720, and the resultant poloidal
field traces in figure \ref{fig:Bpol_meas_cf_sim29089_1720}(b) are
a reasonable match to those in shot  29089, which has compression
parameter $\widetilde{\tau}_{c}=0.5$, helping to confirm the hypothesis. 

\subsection{Results from additional simulated diagnostics\label{subsec:Results-from-additional}}

In this section, results from simulated diagnostics that don't have
experimental counterparts will be presented and discussed.

\subsubsection{Simulated diagnostics\protect \\
for $\psi_{CT}(t),\,T_{max}(t),\,r_{s}(t),\,r_{axis}(t),\,V_{CT}(t),\,\mbox{ and }\beta(t)$\label{subsec:SimDiagMaxTetc}}

\begin{figure}[H]
\subfloat[]{\raggedright{}\includegraphics[width=8cm,height=5cm]{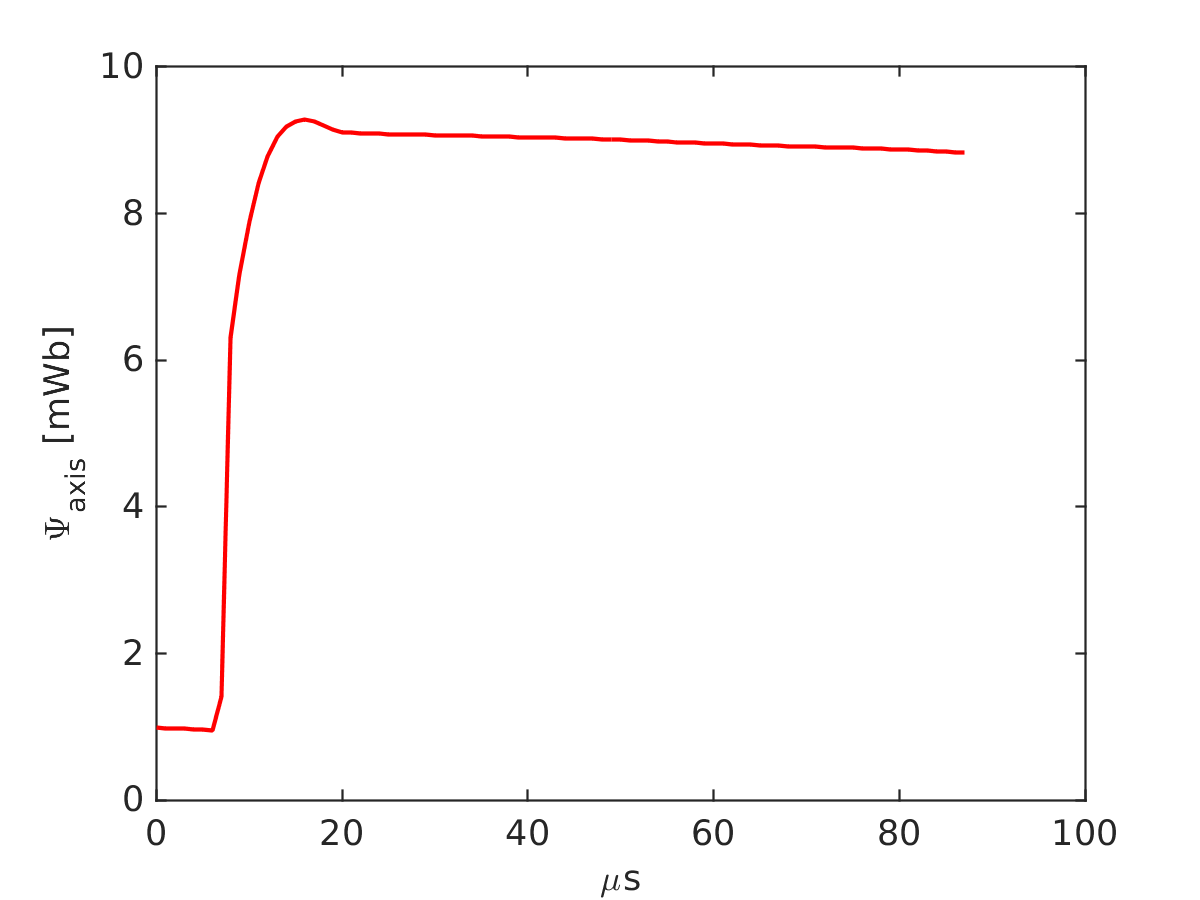}}\hfill{}\subfloat[]{\raggedleft{}\includegraphics[width=8cm,height=5cm]{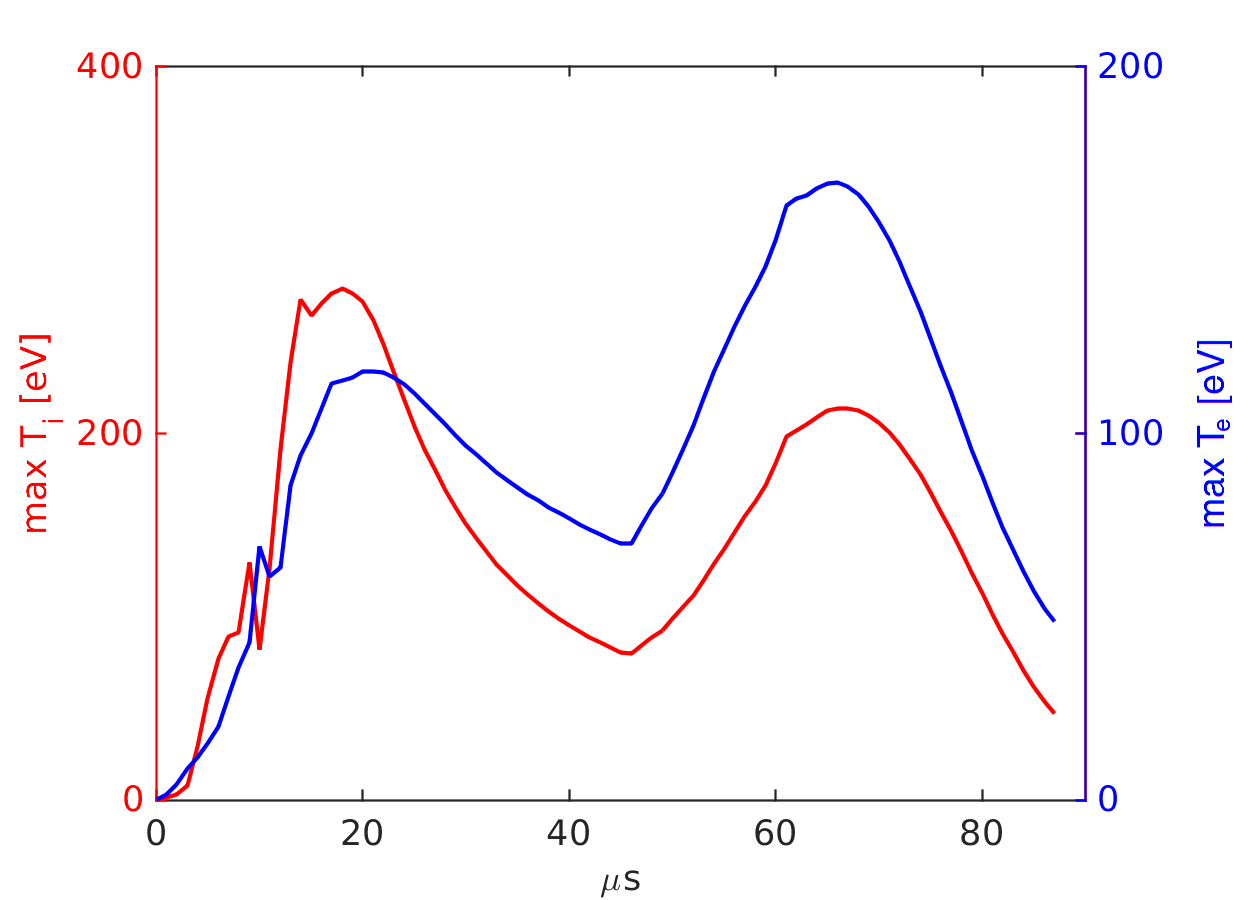}}

\subfloat[]{\raggedright{}\includegraphics[width=8cm,height=5cm]{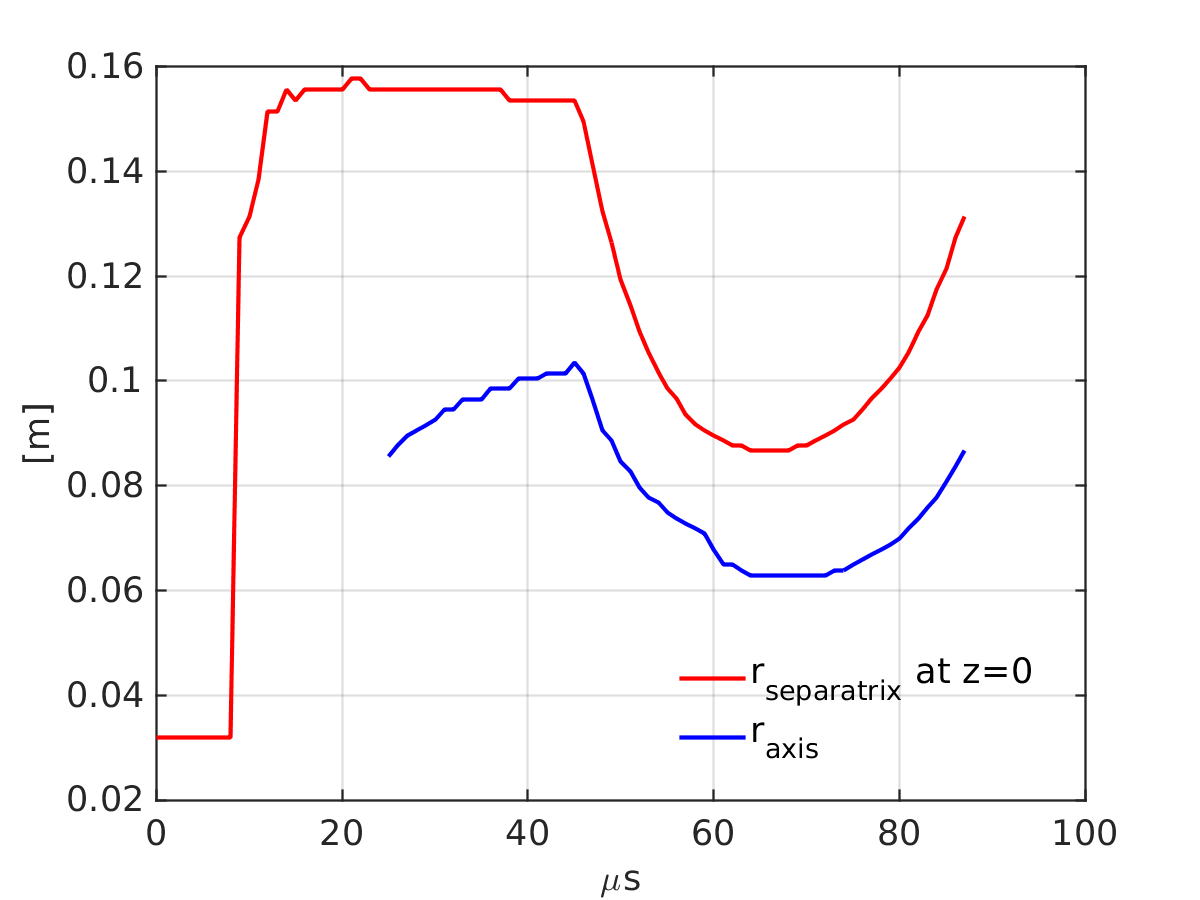}}\hfill{}\subfloat[]{\raggedleft{}\includegraphics[width=8cm,height=5cm]{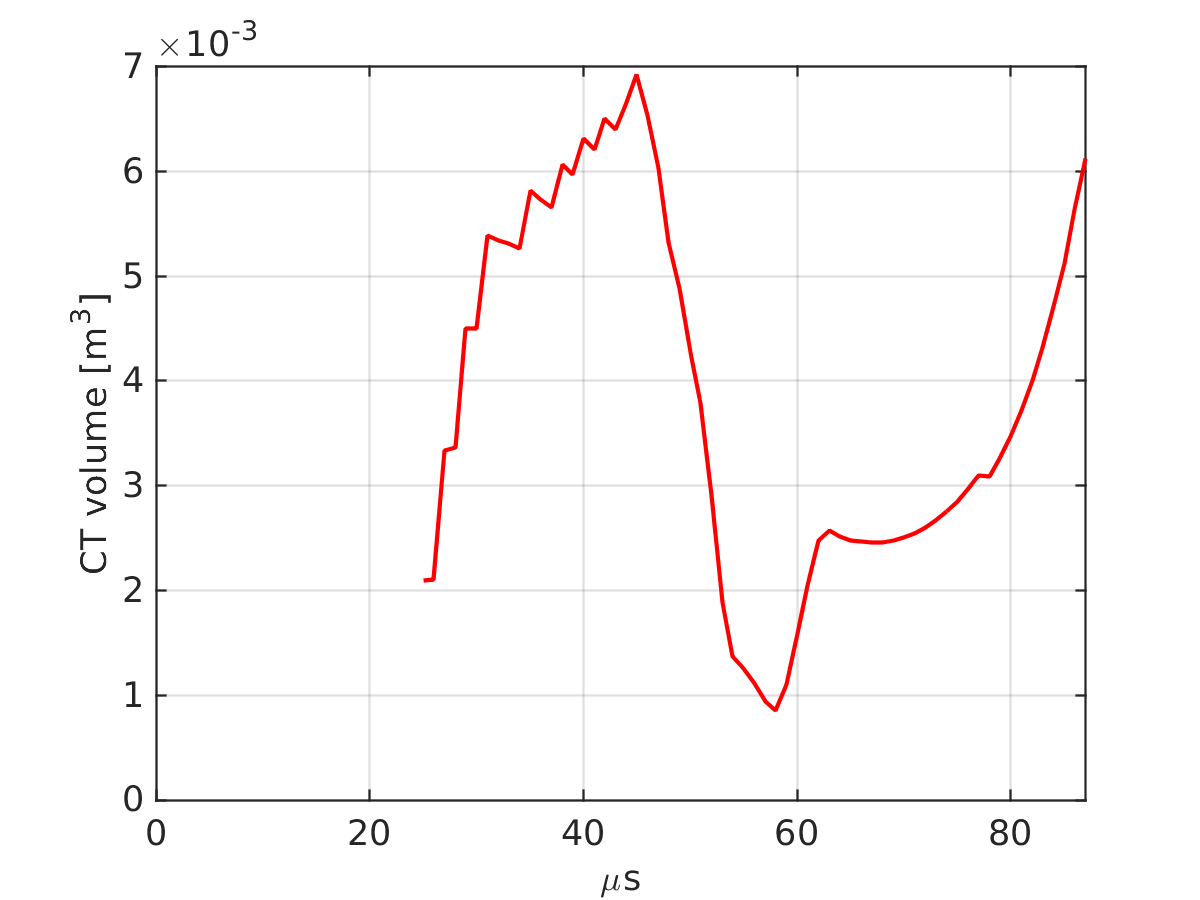}}

\subfloat[]{\raggedright{}\includegraphics[width=8cm,height=5cm]{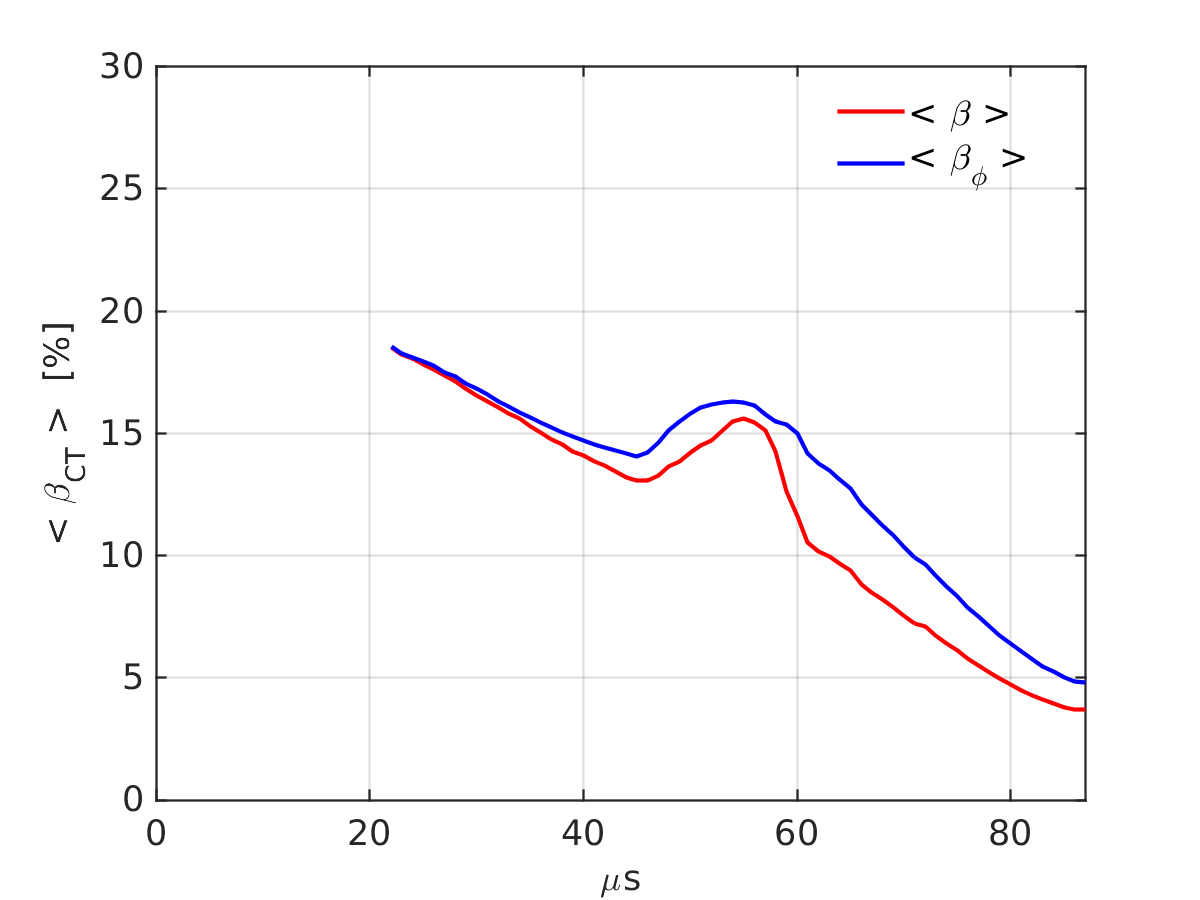}}\hfill{}\subfloat[]{\raggedleft{}\includegraphics[width=8cm,height=5cm]{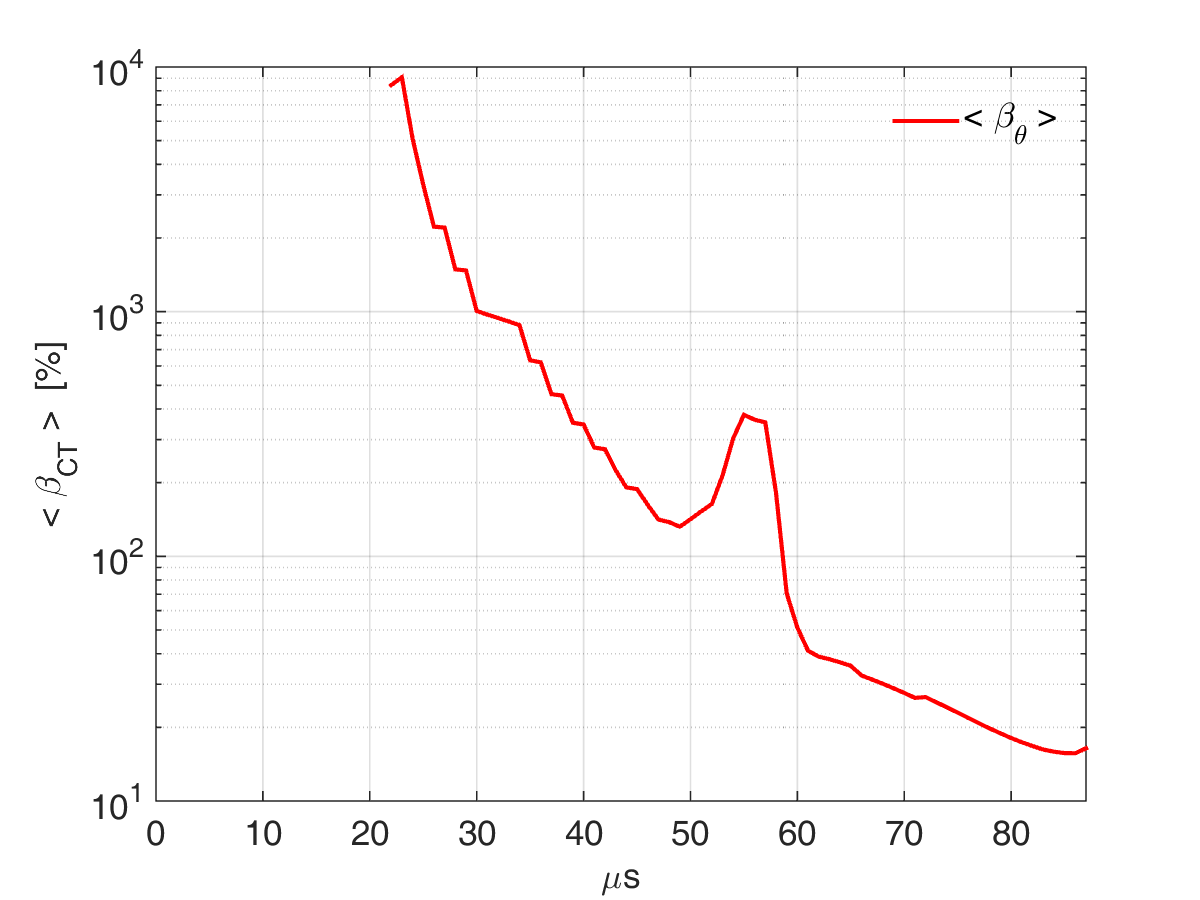}}

\caption{\label{fig:varSIM2353}$\,\,\,\,$Selection of additional simulated
diagnostics for simulation  2353}
\end{figure}
Figure \ref{fig:varSIM2353} indicates the evolution of various parameters
for simulation  2353, which had the code inputs listed in table \ref{tab:Sim parameters}.
Figure \ref{fig:varSIM2353}(a) indicates the evolution of $\psi_{CT}$
, the value of $\psi$ at the magnetic axis of the CT, and is calculated
as the maximum value of $\psi$ in the part of the simulation domain
representing the CT confinement region. Poloidal flux is approximately
conserved throughout the simulation, also at compression. The gradual
drop in $\psi_{CT}$ over time is due to resistive decay of the CT
toroidal current. 

Figure \ref{fig:varSIM2353}(b) shows the maximum ion and electron
temperatures in the computational domain. Ions attains their maximum
temperature around the entrance to the CT confinement region, due
to viscous heating during the formation process, when they are accelerated
to extreme velocities by the action of the formation $\mathbf{J}_{r}\times\mathbf{B}_{\phi}$
force, and by the jets associated with magnetic reconnection of the
poloidal stuffing flux as closed CT flux surfaces are formed (see
figure \ref{fig: Ti_11coils}(d)). The electron fluid attains its
maximum temperature later around peak compression, in the CT core,
due to compressional heating and enhanced ohmic heating.

Simulated CT outer separatrix, $r_{s}(t)$, also indicated in figure
\ref{fig:Rsep_comp_comparison}, is plotted in figure \ref{fig:varSIM2353}(c),
along with $r_{axis}(t)$, which the $r$ coordinate of the magnetic
axis of the CT. $r_{axis}(t)$ is calculated by finding the $r$ coordinate
of the location of $\psi_{CT}$ over time. Note that the CT, and $r_{axis}$,
is defined only after the first closed poloidal flux surface is formed,
at around 25$\,\upmu$s. It can be seen that the simulation indicates
that the aspect ratio $\left(r_{s}-r_{axis}\right)/r_{axis}$ is approximately
constant over compression. 

Figure \ref{fig:varSIM2353}(d) indicates the evolution of the CT
volume, calculated using equation \ref{eq:300}, over time. CT volume
decreases by a factor of over three during compression. As shown in
figures \ref{fig: psi_11coils}(d) and (e), simulations indicates
that open field lines surround the closed flux surfaces until shortly
after halfway through the primary compression cycle. These are pinched
off and magnetically reconnect to form additional closed CT flux surfaces
at compression, explaining the sharp increase in CT volume at $t\sim59\,\upmu$s.
This increase is absent when compression initiation is delayed to
a time when there are no open lines surrounding the CT. 

Figures \ref{fig:varSIM2353}(e) and (f) indicate how magnetic energy
associated with CT toroidal field is greater than that associated
with CT poloidal field. $<\beta>,\,<\beta_{\theta}>$, and $<\beta_{\phi}>$
are calculated using equation \ref{eq:301}. The volume-averaged magnetic
pressure associated with poloidal field is greater than volume-averaged
plasma pressure only after $\sim57\,\upmu$s (figure \ref{fig:varSIM2353}(f)).
At this time, open field lines surrounding the CT are pinched off
during magnetic compression, to form additional closed field lines
that are then associated with the exterior of the CT. When this occurs,
high levels of toroidal current flowing along the originally open
field lines results in a sudden increase of poloidal field associated
with the CT exterior. Note that because the simulated CT toroidal
current profiles are extremely hollow (see figure \ref{fig: Jp_11coils}),
the poloidal magnetic field is particularly weak in the CT interior.
Simulated volume-averaged total beta increases from $\sim13\%$ to
$\sim16\%$ over compression. 

\subsubsection{Simulated internal magnetic field\label{subsec:SimDiagBint} }

\begin{figure}[H]
\subfloat[]{\raggedright{}\includegraphics[width=8cm,height=5cm]{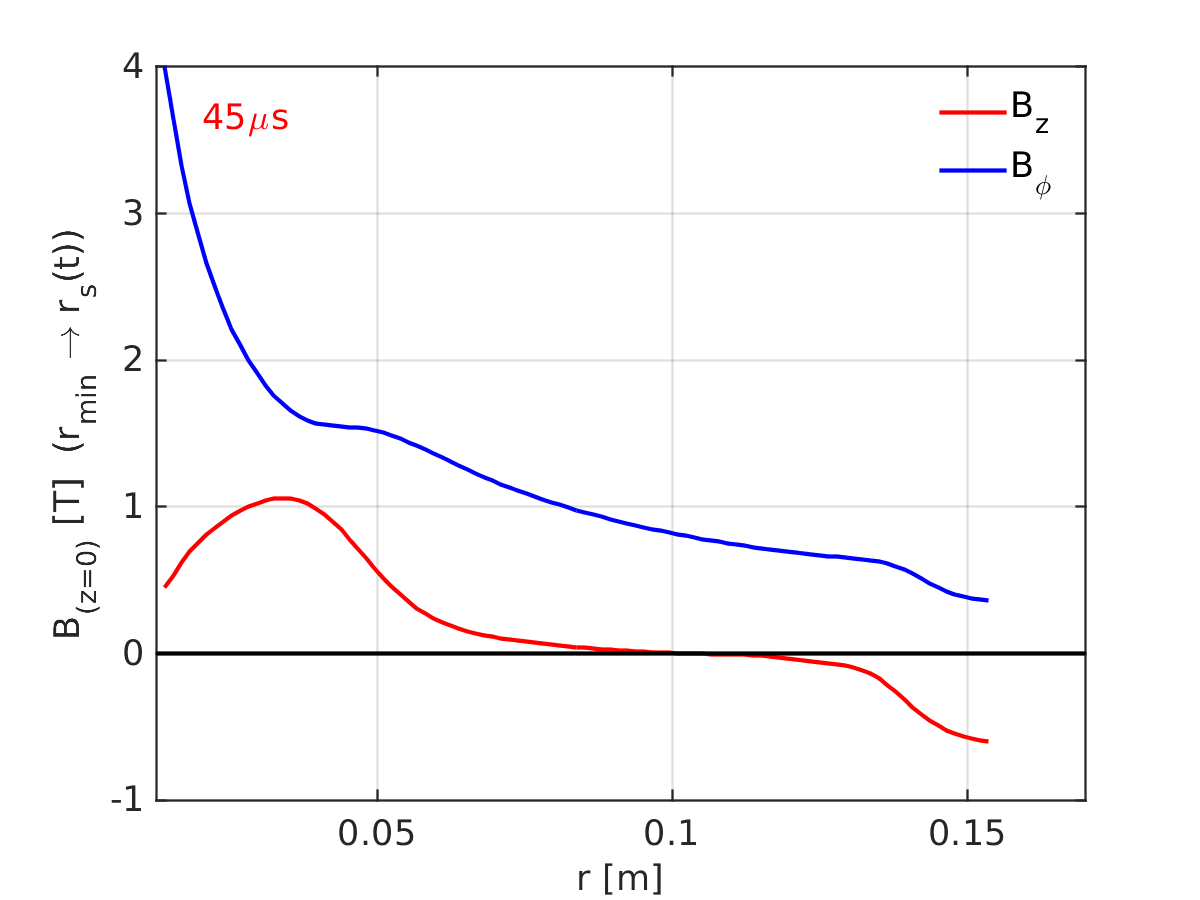}}\hfill{}\subfloat[]{\raggedleft{}\includegraphics[width=8cm,height=5cm]{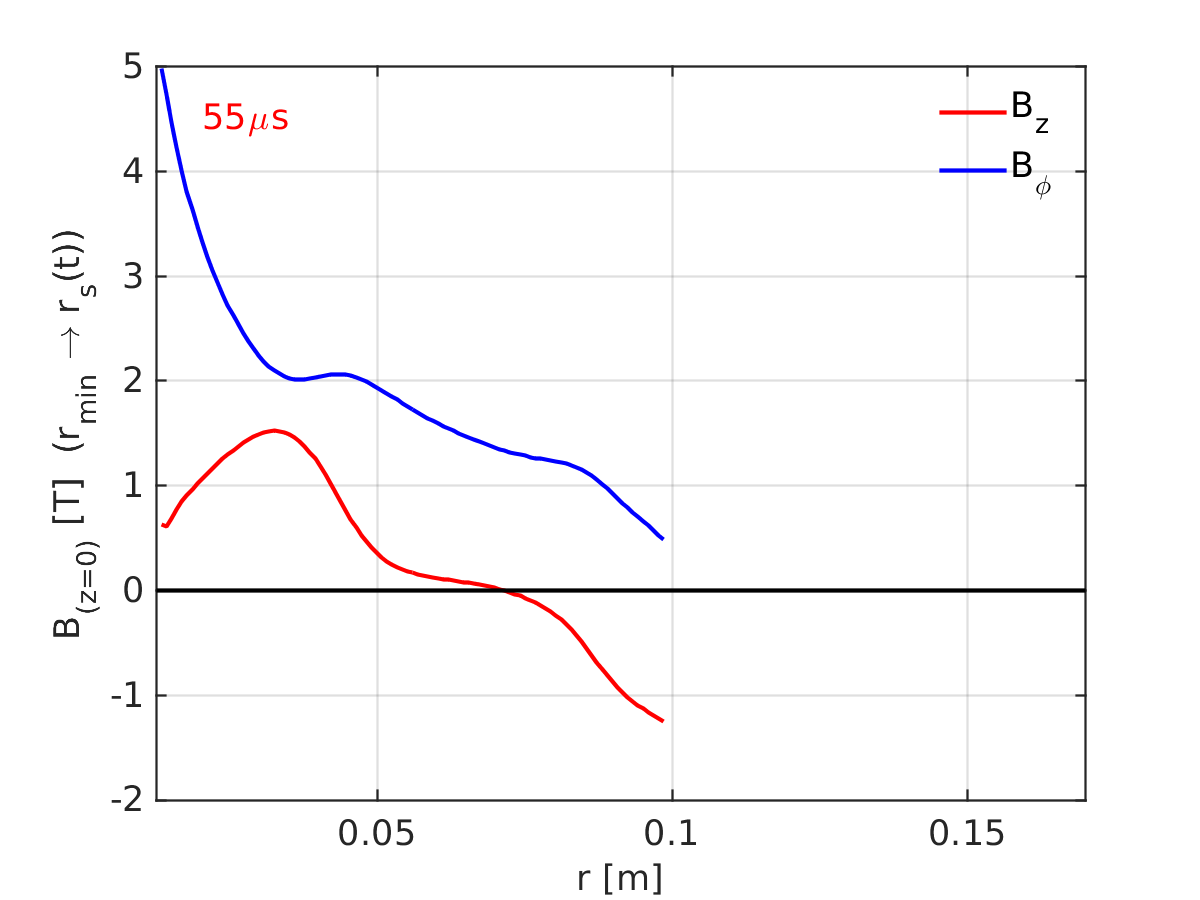}}

\subfloat[]{\raggedright{}\includegraphics[width=8cm,height=5cm]{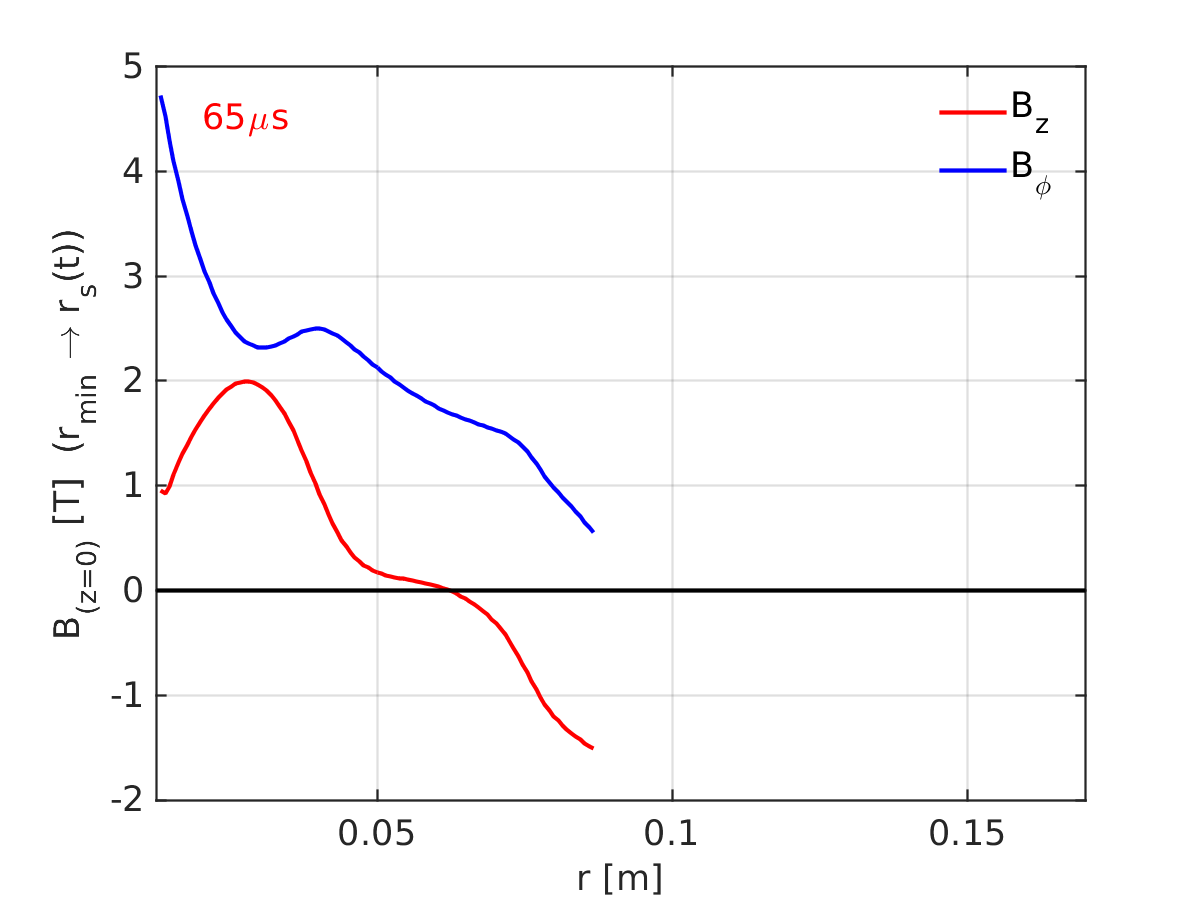}}\hfill{}\subfloat[]{\raggedleft{}\includegraphics[width=8cm,height=5cm]{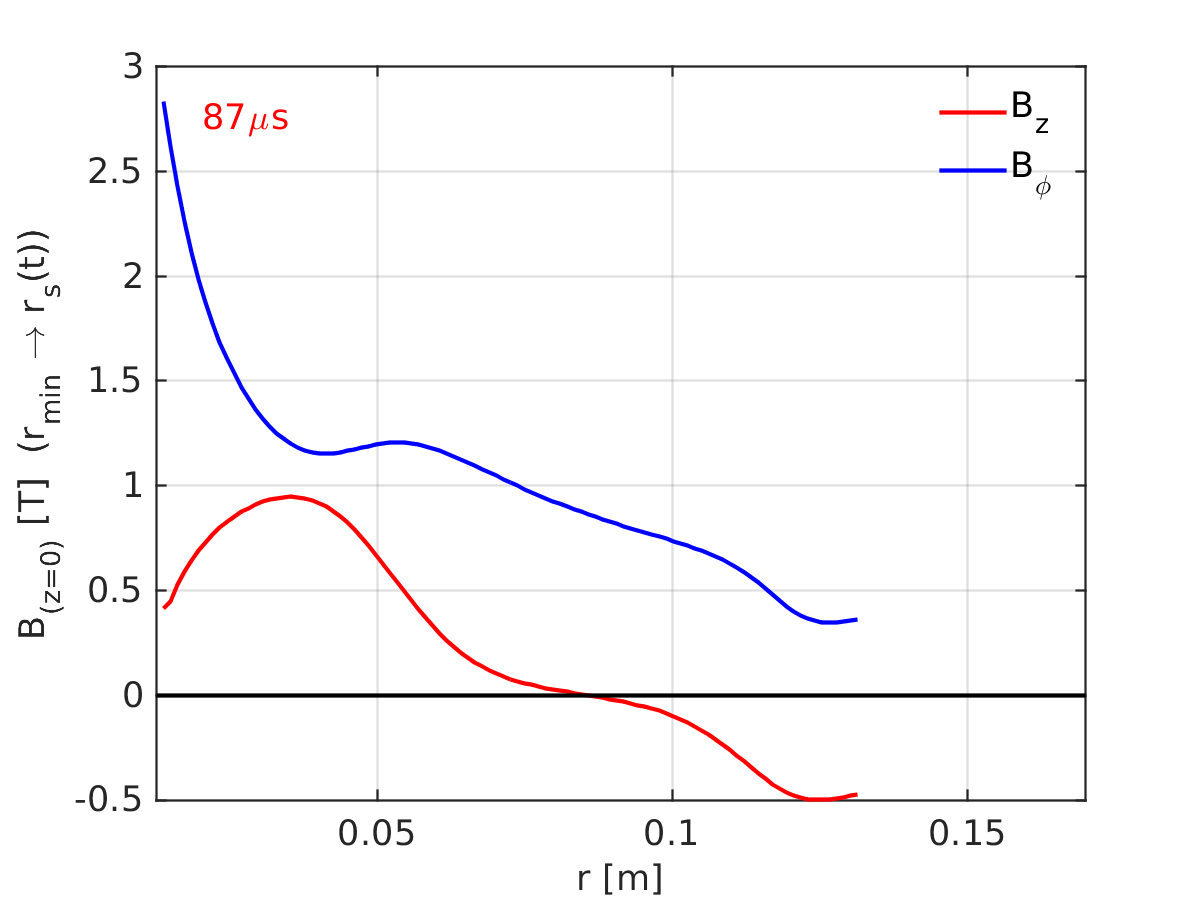}}

\caption{\label{fig:BzBp_SIM2353}$\,\,\,\,$Internal CT $B_{z}$ and $B_{\phi}$
profiles from simulation  2353}
\end{figure}
For the same simulation  2353 ($t_{comp}=45\,\upmu$s), figure \ref{fig:BzBp_SIM2353}
indicates the simulated values, at $t=45,\,55,\,65$, and $87\,\upmu$s,
of $B_{z}$ and $B_{\phi}$ along the horizontal line spanning from
the chalice waist at $(r,\,z)=(r_{min}\sim1.4$ cm, 0 cm) to the CT
outboard equatorial separatrix at $(r,\,z)=(r_{s}(t)$, 0 cm). Although
the CT magnetic axis is not located precisely at $z$=0, especially
at earlier times, and the closed poloidal field\textbf{ }lines are
not exactly circular,\textbf{ $B_{z}$ }is still a good approximation
for poloidal field at $z$=0.\textbf{ }In each of these graphs, the
$r$ component of the magnetic axis location, also indicated in figure
\ref{fig:varSIM2353}(c), is the $r$ coordinate at which $B_{z}$
crosses through zero. Note that the angular shape of the chalice wall,
at the waist at $(r,\,z)=(r_{min}\sim1.4$ cm, 0 cm), results in expansion
and weakening of the poloidal field in that region from $r_{min}\sim1.4$
cm to $r\sim3$ cm at $z=0$ cm, as can be seen in figures \ref{fig:BzBp_SIM2353}(a)
to (d), and also in figures \ref{fig: psiC_11coils}(c) to (f). Also
evident from \ref{fig:BzBp_SIM2353}(a) to (d) is that the CT toroidal
field falls especially rapidly with increasing $r$ in that region,
due to the relatively low level of poloidal current there (poloidal
currents sustain the externally imposed $B_{\phi})$, as can also
be seen in figures \ref{fig: FC_11coils}(c) to (f), and in figures
\ref{fig: Jpol_11coils}(c) to (f).

Comparing the profiles with the depiction of the field profiles for
the typical tokamak and spheromak configurations in figure \ref{fig:Bprof_toka_sphero},
it can be seen how the CT field profile has more in common with that
of a tokamak. Crow-barred shaft current flowing around the confinement
region produces a tokamak-like toroidal field profile that, in a tokamak,
is due to current in the toroidal field coils (figure \ref{fig:Tok_icf}(a)).
However, as shown in section \ref{subsec:SimDiagQprofile}, the simulated
CT $q$ profile indicates that the shaft current is insufficient to
reproduce a typical tokamak $q$ profile. 

The magnetic compression experiment did not have internal magnetic
measurements that would enable confirmation of these simulated diagnostics
for internal field profiles. It is likely that the 2D simulations
overestimates the level of hollowness of the current profiles. On
the other hand, the magnetic measurements that do exist (magnetic
field at the chalice walls) are well matched by the simulated diagnostics,
so it seems reasonable to have some level of trust in the simulated
field diagnostics.

\subsubsection{Simulated $q$ profile\label{subsec:SimDiagQprofile}}

\begin{figure}[H]
\subfloat[]{\raggedright{}\includegraphics[width=7cm,height=5cm]{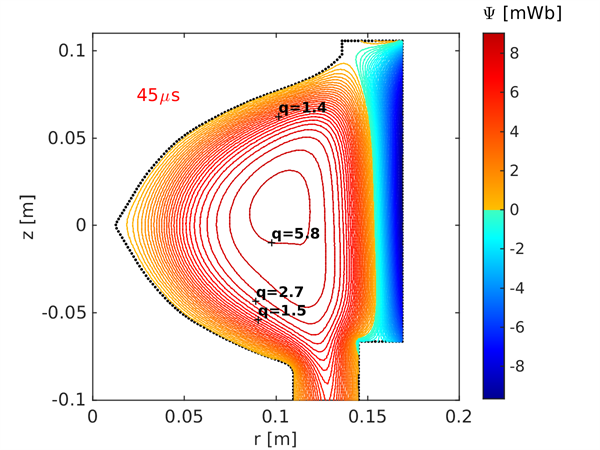}}\hfill{}\subfloat[]{\raggedleft{}\includegraphics[width=7cm,height=5cm]{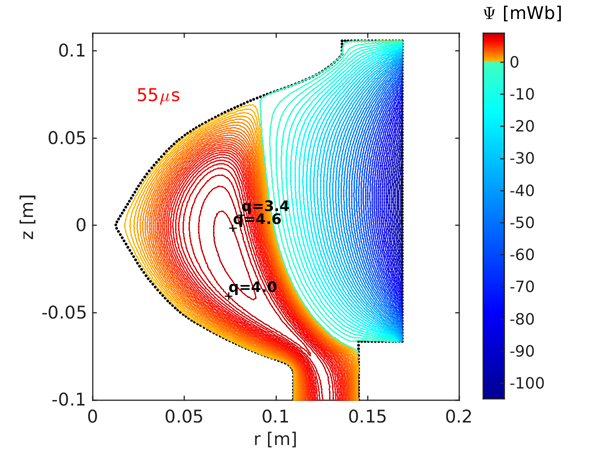}}

\subfloat[]{\raggedright{}\includegraphics[width=7cm,height=5cm]{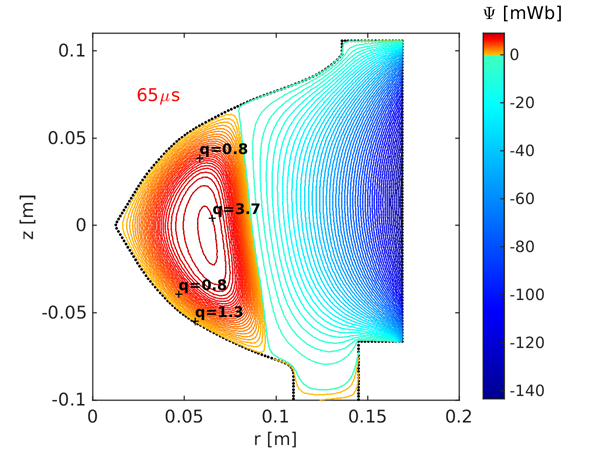}}\hfill{}\subfloat[]{\raggedleft{}\includegraphics[width=7cm,height=5cm]{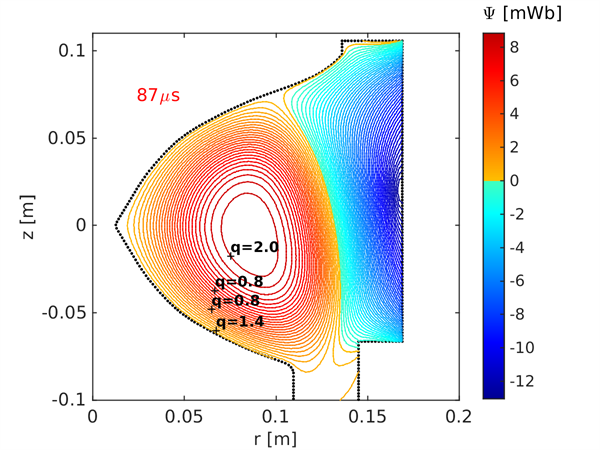}}

\subfloat[]{\raggedright{}\includegraphics[width=7cm,height=5cm]{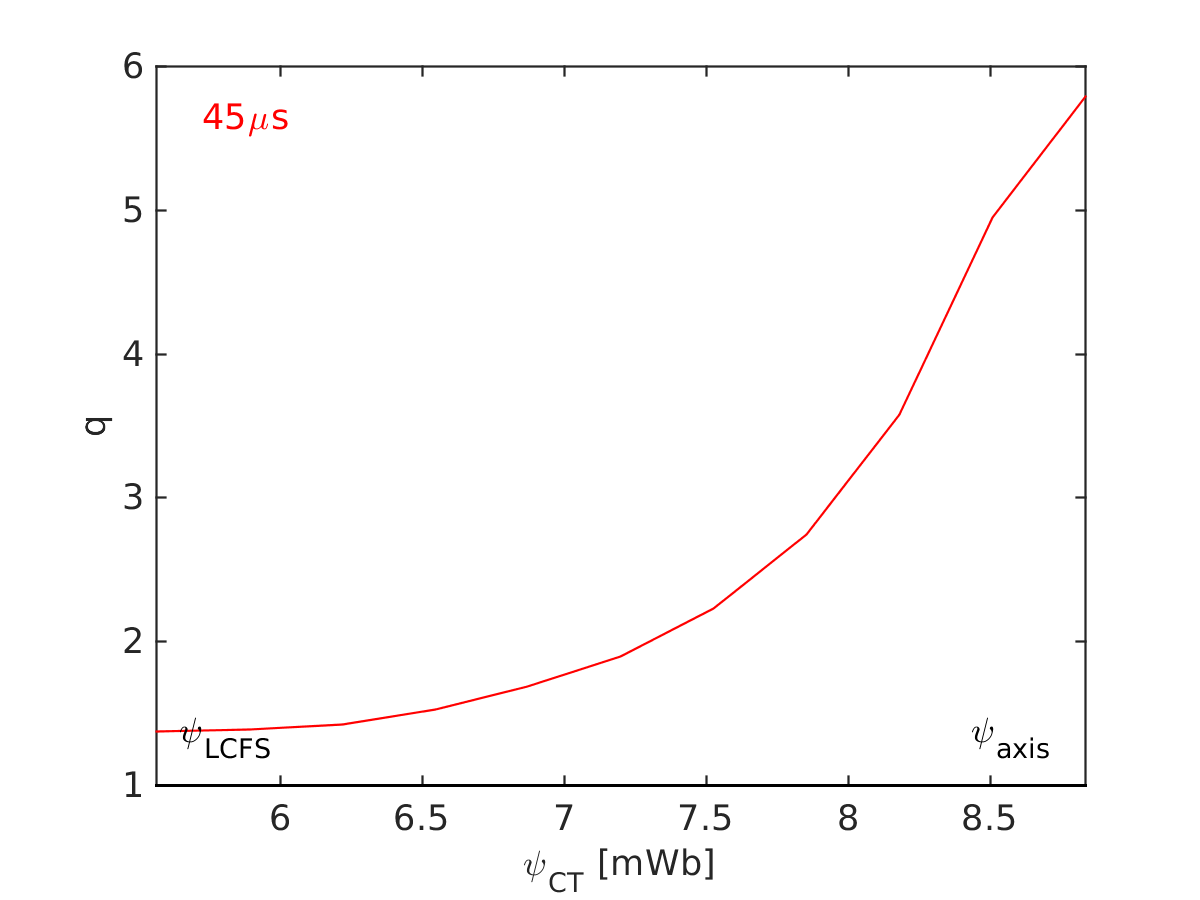}}\hfill{}\subfloat[]{\raggedleft{}\includegraphics[width=7cm,height=5cm]{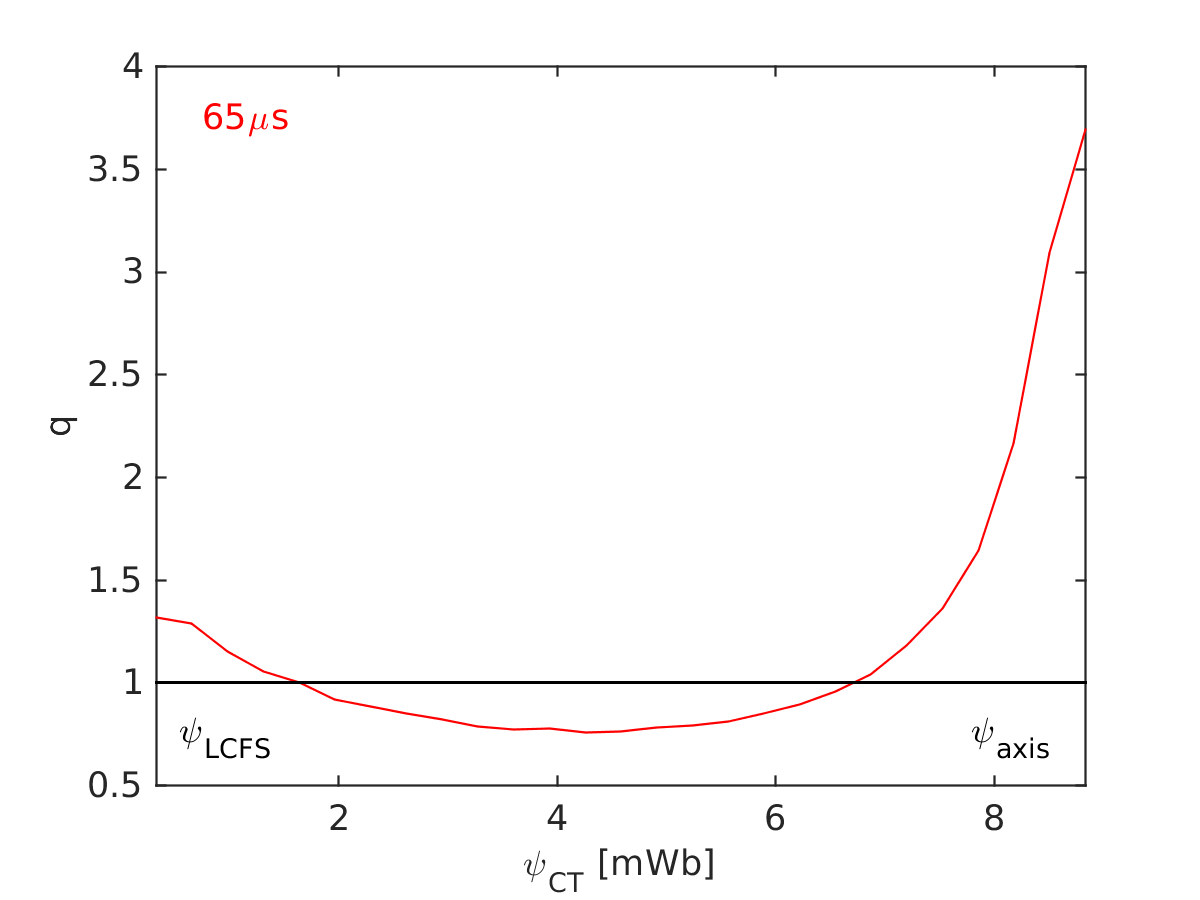}}

\caption{\label{fig:Qpsi_2353}$\,\,\,\,$$q$ profile, simulation  2353}
\end{figure}
Figure \ref{fig:Qpsi_2353} shows the profile of safety factor $q(\psi)$
for simulation  2353 at various times. The method developed to find
the $q$ profile from the simulated magnetic fields has been described
in section \ref{subsec:q-profile}. Simulated $q(\psi)$ shows two
trends at compression, depending on the value of $t_{comp}$. When
the compression banks are fired early in the CT's life, for example
at $t_{comp}=45\,\upmu$s as in simulation  2353, the CT, defined
by regions of closed $\psi$ contours, is still, at $t=t_{comp}$,
surrounded by open field lines that are pinned to the inner and outer
electrodes (figure \ref{fig:Qpsi_2353}(a)), and $q(\psi)$ ranges
from $q\sim5.8$ near the magnetic axis (at $\sim8.75\mbox{ mWb}$)
to $q\sim1.4$ at the last closed flux surface ($\mbox{LCFS}$) at
$\sim5.5\mbox{ mWb}$ (figure \ref{fig:Qpsi_2353}(e)). At compression,
the open field lines surrounding the CT are pinched off, to form additional
closed field lines that are then associated with the exterior of the
CT, as indicated in figure \ref{fig:Qpsi_2353}(c). High levels of
toroidal current flowing along the originally open field lines results
in these field lines being associated with low $q$ when they are
pinched off. At $65\,\upmu$s, $q(\psi)$ ranges from $q\sim3.7$
near the magnetic axis (at $\sim8.7\mbox{ mWb}$) to $q\sim1.3$ at
the $\mbox{LCFS}$ at $\sim0.3\mbox{ mWb}$ (figure \ref{fig:Qpsi_2353}(f)),
while dipping below $q=1$ over a large extent between the magnetic
axis and the $\mbox{LCFS}$. 
\begin{figure}[H]
\subfloat[]{\raggedright{}\includegraphics[width=7cm,height=5cm]{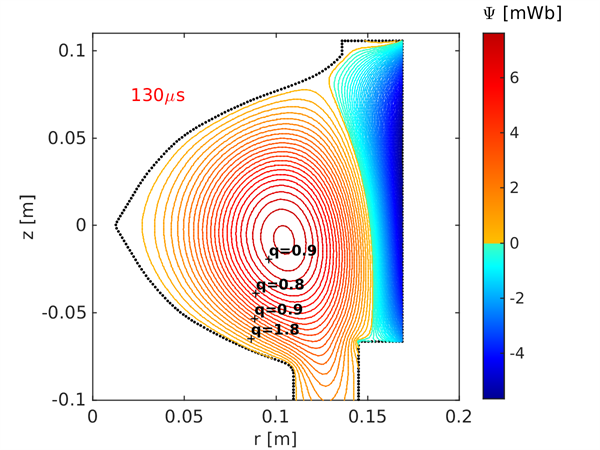}}\hfill{}\subfloat[]{\raggedleft{}\includegraphics[width=7cm,height=5cm]{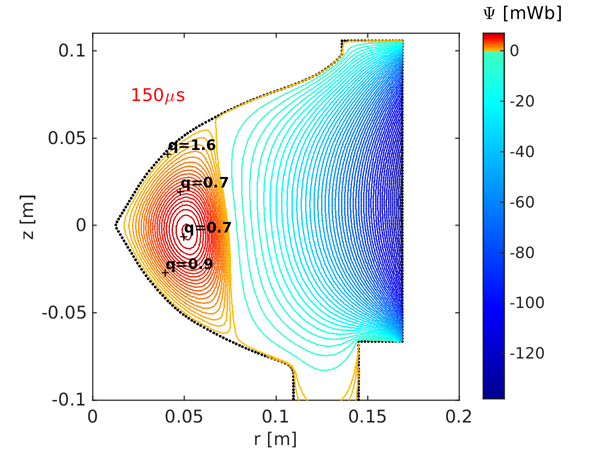}}

\subfloat[]{\raggedright{}\includegraphics[width=7cm,height=5cm]{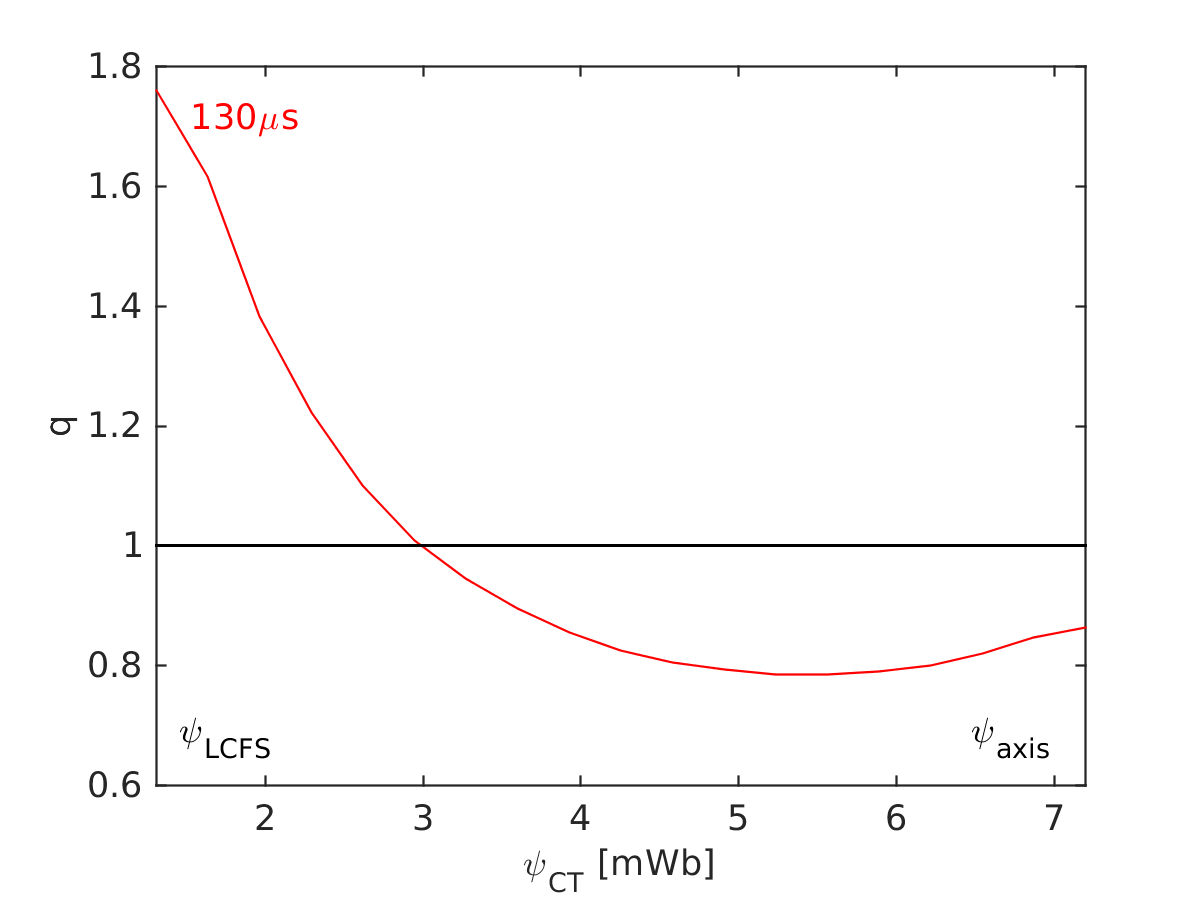}}\hfill{}\subfloat[]{\raggedleft{}\includegraphics[width=7cm,height=5cm]{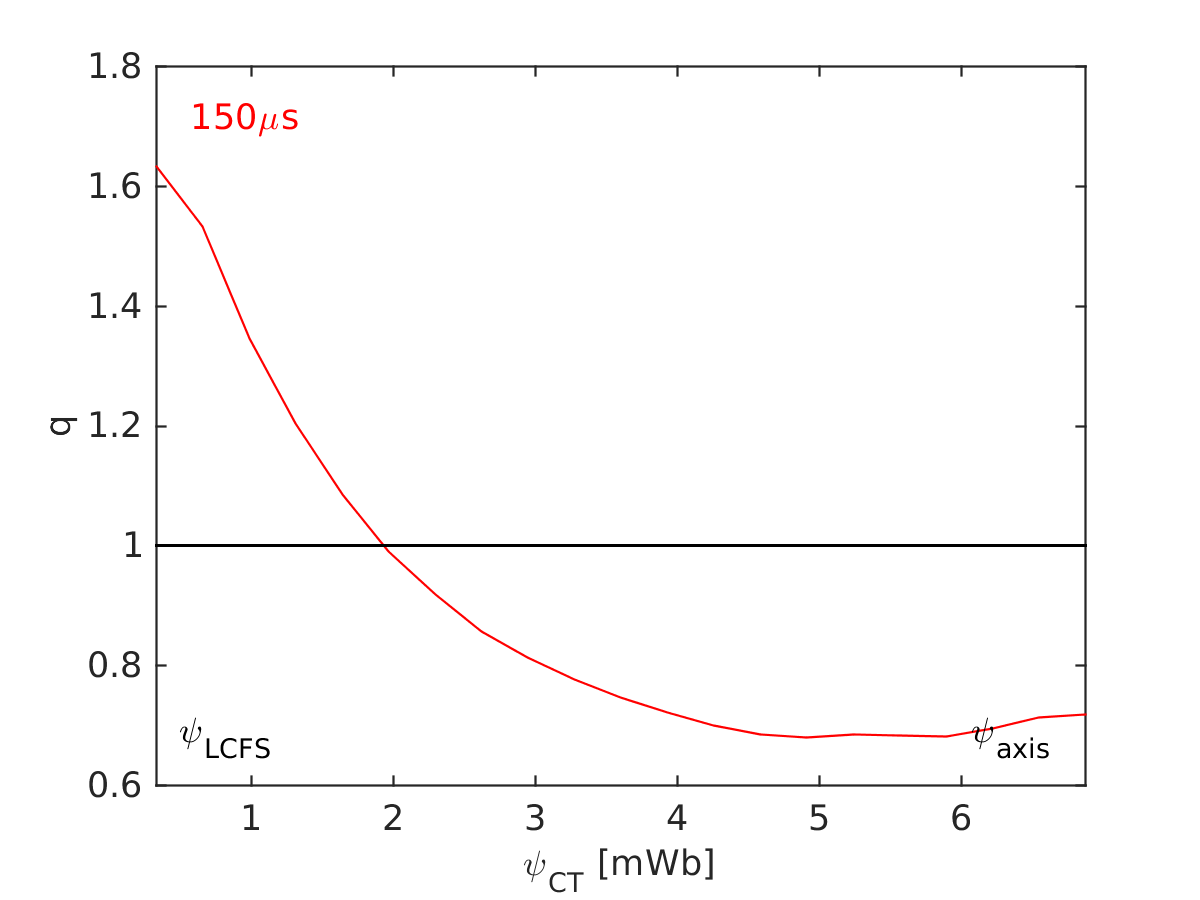}}

\caption{\label{fig:Qpsi_2287}$\,\,\,\,$$q$ profile, simulation 2287}
\end{figure}
When compression is started late in the CT's life, for example at
$t_{comp}=130\,\upmu$s in simulation  2287, most of the open poloidal
field lines that previously surrounded the CT have already reconnected
because $T_{e}$ has dropped and $\eta$ has increased. Then, MHD
simulations typically show $q>1$ at the $\mbox{LCFS}$, while the
region with $q<1$ extends all the way to the magnetic axis prior
to compression and also at peak compression (figures \ref{fig:Qpsi_2287}(a)
- (d)).

For both early and late magnetic compression, simulations indicate
that the $q$ profile is not contingent to magnetohydrodynamic stability
\cite{WessonTokamaks}, as $q<2$ at the LCFS in either case. Also,
for both early and late compression, $q$ drops below one over extensive
spans between the magnetic axis and the LCFS. The 2D simulations,
which neglect inherently three dimensional turbulent transport and
flux conversion, are likely to overestimate the level of hollowness
of the current profiles, and lead to an underestimation of $q$ towards
the CT edge, but without further internal experimental diagnostics
or 3D simulations, the level of underestimation remains uncertain.
The Kruskal-Shafranov limit determines that magnetically confined
plasmas are unstable to external kink modes when $q<1$. This condition
for instability doesn't pertain to toroidal sausage modes. As discussed
in section \ref{subsec:Compressional-Instability}, the external kink
or toroidal sausage modes are the most obvious candidates that could
lead to the toroidal field behavior that was observed on most compression
shots. Both the kink and sausage instabilities can be stabilized with
addition toroidal field, so the implementation of additional shaft
current should be considered if the magnetic compression experiment
was repeated. 

\subsubsection{Adiabatic compression scalings\label{subsec:SimDiagCompression-scalings}}

As discussed in section \ref{subsec:Scaling-laws-for}, if a magnetically
confined plasma is compressed on a time-scale that is short compared
with the resistive magnetic decay time and particle confinement time
of the plasma, the adiabatic compression scaling laws presented in
table \ref{tab:Scaling-parameters-for} should apply. Diagnostics
internal to the CT that would enable assessment of the scalings are
not available, but it is possible to estimate them using outputs from
simulations that match the available fixed-point external diagnostics
for magnetic field, and internal line-averaged diagnostics for density
and ion temperature along fixed chords. The CT cross-sectional area
in the poloidal plane is irregular, so the scalings in the first row
of table \ref{tab:Scaling-parameters-for} are relevant. $L(t)$ and
$S(t)$, the perimeter-length and area of the poloidal CT cross-section,
and $V(t)$, the CT volume, can be calculated using the coordinates
of the points that define the $\psi$ contour pertaining to the LCFS.
As discussed in reference to figures \ref{fig: psi_11coils}(d) and
(e), poloidal field lines that remain open surrounding closed CT poloidal
field lines are pinched off during magnetic compression, and reconnect
to form additional closed CT field lines. This affects the definition
of $\psi_{LCFS}$, and therefore of the values of the geometric scalings
$a_{0}(t),\,L(t),\,S(t)\mbox{ and }V(t)$ that are defined by the
location of the LCFS and are required to determine the predicted adiabatic
scalings. Hence, compression scalings are best assessed from simulations
in which compression is initiated relatively late in time when $\psi_{LCFS}$
is close to zero, and few open poloidal field lines surround the CT.
In addition, simulations indicate that closed poloidal CT field lines
that extend partially down into the gun barrel entrance can be pinched
off, and reconnect at compression (figures \ref{fig: psi_11coils}(d)
and (e)), which also affects the geometric scalings. A solution is
to assess the parameters of interest, including the geometric parameters,
relevant to a $\psi$ contour, defined by a fixed value of $\psi=\psi^{0}$,
that is internal to the pre-compression LCFS, and doesn't extend partially
into the gun barrel, a strategy that naturally does not affect the
compression scalings.
\begin{figure}[H]
\subfloat[]{\raggedright{}\includegraphics[width=7cm,height=5cm]{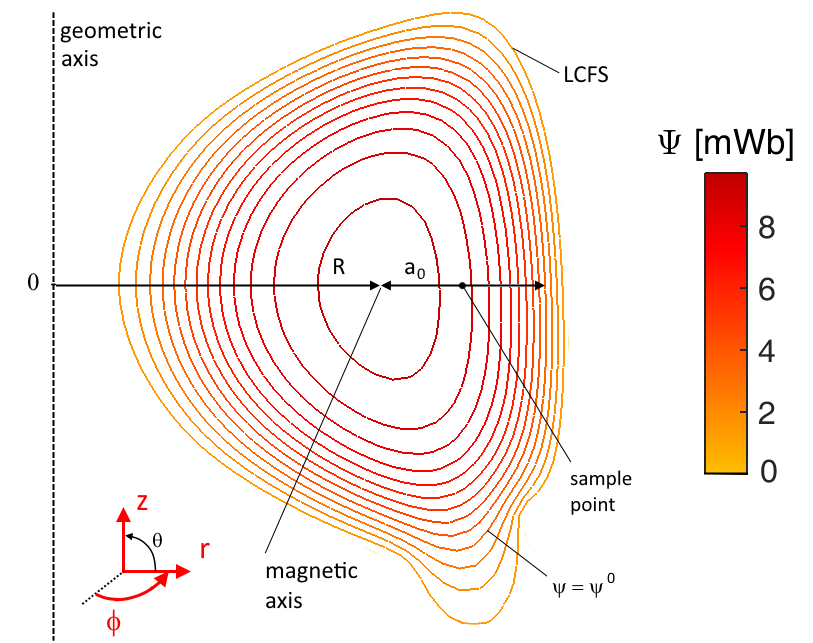}}\hfill{}\subfloat[]{\raggedleft{}\includegraphics[width=7cm,height=5cm]{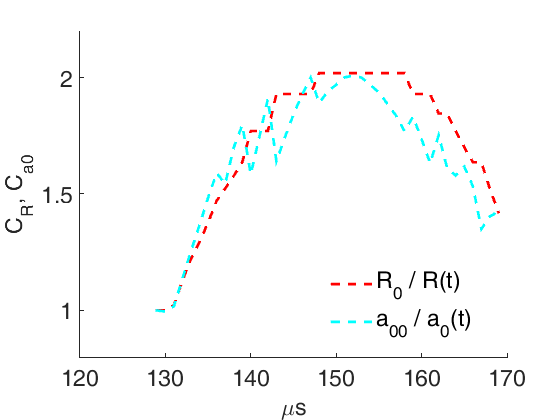}}

\subfloat[]{\raggedright{}\includegraphics[width=7cm,height=5cm]{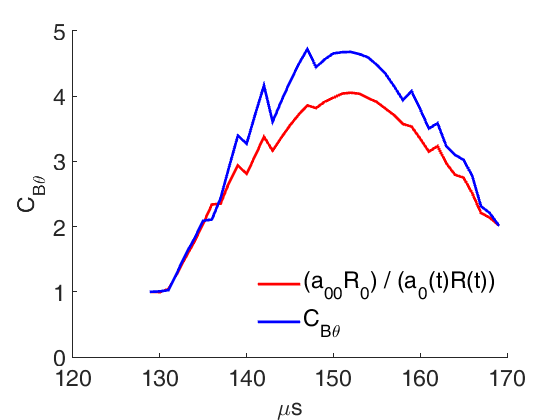}}\hfill{}\subfloat[]{\raggedleft{}\includegraphics[width=7cm,height=5cm]{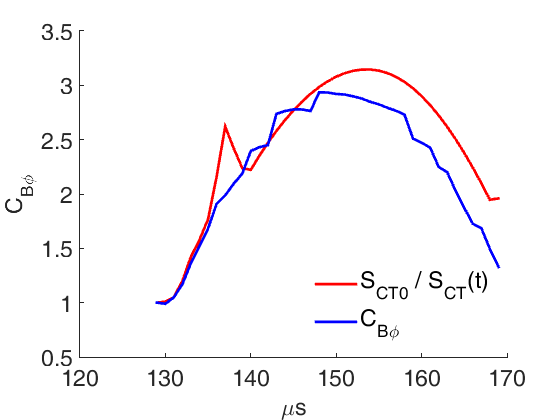}}

\subfloat[]{\raggedright{}\includegraphics[width=7cm,height=5cm]{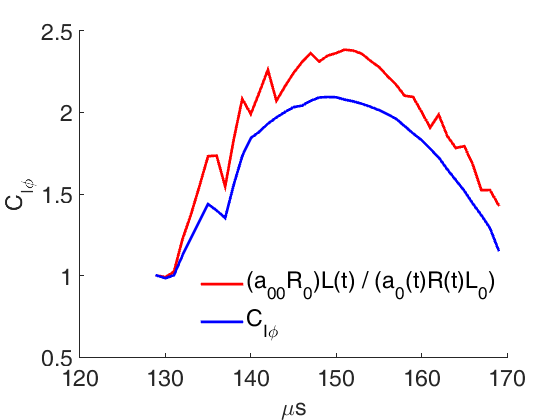}}\hfill{}\subfloat[]{\raggedright{}\includegraphics[width=7cm,height=5cm]{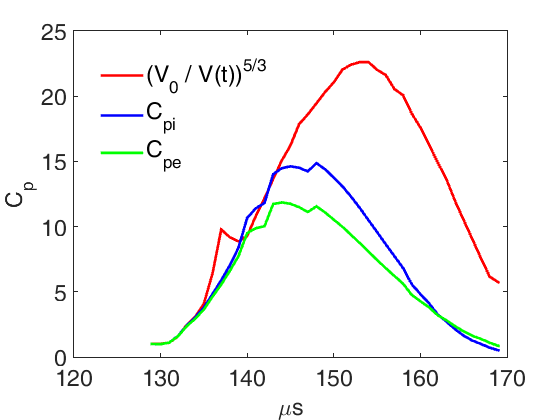}}

\caption{\label{fig:Compscalings2275}$\,\,\,\,$Compression scalings, simulation
 2287}
\end{figure}
Figure \ref{fig:Compscalings2275}(a) shows pre-compression CT $\psi$
contours from an MHD simulation at relatively late time, when most
open poloidal field lines have reconnected to form closed CT field
lines, and $\psi_{LCFS}\sim0$. A flux contour with $\Psi=\Psi^{0}=2$
mWb, that is suitable for assessing compression scaling parameters,
and definitions of $R(t)$ and (modified) $a_{0}(t)$, are also indicated.
Note that $a_{0}(t)$ has been re-defined as the distance from the
CT magnetic axis radially outwards at poloidal angle $\theta=0$ to
the closed flux surface defined by $\Psi=\Psi^{0}$. Simulated geometric
compression scalings for $R(t)$ and $a_{0}(t)$ from simulation 2287
are shown in figure \ref{fig:Compscalings2275}(b), where the subscript
0 denotes pre-compression values. This indicates approximately constant
aspect ratio (in irregular geometry), and that the compression is
close to the \textquotedbl Type A\textquotedbl{} compression regime
defined in \cite{Furth}. With constant aspect ratio, this indicates
a geometric compression factor, in terms of equatorial outboard CT
separatrix, of $C_{s}\sim C_{R}\sim C_{a0}\sim2$. As described in
section \ref{subsec:Rsep_comp}, a geometric compression factor $C_{s}\sim1.7$
was determined experimentally, and confirmed by MHD simulation, for
shot 39738 ($V_{comp}=18$ kV  and $t_{comp}=45\,\upmu$s), and that
more extreme compression in $r_{s}$ cannot be experimentally evaluated
due to limitations on the technique. More extreme compression would
be expected for shots at comparable $V_{comp}$, with $t_{comp}$
delayed to when pre-compression CT flux has decayed to lower levels.
Results from simulation 2287 have also been presented in figures \ref{fig: ring_psi_f},
\ref{fig: ring_psi_f-1}, and \ref{fig:Bpol_Btor_meas_cf_sim39735_2287}.
Simulation 2287 pertains to shot 39735 ($V_{comp}=18$ kV  and $t_{comp}=130\,\upmu$s),
so the increased estimate for $C_{s}$ is consistent with the shot
parameters. Note that, as outlined in chapter \ref{Chap:Magnetic-Compression},
with $\widetilde{\tau}_{c}=0.6$, shot 39735 is not classified as
a flux-conserving shot, so the adiabatic compression scalings evaluated
here pertain to the shot only up until the time when flux started
to be lost, just before peak compression. As outlined above, the technique
described here cannot practically be applied to simulations with compression
initiated early, and flux-conserving compressed shots were generally
taken with $t_{comp}=45\,\upmu$s. Only a few shots were taken with
late compression, none of which conserved flux very well over compression,
as determined by the $\widetilde{\tau}_{c}$ metric.

Figure \ref{fig:Compscalings2275}(c) shows how, for simulation  2287,
poloidal field scales approximately adiabatically as $B_{\theta}\rightarrow a_{0}^{-1}R^{-1}$,
where the sample point used to determine the scalings, indicated in
figure \ref{fig:Compscalings2275}(a), is located halfway between
the magnetic axis and the outboard point where $\psi=\psi^{0}$ at
the same axial coordinate as the magnetic axis. The notation $C_{B\theta}$
denotes the scaling of poloidal field as $C_{B\theta}(t)=\frac{B_{\theta}(t)}{B_{\theta0}}$
where $B_{\theta0}$ is the pre-compression magnetic field at the
sample point. Similarly, figure \ref{fig:Compscalings2275}(d) shows
how toroidal field at the same sample point also scales adiabatically,
as $B_{\phi}\rightarrow S^{-1}$. Figure \ref{fig:Compscalings2275}(e)
shows how plasma current, calculated as the integral of toroidal current
density over the area inside the closed flux surface at $\psi=\psi^{0}$,
evolves approximately according to the adiabatic scaling for plasma
current. 

As indicated in \ref{fig:Compscalings2275}(f), the scaling for pressure
(and hence also for the $\beta$ scalings) does not follow the adiabatic
prediction of $p\rightarrow V^{-\frac{5}{3}}$, due to the presence
of artificial density diffusion, which effectively relocates particles
from high density to low density regions. For this simulation, density
diffusion was $\zeta=50$ m$^{2}$/s, which is close to the minimum
value required for numerical stability at moderate timestep and mesh
resolution for simulations including magnetic compression, and $n_{e}$
follows the adiabatic scaling $n_{e}\rightarrow V^{-1}$ for only
the first $5\,\upmu$s after compression initiation. Ion and electron
pressures follow the adiabatic predictions for $15\,\upmu$s - the
extension is due to approximate internal force balance during this
portion of the compression cycle, which leads to increased temperature
in regions of low density. Temperatures at compression increase more,
while density increases less, than the predicted increases based on
the adiabatic scalings. When $\zeta$ is increased to $150$ m$^{2}$/s,
the duration over which $n_{e}$ follows the adiabatic scaling is
reduced further to around $2\,\upmu$s. 

This simulation, which produces results that closely match the available
experimental measurements for shot 39735 over most of the compression
cycle, indicates that CT aspect ratio is approximately constant over
compression, with $C_{s}\sim C_{R}\sim C_{a0}\sim2$, and that internal
CT poloidal and toroidal fields, and CT toroidal current, scale approximately
adiabatically, increasing over the main compression cycle by factors
of approximately four, three and two respectively. 

\subsubsection{System energy components\label{subsec:SimDiag_Energy}}

\begin{figure}[H]
\begin{centering}
\subfloat{\raggedleft{}\includegraphics[width=15cm,height=8cm]{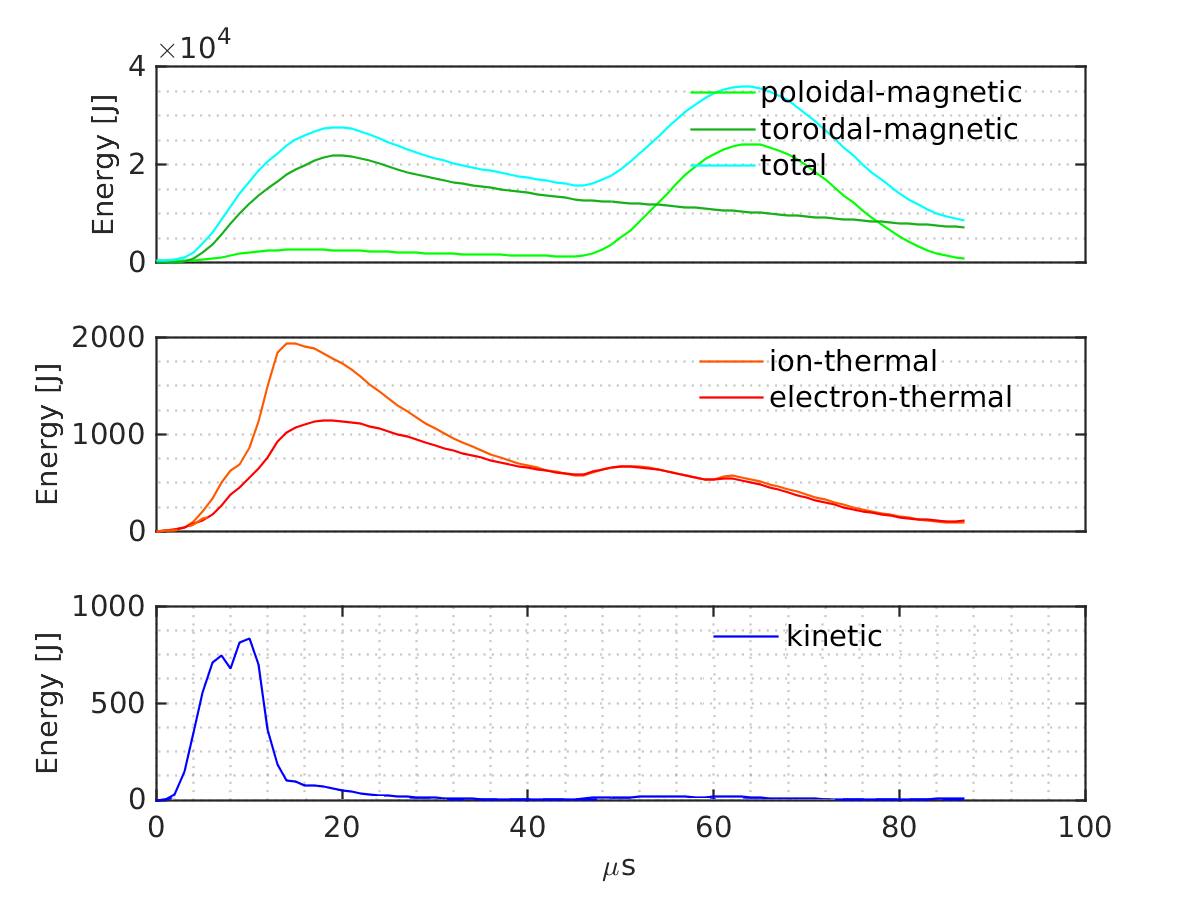}}\caption{\label{fig:Energy2353}$\,\,\,\,$System energy evolution for simulation
 2353}
\par\end{centering}
\end{figure}
Figure \ref{fig:Energy2353} indicates the evolution of the components
of system energy for simulation  2353. Boundary conditions are explicitly
applied to the pressure fields to enable thermal losses in accordance
with the thermal conduction model, and time-dependent external sources
of magnetic energy associated with formation, levitation and compression
are imposed. Finite boundary conditions for $\psi$ enable outward
electromagnetic energy losses (Poynting flux). Total system energy
is \emph{not} conserved, in contrast to the energy conservation in
cases with no thermal diffusion through the boundary and perfectly
electrically conducting boundaries, as presented in figure \ref{fig:Energy-partition-of-},
and in figure \ref{fig:Energy_N_cons} in chapter \ref{chap:Neutral-models}.
Figure \ref{fig:Energy2353} shows how total energy and toroidal magnetic
energy ($i.e.,$ magnetic energy associated with toroidal field) rise
at formation until $\sim20\,\upmu$s, as toroidal magnetic flux associated
with formation is added to the region below the physical location
of the gas valves. The axial gradient in $f$ associated with the
formation flux input profile (figure \ref{fig:gform_Phiform_Vform}(a))
is associated with radial formation current between the electrodes,
which causes ohmic heating of the electrons, and leads to a $\mathbf{J}_{r}\times\mathbf{B}_{\phi}$
force which advects the plasma rapidly upwards. This lead to an increase
in kinetic energy, and to ion viscous heating, especially over the
first $15\,\upmu$s of the simulation. Poloidal magnetic energy also
rises until $15\,\upmu$s as toroidal plasma currents are induced
to flow, leading to conversion of toroidal flux to poloidal flux during
the bubble-in process. As the plasma currents decay resistively, the
magnetic energy associated with internal currents decreases and is
initially converted to electron thermal energy. Recall that $\dot{U}{}_{M\theta\eta}=-\dot{U}{}_{Th\phi\eta}$,
equation \ref{eq:376}, and $\dot{U}{}_{M\phi_{\eta}}=-\dot{U}{}_{Th\theta\eta}$,
equation \ref{eq:127.5}. Electrons are ohmically heated and impart
energy to ions by collisions (term $Q_{ie}$, equation \ref{eq:517.61}).
Rapid diffusion of heat through the boundary means that thermal energy
is lost faster than it is sourced from magnetic energy after the initial
$\sim20\,\upmu$s period of maximum thermal energy gain, except during
the time when the CT is being magnetically compressed. Poloidal magnetic
energy associated with the external levitation field decreases as
the currents in the levitation coils decay. Further electromagnetic
energy is lost from the system due to Poynting flux through the boundary.
Note that, as described at the end of section \ref{subsec:Continuous_eqns_conservation_props},
Poynting flux out of the system is finite unless appropriate boundary
conditions, for example $\mathbf{v}|_{\Gamma}=\mathbf{0}$, $\psi|_{\Gamma}=0$
and $\left(\nabla_{\perp}\,f\right)|_{\Gamma}=0$ are imposed. In
this simulation with an external source of time-dependent poloidal
magnetic energy ($i.e.,$ finite $\psi(t)|_{\Gamma})$, the second
condition is not satisfied. The third condition is satisfied only
on the parts of the boundary that represent electrically conducting
walls; boundary conditions for $f$ consistent with conservation of
system toroidal flux are applied explicitly on the part of the boundary
that represents an electrical insulator, which breaks the third condition
on that part of the boundary.

Ion and electron thermal energy, magnetic energy, kinetic energy and
total energy continue to fall until $t_{comp}=45\,\upmu$s, when around
25 kJ of poloidal magnetic energy, due to over 1 MA toroidal current
in the levitation/compression coils, is added to the system over around
20$\,\upmu$s. The rapid inward compression of the CT leads to a slight
increase in kinetic energy. Ion and electron thermal energies increase
due to compressional heating. Toroidal plasma current increases as
the CT is compressed, leading to increased ohmic heating and an additional
rise in electron thermal energy. Again, electrons transfer some of
this energy to the ion fluid, so ion thermal energy also rises due
to enhanced ohmic heating. 

\section{Summary\label{sec:SummarySIMresults}}

The inclusion of a model for the insulating wall enables confirmation
that the level of penetration of poloidal field, that is advected
into the containment region during CT formation, into the insulating
region is reduced in the six coil configuration when $|t_{lev}|$
is increased, and that field penetration is avoided in the eleven
coil configuration. Simulated diagnostics from the two dimensional
MHD simulations show generally good agreement with experimental measurements.
Simulated poloidal and toroidal field at the magnetic probe locations
are very close to experimental measurements, for compression and levitation-only
shots. Levitation circuit resistance modification was a key strategy
developed to improve the performance of levitated CTs and avoid unintentional
magnetic compression and the associated instability. The modifications
were modelled by varying the rate of change of the amplitudes of boundary
conditions for $\psi_{lev}$ in proportion to the corresponding experimentally
measured levitation current signals, and the method works well to
approximately reproduce the experimentally measured poloidal field
in the two principal circuit configurations investigated. The time
evolutions of simulated outboard equatorial separatrix radius are
a close reproduction of the experimentally measured profiles, for
compression shots and levitation-only shots in both levitation circuit
configurations. 

Simulated ion temperature and electron density are also qualitatively
close matches to experimental measurements. The level to which artificial
density diffusion affects the simulated density diagnostic, and the
evolution of the other fields, should be investigated further. Simulations
do not capture the extremely high density that is measured experimentally
when plasma first enters the CT containment area, perhaps due to artificial
diffusion, or due to the inability of the model to capture the effect
of sputtering of high $Z$ ions during formation, and their subsequent
recombination as the plasma cools. Ion temperature is measured experimentally
using the ion-Doppler diagnostic, while the corresponding simulated
diagnostic works by obtaining the line averaged ion temperature along
the ion-Doppler chords, which seems reasonable as a crude first approximation.
The simulated ion temperature model could be developed further to
include additional physics, including density dependence. Qualitatively,
the simulated diagnostic for ion temperature approximately reproduces
the experimental measurement with the current model, and a suitable
choice of $\nu_{phys}$ enables scaling to match the experimental
measurement. 

The poloidal field profiles experimentally measured during magnetic
compression routinely indicated loss of CT poloidal flux during compression,
suggesting the action of a disruptive instability. It was found to
be possible to approximately reproduce the profiles by forcing CT
poloidal flux loss during simulated magnetic compression, thereby
verifying that a flux loss mechanism that was involved in the instability. 

A range of simulated diagnostics looking at physics internal to the
CT were developed. Without experimental counterparts to verify these
diagnostics, there is some uncertainty about the validity of the findings.
The simulated diagnostics that do have experimental counterparts match
experimental measurements very well in general, and this lends some
confidence to the internal simulated diagnostics. Simulations confirmed,
as also indicated with scintillator and ion-Doppler data, that ion
viscous heating is extreme during the formation process, while confirming
that electrons are heated more than ions at compression, likely due
to enhanced ohmic heating in combination with compressional heating.
Plasma jets associated with magnetic reconnection lead to high levels
of ion viscous heating, particularly around the entrance to the CT
containment region during formation of closed poloidal CT flux surfaces.
Simulations show that the pinching off of closed field lines that
extend partially down the gun when magnetic compression is initiated
early in the life of the CT leads to the formation of a smaller CT
located below the entrance to the containment region. Field lines
that remain open surrounding the closed CT flux surfaces are then
also pinched and reconnect to form additional closed field lines around
the main CT. The newly reconnected open field lines below the main
CT act like a slingshot that advects the smaller CT down the gun.
This effect cannot be verified experimentally, as the magnetic probes
located along the gun barrel were not functioning properly at any
time. Simulations of magnetic compression indicate that CT volume
decreases by a factor of around three over the primary compression
cycle - this observation is partially supported in that the time evolution
of simulated outboard equatorial separatrix matches the experimental
observation. It is shown that the oscillating compression field resulted
in the inductive formation, magnetic compression, and subsequent extinguishing
(magnetic reconnection of CT poloidal field with the compression field)
of a second and third CT. This is supported experimentally in that
experimental poloidal field measurements closely match the corresponding
simulated diagnostic, and that Xray-phosphor imaging indicates the
compressional heating of up to three distinct plasmoids on many compression
shots. With extremely high values for poloidal beta, and toroidal
beta less than one, it was indicated that magnetic energy associated
with CT toroidal field greatly exceeds that associated with CT poloidal
field. Simulated volume-averaged total beta increases from $\sim13\%$
to $\sim16\%$ over compression. Simulated diagnostics for magnetic
field profiles internal to the CT indicate that the CT has a field
profile with more in common with a tokamak plasma than a spheromak
plasma, but with the amplitude of the toroidal field far lower than
that of a tokamak plasma. This is verified by the simulated $q(\psi)$
profile. Tokamaks typically operate at $q\gg1$ at all points in the
plasma, increasing MHD stability. Simulations indicates that, at early
times ($\sim45\,\upmu$s) in the CT life, while $q\sim6$ around the
magnetic axis, it falls to a value not much higher than one towards
the LCFS. Over time, as the CT current profile becomes less hollow,
$q$ drops to below one at the axis, and rises to around two at the
LCFS (figure \ref{fig:Qpsi_2287}(c)). It was found that $q<1$ over
extended regions between the magnetic axis and the LCFS at peak magnetic
compression, regardless of whether compression was initiated at early
or at late times in the CT life. Internal simulated diagnostics were
used to compare theoretical adiabatic compression scalings with simulated
compression scalings. It was found that while the magnetic fields,
and toroidal current at a point located halfway between the magnetic
axis and the outboard equatorial separatrix, did rise at compression
approximately according to the adiabatic scalings, that the scalings
for the other quantities of interest ($n_{e},\,T_{e},\,T_{i},\,\beta_{\theta}$
and $\beta_{\phi}$) did not, due to the effect of artificial density
diffusion. Approximately constant aspect ratio compression (in irregular
geometry) was indicated, so that compression is close to the \textquotedbl Type
A\textquotedbl{} regime defined in \cite{Furth}. With constant aspect
ratio, this indicates a geometric compression factor, in terms of
outboard equatorial CT separatrix, of $C_{s}\sim C_{R}\sim C_{a0}\sim2$.
As described in section \ref{subsec:Rsep_comp}, an outboard equatorial
separatrix radius at peak compression of $r_{s}\sim9$ cm, and a geometric
compression factor $C_{s}\sim1.7$, was determined experimentally,
and confirmed by MHD simulation, for a shot which was compressed relatively
early in time ($V_{comp}=18$ kV  and $t_{comp}=45\,\upmu$s). As
noted in section \ref{subsec:Rsep_comp}, when $r_{s}\lesssim9$ cm,
the slope of the functional fit in \ref{Bz_rsep39650_1}(b) is too
flat to be successfully inverted with good accuracy. For this reason,
$C_{s}$ cannot be experimentally evaluated for shots at comparable
$V_{comp}$ if compression is fired relatively late in time, when
pre-compression CT flux has decayed to lower levels and compression
is more extreme.

The components of system energy evolve as expected with conversions
initially from toroidal magnetic energy associated with formation
current to thermal energy, kinetic energy and poloidal magnetic energy.
Magnetic energy is converted to electron thermal energy via ohmic
heating, with kinetic energy being converted to ion thermal energy
due to viscous effects. Rapid diffusion of heat through the boundary
means that thermal energy is lost faster than it is sourced from magnetic
energy after the initial $\sim20\,\upmu$s period of maximum thermal
energy gain, except during the time when the CT is being magnetically
compressed. At CT magnetic compression, the principal energy conversion
mechanisms are that poloidal magnetic energy does work to compress
the CT, leading to compressional heating of ions and electrons. Toroidal
plasma current increases as the CT is compressed, leading to increased
ohmic heating and an additional rise in electron thermal energy. Again,
electrons transfer some of this energy to the ion fluid, so ion thermal
energy also rises due to enhanced ohmic heating. \newpage{}

\chapter{Model for interaction between plasma and neutral fluids\label{chap:Neutral-models}}

This chapter focuses on the development and implementation of a model
for simulating interaction between plasma fluid and neutral fluid
over the process of CT formation into a levitation field and magnetic
compression. A paper \cite{Neut_paper}, based on the material presented
this chapter, has been submitted for publication. 

A primary motivation for including plasma / neutral fluid interaction
in the MHD model was to reduce simulated ion temperature to levels
corresponding to the ion-Doppler diagnostic measurements while maintaining
energy conservation, $i.e.,$ while keeping $\nu_{num}=\nu_{phys}$
(see appendix \ref{subsec:Viscous-diffusion}). Charge exchange collisions
are an important mechanism by which ion temperature is reduced - in
a charge exchange reaction, a hot ion takes an electron from a cold
neutral particle, resulting in a hot neutral particle and a cold ion.
Secondly, it was expected that the presence of a neutral fluid in
the gun around the gas valves, which is also a physical phenomenon,
would slow down the axially directed plasma during the formation process,
leading to reduced viscous heating. In the actual experiment, the
formation process is initiated when an electric field is applied across
a cloud of cold gas between the inner and outer machine electrodes.
The gas cloud is concentrated near the gas-puff valves, at $z=-0.43$
m (see figure \ref{fig:Machine-Schematic-1}). The gas valves are
typically opened approximately $400\,\upmu$s before the formation
voltage is applied across the electrodes, so that a neutral gas cloud
has ample time to diffuse away from the valves. The ionization action
of the applied formation field results in seed electrons that lead
to further impact ionization. The resultant plasma that is advected
up the gun by the $\mathbf{J}_{r}\times\mathbf{B}_{\phi}$ force is
initially surrounded by a relatively static neutral gas cloud that
it ionizes or displaces. The neutral cloud slows down the advected
plasma through collisions, resulting in a reduction of ion viscous
heating as $\mathbf{v}$ and $\nabla\mathbf{v}$ are reduced. 

The first attempt at modelling plasma-neutral interactions, in order
to reduce simulated ion temperature while maintaining energy conservation,
included charge exchange reactions only. This approach succeeds in
reducing simulated ion temperature, but is unphysical because the
rate of change of the neutral fluid density, which depends on the
rates of ionization and recombination, and therefore on the electron
temperature, cannot be determined and is held constant. If the neutral
fluid density is kept constant, ion cooling due to charge exchange
continues even in situations where the electron temperature is so
high that physically the neutral fluid density would be minuscule.

Finally, a plasma-neutral model, in which a single-fluid MHD plasma
interacts with a neutral fluid was developed and implemented in the
MHD code. The part of the dynamics of both the plasma and neutral
fluid that does not depend on inelastic collisions is based on the
set of MHD equations presented in section \ref{sec:Axisymmetric-two-temperature-MHD}.
For the neutral fluid equations, terms in the MHD equations relating
to electromagnetic phenomena are omitted. Thus, total system energy
conservation is maintained when the energy lost by electrons due to
impact ionization and radiative recombination is accounted for. It
turned out that, while the inclusion of a neutral fluid in the CT
formation simulations did reduce simulated ion temperature, the cause
of the temperature reduction was due to bulk inertial effects, which
reduces axial velocities and therefore reduces viscous heating, rather
than direct interaction between the fluids. It was found that the
same level of ion temperature reduction can be achieved in simulations
without a neutral fluid, simply by increasing the initial plasma fluid
density until it is equal to the combined neutral and plasma densities
for a simulation with the same initial spatial distributions of the
neutral and plasma fluids. Due to equilibration between the ion and
electron temperatures, ionization rates are high in regions where
ions are hot, so that neutral fluid density, and hence the level of
charge-exchange related ion cooling, is approximately negligible in
those key regions. The model proved to be useful because it helped
clarify the mechanisms behind the significant increases in electron
density that are routinely observed at $\sim500\,\upmu$s with CTs
formed in the SPECTOR plasma injector (figure \ref{fig:GFinjectors}(b)).
Neutral gas, which remains concentrated below the gas valves after
CT formation, diffuses up the gun barrel to the CT containment region
where it is ionized, leading to the observed electron density increases. 

This chapter is arranged as follows. A model overview is presented
in section \ref{sec:Model-overview}. Development of the scattering
terms, and the reactive terms pertaining to ionization, recombination
and charge exchange collisions, that are included in the plasma and
neutral fluid equations, is presented in sections \ref{subsec:Scattering-collision-terms}
and \ref{subsec:Reactive-collision-terms} respectively. The resultant
fluid equations for ions, electrons and neutral particles are derived
in section \ref{sec:3-fluid-MHD-equations}. The ion and electron
MHD equations are reduced to a set of single plasma fluid equations,
that include terms pertaining to interaction with the neutral fluid,
in section \ref{sec:2-fluid-MHD-equations}. A demonstration of total
system energy and particle count conservation with inclusion of a
neutral fluid in the MHD model is presented and discussed in section
\ref{sec:Energy-conservation-with}. Implementation of diffusion terms
for the neutral fluid is discussed in section \ref{sec:Neutral-thermal-diffusion}.
A simple model for including effects of charge-exchange reactions
only without evolving a neutral fluid is presented in section \ref{sec:Simple-CX-model}.
Simulation results are presented and discussed in section \ref{sec:Simulation-results-with}.
Section \ref{sec:SummaryNeut} concludes the chapter with an overview
of the principal findings.

\section{Model overview\label{sec:Model-overview}}

Based on the work presented in \cite{Meier,MeierPhd}, the model includes
resonant charge exchange, electron impact ionization, and radiative
recombination reactions:
\begin{align}
i^{+}+n & \rightarrow n+i^{+}\nonumber \\
e^{-}+n & \rightarrow i^{+}+2e^{-}-\phi_{ion}\label{eq:520}\\
e^{-}+i^{+} & \rightarrow n+h\nu_{p}\nonumber 
\end{align}
Here, $i^{+},\,e^{-}$ and $n$ respectively represent singly charged
ions and electrons, and neutral particles, $h\,[\mbox{m}^{2}\mbox{kg/s}]$
is Planck's constant, and $\nu_{p}$ is the wave frequency associated
with the photon emitted in the recombination reaction. The charge
exchange process is \emph{resonant} because the initial and final
states have the same quantum mechanical energy - the exchanged electron's
initial and final energy states are the same so that the combined
kinetic energy and momentum of the ion and neutral is unchanged \cite{Goldston,HazeltineWaelbroeck}.
In the following derivations, singly charged ions and a single neutral
species will be considered. The plasma is assumed to be optically
thin, so that radiation energy $h\nu_{p}$ associated with radiative
recombination is lost from the system. Following from \cite{Meier,MeierPhd},
excited states are not tracked in order to simplify the model. Instead,
an effective ionization potential, $\phi_{ion}$, includes the excitation
energy that is expended on average for each ionization event, as well
as the electron binding energy. \\
\\
The Boltzmann equation (\ref{eq:360}) is 
\[
\frac{\partial f_{\alpha}}{\partial t}+\nabla\cdot(\mathbf{V}\,f_{\alpha})+\nabla_{v}\cdot\left(\left(\frac{q_{\alpha}}{m_{\alpha}}(\mathbf{E}(\mathbf{r},t)\mathbf{+V}\mathbf{\times B}(\mathbf{r},t))\right)\,f_{\alpha}\right)=\frac{\partial f_{\alpha}}{\partial t}|_{collisions}=C_{\alpha}(f)
\]
where $\alpha=i,\,e,$ or $n$ denote ions, electrons, or neutral
particles. If $\alpha=n$, then the acceleration term vanishes, because
$q_{n}=0$. \\
\\
As mentioned in appendix \ref{subsec:MomentsCollision}, the collision
operator can be split into parts pertaining to elastic \emph{scattering}
collisions and \emph{reacting} collisions: $C_{\alpha}(f)=C_{\alpha}^{scatt.}(f)+C_{\alpha}^{react.}(f)$.
Electron impact ionization, and radiative recombination are inelastic
reacting collisions, whereas the resonant charge exchange process
is an elastic reacting collision because the initial and final states
are degenerate. Moments of $C_{\alpha}^{scatt.}$ were taken in appendix
\ref{subsec:MomentsCollision}, and used when deriving the two plasma-fluid
Braginskii equations in appendix \ref{subsec:Braginski-equations-from}.
In section \ref{subsec:Scattering-collision-terms}, moments of the
scattering collision operators when a neutral species is included
will be considered. 

In order to derive the collisional terms in a three-component model
(ion, electron and neutral fluids), which will then be reduced to
a two-component model (single plasma-fluid and neutral fluid), moments
of the part of the collision operator that pertains to reacting collisions
can be taken, as is done in \cite{Meier,MeierPhd}. In this work,
we show how, for the inelastic reacting collisions of ionization and
recombination, the results for the collisional terms in a three-component
model that are presented in \cite{Meier,MeierPhd} can be obtained
without taking moments. Instead, the sources or sinks that arise in
the species continuity equations due to the reacting collisions of
ionization and recombination are used directly to determine the consequent
terms that arise in the species momentum and energy equations. This
approach, described in section \ref{subsec:Ionization-and-recombination},
is intuitive and also allows for determination of the terms that must
be included in the MHD equations when external particle sources are
present. In addition, this approach allows for the determination of
terms in the plasma and neutral fluid energy equations that represent
thermal energy lost by electrons and transferred to photons and neutral
particles in radiative recombination processes. These terms cannot
be evaluated using the moment-taking approach and have been neglected
in studies based on the model presented in \cite{Meier}. 

The terms in the MHD equations that correspond to charge exchange
collisions cannot be evaluated using this approach because charge
exchange collisions do not give rise to sources or sinks in the species
continuity equations. Consequently, the terms in the MHD equations
that correspond to reactive charge exchange collisions must be evaluated
by taking moments of the part of the collision operator associated
with charge-exchange collisions, as is done in \cite{Meier,MeierPhd}
- the procedure is briefly outlined in section \ref{subsec:Charge-exchange}.

\section{Scattering collision terms\label{subsec:Scattering-collision-terms}}

Scattering collisions have no $0^{th}$ moment effect, as was indicated
in appendix \ref{subsec:MomentsCollision}, equation \ref{eq:466.1}.
To find the contribution of scattering collisions to the rates of
change of momentum of the ions, electrons and neutral particles, the
first moments of the operators for scattering collisions can be taken.
For convenience, equation \ref{eq:467-1} is repeated: $\mathbf{R}_{\alpha}=\underset{\sigma}{\Sigma}\mathbf{R}_{\alpha\sigma}=\int m_{\alpha}\underset{\sigma}{\Sigma}\,C_{\alpha\sigma}^{scatt.}\mathbf{V}d\mathbf{V}=\int m_{\alpha}C_{\alpha}^{scatt.}\mathbf{V}d\mathbf{V}$.
Here, $\mathbf{R}_{\alpha}$ is the total friction force acting on
species $\alpha$ due to the net effect of the frictional interaction
with each of species $\sigma$. Note that, using equations \ref{eq:354},
\ref{eq:355-2} and \ref{eq:466.1}, we can write: $\mathbf{R}_{\alpha}=\int m_{\alpha}C_{\alpha}^{scatt.}\mathbf{c}_{\alpha}d\mathbf{V}$.
In a system with just two species, ions and electrons, $\mathbf{R}_{\alpha}$
is given by the Chapman-Enskog closures (equation \ref{eq:472.471}).
Assuming isotropic resistivity, and ignoring the thermal force terms
for simplicity, equation \ref{eq:472.471} gives $\mathbf{R}_{e}=-\mathbf{R}_{i}=\eta'ne\mathbf{J}=\nu_{ei}\rho_{e}(\mathbf{v}_{i}-\mathbf{v}_{e})$.
Recall, from appendix \ref{subsec:MomentsCollision}, that $\nu_{ei}=\frac{1}{\tau_{ei}}$
is the electron-ion collision frequency where $\tau_{ei}$ (equation
\ref{eq:472.38}) is the electron-ion collision time, the time it
takes for an electron to be scattered by $90^{o}$ due to collisions
with ions. When there are just two species, then, since $\mathbf{R}_{\alpha}=\underset{\sigma}{\Sigma}\mathbf{R}_{\alpha\sigma}$
(equation \ref{eq:467-1}) and $\mathbf{R}_{\sigma\sigma}=0$ (a fluid
does not exert friction on itself), the equivalent notation $\mathbf{R}_{e}\equiv\mathbf{R}_{ei}$
and $\mathbf{R}_{i}\equiv\mathbf{R}_{ie}$ can be introduced for the
frictional forces due to scattering collisions. 

The notation $(X)_{scatt.}$ is introduced to represent the part of
$X$ that pertains to scattering collisions. Recall that $m_{\alpha}$
times the first moment of the first term of the Boltzmann equation
is $\frac{\partial\left(\rho_{\alpha}\mathbf{v}_{\alpha}\right)}{\partial t}$
(appendix \ref{subsec:Braginski-equations-from}). Since $\left(\frac{\partial\left(\rho_{\alpha}\mathbf{v}_{\alpha}\right)}{\partial t}\right)_{scatt.}=\rho_{\alpha}\left(\frac{\partial\mathbf{v}_{\alpha}}{\partial t}\right)_{scatt.}+\mathbf{v}_{\alpha}\left(\frac{\partial\rho_{\alpha}}{\partial t}\right)_{scatt.}$,
and scattering collisions are not a source of particles $\left(\Rightarrow\left(\frac{\partial\rho_{\alpha}}{\partial t}\right)_{scatt.}=0\right)$,
the terms $\rho_{\alpha}\left(\frac{\partial\mathbf{v}_{\alpha}}{\partial t}\right)_{scatt.}$
can be expressed as: 
\begin{equation}
\rho_{\alpha}\left(\frac{\partial\mathbf{v}_{\alpha}}{\partial t}\right)_{scatt.}=\mathbf{R}_{\alpha}=\underset{\sigma}{\Sigma}\mathbf{R}_{\alpha\sigma}\label{eq:520.1}
\end{equation}
With the inclusion of a neutral fluid, this leads to the following
relations: 
\begin{align}
\rho_{i}\left(\frac{\partial\mathbf{v}_{i}}{\partial t}\right)_{scatt.}=\mathbf{R}_{i} & =\mathbf{R}_{ie}+\mathbf{R}_{in}\nonumber \\
\rho_{e}\left(\frac{\partial\mathbf{v}_{e}}{\partial t}\right)_{scatt.}=\mathbf{R}_{e} & =\mathbf{R}_{ei}+\mathbf{R}_{en}=-\mathbf{R}_{ie}+\mathbf{R}_{en}\label{eq:521}\\
\rho_{n}\left(\frac{\partial\mathbf{v}_{n}}{\partial t}\right)_{scatt.}=\mathbf{R}_{n} & =\mathbf{R}_{ni}+\mathbf{R}_{ne}=-\mathbf{R}_{in}-\mathbf{R}_{en}\nonumber 
\end{align}
where the identities $\mathbf{R}_{\alpha\sigma}=-\mathbf{R}_{\sigma\alpha}$
(implying that the frictional force exerted by species $\alpha$ on
species $\sigma$ is balanced by the frictional force exerted by species
$\sigma$ on species $\alpha$) have been used. Note that $\mathbf{R}_{ei}=\nu_{ei}\rho_{e}(\mathbf{v}_{i}-\mathbf{v}_{e})$
and $\mathbf{R}_{ie}=\nu_{ie}\rho_{i}(\mathbf{v}_{e}-\mathbf{v}_{i})$,
where $\nu_{ie}=\sim\frac{m_{e}}{m_{i}}\nu_{ei}=\frac{1}{\tau_{ie}}$
is the ion-electron collision frequency and $\tau_{ie}$ is the ion-electron
collision time, the time it takes for an ion to be scattered $90^{o}$
due to collisions with electrons. Similarly, the forms for the charged-neutral
friction forces are $\mathbf{R}_{in}=\nu_{in}\rho_{i}(\mathbf{v}_{n}-\mathbf{v}_{i})$
and $\mathbf{R}_{en}=\nu_{en}\rho_{e}(\mathbf{v}_{n}-\mathbf{v}_{e})$,
where $\nu_{\alpha\sigma}\sim\frac{m_{\alpha}}{m_{\sigma}}\nu_{\sigma\alpha}$
is the frequency for scattering of particles of species $\alpha$
from particles of species $\sigma$ \cite{bellan fundamentals}. The
terms $\mathbf{R}_{in}\mbox{ and }\mathbf{R}_{en}$ can be neglected
in many cases - in general, neutral-charged particle scattering collisions
are unimportant when the plasma is ionized by even a few percent \cite{Goldston,Meier}. 

To find the contribution of scattering collisions to the rates of
change of energy of the ions, electrons and neutral particles, the
second moments of the operators for scattering collisions can be taken.
Since $\mathbf{R}_{\alpha}=\underset{\sigma}{\Sigma}\mathbf{R}_{\alpha\sigma}$
and $Q_{\alpha}=\underset{\sigma}{\Sigma}Q_{\alpha\sigma}$ (equation
\ref{eq:467-2}), equation \ref{eq:468-1} becomes:\\
\begin{align}
\int C_{\alpha}^{scatt.}(\frac{1}{2}m_{\alpha}V^{2})\,d\mathbf{V} & =Q_{\alpha}+\mathbf{v}_{\alpha}\cdot\mathbf{R}_{\alpha}=\underset{\sigma}{\Sigma}Q_{\alpha\sigma}+\mathbf{v}_{\alpha}\cdot\underset{\sigma}{\Sigma}\mathbf{R}_{\alpha\sigma}\label{eq:521.1}
\end{align}
From appendix \ref{subsec:Braginski-equations-from}, the second moment
of the first term of the Boltzmann equation is
\[
\frac{\partial}{\partial t}\left(\frac{1}{2}\rho_{\alpha}v_{\alpha}^{2}+\frac{p_{\alpha}}{\gamma-1}\right)=\frac{v^{2}}{2}\frac{\partial\rho_{\alpha}}{\partial t}+\rho_{\alpha}\mathbf{v}_{\alpha}\cdot\frac{\partial\mathbf{v}_{\alpha}}{\partial t}+\frac{1}{\gamma-1}\frac{\partial p_{\alpha}}{\partial t}
\]
where $\gamma=\frac{5}{3}$. Since $\left(\frac{\partial\rho_{\alpha}}{\partial t}\right)_{scatt.}=0$,
this leads to
\[
\frac{1}{\gamma-1}\left(\frac{\partial p_{\alpha}}{\partial t}\right)_{scatt.}=\left(\frac{\partial}{\partial t}\left(\frac{1}{2}\rho_{\alpha}v_{\alpha}^{2}+\frac{p_{\alpha}}{\gamma-1}\right)\right)_{scatt.}-\rho_{\alpha}\mathbf{v}_{\alpha}\cdot\left(\frac{\partial\mathbf{v}_{\alpha}}{\partial t}\right)_{scatt.}
\]
Using equations \ref{eq:521.1} and \ref{eq:520.1}, this implies
that 
\[
\frac{1}{\gamma-1}\left(\frac{\partial p_{\alpha}}{\partial t}\right)_{scatt.}=\underset{\sigma}{\Sigma}Q_{\alpha\sigma}+\cancel{\mathbf{v}_{\alpha}\cdot\underset{\sigma}{\Sigma}\mathbf{R}_{\alpha\sigma}}-\cancel{\mathbf{v}_{\alpha}\cdot\underset{\sigma}{\Sigma}\mathbf{R}_{\alpha\sigma}}
\]
\begin{align}
\Rightarrow\frac{1}{\gamma-1}\left(\frac{\partial p_{i}}{\partial t}\right)_{scatt.} & =Q_{ie}+Q_{in}\nonumber \\
\frac{1}{\gamma-1}\left(\frac{\partial p_{e}}{\partial t}\right)_{scatt.} & =Q_{ei}+Q_{en}\label{eq:522}\\
\frac{1}{\gamma-1}\left(\frac{\partial p_{n}}{\partial t}\right)_{scatt.} & =Q_{ni}+Q_{ne}\nonumber 
\end{align}
Here, $Q_{\alpha\sigma}$ represents the heat gained by species $\alpha$
due to scattering interactions with species $\sigma$. Again, due
to the relative unimportance of neutral-charged particle scattering
collisions \cite{Goldston,Meier}, the terms $Q_{in}$, $Q_{en}$,
$Q_{ni}$, and $Q_{ne}$ can usually be neglected.

\section{Reactive collision terms\label{subsec:Reactive-collision-terms}}

In this section, the terms in the species (ions, electrons and neutral
fluids) mass, momentum and energy conservation equations that pertain
to collisions associated with ionization, recombination, and charge
exchange reactions will be assessed. 

\subsection{Ionization and recombination\label{subsec:Ionization-and-recombination}}

\subsubsection{Mass Conservation}

For electron impact ionization, the collision operator defining the
time-rate of change of $f_{n}$ due to \emph{reacting} collisions
between neutral particles and electrons that result in electron impact
ionization is \cite{Meier}: 
\begin{align}
C_{n}^{ion} & =-f_{n}(\mathbf{r},\mathbf{V},t)\int f_{e}(\mathbf{r},\mathbf{V},t)\,\sigma_{ion}(V_{rel})\,V_{rel}\,d\mathbf{V}\label{eq:528}
\end{align}
Here, $\sigma_{ion}(V_{rel})\,\,[\mbox{m}^{2}]$ is the cross-section
for electron impact ionization, and the relative particle speed for
the ionization reaction is $V_{rel}=|\mathbf{V}_{e}-\mathbf{V}_{n}|,$
where $\mathbf{V}_{e}$ is the electron particle velocity and $\mathbf{V}_{n}$
is the neutral particle velocity. To find the source rate for neutral
particles due to electron impact ionization, the $0^{th}$ moment
of this operator is taken: $\Gamma_{n}^{ion}=-\int f_{n}(\mathbf{r},\mathbf{V'},t)\left(\int f_{e}(\mathbf{r},\mathbf{V},t)\,\sigma_{ion}(V_{rel})\,V_{rel}\,d\mathbf{V}\right)\,d\mathbf{V'}$.
Referring to equation \ref{eq:352.0}, the inner integral is $\int f_{e}\,\sigma_{ion}\,V_{rel}d\mathbf{V}=n_{e}(\mathbf{r},t)<\sigma_{ion}\,V_{rel}>$,
so that, using equation \ref{eq:352-1}, $\Gamma_{n}^{ion}=-n_{n}n_{e}<\sigma_{ion}\,V_{rel}>$.
Since the electron mass is much less than the neutral particle mass,
it can be assumed that the electron particle ($i.e.,$ thermal) velocity
is much greater than the neutral particle velocity, so that $V_{rel}\sim V_{e}$:
\begin{equation}
\Gamma_{n}^{ion}=\int C_{n}^{ion}d\mathbf{V}=-n_{n}n_{e}<\sigma_{ion}\,V_{e}>\label{eq:529}
\end{equation}
$\Gamma_{n}^{ion}[\mbox{m}^{-3}\mbox{s}^{-1}]$ (units of particles
per metre cubed per second) will appear on the right hand side of
the neutral fluid mass continuity equation, constituting a sink of
neutral particles due to impact ionization. For each neutral particle
lost to ionization, an ion and electron are added to the system. The
corresponding collision operators and source rates are 
\begin{equation}
C_{i}^{ion}=C_{e}^{ion}=-C_{n}^{ion}\label{eq:529.01}
\end{equation}
and 
\begin{equation}
\Gamma_{i}^{ion}=\Gamma_{e}^{ion}=-\Gamma_{n}^{ion}\label{eq:529.1}
\end{equation}
Equation \ref{eq:529} can also be derived with a more intuitive method
that doesn't use the collision operator \cite{Goldston}. The mean
free path for impact ionization collisions is defined by considering
that there is one particle in the volume swept out by the cross-sectional
area for impact ionization collisions over one mean free path: $n_{n}\sigma_{ion}(V_{rel})\lambda_{mfp}^{ion}=1\Rightarrow\lambda_{mfp}^{ion}=\frac{1}{n_{n}\sigma_{ion}(V_{rel})}$.
The frequency for electron impact ionization collisions is defined
as an average over all velocities in the Maxwellian distribution:
\[
\nu_{ion}=<\frac{V_{rel}}{\lambda_{mfp}^{ion}}>=n_{n}<\sigma_{ion}(V_{rel})V_{rel}>
\]
so that, with $V_{e}\gg V_{n}$, and $V_{rel}=|\mathbf{V}_{e}-\mathbf{V}_{n}|\approx V_{e}$,
$\nu_{ion}=n_{n}<\sigma_{ion}V_{e}>$. From this, the rate of increase
of ions per unit volume is again $\Gamma_{i}^{ion}=\Gamma_{e}^{ion}=-\Gamma_{n}^{ion}=n_{n}n_{e}<\sigma_{ion}\,V_{e}>$.
The velocity integrated quantity $<\sigma_{ion}\,V_{e}>\,[\mbox{m}^{3}/\mbox{s}]$
is the\emph{ ionization rate parameter - }its value can be found as
a function of temperature from the fitting formula given by Voronov
\cite{Voronov}, as 
\begin{equation}
<\sigma_{ion}\,V_{e}>(\mathbf{r},t)=\frac{A\left(1+P\sqrt{U}\right)U^{K}exp\left(-U\right)}{U+X}\label{eq:530}
\end{equation}
where $U(\mathbf{r},t)=\frac{\phi_{ion}}{T_{e}(\mathbf{r},t)}$, with
$\phi_{ion}$ being the effective ionization potential in the same
units as $T_{e}$. As mentioned previously, an effective ionization
potential including the excitation energy that is expended on average
for each ionization event, as well as the electron binding energy,
is used instead of the regular ionization energy, because, for simplicity,
excited states are not tracked. An estimate of the validity of the
formula for the ionization rate parameter is given by Voronov for
each element. For example, for hydrogen, accuracy is to within 5\%
for electron temperatures from 1 $\mbox{eV}$ to 20 $\mbox{keV}$.
The MHD code has the option of either hydrogen, deuterium or helium
as the neutral gas and plasma source. From \cite{Voronov}, the coefficients
required for equation \ref{eq:530} for these options are shown in
table \ref{tab:coef_ionization rate parameter}. Also included is
the atomic diameter ($d_{atom}$) for each atom, which is used to
calculate the viscous and thermal diffusion coefficients for the neutral
fluid, using equations \ref{eq:472.35} and \ref{eq:472.36}, where
the mean free path for neutral-neutral collisions is defined in equation
\ref{eq:472.310}. The values for effective ionization potentials
are taken from \cite{Yusupaliev}, the atomic diameters from \cite{MeierPhd}
and \cite{wikiAradius}.
\begin{table}[H]
\centering{}%
\begin{tabular}{|c|c|c|c|}
\hline 
$\mbox{ion\,\,type}:$ & H & $\mbox{D}$ & $\mbox{He}$\tabularnewline
\hline 
\hline 
$\phi_{ion}\,[\mbox{eV}]$ & $13.6$ & $33$ & $28$\tabularnewline
\hline 
$A$ & $2.91\times10^{-14}\,[\mbox{m}^{3}/\mbox{s}]$ & $2.91\times10^{-14}\,[\mbox{m}^{3}/\mbox{s}]$ & $1.75\times10^{-14}\,[\mbox{m}^{3}/\mbox{s}]$\tabularnewline
\hline 
$P$ & $0$ & $0$ & $0$\tabularnewline
\hline 
$K$ & $0.39$ & $0.39$ & $0.35$\tabularnewline
\hline 
$X$ & $0.232$ & $0.232$ & $0.18$\tabularnewline
\hline 
$d_{atom}[\mbox{m}]$ & $1.06\times10^{-10}$ & $2.4\times10^{-10}$ & $2.8\times10^{-10}$\tabularnewline
\hline 
\end{tabular}\caption{\label{tab:coef_ionization rate parameter}$\,\,\,\,$Coefficients
for calculating ionization rate parameters}
\end{table}

The collision operators defining the time-rates of change of $f_{\alpha}$
due to reacting collisions between ions and electrons that result
in radiative recombination are \cite{Meier}: 
\begin{align}
C_{n}^{rec} & =\frac{m_{e}}{m_{n}}f_{e}\int f_{i}\,\sigma_{rec}(V_{rel})\,V_{rel}\,d\mathbf{V}+\frac{m_{i}}{m_{n}}f_{i}\int f_{e}\,\sigma_{rec}(V_{rel})\,V_{rel}\,d\mathbf{V}\nonumber \\
C_{i}^{rec} & =-f_{i}\int f_{e}\,\sigma_{rec}(V_{rel})\,V_{rel}\,d\mathbf{V}\nonumber \\
C_{e}^{rec} & =-f_{e}\int f_{i}\,\sigma_{rec}(V_{rel})\,V_{rel}\,d\mathbf{V}\label{eq:530.1}
\end{align}
Here, $V_{rel}=|\mathbf{V}_{e}-\mathbf{V}_{i}|$ is the relative particle
speed, and $\sigma_{rec}(V_{rel})\,\,[\mbox{m}^{2}]$ is the cross-section
for radiative recombination. Since $m_{e}\ll m_{n}$, and $m_{i}\approx m_{n}$,
$C_{n}^{rec}\approx-C_{i}^{rec}$. With $V_{e}\gg V_{i}$ , and following
the process of the derivation of equation \ref{eq:529}, we arrive
at 

\begin{equation}
-\Gamma_{n}^{rec}\approx\Gamma_{i}^{rec}=\Gamma_{e}^{rec}=\int C_{e}^{rec}d\mathbf{V}=-n_{i}n_{e}<\sigma_{rec}\,V_{e}>\label{eq:531}
\end{equation}
As was shown above for the ionization rate parameter, this result
can also be obtained, without taking moments of the collision operator,
by considering the mean free path and collision frequency associated
with recombination \cite{Goldston}. The velocity integrated quantity
$<\sigma_{rec}\,V_{e}>\,[\mbox{m}^{3}/\mbox{s}]$ is the\emph{ recombination
rate parameter - }its value can be estimated as a function of electron
temperature as \cite{McWhirter,MeierPhd,Goldston} 
\begin{equation}
<\sigma_{rec}\,V_{e}>(\mathbf{r},t)=2.6\times10^{-19}\frac{\,Z_{eff}^{2}}{\sqrt{T_{e}(\mathbf{r},t)\,[eV]}}\label{eq:531.0}
\end{equation}
Here, the recombination rate is for recombination to charge state
$Z_{eff}-1$. Referring to the expressions for two-plasma-fluid mass
continuity in the absence of neutral particles and reactive collisions
(equation \ref{eq:469}), when the particle source rates due to reactive
collisions defined in equations \ref{eq:529}, \ref{eq:529.1}, and
\ref{eq:531} are included, we arrive at the mass continuity equations
for the three-fluid system:

\begin{align}
\dot{n}_{i} & =-\nabla\cdot(n_{i}\mathbf{v}_{i})+\Gamma_{i}^{ion}-\Gamma_{n}^{rec}\nonumber \\
\dot{n}_{e} & =-\nabla\cdot(n_{e}\mathbf{v}_{e})+\Gamma_{i}^{ion}-\Gamma_{n}^{rec}\label{eq:531.2}\\
\dot{n}_{n} & =-\nabla\cdot(n_{n}\mathbf{v}_{n})+\Gamma_{n}^{rec}-\Gamma_{i}^{ion}+\Gamma_{n}^{ext}\nonumber 
\end{align}
Note that all source terms here, as well as each of $n_{\alpha}$
and $\mathbf{v}_{\alpha}$, are functions of $\mathbf{r}$ and $t$.
In the experiment, the gas puff valves take time to shut, and remain
open for several hundred microseconds after the formation banks are
fired, so neutral particles are being added to the system near the
valves. Additionally, neutral particles are sourced through recycling
processes at the vacuum vessel walls, although this recycling effect
hasn't yet been implemented in the code. An additional neutral \emph{external}
source term, $\Gamma_{n}^{ext}(\mathbf{r},t)\,[\mbox{m}^{-3}\mbox{ s}^{-1}]$,
has been included on the right side of the expression for $\dot{n}_{n}$,
in order to be able to simulate neutral particle injection.

\subsubsection{Momentum Conservation}

To find the contributions of ionization and recombination reactions
to the rates of change of momentum of the ions, electrons, and neutral
particles, the first moments of the reacting collision operators defined
in equations \ref{eq:528}, \ref{eq:529.01}, and \ref{eq:530.1}
can be taken. This can be done explicitly, as is shown in \cite{Meier,MeierPhd}.
However, it is instructive to demonstrate that the formal process
of taking first moments (and second moments for the contributions
to the rates of change of energy) can be skipped, to arrive at results
equivalent to those presented in \cite{Meier,MeierPhd}, but with
evaluation of the terms which determine the volumetric rate of thermal
energy transfer from electrons to photons and neutral particles due
to radiative recombination.

Referring to equation \ref{eq:531.2}, the general form of the expression
for the species rates of change of number density that correspond
to the reactive collisions of ionization and recombination, and also
to any external particle sources, is 
\begin{equation}
\left(\frac{\partial n_{\alpha}}{\partial t}\right)_{ire}=\underset{k}{\Sigma}S_{\alpha k}\label{eq:531.40}
\end{equation}
Here, $\left(X\right)_{ire}$ denotes the part of $X$ that pertains
to the reactive collisions of \emph{ionization} and \emph{recombination,}
and to any \emph{external} particle sources. $S_{\alpha k}\,[\mbox{m}^{-3}\mbox{s}^{-1}]$
represents the $k^{th}$ source (in units of particles per metres
cubed per second) for particles of type $\alpha,$ as determined from
equation \ref{eq:531.2}. Here, $S_{i1}=S_{e1}=-S_{n2}=\Gamma_{i}^{ion},\,S_{i2}=S_{e2}=-S_{n1}=-\Gamma_{n}^{rec}$,
and $S_{n3}=\Gamma_{n}^{ext}$. Note that particle \textquotedbl sources\textquotedbl{}
with a negative sign such as $S_{i2}=-\Gamma_{n}^{rec}$ in the expression
for $\dot{n}_{i}$ in equation \ref{eq:531.2}, are actually particle
sinks. \\
\\
Species momentum conservation is described, in the absence of reactive
collisions, by equation \ref{eq:469.1}: 
\[
\frac{\partial(\rho_{\alpha}\mathbf{v}_{\alpha})}{\partial t}=-\nabla\cdot\underline{\mathbf{p}}_{\alpha}-\nabla\cdot(\rho_{\alpha}\mathbf{v}_{\alpha}\mathbf{v}_{\alpha})+q_{\alpha}n_{\alpha}\left(\mathbf{E}+\mathbf{v}_{\alpha}\times\mathbf{B}\right)+\mathbf{R}_{\alpha}
\]
To include the terms that correspond to the reactive collisions of
ionization and recombination, and to any external particle sources,
this can be written as\\
 $\frac{\partial(m_{\alpha}n_{\alpha}\mathbf{v}_{\alpha})}{\partial t}=...+\left(\frac{\partial(m_{\alpha}n_{\alpha}\mathbf{v}_{\alpha})}{\partial t}\right)_{ire}$,
where \textquotedbl$...$\textquotedbl{} represents the right side
of equation \ref{eq:469.1}. Particles sourced by $S_{\alpha k}$
add, or (for sources with negative sign), remove, species $\alpha$
momentum $\underset{j}{\Sigma}m_{0jk}\mathbf{v}_{0jk}$, where $m_{0jk}$
and $\mathbf{v}_{0jk}$ are the mass and fluid velocity of the particles
of type $j$ which have their momentum introduced or taken away. The
summation over sourced particles of type $j$ is relevant only for
$S_{\alpha k}=S_{n1}=\Gamma_{n}^{rec}$; recombination is a source
for total neutral particle momentum, and each neutral particle added
to the neutral population through recombination initially has momentum
$m_{i}\mathbf{v}_{i}+m_{e}\mathbf{v}_{e}$. The general form of the
expression for the species rates of change of momentum that correspond
to the reactive collisions of ionization and recombination, and also
to any external particle sources, is 
\begin{equation}
\left(\frac{\partial(m_{\alpha}n_{\alpha}\mathbf{v}_{\alpha})}{\partial t}\right)_{ire}=\underset{k}{\Sigma}\left(S_{\alpha k}\underset{j}{\Sigma}\left(m_{0jk}\mathbf{v}_{0jk}\right)\right)\label{eq:531.41}
\end{equation}
This expression must be retained for the neutral recombination source
term $S_{n1}$. However, for all other source terms, $\underset{j}{\Sigma}\left(m_{0jk}\mathbf{v}_{0jk}\right)\rightarrow m_{\alpha}\mathbf{v}_{0k}$,
where $\mathbf{v}_{0k}$ is the \textquotedbl initial\textquotedbl{}
fluid velocity ($i.e.,$ the fluid velocity at the time when the ionization
or recombination reaction occurs) of the particles of type $\alpha$
which are introduced or taken away due to source $S_{\alpha k}$,
and the general expression can be simplified to\\
\[
\left(\frac{\partial(m_{\alpha}n_{\alpha}\mathbf{v}_{\alpha})}{\partial t}\right)_{ire}=m_{\alpha}\underset{k}{\Sigma}\left(S_{\alpha k}\mathbf{v}_{0k}\right)
\]
The corresponding additional terms on the right side of the momentum
equations are: 
\begin{align}
\left(\frac{\partial(m_{i}n_{i}\mathbf{v}_{i})}{\partial t}\right)_{ire} & =\Gamma_{i}^{ion}m_{i}\mathbf{v}_{n}-\Gamma_{n}^{rec}m_{i}\mathbf{v}_{i}\nonumber \\
\left(\frac{\partial(m_{e}n_{e}\mathbf{v}_{e})}{\partial t}\right)_{ire} & =\Gamma_{i}^{ion}m_{e}\mathbf{v}_{n}-\Gamma_{n}^{rec}m_{e}\mathbf{v}_{e}\label{eq:531.5}\\
\left(\frac{\partial(m_{n}n_{n}\mathbf{v}_{n})}{\partial t}\right)_{ire} & =\Gamma_{n}^{rec}(m_{i}\mathbf{v}_{i}+m_{e}\mathbf{v}_{e})-\Gamma_{i}^{ion}m_{n}\mathbf{v}_{n}+\Gamma_{n}^{ext}m_{n}\mathbf{v}_{n0}\nonumber 
\end{align}
For example, in the expression above for $\left(\frac{\partial(m_{i}n_{i}\mathbf{v}_{i})}{\partial t}\right)_{ire}$,
ions that are sourced from neutral particles through ionization add
to the total ion momentum, and newly ionized particles are introduced
with velocity $\mathbf{v}_{n}$. Meanwhile, ions with velocity $\mathbf{v}_{i}$,
that are lost to recombination, take away from the total ion momentum.
Each neutral particle introduced by recombination initially has momentum
$m_{i}\mathbf{v}_{i}+m_{e}\mathbf{v}_{e}$. Neutral particles introduced
by external sources such as gas puffing also add neutral particle
momentum - each externally sourced neutral has initial momentum $m_{n}\mathbf{v}_{n0}$,
where $\mathbf{v}_{n0}$ is its initial velocity.\\
\\
Using equation \ref{eq:531.40}, equation \ref{eq:531.41} can be
recast as 
\begin{align}
\left(\frac{\partial(m_{\alpha}n_{\alpha}\mathbf{v}_{\alpha})}{\partial t}\right)_{ire} & =m_{\alpha}n_{\alpha}\left(\frac{\partial\mathbf{v}_{\alpha}}{\partial t}\right)_{ire}+m_{\alpha}\mathbf{v}_{\alpha}\left(\frac{\partial n_{\alpha}}{\partial t}\right)_{ire}=\underset{k}{\Sigma}\left(S_{\alpha k}\underset{j}{\Sigma}\left(m_{0jk}\mathbf{v}_{0jk}\right)\right)\nonumber \\
 & =m_{\alpha}n_{\alpha}\left(\frac{\partial\mathbf{v}_{\alpha}}{\partial t}\right)_{ire}+m_{\alpha}\mathbf{v}_{\alpha}\left(\underset{k}{\Sigma}S_{\alpha k}\right)\nonumber \\
\Rightarrow m_{\alpha}n_{\alpha}\left(\frac{\partial\mathbf{v}_{\alpha}}{\partial t}\right)_{ire} & =\underset{k}{\Sigma}\left(S_{\alpha k}\underset{j}{\Sigma}\left(m_{0jk}\mathbf{v}_{0jk}\right)\right)-m_{\alpha}\mathbf{v}_{\alpha}\left(\underset{k}{\Sigma}S_{\alpha k}\right)\label{eq:531.51}
\end{align}
 For the ions and electrons (all sources), and for the neutral source
terms corresponding to ionization and external sources, where $\underset{j}{\Sigma}\left(m_{0jk}\mathbf{v}_{0jk}\right)\rightarrow m_{\alpha}\mathbf{v}_{0k}$,
this general expression can be simplified to 
\begin{equation}
m_{\alpha}n_{\alpha}\left(\frac{\partial\mathbf{v}_{\alpha}}{\partial t}\right)_{ire}=m_{\alpha}\underset{k}{\Sigma}\left(S_{\alpha k}\left(\mathbf{v}_{0k}-\mathbf{v}_{\alpha}\right)\right)\label{eq:531.52}
\end{equation}
Equation \ref{eq:531.52} and (for neutral recombination only) equation
\ref{eq:531.51} lead to:\\
\begin{align}
\rho_{i}\left(\frac{\partial\mathbf{v}_{i}}{\partial t}\right)_{ire} & =\Gamma_{i}^{ion}m_{i}(\mathbf{v}_{n}-\mathbf{v}_{i})\nonumber \\
\rho_{e}\left(\frac{\partial\mathbf{v}_{e}}{\partial t}\right)_{ire} & =\Gamma_{i}^{ion}m_{e}(\mathbf{v}_{n}-\mathbf{v}_{e})\label{eq:531.6}\\
\rho_{n}\left(\frac{\partial\mathbf{v}_{n}}{\partial t}\right)_{ire} & =\Gamma_{n}^{rec}(m_{i}\mathbf{v}_{i}+m_{e}\mathbf{v}_{e}-m_{n}\mathbf{v}_{n})+\Gamma_{n}^{ext}m_{n}(\mathbf{v}_{n0}-\mathbf{v}_{n})\nonumber 
\end{align}

\subsubsection{Energy Conservation\label{subsec:Econ_ionrecomb}}

Assuming Maxwellian distributions for each of species $\alpha$, so
that $p_{\alpha}=n_{\alpha}T_{\alpha}$, the part of the species energy
equation that corresponds to particle sources due to the reactive
collisions of ionization and recombination, and to external particle
sources, can be written as
\begin{equation}
\left(\frac{\partial}{\partial t}\left(\frac{1}{2}m_{\alpha}n_{\alpha}v_{\alpha}^{2}+\frac{p_{\alpha}}{\gamma-1}\right)\right)_{ire}=\frac{1}{2}\underset{k}{\Sigma}\left(S_{\alpha k}\underset{j}{\Sigma}\left(m_{0jk}v_{0jk}^{2}\right)\right)+\frac{1}{\gamma-1}\underset{k}{\Sigma}\left(\xi_{\alpha k}S_{\alpha k}\underset{j}{\Sigma}T_{0jk}\right)\label{eq:531.39}
\end{equation}
Here, $\xi_{\alpha k}$ is a particle\emph{ mass ratio} that must
be considered for the ionization source that introduces ions ($S_{i1}=\Gamma_{i}^{ion}$)
and electrons ($S_{e1}=\Gamma_{i}^{ion}$), and for the recombination
source that introduces neutral particles $(S_{n1}=\Gamma_{n}^{rec})$.
To clarify this for the ionization source, when a neutral particle
with thermal energy $\frac{1}{\gamma-1}T_{n}$ is ionized, the resultant
ion and electron have thermal energies $\frac{m_{i}}{m_{n}}\frac{1}{\gamma-1}T_{n}$
and $\frac{m_{e}}{m_{n}}\frac{1}{\gamma-1}T_{n}$ respectively, so
that their combined thermal energy is equal to that of the original
neutral. 

To clarify the relevance of $\xi_{\alpha k}$ for the recombination
source, simple analysis of the kinematics of the radiative recombination
reaction indicates that the bulk of the electron thermal energy, and
a fraction of the ion thermal energy, is transferred to the emitted
photon. Noting that $m_{i}\sim m_{n}\gg m_{e}$, then, in the rest
frame of the neutral particle (post reaction), the ion (prior to the
reaction) has negligible kinetic energy ($ie.,$ thermal energy, since
we are considering single particles with random velocities), the electron
has approximately the same energy that it has in the laboratory frame,
and the neutral particle has no kinetic energy. Consequently, as an
approximation, the bulk of the electron thermal energy (of the order
$\sim(m_{i}/m_{n})\,T_{e}/(\gamma-1)$) and a negligible portion of
the ion thermal energy ($\sim(m_{e}/m_{n})\,T_{i}/(\gamma-1)$) is
transferred to the photon emitted in the radiative recombination reaction,
while a negligible portion of the electron thermal energy (of the
order $\sim(m_{e}/m_{n})\,T_{e}/(\gamma-1)$) and the bulk of the
ion thermal energy ($\sim(m_{i}/m_{n})\,T_{i}/(\gamma-1)$) is transferred
to the neutral particle. The combined thermal energy of the neutral
particle and photon is equal to the combined thermal energy of the
ion and electron. Note that for $\Gamma_{n}^{ext}$, the external
neutral particle source, $\xi_{\alpha k}=1$. 

Once again, the summation over $j$ is relevant only for $S_{\alpha k}=S_{n1}=\Gamma_{n}^{rec}$;
recombination is a source for neutral particle energy, and it is assumed
that each neutral particle added to the neutral particle population
through recombination initially has energy $\frac{1}{2}\left(m_{i}v_{i}^{2}+m_{e}v_{e}^{2}\right)+\frac{1}{\gamma-1}\left(\frac{m_{i}}{m_{n}}\,T_{i}+\frac{m_{e}}{m_{n}}\,T_{e}\right)$.
Note that for all other sources (apart from recombination) the summation
over $j$ in equation \ref{eq:531.39} can be neglected; $\underset{j}{\Sigma}T_{0jk}\rightarrow T_{0k}$,
where $T_{0k}$ is the initial temperature of the sourced particle,
and $\underset{j}{\Sigma}\left(m_{0jk}v_{0jk}^{2}\right)\rightarrow m_{\alpha}v_{0k}^{2}.$
We want to obtain an expression for $\left(\frac{\partial p_{\alpha}}{\partial t}\right)_{ire}$,
which will be included on the right side of the species energy equations
which have the form $\frac{\partial p_{\alpha}}{\partial t}=.....\,$.
The first step is to use equation \ref{eq:531.40} to expand the partial
derivative in equation \ref{eq:531.39}:{\footnotesize{}
\begin{align*}
\frac{1}{2}m_{\alpha}v_{\alpha}^{2}\left(\frac{\partial n_{\alpha}}{\partial t}\right)_{ire}+m_{\alpha}n_{\alpha}\mathbf{v}_{\alpha}\cdot\left(\frac{\partial\mathbf{v}_{\alpha}}{\partial t}\right)_{ire}+\frac{1}{\gamma-1}\left(\frac{\partial p_{\alpha}}{\partial t}\right)_{ire}=\frac{1}{2}\underset{k}{\Sigma}\left(S_{\alpha k}\underset{j}{\Sigma}\left(m_{0jk}v_{0jk}^{2}\right)\right)+\frac{1}{\gamma-1}\underset{k}{\Sigma}\left(\xi_{\alpha k}S_{\alpha k}\underset{j}{\Sigma}T_{0jk}\right)\\
\Rightarrow\frac{1}{\gamma-1}\left(\frac{\partial p_{\alpha}}{\partial t}\right)_{ire}=\frac{1}{\gamma-1}\underset{k}{\Sigma}\left(\xi_{\alpha k}S_{\alpha k}\underset{j}{\Sigma}T_{0jk}\right)+\frac{1}{2}\underset{k}{\Sigma}\left(S_{\alpha k}\underset{j}{\Sigma}\left(m_{0jk}v_{0jk}^{2}\right)\right)-\frac{1}{2}m_{\alpha}v_{\alpha}^{2}\underset{k}{\Sigma}\left(S_{\alpha k}\right)-m_{\alpha}n_{\alpha}\mathbf{v}_{\alpha}\cdot\left(\frac{\partial\mathbf{v}_{\alpha}}{\partial t}\right)_{ire}
\end{align*}
}Using equation \ref{eq:531.51}, this implies that {\small{}
\begin{align}
\left(\frac{\partial p_{\alpha}}{\partial t}\right)_{ire} & =\underset{k}{\Sigma}\left(\xi_{\alpha k}S_{\alpha k}\underset{j}{\Sigma}T_{0jk}\right)+(\gamma-1)\biggl[\frac{1}{2}\underset{k}{\Sigma}\left(S_{\alpha k}\underset{j}{\Sigma}\left(m_{0jk}v_{0jk}^{2}\right)\right)\nonumber \\
 & -\frac{1}{2}m_{\alpha}v_{\alpha}^{2}\underset{k}{\Sigma}\left(S_{\alpha k}\right)-\mathbf{v}_{\alpha}\cdot\left(\underset{k}{\Sigma}\left(S_{\alpha k}\underset{j}{\Sigma}\left(m_{0jk}\mathbf{v}_{0jk}\right)\right)-m_{\alpha}\mathbf{v}_{\alpha}\left(\underset{k}{\Sigma}S_{\alpha k}\right)\right)\biggr]\nonumber \\
\Rightarrow\left(\frac{\partial p_{\alpha}}{\partial t}\right)_{ire} & =\underset{k}{\Sigma}\left(\xi_{\alpha k}S_{\alpha k}\underset{j}{\Sigma}T_{0jk}\right)+(\gamma-1)\left(\frac{1}{2}m_{\alpha}v_{\alpha}^{2}\underset{k}{\Sigma}\left(S_{\alpha k}\right)+\underset{k}{\Sigma}\left(S_{\alpha k}\underset{j}{\Sigma}\left(m_{0jk}\left(\frac{1}{2}v_{0jk}^{2}-\mathbf{v}_{\alpha}\cdot\mathbf{v}_{0jk}\right)\right)\right)\right)\nonumber \\
\Rightarrow\left(\frac{\partial p_{\alpha}}{\partial t}\right)_{ire} & =\underset{k}{\Sigma}\left(\xi_{\alpha k}S_{\alpha k}\underset{j}{\Sigma}T_{0jk}\right)+(\gamma-1)\left(\underset{k}{\Sigma}\left(S_{\alpha k}\left(\frac{1}{2}m_{\alpha}v_{\alpha}^{2}+\underset{j}{\Sigma}\left(m_{0jk}\left(\frac{1}{2}v_{0jk}^{2}-\mathbf{v}_{\alpha}\cdot\mathbf{v}_{0jk}\right)\right)\right)\right)\right)\nonumber \\
\nonumber \\
\label{eq:532}
\end{align}
}For the ions and electrons (all sources), and for the neutral source
terms corresponding to ionization and external sources, where $\underset{j}{\Sigma}m_{0jk}\mathbf{v}_{0jk}\rightarrow m_{\alpha}\mathbf{v}_{0k}$,
$\underset{j}{\Sigma}\left(m_{0jk}v_{0jk}^{2}\right)\rightarrow m_{\alpha}v_{0k}^{2}$,
and $\underset{j}{\Sigma}T_{0jk}\rightarrow T_{0k}$, this general
expression can be simplified to
\begin{align*}
\left(\frac{\partial p_{\alpha}}{\partial t}\right)_{ire} & =\underset{k}{\Sigma}\left(\xi_{\alpha k}S_{\alpha k}T_{0k}\right)+(\gamma-1)\left(\frac{1}{2}m_{\alpha}\underset{k}{\Sigma}\left(S_{\alpha k}\left(v_{\alpha}^{2}-2\mathbf{v}_{\alpha}\cdot\mathbf{v}_{0k}+v_{0k}^{2}\right)\right)\right)\\
\Rightarrow\left(\frac{\partial p_{\alpha}}{\partial t}\right)_{ire} & =\underset{k}{\Sigma}\left(\xi_{\alpha k}S_{\alpha k}T_{0k}\right)+(\gamma-1)\left(\frac{1}{2}m_{\alpha}\underset{k}{\Sigma}\left(S_{\alpha k}\left(\mathbf{v}_{\alpha}-\mathbf{v}_{0k}\right)^{2}\right)\right)
\end{align*}
However, the more complicated form of equation \ref{eq:532} must
be retained for the recombination neutral source term.\\
\\
The resultant forms for the individual species are:

\begin{align}
\left(\frac{\partial p_{i}}{\partial t}\right)_{ire} & =\Gamma_{i}^{ion}\frac{m_{i}}{m_{n}}\,T_{n}-\Gamma_{n}^{rec}\,T_{i}+(\gamma-1)\frac{1}{2}m_{i}\left(\Gamma_{i}^{ion}\left(\mathbf{v}_{i}-\mathbf{v}_{n}\right)^{2}\right)\nonumber \\
\left(\frac{\partial p_{e}}{\partial t}\right)_{ire} & =\Gamma_{i}^{ion}\frac{m_{e}}{m_{n}}\,T_{n}-\Gamma_{n}^{rec}\,T_{e}+(\gamma-1)\left(\Gamma_{i}^{ion}\left(\frac{1}{2}m_{e}\left(\mathbf{v}_{e}-\mathbf{v}_{n}\right)^{2}-\phi_{ion}\right)\right)\nonumber \\
\left(\frac{\partial p_{n}}{\partial t}\right)_{ire} & =\Gamma_{n}^{rec}\left(\frac{m_{i}}{m_{n}}\,T_{i}+\frac{m_{e}}{m_{n}}\,T_{e}\right)-\Gamma_{i}^{ion}\,T_{n}+\Gamma_{n}^{ext}\,T_{n0}+(\gamma-1)\biggl[\Gamma_{n}^{rec}\biggl(\frac{1}{2}m_{n}v_{n}^{2}+\frac{1}{2}m_{i}v_{i}^{2}\nonumber \\
 & \,\,\,\,\,\,\,+\frac{1}{2}m_{e}v_{e}^{2}-m_{i}\mathbf{v}_{n}\cdot\mathbf{v}_{i}-m_{e}\mathbf{v}_{n}\cdot\mathbf{v}_{e}\biggr)+\Gamma_{n}^{ext}\,\frac{1}{2}m_{n}\left(\mathbf{v}_{n}-\mathbf{v}_{n0}\right)^{2}\biggr]\label{eq:533}
\end{align}
Here, $T_{n0}$ is the initial temperature of the externally sourced
neutral particles. The effective ionization energy has been included
as a sink of electron energy - recalling that $U_{Th}=\frac{p}{\gamma-1}$,
for each electron with energy $\left(\frac{1}{\gamma-1}\frac{m_{e}}{m_{n}}\,T_{n}+\frac{1}{2}m_{e}\left(\mathbf{v}_{e}-\mathbf{v}_{n}\right)^{2}\right)$
Joules that is sourced by ionization, another electron has expended
$\phi_{ion}$ Joules to initiate the ionization process.\\
\\
The following definitions are made, representing the thermal energy
per unit volume per second transferred between species due to ionization
and recombination processes: 
\begin{flalign}
Q_{n}^{ion} & =\Gamma_{i}^{ion}\frac{1}{\gamma-1}T_{n} &  & \mbox{(neutral particles\ensuremath{\rightarrow} ions and electrons, due to ionization)}\nonumber \\
Q_{i}^{rec} & =\Gamma_{n}^{rec}\frac{1}{\gamma-1}T_{i} &  & \mbox{(ions \ensuremath{\rightarrow} neutral particles (and photons), due to recombination)}\nonumber \\
Q_{e}^{rec} & =\Gamma_{n}^{rec}\frac{1}{\gamma-1}T_{e} &  & \mbox{(electrons \ensuremath{\rightarrow} photons (and neutral particles), due to recombination)}\label{eq:533.0}
\end{flalign}
Hence, equation \ref{eq:533} can be re-expressed as:
\begin{align}
\left(\frac{\partial p_{i}}{\partial t}\right)_{ire} & =(\gamma-1)\left(\frac{m_{i}}{m_{n}}Q_{n}^{ion}-Q_{i}^{rec}+\frac{1}{2}m_{i}\left(\Gamma_{i}^{ion}\left(\mathbf{v}_{i}-\mathbf{v}_{n}\right)^{2}\right)\right)\nonumber \\
\left(\frac{\partial p_{e}}{\partial t}\right)_{ire} & =(\gamma-1)\left(\frac{m_{e}}{m_{n}}Q_{n}^{ion}-Q_{e}^{rec}+\Gamma_{i}^{ion}\left(\frac{1}{2}m_{e}\left(\mathbf{v}_{e}-\mathbf{v}_{n}\right)^{2}-\phi_{ion}\right)\right)\nonumber \\
\left(\frac{\partial p_{n}}{\partial t}\right)_{ire} & =(\gamma-1)\biggl[\frac{m_{i}}{m_{n}}\,Q_{i}^{rec}+\frac{m_{e}}{m_{n}}\,Q_{e}^{rec}-Q_{n}^{ion}\nonumber \\
 & \,\,\,\,\,\,\,+\Gamma_{n}^{rec}\biggl(\frac{1}{2}m_{n}v_{n}^{2}+\frac{1}{2}m_{i}v_{i}^{2}+\frac{1}{2}m_{e}v_{e}^{2}-m_{i}\mathbf{v}_{n}\cdot\mathbf{v}_{i}-m_{e}\mathbf{v}_{n}\cdot\mathbf{v}_{e}\biggr)\nonumber \\
 & \,\,\,\,\,\,\,+\Gamma_{n}^{ext}\frac{1}{2}m_{n}\left(\mathbf{v}_{n}-\mathbf{v}_{n0}\right)^{2}\biggr]+\Gamma_{n}^{ext}T_{n0}\label{eq:533.1}
\end{align}
It is worth noting that in the derivations presented in \cite{Meier,MeierPhd},
that $Q_{e}^{rec}$ (thermal energy transferred from electrons to
neutral particles due to recombination, per $\mbox{m}^{3}$ per second)
is not evaluated because its derivation with the moment-taking method
leads to an integral that cannot be solved easily. In particular,
the second moment of $C_{e}^{rec}$, where $C_{e}^{rec}$ is defined
in equation \ref{eq:530.1}, is 
\begin{align*}
\int\frac{1}{2}m_{e}V^{2}C_{e}^{rec}\,d\mathbf{V} & =-\frac{1}{2}m_{e}\int V'^{2}\left(f_{e}\int f_{i}\,\sigma_{rec}\,(V_{rel})\,V_{rel}\,d\mathbf{V}\right)\,d\mathbf{V'}\\
 & =-\frac{1}{2}m_{e}n_{i}\left(v_{e}^{2}\int f_{e}\,\sigma_{rec}\,(V_{rel})\,V_{rel}\,d\mathbf{V}+\int c_{e}^{2}\,f_{e}\,\sigma_{rec}(V_{rel})\,V_{rel}\,d\mathbf{V}\right)
\end{align*}
Here the substitution $\mathbf{V=}\mathbf{c}_{e}(\mathbf{r},t)+\mathbf{\mathbf{v}}_{e}(\mathbf{r},t)$
(equation \ref{eq:354}) has been made, and equations \ref{eq:355-1}
and \ref{eq:355-2} are used. Following the procedure for the example
derivation at the beginning of this section ($0^{th}$ moment of $C_{n}^{ion}$),
the first integral is simply\\
 $-\frac{1}{2}m_{e}n_{i}\left(v_{e}^{2}\int f_{e}\,\sigma_{rec}\,(V_{rel})\,V_{rel}\,d\mathbf{V}\right)=-\Gamma_{n}^{rec}\frac{1}{2}m_{e}v_{e}^{2}$.
It is assumed that the electron thermal speed is much greater than
the ion thermal speed $V_{the}\gg V_{thi}$, and also much greater
than the relative ion-electron fluid speed $V_{the}\gg|\mathbf{v}_{i}-\mathbf{v}_{e}|$,
so that $V_{rel}\approx c_{e}$, and the second integral reduces to
$-\frac{1}{2}m_{e}n_{i}\left(\int c_{e}^{3}f_{e}\sigma_{rec}(c_{e})\,d\mathbf{V}\right)$.
The definition $Q_{e}^{rec}=\frac{1}{2}m_{e}n_{i}\left(\int c_{e}^{3}f_{e}\sigma_{rec}(c_{e})\,d\mathbf{V}\right)$
is made in \cite{Meier,MeierPhd}, but the integral cannot be easily
evaluated, and it is suggested that this term can be dropped if the
loss of electron thermal energy due to recombination is not expected
to play an important role in the energy balance. However, from equation
\ref{eq:533.0}, it can be seen that $Q_{e}^{rec}=\frac{T_{e}}{T_{i}}Q_{i}^{rec}$,
so that in cases where $T_{i}\sim T_{e},$ it would be unreasonable
to neglect $Q_{e}^{rec}$ while retaining $Q_{i}^{rec}$. The $Q_{e}^{rec}$
term is included as an undetermined energy sink/source for the electron/neutral
fluids respectively in \cite{Meier,MeierPhd}, without scaling by
the factor $m_{e}/m_{n}$ in the neutral fluid energy equation, and
is ignored when the equations are implemented to code. When implemented
to the DELiTE framework MHD code, it was found that inclusion of the
$Q_{e}^{rec}$ term as an energy source for the neutral fluid (without
the scaling factor $m_{e}/m_{n}$ in equation \ref{eq:533.1}) leads
to significant increases in neutral particle temperature, as shown
in section \ref{subsec:Effect-of-inclusion}. 

However, as discussed above, from looking at the kinematics of the
radiative recombination reaction, it is more physical to neglect $Q_{e}^{rec}$
as an energy source for the neutral fluid, but include it as an energy
sink for the electron fluid, while $Q_{i}^{rec}$ may be included
as both an energy source for the neutral fluid and an energy sink
for the ion fluid. As discussed in section \ref{subsec:Effect-of-inclusion},
peak electron temperature falls by around just one percent when the
$Q_{e}^{rec}$ term is included as energy sink for the electron fluid.
Overall, in the regimes studied, it turns out that the $Q_{e}^{rec}$
term can be neglected without significantly affecting electron temperature.

\subsection{Charge exchange\label{subsec:Charge-exchange}}

In order to find the terms in the MHD equations that correspond to
the charge exchange reactions, the process of taking moments of the
charge exchange collision operators can't be avoided due to, and is
complicated by, the degeneracy associated with the charge exchange
reaction. In this work, the details of the process won't be reproduced
- only the required results that were originally achieved in \cite{Pauls},
and then very well detailed in \cite{Meier,MeierPhd}, will be presented. 

The collision operators defining the time-rates of change of $f_{n}$
and $f_{i}$ due to reacting collisions between ions and neutral particles
that result in charge exchange are \cite{Meier}: 
\begin{align*}
C_{n}^{cx} & =\frac{m_{i}}{m_{n}}f_{i}\int f_{n}\sigma_{cx}(V_{rel})\,V_{rel}\,d\mathbf{V}-\frac{m_{i}}{m_{n}}f_{n}\int f_{i}\sigma_{cx}(V_{rel})\,V_{rel}\,d\mathbf{V}\\
C_{i}^{cx} & =f_{n}\int f_{i}\sigma_{cx}(V_{rel})\,V_{rel}\,d\mathbf{V}-f_{i}\int f_{n}\sigma_{cx}(V_{rel})\,V_{rel}\,d\mathbf{V}
\end{align*}
$\sigma_{cx}(V_{rel})\,[\mbox{m}^{2}]$ is the cross-section for charge
exchange collisions. In this case, $V_{rel}=|\mathbf{V}_{i}-\mathbf{V}_{n}|,$
is the relative particle speed for the reaction, where $\mathbf{V}_{i}$
is the ion particle velocity and $\mathbf{V}_{n}$ is the neutral
particle velocity. The first term in the expression for $C_{n}^{cx}$
and the second term in the expression for $C_{i}^{cx}$ represent
the conversion of ions to neutral particles, the other two terms represent
the conversion of neutral particles to ions. 

\subsubsection{$\mathbf{0}^{\mathbf{th}}$ moment of the charge exchange collision
operator}

Charge exchange collisions do not change the number of ions, electrons
or neutral particles, but the $0^{th}$ moment will be taken here
to help illustrate some of the details of the charge exchange collision
terms. Using the method outlined in \cite{Pauls}, and detailed in
\cite{MeierPhd}, a good approximation for $C_{i}^{cx}$ is 
\[
C_{i}^{cx}\approx\sigma_{cx}(V_{i}^{*}n_{i}f_{n}-V_{n}^{*}n_{n}f_{i})
\]
where $V_{\alpha}^{*}=V_{th\alpha}\sqrt{4/\pi+(|\mathbf{V}-\mathbf{v}_{\alpha}|/V_{th\alpha})^{2}}$.
The $0^{th}$ moments of $C_{i}^{cx}$ and $C_{n}^{cx}$ can then
be evaluated as\- $\int C_{i}^{cx}d\mathbf{V}=(\Gamma^{cx}-\Gamma^{cx})=0$,
and $\int C_{n}^{cx}d\mathbf{V}=\frac{m_{i}}{m_{n}}(\Gamma^{cx}-\Gamma^{cx})=0$,
where $\Gamma^{cx}\,[\mbox{m}^{-3}\mbox{s}^{-1}]$ is the source rate
of neutral particles, equal to the source rate of ions, for the charge
exchange reaction: 
\begin{equation}
\Gamma^{cx}=n_{i}n_{n}\sigma_{cx}(V_{cx})V_{cx}\label{eq:535}
\end{equation}
Note $\sigma_{cx}$ is evaluated at $V_{cx,}$where $V_{cx}$ is a
representative speed for charge exchange collisions \cite{Pauls,Meier}:
\begin{equation}
V_{cx}=\sqrt{\frac{4}{\pi}V_{thi}^{2}+\frac{4}{\pi}V_{thn}^{2}+v_{in}^{2}}\label{eq:536}
\end{equation}
where $v_{in}=|\mathbf{v}_{in}|=|\mathbf{v}_{i}-\mathbf{v}_{n}|$.
A formula for $\sigma_{cx}(V_{cx})\,[\mbox{m}^{2}]$ can be found
based on charge exchange data from Barnett \cite{Barnett,MeierPhd}.
For hydrogen and deuterium the fitting formulae are 
\begin{align}
\sigma_{cx-H}(V_{cx}) & =1.12\times10^{-18}-7.15\times10^{-20}\mbox{ln}(V_{cx})\nonumber \\
\sigma_{cx-D}(V_{cx}) & =1.09\times10^{-18}-7.15\times10^{-20}\mbox{ln}(V_{cx})\label{eq:536.1}
\end{align}

\subsubsection{$\mathbf{1}^{\mathbf{st}}$ moment of the charge exchange collision
operator}

The first moments of the charge exchange collision operators are \cite{Pauls,Meier}:

\begin{align}
\int m_{i}\mathbf{V}C_{i}^{cx}d\mathbf{V} & \approx-m_{i}\mathbf{v}_{in}\Gamma^{cx}-\mathbf{R}_{ni}^{cx}+\mathbf{R}_{in}^{cx}\nonumber \\
\int m_{n}\mathbf{V}C_{n}^{cx}d\mathbf{V} & \approx m_{i}\mathbf{v}_{in}\Gamma^{cx}+\mathbf{R}_{ni}^{cx}-\mathbf{R}_{in}^{cx}\label{eq:536.2}
\end{align}
where 

\begin{align}
\mathbf{R}_{in}^{cx}\approx-m_{i}\sigma_{cx}(V_{cx})n_{i}n_{n}\mathbf{v}_{in}V_{thn}^{2}\left(4\left(\frac{4}{\pi}V_{thi}^{2}+v_{in}^{2}\right)+\frac{9\pi}{4}V_{thn}^{2}\right)^{-\frac{1}{2}}\nonumber \\
\mathbf{R}_{ni}^{cx}\approx m_{i}\sigma_{cx}(V_{cx})n_{i}n_{n}\mathbf{v}_{in}V_{thi}^{2}\left(4\left(\frac{4}{\pi}V_{thn}^{2}+v_{in}^{2}\right)+\frac{9\pi}{4}V_{thi}^{2}\right)^{-\frac{1}{2}}\label{eq:538}
\end{align}
The term $m_{i}\mathbf{v}_{in}\Gamma^{cx}$ represents the transfer
of momentum due to bulk fluid effects \cite{Meier,MeierPhd}. $\mathbf{R}_{ni}^{cx}\mbox{ and }\mathbf{R}_{in}^{cx}$
represent frictional drag forces that arise due to charge exchange,
with $\mathbf{v}_{in}=\mathbf{v}_{i}-\mathbf{v}_{n}$. Such frictional
terms do not arise for the ionization and recombination processes,
in which the electron thermal speed is assumed to be far higher than
the relative particle motion \cite{Meier,MeierPhd}. The derivation
of the moments of the reacting collision operators assumes that the
reacting species are described by Maxwellian distributions. As a result,
non-Maxwellian effects due to thermal gradients are neglected and
don't appear in the expressions for $\mathbf{R}_{ni}^{cx}\mbox{ and }\mathbf{R}_{in}^{cx}$
\cite{Meier}, as they did in the expressions for $\mathbf{R}_{ei}$
and $\mathbf{R}_{ie}$ (equation \ref{eq:472.471}). 

The notation $(X)_{cx}$ is introduced to represent the part of $X$
that pertains to charge exchange collisions, in the same way that
$(X)_{ire}$ is the part of $X$ relating to ionization, recombination
and external sources (section \ref{subsec:Ionization-and-recombination}).
Recall that $m_{\alpha}$ times the first moment of the first term
of the Boltzmann equation is $\frac{\partial\left(\rho_{\alpha}\mathbf{v}_{\alpha}\right)}{\partial t}$
(appendix \ref{subsec:Braginski-equations-from}). Since $\left(\frac{\partial\left(\rho_{\alpha}\mathbf{v}_{\alpha}\right)}{\partial t}\right)_{cx}=\rho_{\alpha}\left(\frac{\partial\mathbf{v}_{\alpha}}{\partial t}\right)_{cx}+\mathbf{v}_{\alpha}\left(\frac{\partial\rho_{\alpha}}{\partial t}\right)_{cx}$,
and charge exchange is not a net source of particles $\left(\Rightarrow\left(\frac{\partial\rho_{\alpha}}{\partial t}\right)_{cx}=0\right)$,
the terms $\rho_{\alpha}\left(\frac{\partial\mathbf{v}_{\alpha}}{\partial t}\right)_{cx}$
can be expressed, using equation \ref{eq:536.2}, as: 
\begin{align}
\rho_{i}\left(\frac{\partial\mathbf{v}_{i}}{\partial t}\right)_{cx} & \approx-m_{i}\mathbf{v}_{in}\Gamma^{cx}-\mathbf{R}_{ni}^{cx}+\mathbf{R}_{in}^{cx}\nonumber \\
\rho_{n}\left(\frac{\partial\mathbf{v}_{n}}{\partial t}\right)_{cx} & \approx m_{i}\mathbf{v}_{in}\Gamma^{cx}+\mathbf{R}_{ni}^{cx}-\mathbf{R}_{in}^{cx}\label{eq:537}
\end{align}

\subsubsection{$\mathbf{2}^{\mathbf{nd}}$ moment of the charge exchange collision
operator}

The second moments of the charge exchange collision operators are
\cite{Pauls,Meier}:

\begin{align}
\int\frac{1}{2}m_{i}V^{2}C_{i}^{cx}d\mathbf{V} & \approx\Gamma^{cx}\frac{1}{2}m_{i}(v_{n}^{2}-v_{i}^{2})+\mathbf{v}_{n}\cdot\mathbf{R}_{in}^{cx}-\mathbf{v}_{i}\cdot\mathbf{R}_{ni}^{cx}+Q_{in}^{cx}-Q_{ni}^{cx}\nonumber \\
\int\frac{1}{2}m_{n}V^{2}C_{n}^{cx}d\mathbf{V} & \approx\Gamma^{cx}\frac{1}{2}m_{i}(v_{i}^{2}-v_{n}^{2})-\mathbf{v}_{n}\cdot\mathbf{R}_{in}^{cx}+\mathbf{v}_{i}\cdot\mathbf{R}_{ni}^{cx}-Q_{in}^{cx}+Q_{ni}^{cx}\label{eq:539}
\end{align}
where 

\begin{align}
Q_{in}^{cx}\approx m_{i}\sigma_{cx}(V_{cx})n_{i}n_{n}\frac{3}{4}V_{thn}^{2}\sqrt{\frac{4}{\pi}V_{thi}^{2}+\frac{64}{9\pi}V_{thn}^{2}+v_{in}^{2}}\nonumber \\
Q_{ni}^{cx}\approx m_{i}\sigma_{cx}(V_{cx})n_{i}n_{n}\frac{3}{4}V_{thi}^{2}\sqrt{\frac{4}{\pi}V_{thn}^{2}+\frac{64}{9\pi}V_{thi}^{2}+v_{in}^{2}}\label{eq:540}
\end{align}
Here, $Q_{in}^{cx}$ and $Q_{ni}^{cx}$ represent the transfer of
thermal energy \cite{Meier,MeierPhd} associated with charge exchange
reactions. As mentioned above, non-Maxwellian effects due to thermal
gradients are neglected; therefore they don't appear in the expressions
for $Q_{in}^{cx}\mbox{ and }Q_{ni}^{cx}$ \cite{Meier}, as they did
in the expressions for $Q_{ei}$ (equation \ref{eq:472.473}).

From appendix \ref{subsec:Braginski-equations-from}, the second moment
of the first term of the Boltzmann equation is $\frac{\partial}{\partial t}\left(\frac{1}{2}\rho_{\alpha}v_{\alpha}^{2}+\frac{p_{\alpha}}{\gamma-1}\right)=\rho_{\alpha}\mathbf{v}_{\alpha}\cdot\frac{\partial\mathbf{v}_{\alpha}}{\partial t}+\frac{1}{\gamma-1}\frac{\partial p_{\alpha}}{\partial t}$
so that: 
\[
\frac{1}{\gamma-1}\left(\frac{\partial p_{\alpha}}{\partial t}\right)_{cx}=\left(\frac{\partial}{\partial t}\left(\frac{1}{2}\rho_{\alpha}v_{\alpha}^{2}+\frac{p_{\alpha}}{\gamma-1}\right)\right)_{cx}-\rho_{\alpha}\mathbf{v}_{\alpha}\cdot\left(\frac{\partial\mathbf{v}_{\alpha}}{\partial t}\right)_{cx}
\]
Using equations \ref{eq:537} and \ref{eq:539}, this implies that
\begin{align}
\frac{1}{\gamma-1}\left(\frac{\partial p_{i}}{\partial t}\right)_{cx} & \approx\left(\Gamma^{cx}\frac{1}{2}m_{i}(v_{n}^{2}-v_{i}^{2})+\mathbf{v}_{n}\cdot\mathbf{R}_{in}^{cx}-\mathbf{v}_{i}\cdot\mathbf{R}_{ni}^{cx}+Q_{in}^{cx}-Q_{ni}^{cx}\right)\nonumber \\
 & -\mathbf{v}_{i}\cdot\left(-m_{i}(\mathbf{v}_{i}-\mathbf{v}_{n})\Gamma^{cx}-\mathbf{R}_{ni}^{cx}+\mathbf{R}_{in}^{cx}\right)\nonumber \\
 & =\Gamma^{cx}m_{i}(\frac{1}{2}v_{n}^{2}-\mathbf{v}_{i}\cdot\mathbf{v}_{n}+\frac{1}{2}v_{i}^{2})+(\mathbf{v}_{n}-\mathbf{v}_{i})\cdot\mathbf{R}_{in}^{cx}+Q_{in}^{cx}-Q_{ni}^{cx}\nonumber \\
\Rightarrow\frac{1}{\gamma-1}\left(\frac{\partial p_{i}}{\partial t}\right)_{cx} & \approx\Gamma^{cx}\frac{1}{2}m_{i}(\mathbf{v}_{n}-\mathbf{v}_{i})^{2}+(\mathbf{v}_{n}-\mathbf{v}_{i})\cdot\mathbf{R}_{in}^{cx}+Q_{in}^{cx}-Q_{ni}^{cx}\label{eq:540.1}
\end{align}
and 
\begin{align}
\frac{1}{\gamma-1}\left(\frac{\partial p_{n}}{\partial t}\right)_{cx} & \approx\left(\Gamma^{cx}\frac{1}{2}m_{i}(v_{i}^{2}-v_{n}^{2})-\mathbf{v}_{n}\cdot\mathbf{R}_{in}^{cx}+\mathbf{v}_{i}\cdot\mathbf{R}_{ni}^{cx}-Q_{in}^{cx}+Q_{ni}^{cx}\right)\nonumber \\
 & -\mathbf{v}_{n}\cdot\left(m_{i}(\mathbf{v}_{i}-\mathbf{v}_{n})\Gamma^{cx}+\mathbf{R}_{ni}^{cx}-\mathbf{R}_{in}^{cx}\right)\nonumber \\
 & =\Gamma^{cx}m_{i}(\frac{1}{2}v_{i}^{2}-\mathbf{v}_{n}\cdot\mathbf{v}_{i}+\frac{1}{2}v_{n}^{2})+(\mathbf{v}_{i}-\mathbf{v}_{n})\cdot\mathbf{R}_{ni}^{cx}-Q_{in}^{cx}+Q_{ni}^{cx}\nonumber \\
\Rightarrow\frac{1}{\gamma-1}\left(\frac{\partial p_{n}}{\partial t}\right)_{cx} & \approx\Gamma^{cx}\frac{1}{2}m_{i}(\mathbf{v}_{i}-\mathbf{v}_{n})^{2}+(\mathbf{v}_{i}-\mathbf{v}_{n})\cdot\mathbf{R}_{ni}^{cx}-Q_{in}^{cx}+Q_{ni}^{cx}\label{eq:540.2}
\end{align}
Note that the term $(\mathbf{v}_{n}-\mathbf{v}_{i})\cdot\mathbf{R}_{in}^{cx}=-\mathbf{v}_{in}\cdot\mathbf{R}_{in}^{cx}$
in the ion energy equation represents the rate of frictional work
done by neutral fluid on the ion fluid as a result of charge exchange
reactions, and the similar term $\mathbf{v}_{in}\cdot\mathbf{R}_{ni}^{cx}$
in the neutral energy equation represents the rate of frictional work
done by $\mathbf{R}_{ni}^{cx}$, which acts on the neutral fluid with
relative velocity $\mathbf{v}_{in}$.\\
\\

\section{3-fluid MHD equations\label{sec:3-fluid-MHD-equations}}

$\left(X\right)_{CE},\,\left(X\right)_{scatt.},\,\left(X\right)_{react.},\,\left(X\right)_{ext.},\,\left(X\right)_{ire},\,\mbox{ and }\left(X\right)_{cx}$
are, respectively, the parts of $X$ that pertain to the combination
of \emph{collisions} and \emph{external} sources, to \emph{scattering}
collisions, to \emph{reacting} collisions\emph{, }to\emph{ external}
sources, to the combination of \emph{ionization}, \emph{recombination}
and \emph{external} sources, and to \emph{charge} \emph{exchange}
collisions. These quantities are related as 
\[
\left(X\right)_{CE}=\left(X\right)_{scatt.}+\left(X\right)_{react.}+\left(X\right)_{ext.}=\left(X\right)_{scatt.}+\left(X\right)_{ire}+\left(X\right)_{cx}
\]
Combining equations \ref{eq:521}, \ref{eq:531.6}, and \ref{eq:537},
and using the identity $\mathbf{v}_{in}=\mathbf{v}_{i}-\mathbf{v}_{n}$,
the complete set of terms that arise in the species momentum equations
due to scattering and reactive collisions, and an external neutral
particle source can be assembled:

\begin{align}
\left(\frac{\partial\mathbf{v}_{i}}{\partial t}\right)_{CE} & =\frac{1}{\rho_{i}}\left(\mathbf{R}_{ie}+\mathbf{R}_{in}-\Gamma_{i}^{ion}m_{i}\mathbf{v}_{in}-\Gamma^{cx}m_{i}\mathbf{v}_{in}-\mathbf{R}_{ni}^{cx}+\mathbf{R}_{in}^{cx}\right)\nonumber \\
\left(\frac{\partial\mathbf{v}_{e}}{\partial t}\right)_{CE} & =\frac{1}{\rho_{e}}\left(\mathbf{R}_{ei}+\mathbf{R}_{en}+\Gamma_{i}^{ion}m_{e}(\mathbf{v}_{n}-\mathbf{v}_{e})\right)\label{eq:541}\\
\left(\frac{\partial\mathbf{v}_{n}}{\partial t}\right)_{CE} & =\frac{1}{\rho_{n}}\biggl(\mathbf{R}_{ni}+\mathbf{R}_{ne}+\Gamma_{n}^{rec}(m_{i}\mathbf{v}_{i}+m_{e}\mathbf{v}_{e}-m_{n}\mathbf{v}_{n})\nonumber \\
 & \,\,\,\,\,\,\,+\Gamma_{n}^{ext}m_{n}(\mathbf{v}_{n0}-\mathbf{v}_{n})+\Gamma^{cx}m_{i}\mathbf{v}_{in}+\mathbf{R}_{ni}^{cx}-\mathbf{R}_{in}^{cx}\biggr)\nonumber 
\end{align}
Similarly, combining equations \ref{eq:522}, \ref{eq:533.1}, \ref{eq:540.1},
and \ref{eq:540.2}, the equivalent set of terms in the species energy
equations are:{\small{}
\begin{align}
\left(\frac{\partial p_{i}}{\partial t}\right)_{CE} & =\left(\gamma-1\right)\biggl[Q_{ie}+Q_{in}+\frac{m_{i}}{m_{n}}\,Q_{n}^{ion}-Q_{i}^{rec}+\frac{1}{2}m_{i}(\Gamma_{i}^{ion}+\Gamma^{cx})v_{in}^{2}\nonumber \\
 & -\mathbf{v}_{in}\cdot\mathbf{R}_{in}^{cx}+Q_{in}^{cx}-Q_{ni}^{cx}\biggr]\nonumber \\
\left(\frac{\partial p_{e}}{\partial t}\right)_{CE} & =\left(\gamma-1\right)\left(Q_{ei}+Q_{en}+\frac{m_{e}}{m_{n}}\,Q_{n}^{ion}-Q_{e}^{rec}+\Gamma_{i}^{ion}\left(\frac{1}{2}m_{e}\left(\mathbf{v}_{e}-\mathbf{v}_{n}\right)^{2}-\phi_{ion}\right)\right)\label{eq:542}\\
\left(\frac{\partial p_{n}}{\partial t}\right)_{CE} & =\left(\gamma-1\right)\biggl[Q_{ni}+Q_{ne}+\frac{m_{i}}{m_{n}}\,Q_{i}^{rec}+\frac{m_{e}}{m_{n}}\,Q_{e}^{rec}-Q_{n}^{ion}+\Gamma_{n}^{rec}\biggl(\frac{1}{2}m_{n}v_{n}^{2}+\frac{1}{2}m_{i}v_{i}^{2}+\frac{1}{2}m_{e}v_{e}^{2}-m_{i}\mathbf{v}_{n}\cdot\mathbf{v}_{i}\nonumber \\
 & -m_{e}\mathbf{v}_{n}\cdot\mathbf{v}_{e}\biggr)+\Gamma_{n}^{ext}\frac{1}{2}m_{n}\left(\mathbf{v}_{n}-\mathbf{v}_{n0}\right)^{2}+\Gamma^{cx}\frac{1}{2}m_{i}v_{in}^{2}+\mathbf{v}_{in}\cdot\mathbf{R}_{ni}^{cx}-Q_{in}^{cx}+Q_{ni}^{cx}\biggr]+\Gamma_{n}^{ext}T_{n0}\nonumber 
\end{align}
}{\small\par}

\subsection{Mass conservation}

Mass conservation is expressed by equation \ref{eq:531.2}:

\begin{align}
\dot{n}_{i} & =-\nabla\cdot(n_{i}\mathbf{v}_{i})+\Gamma_{i}^{ion}-\Gamma_{n}^{rec}\nonumber \\
\dot{n}_{e} & =-\nabla\cdot(n_{e}\mathbf{v}_{e})+\Gamma_{i}^{ion}-\Gamma_{n}^{rec}\label{eq:543}\\
\dot{n}_{n} & =-\nabla\cdot(n_{n}\mathbf{v}_{n})+\Gamma_{n}^{rec}-\Gamma_{i}^{ion}+\Gamma_{n}^{ext}\nonumber 
\end{align}

\subsection{Momentum conservation}

The expression for species momentum conservation (equation \ref{eq:472-2})
can be combined with equation \ref{eq:541} for the species momentum
equations in the three-fluid system. The term $\frac{1}{\rho_{\alpha}}\mathbf{R}_{\alpha}$
in equation \ref{eq:472-2}, which arose by taking the first moment
of the collision operator for scattering collisions only, is replaced
with the terms in equation \ref{eq:541} to yield:
\begin{align}
\frac{\partial\mathbf{v}_{i}}{\partial t} & =-(\mathbf{v}_{i}\cdot\nabla)\mathbf{v}_{i}+\frac{1}{\rho_{i}}\biggl[-\nabla p_{i}-\nabla\cdot\overline{\boldsymbol{\pi}}_{i}+q_{i}n_{i}\left(\mathbf{E}+\mathbf{v}_{i}\times\mathbf{B}\right)+\mathbf{R}_{ie}+\mathbf{R}_{in}\nonumber \\
 & -\Gamma_{i}^{ion}m_{i}\mathbf{v}_{in}-\Gamma^{cx}m_{i}\mathbf{v}_{in}-\mathbf{R}_{ni}^{cx}+\mathbf{R}_{in}^{cx}\biggr]\nonumber \\
\frac{\partial\mathbf{v}_{e}}{\partial t} & =-(\mathbf{v}_{e}\cdot\nabla)\mathbf{v}_{e}+\frac{1}{\rho_{e}}\biggl[-\nabla p_{e}-\nabla\cdot\overline{\boldsymbol{\pi}}_{e}+q_{e}n_{e}\left(\mathbf{E}+\mathbf{v}_{e}\times\mathbf{B}\right)+\mathbf{R}_{ei}+\mathbf{R}_{en}\nonumber \\
 & +\Gamma_{i}^{ion}m_{e}(\mathbf{v}_{n}-\mathbf{v}_{e})\biggr]\nonumber \\
\frac{\partial\mathbf{v}_{n}}{\partial t} & =-(\mathbf{v}_{n}\cdot\nabla)\mathbf{v}_{n}+\frac{1}{\rho_{n}}\biggl[-\nabla p_{n}-\nabla\cdot\overline{\boldsymbol{\pi}}_{n}+\mathbf{R}_{ni}+\mathbf{R}_{ne}+\Gamma_{n}^{rec}(m_{i}\mathbf{v}_{i}+m_{e}\mathbf{v}_{e}-m_{n}\mathbf{v}_{n})\nonumber \\
 & +\Gamma^{cx}m_{i}\mathbf{v}_{in}+\mathbf{R}_{ni}^{cx}-\mathbf{R}_{in}^{cx}+\Gamma_{n}^{ext}m_{n}(\mathbf{v}_{n0}-\mathbf{v}_{n})\biggr]\label{eq:544}
\end{align}

\subsection{Energy conservation}

The expression for species energy conservation (equation \ref{eq:472.31})
can be combined with equation \ref{eq:542} for the species energy
equations in the three-fluid system. The term $\left(\gamma-1\right)Q_{\alpha}$
in equation \ref{eq:472.31}, which arose by taking the second moment
of the collision operator for scattering collisions only, is replaced
with the terms in equation \ref{eq:542}. The resultant species energy
equations for the three-fluid system are: 
\begin{align*}
\frac{\partial p_{i}}{\partial t} & =-\mathbf{v}_{i}\cdot\nabla p_{i}-\gamma p_{i}\nabla\cdot\mathbf{v}_{i}+(\gamma-1)\biggl[-\overline{\boldsymbol{\pi}}_{i}:\nabla\mathbf{v}_{i}-\nabla\cdot\mathbf{q}_{i}+Q_{ie}+Q_{in}+\frac{m_{i}}{m_{n}}Q_{n}^{ion}\\
 & \,\,\,\,\,\,\,-Q_{i}^{rec}+\frac{1}{2}m_{i}(\Gamma_{i}^{ion}+\Gamma^{cx})v_{in}^{2}-\mathbf{v}_{in}\cdot\mathbf{R}_{in}^{cx}+Q_{in}^{cx}-Q_{ni}^{cx}\biggr]\\
\frac{\partial p_{e}}{\partial t} & =-\mathbf{v}_{e}\cdot\nabla p_{e}-\gamma p_{e}\nabla\cdot\mathbf{v}_{e}+(\gamma-1)\biggl[-\overline{\boldsymbol{\pi}}_{e}:\nabla\mathbf{v}_{e}-\nabla\cdot\mathbf{q}_{e}+Q_{ei}+Q_{en}\\
 & \,\,\,\,\,\,\,+\frac{m_{e}}{m_{n}}Q_{n}^{ion}-Q_{e}^{rec}+\Gamma_{i}^{ion}\left(\frac{1}{2}m_{e}\left(\mathbf{v}_{e}-\mathbf{v}_{n}\right)^{2}-\phi_{ion}\right)\biggr]\\
\frac{\partial p_{n}}{\partial t} & =-\mathbf{v}_{n}\cdot\nabla p_{n}-\gamma p_{n}\nabla\cdot\mathbf{v}_{n}+(\gamma-1)\biggl[-\overline{\boldsymbol{\pi}}_{n}:\nabla\mathbf{v}_{n}-\nabla\cdot\mathbf{q}_{n}+Q_{ni}+Q_{ne}+\frac{m_{i}}{m_{n}}\,Q_{i}^{rec}+\frac{m_{e}}{m_{n}}\,Q_{e}^{rec}\\
 & \,\,\,\,\,\,\,-Q_{n}^{ion}+\Gamma_{n}^{rec}\biggl(\frac{1}{2}m_{n}v_{n}^{2}+\frac{1}{2}m_{i}v_{i}^{2}+\frac{1}{2}m_{e}v_{e}^{2}-m_{i}\mathbf{v}_{n}\cdot\mathbf{v}_{i}-m_{e}\mathbf{v}_{n}\cdot\mathbf{v}_{e}\biggr)+\Gamma^{cx}\frac{1}{2}m_{i}v_{in}^{2}\\
 & \,\,\,\,\,\,\,+\mathbf{v}_{in}\cdot\mathbf{R}_{ni}^{cx}-Q_{in}^{cx}+Q_{ni}^{cx}+\Gamma_{n}^{ext}\frac{1}{2}m_{n}\left(\mathbf{v}_{n}-\mathbf{v}_{n0}\right)^{2}\biggr]+\Gamma_{n}^{ext}T_{n0}
\end{align*}

\section{2-fluid MHD equations\label{sec:2-fluid-MHD-equations}}

Applying the same procedure described in appendix \ref{subsec:Single-fluid-Magnetohydrodynamic}
for the terms in the 3-fluid MHD equations that don't correspond to
interspecies collisions, and summing terms corresponding to interspecies
collisions, with the limit $m_{e}\rightarrow0$, the ion and electron
conservation equations can be reduced to a single plasma-fluid description.
In the limit $m_{e}\rightarrow0$, the single plasma-fluid velocity
is approximately the ion fluid velocity as is clarified by the identities
$\mathbf{v}=\underset{\alpha}{\frac{1}{\rho}\Sigma}\rho_{\alpha}\mathbf{v}_{\alpha}\mbox{ and }\rho=\underset{\alpha}{\Sigma}\rho_{\alpha}$,
so that $\mathbf{v}_{in}\rightarrow(\mathbf{v}-\mathbf{v}_{n})$.
The charged-neutral particle frictional forces $\mathbf{R}_{in}=-\mathbf{R}_{ni}\mbox{ and }\mathbf{R}_{en}=-\mathbf{R}_{ne}$,
and heat exchange terms $Q_{in},\,Q_{en},\,Q_{ni}$, and $Q_{ne}$
can be neglected, as mentioned in section \ref{subsec:Scattering-collision-terms}.
Following from \cite{MeierPhd}, the parameter $\lambda$, where $0\leq\lambda\leq1$,
is introduced. If the mean free paths of the charge-exchanged neutral
particles are expected to be large, then $\lambda$ is set between
$0$ and $1$, so that charge-exchanged neutral particles leave the
system without reacting again, taking their energy and momentum with
them. If the mean free paths of the charge-exchanged neutral particles
are expected to be short, then $\lambda$ is set to zero. By default,
$\lambda$ is set to zero in MHD simulations, consistent with conservation
of energy associated with the neutral fluid. The resulting set of
conservation equations, in continuous form, for the two-fluid system,
including ${\color{red}\ensuremath{\mbox{ionization \& recombination}}}$
and ${\color{blue}\mbox{charge exchange}}$ terms, as well as \textcolor{cyan}{neutral
source} terms, and \textcolor{red}{${\color{green}\mbox{density diffusion}}$}
(correction terms ${\color{green}\mathbf{f}_{\zeta}\ensuremath{/\rho}}$
and ${\color{green}\mathbf{f}_{\zeta_{n}}\ensuremath{/\rho_{n}}}$
are included to maintain energy and (in some cases) angular momentum
conservation, as described in section \ref{subsec:Maintenance-of-momentum}),
is

\selectlanguage{german}%
\texttt{
\begin{align*}
\dot{n} & =-\nabla\cdot(n\mathbf{v}){\color{red}+\Gamma_{i}^{ion}-\Gamma_{n}^{rec}{\color{green}+\ensuremath{\nabla\cdot}\left(\zeta\nabla n\right)}}\\
\dot{\mathbf{v}} & =-\mathbf{v}\cdot\nabla\mathbf{v}+\frac{1}{\rho}\left(-\nabla p-\nabla\cdot\overline{\boldsymbol{\pi}}+\mathbf{J\times}\mathbf{B}{\color{red}-\Gamma_{i}^{ion}m_{i}\mathbf{v}_{in}}{\color{blue}\ensuremath{-\Gamma^{cx}m_{i}\mathbf{v}_{in}\mathbf{-R}_{ni}^{cx}+\mathbf{R}_{in}^{cx}}}{\color{green}+\mathbf{f}_{\zeta}}\right)\\
\dot{p} & =-\mathbf{v}\cdot\nabla p-\gamma p\,\nabla\cdot\mathbf{v}+(\gamma-1)\biggl(-\overline{\boldsymbol{\pi}}:\nabla\mathbf{v}-\nabla\cdot\mathbf{q}+\eta'J^{2}{\color{red}+\Gamma_{i}^{ion}}{\color{red}\frac{1}{2}m_{i}v_{in}^{2}+Q_{n}^{ion}-\Gamma_{i}^{ion}\phi_{ion}}\\
 & {\color{blue}\,\,\,\,\,\,\,{\color{red}-Q_{i}^{rec}-Q_{e}^{rec}}-\mathbf{R}_{in}^{cx}\cdot\mathbf{v}_{in}}{\color{blue}+Q_{in}^{cx}-Q_{ni}^{cx}}{\color{blue}+\Gamma^{cx}\frac{1}{2}m_{i}v_{in}^{2}}\biggr)\\
\dot{n}_{n} & =-\nabla\cdot(n_{n}\mathbf{v}_{n}){\color{red}{\color{blue}{\color{blue}-}\lambda\Gamma^{cx}}-\Gamma_{i}^{ion}+\Gamma_{n}^{rec}}{\color{green}{\color{cyan}{\color{magenta}{\color{cyan}+\Gamma_{n}^{ext}}}}\ensuremath{+\ensuremath{\nabla\cdot}\left(\zeta_{n}\nabla n_{n}\right)}}\\
\dot{\mathbf{v}}_{n} & =-\mathbf{v}_{n}\cdot\nabla\mathbf{v}_{n}+\frac{1}{\rho_{n}}\biggl(-\nabla p_{n}-\nabla\cdot\overline{\boldsymbol{\pi}}_{n}{\color{red}+\Gamma_{n}^{rec}m_{i}\mathbf{v}_{in}}{\color{blue}{\color{blue}-\mathbf{R}_{in}^{cx}+(1-\lambda)(\mathbf{R}_{ni}^{cx}}+\Gamma^{cx}m_{i}\mathbf{v}_{in})}\\
 & \,\,\,\,\,\,\,{\color{cyan}+\Gamma_{n}^{ext}m_{n}(\mathbf{v}_{n0}-\mathbf{v}_{n})}{\color{green}+\mathbf{f}_{\zeta_{n}}\biggr)}\\
\dot{p}_{n} & =-\mathbf{v}_{n}\cdot\nabla p_{n}-\gamma p_{n}\,\nabla\cdot\mathbf{v}_{n}+(\gamma-1)\biggl(-\overline{\boldsymbol{\pi}}_{n}:\nabla\mathbf{v}_{n}-\nabla\cdot\mathbf{q}_{n}{\color{red}-Q_{n}^{ion}+\Gamma_{n}^{rec}\frac{1}{2}m_{i}v_{in}^{2}+Q_{i}^{rec}}\\
 & \,\,\,\,\,\,\,{\color{blue}+(1-\lambda)(\mathbf{R}_{ni}^{cx}\cdot\mathbf{v}_{in}+\Gamma^{cx}\,\frac{1}{2}m_{i}v_{in}^{2}+Q_{ni}^{cx})-Q_{in}^{cx}}{\color{cyan}+\Gamma_{n}^{ext}\frac{1}{2}m_{n}(\mathbf{v}_{n}-\mathbf{v}_{n0})^{2}}\biggr){\color{cyan}+\Gamma_{n}^{ext}T_{n0}}
\end{align*}
}\foreignlanguage{english}{As mentioned in appendix \ref{sec:SummaryKin_MHD_EQ},
the code has the option to evolve the single plasma-fluid energy equation
or to evolve separate energy equations for the ions and electrons.
For the latter option, when plasma-neutral interaction is included,
the ion and electron energy equations are}\texttt{
\begin{align*}
\dot{p}_{i} & =-\mathbf{v}\cdot\nabla p_{i}-\gamma p_{i}\,\nabla\cdot\mathbf{v}+(\gamma-1)\biggl(-\overline{\boldsymbol{\pi}}:\nabla\mathbf{v}-\nabla\cdot\mathbf{q}_{i}+Q_{ie}{\color{red}+\Gamma_{i}^{ion}}{\color{red}\frac{1}{2}m_{i}v_{in}^{2}+Q_{n}^{ion}-Q_{i}^{rec}}\\
 & {\color{blue}-\mathbf{R}_{in}^{cx}\cdot\mathbf{v}_{in}}{\color{blue}+Q_{in}^{cx}-Q_{ni}^{cx}}{\color{blue}+\Gamma^{cx}\frac{1}{2}m_{i}v_{in}^{2}}\biggr)\\
\dot{p}_{e} & =-\mathbf{v}\cdot\nabla p_{e}-\gamma p_{e}\,\nabla\cdot\mathbf{v}+(\gamma-1)\left(+\eta'J^{2}-\nabla\cdot\mathbf{q}_{e}-Q_{ie}{\color{red}-\Gamma_{i}^{ion}\phi_{ion}-Q_{e}^{rec}}\right)
\end{align*}
}\foreignlanguage{english}{where $Q_{ie}$ is defined in equation
\ref{eq:517.61}. }
\selectlanguage{english}%

\section{Conservation properties with inclusion of neutral fluid\label{sec:Energy-conservation-with}}

\begin{figure}[H]
\centering{}\subfloat[]{\includegraphics[width=10cm,height=6.5cm]{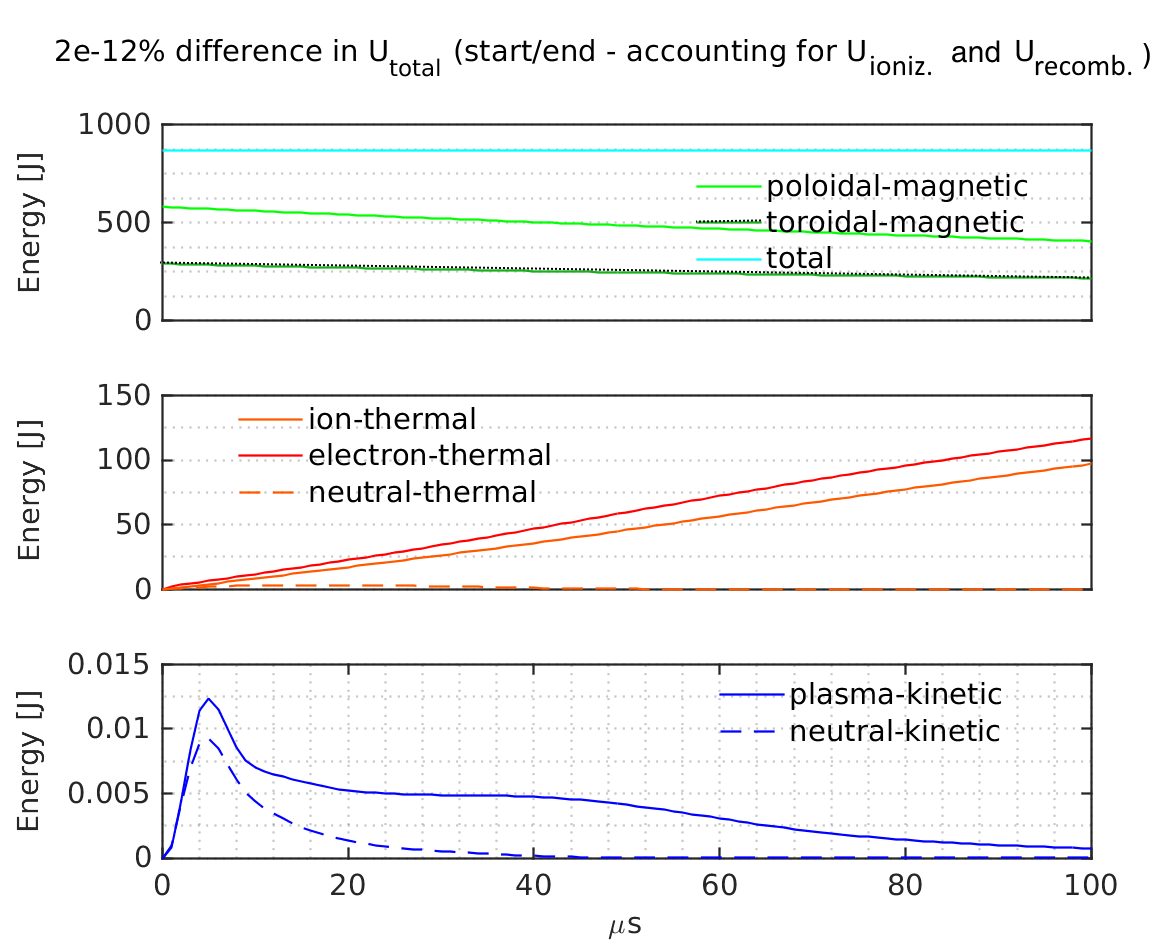}

}\hfill{}\subfloat[]{\includegraphics[width=6cm,height=5cm]{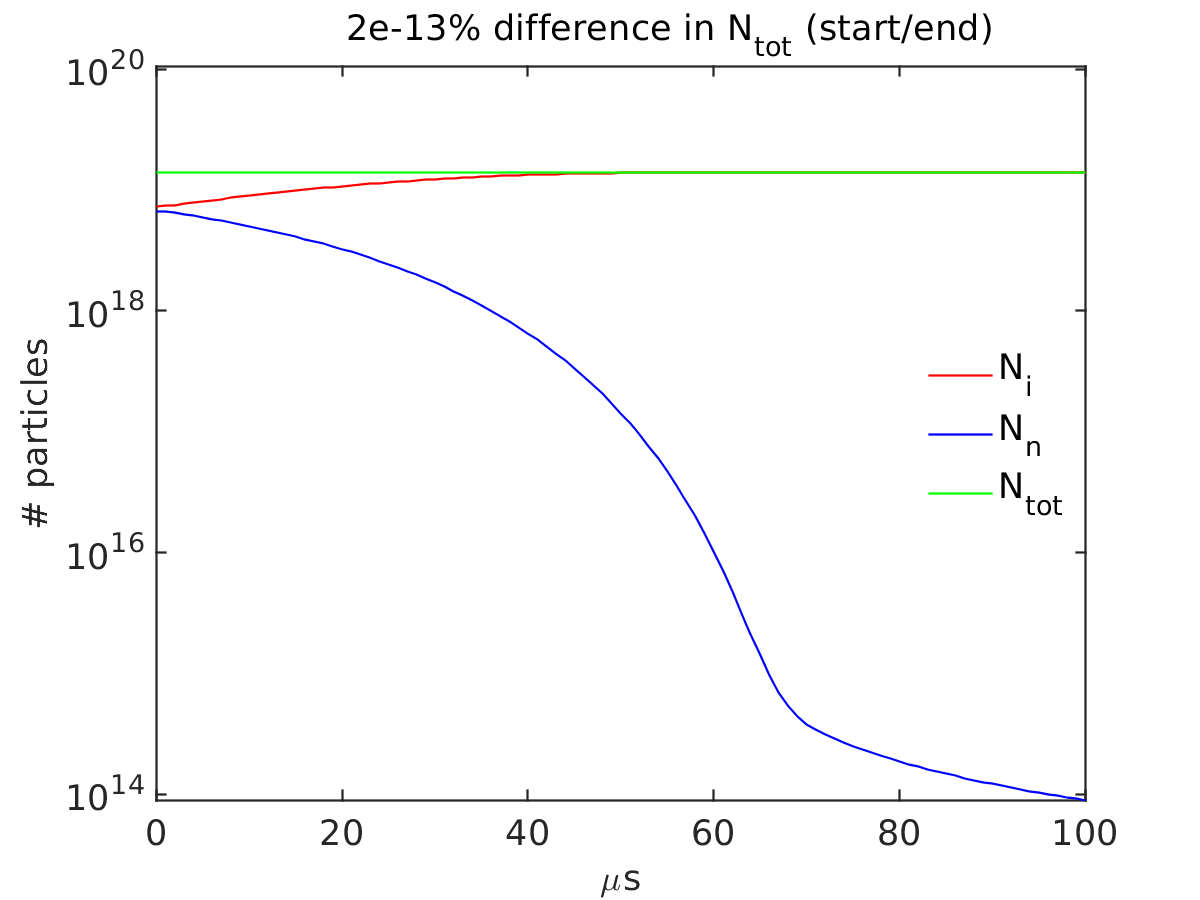}}\caption{\label{fig:Energy_N_cons}$\,\,\,\,$Illustration of energy and particle
conservation for MHD simulation with neutral fluid}
\end{figure}
Figure \ref{fig:Energy_N_cons}(a) indicates the partition of energy,
and how total system energy is conserved for a simulation where a
neutral fluid is evolved along with the plasma fluid. In this case,
the initial neutral fluid number density  was approximately equal
to the initial plasma fluid number density, with spatially uniform
initial distributions. The simulation started from a Grad-Shafranov
equilibrium. The only explicitly applied boundary conditions were
$v_{r}|_{\Gamma}=v_{z}|_{\Gamma}=v_{nr}|_{\Gamma}=v_{nz}|_{\Gamma}=0\mbox{ and }\psi|_{\Gamma}=0$,
so that the thermal and Poynting fluxes are zero through the boundary,
as outlined in section \ref{subsec:Conservation-properties}. Because
the equations for the neutral fluid are analogous, in terms of energy
conserving properties, to the equations for the plasma fluid, it is
to be expected that energy conservation will be maintained when the
evolution of a neutral fluid is simulated along with that of a plasma
fluid. When the electron fluid energy lost through the ionization
and recombination (the model assumes that in the radiative recombination
reaction, the electron thermal energy is lost to the photon emitted,
which leaves the optically thin system without further interaction)
processes are accounted for, it is evident that total energy is conserved
to numerical precision. The partitions of magnetic energy and the
thermal and kinetic plasma fluid energies follow the trends outlined
in section \ref{subsec:Conservation-properties} for the case without
neutral fluid evolution. 

Note that the neutral fluid thermal energy is negligible compared
with the plasma thermal energy. Initial spatially uniform temperatures
were $T_{i}=T_{e}=T_{n}=0.02$ eV. Due to ionization, neutral fluid
density is low in regions in which electrons are hot. In these regions,
ions are also generally hot due to heat exchange with electrons. The
hot ions result, through the charge exchange process, in hot neutral
particles, so regions with hot neutral particles are regions of relatively
low neutral fluid density. 

Neutral fluid kinetic energy is small compared with the kinetic energy
of the ions and electrons. Charged particle acceleration leads to
neutral particle acceleration, largely due to frictional forces between
the plasma and neutral fluids associated with charge-exchange reactions,
and due to momentum exchange arising from recombination processes.
Plasma is accelerated by $\mathbf{J}\times\mathbf{B}$ forces, and
is heated in regions where it is accelerated, either by ohmic heating
(electrons) or viscous heating (ions). Thus, neutral fluid density
tends to be low in regions where neutral fluid is accelerated due
to high ionization rates in those regions, so net neutral fluid kinetic
energy remains low. Figure \ref{fig:Energy_N_cons}(b) shows how total
particle count ($N_{tot})$ is conserved to numerical precision for
the same simulation. Initial neutral particle count ($N_{n}$) was
around equal to the ion inventory ($N_{i}$), and by the end of the
simulation, neutral particles account for around one in 100,000 of
the total number of particles, due to ionization.

Net angular momentum of the plasma and neutral fluids is also conserved
to numerical precision. No boundary conditions are explicitly applied
to $v_{\phi}$ or to $v_{n\phi}$, so the natural boundary conditions
$\left(\nabla_{\perp}\,\omega\,\right)|_{\Gamma}=\left(\nabla_{\perp}\,\omega_{n}\,\right)|_{\Gamma}=0$
are automatically imposed (see section \ref{subsec:Angular-momentum-conservation}).

\section{Neutral fluid thermal diffusion\label{sec:Neutral-thermal-diffusion}}

The thermal diffusion coefficient for the neutral fluid, given by
the Chapman-Cowling closures (equation \ref{eq:472.36}), can be expressed,
with the definition of the neutral-neutral scattering collision frequency,
$\nu_{cn}=\frac{V_{thn}}{\lambda_{mfp}}$, as:

\begin{equation}
\chi_{n}\approx\frac{75\sqrt{\pi}}{64}\frac{V_{thn}^{2}}{\nu_{cn}}\label{eq:550}
\end{equation}
The expression for $\chi_{n}$ was derived with a Chapman-Enskog expansion
using the ratio of the mean free path for scattering collisions to
the characteristic system length scale as the small ordering parameter
(the principle is outlined in appendix \ref{subsec:Chapman-Enskog-closures}).
However, if the charge exchange collision frequency is higher than
the frequency for neutral-neutral scattering collisions, neutral thermal
conductivity should be reduced. As outlined in \cite{Meier,MeierPhd},
an approximation to include the effect of charge exchange collisions
on neutral thermal conduction is to replace the term $\nu_{cn}$ in
the definition for $\chi_{n}$ with $\nu_{cn}+\nu_{cx}$ where $\nu_{cx}=\frac{V_{thn}}{\lambda_{cx}}=V_{thn}\sigma_{cx}n_{n}$
is the charge exchange frequency, and $\lambda_{cx}\approx\frac{1}{\sigma_{cx}n_{n}}$
is the mean free path for charge exchange collisions. In this way,
$\chi_{n}$ will be limited by whichever frequency dominates. The
resulting expression, which is implemented to the code as an option,
is 
\begin{equation}
\chi_{n}\approx\frac{75\sqrt{\pi}}{64}\frac{V_{thn}^{2}}{\nu_{cn}+\nu_{cx}}=\frac{75\sqrt{\pi}}{64}\frac{V_{thn}}{\frac{1}{\lambda_{mfp}}+\sigma_{cx}n_{n}}\label{eq:550-1}
\end{equation}
As described in \cite{MeierPhd}, when charge exchange dominates over
scattering, the model compares well with an established model for
neutral thermal diffusion \cite{Helender}, which was derived rigorously
with a formal Chapman-Enskog-like procedure. A similar modification
to the neutral viscous diffusion coefficient (equation \ref{eq:472.35})
could be made but hasn't been experimented with in this work.

\section{Simple model for charge exchange reactions\label{sec:Simple-CX-model}}

If evolving the plasma fluid only, the terms $\frac{1}{\rho}\left({\color{red}{\color{blue}\ensuremath{-\mathbf{R}_{ni}^{cx}-\Gamma^{cx}m_{i}\mathbf{v}_{in}}}}\right)$
and\\
 $(\gamma-1)\left({\color{blue}-Q_{ni}^{cx}+{\color{blue}{\color{blue}\Gamma^{cx}\frac{m_{i}}{2}v_{in}^{2}}}}\right)$
may be retained in the expressions for $\dot{\mathbf{v}}$ and $\dot{p}_{i}$.
This represents that hot ions that react with cold neutral particles
(constant $\rho_{n}$) ionize and impart half their energy to the
neutral particles and leave the system without reacting again. The
simple model is not physically representative because the quantities
$\mathbf{R}_{ni}^{cx},\,\Gamma^{cx},\,Q_{ni}^{cx}$ and $\mathbf{v}_{in}$
depend on local neutral-fluid-related variables such as $n_{n}(\mathbf{r},\,t)\mbox{ and }T_{n}(\mathbf{r},\,t)$
that would need to be estimated, but may be useful to study particular
scenarios.\newpage{}

\section{Simulation results with neutral fluid\label{sec:Simulation-results-with}}

\begin{table}[H]
\centering{}%
\begin{tabular}{|c|c|c|c|c|}
\hline 
\textbf{\small{}$\mathbf{neutralfluid}$} & {\small{}$\mathbf{N_{0}\,[\mbox{\textbf{m}}^{-3}]}$} & \textbf{\small{}$\mathbf{\boldsymbol{\sigma}}_{\mathbf{N}}^{\mathbf{2}}\,[\mbox{\textbf{m}}^{2}]$} & \textbf{\small{}$\mathbf{\boldsymbol{\zeta}_{n}\,[\mbox{\textbf{m\ensuremath{\mathbf{^{\mathbf{2}}}}/s}}]}$} & \textbf{\small{}$\mathbf{add_{N}}$}\tabularnewline
\hline 
{\small{}1} & {\small{}$4.5\times10^{20}$} & {\small{}0.01} & {\small{}90} & {\small{}1}\tabularnewline
\hline 
 &  &  &  & \tabularnewline
\hline 
{\small{}$\mathbf{vary_{\chi_{N}}}$} & \textbf{\small{}$\mathbf{vary_{\nu_{N}}}$} & \textbf{$\boldsymbol{\chi_{Nmax}}$$\,[\mbox{\textbf{m\ensuremath{\mathbf{^{\mathbf{2}}}}/s}}]$} & \textbf{$\boldsymbol{\nu_{Nmax}}$$\,[\mbox{\textbf{m\ensuremath{\mathbf{^{\mathbf{2}}}}/s}}]$} & {\small{}$\boldsymbol{\chi_{CX}}$}\tabularnewline
\hline 
{\small{}1} & 1 & $5\times10^{4}$ & $1\times10^{4}$ & {\small{}1}\tabularnewline
\hline 
\end{tabular}\caption{\label{tab:Sim parametersNeut} $\,\,\,\,$Neutral-relevant code input
parameters for simulation  2353 }
\end{table}
The code inputs in table \ref{tab:Sim parametersNeut} are related
to the neutral fluid dynamics for simulation  2353, for which the
principal code input parameters were presented in table \ref{tab:Sim parameters}.
When interaction between the plasma and a neutral fluid is evolved,
code input parameter $neutralfluid$ is set equal to one. Analogous
to the case described in equation \ref{eq:900} for the initial plasma
distribution, the initial static neutral fluid distribution is determined
by a Gaussian profile with variance $\sigma_{N}^{2}$ determining
the degree of neutral fluid spread around the gas valves, and neutral
number density scaling $N_{0}.$ $\zeta_{n}\,[\mbox{m}^{2}\mbox{/s}]$
is the coefficient of neutral fluid density diffusion. $add_{N}=1$
implies that neutral fluid is added to the simulation domain at the
location of the gas valves throughout the simulation. Physically,
the gas valves remain open for up to around a millisecond after they
are first opened. In general, simulations including neutral dynamics
have $vary_{\nu_{N}}$ and $vary_{\chi_{N}}$ set to one, so that
the analytical closures given by the Chapman-Enskog formulae (equation
\ref{eq:472.35} and \ref{eq:472.36}) for the neutral fluid viscous
and thermal diffusion coefficients are used. However, if code input
$\chi_{CX}$ is also set to one, as it is for this simulation, the
modified expression for $\chi_{N}$, equation \ref{eq:550-1}, is
used to determine thermal diffusion for the neutral fluid. It is found
that this expression results in an increase of maximum $T_{n}$ of
around 10\% compared with cases where $\chi_{CX}$ is set to zero
and $vary_{\chi_{N}}$ is set to one, in which case equation \ref{eq:472.36}
is used to determine $\chi_{N}$. Constant coefficients are used if
$vary_{\nu_{N}}$ and $vary_{\chi_{N}}$ are set to zero. $\chi_{Nmax}$
and $\nu_{Nmax}$ determine the upper limits applied to the neutral
fluid viscous and thermal diffusion coefficients. 
\begin{figure}[H]
\subfloat[]{\raggedright{}\includegraphics[scale=0.5]{fig_190.png}}\hfill{}\subfloat[]{\raggedright{}\includegraphics[scale=0.5]{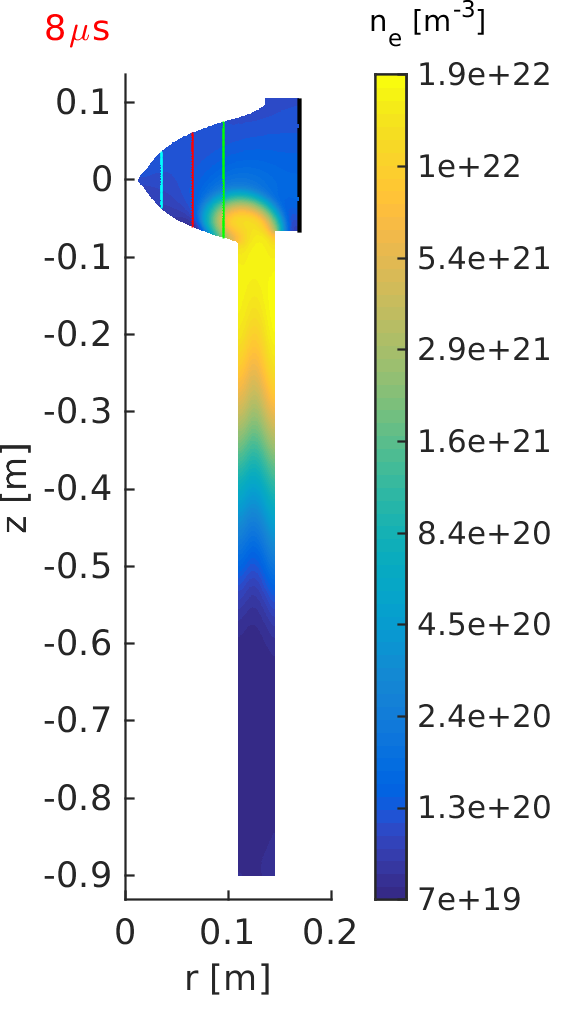}}\hfill{}\subfloat[]{\raggedright{}\includegraphics[scale=0.5]{fig_191.png}}

\subfloat[]{\raggedright{}\includegraphics[scale=0.5]{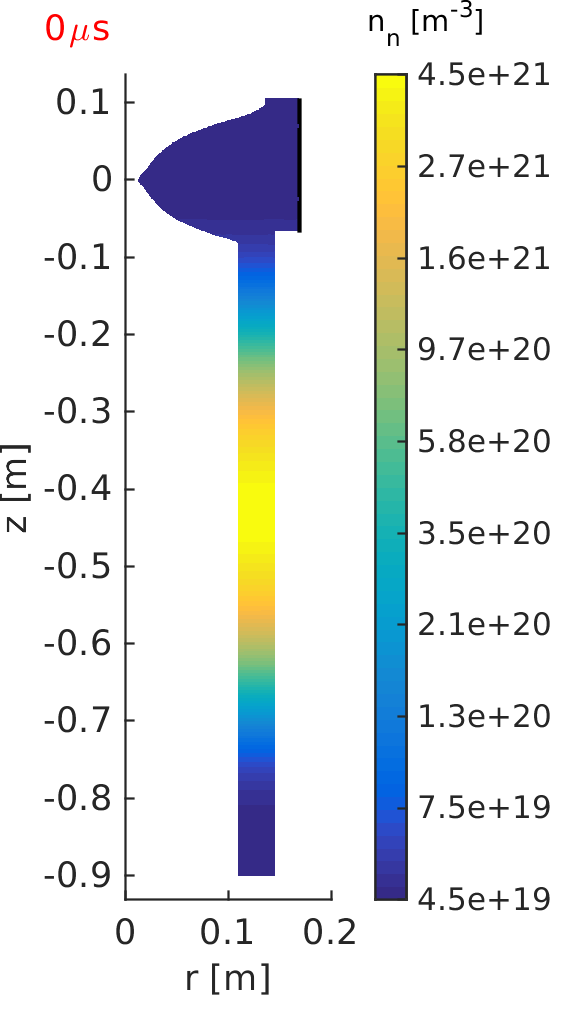}}\hfill{}\subfloat[]{\raggedright{}\includegraphics[scale=0.5]{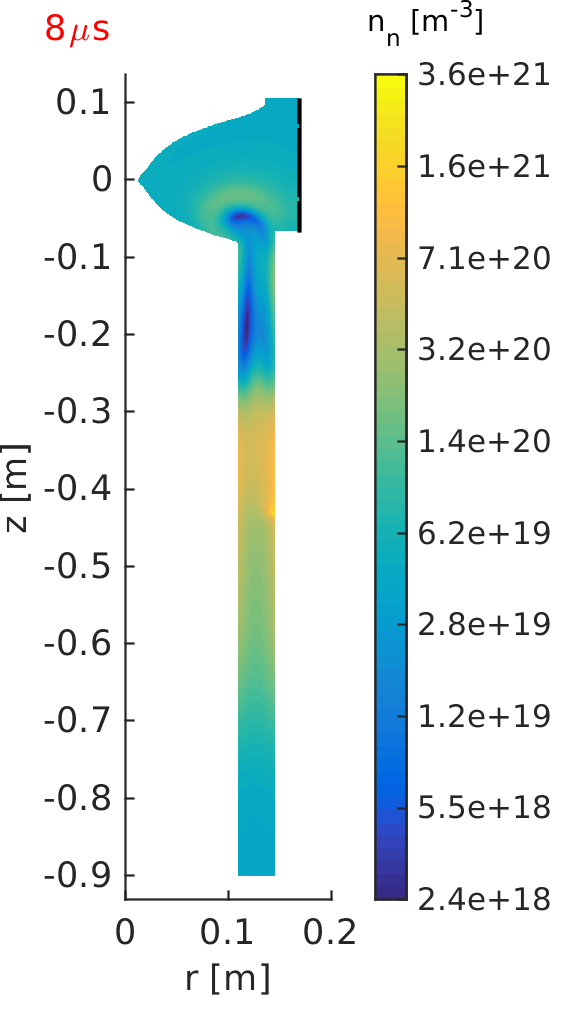}}\hfill{}\subfloat[]{\raggedright{}\includegraphics[scale=0.5]{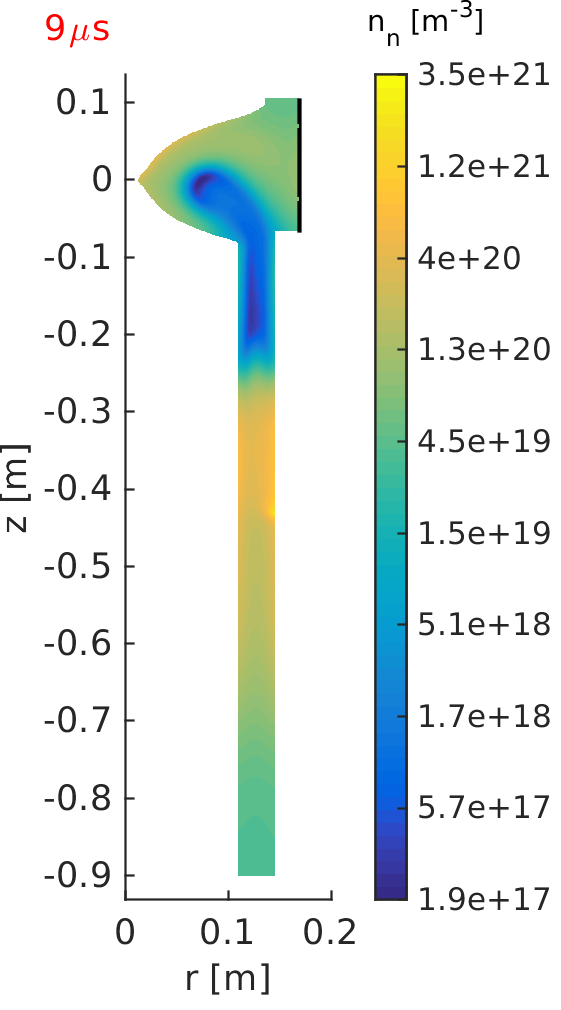}}

\caption{\label{fig: neut_bub_0}$\,\,\,\,$Neutral fluid dynamics at bubble-in
(1)}
\end{figure}
Figures \ref{fig: neut_bub_0}(a) and (d) show the initial distributions
for plasma density, represented by $n_{e}$, and neutral fluid density
$n_{n}$. The initial density distributions are Gaussian profiles,
centered around the gas puff locations at $z=-0.43$ m, with a higher
variance for the neutral fluid distribution, representing that the
neutral gas has diffused around the gas puff valve locations, while
the initial plasma distribution, is more localised to the gas puff
locations. The initial neutral particle inventory, determined by $\sigma_{N}^{2}$
and $N_{0}$, was over half the initial plasma particle inventory
for this simulation with $\sigma_{n}^{2}<\sigma_{N}^{2}$ and $N_{0}=n_{0}/2$.
Note that $n_{e}=Z_{eff}\,n_{i}$, where $Z_{eff}=1.3$ for simulation
2353. As shown in figures \ref{fig: neut_bub_0}(b) and (c), plasma
is starting to enter the CT containment region at 8$\,\upmu$s and
$9\,\upmu$s. A front of neutral fluid precedes the plasma as it is
advected upwards (figures \ref{fig: neut_bub_0}(e) and (f)). Note
that neutral particles are being added at the gas puff valve locations
by the outer boundary at $z=-0.43$ m. In the experiment, the gas
valves are opened at $t\sim-400\,\upmu$s, and remain open for $\sim1\mbox{ ms}$,
so that cold neutral gas is being added to the vacuum vessel throughout
the simulation, at a rate that can be estimated and assigned to the
simulated neutral particle source terms. 
\begin{figure}[H]
\subfloat[]{\raggedright{}\includegraphics[width=7cm,height=5cm]{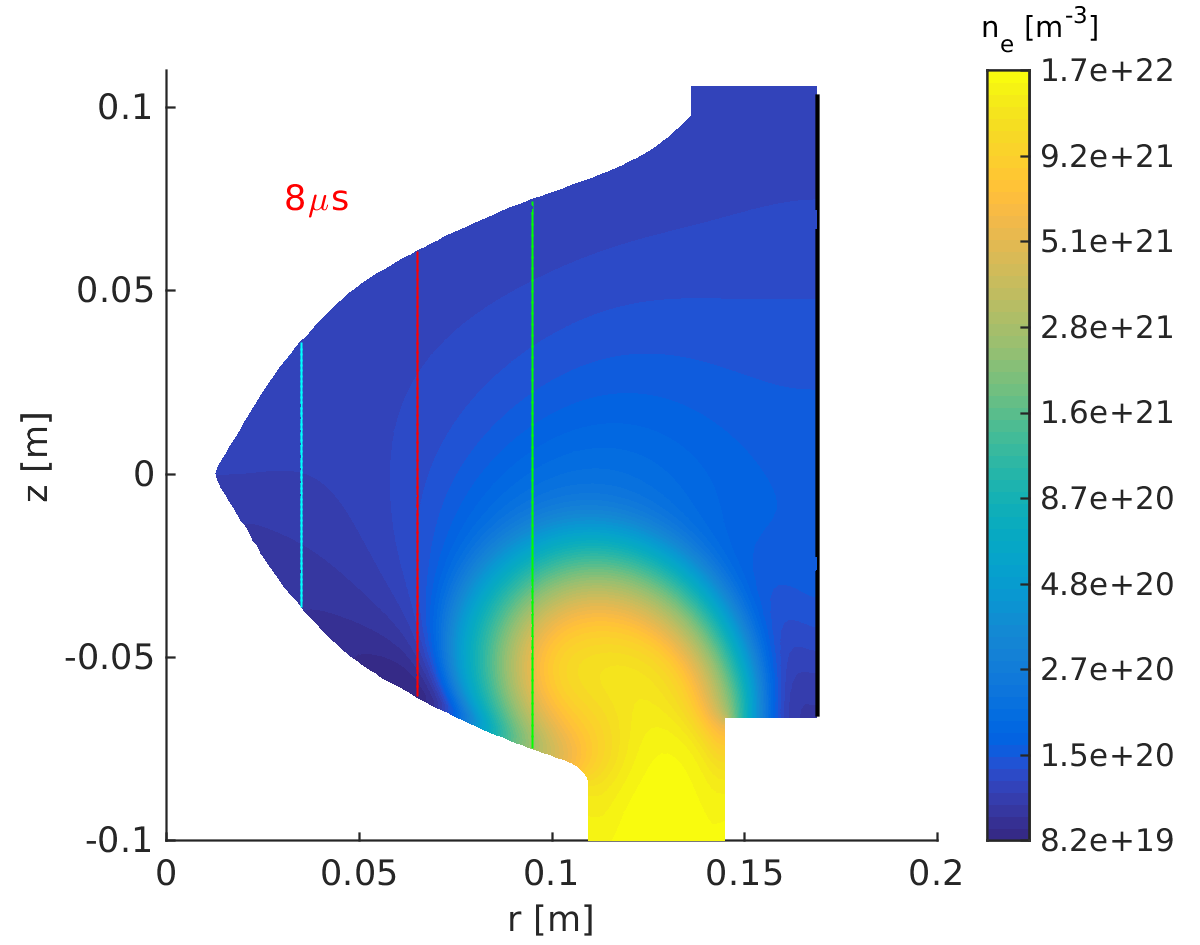}}\hfill{}\subfloat[]{\raggedright{}\includegraphics[width=7cm,height=5cm]{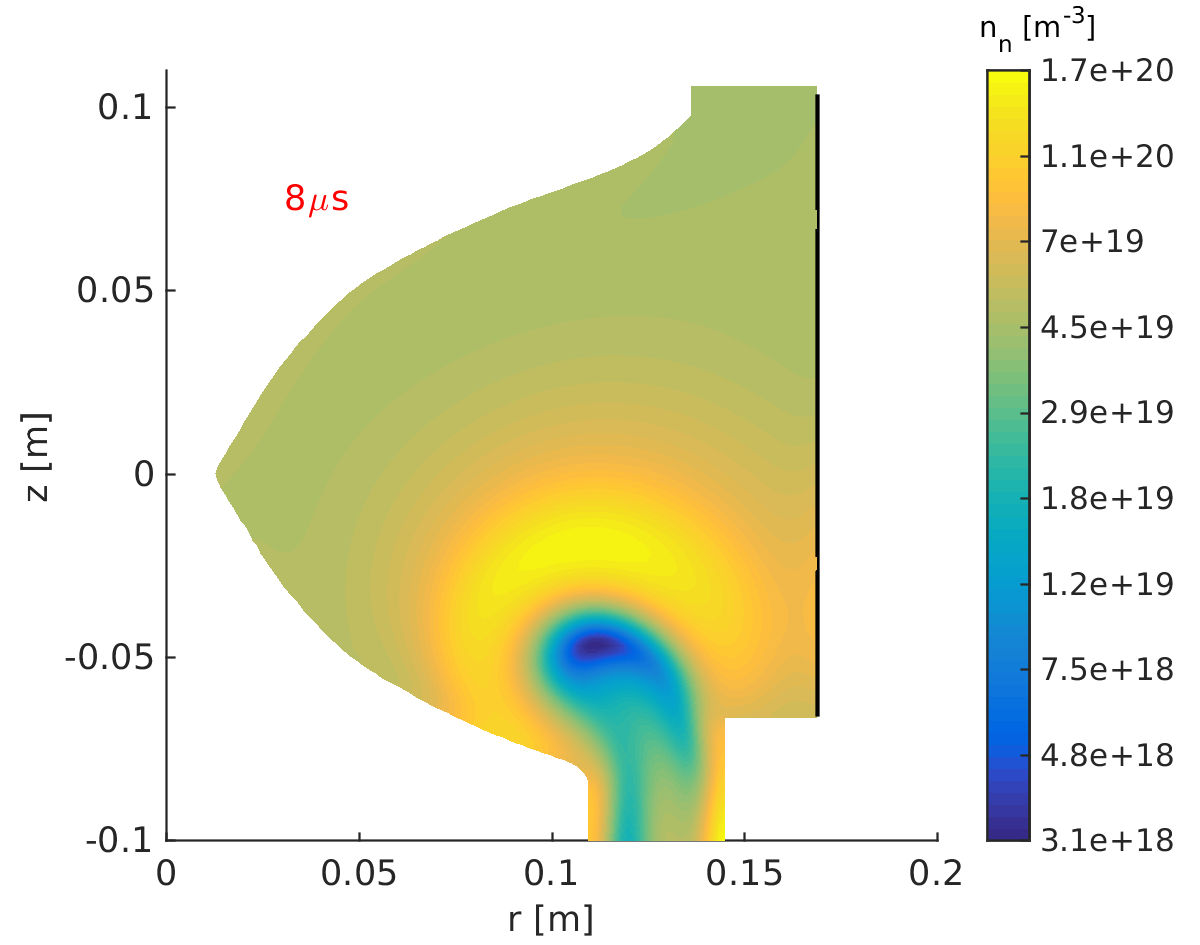}}

\subfloat[]{\raggedright{}\includegraphics[width=7cm,height=5cm]{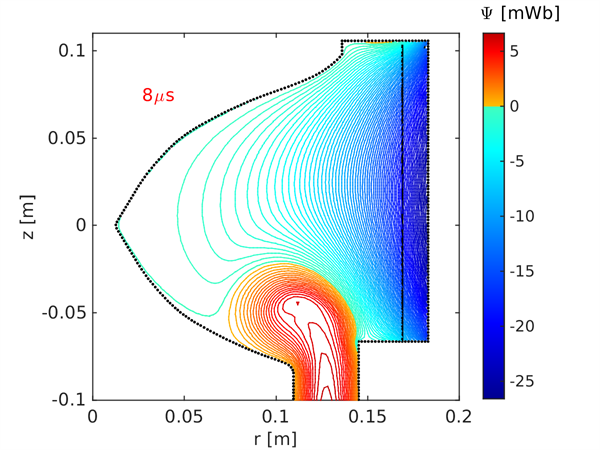}}\hfill{}\subfloat[]{\raggedright{}\includegraphics[width=7cm,height=5cm]{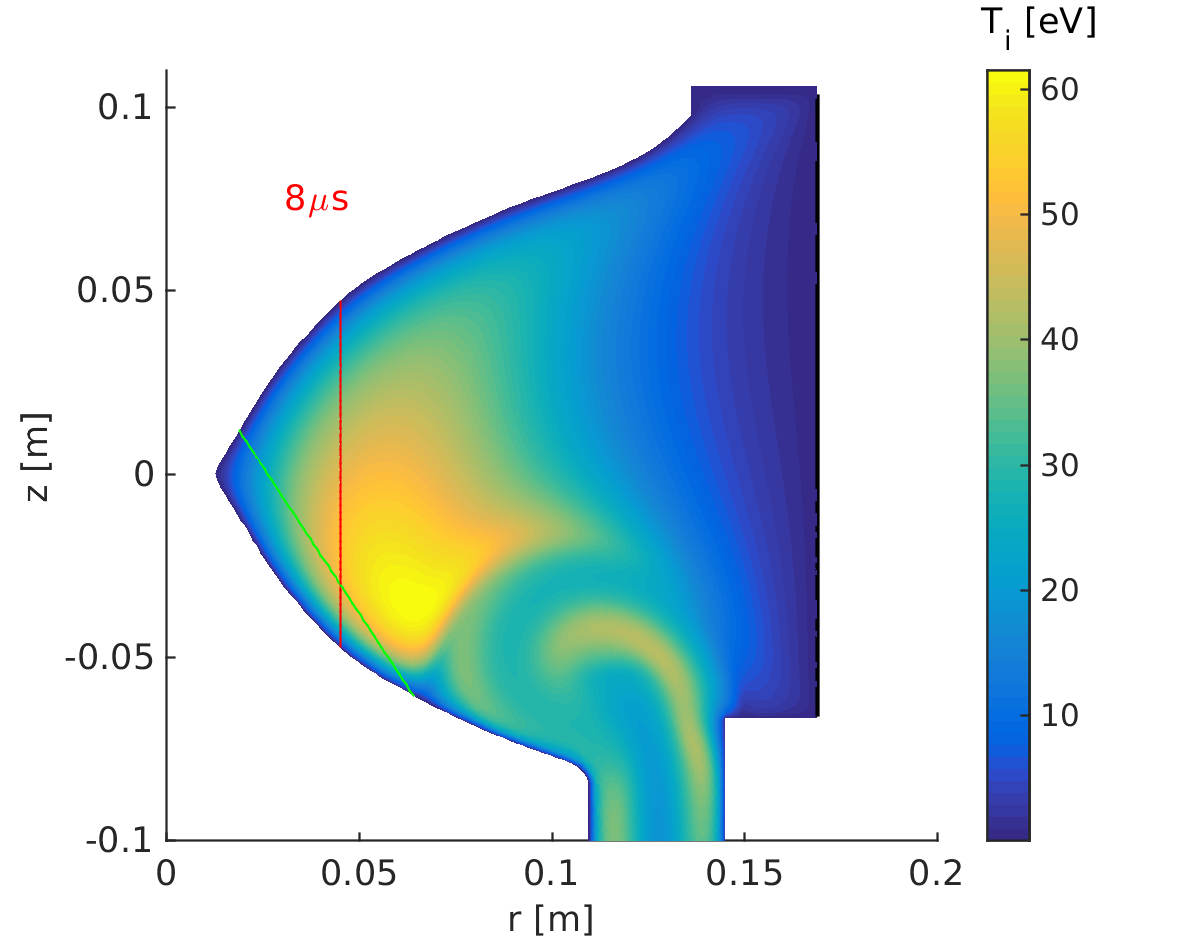}}

\subfloat[]{\raggedright{}\includegraphics[width=7cm,height=5cm]{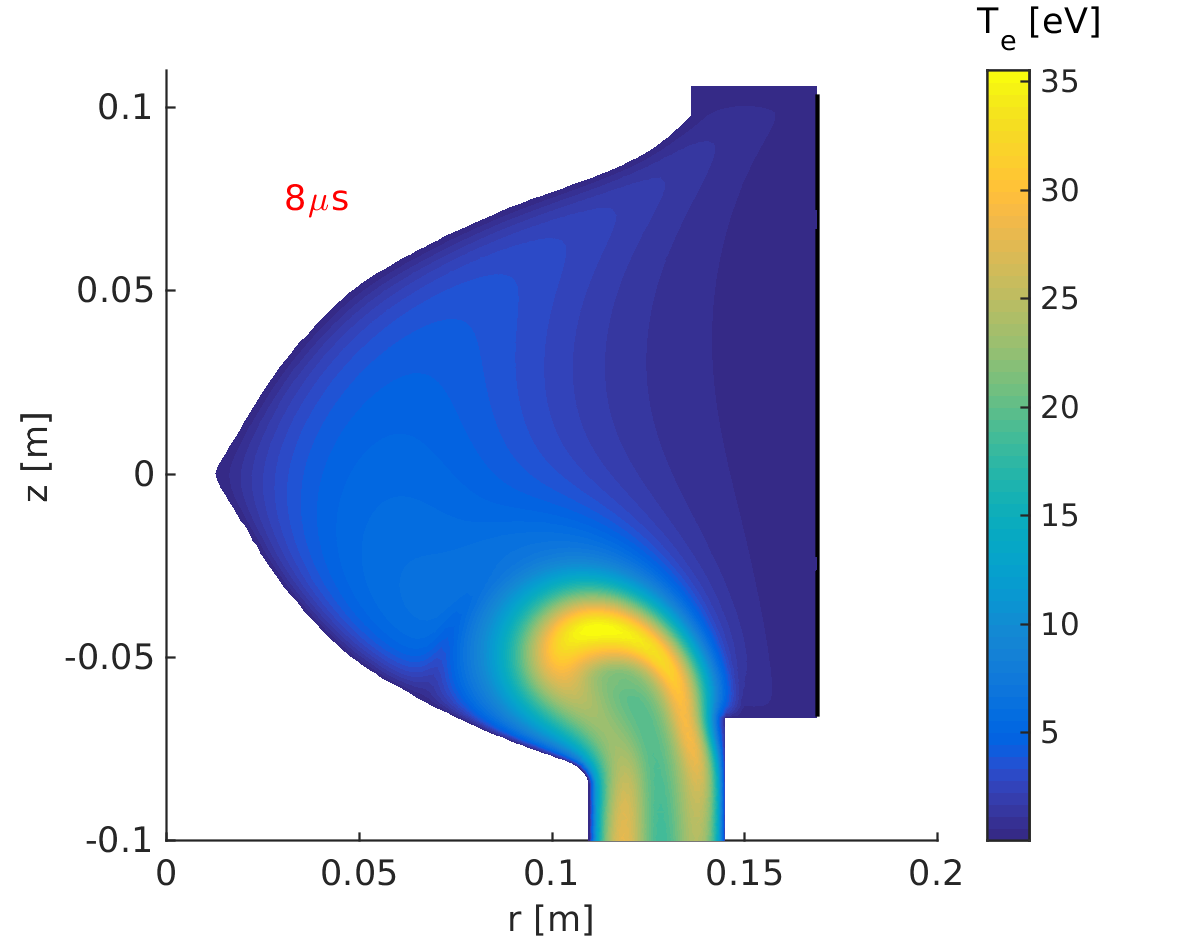}}\hfill{}\subfloat[]{\raggedright{}\includegraphics[width=7cm,height=5cm]{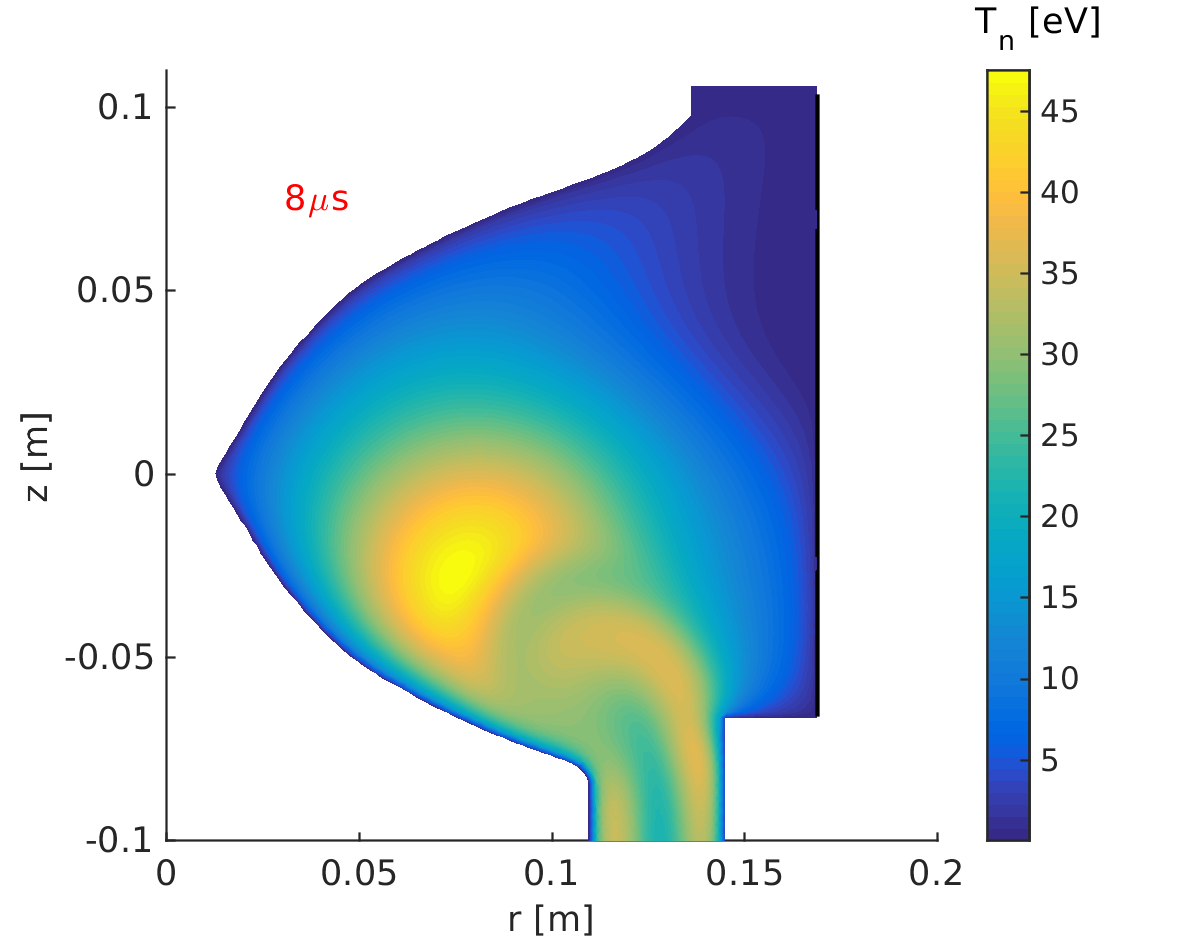}}

\caption{\label{fig: neut_bub_1}$\,\,\,\,$Neutral fluid dynamics at bubble-in
(2) }
\end{figure}
Figures \ref{fig: neut_bub_1}(a) and (b) show close-up views of $n_{e}$
and $n_{n}$ at 8$\,\upmu$s. Figures \ref{fig: neut_bub_1}(c) and
(d) show $\psi$ contours and the distribution of $T_{i}$ at the
same time. Ions are hot due to viscous heating. Ohmic heating in combination
with heat exchange with ions results in hot electrons (figure \ref{fig: neut_bub_1}(e)).
Note that neutral fluid density is low where $T_{e}$ is high due
ionization (figure \ref{fig: neut_bub_1}(b)). Due to charge exchange
reactions, neutral fluid temperature tends to equilibrise with ion
temperature (figures \ref{fig: neut_bub_1} (d) and (f)), and can
become hotter than ions if the thermal diffusion for neutral fluid
is set to be lower than ion thermal diffusion. In general, when $\chi_{\parallel i}$
and $\chi_{\parallel e}$ are fixed at moderate experimentally relevant
values such as those in table \ref{tab:Sim parameters}, and $\chi_{N}$
is determined by equation \ref{eq:550} or \ref{eq:550-1}, with $\chi_{Nmax}\gtrsim5\times10^{4}\,[\mbox{m}^{2}\mbox{/s}]$,
as is the case for this simulation, it is found that $T_{N}<T_{i}$.
\begin{figure}[H]
\subfloat[]{\raggedright{}\includegraphics[width=7cm,height=5cm]{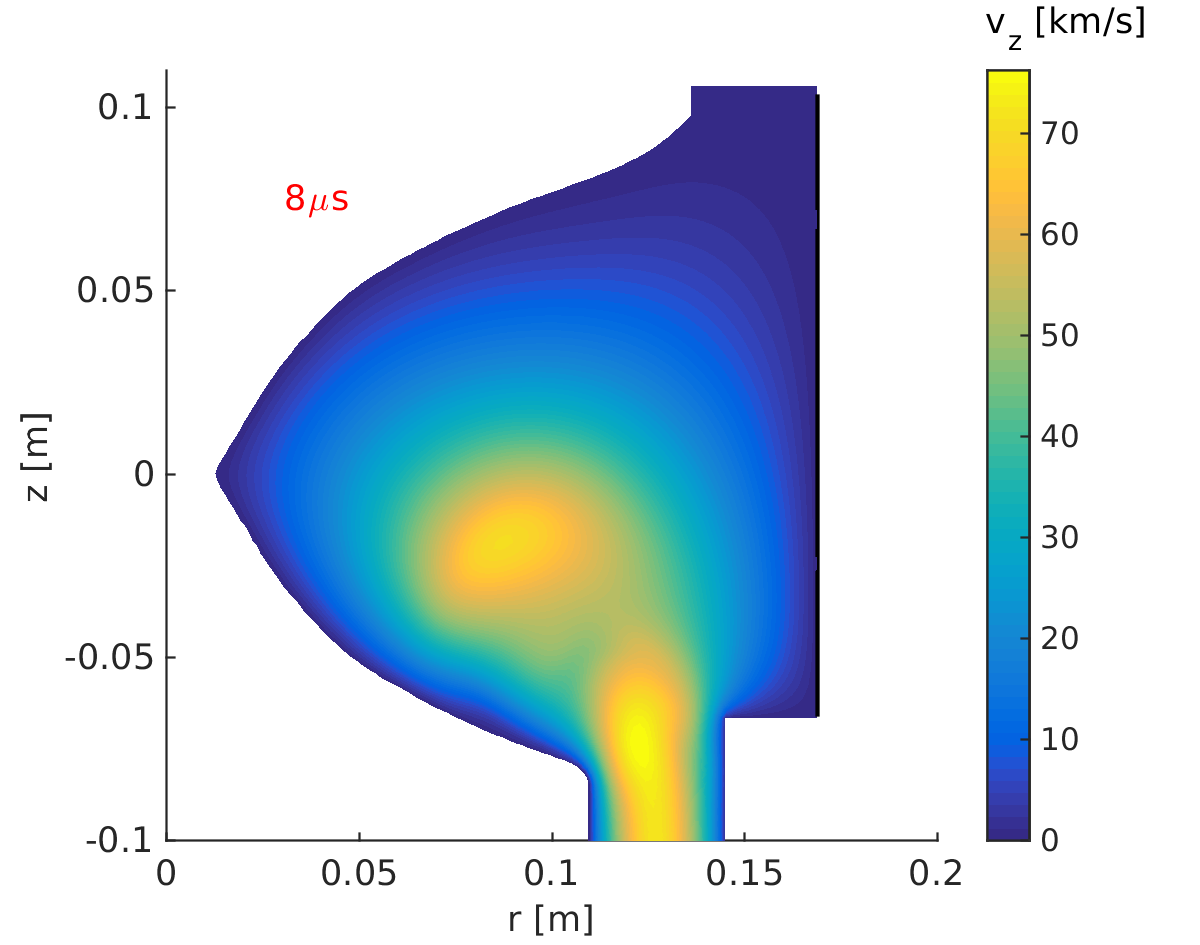}}\hfill{}\subfloat[]{\raggedright{}\includegraphics[width=7cm,height=5cm]{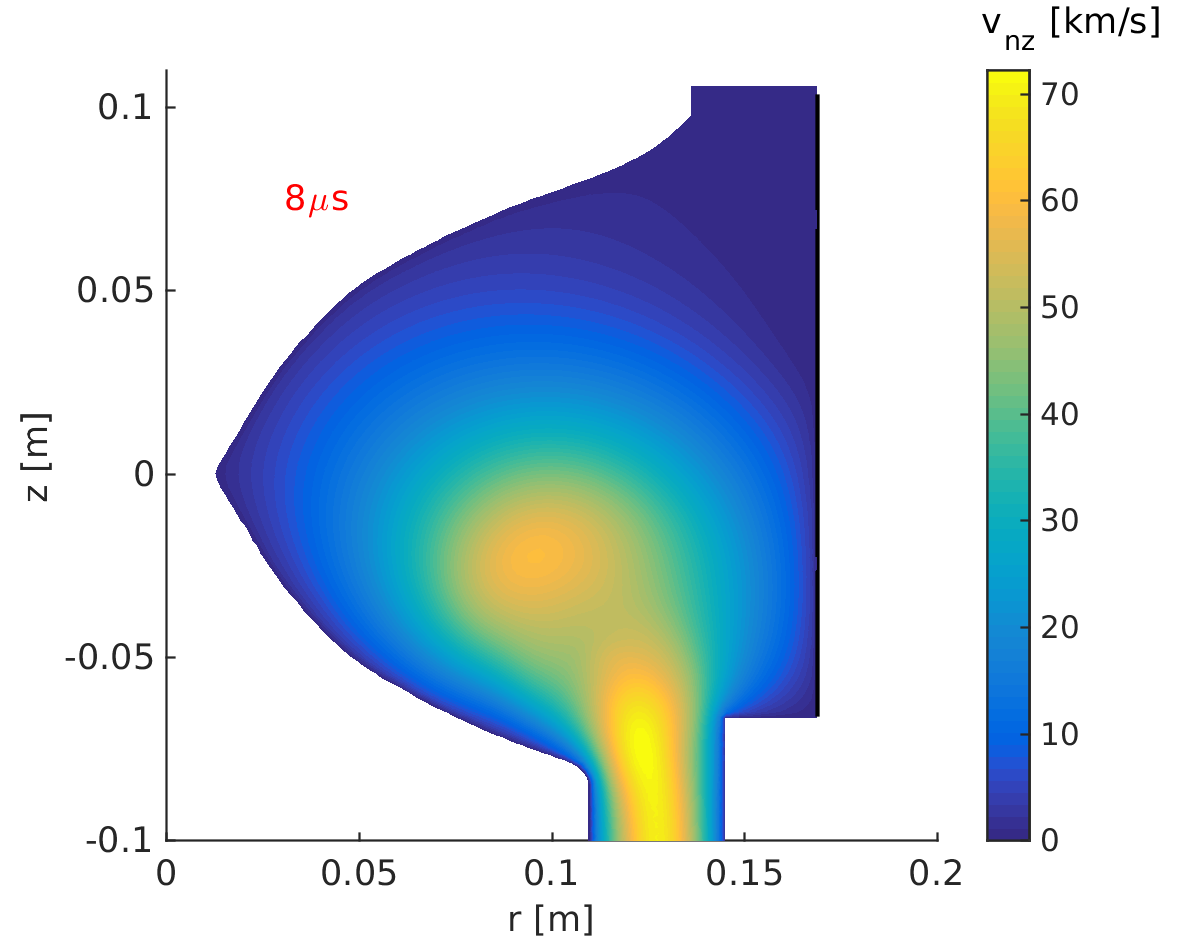}}

\subfloat[]{\raggedright{}\includegraphics[width=7cm,height=5cm]{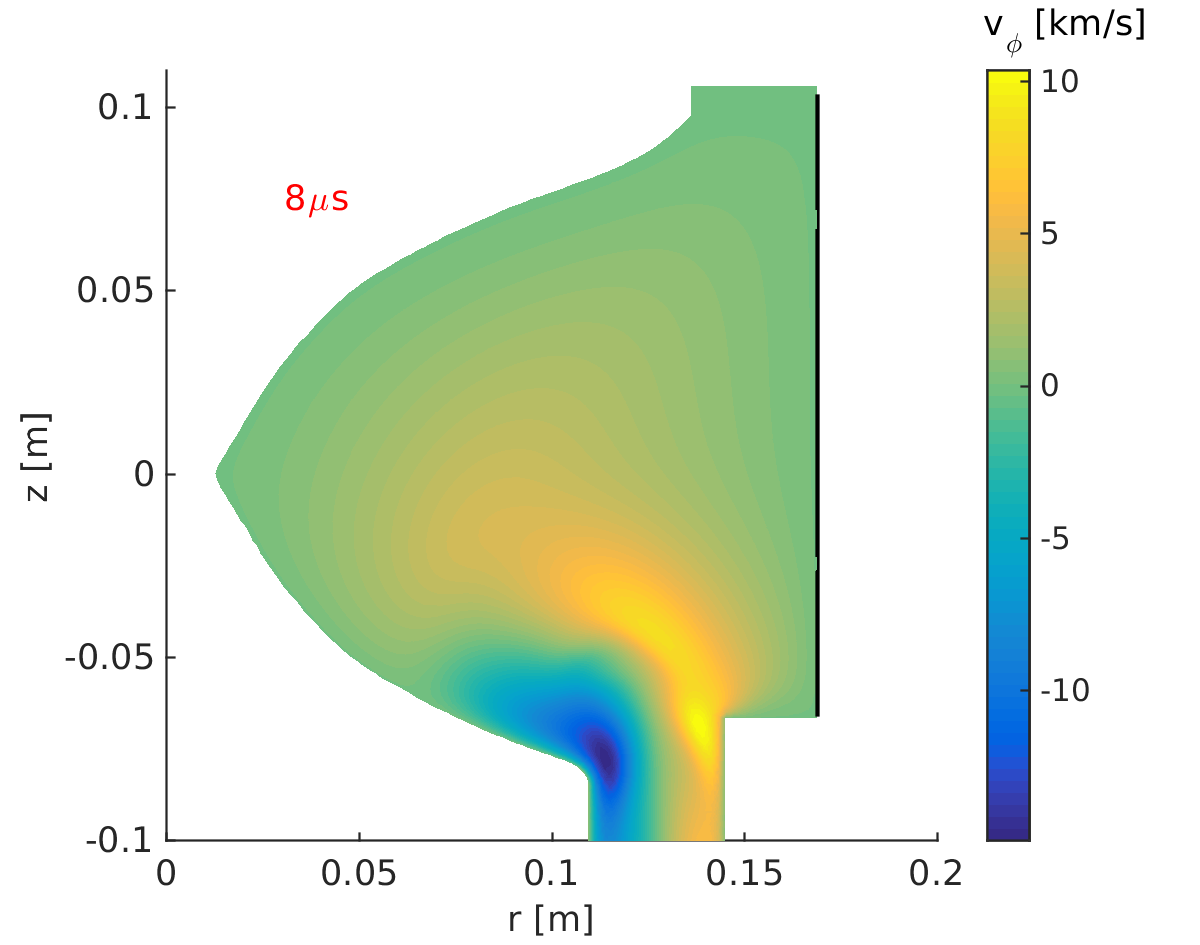}}\hfill{}\subfloat[]{\raggedright{}\includegraphics[width=7cm,height=5cm]{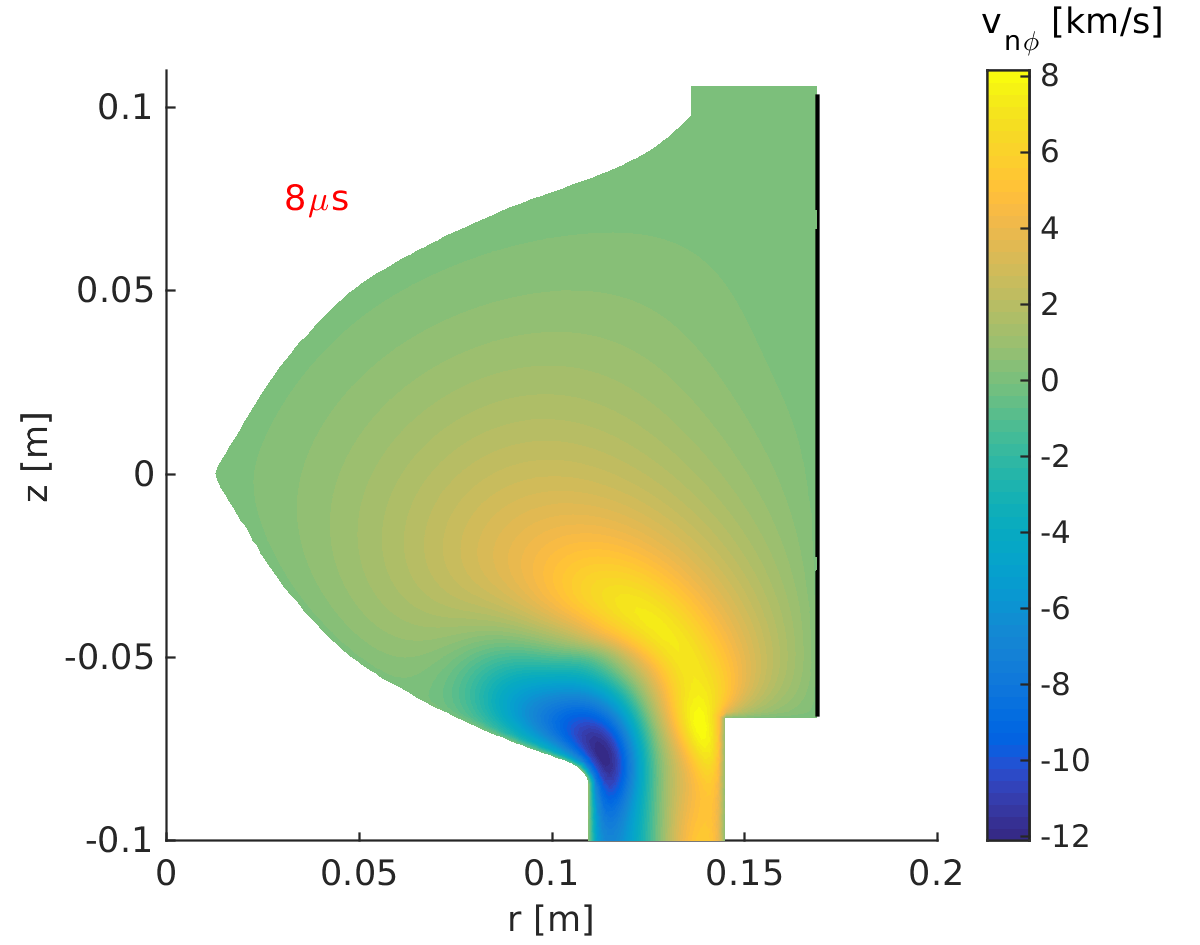}}

\caption{\label{fig: neut_bub_2}$\,\,\,\,$Neutral fluid dynamics at bubble-in
(3) }
\end{figure}
Figures \ref{fig: neut_bub_2}(a) and (b) shows profiles of axial
velocity at $8\,\upmu$s for the plasma fluid and neutral fluid respectively,
while azimuthal velocity profiles are presented at the same output
time in figures \ref{fig: neut_bub_2}(c) and (d). Plasma acceleration
leads to neutral fluid acceleration, due to frictional forces associated
with charge-exchange reactions, and due to momentum exchange arising
from recombination processes. It can be seen how the neutral fluid
attains nearly the same velocity magnitudes as the plasma fluid.

\subsection{Effect of inclusion of the $Q_{e}^{rec}$ term\label{subsec:Effect-of-inclusion}}

$Q_{e}^{rec}\,[\mbox{J m}^{-3}\,\mbox{s}^{-1}]$ determines the volumetric
rate of thermal energy transfer from electrons to photons and neutral
particles due to radiative recombination. As discussed in section
\ref{subsec:Econ_ionrecomb}, the method described in this chapter
to evaluate the terms determining plasma-neutral interactions allows
for the evaluation of $Q_{e}^{rec}$, which could not be evaluated
by the formal moment-taking process described in \cite{Meier}, and
has been neglected in studies \cite{MeierPhd,Leake} based on the
model. From looking at the kinematics of the radiative recombination
reaction (section \ref{subsec:Econ_ionrecomb}), it is more physical
to neglect $Q_{e}^{rec}$ as an energy source for the neutral fluid
(most of the electron thermal energy is transferred to the photon),
but include it as an energy sink for the electron fluid. It is interesting
however to note the effect of including the term as an energy source
for the neutral fluid - in this (unphysical) scenario it is assumed,
as presented in \cite{Meier}, that all the electron thermal energy
lost during radiative recombination is transferred to the neutral
particle.\\
\begin{figure}[H]
\subfloat[]{\raggedright{}\includegraphics[width=7cm,height=5cm]{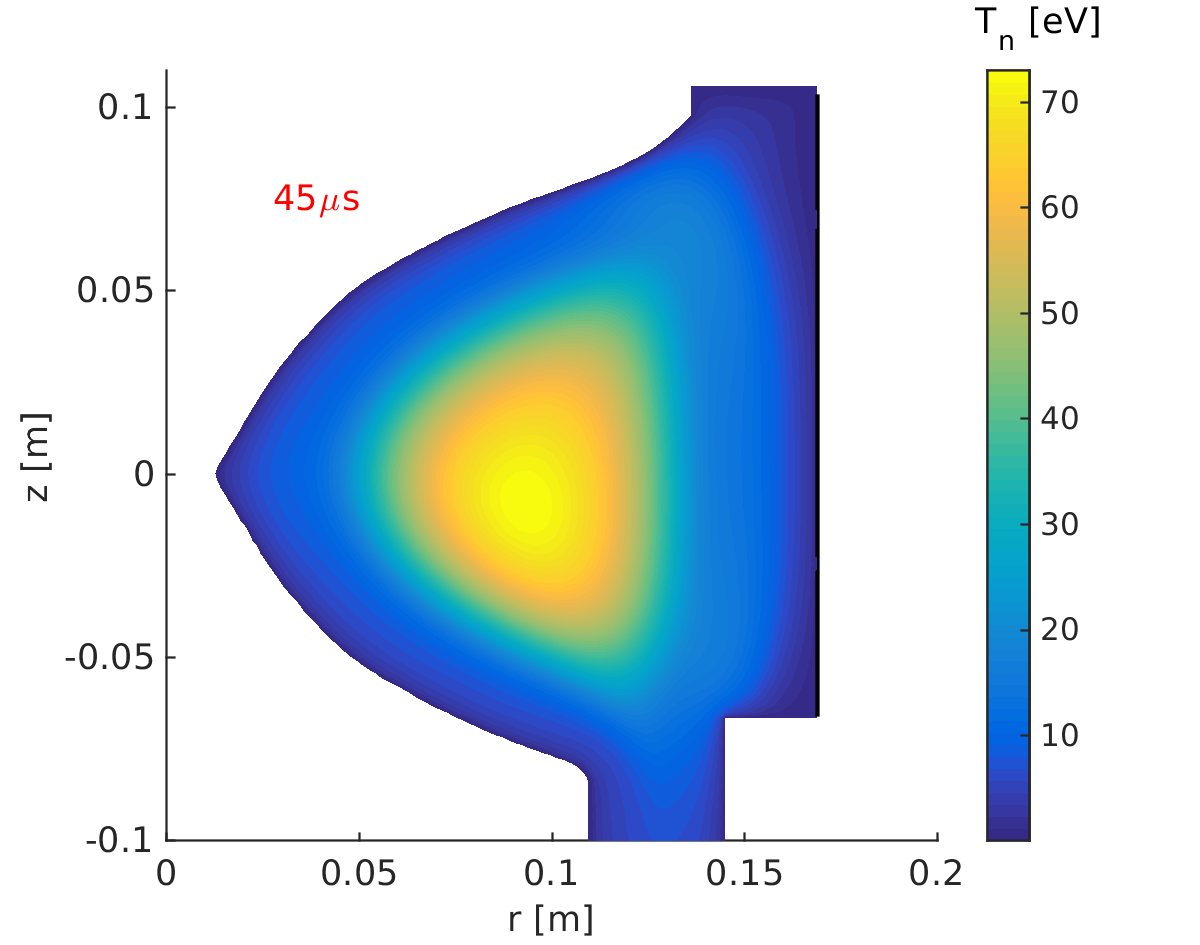}}\hfill{}\subfloat[]{\raggedright{}\includegraphics[width=7cm,height=5cm]{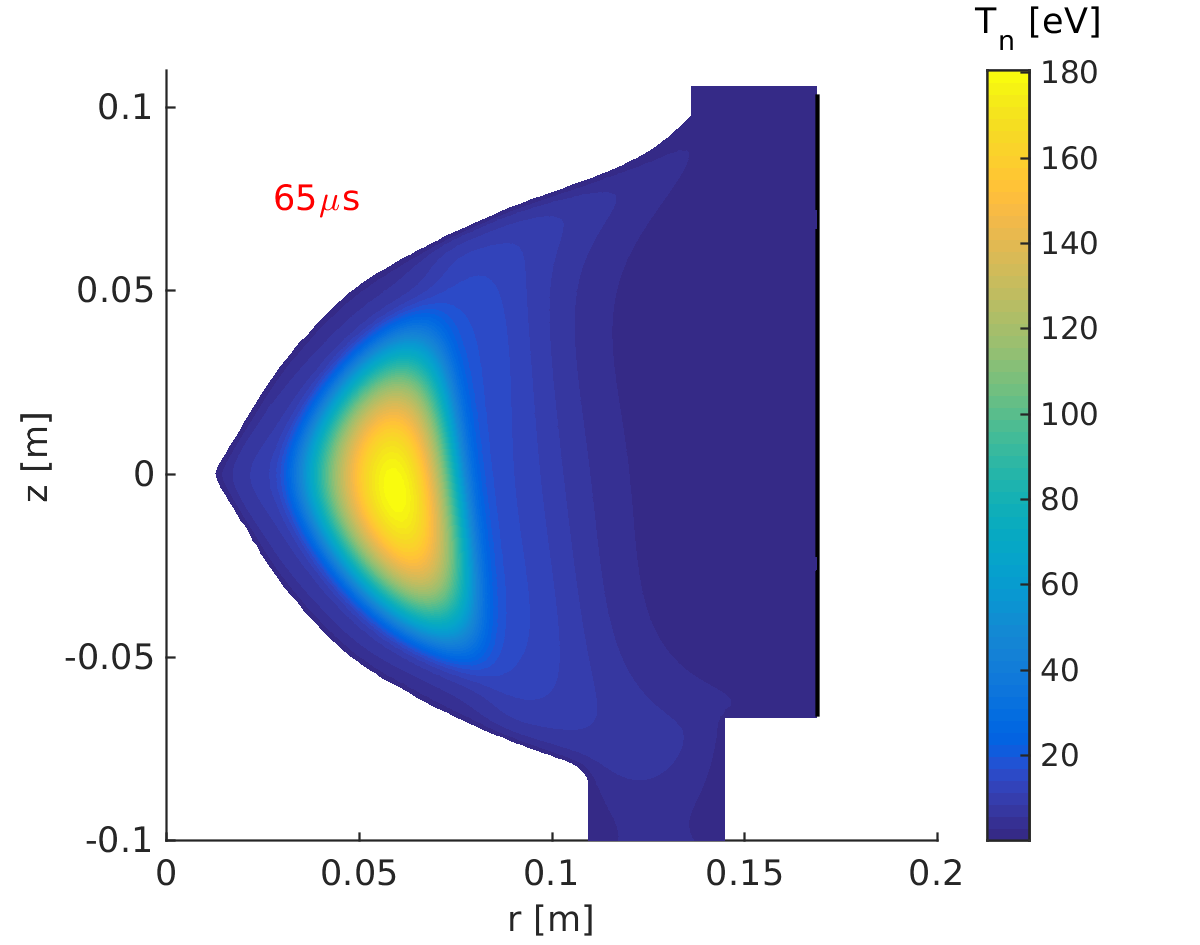}}

\subfloat[]{\raggedright{}\includegraphics[width=7cm,height=5cm]{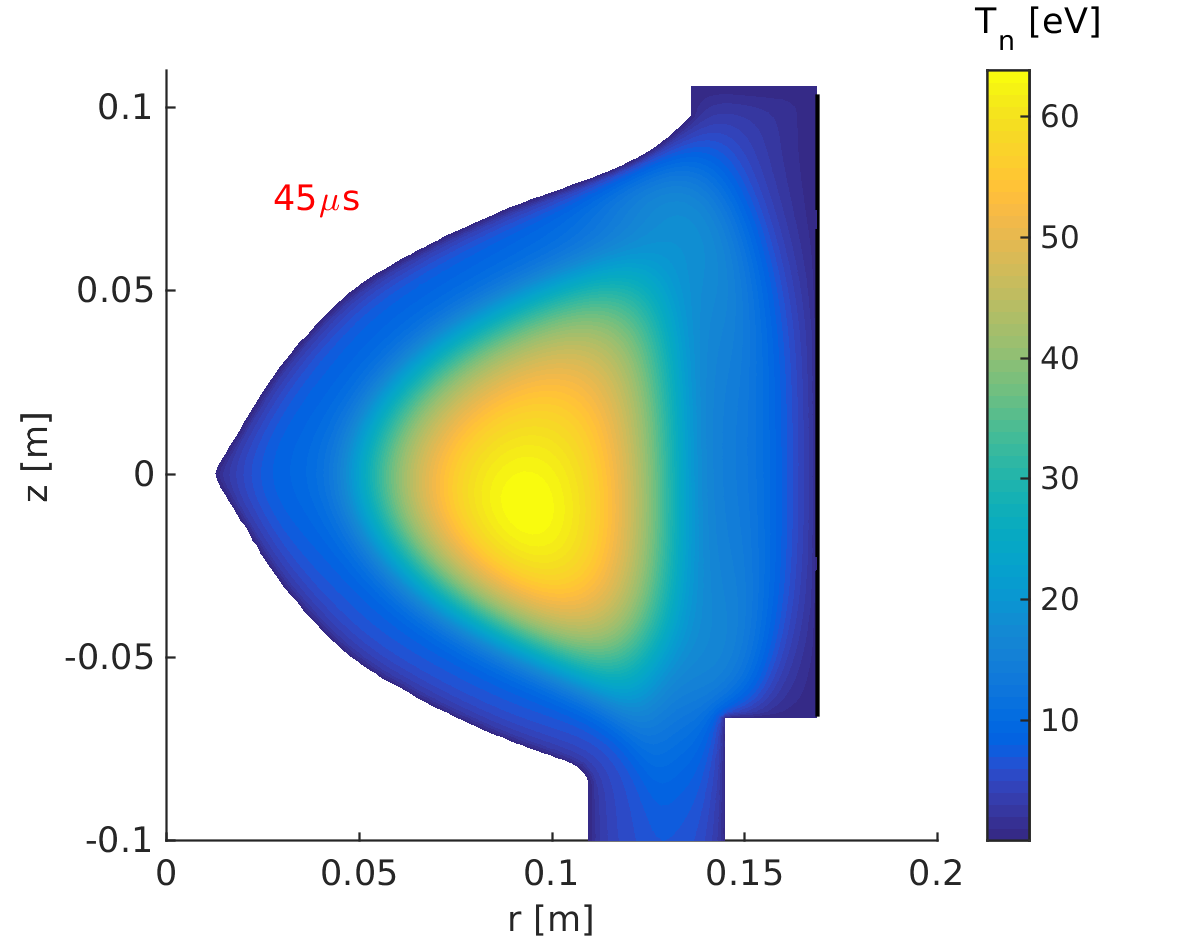}}\hfill{}\subfloat[]{\raggedright{}\includegraphics[width=7cm,height=5cm]{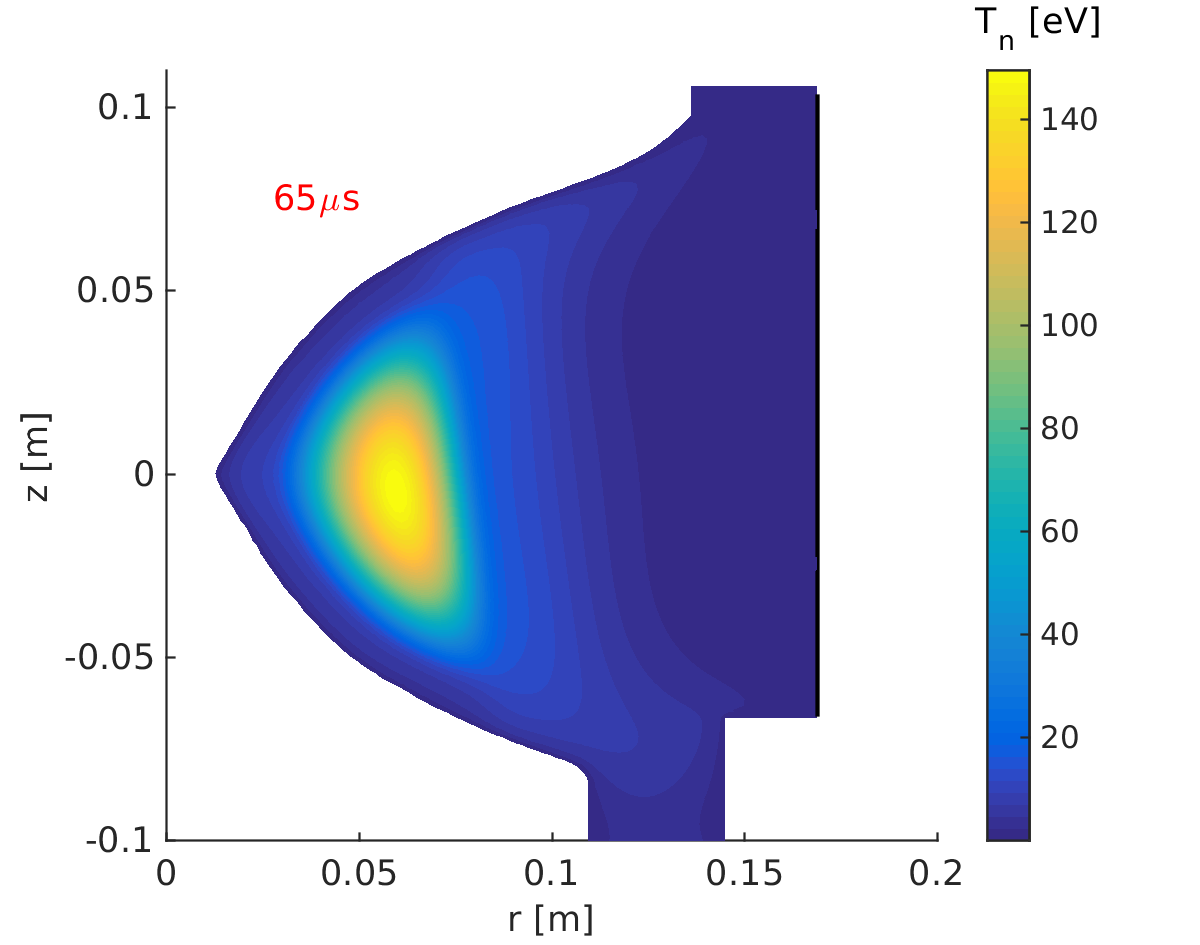}}

\caption{\label{fig: Qerec}$\,\,\,\,$Effect of inclusion of the $Q_{e}^{rec}$
term as an energy source for the neutral fluid}
\end{figure}
Figures \ref{fig: Qerec}(a) and (b) show profiles of $T_{n}$ from
simulation 2353, in which the $Q_{e}^{rec}$ term is included in the
energy equations for the electron and neutral fluids, at $45\,\upmu$s
just prior to magnetic compression, and at peak compression at $65\,\upmu$s.
Note that electron and ion temperature profiles are presented at the
same times, from the same simulation, in figures \ref{fig: Te_11coils}(e)
and (f) and in figures \ref{fig: Ti_11coils}(e) and (f). Figures
\ref{fig: Qerec}(c) and (d) show $T_{n}$ profiles at the same times
from a simulation that is identical except that the $Q_{e}^{rec}$
term is not included in the electron and neutral fluid energy equations.
It can be seen how peak $T_{n}$ increases by around 20\% at $45\,\upmu$s,
and by around 30\% at $65\,\upmu$s, when the $Q_{e}^{rec}$ term
is included. Note that if $\chi_{Nmax}$ is increased from $5\times10^{4}$
{[}m$^{2}$/s{]} to $1\times10^{5}$ {[}m$^{2}$/s{]}, that peak $T_{n}$
increases by around 30\% at $45\,\upmu$s, and by around 80\% at $65\,\upmu$s
when the $Q_{e}^{rec}$ term is included. From equations \ref{eq:531},
\ref{eq:531.0}, and \ref{eq:533.0}, it can be seen that 
\[
Q_{e}^{rec}\propto Z_{eff}^{3}\,n_{i}^{2}\sqrt{T_{e}}
\]
Hence, $Q_{e}^{rec}$ is high in regions where plasma density and
electron temperature are high. The increase in $T_{n}$ when $Q_{e}^{rec}$
is included is especially large in such regions, for example near
the CT core at peak magnetic compression, where the rate of ionization
is high and hence $n_{n}$, and thermal energy associated with neutral
particles, is low. 

Not shown here, peak electron temperature falls by around 1\% when
the $Q_{e}^{rec}$ term is included in the electron fluid energy equation.
$Q_{e}^{rec}$ appears as a (physical) thermal energy sink in the
electron fluid energy equation, but the reduction in $T_{e}$ when
$Q_{e}^{rec}$ is included is negligible, even in regions where $Q_{e}^{rec}$
is high, due to the high levels of electron thermal energy in such
regions. 

\subsection{Neutral fluid interaction in SPECTOR geometry\label{sec:Neutral_SPECTOR} }

A schematic of the SPECTOR machine \cite{spectPoster} is presented
in figure \ref{fig:GFinjectors}(b). SPECTOR has a separate circuit
to drive up to 1 MA shaft current, as indicated in figure \ref{fig:GFinjectors}(b),
and produces CTs with typical lifetimes of around two milliseconds.
It is usual to observe a significant rise in electron density at around
$500\,\upmu$s on the SPECTOR machine, and it was thought that this
may be a result of neutral gas, that remains concentrated around the
gas valve locations after CT formation, diffusing up the gun. Ionization
of the neutral particles would lead to CT fueling and an increase
in observed electron density. The model for interaction between plasma
and neutral fluids was applied to study the issue. 

\begin{figure}[H]
\subfloat[]{\raggedright{}\includegraphics[scale=0.5]{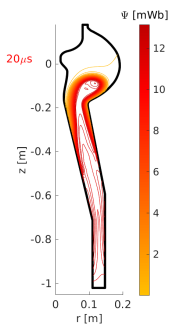}}\hfill{}\subfloat[]{\raggedright{}\includegraphics[scale=0.5]{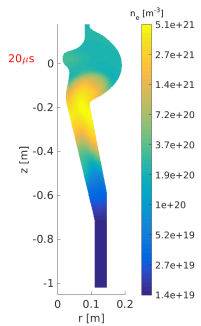}}\hfill{}\subfloat[]{\raggedright{}\includegraphics[scale=0.5]{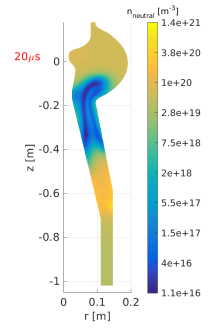}}

\subfloat[]{\raggedright{}\includegraphics[scale=0.5]{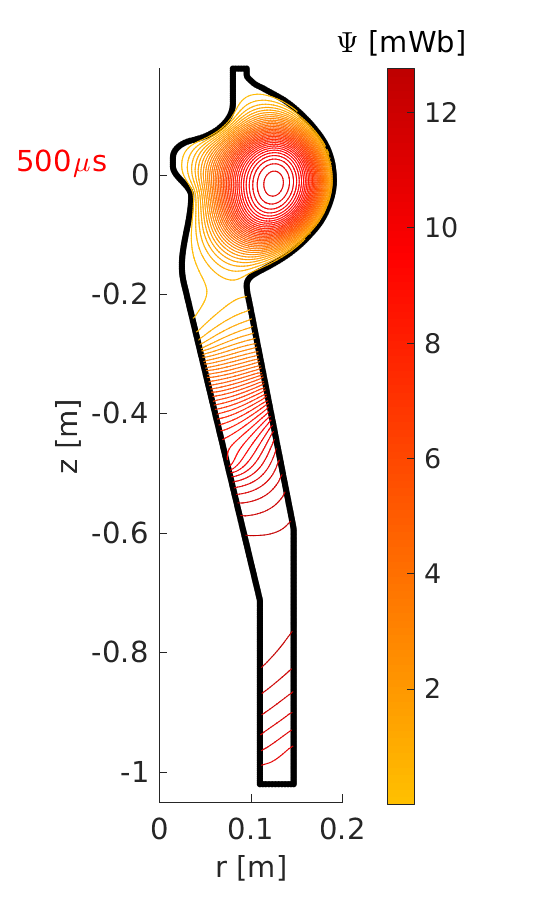}}\hfill{}\subfloat[]{\raggedright{}\includegraphics[scale=0.5]{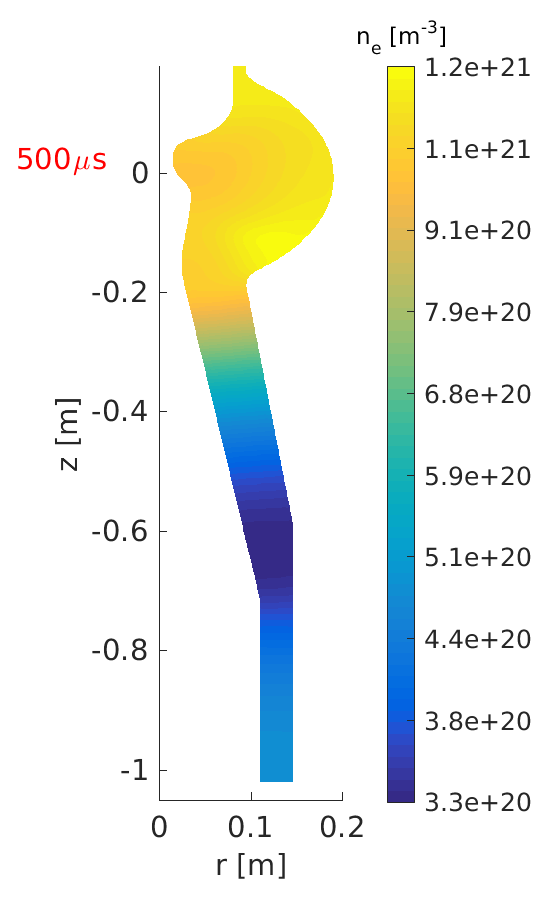}}\hfill{}\subfloat[]{\raggedright{}\includegraphics[scale=0.5]{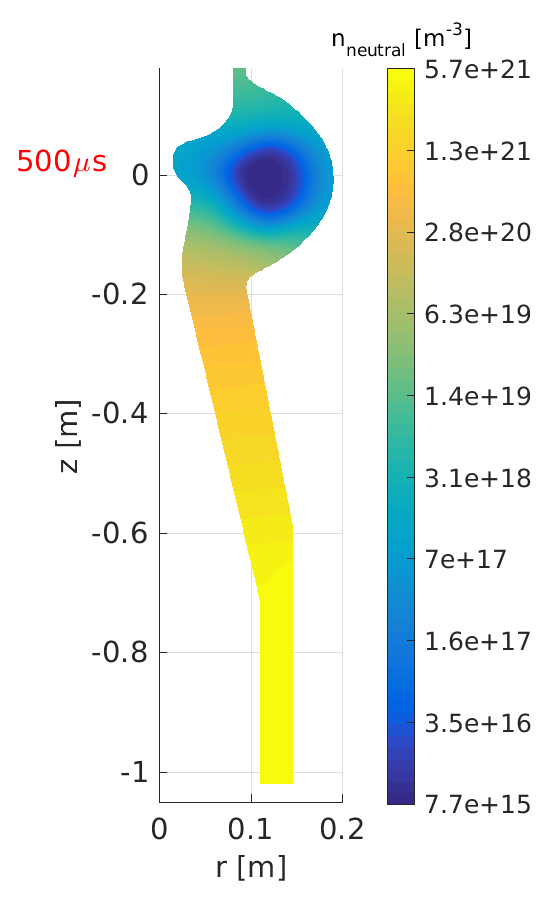}}

\caption{\label{fig:SPEC_psi_n_nN}$\,\,\,\,$Poloidal flux contours and profiles
of electron and neutral fluid densities in SPECTOR geometry}
\end{figure}
Figures \ref{fig:SPEC_psi_n_nN}(a), (b) and (c) show $\psi$ contours
and profiles of $n_{e}$ and $n_{n}$ at $20\,\upmu$s, as plasma
enters the CT containment region. Profiles of the same quantities
are shown in figures \ref{fig:SPEC_psi_n_nN}(d), (e) and (f) at $500\,\upmu$s,
around the time when the density rise is usually observed. It can
be seen how neutral fluid density is highest at the bottom of the
gun barrel (figure \ref{fig:SPEC_psi_n_nN}(f)) - any neutral gas
advected or diffusing upwards is ionized. A region of particularly
high electron density is apparent just above, and outboard of, the
entrance to the containment region (figure \ref{fig:SPEC_psi_n_nN}(e))
- this is due to the fueling effect arising from neutral gas diffusion.
\\
\begin{figure}[H]
\subfloat[]{\raggedright{}\includegraphics[width=7cm,height=6.3cm]{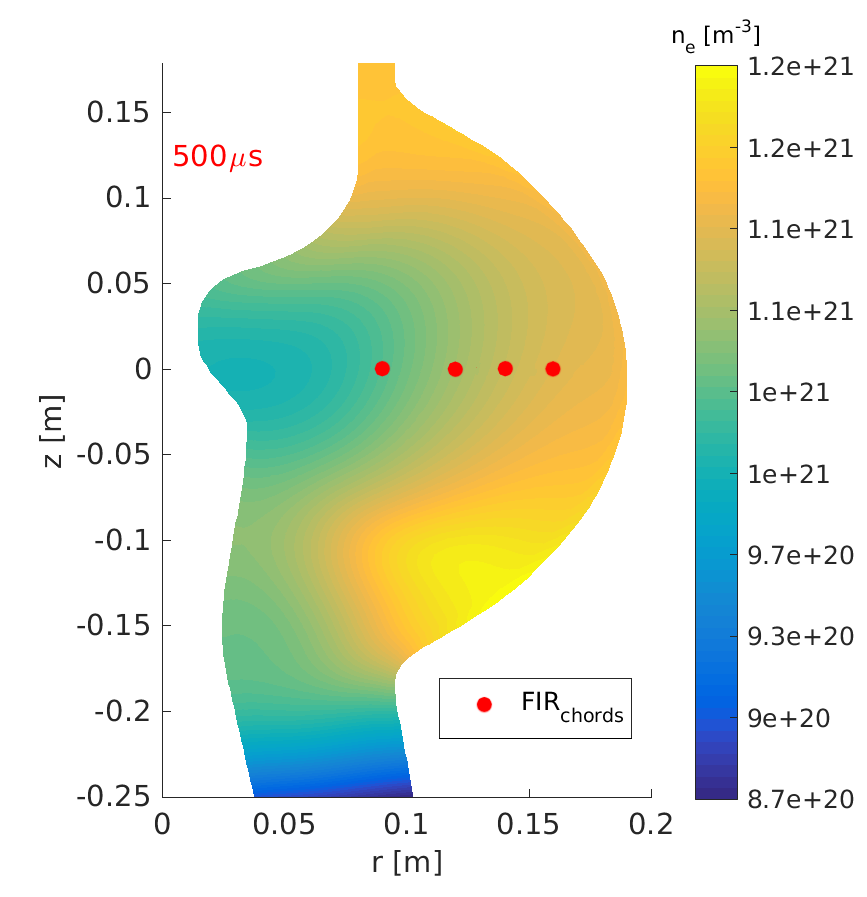}}\hfill{}\subfloat[]{\raggedright{}\includegraphics[width=7cm,height=6.3cm]{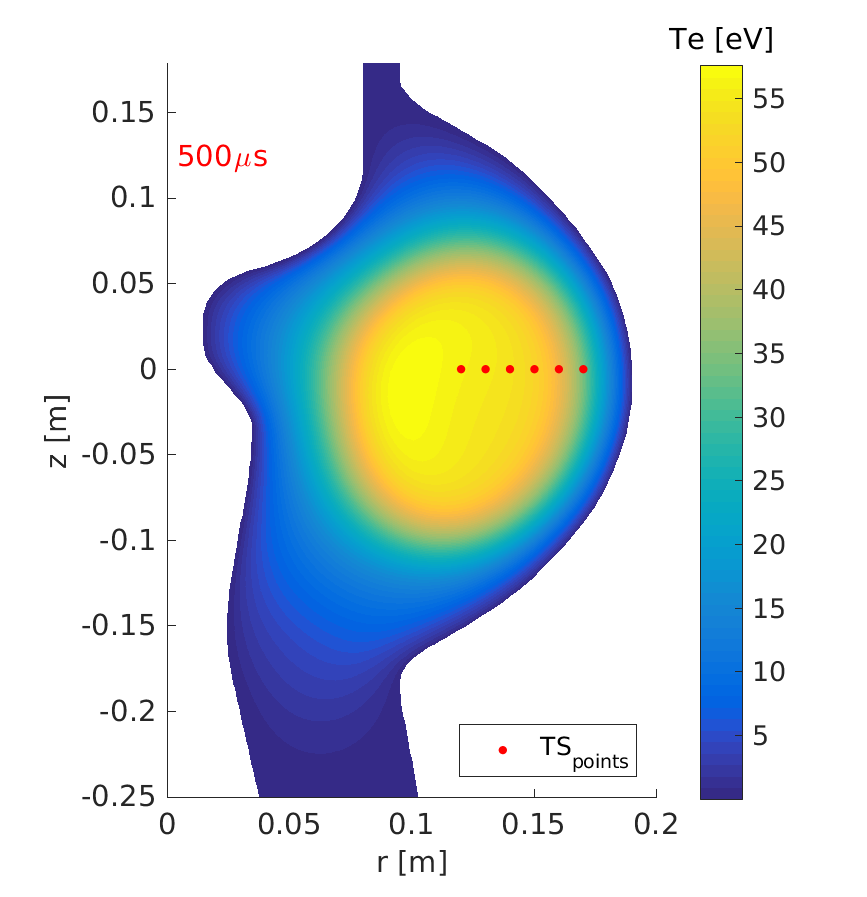}}

\caption{\label{fig: Spect_n_Te}$\,\,\,\,$Profiles of electron density and
temperature in SPECTOR geometry}
\end{figure}
The region of particularly high electron density is more defined in
figure \ref{fig: Spect_n_Te}(a), in which cross-sections of the horizontal
chords representing the lines of sight of the FIR (far-infrared) interferometer
\cite{polarimetryGF} are also depicted. The electron temperature
profile at 500$\,\upmu$s is shown in figure \ref{fig: Spect_n_Te}(b).
Referring to figure \ref{fig:SPEC_psi_n_nN}(f), it can be seen how
neutral fluid density is low in regions of high $T_{e}$ as a result
of ionization.

\begin{figure}[H]
\subfloat[]{\raggedright{}\includegraphics[width=7cm,height=5cm]{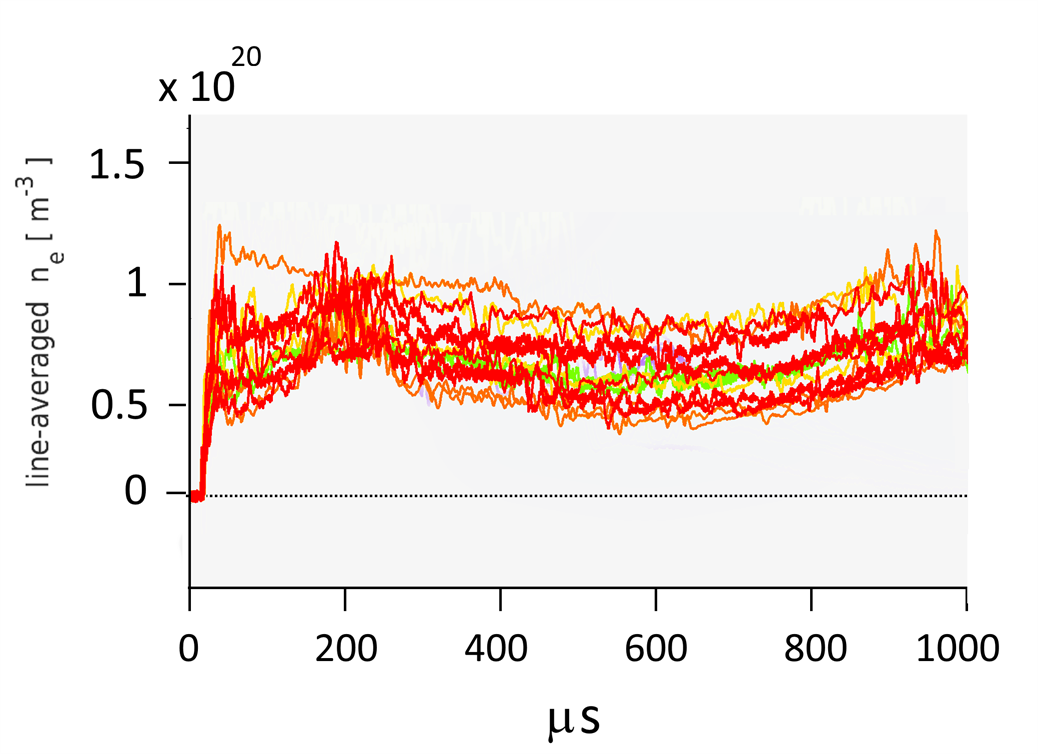}}\hfill{}\subfloat[]{\raggedright{}\includegraphics[width=7cm,height=5cm]{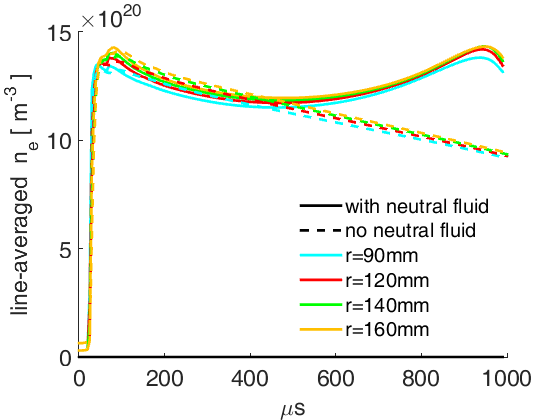}}

\caption{\label{fig:Spector_nN}$\,\,\,\,$Effect of neutral fluid dynamics
in SPECTOR geometry}
\end{figure}
Figure \ref{fig:Spector_nN}(a) shows line-averaged electron density
measured along the interferometer chord at $r=140$ mm from a selection
of several shots in SPECTOR. It can be seen how density starts to
rise at around 500 to $600\,\upmu$s. Figure \ref{fig:Spector_nN}(b)
shows the simulated diagnostic for line-averaged electron density
along the chords indicated in figure \ref{fig: Spect_n_Te}(a). The
density rise is qualitatively reproduced when a neutral fluid is included
in the simulation. Similar simulations without the inclusion of neutral
fluid do not indicate this density rise (dashed lines in figure \ref{fig:Spector_nN}(b)).
The density rise was not observed in the magnetic compression experiment
because CT lifetimes were shorter than the time it takes for a sufficient
amount of neutral gas to diffuse upwards toward the containment region.
Note that plasma densities in SPECTOR are far lower than in the magnetic
compression device, so that, as discussed in appendix \ref{sec:Timestepping-methods},
the timestep must be reduced for simulations with correspondingly
low densities. The simulations presented in figure \ref{fig:Spector_nN}(b)
were run with artificially high plasma density in order to allow for
an increased timestep and moderately short simulation run-times. Note
that the electron temperatures indicated in figure \ref{fig: Spect_n_Te}(b)
are underestimations of the actual temperatures due to the overestimation
of density in the simulation. The main goal of these simulations was
to demonstrate that the inclusion of neutral fluid interaction can
qualitatively model the observed electron density increase.

The CT fueling and cooling effect of neutral gas diffusing up the
gun is thought to be related to the unusually significant increases
in CT lifetime and electron temperature observed when a biased electrode
was inserted 11 mm into the CT edge on the SPECTOR plasma injector
\cite{Spector_biasing}. Electrode biasing involves the insertion
of an electrode, that is biased relative to the vessel wall near the
point of insertion, into the edge of a magnetized plasma. This leads
to a radially directed electric field between the probe and the wall.
The resultant $\mathbf{J}_{r}\times\mathbf{B}$ force imposed on the
plasma at the edge of the plasma confinement region varies with distance
between the probe and the wall, because $E_{r}$, as well as the magnetic
field, vary in that region. The associated torque overcomes viscous
forces, spinning up the edge plasma, and results in shearing of the
particle velocities between the probe and the wall. The sheared velocity
profile is thought to suppress the growth of turbulent eddies that
advect hot plasma particles to the wall, thereby reducing this plasma
cooling mechanism. In general, high confinement modes induced by probe
biasing share features of those initiated by various methods of heating,
including a density pedestal near the wall (near the probe radius
for probe biasing), diminished levels of recycling as evidenced by
reduced $\mbox{H}_{\alpha}$ emission intensity, and increased particle
and energy confinement times. Note the $\mbox{H}_{\alpha}$ line is
located in the deep-red visible spectrum, in the Balmer series (one
the six named series that describe spectral line emission of hydrogen
atoms), with a wavelength of $\lambda_{H_{\alpha}}=656.28$ nm in
air. A photon with energy $E_{H_{\alpha}}=h\nu_{H_{\alpha}}=\frac{hc}{\lambda_{H_{\alpha}}}$
is emitted when an electron in a neutral hydrogen atom falls from
its third to second lowest energy level. Here, $h=6.63\times10^{-34}$
{[}J-s{]} is Planck's constant, $\nu_{H_{\alpha}}\,[\mbox{s}^{-1}]$
is the frequency of the photon emitted, and $c$ is the speed of light.
Ions and electrons, that diffuse out of the plasma to the vessel walls,
cool and recombine at the walls to form neutral atoms, which can then
exhibit $\mbox{H}_{\alpha}$ emission, before being re-ionized and
$\mbox{recycled}$ back into the plasma as cold ions and electrons.
The presence of a transport barrier reduces the level of $\mbox{H}_{\alpha}$
emission by reducing the flux of charged particles to the vessel walls.

In contrast to most of the biasing experiments conducted on tokamak
plasmas, where electron density is found to increase as a consequence
of biasing, electron density was markedly reduced in the SPECTOR edge
biasing experiment. This density reduction is thought to be due to
the effect of the transport barrier impeding the CT fueling associated
with neutral gas diffusing up the gun. CT lifetimes and temperatures
were found to increase by factors of around 2.4 with edge biasing.
The scale of the improvements observed is significantly greater than
that associated with prior biasing experiments, and it is thought
that this result is associated with a reduction of the cooling effect
associated with CT fueling. A synopsis of the experiment, based on
\cite{Spector_biasing}, is presented in appendix \ref{chap:First-results-from}.

\section{Summary\label{sec:SummaryNeut}}

It has been shown how the terms that determine the source rates of
species momentum and energy due to ionization and recombination can
be derived from basic principles rather than the more formal and involved
process of taking successive moments of the collision operators pertaining
to the reactions. The moment-taking method must be used to determine
terms relating to charge exchange reactions. Only the simple method
enables determination of $Q_{e}^{rec}$, which prescribes the volumetric
rate of thermal energy transfer from electrons to photons and neutral
particles due to radiative recombination. From looking at the kinematics
of the radiative recombination reaction, it appears reasonable to
neglect $Q_{e}^{rec}$ as an energy source for the neutral fluid (most
of the electron thermal energy is transferred to the photon), but
include it as an energy sink for the electron fluid. Inclusion of
this term in the electron energy equation leads to an insignificant
reduction in electron fluid temperature in the regime examined. The
source rate terms have been used to develop the three fluid, and two
fluid equations for plasma and neutral gas, which have been implemented
to the code.

Energy conservation is maintained with the inclusion of the neutral
fluid model if the electron thermal energy expended on ionization
and recombination processes is accounted for. The initial motivation
for the study of plasma/neutral interaction was that inclusion of
a neutral model was expected to lead to net ion cooling; this concept
has been disproved. Increased net initial particle inventory results
in lower ion temperatures, regardless of whether part of the initial
inventory includes neutral particles. Charge exchange reactions lead
to ion-neutral heat exchange, but the ion temperature is relatively
unchanged by charge exchange reactions in hot-ion regions where electron
temperature and ionization rates are also high and neutral fluid density
is consequently low. The effect of residual neutral gas, that is concentrated
around the gas valves after the formation process, diffusing up the
gun to the CT containment area at relatively late times in SPECTOR
geometry, leading to increases in measured CT electron density, can
be captured by the model. This insight helps account for the exceptionally
significant increase in electron temperature, and markedly reduced
electron density, observed during the electrode edge biasing experiment
conducted on SPECTOR (see appendix \ref{chap:First-results-from}).
The model implementation is a good testbed for further studies and
improvements.

\newpage{}

\part*{PART 3: CONCLUSIONS\label{part:Conclusions}\addcontentsline{toc}{part}{PART 3: CONCLUSIONS} }

\chapter{Concluding remarks\label{chap:Concluding-remarks} }

In the study of CT formation into a levitation field, interaction
between plasma and the insulating wall surrounding the containment
region during the CT formation process led to high levels of plasma
impurities and consequent radiative cooling. The longest levitated
CT lifetimes were up to $\sim270\,\upmu$s with the 25-turn coil configuration,
despite the presence of the quartz wall. This was almost double the
maximum of $\sim150\,\upmu$s lifetimes seen with six coils around
quartz wall, but still less than the $\sim340\,\upmu$s lifetimes
observed without levitation with an aluminum flux conserver. The ceramic
alumina wall was far less contaminating than the quartz wall. In the
six coil configuration, best levitated CT lifetimes decreased significantly
when the ceramic wall was replaced with a quartz wall, despite the
larger inner radius of the quartz wall, which should have allowed
for a 50\% increase in lifetime if the material had not also been
changed. A revised wall design, such as one implementing a thin-walled
tube of pyrolytic boron nitride located inside an alumina tube for
vacuum support, would likely be beneficial. Future designs may ideally
use levitation/compression coils internal to the vacuum vessel, eliminating
the need for an insulating wall, but that would introduce alternative
complications.

In the original six-coil configurations, plasma being rapidly advected
into the containment region during the formation process was able
to displace the levitation field into the large gaps above the coil
stack, and come into contact with the insulating wall. Some mitigation
of this effect was achieved by firing the levitation banks earlier,
allowing the levitation field to soak through and become resistively
pinned in the steel above and below the insulating wall. As supported
by MHD simulation, this line-tying effect reduced the level of penetration
of stuffing field into the insulating wall during CT formation, resulting
in up to $20\,\upmu$s increase in CT lifetime. Also supported by
MHD simulation, interaction between plasma and the insulating wall
during formation was reduced with the modified levitation field profiles
of the 25-turn coil and 11-coil configurations, in which current carrying
coils extended along the entirety of the outer surface of the insulating
wall. These configurations were implemented towards the end of the
experimental phase. Spectrometer data and observations of CT lifetime
confirm that the improved design led to reduced levels of plasma impurities
and radiative cooling. Consistent with this explanation for the improvement,
at the same initial CT poloidal flux, as determined by the voltage
on the formation capacitors and the current in the main coil, CT lifetimes
were around the same for the six-coil, 25-turn coil, and eleven-coil
configurations. However the setups with the 25-turn coil and with
eleven coils allowed for the successful formation of higher flux,
physically larger, CTs - formation voltage could be increased from
$12$ to $16\mbox{ kV}$ and main coil current could be increased
from 45 to 70A, corresponding to an increase in $\Psi_{gun}$ from
around 8 to 12  mWb. In contrast, the benefit of increased initial
CT flux was surpassed by the performance degradation due to increased
wall interaction in the six coil setup. Although the recurrence rate
of good shots in the 25-turn coil configuration was significantly
worse than that in the 11-coil configuration, the longest lived CTs
produced in the former configuration endured for noticeably longer
than those produced in the latter configuration. The stronger levitation
field at the top and bottom of the insulating wall in the 25-turn
coil configuration, and consequent reduction in plasma-wall interaction
and radiative cooling may partially account for this. In addition,
the increased ratio of the levitation coil to levitation circuit holding
inductance associated with the 25-turn coil configuration, which led
to more levitation flux increase upon plasma entry to the containment
region, may have played a role. The longer rise time of the levitation
current associated with the 25-turn configuration required an increase
in the delay between the firing of levitation and formation banks,
which can lead to impediment, through the line-tying effect, of plasma
entry to the CT containment region. This is likely to have been the
cause of the poor repeatability of good shots in the 25-turn configuration.
The benefit of slightly reducing plasma-wall interaction by increasing
the delay, and the line-tying effect, outweighed the detrimental effect
of impeding plasma entry to the CT containment region in the 6-coil
configuration only. Future designs should optimise between the ideals
of low coil inductances and high coil to levitation circuit inductance
ratios.

Compared with the aluminum flux conserver, the more resistive stainless
steel wall led to increased impurity levels and shorter CT lifetimes,
likely due to more CT field-diffusion in the material, leading to
enhanced impurity sputtering. CT lifetime in the configuration with
the stainless steel wall was increased by allowing a levitation field
of moderate strength to soak through the wall and partially hold the
CT off it. Magnetic perturbations with toroidal mode number $n=2$
were observed on CTs produced with both stainless steel and aluminum
outer flux conservers, and remained even when a moderate levitation
field was allowed to soak through the stainless steel wall, but were
absent in all configurations tested in which a CT was held off an
outer insulating wall by a levitation field. It is known that $n=2$
fluctuations are a sign of internal MHD activity associated with increased
electron temperature. However, the longest-lived CTs produced with
the 25-turn configuration endured for up to 10\% longer than, and
may therefore be reasonably assumed to be hotter than the CTs produced
with the stainless steel outer flux conserver. It is possible that
the levitation field acts to damp out helically propagating magnetic
fluctuations at the outboard CT edge and that internal MHD activity
is relatively unchanged. The $n=1$ magnetic fluctuations observed
when 80 kA additional crowbarred shaft current was applied to the
machine in the eleven coil configuration confirmed coherent toroidal
CT rotation, and may have been a result of more vigorous MHD activity
that remained apparent despite damping. 

Indications of an instability, thought to be an external kink, occurred
very frequently during magnetic compression and during under-damped
magnetic levitation. Levitation circuit modification to match the
decay rates of the levitation and plasma currents led to more stable,
longer lived plasmas, and a greatly increased the recurrence rate
of good shots, by avoiding unintentional magnetic compression during
CT levitation. MHD simulation results, which closely match the available
experimental measurements, indicate that $q<1$ over extensive regions
between the CT magnetic axis and LCFS. An obvious improvement to the
experiment design would be to drive additional shaft current and raise
the $q$ profile to MHD stable regimes. 

The recurrence rate of shots in which CT poloidal flux was conserved
during magnetic compression is an indication of resilience against
a disruption-inducing instability during compression, and was increased
from around 10\% to 70\% with the transition to the levitation/compression
field profile of the eleven-coil configuration. The improvement is
likely to be largely due to the compression field profile itself,
which led to more uniform outboard compression, as opposed to the
largely equatorial outboard compression associated with the six coil
configuration. The effect of having a reduced impurity concentration
and increased CT plasma temperature prior to compression initiation,
as a consequence of the improved levitation field profile, may also
have played a role. Due to improved flux conservation at compression,
magnetic compression ratios increased significantly with the eleven
coil configuration. Magnetic compression usually did not exhibit good
toroidal symmetry.

In the eleven coil configuration, poloidal field at the CT edge, at
fixed $r=26$ mm, increased by a factor of up to six at compression,
while line averaged electron density at fixed $r=65$ mm was observed
to increase by a factor of up to four, with the electron density front
moving inwards at up to 10 km/s. Ion Doppler measurements, at fixed
$r=45$ mm indicated increases in ion temperature by factors of up
to four. Increases in poloidal field, density, and ion temperature
at compression were significant only in the eleven coil configuration.
The experimental technique developed to measure the CT outboard separatrix
confirmed that increasing the damping of the levitation field over
time led to CTs that remained physically larger over extended times.
Separatrix radii trajectories from MHD simulations matched those obtained
experimentally for various magnetic levitation and compression scenarios,
and indicated a radial compression factor, in terms of equatorial
outboard CT separatrix, of up to $1.7$. MHD simulation results indicate
that CT aspect ratio is approximately constant over compression, and
that internal CT poloidal and toroidal fields, and CT toroidal current,
scale approximately adiabatically. Due to the effect of artificial
density diffusion, simulated density and temperature do not scale
according to the theoretical adiabatic compression scalings. 

Based on a linear finite element method, various differential matrix
operators with useful mimetic properties have been developed, and
used to ensure conservation of total energy, particle count, toroidal
flux, and (in some scenarios) total angular momentum in an axisymmetric
numerical scheme that implements the non-linear single fluid, two
temperature MHD equations. The principal mimetic qualities of the
operators are that they satisfy discrete forms of the differential
product rule and the divergence theorem. A novelty of the code is
that all discrete spatial differential operators are represented as
matrices, and the discretized forms of continuous differential equations
may be obtained by simply replacing the original continuous differentiations
with the corresponding matrix operators. The resultant DELiTE framework
may be applied to solve a wide range of systems of differential equations,
and may serve as an educational tool. 

A freely available mesh generator has been modified to produce the
computational grid. Forward Euler timestepping is found to be adequate
for good solution convergence for the short physical timespans associated
with the magnetic compression experiment, but the options of the higher
order Runge-Kutta 2 and Runge-Kutta 4 timestepping schemes, which
have been implemented to the code, may be advantageous for simulations
with extended timespans. Methods were developed to vary the simulation
timestep according to the time-evolving conditions of the various
field solutions, which determine the maximum allowable timestep for
stable time-advance, while managing simulation data preservation. 

A numerical method to solve for the Grad-Shafranov equilibrium solution
has been developed. MHD simulations can begin from an initial equilibrium
state, or with the CT formation process. Various simulated diagnostics
have been implemented to the code. Some of these are counterparts
to the available experimental diagnostics. Several other simulated
diagnostics have been developed, including the time evolutions of
the $q(\psi)$ profile, magnetic axis location, volume-averaged $\beta$,
CT volume and magnetic fluxes, system energy components, and maximum
ion and electron temperatures. The Grad-Shafranov equilibrium and
$q(\psi)$ profile solutions have been benchmarked against solutions
from a well-established code, and tests that verified convergence
of MHD solutions with increasing mesh resolution were conducted. Two
dimensional simulations neglect inherently three dimensional turbulent
transport and flux conversion, and are likely to overestimate the
level of hollowness of the current profiles, and lead to an underestimation
of $q$ towards the CT edge, but without further internal experimental
diagnostics or 3D simulations, the level of underestimation remains
uncertain.

A model for anisotropic thermal diffusion has been formulated and
implemented to the code. Viscosity and resistivity are modelled as
being isotropic. The Spitzer resistvity formula is implemented for
resistive diffusion coefficients. Constant viscous and thermal diffusion
coefficients are generally implemented although the code has the options
of implementing thermal diffusion coefficients based on the Chapmnn-Enskog
closures, with the option also for Bohm scalings for the perpendicular
thermal diffusion coefficients. Thermal diffusion coefficients were
set to constant values for the simulations presented in this work,
with the high parallel coefficients representing the physical case
where particles are free to stream unimpeded along field lines. Perpendicular
thermal diffusion coefficients are chosen such that simulated CT lifetimes
are comparable to the experimentally observed lifetimes. A model for
radiative cooling has not been included - relatively high perpendicular
thermal diffusion coefficients act as a proxy to include the effects
of this cooling mechanism.

Methods for simulating CT formation, magnetic levitation and magnetic
compression to study this novel magnetic compression experiment have
been developed. The CT formation process is simulated by adding toroidal
flux associated with the measured formation voltage to the domain.
Even with the simplifying assumption that radial formation current
between the electrodes flows at a fixed axial location, the model
is able to reproduce experimental measurements to a very acceptable
level. Tests to allow variation of the axial location of intra-electrode
formation current, with upper and lower bounds, were able to precisely
reproduce the experimentally measured formation current signals over
the times when the axial location was unrestricted. Boundary conditions
for poloidal flux are evaluated and varied over time to model the
levitation and compression fields, and superimposed on boundary conditions
for poloidal flux that correspond to the stuffing field, which are
held constant over time. In combination with explicitly applied boundary
conditions for velocity components, properties of the discrete differential
operators lead to the natural imposition of boundary conditions on
$f$, resulting in a model for perfectly electrically conducting boundaries.
Boundary conditions for $f$ on the part of the boundary that represents
the insulating wall surrounding the CT containment region are overwritten
with a constant value, which is updated at each timestep. The value
for this constant is evaluated to be consistent with the maintenance
of system toroidal flux conservation in accordance with the model
in which crow-barred poloidal shaft current flows in the walls of
the machine and in the aluminum bars surrounding the insulating wall.
Conservation of system toroidal flux enables simulations to reproduce
poloidal currents that, in the experiment, are induced to flow between
the conducting walls across various regions of ambient plasma external
to the CT. In particular, simulations indicate that wall-to-wall currents
flow outboard of the CT during magnetic compression. This is thought
to be a representation of real physical phenomenon which led to experimentally
measured toroidal field signals that suggested the appearance of an
external kink instability at magnetic compression.

Special care has been taken to simulate the poloidal vacuum field
in the insulating region between the inner radius of the insulating
tube and the levitation/compression coils, and to couple this solution
to the evaluation of poloidal flux in the plasma domain, while maintaining
toroidal flux conservation. This enables a quantitative model of plasma/wall
interaction in various coil configurations. 

Simulated diagnostics from the two dimensional MHD simulations show
generally good agreement with experimental measurements. Simulations
do not capture the extremely high density that is measured experimentally
when plasma first enters the CT containment area, perhaps due to artificial
diffusion, or due to the inability of the model to capture the effect
of sputtering of high $Z$ ions during formation, and their subsequent
recombination as the plasma cools. Simulations confirmed, as also
indicated with scintillator and ion-Doppler data, that ion viscous
heating is extreme during the formation process. The simulated diagnostic
for ion temperature works by obtaining the line averaged ion temperature
along the ion-Doppler chords, which seems reasonable as a crude first
approximation, but could be developed further to include additional
physics. Qualitatively, the simulated diagnostic for ion temperature
approximately reproduces the experimental measurement with the current
model, and a suitable choice for the viscous diffusion coefficient
in the ion energy equation enables scaling to match the experimental
measurement. Plasma jets associated with magnetic reconnection lead
to high levels of ion viscous heating, particularly around the entrance
to the CT containment region during formation of closed poloidal CT
flux surfaces. Simulations show that the pinching off of closed field
lines that extend partially down the gun when magnetic compression
is initiated early in the life of the CT leads to the formation of
a smaller CT located below the entrance to the containment region.
Field lines that remain open surrounding the closed CT flux surfaces
are then also pinched and reconnect to form additional closed field
lines around the main CT. The newly reconnected open field lines below
the main CT act like a slingshot that advects the smaller CT down
the gun. This effect cannot be verified experimentally, as the magnetic
probes located along the gun barrel were not functioning properly
at any time. The poloidal field profiles experimentally measured during
magnetic compression routinely indicated loss of CT poloidal flux
during compression, particularly in the six coil configuration, suggesting
the action of a disruptive instability. It was found to be possible
to approximately reproduce the profiles by forcing CT poloidal flux
loss during simulated magnetic compression, thereby verifying that
a flux loss mechanism was involved in the instability. Simulations
of magnetic compression indicate that the oscillating compression
field resulted in the inductive formation, magnetic compression, and
subsequent extinguishing (magnetic reconnection of CT poloidal field
with the compression field) of a second and third CT. This is supported
experimentally in that experimental poloidal field measurements closely
match the corresponding simulated diagnostic, and that Xray-phosphor
imaging indicates the compressional heating of up to three distinct
plasmoids on many compression shots. 

The components of system energy for MHD simulations involving CT formation
and magnetic compression evolve as expected, with conversions initially
from toroidal magnetic energy associated with formation current to
thermal energy, kinetic energy and poloidal magnetic energy. Magnetic
energy is converted to electron thermal energy via ohmic heating,
with kinetic energy being converted to ion thermal energy due to viscous
effects. Rapid diffusion of heat through the boundary means that thermal
energy is lost faster than it is sourced from magnetic energy after
the initial period of maximum thermal energy gain, except during the
time when the CT is being magnetically compressed. At CT magnetic
compression, the principal energy conversion mechanisms are that poloidal
magnetic energy does work to compress the CT, leading to compressional
heating of ions and electrons. Toroidal plasma current increases as
the CT is compressed, leading to increased ohmic heating and an additional
rise in electron thermal energy. Again, electrons transfer some of
this energy to the ion fluid, so ion thermal energy also rises due
to enhanced ohmic heating.

A model of plasma/neutral fluid interaction was developed and included
in the DELiTE framework. The terms that determine the source rates
of species momentum and energy due to ionization and recombination
were derived from basic principles rather than applying the involved
process of taking successive moments of the collision operators pertaining
to the reactions. This method enables determination of the volumetric
rate of thermal energy transfer from electrons to photons and neutral
particles due to radiative recombination, which has been neglected
in other studies. The implementation of the model has enabled clarification
of the mechanisms behind the significant increases in CT electron
density that are routinely observed on the SPECTOR plasma injector;
simulations confirm that neutral gas, which remains concentrated below
the gas valves after CT formation, diffuses up the gun barrel to the
CT containment region where it is ionized, leading to the observed
electron density increases. This understanding helps account for the
exceptionally significant increase in electron temperature, and markedly
reduced electron density, observed during the electrode edge biasing
experiment conducted on SPECTOR. It is thought that edge fueling impediment,
a consequence of a biasing-induced transport barrier, is largely responsible
for the temperature increase and density decrease. System energy conservation
is maintained with the inclusion of the neutral fluid model if the
electron thermal energy expended on ionization and recombination processes
is accounted for. The initial motivation for the study of plasma/neutral
fluid interaction was that inclusion of a neutral fluid model was
expected to lead to net ion cooling. However simulations indicated
that ion temperature is relatively unchanged by charge exchange reactions
in hot-ion regions, where electron temperature and ionization rates
are also high and neutral fluid density is particularly low. The model
implementation is a good testbed for further studies and improvements.

Further code development may include a more generally applicable model
for density diffusion with correction terms that maintain angular
momentum conservation even in simulation scenarios that include CT
formation and magnetic compression. The level to which artificial
density diffusion affects the simulated density diagnostic, and the
evolution of the other fields, should be investigated further. An
implicit timestepping scheme may be implemented as this would enable
faster simulation times while enabling the use of a greater range
of values for the various diffusion coefficients. The ability to model
part of the domain as a plasma-free material is useful because it
expands the code's range of applicability as a problem-solving tool.
For example, the feature can easily be adapted to model the penetration
of field associated with magnetically confined plasma into surrounding
conductors. Even a much simpler model, such as the isothermal version
of the code, can qualitatively reproduce experimental diagnostics
for magnetic field and density, and with resistivity simply set artificially
high in the regions representing the insulating wall, is able to illustrate
intersection of stuffing field with the wall during CT formation,
and the mitigation of the effect in the eleven coil configuration.
However, the effort invested in developing a conservative scheme with
many extended features has resulted in a code framework with a solid
foundation for further development and improvement. 

The magnetic compression experiment would certainly be worth attempting
again. It is unclear if it can work to meet the original goal of supporting
the PCS program, as the dynamics of a CT being compressed by a moving
metal wall may be too different from those associated with a compressing
external magnetic field. However, the goal of achieving levitated
CT temperatures and lifetimes comparable to that of CTs formed into
a containment region surrounded by a flux conserver appears within
reach, if a suitable choice of material is made for the insulating
wall, or if compression coils are integrated within the vacuum vessel,
and if the levitation circuit parameters are optimised. Furthermore,
it has been inferred that CT stability needs to be improved; a first
step towards improvement would be to add toroidal field through the
action of externally driven shaft current, as is currently done on
SPECTOR machines. Having already demonstrated reasonably significant
temperature increases at compression, rapid further progress with
some design changes based on lessons learned seems likely.\\

\chapter*{$ $\label{bib}\addcontentsline{toc}{chapter}{BIBLIOGRAPHY}$ $}

\part*{PART 4: APPENDICES\label{part:apx__________________}\addcontentsline{toc}{part}{PART 4: APPENDICES} }

\appendix

\chapter{Plasma kinetic theory, MHD equations, and equilibrium models \label{chap:Kinetic,MhD,GS}}

This appendix is largely comprised of background material, assembled
mostly from material found in \cite{HazeltineWaelbroeck,farside braginskii,bellan fundamentals,bittencourt,Bellan_Spheromaks}.
An overview is presented of the well-established methods used to derive
the MHD equations and Grad-Shafranov equation, which are solved for
in the code. The overview is included mainly because it lays the groundwork
for chapter \ref{chap:Neutral-models}, in which the set of equations
that describes co-interacting plasma and neutral fluids are derived,
in a process that involves taking moments of collision operators that
pertain to the scattering and reactive collisions associated with
plasma-neutral interactions. Readers familiar with this material may
wish to skip directly to section \ref{sec:SummaryKin_MHD_EQ}, where
the equations which are solved for in the MHD code (without plasma-neutral
interaction) are collected. This appendix begins with a brief overview
of plasma kinetic theory in section \ref{subsec:DistFnc_BoltzEqn}.
In section \ref{subsec:MomentsDist}, it will be shown how various
important plasma fluid parameters can be obtained by taking successive
moments of the distribution function. For example, the plasma fluid
number density, fluid velocity, total stress tensor, and energy flux
density are derived by taking the zeroth, first, second and third
moments respectively. In section \ref{subsec:MomentsCollision}, it
will be shown how taking moments of the part of the collision operator
pertaining to scattering collisions results in finite terms that will
be included in the ion and electron plasma fluid momentum and energy
conservation equations. Using the derivations presented in sections
\ref{subsec:MomentsDist} and \ref{subsec:MomentsCollision}, the
process of taking moments of the Boltzmann equation results in the
two-fluid Braginskii MHD equations, as shown in section \ref{subsec:Braginski-equations-from}.
A brief overview of the Chapmann-Enskog scheme for obtaining fluid
closures, which yield expressions for the ion and electron heat flux
densities and viscous diffusion coefficients, is presented in section
\ref{subsec:Chapman-Enskog-closures}. The process for obtaining the
single-fluid MHD equations from the Boltzmann equation and the two-fluid
MHD equations is outlined in section \ref{subsec:Single-fluid-Magnetohydrodynamic}.
Derivations of the standard form and a simplified form of the Grad-Shafranov
equation, which describes the force balance equilibrium in ideal MHD
with toroidal symmetry, are presented in section \ref{sec:Equilibrium-models}.
The appendix will conclude with section \ref{sec:SummaryKin_MHD_EQ},
where the core set of equations implemented to the DELiTE code framework
are collected. 

\section{Distribution functions and the Boltzmann equation\label{subsec:DistFnc_BoltzEqn}}

We consider a system of point particles with a given charge and mass.
The system has two species of particles, electrons and singly charged
ions. In six-dimensional phase space $(\mathbf{r,V})$, each particle
has its own trajectory - the trajectory of the $i^{th}$ particle
is $\mathbf{r}_{i}(t),\,\mathbf{V}_{i}(t)$. Since $\mathbf{V}_{i}(t)=\dot{\mathbf{r}}_{i}(t)$,
if $\mathbf{r}_{i}(t)\mbox{ and }\mathbf{V}_{i}(t)$ are known at
one time, they are known at all times. The instantaneous dynamic state
of each particle is given by a point in phase space that's defined
by the six coordinates $x,\,y,\,z,\,V_{x}\,,V_{y},\,V_{z}$. The microscopic
number density $[\mbox{m}^{-3}]$ of particles of species $\alpha$
in phase space is described by the Klimontovich density distribution
function \cite{AbhayRam}: 
\begin{equation}
F_{\alpha}(\mathbf{r},\mathbf{V},t)=\stackrel[i=1]{N_{0}}{\Sigma}\delta(\mathbf{r}-\mathbf{r}_{i}(t))\,\delta(\mathbf{V}-\mathbf{V}_{i}(t))\label{eq:344}
\end{equation}
where $N_{0}$ is the number of particles of each species, and $\delta(x)$
is the Dirac-delta function: 
\[
\delta(x)=\{_{0,\,if\,x\neq0}^{\infty,\,if\,x=0}
\]
 
\[
\int_{-\infty}^{\infty}\delta(x)\,dx=1
\]
Note that for point particles, number density is infinite at singular
points where there are particles. $F_{\alpha}$ is normalised so that
its velocity integral is the regular particle number density \cite{HazeltineWaelbroeck}:
\[
\int F_{\alpha}(\mathbf{r},\mathbf{V},t)\,d\mathbf{V}=n_{\alpha}(\mathbf{r},t)
\]
The distribution function satisfies the continuity equation in phase
space: 
\[
\frac{dF_{\alpha}}{dt}=\left(\frac{\partial}{\partial t}+\frac{\partial\mathbf{r}}{\partial t}\cdot\frac{\partial}{\partial\mathbf{r}}+\frac{\partial\mathbf{V}}{\partial t}\cdot\frac{\partial}{\partial\mathbf{V}}\right)F_{\alpha}=\frac{\partial F_{\alpha}}{\partial t}+\mathbf{V\cdot}\nabla F_{\alpha}+\mathbf{a_{\alpha}\cdot}\nabla_{v}F_{\alpha}=0
\]
 where $\mathbf{a}_{\alpha}(\mathbf{r},t)=\frac{q_{\alpha}}{m_{\alpha}}(\mathbf{E}(\mathbf{r},t)\mathbf{+V}(t)\mathbf{\times B}(\mathbf{r},t))$
is the species particle acceleration determined by the Lorentz force.
The Klimontovich equation is:
\begin{equation}
\frac{\partial F_{\alpha}}{\partial t}+\mathbf{V\cdot}\nabla F_{\alpha}+\frac{q_{\alpha}}{m_{\alpha}}(\mathbf{E}(\mathbf{r},t)\mathbf{+V}\mathbf{\times B}(\mathbf{r},t))\mathbf{\cdot}\nabla_{v}F_{\alpha}=0\label{eq:345}
\end{equation}
The fields are due to external sources in combination with microscopic
fields self-consistently produced by the point particles: $\mathbf{E}(\mathbf{r},t)=\mathbf{E}_{ext}(\mathbf{r},t)+\mathbf{E}_{m}(\mathbf{r},t)$,
and \\
$\mathbf{B}(\mathbf{r},t)=\mathbf{B}_{ext}(\mathbf{r},t)+\mathbf{B}_{m}(\mathbf{r},t)$.
The microscopic fields are determined by Maxwell's equations:\\
\begin{align}
\nabla\cdot\mathbf{E}_{m}(\mathbf{r},t) & =\frac{\rho_{c-m}(\mathbf{r},t)}{\epsilon_{0}}\nonumber \\
\nabla\cdot\mathbf{B}_{m}(\mathbf{r},t) & =0\nonumber \\
\nabla\times\mathbf{E}_{m}(\mathbf{r},t) & =-\mathbf{\dot{B}}_{m}(\mathbf{r},t)\label{eq:346}\\
\nabla\times\mathbf{B}_{m}(\mathbf{r},t) & =\mu_{0}(\mathbf{J}_{m}(\mathbf{r},t)+\epsilon_{0}\mathbf{\dot{E}}_{m}(\mathbf{r},t))\nonumber 
\end{align}
The microscopic charge density is 
\begin{equation}
\rho_{c-m}(\mathbf{r},t)=\stackrel[\alpha]{}{\Sigma}q_{\alpha}\int F_{\alpha}(\mathbf{r},\mathbf{V},t)\,d\mathbf{V}\label{eq:347}
\end{equation}
and the microscopic current density is 
\begin{equation}
\mathbf{J}_{m}(\mathbf{r},t)=\stackrel[\alpha]{}{\Sigma}q_{\alpha}\int\mathbf{V}F_{\alpha}(\mathbf{r},\mathbf{V},t)\,d\mathbf{V}\label{eq:348}
\end{equation}
It can be said that plasma physics can be viewed formally as a closure
of Maxwell's equations by means of constitutive relations - expressions
for charge density and current density in terms of the electric and
magnetic fields \cite{HazeltineWaelbroeck}. Given initial fields
that are consistent with Maxwell's equations, equations \ref{eq:347}
and \ref{eq:348}, in combination with equations \ref{eq:344} to
\ref{eq:346} make up the desired constitutive relations. The problem
is completely deterministic and densities and fields are exactly determined
at all times. Although the Klimontovich formalism is exact, it is
far too detailed to be of practical use \cite{AbhayRam}. The fine-grained,
spiky distribution function is microscopically exact but does not
correspond to the macroscopically smooth quantities observed in experiment.
A more useful description is obtained by an averaging process. One
method to achieve this is based on the Liouville distribution and
the BBGKY hierarchy equations \cite{LandauLifshitz,AbhayRam}. An
equivalent, more straightforward method, is based on taking the ensemble
averages of the distribution functions and fields. Splitting the distribution
functions and the microscopic parts of the fields into averaged and
fluctuating contributions, we define:
\begin{align}
F_{\alpha}(\mathbf{r},\mathbf{V},t) & =\ll F_{\alpha}(\mathbf{r},\mathbf{V},t)\gg+\delta F_{\alpha}(\mathbf{r},\mathbf{V},t)\nonumber \\
\mathbf{E}(\mathbf{r},t) & =\mathbf{E}_{ext}(\mathbf{r},t)+\ll\mathbf{E}_{m}(\mathbf{r},t)\gg+\delta\mathbf{E}_{m}(\mathbf{r},t)\label{eq:349}\\
\mathbf{B}(\mathbf{r},t) & =\mathbf{B}_{ext}(\mathbf{r},t)+\ll\mathbf{B}_{m}(\mathbf{r},t)\gg+\delta\mathbf{B}_{m}(\mathbf{r},t)\nonumber 
\end{align}
where $\ll...\gg$ indicates ensemble averaging ($e.g.,$ $f_{\alpha}(\mathbf{r},\mathbf{V},t)=\ll F_{\alpha}(\mathbf{r},\mathbf{V},t)\gg$
is a smoothed distribution function for species $\alpha$ that is
obtained from a statistical average of the microscopic distribution
function), and $\delta$ indicates fluctuating contributions. Using
the evolution of $f_{\alpha}$ to characterise the system does not
track individual particles (as is done when using $F_{\alpha}$) -
instead, classes of particles having approximately the same $\mathbf{r}$
and $\mathbf{v}$ are tracked \cite{bellan fundamentals}.\textbf{
}$f_{\alpha}(\mathbf{r},\mathbf{V},t)$ represents the average of
the number density $[\mbox{m}^{-6}\,\mbox{s}^{3}]$ of particles of
species $\alpha$ in a 6D volume element in phase space, centered
around $\mathbf{r},\mathbf{V}$, with dimensions $dx.dy.dz.dV_{x}.dV_{y}.dV_{z}$,
where the average is taken in the $6D$ volume. Consequently, \\
$f_{\alpha}(\mathbf{r,V},t)\,d\mathbf{r}\,d\mathbf{V}=f_{\alpha}(\mathbf{r,V},t)\,dx.dy.dz.dV_{x}.dV_{y}.dV_{z}=$
  number of particles of species $\alpha$ in the same 6D volume element. 

Using equation \ref{eq:349} in equation \ref{eq:345}, and ensemble-averaging,
noting that\\
$\ll\delta F_{\alpha}\gg=\ll\delta\mathbf{E}_{m}\gg=$$\ll\delta\mathbf{B}_{m}\gg=0$
(statistical average of a random fluctuation is zero), \\
$\ll\ll a\gg\gg=\ll a\gg$, and that $\ll a+b\gg=\ll a\gg+\ll b\gg$,
we obtain:
\begin{align}
 & \frac{\partial f_{\alpha}}{\partial t}+\mathbf{V\cdot}\nabla f_{\alpha}+\frac{q_{\alpha}}{m_{\alpha}}(\ll\mathbf{E}(\mathbf{r},t)\gg\mathbf{+V}\mathbf{\times\ll B}(\mathbf{r},t)\gg)\mathbf{\cdot}\nabla_{v\,}f_{\alpha}\nonumber \\
 & =-\frac{q_{\alpha}}{m_{\alpha}}\ll(\mathbf{\delta E}_{m}(\mathbf{r},t)\mathbf{+V}\mathbf{\times\delta B}_{m}(\mathbf{r},t))\mathbf{\cdot}\nabla_{v}\delta F_{\alpha}\gg\label{eq:350}
\end{align}
where $\ll\mathbf{E}(\mathbf{r},t)\gg=\mathbf{E}_{ext}(\mathbf{r},t)+\ll\mathbf{E}_{m}(\mathbf{r},t)\gg$,
and $\ll\mathbf{B}(\mathbf{r},t)\gg=\mathbf{B}_{ext}(\mathbf{r},t)+\ll\mathbf{B}_{m}(\mathbf{r},t)\gg$.
The left side of equation \ref{eq:350} represents collective effects,
the right side represents collisional effects. Changing to the standard
notations $\ll\mathbf{E}(\mathbf{r},t)\gg\rightarrow\mathbf{E}(\mathbf{r},t)$,
$\ll\mathbf{B}(\mathbf{r},t)\gg\rightarrow\mathbf{B}(\mathbf{r},t)$,
and $-\frac{q_{\alpha}}{m_{\alpha}}\ll(\mathbf{\delta E}_{m}(\mathbf{r},t)\mathbf{+V}\mathbf{\times\delta B}(\mathbf{r},t))\mathbf{\cdot}\nabla_{v}\delta F_{\alpha}\gg\rightarrow\frac{\partial f_{\alpha}}{\partial t}|_{collisions}=C_{\alpha}(f)$,
we obtain the standard form of the Boltzmann equation:
\begin{equation}
\frac{\partial f_{\alpha}}{\partial t}+\mathbf{V\cdot}\nabla f_{\alpha}+\frac{q_{\alpha}}{m_{\alpha}}(\mathbf{E}(\mathbf{r},t)\mathbf{+V}\mathbf{\times B}(\mathbf{r},t))\mathbf{\cdot}\nabla_{v}f_{\alpha}=C_{\alpha}(f)\label{eq:351}
\end{equation}
Here, $C_{\alpha}(f)$ is the collision operator, which is not necessarily
linear, and generally involves the distribution functions of all the
species in the system \cite{HazeltineWaelbroeck} - for this reason
the subscript on the argument of $C_{\alpha}$ is not included. When
collisions are neglected, we have the Vlasov equation:
\begin{equation}
\frac{\partial f_{\alpha}}{\partial t}+\mathbf{V\cdot}\nabla f_{\alpha}+\frac{q_{\alpha}}{m_{\alpha}}(\mathbf{E}(\mathbf{r},t)\mathbf{+V}\mathbf{\times B}(\mathbf{r},t))\mathbf{\cdot}\nabla_{v}f_{\alpha}=0\label{eq:352}
\end{equation}

\section{Moments of the distribution function\label{subsec:MomentsDist} }

Each particle in the plasma can be associated with some molecular
property $\zeta(\mathbf{r,V},t)$, such as the mass, velocity, momentum
or energy of the particle. Since $f_{\alpha}(\mathbf{r,V},t)\,d\mathbf{r}\,d\mathbf{V}$
is the number of particles of species $\alpha$ in the 6D volume element
in phase space, it follows that the total combined value of $\zeta$
for all the particles of type $\alpha$ in the volume element is $\zeta(\mathbf{r,V},t)\,f_{\alpha}(\mathbf{r,V},t)\,d\mathbf{r}\,d\mathbf{V}$.
Therefore, the total combined value of $\zeta$, irrespective of particle
velocity, for all particles of type $\alpha$ in the $3D$ volume
element $d\mathbf{r}=dx.dy.dz$ in configuration space is $d\mathbf{r}\int\zeta(\mathbf{r,V},t)\,f_{\alpha}(\mathbf{r,V},t)\,d\mathbf{V}$,
where the integral is over all velocity space. The average value of
$\zeta$ can then be found by dividing this by $n_{\alpha}(\mathbf{r},t)d\mathbf{r},$
which is the number of particles of type $\alpha$ in $d\mathbf{r}$
\cite{bittencourt}:
\begin{equation}
<\zeta(\mathbf{r,V},t)>_{\alpha}=\zeta_{\alpha-ave}(\mathbf{r},t)=\frac{1}{n_{\alpha}(\mathbf{r},t)}\int\zeta(\mathbf{r,V},t)\,f_{\alpha}(\mathbf{r,V},t)\,d\mathbf{V}\label{eq:352.0}
\end{equation}
Considering the cases $\zeta(\mathbf{r,V},t)=1$ and $\zeta(\mathbf{r,V},t)=\mathbf{V}$,
the first two moments of the distribution function can be defined
as follows. The number density $[\mbox{m}^{-3}]$ for species $\alpha$
is defined using the distribution function as 
\begin{equation}
n_{\alpha}(\mathbf{r},t)=\int f_{\alpha}(\mathbf{r,V},t)\,d\mathbf{V}\label{eq:352-1}
\end{equation}
The species fluid velocity is defined as \textbf{
\begin{equation}
\mathbf{v}_{\alpha}(\mathbf{r},t)=\frac{1}{n_{\alpha}(\mathbf{r},t)}\int\mathbf{V}\,f_{\alpha}\,d\mathbf{V}\label{eq:352-2}
\end{equation}
}The quantities $n_{\alpha}$ and \textbf{$\Gamma_{\alpha}=n_{\alpha}\mathbf{v}_{\alpha}$,
}are the first two \emph{moments }of the distribution function ($0^{th}$
and $1^{st}$ order moments). Successive moments are of the form $M_{k}=\int f\mathbf{\,VVV}...\mathbf{V}\,d\mathbf{V}$,
for $k=2,3,...,\infty$, with $k$ factors of $\mathbf{V}$ - in general,
$M_{k}$ is a tensor of rank $k$, although it is often contracted
to lower rank \cite{HazeltineWaelbroeck}. For example, the $2^{nd}$
order moment, describing the flow of momentum in the laboratory frame
\cite{farside braginskii}, is the total stress tensor given by 
\begin{equation}
\underline{\mathbf{P}}_{\alpha}(\mathbf{r},t)=m_{\alpha}\int\mathbf{VV}\,f_{\alpha}\,d\mathbf{V}\label{eq:353}
\end{equation}
where $m_{\alpha}$ is the particle's mass. We define 
\begin{equation}
\mathbf{c}_{\alpha}(\mathbf{r},t)=\mathbf{V-\mathbf{v}_{\alpha}}(\mathbf{r},t)\label{eq:354}
\end{equation}
as the species random particle velocity relative to the species fluid
velocity \cite{bittencourt}. Note that $\mathbf{V}$ is independent
of\textbf{ $\mathbf{r}$ }and\textbf{ $t$.} By definition, 
\begin{equation}
\int\mathbf{c}_{\alpha}\,f_{\alpha}\,d\mathbf{V}=\int c_{\alpha}\,f_{\alpha}\,d\mathbf{V}=0\label{eq:355-1}
\end{equation}
because the statistical average of a random quantity is zero, and,
since the average of an averaged value is unchanged: 
\begin{equation}
\int\mathbf{v}_{\alpha}\,f_{\alpha}\,d\mathbf{V}=\mathbf{v}_{\alpha}\label{eq:355-2}
\end{equation}
The pressure tensor ($2^{nd}$ order moment in the rest frame of species
$\alpha$) is defined as 
\begin{equation}
\underline{\mathbf{p}}_{\alpha}(\mathbf{r},t)=m_{\alpha}\int\mathbf{c_{\alpha}c_{\alpha}}\,f_{\alpha}\,d\mathbf{V}\label{eq:355}
\end{equation}
Scalar pressure is given by the trace of the pressure tensor: $p_{\alpha}=\frac{1}{3}Tr(\underline{\mathbf{p}}_{\alpha})$.
Expressing $\mathbf{c}_{\alpha}$ in terms of Cartesian coordinates
$x,\,y,\,z$ as $(c_{\alpha x},\,c_{\alpha y},\,c_{\alpha z})$, then
with the definition\\
$\left(\underline{\mathbf{p}}_{\alpha}\right)_{ij}=m_{\alpha}\int c_{\alpha i}\,c_{\alpha j}\,f_{\alpha}\,d\mathbf{V}$,
the scalar pressure is: 
\begin{align}
p_{\alpha} & =\frac{1}{3}\left(\left(\underline{\mathbf{p}}_{\alpha}\right)_{xx}+\left(\underline{\mathbf{p}}_{\alpha}\right)_{yy}+\left(\underline{\mathbf{p}}_{\alpha}\right)_{zz}\right)=\frac{m_{\alpha}}{3}\int\left(c_{\alpha x}^{2}+c_{\alpha y}^{2}+c_{\alpha z}^{2}\right)\,f_{\alpha}\,d\mathbf{V}\nonumber \\
\Rightarrow p_{\alpha} & =\frac{m_{\alpha}}{3}\int c_{\alpha}^{2}\,f_{\alpha}\,d\mathbf{V}\label{eq:355-3}
\end{align}
Substituting equation \ref{eq:354} into equation \ref{eq:353}, we
have:

\begin{align}
\underline{\mathbf{P}}_{\alpha} & =m_{\alpha}\left(\int\mathbf{c_{\alpha}c_{\alpha}}\,f_{\alpha}\,d\mathbf{V}+\int\mathbf{c_{\alpha}v_{\alpha}}\,f_{\alpha}\,d\mathbf{V}+\int\mathbf{v_{\alpha}c_{\alpha}}\,f_{\alpha}\,d\mathbf{V}+\int\mathbf{v_{\alpha}v_{\alpha}}\,f_{\alpha}\,d\mathbf{V}\right)\nonumber \\
 & =m_{\alpha}\left(\frac{1}{m_{\alpha}}\underline{\mathbf{p}}_{\alpha}+\cancelto{0}{\left(\int\mathbf{c_{\alpha}}\,f_{\alpha}\,d\mathbf{V}\right)}\mathbf{v_{\alpha}}+\mathbf{v}_{\alpha}\cancelto{0}{\left(\int\mathbf{c_{\alpha}}\,f_{\alpha}\,d\mathbf{V}\right)}+\mathbf{v_{\alpha}v_{\alpha}}n_{\alpha}\right)\nonumber \\
 & \,\,\,\,\,\,\,\,\,\,\,\,\,\,\,\,\,\,\,\,\,\,\,\,\,\,\,\,\,\,\,\,\,\,\,\,\,\mbox{(use eqns. \ref{eq:355}, \ref{eq:355-1}, \ref{eq:355-2}, and \ref{eq:352-1})}\nonumber \\
\Rightarrow\underline{\mathbf{P}}_{\alpha} & =\underline{\mathbf{p}}_{\alpha}+\rho_{\alpha}\mathbf{v}_{\alpha}\mathbf{v}_{\alpha}\label{eq:356-1}
\end{align}
For convenience, $\underline{\mathbf{p}}_{\alpha}$ can be separated:
\begin{equation}
\underline{\mathbf{p}}_{\alpha}=p_{\alpha}\underline{\mathbf{I}}+\overline{\boldsymbol{\pi}}_{\alpha}\label{eq:357}
\end{equation}
where $\underline{\mathbf{I}}$ is the identity tensor, and $\overline{\boldsymbol{\pi}}$
is the generalised viscosity tensor, which is symmetric and has zero
trace \cite{farside braginskii}. An important contracted $3^{rd}$
order moment is the energy flux density:

\begin{equation}
\mathbf{Q'}_{\alpha}(\mathbf{r},t)=\int\left(\frac{1}{2}m_{\alpha}V^{2}\right)\mathbf{V}\,f_{\alpha}\,d\mathbf{V}\label{eq:358}
\end{equation}
The heat flux density (contracted $3^{rd}$ order moment in the rest
frame of species $\alpha$), defined as the flux of random ($i.e.,$
thermal) energy \cite{bittencourt}, is:

\begin{equation}
\mathbf{q}_{\alpha}(\mathbf{r},t)=\int\left(\frac{1}{2}m_{\alpha}c_{\alpha}^{2}\right)\mathbf{c_{\alpha}}\,f_{\alpha}\,d\mathbf{V}\label{eq:359}
\end{equation}
Using equation \ref{eq:354} , $\mathbf{Q'}_{\alpha}$ can be expressed
in terms of $\mathbf{q}_{\alpha}$:
\begin{align*}
\mathbf{Q'_{\alpha}} & =\frac{1}{2}m_{\alpha}\int\left(v_{\alpha}^{2}+2\mathbf{v_{\alpha}}\cdot\mathbf{c_{\alpha}}+c_{\alpha}^{2}\right)\left(\mathbf{v_{\alpha}+c_{\alpha}}\right)\,f_{\alpha}\,d\mathbf{V}\\
 & =\frac{1}{2}m_{\alpha}\biggl(\int v_{\alpha}^{2}\mathbf{v_{\alpha}}\,f_{\alpha}\,d\mathbf{V}+2\int\mathbf{c_{\alpha}\cdot\left(v_{\alpha}v_{\alpha}\right)}\,f_{\alpha}\,d\mathbf{V}+\int c_{\alpha}^{2}\mathbf{v_{\alpha}}\,f_{\alpha}\,d\mathbf{V}\\
 & +\int v_{\alpha}^{2}\mathbf{c_{\alpha}}\,f_{\alpha}\,d\mathbf{V}+2\int\mathbf{v_{\alpha}\cdot\left(c_{\alpha}c_{\alpha}\right)}\,f_{\alpha}\,d\mathbf{V}+\int c_{\alpha}^{2}\mathbf{c_{\alpha}}\,f_{\alpha}\,d\mathbf{V}\biggr)\\
 & =\frac{1}{2}m_{\alpha}\biggl[v_{\alpha}^{2}\mathbf{v_{\alpha}}\cancelto{n_{\alpha}(\mathbf{r},t)}{\left(\int f_{\alpha}\,d\mathbf{V}\right)}+2\cancelto{0}{\left(\int\mathbf{c_{\alpha}}\,f_{\alpha}\,d\mathbf{V}\right)}\cdot\mathbf{v_{\alpha}v_{\alpha}}+\left(\int c_{\alpha}^{2}f_{\alpha}\,d\mathbf{V}\right)\mathbf{v_{\alpha}}\\
 & +v_{\alpha}^{2}\cancelto{0}{\left(\int\mathbf{c_{\alpha}}\,f_{\alpha}\,d\mathbf{V}\right)}+2\mathbf{v_{\alpha}}\cdot\left(\int\mathbf{c_{\alpha}c_{\alpha}}\,f_{\alpha}\,d\mathbf{V}\right)+\int c_{\alpha}^{2}\mathbf{c_{\alpha}}\,f_{\alpha}\,d\mathbf{V}\biggr]\\
 & \,\,\,\,\,\,\,\,\,\,\,\,\,\,\,\,\,\,\,\,\,\,\,\,\,\,\,\,\,\mbox{(use eqns. \ref{eq:352-1}, \ref{eq:355-1}, and \ref{eq:355-2})}
\end{align*}

\begin{align}
\Rightarrow\mathbf{Q'_{\alpha}} & =\frac{1}{2}\rho_{\alpha}v_{\alpha}^{2}\mathbf{v_{\alpha}}+\frac{3}{2}p_{\alpha}\mathbf{v_{\alpha}}+\mathbf{v_{\alpha}\cdot}\underline{\mathbf{p}}_{\alpha}+\mathbf{q}_{\alpha}\,\,\,\,\mbox{(use eqns. \ref{eq:355-3}, \ref{eq:355}, \& \ref{eq:359})}\label{eq:465}
\end{align}

\section{Moments of the collision operator \label{subsec:MomentsCollision}}

At the atomic level, collisions are interactions between the force
fields associated with the interacting particles. The collision operator
can be split into parts pertaining to elastic \emph{scattering} collisions
and \emph{reacting} collisions as $C_{\alpha}(f)=C_{\alpha}^{scatt.}(f)+C_{\alpha}^{react.}(f)$.
Electron impact ionization and radiative recombination are examples
of inelastic reacting collisions, while resonant charge exchange is
an elastic reacting collision \cite{Goldston,Meier}. Reacting collisions
will be included in a plasma/neutral fluid model in chapter \ref{chap:Neutral-models}.
Mass, momentum and energy are conserved and particles are not created
or destroyed in elastic scattering collisions.

The Coulomb interaction of charged particles is a long range one characterised
by multiple simultaneous interactions. The short range fields (within
the electronic shells) of neutral particles result, by contrast, in
binary collisions. Because of the long range nature of the Coulomb
force, small-angle deflections associated with Coulomb collisions
are much more frequent than the large angle deflections associated
with binary collisions. The cumulative effect of many small-angle
collisions is much larger than that of relatively fewer large-angle
collisions \cite{Goldston}. It is possible to deal with multiple
collisions by approximating them as a number of simultaneous binary
collisions \cite{bittencourt}. Boltzmann's collision operator for
neutral gas is:

\begin{equation}
C_{\alpha}^{scatt.}(f)=\underset{\sigma}{\Sigma}\,C_{\alpha\sigma}^{scatt.}(f_{\alpha},\,f_{\sigma})\label{eq:465.1}
\end{equation}
Here, $C_{\alpha\sigma}^{scatt.}(f_{\alpha},\,f_{\sigma})$, the rate
of change of $f_{\alpha}$ due to collisions of species $\alpha$
with species $\sigma$, considers only binary collisions and is therefore
bilinear because $C_{\alpha\sigma}^{scatt.}$ is a linear function
of both its arguments \cite{bellan fundamentals,HazeltineWaelbroeck}.
In plasmas, where long-range Coulomb interactions lead to Debye shielding,
a many-body effect, collisions are not strictly binary. However, in
a weakly coupled plasma, the departure from bilinearity is logarithmic,
and can be neglected to a good approximation since the logarithm is
a relatively weakly varying function \cite{farside braginskii}. The
collisional process for the elastic collisions described by $C_{\alpha}^{scatt.}$
conserves particles, momentum and energy at each point \cite{HazeltineWaelbroeck}.
\\
\\
Particle conservation is expressed by: 
\begin{equation}
\int C_{\alpha\sigma}^{scatt.}d\mathbf{V}=0\label{eq:466}
\end{equation}
In combination with equation \ref{eq:465.1}, this yields 
\begin{equation}
\int C_{\alpha}^{scatt.}d\mathbf{V}=0\label{eq:466.1}
\end{equation}
Momentum conservation requires that collisions between different species
conserve the total momentum: 
\[
\int m_{\alpha}C_{\alpha\sigma}^{scatt.}\mathbf{V}d\mathbf{V}=-\int m_{\sigma}C_{\sigma\alpha}^{scatt.}\mathbf{V}d\mathbf{V}
\]
The rate of collisional momentum exchange, or the collisional friction
force, is defined as

\[
\mathbf{R}_{\alpha\sigma}=\int m_{\alpha}C_{\alpha\sigma}^{scatt.}\mathbf{V}d\mathbf{V}
\]
so that the total collisional friction force experienced by species
$\alpha$ is:

\begin{equation}
\mathbf{R}_{\alpha}=\underset{\sigma}{\Sigma}\mathbf{R}_{\alpha\sigma}=\int m_{\alpha}\underset{\sigma}{\Sigma}\,C_{\alpha\sigma}^{scatt.}\mathbf{V}d\mathbf{V}=\int m_{\alpha}C_{\alpha}^{scatt.}\mathbf{V}d\mathbf{V}\label{eq:467-1}
\end{equation}
With these definitions, collisional momentum conservation (the total
plasma does not exert a friction on itself) can be re-expressed as
\cite{bellan fundamentals,HazeltineWaelbroeck}: 
\begin{equation}
\underset{\alpha}{\Sigma}\mathbf{R}_{\alpha}=0\label{eq:467.2}
\end{equation}
Collisional energy conservation requires that $Q_{L\alpha\sigma}+Q_{L\sigma\alpha}=0$,
where\\
$Q_{L\alpha\sigma}=\int C_{\alpha\sigma}^{scatt.}(\frac{1}{2}m_{\alpha}V^{2})\,d\mathbf{V}$
is the rate at which species $\sigma$ transfers energy to species
$\alpha$ via collisions. The $L$ subscript is to indicate that the
kinetic energy of both species is measured in the same ($e.g.,$ Laboratory)
frame. $Q_{\alpha\sigma}=\int C_{\alpha\sigma}^{scatt.}(\frac{1}{2}m_{\alpha}c_{\alpha}^{2})\,d\mathbf{V}$
is the rate at which species $\sigma$ collisionally transfers energy
to species $\alpha$ in the rest frame of species $\alpha$ (the frame
moving at $\mathbf{v}_{\alpha}$, the fluid velocity of species $\alpha$)
\cite{HazeltineWaelbroeck,farside braginskii}. The total rate of
collisional energy transfer to species $\alpha$ in the rest frame
of species $\alpha$ is 
\begin{equation}
Q_{\alpha}=\underset{\sigma}{\Sigma}Q_{\alpha\sigma}\label{eq:467-2}
\end{equation}
With the relation $\mathbf{c}_{\alpha}(\mathbf{r},t)=\mathbf{V-\mathbf{v}_{\alpha}}(\mathbf{r},t)$,
it can be seen, following the procedure used for the derivation of
equations \ref{eq:356-1} and \ref{eq:465}, that\\
$Q_{L\alpha\sigma}=\int C_{\alpha\sigma}^{scatt.}(\frac{1}{2}m_{\alpha}(\mathbf{c}_{\alpha}+\mathbf{v}_{\alpha})^{2})\,d\mathbf{V}=Q_{\alpha\sigma}+\mathbf{v}_{\alpha}\cdot\mathbf{R}_{\alpha\sigma}$,
so that: 
\begin{align}
\underset{\sigma}{\Sigma}Q_{L\alpha\sigma}=\int C_{\alpha}^{scatt.}(\frac{1}{2}m_{\alpha}V^{2})\,d\mathbf{V} & =Q_{\alpha}+\mathbf{v}_{\alpha}\cdot\mathbf{R}_{\alpha} & \mbox{ (use eqns. \ref{eq:465.1}, \ref{eq:467-2} and \ref{eq:467-1})}\label{eq:468-1}
\end{align}
With this, collisional energy conservation can be re-expressed as
\cite{HazeltineWaelbroeck}: 
\begin{align}
\underset{\sigma}{\Sigma}\left(Q_{L\alpha\sigma}+Q_{L\sigma\alpha}\right) & =0\nonumber \\
\Rightarrow\underset{\alpha}{\Sigma}\left(Q_{\alpha}+\mathbf{v}_{\alpha}\cdot\mathbf{R}_{\alpha}\right) & =0\label{eq:468.2}
\end{align}

\section{Braginskii equations from moments of the Boltzmann equation\label{subsec:Braginski-equations-from}}
\begin{flushleft}
Noting that $\mathbf{V}$ is independent of $\mathbf{r}$ (so that
$\nabla\cdot(\mathbf{V}\,f_{\alpha})=\mathbf{V\cdot}\nabla f_{\alpha})$,
and that $\mathbf{V}\perp\mathbf{V\times B}$ (so that $\nabla_{v}\cdot(\mathbf{V}\mathbf{\times B}(\mathbf{r},t)\,f_{\alpha})=\left(\mathbf{V}\mathbf{\times B}(\mathbf{r},t)\right)\mathbf{\cdot}\nabla_{v}\,f_{\alpha})$,
it is convenient to rewrite the Boltzmann equation (equation \ref{eq:351})
as: 
\begin{equation}
\frac{\partial f_{\alpha}}{\partial t}+\nabla\cdot(\mathbf{V}\,f_{\alpha})+\nabla_{v}\cdot\left(\left(\frac{q_{\alpha}}{m_{\alpha}}(\mathbf{E}(\mathbf{r},t)\mathbf{+V}\mathbf{\times B}(\mathbf{r},t))\right)\,f_{\alpha}\right)=C_{\alpha}^{scatt.}(f_{\alpha})\label{eq:360}
\end{equation}
where only scattering collisions will be considered, $i.e.,$ $C_{\alpha}\rightarrow C_{\alpha}^{scatt.}.$
Taking the $0^{th},\,1^{st}$ and $2^{nd}$ moments of the Boltzmann
equation we will arrive at the species conservation equations for
mass, momentum and energy. The following identities will be used:
\begin{align}
\int\nabla_{v}\cdot\left(\frac{q_{\alpha}}{m_{\alpha}}(\mathbf{E}(\mathbf{r},t)\mathbf{+V}\mathbf{\times B}(\mathbf{r},t))\,f_{\alpha}\right)d\mathbf{V} & =\int\nabla_{v}\cdot\left(\mathbf{a}_{\alpha}\,f_{\alpha}\right)d\mathbf{V}=\int\left(\mathbf{a}_{\alpha}\,f_{\alpha}\right)\cdot d\mathbf{S}_{v}=0\nonumber \\
\nonumber \\
\label{eq:467}\\
\nonumber \\
 & \mbox{(Using Gauss' law, the integral is over a surface }\nonumber \\
 & \mbox{in velocity space at \ensuremath{V}\ensuremath{\rightarrow\infty}}.\nonumber \\
 & \mbox{Since \mbox{\ensuremath{f_{\alpha}(\mathbf{r},\mathbf{V},t)\rightarrow0}}\mbox{ as }\ensuremath{V}\ensuremath{\rightarrow\infty}, the integral \ensuremath{\rightarrow}0})\nonumber 
\end{align}
 
\begin{equation}
\nabla_{v}\mathbf{V=}\left(\frac{\partial\mathbf{V}}{\partial\mathbf{V}}\right)_{ij}=\delta_{ij}\label{eq:468}
\end{equation}
\\
\par\end{flushleft}

\subsection{$\mathbf{0}^{\mathbf{th}}$ moment (mass continuity equation)}
\begin{flushleft}
\begin{align*}
1^{st}\mbox{ term}: & \int\frac{\partial f_{\alpha}}{\partial t}d\mathbf{V}=\frac{\partial}{\partial t}\left(\int f_{\alpha}d\mathbf{V}\right)=\frac{\partial n_{\alpha}}{\partial t} & \mbox{(use eqn. \ref{eq:352-1})}\\
2^{nd}\mbox{ term}: & \int\nabla\cdot(\mathbf{V}\,f_{\alpha})d\mathbf{V}=\nabla\cdot(\int\mathbf{V}\,f_{\alpha}d\mathbf{V})=\nabla\cdot(n_{\alpha}\mathbf{v}_{\alpha}) & \mbox{(use eqn. \ref{eq:352-2})}\\
3^{rd}\mbox{ term}: & \int\nabla_{v}\cdot\left(\mathbf{a}_{\alpha}\,f_{\alpha}\right)d\mathbf{V}=0 & \mbox{(use eqn. \ref{eq:467})}\\
4^{th}\mbox{ term}: & \int C_{\alpha}^{scatt.}d\mathbf{V}=0 & \mbox{(use eqn. \ref{eq:466.1})}
\end{align*}
\begin{align}
\Rightarrow\frac{\partial n_{\alpha}(\mathbf{r},t)}{\partial t}+\nabla\cdot(n_{\alpha}(\mathbf{r},t)\,\mathbf{v}_{\alpha}(\mathbf{r},t))=0\nonumber \\
\Rightarrow\frac{\partial\rho_{\alpha}(\mathbf{r},t)}{\partial t}+\nabla\cdot(\rho_{\alpha}(\mathbf{r},t)\,\mathbf{v}_{\alpha}(\mathbf{r},t))=0 & \,\,\,\,\,\,\,\mbox{(use \ensuremath{\rho_{\alpha}=m_{\alpha}n_{\alpha})}}\label{eq:469}
\end{align}
\par\end{flushleft}

\subsection{$\mathbf{1}^{\mathbf{st}}$ moment (momentum equation)}
\begin{flushleft}
$1^{st}$ term: $\int\mathbf{V}\frac{\partial f_{\alpha}}{\partial t}d\mathbf{V}=\frac{\partial}{\partial t}\left(\int\mathbf{V}f_{\alpha}d\mathbf{V}\right)=\frac{\partial(n_{\alpha}\mathbf{v}_{\alpha})}{\partial t}$
(using equation \ref{eq:352-2})\\
$ $\\
$ $\\
$2^{nd}$ term:\\
$ $\\
$\int\nabla\cdot\left(\mathbf{V}f_{\alpha}\right)\mathbf{V}d\mathbf{V}=\int\nabla\cdot\left(\mathbf{VV}\,f_{\alpha}\right)d\mathbf{V}-\int\mathbf{V}\,f_{\alpha}\left(\cancelto{0}{\nabla\cdot\mathbf{V}}\right)d\mathbf{V}=\frac{1}{m_{\alpha}}\nabla\cdot\underline{\mathbf{P}}_{\alpha}$
\\
(note $\mathbf{V}$ is independent of $\mathbf{r},$ and use equation
\ref{eq:353})\\
$ $\\
$3^{rd}$ term: 
\begin{align*}
\int\nabla_{v}\cdot(\mathbf{a}_{\alpha}\,f_{\alpha})\mathbf{V}\,d\mathbf{V} & =\cancelto{0}{\int\nabla_{v}\cdot(\mathbf{a}_{\alpha}\,f_{\alpha}\mathbf{V})\,d\mathbf{V}}-\int\mathbf{a}_{\alpha}\,f_{\alpha}\cdot\left(\cancelto{\underline{\mathbf{I}}}{\mathbf{\nabla}_{v}\mathbf{V}}\right)\,d\mathbf{V}) & \mbox{ (use eqns. \ref{eq:467}, \ref{eq:468})}\\
 & =-\frac{q_{\alpha}}{m_{\alpha}}\left(\mathbf{E}\cancelto{n_{\alpha}}{\int f_{\alpha}d\mathbf{V}}+\left(\cancelto{n_{\alpha}\mathbf{v}_{\alpha}}{\int\mathbf{V}f_{\alpha}d\mathbf{V}}\right)\times\mathbf{B}\right) & \mbox{ (use eqns. \ref{eq:352-1}, \ref{eq:352-2})}\\
 & =-\frac{q_{\alpha}n_{\alpha}}{m_{\alpha}}\left(\mathbf{E}+\mathbf{v}_{\alpha}\times\mathbf{B}\right) & \mbox{}
\end{align*}
$ $\\
$4^{th}$ term: $\int C_{\alpha}^{scatt.}\mathbf{V}\,d\mathbf{V}=\frac{1}{m_{\alpha}}\mathbf{R}_{\alpha}$
(using equation \ref{eq:467-1}). \\
$ $\\
Collecting all four terms, the momentum equation is:
\[
\frac{\partial(\rho_{\alpha}\mathbf{v}_{\alpha})}{\partial t}+\nabla\cdot\underline{\mathbf{P}}_{\alpha}-q_{\alpha}n_{\alpha}\left(\mathbf{E}+\mathbf{v}_{\alpha}\times\mathbf{B}\right)=\mathbf{R}_{\alpha}
\]
\par\end{flushleft}

\begin{align}
\Rightarrow\frac{\partial(\rho_{\alpha}\mathbf{v}_{\alpha})}{\partial t} & =-\nabla\cdot\underline{\mathbf{p}}_{\alpha}-\nabla\cdot(\rho_{\alpha}\mathbf{v}_{\alpha}\mathbf{v}_{\alpha})+q_{\alpha}n_{\alpha}\left(\mathbf{E}+\mathbf{v}_{\alpha}\times\mathbf{B}\right)+\mathbf{R}_{\alpha} & \mbox{(use eqn. \ref{eq:356-1})}\nonumber \\
\label{eq:469.1}
\end{align}
Taking the $2^{nd}$ term on the right over to the left, the momentum
equation can be re-expressed using the following tensor relations:
\begin{equation}
\nabla\cdot\left(d\underline{\mathbf{T}}\right)=\nabla d\cdot\underline{\mathbf{T}}+d\nabla\cdot\underline{\mathbf{T}}\label{eq:470}
\end{equation}
\begin{equation}
\nabla\cdot\left(\mathbf{ab}\right)=(\nabla\cdot\mathbf{a})\mathbf{b}+\mathbf{a}\cdot\nabla\mathbf{b}\label{eq:471}
\end{equation}

\begin{align*}
\frac{\partial(\rho_{\alpha}\mathbf{v}_{\alpha})}{\partial t}+\nabla\cdot(\rho_{\alpha}\mathbf{v}_{\alpha}\mathbf{v}_{\alpha}) & =\rho_{\alpha}\frac{\partial\mathbf{v}_{\alpha}}{\partial t}-\mathbf{v}_{\alpha}\nabla\cdot(\rho_{\alpha}\mathbf{v}_{\alpha}) & \mbox{}\\
 & +\nabla\rho_{\alpha}\cdot(\mathbf{v}_{\alpha}\mathbf{v}_{\alpha})+\rho_{\alpha}\nabla\cdot(\mathbf{v}_{\alpha}\mathbf{v}_{\alpha}) & \mbox{(use eqns. \ref{eq:469}, \ref{eq:470})}\\
 & =\rho_{\alpha}\frac{\partial\mathbf{v}_{\alpha}}{\partial t}-\bcancel{\mathbf{v}_{\alpha}\rho_{\alpha}\nabla\cdot\mathbf{v}_{\alpha}}-\cancel{\left(\mathbf{v}_{\alpha}\mathbf{v}_{\alpha}\right)\cdot\nabla\rho_{\alpha}} & \mbox{\mbox{}}\\
 & +\cancel{\nabla\rho_{\alpha}\cdot(\mathbf{v}_{\alpha}\mathbf{v}_{\alpha})}+\bcancel{\rho_{\alpha}(\nabla\cdot\mathbf{v}_{\alpha})\mathbf{v}_{\alpha}}+\rho_{\alpha}(\mathbf{v}_{\alpha}\cdot\nabla)\mathbf{v}_{\alpha} & \mbox{(use eqn. \ref{eq:471})}
\end{align*}
This is used to re-arrange equation \ref{eq:469.1} to obtain the
final form of the momentum equation:
\begin{align}
\rho_{\alpha}\frac{\partial\mathbf{v}_{\alpha}}{\partial t}+\rho_{\alpha}(\mathbf{v}_{\alpha}\cdot\nabla)\mathbf{v}_{\alpha}=-\nabla\cdot\underline{\mathbf{p}}_{\alpha}+q_{\alpha}n_{\alpha}\left(\mathbf{E}+\mathbf{v}_{\alpha}\times\mathbf{B}\right)+\mathbf{R}_{\alpha} & \mbox{\mbox{}}\label{eq:472}
\end{align}

\begin{align}
\Rightarrow\frac{\partial\mathbf{v}_{\alpha}}{\partial t}=-(\mathbf{v}_{\alpha}\cdot\nabla)\mathbf{v}_{\alpha}+\frac{1}{\rho_{\alpha}}\left(-\nabla p_{\alpha}-\nabla\cdot\overline{\boldsymbol{\pi}}_{\alpha}+q_{\alpha}n_{\alpha}\left(\mathbf{E}+\mathbf{v}_{\alpha}\times\mathbf{B}\right)+\mathbf{R}_{\alpha}\right) & \mbox{\,\,\,\,\mbox{ (use eqn. \ref{eq:357})}}\label{eq:472-2}
\end{align}

\subsection{$\mathbf{2}^{\mathbf{nd}}$ moment (energy equation)}

$1^{st}$ term: 
\begin{align*}
\int\left(\frac{1}{2}m_{\alpha}V^{2}\right)\frac{\partial f_{\alpha}}{\partial t}d\mathbf{V} & =\frac{1}{2}m_{\alpha}\frac{\partial}{\partial t}\left(\int v_{\alpha}^{2}f_{\alpha}d\mathbf{V}+2\int\mathbf{v_{\alpha}}\cdot\mathbf{c_{\alpha}}f_{\alpha}d\mathbf{V}+\int c_{\alpha}^{2}f_{\alpha}d\mathbf{V}\right) & \mbox{(use eqn. \ref{eq:354})}\\
 & =\frac{1}{2}m_{\alpha}\frac{\partial}{\partial t}\left(v_{\alpha}^{2}\cancelto{n_{\alpha}}{\int f_{\alpha}d\mathbf{V}}+2\mathbf{v_{\alpha}}\cdot\cancelto{0}{\int\mathbf{c_{\alpha}}f_{\alpha}d\mathbf{V}}+\cancelto{\frac{3p_{\alpha}}{m_{\alpha}}}{\int c_{\alpha}^{2}f_{\alpha}d\mathbf{V}}\right)\\
 & \,\,\,\,\,\,\,\,\,\,\,\,\,\,\,\,\,\,\,\,\,\,\,\,\,\,\,\,\,\,\mbox{(use eqns. \ref{eq:355-2}, \ref{eq:352-1}, \ref{eq:355-1} and \ref{eq:355-3})}\\
 & =\frac{\partial}{\partial t}\left(\frac{1}{2}\rho_{\alpha}v_{\alpha}^{2}+\frac{3}{2}p_{\alpha}\right)
\end{align*}
 $2^{nd}$ term: 
\begin{align*}
\int\nabla\cdot\left(\mathbf{V}f_{\alpha}\right)\left(\frac{1}{2}m_{\alpha}V^{2}\right)d\mathbf{V} & =\int\nabla\cdot\left(\frac{1}{2}m_{\alpha}V^{2}\mathbf{V}f_{\alpha}\right)d\mathbf{V}\\
 & \,\,\,-\int f_{\alpha}\mathbf{V}\cdot\cancelto{0}{\nabla\left(\frac{1}{2}m_{\alpha}V^{2}\right)}d\mathbf{V} & \mbox{\mbox{(\ensuremath{\mathbf{V}} is independent of \ensuremath{\mathbf{r}})}}\\
 & =\nabla\cdot\int\left(\frac{1}{2}m_{\alpha}V^{2}\mathbf{V}f_{\alpha}\right)d\mathbf{V} & \mbox{}\mbox{(\ensuremath{\mathbf{V}} is independent of \ensuremath{\mathbf{r}})}\\
 & =\nabla\cdot\mathbf{Q'}_{\alpha} & \mbox{(use eqn. \ref{eq:358})}
\end{align*}
$3^{rd}$ term: 
\begin{align}
\int\nabla_{v}\cdot(\mathbf{a}_{\alpha}\,f_{\alpha})\left(\frac{1}{2}m_{\alpha}V^{2}\right)\,d\mathbf{V} & =\cancelto{0}{\int\nabla_{v}\cdot\left(\mathbf{a}_{\alpha}\,f_{\alpha}\left(\frac{1}{2}m_{\alpha}V^{2}\right)\right)\,d\mathbf{V}} & \mbox{(use eqn. \ref{eq:467})}\nonumber \\
 & -\int f_{\alpha}\mathbf{a}_{\alpha}\cdot\mathbf{\nabla}_{v}\left(\frac{1}{2}m_{\alpha}V^{2}\right)\,d\mathbf{V}\nonumber \\
 & =-\int f_{\alpha}\mathbf{a}_{\alpha}\cdot\frac{\partial}{\partial\mathbf{V}}\left(\frac{1}{2}m_{\alpha}\mathbf{V}\cdot\mathbf{V}\right)\,d\mathbf{V} & \mbox{}\nonumber \\
 & =-\int f_{\alpha}\mathbf{a}_{\alpha}\cdot\left(m_{\alpha}\underline{\mathbf{I}}\cdot\mathbf{V}\right)\,d\mathbf{V} & \mbox{}\nonumber \\
 & =-q_{\alpha}\int f_{\alpha}\left(\mathbf{E}+\mathbf{V}\mathbf{\times B}\right)\cdot\mathbf{V}\,d\mathbf{V}\nonumber \\
 & =-q_{\alpha}\mathbf{E}(\mathbf{r},t)\cdot\cancelto{n_{\alpha}\mathbf{v}_{\alpha}}{\int f_{\alpha}\mathbf{V}\,d\mathbf{V}} & \mbox{ (use eqn. \ref{eq:352-1})}\nonumber \\
 & =-q_{\alpha}n_{\alpha}\mathbf{E}\cdot\mathbf{v}_{\alpha}\label{eq:472.3}
\end{align}
\\
$4^{th}$ term: $\int C_{\alpha}^{scatt.}\left(\frac{1}{2}m_{\alpha}V^{2}\right)\,d\mathbf{V}=Q_{\alpha}+\mathbf{v}_{\alpha}\cdot\mathbf{R}_{\alpha}$
(use equation \ref{eq:468-1})
\begin{align*}
\Rightarrow\frac{\partial}{\partial t}\left(\frac{1}{2}\rho_{\alpha}v_{\alpha}^{2}+\frac{3}{2}p_{\alpha}\right)+\nabla\cdot\mathbf{Q'}_{\alpha}-q_{\alpha}n_{\alpha}\mathbf{E}\cdot\mathbf{v}_{\alpha}=Q_{\alpha}+\mathbf{v}_{\alpha}\cdot\mathbf{R}_{\alpha}\\
\Rightarrow\frac{1}{2}v_{\alpha}^{2}\frac{\partial\rho_{\alpha}}{\partial t}+\frac{1}{2}\rho_{\alpha}\frac{\partial(v_{\alpha}^{2})}{\partial t}+\frac{3}{2}\frac{\partial p_{\alpha}}{\partial t}+\frac{1}{2}\nabla\cdot\left(\rho_{\alpha}v_{\alpha}^{2}\mathbf{v}_{\alpha}\right)+\frac{3}{2}\nabla\cdot\left(p_{\alpha}\mathbf{v}_{\alpha}\right) & \mbox{}\\
+\nabla\cdot\left(\mathbf{v}_{\alpha}\cdot\underline{\mathbf{p}}_{\alpha}\right)+\nabla\cdot\mathbf{q}_{\alpha}-q_{\alpha}n_{\alpha}\mathbf{E}\cdot\mathbf{v}_{\alpha}=Q_{\alpha}+\mathbf{v}_{\alpha}\cdot\mathbf{R}_{\alpha} & \mbox{ (use eqn. \ref{eq:465})}
\end{align*}
The first two terms can be rearranged using the mass and momentum
conservation equations (\ref{eq:469} \& \ref{eq:472}):

\begin{align*}
\frac{1}{2}v_{\alpha}^{2}\frac{\partial\rho_{\alpha}}{\partial t}+\frac{1}{2}\rho_{\alpha}\frac{\partial(v_{\alpha}^{2})}{\partial t} & =-\frac{1}{2}v_{\alpha}^{2}\nabla\cdot(\rho_{\alpha}\mathbf{v}_{\alpha})+\rho_{\alpha}v_{\alpha}\left(\frac{\partial v_{\alpha}}{\partial t}\right)\\
 & =-\frac{1}{2}v_{\alpha}^{2}\nabla\cdot(\rho_{\alpha}\mathbf{v}_{\alpha})+\mathbf{v}_{\alpha}\cdot\left(\rho_{\alpha}\frac{\partial\mathbf{v}_{\alpha}}{\partial t}\right)\\
 & =-\frac{1}{2}v_{\alpha}^{2}\nabla\cdot(\rho_{\alpha}\mathbf{v}_{\alpha})+\mathbf{v}_{\alpha}\cdot\biggl(-\rho_{\alpha}(\mathbf{v}_{\alpha}\cdot\nabla)\mathbf{v}_{\alpha}-\nabla\cdot\underline{\mathbf{p}}_{\alpha}\\
 & \,\,\,\,\,\,\,+q_{\alpha}n_{\alpha}\left(\mathbf{E}+\mathbf{v}_{\alpha}\times\mathbf{B}\right)+\mathbf{R}_{\alpha}\biggr)
\end{align*}
Assemble all four terms:

\begin{align*}
\Rightarrow\cancel{-\frac{1}{2}v_{\alpha}^{2}\nabla\cdot(\rho_{\alpha}\mathbf{v}_{\alpha})}-\bcancel{\frac{1}{2}\rho_{\alpha}\mathbf{v}_{\alpha}\cdot\nabla\left(v_{\alpha}^{2}\right)}-\mathbf{v}_{\alpha}\cdot\left(\nabla\cdot\underline{\mathbf{p}}_{\alpha}\right)+\cancel{q_{\alpha}n_{\alpha}\mathbf{v}_{\alpha}\cdot\mathbf{E}}+\bcancel{\mathbf{v}_{\alpha}\cdot\mathbf{R}_{\alpha}}\\
+\frac{3}{2}\frac{\partial p_{\alpha}}{\partial t}+\cancel{\frac{1}{2}v_{\alpha}^{2}\nabla\cdot\left(\rho_{\alpha}\mathbf{v}_{\alpha}\right)}+\bcancel{\frac{1}{2}\rho_{\alpha}\mathbf{v}_{\alpha}\cdot\nabla\left(v_{\alpha}^{2}\right)}+\frac{3}{2}p_{\alpha}\nabla\cdot\mathbf{v}_{\alpha}+\frac{3}{2}\mathbf{v}_{\alpha}\cdot\nabla p_{\alpha}\\
+\nabla\cdot\left(\mathbf{v}_{\alpha}\cdot\underline{\mathbf{p}}_{\alpha}\right)+\nabla\cdot\mathbf{q}_{\alpha}-\cancel{q_{\alpha}n_{\alpha}\mathbf{E}\cdot\mathbf{v}_{\alpha}}=Q_{\alpha}+\bcancel{\mathbf{v}_{\alpha}\cdot\mathbf{R}_{\alpha}}
\end{align*}
Using the tensor relationship 
\begin{equation}
\nabla\cdot\left(\mathbf{a}\cdot\underline{\mathbf{T}}\right)-\mathbf{a}\cdot\left(\nabla\cdot\underline{\mathbf{T}}\right)=\underline{\mathbf{T}}:\nabla\mathbf{a}\label{eq:472.301}
\end{equation}
where the contraction (inner product) of two second order tensors
\cite{outerproduct} is defined as $\underline{\mathbf{T}}:\underline{\boldsymbol{U}}=\underset{i}{\Sigma}\underset{j}{\Sigma}T_{ij}U_{ij}$
so that $\underline{\mathbf{T}}:\nabla\mathbf{a}=T_{ij}\frac{\partial a_{j}}{\partial r_{i}},$
this can be expressed as:

\begin{align}
\frac{3}{2}\frac{\partial p_{\alpha}}{\partial t}+\frac{3}{2}\mathbf{v}_{\alpha}\cdot\nabla p_{\alpha}+\frac{3}{2}p_{\alpha}\nabla\cdot\mathbf{v}_{\alpha}+\underline{\mathbf{p}}_{\alpha}:\nabla\mathbf{v}_{\alpha}+\nabla\cdot\mathbf{q}_{\alpha}=Q_{\alpha} & \mbox{\mbox{}}\nonumber \\
\Rightarrow\frac{3}{2}\frac{\partial p_{\alpha}}{\partial t}+\frac{3}{2}\mathbf{v}_{\alpha}\cdot\nabla p_{\alpha}+\frac{3}{2}p_{\alpha}\nabla\cdot\mathbf{v}_{\alpha}+p_{\alpha}\left(\cancelto{\nabla\cdot\mathbf{v}_{\alpha}}{\underline{\mathbf{I}}:\nabla\mathbf{v}_{\alpha}}\right)+\overline{\boldsymbol{\pi}}_{\alpha}:\nabla\mathbf{v}_{\alpha}+\nabla\cdot\mathbf{q}_{\alpha}=Q_{\alpha} & \mbox{ (use eqn. \ref{eq:357})}\nonumber \\
\Rightarrow\frac{\partial p_{\alpha}}{\partial t}=-\mathbf{v}_{\alpha}\cdot\nabla p_{\alpha}-\gamma p_{\alpha}\nabla\cdot\mathbf{v}_{\alpha}+(\gamma-1)\left(-\overline{\boldsymbol{\pi}}_{\alpha}:\nabla\mathbf{v}_{\alpha}-\nabla\cdot\mathbf{q}_{\alpha}+Q_{\alpha}\right)\label{eq:472.31}
\end{align}
Here, $\gamma=\frac{c_{p}}{c_{v}}$ is the adiabatic gas constant,
where $c_{p}\,${[}J$\,$kg$^{-1}$K$^{-1}${]} is the specific heat
capacity at constant pressure, and $c_{v}\,${[}J$\,$kg$^{-1}$K$^{-1}${]}
is the specific heat capacity at constant volume. $\gamma=\frac{N+2}{N}$,
where $N$ is the number of degrees of freedom of the bulk species.
For a monatomic gas, $N=3,$ so that $\gamma=\frac{5}{3}.$ For a
given material, the availability of translational, rotational and
intra-molecular exchanges characterise the value of $N$. 

\section{Chapman-Enskog closures\label{subsec:Chapman-Enskog-closures}}

Lack of closure is an intrinsic property of fluid theory because the
evolution equation for each moment involves the next higher order
moment, so the system cannot be closed without losing some information.
For example, taking the $0^{th}$ moment of the Boltzmann equation
yields the mass conservation equation which involves $\mathbf{v}_{\alpha}$,
the $1^{st}$ moment of the distribution function, and taking the
$1^{st}$ moment of the Boltzmann equation yields the momentum conservation
equation which requires $\overline{\boldsymbol{\pi}}_{\alpha}$, the
$2^{nd}$ moment of the distribution function. Truncation closure
schemes set higher order moments to zero or describe them in terms
of lower order moments, and can be useful to provide insight, but
are associated with uncontrolled approximation \cite{farside braginskii}.
Asymptotic schemes work with expansions in terms of some small ordering
parameter and have the advantage of providing an estimate of the error
associated with the closure, but involve direct interaction with the
kinetic equations and so are mathematically demanding \cite{farside braginskii}. 

\subsection{Closures for a neutral gas\label{subsec:Closures_neutral}}

The Chapman-Enskog asymptotic closure scheme for a neutral gas dominated
by collisions uses the ratio of the mean free path to the characteristic
system length scale as the small ordering parameter \cite{farside braginskii}.
The fluid equations for a neutral gas are similar to those derived
above for the two-fluid plasma, but without electromagnetic forces
as the fluid isn't conducting, and also without friction forces since
there is only one species:
\begin{align*}
\dot{\rho}_{n} & =-\nabla\cdot(\rho_{n}\mathbf{v}_{n})\\
\dot{\mathbf{v}}_{n} & =-\mathbf{v}_{n}\cdot\nabla\mathbf{v}_{n}+\frac{1}{\rho_{n}}\left(-\nabla p_{n}-\nabla\cdot\overline{\boldsymbol{\pi}}_{n}\right)\\
\dot{p}_{n} & =-\mathbf{v}_{n}\cdot\nabla p_{n}-\gamma p_{n}\,\nabla\cdot\mathbf{v}_{n}+(\gamma-1)\left(-\overline{\boldsymbol{\pi}}_{n}:\nabla\mathbf{v}_{n}-\nabla\cdot\mathbf{q}_{n}\right)
\end{align*}
Here, the subscript $n$ implies parameters pertaining to the neutral
fluid. The mean free path for a neutral gas is defined by considering
that there is, on average, approximately one particle in the volume
swept out by another particle over one mean free path: 
\begin{equation}
\lambda_{mfp}=\frac{1}{\sqrt{2}n_{n}\pi d^{2}}\label{eq:472.310}
\end{equation}
where $d$ is the particle diameter. Denoting $L$ as the characteristic
system length scale, then $\epsilon=\frac{\lambda_{mfp}}{L}\ll1$,
and the distribution function is expanded in orders of $\epsilon$:
\[
f(\mathbf{r},\mathbf{V},t)=f_{0}(\mathbf{r},\mathbf{V},t)+\epsilon f_{1}(\mathbf{r},\mathbf{V},t)+\epsilon^{2}f_{2}(\mathbf{r},\mathbf{V},t)+....
\]
To zeroth order in $\epsilon$, the Boltzmann equation requires that
$f_{0}$ is a Maxwellian distribution \cite{farside braginskii}:\\
 $f_{0}(\mathbf{r},\mathbf{V},t)=n_{n}(\mathbf{r},t)\left(\frac{m_{n}}{2\pi T_{n}(\mathbf{r},t)}\right)^{\frac{3}{2}}\mbox{exp}\left(-\frac{m_{n}(\mathbf{V}-\mathbf{v}_{n}(\mathbf{r},t))^{2}}{2T_{n}(\mathbf{r},t)}\right)$,
where $T_{n}\,[\mbox{J}]$ is the neutral fluid temperature. In a
gas at rest in uniform conditions, the velocity distribution tends
towards the Maxwellian distribution. Actually, this would occur very
rapidly, in a few collision times \cite{Degrez}. In a flowing gas
in a non-uniform state, molecules moving into a neighboring region
cause deviations from the Maxwellian distribution. However, if macroscopic
gradients are small, it is to be expected that the deviations from
the Maxwellian distribution will also be small and that collisions
driving the gas towards equilibrium will remain the dominant effect.
One interpretation of the principal steps in the Chapman-Enskog method
is that the small ordering parameter is introduced into the kinetic
equation to give more weight to the collision term, emphasising the
strength of the collision-driven drive towards equilibrium \cite{Degrez}. 

There is no heat flow or viscous stress associated with a Maxwellian
distribution function, so $\mathbf{q}_{n}$ and $\overline{\boldsymbol{\pi}}_{n}$
depend on the higher order non-Maxwellian corrections to the distribution
function \cite{farside braginskii}. The Boltzmann equation can be
linearised ($i.e.,$ neglect term of order $\epsilon^{2}$ and higher)
and rearranged into an integral equation for $f_{1}$ in terms of
$f_{0}$. The equation is integral because the collision operator
is an integral one. The equation is solved by expanding $f_{1}$ in
velocity space using Laguerre (Sonine) polynomials, so that it reduces
to an infinite set of simultaneous algebraic equations (which can
be truncated) for the coefficients in the expansion. The Laguerre
polynomial expansion converges rapidly and it is convenient to keep
only the first two terms in the expansion without much loss of accuracy
in the solutions for the coefficients that define $f_{1}$ \cite{farside braginskii}.
Moments of $f_{1}$ are taken to obtain expressions for $\mathbf{q}_{n}$
and $\overline{\boldsymbol{\pi}}_{n}$:
\begin{align}
\mathbf{q}_{n} & =-\kappa_{n}\nabla T_{n}\label{eq:472.33}\\
\left(\overline{\boldsymbol{\pi}}_{n}\right)_{ij} & =-\mu_{n}\left(\frac{\partial v_{ni}}{\partial r_{j}}+\frac{\partial v_{nj}}{\partial r_{i}}-\delta_{ij}\frac{2}{3}\nabla\cdot\mathbf{v}_{n}\right)\label{eq:472.34}
\end{align}
Here, $\kappa_{n}$ {[}(m-s)$^{-1}${]} is the thermal conductivity
for the neutral fluid, $\mu_{n}\:[\mbox{kg\,m\ensuremath{^{-1}}s\ensuremath{^{-1}}}]$
is the dynamic viscosity, $\delta_{ij}$ is the Kronecker delta, and
$v_{ni}$ refers to the $i^{th}$ component of $\mathbf{v}_{n}$.
$\kappa_{n}$ and $\mu_{n}$ can be expressed in terms of diffusion
coefficients as $\kappa_{n}=n_{n}\chi_{n}\,[\mbox{m}^{2}/\mbox{s}]$
and $\mu_{n}=\rho_{n}\nu_{n}\,[\mbox{m}^{2}/\mbox{s}]$, where $\chi_{n}$
is the thermal diffusivity and $\nu_{n}$ is the kinematic viscosity.
For hard sphere molecules, these diffusion coefficients, (from Chapman
and Cowling \cite{Chapman}) are: 
\begin{equation}
\nu_{n}=\frac{5\sqrt{\pi}}{16}\left(1+\frac{1}{44}+...\right)\nu_{cn}\lambda_{mfp}^{2}\approx\frac{5\sqrt{\pi}\,V_{thn}\lambda_{mfp}}{16}\label{eq:472.35}
\end{equation}

\begin{equation}
\chi_{n}=\frac{75\sqrt{\pi}}{64}\left(1+\frac{3}{202}+...\right)\nu_{cn}\lambda_{mfp}^{2}\approx\frac{75\sqrt{\pi}\,V_{thn}\lambda_{mfp}}{64}\label{eq:472.36}
\end{equation}
Here, the first two terms in the Laguerre polynomial expansion are
shown explicitly in the brackets. $\nu_{cn}=\frac{V_{thn}}{\lambda_{mfp}}$
is the neutral-neutral scattering collision frequency where the neutral
particle thermal speed is $V_{thn}=\sqrt{\frac{2T_{n}}{m_{n}}}$.
Viscous and thermal diffusivities for a neutral gas may be interpreted
in terms of the random-walk diffusion of molecules with above average
momentum and energy, respectively \cite{farside braginskii}. From
classical stochastic theory, if particles move a step length $l$
in a random direction (for $\nu_{n}$ and $\chi_{n}$, $l=\lambda_{mfp}$,
the mean free path for neutral-neutral scattering collisions), with
a frequency $\nu$, the motion is associated with a diffusivity of
$\nu l^{2}$. Chapman-Enskog theory enables, for a given force law
between molecules, the determination of the constants that precede
this classical diffusivity. The constants, as well as the expression
for the mean free path, will vary depending on the force law in question
\cite{farside braginskii}.

\subsection{Closures for an unmagnetized collisional plasma\label{subsec:Closures_unmagnetised}}

For a plasma, asymptotic closure schemes can be based on one of two
possible small parameters. For collisional plasmas, the small parameter
has the same form as that for a neutral gas: $\epsilon=\frac{\lambda_{mfp}}{L}\ll1$,
where $\lambda_{mfp}$ and $L$ are the mean free path and characteristic
length scale for the plasma. The ion-ion collision time is 
\begin{equation}
\tau_{ii}=\frac{12\pi^{1.5}\epsilon_{0}^{2}\sqrt{m_{i}}\,T_{i}^{1.5}}{\varLambda\,e^{4}n}\label{eq:472.37}
\end{equation}
and 
\begin{equation}
\tau_{ei}=\frac{6\sqrt{2}\pi^{1.5}\epsilon_{0}^{2}\sqrt{m_{e}}\,T_{e}^{1.5}}{\varLambda\,e^{4}n\,Z_{eff}^{2}}\label{eq:472.38}
\end{equation}
is the electron-ion collision time. Collision times are the average
times between $90^{o}$ scattering events. Here, $e$ is the electron
charge, $\varLambda$ is the Coulomb logarithm, $\epsilon_{0}$ is
vacuum permittivity, and $T_{e}$, $T_{i}$ are the electron and ion
temperatures in Joules. Note that the collision times increase with
temperature - as plasma temperature increases, the cross-section for
Coulomb collisions decreases because the distance of closest approach
become smaller. 

Unmagnetized plasmas are defined by $\omega_{ci}\tau_{ii},\,\omega_{ce}\tau_{ei}\ll1$,
where $\omega_{c\alpha}=\frac{q_{\alpha}B}{m_{\alpha}}$ is the species
cyclotron frequency. The definition implies that, in an unmagnetized
plasma, many collisions occur during one gyro-rotation. For collisional,
unmagnetized plasma with $Z_{eff}=1$, the two-Laguerre-polynomial
Chapman-Enskog closure scheme gives 

\begin{align}
\mathbf{R}_{e}=-\mathbf{R}_{i} & =\eta'ne\mathbf{J}-0.71n\nabla T_{e}\label{eq:472.39}\\
Q_{i} & =\frac{3m_{e}}{m_{i}}\frac{n(T_{e}-T_{i})}{\tau_{ei}}\label{eq:472.40}\\
Q_{e} & =-Q_{i}+\frac{\mathbf{J\cdot R}_{e}}{ne}=-Q_{i}+\eta'J^{2}-0.71\frac{\mathbf{J}\cdot\nabla T_{e}}{e}\label{eq:472.41}
\end{align}
Here, $\eta'\,[\Omega-\mbox{m}]=\frac{m_{e}\nu_{ei}}{1.96ne^{2}}$
is the plasma resistivity, where $\nu_{ei}=\frac{1}{\tau_{ei}}$ is
the electron-ion collision frequency. Assuming quasineutrality and
singly charged ions, ($i.e.$, charge density, $\rho_{c}=\underset{\alpha}{\Sigma}n_{\alpha}q_{\alpha}=0$,
$q_{i}=e\mbox{, and }q_{e}=-e$, so that $n_{i}=n_{e}=n$), the current
density is given by $\mathbf{J}=\underset{\alpha}{\Sigma}n_{\alpha}q_{\alpha}\mathbf{v}_{\alpha}=ne(\mathbf{v}_{i}-\mathbf{v}_{e})$,
so that the first term in the expression for $\mathbf{R}_{e}$ is
$\nu_{ei}\rho_{e}(\mathbf{v}_{i}-\mathbf{v}_{e})$. In a collision
time $\tau_{ei}$, the electrons lose momentum $-m_{e}(\mathbf{v}_{i}-\mathbf{v}_{e})$
as a result of losing their ordered velocity with respect to that
of the ions, so the frictional force on the electrons, which is the
rate of change of their momentum due to interaction with the ions,
is $\nu_{ei}\rho_{e}(\mathbf{v}_{i}-\mathbf{v}_{e})$. Since $\eta'\propto\nu_{ei}$
which is inversely proportional to $T_{e}^{1.5}$, the faster (hotter)
electrons in the distribution are the most conductive - current in
plasma is carried primarily by fast electrons. This phenomenon gives
rise to the thermal force which is the second term in the expression
for $\mathbf{R}_{e}$ \cite{farside braginskii}. The origin of the
thermal force can be clarified as follows. Assuming, for simplicity,
that the ion and electron fluids are at rest ($\mathbf{v}_{e}=\mathbf{v}_{i}=\mathbf{0})$,
then, at any point $x_{0}$ ($e.g.,$ on the $x$ axis), the flux
of electrons directed to the left is balanced by the flux of electrons
directed to the right. Due to electron-ion collisions, these fluxes
experience frictional forces of order $m_{e}nV_{the}/\tau_{ei}$,
where $V_{the}$ is the electron thermal speed. Since $V_{the}/\tau_{ei}\sim T_{e}^{-1}$,
if electrons coming from the right are hotter on average than those
coming from the left, there will be a net force directed to the left,
against the electron temperature gradient.

$Q_{i}$ gives the rate at which energy is imparted from the electrons
to the ions due to collisions between the ion and electron fluids.
The first term in the expression for $Q_{e}$ is simply $-Q_{i}$.
The second term on the extreme right is the ohmic heating term, representing
the change from ordered electronl motion relative to the ions into
random motion (heat) due to collisions with the ions. The third term
is the work done by the electrons against the thermal force \cite{farside braginskii}.

The closure scheme gives the ion and electron heat flux densities
as

\begin{align}
\mathbf{q}_{i} & =-n\chi_{\parallel i}\nabla T_{i}\nonumber \\
\mathbf{q}_{e} & =-n\chi_{\parallel e}\nabla T_{e}-0.71\frac{T_{e}}{e}\mathbf{J}\label{eq:472.43}
\end{align}
where the ion and electron thermal diffusion coefficients are

\begin{align}
\chi_{\parallel i} & =3.9\,\tau_{ii}\frac{T_{i}}{m_{i}}\nonumber \\
\chi_{\parallel e} & =3.2\,\tau_{ei}\frac{T_{e}}{m_{e}}\label{eq:472.45}
\end{align}
$\mathbf{q}_{i}$ and the first term in the expression for $\mathbf{q}_{e}$
are analogous to the corresponding expressions for $\mathbf{q}_{n}$
(equation \ref{eq:472.33}) (recall $\kappa=n\chi$), while $\chi_{\parallel i}$
and $\chi_{\parallel e}$ can be compared with the neutral conductivity
(equation \ref{eq:472.36}). The parallel subscripts on $\chi_{\alpha}$
are included because it turns out that the values of $\chi_{\alpha}$
in the unmagnetized case are the same as the values of $\chi_{\parallel\alpha}$
in the magnetized case (in the magnetized case, the parallel subscripts
indicate that the coefficients pertain to thermal diffusion in the
direction parallel to $\mathbf{B})$. As with the closures for a neutral
fluid, the ion and electron thermal diffusivities $\chi_{\parallel i}$
and $\chi_{\parallel e}$ represent random-walk heat diffusion, and
are given by expressions of the form $\chi=C(\nu l^{2})=C(\frac{1}{\nu}V_{th}^{2})$,
for some constant $C$, and with $l=\frac{V_{th}}{\nu}$. For example,
in equation \ref{eq:472.45}, we have $\chi_{\parallel i}\sim\frac{\tau_{ii}T_{i}}{m_{i}}\sim\frac{V_{thi}^{2}}{\nu_{ii}}$
where $\nu_{ii}=\frac{1}{\tau_{ii}}$ is the ion-ion collision frequency
and $V_{thi}=\sqrt{\frac{2T_{i}}{m_{i}}}$ is the ion thermal speed.
Note that, for $T_{i}\sim T_{e}$, that $\frac{\chi_{\parallel e}}{\chi_{\parallel i}}\sim\frac{m_{i}}{m_{e}}\frac{\tau_{ei}}{\tau_{ii}}\sim\sqrt{\frac{m_{i}}{m_{e}}}\sim43$,
so electron heat diffusivity is much greater than that of the ions
in the unmagnetized case, and in the magnetized case, parallel heat
diffusivity is much greater for electrons than for ions.

The second term in the expression for $\mathbf{q}_{e}$ represents
additional convection of heat due to relative electron-ion motion.
If $\mathbf{v}_{rel}=\mathbf{v}_{e}-\mathbf{v}_{i}$, then in the
frame moving at $\mathbf{v}_{e}$, there are more fast electrons than
slow electrons moving in the same direction as $\mathbf{v}_{rel}$
and more slow electrons than fast electrons moving in the opposite
direction. Fast electrons carry more energy than slow electrons so
additional heat is convected in the direction of $\mathbf{v}_{rel}$.
Since $\mathbf{J}\sim(\mathbf{v}_{i}-\mathbf{v}_{e})$, the additional
heat is convected in the direction opposite that of $\mathbf{J}$.\\
\\
The ion and electron viscosity tensors have the same form as $\overline{\boldsymbol{\pi}}_{n}$
(equation \ref{eq:472.34}): 
\begin{align}
\left(\overline{\boldsymbol{\pi}}_{\alpha}\right)_{ij} & =-\rho_{\alpha}\nu_{\alpha}\left(\frac{\partial v_{\alpha i}}{\partial r_{j}}+\frac{\partial v_{\alpha j}}{\partial r_{i}}-\delta_{ij}\frac{2}{3}\nabla\cdot\mathbf{v}_{\alpha}\right)\label{eq:472.46}
\end{align}
where the coefficients of kinematic viscosity (viscous diffusion coefficients)
are given as 
\begin{align}
\nu_{i} & =0.96\,\tau_{ii}\,\frac{T_{i}}{m_{i}}\nonumber \\
\nu_{e} & =0.73\,\tau_{ei}\,\frac{T_{e}}{m_{e}}\label{eq:472.47}
\end{align}
As with the closure for neutral viscosity, the ion and electron viscous
diffusivities $\nu_{i}$ and $\nu_{e}$ represent random-walk diffusion,
given by expressions of the form $\nu=C(\nu\lambda^{2})=C(\frac{1}{\nu}V_{th}^{2})$.
It can be seen how, for $T_{i}\sim T_{e}$, that $\frac{\mu_{i}}{\mu_{e}}=\frac{\rho_{i}\nu_{i}}{\rho_{e}\nu_{e}}\sim\frac{\tau_{ii}}{\tau_{ei}}\sim\sqrt{\frac{m_{i}}{m_{e}}}\sim43$,
so that the viscosity (diffusion of momentum) of a plasma is mostly
due to the ions \cite{farside braginskii}. 

\subsection{Closures for a magnetized plasma\label{subsec:Closures_magnetised}}

Magnetized plasmas are defined by $\omega_{ci}\tau_{ii},\,\omega_{ce}\tau_{ei}\gg1$,
which implies that many gyro-rotations occur in one collision time,
and that the electron and ion Larmor radii are much smaller than the
mean free path \cite{farside braginskii}. In the magnetized limit,
the small ordering parameter is $\epsilon=\frac{r_{L}}{L}\ll1$, where
$r_{L}=\frac{V_{\perp}}{\omega_{c}}$ is the Larmor radius, where
$V_{\perp}$ is the particle speed perpendicular to $\mathbf{B}$.
The two-Laguerre-polynomial Chapman-Enskog closure scheme gives 
\begin{align}
\mathbf{R}_{e} & =-\mathbf{R}_{i}=ne(\eta'_{\parallel}\mathbf{J}_{\parallel}+\eta'_{\perp}\mathbf{J}_{\perp})-0.71n\nabla_{\parallel}T_{e}-\frac{3n}{2\omega_{ce}\tau_{ei}}\hat{\mathbf{b}}\times\nabla_{\perp}T_{e}\label{eq:472.471}\\
Q_{i} & =\frac{3m_{e}}{m_{i}}\frac{n(T_{e}-T_{i})}{\tau_{ei}}\label{eq:472.472}\\
Q_{e} & =-Q_{i}+\frac{\mathbf{J\cdot R}_{e}}{ne}\label{eq:472.473}
\end{align}
where $\eta'_{\parallel}\,[\Omega-\mbox{m}]=\eta'=\frac{m_{e}\nu_{ei}}{1.96ne^{2}}$
is the plasma resistivity effective parallel to $\mathbf{B}$, $\eta'_{\perp}\,[\Omega-\mbox{m}]=1.96\,\eta'_{\parallel}$
is the resistivity effective perpendicular to $\mathbf{B}$, $\mathbf{J}_{\parallel}$
and $\mathbf{J}_{\perp}$ are the current densities parallel and perpendicular
to $\mathbf{B}$, and $\hat{\mathbf{b}}=\frac{\mathbf{B}}{B}$. The
third and fourth terms in the expression for $\mathbf{R}_{e}$ constitute
the thermal force in the magnetized limit. The MHD model we developed
uses the approximation of isotropic resistivity - this was considered
a reasonable simplification since $\eta'_{\parallel}\sim\eta'_{\perp}$.
A further additional simplification is the neglection of last two
terms in equation \ref{eq:472.473}, representing work by the electron
fluid done against the thermal force. The expressions for the ion
and electron viscosity tensors are complicated in the magnetized limit
and the details wont be looked here because our MHD model presently
also approximates the viscosity as being isotropic. 

In the magnetized limit, the ion and electron heat flux densities
are given by 
\begin{align}
\mathbf{q}_{i} & =-n\chi_{\parallel i}\nabla_{\parallel}T_{i}-n\chi_{\perp i}\nabla_{\perp}T_{i}-\left(n\chi_{\times i}\hat{\mathbf{b}}\times\nabla_{\perp}T_{i}\right)\label{eq:472.48}\\
\mathbf{q}_{e} & =-n\chi_{\parallel e}\nabla_{\parallel}T_{e}-n\chi_{\perp e}\nabla_{\perp}T_{e}-\left(n\chi_{\times e}\hat{\mathbf{b}}\times\nabla_{\perp}T_{e}+0.71\frac{T_{e}}{e}\mathbf{J}_{\parallel}+\frac{3T_{e}}{2\omega_{ce}\tau_{ei}e}\hat{\mathbf{b}}\times\mathbf{J}_{\perp}\right)\label{eq:472.49}
\end{align}
The $1^{st}$ and $2^{nd}$ terms in the expressions for $\mathbf{q}_{i}$
and $\mathbf{q}_{e}$ represent diffusive heat transport parallel
to and perpendicular to the magnetic field. For simplicity, only these
terms are retained in our MHD model. The $3^{rd}$ terms in the two
expressions represent transport that is perpendicular to both the
field and to the perpendicular temperature gradients. The $4^{th}$
and $5^{th}$ terms in the expression for $\mathbf{q}_{e}$ correspond
to additional convection of heat due to relative electron-ion motion
- details can be found in \cite{farside braginskii}. The Chapman-Enskog
scheme gives the ion and electron parallel thermal diffusion coefficients
as in equation \ref{eq:472.45}, and the perpendicular thermal diffusion
coefficients as

\begin{align}
\chi_{\perp i} & =2\,\frac{1}{\tau_{ii}}\frac{T_{i}}{m_{i}\,\omega_{ci}^{2}}\nonumber \\
\chi_{\perp e} & =4.7\,\frac{1}{\tau_{ei}}\frac{T_{e}}{m_{e}\,\omega_{ce}^{2}}\label{eq:472.50}
\end{align}
Once again, the perpendicular ion and electron thermal diffusivities
$\chi_{\perp i}$ and $\chi_{\perp e}$ represent random-walk heat
diffusion, and are given by expressions of the form $\chi=C(\nu l^{2})$,
for some constant $C$. Here, the step length $l$ is the Larmor radius:
$l_{\alpha}=r_{L\alpha}=\frac{V_{\perp\alpha}}{\omega_{c\alpha}}=\frac{V_{th\alpha}}{\omega_{c\alpha}}=\frac{1}{\omega_{c\alpha}}\sqrt{\frac{2T_{\alpha}}{m_{\alpha}}}$.
For example, $\chi_{\perp i}\sim\nu_{ii}r_{Li}^{2}$, where $\nu_{ii}=\frac{1}{\tau_{ii}}$
is the ion-ion collision frequency and $r_{Li}=\frac{1}{\omega_{ci}}\sqrt{\frac{2T_{i}}{m_{i}}}$.
Note that, for $T_{i}\sim T_{e}$, that $\frac{\chi_{\perp i}}{\chi_{\perp e}}\sim\sqrt{\frac{m_{e}}{m_{i}}}\left(\frac{m_{i}}{m_{e}}\right)^{2}\frac{m_{e}}{m_{i}}\sim\sqrt{\frac{m_{i}}{m_{e}}}\sim43$,
so perpendicular heat diffusivity is much greater for ions than for
electrons. Since $\chi\sim\nu l^{2}\sim V_{th}l$, the ratios of the
parallel to perpendicular thermal diffusion coefficients are given
by the ratios of the diffusion random walk step lengths: $\frac{\chi_{\parallel\alpha}}{\chi_{\perp\alpha}}=\frac{\lambda_{mfp\alpha}}{r_{L\alpha}}=\frac{V_{th\alpha}\tau_{\alpha}}{(V_{th\alpha}/\omega_{c\alpha})}=\omega_{c\alpha}\tau_{\alpha}$
(here, $\tau_{i}\Rightarrow\tau_{ie}$, and $\tau_{e}\Rightarrow\tau_{ei}$).
By definition, plasmas are considered to be magnetized when $\omega_{c\alpha}\tau_{\alpha}\gg1$,
so the parallel thermal diffusion is far higher than perpendicular
thermal diffusion. For example, at $T_{i}=T_{e}=100\mbox{ eV}$, $n=1\times10^{20}\mbox{ m}^{-3}$,
and $B=1$ T, equations \ref{eq:472.45} and \ref{eq:472.50} yield
$\chi_{\parallel e}\sim3\times10^{8}\mbox{ m}^{2}/\mbox{s},\,\chi_{\parallel i}\sim6\times10^{6}\mbox{ m}^{2}/\mbox{s},\,\chi_{\perp i}\sim1\mbox{ m}^{2}/\mbox{s},\mbox{ and }\chi_{\perp e}\sim0.1\mbox{ m}^{2}/\mbox{s}$. 

\section{\label{subsec:Single-fluid-Magnetohydrodynamic}Single fluid magnetohydrodynamic
equations}

The two-fluid equations describe particle motion with the species
(ions and electrons) mean velocities $\mathbf{v}_{\alpha}$, and the
pressures $\underline{\mathbf{p}}_{\alpha}$which describe the random
deviation of the particle velocities from the average values \cite{bellan fundamentals}.
The single fluid MHD equations use single fluid center of mass velocity
$\mathbf{v}$, and current density $\mathbf{J}$, to describe average
motion, where $\mathbf{v}=\frac{1}{\rho}\underset{\alpha}{\Sigma}\rho_{\alpha}\mathbf{v}_{\alpha}$
(with $\rho=\underset{\alpha}{\Sigma}\rho_{\alpha}$), and $\mathbf{J}=\underset{\alpha}{\Sigma}n_{\alpha}q_{\alpha}\mathbf{v}_{\alpha}$.\\

\subsection{$\mathbf{0}^{\mathbf{th}}$ moment (mass continuity equation)}

The MHD mass continuity equation can be obtained by summing the two-fluid
mass continuity equations (\ref{eq:469}) over species:
\begin{equation}
\frac{\partial\rho(\mathbf{r},t)}{\partial t}+\nabla\cdot(\rho(\mathbf{r},t)\mathbf{v}(\mathbf{r},t))=0\label{eq:472.32}
\end{equation}

\subsection{$\mathbf{1}^{\mathbf{st}}$ moment (momentum equation)}

Formally, the MHD momentum equation is found by taking the first moment
of the Boltzmann equation (\ref{eq:360}), multiplying by $m_{\alpha}$,
and summing over species \cite{bellan fundamentals}:{\small{}
\begin{align}
\underset{\alpha}{\Sigma}\left(m_{\alpha}\int\left[\frac{\partial f_{\alpha}}{\partial t}+\nabla\cdot(\mathbf{V}\,f_{\alpha})+\nabla_{v}\cdot\left(\left(\frac{q_{\alpha}}{m_{\alpha}}(\mathbf{E}(\mathbf{r},t)\mathbf{+V}\mathbf{\times B}(\mathbf{r},t))\right)\,f_{\alpha}\right)\right]\mathbf{V}\,d\mathbf{V}\right) & =\underset{\alpha}{\Sigma}m_{\alpha}\int C_{\alpha}^{scatt.}\mathbf{V}\,d\mathbf{V}\nonumber \\
\nonumber \\
\label{eq:472.4}
\end{align}
}The subsequent derivation follows the procedure detailed earlier
in the derivation of the expression for two-fluid momentum conservation
(equation \ref{eq:472-2}). The $1^{st}$ term is\\
\[
\underset{\alpha}{\Sigma}\left(m_{\alpha}\int\frac{\partial f_{\alpha}}{\partial t}\mathbf{V}\,d\mathbf{V}\right)=\frac{\partial}{\partial t}\left(\underset{\alpha}{\Sigma}\left(m_{\alpha}\int f_{\alpha}\mathbf{V}\,d\mathbf{V}\right)\right)=\frac{\partial}{\partial t}\left(\underset{\alpha}{\Sigma}\left(m_{\alpha}n_{\alpha}\mathbf{v}_{\alpha}\right)\right)=\frac{\partial(\rho\mathbf{v})}{\partial t}
\]
\\
In particular, the $2^{nd}$ term is\\
 
\[
\underset{\alpha}{\Sigma}\left(m_{\alpha}\int\nabla\cdot\left(\mathbf{V}f_{\alpha}\right)\mathbf{V}d\mathbf{V}\right)=\underset{\alpha}{\Sigma}\left(m_{\alpha}\nabla\cdot\left(\int\mathbf{VV}\,f_{\alpha}d\mathbf{V}\right)-m_{\alpha}\int\mathbf{V}\,f_{\alpha}\cancelto{0}{\nabla\cdot\mathbf{V}}d\mathbf{V}\right)=\nabla\cdot\underline{\mathbf{P}}
\]
where $\underline{\mathbf{P}}=\underset{\alpha}{\Sigma}\left(m_{\alpha}\int\mathbf{VV}\,f_{\alpha}d\mathbf{V}\right)$
is the total stress tensor for the single plasma fluid. Instead of
defining random particle velocities relative to $\mathbf{v}_{\alpha}$,
as was done in the derivation of the two-fluid momentum equation,
random velocities are defined relative to the single fluid velocity
$\mathbf{v}$: 
\begin{equation}
\mathbf{c}_{0\alpha}(\mathbf{r},t)=\mathbf{V-\mathbf{v}}(\mathbf{r},t)\label{eq:473}
\end{equation}
Then, through manipulation analogous to that in the derivation of
equation \ref{eq:356-1}, $\underline{\mathbf{P}}$ can be expressed
as $\underline{\mathbf{P}}=\underline{\mathbf{p}}+\rho\mathbf{v}\mathbf{v}$
\cite{bellan fundamentals}, where the single fluid pressure tensor
is defined as: 
\[
\underline{\mathbf{p}}(\mathbf{r},t)=\underset{\alpha}{\Sigma}\left(m_{\alpha}\int\mathbf{c}_{0\alpha}\mathbf{c}_{0\alpha}\,f_{\alpha}\,d\mathbf{V}\right)
\]
$\underline{\mathbf{p}}$ can be related to $\underline{\mathbf{p}}_{\alpha}$
- firstly the species diffusion velocity is defined as 
\begin{equation}
\mathbf{w}_{\alpha}(\mathbf{r},t)=\mathbf{\mathbf{\mathbf{v}}_{\alpha}}(\mathbf{r},t)\mathbf{-\mathbf{v}}(\mathbf{r},t)\label{eq:473.01}
\end{equation}
Using this definition and equation \ref{eq:354}, the species random
particle velocities in the frame moving at velocity $\mathbf{v}$
can be expressed as \cite{bittencourt}: 
\begin{equation}
\mathbf{c}_{0\alpha}=\mathbf{V-\mathbf{v}}=(\mathbf{c}_{\alpha}+\mathbf{\mathbf{v}}_{\alpha})-(\mathbf{\mathbf{\mathbf{v}}_{\alpha}}-\mathbf{w}_{\alpha})=\mathbf{c}_{\alpha}+\mathbf{w}_{\alpha}\label{eq:473.1}
\end{equation}
Note that the average of an averaged value is unchanged: 
\begin{equation}
\int\mathbf{w}_{\alpha}\,f_{\alpha}\,d\mathbf{V}=\mathbf{w}_{\alpha}\label{eq:475}
\end{equation}
Using these relations, the single fluid pressure tensor can be expressed
as:

\begin{align}
\underline{\mathbf{p}}(\mathbf{r},t) & =\underset{\alpha}{\Sigma}\left(m_{\alpha}\int(\mathbf{c}_{\alpha}+\mathbf{w}_{\alpha})(\mathbf{c}_{\alpha}+\mathbf{w}_{\alpha})\,f_{\alpha}\,d\mathbf{V}\right) & \mbox{(use eqn. \ref{eq:473.1})}\nonumber \\
 & =\underset{\alpha}{\Sigma}\left(m_{\alpha}\int(\mathbf{c}_{\alpha}\mathbf{c}_{\alpha}+\mathbf{c}_{\alpha}\mathbf{w}_{\alpha}+\mathbf{w}_{\alpha}\mathbf{c}_{\alpha}+\mathbf{w}_{\alpha}\mathbf{w}_{\alpha})\,f_{\alpha}\,d\mathbf{V}\right)\nonumber \\
 & =\underset{\alpha}{\Sigma}\left(m_{\alpha}\left(\frac{1}{m_{\alpha}}\underline{\mathbf{p}}_{\alpha}+\cancelto{0}{\left(\int\mathbf{c}_{\alpha}\,f_{\alpha}\,d\mathbf{V}\right)\mathbf{w}_{\alpha}}+\mathbf{w}_{\alpha}\cancelto{0}{\left(\int\mathbf{c}_{\alpha}\,f_{\alpha}\,d\mathbf{V}\right)}+\mathbf{w}_{\alpha}\mathbf{w}_{\alpha}n_{\alpha}\right)\right)\nonumber \\
 & \mbox{\mbox{(use eqns. \ref{eq:355}, \ref{eq:475}, \ref{eq:355-1} and \ref{eq:352-1})}}\nonumber \\
\Rightarrow\underline{\mathbf{p}} & =\underset{\alpha}{\Sigma}\underline{\mathbf{p}}_{\alpha}+\underset{\alpha}{\Sigma}\rho_{\alpha}\mathbf{w}_{\alpha}\mathbf{w}_{\alpha}\label{eq:475.1}
\end{align}
Scalar pressure is given by the trace of the pressure tensor: $p=\frac{1}{3}Tr(\underline{\mathbf{p}})$
\cite{bittencourt}. Expressing $\mathbf{c}_{0\alpha}$ in terms of
Cartesian coordinates $x,\,y,\,z$ as $(c_{0\alpha x},\,c_{0\alpha y},\,c_{0\alpha z})$,
then with the definition $\left(\underline{\mathbf{p}}\right)_{ij}=\underset{\alpha}{\Sigma}\left(m_{\alpha}\int c_{0\alpha i}c_{0\alpha j}\,f_{\alpha}\,d\mathbf{V}\right)$,
the scalar pressure is: 
\begin{align}
p & =\frac{1}{3}\left(\left(\underline{\mathbf{p}}\right)_{xx}+\left(\underline{\mathbf{p}}\right)_{yy}+\left(\underline{\mathbf{p}}\right)_{zz}\right)\nonumber \\
 & =\underset{\alpha}{\Sigma}\left(\frac{m_{\alpha}}{3}\int\left(c_{0\alpha x}^{2}+c_{0\alpha y}^{2}+c_{0\alpha z}^{2}\right)\,f_{\alpha}\,d\mathbf{V}\right)\nonumber \\
 & =\underset{\alpha}{\Sigma}\left(\frac{m_{\alpha}}{3}\int c_{0\alpha}^{2}\,f_{\alpha}\,d\mathbf{V}\right)\nonumber \\
 & =\underset{\alpha}{\Sigma}\left(\frac{m_{\alpha}}{3}\int(\mathbf{c}_{\alpha}+\mathbf{w}_{\alpha})\cdot(\mathbf{c}_{\alpha}+\mathbf{w}_{\alpha})\,f_{\alpha}\,d\mathbf{V}\right) & \mbox{(use eqn. \ref{eq:473.1})}\nonumber \\
 & =\underset{\alpha}{\Sigma}\left(\frac{m_{\alpha}}{3}\int(c_{\alpha}^{2}+2\mathbf{w}_{\alpha}\cdot\mathbf{c}_{\alpha}+w_{\alpha}^{2})\,f_{\alpha}\,d\mathbf{V}\right)\nonumber \\
 & =\underset{\alpha}{\Sigma}\left(\frac{m_{\alpha}}{3}\left(\cancelto{\frac{3p_{\alpha}}{m_{\alpha}}}{\int c_{\alpha}^{2}\,f_{\alpha}\,d\mathbf{V}}+2\mathbf{w}_{\alpha}\cdot\cancelto{0}{\int\mathbf{c}_{\alpha}\,f_{\alpha}\,d\mathbf{V}}+w_{\alpha}^{2}\cancelto{n_{\alpha}}{\int\,f_{\alpha}\,d\mathbf{V}}\,\,\,\,\,\,\,\,\,\,\right)\right)\nonumber \\
 & \mbox{\mbox{(use eqns. \ref{eq:355-3}, \ref{eq:475}, \ref{eq:355-1} and \ref{eq:352-1})}}\nonumber \\
\Rightarrow p & =\underset{\alpha}{\Sigma}p_{\alpha}+\frac{1}{3}\underset{\alpha}{\Sigma}\rho_{\alpha}w_{\alpha}^{2} & \mbox{}\label{eq:476}
\end{align}
 Analogous to equation \ref{eq:357}, we have: 
\begin{equation}
\underline{\mathbf{p}}(\mathbf{r},t)=p(\mathbf{r},t)\underline{\mathbf{I}}+\overline{\boldsymbol{\pi}}(\mathbf{r},t)\label{eq:476.1}
\end{equation}
with $\overline{\boldsymbol{\pi}}$ being the single fluid generalised
stress tensor, so starting with equation \ref{eq:475.1}, we can define
$\overline{\boldsymbol{\pi}}$ in terms of $\overline{\boldsymbol{\pi}}_{\alpha}$:

\begin{align}
\underline{\mathbf{p}} & =\underset{\alpha}{\Sigma}\underline{\mathbf{p}}_{\alpha}+\underset{\alpha}{\Sigma}\rho_{\alpha}\mathbf{w}_{\alpha}\mathbf{w}_{\alpha}\nonumber \\
\Rightarrow p\underline{\mathbf{I}}+\overline{\boldsymbol{\pi}} & =\underset{\alpha}{\Sigma}\left(p_{\alpha}\underline{\mathbf{I}}+\overline{\boldsymbol{\pi}}_{\alpha}+\rho_{\alpha}\mathbf{w}_{\alpha}\mathbf{w}_{\alpha}\right) & \mbox{(use eqns. \ref{eq:476.1} \& \ref{eq:357})}\nonumber \\
\Rightarrow\cancel{\underset{\alpha}{\Sigma}p_{\alpha}\underline{\mathbf{I}}}+\frac{1}{3}\underset{\alpha}{\Sigma}\rho_{\alpha}w_{\alpha}^{2}\underline{\mathbf{I}}+\overline{\boldsymbol{\pi}} & =\underset{\alpha}{\Sigma}\left(\cancel{p_{\alpha}\underline{\mathbf{I}}}+\overline{\boldsymbol{\pi}}_{\alpha}+\rho_{\alpha}\mathbf{w}_{\alpha}\mathbf{w}_{\alpha}\right) & \mbox{(use eqn. \ref{eq:476})}\nonumber \\
\Rightarrow\overline{\boldsymbol{\pi}} & =\underset{\alpha}{\Sigma}\left(\overline{\boldsymbol{\pi}}_{\alpha}+\rho_{\alpha}\left(\mathbf{w}_{\alpha}\mathbf{w}_{\alpha}-\frac{1}{3}w_{\alpha}^{2}\underline{\mathbf{I}}\right)\right) & \mbox{}
\end{align}
The $3^{rd}$ term in equation \ref{eq:472.4} is 
\[
\underset{\alpha}{\Sigma}\left(m_{\alpha}\int\nabla_{v}\cdot\left(\frac{q_{\alpha}}{m_{\alpha}}(\mathbf{E}\mathbf{+V}\mathbf{\times B})\,f_{\alpha}\right)\,\mathbf{V}\,d\mathbf{V}\right)=-\underset{\alpha}{\Sigma}\left(q_{\alpha}n_{\alpha}\left(\mathbf{E}+\mathbf{v}_{\alpha}\times\mathbf{B}\right)\right)=-\rho_{c}\mathbf{E}-\mathbf{J}\times\mathbf{B}
\]
where $\rho_{c}(\mathbf{r},t\mathbf{)}=\underset{\alpha}{\Sigma}q_{\alpha}n_{\alpha}$
is the charge density. The $4^{th}$ term is\\
$\underset{\alpha}{\Sigma}\left(m_{\alpha}\int C_{\alpha}^{scatt.}\mathbf{V}\,d\mathbf{V}\right)=\underset{\alpha}{\Sigma}\mathbf{R}_{\alpha}=0$
since $\mathbf{R}_{i}+\mathbf{R}_{e}=0$ (equation \ref{eq:467.2}
- the total plasma does not exert a friction on itself). In the MHD
description, the plasma is assumed to be quasineutral because we are
interested in plasma behavior at frequencies low compared with the
plasma frequency $\left(\omega\ll\omega_{p}=\sqrt{\frac{ne^{2}}{m_{e}\epsilon_{0}}}\right)$,
and at length scales long compared to the Debye length $\left(L\gg\lambda_{D}=\sqrt{\frac{\epsilon_{0}k_{b}T}{ne^{2}}}\right)$
\cite{Callen}. Assembling the four terms, in the limit of quasineutrality,
the single fluid momentum equation is:
\begin{equation}
\frac{\partial\mathbf{v}}{\partial t}=-(\mathbf{v}\cdot\nabla)\mathbf{v}+\frac{1}{\rho}\left(-\nabla p-\nabla\cdot\overline{\boldsymbol{\pi}}+\mathbf{J}\times\mathbf{B}\right)\label{eq:476.3}
\end{equation}

\subsection{Ohm's law}

The single fluid momentum equation gives one relationship between
$\mathbf{J}$ and $\mathbf{v}$. Another relationship, Ohm's law,
can be found starting with the expression for electron momentum conservation
(equation \ref{eq:472-2}), which can be written as:

\textbf{
\[
\rho_{e}\frac{d\mathbf{v}_{e}}{dt}=-\nabla\cdot\underline{\mathbf{p}}_{e}-en_{e}\left(\mathbf{E}+\mathbf{v}_{e}\times\mathbf{B}\right)+\mathbf{R}_{e}
\]
}Here, \textbf{$\frac{d\mathbf{v}_{e}}{dt}=\frac{\partial\mathbf{v}_{e}}{\partial t}+\mathbf{v}_{e}\cdot\nabla\mathbf{v}_{e}$
}is the convective derivative of $\mathbf{v}_{e}$, and $e=-q_{e}$
is the electron charge. Neglecting the thermal force, equation \ref{eq:472.39}
defines the friction force that the ions exert on the electrons as
$\mathbf{R}_{e}=\eta'ne\mathbf{J}$. Along with the assumption of
quasineutrality, this implies the electron momentum equation is:\textbf{
\[
m_{e}\frac{d\mathbf{v}_{e}}{dt}=-\frac{1}{n}\nabla\cdot\underline{\mathbf{p}}_{e}-e\mathbf{E}-e\mathbf{v}_{e}\times\mathbf{B}+\eta'e\mathbf{J}
\]
}Comparing the magnitudes of the electron inertia and the magnetic
force terms, we see that\\
$|m_{e}\frac{d\mathbf{v}_{e}}{dt}|\,/\,|e\mathbf{v}_{e}\times\mathbf{B}|=\frac{m_{e}}{\tau eB}\sim\frac{\omega}{\omega_{ce}}$
, where $\omega_{ce}=\frac{eB}{m_{e}}$ is the electron cyclotron
frequency and $\omega$ is the frequency of the plasma behaviour that
we wish to study. Since MHD is concerned with low frequency phenomena,
$\frac{\omega}{\omega_{ce}}\ll1$, so the electron inertia term can
be neglected. It's worth noting that this assumption is good for velocities
perpendicular to $\mathbf{B}$, but not necessarily good for parallel
velocities since there are no magnetic forces parallel to $\mathbf{B}$
\cite{bellan fundamentals}. The center of mass velocity is $\mathbf{v}=\underset{\alpha}{\frac{1}{\rho}\Sigma}\rho_{\alpha}\mathbf{v}_{\alpha},$
so with $m_{e}\ll m_{i}$, $\mathbf{v\sim v}_{i}$. Since $\mathbf{J}=ne(\mathbf{v}_{i}-\mathbf{v}_{e})$,
this implies that $\mathbf{v}_{e}\sim-\frac{\mathbf{J}}{ne}+\mathbf{v}$,
so the electron momentum equation reduces to:\textbf{
\begin{align*}
0 & =-\frac{1}{n}\nabla\cdot\underline{\mathbf{p}}_{e}-e\mathbf{E}-e\left(-\frac{\mathbf{J}}{ne}+\mathbf{v}\right)\times\mathbf{B}+\eta'e\mathbf{J}\\
\Rightarrow0 & =-\frac{1}{ne}\nabla\cdot\underline{\mathbf{p}}_{e}-\mathbf{E}+\frac{1}{ne}\mathbf{J}\times\mathbf{B}-\mathbf{v}\times\mathbf{B}+\eta'\mathbf{J}\\
\Rightarrow\mathbf{E}+\mathbf{v}\times\mathbf{B} & +\frac{1}{ne}\nabla\cdot\underline{\mathbf{p}}_{e}-\frac{1}{ne}\mathbf{J}\times\mathbf{B}=\eta'\mathbf{J}
\end{align*}
}The term $-\frac{1}{ne}\mathbf{J}\times\mathbf{B}$ is the Hall term.
Comparing the magnitudes of the Hall term and the resistive term,
we see that $|\frac{1}{ne}\mathbf{J}\times\mathbf{B}|\,/\,|\eta'\mathbf{J}|=\frac{eB}{m_{e}\nu_{ei}}\sim\frac{\omega_{ce}}{\nu_{ei}}$,
so the Hall term can be neglected if the electron-ion collision frequency
is large compared with the electron cyclotron frequency. Note however,
that magnetized plasmas are defined by $\omega_{ci}\tau_{ii},\,\omega_{ce}\tau_{ei}\gg1$
. Even when the collision frequency is not large, it is often reasonable
to neglect both the Hall term and the pressure tensor term \cite{bellan fundamentals,Schunk_Nagy},
leading to the reduced Ohm's law:
\begin{equation}
\mathbf{E}+\mathbf{v}\times\mathbf{B}=\eta'\mathbf{J}\label{eq:476.4}
\end{equation}

\subsection{$\mathbf{2}^{\mathbf{nd}}$ moment (energy equation)}

The MHD energy equation is found by taking the contracted second moment
of the Boltzmann equation (\ref{eq:360}), and summing over species:

{\footnotesize{}
\[
\underset{\alpha}{\Sigma}\left(\int\left[\frac{\partial f_{\alpha}}{\partial t}+\nabla\cdot(\mathbf{V}\,f_{\alpha})+\nabla_{v}\cdot\left(\left(\frac{q_{\alpha}}{m_{\alpha}}(\mathbf{E}(\mathbf{r},t)\mathbf{+V}\mathbf{\times B}(\mathbf{r},t))\right)\,f_{\alpha}\right)\right]\left(\frac{1}{2}m_{\alpha}V^{2}\right)\,d\mathbf{V}\right)=\underset{\alpha}{\Sigma}\int C_{\alpha}^{scatt.}\frac{1}{2}m_{\alpha}V^{2}\,d\mathbf{V}
\]
}Again, the subsequent derivation follows the procedure outlined earlier
in the derivation of the expression for two-fluid energy conservation
(equation \ref{eq:472.31}). However, the $1^{st}$ term is expanded
using random particle velocities relative to $\mathbf{v}$ instead
of to $\mathbf{v}_{\alpha}$, resulting in $\frac{\partial}{\partial t}\left(\frac{1}{2}\rho u^{2}+\frac{3}{2}p\right)$,
which is then manipulated using the expressions for single fluid mass
and momentum conservation (equations \ref{eq:472.32} and \ref{eq:476.3}),
instead of using the two-fluid mass and momentum conservation equations,
as was done to derive the two-fluid energy equation. In particular,
the $2^{nd}$ term is 
\begin{align*}
\underset{\alpha}{\Sigma}\left(\int\nabla\cdot\left(\mathbf{V}f_{\alpha}\right)\left(\frac{1}{2}m_{\alpha}V^{2}\right)d\mathbf{V}\right) & =\underset{\alpha}{\Sigma}\left(\int\nabla\cdot\left(\frac{1}{2}m_{\alpha}V^{2}\mathbf{V}f_{\alpha}\right)d\mathbf{V}-\int f_{\alpha}\mathbf{V}\cdot\cancelto{0}{\nabla\left(\frac{1}{2}m_{\alpha}V^{2}\right)}d\mathbf{V}\right)\\
 & =\nabla\cdot\left(\underset{\alpha}{\Sigma}\int\left(\frac{1}{2}m_{\alpha}V^{2}\mathbf{V}f_{\alpha}\right)d\mathbf{V}\right)=\nabla\cdot\mathbf{Q'}
\end{align*}
where $\mathbf{Q'}(\mathbf{r},t)=\underset{\alpha}{\Sigma}\left(\int(\frac{1}{2}m_{\alpha}V^{2})\mathbf{V}\,f_{\alpha}\,d\mathbf{V}\right)$
is the energy flux density for the single plasma fluid. Analogous
to equation \ref{eq:465}, but using random particle velocities relative
to $\mathbf{v}$ instead of to $\mathbf{v}_{\alpha}$, the energy
flux density for the single plasma fluid can be expressed as $\mathbf{Q'}=\frac{1}{2}\rho v^{2}\mathbf{v}+\frac{3}{2}p\mathbf{v}+\mathbf{v\cdot}\underline{\mathbf{p}}+\mathbf{q}$,
where $\mathbf{q}(\mathbf{r},t)=\underset{\alpha}{\Sigma}\left(\int(\frac{1}{2}m_{\alpha}c_{0\alpha}^{2})\mathbf{c}_{0\alpha}\,f_{\alpha}\,d\mathbf{V}\right)$
is the heat flux density for the single plasma fluid \cite{bittencourt}
(flux of random energy, contracted $3^{rd}$ order moment in the frame
moving with the single fluid at velocity $\mathbf{v}$). Following
from equation \ref{eq:472.3}, the $3^{rd}$ term is $\underset{\alpha}{\Sigma}\left(-q_{\alpha}n_{\alpha}\mathbf{E}\cdot\mathbf{v}_{\alpha}\right)=-\mathbf{E\cdot J}$,
since $\mathbf{J}=\underset{\alpha}{\Sigma}\left(n_{\alpha}q_{\alpha}\mathbf{v}_{\alpha}\right)$.
The $4^{th}$ term vanishes as collisional energy conservation implies
that $\underset{\alpha}{\Sigma}\int C_{\alpha}^{scatt.}\frac{1}{2}m_{\alpha}V^{2}\,d\mathbf{V}=\underset{\alpha}{\Sigma}\left(Q_{\alpha}+\mathbf{v}_{\alpha}\cdot\mathbf{R}_{\alpha}\right)=0$
(equation \ref{eq:468.2}). \\
\\
Assembling the four terms and simplifying the expression in much the
same manner as was done for the derivation of equation \ref{eq:472.31},
the single fluid energy equation is: 
\[
\frac{\partial p}{\partial t}=-\mathbf{v}\cdot\nabla p-\gamma p\,\nabla\cdot\mathbf{v}+(\gamma-1)\left(-\overline{\boldsymbol{\pi}}:\nabla\mathbf{v}-\nabla\cdot\mathbf{q}+\mathbf{E\cdot J}-\mathbf{v}\cdot\mathbf{J}\times\mathbf{B}\right)
\]
The last two terms can be combined as follows: The reduced Ohm's law
(equation \ref{eq:476.4}) gives:\\
$\mathbf{E=\eta'}\mathbf{J}-\mathbf{v\times B}\Rightarrow\mathbf{E\cdot J}=\eta'J^{2}-\mathbf{J\cdot}\mathbf{v\times B}=\eta'J^{2}+\mathbf{J\cdot}\mathbf{B\times v}=\eta'J^{2}+\mathbf{v\cdot}\mathbf{J\times B}$,
so that the single fluid energy equation is: 
\begin{equation}
\frac{\partial p}{\partial t}=-\mathbf{v}\cdot\nabla p-\gamma p\,\nabla\cdot\mathbf{v}+(\gamma-1)\left(\eta'J^{2}-\overline{\boldsymbol{\pi}}:\nabla\mathbf{v}-\nabla\cdot\mathbf{q}\right)\label{eq:477}
\end{equation}

\section{Equilibrium models\label{sec:Equilibrium-models}}

\subsection{Grad-Shafranov equation\label{subsec:Grad-Shafranov-equation}}

In cylindrical coordinates with azimuthal symmetry, the divergence
of the magnetic field is 
\begin{equation}
\nabla\cdot\mathbf{B}=0=\frac{1}{r}\frac{\partial(rB_{r})}{\partial r}+\frac{\partial B_{z}}{\partial z}\label{eq9}
\end{equation}
Poloidal flux through a circular area of radius $r$ in a plane of
constant $z$ is defined as $\Psi(r,z)$ {[}Wb{]}$=\int_{0}^{r}\mathbf{B}_{\theta}(r,z)\cdot d\mathbf{s}$,
where $d\mathbf{s}=ds\widehat{\mathbf{z}}$, and $ds=2\pi r\,dr$
is an annular elemental area of radius $r$ centered at $(0,\,z)$.
Note that, in cylindrical geometry with azimuthal symmetry, the curl
of a toroidal vector has only poloidal components. Hence, using Stokes
theorem 
\[
\Psi(r,z)=\int_{0}^{r}\nabla\times\mathbf{A_{\phi}}(r,z)\cdot d\mathbf{s}=2\pi r\,A_{\phi}(r,z)
\]
Defining $\psi=\frac{1}{2\pi}\Psi$ as the poloidal flux per radian
in toroidal angle $\phi$, this implies that 
\begin{equation}
\psi=rA_{\phi}\label{eq:9.001}
\end{equation}
The poloidal field is defined as 
\[
\mathbf{B}_{\theta}=\nabla\times\mathbf{A}_{\phi}=-\frac{\partial A_{\phi}}{\partial z}\widehat{\mathbf{r}}+\frac{1}{r}\frac{\partial(rA_{\phi})}{\partial r}\widehat{\mathbf{z}}
\]
Using the definition for $\psi,$ this implies that 
\begin{equation}
B_{r}=-\frac{1}{r}\frac{\partial\psi}{\partial z},\,\,\,\,\,B_{z}=\frac{1}{r}\frac{\partial\psi}{\partial r}\label{eq:9.01}
\end{equation}
which satisfies equation \ref{eq9}. In cylindrical coordinates, $\nabla\phi=\frac{\widehat{\boldsymbol{\phi}}}{r}$,
so that $\mathbf{B}_{\theta}=\nabla\times\mathbf{A}_{\phi}=\nabla\times\frac{\psi}{r}\widehat{\boldsymbol{\phi}}=\nabla\times\left(\psi\nabla\phi\right)=\mathbf{\nabla}\psi\times\nabla\phi$.
Using Ampere's law and Stoke's theorem, $\nabla\times\mathbf{B}=\mu_{0}\mathbf{J}\Rightarrow2\pi r\,B_{\phi}=\mu_{0}I_{\theta}\Rightarrow B_{\phi}\widehat{\boldsymbol{\phi}}=\frac{\mu_{0}I_{\theta}}{2\pi r}\,r\nabla\phi$
\[
\Rightarrow\mathbf{B}_{\phi}=\frac{\mu_{0}I_{\theta}}{2\pi}\nabla\phi,
\]
where $I_{\theta}(r,z)$ is the poloidal current linked by a circle
of radius $r$ with center on the z-axis at axial location $z$. Hence,
equivalent forms of the general axisymmetric field are:

\begin{equation}
\mathbf{B}=\mathbf{\nabla}\psi\times\nabla\phi+f(\psi)\nabla\phi=\frac{1}{2\pi}\left(\mathbf{\nabla}\Psi\times\nabla\phi+\mu_{0}I_{\theta}\nabla\phi\right),\label{eq:9.1}
\end{equation}
where 
\begin{equation}
f(\psi)=rB_{\phi}\label{eq:9.2}
\end{equation}
Here, $r$ is the radius of curvature of $\mathbf{B}_{\phi}$. Note
that $\nabla\times\mathbf{B}_{\phi}=\frac{\mu_{0}}{2\pi}\nabla\times(I_{\theta}(r,z)\,\nabla\phi)=\frac{\mu_{0}}{2\pi}\left(\nabla I_{\theta}\times\nabla\phi\right)$,
a poloidal vector with $r$ and $z$ dependence. Referring to Ampere's
law, this implies that
\begin{equation}
\mathbf{J}_{\theta}=\frac{1}{2\pi}\nabla I_{\theta}(r,z)\times\nabla\phi\label{eq:10}
\end{equation}
Equivalently, $\mu_{0}\mathbf{J_{\theta}}=\mathbf{\nabla\times\mathbf{B}_{\phi}=}\nabla\times\left(\frac{f(\psi)}{r}\widehat{\boldsymbol{\phi}}\right)=\left(-\frac{\partial}{\partial z}\left(\frac{f(\psi)}{r}\right)\right)\hat{\boldsymbol{r}}+\left(\frac{1}{r}\frac{\partial}{\partial r}\left(\frac{rf(\psi)}{r}\right)\right)\hat{\boldsymbol{z}}$,
so that
\begin{equation}
J_{r}=-\frac{1}{\mu_{0}r}\frac{\partial f(\psi)}{\partial z},\;\;\;\;\;J_{z}=\frac{1}{\mu_{0}r}\frac{\partial f(\psi)}{\partial r}\label{eq:9.3}
\end{equation}
The curl of a poloidal vector is toroidal in cylindrical coordinates
with azimuthal symmetry, for example $\nabla\times\mathbf{B}_{\theta}=\widehat{\boldsymbol{\phi}}\left(\frac{\partial B_{r}}{\partial z}-\frac{\partial B_{z}}{\partial r}\right)$.
Together with equation \ref{eq:9.01}, this indicates, as shown in
\cite{bellan fundamentals}, that the magnitude of the curl of the
poloidal field is given by a Laplacian-like operator on $\psi$. 
\begin{align}
\nabla\times\mathbf{B}_{\theta} & =|\nabla\times\mathbf{B}_{\theta}|\widehat{\boldsymbol{\phi}}=\left(\widehat{\boldsymbol{\phi}}\cdot\nabla\times\mathbf{B}_{\theta}\right)\widehat{\boldsymbol{\phi}}\nonumber \\
 & =\left(r\nabla\phi\cdot\nabla\times\mathbf{B}_{\theta}\right)\widehat{\boldsymbol{\phi}}\nonumber \\
 & =r\widehat{\boldsymbol{\phi}}\left(\mathbf{B_{\mathbf{\theta}}\cdot}\nabla\times\nabla\phi-\nabla\cdot(\nabla\phi\times\mathbf{B}_{\theta})\right)\nonumber \\
 & =r\widehat{\boldsymbol{\phi}}\left(\nabla\cdot(\mathbf{B}_{\theta}\times\nabla\phi)=\nabla\cdot(\mathbf{(\nabla}\psi\times\nabla\phi)\times\nabla\phi)\right)\nonumber \\
 & =r\widehat{\boldsymbol{\phi}}\left(\nabla\cdot(\mathbf{-\nabla}\psi(\nabla\phi\cdot\nabla\phi)+\nabla\phi(\nabla\phi\cdot\nabla\psi))\right) & \,\,\mbox{(note that }\nabla\phi\perp\nabla\psi)\nonumber \\
 & =-r^{2}\nabla\cdot\left(\frac{\nabla\psi}{r^{2}}\right)\nabla\phi\label{eq:10.1}
\end{align}
With Ampere's law, this can be expressed as 
\begin{equation}
\mathbf{J}_{\phi}=-\frac{r^{2}}{\mu_{0}}\nabla\cdot\left(\frac{\nabla\psi}{r^{2}}\right)\nabla\phi\label{eq:11}
\end{equation}
The elliptic operator $\Delta^{*}$ is defined as: 
\begin{eqnarray}
\Delta^{*}\psi=r^{2}\nabla\cdot\left(\frac{\nabla\psi(\mathbf{r})}{r^{2}}\right) & = & r^{2}\left(\nabla\cdot\left(\frac{1}{r^{2}}\frac{\partial\psi}{\partial r}\hat{\mathbf{r}}+\frac{1}{r^{2}}\frac{\partial\psi}{\partial z}\hat{\mathbf{z}}\right)\right)\nonumber \\
 & = & r^{2}\left(\frac{1}{r}\frac{\partial}{\partial r}\left(\frac{1}{r}\frac{\partial\psi}{\partial r}\right)+\frac{\partial}{\partial z}\left(\frac{1}{r^{2}}\frac{\partial\psi}{\partial z}\right)\right)\nonumber \\
 & = & r\frac{\partial}{\partial r}\left(\frac{1}{r}\frac{\partial\psi}{\partial r}\right)+\frac{\partial^{2}\psi}{\partial z^{2}}\label{eq:19}
\end{eqnarray}
 
\begin{equation}
\Rightarrow\mathbf{J}_{\phi}=-\frac{1}{r\mu_{0}}\Delta^{^{*}}\psi\,\widehat{\boldsymbol{\phi}}\label{eq:20-1}
\end{equation}
In an equilibrium state with no fluid velocity, the expression for
single plasma fluid momentum conservation, equation \ref{eq:476.3},
reduces to the force balance equation
\begin{equation}
\mathbf{J}\times\mathbf{B}=\nabla p\label{eq:19.0}
\end{equation}
which can be expressed as
\begin{equation}
\mathbf{J_{\theta}}\times\mathbf{B_{\phi}}+\mathbf{\mathbf{J_{\phi}}\times\mathbf{B_{\theta}}+\mathbf{\mathbf{J_{\theta}}\times\mathbf{B_{\theta}}}}=\nabla p\label{eq:13}
\end{equation}
Since poloidal terms can be decomposed to radial and axial terms,
$\mathbf{J_{\theta}}\times\mathbf{B_{\theta}}$ is toroidally directed
and can be finite in non-axisymmetric cases. With axisymmetry, $\frac{\partial p}{\partial\phi}=0$\\
so $\mathbf{J_{\theta}}\times\mathbf{B_{\theta}}=0$. Referring to
equations \ref{eq:10} and \ref{eq:9.1}, this implies that $\nabla I_{\theta}\times\nabla\phi\parallel\nabla\psi\times\nabla\phi\,\Rightarrow\,\nabla I_{\theta}\parallel\nabla\psi$.
An arbitrary displacement $d\mathbf{r}$ results in changes in current
and poloidal flux $dI_{\theta}=d\mathbf{r}\cdot\nabla I_{\theta}$
and $d\psi=d\mathbf{r}\cdot\nabla\psi$, so that $\frac{dI_{\theta}}{d\psi}$
can be expressed as $\frac{dI_{\theta}}{d\psi}=\frac{d\mathbf{r}\cdot\nabla I_{\theta}}{d\mathbf{r}\cdot\nabla\psi}$.
Since $\nabla I_{\theta}$ is parallel to $\nabla\psi$, the derivative
is always defined and can be integrated $\Rightarrow I_{\theta}=I_{\theta}(\psi)$,
so that $\nabla I_{\theta}(\psi)=I_{\theta}'(\psi)\nabla\psi$ , where
$I_{\theta}'=\frac{dI_{\theta}}{d\psi}$\cite{bellan fundamentals}.
With this, we can redefine the poloidal current density (equation
\ref{eq:10})
\begin{equation}
\mathbf{J}_{\theta}=\frac{I_{\theta}'(\psi)}{2\pi}\nabla\psi\times\nabla\phi\label{eq:14}
\end{equation}
Using equations \ref{eq:9.1}, \ref{eq:11} and \ref{eq:14} in equation
\ref{eq:13}, we obtain
\begin{eqnarray}
\nabla p & = & \frac{\mu_{0}I_{\theta}I_{\theta}'(\psi)}{4\pi^{2}}(\nabla\psi\times\nabla\phi)\times\nabla\phi-\frac{r^{2}}{\mu_{0}}\nabla\cdot\left(\frac{\nabla\psi}{r^{2}}\right)\nabla\phi\times\mathbf{(\nabla}\psi\times\nabla\phi)\nonumber \\
 & = & \left((\nabla\psi\times\nabla\phi)\times\nabla\phi\right)\left(\frac{\mu_{0}I_{\theta}I_{\theta}'(\psi)}{4\pi^{2}}+\frac{r^{2}}{\mu_{0}}\nabla\cdot\left(\frac{\nabla\psi}{r^{2}}\right)\right)\nonumber \\
 & = & -\nabla\psi\left(\frac{\mu_{0}I_{\theta}I_{\theta}'(\psi)}{4\pi^{2}r^{2}}+\frac{1}{\mu_{0}}\nabla\cdot\left(\frac{\nabla\psi}{r^{2}}\right)\right)\label{eq:16}
\end{eqnarray}
Equation \ref{eq:16} shows that $\nabla p\parallel\nabla\psi\Rightarrow p=p(\psi)$
and $\nabla p=p'\nabla\psi$, following the argument above for $I_{\theta}(\psi)$.
With this, the Grad-Shafranov equation can be expressed as
\begin{align}
\nabla\cdot\left(\frac{\nabla\psi}{r^{2}}\right)+\mu_{0}^{2}\frac{I_{\theta}I_{\theta}'(\psi)}{4\pi^{2}r^{2}}+\mu_{0}p' & =0\nonumber \\
\Rightarrow\Delta^{*}\psi=-\left(\mu_{0}r^{2}\frac{dp}{d\psi}+\frac{\mu_{0}^{2}I_{\theta}}{4\pi^{2}}\frac{dI_{\theta}}{d\psi}\right)\label{eq:20.1}
\end{align}
Since $I_{\theta}=\frac{2\pi f(\psi)}{\mu_{0}}$, $I_{\theta}I_{\theta}'=\frac{4\pi^{2}ff'}{\mu_{0}^{2}}$,
leading to the more usual form of the Grad-Shafranov equation: 
\begin{equation}
\Delta^{*}\psi=-\left(\mu_{0}r^{2}\frac{dp}{d\psi}+f\frac{df}{d\psi}\right)\label{eq:20}
\end{equation}
For convenience, the expressions for the magnetic field and current
density components (equations \ref{eq:9.01}, \ref{eq:9.2} \ref{eq:9.3},
and \ref{eq:20-1}) in the case with toroidal symmetry are reproduced
here:

\begin{align}
B_{r} & =-\frac{1}{r}\frac{\partial\psi}{\partial z} &  & B_{\phi}=\frac{f(\psi)}{r} &  & B_{z}=\frac{1}{r}\frac{\partial\psi}{\partial r}\nonumber \\
J_{r} & =-\frac{1}{\mu_{0}r}\frac{\partial f(\psi)}{\partial z} &  & J_{\phi}=-\frac{1}{\mu_{0}r}\Delta^{^{*}}\psi &  & J_{z}=\frac{1}{\mu_{0}r}\frac{\partial f(\psi)}{\partial r}\label{eq:20.2}
\end{align}

\subsection{Grad-Shafranov equation for case with linear $\lambda(\psi)$ profile\label{subsec:GS_linear_lambda}}

\begin{figure}[H]
\centering{}\includegraphics[scale=0.8]{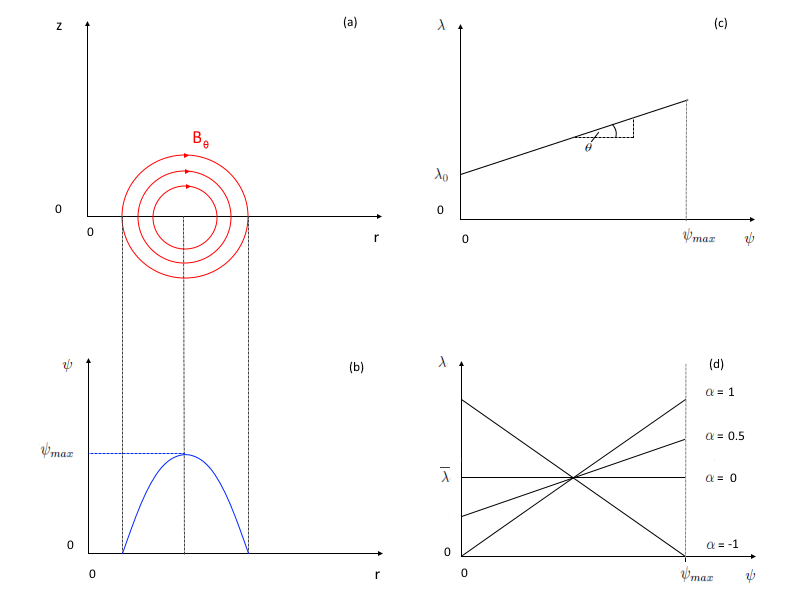}\caption{$\,\,\,\,$Linear $\lambda(\psi)$ profile\label{fig:Lambda-Profile}}
\end{figure}

In equilibrium, when $\mathbf{J}$ is parallel to $\mathbf{B}$, the
magnetic field topology is described by equation \ref{eq:0.01}: $\nabla\times\mathbf{B=}\lambda\mathbf{B}$.
Taking the divergence gives $\mathbf{B}\cdot\nabla\lambda=0$. Referring
to equation \ref{eq:9.1}, this implies that $\left(\mathbf{\nabla}\Psi\times\nabla\phi+\mu_{0}I_{\theta}\nabla\phi\right)\cdot\nabla\lambda=\left(\mathbf{\nabla}\Psi\times\nabla\phi\right)\cdot\nabla\lambda+\mu_{0}I_{\theta}\nabla\phi\cdot\nabla\lambda=0$.
All quantities, including $\lambda$, are symmetric in $\phi$, so
$\nabla\phi\cdot\nabla\lambda=0$. $\mathbf{\nabla}\Psi\times\nabla\phi$
is perpendicular to $\nabla\Psi$, so it is parallel to a flux surface.
This implies that $\nabla\lambda$ is perpendicular to flux surfaces,
so that $\lambda=\lambda(\psi)$, with the consequence that general
axisymmetric force free states must have $\lambda$ constant on a
flux surface. Unlike the Taylor state, in which fields satisfy $\nabla\times\mathbf{B=}\lambda\mathbf{B}$
with constant $\lambda$, $\lambda$ can vary across flux surfaces
in general axisymmetric force free states. A simple model for the
dependence of $\lambda$ on $\psi$ is to assume a linear profile
\cite{Bellan_Spheromaks,Knox}: 
\begin{equation}
\lambda(\psi)=\lambda_{0}+m\psi\label{eq:21}
\end{equation}
where 
\begin{equation}
m=tan\,\theta=(\lambda(\psi_{max})-\lambda_{0})/\psi_{max}\label{eq:22}
\end{equation}
is the slope of the $\lambda(\psi)$ profile, and $\lambda_{0}$ is
the value of $\lambda$ where $\psi=0$, as shown in figure \ref{fig:Lambda-Profile}(c).
The average value of $\lambda$ is 
\begin{equation}
\bar{\lambda}=\frac{\lambda(\psi_{max})-\lambda_{0}}{2}+\lambda_{0}\label{eq:23}
\end{equation}
Equations \ref{eq:22} and \ref{eq:23} imply that 
\begin{equation}
\bar{\lambda}=\frac{m\psi_{max}}{2}+\lambda_{0}\label{eq:24}
\end{equation}
Defining 
\begin{equation}
\lambda_{0}=\bar{\lambda}(1-\alpha)\label{eq:25}
\end{equation}
with $-1\leq\alpha\leq1$ so that $0\leq\lambda_{0}\leq2\bar{\lambda}$,
equation \ref{eq:24} implies that 
\begin{equation}
\bar{\lambda}=\frac{m\psi_{max}}{2}+\bar{\lambda}-\bar{\lambda}\alpha\Rightarrow m=\frac{2\bar{\lambda}\alpha}{\psi_{max}}\label{eq:26}
\end{equation}
Using equations \ref{eq:25} and \ref{eq:26} in equation \ref{eq:21}
implies that 
\begin{eqnarray}
\lambda(\psi) & = & \bar{\lambda}-\bar{\lambda}\alpha+2\bar{\lambda}\alpha\widetilde{\psi}\nonumber \\
\Rightarrow\lambda(\psi) & = & \bar{\lambda}(1+\alpha(2\widetilde{\psi}-1))\label{eq:27}
\end{eqnarray}
where $\psi$ has been normalized: $\widetilde{\psi}=\frac{\psi}{\psi_{max}}$.
From equations \ref{eq:9.1} and \ref{eq:14} it can be seen that
$\frac{J_{\theta}}{B_{\theta}}=\frac{1}{2\pi}I_{\theta}'$. The force
free equation $\nabla\times\mathbf{B=}\lambda\mathbf{B}\Rightarrow$$\mu_{0}J_{\theta}=\lambda B_{\theta}$,
so that:
\begin{align}
\mu_{0}\frac{J_{\theta}}{B_{\theta}} & =\lambda\nonumber \\
\Rightarrow\mu_{0}\frac{dI_{\theta}}{d\psi} & =2\pi\,\bar{\lambda}(1+\alpha(2\widetilde{\psi}-1))\label{eq:28}
\end{align}
This can be integrated to give $I_{\theta}(\psi)$:
\begin{eqnarray}
I_{\theta} & = & 2\pi\,\frac{\bar{\lambda}}{\mu_{0}}\int\left(1+\alpha(2\widetilde{\psi}-1)\right)d\psi\nonumber \\
 & = & 2\pi\,\frac{\bar{\lambda}}{\mu_{0}}\left(\psi+\frac{\alpha\psi^{2}}{\psi_{max}}-\alpha\psi\right)+C\nonumber \\
\Rightarrow\mu_{0}I_{\theta}(\psi) & = & 2\pi\,\bar{\lambda}\psi(1+\alpha(\widetilde{\psi}-1))\label{eq:29}
\end{eqnarray}
Note that with the boundary condition $I_{\theta}(\psi=0)=0$, $C$,
the constant of integration, is zero. Inserting equations \ref{eq:28}
and \ref{eq:29} in equation \ref{eq:20.1}, gives an expression for
the Grad-Shafranov equation with a linear dependence of $\lambda$
on $\psi$: 

\begin{equation}
\Delta^{*}\psi=-\left(\mu_{0}r^{2}\frac{dp}{d\psi}+\bar{\lambda}^{2}\psi(1+\alpha(2\widetilde{\psi}-1))(1+\alpha(\widetilde{\psi}-1))\right)\label{eq:29.2}
\end{equation}
As indicated in figure \ref{fig:Lambda-Profile}(d), the slope of
the linear $\lambda(\psi)$ profile is determined by $\alpha.$ Equilibria
with $\alpha<0$ tend to have \textquotedbl hollow\textquotedbl{}
current profiles, where $\lambda\sim J/B$ decreases towards the magnetic
axis, so that most of the current flows near the exterior of the magnetically
confined plasma, while \textquotedbl peaked\textquotedbl{} current
profiles, where current flows mainly around the center of the plasma,
are associated with $\alpha>0$ . In studies of spheromak formation
on the CTX magnetized Marshall gun \cite{Knox}, it was found that
$\alpha<0$ during CT formation and $\alpha<0$ after formation, during
the CT decay phase. This is consistent with the concept that helicity
flows in the direction of decreasing $\lambda$\cite{Bellan_Spheromaks}.
During formation, helicity is injected into the CT, and when the CT
is decaying, helicity flows outwards from the magnetic axis towards
the wall.

\section{Summary\label{sec:SummaryKin_MHD_EQ}}

It has been shown how the MHD equations are derived using basic principles
of kinetic theory. The techniques outlined, and equations obtained
in this chapter, will be referred to in chapter \ref{chap:Neutral-models},
where the set of equations that describes co-interacting plasma and
neutral fluids will be developed. The single-fluid MHD equations (equations
\ref{eq:472.32}, \ref{eq:476.3}, and \ref{eq:477}), and equations
defining the time-rates of change of the magnetic field with toroidal
symmetry (equations \ref{eq:211} and \ref{eq:214}), which are solved
in the MHD code, are collected here:

\begin{equation}
\dot{\rho}=-\nabla\cdot(\rho\mathbf{v})\label{eq: 478}
\end{equation}

\begin{equation}
\dot{\mathbf{v}}=-\mathbf{v}\cdot\nabla\mathbf{v}+\frac{1}{\rho}\left(-\nabla p-\nabla\cdot\overline{\boldsymbol{\pi}}+\mathbf{J\times}\mathbf{B}\right)\label{eq:479}
\end{equation}

\begin{equation}
\dot{p}=-\mathbf{v}\cdot\nabla p-\gamma p\,\nabla\cdot\mathbf{v}+(\gamma-1)\left(\eta'J^{2}-\overline{\boldsymbol{\pi}}:\nabla\mathbf{v}-\nabla\cdot\mathbf{q}\right)\label{eq:481}
\end{equation}
The code has option of evolving separate energy equations for the
ion and electron fluids. If this option is chosen, then noting that
total plasma fluid pressure is $p=p_{i}+p_{e}$, where $p_{i}=nT_{i}$,
and $p_{e}=Z_{eff}nT_{e}$, and that $\mathbf{q}=\mathbf{q}_{i}+\mathbf{q}_{e}$,
equation \ref{eq:481} is partitioned as

\begin{equation}
\dot{p}_{i}=-\mathbf{v}\cdot\nabla p_{i}-\gamma p_{i}\,\nabla\cdot\mathbf{v}+(\gamma-1)\left(-\overline{\boldsymbol{\pi}}:\nabla\mathbf{v}-\nabla\cdot\mathbf{q}_{i}+Q_{ie}\right)\label{eq:481.3}
\end{equation}

\begin{equation}
\dot{p}_{e}=-\mathbf{v}\cdot\nabla p_{e}-\gamma p_{e}\,\nabla\cdot\mathbf{v}+(\gamma-1)\left(\eta'J^{2}-\nabla\cdot\mathbf{q}_{e}-Q_{ie}\right)\label{eq:481.4}
\end{equation}
Consistent with the model simplifications outlined in section \ref{subsec:Closures_magnetised},
\begin{equation}
\mathbf{q}_{\alpha}=-n_{\alpha}\chi_{\parallel\alpha}\nabla_{\parallel}T_{\alpha}-n_{\alpha}\chi_{\perp\alpha}\nabla_{\perp}T_{\alpha}\label{eq:481.41}
\end{equation}
and

\begin{equation}
Q_{ie}=-Q_{ei}+\eta'J^{2}=\frac{3m_{e}}{m_{i}}\frac{Z_{eff}\,n(T_{e}-T_{i})}{\tau_{ei}}\label{eq:481.5}
\end{equation}
Note that the expressions presented in section \ref{subsec:Chapman-Enskog-closures}
were developed with the assumption that $Z_{eff}=1$. The dependence
of $\mathbf{q}_{\alpha}$(recall that $n_{e}=Z_{eff}\,n)$ and $Q_{ie}$
on $Z_{eff}$ has been reintroduced here. Recall that $Q_{\alpha}=\underset{\sigma}{\Sigma}Q_{\alpha\sigma}$
(equation \ref{eq:467-2}), and $Q_{\alpha\alpha}=0$, so that in
equations \ref{eq:472.472} and \ref{eq:472.473}, which are expressed
in the context of a two fluid system (ion and electron fluids), $Q_{i}\equiv Q_{ie}$,
and $Q_{e}\equiv Q_{ei}$. As mentioned, the MHD model we developed
neglects the terms in the expression for $Q_{ei}$ (equation \ref{eq:472.473})
that represent work done by the electron fluid against the thermal
force. Also, the model approximates resistivity and viscosity as being
isotropic - only the anisotropy pertaining to thermal diffusion is
included. The viscous heating term has been dropped from the electron
energy equation because the magnitudes of species viscous tensor components
$\left(\overline{\boldsymbol{\pi}}_{\alpha}\right)_{ij}$ (equation
\ref{eq:238}) are proportional to $\mu_{\alpha}=\rho_{\alpha}\nu_{\alpha}$.
Since $n_{i}\sim n_{e},$ $T_{i}\sim T_{e}$ (recall $\nu_{\alpha}$
scales with $T_{\alpha})$, and $m_{e}\ll m_{i}$, it is reasonable
to drop this term.

The Grad-Shafranov equation describes MHD equilibrium in the case
with toroidal symmetry:
\[
\Delta^{*}\psi=-\left(\mu_{0}r^{2}\frac{dp}{d\psi}+f\frac{df}{d\psi}\right)
\]
As shown in section \ref{subsec:GS_linear_lambda}, a simple model
assuming a linear dependence of $\lambda$ on $\psi$ leads to: 
\[
\Delta^{*}\psi=-\left(\mu_{0}r^{2}\frac{dp}{d\psi}+\bar{\lambda}^{2}\psi(1+\alpha(2\widetilde{\psi}-1))(1+\alpha(\widetilde{\psi}-1))\right)
\]
The numerical method implemented to code to solve the Grad-Shafranov
equation, using the discrete form of the $\Delta^{*}$ operator (section
\ref{subsec:Lapl_delstar}), is presented in appendix \ref{sec:Numerical-solution-of}.
Example solutions are presented in appendices \ref{sec:Numerical-solution-of}
and \ref{subsec:Equilibrium-solution-comparison}.

\newpage{}

\chapter{Formulation of discretized MHD equations for\protect \\
energy conservation\label{sec:Formulation-of-discretized}}

\section{Overview\label{subsec:Overview}}

The total energy of any physical system is conserved if there are
no sources or sinks of energy into or out of the system. Energy can
be distributed between components of the system, and can be converted
from one type of energy to another type, but the total energy will
remain constant if the system is a completely isolated one. In a plasma
system, the energy is composed of kinetic energy, thermal energy and
magnetic energy (in this discussion, internal energy due to excited
states, electron energy converted to potential energy in ionization
processes, and photon energy lost during radiative recombination reactions,
is ignored). Moving charged particles constitute kinetic energy, and
relative motion between the ion and electron fluids constitute currents,
which cause the magnetic fields associated with magnetic energy. The
frictional force between ions and electrons leads to resistive (ohmic)
heating, and tends to reduce the relative motion between the ion and
electron fluids, so that currents are resistively dissipated. The
resistive contribution to the reduction of the system\textquoteright s
magnetic energy is balanced by the increase in thermal energy due
to ohmic heating. Kinetic energy is dissipated by viscous forces between
particles of the same species, which leads to viscous heating. The
viscous contribution to the reduction of the system\textquoteright s
kinetic energy is balanced by an increase in thermal energy due to
viscous heating. In general, total energy in a real system, for example
a plasma contained within a vacuum vessel, is \emph{not} conserved
because heat is lost from the system by thermal conduction through
the vessel walls, and there is a radiative flux of electromagnetic
energy out of the system. However, energy loss due to thermal and
Poynting fluxes through the system boundary can be eliminated in the
physical system if the vessel wall is perfectly thermally insulating
and perfectly electrically conducting. 

In this appendix, a discrete form of the momentum equation is developed
which, on account of the inherent properties of the discrete differential
matrix operators, ensures that total energy of the system described
by the complete set of discretised MHD equations, in cylindrical geometry
with azimuthal symmetry, is conserved with appropriate boundary conditions.
In the process of developing the discrete form of the momentum equation,
discrete forms of the continuity and energy equations, and expressions
defining the evolution of the axisymmetric magnetic field, will be
defined. Note that the forms of the element-to-node differential operators
(section \ref{subsec:Drn}) were unknown at the time of development
of the discrete momentum equation for system energy conservation.
At that time, only the node-to-element and node-to-node operators
were defined - the process outlined in this section helped to define
the element-to-node operators and their properties. The rate of change
of total system energy can be can be expressed in discretized form
in terms of the discrete differential operators and the discretised
fields quantities. Leaving the discrete forms for $\dot{v}_{\beta}$
as the unknowns to be solved for, discretised expressions for $\dot{\rho},\,\dot{p},\,\dot{\psi}$
and $\dot{f}$ will be substituted in the discrete form of equation
\ref{eq:480.61}, which is repeated here for convenience:{\small{}
\begin{align*}
\dot{U}_{Total} & =\dot{U}_{K}+\dot{U}_{Th}+\dot{U}_{M}=\int\left[\left(\frac{1}{2}\dot{\rho}v^{2}+\rho\mathbf{v}\cdot\dot{\mathbf{v}}\right)+\frac{\dot{p}}{\gamma-1}+\frac{1}{2\mu{}_{0}}\left(\frac{\partial}{\partial t}\left(\left(\frac{\nabla\psi}{r}\right)^{2}+\left(\frac{f}{r}\right)^{2}\right)\right)\right]\:dV
\end{align*}
}The thermal diffusion term is temporarily dropped from the expression
for $\dot{p}$, but will be re-included later. It will turn out that
we will be able to transmute the terms in the resultant discrete expression
for $\dot{U}_{Total}$ as the sum of three collections of terms that
are coefficients of the three velocity components. The form of the
expression will be:{\footnotesize{}
\begin{equation}
\dot{U}_{Total}=\underline{dV}^{T}*\left\{ \Sigma_{\beta}\left(\underline{v_{\beta}}\circ\left(\underline{\rho}\circ\underline{\dot{v}_{\beta}}+\underline{X_{1\beta}}+\underline{X_{2\beta}}+....+\underline{X_{n\beta}}\right)\right)\right\} =0\label{eq:223.2}
\end{equation}
}The $n$ quantities $\underline{X_{j\beta}}\,\,[N_{n}\times1]$ are
complex expressions involving various physical parameters and mesh-based
fields. $n$ will have the same value for the $r$ and $z$ coordinates,
but on account of the assumption of axisymmetry, $n\phi<nr=nz$. The
identity expressed in equation \ref{eq:223.2} can hold only if, for
each coordinate $\beta$, either $v_{\beta}=0$ at every node (a useless
solution), or if $\left(\underline{\rho}\circ\underline{\dot{v}_{\beta}}+\underline{X_{1\beta}}+\underline{X_{2\beta}}+....+\underline{X_{n\beta}}\right)=0$,
for each coordinate $\beta$. The latter expressions lead to discrete
forms, that ensure total system energy conservation, for the three
components of the momentum equation:
\begin{equation}
\underline{\dot{v}_{\beta}}=-\left(\underline{X_{1\beta}}+\underline{X_{2\beta}}+....+\underline{X_{n\beta}}\right)\oslash\underline{\rho}\label{eq:223.3}
\end{equation}
Naturally, the resultant expressions will be the discrete analogs
of the continuous form of the momentum equation components. Various
transpose identities, and some of the inherent properties of the discrete
differential operators, will be used in the rearrangement of equation
\ref{eq:480.61} in terms of discrete quantities. Along with the matrix
transpose operation defined with equation \ref{eq:516.01}, note that
the following relationships hold for any column vectors $\underline{a},\:\underline{b}$,
and $\underline{c}$, with dimensions $N_{n}\times1$, where $\underline{\underline{A}}$
$[N_{n}\times N_{n}]$ is a diagonal array with its main diagonal
composed of the elements of $\underline{a}$: 
\begin{equation}
\underline{a}{}^{T}*(\underline{b}\circ\underline{c})=\underline{b}{}^{T}*\underline{\underline{A}}*\underline{c}=\underline{c}{}^{T}*\underline{\underline{A}}*\underline{b}\label{eq:116-2}
\end{equation}

\begin{equation}
\underline{a}{}^{T}*\underline{b}=\underline{c}{}^{T}*(\underline{a}\circ\underline{b}\oslash\underline{c})\label{eq:116.1}
\end{equation}

This appendix is arranged as follows. A discrete expression for $\dot{U}_{K}$,
the rate of change of system kinetic energy, is derived in section
\ref{subsec:KineticEnergy}. A discrete form of the expression for
$\dot{U}_{Th}$, the rate of change of system thermal energy, is derived
in section \ref{sec:Thermal-Energy}. The focus of section \ref{subsec:Viscous-heating}
is the part of $\dot{U}_{Th}$ that is due to viscosity, leading to
discrete forms of the viscous terms in the energy and momentum equations.
A discrete form of the expression for $\dot{U}_{M}$, the rate of
change of system magnetic energy, is derived in section \ref{subsec:Magnetic-energy},
along with the discrete form of the ohmic heating term in the energy
equation, and discrete forms of the expressions for $\dot{\psi}$
and $\dot{f}$. The discrete form of $\dot{U}_{Total}$ ($i.e.,$
equation \ref{eq:223.2}) is assembled to yield the discrete forms
of the components of the momentum equation in section \ref{sec:Assembly-of-discretised}.
Finally, the complete set of discretised equations for the axisymmetric
MHD model, in a form that ensures conservation of total system energy,
are collected in appendix summary \ref{sec:SummaryB}.

\section{\label{subsec:KineticEnergy}Kinetic Energy}

The rate of change of system kinetic energy can be expressed as
\begin{align}
\dot{U}_{K} & =\int\left(\frac{1}{2}\dot{\rho}v^{2}+\rho\mathbf{v}\cdot\dot{\mathbf{v}}\right)dV\label{eq:212.0}
\end{align}
In discretised form, this is:

\begin{equation}
\dot{U}_{K}=\underline{dV}^{T}*\left\{ \underset{1}{\frac{1}{2}\underline{\dot{\rho}}\circ\underline{v}^{2}}+\underset{2}{\underline{\rho}\circ\underline{v_{r}}\circ\underline{\dot{v}_{r}}}+\underset{3}{\underline{\rho}\circ\underline{v_{\phi}}\circ\underline{\dot{v}_{\phi}}}+\underset{4}{\underline{\rho}\circ\underline{v_{z}}\circ\underline{\dot{v}_{z}}}\right\} \label{eq:212.1}
\end{equation}
From equation \ref{eq: 479.01}, the discrete form of the mass continuity
equation is 
\begin{equation}
\underline{\dot{\rho}}=-\underline{\underline{\nabla}}\cdot(\underline{\rho}\circ\underline{\mathbf{v}})\label{eq:212.2}
\end{equation}
Hence, the first term in equation \ref{eq:212.1}, a scalar expression,
is:
\begin{equation}
\underset{1}{\dot{U}_{K}}=\underline{dV}\,{}^{T}*\left\{ \underset{1a}{-\frac{1}{2}\left(\underline{\underline{Dr}}*\left(\underline{r}\circ\underline{\rho}\circ\underline{v_{r}}\right)\right)\circ\underline{v}^{2}\oslash\underline{r}}-\underset{1b}{\frac{1}{2}\left(\underline{\underline{Dz}}*\left(\underline{r}\circ\underline{\rho}\circ\underline{v_{z}}\right)\right)\circ\underline{v}^{2}\oslash\underline{r}}\right\} \label{eq:229}
\end{equation}
The first term here is:

\begin{align*}
\underset{1a}{\dot{U}_{K}} & =\frac{2\pi}{3}\left(\underline{s}\circ\underline{r}\right){}^{T}*\left\{ -\frac{1}{2}\left(\underline{\underline{Dr}}*\left(\underline{r}\circ\underline{\rho}\circ\underline{v_{r}}\right)\right)\circ\underline{v}^{2}\oslash\underline{r}\right\} \\
 & =-\frac{\pi}{3}\underline{s}{}^{T}*\left\{ \left(\underline{\underline{Dr}}*\left(\underline{r}\circ\underline{\rho}\circ\underline{v_{r}}\right)\right)\circ\underline{v}^{2}\right\} \\
 & =-\frac{\pi}{3}\left(\underline{v}^{2}\right){}^{T}*\left\{ \underline{\underline{S}}*\underline{\underline{Dr}}*\left(\underline{r}\circ\underline{\rho}\circ\underline{v_{r}}\right)\right\}  & \mbox{(use eqn. \ref{eq:116-2})}\\
 & =-\frac{\pi}{3}\left(\underline{r}\circ\underline{\rho}\circ\underline{v_{r}}\right){}^{T}*\underline{\underline{Dr}}{}^{T}*\underline{\underline{S}}{}^{T}*\underline{v}^{2} & (\mbox{transpose the scalar})
\end{align*}
The node-to-node differential operators have the property (equations
\ref{eq:510.23} and \ref{eq:510.24}) that 
\begin{equation}
\left(\underline{\underline{S}}*\underline{\underline{Dr}}+\underline{\underline{Dr}}^{T}*\underline{\underline{S}}\right)_{int}=\left(\underline{\underline{S}}*\underline{\underline{Dz}}+\underline{\underline{Dz}}^{T}*\underline{\underline{S}}\right)_{int}=0\label{eq:230}
\end{equation}
where $\underline{\underline{S}}^{T}=\underline{\underline{S}}$.
Since we will set $v_{r}=0$ on the boundary nodes, we can use the
property to re-express $\underset{1a}{\dot{U}_{K}}$:

\begin{align*}
\underset{1a}{\dot{U}_{K}} & =+\frac{\pi}{3}\left(\underline{r}\circ\underline{\rho}\circ\underline{v_{r}}\right){}^{T}*\underline{\underline{S}}*\underline{\underline{Dr}}*\underline{v}^{2} & \mbox{(use eqn. \ref{eq:230})}\\
 & =\frac{\pi}{3}\underline{s}{}^{T}*\left\{ \left(\underline{r}\circ\underline{\rho}\circ\underline{v_{r}}\right)\circ\left(\underline{\underline{Dr}}*\underline{v}^{2}\right)\right\}  & \mbox{(use eqn. \ref{eq:116-2})}\\
 & =\frac{2\pi}{3}\left(\underline{s}\circ\underline{r}\right){}^{T}*\left\{ \left(\underline{\rho}\circ\underline{v_{r}}\right)\circ\left(\underline{\underline{Dr}}*\left(\frac{\underline{v}^{2}}{2}\right)\right)\right\} \\
 & =\underline{dV}^{T}*\left\{ \underline{v_{r}}\circ\left(\underline{\rho}\circ\left(\underline{\underline{Dr}}*\left(\frac{\underline{v}^{2}}{2}\right)\right)\right)\right\} 
\end{align*}
Repeating the process for the second term in equation \ref{eq:229},
we arrive at the required form of $\dot{U}_{K}$:{\footnotesize{}
\begin{align}
\dot{U}_{K} & =\underline{dV}\,{}^{T}*\left\{ \underline{v_{r}}\circ\left(\underline{\rho}\circ\left(\underline{\underline{Dr}}*\left(\frac{\underline{v}^{2}}{2}\right)\right)\right)+\underline{v_{z}}\circ\left(\underline{\rho}\circ\left(\underline{\underline{Dz}}*\left(\frac{\underline{v}^{2}}{2}\right)\right)\right)+\underline{\rho}\circ\underline{v_{r}}\circ\underline{\dot{v}_{r}}+\underline{\rho}\circ\underline{v_{\phi}}\circ\underline{\dot{v}_{\phi}}+\underline{\rho}\circ\underline{v_{z}}\circ\underline{\dot{v}_{z}}\right\} \label{eq:230.1}
\end{align}
}We want to reproduce the advective term $-\mathbf{v\cdot}\nabla\mathbf{v}$
in the RHS of the momentum equation. Using the vector identity $\mathbf{A}\times(\nabla\times\mathbf{B})=(\nabla\mathbf{B})\cdot\mathbf{A}-(\mathbf{A}\cdot\nabla)\mathbf{B}$,
it can be seen that $-(\mathbf{v}\cdot\nabla)\mathbf{v}=\mathbf{v}\times(\nabla\times\mathbf{v})-\nabla(v^{2}/2)$.
From equation \ref{eq:230.1}, it is evident that the term $-\nabla(v^{2}/2)$
has arisen naturally in the formulation for a momentum equation, so
following from the procedure outlined above, we need to add the term
$\int(-\rho\mathbf{v}\cdot(\mathbf{v}\times(\nabla\times\mathbf{v})))\,dV$
to $\dot{U}_{K}$ in order to reproduce the term $-\mathbf{v\cdot}\nabla\mathbf{v}$.
This addition will not affect the net value of $\dot{U}_{K}$, because
$-\rho\mathbf{v}\cdot(\mathbf{v}\times(\nabla\times\mathbf{v}))=0$,
since $\mathbf{v}\perp(\mathbf{v}\times(\nabla\times\mathbf{v}))$.
In cylindrical coordinates with azimuthal symmetry, $\mathbf{v}\times(\nabla\times\mathbf{v})$
may be expanded as:

\begin{align*}
\mathbf{v}\times(\nabla\times\mathbf{v}) & =\nabla(v^{2}/2)-(\mathbf{v}\cdot\nabla)\mathbf{v}\\
 & =\hat{\mathbf{r}}\left(\frac{\partial}{\partial r}(v^{2}/2)-v_{r}\frac{\partial v_{r}}{\partial r}-v_{z}\frac{\partial v_{r}}{\partial z}+\frac{v_{\phi}^{2}}{r}\right)+\widehat{\boldsymbol{\phi}}\left(-v_{r}\frac{\partial v_{\phi}}{\partial r}-v_{z}\frac{\partial v_{\phi}}{\partial z}-\frac{v_{\phi}v_{r}}{r}\right)\\
 & +\hat{\mathbf{z}}\left(\frac{\partial}{\partial z}(v^{2}/2)-v_{r}\frac{\partial v_{z}}{\partial r}-v_{z}\frac{\partial v_{z}}{\partial z}\right)
\end{align*}
Noting that $\frac{\partial}{\partial r}(v^{2}/2)=\mathbf{v}\cdot\frac{\partial\mathbf{v}}{\partial r}$
and $\frac{\partial}{\partial z}(v^{2}/2)=\mathbf{v}\cdot\frac{\partial\mathbf{v}}{\partial z}$,
this can be written as:

{\small{}
\begin{align*}
\mathbf{v}\times(\nabla\times\mathbf{v}) & =\hat{\mathbf{r}}\left(v_{r}\bcancel{\left(\frac{\partial v_{r}}{\partial r}-\frac{\partial v_{r}}{\partial r}\right)}+v_{z}\left(\frac{\partial v_{z}}{\partial r}-\frac{\partial v_{r}}{\partial z}\right)+v_{\phi}\left(\frac{\partial v_{\phi}}{\partial r}+\frac{v_{\phi}}{r}\right)\right)\\
 & +\widehat{\boldsymbol{\phi}}\left(-v_{r}\left(\frac{\partial v_{\phi}}{\partial r}+\frac{v_{\phi}}{r}\right)-v_{z}\frac{\partial v_{\phi}}{\partial z}\right)+\hat{\mathbf{z}}\left(v_{r}\left(\frac{\partial v_{r}}{\partial z}-\frac{\partial v_{z}}{\partial r}\right)+v_{z}\bcancel{\left(\frac{\partial v_{z}}{\partial z}-\frac{\partial v_{z}}{\partial z}\right)}+v_{\phi}\frac{\partial v_{\phi}}{\partial z}\right)
\end{align*}
}Since $\frac{\partial v_{\phi}}{\partial r}+\frac{v_{\phi}}{r}=\frac{1}{r}\frac{\partial(rv_{\phi})}{\partial r}$
and $\frac{\partial r}{\partial z}=0$, this becomes 

{\small{}
\begin{align}
\mathbf{v}\times(\nabla\times\mathbf{v}) & =\hat{\mathbf{r}}\left(-v_{z}\left(\frac{\partial v_{r}}{\partial z}-\frac{\partial v_{z}}{\partial r}\right)+\frac{v_{\phi}}{r}\frac{\partial(rv_{\phi})}{\partial r}\right)+\widehat{\boldsymbol{\phi}}\left(-\frac{v_{r}}{r}\frac{\partial(rv_{\phi})}{\partial r}-\frac{v_{z}}{r}\frac{\partial(rv_{\phi})}{\partial z}\right)\nonumber \\
 & \,\,\,+\hat{\mathbf{z}}\left(v_{r}\left(\frac{\partial v_{r}}{\partial z}-\frac{\partial v_{z}}{\partial r}\right)+\frac{v_{\phi}}{r}\frac{\partial(rv_{\phi})}{\partial z}\right)\nonumber \\
\label{eq:120.2}
\end{align}
}The discretised form of this expression may be used to add $\int(-\rho\mathbf{v}\cdot(\mathbf{v}\times(\nabla\times\mathbf{v})))\,dV$
to the RHS of equation \ref{eq:230.1} as follows

\begin{align}
\dot{U}_{K} & =\underline{dV}\,^{T}*\biggl\{\underline{v_{r}}\circ\left(\underline{\rho}\circ\left(\underline{\underline{Dr}}*\left(\frac{\underline{v}^{2}}{2}\right)\right)\right)+\underline{v_{z}}\circ\left(\underline{\rho}\circ\left(\underline{\underline{Dz}}*\left(\frac{\underline{v}^{2}}{2}\right)\right)\right)\nonumber \\
 & +\underline{\rho}\circ\underline{v_{r}}\circ\underline{\dot{v}_{r}}+\underline{\rho}\circ\underline{v_{\phi}}\circ\underline{\dot{v}_{\phi}}+\underline{\rho}\circ\underline{v_{z}}\circ\underline{\dot{v}_{z}}\nonumber \\
 & -\underline{\rho}\circ\underline{v_{r}}\circ\left(-\underline{v_{z}}\circ\left(\underline{\underline{Dz}}*\underline{v_{r}}-\underline{\underline{Dr}}*\underline{v_{z}}\right)+\underline{v_{\phi}}\circ\left(\underline{\underline{Dr}}*\left(\underline{r}\circ\underline{v_{\phi}}\right)\right)\oslash\underline{r}\right)\label{eq:234}\\
 & -\underline{\rho}\circ\underline{v_{\phi}}\circ\left(-\underline{\mathbf{v}}\cdot\left(\underline{\underline{\nabla}}\,\,\left(\underline{r}\circ\underline{v_{\phi}}\right)\right)\oslash\underline{r}\right)\nonumber \\
 & -\underline{\rho}\circ\underline{v_{z}}\circ\left(\underline{v_{r}}\circ\left(\underline{\underline{Dz}}*\underline{v_{r}}-\underline{\underline{Dr}}*\underline{v_{z}}\right)+\underline{v_{\phi}}\circ\left(\underline{\underline{Dz}}*\left(\underline{r}\circ\underline{v_{\phi}}\right)\right)\oslash\underline{r}\right)\biggr\}\nonumber 
\end{align}
Equation \ref{eq:234} is the final discretized form for the rate
of change of kinetic energy. A similar technique will be applied to
the terms in $\dot{U}_{M}$ and $\dot{U}_{Th}$, leading to expressions
in the form of equations \ref{eq:223.2} and \ref{eq:223.3}. Note
that the $3^{rd},\,4^{th},$ and $5^{th}$ terms in equation \ref{eq:234}
will be the only terms containing time-derivatives of the velocity
components in the discretised form of $\dot{U}_{Total}$. 

\section{Thermal Energy\label{sec:Thermal-Energy}}

The rate of change of system thermal energy can be expressed as
\begin{equation}
\dot{U}_{Th}=\int\frac{\dot{p}}{\gamma-1}\:dV\label{eq:235}
\end{equation}
Ignoring the thermal diffusion term for now, the single fluid energy
equation (\ref{eq:481}) is\\
$\dot{p}=-\mathbf{v}\cdot\nabla p-\gamma p\nabla\cdot\mathbf{v}+(\gamma-1)\left(\eta'J^{2}-\overline{\boldsymbol{\pi}}:\nabla\mathbf{v}\right)$.
The corresponding discrete form can be expressed as 
\begin{equation}
\underline{\dot{p}}=-\underline{\mathbf{v}}\cdot\left(\underline{\underline{\nabla}}\,\,\underline{p}\right)-\gamma\,\underline{p}\circ\left(\underline{\underline{\nabla}}\cdot\underline{\mathbf{v}}\right)+(\gamma-1)\,\left(\eta'J^{2}-\overline{\boldsymbol{\pi}}:\nabla\mathbf{v}\right)_{disc}\label{eq:235.01}
\end{equation}
where $\left(\eta'J^{2}-\overline{\boldsymbol{\pi}}:\nabla\mathbf{v}\right)_{disc}$
is the discrete form of $\eta'J^{2}-\overline{\boldsymbol{\pi}}:\nabla\mathbf{v}$;
explicit forms of these terms will be developed. The rate of change
of system thermal energy in discrete form is 
\begin{align}
\dot{U}_{Th} & =\underline{dV}\,{}^{T}*\biggl\{\frac{1}{\gamma-1}\biggl(\underset{1}{\underbrace{-\underline{v_{r}}\circ\left(\underline{\underline{Dr}}*\underline{p}\right)}}\underset{2}{\underbrace{-\underline{v_{z}}\circ\left(\underline{\underline{Dz}}*\underline{p}\right)}}\nonumber \\
 & -\gamma\left(\underset{3}{\underbrace{\underline{p}\circ\left(\underline{\underline{Dr}}*\left(\underline{r}\circ\underline{v_{r}}\right)\right)\oslash\underline{r}}}+\underset{4}{\underbrace{\underline{p}\circ\left(\underline{\underline{Dz}}*\left(\underline{r}\circ\underline{v_{z}}\right)\right)\oslash\underline{r}}}\right)\biggr)\biggr\}\label{eq:235.1}\\
 & +\underset{5}{\dot{U}_{Th\pi}}+\underset{6}{\dot{U}{}_{Th\phi\eta}}+\underset{7}{\dot{U}{}_{Th\theta\eta}}\nonumber 
\end{align}
Here, $\dot{U}{}_{Th\pi}=\int\left(-\overline{\boldsymbol{\pi}}:\nabla\mathbf{v}\right)dV$
is the rate of change of thermal energy due to viscous heating, $\dot{U}{}_{Th\phi\eta}=\int\eta'J_{\phi}^{2}\,dV$
is the rate of change of thermal energy in the volume associated with
ohmic heating due to toroidal currents, and $\dot{U}{}_{Th\theta\eta}=\int\eta'J_{\theta}^{2}\,dV$
is the rate of change of thermal energy in the volume associated with
ohmic heating due to poloidal currents.

The first two terms in equation \ref{eq:235.1} are already in the
required form. The $3^{rd}$ term can be expressed as follows:

\begin{align*}
\underset{3}{\dot{U}_{Th}} & =\frac{\gamma}{\gamma-1}\,\underline{dV}^{T}*\left\{ -\underline{p}\circ\left(\underline{\underline{Dr}}*\left(\underline{r}\circ\underline{v_{r}}\right)\right)\oslash\underline{r}\right\} \\
 & =\frac{2\pi\gamma}{3(\gamma-1)}\,\underline{s}{}^{T}*\left\{ -\underline{p}\circ\left(\underline{\underline{Dr}}*\left(\underline{r}\circ\underline{v_{r}}\right)\right)\right\} \\
 & =-\frac{2\pi\gamma}{3(\gamma-1)}\,\left(\underline{p}{}^{T}*\underline{\underline{S}}*\left(\underline{\underline{Dr}}*\left(\underline{r}\circ\underline{v_{r}}\right)\right)\right) & \mbox{(use eqn. \ref{eq:116-2})}\\
 & =-\frac{2\pi\gamma}{3(\gamma-1)}\,\left(\left(\underline{r}\circ\underline{v_{r}}\right){}^{T}*\underline{\underline{Dr}}{}^{T}*\underline{\underline{S}}*\underline{p}\right) & \mbox{\ensuremath{\mbox{(transpose the scalar)}}}\\
 & =+\frac{2\pi\gamma}{3(\gamma-1)}\,\left(\left(\underline{r}\circ\underline{v_{r}}\right){}^{T}*\underline{\underline{S}}*\underline{\underline{Dr}}*\underline{p}\right) & \mbox{(use eqn. \ref{eq:230})}\\
 & =\frac{2\pi\gamma}{3(\gamma-1)}\,\underline{s}{}^{T}*\left(\left(\underline{r}\circ\underline{v_{r}}\right)\circ\left(\underline{\underline{Dr}}*\underline{p}\right)\right) & \mbox{(use eqn. \ref{eq:116-2})}\\
 & =\frac{\gamma}{\gamma-1}\,\underline{dV}^{T}*\left\{ \underline{v_{r}}\circ\left(\underline{\underline{Dr}}*\underline{p}\right)\right\} 
\end{align*}
Similarly, the $4^{th}$ term in equation \ref{eq:235.1} can be re-expressed
as :
\begin{equation}
\underset{4}{\dot{U}_{Th}}=\frac{\gamma}{\gamma-1}\,\underline{dV}^{T}*\left(\underline{v_{z}}\circ\left(\underline{\underline{Dz}}*\underline{p}\right)\right)\label{eq:235.3}
\end{equation}
From this, the first four terms in equation \ref{eq:235.1} can be
written as:

\begin{align}
\underset{1\rightarrow4}{\dot{U}_{Th}} & =\underline{dV}^{T}*\left\{ \underline{v_{r}}\circ\left(\underline{\underline{Dr}}*\underline{p}\right)+\underline{v_{z}}\circ\left(\underline{\underline{Dz}}*\underline{p}\right)\right\} \label{eq:235.4}
\end{align}

\subsection{\label{subsec:Viscous-heating}Viscous heating}

The $5^{th}$ term in equation \ref{eq:235.1} represents the part
of the rate of change thermal energy that is due to viscous heating:
\begin{equation}
\dot{U}{}_{Th\pi}=\int-\overline{\boldsymbol{\pi}}:\nabla\mathbf{v}\:dV\label{eq:237}
\end{equation}
For the viscous terms, we use, for simplicity, the unmagnetised version
of the viscous stress tensor in axisymmetric cylindrical coordinates.
In cylindrical coordinates, with azimuthal symmetry, the components
of the viscosity tensor $\overline{\boldsymbol{\pi}}$ are \cite{Cloutman}:
\begin{equation}
\overline{\boldsymbol{\pi}}=-\mu\left(\begin{array}{ccc}
2\frac{\partial v_{r}}{\partial r}-\frac{2}{3}\nabla\cdot\mathbf{v} & r\frac{\partial}{\partial r}\left(\frac{v_{\phi}}{r}\right) & \frac{\partial v_{z}}{\partial r}+\frac{\partial v_{r}}{\partial z}\\
r\frac{\partial}{\partial r}\left(\frac{v_{\phi}}{r}\right) & 2\frac{v_{r}}{r}-\frac{2}{3}\nabla\cdot\mathbf{v} & \frac{\partial v_{\phi}}{\partial z}\\
\frac{\partial v_{z}}{\partial r}+\frac{\partial v_{r}}{\partial z} & \frac{\partial v_{\phi}}{\partial z} & 2\frac{\partial v_{z}}{\partial z}-\frac{2}{3}\nabla\cdot\mathbf{v}
\end{array}\right)\label{eq:238}
\end{equation}
where $\mu\,${[}kg$\,$m$^{-1}\mbox{s}{}^{-1}${]} is dynamic viscosity.
In the following it is assumed for generality that $\mu$ is spatially
dependent, $\mu=\mu(\mathbf{r})$. The contraction (inner product)
of two second order tensors \cite{outerproduct} is defined as $\underline{\boldsymbol{T}}:\underline{\boldsymbol{U}}=\underset{i}{\Sigma}\underset{j}{\Sigma}T_{ij}U_{ij}$,
and $\nabla\mathbf{v}$, in cylindrical coordinates with azimuthal
symmetry is

\[
\nabla\mathbf{v}=\left(\begin{array}{ccc}
\frac{\partial v_{r}}{\partial r} & -\frac{v_{\phi}}{r} & \frac{\partial v_{r}}{\partial z}\\
\frac{\partial v_{\phi}}{\partial r} & \frac{v_{r}}{r} & \frac{\partial v_{\phi}}{\partial z}\\
\frac{\partial v_{z}}{\partial r} & 0 & \frac{\partial v_{z}}{\partial z}
\end{array}\right)
\]
{\footnotesize{}
\begin{align*}
\Rightarrow\overline{\boldsymbol{\pi}}:\nabla\mathbf{v} & =-\mu\biggl[\underset{1}{\underbrace{\left(2\frac{\partial v_{r}}{\partial r}-\frac{2}{3}\nabla\cdot\mathbf{v}\right)\frac{\partial v_{r}}{\partial r}}}+\underset{2}{\underbrace{\left(r\frac{\partial}{\partial r}\left(\frac{v_{\phi}}{r}\right)\right)\left(-\frac{v_{\phi}}{r}\right)}}+\underset{3}{\underbrace{\left(\frac{\partial v_{z}}{\partial r}+\frac{\partial v_{r}}{\partial z}\right)\left(\frac{\partial v_{r}}{\partial z}\right)}}+\underset{4}{\underbrace{r\frac{\partial}{\partial r}\left(\frac{v_{\phi}}{r}\right)\left(\frac{\partial v_{\phi}}{\partial r}\right)}}\\
 & +\underset{5}{\underbrace{\left(2\frac{v_{r}}{r}-\frac{2}{3}\nabla\cdot\mathbf{v}\right)\left(\frac{v_{r}}{r}\right)}}+\underset{6}{\underbrace{\left(\frac{\partial v_{\phi}}{\partial z}\right)^{2}}}+\underset{7}{\underbrace{\left(\frac{\partial v_{z}}{\partial r}+\frac{\partial v_{r}}{\partial z}\right)\left(\frac{\partial v_{z}}{\partial r}\right)}}+\underset{8}{\underbrace{\left(2\frac{\partial v_{z}}{\partial z}-\frac{2}{3}\nabla\cdot\mathbf{v}\right)\left(\frac{\partial v_{z}}{\partial z}\right)}}\biggr]
\end{align*}
}The $2^{nd}$ and $4^{th}$ terms can be combined: $\underset{2+4}{\overline{\boldsymbol{\pi}}:\nabla\mathbf{v}}=-\mu\left(r\frac{\partial}{\partial r}\left(\frac{v_{\phi}}{r}\right)\right)^{2}$,
as can the $3^{rd}$ and $7^{th}$ terms:\\
 $\underset{3+7}{\overline{\boldsymbol{\pi}}:\nabla\mathbf{v}}=-\mu\left(\frac{\partial v_{z}}{\partial r}+\frac{\partial v_{r}}{\partial z}\right)^{2}$.
The sum of the $1^{st}$ , $5^{th}$ and $8^{th}$ terms is:\\
$\underset{1+5+8}{\overline{\boldsymbol{\pi}}:\nabla\mathbf{v}}=-2\mu\left(\left(\frac{\partial v_{r}}{\partial r}\right)^{2}+\left(\frac{v_{r}}{r}\right)^{2}+\left(\frac{\partial v_{z}}{\partial z}\right)^{2}\right)+\frac{2}{3}\mu\left(\frac{1}{r}\frac{\partial(rv_{r})}{\partial r}+\frac{\partial v_{z}}{\partial z}\right)^{2}$.
Combining terms, and noting that $\frac{\partial r}{\partial z}\rightarrow0$,
and defining the angular speed $\omega=\frac{v_{\phi}}{r}$, this
can be expressed as 

\begin{align}
\overline{\boldsymbol{\pi}}:\nabla\mathbf{v} & =-\mu\biggl[2\left(\left(\frac{\partial v_{r}}{\partial r}\right)^{2}+\left(\frac{v_{r}}{r}\right)^{2}+\left(\frac{\partial v_{z}}{\partial z}\right)^{2}\right)\nonumber \\
 & +\left(r\,\nabla\omega\right)^{2}+\left(\frac{\partial v_{z}}{\partial r}+\frac{\partial v_{r}}{\partial z}\right)^{2}-\frac{2}{3}\left(\frac{1}{r}\frac{\partial(rv_{r})}{\partial r}+\frac{1}{r}\frac{\partial(rv_{z})}{\partial z}\right)^{2}\biggr]\label{eq:239}
\end{align}
In cylindrical coordinates, the divergence of a second order tensor
\cite{tensorwiki} is
\begin{align*}
\nabla\cdot\overline{\boldsymbol{\pi}} & =\left(\frac{\partial\pi_{rr}}{\partial r}+\frac{1}{r}\left(\frac{\partial\pi_{\phi r}}{\partial\phi}+\pi_{rr}-\pi_{\phi\phi}\right)+\frac{\partial\pi_{zr}}{\partial z}\right)\hat{\mathbf{r}}\\
 & +\left(\frac{\partial\pi_{r\phi}}{\partial r}+\frac{1}{r}\left(\frac{\partial\pi_{\phi\phi}}{\partial\phi}+\pi_{r\phi}+\pi_{\phi r}\right)+\frac{\partial\pi_{z\phi}}{\partial z}\right)\boldsymbol{\hat{\mathbf{\phi}}}\\
 & +\left(\frac{\partial\pi_{rz}}{\partial r}+\frac{1}{r}\left(\frac{\partial\pi_{\phi z}}{\partial\phi}+\pi_{rz}\right)+\frac{\partial\pi_{zz}}{\partial z}\right)\hat{\mathbf{z}}
\end{align*}
For a symmetric tensor, and with azimuthal symmetry, this becomes
\begin{align}
\nabla\cdot\overline{\boldsymbol{\pi}} & =\left(\frac{1}{r}\frac{\partial}{\partial r}(r\pi_{rr})-\frac{\pi_{\phi\phi}}{r}+\frac{\partial\pi_{rz}}{\partial z}\right)\hat{\mathbf{r}}+\left(\frac{\partial\pi_{r\phi}}{\partial r}+\frac{2\pi_{r\phi}}{r}+\frac{\partial\pi_{\phi z}}{\partial z}\right)\widehat{\boldsymbol{\phi}}+\left(\frac{1}{r}\frac{\partial}{\partial r}(r\pi_{rz})+\frac{\partial\pi_{zz}}{\partial z}\right)\hat{\mathbf{z}}\nonumber \\
\nonumber \\
\label{eq:241}
\end{align}
Note that the reduction of the system's kinetic energy due to viscous
dissipation is balanced by an increase in thermal energy due to viscous
heating. Referring to equations \ref{eq:479} and \ref{eq:481}, the
rate of change of the energy associated with viscous heating in the
system is
\begin{align*}
\dot{U}_{vh} & =\int\left(\frac{\dot{p}_{vh}}{\gamma-1}+\rho\dot{\mathbf{v}}_{vh}\cdot\mathbf{v}\right)\:dV=-\int\left(\overline{\boldsymbol{\pi}}:\nabla\mathbf{v}+\mathbf{v}\cdot(\nabla\cdot\overline{\boldsymbol{\pi}})\right)\:dV\\
 & =-\int\nabla\cdot\left(\mathbf{v}\cdot\overline{\boldsymbol{\pi}}\right)\:dV & (\mbox{use eqn. \ref{eq:472.301}})\\
 & =-\int\left(\mathbf{v}\cdot\overline{\boldsymbol{\pi}}\right)\cdot d\boldsymbol{\Gamma}
\end{align*}
Thus, these terms balance if $\mathbf{v}_{\perp}|_{\Gamma}=0.$ It
can be shown that $\dot{U}_{vh\phi}$, the part of $\dot{U}_{vh}$
that contains terms with $v_{\phi}$, can be expressed as the integral
over the system's volume of the divergence of a vector that has the
quality that each component is a multiple of $v_{\phi}$. With the
boundary condition $v_{\phi}|_{\Gamma}=0$, the divergence theorem
implies that $\dot{U}_{vh\phi}=0.$ By the same technique, it can
be shown that $\dot{U}_{vhrz}$, the part of $\dot{U}_{vh}$ that
contains terms with $v_{r}$ and $v_{z}$ is zero with boundary conditions
$v_{r}|_{\Gamma}=v_{z}|_{\Gamma}=0$, so that the kinetic energy loss
due to viscous dissipation of toroidal/poloidal fluid motion is balanced
by an increase in thermal energy due to viscous heating arising from
toroidal/poloidal fluid motion.

With this in mind, the next step is to use the discrete expressions
for the viscous contribution to the system's thermal energy to derive
discrete expressions for the viscosity term in the momentum equation.

\subsubsection{Discretisation of $\dot{U}{}_{Th\pi}$\label{par:Disc_Uthpi}}

$ $\\
$ $\\
$\overline{\boldsymbol{\pi}}$ contains terms with squared derivatives,
so when we expand the discretized expression for $\dot{U}{}_{Th\pi}=\int-\overline{\boldsymbol{\pi}}:\nabla\mathbf{v}\:dV$
to get the parts of the expressions, pertaining to the viscous heating
term, that make up the coefficients of the velocity components, in
the same way as was done in section \ref{subsec:KineticEnergy}, we
will obtain terms with 2nd derivatives. For 2nd derivatives, we want
to apply consecutive applications of the node-to-element and element-to-node
differential operators for consistency with the three point stencil
of the numerical scheme. Therefore, the node-to-element operators
will be implemented whenever squared derivatives appear in the following
equation development process. 

Note that equation \ref{eq:239} leads naturally to the discrete volume-averaged
form of $-\overline{\boldsymbol{\pi}}:\nabla\mathbf{v}$, which will
be used in the discretised energy equation:
\begin{align}
\left(-\overline{\boldsymbol{\pi}}:\nabla\mathbf{v}\right)_{disc}=\underline{Q_{\pi}} & =\underline{\underline{W_{n}}}*\biggl[\underline{\mu^{e}}\circ\biggl\{2\left(\underline{\underline{Dr^{e}}}*\underline{v_{r}}\right)^{2}+2\left(\underline{\underline{Dz^{e}}}*\underline{v_{z}}\right)^{2}+\left(\underline{r^{e}}\circ\left(\underline{\underline{\nabla^{e}}}\,\,\underline{\omega}\right)\right)^{2}\nonumber \\
 & +\left(\underline{\underline{Dr^{e}}}*\underline{v_{z}}+\underline{\underline{Dz^{e}}}*\underline{v_{r}}\right)^{2}-\frac{2}{3}\left(\underline{\underline{\nabla^{e}}}\cdot\underline{\mathbf{v}}\right)^{2}\biggr\}\biggr]+2\,\underline{\mu}\circ\left(\underline{v_{r}}\oslash\underline{r}\right)^{2}\label{eq:242}
\end{align}
where $\underline{\underline{W_{n}}}$ is the volume-averaging operator
(equations \ref{eq:515} and \ref{eq:515.1}). Recalling that the
vector of elemental volumes is $\underline{dV^{e}}=2\pi\underline{s^{e}}\circ\underline{r^{e}}$,
then using equations \ref{eq:237} and \ref{eq:239}, the discrete
form of the part of the thermal energy that is due to viscous heating
is

{\scriptsize{}
\begin{align}
\dot{U}{}_{Th\pi} & =2\pi(\underline{s^{e}}\circ\underline{r^{e}}\circ\underline{\mu^{e}}){}^{T}*\biggl\{\underset{1}{\underbrace{2\left(\underline{\underline{Dr^{e}}}*\underline{v_{r}}\right)^{2}}}+\underset{2}{\underbrace{2\left(\underline{\underline{Dz^{e}}}*\underline{v_{z}}\right)^{2}}}\nonumber \\
 & \underset{3}{\underbrace{+\left(\underline{r^{e}}\circ\left(\underline{\underline{Dr^{e}}}*\underline{\omega}\right)\right)^{2}}}\underset{4}{\underbrace{+\left(\underline{r^{e}}\circ\left(\underline{\underline{Dz^{e}}}*\underline{\omega}\right)\right)^{2}}}+\underset{5}{\underbrace{\left(\underline{\underline{Dr^{e}}}*\underline{v_{z}}\right)^{2}}}+\underset{6}{\underbrace{2\left(\underline{\underline{Dr^{e}}}*\underline{v_{z}}\right)\circ\left(\underline{\underline{Dz^{e}}}*\underline{v_{r}}\right)}}+\underset{7}{\underbrace{\left(\underline{\underline{Dz^{e}}}*\underline{v_{r}}\right)^{2}}}\nonumber \\
 & -\frac{2}{3}\bigg[\underset{8}{\underbrace{\left(\left(\underline{\underline{Dr^{e}}}*\left(\underline{r}\circ\underline{v_{r}}\right)\right)\oslash\underline{r^{e}}\right)^{2}}}+\underset{9}{\underbrace{2\left(\left(\underline{\underline{Dr^{e}}}*\left(\underline{r}\circ\underline{v_{r}}\right)\right)\oslash\underline{r^{e}}\right)\circ\left(\left(\underline{\underline{Dz^{e}}}*\left(\underline{r}\circ\underline{v_{z}}\right)\right)\oslash\underline{r^{e}}\right)}}+\underset{10}{\underbrace{\left(\left(\underline{\underline{Dz^{e}}}*\left(\underline{r}\circ\underline{v_{z}}\right)\right)\oslash\underline{r^{e}}\right)^{2}}}\bigg]\biggr\}\nonumber \\
 & +\underset{11}{\underbrace{\frac{2\pi}{3}(\underline{s}\circ\underline{r}){}^{T}*\left(2\underline{\mu}\circ\left(\underline{v_{r}}\oslash\underline{r}\right)^{2}\right)}}\label{eq:254}
\end{align}
}Here, $\underline{\mu^{e}}$ and $\underline{\mu}$ are the vectors
of element-based and node-based dynamic viscosities, and are related
as $\underline{\mu^{e}}=<\underline{\mu}>^{e}$ (equation \ref{eq:502.31}).
Note that each term in equation \ref{eq:254} is a scalar. The $11^{th}$
term, corresponding to the integral over volume of the second term
in equation \ref{eq:239}, does not involve squared derivatives, so
this term can be evaluated using node-based quantities. Recall that
we want to express each scalar term in equation \ref{eq:254} in the
form: $\underline{dV}^{T}*\left(\underline{v_{\beta}}\circ\underline{X_{j\beta}}\right)$,
so that we can use the axiom $\dot{U}_{Total}=0$ to construct the
discretised form of the momentum equation. Only the $3^{rd}$ and
$4^{th}$ terms contain $\underline{v_{\phi}}$. Defining $\underline{w_{3}^{e}}=\underline{\mu^{e}}\circ\underline{s^{e}}\circ\underline{r^{e}}^{3}$,
the $3^{rd}$ term in equation \ref{eq:254} can be expressed as:

\begin{alignat}{2}
\underset{3}{\dot{U}{}_{Th\pi}} & =2\pi\left(\underline{w_{3}^{e}}\circ\left(\underline{\underline{Dr^{e}}}*\underline{\omega}\right)\right){}^{T}*\left(\underline{\underline{Dr^{e}}}*\underline{\omega}\right)\nonumber \\
 & =2\pi\,\underline{\omega}{}^{T}*\underline{\underline{Dr^{e}}}{}^{T}*\left(\underline{w_{3}^{e}}\circ\left(\underline{\underline{Dr^{e}}}*\underline{\omega}\right)\right) & \mbox{(transpose the scalar)}\nonumber \\
 & =\frac{2\pi}{3}(\underline{s}\circ\underline{r}){}^{T}*\left\{ \underline{\omega}\circ\left(\underline{\underline{Dr^{e}}}{}^{T}*\left(\underline{w_{3}^{e}}\circ\left(\underline{\underline{Dr^{e}}}*\underline{\omega}\right)\right)\right)\oslash(\underline{s}\circ\underline{r}/3)\right\}  & (\mbox{use eqn. \ref{eq:116.1}})\nonumber \\
 & =\underline{dV}^{T}*\left\{ \underline{v_{\phi}}\circ\left(3\underline{\underline{Dr^{e}}}{}^{T}*\left(\underline{w_{3}^{e}}\circ\left(\underline{\underline{Dr^{e}}}*\underline{\omega}\right)\right)\right)\oslash(\underline{s}\circ\underline{r}^{2})\right\} \label{eq:255}
\end{alignat}
Similarly, the $4^{th}$ term in equation \ref{eq:254} is 
\begin{equation}
\underset{4}{\dot{U}{}_{Th\pi}}=\underline{dV}^{T}*\left\{ \underline{v_{\phi}}\circ\left(3\underline{\underline{Dz^{e}}}{}^{T}*\left(\underline{w_{3}^{e}}\circ\left(\underline{\underline{Dz^{e}}}*\underline{\omega}\right)\right)\right)\oslash(\underline{s}\circ\underline{r}^{2})\right\} \label{eq:256}
\end{equation}
For convenience, we define:
\begin{equation}
\dot{U}{}_{Th\pi}=\dot{U}{}_{Th\pi r}+\dot{U}{}_{Th\pi\phi}+\dot{U}{}_{Th\pi z}\label{eq:258.1}
\end{equation}
where $\dot{U}{}_{Th\pi\beta}$ is the part of $\dot{U}{}_{Th\pi}$
that contains $v_{\beta}$. Adding equations \ref{eq:255} and \ref{eq:256},
we arrive at :
\begin{equation}
\dot{U}{}_{Th\pi\phi}=\underline{dV}^{T}*\left\{ \underline{v_{\phi}}\circ\underline{\varPi_{\phi}}\right\} \label{eq:258.2}
\end{equation}
where, with reference to the definitions of the element-to node differential
operators (equations \ref{eq:515.02} and \ref{eq:515.0})
\begin{equation}
\underline{\varPi_{\phi}}=-\left(\underline{\underline{\nabla_{n}}}\cdot\left(\underline{\mu^{e}}\circ\underline{r^{e}}^{2}\circ\left(\underline{\underline{\nabla^{e}}}\,\,\underline{\omega}\right)\right)\right)\oslash\underline{r}\label{eq:258.3}
\end{equation}
This term represents the discrete form of $\left(\nabla\cdot\overline{\boldsymbol{\pi}}\right)_{\phi}$,
and will appear in the $\phi$ component of the discretised momentum
equation. 

The $1^{st},\;6^{th},\;7^{th},\;8^{th},\;9^{th}$ and $11^{th}$ terms
in equation \ref{eq:254} include $r$ components of velocity and
will contribute to $\dot{U}{}_{Th\pi r}$. Defining $\underline{w_{1}^{e}}=\underline{\mu^{e}}\circ\underline{s^{e}}\circ\underline{r^{e}}$,
the $1^{st}$ term can be expressed as

\begin{alignat}{2}
\underset{1}{\dot{U}_{Th\pi}} & =4\pi\left(\underline{w_{1}^{e}}\circ\left(\underline{\underline{Dr^{e}}}*\underline{v_{r}}\right)\right){}^{T}*\left(\underline{\underline{Dr^{e}}}*\underline{v_{r}}\right)\nonumber \\
 & =4\pi\,\underline{v_{r}}{}^{T}*\underline{\underline{Dr^{e}}}{}^{T}*\left(\underline{w_{1}^{e}}\circ\left(\underline{\underline{Dr^{e}}}*\underline{v_{r}}\right)\right) & \mbox{(transpose the scalar)}\nonumber \\
 & =\frac{4\pi}{3}(\underline{s}\circ\underline{r}){}^{T}*\left\{ \underline{v_{r}}\circ\left(3\underline{\underline{Dr^{e}}}{}^{T}*\left(\underline{w_{1}^{e}}\circ\left(\underline{\underline{Dr^{e}}}*\underline{v_{r}}\right)\right)\right)\oslash(\underline{s}\circ\underline{r})\right\}  & (\mbox{use eqn. \ref{eq:116.1}})\nonumber \\
 & =\underline{dV}^{T}*\left\{ \underline{v_{r}}\circ\left(6\underline{\underline{Dr^{e}}}{}^{T}*\left(\underline{w_{1}^{e}}\circ\left(\underline{\underline{Dr^{e}}}*\underline{v_{r}}\right)\right)\right)\oslash(\underline{s}\circ\underline{r})\right\} \nonumber \\
 & =\underline{dV}^{T}*\left\{ \underline{v_{r}}\circ\left(-2\underline{\underline{Dr_{n}}}*\left(\underline{\mu^{e}}\circ\underline{r^{e}}\circ\left(\underline{\underline{Dr^{e}}}*\underline{v_{r}}\right)\right)\right)\oslash\underline{r}\right\}  & \mbox{(use eqn. \ref{eq:515.0})}\label{eq:259}
\end{alignat}
Similarly, the $7^{th}$ term becomes 
\begin{equation}
\underset{7}{\dot{U}{}_{Th\pi}}=\underline{dV}^{T}*\left\{ \underline{v_{r}}\circ\left(-\underline{\underline{Dz_{n}}}*\left(\underline{\mu^{e}}\circ\underline{r^{e}}\circ\left(\underline{\underline{Dz^{e}}}*\underline{v_{r}}\right)\right)\right)\oslash\underline{r}\right\} \label{eq:259.1}
\end{equation}
The $6^{th}$ term in equation \ref{eq:254} is: 

\begin{alignat*}{2}
\underset{6}{\dot{U}{}_{Th\pi}} & =4\pi\left(\underline{w_{1}^{e}}\circ\left(\underline{\underline{Dr^{e}}}*\underline{v_{z}}\right)\right){}^{T}*\left(\underline{\underline{Dz^{e}}}*\underline{v_{r}}\right)\\
 & =4\pi\,\underline{v_{r}}{}^{T}*\underline{\underline{Dz^{e}}}{}^{T}*\left(\underline{w_{1}^{e}}\circ\left(\underline{\underline{Dr^{e}}}*\underline{v_{z}}\right)\right) & \mbox{(transpose the scalar)}\\
 & =\frac{4\pi}{3}(\underline{s}\circ\underline{r}){}^{T}*\left\{ \underline{v_{r}}\circ\left(3\underline{\underline{Dz^{e}}}{}^{T}*\left(\underline{w_{1}^{e}}\circ\left(\underline{\underline{Dr^{e}}}*\underline{v_{z}}\right)\right)\right)\oslash(\underline{s}\circ\underline{r})\right\}  & (\mbox{use eqn. \ref{eq:116.1}})\\
 & =\underline{dV}^{T}*\left\{ \underline{v_{r}}\circ\left(-2\underline{\underline{Dz_{n}}}*\left(\underline{\mu^{e}}\circ\underline{r^{e}}\circ\left(\underline{\underline{Dr^{e}}}*\underline{v_{z}}\right)\right)\right)\oslash\underline{r}\right\}  & \mbox{\mbox{(use eqn. \ref{eq:515.0})}}
\end{alignat*}
In the first step above, we could equally have chosen\\
$\underset{6}{\dot{U}{}_{Th\pi}}=4\pi\left(\underline{w_{1}^{e}}\circ\left(\underline{\underline{Dz^{e}}}*\underline{v_{r}}\right)\right){}^{T}*\left(\underline{\underline{Dr^{e}}}*\underline{v_{z}}\right)$
which would have resulted in the expression\\
\[
\underset{6}{\dot{U}{}_{Th\pi}}=\underline{dV}^{T}*\left\{ \underline{v_{z}}\circ\left(-2\underline{\underline{Dr_{n}}}*\left(\underline{\mu^{e}}\circ\underline{r^{e}}\circ\left(\underline{\underline{Dz^{e}}}*\underline{v_{r}}\right)\right)\right)\oslash\underline{r}\right\} 
\]
so it is valid to express the $6^{th}$ term as:

\begin{align}
\underset{6}{\dot{U}{}_{Th\pi}} & =\underline{dV}^{T}*\biggl\{\underline{v_{r}}\circ\left(-\underline{\underline{Dz_{n}}}*\left(\underline{\mu^{e}}\circ\underline{r^{e}}\circ\left(\underline{\underline{Dr^{e}}}*\underline{v_{z}}\right)\right)\right)\oslash\underline{r}\nonumber \\
 & +\underline{v_{z}}\circ\left(-\underline{\underline{Dr_{n}}}*\left(\underline{\mu^{e}}\circ\underline{r^{e}}\circ\left(\underline{\underline{Dz^{e}}}*\underline{v_{r}}\right)\right)\right)\oslash\underline{r}\biggr\}\label{eq:260}
\end{align}
Defining $\underline{w_{0}^{e}}=\underline{\mu^{e}}\circ\underline{s^{e}}\oslash\underline{r^{e}}$,
the $8^{th}$ term in equation \ref{eq:254} is: 

\begin{alignat}{2}
\underset{8}{\dot{U}{}_{Th\pi}} & =-\frac{4}{3}\pi\left(\underline{w_{0}^{e}}\circ\left(\underline{\underline{Dr^{e}}}*\left(\underline{r}\circ\underline{v_{r}}\right)\right)\right){}^{T}*\left(\underline{\underline{Dr^{e}}}*\left(\underline{r}\circ\underline{v_{r}}\right)\right)\nonumber \\
 & =-\frac{4}{3}\pi\left(\underline{r}\circ\underline{v_{r}}\right)^{T}*\underline{\underline{Dr^{e}}}{}^{T}*\left(\underline{w_{0}^{e}}\circ\left(\underline{\underline{Dr^{e}}}*\left(\underline{r}\circ\underline{v_{r}}\right)\right)\right) & \mbox{(transpose the scalar)}\nonumber \\
 & =\underline{dV}^{T}*\left\{ \left(\underline{r}\circ\underline{v_{r}}\right)\circ\left(-2\underline{\underline{Dr^{e}}}{}^{T}*\left(\underline{w_{0}^{e}}\circ\left(\underline{\underline{Dr^{e}}}*\left(\underline{r}\circ\underline{v_{r}}\right)\right)\right)\right)\oslash(\underline{s}\circ\underline{r})\right\}  & \mbox{\ensuremath{(\mbox{use eqn. \ref{eq:116.1}})}}\nonumber \\
 & =\underline{dV}^{T}*\left\{ \underline{v_{r}}\circ\left(-2\underline{\underline{Dr^{e}}}{}^{T}*\left(\underline{w_{0}^{e}}\circ\left(\underline{\underline{Dr^{e}}}*\left(\underline{r}\circ\underline{v_{r}}\right)\right)\right)\right)\oslash\underline{s}\right\} \nonumber \\
 & =\underline{dV}^{T}*\left\{ \underline{v_{r}}\circ\left(\frac{2}{3}\underline{\underline{Dr_{n}}}*\left(\underline{\mu^{e}}\circ\left(\underline{\underline{Dr^{e}}}*\left(\underline{r}\circ\underline{v_{r}}\right)\right)\oslash\underline{r^{e}}\right)\right)\right\}  & \mbox{(use eqn. \ref{eq:515.0})}\label{eq:261}
\end{alignat}
Similarly, the $9^{th}$ term in equation \ref{eq:254} is: 

\begin{alignat*}{2}
\underset{9}{\dot{U}{}_{Th\pi}} & =-\frac{8}{3}\pi\left(\underline{w_{0}^{e}}\circ\left(\underline{\underline{Dr^{e}}}*\left(\underline{r}\circ\underline{v_{r}}\right)\right)\right){}^{T}*\left(\underline{\underline{Dz^{e}}}*\left(\underline{r}\circ\underline{v_{z}}\right)\right)\\
 & =\underline{dV}^{T}*\left\{ \underline{v_{z}}\circ\left(-4\underline{\underline{Dz^{e}}}{}^{T}*\left(\underline{w_{0}^{e}}\circ\left(\underline{\underline{Dr^{e}}}*\left(\underline{r}\circ\underline{v_{r}}\right)\right)\right)\right)\oslash\underline{s}\right\}  & \mbox{}\\
 & =\underline{dV}^{T}*\left\{ \underline{v_{z}}\circ\left(\frac{4}{3}\underline{\underline{Dz_{n}}}*\left(\underline{\mu^{e}}\circ\left(\underline{\underline{Dr^{e}}}*\left(\underline{r}\circ\underline{v_{r}}\right)\right)\oslash\underline{r^{e}}\right)\right)\right\} 
\end{alignat*}
In the first step above, we could equally have chosen: 
\[
\underset{9}{\dot{U}{}_{Th\pi}}=-\frac{8}{3}\pi\left(\underline{w_{0}^{e}}\circ\left(\underline{\underline{Dz^{e}}}*\left(\underline{r}\circ\underline{v_{z}}\right)\right)\right){}^{T}*\left(\underline{\underline{Dr^{e}}}*\left(\underline{r}\circ\underline{v_{r}}\right)\right)
\]
which would have resulted in the expression: 
\[
\underset{9}{\dot{U}{}_{Th\pi}}=\underline{dV}^{T}*\left\{ \underline{v_{r}}\circ\left(\frac{4}{3}\underline{\underline{Dr_{n}}}*\left(\underline{\mu^{e}}\circ\left(\underline{\underline{Dz^{e}}}*\left(\underline{r}\circ\underline{v_{z}}\right)\right)\oslash\underline{r^{e}}\right)\right)\right\} 
\]
so it is valid to express the $9^{th}$ term as:

\begin{align}
\underset{9}{\dot{U}{}_{Th\pi}} & =\underline{dV}^{T}*\biggl\{\underline{v_{z}}\circ\left(\frac{2}{3}\underline{\underline{Dz_{n}}}*\left(\underline{\mu^{e}}\circ\left(\underline{\underline{Dr^{e}}}*\left(\underline{r}\circ\underline{v_{r}}\right)\right)\oslash\underline{r^{e}}\right)\right)\nonumber \\
 & +\underline{v_{r}}\circ\left(\frac{2}{3}\underline{\underline{Dr_{n}}}*\left(\underline{\mu^{e}}\circ\left(\underline{\underline{Dz^{e}}}*\left(\underline{r}\circ\underline{v_{z}}\right)\right)\oslash\underline{r^{e}}\right)\right)\biggr\}\label{eq:264}
\end{align}
Transforming the $11^{th}$ term in equation \ref{eq:254} is straightforward: 

\begin{alignat}{1}
\underset{11}{\dot{U}{}_{Th\pi}} & =\underline{dV}^{T}*\left\{ 2\underline{\mu}\circ\left(\underline{v_{r}}\oslash\underline{r}\right)^{2}\right\} =\underline{dV}^{T}*\left\{ \underline{v_{r}}\circ\left(2\underline{\mu}\circ\underline{v_{r}}\oslash\underline{r}^{2}\right)\right\} \label{eq:265}
\end{alignat}
Adding equations \ref{eq:259}, \ref{eq:259.1}, \ref{eq:261}, \ref{eq:265},
and the parts of equations \ref{eq:260} and \ref{eq:264} that pertain
to $v_{r}$, we arrive at:
\begin{equation}
\dot{U}{}_{Th\pi r}=\underline{dV}^{T}*\left\{ \underline{v_{r}}\circ\underline{\varPi_{r}}\right\} \label{eq:257-1}
\end{equation}
where
\begin{align*}
\underline{\varPi_{r}} & =\biggl[-\left(2\underline{\underline{Dr_{n}}}*\left(\underline{\mu^{e}}\circ\underline{r^{e}}\circ\left(\underline{\underline{Dr^{e}}}*\underline{v_{r}}\right)\right)\right)\oslash\underline{r}-\left(\underline{\underline{Dz_{n}}}*\left(\underline{\mu^{e}}\circ\underline{r^{e}}\circ\left(\underline{\underline{Dz^{e}}}*\underline{v_{r}}\right)\right)\right)\oslash\underline{r}\\
 & \,\,\,\,\,\,\,-\left(\underline{\underline{Dz_{n}}}*\left(\underline{\mu^{e}}\circ\underline{r^{e}}\circ\left(\underline{\underline{Dr^{e}}}*\underline{v_{z}}\right)\right)\right)\oslash\underline{r}+\frac{2}{3}\underline{\underline{Dr_{n}}}*\left(\underline{\mu^{e}}\circ\left(\underline{\underline{Dr^{e}}}*\left(\underline{r}\circ\underline{v_{r}}\right)\right)\oslash\underline{r^{e}}\right)\\
 & \,\,\,\,\,\,\,+\frac{2}{3}\underline{\underline{Dr_{n}}}*\left(\underline{\mu^{e}}\circ\left(\underline{\underline{Dz^{e}}}*\left(\underline{r}\circ\underline{v_{z}}\right)\right)\oslash\underline{r^{e}}\right)\biggr]+2\underline{\mu}\circ\underline{v_{r}}\oslash\underline{r}^{2}
\end{align*}
Note that the $2^{nd}$ and $3^{rd}$ terms can be combined. With
reference to the definition of the node-to-element divergence operation
(equation \ref{eq:505.02}), the $4^{th}$ and $5^{th}$ terms can
also be combined, leading to the discrete form of $\left(\nabla\cdot\overline{\boldsymbol{\pi}}\right)_{r}$:

\begin{align}
\underline{\varPi_{r}} & =\left[-\left(2\underline{\underline{Dr_{n}}}*\left(\underline{\mu^{e}}\circ\underline{r^{e}}\circ\left(\underline{\underline{Dr^{e}}}*\underline{v_{r}}\right)\right)\right)-\left(\underline{\underline{Dz_{n}}}*\left(\underline{\mu^{e}}\circ\underline{r^{e}}\circ\left(\underline{\underline{Dr^{e}}}*\underline{v_{z}}+\underline{\underline{Dz^{e}}}*\underline{v_{r}}\right)\right)\right)\right]\oslash\underline{r}\nonumber \\
 & \,\,\,+\frac{2}{3}\underline{\underline{Dr_{n}}}*\left(\underline{\mu^{e}}\circ\left(\underline{\underline{\nabla^{e}}}\cdot\underline{\mathbf{v}}\right)\right)+2\underline{\mu}\circ\underline{v_{r}}\oslash\underline{r}^{2}\label{eq:266}
\end{align}
The $2^{nd},\;5^{th},\;6^{th},\;9^{th}$ and $10^{th}$ terms in equation
\ref{eq:254} include $z$ components of velocity and will contribute
to $\dot{U}{}_{Th\pi z}$. The re-expression for the $2^{nd}$, $5^{th}$
and $10^{th}$ terms follow from the re-expressions of the $1^{st}$
and $8^{th}$ terms:
\begin{equation}
\underset{2}{\dot{U}{}_{Th\pi}}=\underline{dV}^{T}*\left\{ \underline{v_{z}}\circ\left(-2\underline{\underline{Dz_{n}}}*\left(\underline{\mu^{e}}\circ\underline{r^{e}}\circ\left(\underline{\underline{Dz^{e}}}*\underline{v_{z}}\right)\right)\right)\oslash\underline{r}\right\} \label{eq:267}
\end{equation}
\begin{equation}
\underset{5}{\dot{U}{}_{Th\pi}}=\underline{dV}^{T}*\left\{ \underline{v_{z}}\circ\left(-\underline{\underline{Dr_{n}}}*\left(\underline{\mu^{e}}\circ\underline{r^{e}}\circ\left(\underline{\underline{Dr^{e}}}*\underline{v_{z}}\right)\right)\right)\oslash\underline{r}\right\} \label{eq:268}
\end{equation}

\begin{equation}
\underset{10}{\dot{U}{}_{Th\pi}}=\underline{dV}^{T}*\left\{ \underline{v_{z}}\circ\left(\frac{2}{3}\underline{\underline{Dz_{n}}}*\left(\underline{\mu^{e}}\circ\left(\underline{\underline{Dz^{e}}}*\left(\underline{r}\circ\underline{v_{z}}\right)\right)\oslash\underline{r^{e}}\right)\right)\right\} \label{eq:269}
\end{equation}
Adding equations \ref{eq:267}, \ref{eq:268}, \ref{eq:269}, and
the parts of equations \ref{eq:260} and \ref{eq:264} that pertain
to $v_{z}$, leads to 
\begin{equation}
\dot{U}{}_{Th\pi z}=\underline{dV}^{T}*\left\{ \underline{v_{z}}\circ\underline{\varPi_{z}}\right\} \label{eq:269.1}
\end{equation}
where the discrete form of $\left(\nabla\cdot\overline{\boldsymbol{\pi}}\right)_{z}$
is
\begin{align}
\underline{\varPi_{z}} & =\left[-\left(2\underline{\underline{Dz_{n}}}*\left(\underline{\mu^{e}}\circ\underline{r^{e}}\circ\left(\underline{\underline{Dz^{e}}}*\underline{v_{z}}\right)\right)\right)-\left(\underline{\underline{Dr_{n}}}*\left(\underline{\mu^{e}}\circ\underline{r^{e}}\circ\left(\underline{\underline{Dr^{e}}}*\underline{v_{z}}+\underline{\underline{Dz^{e}}}*\underline{v_{r}}\right)\right)\right)\right]\oslash\underline{r}\nonumber \\
 & \,\,\,+\frac{2}{3}\underline{\underline{Dz_{n}}}*\left(\underline{\mu^{e}}\circ\left(\underline{\underline{\nabla^{e}}}\cdot\underline{\mathbf{v}}\right)\right)\label{eq:269.2}
\end{align}

\subsection{Final expression for discretised form of $\dot{U}_{Th}$}

Combining equations \ref{eq:235.1}, \ref{eq:235.4}, \ref{eq:258.1},
\ref{eq:258.2}, \ref{eq:257-1} and \ref{eq:269.1}, we arrive at
the final expression for discretised form of $\dot{U}_{Th}$:

\begin{align}
\dot{U}_{Th} & =\underbrace{\underline{dV}\,{}^{T}*\left\{ \underline{v_{r}}\circ\left(\underline{\underline{Dr}}*\underline{p}+\underline{\varPi_{r}}\right)+\underline{v_{\phi}}\circ\underline{\varPi_{\phi}}+\underline{v_{z}}\circ\left(\underline{\underline{Dz}}*\underline{p}+\underline{\varPi_{z}}\right)\right\} }_{\dot{U}_{Th\,ideal}}+\dot{U}{}_{Th\theta\eta}+\dot{U}{}_{Th\phi\eta}\label{eq:269.3}
\end{align}
where $\dot{U}_{Th\,ideal}$ refers to the part of the rate of change
of system thermal energy that is not related to ohmic heating. The
expressions for $\dot{U}{}_{Th\phi\eta}$ and $\dot{U}{}_{Th\theta\eta}$,
the rates of change of thermal energy in the volume associated with
ohmic heating due to toroidal and poloidal currents respectively,
will be defined in section \ref{subsec:Magnetic-energy}.

\section{\label{subsec:Magnetic-energy}Magnetic energy}

\subsection{\label{subsec:poloidalMagnetic-energy}Discretisation of $\dot{U}_{M\theta}$}

Referring to equation \ref{eq:480.61} , the rate of change of the
magnetic energy associated with poloidal magnetic field is:

\begin{equation}
\dot{U}_{M\theta}=\frac{1}{2\mu_{0}}\int\frac{\partial}{\partial t}\left(\left(\frac{\nabla\psi}{r}\right)^{2}\right)\:dV=\frac{1}{\mu_{0}}\int\frac{1}{r^{2}}\left(\frac{\partial\psi}{\partial r}\frac{\partial\dot{\psi}}{\partial r}+\frac{\partial\psi}{\partial z}\frac{\partial\dot{\psi}}{\partial z}\right)\:dV\label{eq:367}
\end{equation}
As described in section \ref{par:Disc_Uthpi}, we need to we use the
node-to-element differential operators to expand terms with squared
derivatives. In terms of the discrete quantities, equation \ref{eq:367}
is

{\small{}
\begin{align}
\dot{U}_{M\theta} & =\frac{1}{\mu_{0}}\,\underline{dV^{e}}^{T}*\left\{ \left(\underline{\underline{Dr^{e}}}*\underline{\psi}\right)\circ\left(\underline{\underline{Dr^{e}}}*\dot{\underline{\psi}}\right)\oslash\underline{r^{e}}^{2}\right\} +\frac{1}{\mu_{0}}\underline{dV^{e}}^{T}*\left\{ \left(\underline{\underline{Dz^{e}}}*\underline{\psi}\right)\circ\left(\underline{\underline{Dz^{e}}}*\dot{\underline{\psi}}\right)\oslash\underline{r^{e}}^{2}\right\} \nonumber \\
 & =\frac{2\pi}{\mu_{0}}\,(\underline{s^{e}}\circ\underline{r^{e}}){}^{T}*\left\{ \left(\underline{\underline{Dr^{e}}}*\underline{\psi}\right)\circ\left(\underline{\underline{Dr^{e}}}*\dot{\underline{\psi}}\right)\oslash\underline{r^{e}}^{2}\right\} +\frac{2\pi}{\mu_{0}}(\underline{s^{e}}\circ\underline{r^{e}}){}^{T}*\left\{ \left(\underline{\underline{Dz^{e}}}*\underline{\psi}\right)\circ\left(\underline{\underline{Dz^{e}}}*\dot{\underline{\psi}}\right)\oslash\underline{r^{e}}^{2}\right\} \nonumber \\
 & =\frac{2\pi}{\mu_{0}}\left(\underline{s^{e}}\circ\left(\underline{\underline{Dr^{e}}}*\underline{\psi}\right)\oslash\underline{r^{e}}\right){}^{T}*\left(\underline{\underline{Dr^{e}}}*\dot{\underline{\psi}}\right)+\frac{2\pi}{\mu_{0}}\left(\underline{s^{e}}\circ\left(\underline{\underline{Dz^{e}}}*\underline{\psi}\right)\oslash\underline{r^{e}}\right){}^{T}*\left(\underline{\underline{Dz^{e}}}*\dot{\underline{\psi}}\right)\nonumber \\
 & =\frac{2\pi}{\mu_{0}}\,\dot{\underbar{\ensuremath{\psi}}}{}^{T}*\left(\underline{\underline{Dr^{e}}}{}^{T}*\left(\underline{s^{e}}\circ\left(\underline{\underline{Dr^{e}}}*\underline{\psi}\right)\oslash\underline{r^{e}}\right)+\underline{\underline{Dz^{e}}}{}^{T}*\left(\underline{s^{e}}\circ\left(\underline{\underline{Dz^{e}}}*\underline{\psi}\right)\oslash\underline{r^{e}}\right)\right)\nonumber \\
 & =\frac{2\pi}{\mu_{0}}\,\dot{\underbar{\ensuremath{\psi}}}{}^{T}*\left(\underline{\underline{K}}*\underline{\psi}\right)\label{eq:369}
\end{align}
}where $\underline{\underline{K}}=\underline{\underline{Dr^{e}}}{}^{T}*\underline{\underline{\widehat{S}}}*\underline{\underline{\widehat{R}}}^{-1}*\underline{\underline{Dr^{e}}}+\underline{\underline{Dz^{e}}}{}^{T}*\underline{\underline{\widehat{S}}}*\underline{\underline{\widehat{R}}}^{-1}*\underline{\underline{Dz^{e}}}$.
Using identity \ref{eq:116.1}, equation \ref{eq:369} can be re-expressed
as

\begin{align}
\dot{U}{}_{M\theta} & =\frac{2\pi}{3}(\underline{s}\circ\underline{r}){}^{T}*\left(\left(3\,\dot{\underbar{\ensuremath{\psi}}}\circ\left(\underline{\underline{K}}*\underline{\psi}\right)\right)\oslash\left(\mu_{0}\,\underline{s}\circ\underline{r}\right)\right)\nonumber \\
 & =\underline{dV}^{T}*\left\{ \left(\dot{\underbar{\ensuremath{\psi}}}\circ\left(3\,\underline{\underline{R}}*\underline{\underline{S}}^{-1}*\underline{\underline{K}}\right)*\underline{\psi}\right)\oslash\left(\mu_{0}\,\underline{r}^{2}\right)\right\} \nonumber \\
 & =\underline{dV}^{T}*\left\{ \left(-\dot{\underbar{\ensuremath{\psi}}}\circ\left(\underline{\underline{\Delta^{^{*}}}}\,\,\underline{\psi}\right)\right)\oslash\left(\mu_{0}\,\underline{r}^{2}\right)\right\} \label{eq:370}
\end{align}
In the last step, the definition $\underline{\underline{\Delta^{^{*}}}}=-3\underline{\underline{R}}*\underline{\underline{S}}^{-1}*\underline{\underline{K}}$
was used (refer to equations \ref{eq:515.0} and \ref{eq:515.72}).
Expressed in terms of discrete quantities and operators, equation
\ref{eq:479.3} is:
\begin{equation}
\dot{\underbar{\ensuremath{\psi}}}=\dot{\underbar{\ensuremath{\psi}}}_{ideal}+\dot{\underbar{\ensuremath{\psi}}}_{\eta}=-\underline{\mathbf{v}}\circ\left(\underline{\underline{\nabla}}\,\,\underbar{\ensuremath{\psi}}\right)+\underline{\eta}\circ\left(\underline{\underline{\Delta^{^{*}}}}\,\,\underline{\psi}\right)\label{eq:374}
\end{equation}
The combination of equations \ref{eq:370} and \ref{eq:374} leads
to
\begin{align}
\dot{U}{}_{M\theta ideal} & =\underline{dV}^{T}*\left\{ \left(\left(\underline{v_{r}}\circ(\underline{\underline{Dr}}*\underbar{\ensuremath{\psi}})+\underline{v_{z}}\circ(\underline{\underline{Dz}}*\underbar{\ensuremath{\psi}})\right)\circ\left(\underline{\underline{\Delta^{^{*}}}}\,\,\underline{\psi}\right)\right)\oslash\left(\mu_{0}\,\underline{r}^{2}\right)\right\} \label{eq:374.1}
\end{align}
and

\begin{align}
\dot{U}{}_{M\theta\eta} & =\underline{dV}^{T}*\left\{ \left(-\underline{\eta}\circ\left(\underline{\underline{\Delta^{^{*}}}}\,\,\underline{\psi}\right)^{2}\right)\oslash\left(\mu_{0}\,\underline{r}^{2}\right)\right\} \label{eq:374.2}
\end{align}
Here, $\dot{U}_{M\theta ideal}$ and $\dot{U}_{M\theta\eta}$ are
the parts of $\dot{U}_{M\theta}$ due to advective and diffusive ($i.e.,$
resistive) effects respectively. The two terms that define the coefficients
of the velocity components in the expression for $\dot{U}{}_{M\theta ideal}$
will constitute part of the formulation for the $\mathbf{J\times B}$
force in the discretised momentum equation. This can be verified as
follows: 
\begin{equation}
\mathbf{J\times B}=(J_{\phi}B_{z}-J_{z}B_{\phi}))\hat{\mathbf{r}}+(J_{z}B_{r}-J_{r}B_{z}))\widehat{\boldsymbol{\phi}}+(J_{r}B_{\phi}-J_{\phi}B_{r}))\hat{\mathbf{z}}\label{eq:375.0}
\end{equation}
Equation \ref{eq:20.2} defines the components of $\mathbf{B}$ and
$\mathbf{J}$:

\begin{align}
B_{r} & =-\frac{1}{r}\frac{\partial\psi}{\partial z} & B_{\phi} & =\frac{f(\psi)}{r} & B_{z} & =\frac{1}{r}\frac{\partial\psi}{\partial r}\nonumber \\
J_{r} & =-\frac{1}{\mu_{0}r}\frac{\partial f(\psi)}{\partial z} & J_{\phi} & =-\frac{1}{\mu_{0}r}\Delta^{^{*}}\psi & J_{z} & =\frac{1}{\mu_{0}r}\frac{\partial f(\psi)}{\partial r}\label{eq:375.1}
\end{align}
It can be seen that the terms in equation \ref{eq:374.1} will contribute
to the $r$ and $z$ components of the $\mathbf{J\times B}$ force
in the discretised momentum equation, making up the\\
$1^{st}$ $\left(J_{\phi}B_{z}=-\frac{1}{\mu_{0}r^{2}}\frac{\partial\psi}{\partial r}\Delta^{^{*}}\psi\right)$
and $6^{th}$ $\left(-J_{\phi}B_{r}=-\frac{1}{\mu_{0}r^{2}}\frac{\partial\psi}{\partial z}\Delta^{^{*}}\psi\right)$
terms in equation \ref{eq:375.0}. The remaining terms in equation
\ref{eq:375.0} will arise from the rearrangement of terms in the
expression for the rate of change of the part of magnetic energy associated
with toroidal magnetic field.

Since toroidal currents in the plasma decay resistively with time,
$\dot{U}{}_{M\theta\eta}<0$ (equation \ref{eq:374.2}), as expected.
It is also expected that the reduction in the magnetic energy associated
with poloidal magnetic field would be balanced by the increase in
thermal energy associated with ohmic heating due to toroidal currents.
Referring to equation \ref{eq:481}, and recalling that $\dot{U}_{Th}=\int\frac{\dot{p}}{\gamma-1}\:dV$
(equation \ref{eq:235}), the expression for $\dot{U}{}_{Th\phi\eta}$,
the part of the rate of change of thermal energy in the volume due
to toroidal currents, is $\dot{U}{}_{Th\phi\eta}=\int\eta'J_{\phi}^{2}\,dV$.
From equation \ref{eq:375.1}, $J_{\phi}=-\frac{1}{\mu_{0}r}\Delta^{^{*}}\psi$,
and $\eta'=\mu_{0}\eta$, so that the discrete form of $\eta'\,J_{\phi}^{2}$,
which will appear in the discrete form of the energy equation, is
\begin{equation}
\left(\eta'\,J_{\phi}^{2}\right)_{disc}=\left(\underline{\eta}/\mu_{0}\right)\circ\left(\left(\underline{\underline{\Delta^{^{*}}}}\,\,\underline{\psi}\right)\oslash\underline{r}\right)^{2}\label{eq:375.2}
\end{equation}
Hence, referring to equation \ref{eq:374.2}, $\dot{U}{}_{M\theta\eta}=-\dot{U}{}_{Th\phi\eta}$,
as expected:

\begin{equation}
\dot{U}{}_{M\theta\eta}=-\dot{U}{}_{Th\phi\eta}=\underline{dV}^{T}*\left\{ -\underline{\eta}\circ\left(\underline{\underline{\Delta^{^{*}}}}\,\,\underline{\psi}\right)^{2}\oslash\left(\mu_{0}\,\underline{r}^{2}\right)\right\} \label{eq:376}
\end{equation}
Finally, from equations \ref{eq:374.1} and \ref{eq:376}, the required
expression for $\dot{U}{}_{M\phi_{\eta}}$ is 
\begin{align}
\dot{U}{}_{M\theta} & =\dot{U}{}_{M\theta ideal}+\dot{U}{}_{M\theta\eta}\nonumber \\
 & =\underline{dV}^{T}*\left\{ \left(\left(\underline{v_{r}}\circ(\underline{\underline{Dr}}*\underbar{\ensuremath{\psi}})+\underline{v_{z}}\circ(\underline{\underline{Dz}}*\underbar{\ensuremath{\psi}})\right)\circ\left(\underline{\underline{\Delta^{^{*}}}}\,\,\underline{\psi}\right)\right)\oslash\left(\mu_{0}\,\underline{r}^{2}\right)\right\} -\dot{U}{}_{Th\phi\eta}\label{eq:376.1}
\end{align}

\subsection{\label{subsec:toroidalmagEnergy}Discretisation of $\dot{U}_{M\phi}$}

In terms of discrete quantities, the rate of change of the part of
magnetic energy associated with toroidal magnetic field is

\begin{equation}
\dot{U}_{M\phi}=\frac{1}{\mu_{0}}\underline{dV}^{T}*\left\{ \underline{f}\circ\dot{\underline{f}}\oslash\underline{r}^{2}\right\} \label{eq:376.2}
\end{equation}
Referring to equation \ref{eq:479.4}, the discrete form for $\dot{f}_{ideal}$
is 
\begin{equation}
\dot{\underline{f}}_{ideal}=\underline{r^{2}}\circ\left[-\underline{\underline{\nabla}}\cdot\left(\underline{f}\circ\underline{\mathbf{v}}\oslash\underline{r}^{2}\right)+\underline{\underline{\nabla_{n}}}\cdot\left(\underline{\mathbf{B}_{\theta}^{e}}\circ\underline{\omega^{e}}\right)\right]\label{eq:376.3}
\end{equation}
where, with reference to equation \ref{eq:375.1}, $\underline{\mathbf{B}_{\theta}^{e}}=\left(-\left(\underline{\underline{Dz^{e}}}*\underline{\psi}\right)\hat{\mathbf{r}}+\left(\underline{\underline{Dr^{e}}}*\underline{\psi}\right)\hat{\mathbf{z}}\right)\oslash\underline{r^{e}}$.
Note that, with azimuthal symmetry, the toroidal component of $\mathbf{B}$
can be dropped from the second divergence term in the expression for
$\dot{\underline{f}}_{ideal}$. Note that, for consistency with the
numerical scheme, the second derivatives in the $2^{nd}$ term in
equation \ref{eq:376.3} are expressed as successive applications
of the node-to-element and element-to-node differential operators.
Hence, the discrete form for $\dot{U}_{M\phi ideal}$ is
\begin{align}
\dot{U}_{M\phi ideal} & =\frac{1}{\mu_{0}}\,\underline{dV}^{T}*\left\{ \underline{f}\circ\dot{\underline{f}}_{ideal}\oslash\underline{r}^{2}\right\} \nonumber \\
 & =\underline{dV}^{T}*\biggl\{\biggl[-\underline{f}\circ\left(\underline{\underline{Dr}}*\left(\underline{v_{r}}\circ\underline{f}\oslash\underline{r}\right)\right)-\underline{f}\circ\left(\underline{\underline{Dz}}*\left(\underline{v_{z}}\circ\underline{f}\oslash\underline{r}\right)\right)\nonumber \\
 & -\underline{f}\circ\left(\underline{\underline{Dr_{n}}}*\left(\underline{\omega^{e}}\circ\left(\underline{\underline{Dz^{e}}}*\underline{\psi}\right)\right)\right)+\underline{f}\circ\left(\underline{\underline{Dz_{n}}}*\left(\underline{\omega^{e}}\circ\left(\underline{\underline{Dr^{e}}}*\underline{\psi}\right)\right)\right)\biggr]\oslash\left(\mu_{0}\,\underline{r}\right)\biggr\}\label{eq:127.1}
\end{align}
Here, the vector of element-based angular speeds is $\underline{\omega^{e}}=<\underline{v_{\phi}}\oslash\underline{r}>^{e}$
(equation \ref{eq:502.31}). The first term is

\begin{align*}
\underset{1}{\dot{U}_{M\phi ideal}} & =-\frac{2\pi}{3}\left(\underline{s}\circ\underline{r}\right){}^{T}*\left\{ \underline{f}\circ\left(\underline{\underline{Dr}}*\left(\underline{v_{r}}\circ\underline{f}\oslash\underline{r}\right)\right)\oslash\left(\mu_{0}\,\underline{r}\right)\right\} \\
 & =-\frac{2\pi}{3\mu_{0}}\,\underline{s}{}^{T}*\left\{ \underline{f}\circ\left(\underline{\underline{Dr}}*\left(\underline{v_{r}}\circ\underline{f}\oslash\underline{r}\right)\right)\right\} \\
 & =-\frac{2\pi}{3\mu_{0}}\,\underline{f}{}^{T}*\underline{\underline{S}}*\left(\underline{\underline{Dr}}*\left(\underline{v_{r}}\circ\underline{f}\oslash\underline{r}\right)\right) & \mbox{(use eqn. \ref{eq:116-2})}\\
 & =-\frac{2\pi}{3\mu_{0}}\,\left(\underline{v_{r}}\circ\underline{f}\oslash\underline{r}\right){}^{T}*\underline{\underline{Dr}}{}^{T}*\underline{\underline{S}}*\underline{f} & \mbox{(transpose the scalar)}\\
 & =+\frac{2\pi}{3\mu_{0}}\,\left(\underline{v_{r}}\circ\underline{f}\oslash\underline{r}\right){}^{T}*\underline{\underline{S}}*\underline{\underline{Dr}}*\underline{f} & \mbox{(use eqn. \ref{eq:230})}\\
 & =\:\;\frac{2\pi}{3\mu_{0}}\,\underline{s}{}^{T}*\left(\left(\underline{v_{r}}\circ\underline{f}\oslash\underline{r}\right)\circ\left(\underline{\underline{Dr}}*\underline{f}\right)\right) & \mbox{(use eqn. \ref{eq:116-2})}\\
 & =\;\;\frac{2\pi}{3\mu_{0}}\left(\underline{s}\circ\underline{r}\right){}^{T}*\left(\left(\underline{v_{r}}\circ\underline{f}\oslash\underline{r}^{2}\right)\circ\left(\underline{\underline{Dr}}*\underline{f}\right)\right)\\
 & =\;\;\underline{dV}^{T}*\left\{ \underline{v_{r}}\circ\left(\underline{f}\circ\left(\underline{\underline{Dr}}*\underline{f}\right)\oslash\left(\mu_{0}\,\underline{r}^{2}\right)\right)\right\} 
\end{align*}
Note that in the $5^{th}$ step, the differential operator property
defined in equation \ref{eq:230} was used, with the assumption of
boundary condition $v_{r}|_{\Gamma}=0.$ Similarly, the second term
is 
\[
\underset{2}{\dot{U}_{M\phi ideal}}=\underline{dV}^{T}*\left\{ \underline{v_{z}}\circ\left(\underline{f}\circ\left(\underline{\underline{Dz}}*\underline{f}\right)\oslash\left(\mu_{0}\,\underline{r}^{2}\right)\right)\right\} 
\]
The third term is 
\begin{align*}
\underset{3}{\dot{U}_{M\phi ideal}} & =-\frac{2\pi}{3}\left(\underline{s}\circ\underline{r}\right){}^{T}*\left\{ \underline{f}\circ\left(\underline{\underline{Dr_{n}}}*\left(\underline{\omega^{e}}\circ\left(\underline{\underline{Dz^{e}}}*\underline{\psi}\right)\right)\right)\oslash\left(\mu_{0}\,\underline{r}\right)\right\} \\
 & =-\frac{2\pi}{3\mu_{0}}\,\underline{s}{}^{T}*\left\{ \underline{f}\circ\left(\underline{\underline{Dr_{n}}}*\left(\underline{\omega^{e}}\circ\left(\underline{\underline{Dz^{e}}}*\underline{\psi}\right)\right)\right)\right\} \\
 & =-\frac{2\pi}{3\mu_{0}}\,\underline{f}{}^{T}*\underline{\underline{S}}*\left(\underline{\underline{Dr_{n}}}*\left(\underline{\omega^{e}}\circ\left(\underline{\underline{Dz^{e}}}*\underline{\psi}\right)\right)\right) & \mbox{(use eqn. \ref{eq:116-2})}\\
 & =-\frac{2\pi}{3\mu_{0}}\,\left(\underline{\omega^{e}}\circ\left(\underline{\underline{Dz^{e}}}*\underline{\psi}\right)\right){}^{T}*\underline{\underline{Dr_{n}}}{}^{T}*\underline{\underline{S}}*\underline{f} & \mbox{(transpose the scalar)}\\
 & =-\frac{2\pi}{3\mu_{0}}\,\left(\underline{\omega^{e}}\circ\left(\underline{\underline{Dz^{e}}}*\underline{\psi}\right)\right){}^{T}*\left(-3\,\underline{\underline{S}}^{-1}*\left(\underline{\underline{Dr^{e}}}^{T}*\widehat{\underline{\underline{S}}}\right)\right){}^{T}*\underline{\underline{S}}*\underline{f} & \mbox{(use eqn. \ref{eq:515.0})}\\
 & =+\frac{2\pi}{\mu_{0}}\,\left(\underline{\omega^{e}}\circ\left(\underline{\underline{Dz^{e}}}*\underline{\psi}\right)\right){}^{T}*\left(\underline{\underline{\widehat{S}}}*\underline{\underline{Dr^{e}}}*\underline{f}\right) & \mbox{(transpose)}\\
 & =\:\;\frac{2\pi}{\mu_{0}}\,\underline{s^{e}}{}^{T}*\left\{ \left(\underline{\omega^{e}}\circ\left(\underline{\underline{Dz^{e}}}*\underline{\psi}\right)\right)\circ\left(\underline{\underline{Dr^{e}}}*\underline{f}\right)\right\}  & \mbox{(use eqn. \ref{eq:116-2})}\\
 & =\:\;\frac{2\pi}{\mu_{0}}\,\left(\underline{s^{e}}\circ\underline{r^{e}}\right){}^{T}*\left\{ \left(\underline{\omega^{e}}\circ\left(\underline{\underline{Dz^{e}}}*\underline{\psi}\right)\oslash\underline{r^{e}}\right)\circ\left(\underline{\underline{Dr^{e}}}*\underline{f}\right)\right\} \\
 & =\:\;\underline{dV^{e}}{}^{T}*\left\{ \underline{\omega^{e}}\circ\left(\left(\underline{\underline{Dz^{e}}}*\underline{\psi}\right)\circ\left(\underline{\underline{Dr^{e}}}*\underline{f}\right)\oslash\underline{r^{e}}\right)/\mu_{0}\right\} \\
 & =\:\;\underline{dV}{}^{T}*\left\{ \underline{\omega}\circ\left(\underline{\underline{W_{n}}}*\left(\left(\underline{\underline{Dz^{e}}}*\underline{\psi}\right)\circ\left(\underline{\underline{Dr^{e}}}*\underline{f}\right)\oslash\underline{r^{e}}\right)\right)/\mu_{0}\right\}  & \mbox{(use eqn. \ref{eq:516})}\\
 & =\:\;\underline{dV}{}^{T}*\left\{ \underline{v_{\phi}}\circ\left(\underline{\underline{W_{n}}}*\left(\left(\underline{\underline{Dz^{e}}}*\underline{\psi}\right)\circ\left(\underline{\underline{Dr^{e}}}*\underline{f}\right)\oslash\underline{r^{e}}\right)\right)\oslash\left(\mu_{0}\,\underline{r}\right)\right\} 
\end{align*}
Note that in the second last step, the volume-averaging operation,
defined by equation \ref{eq:516}, where $\underline{Q^{e}}=\underline{\omega^{e}}$
and $\underline{U^{e}}=\left(\left(\underline{\underline{Dz^{e}}}*\underline{\psi}\right)\circ\left(\underline{\underline{Dr^{e}}}*\underline{f}\right)\oslash\underline{r^{e}}\right)$,
has been implemented. By the same procedure, the $4^{th}$ term in
equation \ref{eq:127.1} is 
\[
\underset{4}{\dot{U}_{M\phi ideal}}=\underline{dV}{}^{T}*\left\{ -\underline{v_{\phi}}\circ\left(\underline{\underline{W_{n}}}*\left(\left(\underline{\underline{Dr^{e}}}*\underline{\psi}\right)\circ\left(\underline{\underline{Dz^{e}}}*\underline{f}\right)\oslash\underline{r^{e}}\right)\right)\oslash\left(\mu_{0}\,\underline{r}\right)\right\} \ 
\]
Collecting terms, the resultant expression for $\dot{U}_{M\phi ideal}$
is 

\begin{align}
\dot{U}_{M\phi ideal} & =\underline{dV}^{T}*\biggl\{\biggl[\underline{v_{r}}\circ\left(\underline{f}\circ\left(\underline{\underline{Dr}}*\underline{f}\right)\right)+\underline{v_{z}}\circ\left(\underline{f}\circ\left(\underline{\underline{Dz}}*\underline{f}\right)\right)\nonumber \\
 & -\underline{v_{\phi}}\circ\left(\underline{r}\circ\left(\underline{\underline{W_{n}}}*\left(\underline{\mathbf{B}_{\theta}^{e}}\cdot\left(\underline{\underline{\nabla^{e}}}\,\,\underline{f}\right)\right)\right)\right)\biggr]\oslash\left(\mu_{0}\,\underline{r^{2}}\right)\biggr\}\label{eq:127.3}
\end{align}
Recall that the $1^{st}$ and $6^{th}$ terms in the expression for
the $\mathbf{J\times}\mathbf{B}$ force (equation \ref{eq:375.0})
arose from the rearrangement of terms in the expression for the rate
of change of the part of magnetic energy associated with poloidal
magnetic field, and will contribute to the $r$ and $z$ components
of the $\mathbf{J\times B}$ force in the discretised momentum equation.
Referring to equations \ref{eq:375.0} and \ref{eq:375.1}, which
define the components of the $\mathbf{J\times}\mathbf{B}$ force,
it can be seen that the remaining four terms that will be included
in the discretised momentum equation are contained in equation \ref{eq:127.3}. 

The discrete form of the part of $\dot{U}_{M\phi}$ pertaining to
diffusive effects is 
\begin{equation}
\dot{U}_{M\phi\eta}=\frac{1}{\mu_{0}}\,\underline{dV}^{T}*\left\{ \underline{f}\circ\dot{\underline{f}}_{\eta}\oslash\underline{r}^{2}\right\} \label{eq:127.0}
\end{equation}
We expect that the reduction in the magnetic energy associated with
toroidal magnetic field would be balanced by the increase in thermal
energy due ohmic heating associated with poloidal currents, $i.e.,$
$\dot{U}{}_{M\phi_{\eta}}=-\dot{U}{}_{Th\theta\eta}$. The part of
the rate of change of thermal energy in the volume due to ohmic heating
with toroidal currents, is $\dot{U}{}_{Th\theta\eta}=\int\eta'J_{\theta}^{2}\,dV$.
Referring to equation \ref{eq:375.1}, the discrete form of $\eta'\,J_{\theta}^{2}=\eta'\left(\frac{1}{\mu_{0}r}\nabla f\right)^{2}$,
is 
\[
\left(\eta'\,J_{\theta}^{2}\right)_{disc}=\left(\underline{\eta^{e}}/\mu_{0}\right)\circ\left(\left(\underline{\underline{\nabla^{e}}}\,\,\underline{f}\right)\oslash\underline{r^{e}}\right){}^{2}
\]
Once again, for consistency with the numerical scheme, derivatives
squared have been expressed with node-to-element differential operators.
$\underline{\eta^{e}}=<\underline{\eta}>^{e}$ (equation \ref{eq:502.31})
is the vector of element-based resistive diffusivities. Referring
to equation \ref{eq:515.1}, note that $\left(\eta'\,J_{\theta}^{2}\right)_{disc}$
can be expressed as a node-based volume-average, which will appear
in the discrete form of the energy equation:

\begin{equation}
\left(\eta'\,J_{\theta}^{2}\right)_{disc}=\underline{\underline{W_{n}}}*\left(\left(\underline{\eta^{e}}/\mu_{0}\right)\circ\left(\left(\underline{\underline{\nabla^{e}}}\,\,\underline{f}\right)\oslash\underline{r^{e}}\right){}^{2}\right)\label{eq:127.4}
\end{equation}
The identity $\dot{U}{}_{M\phi_{\eta}}=-\dot{U}{}_{Th\theta\eta}$
implies that 

\begin{equation}
\dot{U}{}_{M\phi_{\eta}}=-\dot{U}{}_{Th\theta\eta}=-\frac{1}{\mu_{0}}\,\underline{dV^{e}}^{T}*\left\{ \left(\left(\underline{\underline{Dr^{e}}}*\underline{f}\right){}^{2}+\left(\underline{\underline{Dz^{e}}}*\underline{f}\right){}^{2}\right)\circ\underline{\eta^{e}}\oslash\underline{r^{e}}{}^{2}\right\} \label{eq:127.5}
\end{equation}
In order to find the discrete form of $\dot{f}_{\eta}$, we can start
by defining 
\begin{equation}
\dot{\underline{f}}_{\eta}=\underline{\underline{D}}*\underline{f}\label{eq:127.51}
\end{equation}
Equations \ref{eq:127.0} and \ref{eq:127.5} imply that
\begin{align*}
 & \frac{1}{\mu_{0}}\,\underline{dV}^{T}*\left\{ \underline{f}\circ\dot{\underline{f}}_{\eta}\oslash\underline{r}^{2}\right\} =-\frac{1}{\mu_{0}}\,\underline{dV^{e}}^{T}*\left\{ \left(\left(\underline{\underline{Dr^{e}}}*\underline{f}\right){}^{2}+\left(\underline{\underline{Dz^{e}}}*\underline{f}\right){}^{2}\right)\circ\underline{\eta^{e}}\oslash\underline{r^{e}}{}^{2}\right\} \\
 & \Rightarrow\frac{1}{3}(\underline{s}\circ\underline{r}){}^{T}*\left\{ \underline{f}\circ(\underline{\underline{D}}*\underline{f})\oslash\underline{r}^{2}\right\} =-(\underline{s^{e}}\circ\underline{r^{e}}){}^{T}*\left\{ \left(\left(\underline{\underline{Dr^{e}}}*\underline{f}\right){}^{2}+\left(\underline{\underline{Dz^{e}}}*\underline{f}\right){}^{2}\right)\circ\underline{\eta^{e}}\oslash\underline{r^{e}}^{2}\right\} \\
 & \Rightarrow\frac{1}{3}\,\underline{s}{}^{T}*\left\{ \underline{f}\circ\left(\underline{\underline{D}}*\underline{f}\right)\oslash\underline{r}\right\} =-\underline{s^{e}}{}^{T}*\left\{ \left(\left(\underline{\underline{Dr^{e}}}*\underline{f}\right){}^{2}+\left(\underline{\underline{Dz^{e}}}*\underline{f}\right){}^{2}\right)\circ\underline{\eta^{e}}\oslash\underline{r^{e}}\right\} \\
 & \Rightarrow\frac{1}{3}\,(\underline{s}\oslash\underline{r}){}^{T}*\left\{ \underline{f}\circ\left(\underline{\underline{D}}*\underline{f}\right)\right\} =-(\underline{s^{e}}\circ\underline{\eta^{e}}\oslash\underline{r^{e}}){}^{T}*\left\{ \left(\underline{\underline{Dr^{e}}}*\underline{f}\right){}^{2}+\left(\underline{\underline{Dz^{e}}}*\underline{f}\right){}^{2}\right\} \\
 & \Rightarrow\frac{1}{3}\,\underline{f\,}{}^{T}*\underline{\underline{S}}*\underline{\underline{R}}^{-1}*\left(\underline{\underline{D}}*\underline{f}\right)=-\left(\underline{\underline{Dr^{e}}}*\underline{f}\right){}^{T}*\widehat{\underline{\underline{S}}}*\widehat{\underline{\underline{\eta}}}*\widehat{\underline{\underline{R}}}^{-1}*\left(\underline{\underline{Dr^{e}}}*\underline{f}\right)\,\,\,\,\,\,\,\,\mbox{(use eqn. \ref{eq:116-2})} & \mbox{\mbox{}}\\
 & \,\,\,\,\,\,\,-\left(\underline{\underline{Dz^{e}}}*\underline{f}\right){}^{T}*\widehat{\underline{\underline{S}}}*\widehat{\underline{\underline{\eta}}}*\widehat{\underline{\underline{R}}}^{-1}*\left(\underline{\underline{Dz^{e}}}*\underline{f}\right)\\
 & \Rightarrow\frac{1}{3}\,\underline{f\,}{}^{T}*\underline{\underline{S}}*\underline{\underline{R}}^{-1}*\left(\underline{\underline{D}}*\underline{f}\right)=-\left(\underline{f}{}^{T}*\underline{\underline{Dr^{e}}}{}^{T}\right)*\widehat{\underline{\underline{S}}}*\widehat{\underline{\underline{\eta}}}*\widehat{\underline{\underline{R}}}^{-1}*\left(\underline{\underline{Dr^{e}}}*\underline{f}\right)\\
 & \,\,\,\,\,\,\,-\left(\underline{f}{}^{T}*\underline{\underline{Dz^{e}}}{}^{T}\right)*\widehat{\underline{\underline{S}}}*\widehat{\underline{\underline{\eta}}}*\widehat{\underline{\underline{R}}}^{-1}*\left(\underline{\underline{Dz^{e}}}*\underline{f}\right)\\
 & \Rightarrow\frac{1}{3}\,\underline{f\,}{}^{T}*\underline{\underline{S}}*\underline{\underline{R}}^{-1}*\left(\underline{\underline{D}}*\underline{f}\right)=-\underline{f}\,{}^{T}*\underline{\underline{K1}}*\underline{f}
\end{align*}
where $\widehat{\underline{\underline{\eta}}}$ is the diagonal matrix
constructed from $\underline{\eta^{e}}$, and 
\begin{equation}
\underline{\underline{K1}}=\underline{\underline{Dr^{e}}}{}^{T}*\widehat{\underline{\underline{S}}}*\widehat{\underline{\underline{\eta}}}*\widehat{\underline{\underline{R}}}^{-1}*\underline{\underline{Dr^{e}}}+\underline{\underline{Dz^{e}}}{}^{T}*\widehat{\underline{\underline{S}}}*\widehat{\underline{\underline{\eta}}}*\widehat{\underline{\underline{R}}}^{-1}*\underline{\underline{Dz^{e}}}\label{eq:127.7}
\end{equation}
so that 
\begin{equation}
\underline{\underline{D}}=-3\,\underline{\underline{R}}*\underline{\underline{S}}^{-1}*\underline{\underline{K1}}\label{eq:127.8}
\end{equation}
With reference to equations \ref{eq:515.02} and \ref{eq:515.0},
which define the element-to-node divergence operation and differential
operators, equations \ref{eq:127.51} and \ref{eq:127.8} imply that
\begin{align}
\dot{\underline{f}}_{\eta} & =\underline{\underline{D}}*\underline{f}\nonumber \\
 & =\underline{r^{2}}\circ\,\underline{\underline{\nabla_{n}}}\cdot\left(\underline{\eta^{e}}\circ\left(\underline{\underline{\nabla^{e}}}\,\,\underline{f}\right)\oslash\underline{r^{e}}{}^{2}\right)\label{eq:127.81}
\end{align}
Note that the expressions for the element-to-node differential operators
were undefined at the time of the development of the discrete momentum
equation for system energy conservation. Back then, only the forms
of the element-to-node and node-to-node operators were known - the
process outlined here to construct an expression for $\dot{\underline{f}}_{\eta}$
helped to define the element-to-node operators, and confirms that
$\dot{U}{}_{M\phi_{\eta}}=-\dot{U}{}_{Th\theta\eta}$ in discrete
form. Now, with the operators and their properties pre-defined, equation
\ref{eq:127.81} is obviously just the discrete form of $\dot{f}_{\eta}$
as defined by equation \ref{eq:479.4}, with continuous differential
operators simply replaced with the discrete counterparts. 

From equation \ref{eq:127.0}, the required expression for $\dot{U}{}_{M\phi_{\eta}}$
is 
\begin{equation}
\dot{U}{}_{M\phi_{\eta}}=-\dot{U}{}_{Th\theta\eta}=-\underline{dV}^{T}*\left\{ \underline{f}\circ\left(\,\underline{\underline{\nabla_{n}}}\cdot\left(\underline{\eta^{e}}\circ\left(\underline{\underline{\nabla^{e}}}\,\,\underline{f}\right)\oslash\underline{r^{e}}{}^{2}\right)\right)/\mu_{0}\right\} \label{eq:127.9}
\end{equation}
In combination with equation \ref{eq:127.3}, this implies that 
\begin{align}
\dot{U}{}_{M\phi} & =\dot{U}{}_{M\phi_{ideal}}+\dot{U}{}_{M\phi_{\eta}}\nonumber \\
 & =\dot{U}{}_{M\phi_{ideal}}-\dot{U}{}_{Th\theta\eta}\nonumber \\
 & =\underline{dV}^{T}*\biggl\{\biggl[\underline{v_{r}}\circ\left(\underline{f}\circ\left(\underline{\underline{Dr}}*\underline{f}\right)\right)+\underline{v_{z}}\circ\left(\underline{f}\circ\left(\underline{\underline{Dz}}*\underline{f}\right)\right)\nonumber \\
 & -\underline{v_{\phi}}\circ\left(\underline{r}\circ\left(\underline{\underline{W_{n}}}*\left(\underline{\mathbf{B}_{\theta}^{e}}\cdot\left(\underline{\underline{\nabla^{e}}}\,\,\underline{f}\right)\right)\right)\right)\biggr]\oslash\left(\mu_{0}\,\underline{r}^{2}\right)\biggr\}-\dot{U}{}_{Th\theta\eta}\label{eq:127.10}
\end{align}

\subsection{Final expression for discretised form of $\dot{U}_{M}$}

$ $\\
$ $\\
Combining equations \ref{eq:376.1} and \ref{eq:127.10}, we arrive
at the final expression for the discretised form of $\dot{U}_{M}$:

\begin{align}
\dot{U}{}_{M} & =\dot{U}{}_{M\theta}+\dot{U}{}_{M\phi}\nonumber \\
 & =\underline{dV}^{T}*\biggl\{\biggl[\left(\underline{v_{r}}\circ\left(\underline{\underline{Dr}}*\underbar{\ensuremath{\psi}}\right)+\underline{v_{z}}\circ\left(\underline{\underline{Dz}}*\underbar{\ensuremath{\psi}}\right)\right)\circ\left(\underline{\underline{\Delta^{^{*}}}}\,\,\underline{\psi}\right) & \left(\dot{U}{}_{M\theta ideal}\right)\nonumber \\
 & +\underline{v_{r}}\circ\left(\underline{f}\circ\left(\underline{\underline{Dr}}*\underline{f}\right)\right)+\underline{v_{z}}\circ\left(\underline{f}\circ\left(\underline{\underline{Dz}}*\underline{f}\right)\right) & \left(\dot{U}{}_{M\phi ideal}\right)\nonumber \\
 & -\underline{v_{\phi}}\circ\left(\underline{r}\circ\left(\underline{\underline{W_{n}}}*\left(\underline{\mathbf{B}_{\theta}^{e}}\cdot\left(\underline{\underline{\nabla^{e}}}\,\,\underline{f}\right)\right)\right)\right)\biggr]\oslash\left(\mu_{0}\,\underline{r}^{2}\right)\biggr\} & \left(\dot{U}{}_{M\phi ideal}\right)\nonumber \\
 & -\dot{U}{}_{Th\phi\eta} & \left(\dot{U}{}_{M\theta\eta}\right)\nonumber \\
 & -\dot{U}{}_{Th\theta\eta} & \left(\dot{U}{}_{M\phi\eta}\right)\nonumber \\
\nonumber \\
\label{eq:127.11}
\end{align}

\section{Assembly of discretised momentum equation\label{sec:Assembly-of-discretised}}

Combining equations \ref{eq:234}, \ref{eq:269.3}, and \ref{eq:127.11},
we obtain the final expression for $\dot{U}_{Total}$:

{\small{}
\begin{align*}
\dot{U}_{Total} & =\dot{U}_{K}+\dot{U}_{Th}+\dot{U}_{M}\\
 & =\underline{dV}\,{}^{T}*\biggl\{\underline{v_{r}}\circ\left(\underline{\rho}\circ\left(\underline{\underline{Dr}}*\left(\frac{\underline{v}^{2}}{2}\right)\right)\right)+\underline{v_{z}}\circ\left(\underline{\rho}\circ\left(\underline{\underline{Dz}}*\left(\frac{\underline{v}^{2}}{2}\right)\right)\right) & \left(\dot{U}_{K}\right)\\
 & +\underline{\rho}\circ\underline{v_{r}}\circ\underline{\dot{v}_{r}}+\underline{\rho}\circ\underline{v_{\phi}}\circ\underline{\dot{v}_{\phi}}+\underline{\rho}\circ\underline{v_{z}}\circ\underline{\dot{v}_{z}} & \left(\dot{U}_{K}\right)\\
 & -\underline{\rho}\circ\underline{v_{r}}\circ\left(-\underline{v_{z}}\circ\left(\underline{\underline{Dz}}*\underline{v_{r}}-\underline{\underline{Dr}}*\underline{v_{z}}\right)+\underline{v_{\phi}}\circ\left(\underline{\underline{Dr}}*\left(\underline{r}\circ\underline{v_{\phi}}\right)\right)\oslash\underline{r}\right) & \left(\dot{U}_{K}\right)\\
 & -\underline{\rho}\circ\underline{v_{\phi}}\circ\left(-\underline{\mathbf{v}}\cdot\left(\underline{\underline{\nabla}}\,\,\left(\underline{r}\circ\underline{v_{\phi}}\right)\right)\oslash\underline{r}\right) & \left(\dot{U}_{K}\right)\\
 & -\underline{\rho}\circ\underline{v_{z}}\circ\left(\underline{v_{r}}\circ\left(\underline{\underline{Dz}}*\underline{v_{r}}-\underline{\underline{Dr}}*\underline{v_{z}}\right)+\underline{v_{\phi}}\circ\left(\underline{\underline{Dz}}*\left(\underline{r}\circ\underline{v_{\phi}}\right)\right)\oslash\underline{r}\right) & \left(\dot{U}_{K}\right)\\
 & +\underline{v_{r}}\circ\left(\underline{\underline{Dr}}*\underline{p}+\underline{\varPi_{r}}\right)+\underline{v_{\phi}}\circ\underline{\varPi_{\phi}}+\underline{v_{z}}\circ\left(\underline{\underline{Dz}}*\underline{p}+\underline{\varPi_{z}}\right) & \left(\dot{U}_{Th\,ideal}\right)\\
 & \cancel{+\dot{U}{}_{Th\phi\eta}} & \left(\dot{U}{}_{Th\phi\eta}\right)\\
 & \bcancel{+\dot{U}{}_{Th\theta\eta}} & \left(\dot{U}{}_{Th\theta\eta}\right)\\
 & +\biggl[\left(\underset{}{\underline{v_{r}}\circ\left(\underline{\underline{Dr}}*\underbar{\ensuremath{\psi}}\right)+}\underset{}{\underline{v_{z}}\circ\left(\underline{\underline{Dz}}*\underbar{\ensuremath{\psi}}\right)}\right)\circ\left(\underline{\underline{\Delta^{^{*}}}}\,\,\underline{\psi}\right) & \left(\dot{U}{}_{M\theta ideal}\right)\\
 & +\underline{v_{r}}\circ\left(\underline{f}\circ\left(\underline{\underline{Dr}}*\underline{f}\right)\right)+\underline{v_{z}}\circ\left(\underline{f}\circ\left(\underline{\underline{Dz}}*\underline{f}\right)\right) & \left(\dot{U}{}_{M\phi ideal}\right)\\
 & -\underline{v_{\phi}}\circ\left(\underline{r}\circ\left(\underline{\underline{W_{n}}}*\left(\underline{\mathbf{B}_{\theta}^{e}}\cdot\left(\underline{\underline{\nabla^{e}}}\,\,\underline{f}\right)\right)\right)\right)\biggr]\oslash\left(\mu_{0}\,\underline{r}^{2}\right)\biggr\} & \left(\dot{U}{}_{M\phi ideal}\right)\\
 & \cancel{-\dot{U}{}_{Th\phi\eta}} & \left(\dot{U}{}_{M\theta\eta}\right)\\
 & \bcancel{-\dot{U}{}_{Th\theta\eta}} & \left(\dot{U}{}_{M\phi\eta}\right)
\end{align*}
}Collecting the terms that make up the coefficients of each velocity
component, we obtain:

\begin{align}
\dot{U}_{Total} & =\underline{dV}\,{}^{T}*\biggl\{\underline{v_{r}}\circ\biggl[\underline{\rho}\circ\underline{\dot{v}_{r}}+\underline{\rho}\circ\left(\underline{\underline{Dr}}*\left(\frac{\underline{v}^{2}}{2}\right)\right)\nonumber \\
 & -\underline{\rho}\circ\left(-\underline{v_{z}}\circ\left(\underline{\underline{Dz}}*\underline{v_{r}}-\underline{\underline{Dr}}*\underline{v_{z}}\right)+\underline{v_{\phi}}\circ\left(\underline{\underline{Dr}}*\left(\underline{r}\circ\underline{v_{\phi}}\right)\right)\oslash\underline{r}\right)\nonumber \\
 & +\underline{\underline{Dr}}*\underline{p}+\underline{\varPi_{r}}+\left[\underset{}{\left(\underline{\underline{Dr}}*\underbar{\ensuremath{\psi}}\right)}\circ\left(\underline{\underline{\Delta^{^{*}}}}\,\,\underline{\psi}\right)+\left(\underline{f}\circ\left(\underline{\underline{Dr}}*\underline{f}\right)\right)\right]\oslash\left(\mu_{0}\,\underline{r}^{2}\right)\biggr]\nonumber \\
 & +\underline{v_{\phi}}\circ\biggl[\underline{\rho}\circ\underline{\dot{v}_{\phi}}-\underline{\rho}\circ\left(-\underline{\mathbf{v}}\cdot\left(\underline{\underline{\nabla}}\,\,\left(\underline{r}\circ\underline{v_{\phi}}\right)\right)\oslash\underline{r}\right)\nonumber \\
 & +\underline{\varPi_{\phi}}-\left(\underline{\underline{W_{n}}}*\left(\underline{\mathbf{B}_{\theta}^{e}}\cdot\left(\underline{\underline{\nabla^{e}}}\,\,\underline{f}\right)\right)\right)\oslash\left(\mu_{0}\,\underline{r}\right)\biggr]\nonumber \\
 & +\underline{v_{z}}\circ\biggl[\underline{\rho}\circ\underline{\dot{v}_{z}}+\underline{\rho}\circ\left(\underline{\underline{Dz}}*\left(\frac{\underline{v}^{2}}{2}\right)\right)\nonumber \\
 & -\underline{\rho}\circ\left(\underline{v_{r}}\circ\left(\underline{\underline{Dz}}*\underline{v_{r}}-\underline{\underline{Dr}}*\underline{v_{z}}\right)+\underline{v_{\phi}}\circ\left(\underline{\underline{Dz}}*\left(\underline{r}\circ\underline{v_{\phi}}\right)\right)\oslash\underline{r}\right)\nonumber \\
 & +\underline{\underline{Dz}}*\underline{p}+\underline{\varPi_{z}}+\left[\left(\underline{\underline{Dz}}*\underbar{\ensuremath{\psi}}\right)\circ\left(\underline{\underline{\Delta^{^{*}}}}\,\,\underline{\psi}\right)+\underline{f}\circ\left(\underline{\underline{Dz}}*\underline{f}\right)\right]\oslash\left(\mu_{0}\,\underline{r}^{2}\right)\biggr]\biggr\}\label{eq:127.111}
\end{align}
Now, the only way to achieve $\dot{U}_{Total}=0$, apart from having
all velocity components equal to zero everywhere, is to set the expressions
that make up the coefficients of the velocity components individually
to zero, leading to the components of the discretised momentum equation,
which are presented along with the discrete forms of the continuity
and energy equations, and discretised expressions for $\dot{\psi}$
and $\dot{f}$, in the following appendix summary.

\section{Summary\label{sec:SummaryB}}

Here, the complete set of discretised equations for the axisymmetric
MHD model, in a form that ensures conservation of total system energy,
are collected. The discrete form of the expression for mass conservation
is given by equation \ref{eq:212.2}: 
\[
\underline{\dot{n}}=-\underline{\underline{\nabla}}\cdot(\underline{n}\circ\underline{\mathbf{v}})
\]
From equation \ref{eq:127.111}, the discrete forms of the components
of the momentum conservation equation are 
\begin{align*}
\underline{\dot{v}_{r}} & =\underbrace{-\underline{\underline{Dr}}*\left(\frac{\underline{v}^{2}}{2}\right)-\underline{v_{z}}\circ\left(\underline{\underline{Dz}}*\underline{v_{r}}-\underline{\underline{Dr}}*\underline{v_{z}}\right)+\underline{v_{\phi}}\circ\left(\underline{\underline{Dr}}*\left(\underline{r}\circ\underline{v_{\phi}}\right)\right)\oslash\underline{r}}_{\mathclap{{-(\mathbf{v\cdot\nabla}\mathbf{v})_{r}=\left(-\nabla(v^{2}/2)+\mathbf{v}\times(\nabla\times\mathbf{v})\right)_{r}}}}\,\,\underbrace{-\left(\underline{\underline{Dr}}*\underline{p}\right)\oslash\underline{\rho}}_{{-\frac{1}{\rho}(\nabla p)_{r}}}\\
 & \,\,\,\,\,\,\,\underbrace{-\underline{\varPi_{r}}\oslash\underline{\rho}}_{-\frac{1}{\rho}(\nabla\cdot\overline{\boldsymbol{\pi}})_{r}}\,\,+\underbrace{\left(\underset{}{-(\underline{\underline{Dr}}*\underbar{\ensuremath{\psi}})}\circ\left(\underline{\underline{\Delta^{^{*}}}}\,\,\underline{\psi}\right)-\left(\underline{f}\circ\left(\underline{\underline{Dr}}*\underline{f}\right)\right)\right)\oslash\left(\mu_{0}\,\underline{r}{}^{2}\circ\underline{\rho}\right)}_{\frac{1}{\rho}(\mathbf{J\times}\mathbf{B})_{r}}\\
\underline{\dot{v}_{\phi}} & =\underbrace{-\underline{\mathbf{v}}\cdot\left(\underline{\underline{\nabla}}\,\,\left(\underline{r}\circ\underline{v_{\phi}}\right)\right)\oslash\underline{r}}_{\mathclap{{-(\mathbf{v\cdot\nabla}\mathbf{v})_{\phi}=\left(-\nabla(v^{2}/2)+\mathbf{v}\times(\nabla\times\mathbf{v})\right)_{\phi}}}}\,\,\,\,\,\,\,\,\,\,\underbrace{-\underline{\varPi_{\phi}}\oslash\underline{\rho}}_{-\frac{1}{\rho}(\nabla\cdot\overline{\boldsymbol{\pi}})_{\phi}}+\underbrace{\left(\underline{\underline{W_{n}}}*\left(\underline{\mathbf{B}_{\theta}^{e}}\cdot\left(\underline{\underline{\nabla^{e}}}\,\,\underline{f}\right)\right)\right)\oslash\left(\mu_{0}\,\underline{r}\circ\underline{\rho}\right)}_{\frac{1}{\rho}(\mathbf{J\times}\mathbf{B})_{\phi}}\\
\underline{\dot{v}_{z}} & =\underbrace{-\underline{\underline{Dz}}*\left(\frac{\underline{v}^{2}}{2}\right)+\underline{v_{r}}\circ\left(\underline{\underline{Dz}}*\underline{v_{r}}-\underline{\underline{Dr}}*\underline{v_{z}}\right)+\underline{v_{\phi}}\circ\left(\underline{\underline{Dz}}*\left(\underline{r}\circ\underline{v_{\phi}}\right)\right)\oslash\underline{r}}_{\mathclap{{-(\mathbf{v\cdot\nabla}\mathbf{v})_{z}=\left(-\nabla(v^{2}/2)+\mathbf{v}\times(\nabla\times\mathbf{v})\right)_{z}}}}\,\,\underbrace{-\left(\underline{\underline{Dz}}*\underline{p}\right)\oslash\underline{\rho}}_{{-\frac{1}{\rho}(\nabla p)_{z}}}\,\,\\
 & \,\,\,\,\,\,\,\underbrace{-\underline{\varPi_{z}}\oslash\underline{\rho}}_{-\frac{1}{\rho}(\nabla\cdot\overline{\boldsymbol{\pi}})_{z}}\,\,+\underbrace{\left(-\left(\underline{\underline{Dz}}*\underbar{\ensuremath{\psi}}\right)\circ\left(\underline{\underline{\Delta^{^{*}}}}\,\,\underline{\psi}\right)-\underline{f}\circ\left(\underline{\underline{Dz}}*\underline{f}\right)\right)\oslash\left(\mu_{0}\,\underline{r}{}^{2}\circ\underline{\rho}\right)}_{\frac{1}{\rho}(\mathbf{J\times}\mathbf{B})_{z}}
\end{align*}
Here, the components of $\underline{\boldsymbol{\varPi}}$ represent
the discrete forms of the components of $\nabla\cdot\boldsymbol{\underline{\pi}}$,
and are defined by equations \ref{eq:258.3}, \ref{eq:266} and \ref{eq:269.2}:
\begin{align}
\underline{\varPi_{r}} & =\biggl[\underset{}{-2\underline{\underline{Dr_{n}}}*\left(\underline{\mu^{e}}\circ\underline{r^{e}}\circ\left(\underline{\underline{Dr^{e}}}*\underline{v_{r}}\right)\right)}-\underset{}{\underline{\underline{Dz_{n}}}*\left(\underline{\mu^{e}}\circ\underline{r^{e}}\circ\left(\underline{\underline{Dr^{e}}}*\underline{v_{z}}+\underline{\underline{Dz^{e}}}*\underline{v_{r}}\right)\right)}\biggr]\oslash\underline{r}\nonumber \\
 & \underset{}{+\frac{2}{3}\left(\underline{\underline{Dr_{n}}}*\left(\underline{\mu^{e}}\circ\left(\underline{\underline{\nabla^{e}}}\cdot\underline{\mathbf{v}}\right)\right)\right)}+\underset{}{2\,\underline{\mu}\circ\underline{v_{r}}\oslash\underline{r}^{2}}\nonumber \\
\underline{\varPi_{\phi}} & =-\left(\underline{\underline{\nabla_{n}}}\cdot\left(\underline{\mu^{e}}\circ\underline{r^{e}}^{2}\circ\left(\underline{\underline{\nabla^{e}}}\,\,\underline{\omega}\right)\right)\right)\oslash\underline{r}\nonumber \\
\underline{\varPi_{z}} & =\biggl[\underset{}{-2\underline{\underline{Dz_{n}}}*\left(\underline{\mu^{e}}\circ\underline{r^{e}}\circ\left(\underline{\underline{Dz^{e}}}*\underline{v_{z}}\right)\right)}-\underset{}{\underline{\underline{Dr_{n}}}*\left(\underline{\mu^{e}}\circ\underline{r^{e}}\circ\left(\underline{\underline{Dr^{e}}}*\underline{v_{z}}+\underline{\underline{Dz^{e}}}*\underline{v_{r}}\right)\right)}\biggr]\oslash\underline{r}\nonumber \\
 & \underset{}{+\frac{2}{3}\left(\underline{\underline{Dz_{n}}}*\left(\underline{\mu^{e}}\circ\left(\underline{\underline{\nabla^{e}}}\cdot\underline{\mathbf{v}}\right)\right)\right)}\label{eq:127.1111}
\end{align}
The discrete form of the single fluid energy equation, where thermal
diffusion is temporarily neglected, is given by the combination of
equations \ref{eq:235.01}, \ref{eq:242}, \ref{eq:375.2}, and \ref{eq:127.4}:
\begin{align}
\underline{\dot{p}} & =-\underline{\mathbf{v}}\cdot\left(\underline{\underline{\nabla}}\,\,\underline{p}\right)-\gamma\,\underline{p}\circ\left(\underline{\underline{\nabla}}\cdot\underline{\mathbf{v}}\right)+(\gamma-1)\,\biggl[\underbrace{\underline{Q_{\pi}}}_{-\overline{\boldsymbol{\pi}}:\nabla\mathbf{v}}+\underset{\eta'J_{\phi}^{2}}{\underbrace{\left(\underline{\eta}/\mu_{0}\right)\circ\left(\left(\underline{\underline{\Delta^{^{*}}}}\,\,\underline{\psi}\right)\oslash\underline{r}\right)^{2}}}\nonumber \\
 & \,\,\,\,+\underset{\eta'J_{\theta}^{2}}{\underbrace{\underline{\underline{W_{n}}}*\left(\left(\underline{\eta^{e}}/\mu_{0}\right)\circ\left(\left(\underline{\underline{\nabla^{e}}}\,\,\underline{f}\right)\oslash\underline{r^{e}}\right)^{2}\right)}}\biggr]\label{eq:127.112}
\end{align}
Here, $\underline{Q_{\pi}}$, the discrete form of $-\overline{\boldsymbol{\pi}}:\nabla\mathbf{v}$,
is given by equation \ref{eq:242}: 
\begin{align}
\underline{Q_{\pi}} & =\underline{\underline{W_{n}}}*\biggl[\underline{\mu^{e}}\circ\biggl\{2\left(\underline{\underline{Dr^{e}}}*\underline{v_{r}}\right)^{2}+2\left(\underline{\underline{Dz^{e}}}*\underline{v_{z}}\right)^{2}+\left(\underline{r^{e}}\circ\left(\underline{\underline{\nabla^{e}}}\,\,\underline{\omega}\right)\right)^{2}\nonumber \\
 & +\left(\underline{\underline{Dr^{e}}}*\underline{v_{z}}+\underline{\underline{Dz^{e}}}*\underline{v_{r}}\right)^{2}-\frac{2}{3}\left(\underline{\underline{\nabla^{e}}}\cdot\underline{\mathbf{v}}\right)^{2}\biggr\}\biggr]+2\,\underline{\mu}\circ\left(\underline{v_{r}}\oslash\underline{r}\right)^{2}\label{eq:127.113}
\end{align}
\\
The discretised single fluid energy equation can be partitioned into
parts pertaining to the ions and electrons, as shown in section \ref{subsec:Discretised-MHD-model}.
Discretised expressions for $\dot{\psi}$ and $\dot{f}$ are given
by equations \ref{eq:374}, and the combination of equations \ref{eq:376.3}
and \ref{eq:127.81}:

\begin{align*}
\underline{\dot{\psi}} & =-\underline{\mathbf{v}}\cdot\left(\underline{\underline{\nabla}}\,\,\underline{\psi}\right)+\underline{\eta}\circ\left(\underline{\underline{\Delta^{^{*}}}}\,\,\underline{\psi}\right)\\
\underline{\dot{f}} & =\underline{r^{2}}\circ\left[-\underline{\underline{\nabla}}\cdot\left(\underline{f}\circ\underline{\mathbf{v}}\oslash\underline{r}^{2}\right)+\underline{\underline{\nabla_{n}}}\cdot\left(\underline{\mathbf{B}_{\theta}^{e}}\circ\underline{\omega^{e}}\right)+\underline{\underline{\nabla_{n}}}\cdot\left(\underline{\eta^{e}}\circ\left(\underline{\underline{\nabla^{e}}}\,\,\underline{f}\right)\oslash\underline{r^{e}}^{2}\right)\right]
\end{align*}

\newpage{}

\chapter{General code setup\label{chap:Gen code setup}}

In this appendix, general aspects of the code setup will be discussed.
Section \ref{subsec:Computational-grid} describes the computation
mesh, and the methods developed to repair damaged meshes and introduce
non-uniform cell size. Section \ref{sec:Timestepping-methods} outlines
the options implemented for timestepping, and describes the protocols
implemented for timestep correction and data preservation. Section
\ref{subsec:Diffcoefs} presents a general discussion of diffusion
coefficients, and how they are implemented to the code. The numerical
methods developed to solve the Grad-Shafranov equation are presented
in section \ref{sec:Numerical-solution-of}. Section \ref{sec:Code-validation}
describes the analysis used to verify and validate the code. This
appendix concludes with a summary in section \ref{sec:SummaryGencode setup}.

\section{Computational grid\label{subsec:Computational-grid} }

We adapted the freely-available DistMesh MATLAB mesh generator algorithms
described in \cite{MeshPersson1,MeshPersson2,MeshPersson3} to provide
the computational grid. DistMesh generates unstructured triangular
and tetrahedral meshes, and employs Delaunay triangulation to optimize
node locations with a force-based smoothing procedure, regularly updating
the topology with the Delaunay routine. The boundary points are restrained
and can only move tangentially to the boundary. The $r$ and $z$
coordinates of points that denote the boundary of the computational
domain are required as inputs to the meshing program. These points,
which should have an interspacing that is sufficiently small for definition
of any curved features, can be extracted from Solidworks models of
the machine geometry. The points at the corners of the computational
domain are assigned as fixed nodes that remain fixed in space during
the mesh generation. For simulations where the MHD solutions are coupled
to the levitation/compression vacuum field solution (see section \ref{sec:Vacuum-field-in})
in electrically insulating areas, fixed nodes are also assigned along
the interface between the insulating region and the plasma-domain.

The mesh generator code was found to be not entirely robust, and some
repairs had to be implemented in order to obtain usable meshes. 
\begin{figure}[H]
\subfloat[Mesh over CT confinement region indicating smallest element location]{\begin{centering}
\includegraphics[width=5cm,height=4.7cm]{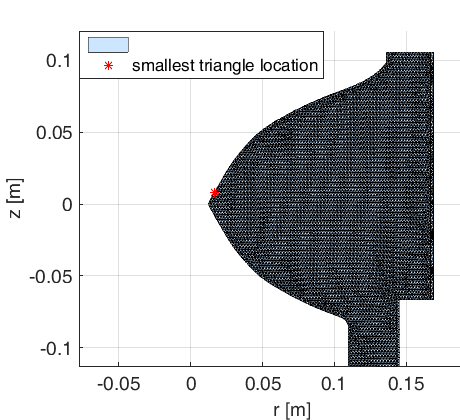}
\par\end{centering}
}\hfill{}\subfloat[Close-up of smallest triangle]{\centering{}\includegraphics[width=5.5cm,height=4.7cm]{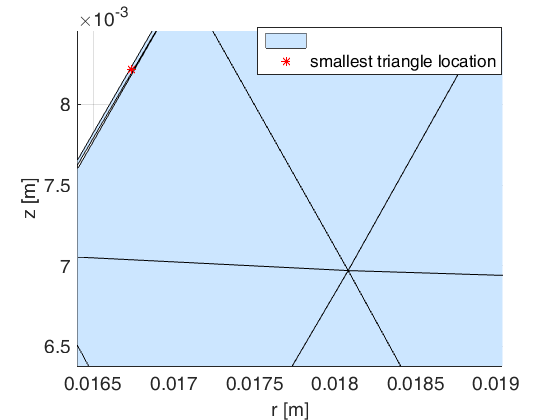}}\hfill{}\subfloat[Element size distribution]{\centering{}\includegraphics[width=5.5cm,height=4.7cm]{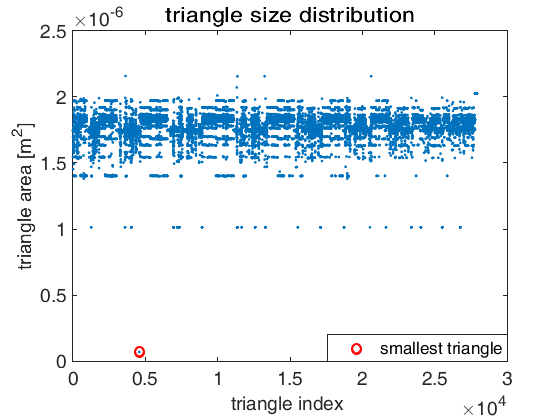}}\caption{$\,\,\,\,$Mesh repairs\label{fig:smallest triangle}}
\end{figure}
In particular, meshes produced with DistMesh would often include exceptionally
small elongated triangle elements, usually located near curved boundaries,
as indicated in figure \ref{fig:smallest triangle}. Figures \ref{fig:smallest triangle}(a)
and (b) show the part of a mesh that represents the CT confinement
region, with the smallest triangle highlighted. The figures here are
from a Matlab program that was made to diagnose and repair the meshes
produced by DistMesh. The mesh shown here is intended to have approximately
uniform triangle sizes, $h_{e}=2\mbox{\mbox{ mm}}$. Figure \ref{fig:smallest triangle}(c)
indicates the triangle size distribution. 

Arrays $\mathbb{P}$ and $\mathbb{T}$ are the outputs of the DistMesh
program. $\mathbb{P}$ has dimensions $[N_{n}\times2]$, where $N_{n}$
is the number of nodes in the mesh; each row contains the $r$ and
$z$ coordinates of an individual node. $\mathbb{T}$ has dimensions
$[N_{e}\times3]$, where $N_{e}$ is the number of triangles in the
mesh, and each row in $\mathbb{T}$ contains the indexes of the nodes
that define the vertices of an individual triangle. The node index
is defined as the number of the node-corresponding row in $\mathbb{P}$.

To allow the MHD code to run, any erroneously small triangles must
be removed, by deleting the row(s) corresponding to the small triangle(s)
in array $\mathbb{T}$. In cases where the exceptionally small triangle
is located exterior to the nominal boundary, for example, as indicated
in figures \ref{fig:smallest triangle}(a) and (b), this repair is
sufficient. When several tiny triangles are included in the original
mesh, removing the corresponding rows from $\mathbb{T}$ may results
in mapping errors, leading to holes (missing triangles) in the mesh.
This can be addressed by adding new row(s) to $\mathbb{T}$ that include
the indexes of the nodes that make up the vertices of the missing
element(s). 

The MHD code has the option of running with a mesh with uniformly
sized triangular elements, or with a mesh with increased resolution
near the boundaries. To implement the variable resolution option,
the $dpoly$ function, that is part of the DistMesh package, is used
calculate the distances from mesh-interior points to the nearest boundary,
so that triangle size can be scaled inversely to the distance from
the nearest boundary. The principle mesh options for MHD code, characterised
by code-parameter $meshtype$, are:
\begin{table}[H]
\raggedright{}%
\begin{tabular}{cl}
$meshtype=0$ & Uniform triangles of size $h_{e}$\tabularnewline
$meshtype=1$ & Triangles have size $h_{e}$ at boundaries, and gradually get bigger
towards \tabularnewline
 & the interior, to maximum size $h_{max}$. Triangles \tabularnewline
 & in the insulating areas have size $h_{max}$.\tabularnewline
$meshtype=2$ & Triangles have size $h_{e}$ at boundaries, and gradually get bigger
towards \tabularnewline
 & the interior, to maximum size $h_{max}$ wherever $d\geq d_{electrode}$($i.e.,$
distance \tabularnewline
 & to nearest boundary is greater or equal to the distance between the
inner \tabularnewline
 & and outer formation electrodes).\tabularnewline
\end{tabular}
\end{table}
\begin{figure}[H]
\subfloat[$meshtype=1$ ]{\begin{centering}
\includegraphics[width=7cm,height=5cm]{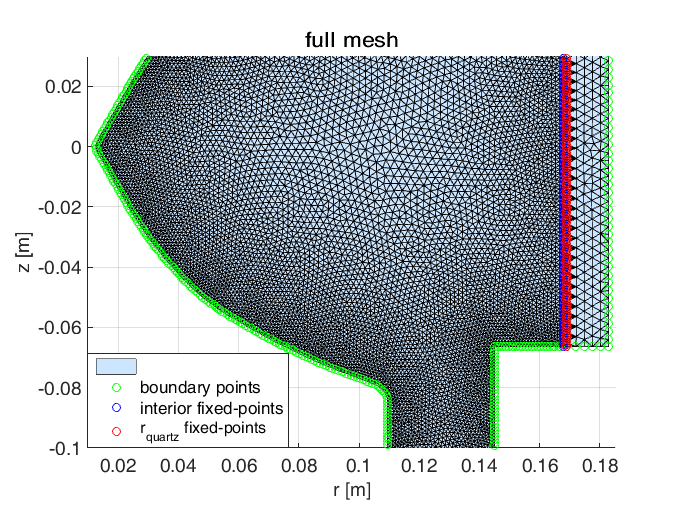}
\par\end{centering}
}\hfill{}\subfloat[$meshtype=2$ ]{\begin{centering}
\includegraphics[width=7cm,height=5cm]{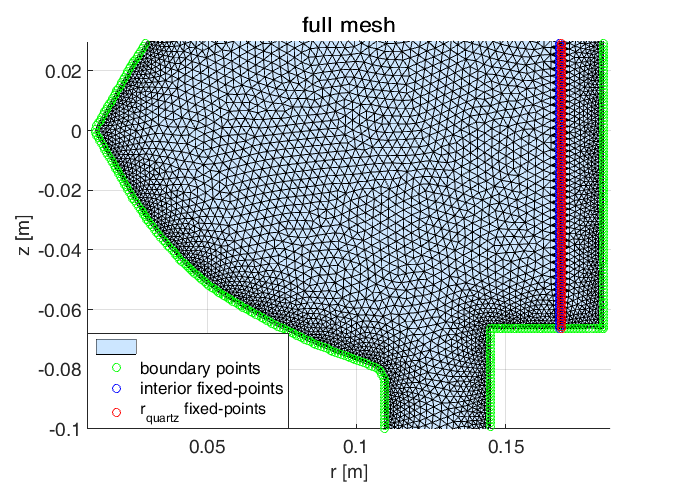}
\par\end{centering}
}\caption{\label{fig:Variable-resolution-meshes-with}$\,\,\,\,$Variable-resolution
meshes}
\end{figure}
Figures \ref{fig:Variable-resolution-meshes-with}(a) and (b) indicate
sections of meshes produced for cases with $meshtype=1$ and $meshtype=2.$
The functionality of the indicated interior and $r_{quartz}$ fixed
points, which are required to solve different sets of equations in
different parts of the domain, is described in section \ref{sec:Vacuum-field-in}.
In general, as shown in section \ref{subsec:Convergence-Study}, a
good degree of solution accuracy can be obtained with meshes with
resolution, defined by $h_{e}$, of around 5 mm. Mesh resolution should
be increased, to $h_{e}\sim2\mbox{\mbox{ mm}}$, for high quality
contour plots of output fields. The main benefit of increasing resolution
near the boundaries is that simulation times can be reduced (fewer
elements) while maintaining resolution at boundaries, where the imposition
of temperature and velocity boundary conditions in particular is more
likely to lead to solution overshooting with coarser meshes. 

\section{Timestepping methods\label{sec:Timestepping-methods}}

The code has the option of using either the forward Euler, Runge-Kutta
2, or Runge-Kutta 4 timestepping scheme. The forward Euler method
is adequate for the short physical timespans ($\lesssim300\,\upmu\mbox{s)}$
required to model the magnetic compression experiments. For these
short simulations, the Runge-Kutta schemes don't produce different
results but increase simulation runtimes by a factor of approximately
two (Runge-Kutta 2) or four (Runge-Kutta 4). The error associated
with the forward Euler scheme scales with $N\,dt$, where $dt$ is
the timestep and $N$ is the number of iterations, and so it increases
with the physical timespan associated with the simulation. For extended
simulations, the Runge-Kutta schemes, where error scales with $N\,dt^{2}$
(Runge-Kutta 2) or $N\,dt^{4}$ (Runge-Kutta 4) are more suitable.
With the forward Euler method, the field $X(t)$ being evolved is
updated at each timestep using the value at the previous timestep
according to:
\begin{equation}
X(t+dt)=X(t)+dt\,\dot{X}\label{eq:143.2}
\end{equation}

\subsection{Timestep correction and data preservation}

The code saves the evolved fields to file every $t_{out}$ seconds
of simulation time, where $t_{out}$ is a code input parameter, typically
set to $t_{out}=1\,\upmu$s. If code input parameter $Saveplots$
is set to one, plots of various interesting quantities against simulation
time are saved to file each time the evolved fields are saved to file.
For example, time-evolution of simulated diagnostics such as \textbf{$B_{\theta}$
}and\textbf{ $B_{\phi}$} at the magnetic probe locations, and line-averaged
electron density and ion temperature along the interferometer and
ion-Doppler chords are produced. In addition, graphs are saved to
record the time-evolution of $\psi_{CT}$, total system toroidal flux,
CT $q(\psi)$ (safety factor) profile, maximum ion and electron temperatures
in the system, system energy components, CT volume, CT outer separatrix,
and magnetic field components at the locations of the magnetic probes
on the Marshall gun wall below the containment region. If the code
input parameter $Savecontourplots$ is also set to one, contour plots
of the evolved fields ($\psi,\,f,\,n_{e},\,p_{i},\,p_{e},\,v_{r},\,v_{\phi}$
and $v_{z}$, as well as $T_{i},\,T_{e}$, $J_{r},\,J_{\phi}$ and
$J_{z}$ are saved to file every $t_{out}\,\upmu$s. When code input
parameter $neutralfluid$ is set equal to one, and a neutral fluid
is also being evolved (see chapter \ref{chap:Neutral-models}), contour
plots of $n_{n},\,p_{n},\,T_{n},\,v_{nr},\,v_{n\phi}$ and $v_{nz}$
(the neutral fluid number density, pressure, temperature and velocity
components) are also saved to file when $Savecontourplots$ is also
set to one.

If the simulation timestep exceeds the minimum required for numerical
stability, unphysical field values, such as negative densities or
pressures, can be calculated. These errors ultimately lead to imaginary
values in all the evolved fields. Throughout the simulations, after
each timestep, a check is performed to see if any of the values of
the $\psi$ field (which is coupled to the other field solutions)
are imaginary. If so, the timestep is reduced by a factor of $tsrf\sim2$
(the \textquotedbl timestep reduction factor\textquotedbl , a code
input parameter is typically set to \textasciitilde 2). When this
occurs, the simulation time and evolved fields are overwritten with
the last saved values, and the simulation continues with the reduced
timestep.

Although extremely low or high diffusion coefficient values can be
used while maintaining numerical stability if the timestep is reduced,
lower and upper limits need to be imposed on diffusion coefficients
in order to achieve numerical stability in combination with acceptably
short simulation runtimes. This is consistent with the standard stability
criteria \cite{JARDIN} that timestep is limited by the constraints
$dt\lesssim D_{min}/V_{max}^{2}$ and $dt\lesssim h_{e}^{2}/D_{max}$,
where $D_{min}$ and $D_{max}$$\,[\mbox{m}^{2}\mbox{/s}]$ are the
minimum and maximum diffusion coefficients, and $V_{max}$ is the
maximum speed associated with the system.

The code can also calculate, at each timestep, the maximum allowed
timestep for the subsequent timestep, based on the mesh resolution,
the ion sound speed and Alfven speed, and the diffusion coefficients.
The limit applied on the maximum allowed timestep by the Alfven speed,
$V_{A}=\frac{B}{\sqrt{\mu_{0}\rho}}$, is $dt_{V_{A}}=\frac{h_{e}}{V_{A}}$,
where $h_{e}$ is the dimension associated with the smallest triangle
element in the mesh. Similarly, the ion acoustic speed $c_{s}=\sqrt{\frac{T_{i}\mbox{[J]}+T_{e}\mbox{[J]}}{m_{i}}}$
imposes a limit of $dt_{c_{s}}=\frac{h_{e}}{c_{s}}$. The limit imposed
by the diffusion coefficients is $dt_{D}\sim\frac{h_{e}^{2}}{4D_{max}}$.
Here, $D_{max}\,[\mbox{m}^{2}/\mbox{s}]$ is the maximum of all the
active diffusion coefficients, $D_{max}=max\,(\chi_{\parallel i},\,\chi_{\parallel e},\,\chi_{\perp i},\,\chi_{\perp e},\,\eta,\,\nu,\,\zeta)$.
Note that all these diffusion coefficients may vary, depending on
other code input parameters, from node to node as a function of plasma
parameters. The factor of four in the expression for $dt_{D}$ arises
when the system is two-dimensional \cite{wwwDiffDT}. It was found
that, for reasonably high $D_{max}$ ($e.g.,$ $D_{max}=\chi_{\parallel e}\sim16000\,\mbox{m}^{2}/\mbox{s}$,
that $dt_{D}$ places the limit on the maximum allowed timestep over
the early part of the simulation, but that later on in the simulation
$dt_{v_{A}}<dt_{D}$, and so $dt_{v_{A}}$ imposes the limit as the
magnetic field strength increases after the bubble-in process. At
typical magnetic compression settings, $dt_{v_{A}}$ decreases further
near peak magnetic compression, and then increases again after peak
compression. 

In practice, it was found that no significant reduction in overall
simulation time was found with the more sophisticated method - the
simpler timestep adjustment method using the timestep reduction factor
seems quite adequate. It is worth noting that artificially increasing
the ion mass $m_{i}$, allows for an increased timestep, as it reduces
the Alfven and thermal speeds, but also leads to increased plasma
fluid inertia, with a consequently delayed and weakened formation
process for simulations that include CT formation. Reducing Alfven
speed by increasing $\mu_{0}$ also enables increased timestep, but
leads to weakened formation and compression, as it reduces advective
$\mathbf{J}\times\mathbf{B}$ forces. Overall, these methods are not
practical when simulating CT formation as they alter the physics,
but can be useful for code testing.

\section{General discussion of diffusion coefficients\label{subsec:Diffcoefs}}

Care must be taken when choosing the values of the various diffusion
coefficients, which are inputs to the code. The primary diffusion
coefficients, $\nu,\,\eta$, and $\chi$, which determine viscosity,
electrical resistivity, and thermal conduction respectively, are physical
quantities that depend on inter-particle collisions on the microscopic
level, and determination of the physically correct values is not trivial. 

On the other hand, minimum levels of diffusion are required for numerical
stability. Excessive spatial gradients in the various fields being
evolved can result, at the timestep update, in overshooting of the
field solutions, leading to spurious results at the regions of steep
gradients. For example, viscosity, which appears in the momentum equation,
acts to smooth out the velocity field. Resistivity appears in the
expressions for $\dot{\psi}$ and $\dot{f}$, and smooths out gradients
in the magnetic field, and thermal conduction in the species energy
equations acts to smooth the pressure fields. In addition, density
diffusion, which is artificial, is required to smooth out the density
field, and also contributes to smoothing of the pressure fields. Excessive
smoothing can blur the physics. Ideally, the \textquotedbl physically
correct\textquotedbl{} diffusion coefficients would be higher than
the coefficients required for stability, but this may not the case.
Increasing mesh resolution in regions of high gradients, and reducing
the timestep, can reduce the minimum required levels of diffusion,
but increasing resolution or reducing the timestep indefinitely is
computationally impractical as it leads to excessive simulation run-times.

\subsection{Thermal diffusion}

Code input parameter $vary_{\chi}$ determines whether constant values,
or values that vary with plasma parameters according to established
relationships, are used as thermal diffusion coefficients. If $vary_{\chi}$
is set equal to one, then the thermal diffusion coefficients\\
 $\chi_{\parallel e}(T_{e}(\mathbf{r},t),\,n(\mathbf{r},t)),\,\,\,\chi_{\parallel i}(T_{i}(\mathbf{r},t),\,n(\mathbf{r},t)),\,\,\,\chi_{\perp e}(T_{e}(\mathbf{r},t),\,n(\mathbf{r},t),\,B(\mathbf{r},t))$,
and\\
 $\chi_{\perp i}(T_{i}(\mathbf{r},t),\,n(\mathbf{r},t),\,B(\mathbf{r},t))$
are updated at each timestep according to the Braginskii expressions
given in equations \ref{eq:472.45} and \ref{eq:472.50}. 

As outlined in section \ref{sec:Timestepping-methods}, the maximum
acceptable timestep for explicit timestepping schemes can be limited
by the maximum diffusion coefficient, so inconveniently small timesteps
are required for very large diffusion coefficients. In addition, in
order to have sufficient smoothing of temperature gradients, mesh
resolution needs to be increased, or time-step reduced, as diffusion
coefficients are reduced to extremely small values. As an example,
at $T_{i}=T_{e}=100\mbox{ eV}$, $n=1\times10^{20}\mbox{ m}^{-3}$,
$Z_{eff}=1$, $m_{i}=4m_{p},$ where $m_{p}$ is the proton mass,
and $B=1$ Tesla, the Braginskii coefficients give $\chi_{\parallel e}\sim2\times10^{7},\,\,\,\chi_{\parallel i}\sim4\times10^{5},\,\,\,\chi_{\perp e}\sim0.01,\mbox{ and }\chi_{\perp i}\sim0.05\,\,[\mbox{m}^{2}/\mbox{s}]$.
Hence, if the Braginskii expressions for the thermal diffusion coefficients
are used, it is necessary to place upper and lower limits on the coefficients
in order to have a conveniently large timestep ($e.g.,\,dt\geq1\times10^{-11}$
s), moderate mesh resolution ($e.g.,\,h_{e}\geq2$ mm), and consequently
moderate simulation run times ($e.g.,$ \textasciitilde 1-2 days
for a 90$\,\upmu$s simulation with $h_{e}=2$ mm). For these practical
values of $h_{e}$ and $dt,$ it was found that, for formation simulations,
that the thermal diffusion coefficients should lie between around
25 m$^{2}/$s and 20000 m$^{2}/$s. Implementation of an implicit
timestepping scheme would extend the range of acceptable thermal diffusion
coefficients. In any case, the Braginskii formulae for diffusion coefficients
are based on a model that doesn't include turbulence and other anomalous
effects, and wouldn't be expected to be completely suitable, especially
for CT formation simulations which involve extreme accelerations and
turbulence. 

If $vary_{\chi}$ is set equal to two, the empirical Bohm scaling
\cite{bittencourt}, expressed as 
\[
\chi_{\perp\alpha}(\mathbf{r},t)=(T_{\alpha}(\mathbf{r},t)\,[\mbox{eV}])/(16\,B(\mathbf{r},t)\,[\mbox{T}])
\]
is used to update the perpendicular thermal diffusion coefficients,
again constrained by upper and lower bounds, at each timestep, and
constant values are chosen for the field-parallel coefficients.

In general, it is found that moderate run times and a good agreement
to experimental magnetic compression data can be had by keeping $vary_{\chi}=0$,
and using constant values for $\chi_{\parallel\alpha}$ and $\chi_{\perp\alpha}$.

\subsection{Viscous diffusion\label{subsec:Viscous-diffusion}}

The Braginskii formulae give expressions for species kinematic viscosity:\\
$\nu_{\alpha}\,[\mbox{m}^{2}/\mbox{s}]=\mu_{\alpha}\,[\mbox{kg/m-s}]/\rho_{\alpha}\,[\mbox{kg/m}^{3}]$
in an unmagnetized, turbulence-free plasma. When single fluid MHD
is implemented, only the ion viscosity is relevant. For a magnetized
plasma, the Braginskii formulae give expressions for $\nu_{\parallel\alpha}$
and $\nu_{\perp\alpha}$, but again these expressions are only relevant
for cases without turbulence. Hence, the physically correct values
are basically undetermined. Equation \ref{eq:472.47} gives the expression
for the parallel ion viscosity as $\nu_{\parallel i}=0.96\,\tau_{ii}\,T_{i}\,[\mbox{J}]/m_{i}$,
while perpendicular ion viscosity is given by $\nu_{\perp i}=3\,T_{i}\,[\mbox{J}]/10\,\omega_{ci}^{2}\,m_{i}\,\tau_{ii}$\cite{farside braginskii},
where $\tau_{ii}$ is the ion-ion collision time is given by equation
\ref{eq:472.37}, and $\omega_{ci}$ is the ion cyclotron frequency.
For $n=1\times10^{20}\,[\mbox{m}^{-3}]$, $T_{i}=100$ eV, $B=1$
T, and $m_{i}=4m_{p}$, this gives the parallel and perpendicular
components of the ion viscosity as $\nu_{\parallel i}\sim1\times10^{5}\,[\mbox{m\ensuremath{^{2}}/s}]$,
and $\nu_{\perp i}\sim0.03\,[\mbox{m\ensuremath{^{2}}/s}]$. As mentioned
in appendix \ref{sec:SummaryKin_MHD_EQ}, electron viscous heating
is neglected in the two-temperature, single fluid MHD model. Including
anisotropic viscosity would add complexity to the expression for the
viscous tensor \cite{farside braginskii}, so isotropic viscosity
is implemented in the code for simplicity.

In practice, for formation simulations using a moderate timestep,
and with a mesh of practical resolution ($h_{e}\sim2\mbox{\mbox{\mbox{ mm}}})$,
it is found that we need to use a value of $\nu_{num}\gtrsim500\;\mbox{m}^{2}/\mbox{s}$
to maintain sufficiently smooth velocity fields for numerical stability.
This level appears excessively high, as it leads to excessive viscous
heating of the ions, especially for simulations with low density,
when plasma velocity is high, such as during CT formation. Increased
number density reduces the maximum observed ion temperature. Some
researchers \cite{Hooper,Hooper2} remove the viscous heating term
entirely from the ion energy equation, but this is not considered
to be a good choice when CT formation is being modelled, when the
high ion speeds do in fact lead to extreme viscous heating. When deuterium
is used for gas-fill, fusion neutrons are routinely detected during
CT formation on GF plasma injectors, and have also been detected during
formation on the magnetic compression experiment on the last night
before the experiment was disassembled, when scintillators were installed
on the machine (see section \ref{subsec:Scintillator-data}). Viscous
heating is thought to be the mechanism behind this ion heating. We
want to reduce viscous heating to acceptable levels, but do not want
to eliminate it entirely. 

To reduce ion viscous heating for low density simulations, so that
simulated ion temperature is close to the levels indicated by the
ion-Doppler diagnostic, $\nu_{num}$ is used in the viscous tensor
in the expression for $-\nabla\cdot\overline{\boldsymbol{\pi}}$ in
the momentum equation, and a reduced value, a crude estimate of the
\textquotedbl physically correct\textquotedbl{} value, $\nu_{phys}<\nu_{num}$
may be used in the expression for $-\overline{\boldsymbol{\pi}}:\nabla\mathbf{v}$
in the ion energy equation. Other codes, $e.g.,$ HiFi \cite{MeierPhd}
also adopt this approach. In practice, determining $\nu_{phys}$ is
not straightforward, $\nu_{phys}$ is always an approximation at best.
For the relatively high densities associated with the magnetic compression
experiment, it is possible to set $\nu_{phys}\sim\nu_{num}$, or even
equal to $\nu_{num}$, and still achieve reasonable simulated ion
temperatures. However, it was necessary to increase numerical viscosity
to $\nu_{num}>\sim1500\;\mbox{m}^{2}/\mbox{s}$, and to set $\nu_{phys}\sim\nu_{num}/3$,
for numerical stability and reasonable simulated ion temperatures
when simulating CT formation in SPECTOR (see section \ref{sec:Neutral_SPECTOR})
geometry. SPECTOR CTs have densities around five to ten times lower
than those associated with the magnetic compression experiment. Ideally,
the strategy of having different values for $\nu_{num}$ and $\nu_{phys}$
would be avoided because it breaks the conservation of system energy
pertaining to the viscous terms (the terms in the last set of square
brackets in equation \ref{eq:518}). However, it was found that the
discrepancy does not, apart from a reduction in ion viscous heating,
significantly alter these particular simulation results, and appears
at this stage to be an acceptable method for reducing ion heating
in simulation scenarios with low density and high plasma velocities.

\subsection{Resistive diffusion}

Code input parameter $vary_{\eta}$ determines whether fixed values,
or values that vary in space and time according to the Spitzer formula,
are used for the isotropic plasma resistive diffusion coefficient.
If $vary_{\eta}=1$, equation \ref{eq:517.6} is used to calculate
the resistive diffusion coefficient at each timestep. To limit the
timestep to acceptably high values, an upper limit is placed on $\eta$,
typically $\eta_{max}=5000\,[\mbox{m}^{2}/\mbox{s}]$, corresponding
to constant $\eta=\eta_{max}$ at nodes where $T_{e}\apprle0.5$ eV.
It would be reasonably straightforward to implement anisotropic resistivity
(much simpler than implementing anisotropic viscosity), but since
the parallel and perpendicular coefficients are of the same order
($\eta_{\parallel}\sim\frac{1}{2}\eta_{\perp}$\cite{farside braginskii}),
isotropic resistivity was considered to be a good approximation of
the physical situation. 

\subsection{Density diffusion}

An artificial density diffusion term $\underline{\underline{\nabla_{n}}}\cdot\left(\underline{\zeta}\circ\underline{\underline{\nabla^{e}}}\,\,\underline{n}\right)$
is included in the mass continuity equation (see equation \ref{eq:517.3}).
A certain minimum level of density diffusion is required for numerical
stability, as density diffusion acts to smooth out gradients in the
density field. Increasing mesh resolution in regions of high gradients
can reduce the minimum required level of density diffusion, but this
is often computationally impractical. The code input parameter $\zeta$
gives the coefficient of density diffusion, and, for simulations including
magnetic compression, is typically required to be at least around
30 m$^{2}$/s for numerical stability in combination with an acceptably
high timestep and consequently short runtimes. As shown in section
\ref{subsec:Maintenance-of-momentum}, correction terms have been
developed and included in the discrete momentum and energy equations
in order to maintain energy conservation, and angular momentum conservation
in some scenarios, when density diffusion is included in the model.

\section{Numerical solution of the Grad-Shafranov equation\label{sec:Numerical-solution-of}}

The MHD code has the option of using an equilibrium described by the
Grad-Shafranov equation (section \ref{sec:Equilibrium-models}) as
starting point from which the dynamical solutions to the various fields
are evolved. The Grad-Shafranov equation (equation \ref{eq:20}) can
be written as: 
\begin{equation}
\Delta^{*}\psi+w(\psi)=0\label{eq:35.1}
\end{equation}
where $w(\psi)=\mu_{0}r^{2}\frac{dp(\psi)}{d\psi}+f(\psi)\frac{df(\psi)}{d\psi}$.
In general, $w(\psi)$ has non-linear dependence on $\psi$, while
$\Delta^{*}\psi$ has linear dependence on $\psi$. 

An iterative pseudo time-stepping scheme method was developed to solve
equation \ref{eq:35.1}. Over artificial time $\widetilde{t}$, an
initial guess for the solution $\psi$ converges to the correct value
that satisfies equation \ref{eq:35.1}. Using the forward Euler method,
with pseudo timestep $\Delta\widetilde{t}$, the scheme is
\begin{align}
\frac{\psi{}^{n+1}-\psi^{n}}{\Delta\widetilde{t}} & =\Delta^{*}\psi^{n}+w(\psi^{n})\nonumber \\
\Rightarrow\psi{}^{n+1} & =\Delta\widetilde{t}(\Delta^{*}\psi^{n}+w(\psi^{n}))+\psi^{n}\label{eq:35.11}
\end{align}
This explicit scheme results in unstable solutions for $\psi$ without
a suitably small timestep. A better approach for faster convergence
is to use an implicit scheme for the iterations. We keep $w(\psi^{n})$,
the non-linear part of equation \ref{eq:35.1}, as an explicit term,
and solve implicitly for the linear term $\Delta^{*}\psi$:
\begin{align}
\frac{\psi{}^{n+1}-\psi^{n}}{\Delta\widetilde{t}} & =\Delta^{*}\psi^{n+1}+w(\psi^{n})\nonumber \\
\Rightarrow\psi{}^{n+1} & =\Delta\widetilde{t}(\Delta^{*}\psi^{n+1}+w(\psi^{n}))+\psi^{n}\nonumber \\
 & =\Delta\widetilde{t}(\Delta^{*}\psi^{n+1}+w(\psi^{n}))+\psi^{n}+\Delta\widetilde{t}(\Delta^{*}\psi^{n}-\Delta^{*}\psi^{n})\nonumber \\
 & =\Delta\widetilde{t}\;\Delta^{*}\psi^{n+1}+\Delta\widetilde{t}(\Delta^{*}\psi^{n}+w(\psi^{n}))+\psi^{n}-\Delta\widetilde{t}\;\Delta^{*}\psi^{n}\nonumber \\
\Rightarrow\left(1-\Delta\widetilde{t}\;\Delta^{*}\right)\psi^{n+1} & =\Delta\widetilde{t}(\Delta^{*}\psi^{n}+w(\psi^{n}))+\left(1-\Delta\widetilde{t}\;\Delta^{*}\right)\psi^{n}\nonumber \\
\Rightarrow(1-\Delta\widetilde{t}\;\Delta^{*})(\psi^{n+1}-\psi_{n}) & =\Delta\widetilde{t}(\Delta^{*}\psi^{n}+w(\psi^{n}))\nonumber \\
\Rightarrow\psi^{n+1} & =(1-\Delta\widetilde{t}\;\Delta^{*})^{-1}\times(\Delta\widetilde{t}(\Delta^{*}\psi^{n}+w(\psi^{n})))+\psi_{n}\label{eq:35.12}
\end{align}
To find the discrete expression for equation \ref{eq:35.12}, note
that $\left(\nabla_{\perp}\psi\right)|_{\Gamma}\neq0$, while the
differential matrix operator $\underline{\underline{\Delta^{^{*}}}}$,
described in section \ref{subsec:Lapl_delstar}, produces accurate
results at the boundary nodes only when the boundary-normal component
of the gradient of the operand field is zero at the boundary. 

Defining $\underline{\psi}=\underline{\psi_{\Gamma}}+\underline{\psi_{int}}$,
where $\underline{\psi}$ is the vector containing the values of $\psi$
for the Grad-Shafranov solution at the nodes, while $\underline{\psi_{\Gamma}}$
and $\underline{\psi_{int}}$ have the values of $\psi$ set to zero
at internal / boundary nodes respectively, the discrete form of $\Delta^{*}\psi$
is 
\begin{equation}
\underline{\underline{\Delta^{*}}}\,\,\underline{\psi}=\underline{\underline{\Delta^{*}}}\,\,\underline{\psi_{\Gamma}}+\underline{\underline{\Delta_{0}^{*}}}\,\,\underline{\psi_{int}}\label{eq:35.13}
\end{equation}
The operator $\underline{\underline{\Delta_{0}^{^{*}}}}$ is designed
to produce the same results as $\underline{\underline{\Delta^{^{*}}}}$
at internal nodes if the boundary values of the operand field are
set to zero. The boundary values of $\underline{a}$ are preserved
in the operation $\underline{c}=\underline{\underline{\Delta_{0}^{^{*}}}}\,\,\underline{a}$,
and do not contribute to the values calculated for elements corresponding
to internal nodes in \textbf{$\underline{c}$}. Defining $\underline{b}\,[N_{n}\times1]$
as the logical vector defining the indexes of the boundary nodes ($i.e.,$
$\underline{b}(i)=1$ if node $i$ is a boundary node, otherwise $\underline{b}(i)=0$),
$\underline{\underline{\Delta_{0}^{^{*}}}}$ is defined as $\underline{\underline{\Delta_{0}^{^{*}}}}=\underline{\underline{\Delta^{^{*}}}},$
with $\underline{\underline{\Delta_{0}^{^{*}}}}(\underline{b},\,:)=\underline{\underline{\Delta_{0}^{^{*}}}}(:,\,\underline{b})=0$,
and $\underline{\underline{\Delta_{0}^{^{*}}}}(\underline{b},\,\underline{b})=1$.
The Grad-Shafranov solution for $\psi$ is defined as having values
of zero at the boundary nodes, $i.e.,$ $\underline{\psi_{\Gamma}}=\underline{0}$,
so that 
\[
\underline{\underline{\Delta^{*}}}\,\,\underline{\psi}=\underline{\underline{\Delta_{0}^{*}}}\,\,\underline{\psi_{int}}
\]
Hence, referring to equation \ref{eq:35.12}, the Grad-Shafranov solution
for $\psi$ can be evaluated at internal nodes, as 
\begin{align*}
\underline{\psi_{int}}{}^{n+1} & =\left(\underline{\underline{I}}-\Delta\widetilde{t}\;\underline{\underline{\Delta_{0}^{*}}}\right)^{-1}*\Delta\widetilde{t}\;\underline{\Lambda{}_{int}}^{n}+\underline{\psi_{int}}^{n}
\end{align*}
where $\underline{\underline{I}}$ is the identity matrix with dimensions
$N_{n}\times N_{n}$. For the explicit scheme (equation \ref{eq:35.11}),
the Grad-Shafranov solution can be evaluated at internal nodes as
\begin{align*}
\underline{\psi_{int}}{}^{n+1} & =\Delta\widetilde{t}\,\underline{\Lambda{}_{int}}^{n}+\underline{\psi_{int}}^{n}
\end{align*}
Introducing the $\underline{\underline{\Delta_{0}^{*}}}$ operator
and solving equation \ref{eq:35.12} at internal nodes eliminates,
for the implicit scheme, the requirements for boundary values associated
with the regular $\underline{\underline{\Delta^{*}}}$ operator. $\underline{\Lambda{}_{int}}^{n}=\left(\underline{\underline{\Delta^{*}}}\,\,\underline{\psi}+\underline{f}\circ\underline{f'}+\mu_{0}\underline{r}^{2}\circ\underline{p'}\right)_{int}^{n}$is
the discrete form of the left side of equation \ref{eq:35.1}, where
the values at the boundary nodes have been set to zero, and is the
term that converges to zero. At each iteration, the expression for
$\underline{\Lambda}^{n}$ is found and then the boundary values of
$\underline{\Lambda}^{n}$ are set to zero. The $L_{2}$ norm of $\underline{\Lambda_{int}}^{n}$
is then determined as $d\Lambda^{n}=\left(\underline{\Lambda_{int}}^{n}\right)^{T}*\left(\underline{\Lambda_{int}}^{n}\right)$;
if $d\Lambda^{n}<\epsilon\ll1$, the solution $\underline{\psi}$
is considered to have converged. Otherwise, the boundary values of
$\underline{\psi}^{n+1}$ are set to zero for the next iteration (in
which $\underline{\psi}^{n+1}\rightarrow\underline{\psi}^{n})$. At
convergence, the solution fields are evaluated as $\underline{\psi}=\underline{\psi}^{n}$,
with $\underline{\psi}(\underline{b})=0,$ $\underline{f}=\underline{f}(\underline{\psi})$,
and $\underline{p}=\underline{p}(\underline{\psi})$.

$\underline{f}(\underline{\psi})$, $\underline{f'}(\underline{\psi})$,
$\underline{p}(\underline{\psi})$ and $\underline{p'}(\underline{\psi})$
are the discretised functional forms of $f(\psi),\;\frac{\partial f(\psi)}{\partial\psi}$,
$p(\psi)$ and $\frac{\partial p(\psi)}{\partial\psi}$, and are defined
as code inputs. As described in appendix \ref{subsec:GS_linear_lambda},
possible functional forms for $f(\psi)$ and $f'(\psi)$ are $f(\psi)=\bar{\lambda}\psi(1+\alpha(\widetilde{\psi}-1))$
and $f'(\psi)=\bar{\lambda}(1+\alpha(2\widetilde{\psi}-1))$, but
any functional form can be chosen. When external shaft current is
included in this particular model, the expression for $f$ is modified
to $f(\psi)=f_{external}+\bar{\lambda}\psi(1+\alpha(\widetilde{\psi}-1))$,
where $f_{external}=\frac{\mu_{0}I_{external}}{2\pi}=0.2\,I_{shaft}\,[\mbox{MA}]$.
For pressure, a reasonable choice of input functional form is the
parabolic profile $p(\psi(\mathbf{r}))=p_{\Gamma}+p_{axis}\left(\psi(\mathbf{r})/\psi_{axis}\right)$,
so that $p'=p_{axis}/\psi_{axis}$. Here, $p_{\Gamma}$ is the choice
of boundary condition for pressure, $p_{axis}$ is the prescribed
peak pressure at the magnetic axis, and $\psi_{axis}$ is the value
of $\psi$ at the magnetic axis. Setting $p=0\mbox{ and }p'=0$ is
also possible - in general, finite pressure has little effect on the
Grad-Shafranov equilibrium, apart from a slight outward translation
of the magnetic axis (known as the \textquotedbl Shafranov shift\textquotedbl ),
and some upward scaling of $\psi_{axis}$. Of course, pressure should
be finite for equilibria that are used as an initial condition for
MHD simulations. 

With an initial guess of uniform $\underline{\psi}^{0}=1\times10^{-3}*\underline{1},$
with $\underline{\psi^{0}}\,(\underline{b})=0$, and mesh resolution
$h_{e}=5$ mm, the number of iterations required for convergence to
$\epsilon=5\times10^{-22}$ with the implicit scheme varies non-linearly
with $\Delta\widetilde{t}$. Optimal $\Delta\widetilde{t}$ is around
$1\times10^{-3}$, for convergence within $N_{iter}=$200 iterations.
$N_{iter}$ increases as $\Delta\widetilde{t}$ is increased or reduced
around the optimal value. In contrast, with the explicit scheme, the
number of iterations required for convergence varies linearly ($i.e.,$
constant product $\Delta\widetilde{t}\times N_{iter})$ with $N_{iter}$,
but only within a particular range. The optimal point for the explicit
scheme is $\Delta\widetilde{t}\sim4\times10^{-6}$, for convergence
after $\sim13,000$ iterations. The solution for the explicit scheme
doesn't converge with $\Delta\widetilde{t}\gtrsim4\times10^{-6}$,
and the range over which the number of iterations required for convergence
varies linearly with $\Delta\widetilde{t}$ extends from $\Delta\widetilde{t}\sim4\times10^{-6}$
to $\Delta\widetilde{t}\sim2\times10^{-7}$.

The explicit scheme is not very sensitive to the initial estimate
for $\psi^{0}$. The number of iterations for convergence ranges from
$\sim20,000$ at $\psi^{0}=1\times10^{10}$, to $\sim13,000$ at $\psi^{0}=1\times10^{-3}$
and $\sim16,000$ at $\psi^{0}=1\times10^{-10}$. The implicit scheme
is even less sensitive to the initial estimate - the number of iterations
for convergence ranges from $\sim190$ at $\psi^{0}=1\times10^{-3}$,
to $\sim210$ at $\psi^{0}=1\times10^{\pm10}$.
\begin{figure}[H]
\subfloat[]{\includegraphics[width=7cm,height=6cm]{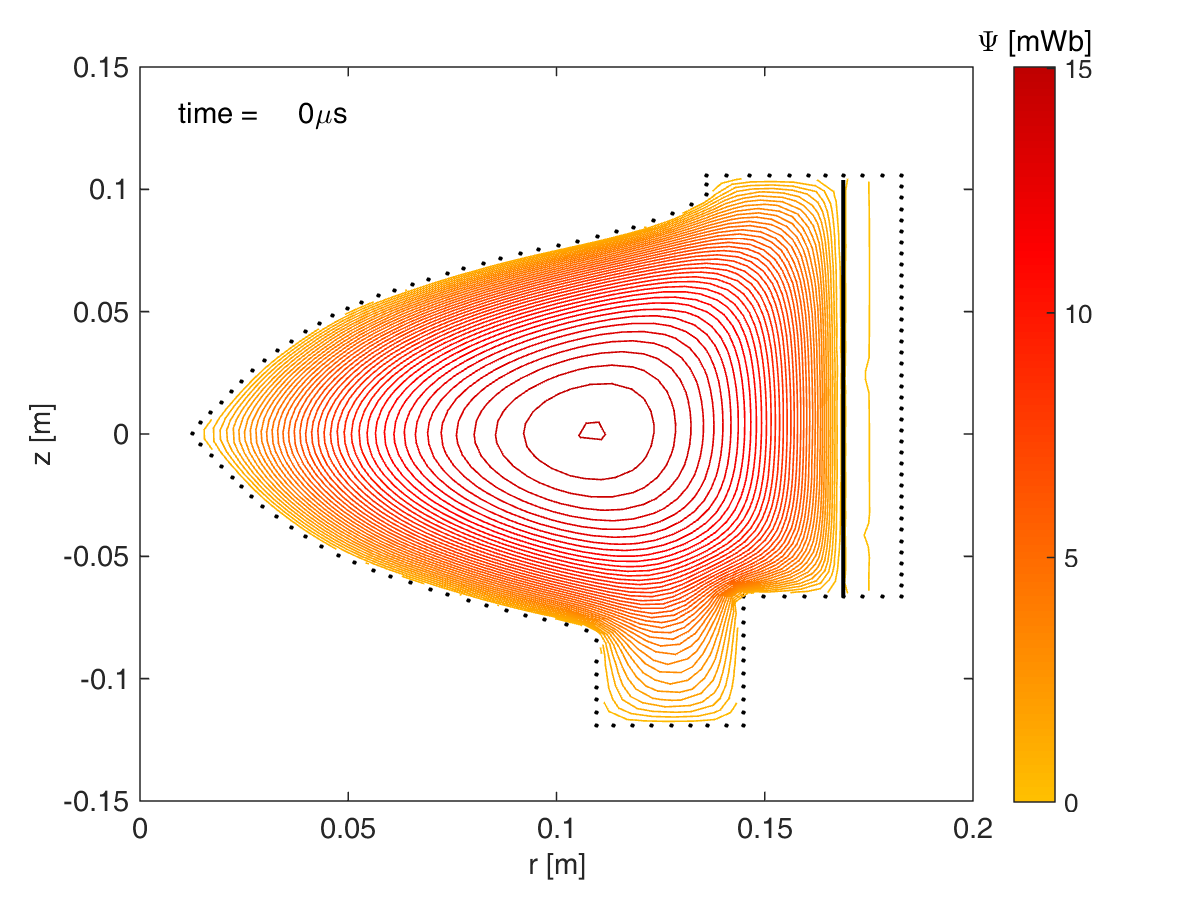}}\hfill{}\subfloat[]{\includegraphics[width=7cm,height=6cm]{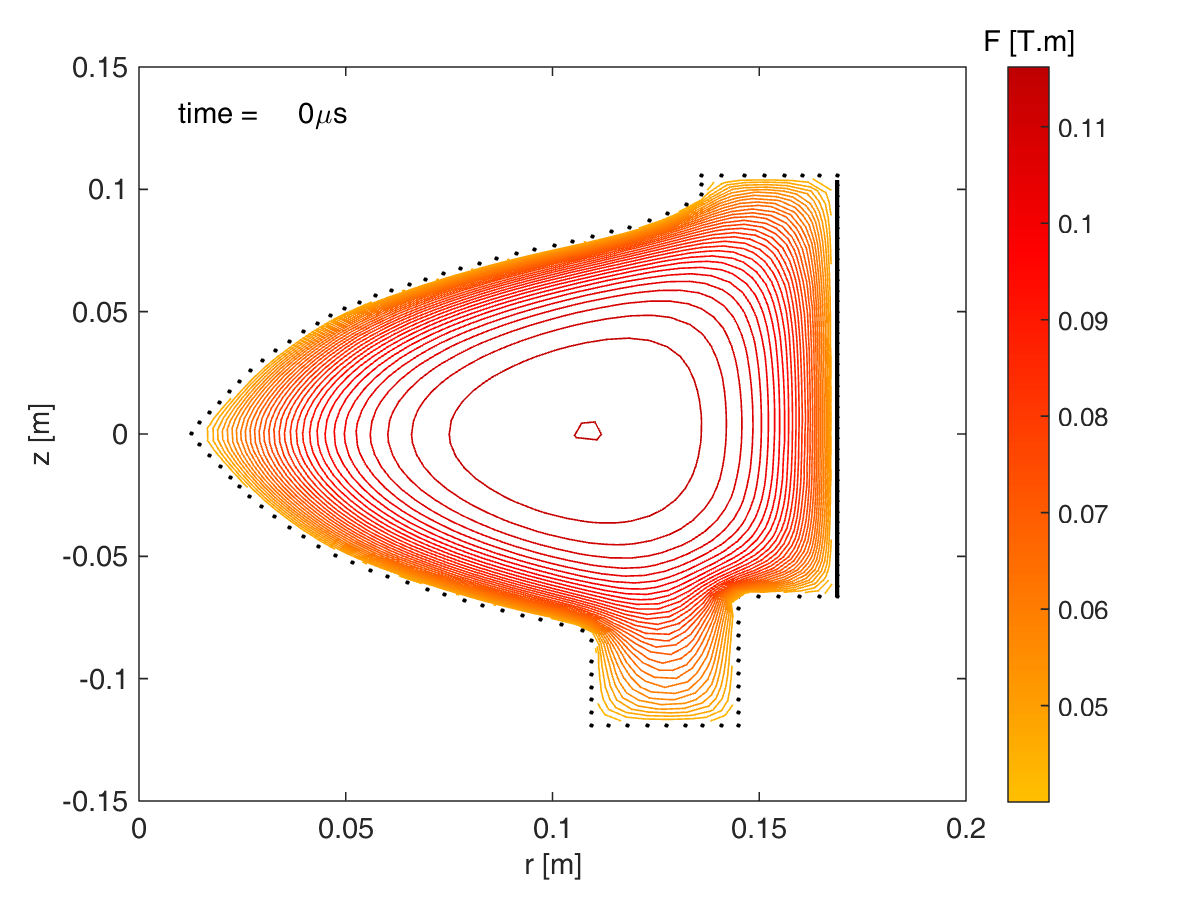}

}
\begin{centering}
\subfloat[]{\includegraphics[width=7cm,height=6cm]{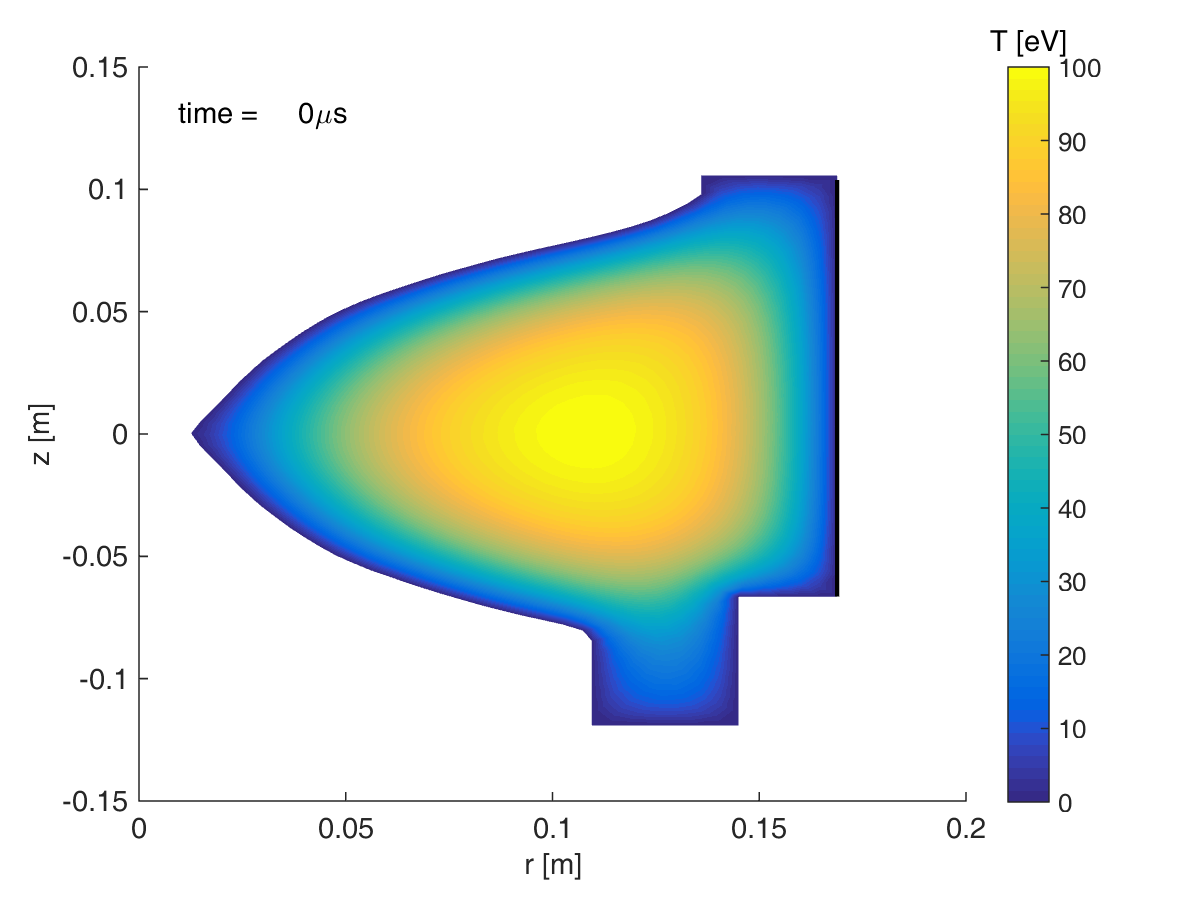}}
\par\end{centering}
\caption{\label{fig:GSsolns}$\,\,\,\,$Grad-Shafranov equilibrium solution
example}
\end{figure}
Figure \ref{fig:GSsolns} shows contours of $\psi$, $f$, and $T$
for a Grad-Shafranov solution run using the model for $f(\psi)$ described
in appendix \ref{subsec:GS_linear_lambda}, with code input parameters
$\bar{\lambda}=31.85$ {[}m$^{-1}${]}, $\alpha=-0.8$, $I_{shaft}=200\,\mbox{ kA}$,
$n_{0}=5\times10^{20}$ m$^{-3}$, $Z_{eff}=1$, peak temperature
$T_{0}=100$ eV, and $T|_{\Gamma}=0.02$ eV. These input parameters
determine the value of $\psi_{axis}$, 15 mWb in this case. The pressure
profile is described by $p(\psi(\mathbf{r}))=p_{\Gamma}+p_{axis}\left(\psi(\mathbf{r})/\psi_{axis}\right)$,
where $p=n_{0}T\,[\mbox{J}](1+Z_{eff})$. As discussed in appendix
\ref{subsec:GS_linear_lambda}, having negative $\alpha$ for this
prescription for $f$ results in \textquotedbl hollow\textquotedbl{}
current profiles where the magnetic field and current is concentrated
towards the CT exterior. A solution with a less hollow profile is
presented, and compared with the solution from a well-established
equilibrium code, in section \ref{sec:Code-validation}. The MHD simulation
results presented in chapter \ref{chap:Simulation-results} are generally
from simulations that include CT formation. Evolution of simulated
CT toroidal field, and comparison with experiment, is presented for
an MHD simulation that started with a Grad-Shafranov equilibrium in
section \ref{subsec:Simulations-starting-withGS11}.

\section{Code verification and validation\label{sec:Code-validation}}

\subsection{Convergence studies (verification)\label{subsec:Convergence-Study}}

\begin{figure}[H]
\subfloat[]{\includegraphics[width=5.3cm,height=5.3cm]{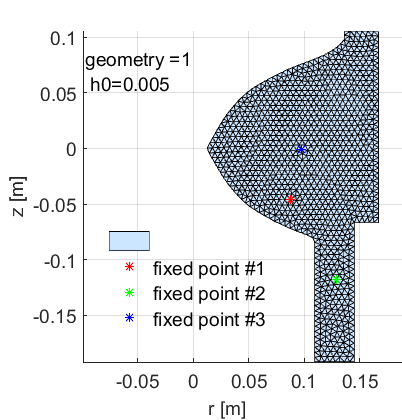}}\hfill{}\subfloat[]{\includegraphics[width=4.8cm,height=5.5cm]{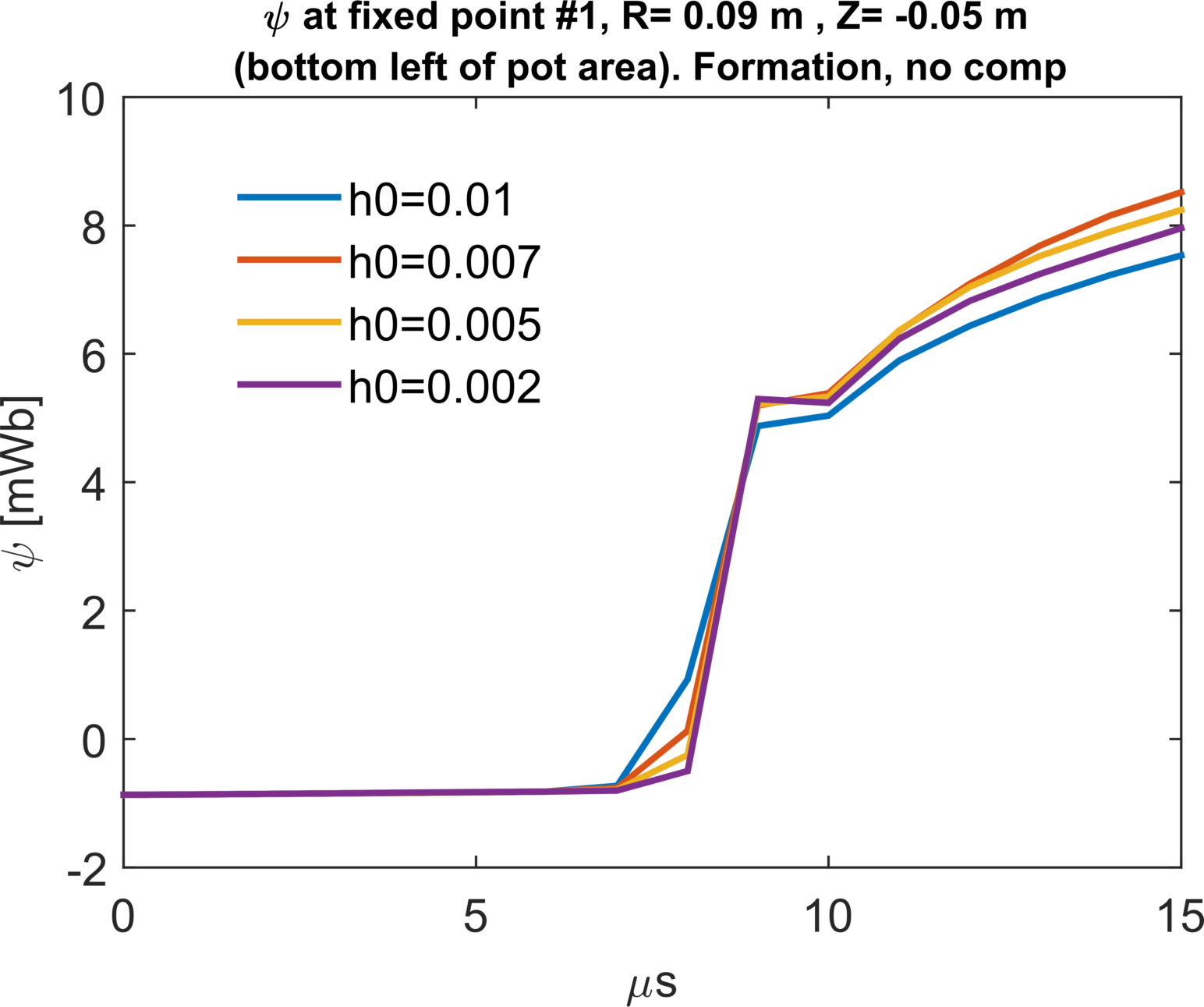}

}\hfill{}\subfloat[]{\includegraphics[width=5.4cm,height=5.7cm]{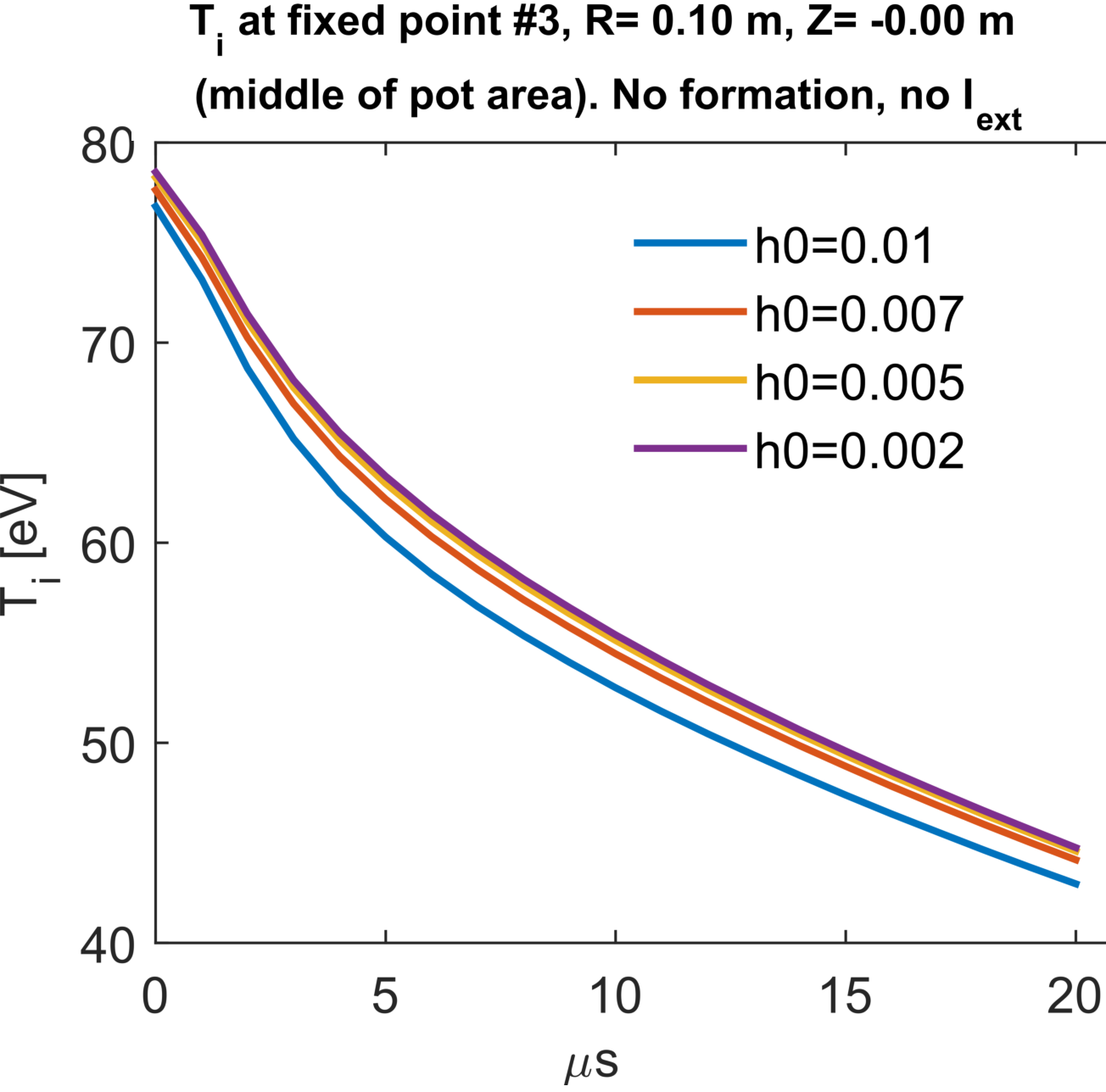}}

\caption{\label{fig: convergence_studies}$\,\,\,\,$Example convergence studies
with varying $h_{e}$ (triangular element size $[\mbox{m}]$)}
\end{figure}
To study and verify the convergence of the equilibrium and MHD solutions
with increasing mesh resolution, we ran a series of tests to record
the values of various solution fields at some fixed points in the
domain for various values of $h_{e}$, which quantifies triangular
element size. Figure \ref{fig: convergence_studies}(a) indicates
the fixed point locations for a mesh with $h_{e}=5\mbox{\mbox{\mbox{ mm}}}$
- the mesh generator was forced to include a solution node at the
fixed points for the test meshes with $h_{e}=2,\,5,\,7$ and 10 mm.
The truncated mesh shown here was used for non-formation tests - a
mesh with the full extent of the gun included was used for tests that
included simulated CT formation, and has additional fixed points down
the gun. Convergence with increasing mesh resolution was verified
for all solution fields at the various fixed points. Example convergence
study results for $\psi$ and $T_{i}$ are shown in figures \ref{fig: convergence_studies}(b)
(formation simulation, with levitation, without compression) and (c)
(no formation, no levitation or compression - the dynamics of this
simulation started from a Grad-Shafranov equilibrium, with $\psi=0$
maintained on the boundaries). 

\subsection{Equilibrium solution comparison with Corsica solution (validation)\label{subsec:Equilibrium-solution-comparison} }

The code's Grad-Shafranov solution was benchmarked against a solution
obtained using the well-developed Corsica code \cite{corsica code}.
The simple model \cite{Knox} that assumes a linear dependence of
$\lambda(\psi)$ on $\psi$ was used to define $f(\psi)$ (see appendix
\ref{subsec:GS_linear_lambda}). Corsica's Grad-Shafranov solution
is solved using a finite difference method on a structured grid. 
\begin{figure}[H]
\subfloat[]{\includegraphics[width=7cm,height=5cm]{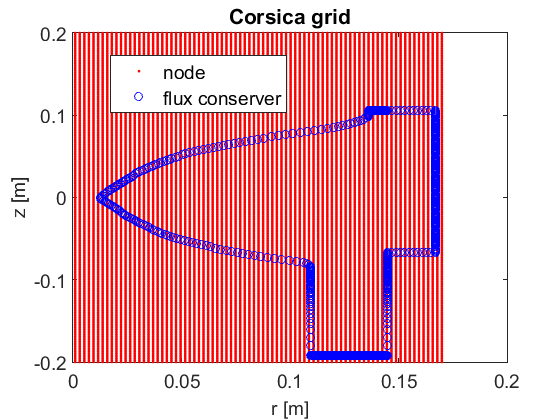}}\hfill{}\subfloat[]{\includegraphics[width=7cm,height=5cm]{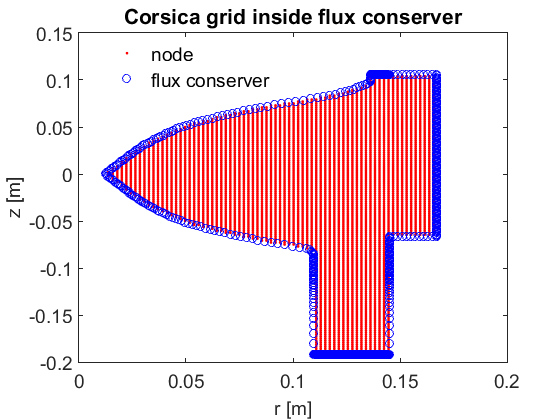}

}

\subfloat[]{\includegraphics[width=7cm,height=5cm]{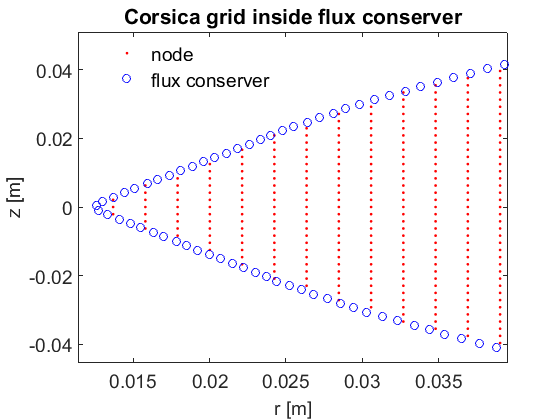}}\hfill{}\subfloat[]{\includegraphics[width=7cm,height=5cm]{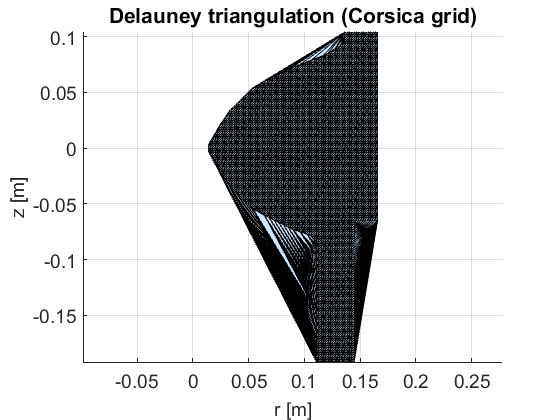}}

\subfloat[]{\includegraphics[width=7cm,height=5cm]{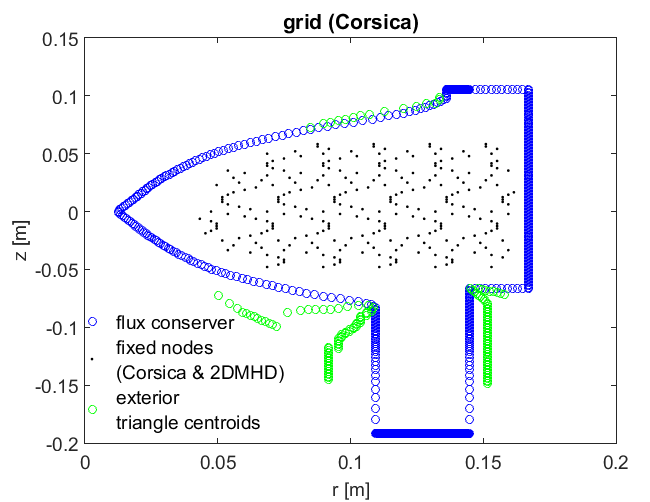}

}\hfill{}\subfloat[]{\includegraphics[width=7cm,height=5cm]{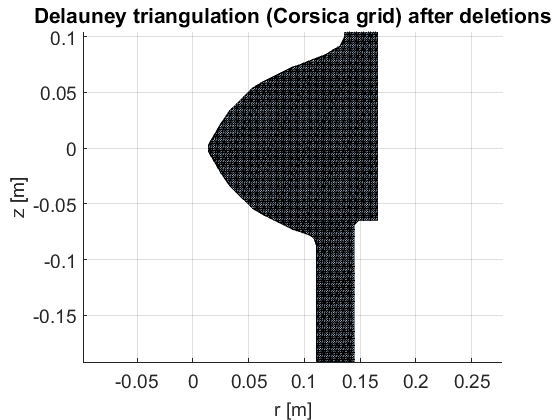}}

\caption{\label{fig:Corsica-grid-with-1}$\,\,\,\,$Corsica grid with triangulation}
\end{figure}
Figure \ref{fig:Corsica-grid-with-1}(a) indicates the locations of
the computational nodes, and the positions that define the flux conserver
for the full Corsica grid. Figures \ref{fig:Corsica-grid-with-1}(b)
and (c) indicate the locations of the nodes inside the flux conserver.
As shown in figure \ref{fig:Corsica-grid-with-1}(d), the vectors
defining the $r$ and $z$ coordinates of these interior nodes were
used to produce a grid of triangular elements using Matlab's Delaunay
triangulation function, for plotting contours of $\psi$ from the
Corsica solution. Figure \ref{fig:Corsica-grid-with-1}(e) shows the
locations of the ($\sim300$) nodes in the Corsica grid that are assigned
as fixed nodes in the grid for our code, so that a direct (no interpolation)
comparison of solutions for $\psi$ at these nodes can be made. Also
shown in figure \ref{fig:Corsica-grid-with-1}(e) are the locations
of the centroids of triangles that lie outside the flux conserver
- these triangles were deleted from the mesh.

\begin{figure}[H]
\subfloat[]{\includegraphics[width=6.5cm,height=7cm]{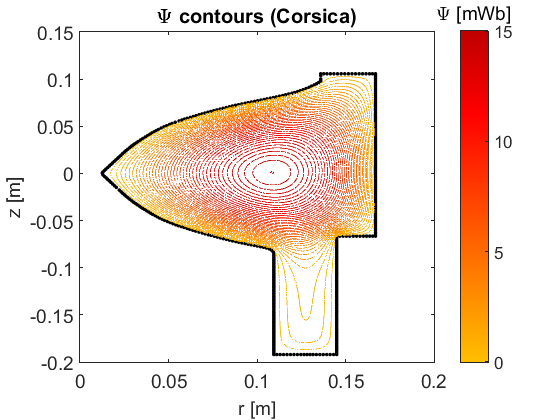}}\hfill{}\subfloat[]{\includegraphics[width=6.5cm,height=7cm]{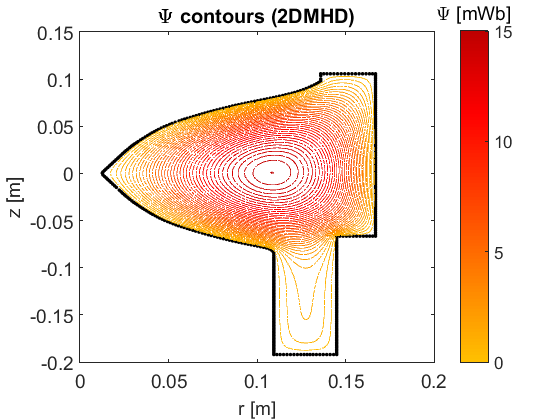}

}

\subfloat[]{\includegraphics[scale=0.55]{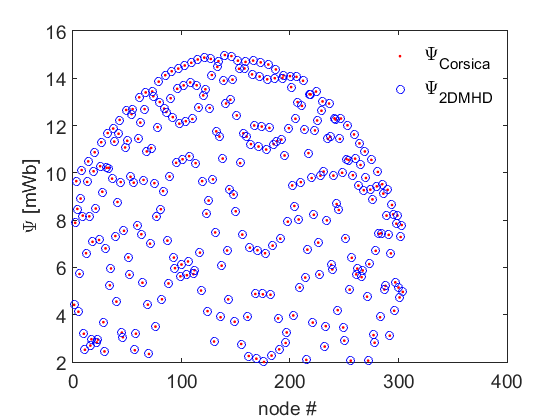}}\hfill{}\subfloat[]{\includegraphics[scale=0.55]{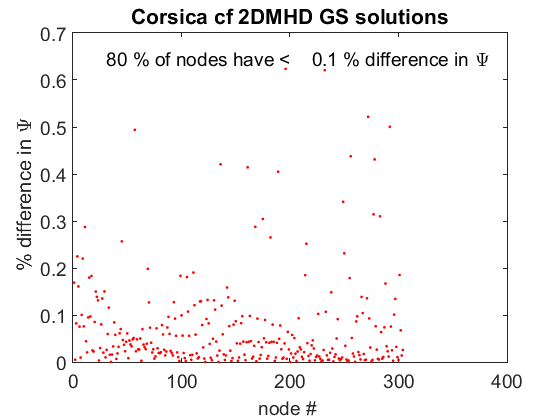}

}
\begin{centering}
\subfloat[]{\includegraphics[scale=0.55]{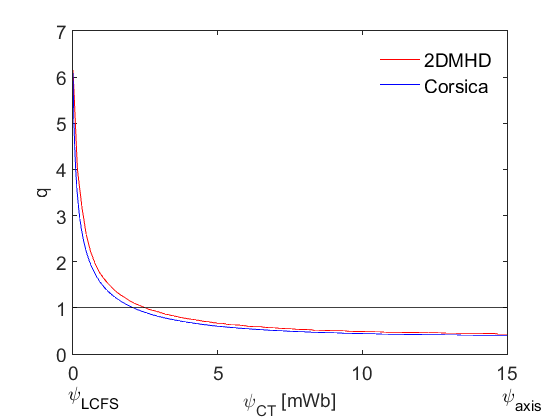}

}
\par\end{centering}
\caption{\label{fig:Corsicasolns-1}$\,\,\,\,$Equilibrium solution comparison}
\end{figure}
Figures \ref{fig:Corsicasolns-1}(a) and (b) show $\psi$ contours
from the Corsica and DELiTE MHD code respectively, for solutions with
comparable grid resolutions, and with $\bar{\lambda}=23.18$ {[}m$^{-1}${]},
$I_{shaft}=200$ kA, $p=0$, and $\alpha=-0.2$. The calculated values
of $\psi$ at the fixed nodes are shown in figure \ref{fig:Corsicasolns-1}(c).
As indicated in figure \ref{fig:Corsicasolns-1}(d), the difference
between the two solutions, expressed as a percentage of $\psi_{corsica}$,
is less than $0.1\%$ at $80\%$ of the fixed nodes, and less than
$\sim0.5\%$ at all fixed nodes. Figure \ref{fig:Corsicasolns-1}(e)
indicates the excellent match between the profiles of $q(\psi)=\frac{\partial\phi(\psi)}{\partial\psi}$
for the two codes. Note that $\psi_{axis}$ and $\psi_{LCFS}$ denote
$\psi$ at the magnetic axis and at the last closed flux surface.
See section \ref{subsec:q-profile} for details of the method developed
to evaluate the CT $q$ profile. Additional tests were done that indicate
similar levels of agreement between the calculated $q$ and $\psi$
profiles for the codes with various inputs parameters $\bar{\lambda},\;I_{shaft},\;p$,
and $\alpha$. 

To date, no benchmark code comparison tests have been done to validate
the MHD part of the code. As presented in chapter \ref{chap:Simulation-results},
the code reproduces many features of the experimental observations
- this is a form of validation of the MHD model.

\section{Summary\label{sec:SummaryGencode setup}}

A freely available mesh generator has been modified to produce the
computational grid. Forward Euler timestepping is found to be adequate
for good solution convergence for the short physical timespans associated
with the magnetic compression experiment, but the options of the higher
order Runge-Kutta 2 and Runge-Kutta 4 timestepping schemes, which
have been implemented to the code, may be advantageous for simulations
with extended timespans. Methods were developed to vary the simulation
timestep according to the time-evolving conditions of the various
field solutions, which determine the maximum allowable timestep for
stable time-advance, while managing simulation data preservation.
Many approximations and simplifications are made when determining
the various diffusion coefficients, which are generally representations
of their physical counterparts, but are also required for numerical
stability. Density diffusion is an artificial term that is required
purely for numerical stability. For turbulent plasmas, there is no
widely applicable formula that can be applied for determination of
the thermal and viscous diffusion coefficients. When setting the viscous
diffusion coefficient, care must be taken, especially when simulating
CT formation, which involves regions of high velocity and low density,
to have adequate viscous diffusion for numerical stability, while
avoiding excessive ion viscous heating. The timestepping scheme is
explicit, leading to restrictions imposed by the minimum acceptable
timestep on the values of the diffusion coefficients. A numerical
method to solve for the Grad-Shafranov equilibrium solution has been
developed. The Grad-Shafranov equilibrium and $q(\psi)$ profile solutions
have been benchmarked against solutions from a well-established code,
and demonstrate an excellent match. Ideally, tests would be conducted
to validate the MHD solutions against a recognised code, this has
not yet been done. Tests that verified convergence of code solutions
with increasing mesh resolution were successfully conducted. As shown
in chapter \ref{chap:Simulation-results}, experimental diagnostics
are closely matched by the counterpart simulated diagnostics; this
is also a form of code validation.\newpage{}

\chapter{First results from plasma edge biasing on SPECTOR\label{chap:First-results-from}}

\section{Introduction}

In this appendix, a description of an edge-biasing experiment conducted
on the SPECTOR plasma injector \cite{Spector_biasing}, and initial
results, are presented. The insertion of a disc-shaped molybdenum
electrode (probe), biased at up to $+100$V, into the edge of the
CT resulted in up to 1 kA radial current being drawn. Core electron
temperature, as measured with a Thomson-scattering diagnostic, was
found to increase by a factor of up to 2.4 in the optimal configuration
tested. $\mbox{H}_{\alpha}$ intensity was observed to decrease, and
CT lifetimes increased by a factor of up to 2.3. A significant reduction
in electron density was observed; this is thought to be due to the
effect of a transport barrier impeding CT fueling, where, as described
in section \ref{sec:Neutral_SPECTOR}, the fueling source is neutral
gas that remains concentrated around the gas valves after CT formation. 

High confinement mode (H-mode) has been implemented by various means
($e.g.,$ edge biasing, neutral beams, ion or electron cyclotron heating,
lower hybrid heating, and ohmic heating) on a range of magnetic confinement
configurations including tokamaks, reversed field pinches, stellarators,
and mirror machines. The first H-mode was produced in the ASDEX tokamak
by neutral beam injection in 1982 \cite{wagnerASDEX}. In 1989, H-mode
was first produced by electrode edge biasing on the CCT tokamak \cite{Taylor_CCT,WEynants_Taylor_CCT}.
In 1990, it was observed that edge impurity ion poloidal speed is
modified abruptly during transitions from low to high confinement
modes on the DIII-D tokamak \cite{Groebner}. H-mode has subsequently
been produced routinely on many toroidal magnetic-fusion experiments,
including practically all the large tokamaks including JET, TFTR,
and JT-60. Since the initial electrode-biasing experiments on CCT,
H-mode has been produced by edge biasing on many tokamaks, for example
CASTOR \cite{CAstor,OOst}, T-10 \cite{OOst,T10}, STOR-M \cite{StorM},
ISTTOK \cite{Figueiredo_ISTTOK}, TEXTOR \cite{OOst,Jachmich}, and
J-TEXT \cite{JTEXT}. 

Electrode biasing involves the insertion of an electrode, that is
biased relative to the vessel wall near the point of insertion, into
the edge of a magnetized plasma. This leads to a radially directed
electric field between the probe and the wall. The resultant $\mathbf{J}_{r}\times\mathbf{B}$
force imposed on the plasma at the edge of the plasma confinement
region varies with distance between the probe and the wall, because
$E_{r}$, as well as the magnetic field, vary in that region. The
associated torque overcomes viscous forces, spinning up the edge plasma,
and results in shearing of the particle velocities between the probe
and the wall. The sheared velocity profile is thought to suppress
the growth of turbulent eddies that advect hot plasma particles to
the wall, thereby reducing this plasma cooling mechanism. In general,
H-modes induced by probe biasing share features of those initiated
by various methods of heating, including a density pedestal near the
wall (near the probe radius for probe biasing), diminished levels
of recycling as evidenced by reduced $\mbox{H}_{\alpha}$ emission
intensity, and increased particle and energy confinement times. For
example, increases in energy confinement times by factors of 1.5,
1.5, 1.2, and 1.8 were reported for CCT, STOR-M, TEXTOR and T-10 respectively.
Core electron density increased by a factor of four on CCT, while
line-averaged electron density increased by factors of 2, 2, 1.5,
and 1.8 on STOR-M, TEXTOR, T-10, and CASTOR respectively. Of these
five examples, a biasing-induced temperature increase was noted only
for the T-10 experiment, with an increase in core ion temperature
by a factor of 1.4 reported, while reduced $\mbox{H}_{\alpha}$ emission
intensity was recorded in each case.

Positive as well as negative electrode biasing works well on some
machines; in other instances only one biasing polarity has the desired
effect. Most biasing experiments have used passive electrodes, while
some have implemented electron-emitting electrodes. Emissive electrodes
have, in addition to a circuit to bias the electrode relative to the
vacuum vessel, a separate heating circuit, and are heated until they
emit electrons. Materials traditionally used for emissive electrodes
include lanthanum hexaboride (LaB6) and tungsten (W). Generally speaking,
emissive electrodes add complexity to an experiment, but may be beneficial
when the edge plasma electron density is so low that dangerously high
voltages (which could initiate a current arc that could damage the
electrode and vessel) would be required in order to draw an edge current
sufficiently high enough for the $\mathbf{J}_{r}\times\mathbf{B}$
force to overcome inertial effects (viscosity, friction) and drive
edge rotation. In the CCT tokamak \cite{Taylor_CCT}, LaB6 cathodes
heated by carbon rods drew edge current up to $40\mbox{ A}$ when
the voltage measured between the electrode (probe) and vessel wall
was $V_{probe}\sim-250\mbox{ V}$. On CCT, for negative bias, it was
found that both electron-emissive electrodes and passive graphite
electrodes produced similar results, as long as the electrode was
large enough to draw sufficient current ($\sim20\,\mbox{A}$), and
small enough not to form a limiter \cite{Taylor_CCT}. For negative
biasing on ISTTOK, it was not possible to draw more than $2\mbox{ to }3\mbox{ A}$
with a passive electrode, $cf.\sim20\mbox{ A}$ with an emissive electrode,
while the current drawn with positive biasing was the same for emissive
and non-emissive electrode ($I_{probe}\sim28\mbox{ A}$ at $V_{probe}\sim+130\mbox{ V}$). 

Pre-biasing conditions of radial electric field and an extensive range
of plasma parameters play roles in determining the beneficial polarity
and the level of bias-induced plasma confinement improvement \cite{Tendler}.
A reduction of radial transport at the edge would be beneficial for
confinement not only because of reduced outward thermal transport,
but also due to reduced inward transport of cold wall-recycled particles
to the core. This latter effect is especially relevant on small machines
for which the surface area to volume ratio of the magnetically confined
plasma is large, particularly in the absence of a limiter or divertor,
where the recycling process is more important. It may be partly due
to this effect, in combination with the effect of a transport barrier
impeding the CT fueling process that arises due to neutral particles
diffusing up the gun after the CT formation process (section \ref{sec:Neutral_SPECTOR}),
that the improvements in confinement times and electron temperatures
observed with the initial edge biasing tests on SPECTOR ($R\sim11\mbox{ cm},\,a\sim8\mbox{ cm},$
no limiter or divertor) appear especially significant. 

An overview of the experiment setup with a description of the biasing
electrode assembly is presented in section \ref{sec:Experiment-setup}.
Circuit analysis, leading to an estimate for the resistance of the
plasma between the electrode and flux conserver, which was useful
for optimising the circuit, is the focus of section \ref{sec:Circuit-analysis}.
Main results are presented in section \ref{sec:Main-results}. The
appendix ends with a discussion of principal findings, conclusions
and possible further improvements to the experiment in section \ref{sec:Summary_biasing_exp}

\section{Experiment setup\label{sec:Experiment-setup}}

A schematic of the SPECTOR \cite{spectPoster} plasma injector is
depicted in figure \ref{fig:GFinjectors}(b), where the red dots represent
the locations of magnetic probes. SPECTOR is a magnetized Marshall
gun that produces compact tori (CTs). It has, in addition to the formation
circuit that drives up to $0.8\mbox{ MA}$ formation current over
around $80\,\upmu$s, a separate circuit to produce an approximately
constant shaft current of up to $0.5\mbox{ MA}$, which flows up the
outer walls of the machine and down the central shaft, increasing
CT toroidal field and making the CT more robust against MHD instability.
Shaft current duration is extended to around 3 ms with a crowbar inductor/diode
circuit, which is indicated schematically in figure \ref{fig:GFinjectors}(b).
Toroidal field at the CT core is typically around $0.5\mbox{ T}$.
The high CT aspect ratio, and the $q$ profile, define the CTs as
spherical tokamaks. Coaxial helicity injection produces plasma currents
in the range $300-800\mbox{ kA}$. A selection of Thomson-scattering
(TS) system-produced electron temperature and electron density measurements
\cite{TS_GF} (both taken at $300$$\,\upmu\mbox{s}$ after CT formation),
electron density measurements obtained with a far-infrared (FIR) interferometer
\cite{polarimetryGF}, spectral data, and magnetic probe data, will
be presented in the following. 
\begin{figure}[H]
\centering{}\subfloat{\includegraphics[width=14cm,height=7cm]{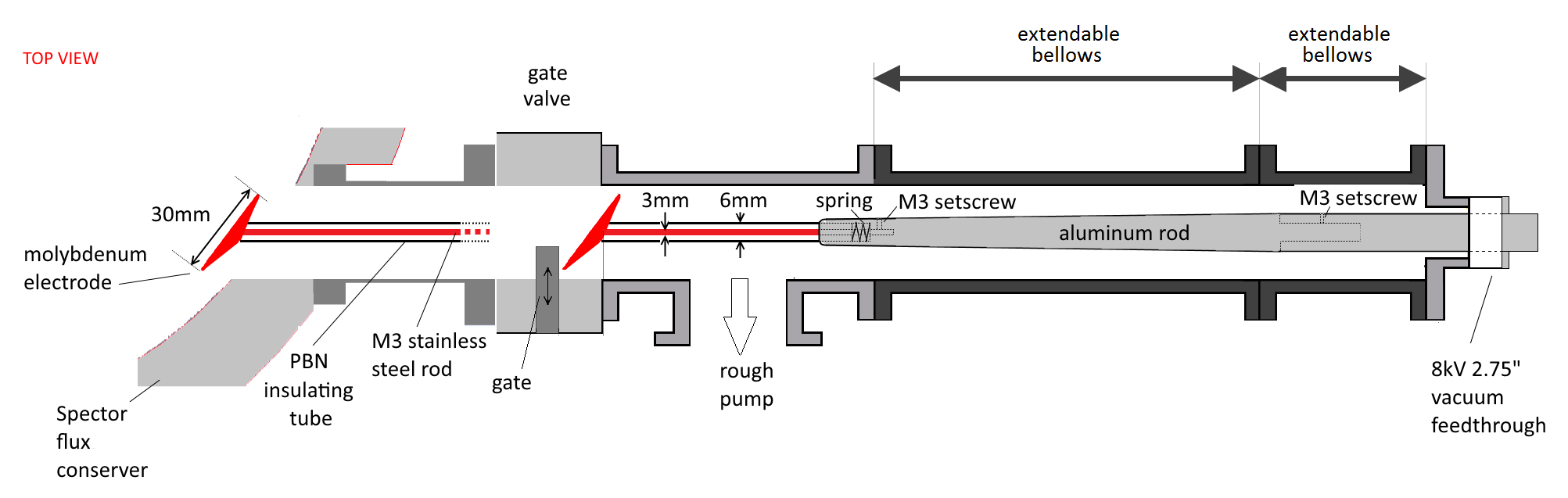}}\caption{\label{fig:bias probe assembly}$\,\,\,\,$Biasing probe assembly}
\end{figure}
Figure \ref{fig:bias probe assembly} indicates a top-view of the
electrode (probe) assembly with extendable vacuum bellows. The biasing
electrode can be retracted behind the gate valve and isolated from
the machine vacuum. The approximately disc-shaped electrode is machined
from molybdenum and has a 30 mm diameter. Molybdenum was chosen for
its high work function against sputtering, high melting point, and
its resilience against the corrosive action of lithium, which is used
as a gettering agent on SPECTOR. A pyrolytic boron nitride (PBN) tube
is used as a plasma-compatible insulator around the M3 stainless steel
rod that connects the electrode to a tapered aluminum rod, which is
in turn connected to the $0.5"$ diameter copper rod that forms part
of the $8\mbox{ kV}$ $2.75"$ CF vacuum feedthrough. The electrode
can be inserted up to 45 mm into the vacuum vessel; insertion depth
was 11 mm for the results presented here.

\section{Circuit analysis\label{sec:Circuit-analysis}}

\begin{figure}[H]
\centering{}\subfloat{\includegraphics[width=16cm,height=7cm]{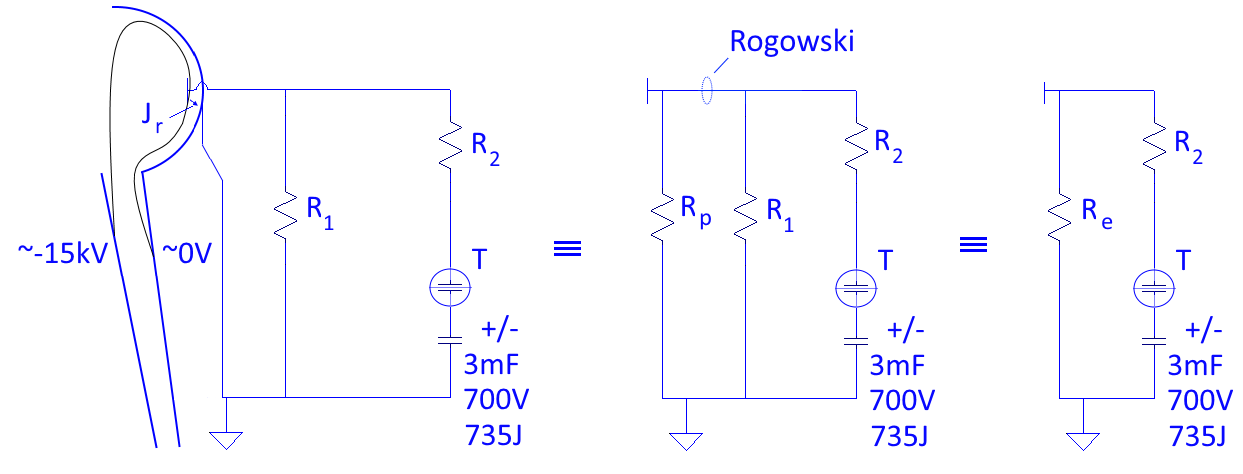}}\caption{\label{fig:Biasing-probe-circuit}$\,\,\,\,$Biasing probe circuit
diagram}
\end{figure}
Figure \ref{fig:Biasing-probe-circuit} indicates the most optimal
of the biasing probe circuit configurations tested. The biasing circuit
was kept open-circuited until well after CT formation, in order to
protect biasing circuit components. A thyratron switch (indicated
in the figure) is robust against large amplitude negative voltage
spikes that can appear on the probe during CT formation when initially
open stuffing-field lines, that are resistively pinned to the injector
inner and outer electrodes, intersect the probe (see left subfigure).
These thyratron switches are designed to operate at several kilovolts,
and usually require several kiloamps of current to remain closed,
but careful setting of switch temperature enabled operation at moderate
voltages and currents. The biasing capacitor voltage setting $V_{bc0}$,
parallel and series resistors $R_{1}$ and $R_{2}$, and $R_{p}$,
the plasma resistance between the electrode ($i.e.,$ probe) and flux
conserver, determine $V_{probe}$, the voltage measured between the
probe and flux conserver, and $I_{probe},$ the radial current drawn
through the plasma edge. The radial current leads, in the classical
edge biasing scenario, to $\mathbf{J}_{r}\times\mathbf{B}$ driven
edge velocity shearing and consequential decorrelation of turbulence
cells and confinement improvement. For the circuit with the $3\mbox{ mF}$
capacitor depicted in figure \ref{fig:Biasing-probe-circuit}, optimal
circuit resistances were found to be $R_{1}\sim0.2\,\Omega$, and
$R_{2}\sim0.5\,\Omega$. Negative electrode biasing was briefly tested;
the results presented here were obtained with positive biasing. The
effective resistance $R_{e}$, comprised of $R_{p}$ and $R_{1}$
in parallel (see figure \ref{fig:Biasing-probe-circuit}, right subfigure),
is given by 
\begin{equation}
R_{e}(t)=\frac{R_{p}(t)\,R_{1}}{R_{p}(t)+R_{1}}\label{eq:1-1}
\end{equation}
The voltage applied by the capacitor on the probe is 
\begin{equation}
V_{applied}(t)=V_{bc}(t)\left(\frac{R_{e}(t)}{R_{e}(t)+R_{2}}\right)\label{eq:2}
\end{equation}
where $V_{bc}(t)$ is the voltage across the biasing capacitor. Equations
\ref{eq:1-1} and \ref{eq:2} can be combined to provide an expression
for $R_{p}$:
\begin{equation}
R_{p}(t)=\frac{R_{1}R_{2}V_{applied}(t)}{R_{1}(V_{bc}(t)-V_{applied}(t))-V_{applied}(t)\,R_{2}}\label{eq:3}
\end{equation}
\begin{figure}[H]
\centering{}\subfloat[]{\includegraphics[width=8cm,height=5cm]{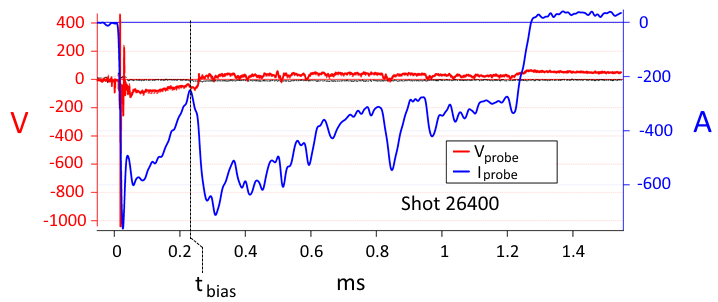}}\hfill{}\subfloat[]{\includegraphics[width=8cm,height=5cm]{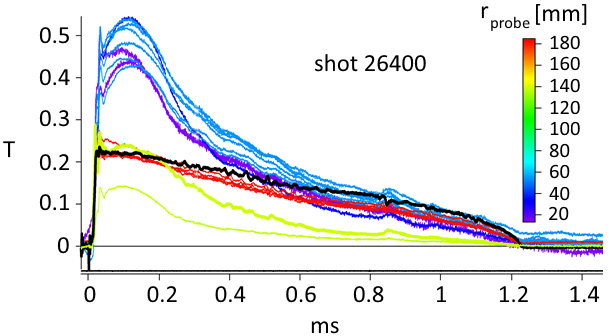}}\caption{\label{26400}$\,\,\,\,$Measured bias probe voltage and current (a),
and poloidal field (b) for shot 26400, which had $V_{bc0}=700$ V
(3 mF capacitor)}
\end{figure}
Figure \ref{26400}(a) shows the voltage measured between the probe
and the vacuum vessel, and the current drawn through the plasma edge,
as measured with the Rogowski coil indicated in figure \ref{fig:Biasing-probe-circuit},
for shot 26400. At the biasing capacitor voltage found to be most
optimal for CT lifetime and electron temperature (as obtained with
the TS system), the voltage measured between the probe and vacuum
vessel was typically $V_{probe}\sim+50\mbox{ V}$ to $+80\mbox{ V}$,
and the maximum radial current drawn to the probe from the wall was
$I_{probe}\sim700\mbox{ A}$ to $\sim1\mbox{ kA}$ shortly after firing
the biasing capacitor(s). For shot 26400, the electrode was inserted
11 mm into the edge plasma, and biased at $t_{bias}=230\,\upmu\mbox{s}$
after firing the formation capacitor banks, as indicated in figure
\ref{26400}(a). Note that current is already flowing through the
plasma edge, and through resistor $R_{1},$ before $t_{bias}$, as
a result of the plasma-imposed potential on the electrode, which typically
led to a measurement of $V_{probe}\sim-100\mbox{ V}$ when magnetized
plasma first enters the CT confinement area at around $20\,\upmu$s.
$V_{probe}$ and $I_{probe}$ decrease over time at a rate that depends
on plasma and circuit parameters. Figure \ref{26400}(b) indicates,
for shot 26400, the poloidal field measured at the magnetic probes
indicated as red dots in figure \ref{fig:GFinjectors}(b). It is interesting
that the fluctuations in $B_{\theta}$, which are thought to be associated
with internal reconnection events, are also manifested on the biasing
voltage and current measurements, $e.g.,$ at $\sim845\,\upmu$s in
figures \ref{26400}(a) and (b). This observation is enabled by the
presence of the small parallel $R_{1}$. As edge plasma impedance
varies, as determined by internal MHD events, the system can divert
varying proportions of capacitor driven current through $R_{1}$.
In future studies, it may be possible to influence the behaviour of
the internal modes that cause the $B_{\theta}$ fluctuations, by driving
an edge current that is resonant with the fluctuations. 
\begin{figure}[H]
\begin{centering}
\subfloat[Biasing probe circuit diagram, including plasma voltage source ]{\includegraphics[width=5.5cm,height=6.5cm]{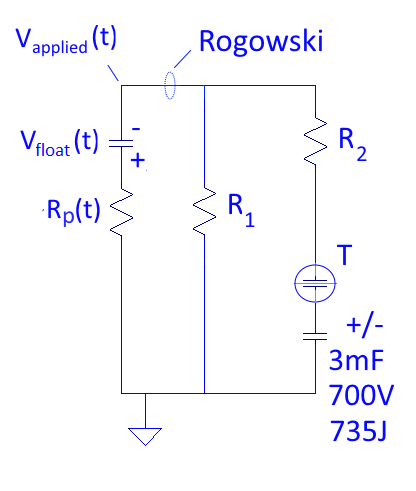}}\hfill{}\subfloat[Current flow schematic for case with switch open]{\includegraphics[width=4.4cm,height=7cm]{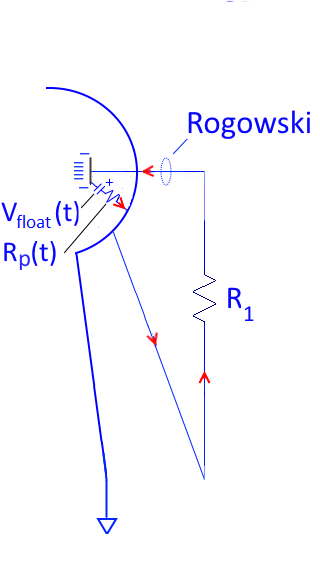}}\hfill{}\subfloat[Current flow schematic for example case with switch closed and $R_{1}=2R_{p}$,
and $V_{bc0}>0$]{\includegraphics[width=5.5cm,height=7cm]{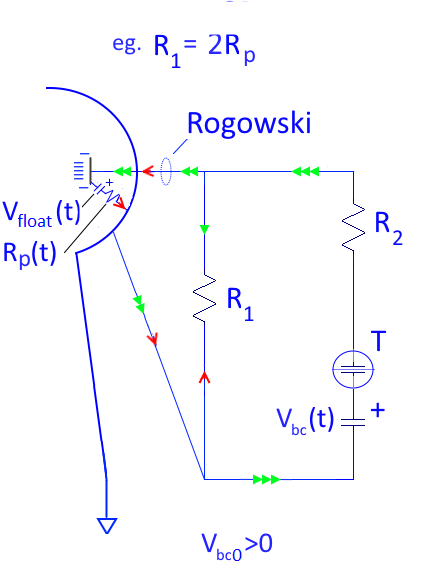}}
\par\end{centering}
\centering{}\caption{\label{fig:Circuit_plasma}$\,\,\,\,$Biasing probe circuit diagram,
including plasma voltage source, with current flow schematics }
\end{figure}
When the plasma is considered as a time-dependent voltage source,
which biases the probe to floating potential $V_{float}(t)$, a more
complete circuit diagram is as depicted in figure \ref{fig:Circuit_plasma}(a).
The inclusion of $R_{1}$, a small external resistance in parallel
with $R_{p}$ (the plasma resistance between the probe and wall),
allows current driven by the floating potential to flow in the circuit
in the case where the thyratron switch is open (see figure \ref{fig:Circuit_plasma}(b)).
When the switch is closed, a proportion of the biasing capacitor-driven
current may divert to flow through $R_{1}$, see figure \ref{fig:Circuit_plasma}(c).
This proportion increases as $R_{p}$ increases with reducing electron
temperature as the CT decays, thereby allowing $I_{probe}$ to decrease
at a rate roughly in proportion to the rate of decrease of the main
CT plasma currents. The presence of an appropriately sized $R_{1}$
also prevents development of a sustained arc, which could damage the
wall and probe, through the ambient plasma that remains between the
probe and wall after the CT has extinguished. In previous edge biasing
studies on tokamaks, the standard is to maintain approximately constant
$V_{applied}$ and $I_{probe}$ for an extended time which is a segment
of the duration over which the approximately constant externally driven
toroidal plasma current flows. On SPECTOR plasmas, the plasma currents
are not driven and are allowed to decay naturally after formation,
so a circuit configuration that establishes constant $V_{applied}$
and $I_{probe}$ would not be compatible.

The differential voltage measured between the probe and flux conserver
is 
\begin{equation}
V_{probe}(t)=V_{applied}(t)+V_{float}(t)\label{eq:2.1}
\end{equation}
If the bias capacitor is not fired, and $R_{1}$ is removed from the
circuit, then in the open circuit condition $V_{probe}(t)=V_{float}(t)$.
Note that $V_{float}$ is not measured directly on each shot, however,
looking at the $V_{probe}$ measurements taken during several open
circuit, probe-in shots, the floating potential can be approximated
as an RC rise of the form 
\begin{equation}
V_{float}(t)=V_{f0}\,e^{-\frac{t}{\tau_{RCf}}}\label{eq:2.11}
\end{equation}
with $V_{f0}\sim-80\mbox{ V}$, and, (depending on CT lifetime) $\tau_{RCf}\sim1\mbox{ ms}$.
$V_{float}(t)$ rises from $\sim-80\mbox{ V}$ at the time when plasma
enters the CT confinement region, to $0\mbox{ V}$ when the CT has
decayed away. With this, an approximation for $V_{applied}$ can be
made using equation \ref{eq:2.1}. $V_{bc}(t)$, the voltage across
the biasing capacitor, was not measured directly in the experiment,
but can be estimated as 
\begin{equation}
V_{bc}(t)=V_{bc0}\,e^{-\frac{t}{\tau_{RCb}}}\label{eq:2.12}
\end{equation}
where, for shot 26400, $V_{bc0}=700\mbox{ V}$ and $\tau_{RCb}=0.5\,\Omega*3\mbox{ mF}=1.5\mbox{ ms}$
(resistance $R_{2}=0.5\,\Omega\gg R_{e})$. With these approximations
for $V_{applied}(t)$ and $V_{bc}(t)$, an estimate of the plasma
resistance along a path that has a principal component along the helical
magnetic field between the probe (with insertion depth $11\mbox{ mm}$)
and flux conserver is evaluated, between $t_{bias}=250\,\upmu\mbox{s}$
until the time when the CT has decayed, using equation \ref{eq:3}:
\begin{figure}[H]
\centering{}\subfloat[]{\includegraphics[width=5.3cm,height=5cm]{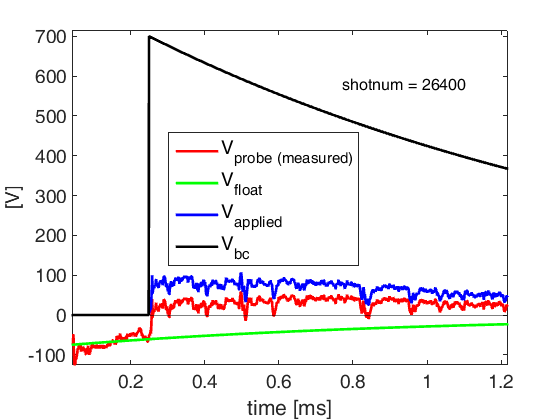}}\hfill{}\subfloat[]{\includegraphics[width=4.7cm,height=5cm]{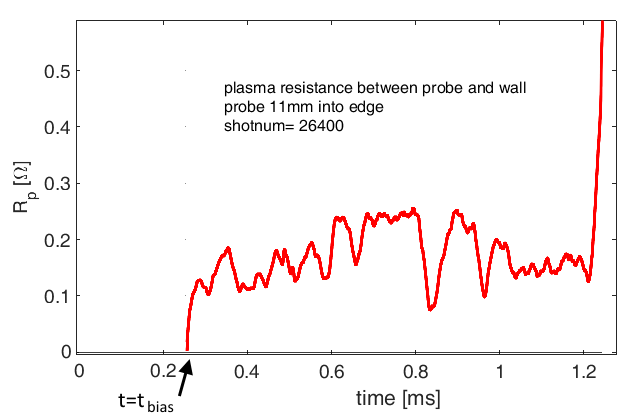}}\hfill{}\subfloat[]{\includegraphics[width=5.5cm,height=5cm]{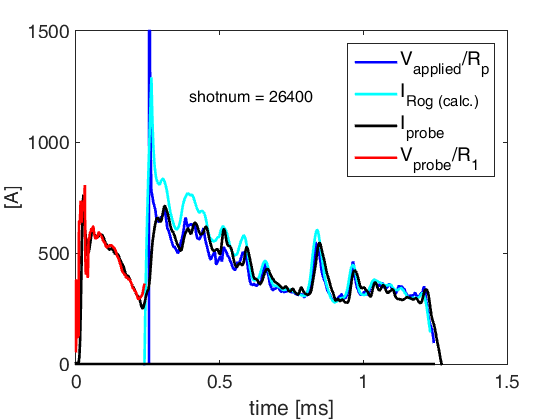}}\caption{\label{26400VRI_bias}$\,\,\,\,$Calculated / measured probe voltages,
edge plasma resistance and bias currents for shot 26400}
\end{figure}
The approximations (from equations \ref{eq:2.1}, \ref{eq:2.11},
and \ref{eq:2.12}) for $V_{applied}(t)$, $V_{float}(t)$, and $V_{bc}(t)$,
and measured $V_{probe}(t)$, for shot 26400, are shown in figure
\ref{26400VRI_bias}(a). Figure \ref{26400VRI_bias}(b) shows the
estimation, from equation \ref{eq:3}, for $R_{p}(t)$. $R_{p}(t)\sim0.15\,\Omega$
to $0.2\,\Omega$, and rises as $T_{e}$ decreases ($\eta_{plasma}$
increases) over CT decay, then drops as the edge current path length
$L$ (recall $R(t)=\eta(t)L(t)/A(t)$) decreases. Path length decreases
because $B_{\theta}$ decreases faster than $B_{\phi}$ (CT toroidal
field is maintained at a relatively constant level by the crow-barred
external shaft current) as the CT decays, $i.e.,$ $q$ increases
- there are fewer poloidal transits for each toroidal transit along
the path which defines $R_{p}$. The sharp dip in $R_{p}$ at $t\sim845\,\upmu\mbox{s}$
coincides with the fluctuations in $I_{probe}(t)$ and $B_{\theta}$
seen in figures \ref{26400}(a) and (b). Note that the current through
the path enclosed by the Rogowski coil depicted in figures \ref{fig:Biasing-probe-circuit}
and \ref{fig:Circuit_plasma} can be calculated using basic circuit
theory as: 
\begin{equation}
I_{rog\,(calc.)}=\frac{1}{R_{p}(t)}\left[\frac{R_{2}\left(V_{bc}(t)\,R_{1}+V_{float}(t)\,R_{1}+V_{bc}(t)\,R_{p}(t)\right)}{R_{1}R_{2}+R_{2}R_{p}(t)+R_{p}(t)\,R_{1}}-V_{bc}(t)-V_{float}(t)\right]\label{eq:4}
\end{equation}
Figure \ref{26400VRI_bias}(c) compares measured $I_{probe}(t)$ (black
trace) with calculated parameters, to verify the calculation of $R_{p}(t)$.
Referring to figure \ref{fig:Circuit_plasma}(c), it is seen that
$V_{applied}(t)/R_{p}(t)$ should, as is confirmed in figure \ref{26400VRI_bias}(c)
(dark blue trace), give the measured $I_{probe}(t)$ current when
the switch is closed after $t=t_{bias}$. Referring to figure \ref{fig:Circuit_plasma}(b),
$V_{probe}(t)/R_{1}\sim I_{probe}(t)$ when the switch is open before
$t=t_{bias}$ (red trace in \ref{26400VRI_bias}(c)). A good match
to measured $I_{probe}(t)$ is found by using calculated $R_{p}(t)$
and the estimated profile of $V_{float}(t)$ in equation \ref{eq:4}
(after $t=t_{bias}$, cyan trace). A good estimate of $R_{p}(t)$
is useful for optimizing external circuit resistances.

\section{Main results\label{sec:Main-results} }

\begin{figure}[H]
\centering{}\subfloat[]{\includegraphics[width=8cm,height=6cm]{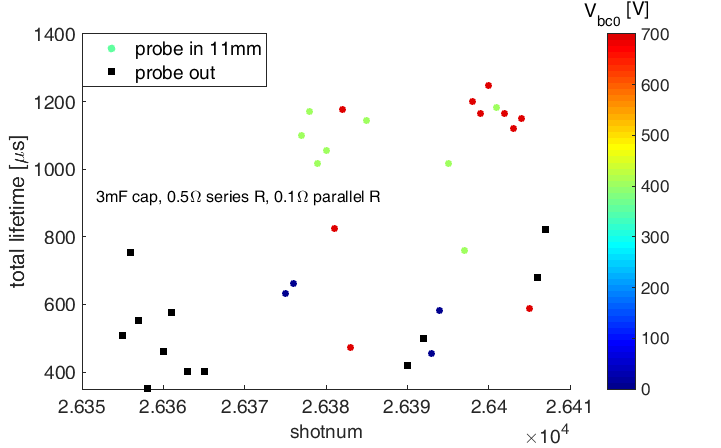}}\hfill{}\subfloat[]{\includegraphics[width=8cm,height=6cm]{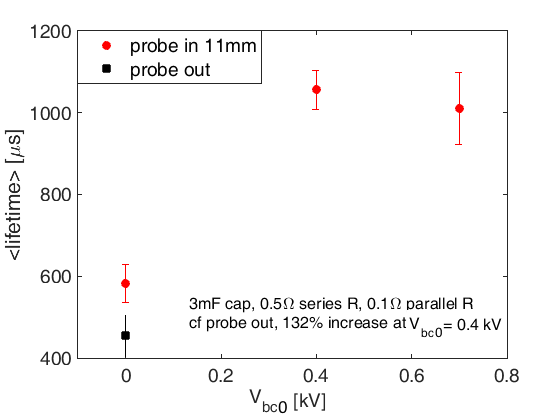}
\raggedleft{}}\caption{\label{fig:life_3mF_se}$\,\,\,\,$$3\mbox{ mF}$ capacitor: total
CT lifetimes $cf.\,\,V_{bc0}$ (bias capacitor voltage setting)}
\end{figure}
Figure \ref{fig:life_3mF_se}(a) indicates how CT lifetimes varied
with $V_{bc0}$ (coloured circles) for shots taken with the biasing
probe inserted $11\mbox{ mm}$ into the plasma edge in the configuration
using the $3\mbox{ mF}$ biasing capacitor circuit, with $R_{1}=0.1\,\Omega$
and $R_{2}=0.5\,\Omega$, compared with shots taken with the probe
removed (black squares). Figure \ref{fig:life_3mF_se}(b) indicates
the average of CT lifetimes for the probe-out configuration (black
squares), and the averages for the probe-in configuration (red circles)
for the setpoints $V_{bc0}=0\mbox{ V},\,400$ V, and $700$ V. It
is indicated that CT lifetime increased from around 450 to 600 eV,
even when the biasing capacitor was not fired - in that case, the
presence of the resistor ($R_{1}$) in parallel with the biasing capacitor
enables current, driven by the potential applied by the plasma, to
flow from the electrode to the wall. At $V_{bc0}=400$ V, CT lifetime
increased by a factor of around 2.3, from $\sim460\,\upmu$s to $\sim1070\,\upmu$s.
Note that TS data is not available for the configuration with the
3 mF capacitor in the biasing circuit.
\begin{figure}[H]
\centering{}\subfloat[]{\includegraphics[width=8cm,height=6cm]{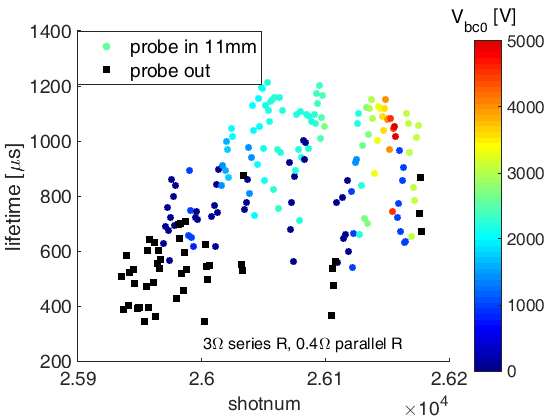}}\hfill{}\subfloat[]{\includegraphics[width=8cm,height=6cm]{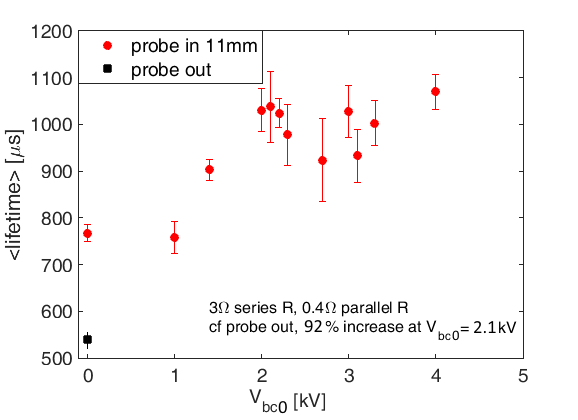}
\raggedleft{}}\caption{\label{fig:life_100uF_se}$\,\,\,\,$$100\,\upmu$F capacitor: CT
lifetimes $cf.\,\,V_{bc0}$ }
\end{figure}
Figure \ref{fig:life_100uF_se} shows equivalent information for shots
taken with a $100\,\upmu$F, 5 kV  capacitor in the biasing circuit,
with $R_{1}=0.4\,\Omega$ and $R_{2}=3\:\Omega$ (TS data is available
for this configuration). CT lifetimes were approximately doubled in
this configuration, with an optimal biasing capacitor setpoint of
$V_{bc0}\sim2$ kV. 
\begin{figure}[H]
\centering{}\subfloat[]{\includegraphics[width=8cm,height=6cm]{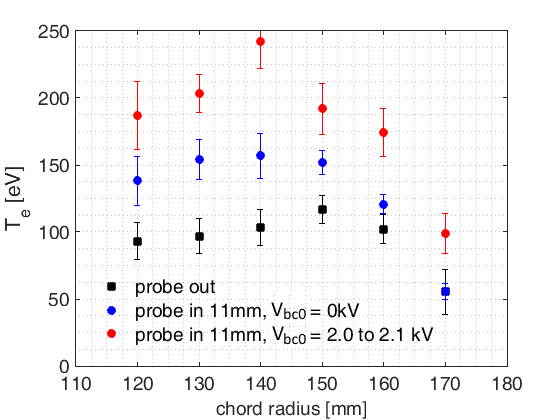}}\hfill{}\subfloat[]{\includegraphics[width=8cm,height=6cm]{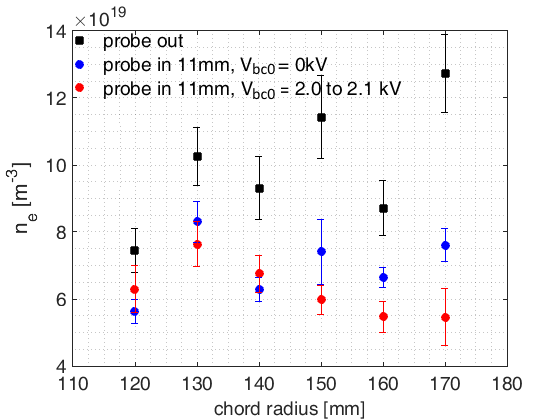}
\raggedleft{}}\caption{\label{fig:TS_2profile}$\,\,\,\,$$100\,\upmu$F capacitor: electron
temperature and density profiles at $300\,\upmu$s, for $V_{bc0}\sim2$$\mbox{ kV}$ }
\end{figure}
Figure \ref{fig:TS_2profile} shows shot data indicating the temperature
and density profiles obtained with the TS system at $300\,\upmu$s
after firing the formation capacitor banks, for the configuration
with the $100\,\upmu$F, 5 kV  biasing capacitor. Note that the TS
sampling points are indicated in figure \ref{fig: Spect_n_Te}(b).
With $V_{bc0}\sim2$ kV, the measurements indicate that temperature
is more than doubled at the inner sampling points, increasing by a
factor of around 2.4 at the sampling point at $r=140$ mm, (black
squares $cf.$ red circles) and the proportional increase in temperature
falls off towards the CT edge. Note that current drawn through the
CT edge leads to a temperature increase even when no voltage is externally
applied to the electrode (black squares $cf.$ blue circles). Referring
to figure \ref{fig:TS_2profile}(b), electron density is markedly
reduced when the electrode is inserted and the reduction is enhanced
when the electrode is externally biased. The proportional decrease
in density is greater towards the CT edge, consistent with the theory
that edge fueling impediment due to an edge transport barrier is largely
responsible for the density reduction. The diagnostic indicates an
electron density reduction by factors of approximately 1.5 and 2.3
at $r=130$ mm and $r=170$ mm respectively.
\begin{figure}[H]
\centering{}\subfloat[]{\includegraphics[width=8cm,height=5cm]{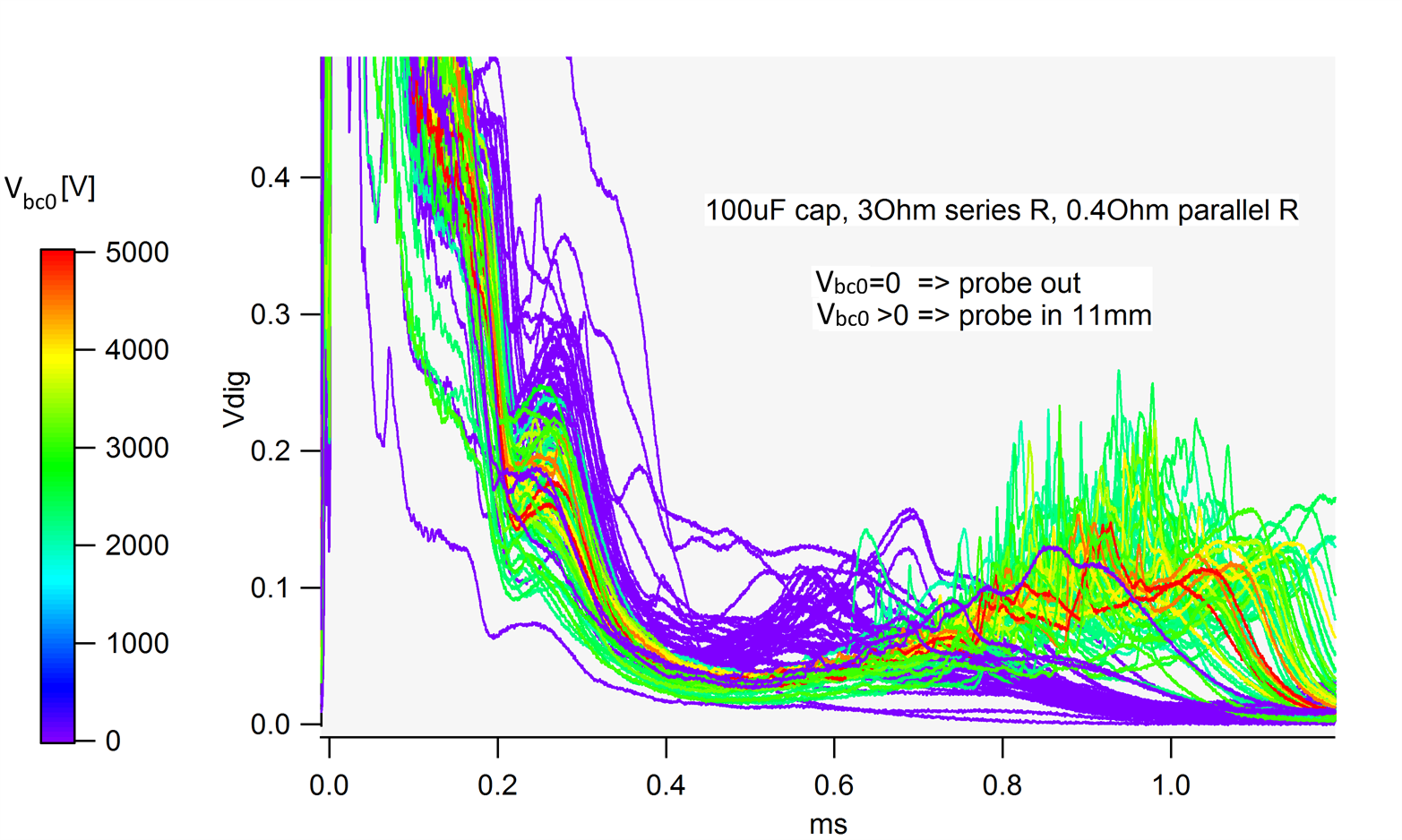}}\hfill{}\subfloat[]{\includegraphics[width=8cm,height=5cm]{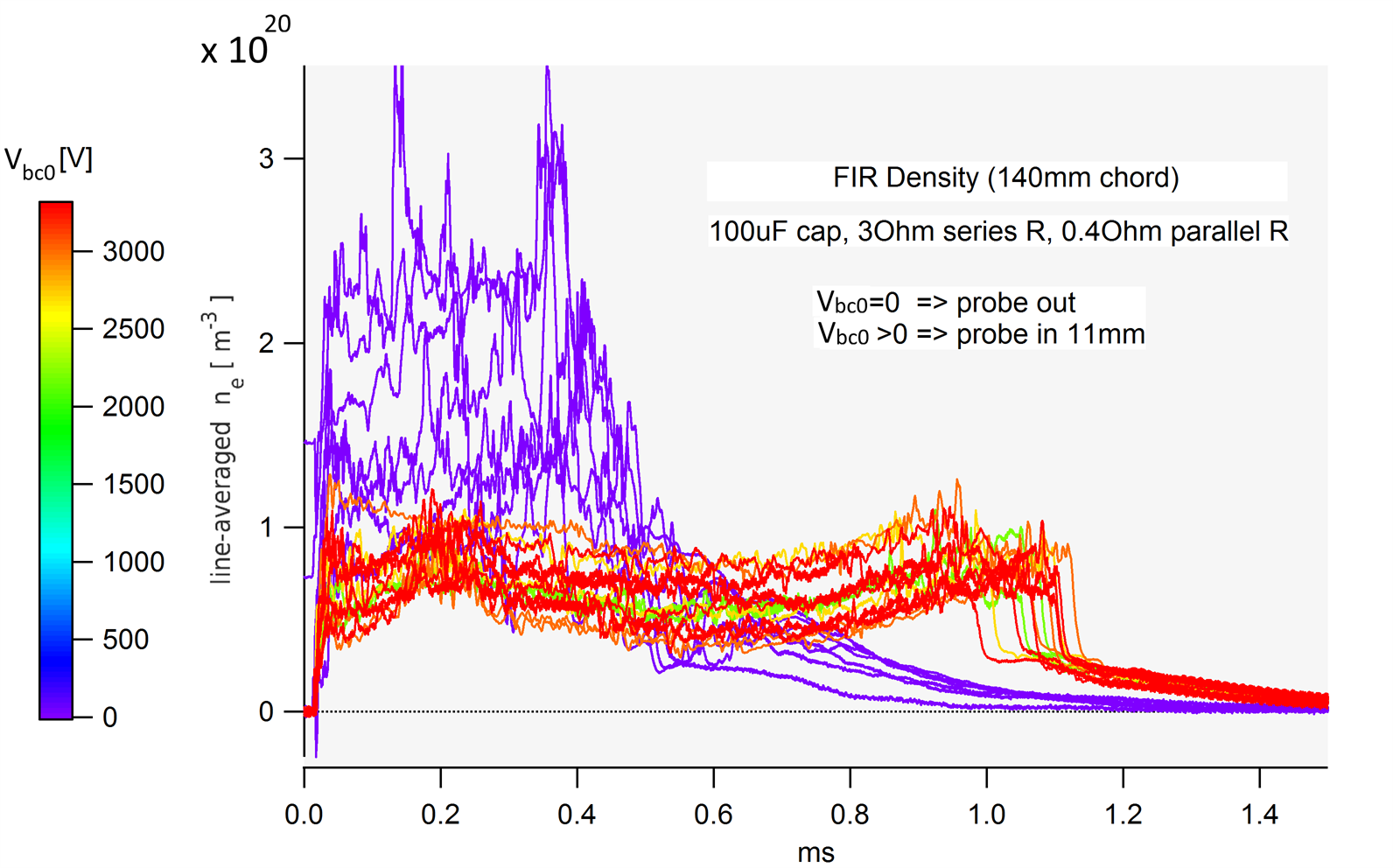}
\raggedleft{}}\caption{\label{fig:Halphavert}$\,\,\,\,$$100\,\upmu$F capacitor: (a) $\mbox{H}_{\alpha}$
intensity, (b) electron density (FIR interferometer)}
\end{figure}
Figure \ref{fig:Halphavert}(a) indicates how $\mbox{H}_{\alpha}$
intensity, along a vertical chord located at $r=88$ mm, is reduced
when the electrode is inserted and biased. $\mbox{H}_{\alpha}$ intensity
reduction is a sign of reduced recombination at the vessel walls,
and is associated with improved confinement. The purple traces are
from shots with the electrode removed from the vacuum vessel. As shown
in figure \ref{fig:Halphavert}(b), time-resolved FIR interferometer
data from the chord at 140 mm (FIR chord locations are indicated in
figure \ref{fig: Spect_n_Te}(a)) confirms the reduction in electron
density when the biased probe is inserted into the plasma edge. Again,
the purple traces are from shots taken with the electrode retracted.
This density reduction is thought to be due to the effect of the transport
barrier impeding the level of CT fueling associated with neutral gas
diffusing up the gun, as discussed in section \ref{sec:Neutral_SPECTOR}.
Note that the fueling effect is not entirely eliminated by the biasing
effect - density starts to increase at around $500$ to 600$\,\upmu$s
($cf.$ figure \ref{fig:Spector_nN}(b)). Note that H$_{\alpha}$
intensity increases dramatically at around the same time (figure \ref{fig:Halphavert}(a)).
The CTs associated with the purple traces in figure \ref{fig:Halphavert}(a)
and (b) do not last for long enough to enable observation of the density
and H$_{\alpha}$ intensity increases at that time. 

\section{Summary\label{sec:Summary_biasing_exp}}

Significant increases in CT lifetime and electron temperature, and
reductions in electron density and $\mbox{H}_{\alpha}$ intensity,
were observed when the electrode was inserted into the plasma edge,
even when the biasing capacitor was not fired. In that case, the presence
of the resistor ($R_{1}$) in parallel with the biasing capacitor
enables current, driven by the potential applied by the plasma, to
flow from the electrode to the wall. Note that in cases where the
biasing capacitor was not fired, the enhanced performance was eliminated
when $R_{1}$ was removed from the circuit. In terms of enhanced CT
lifetime, which was observed to increase by a factor of up to 2.3,
the optimal biasing circuit tested was with the 3 mF capacitor in
place, but TS data was not available in that configuration. CT lifetimes
and electron temperatures were observed to increase by factors of
around 2 and 2.4 (temperature near the CT core) respectively in the
configuration with the 100$\,\upmu$F capacitor charged to 2.1 kV,
while density decreased by a factor of around 2.3 near the CT edge.
This density reduction is thought to be due to the effect of the transport
barrier impeding level of CT fueling associated with neutral gas diffusing
up the gun. The consequent reduction of cool particle influx to the
CT is thought to partially responsible for the particularly significant
increases in observed temperature, as compared with prior edge biasing
experiments. Up to $\sim1200$ A was drawn $11\mbox{\mbox{ mm}}$
through the edge plasma, while improving CT lifetime and temperature. 

Note that the biasing experiment was conducted without a fresh lithium
coating on the inside of the SPECTOR flux conserver. With a fresh
coating, CT lifetimes are typically around 2 ms. The biasing experiment
may be run again with a fresh coating. The experiment was conducted
over a short period (less than two weeks). As the majority of the
probe-out shots were taken at the beginning of each day, there is
likely some data skew due to cleaning effects. The improvement shown
with biasing may be extended with further circuit optimization. Negative
biasing was tested briefly - a slight increase of electron temperature
and a peaking of the electron temperature profile was observed, but
there was no evidence of lifetime increase. It may be that the ion-sputtering
of the probe associated with negative biasing lead to performance
degradation associated with plasma impurities that offset the improvement
associated with the establishment of a transport barrier. Perhaps
more cleaning shots are required to see a significant improvement
with negative biasing - the efficacy of negative biasing hasn't been
confirmed. An IV curve was produced with the electrode biased to a
range of positive and negative voltages on a shot to shot basis. Langmuir
analysis indicated $T_{e}\sim130$$\,\mbox{\mbox{eV}}$ and $n_{e}\sim10^{19}\,[\mbox{m}^{-3}]$
at the probe location at $300$$\,\upmu\mbox{s}$, and $T_{e}\sim85$
$\mbox{\mbox{eV}}$ and $n_{e}\sim5\times10^{18}\,[\mbox{m}^{-3}]$
at $600$$\,\upmu\mbox{s}$. Compared with TS data, the electron temperature
estimates in particular appear too high. The fact that probe biasing
affects electron temperature and electron density makes the Langmuir
analysis results dubious at best, but it may be possible to correct
for this effect. 

It would be worth repeating the experiment with a fresh lithium coating
on the inner flux conserver. Circuit parameters, probe insertion depth,
and machine operation settings should be optimized further. The effects
of biasing on edge conditions should be characterised using Langmuir
and Mach probes, and ion Doppler diagnostics. Negative biasing may
be tested more rigorously. It would be interesting to look at the
effects of driving edge current resonant to the MHD behaviour that
manifests itself in the form of fluctuations on measurements including
CT poloidal field.

The biasing experiment was especially noteworthy because it has generally
been found that insertion of foreign objects, such as thin alumina
tubes containing magnetic probes, into SPECTOR and MRT CTs, leads
to performance degradation associated with plasma impurities. After
the extensive problems encountered relating to plasma/material interaction
and impurities during the magnetic compression experiment, special
care was taken to choose a plasma-compatible material for the biasing
electrode assembly. The pyrolytic boron nitride tube and molybdenum
electrode combination seems to have been a good choice - at least
the benefit due to drawing a current through the CT edge outweighed
any performance degradation that may have been associated with impurities
introduced to the system.
\end{document}